\documentclass[12pt]{iopart}
\makeindex
\usepackage{url}

\usepackage{graphicx}
\usepackage[english]{babel}

\usepackage{aas_macros}

\graphicspath{{Figs/}}

\expandafter\let\csname equation*\endcsname\relax
\expandafter\let\csname endequation*\endcsname\relax

\usepackage{amsmath,mathrsfs,wasysym,amssymb}
\usepackage{textcomp, gensymb}
\usepackage[sectionbib]{chapterbib}
\usepackage[unicode, pdfusetitle]{hyperref}
\usepackage[all]{hypcap}
\usepackage{cite}
\usepackage[dvipsnames]{xcolor}
\usepackage{amsfonts}
\usepackage{mathtools}
\usepackage{float}
\usepackage{tensor}
\usepackage{braket}
\usepackage{appendix}
\usepackage{array}
\usepackage{tabularx}
\usepackage{tabularray}
\usepackage{ifthen}
\usepackage{booktabs}
\usepackage{multirow}
\usepackage{colortbl}
\definecolor{p-r}{RGB}{171, 40, 52}
\hypersetup{colorlinks=true, citecolor=p-r, linkcolor=p-r, urlcolor=p-r}
\usepackage{tikz}
\usetikzlibrary{decorations.pathmorphing}
\usetikzlibrary{decorations.markings}
\usepackage[normalem]{ulem}
\usepackage{xspace}

\usepackage[T1]{fontenc}
\usepackage[utf8]{inputenc}
\usepackage{microtype}
\usepackage{revsymb} %
\usepackage{aas_macros} %

\numberwithin{equation}{section}
\let\Re\relax
\DeclareMathOperator{\Re}{{Re}}
\let\Im\relax
\DeclareMathOperator{\Im}{{Im}}

\newcommand{\cA}{\mathcal{A}}
\newcommand{\f}{\frac}

\newcommand{\sch}{Schwarzschild }
\newcommand\underrel[2]{\mathrel{\mathop{#2}\limits_{#1}}}
\newcommand*{\scri}{\ensuremath{\mathscr{I}}}

\usepackage{wasysym} %
\newcommand{\tho}{\text{\thorn}} %
\newcommand{\edt}{\text{\dh}}

\mathchardef\minus = "002D  %
\newcommand{\swY}[4][]{
	\ifthenelse{ \equal {#1} {} }{ \def\tmp {Y} }{ \ifthenelse{ \equal {#1} {*} }{ \def\tmp {Y^*} }{ \def\tmp {\bar{Y}} } }
	{}_{{}_{#2}}\!\tmp_{#3}(#4)
}
\newcommand{\swSH}[5][]{
	\ifthenelse{ \equal {#1} {} }{ \def\tmp {S} }{ \ifthenelse{ \equal {#1} {*} }{ \def\tmp {S^*} }{ \def\tmp {\bar{S}} } }
	{}_{{}_{#2}}\tmp_{#3}(#4;#5)
}
\newcommand{\swS}[5][]{
	\ifthenelse{ \equal {#1} {} }{ \def\tmp {S} }{ \ifthenelse{ \equal {#1} {*} }{ \def\tmp {S^*} }{ \def\tmp {\bar{S}} } }
	{}_{{}_{#2}}\tmp_{#3}(#4;#5)
}
\newcommand{\scA}[4][]{
	\ifthenelse{ \equal {#1} {} }{ \def\tmp {A} }{ \ifthenelse{ \equal {#1} {*} }{ \def\tmp {A^*} }{ \def\tmp {\bar{A}} } }
	{}_{{}_{#2}}\tmp_{#3}(#4)
}
\newcommand{\YSH}[4][]{
	\ifthenelse{ \equal {#1} {} }{ \def\tmp {\mathcal{A}} }{ \ifthenelse{ \equal {#1} {*} }{ \def\tmp {\mathcal{A}^*} }{\def\tmp {\bar{\mathcal{A}} } } }
	{}_{{}_{#2}}\tmp_{#3}(#4)
}

\DeclareMathOperator{\sign}{sign}

\newcommand{\brak}[2]{ {\langle {#1} \, | \, {#2} \rangle} }
\newcommand{\ketbra}[2]{ \ket{#1}\bra{#2} }

\def\eqn#1{Eq.~(\ref{#1})}

\newcommand{\Eqnsa}[2]{Equations~(\ref{#1}) and (\ref{#2})}

\def\SL#1{Sturm-Liouville}
\def\SLT#1{Sturm-Liouville theory}
\newcommand{\W}{{\mathrm{W}}}

\newcommand{\net}{\ensuremath{\mathrm{net}}\xspace}
\newcommand{\data}{\ensuremath{d}\xspace}
\newcommand{\vdata}{\ensuremath{\mathbf{\data}}\xspace}
\newcommand{\noise}{\ensuremath{n}\xspace}
\newcommand{\vnoise}{\ensuremath{\mathbf{\noise}}\xspace}
\newcommand{\signal}{\ensuremath{h}\xspace}
\newcommand{\vsignal}{\ensuremath{\mathbf{\signal}}\xspace}
\newcommand{\params}{\ensuremath{\boldsymbol{\lambda}}\xspace}
\newcommand{\residual}{\ensuremath{r}\xspace}
\newcommand{\vresidual}{\ensuremath{\mathbf{\residual}}\xspace}
\newcommand{\covmat}{\ensuremath{C}\xspace}
\newcommand{\vcovmat}{\ensuremath{\mathbf{C}}\xspace}
\newcommand{\circmat}{\ensuremath{\Sigma}\xspace}
\newcommand{\vcircmat}{\ensuremath{\boldsymbol{\Sigma}}\xspace}
\newcommand{\dftmat}{\ensuremath{\Omega}\xspace}
\newcommand{\vdftmat}{\ensuremath{\boldsymbol{\Omega}}\xspace}
\newcommand{\transpose}{\ensuremath{\mathsf{T}}\xspace}

\begin{document}

{
\footnotesize
\begin{flushright}
    RUP-25-10, YITP-25-64, RIKEN-iTHEMS-Report-25
    \vspace{-\baselineskip}
\end{flushright}
}

\title{Black hole spectroscopy: from theory to experiment}

\newlength{\savedmathindent}
\setlength{\savedmathindent}{\mathindent}
\setlength{\mathindent}{4pc}

\newcommand{\bham}{School of Physics and Astronomy and Institute for Gravitational Wave Astronomy, University of Birmingham, Edgbaston, Birmingham, B15 9TT, UK}
\newcommand{\nbi}{Center of Gravity, Niels Bohr Institute, Blegdamsvej 17, 2100 Copenhagen, Denmark}
\newcommand{\centra}{CENTRA, Departamento de F\'{\i}sica, Instituto Superior T\'ecnico -- IST, Universidade de Lisboa -- UL, Avenida Rovisco Pais 1, 1049-001 Lisboa, Portugal}
\newcommand{\cit}{TAPIR, California Institute of Technology, Pasadena, CA 91125, USA}
\newcommand{\citb}{Walter Burke Institute for Theoretical Physics, California Institute of Technology, Pasadena, CA 91125, USA}
\newcommand{\cornell}{Cornell Center for Astrophysics and Planetary
  Science, Cornell University, Ithaca, New York 14853, USA}
\newcommand{\JHU}{William H. Miller III Department of Physics and Astronomy, Johns Hopkins University, Baltimore, Maryland 21218, USA}
\newcommand{\UMiss}{Department of Physics and Astronomy,
    University of Mississippi, University, MS 38677, USA}
\newcommand{\WFU}{Department of Physics,
    Wake Forest University, Winston-Salem, North Carolina 27109, USA}
\newcommand{\aei}{Max Planck Institute for Gravitational Physics (Albert Einstein Institute), D-14476 Potsdam, Germany}
\newcommand{\aeihan}{Max Planck Institute for Gravitational Physics (Albert Einstein Institute), D-30167 Hannover, Germany}
\newcommand{\UIUC}{Department of Physics and Illinois Center for Advanced Studies of the Universe, University of Illinois at Urbana-Champaign, Urbana, Illinois 61801, USA}
\newcommand{\yitp}{Center for Gravitational Physics and Quantum Information, Yukawa Institute for Theoretical Physics, Kyoto University, 606-8502, Kyoto, Japan}
\newcommand{\hakubi}{The Hakubi Center for Advanced Research, Kyoto University,
Yoshida Ushinomiyacho, Sakyo-ku, Kyoto 606-8501, Japan}
\newcommand{\ithems}{RIKEN iTHEMS, Wako, Saitama, 351-0198, Japan}
\newcommand{\KGU}{Division of Liberal Arts, Kogakuin University, 2665-1 Nakano-machi, Hachioji, Tokyo, 192-0015, Japan}
\newcommand{\sns}{Scuola Normale Superiore, Piazza dei Cavalieri 7, 56126 Pisa, Italy}
\newcommand{\infnpi}{INFN, Sezione di Pisa, Largo Bruno Pontecorvo 3, 56127 Pisa, Italy}
\newcommand{\pisa}{Dipartimento di Fisica ``Enrico Fermi,'' Universit\`a di Pisa, Largo Bruno Pontecorvo 3, Pisa I-56127, Italy}
\newcommand{\unige}{Department of Theoretical Physics and
 Gravitational Wave Science Center,  
24 quai E. Ansermet, CH-1211 Geneva 4, Switzerland}
\newcommand{\sissa}{SISSA, Via Bonomea 265, 34136 Trieste, Italy}
\newcommand{\infnts}{INFN Sezione di Trieste, Padriciano 99, 34012 Trieste, Italy}
\newcommand{\ifpu}{IFPU - Institute for Fundamental Physics of the Universe, Via Beirut 2, 34014 Trieste, Italy}
\newcommand{\uon}{School of Mathematical Sciences, University of Nottingham, University Park, Nottingham, NG7 2RD, United Kingdom}
\newcommand{\GSSI}{Gran Sasso Science Institute (GSSI), I-67100 L'Aquila, Italy}
\newcommand{\GranSasso}{INFN, Laboratori Nazionali del Gran Sasso, I-67100 Assergi, Italy}
\newcommand{\jena}{Theoretisch-Physikalisches Institut, Friedrich-Schiller-Universit{\"a}t Jena, 07743, Jena, Germany}
\newcommand{\infnto}{INFN sezione di Torino, Torino, 10125, Italy}
\newcommand{\kulitf}{Institute for Theoretical Physics, KU Leuven, Celestijnenlaan 200D, B-3001 Leuven, Belgium}
\newcommand{\lgi}{Leuven Gravity Institute, KU Leuven, Celestijnenlaan 200D, B-3001 Leuven, Belgium}
\newcommand{\umd}{Department of Astronomy, University of Maryland, College Park, MD 20742, USA}
\newcommand{\gsfc}{Astroparticle Physics Laboratory, NASA/GSFC, Greenbelt, MD 20771, USA}
\newcommand{\MIT}{Department of Physics and MIT Kavli Institute, Massachusetts Institute of Technology, Cambridge, MA 02139, USA}
\newcommand{\cresst}{Center for Research and Exploration in Space Science and Technology, NASA/GSFC, Greenbelt, MD 20771, USA}
\newcommand{\ufabc}{Centro de Matem\'atica, Computa\c{c}\~ao e Cogni\c{c}\~ao, Universidade Federal do ABC, 09280-560, Santo Andr\'e, S\~ao Paulo, Brazil}
\newcommand{\ldit}{Laboratoire des 2 Infinis - Toulouse (L2IT-IN2P3), Université de Toulouse, CNRS, UPS, F-31062 Toulouse Cedex 9, France}
\newcommand{\uri}{Department of Physics and Center for Computational Research, University of Rhode Island, Kingston, RI 02881, USA}
\newcommand{\uiuc}{Illinois Center for Advanced Studies of the Universe \& Department of Physics, University of Illinois Urbana-Champaign, Urbana, Illinois 61801, USA}
\newcommand{\cea}{IRFU, CEA, Universit\'{e} Paris-Saclay, F-91191, Gif-sur-Yvette, France}
\newcommand{\pgi}{Princeton Gravity Initiative, Princeton University, Princeton, New Jersey, 08544, USA}
\newcommand{\bimsa}{Beijing Institute of Mathematical Sciences and Applications (BIMSA), Huairou District, Beijing 101408, P. R. China}
\newcommand{\iccub}{Departament de F\'isica Qu\`antica i Astrof\'isica, Institut de Ci\`encies del Cosmos, Universitat de Barcelona, Mart\'i i Franqu\`es 1, E-08028 Barcelona, Spain}
\newcommand{\perimeter}{Perimeter Institute for Theoretical Physics, Waterloo, ON N2L2Y5, Canada}
\newcommand{\stag}{STAG Research Centre and Mathematical Sciences, University of Southampton, University Road Southampton SO17 1BJ, U.K.}
\newcommand{\damtp}{DAMTP, Centre for Mathematical Sciences, University of Cambridge, Wilberforce Road, Cambridge CB3 0WA, U.K.}
\newcommand{\uab}{Instituto de Astrof\'isica, Departamento de F\'isica y Astrnom\'ia, Universidad Andr\'es Bello, Santiago, Chile}
\newcommand{\tsinghua}{Department of Astronomy, Tsinghua University, Beijing 100084, China}
\newcommand{\ditech}{Department of Information, Artificial Intelligence and Data Science, Daiichi Institute of Technology, Tokyo 110-0005, Japan}
\newcommand{\rikkyo}{Department of Physics, Rikkyo University, Toshima, Tokyo 171-8501, Japan}
\newcommand{\sapienza}{Dipartimento di Fisica, Sapienza Università 
	di Roma \& INFN Sezione di Roma, Piazzale Aldo Moro 5, 00185, Roma, Italy}
\newcommand{\cca}{Center for Computational Astrophysics, Flatiron Institute, 162 5th Avenue, New York, NY 10010, USA}
\newcommand{\stonybrook}{Department of Physics and Astronomy, Stony Brook University, Stony Brook, NY 11794, USA}
\newcommand{\uib}{Universitat de les Illes Balears, Departament de Fisica, IAC3 – IEEC, Crta. Valldemossa km 7.5, E-07122 Palma, Spain}
\newcommand{\apc}{Universit\'e Paris Cit\'e, CNRS, Astroparticule et Cosmologie, F-75013 Paris, France}
\newcommand{\ucm}{Departamento de F\'isica Te\'orica and IPARCOS, Facultad de Ciencias F\'isicas, Universidad Complutense de Madrid, 28040 Madrid, Spain}
\newcommand{\old}{Institute of Physics, University of Oldenburg, D-26111 Oldenburg, Germany}
\newcommand{\unimib}{Dipartimento di Fisica ``G. Occhialini,'' Universit´a degli Studi di Milano-Bicocca, Piazza della Scienza 3, 20126 Milano, Italy}
\newcommand{\infnmi}{INFN, Sezione di Milano-Bicocca, Piazza della Scienza 3, 20126 Milano, Italy}
\newcommand{\waterloo}{Department of Physics and Astronomy, University of Waterloo, 200 University Ave W, N2L 3G1, Waterloo, Canada}
\newcommand{\wca}{Waterloo Centre for Astrophysics, University of Waterloo, Waterloo, ON, N2L 3G1, Canada}
\newcommand{\ifae}{Institut de Física d'Altes Energies (IFAE), The Barcelona Institute
of Science and Technology, UAB Campus, E-08193 Barcelona, Spain}
\newcommand{\radboud}{Institute for Mathematics, Astrophysics and Particle Physics, Radboud University, Heyendaalseweg 135, 6525 AJ Nijmegen, The Netherlands}
\newcommand{\kcl}{King's  College  London,  Strand,  London  WC2R  2LS,  United Kingdom}
\newcommand{\columbia}{Department of Astronomy and Columbia Astrophysics Laboratory,
Columbia University, 550 W 120th St, New York, NY 10027, USA}
\newcommand{\tmu}{Department of Physics, Tokyo Metropolitan University, 1-1 Minami-Osawa, Hachioji, Tokyo 192-0397, Japan}
\newcommand{\syr}{Department of Physics, Syracuse University, Syracuse, New York 13244, USA}
\newcommand{\dart}{University of Massachusetts Dartmouth, 285 Old Westport Rd, North Dartmouth, Massachusetts 02747, USA}
\newcommand{\igc}{Institute for Gravitation and the Cosmos, Department of Physics, Penn State University, University Park, PA 16802, USA}
\newcommand{\psu}{Department of Astronomy and Astrophysics, Penn State University, University Park, PA 16802, USA}
\newcommand{\cardiff}{School of Physics and Astronomy, Cardiff University, Cardiff, CF24 3AA, United Kingdom}

\author{%
Emanuele~Berti$^{1}$,
Vitor~Cardoso$^{2,3}$,
Gregorio~Carullo$^{2,4, *}$,
Jahed~Abedi$^{5}$,
Niayesh~Afshordi$^{6,7,8}$,
Simone~Albanesi$^{9,10}$,
Vishal~Baibhav$^{11}$,
Swetha~Bhagwat$^{4}$,
José~Luis~Blázquez-Salcedo$^{12}$,
B\'eatrice~Bonga$^{13}$,
Bruno~Bucciotti$^{14,15}$,
Giada~Caneva~Santoro$^{16}$,
Pablo~A.~Cano$^{17}$,
Collin~Capano$^{18,19}$,
Mark~Ho-Yeuk~Cheung$^{1}$,
Cecilia~Chirenti$^{20,21,22,23}$,
Gregory~B.~Cook$^{24}$,
Adrian~Ka-Wai~Chung$^{25,62}$,
Marina~De~Amicis$^{2}$,
Kyriakos~Destounis$^{3}$,
Oscar~J.~C.~Dias$^{26}$,
Walter~Del~Pozzo$^{27,15}$,
Francisco~Duque$^{28}$,
Will~M.~Farr$^{29,30}$,
Eliot~Finch$^{31}$,
Nicola~Franchini$^{32,3}$,
Kwinten~Fransen$^{33}$,
Vasco~Gennari$^{34}$,
Stephen~R.~Green$^{35}$,
Scott~A.~Hughes$^{36}$,
Maximiliano~Isi$^{30}$,
Xisco~Jimenez~Forteza$^{37}$,
Gaurav~Khanna$^{38,19}$,
Fech~Scen~Khoo$^{12,39}$,
Masashi~Kimura$^{40,41}$,
Badri~Krishnan$^{42,13}$,
Adrien~Kuntz$^{43,44,45,3}$,
Macarena~Lagos$^{46}$,
Rico~K.~L.~Lo$^{2}$,
Lionel~London$^{47}$,
Sizheng~Ma$^{8}$,
Simon~Maenaut$^{48,49}$,
Lorena~Maga\~na~Zertuche$^{2}$,
Elisa~Maggio$^{28}$,
Andrea~Maselli$^{50,51}$,
Keefe~Mitman$^{52}$,
Hayato~Motohashi$^{53}$,
Naritaka~Oshita$^{54,55,56}$,
Costantino~Pacilio$^{57,58}$,
Paolo~Pani$^{59}$,
Rodrigo~Panosso~Macedo$^{2}$,
Chantal~Pitte$^{60,43,44,45}$,
Lorenzo~Pompili$^{28}$,
Jaime~Redondo-Yuste$^{2}$,
Maur\'icio~Richartz$^{23}$,
Antonio~Riotto$^{61}$,
Jorge~E.~Santos$^{62}$,
Bangalore~Sathyaprakash$^{63,64,65}$,
Laura~Sberna$^{35}$,
Hector~O.~Silva$^{25,28}$,
Leo~C.~Stein$^{66}$,
Alexandre~Toubiana$^{57,58}$,
Sebastian~H.~V\"olkel$^{28}$,
Julian~Westerweck$^{4}$,
Huan~Yang$^{67}$,
Sophia~Yi$^{1}$,
Nicolas~Yunes$^{25}$
and
Hengrui~Zhu$^{68}$
}

\newcommand{\affilInfo}{%
\address{$^{1}$~\JHU}
\address{$^{2}$~\nbi}
\address{$^{3}$~\centra}
\address{$^{4}$~\bham}
\address{$^{5}$~\bimsa}
\address{$^{6}$~\wca}
\address{$^{7}$~\waterloo}
\address{$^{8}$~\perimeter}
\address{$^{9}$~\jena}
\address{$^{10}$~\infnto}
\address{$^{11}$~\columbia}
\address{$^{12}$~\ucm}
\address{$^{13}$~\radboud}
\address{$^{14}$~\sns}
\address{$^{15}$~\infnpi}
\address{$^{16}$~\ifae}
\address{$^{17}$~\iccub}
\address{$^{18}$~\syr}
\address{$^{19}$~\dart}
\address{$^{20}$~\umd}
\address{$^{21}$~\gsfc}
\address{$^{22}$~\cresst}
\address{$^{23}$~\ufabc}
\address{$^{24}$~\WFU}
\address{$^{25}$~\UIUC}
\address{$^{26}$~\stag}
\address{$^{27}$~\pisa}
\address{$^{28}$~\aei}
\address{$^{29}$~\stonybrook}
\address{$^{30}$~\cca}
\address{$^{31}$~\cit}
\address{$^{32}$~\apc}
\address{$^{33}$~\citb}
\address{$^{34}$~\ldit}
\address{$^{35}$~\uon}
\address{$^{36}$~\MIT}
\address{$^{37}$~\uib}
\address{$^{38}$~\uri}
\address{$^{39}$~\old}
\address{$^{40}$~\ditech}
\address{$^{41}$~\rikkyo}
\address{$^{42}$~\aeihan}
\address{$^{43}$~\sissa}
\address{$^{44}$~\infnts}
\address{$^{45}$~\ifpu}
\address{$^{46}$~\uab}
\address{$^{47}$~\kcl}
\address{$^{48}$~\kulitf}
\address{$^{49}$~\lgi}
\address{$^{50}$~\GSSI}
\address{$^{51}$~\GranSasso}
\address{$^{52}$~\cornell}
\address{$^{53}$~\tmu}
\address{$^{54}$~\yitp}
\address{$^{55}$~\hakubi}
\address{$^{56}$~\ithems}
\address{$^{57}$~\unimib}
\address{$^{58}$~\infnmi}
\address{$^{59}$~\sapienza}
\address{$^{60}$~\cea}
\address{$^{61}$~\unige}
\address{$^{62}$~\damtp}
\address{$^{63}$~\igc}
\address{$^{64}$~\psu}
\address{$^{65}$~\cardiff}
\address{$^{66}$~\UMiss}
\address{$^{67}$~\tsinghua}
\address{$^{68}$~\pgi}
}

\newcommand{\emaillink}[1]{\href{mailto:#1}{#1}}

\eads{\emaillink{berti@jhu.edu}, \emaillink{vitor.cardoso@nbi.ku.dk} and\\
*\emaillink{g.carullo@bham.ac.uk} (Corresponding author)}
\begin{indented}\item[]{} (Affiliation list at end.)\end{indented}

\hypersetup{pdfauthor={Berti et al.}}

\begin{abstract}
The ``ringdown'' radiation emitted by oscillating black holes has great scientific potential. By carefully predicting the frequencies and amplitudes of black hole quasinormal modes and comparing them with gravitational-wave data from compact binary mergers we can advance our understanding of the two-body problem in general relativity, verify the predictions of the theory in the regime of strong and dynamical gravitational fields, and search for physics beyond the Standard Model or new gravitational degrees of freedom. We summarize the state of the art in our understanding of black hole quasinormal modes in general relativity and modified gravity, their excitation, and the modeling of ringdown waveforms. We also review the status of LIGO-Virgo-KAGRA ringdown observations, data analysis techniques, and the bright prospects of the field in the era of LISA and next-generation ground-based gravitational-wave detectors.
\end{abstract}

\setlength{\mathindent}{\savedmathindent}

\clearpage
\tableofcontents

\renewcommand{\sectionmark}[1]{\markboth{{\scshape\thesection.\ #1}}{} }

\clearpage
\section{Overview}

\noindent
{\em In my entire scientific life, extending over forty-five years, the most shattering experience has been the realization that an exact solution of Einstein's equations of general relativity, discovered by the New Zealand mathematician Roy Kerr, provides the absolute exact representation of untold numbers of massive black holes that populate the universe. This ``shuddering before the beautiful,'' this incredible fact that a discovery motivated by a search after the beautiful in mathematics should find its exact replica in Nature, persuades me to say that beauty is that to which the human mind responds at its deepest and most profound level.}

\vspace{.2cm}

\noindent
\begin{flushright}
Subrahmanyan Chandrasekhar~\cite{Chandrasekhar:1985kt}
\end{flushright}

\vspace{.2cm}

The gravitational physics landscape has evolved rapidly in the last few decades, driven in large part by experiments that probe astrophysical systems in which gravity is strong and dynamical. Black holes (BHs), the most extraordinary macroscopic objects in the Universe, play a special role in these developments. In Einstein's theory of general relativity (GR), the exact solutions describing their exterior are extremely simple, while the analysis of their interior reveals features such as the existence of singularities and mass inflation that imply a failure (or, at least, the incompleteness) of the underlying classical theory.
At the same time, astrophysical BHs can be seen as natural laboratories that amplify the effects of new physics, possibly revealing the presence of new particles or fields.
There are thus strong reasons to perform experimental tests of BH spacetimes and of their dynamics, and to leverage these objects to search for new physics. These efforts are becoming more and more accurate thanks to the experimental revolution enabled by gravitational-wave (GW) astronomy, infrared interferometry, and very large baseline radio interferometry. 

This review focuses on BH spectroscopy, i.e., the idea of treating BHs as ``gravitational atoms''. The hope is that GW measurements of the complex quasinormal mode (QNM) frequencies of BHs in the so-called ``ringdown'' phase can drive our understanding of some of the deepest mysteries in gravitational physics (such as the information loss paradox, strong cosmic censorship, the nature of dark matter and dark energy), very much like atomic spectroscopy shaped our understanding of the laws of quantum mechanics.

There are many excellent reviews and popular articles on GWs and ringdown (see e.g.~\cite{Kokkotas:1999bd,Nollert:1999ji,Ferrari:2007dd,Berti:2009kk,Konoplya:2011qq,Berti:2018vdi,Cardoso:2019rvt,2025PhT..2025d4178B}), so why did we feel the need to write a new one? 

The first observation of GWs from a binary black hole (BBH) merger led to a blossoming of theoretical and experimental studies of BH spectroscopy, and all of the existing reviews have quickly become either outdated or incomplete.
In particular, since the 2009 review by Berti, Cardoso and Starinets~\cite{Berti:2009kk} there have been major theoretical developments in our understanding of GW spectra and of their excitation in BBH mergers. 
The field of BH perturbation theory, the foundation of ringdown modeling, has witnessed a second golden age. 
Our understanding of the QNM spectra of BHs in GR (Chapter~\ref{chap2}) and beyond GR (Chapter~\ref{sec:beyondGR}) has improved enormously.
Multiple groups are pushing BH perturbation theory to higher orders and using insight from numerical simulations (Chapter~\ref{sec:amplitudes}) with the goal of developing analytical models for the post-merger phase from first principles (Chapter~\ref{sec:waveforms}).
In this sense, ringdown modeling is essential to complete the analytical description of the two-body problem in GR -- a major milestone in gravity theory.
The progress in observational techniques has been even more spectacular, leading to an explosion of alternative and complementary formulations of ringdown data analysis that have been applied with great success to GW data (Chapter~\ref{sec:DataAnalysis}). Last but not least, ongoing efforts in the experimental development of more sensitive detectors on the ground and in space will expand our ability to do BH spectroscopy in unprecedented ways (Chapter~\ref{sec:nextgen}).

This review was stimulated by the ``Ringdown Inside and Out'' meeting, held in August 2024 in Copenhagen~\cite{ringdownIO}.
The field is now so broad and vibrant that a community effort was necessary to summarize even only part of the important developments that have occurred in the past years. 
We have deliberately chosen to focus on those developments that are most observationally relevant. Therefore (with a few exceptions) we did not attempt to review the extensive recent literature on perturbations of higher dimensional and nonasympotically flat spacetimes, BH solutions beyond GR, exotic compact objects, and so on.
To keep the review self-contained, we briefly summarize some of these developments in the appendices. In Appendix~\ref{sec:public_codes} we list some pedagogical introductions to BH perturbation theory, as well as public software and repositories related to the theoretical calculations and data analysis tools discussed in the review. In Appendix~\ref{sec:NP_GHP} we present a short introduction to the Newman-Penrose and Geroch-Held-Penrose formalisms. We have also included three appendices on QNM spectra in (Anti-)de Sitter BH spacetimes and in higher dimensions (Appendix~\ref{sec:GRcosmoHighD}), superradiant instabilities (Appendix~\ref{sec:superradiant_instabilities}), and analog BHs (Appendix~\ref{sec:analog}). Research in these areas is so active that it would have been impossible to do it justice here. The main purpose of these appendices is to point the interested reader to additional resources on these topics.

\vspace{1cm}
\begin{flushright}\noindent
Emanuele Berti\\
Vitor Cardoso\\
Gregorio Carullo\\
(Editors)
\end{flushright}

\clearpage

\subsection{Suggested reading pathways}

For students first approaching the subject who may find this review too technical, good starting points are the older review~\cite{Berti:2009kk} and the primers listed in Appendix~\ref{sec:public_codes}.
Readers already acquainted with the necessary background can follow different reading pathways, depending on their interests.

A theory-oriented reader interested in beyond-GR extensions could start from Sections~\ref{sec_21},~\ref{sec:KNqnm} and~\ref{subsec:LR}, followed by Chapter~\ref{sec:beyondGR}; this could be complemented by the introduction to Section~\ref{subsec:WF_modeling_dependence} and then by Sections~\ref{subsec:current_observations}-\ref{subsec:echoes_DS} and Sections~\ref{subsec:future_rates}-\ref{subsec:future_tests} for an overview of current observational constraints and future prospects.

Readers with an inclination towards mathematical relativity may want to start from Chapters~\ref{chap2} and~\ref{sec:amplitudes}, together with Sections~\ref{sec:bms-frames-memory}-\ref{sec:horizons}.

Readers interested in the two-body problem in GR and gravitational waveform modeling can focus on Chapter~\ref{sec:amplitudes} and Sections~\ref{subsec:WF_modeling_dependence}-\ref{sec:effective-one-body}. More advanced topics can be found in the remainder of Chapter~\ref{sec:waveforms}, as well as Sections~\ref{subsec:LR}-\ref{sec:spectral_environmental}.

Observers interested in experimental applications could start with Sections~\ref{subsec:WF_modeling_dependence}-\ref{sec:effective-one-body} and Chapters~\ref{sec:DataAnalysis}-\ref{sec:nextgen}, possibly followed by Chapter~\ref{sec:amplitudes} (if their focus is on nonstationary/nonlinear effects in GR) or Chapter~\ref{sec:beyondGR} (if they are mainly interested in searches for new physics).

\clearpage

\subsection*{Notation and conventions}

Unless otherwise and explicitly stated, we use geometrical units 
($G=c=1$). 
We also adopt the ``mostly plus''
$({}-{}+{}+{}+{})$ convention for the metric.  
For reference, in Tables~\ref{tab:notation} and~\ref{tab:acro} we list symbols and acronyms that are used often throughout the text.

At the beginning of each section we list the names of its initial writers, who usually contributed the majority of the material.
Subsequent iterations involved contributions from the other authors.
The editors contributed to all of the chapters.

\begin{table}[h]
\begin{tabular}{ll}
  $M$ & Black hole mass. \\
  $a$ & Kerr rotation parameter: $a = J/M$, with $J$ the intrinsic angular momentum. \\
  $\chi=a/M$ & Dimensionless angular momentum of BH, ranging within [0,1].\\
  $r_{+/-}$ & Radius of the outer/inner BH event horizon in the chosen coordinates.\\
  $\Omega_H$ & Horizon angular velocity. At extremality, $\Omega_H=\Omega_H^{\rm ext}.$ \\
  $\omega$ & Fourier transform variable. The time dependence of any
  field is $\sim e^{-i\omega t}$.  \\
  & For stable spacetimes, ${\rm Im}(\omega)<0$.\\
  $\omega_R,\,\omega_I$ & Real and imaginary part of the QNM
  frequencies.\\
  $s$ & Spin of the field.\\
  $\ell, m$ & Orbital (polar) and azimuthal angular momentum indices. \\ 
  $n$ & Overtone number, an integer labeling the QNMs by increasing $|{\rm Im}(\omega)|$.\\
  & We conventionally start counting from a ``fundamental mode'' with $n=0$.\\
  $\omega^+,\,\omega^-$ & Frequencies of ordinary and mirror modes, where $\omega^-_{\ell m n}=-(\omega^+_{\ell(-m)n})^*$. \\
  $\Phi_{1,2}$ and $\Psi_{0,1,2,3,4}$ & Complex Weyl scalars of the Newman-Penrose formalism.\\ %
  $\Psi^\pm_2$ & Gravitational Regge-Wheeler-Zerilli functions. \\
  $V^\pm_2$ & Regge-Wheeler-Zerilli gravitational potentials. \\
  $\swY{s}{\ell{m}}{\theta,\phi}$ & Spin-weighted spherical harmonics.\\
  $\scA{s}{\ell{m}}{c}$ & Angular Teukolsky equation separation constant.\\
  $\swSH{s}{\ell{m}}{\theta,\phi}{c}$ & Spin-weighted spheroidal harmonics.\\
  $\swS{s}{\ell{m}}{x}{c}$ & Spin-weighted spheroidal function such that \\
  & $\swSH{s}{\ell{m}}{\theta,\phi}{c}=\frac1{\sqrt{2\pi}}\swS{s}{\ell{m}}{\cos\theta}{c}e^{i m\phi}$ and \\
  & $\swY{s}{\ell{m}}{\theta,\phi}=\frac1{\sqrt{2\pi}}\swS{s}{\ell{m}}{\cos\theta}{0}e^{i m\phi}$.\\
  $\YSH{s}{\acute\ell\ell{m}}{c}$ & Spherical-spheroidal expansion coefficients.\\
  $\mathcal{Q}$ & Radial Starobinsky constant for algebraically-special modes.\\
  $\lambdabar$ & Variable used by Chandrasekhar~\cite{Chandrasekhar:1984mgh} to simplify various expressions.\\
  & Related to the Teukolsky angular separation constant.\\
  $\mu$ & Mass parameter of a scalar field (of mass $\mu \hbar$, with $\hbar$ Planck's constant).\\
\end{tabular}
\caption{Notation and conventions adopted in this review.}
\label{tab:notation}
\end{table}

\begin{table}[h]
\begin{tabular}{ll}
  BBH   & Binary black hole.\\
  BF    & Bayes factor.\\
  BH    & Black hole.\\
  CE    & Cosmic Explorer.\\
  dCS   & Dynamical Chern-Simons.\\
  EdGB  & Einstein-dilaton-Gauss-Bonnet.\\
  EFT   & Effective field theory.\\
  EOB   & Effective one body.\\
  EsGB  & Einstein-scalar-Gauss-Bonnet.\\
  ET    & Einstein Telescope.\\
  FD    & Frequency domain.\\
  GR    & General relativity.\\
  GW    & Gravitational wave.\\
  GWTC  & Gravitational-wave transient catalog.\\
  IMR   & Inspiral-merger-ringdown.\\ %
  LVK   & LIGO-Virgo-KAGRA.\\ %
   MBH  & Massive black hole.\\
   NIA  & Negative imaginary axis.\\
    NR  & Numerical relativity.\\
   ODE  & Ordinary differential equation.\\
   PDE  & Partial differential equation.\\
   QNM  & Quasinormal mode.\\
   QQNM & Quadratic quasinormal mode.\\ 
    RWZ & Regge-Wheeler-Zerilli.\\
    SCC & Strong cosmic censorship.\\
    SNR & Signal-to-noise ratio.\\
    TD  & Time domain.\\
    TDI & Time-delay interferometry.\\
    TGR & Testing general relativity.\\
    TTM & Total-transmission mode.\\
    XG  & Next-generation detector.\\
    ZDM & Zero-damping mode.\\ %
\end{tabular}
\caption{Main acronyms used in this review.}
\label{tab:acro}
\end{table}
\clearpage
\section{The quasinormal mode spectrum in general relativity}
\label{chap2}

\noindent
{\em In this paper we have given the complex frequencies of the free oscillations of rotating BHs for different spherical harmonic indices. After the advent of gravitational wave astronomy, the observation of these resonant frequencies might finally provide direct evidence of BHs with the same certainty as, say, the 21cm line identifies interstellar hydrogen.}

\vspace{.2cm}

\noindent
\begin{flushright}
Steven Detweiler, 1979~\cite{Detweiler:1980gk}
\end{flushright}

\vspace{.2cm}

The spectrum of linearized fluctuations of massless fields -- scalar, electromagnetic or gravitational waves -- in black hole (BH) spacetimes is a rich and interesting topic {\it per se}. The dissipative nature of BH spacetimes makes them ``poor'' oscillators, and introduces novel features relative to normal-mode systems, that require challenging techniques to dissect them in real signals. In this chapter we overview our knowledge of the spectrum of BHs, starting in Section~\ref{sec_21} with well-known and new results concerning the spectrum of rotating Kerr BHs along with a list of publicly available data and tools. Adding electromagnetic charge, abundant in our universe, introduces a number of technical challenges that were recently overcome. These breakthroughs and the spectrum of rotating, charged BHs are the topic of Section~\ref{sec:KNqnm}. One of the novel features apparent in the spectrum is mode avoidance, which we discuss at length in a broader setup in Section~\ref{sec:avoidance}. In recent years, a geometrical picture of the spectrum of BHs in terms of localized states close to the light ring has emerged. This geometrical interpretation, which we review in Section~\ref{subsec:LR}, is both important and useful, as it provides intuition and a characterization of the spectrum in terms of local quantities. The challenges associated with the noncompleteness of the spectrum of BHs and possible ways to extract such modes have recently witnessed a flurry of activity, that we review in Section~\ref{sec:scalar_products}. From an experimental point of view, it is essential to consider the fact that BHs in the cosmos are not isolated and that the presence of matter affects their spectrum, sometimes triggering spectral instabilities. In Section~\ref{sec:spectral_environmental} we review recent work quantifying the impact of matter on the oscillation spectrum of BHs.

Throughout this review, we refer to several frameworks and formalisms developed to study BH perturbations.
For the benefit of the reader, we summarize here the most widely used approaches, highlighting their differences. The Regge-Wheeler-Zerilli (RWZ)~\cite{Regge:1957td,Zerilli:1970wzz}, Teukolsky~\cite{Teukolsky:1973ha} and Sasaki-Nakamura~\cite{Sasaki:1981,Sasaki:1981kj,Sasaki:1981sx} formalisms refer to master equations that describe small fluctuations around a background metric. 
The RWZ formalism, originally developed for the Schwarzschild spacetime, is based on metric perturbations expanded in tensor spherical harmonics. 
It yields master equations governing the odd-parity and even-parity sectors (although the two wave equations are isospectral~\cite{Chandrasekhar:1985kt}). 
In contrast, the Teukolsky master equation describes perturbations of Kerr black holes in terms of curvature fluctuations. 
The Teukolsky equation applies equally well in the Schwarzschild limit, but its potential (unlike the RWZ potentials) is not short-ranged. To improve its asymptotic behavior, Sasaki and Nakamura introduced a transformation of the Teukolsky equation that yields a new master equation with a short-ranged potential. 
At a different level we find the Newman-Penrose (NP)~\cite{Newman:1961qr} and Geroch-Held-Penrose (GHP)~\cite{Geroch:1973am} formalisms.
The NP formalism provides the geometric foundation underlying the Teukolsky and Sasaki–Nakamura equations. 
It describes generic spacetimes using a null tetrad, so that information about the spacetime curvature is encoded in the so-called ``spin coefficients'' and ``Weyl scalars''.
The GHP formalism refines the NP formalism by incorporating part of the local Lorentz symmetry directly into the associated variables and differential operators. 
In Appendix~\ref{sec:NP_GHP} we discuss in more detail the geometric foundation of the NP and GHP formalisms. 
Here we remark that the NP and GHP formalisms are fully generic reformulations of GR. On the other hand, the RWZ and Teukolsky equations describe linear perturbations, which admit systematic, but increasingly intricate, higher-order generalizations.

\subsection{The Kerr spectrum and algebraically special modes}\label{sec_21}

\vspace{-.1cm}

\noindent \textit{Initial contributors: Cook, Stein}

\vspace{.2cm}

This review deals with BH spectroscopy, the study of the complex frequencies of the free oscillations of rotating BHs.
Much of the appeal of BH spectroscopy stems from a very special property of GR: BHs are simple. Uniqueness results show that in vacuum, stationary and asymptotically flat BHs are fully described by their mass and angular momentum only, the so-called Kerr family of spacetimes~\cite{Kerr:1963ud,Newman:1965my,Israel:1967wq,Hawking:1971vc,Robinson:1975bv,Mazur:1982db,Bekenstein:1996pn,Carter:1997im,Mazur:2000pn,Robinson:2004zz,Chrusciel:2012jk,Ginzburg,Zeldovich,Bunting}.

In Boyer-Lindquist coordinates, the line element of the Kerr metric is given by
\begin{subequations} \label{eq:Kerr_Metric}
\begin{align}
  \text{d}s^2 &= -\left(1-\frac{2Mr}{\Sigma}\right)\text{d}t^2
       -\frac{4Mra\sin^2\theta}{\Sigma}\text{d}t\text{d}\phi 
       + \frac{\Sigma}{\Delta}\text{d}r^2 + \Sigma\rm{d}\theta^2
\\ &\mbox{}\hspace{0.2in}
       + \left(r^2+a^2+\frac{2Mra^2\sin^2\theta}{\Sigma}\right)
       \sin^2\theta\text{d}\phi^2, \nonumber \\
\Sigma &\equiv r^2 + a^2\cos^2\theta, \\
\Delta &\equiv r^2-2Mr+a^2.\label{def:delta}
\end{align}
\end{subequations}
The solution depends only on the BH mass $M$ and on the Kerr rotation parameter $a=J/M$, where $J$ is the magnitude of the BH's angular momentum. It is often useful to define a dimensionless spin $\chi\equiv J/M^2$, ranging within [0,1]. Other quantities of relevance are the event horizon radius $r_+=M+\sqrt{M^2-a^2}$, the Cauchy horizon radius $r_-=M-\sqrt{M^2-a^2}$, the angular velocity of the horizon $\Omega_H=a/(2Mr_+)$, and the temperature $T_{\rm H}=(r_+-r_-)/(8 \pi M r_+)$. When $a=M$, the horizons coincide and the BH is said to be extremal. 
If $a>M$, there is no event horizon, and the Kerr metric describes a naked singularity rather than a BH.

Astrophysical BHs do not exist in isolation, but it turns out that the oscillation modes of BHs are localized close to the horizon (more precisely, close to the light ring: see Section~\ref{subsec:LR} below), where the matter content can to some extent be neglected. The validity of this approximation will be discussed in 
Section~\ref{sec:spectral_environmental} below. For simplicity, we will start by considering pure vacuum solutions.

Here we are interested in BH spectroscopy -- the study of spacetimes which are perturbed away from the stationary, equilibrium Kerr metric. A powerful tool to understand BH dynamics is linear perturbation theory. Linear perturbations of the Kerr geometry are most commonly explored in terms of the complex Newman-Penrose scalars (nonlinear perturbations will be considered later on). For gravitational perturbations, these are the ingoing and outgoing radiative parts of the Weyl tensor ($\Psi_0$ and $\Psi_4$, respectively). The radiative components for scalar, electromagnetic and gravitational fields are all governed by the Teukolsky master equation~\cite{Teukolsky:1972my, Teukolsky:1973ha}
\begin{align}\label{eq:Teukolsky_Master}
\left[\frac{(r^2+a^2)^2}{\Delta} - a^2\sin^2\theta\right]
   \frac{\partial^2{}_s\psi}{\partial{t}^2}
+ \frac{4Mar}{\Delta}\frac{\partial^2{}_s\psi}{\partial{t}\partial\phi}
+ \left[\frac{a^2}{\Delta} - \frac1{\sin^2\theta}\right]
   \frac{\partial^2{}_s\psi}{\partial\phi^2}
 &\nonumber \\
- \Delta^{-s}\frac{\partial}{\partial{r}}\left(
    \Delta^{s+1}\frac{\partial{}_s\psi}{\partial{r}}\right)
-  \frac1{\sin\theta}\frac{\partial}{\partial\theta}\left(
   \sin\theta\frac{\partial{}_s\psi}{\partial\theta}\right)
- 2s\left[\frac{a(r-M)}{\Delta} + \frac{i\cos\theta}{\sin^2\theta}\right]
     \frac{\partial{}_s\psi}{\partial\phi}
 & \\
- 2s\left[\frac{M(r^2-a^2)}{\Delta} - r - ia\cos\theta\right]
     \frac{\partial{}_s\psi}{\partial{t}}
 &\nonumber \\
+ (s^2\cot^2\theta - s){}_s\psi
&= 4\pi\Sigma T,\nonumber 
\end{align}
where ${}_s\psi$ is a scalar function representing a field of spin-weight $s=0,\pm 1,\pm 2$ for scalar, electromagnetic and gravitational fields (we will focus mostly on the latter, unless stated otherwise), and $T$ is built out of the matter stress-energy tensor. In vacuum ($T=0$), Eq.~\eqref{eq:Teukolsky_Master} separates if one lets
\begin{equation}\label{eq:Teukolsky_separation_form}
  {}_s\psi(t,r,\theta,\phi) = e^{-i\omega{t}} e^{im\phi}S(\theta)R(r)\,.
\end{equation}
The function $S(\theta)=\swS{s}{\ell{m}}{x}{a\omega}$, with $x\equiv\cos\theta$, is the spin-weighted spheroidal function satisfying
\begin{align}\label{eq:swSF_DiffEqn}
\partial_x \Big[ (1-x^2)\partial_x [\swS{s}{\ell{m}}{x}{c}]\Big]
+ \bigg[(cx)^2 - 2 csx + s + \scA{s}{\ell m}{c}
- \frac{(m+sx)^2}{1-x^2}\bigg]\swS{s}{\ell{m}}{x}{c} = 0,
\end{align}
where $c=a\omega$ is the oblateness parameter, $\ell$ and $m$ are the polar and azimuthal indices, and $\scA{s}{\ell m}{c}$ is the angular separation constant.  
This differential equation is commonly referred to as the angular Teukolsky equation. The radial function $R(r)$ must satisfy
\begin{align}\label{eq:radialR:Diff_Eqn}
\Delta^{-s}\frac{d}{dr}&\left[\Delta^{s+1}\frac{dR(r)}{dr}\right]
+ \left[\frac{K^2 -2is(r-M)K}{\Delta} + 4is\omega{r} - \lambdabar\right]R(r)=0,
\end{align}
where
\begin{subequations}
\begin{align}
  K &\equiv (r^2+a^2)\omega - am, \\
\label{eq:lambdabar def}
  \lambdabar &\equiv \scA{s}{\ell{m}}{a\omega} + a^2\omega^2 - 2am\omega.
\end{align}
\end{subequations}
Equation~(\ref{eq:radialR:Diff_Eqn}) is commonly referred to as the radial Teukolsky equation.  Together, Eqs.~(\ref{eq:swSF_DiffEqn}) and (\ref{eq:radialR:Diff_Eqn}) can be solved as a coupled, nonlinear eigenvalue problem for the mode frequencies $\omega_{\ell{m}n}$ and the angular separation constants $\scA{s}{\ell{m}}{a\omega}$. Here, the ``overtone index'' $n$ enumerates the eigensolutions of the coupled equations.

Linear perturbations of BHs were first explored in the case of nonrotating (Schwarzschild) BHs by Regge and Wheeler~\cite{Regge:1957td} and Zerilli~\cite{Zerilli:1970se,Zerilli:1970wzz} using a decomposition in terms of tensorial spherical harmonics (see~\cite{Nollert:1999ji,Nagar:2005ea,Berti:2009kk,Berti:2014bla} for reviews). The various metric component perturbations have different angular dependencies, but the radial equations reduce to a single, simple Schr\"odinger-like master equation
\begin{align}\label{eq:RWZ master}
    \frac{{\rm d}^2\Psi^\pm_2}{{\rm d}r^2_*} + (\omega^2-V^\pm_2)\Psi^\pm_2 = 0, 
\end{align}
where $r_*$ is the tortoise coordinate defined by ${\rm d}r_*/dr=r/(r-2M)$, $\Psi^\pm_2$ are scalars representing spin-$2$ gravitational perturbations, and $V^\pm_2$ are the respective potentials.  
The two signs refer to the odd (or axial) Regge-Wheeler perturbations $\Psi^-_2$, with parity $(-1)^{\ell+1}$; and to the even (or polar) Zerilli  perturbations $\Psi^+_2$, with parity $(-1)^\ell$.  

The Regge-Wheeler and Zerilli potentials are
\begin{align}
\label{eq:RW-potential}
    V^-_2 &= f(r)\left[\frac{\ell(\ell+1)}{r^2} - \frac{6M}{r^3}\right] \\
\intertext{and}
\label{eq:Zerilli-potential}
    V^+_2 &= \frac{2f(r)}{r^3}\left[\frac{9M^3 +9\lambda M^2r + 3\lambda^2Mr^2+\lambda^2(1+\lambda)r^3}{(3M+\lambda r)^2}\right],
\end{align}
where $f(r)\equiv 1-2M/r$ and $\lambda\equiv(\ell-1)(\ell+2)/2$. The master variables are sometimes called the RWZ wavefunctions. There are important relations between the Regge-Wheeler function, the Zerilli function, and the Teukolsky variables~\cite{Chandrasekhar:1985kt}.
The master equation (\ref{eq:RWZ master}) also applies to perturbations with spins other than 2, each with its own potential~\cite{Berti:2009kk}.

From now on we will mostly focus on the eigenvalues and eigenfunctions of Kerr BHs, because they are the most interesting in an astrophysical context. The Kerr-Newman (charged and rotating) case will be discussed in Section~\ref{sec:KNqnm}.

\subsubsection{Symmetries of the Teukolsky equations}\label{sec:Teukolsky_symms}
By itself, Eq.~\eqref{eq:swSF_DiffEqn} is a linear eigenvalue problem if one demands that solutions are regular at the boundaries.  Solutions of this eigenvalue problem can be used to construct the spin-weighted spheroidal harmonics $\swSH{s}{\ell{m}}{\theta,\phi}{c}$, which are related to the eigenfunctions $\swS{s}{\ell{m}}{x}{c}$ by
\begin{align}\label{eq:swharmtofunct}
  \swSH{s}{\ell{m}}{\theta,\phi}{c} = 
  \frac1{\sqrt{2\pi}}\,\swS{s}{\ell{m}}{\cos\theta}{c}e^{im\phi}.
\end{align}
The spin-weighted spheroidal harmonics are generalizations of the spin-weighted spherical harmonics $\swY{s}{\ell{m}}{\theta,\phi}=\swSH{s}{\ell{m}}{\theta,\phi}{0}$, and the factor of $1/\sqrt{2\pi}$ in Eq.~\eqref{eq:swharmtofunct} guarantees that the functions and harmonics can be individually normalized:
\begin{align}
    \int_{-1}^1{|\swS{s}{\ell{m}}{x}{c}|^2{\rm d}x}=\oint{|\swSH{s}{\ell{m}}{\theta,\phi}{c}|^2{\rm d}\Omega}=1.
\end{align}
For fixed values of $c$, $s$, and $m$, the spin-weighted spherical functions, while they can be normalized, are not orthogonal in general~\cite{Berti:2005gp,London:2020uva}. That is,
\begin{align}\label{eq:non-othogonal}
	\int_{-1}^1{\swS[*]{s}{\acute\ell{m}}{x}{c}\swS{s}{\ell{m}}{x}{c}dx} &= {}_{{}_s}\alpha_{m\acute\ell\ell}(c)\not\propto\delta_{\acute\ell\ell}, \\
\intertext{and when $m$ is allowed to vary,}
\label{eq:non-othogonal-harm}
	\oint{\swSH[*]{s}{\acute\ell\acute{m}}{\theta,\phi}{c}\swSH{s}{\ell{m}}{\theta,\phi}{c}d\Omega} &= {}_{{}_s}\alpha_{m\acute\ell\ell}(c)\delta_{\acute{m}m}.
\end{align}
For the special cases of $c=0$, or $s=0$ with $c^2$ real, the functions and harmonics are orthogonal.  However, for fixed values of $c$, $s$, and $m$, the spin-weighted spheroidal functions are biorthogonal~\cite{ronveaux1995heun,London:2020uva,London:2023aeo,London:2023idh}, which means that
\begin{align}\label{eq:biothogonal}
	\int_{-1}^1{\swS{s}{\acute\ell{m}}{x}{c}\swS{s}{\ell{m}}{x}{c}dx}\propto\delta_{\acute\ell\ell}.
\end{align}

The fundamental symmetries of the angular Teukolsky equation are found by considering three transformations: $\{s\to-s,x\to-x\}$, $\{m\to-m,x\to-x,c\to-c\}$, and complex conjugation. For complex conjugation, if $\alpha$ is a complex quantity, then the conjugate of $\alpha$ is written as $\alpha^*$.  The angular separation constants satisfy the properties \begin{subequations}\label{eq:swSF_all_ident}
\begin{align}
\label{eq:swSF_sA_ident}
\scA{-s}{\ell{m}}{c} &= \scA{s}{\ell{m}}{c} + 2s, \\
\label{eq:swSF_mcA_ident}
\scA{s}{\ell(-m)}{c} &= \scA{s}{\ell{m}}{-c}, \\
\label{eq:swSF_cA_ident}
\scA[*]{s}{\ell{m}}{c} &= \scA{s}{\ell{m}}{c^*},\\
\intertext{and the spin-weighted spheroidal functions can be chosen to satisfy:}
\label{eq:swSF_sx_ident}
\swS{-s}{\ell{m}}{x}{c} &= (-1)^{\ell-m}\swS{s}{\ell{m}}{-x}{c}, \\
\label{eq:swSF_mxc_ident}
\swS{s}{\ell(-m)}{x}{c} &= (-1)^{\ell+s}\swS{s}{\ell{m}}{-x}{-c}, \\
\label{eq:swSF_cc_ident}
\swS[*]{s}{\ell{m}}{x}{c} &= \swS{s}{\ell{m}}{x}{c^*}.
\end{align}
\end{subequations}
With these choices for the spin-weighted spheroidal functions, the spin-weighted spheroidal harmonics of Eq.~\eqref{eq:swharmtofunct} are consistent with the Condon-Shortley phase convention for the spin-weighted spherical harmonics, yielding the more familiar expressions
\begin{subequations}\label{eq:csphase}
\begin{align}
\label{eq:csphase_conj}
\swSH[*]{s}{\ell{m}}{\theta,\phi}{c} &= (-1)^{s-m}\swSH{-s}{\ell(-m)}{\theta,\phi}{-c^*},\\
\label{eq:csphase_parity}
\swSH{s}{\ell{m}}{\pi-\theta,\pi+\phi}{c} &= (-1)^\ell\swSH{-s}{\ell{m}}{\theta,\phi}{c}. 
\end{align}
\end{subequations}
Note that the index combinations in Eqs.~(\ref{eq:swSF_all_ident}) and (\ref{eq:csphase}) have been carefully chosen to be valid for both integer and half-odd integer values of $s$.

The spin-weighted spheroidal harmonics and functions can be expanded in terms of the
spin-weighted spherical harmonics and functions,
\begin{align}
\swSH{s}{\ell{m}}{\theta,\phi}{c} &= \sum_{\acute\ell=\ell_{\text{\tiny min}}}
^\infty{\YSH{s}{\acute\ell\ell{m}}{c}\swY{s}{\acute\ell{m}}{\theta,\phi}},\\
\intertext{and}
\swS{s}{\ell{m}}{x}{c} &= \sum_{\acute\ell=\ell_{\text{\tiny min}}}^\infty
    \YSH{s}{\acute\ell\ell{m}}{c}\swS{s}{\acute\ell{m}}{x}{0},
\end{align}
where $\ell_{\text{min}}\equiv\max(|m|,|s|)$.  Here, the $\YSH{s}{\acute\ell\ell{m}}{c}$ are the spherical-spheroidal expansion coefficients, which have the following symmetries, consistent with Eqs.~(\ref{eq:swSF_all_ident}):
\begin{subequations}\label{eq:SSEC_sym_all}
\begin{align}
\label{eq:SSEC_sym_s}
\YSH{-s}{\acute\ell\ell{m}}{c} &= (-1)^{\acute\ell-\ell}\YSH{s}{\acute\ell\ell{m}}{c},\\
\label{eq:SSEC_sym_m}
\YSH{s}{\acute\ell\ell(-m)}{c} &= (-1)^{\acute\ell-\ell}\YSH{s}{\acute\ell\ell{m}}{-c}, \\
\label{eq:SSEC_sym_c}
\YSH{s}{\acute\ell\ell{m}}{c^*} &= \YSH[*]{s}{\acute\ell\ell{m}}{c}.
\end{align}
\end{subequations}

The coupled equations (\ref{eq:swSF_DiffEqn}) and (\ref{eq:radialR:Diff_Eqn}), taken together, produce an additional symmetry. Combining Eqs.~(\ref{eq:swSF_mcA_ident}) and (\ref{eq:swSF_cA_ident}) yields $\scA{s}{\ell(-m)}{c}=\scA[*]{s}{\ell m}{-c^*}$. Because of this, if $\omega_{\ell{m}n}$ is an eigenvalue of Eq.~\eqref{eq:radialR:Diff_Eqn} with eigenfunction $R_{\ell{m}n}(r)$, then $-\omega^*_{\ell(-m)n}\,(=\omega_{\ell{m}n})$ is also an eigenvalue with eigenfunction $R^*_{\ell(-m)n}(r)\propto R_{\ell{m}n}(r)$. Because of this symmetry, the modes of the Kerr geometry belong to two families of solutions, often referred to as ``mirror'' solutions, because the relation
\begin{align}\label{eq:mirror mode symmetry}
   \omega^-_{\ell{m}n} = -(\omega^+_{\ell(-m)n})^*
\end{align}
is a reflection through the imaginary axis.  One family of solutions is labeled by $\omega^+_{\ell{m}n}$ and the other by $\omega^-_{\ell{m}n}$.  In general, the $\omega^+_{\ell{m}n}$ modes are chosen to have a positive real component.  However, this choice is not always possible, since certain continuous sequences of mode frequencies will cross the imaginary axis: see e.g. the $\omega^+_{2(-2)2}$ sequence in Fig.~\ref{fig:TTMLl2allmn}.
Below, in contexts where the $-$ component is suppressed compared to the $+$, the $\pm$ label will be omitted to avoid clutter, and the $+$ component will be intended.

\subsubsection{Quasinormal and Total-Transmission modes}\label{sec:QNMsTTMs}
The possible mode solutions of the coupled Teukolsky equations depend on the boundary conditions imposed on them. The boundary conditions on the angular solutions are always chosen to produce solutions which are regular on the polar axis. For the radial equation, there are three physically interesting choices for the boundary conditions. The most interesting and useful choice is to consider mode solutions that forbid waves to travel in from infinity or out of the event horizon of the BH. Such modes are referred to as quasinormal modes (QNMs), and represent the natural ringing modes of the BH. They are the subject of most of this review. Alternatively, if one of these two conditions is reversed, then the resulting mode solutions represent total-transmission modes (TTMs).  There are two possibilities for the TTM boundary conditions. If the boundary condition at the event horizon is reversed so that waves are forbidden from traveling into the BH, then one has what is referred to as a ``left'' total-transmission mode (TTM${}_L$). In this case, the mode travels toward spatial infinity without scattering off of the potential barrier. Alternatively, if  the boundary condition at infinity is reversed so that waves are forbidden from traveling out of the domain, then one has what is referred to as a ``right'' total-transmissions modes (TTM${}_R$). In this case, the mode travels toward the event horizon without scattering off of the potential barrier.

The QNM solutions can be determined by many different approaches (see~\cite{Berti:2009kk} for an overview).  The result is a set of complex mode frequencies $\omega_{\ell{m}n}(a)$ that obey the mirror mode symmetry of Eq.~\eqref{eq:mirror mode symmetry}.  The polar $\ell$ and azimuthal $m$ indices fix the angular behavior of the mode, while the overtone index $n$ enumerates the set of modes in the same angular family.  In the Schwarzschild limit $a=0$, the azimuthal modes are all degenerate and the QNMs obtained from Eq.~\eqref{eq:RWZ master} by the Regge-Wheeler and Zerilli potentials (\ref{eq:RW-potential}) and (\ref{eq:Zerilli-potential}) are isospectral, i.e., they yield the same set of frequencies: see Section~\ref{sec:KNqnm} and Appendix~A of~\cite{Berti:2009kk} for additional details.  Each mode also depends on the angular momentum of the BH through the Kerr rotation parameter $a$.  

\begin{figure*}[t]
    \centering
    \includegraphics[width=0.48\linewidth]{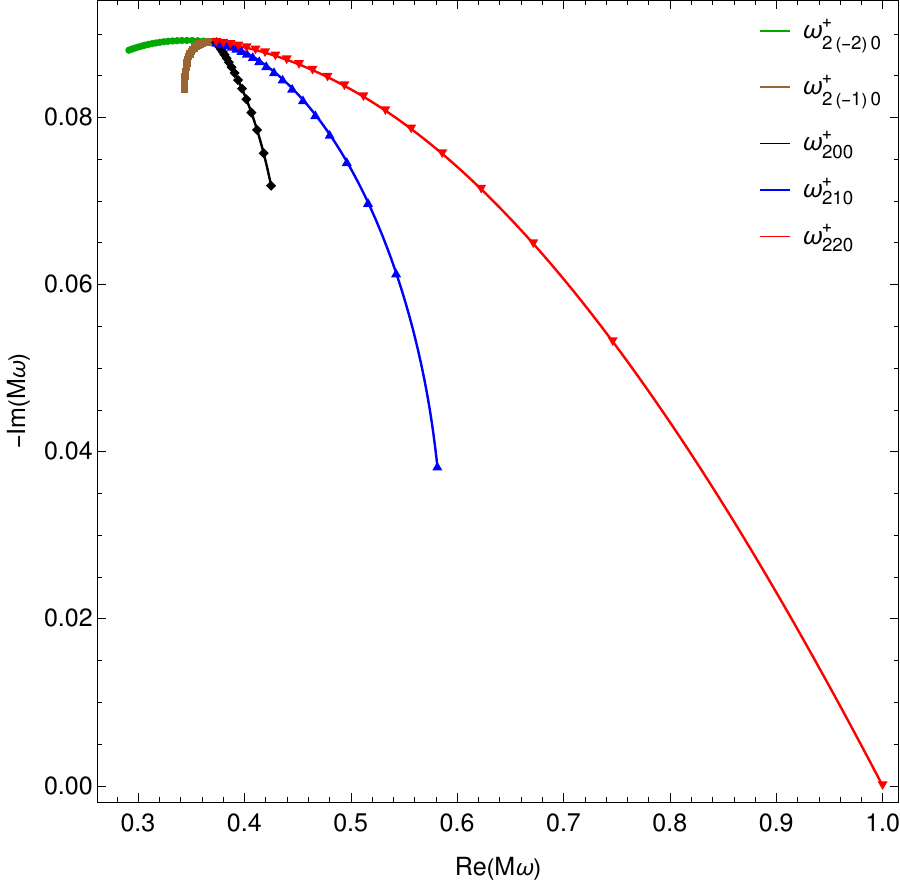}\hspace*{32pt}
    \includegraphics[width=0.48\linewidth]{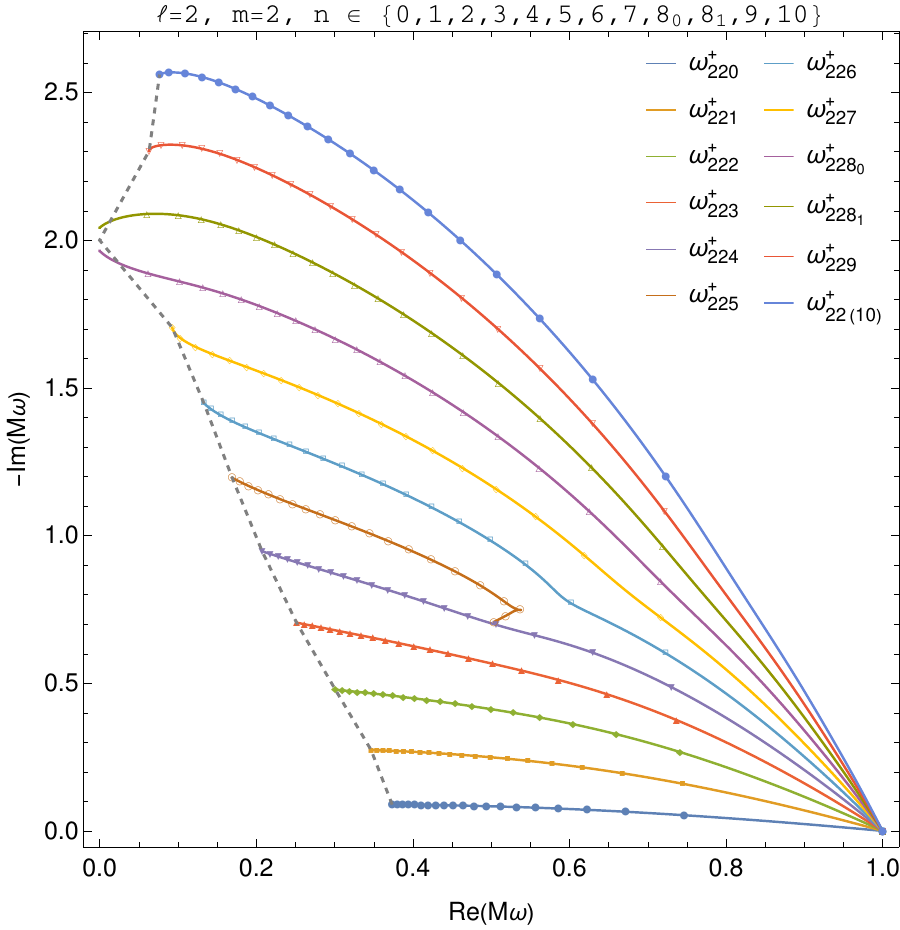}
    \caption{Examples of the complex mode frequencies $M\omega^+$ for the gravitational ($s=-2$), quadrupolar ($\ell=2$) QNMs. Left panel: sequences for all 5 azimuthal modes ($-2\le{m}\le2$) and for the fundamental mode ($n=0$). The Schwarzschild limit ($a=0$) of each sequence is at $M\omega^+\approx0.37-0.089i$. Each sequence is parameterized by $a$ over the range $0\le{a}<M$, and markers along each sequence are at intervals of $\Delta{a}=0.05M$. Right panel: sequences for the first 12 overtones (the bottom line is $n=0$) of the $m=2$ axial mode. Note that $0\le{n}\le10$, but that there are two $n=8$ overtones, labeled as $n=8_0$ and $n=8_1$. This overtone multiplet notation is used to make the overtone notation for the sequences consistent with the labeling of the Schwarzschild modes.}
    \label{fig:QNMl2behavior}
\end{figure*}

\begin{figure*}[t]
    \centering
    \includegraphics[width=0.48\linewidth]{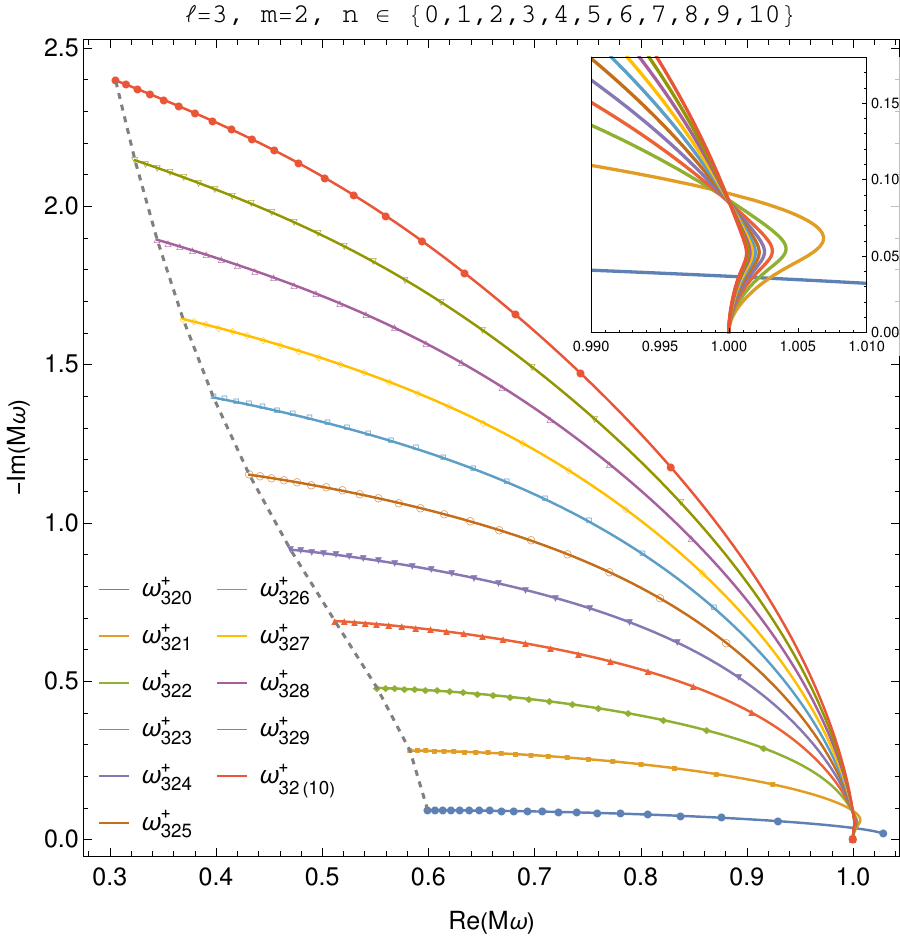}\hspace*{32pt}
    \includegraphics[width=0.48\linewidth]{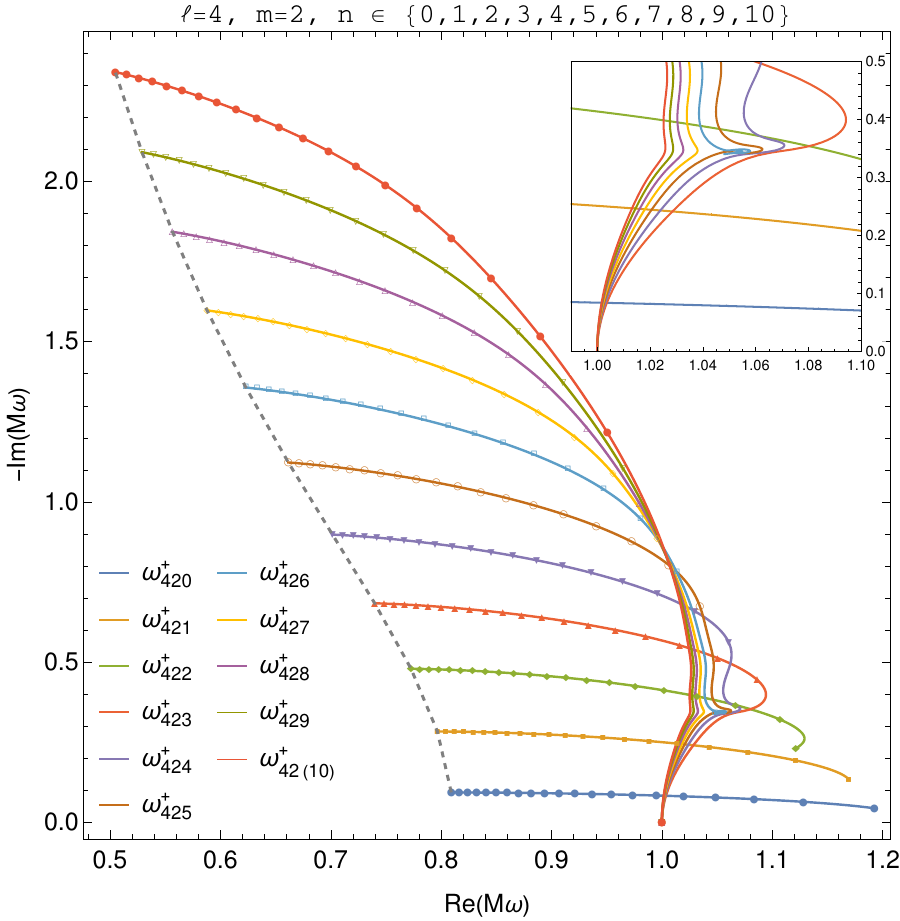}
    \caption{Examples of the complex mode frequency $M\omega$ for the gravitational ($s=-2$) QNMs with multipolar indices $\ell=3$ and $\ell=4$.  Only the $m=2$ azimuthal modes are plotted.  See Fig.~\ref{fig:QNMl2behavior} for additional details.}
    \label{fig:QNMl34}
\end{figure*}

The behavior of the gravitational ($s=-2$) QNM spectrum is illustrated in Figs.~\ref{fig:QNMl2behavior} and~\ref{fig:QNMl34}.  The left panel of Fig.~\ref{fig:QNMl2behavior} illustrates the behavior of the quadrupolar ($\ell=2$), fundamental ($n=0$) gravitational Kerr QNMs.  In the Schwarzschild limit, the sequences with $-2\le m\le2$ all begin with $M\omega^+_{2m0}\approx 0.37-0.089i$.  However, as the Kerr rotation parameter varies over its range $0\le a<M$, the degeneracy is broken, and each azimuthal mode traces out its own sequence of solutions.  Notice that the $m=2$ mode approaches the real axis at $M\omega=1$ in the extremal limit $a/M\to1$.  The right panel of Fig.~\ref{fig:QNMl2behavior} illustrates the behavior of the $\ell=2$, $m=2$ gravitational Kerr QNMs for all overtones in the range $0\le n\le 10$.  In general, the decay rate (magnitude of the imaginary part of $\omega$) increases with increasing $n$.  The dashed line in the figure connects each of the modes at the Schwarzschild limit $a=0$. Each solid line with markers plots the mode frequencies $\omega^+_{22n}(a)$ as $a$ increases from the Schwarzschild limit toward the extremal limit at $a=M$.

The two panels in Fig.~\ref{fig:QNMl34} illustrate the behavior of the $\ell=3$ and $\ell=4$ mode frequencies with $m=2$ and $0\le n\le 10$.  Together, the right panel of Fig.~\ref{fig:QNMl2behavior} and the two panels in Fig.~\ref{fig:QNMl34} illustrate several common features of QNMs, as well as some exceptional behavior.  Consider the mode frequencies in the Schwarzschild limit.  As the polar index $\ell$ increases, each overtone family begins further to the right, at larger values of $\Re(\omega)$.  We also see that many, but not all, of the $m=2$ sequences approach an accumulation point at $M\omega^+_{\ell2n} = 1$ in the extremal limit.  In general, it is found that for $m>0$, many sequences exhibit the behavior $\lim_{a\to{M}}M\omega^+_{\ell{m}n}=m/2$.  By symmetry, similar accumulation points will occur for $m<0$ along the negative real axis, with $\lim_{a\to{M}}M\omega^-_{\ell{m}n}=m/2$.  This behavior was first predicted by Detweiler~\cite{Detweiler:1980gk}, and further studied in Refs.~\cite{Cardoso:2004hh,Hod:2008zz,Yang:2012pj,Yang:2013uba,Cook:2014cta}.

\begin{figure*}[t]
    \centering
    \includegraphics[width=374pt]{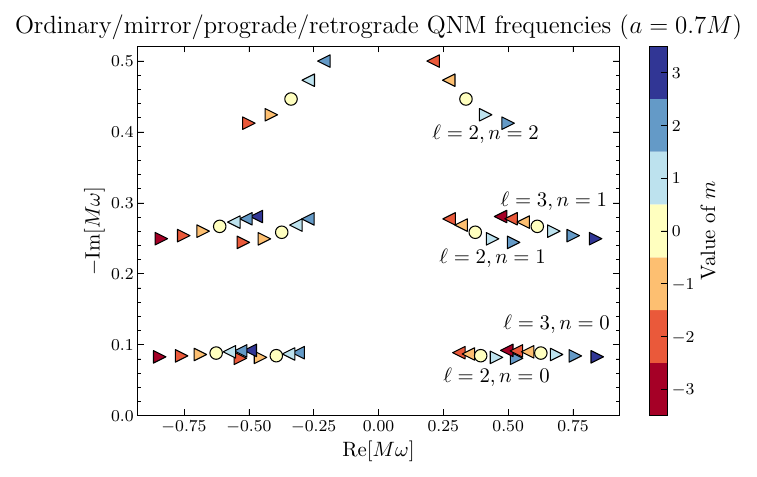}
    \caption{Right-pointing triangles are prograde Kerr QNM  frequencies computed for $a/M=0.7$, while left-pointing triangles are retrograde modes. Note that prograde and retrograde modes are present both in the right half-plane (ordinary modes) and in the left half-plane (so-called ``mirror'' modes).
    Some general trends are visible. As we increase either $\ell$ or $m$, the complex QNM frequencies have larger values of $|\Re[M\omega]|$; as we increase $n$, the QNM frequencies move to larger values of $-\Im[M\omega]$. Figure taken from~\cite{MaganaZertuche:2021syq}.}
    \label{fig:mirror_demo}
\end{figure*}

The ordinary/mirror modes classification adopted above is based on the sign of the modes real frequency.
Mode solutions are sometimes referred to as prograde or retrograde (or equivalently co-/counter-rotating), depending on the direction of rotation of the perturbation relative to the BH spin.
The literature adopts various criteria to define this classification~\cite{Li:2021wgz,MaganaZertuche:2021syq}.  
In this review, these terms will be defined to refer to the propagation of a mode's surfaces of constant phase.  From the decomposition of ${}_s\psi$ in Eq.~\eqref{eq:Teukolsky_separation_form}, surfaces of constant phase evolve as $\exp\{-i(\Re[\omega_{\ell{m}n}]t-m\phi)+\Im[\omega_{\ell{m}n}]t\}$.
For prograde modes, surfaces of constant phase move in the direction of increasing $\phi$, and have the property that $\sign{\Re[\omega_{\ell{m}n}]}=\sign{m}$. Thus, retrograde modes move in the opposite direction, and have the property that $\sign{\Re[\omega_{\ell{m}n}]}=-\sign{m}$.  So long as $\Re[\omega^+_{\ell{m}n}]>0$ (this will be true for quasinormal modes, but may not always be true for total-transmission modes), then prograde modes consist of all $\omega^+_{\ell{m}n}$ modes for $m>0$, and their mirror modes $\omega^-_{\ell(-m)n}$.  Similarly, retrograde modes consist of all $\omega^+_{\ell{m}n}$ modes for $m<0$, and their mirror modes. This notion of prograde/retrograde modes does not apply to modes with $m=0$.
The distinction between prograde, retrograde, ordinary and mirror modes, as defined in Eq.~\eqref{eq:mirror mode symmetry}, is illustrated in Fig.~\ref{fig:mirror_demo}.
In the following we will sometimes exploit the reflection symmetry around $\rm Re [M \omega]$ displayed in this figure, writing $\omega_{\ell m n}$ without specifying the branch to which the frequency belongs; the second branch can be recovered by symmetry.
For an extended discussion of retrograde modes, including their astrophysical excitation, see Sec.~\ref{subsubsec:retrograde_modes_excitations}.

\begin{figure*}[t]
    \centering
    \includegraphics[width=0.8\linewidth]{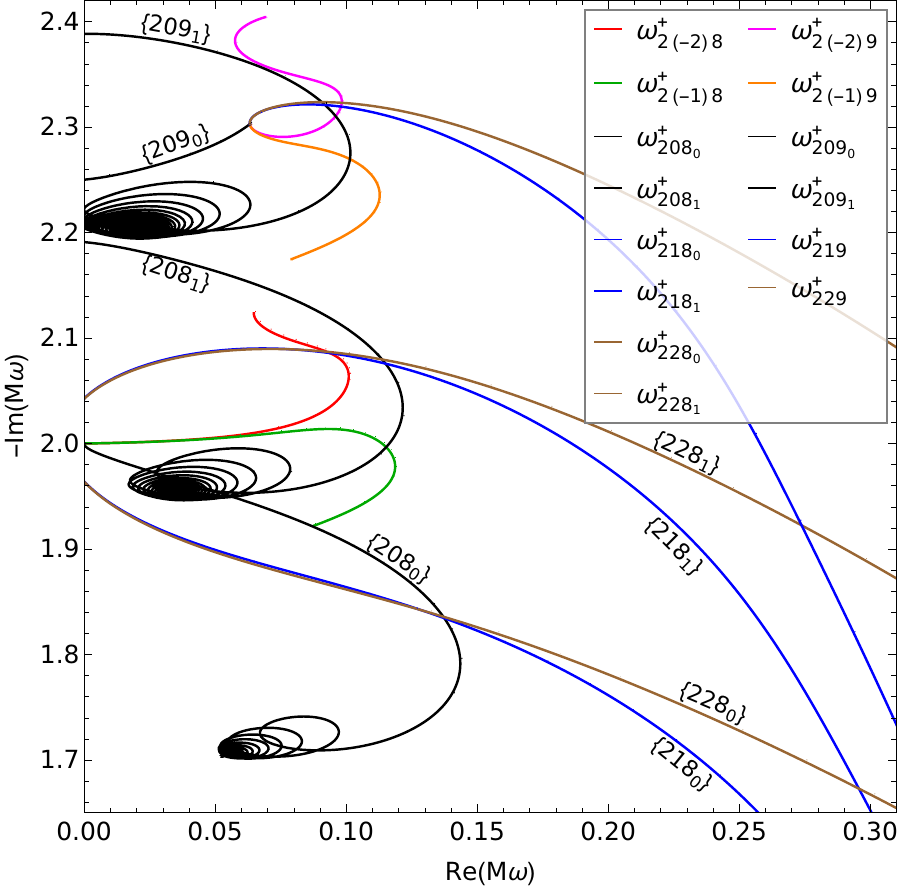}
    \caption{Examples of the exceptional behavior of the complex mode frequency $M\omega$ for the gravitational ($s=-2$), quadrupolar ($\ell=2$) QNM overtones $n=8$ and $9$.  For $n=8$, the $m=1$ and $2$ sequences occur as overtone multiplets, and none of these $4$ sequences terminate at the Schwarzschild limit of $M\omega=-2i$.  There are also two $m=0$ sequences.  The $\omega^+_{208_0}$ sequence does terminate at the Schwarzschild limit, but the $\omega^+_{208_1}$ sequence begins on the NIA with $a\approx0.32M$ as a polynomial mode which is simultaneously a QNM and a TTM${}_L$.  For the $n=9$ sequences, all of the $m\ne0$ sequences are nonexceptional.  However the $m=0$ sequence is noncontinuous and exceptional.  The initial $\omega^+_{209_0}$ portion of the sequence terminates on the NIA as a polynomial QNM at $a\approx0.31M$, then $\omega^+_{209_1}$ re-emerges from the NIA at $a\approx0.40M$ as a polynomial mode which is simultaneously a QNM and a TTM${}_L$.  The $\omega^+_{209_1}$ sequence soon begins to undertake many loops,  the first $7$ of which become tangent to the NIA.  Each of these points of tangency does not represent either a QNM or a TTM. 
    Figure adapted from~\cite{Cook:2016fge}.}
    \label{fig:QNMl2allmn8-9}
\end{figure*}

A first example of exceptional behavior is seen in the right panel of Fig.~\ref{fig:QNMl2behavior}, where the $n=8$ Schwarzschild limit appears to be near $M\omega=-2i$.  There are two sequences displayed in Fig.~\ref{fig:QNMl2behavior} which approach this limit ($\omega^+_{228_0}$ and $\omega^+_{228_1}$), however neither fully extend to the Schwarzschild limit.

The exceptional behavior near $M\omega=-2i$ is further illustrated in Fig.~\ref{fig:QNMl2allmn8-9}, where we see that the $\omega^+_{2m8}$ mode frequencies with $m<0$ do appear to begin at $M\omega=-2i$, as does the $m=0$ mode labeled as $\omega^+_{208_0}$.  However, there are two additional modes, $\omega^+_{218_{0,1}}$ which have exceptional behavior similar to that seen for the two modes $\omega^+_{228_{0,1}}$.  Hints at this behavior were first seen by Leaver~\cite{Leaver:1985ax}, and then with increasing detail in Refs.~\cite{Onozawa:1996ux,Berti:2003jh,Cook:2014cta}.    This exceptional behavior motivated the use of overtone multiplet notation~\cite{Cook:2014cta} where some overtones obtain a subscript.  In this example, the notation allows for two sets of $m\ge0$ sequences to be associated with the $n=8$ Schwarzschild limit.  A second use for multiplet notation arises when a sequence such as $\omega^+_{209_{0,1}}$ has two noncontinuous segments.  These exceptional behaviors will be considered in more detail below, but they are tied to the behavior of the TTMs.
The peculiar behavior of the $\omega^+_{225}$ mode frequency near $a=0.9$ visible in the right panel of Fig.~\ref{fig:QNMl2behavior} deserves a separate discussion (see Section~\ref{sec:avoidance} below).

The gravitational TTMs of Kerr were first considered by Wald~\cite{wald1973perturbations} and in detail by Chandrasekhar~\cite{Chandrasekhar:1984mgh} in the context of ``algebraically special'' perturbations, for which only one of the complex Weyl scalars $\Psi_0$ or $\Psi_4$ can be nonzero. Chandrasekhar showed that the TTM${}_L$s (for which $\Psi_4$ is nonzero) and the TTM${}_R$s (for which $\Psi_0$ is nonzero) exist only if the Starobinsky constant $|\mathcal{Q}|^2$ vanishes, where
\begin{align}\label{eq:Starobinsky_const}
  |\mathcal{Q}|^2 =\lambdabar^2(\lambdabar+2)^2 
  &+ 8\lambdabar{a}\omega\left[
      6(a\omega+m) -5\lambdabar(a\omega-m)\right]
      \nonumber\\ \mbox{}&
      + 144\omega^2\left[M^2+a^2(a\omega-m)^2\right].
\end{align}

For TTM${}_{\rm L}$s, $s=-2$ and
\begin{equation}
  \lambdabar=\lambdabar_-\equiv \scA{-2}{\ell{m}}{a\omega}
     + a^2\omega^2 - 2ma\omega.
\end{equation}

For TTM${}_{\rm R}$s, $s=+2$ and
\begin{equation}
  \lambdabar=\lambdabar_+\equiv \scA{2}{\ell{m}}{a\omega}
     + a^2\omega^2 - 2ma\omega+4.
\end{equation}

Note that $\lambdabar$ in Eq.~\eqref{eq:Starobinsky_const} is the same as Eq.~\eqref{eq:lambdabar def} for $s=-2$, but not for $s=2$.  This is a notational carryover from~\cite{Chandrasekhar:1984mgh}.
However, because of Eq.~\eqref{eq:swSF_sA_ident}, it follows that $\lambdabar_+=\lambdabar_-$, and we find that the TTM${}_{\rm L}$ and TTM${}_{\rm R}$ algebraically special modes share the same frequency spectrum.

\begin{figure*}[t]
    \centering
    \includegraphics[width=0.316\linewidth]{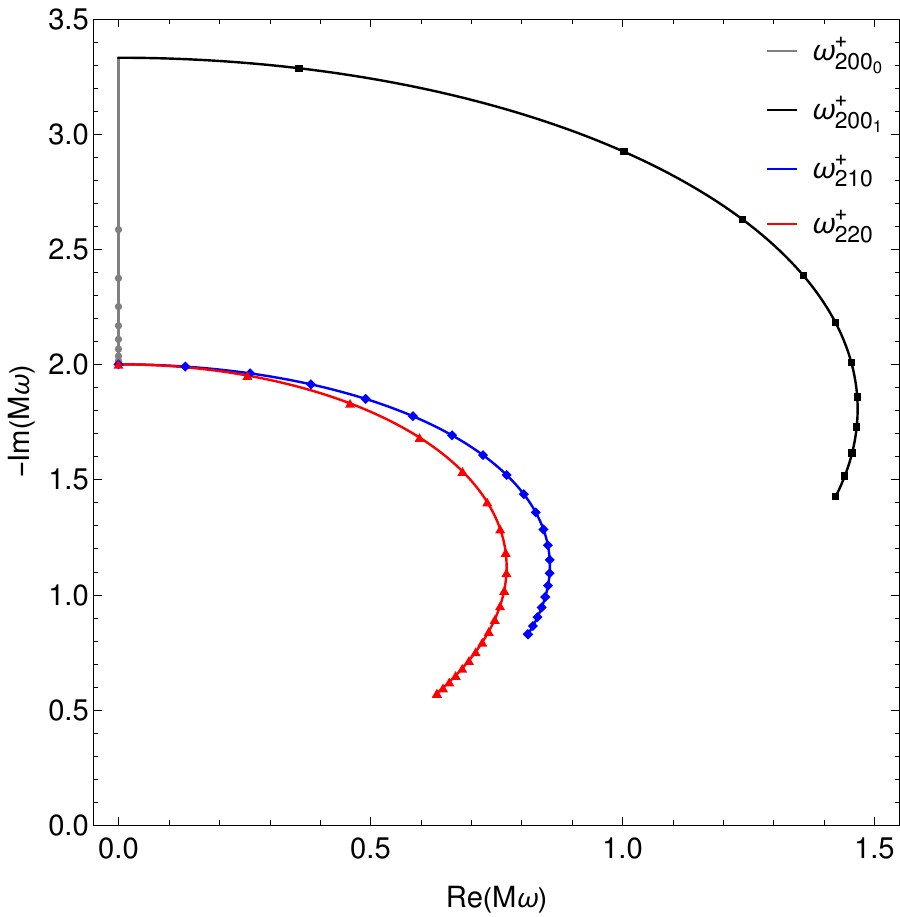} \hspace{4pt}
    \includegraphics[width=0.316\linewidth]{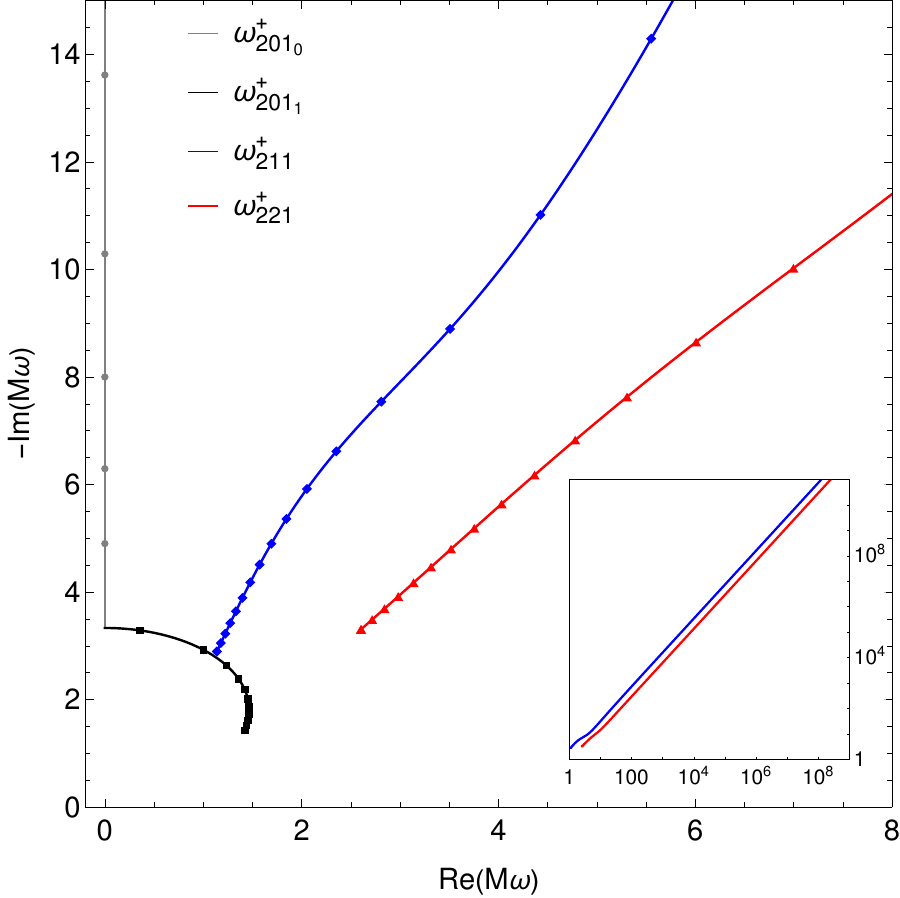}\hspace*{4pt}
    \includegraphics[width=0.316\linewidth]{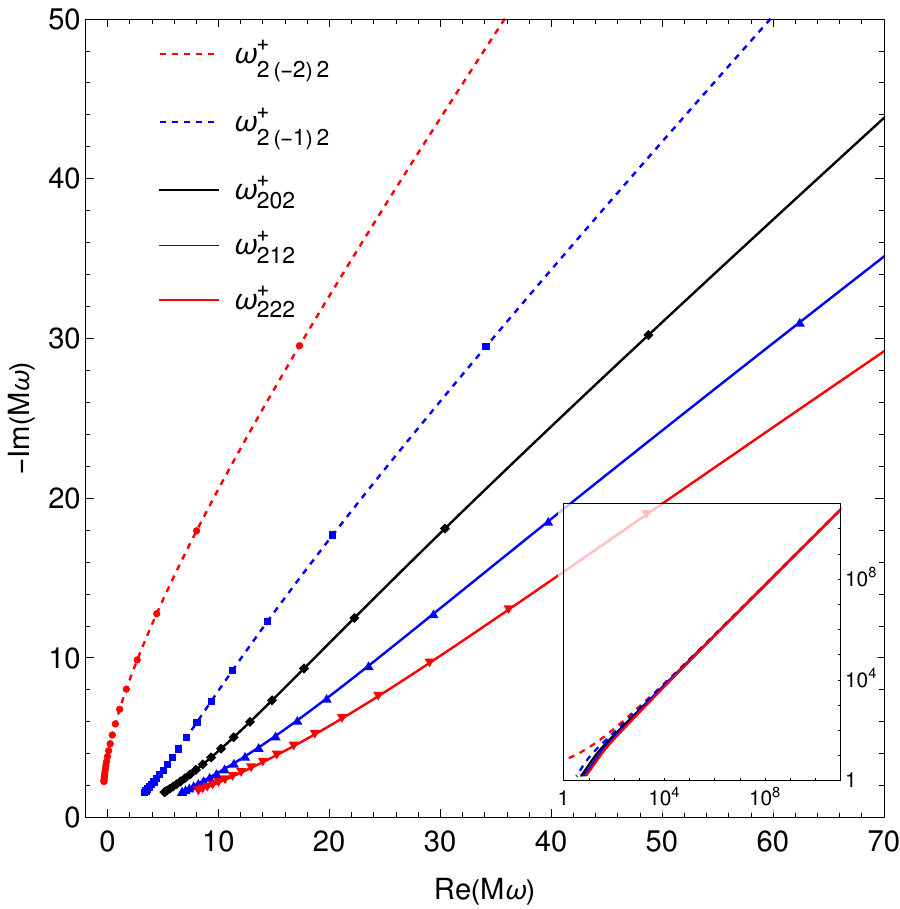}
    \caption{Examples of the complex mode frequency $M\omega$ for the gravitational ($s=-2$) TTM${}_L$s for $\ell=2$.  The left plot with $n=0$ illustrates the family of TTMs which connect to the Schwarzschild TTM frequencies $\Omega_\ell$.  The next two plots with $n=1$ and $n=2$ illustrate the 2 families of TTMs whose Schwarzschild limit frequencies are at complex infinity.  The insets in each of these plots are log-log plots showing the behavior of the sequences for large $M\omega$.  Note that for the first 2 families, only the $m\ge0$ sequences are plotted, because the two mirror-mode families are degenerate.
    Figure adapted from~\cite{Cook:2022kbb}.}
    \label{fig:TTMLl2allmn}
\end{figure*}
In the Schwarzschild limit, the angular separation constant has the form $\scA{s}{\ell{m}}{0}=\ell(\ell+1)-s(s+1)$. Then the TTM frequencies can be computed analytically as 
\begin{align}\label{eq:sch_alg_spec_modes}
    M\omega \equiv M\Omega_\ell = -\frac{i}{12}(\ell-1)\ell(\ell+1)(\ell+2),
\end{align}
and we note that $M\Omega_2=-2i$.  For $a>0$, computing the Starobinsky constant must be coupled with solving the angular equation~\eqref{eq:swSF_DiffEqn}, and the TTM frequencies must be determined numerically.  There are 3 families of TTMs known at this time, and the $\ell=2$ mode frequencies of these 3 families are shown in Fig.~\ref{fig:TTMLl2allmn}.  The left panel displays the TTMs of the family that connects directly to $M\Omega_\ell$ in the Schwarzschild limit.  This family of solutions was first explored by Chandrasekhar~\cite{Chandrasekhar:1984mgh} and then in Refs.~\cite{Onozawa:1996ux,Cook:2014cta}, and we label members of this family with overtone $n=0$.  However, two additional families of solutions, labeled respectively with overtones $n=1$ and $n=2$, were recently discovered~\cite{Cook:2018ses,Cook:2022kbb} where the mode frequency in the Schwarzschild limit resides at complex infinity.  The center and right panels display the $\ell=2$ mode frequencies of these two additional families.  Interestingly, the $n=0$ and $n=1$ TTM families share the property that the mirror-mode solutions are degenerate~\cite{Cook:2022kbb}, i.e., $\omega^-_{\ell{m}(0,1)}=\omega^+_{\ell{m}(0,1)}$.  Because of this, a common perspective is to assume that only the $m\ge0$ TTMs exist for $\omega^+_{\ell{m}(0,1)}$, and only the $m\le0$ TTMs exist for $\omega^-_{\ell{m}(0,1)}$.
\subsubsection{Solving the mode equations}\label{sec:Solving for modes}
A unified framework for solving the mode equations is found by recognizing that both Eqs.~(\ref{eq:radialR:Diff_Eqn}) and (\ref{eq:swSF_DiffEqn}) are confluent Heun equations~\cite{ronveaux1995heun,blandin1983general,Batic:2007it,Fiziev:2009wn}.  This approach is particularly fruitful for understanding solutions of the radial equation~\cite{Cook:2014cta}. All mode solutions for subextremal Kerr BHs are in the form of confluent Heun functions which are simultaneously local Frobenius solutions of the regular singular point at the event horizon, and of the irregular singular point at infinity. Furthermore, some mode solutions are in the form of confluent Heun polynomials, which are simultaneously local solutions at all 3 singular points, including the regular singular point at the Cauchy horizon.

The most fruitful approach to solving the radial equation is known as Leaver's method~\cite{Leaver:1985ax}, based on Pincherle's theorem~\cite{Gautschi:1967cat}, which addresses the convergence of continued fractions. This approach is usually augmented with an improvement due to Nollert~\cite{Nollert:1993zz,Cook:2014cta}. When the BH is extremal, the event horizon becomes an irregular singular point and the radial equation transforms into a double confluent Heun equation. As a result, Leaver's method must be modified to accurately determine the QNMs in this case~\cite{Onozawa:1995vu,Richartz:2015saa}. All of the puzzling behaviors observed in the QNM spectrum, some illustrated in Fig.~\ref{fig:QNMl2allmn8-9}, can be understood from the fact that QNM solutions on the negative imaginary axis (NIA) are only allowed in the form of confluent Heun polynomials, and by understanding the detailed behaviors of these polynomial solutions~\cite{Cook:2016ngj,Ansorg:2016ztf}.
We will discuss this further below. Furthermore, TTMs are found to be a particular type of confluent Heun polynomial solution. The necessary conditions for the existence of these polynomial solutions for TTMs yield exactly the vanishing of the Starobinsky constant, with no consideration given to the algebraically special nature of these modes.

Although the angular equation is also of confluent Heun type and can be solved using continued fraction methods, a spectral method for solving the angular equation~\cite{Cook:2014cta} has proven to be more useful.  The added benefit comes from the fact that the spherical-spheroidal expansion coefficients are computed as part of the solution.  These coefficients can be important when the QNM solutions are used to perform ringdown fitting.

The high-overtone (large-$n$) limit and the large angular number (large-$\ell$) limit of the QNM frequencies are known and discussed in detail in~\cite{Berti:2009kk}. The large-$\ell$ limit is closely connected to null particle geodesics, and it has an elegant geometric interpretation~\cite{Ferrari:1984zz,Cardoso:2008bp,Dolan:2010wr,Yang:2012he}, as discussed in Section~\ref{subsec:LR} below.
The large-$n$ limit has been thoroughly explored for nonspinning BHs, where interesting connections to the BH area quantization program were established.
The asymptotic spectrum of spinning BHs is tightly connected to the asymptotics of the angular spin-weighted spheroidal functions. 
The state of the art on these limits as of 2009 is discussed in~\cite{Berti:2009kk}. For more recent developments on this subject,  see e.g.~\cite{Keshet:2012uq,Hod:2012dtv,Hod:2013sna,Casals:2018cgx,Vickers:2022ivk,Cook:2022kbb}.

\subsubsection{Purely damped modes}
As illustrated in Fig.~\ref{fig:QNMl2allmn8-9}, the sequences of QNM frequencies can exhibit unusual behavior near and on the NIA where the mode frequencies become purely damped. The behavior of QNMs near Schwarzschild TTMs (located at the mode frequencies of $\Omega_\ell$) was resolved by Maassen van den Brink~\cite{MaassenvandenBrink:2000iwh}. Cook and Zalutskiy~\cite{Cook:2016ngj} then combined this approach with confluent Heun theory to extend the analysis to all modes of Kerr. The QNM frequencies can only be purely imaginary if the radial mode function is polynomial. From Heun theory, there are two possible ways for a polynomial QNM solution to exist. A necessary, but not sufficient, condition is satisfied if a mode frequency obeys either
\begin{equation}\label{eq:omega_plus}
M\omega=M\omega_+ \equiv
\frac{am-iN_+\sqrt{M^2-a^2}}{2(M+\sqrt{M^2-a^2})},
\end{equation}
where $N_+\ge s+1$ are either integers or half-odd integers, or 
\begin{equation}\label{eq:omega_minus}
M\omega=M\omega_- \equiv
-i\frac{N_-}4,
\end{equation}
where $N_- \ge1$ is an integer.  If a mode might be polynomial, a necessary and sufficient condition must be checked to determine if the mode is actually polynomial~\cite{Cook:2016ngj}.  
The full analysis is complicated, but the results are straightforward. Polynomial QNMs have only been found for $m=0$ (unless $a=0$, so that the different $m$ modes are degenerate).  Any mode with $m\ne0$ appearing to approach the NIA with $a\ne0$ must terminate, and no QNM along the sequence exists exactly on the NIA.  This follows from the fact that only confluent Heun polynomial QNMs can exist precisely on the NIA~\cite{Cook:2016ngj}.  Examples of this behavior are the $\omega^+_{218_{0,1}}$ and $\omega^+_{228_{0,1}}$ sequences in Fig.~\ref{fig:QNMl2allmn8-9}.  All known $m=0$ QNM sequences that approach (leave) the NIA with a limiting frequency obeying Eq.~\eqref{eq:omega_minus} terminate (begin) on the NIA as polynomial QNMs with some value of $a>0$.  An example of this behavior is the $\omega^+_{209_0}$ sequence in Fig.~\ref{fig:QNMl2allmn8-9}, which terminates at $M\omega=-9i/4$ with $a\approx0.31M$.  The behavior for all known $m=0$ QNM sequences that obey Eq.~\eqref{eq:omega_plus} is slightly more complicated.  In Fig.~\ref{fig:QNMl2allmn8-9}, the $\omega^+_{209_1}$ sequence contains 8 points which satisfy Eq.~\eqref{eq:omega_plus}.  The first leaves the NIA at $M\omega\approx-2.39i$ and $a\approx0.40M$, while the remaining 7 are points where the sequence becomes tangent to the NIA near $M\omega\approx-2.21i$.  All known $m=0$ QNM sequences that approach (leave) the NIA with a limiting frequency obeying Eq.~\eqref{eq:omega_plus} and terminate (begin) on the NIA as polynomial modes are simultaneously QNM and TTM${}_L$, while the corresponding TTM${}_R$ does not exist.  However, all known points of tangency with the NIA along $m=0$ QNM sequences have frequencies which also obey Eq.~\eqref{eq:omega_plus}. In all cases, the points of tangency are neither QNMs nor TTMs.  This more complicated behavior arises from the fact that when Eq.~\eqref{eq:omega_plus} is satisfied, the roots of the indicial equation associated with the singular point at the event horizon differ by an integer.

\subsubsection{Near extremal regime: branching and stability}\label{sec:near_extremal_branching}

As previously discussed in Section~\ref{sec:QNMsTTMs}, many sequences of QNM frequencies converge to the real value $M\omega = m/2$ in the extremal limit. However, other sequences may exist (such as $\omega_{225}$ in Fig.~\ref{fig:QNMl2behavior}) with limiting frequencies that possess finite, nonzero imaginary parts. These two distinct sets of modes are referred to as zero-damping modes (ZDMs) and damped modes, respectively~\cite{Yang:2012pj,Yang:2013uba}. The ZDMs exist for all $\ell$ and $m \neq 0$, with their frequencies accumulating on the real axis as the extremal limit is approached. In contrast, damped modes exhibit nonzero damping for all BH spins. They occur for all retrograde modes and for prograde modes that satisfy $0 \leq  \left|m/(l+1/2)\right| \lesssim 0.74$. When the two families coexist, they merge into a single set of QNMs at sufficiently low BH spins. However, as the spin increases towards extremality, ZDMs and damped modes cross each other in the complex $\omega$ plane, as observed in Figs.~\ref{fig:QNMl2behavior} and \ref{fig:QNMl34}.
ZDMs and damped modes have fundamentally different origins. While ZDMs arise from the near-horizon geometry of nearly extremal BHs, damped modes are associated with the potential barrier surrounding the BH~\cite{Yang:2012pj,Yang:2013uba}. Consequently, ZDMs are also referred to as photon sphere modes, whereas damped modes are known as near-horizon modes (see their generalization for Kerr-Newman BHs in Section~\ref{sec:KNqnm}).

The ZDMs have been associated with parametric instabilities that resemble the inverse energy cascade of turbulent fluids~\cite{Yang:2014tla}. Additionally, for near-extremal Kerr BHs
the collective excitation of an infinite number of ZDMs leads to a transient 
tail 
at null infinity~\cite{Yang:2013uba} (see Section~\ref{sec:tails}) and to a
transient power-law horizon instability~\cite{Gralla:2016sxp}. The transient
behavior turns eternal in the extremal limit, where the limiting ZDM frequencies
$M\omega=m/2$ become branch points in the frequency-domain (FD) Green's
function~\cite{Casals:2016mel, Richartz:2017qep, Casals:2019vdb}. In fact, even
though Kerr BHs have been proven to be mode-stable~\cite{Whiting:1988vc,
  TeixeiradaCosta:2019skg}, extremal Kerr BHs are linearly
unstable~\cite{Aretakis:2011gz, Lucietti:2012sf, Aretakis:2012ei}. While mode
stability relies on the absence of exponentially growing modes, linear stability
concerns the existence of perturbations of generic initial data that grow in
time.  The polynomial, rather than exponential, nature of the instability of
extremal Kerr BHs can be attributed to the ZDM branch points~\cite{Casals:2016mel, Gralla:2016sxp, Richartz:2017qep, Casals:2019vdb}. For
discussions on linear and nonlinear stability in the subextremal Kerr regime,
see~\cite{Klainerman:2022ric, Shlapentokh-Rothman:2023bwo} and the references
therein.

\subsubsection{Public codes and data}

Extensive sets of gravitational Kerr QNM and TTM solutions are freely
available~\cite{GRIT, CoG, JHU, PaniRome, cook_2024_14024959,
  motohashi_2024_12696858, Lo:2025njp}, along with the \texttt{Mathematica}
routines used to compute them~\cite{GRIT, CoG, JHU, PaniRome,
  KerrModes_2024_cook}.  An open-source \texttt{python} package for computing
QNMs is also available~\cite{Stein:2019mop}. In Appendix~\ref{sec:public_codes}
we provide a list of these and other public resources.
\subsection{Kerr-Newman spectrum and eigenvalue repulsion}
\label{sec:KNqnm}

\vspace{-.1cm}

\noindent \textit{Initial contributors: Dias, Santos}

\vspace{.2cm}

As we have seen, the perturbations of rotating (Kerr) BHs are separable, and
they can be reduced to the solution of a pair of angular and radial ordinary
differential equations (ODEs).
The situation is more complex in the general (Kerr-Newman) case where both the charge $Q$ and the angular momentum $J$ are nonzero. This is the subject of the present section.

Astrophysical BHs are expected to discharge very rapidly~\cite{Gibbons:1975kk,Blandford:1977ds,Palenzuela:2011es}, and therefore they should be well approximated by the Kerr solution.
The Kerr-Newman solution is still a useful theoretical laboratory to understand QNMs and ringdown beyond the Kerr metric.
Complex new phenomena that appear in the Kerr-Newman spacetime, like eigenvalue repulsion (Section~\ref{subsubsec:eigen_rep}), have analogues in other systems. 
The efficiency of the discharge mechanism depends on the properties of Standard Model particles~\cite{Cardoso:2016olt}, and potential observational signatures of nonzero charge are a robust pathway to uncover new physics (see Section~\ref{subsec:TGR_theory-specific}).

\subsubsection{Formalism}
So far, it has not been possible to describe the most general
gravito-electromagnetic perturbations of a Kerr-Newman BH by a single partial
differential equation (PDE). {\em A priori}, one expects the perturbed Einstein-Maxwell equation around Kerr-Newman to be described by a system of nine coupled PDEs. However, Chandrasekhar demonstrated that, within the Newman-Penrose formalism~\cite{Newman:1961qr,Geroch:1973am,Chandrasekhar:1985kt} and in the so-called {\it phantom gauge} (whereby two perturbed complex Weyl scalars are set to zero), the perturbation problem ``simply'' reduces to two coupled PDEs~\cite{Chandrasekhar:1985kt,Mark:2014aja}. Such a remarkable complexity reduction is a consequence of the fact that Kerr-Newman is a Petrov type~D solution. More recently, it has been shown that generic gravito-electromagnetic perturbations of Kerr-Newman BHs (except for those that change the mass and angular momentum of the solution~\cite{wald1973perturbations}) are described by a coupled system of two PDEs for two {\it gauge-invariant} Newman-Penrose fields~\cite{Dias:2015wqa,Dias:2022oqm}.
The two gauge-invariant Newman-Penrose fields (invariant under infinitesimal diffeomorphisms and tetrad rotations) are given by~\cite{Dias:2015wqa,Dias:2022oqm} 
\begin{align}\label{KNgaugeInvariant}
\widetilde{\psi}_{-2}= \left(\bar{r}^*\right)^4 \Psi_4^{(1)},  \qquad 
\widetilde{\psi}_{-1}=\frac{\left(\bar{r}^*\right)^3}{2\sqrt{2}\Phi_1^{(0)}} \left(2\Phi_1^{(0)}\Psi_3^{(1)} -3 \Psi_2^{(0)}\Phi_2^{(1)}\right).
\end{align}
where $\Phi_{1,2}$ and $\Psi_{2,3,4}$ are the standard complex Weyl scalars with the superscripts $^{(0)}$ and $^{(1)}$ standing for unperturbed and perturbed quantities, respectively, and $\bar{r} = r+ia\cos \theta$~\cite{Chandrasekhar:1985kt,Dias:2015wqa,Dias:2022oqm}. The two coupled PDEs for these master variables $\widetilde{\psi}_{-2}$ and $\widetilde{\psi}_{-1}$ are~\cite{Dias:2015wqa,Dias:2022oqm} %
\begin{align}\label{GaugeInvEqs}
& \left(\mathcal{F}_{-2}+ Q^2 \mathcal{G}_{-2}\right) \widetilde{\psi}_{-2}  + Q^2 \mathcal{H}_{-2}  \widetilde{\psi}_{-1} =0 \,, \nonumber \\
& \left(\mathcal{F}_{-1} +Q^2 \mathcal{G}_{-1}\right)\widetilde{\psi}_{-1} + Q^2  \mathcal{H}_{-1} \widetilde{\psi}_{-2}=0  \,, 
\end{align}
where the differential operators $\mathcal{F}_s, \mathcal{G}_s, \mathcal{H}_s$ ($s=-2,-1$) are functions of the standard Chandrasekhar radial and angular operators  $\mathcal{D}_j$, $\mathcal{L}_j$ ($j=-2,-1,0,1,2$) and their complex conjugates, as defined in Eqs.~(2.24) and (2.25) of~\cite{Dias:2022oqm}.
After gauge fixing, this pair of PDEs reduces to the coupled PDE system in the phantom gauge originally found by Chandrasekhar~\cite{Chandrasekhar:1985kt,Mark:2014aja}. When $Q=0$, the pair of coupled PDEs {\it decouples} and one gets the familiar Teukolsky equations for the decoupled gravitational and electromagnetic perturbations in Kerr~\cite{Teukolsky:1972my}. Note that  the Newman-Penrose master variables $\widetilde{\psi}_{-2}$ and $\widetilde{\psi}_{-1}$ are relevant for the study of perturbations that are regular at the future horizon and outgoing at future null infinity. There is a companion set of two coupled PDEs for $\widetilde{\psi}_{2}$ and $\widetilde{\psi}_{1}$, which are the positive-spin counterparts of $\widetilde{\psi}_{-2}$ and $\widetilde{\psi}_{-1}$. These are relevant when we are interested in perturbations that are outgoing at past null infinity and regular at the past horizon.

\subsubsection{The spectrum}
The Kerr-Newman PDE system~\eqref{GaugeInvEqs}  allows to understand straightforwardly the {\it isospectrality} property of the QNM spectrum of Schwarzschild and Reissner-Nordstr\"om BHs~\cite{Dias:2015wqa,Dias:2022oqm}, namely the fact that frequency spectra of the Regge-Wheeler (a.k.a. axial or odd)~\cite{Regge:1957td} and Zerilli (a.k.a. polar or even)~\cite{Zerilli:1974ai} QNMs are the same~\cite{Chandrasekhar:1985kt}.  
In the Schwarzschild limit, the Kerr-Newman PDE pair~\eqref{GaugeInvEqs}  decouples, and there is a differential map that gives Regge-Wheeler's  $\Phi_{\hbox{\tiny RW}}$ and Zerilli's $\Phi_{\hbox{\tiny Z}}$ master variables in terms of the gauge invariant field $\widetilde{\psi}_{-2}$ (gravitational perturbations) or $\widetilde{\psi}_{-1}$ (electromagnetic perturbations)~\cite{Chandrasekhar:1975nkd,Sasaki:1981,Dias:2013sdc}. That is, the study of a single decoupled pair of Teukolsky ODEs~\cite{Teukolsky:1972my} in the Schwarzschild BH yields, simultaneously, the Regge-Wheeler and the Zerilli QNMs. Isospectrality of these two last modes is thus inevitable~\cite{Dias:2015wqa,Dias:2022oqm} (a similar discussion and conclusion applies to Reissner-Nordstr\"om).

The coupled pair of PDEs~\eqref{GaugeInvEqs} for gravito-electromagnetic perturbations of Kerr-Newman must be supplemented with appropriate (physical) boundary conditions to have a well-posed boundary value problem. 
As usual for QNM problems, at the future event horizon only regular modes are kept in ingoing Eddington-Finkelstein coordinates and, at spatial infinity, we require only outgoing waves. Moreover, we must require regularity along the axis generated by $\partial/\partial \phi$~\cite{Dias:2015wqa,Dias:2021yju,Dias:2022oqm}. 

The system has a scaling symmetry that reduces the physical parameter space. Namely, the equations of motion are left invariant if we scale the metric and Maxwell field strength as $g_{\mu \nu}\to \lambda^2 g_{\mu \nu}$ and $F_{\mu \nu}\to \lambda F_{\mu \nu}$, for an arbitrary constant $\lambda$.
This means that we can scale out one of the 3 parameters of the solution and, to find the frequency spectrum,  we  ``just'' need to scan a 2-dimensional space; e.g. we find $\omega M$ for each dimensionless pair of parameters $\{a/M,Q/M\}$ (or $\{a/r_+,Q/r_+\}$).
The modes are Fourier decomposed as $\widetilde{\psi}_s=\widetilde{\psi}_s(r,\theta)e^{-i\omega t}e^{i m \phi}$ ($s=-2,-1$), and the $t -\phi$ symmetry of the background allows us to consider only modes with  Re$\,\omega \geq 0$, as long as we study both signs of $m$. The QNM spectrum can be organized in terms of a set of three quantum numbers $\{\ell,m,n \}$, where the harmonic number $\ell$ is related to the number of zeros of the eigenfunctions along the polar direction and $n$ is the radial overtone (related to the number of zeros of the eigenfunctions along the radial direction).
In the limit where $J\to 0$ or $Q\to 0$, $\ell$ matches the quantum number of the spherical (spheroidal) harmonics for Reissner-Nordstr\"om (Kerr).

To solve numerically the eigenvalue problem for $\widetilde{\psi}_s(r,\theta)$
($s=-2,-1$) and $\omega$, one can use a {\it pseudospectral collocation method}
that searches directly for specific QNMs using a {\it Newton-Raphson
  root-finding algorithm}. See the review~\cite{Dias:2015nua} for details on the
use of these methods on linear and nonlinear gravitational systems, while~\cite{Trefethen,Boyd,CanutoBook} are excellent references introducing
pseudospectral methods for generic systems (often, when one wants to find
several radial overtones simultaneously, it is also useful to solve the
eigenfrequency problem using \texttt{Mathematica}'s built-in routine
\texttt{Eigensystem} before using the Newton-Raphson root-finding algorithm to
follow a specific family of modes across the BH phase space). The use of such
pseudospectral methods (supplemented with the Newton-Raphson algorithm and/or
\texttt{Mathematica}'s built-in routine) to find QNMs and unstable modes of BHs
dates back to 2009 and has been proven to be extremely efficient and
accurate~\cite{Dias:2009iu, Dias:2010eu, Dias:2010maa, Dias:2010gk, Dias:2011jg,
  Dias:2010ma, Dias:2011tj, Cardoso:2013pza, Dias:2014eua, Dias:2018etb}.  The
exponential convergence of the method, the fact that derivatives are computed
using all the grid points, and the use of quadruple precision, guarantee that
the results of~\cite{Dias:2015wqa,Dias:2021yju,Dias:2022oqm} are accurate up to,
at least, the eighth decimal place (note that the alternative finite-difference
grid discretization only has polynomial convergence, and derivatives are
computed using only a few nearby grid points).

The main property of the gravito-electromagnetic QNM spectrum of Kerr-Newman BHs is that no unstable linear modes were found~\cite{Dias:2015wqa,Dias:2021yju,Dias:2022oqm}. Conversely, this provides strong evidence that (up to at least 99.999\% of extremality) Kerr-Newman BHs are mode stable at the linear level within Einstein-Maxwell theory. This finding is supported by the nonlinear time evolutions and merger simulations of charged BHs that were carried out so far (see e.g.~\cite{Zilhao:2014wqa,Bozzola:2020mjx,Bozzola:2021elc,Bozzola:2022uqu,Bozzola:2023nzo}),
and further motivates a formal proof that Kerr-Newman BHs are (non)linearly stable.

\begin{figure}[t]
\centering
\includegraphics[width=.70\textwidth]{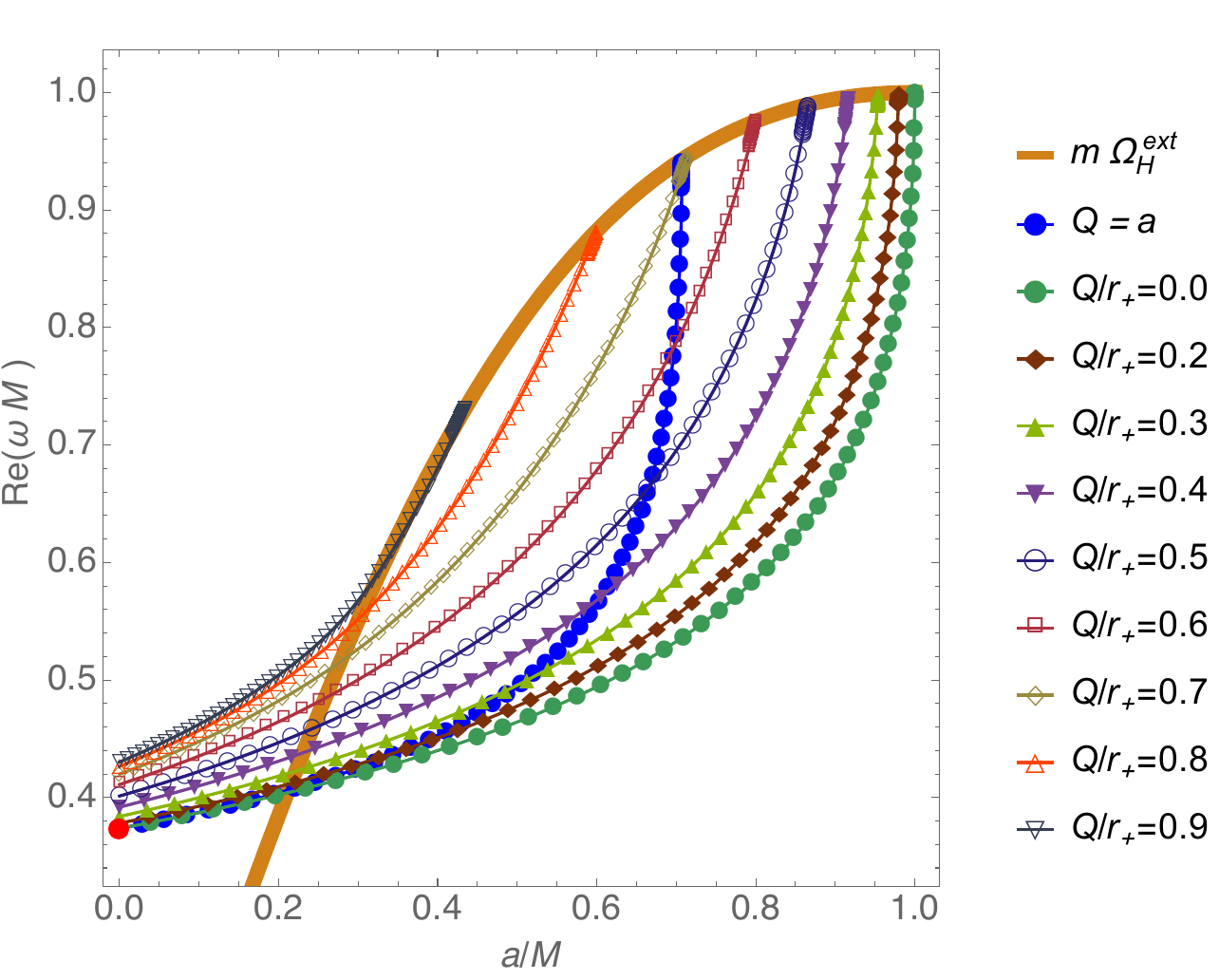}\hspace{0.5cm}
\includegraphics[width=.73\textwidth]{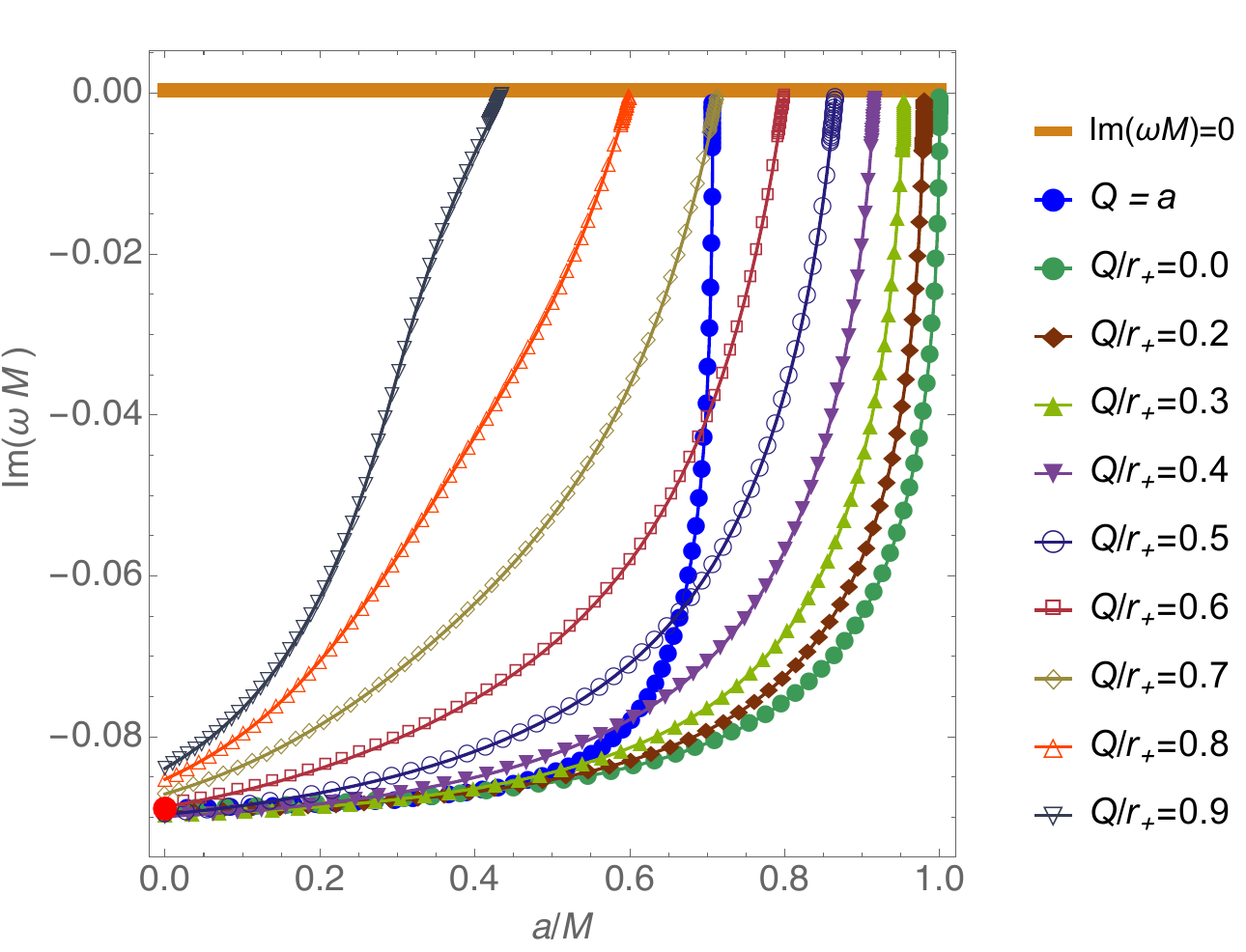}
\caption{Real (top panel) and imaginary (bottom panel) part of the gravito-electromagnetic QNM frequencies with $\ell=m=2, n=0$  of Kerr-Newman BHs with $Q=a$ and fixed $Q/r_+=0.0,0.2,\cdots,0.9$ (see legend) that start at the Schwarzschild gravitational QNM with $\ell=2, n=0$ (red disc). Figure taken from~\cite{Dias:2015wqa}. Data can be downloaded from the arXiv source of~\cite{Dias:2015wqa}.}
\label{Fig:KNl2m2}
\end{figure}

The full 3-dimensional plot ${J/M^2,Q/M,\omega M}$ for the $\ell=m=2, n=0$  gravito-electromagnetic sector, which is  potentially the most unstable because of its longed-lived nature, can be found in Fig.~2 of~\cite{Dias:2021yju} and Fig.~15 of~\cite{Dias:2022oqm};
for other values of $\ell,m$, see Figs.~18--31 of~\cite{Dias:2022oqm}.  
Here, it suffices to illustrate the Kerr-Newman linear stability by selecting a few Kerr-Newman families of BHs with constant dimensionless charge and describe what happens to the frequency of the mode with lowest $|\mathrm{Im}(\omega M)|$ as one increases the rotation $a/M$ from the Reissner-Nordstr\"om limit ($a=0$) all the way up to the extremal Kerr-Newman limit  (where the temperature vanishes and the dimensionless angular velocity is $\Omega_H=\Omega_H^{ext}$; this is $a=M$ only in the Kerr limit). This is done in Fig.~\ref{Fig:KNl2m2}. One can see that $\mathrm{Im}(\omega M)\to 0$ and $\mathrm{Re}(\omega M)\to m\Omega_H^{ext}$ (light brown curve) as the Kerr-Newman extremal solution is approached. 

Prior to the accurate numerical results of~\cite{Dias:2015wqa,Dias:2021yju,Dias:2022oqm}, hints about the gravito-electromagnetic QNM spectrum of Kerr-Newman were obtained within perturbation theory about (i) the Reissner-Nordstr\"om BH in a small rotation parameter $a$ expansion~\cite{Berti:2005eb,Pani:2013ija,Pani:2013wsa} or (ii) the Kerr BH in a small charge parameter $Q$ expansion~\cite{Mark:2014aja} (extended to second order in perturbation theory in~\cite{Blazquez-Salcedo:2022eik}). As expected, these perturbative results match well the numerical results for small rotation and charge, respectively: see Figs.~$4$ and $5$ of~\cite{Dias:2015wqa}.

\subsubsection{Eigenvalue repulsion}\label{subsubsec:eigen_rep}
A second important feature of the QNM spectra of Kerr-Newman is that it contains two main families of QNMs for a given set of quantum numbers $\{\ell,m,n \}$. These two families are denoted as (i) the {\it photon sphere}, and (ii) the {\it near-horizon} families. 
There are windows in the parameter space $\{J/M^2,Q/M \}$ where the frequency of each of these two families can be captured by perturbative expansions (namely, a large-$m$ WKB expansion  and/or a near-horizon matched asymptotic expansion): see Section~3 of~\cite{Dias:2022oqm} for a detailed discussion. This permits to identify the two families of QNMs (thus providing the basis for their nomenclature), and yields approximate analytical formulae to check the frequency results found through numerical methods. These two families are the natural extension to the rotating case ($J\neq 0$) of the photon sphere and near-horizon families of QNMs that are present in the Reissner-Nordstr\"om case (see Fig.~1 of~\cite{Dias:2022oqm}).
Which one of these families dominates the frequency spectra (i.e., which one has lower $|\mathrm{Im}\,\omega |$) depends on the region $\{J/M^2,Q/M \}$ of the parameter space. 

Interestingly, these two families can only be unambiguously distinguished in the Reissner-Nordstr\"om limit, due to a phenomenon called {\it eigenvalue repulsion}, whereby the two families can interact strongly near extremality (so much that the two families lose their original clear identity)~\cite{Dias:2021yju,Dias:2022oqm}.
Typically, two different QNM families of a BH can have eigenfrequencies that may simply cross in the real or imaginary plane (but not in both), but do not interact in any way. However, in Kerr-Newman, an intricate interaction between the photon sphere and near-horizon gravito-electromagnetic modes with the same quantum numbers $\{\ell,m,n \}$ is observed. Namely, in certain parts of the parameter space the frequencies of two QNM families approach in the complex plane very closely, without crossing (i.e., without matching in frequency), before repelling violently and moving apart again: this feature is identified in Figs.~13 and 14 of~\cite{Dias:2022oqm}. These repulsions are very strongly dependent on the BH parameters  $\{J/M^2,Q/M \}$  (a relatively minor change of the BH parameters can cause the repulsion to be absent) and they are a fundamental  structural feature  of the Kerr-Newman QNM spectrum. It helps  understanding how the latter bridges the Reissner-Nordstr\"om and Kerr cases to solve some puzzling properties of the QNM spectra of the latter two~\cite{Yang:2012pj,Yang:2013uba,Zimmerman:2015trm}. 

\begin{figure}[h]
\centering
\includegraphics[width=.49\textwidth]{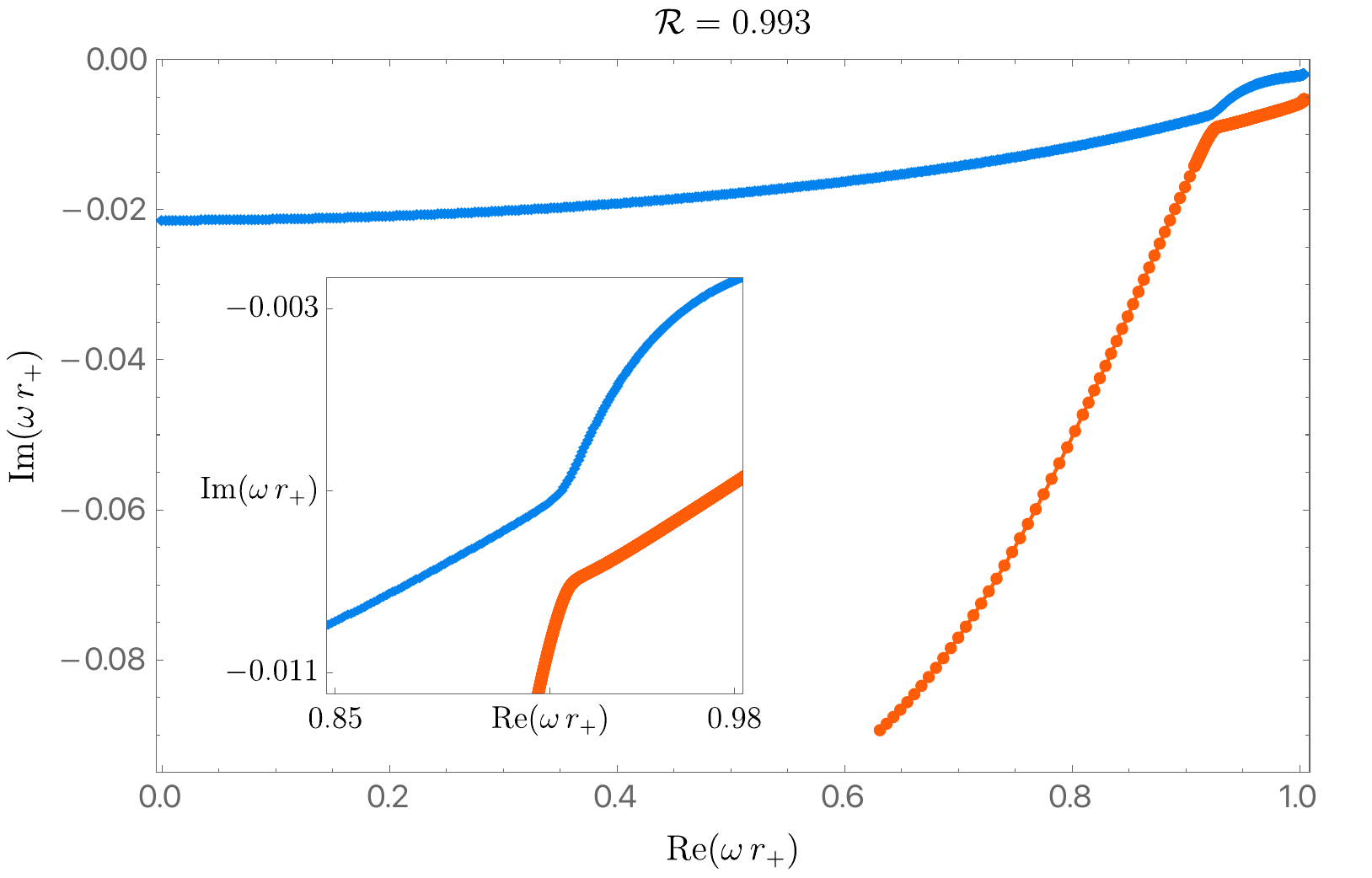}
\hspace{0.0cm}
\includegraphics[width=.49\textwidth]{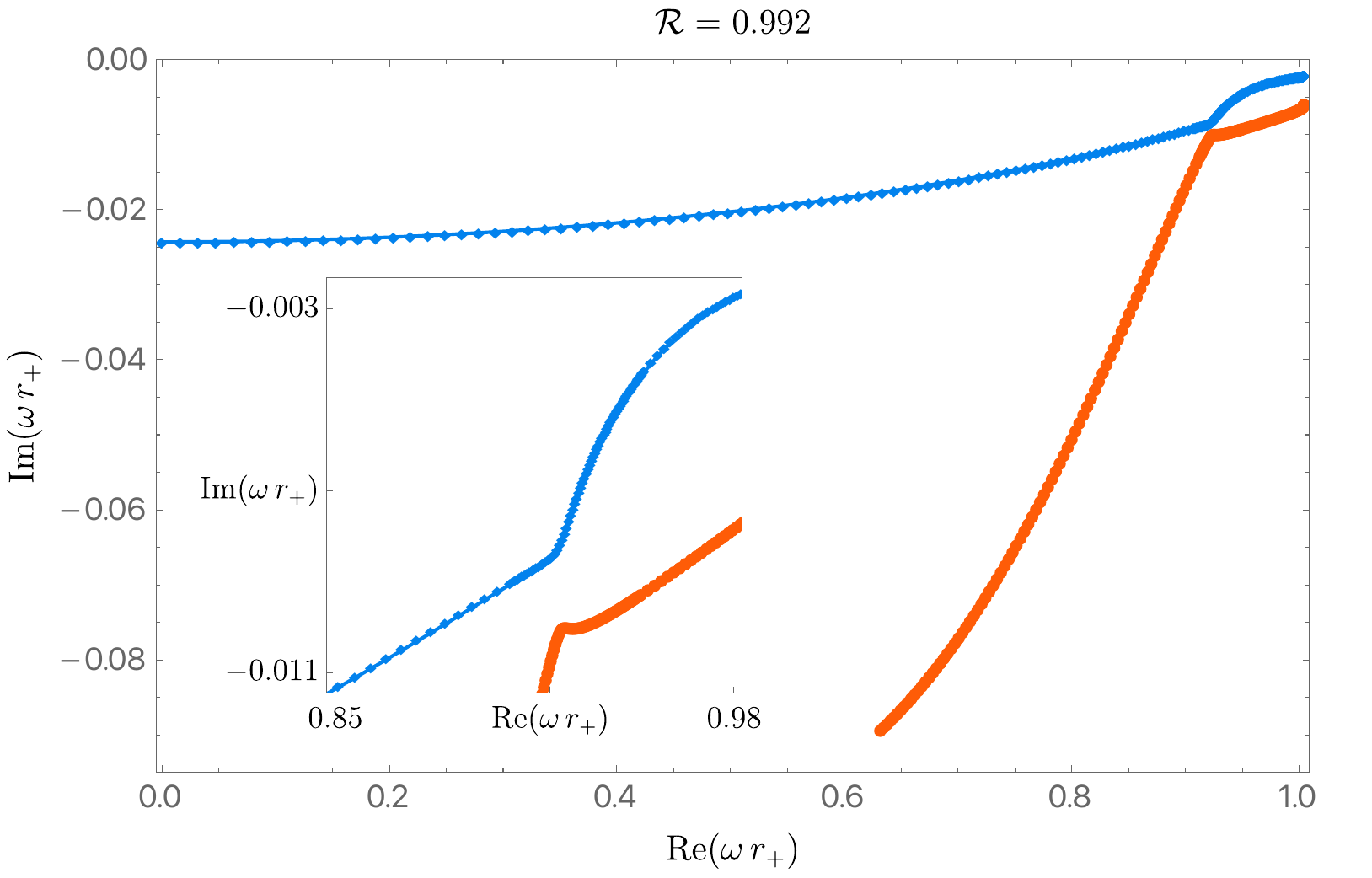}
\vskip 0.2cm
\includegraphics[width=.49\textwidth]{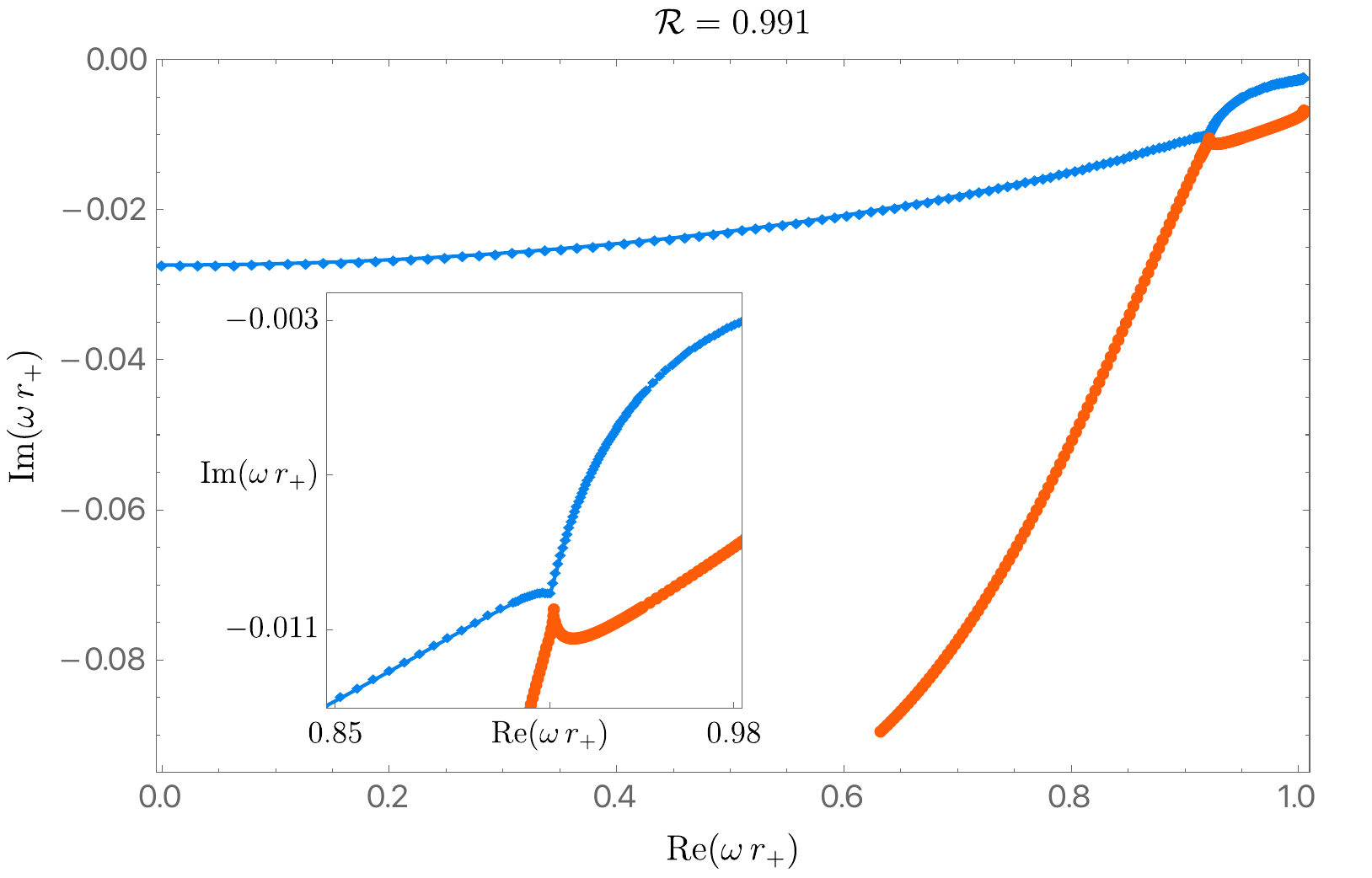}
\hspace{0.0cm}
\includegraphics[width=.49\textwidth]{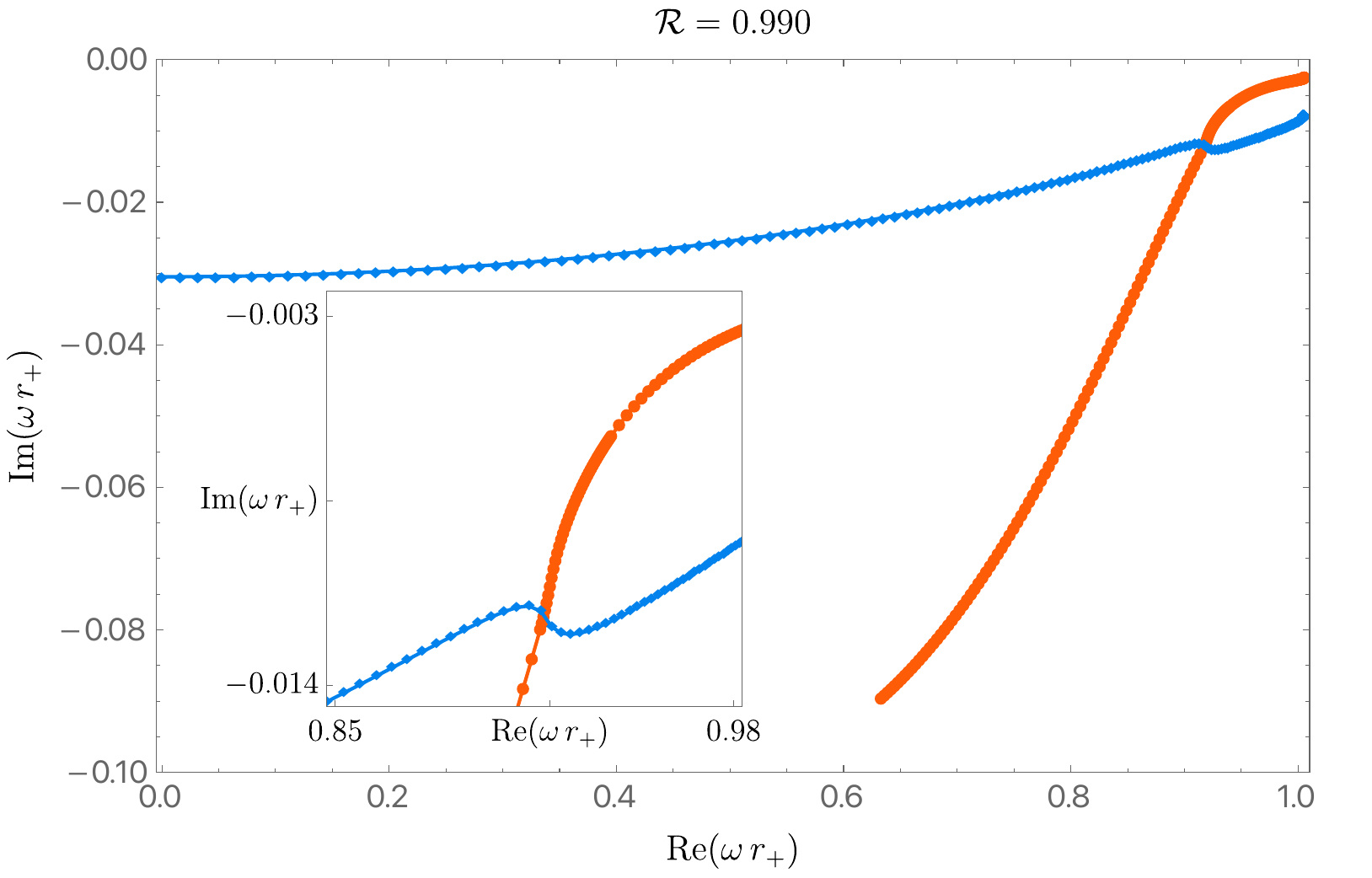}
\vskip 0.2cm
\includegraphics[width=.49\textwidth]{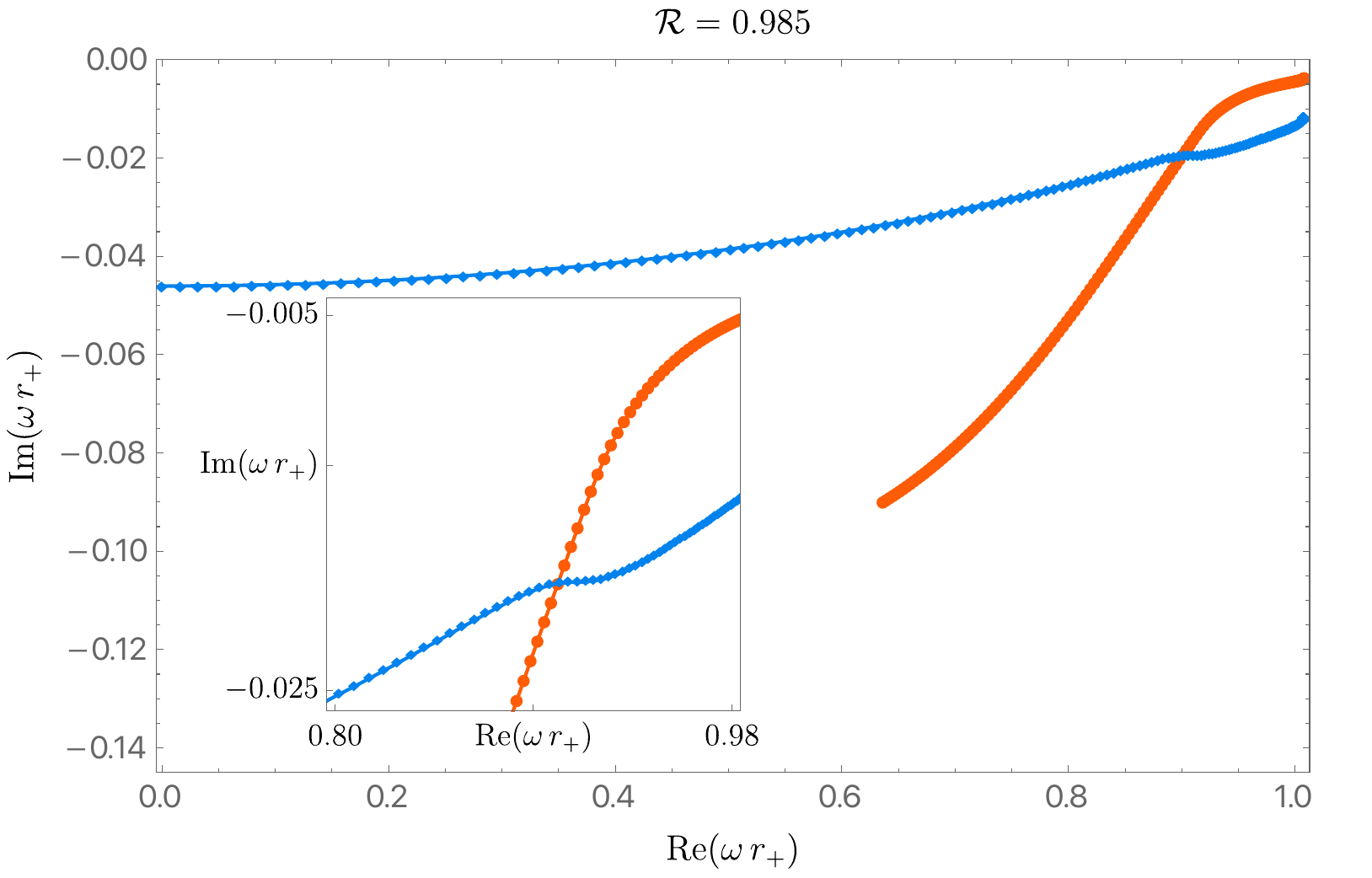}
\hspace{0.0cm}
\includegraphics[width=.49\textwidth]{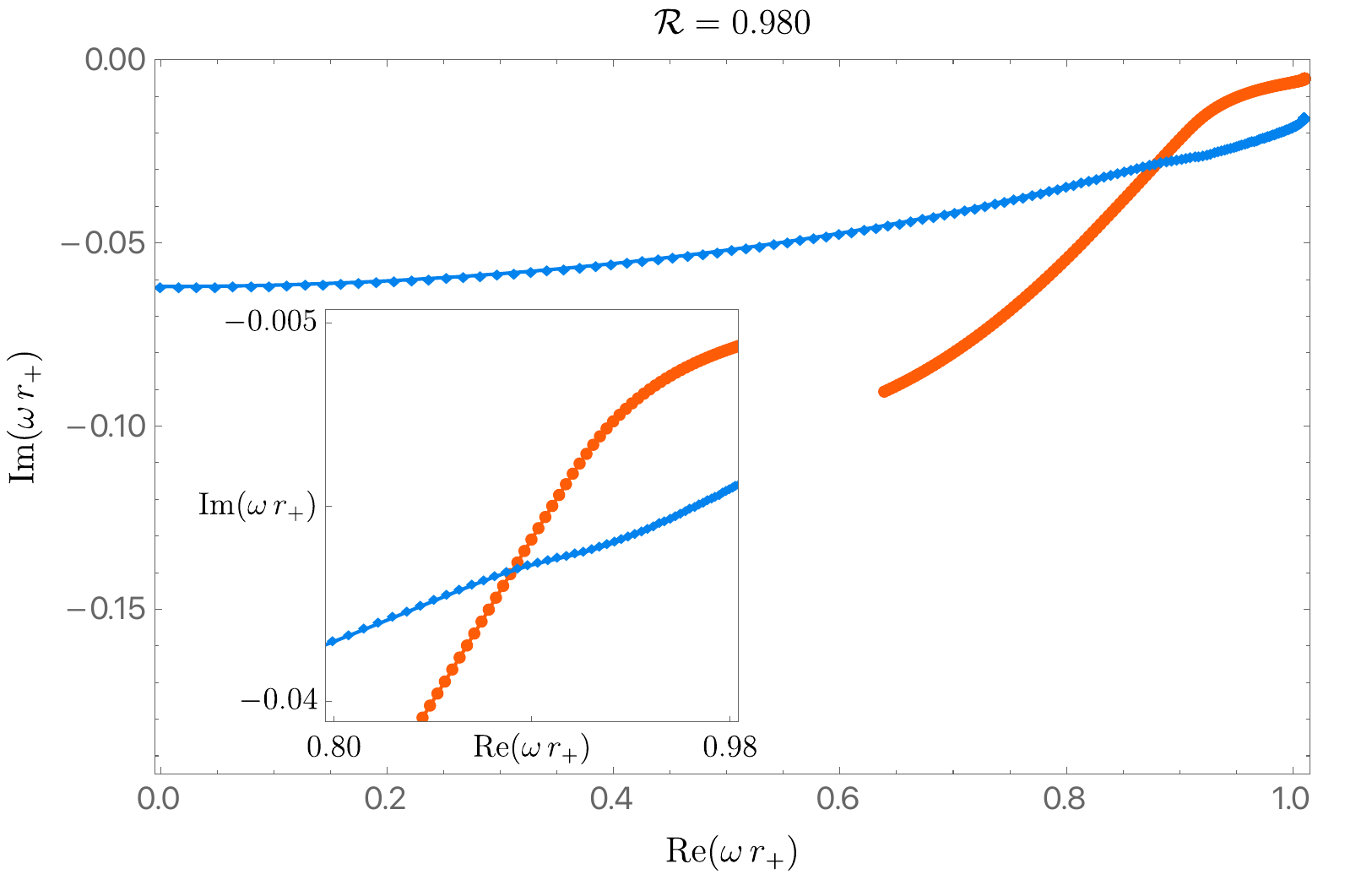}
\caption{Example of eigenvalue repulsion. Imaginary part of the frequency as a function of its real part for the photon sphere (orange disks) and the near-horizon (blue diamonds) families of scalar field QNMs with  $m=\ell=2, n=0$ for a Kerr-Newman family with, following the lexicographic order, $\mathcal{R}=0.993$, $\mathcal{R}=0.992$, $\mathcal{R}=0.991$, $\mathcal{R}=0.990$,  $\mathcal{R}=0.985$ and $\mathcal{R}=0.980$.
Figure taken from~\cite{Davey:2023fin}.}
\label{Fig:KNrepulsion}
\end{figure}  

The phenomenon of \emph{eigenvalue repulsion}  is common in certain eigenvalue
problems of quantum mechanical systems, where it is also known as  \emph{level
  repulsion}, \emph{avoided crossing}, \emph{Wigner-Teller effect} or {\it mode
  avoidance}~\cite{Landau1981Quantum,Cohen-Tannoudji:1977,Hund1927,1929PhyZ...30..467V,LandauQM,Arnold1978,ashcroft1976solid,1932PhyZS...2...46L,Zener:1932ws,Majorana:1932ga,Stu1932,Ivakhnenko:2022sfl,PhysRev.69.674,Herzberg1991,Smirnov:2003da,Wurm:2017cmm,Giganti:2017fhf}. A
first-principles argument that explains why eigenvalue repulsions have a better
chance of occurring in BH families with two (or more) dimensionless parameters
is provided in Section~4 of~\cite{Dias:2022oqm} (see also Section~4.1 of~\cite{Davey:2023fin}), independently of the spin of the perturbations. For
example, the phenomenon occurs not only in Kerr-Newman BHs, but also in
charged~\cite{Dias:2020ncd} or rotating de Sitter~\cite{Davey:2022vyx} BH
solutions, which are also described by 2 dimensionless parameters (of course, it
also occurs in Kerr-Newman de Sitter~\cite{Davey:2024xvd}). Moreover, it is not
restricted to gravito-electromagnetic modes since it also occurs in the QNM
spectrum of a {\it scalar field} in Kerr-Newman~\cite{Davey:2023fin}: see
e.g. Figs.~7--8  and 9--11 of~\cite{Davey:2023fin}.

In fact, the features of eigenvalue repulsions in the gravito-electromagnetic and scalar field  QNM spectra of Kerr-Newman are qualitatively similar, but the latter case allows us to explore more deeply the phenomenon because the system is simply described by an angular/radial pair of ODEs~\cite{Davey:2023fin}, instead of a coupled pair of PDEs~\cite{Dias:2015wqa,Dias:2021yju,Dias:2022oqm}. For this reason, one can borrow results from the scalar field case~\cite{Davey:2023fin} to best illustrate the eigenvalue repulsion phenomenon in Fig.~\ref{Fig:KNrepulsion}. In this figure we use the so-called ``polar parameterization'' $( \mathcal{R}, \Theta)$ of the Kerr-Newman parameter space introduced in~\cite{Davey:2023fin}:
\begin{equation}\label{KN:PolarParameterization}
\frac{a}{r_+}= \mathcal{R}\, \sin \Theta\,, \qquad \frac{Q}{r_+}=\mathcal{R}\, \cos \Theta\,,
\end{equation}
so that $\mathcal{R}=\sqrt{r_-/r_+}$ (where $r_{\pm}$ are the event and Cauchy horizon locations). This parameterization $(\mathcal{R},\Theta)$ has the property that $\mathcal{R}$ is an off-extremality radial measure (since it vanishes in the Schwarzschild limit and attains its maximum value of $\mathcal{R}=1$ at extremality, where $r_-=r_+$), while the polar parameter $\Theta$ ranges between the Reissner-Nordstr\"om solution (where $\Theta=0$ and thus $a=0$) and the Kerr solution (where $\Theta=\pi/2$ and thus $Q=0$).
All curves in Fig.~\ref{Fig:KNrepulsion} follow a family of Kerr-Newman BHs that starts at the Reissner-Nordstr\"om solution ($\Theta=0$; left endpoint) and evolve in $\Theta$ towards the Kerr solution ($\Theta=\pi/2$; right endpoint) while staying always at {\it fixed distance $\mathcal{R}$ from extremality}. The six plots have different $\mathcal{R}$, as displayed in the top label of each plot (note that the blue near-horizon curves always have $\mathrm{Re}\,\omega =0$ at $\Theta=0$). 

The eigenvalue repulsion is clearly observed in the evolution of the first four
plots with $\mathcal{R}\geq 0.990$ (following the lexicographic order). On the
other hand, for $\mathcal{R}\leq 0.990$ (last three plots) we observe a simple
crossover. Although not clear from Fig.~\ref{Fig:KNrepulsion}, in the crossovers
only the imaginary part of frequency (but not the real part) coincides for two
families of QNMs for a given $(\mathcal{R},\Theta)$: see discussion of
Figs.~7--8 of~\cite{Davey:2023fin} (or discussion of Figs.~13--14
of~\cite{Dias:2022oqm}) for full details.  As stated above, eigenvalue
repulsions typically can occur in BH backgrounds described by two or more
dimensionless parameters.  However, they can occur even when the BH background
is parameterized by a single parameter.  In this context, certain peculiar
features of the Kerr QNM spectrum (e.g. the behavior of the fifth radial
overtone of the $\ell = m = 2$ gravitational QNM~\cite{Onozawa:1996ux}) are
ultimately due to the phenomenon of eigenvalue
repulsion~\cite{Motohashi:2024fwt}. This will be reviewed and discussed in
detail in Section~\ref{sec:avoidance} below.

Note that symmetries might induce ``accidental'' (and thus exceptional)
eigenvalue crossings even when the BH background is parameterized by a single
parameter, e.g. for Kerr BHs (see Section~4 of~\cite{Dias:2022oqm} and
Section~4.1 of~\cite{Davey:2023fin}). Indeed, the near horizon geometry of
extremal BHs has an emergent SL$(2,\mathbb{R})$ isometry, responsible for the
clustering of eigenvalues near
extremality~\cite{Dias:2015wqa,Dias:2021yju,Dias:2022oqm,Davey:2023fin}. This
special symmetry can eventually lead to accidental eigenvalue crossings.

The above discussion conveys the intricacies and richness of the spectrum of BHs in GR.
For a more detailed description of the properties of the gravito-electromagnetic and scalar QNM spectrum of Kerr-Newman BHs, we refer the reader to~\cite{Dias:2015wqa,Dias:2021yju,Dias:2022oqm,Davey:2023fin}.
To improve readability, in the Appendices we overview recent developments concerning QNMs in asymptotically (anti-)de Sitter spacetimes (Appendix~\ref{sec:GRcosmoHighD}), perturbations induced by massive fields (Appendix~\ref{sec:superradiant_instabilities}), and the analog BHs program, aiming to reproduce and observe some of this phenomenology in the laboratory (Appendix~\ref{sec:analog}). 
While we have decided to focus this review on four-dimensional asymptotically flat spacetimes and on GW physics, these are independent research fields on their own, with a significant amount of complex and independent content which is foundational to some important questions in physics (such as cosmic censorship, the hierarchy problem, the dark matter puzzle, etc.).

\subsection{Mode avoidance, resonant excitation, and hysteresis}
\label{sec:avoidance}

\vspace{-.1cm}

\noindent \textit{Initial contributors: Motohashi, Richartz}

\vspace{.2cm}

\subsubsection{Mode avoidance and resonant excitation}
\label{sec:avoidance_KerrGWs}

As discussed in Section~\ref{sec:KNqnm}, mode avoidance (also referred to as eigenvalue repulsion) frequently occurs in the QNM spectrum of BHs with two (or more) dimensionless parameters~\cite{Dias:2021yju,Dias:2022oqm,Davey:2022vyx,Davey:2023fin,Davey:2024xvd}.
For example, in Kerr-Newman BHs the phenomenon has been shown to arise regardless of the values of $\ell=m$, $n$, or the spin $s$ of the perturbations, but the phenomenon is expected to be present also for $\ell\neq m$. 
Interestingly, mode avoidance can also occur in systems characterized by a single dimensionless parameter. Apart from sharp avoidance patterns, including those discussed in Section~\ref{sec:KNqnm}, there are instances of mild avoidance, such as those observed in the GW spectrum of Kerr BHs~\cite{Motohashi:2024fwt}. 
Notably, mode avoidance in the spectrum affects in interesting ways the QNM excitation factors and it is associated with resonant amplification of the QNMs. This feature appears to be universal in BH perturbation theory. 

The QNM frequencies $\omega_{\ell mn}$ and excitation factors $B_{\ell mn}$ are formally defined as poles and residues of the Green's function, respectively (see Section~\ref{sec:amplitudes}). 
In terms of a homogeneous solution of the radial Teukolsky equation~\eqref{eq:radialR:Diff_Eqn} satisfying the ingoing boundary condition at the event horizon, 
\begin{align} \label{Rin}
R_{\ell m}(r) &\to
\begin{cases}
\Delta^{-s}e^{-ikr_*} , & r\to r_+, \\
\mathcal{A}_{\rm in} r^{-1} e^{-i\omega r_*} + \mathcal{A}_{\rm out} r^{-1-2s} e^{i\omega r_*} , & r\to \infty,
\end{cases}
\end{align}
where $k=\omega-ma/r_+$, the QNM frequencies $\omega_{\ell mn}$ are defined as roots of 
\begin{align} \label{omegadef}
\mathcal{A}_{\rm in}(\omega_{\ell mn})=0 ,
\end{align} 
and the excitation factors $B_{\ell mn}$ are defined as~\cite{Leaver:1986gd}  
\begin{align} \label{Bdef} 
B_{\ell mn} = \left. \frac{\mathcal{A}_{\rm out}}{2\omega} \left( \frac{\text{d}\mathcal{A}_{\rm in}}{\text{d}\omega}\right)^{-1} \right|_{\omega=\omega_{\ell mn}} ,
\end{align}
where the asymptotic amplitudes $\mathcal{A}_{\rm in}$ and $\mathcal{A}_{\rm out}$ for the Teukolsky equation can be calculated by using the Mano-Suzuki-Takasugi (MST) or the Sasaki-Nakamura (SN) formalism~\cite{Mano:1996gn,Mano:1996vt,Sasaki:2003xr}. 
Both of the QNM frequencies and excitation factors are complex quantities and have the key property of depending solely on the BH parameters, while they are independent of the initial data.
Linear perturbation theory predicts that ringdown GWs are given by a superposition of damped sinusoids, with frequencies and decay times determined by the QNM frequencies, and amplitudes factorized as a product of the (initial-data independent) excitation factors with certain initial-data dependent integrals.
Therefore, in the context of BH spectroscopy, it is essential to understand the nature of QNM frequencies and excitation factors.
\begin{figure}[t]
  \centering
  \includegraphics[width=0.49\columnwidth]{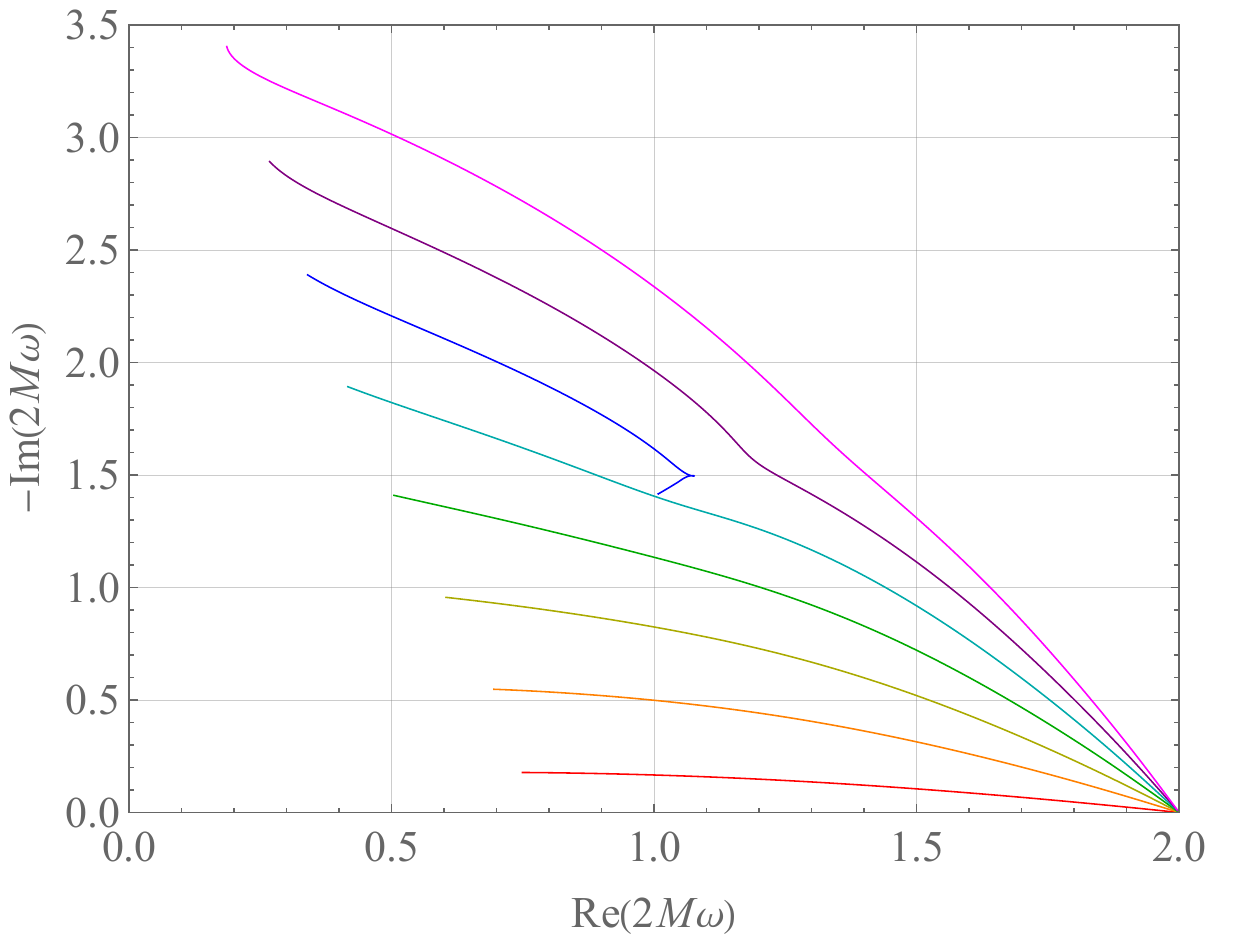}
  \includegraphics[width=0.49\columnwidth]{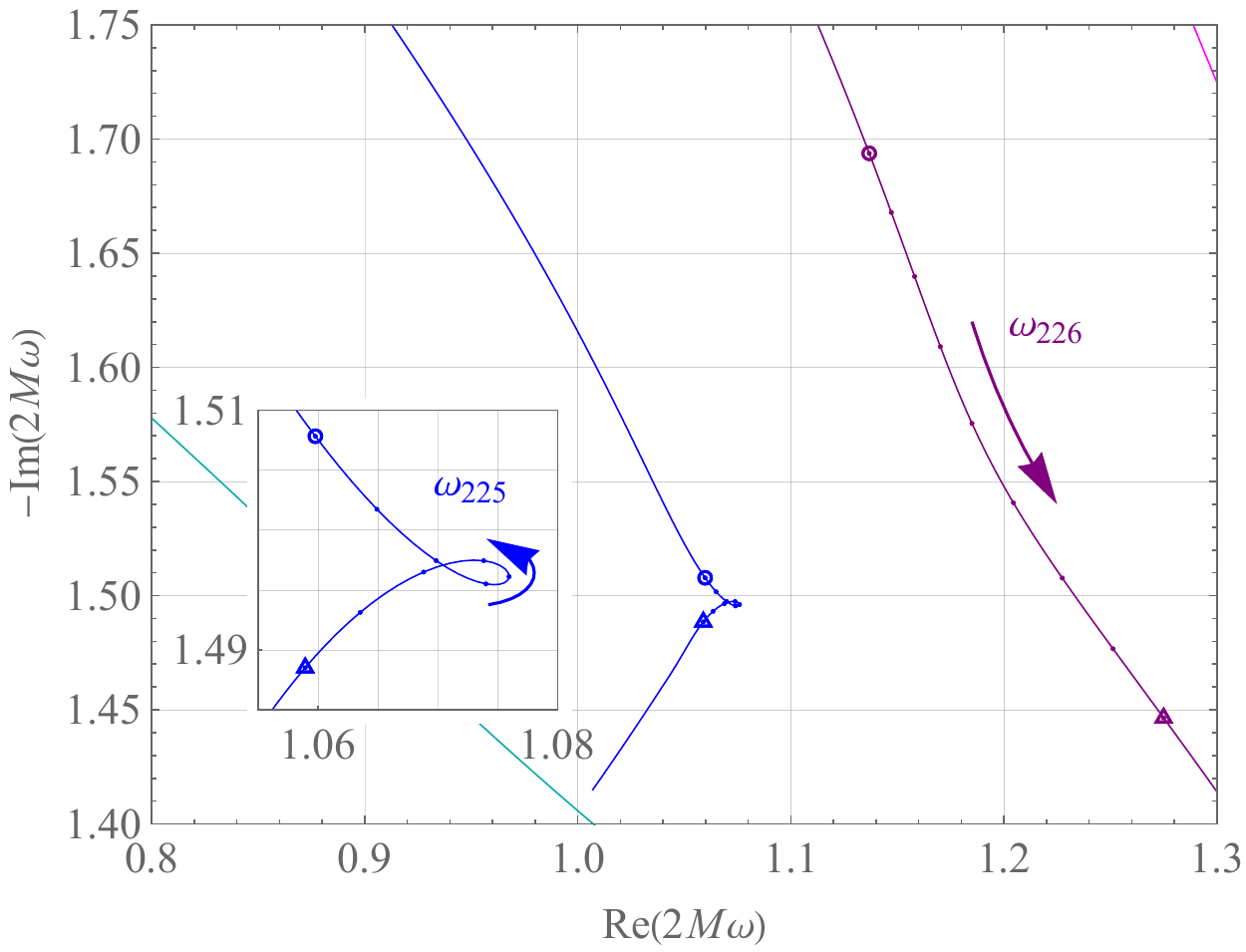}
  \includegraphics[width=0.49\columnwidth]{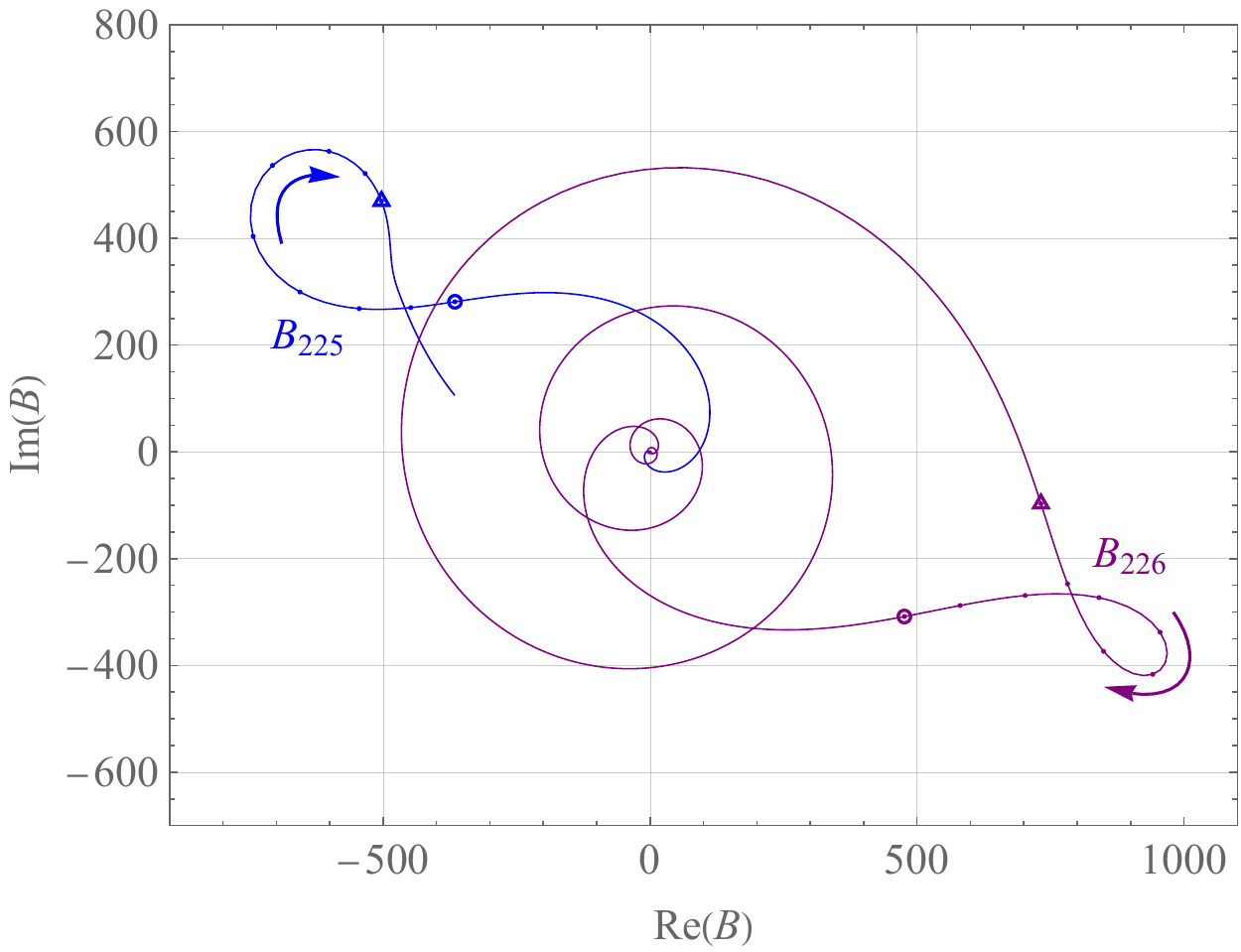}
  \includegraphics[width=0.49\columnwidth]{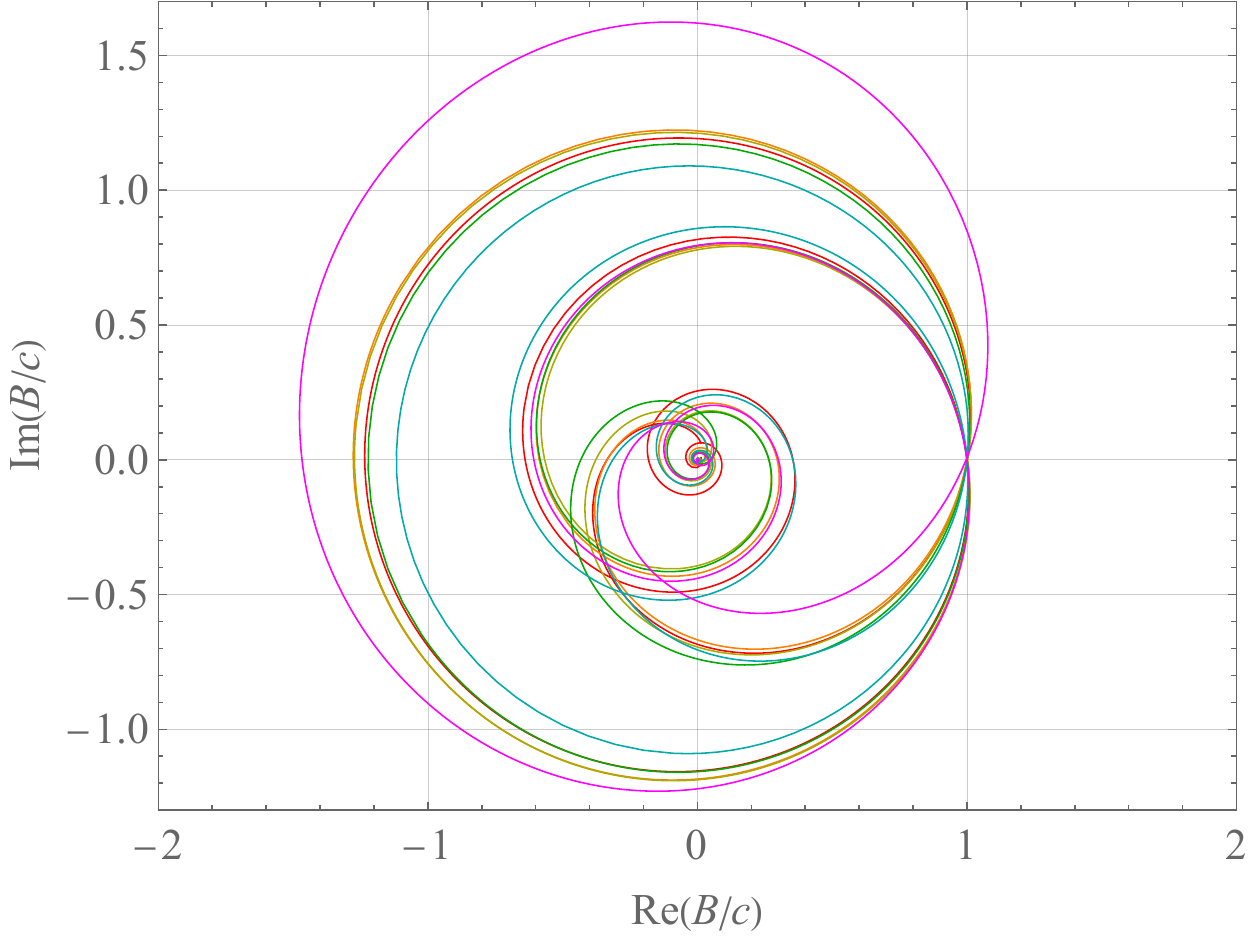}
\caption{
Top left: Kerr QNM frequencies $\omega_{\ell mn}$ (top left) of $(\ell,m)=(2,2)$ from $n=0$ (bottom red) to $7$ (top magenta) for $0\leq a/M\leq 1-10^{-6}$. Top right: zoom-in plot, with arrows indicating the direction of increasing spin.
Bottom: Excitation factors $B_{\ell mn}$ with $n=5,6$ (left) and other overtones (right), the latter of which are normalized by a complex constant $c=c_n$ numerically chosen so that each curve crosses approximately at $1$, to highlight the similarity of the spiral shapes.
In the bottom left panel, the range $a/M=0.875$ (circle) -- $0.915$ (triangle) is highlighted, and the shown small dots have a spacing of $0.005$ in $a/M$.
Figure adapted from~\cite{Motohashi:2024fwt} by using the public data~\cite{motohashi_2024_12696858}.
}
\label{fig:Sec2_4_wB22n}
\end{figure}
As noted in Fig.~\ref{fig:QNMl2behavior}, the $n=5$ overtone of the $\ell=m=2$ Kerr QNM spectrum displays a peculiar behavior near $a/M \simeq 0.9$~\cite{Onozawa:1996ux}.
The top panels of Fig.~\ref{fig:Sec2_4_wB22n} zoom in on this behavior. While the fundamental mode and most overtones migrate smoothly toward the so-called accumulation point at $2M\omega=m$  as the spin parameter increases~\cite{Detweiler:1980gk,Glampedakis:2001js,Cardoso:2004hh,Hod:2008zz,Yang:2012pj,Yang:2013uba}, only the fifth overtone unexpectedly reverses direction around $a/M \simeq 0.9$, creating a knot-shape loop and ultimately isolating itself from the accumulation point. 
This anomaly, first observed in~\cite{Onozawa:1996ux}, was confirmed by subsequent numerical calculations~\cite{Berti:2003jh,Berti:2004md,Cook:2014cta}. Quite remarkably, the absolute values of the excitation factors for the fifth and sixth overtones become unexpectedly large for high spins~\cite{Giesler:2019uxc,Oshita:2021iyn}, even though the Kerr excitation factors typically approach zero at extremality~\cite{Ferrari:1984zz,Berti:2006wq,Zhang:2013ksa} and asymptotically scale as $n^{-1}$ for higher overtones~\cite{Andersson:1996cm,Berti:2006wq}.

The physical origin of such peculiar behavior was recently understood to be mild mode avoidance, and an associated resonance between the fifth and sixth overtones~\cite{Motohashi:2024fwt}.
This mild avoidance is subtle in terms of the QNM frequencies: while leaving a clear imprint on the $n=5$ overtone, it only manifests as a slight distortion of the $n=6$ overtone (see the top right panel of Fig.~\ref{fig:Sec2_4_wB22n}). 
Instead, it is accompanied by noticeable features in the complex excitation factors.
As shown in the bottom panels of Fig.~\ref{fig:Sec2_4_wB22n}, while the fundamental mode and other overtones form spiral trajectories (bottom right), the excitation factors of the fifth and sixth overtones are simultaneously enhanced relative to the ``ordinary'' spiraling behavior near $a/M \simeq 0.9$. In other words, mode avoidance is associated with a resonance.

\begin{figure}[t]
  \centering
  \includegraphics[width=0.49\columnwidth]{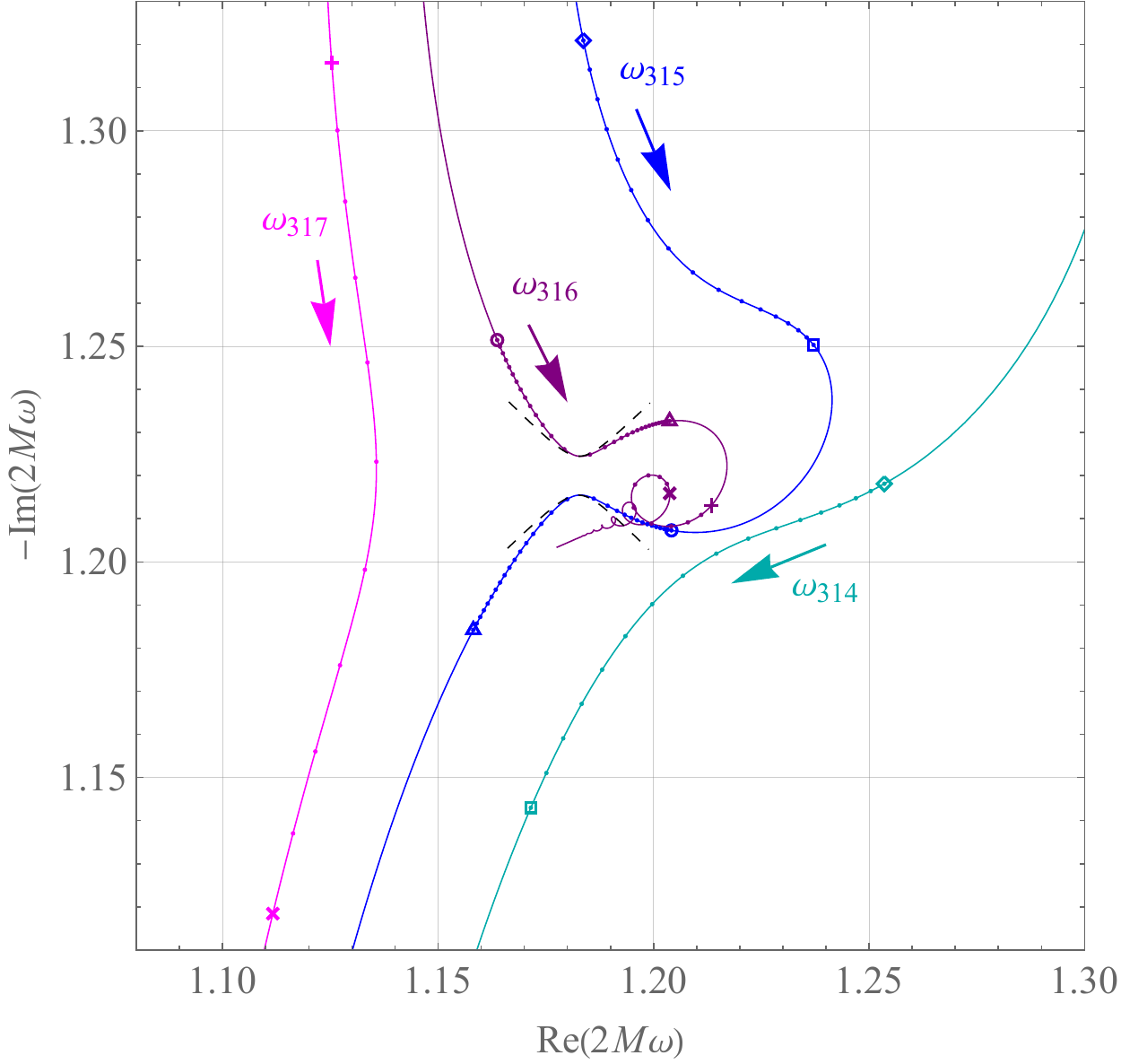}
  \includegraphics[width=0.49\columnwidth]{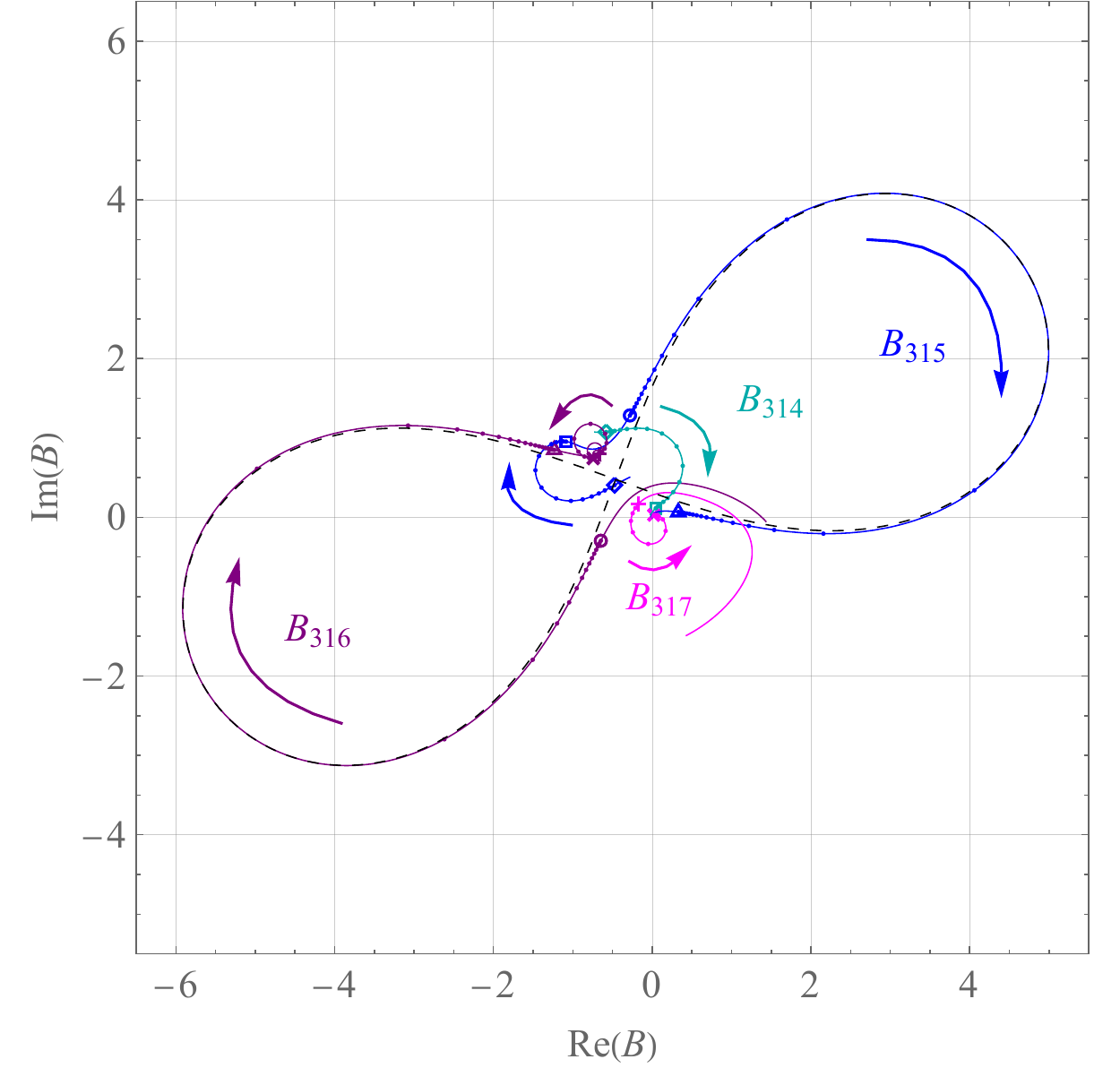}
  \caption{
  Kerr QNM frequencies $\omega_{\ell mn}$ (left) and excitation factors $B_{\ell mn}$ (right) of $(\ell,m)=(3,1)$ from $n=4$ to $7$. The excitation factor curves are shown for $0.95 \leq a/M \leq 1-10^{-6}$. 
  Highlighted are the ranges $a/M=0.952$ (diamond) -- $0.96$ (square), $0.9715$ (circle) -- $0.973$ (triangle), and $0.978$ (plus) -- $0.983$ (cross), with small dots having a spacing of $5\times 10^{-4}$, $5\times 10^{-5}$, and $5\times 10^{-4}$, respectively.
  Black dashed curves represent the hyperbola (left) and lemniscate (right).
  Figure adapted from~\cite{Motohashi:2024fwt} by using the public data~\cite{motohashi_2024_12696858}.
  }
  \label{fig:Sec2_4_wB31n}
\end{figure}

Mode avoidance and the associated resonance are even more pronounced in the $(\ell,m)=(3,1)$ gravitational multipole, as illustrated in Fig.~\ref{fig:Sec2_4_wB31n}. 
Here, as the spin parameter increases, this distinctive behavior appears across successive pairs of overtones. 
Sharper repulsions between two QNM frequencies yield a stronger amplification of the corresponding pair of excitation factors.
The full dataset of QNM frequencies and excitation factors is publicly available~\cite{motohashi_2024_12696858,Lo:2025njp,GRIT,CoG} (see Appendix~\ref{sec:public_codes}).

The resonance associated with mode avoidance is not exclusive to the Kerr GW spectrum, but it is instead a universal phenomenon observed in QNMs near the so-called exceptional points.
The phenomenon has also been observed for scalar and electromagnetic perturbations~\cite{Motohashi:2024fwt,Lo:2025njp}, and it has been demonstrated in the simple case of a double rectangular barrier potential~\cite{Motohashi:2024fwt}, a common toy model used to study spectral instabilities and environmental effects (see Section~\ref{sec:spectral_environmental}). 
An interesting behavior of complex mode avoidance is that the real or the imaginary part of the mode frequencies can cross, and when one part crosses the other repels~\cite{Motohashi:2024fwt,Lo:2025njp}.
In general, sharp resonances share a common feature~\cite{Motohashi:2024fwt}: the QNM frequencies and the excitation factors near resonance have the shape of a hyperbola and of a lemniscate of Bernoulli, respectively, and the resonance peak is characterized by a quarter-power Lorentzian as a function of the spin parameter. 
This topic is discussed in more detail in Section~\ref{sec:EP}.

\subsubsection{Hysteresis}
\label{sec:hysteresis}

Mode avoidance occurs when two or more eigenvalues of the wave equation become nearly degenerate~\cite{Dias:2021yju,Dias:2022oqm,Davey:2023fin,Motohashi:2024fwt}. The occurrence of this phenomenon in the Kerr gravitational spectrum raises the question of whether a configuration exists in which QNMs coincide {\em exactly}, rather than merely approximating one another. Remarkably, for a massive scalar field minimally coupled to a Kerr BH, such exact QNM coincidence is achievable. By exploring the associated parameter space for $\ell=m=1$, the $n=0$ and $n=1$ overtone were shown to coincide when the BH spin $a = a_c  \simeq 0.99946598M$ and when the scalar field mass $\mu M = (\mu M)_c \simeq 0.37049814$~\cite{Cavalcante:2024swt,Cavalcante:2024kmy}.
In this case, for fixed $\mu M$, the overtone number $n$ is defined according to the magnitude
of the imaginary part of the frequency of the QNMs when the spin of the BH is continuously decreased to $a/M=0$.
The degenerate frequency is $M \omega \approx 0.50583048 - 0.02310050 i$, with further precision achievable by fine-tuning the numerical value of the pair $(a/M)_c$, $(\mu M)_c$.
Datasets and numerical codes, based on Leaver's method~\cite{Leaver:1985ax,Nollert:1993zz,Konoplya:2006br,Dolan:2007mj,Konoplya:2013rxa} and on the isomonodromic method~\cite{CarneirodaCunha:2015hzd,daCunha:2021jkm}, are publicly available~\cite{zenodo13961216} (see Appendix~\ref{sec:public_codes}). 

\begin{figure}[t]
  \centering
  \includegraphics[height=0.43\columnwidth]{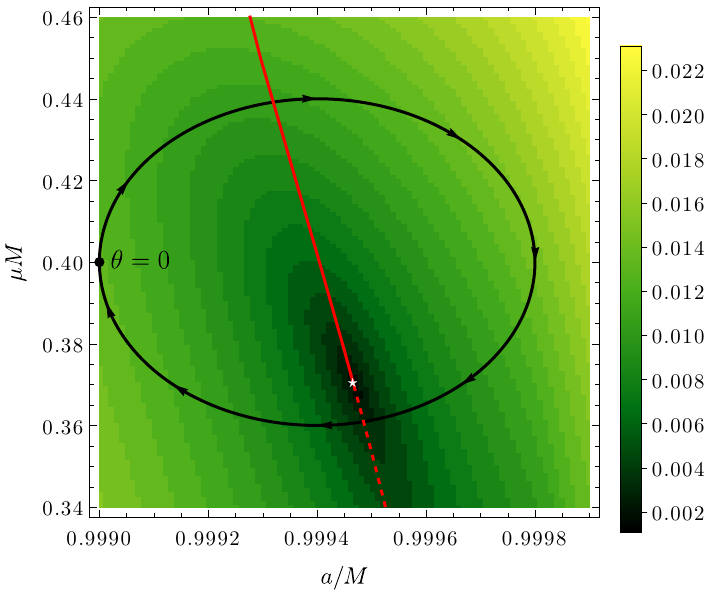} \ \ 
  \includegraphics[height=0.43\columnwidth]{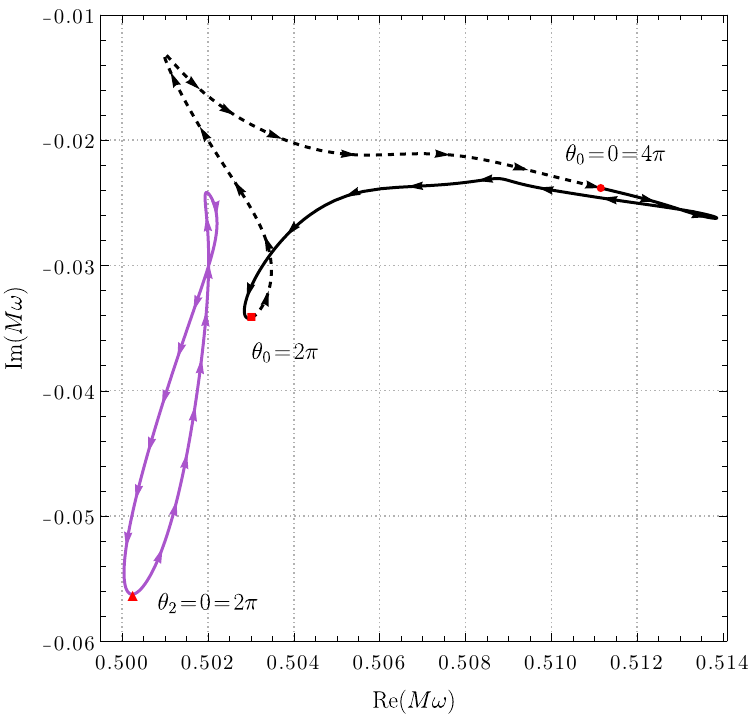}
  \caption{Left panel: parameter space around the degeneracy point (marked by the white star) where the QNM frequencies $\omega_{110}$ and $\omega_{111}$ coincide. Green shades in the background represent the absolute difference between these frequencies, specifically $|M\omega_{111} - M\omega_{110}|$. The solid red line indicates the loci where $\mathrm{Im}(M\omega_{110}) = \mathrm{Im}(M\omega_{111})$, while the dashed red line corresponds to the loci where $\mathrm{Re}(M\omega_{110}) = \mathrm{Re}(M\omega_{111})$. The fundamental mode and its second overtone, initially defined at the black dot ($\theta=0$), are tracked as the parameters change along the closed elliptical path (black curve). 
  Right panel: as the path around the ellipse is traversed, the fundamental mode follows the solid black line, transforming into the first overtone after one full loop. After a second loop, the QNM frequency has traced the dashed black line, returning to its initial value. In contrast, the second overtone follows the purple path, returning to its original value after a single loop around the ellipse.}
  \label{fig:Sec2_4_hysteresis}
\end{figure}

The left panel of Fig.~\ref{fig:Sec2_4_hysteresis} displays the parameter space in the vicinity of the degeneracy point, which is represented by a star. Originating from the degeneracy point, there are two curves. The dashed red curve corresponds to parameter configurations for which the oscillation rates of the $n=0$ and $n=1$ QNMs are equal. In contrast, along the solid red curve the decay times of the $n=0$ and $n=1$ QNMs coincide. As the spin of the BH increases for  $ \mu M > (\mu M)_c$, the red solid curve is crossed, leading to the $n=1$ mode becoming the dominant mode. This phenomenon resembles the behavior of the Kerr ZDMs, which dominate over damped modes as the spin of the BH approaches the extremal limit $a/M=1$~\cite{Andersson:1999wj,Glampedakis:2001js,Cardoso:2004hh,Hod:2011zzd,Yang:2012pj,Yang:2013uba,Cook:2014cta,Richartz:2015saa,daCunha:2021jkm}. It also raises the question of how a QNM would change if a closed path were traced adiabatically in the parameter space, crossing the red solid line once and encircling the degeneracy point.  

Such a configuration exhibits a ``QNM hysteresis'' phenomenon in Kerr BHs~\cite{Cavalcante:2024swt,Cavalcante:2024kmy}: the outcome of adiabatically following a QNM in parameter space can depend on the specific path taken. 
For illustration, consider the evolution of the fundamental mode $M\omega_{110}=0.5112-0.0238i$, initially defined at $(a/M,\mu M) = (0.999,0.4)$, as it follows the elliptical path parameterized by $a/M = 0.9994-0.0004 \cos \theta_0$ and $\mu M = 0.4-0.04 \sin \theta_0$. The angle $\theta_0$ sweeps the ellipse in the clockwise direction, as shown in the left panel of Fig.~\ref{fig:Sec2_4_hysteresis}. After a complete revolution ($\theta_0 = 2\pi$), the spin of the BH and the mass of the scalar field return to their initial values. However, the QNM frequency does not: the fundamental QNM transitions to the first overtone $M\omega_{111}=0.5030-0.0342i$, as shown in the right panel of Fig.~\ref{fig:Sec2_4_hysteresis}. It is only after a second full revolution ($\theta_0 = 4\pi$) that the mode frequency returns to its original value. The right panel of Fig.~\ref{fig:Sec2_4_hysteresis} also shows the evolution of the $n=2$ QNM, $M\omega_{112} = 0.5002-0.0562i$, along the same elliptical path. Since there is no degeneracy point for the $n=2$ QNM within this closed path, the mode frequency returns to its starting value after just one loop ($\theta_2 = 2\pi$).

The QNM hysteresis effect implies that the response of a Kerr BH to massive scalar perturbations is path-dependent, suggesting that the BH retains a form of ``memory'' of previous states, which influences its subsequent behavior. 
This path dependence and the point of degeneracy can be understood in terms of exceptional points and geometric phases of non-Hermitian systems (see Section~\ref{sec:EP}).
A further analogy can be drawn~\cite{Cavalcante:2024swt}, linking the coincidence of QNM frequencies and hysteresis effects in Kerr BHs with the phase space of thermodynamic systems undergoing phase transitions, where phenomena such as critical points and level-crossings are well established.

The existence of additional exceptional points corresponds to degeneracies between other modes~\cite{Cavalcante:2024swt,Cavalcante:2024kmy}, and recent work~\cite{Cavalcante:2025abr} has begun to explore these structures for broader ranges of $\ell$, $m$, and $n$. However, a full characterization of these degeneracies is still lacking. Another avenue for future investigation is the possibility of QNM hysteresis for other types of massive fields and different BH spacetimes. Finally, further research should address whether the QNM hysteresis effect has astrophysical implications. Notably, the critical mass lies in the regime where superradiant instabilities are most efficient~\cite{Cavalcante:2024swt,Cavalcante:2024kmy} (see Appendix~\ref{sec:superradiant_instabilities}). In physical units, this corresponds to 
\[\mu M \simeq \left( \frac{M}{M_{\odot}} \right) \left(
    \frac{\mu} {10^{-10} \, \text{eV}} \right) \sim \mathcal{O}(1). \] 
For astrophysical BHs with masses in the range $\sim (5,
10^{10}) M_{\odot}$, the critical mass
lies within $ \mu \in (10^{-20} - 10^{-11}) \, \text{eV}$,
which is of interest in beyond-Standard Model scenarios and may
have relevance in the context of dark matter searches. In particular, bosonic
condensates~\cite{Hod:2012px,Herdeiro:2014goa} may grow due to superradiant instabilities~\cite{Damour:1976kh,Ternov:1978gq,Zouros:1979iw,Detweiler:1980uk,Dolan:2007mj,Dolan:2012yt} and emit
GWs through various channels -- e.g., direct continuous
emission arising from annihilation of the boson field, or level
transitions between states with different overtone numbers~\cite{Arvanitaki:2009fg,Arvanitaki:2010sy,Brito:2015oca}. Such
processes (superradiant instabilities and GW emission)
will change both the mass and the spin of the BH, drawing
paths in the parameter space (as in the left panel of Fig.~\ref{fig:Sec2_4_hysteresis}) that might encompass an exceptional point. 

\subsubsection{Exceptional points}
\label{sec:EP}
As seen above, QNMs of BHs demonstrate the occurrence of mode avoidance and associated resonances.
This concept was originally developed in quantum mechanics, where two energy eigenvalues cannot coincide unless a certain condition is satisfied~\cite{Hund1927,1929PhyZ...30..467V,LandauQM,Arnold1978}.
Usually, the values of two parameters must be adjusted to satisfy this condition.
Therefore, if only one parameter is varied, the avoided crossing occurs, and the resultant trajectories of eigenvalues correspond to a cross section of the so-called Dirac cone.
Avoided crossings play a pivotal role in experiments and observations across many fields in physics~\cite{ashcroft1976solid,1932PhyZS...2...46L,Zener:1932ws,Majorana:1932ga,Stu1932,Ivakhnenko:2022sfl,PhysRev.69.674,Herzberg1991,Smirnov:2003da,Wurm:2017cmm,Giganti:2017fhf}.

However, a crucial difference between QNMs and bound states in quantum mechanics originates from the boundary conditions.
In this analogy, the QNM frequencies and Weyl scalar perturbations correspond to the energy eigenvalues and wave functions in quantum mechanics, respectively.
Instead of decaying at infinity, the QNM wave function is required to satisfy the ingoing/outgoing wave condition at the event horizon and at spatial infinity.
This leaky boundary condition, known as Siegert boundary condition~\cite{Siegert:1939zz}, implies that the QNM eigenvalues must be complex numbers, in contrast to the real energy eigenvalues of bound sates.
In quantum mechanics, this kind of system is known as the Schr\"odinger problem of open systems, for which the Hamiltonian is not Hermitian~\cite{Gamow:1928zz,LandauQM,Kukulin1989,moiseyev_2011}, utilized for studies of quantum resonances.
More recently, non-Hermitian physics has been extensively studied from both the theoretical and the experimental point of view~\cite{El-Ganainy:2018ksn,Bergholtz:2019deh,Ashida:2020dkc}. 

One of the most important concepts in non-Hermitian system is the {\em exceptional point}~\cite{Kato1995}, 
a branch point singularity in the complex eigenvalue plane~\cite{Ozdemir:2019iqe,Wiersig:2020dgv,Parto2021,Ding:2022juv}.
At the exceptional point, two or more eigenvalues and their corresponding eigenvectors coalesce, and the norm of the eigenvectors becomes zero.
The avoided crossing in non-Hermitian systems occurs in the vicinity of an exceptional point~\cite{PhysRevE.61.929,Heiss:2012dx}.
It is advantageous to study the avoided crossing of BH QNMs in the context of non-Hermitian physics.
However, as we will see in Section~\ref{sec:scalar_products}, the standard inner product used for bound states in quantum mechanics does not extend to QNMs or, more broadly, to open systems. 

In nuclear physics, a wave function for resonances, analogous to the QNM wave function, is known as the Gamow-Siegert state~\cite{Gamow:1928zz,Siegert:1939zz}. An appropriate finite norm is realized by the so-called Zel'dovich regularization~\cite{Zeldovich:1961a} (see also earlier work~\cite{Kapur1938}), where the norm is defined by introducing a regularization factor which decays faster than the wave function itself, maintaining the divergent contribution from the boundary.
This formulation allows one to develop a Rayleigh-Schr\"odinger-like perturbation theory for resonance states~\cite{Zeldovich:1961a,PeZe1998} (see~\cite{Kukulin1989,moiseyev_2011} for more developments in quantum resonances and non-Hermitian physics).

For BH QNMs, a finite scalar product obtained by subtracting the divergent boundary term was introduced in~\cite{Ching:1993gt,Leung:1997was,Leung_1998}.
Several other definitions of the scalar product for QNMs of BHs have been considered, as discussed in Section~\ref{sec:scalar_products} below.
By means of such scalar products, avoided crossings and resonances in the QNM spectrum of BHs can be understood as a universal phenomenon occurring near exceptional points~\cite{Motohashi:2024fwt}. 
Generalizing the scalar product of~\cite{Ching:1993gt,Leung:1997was,Leung_1998} for Schwarzschild BHs, and employing an analogy with the scalar product used in quantum resonances~\cite{Kukulin1989,moiseyev_2011}, a scalar product (or more precisely a ``biorthogonal'' product) can be introduced for the Weyl scalars of Kerr BHs~\cite{Motohashi:2024fwt}.
Using this biorthogonal product, one can develop a Rayleigh-Schr\"odinger-like perturbation theory and predict small changes $\delta\omega$ of the QNM frequencies induced by small changes of the BH parameters. 
Similar results have also been obtained for other products~\cite{Leung:1997was,Leung:1999iq,Yang:2014tla,Zimmerman:2014aha,Mark:2014aja,Yang:2015jja,Cannizzaro:2023jle}.

Consider two overtone QNM frequencies, $\omega_{n_1}$ and $\omega_{n_2}$, that are isolated from other QNMs, and approach closely each other in the complex frequency plane when the BH parameters are varied (here, the subscripts $\ell,m$ are suppressed for simplicity).
An argument analogous to the quantum two-level system can be applied to predict how two QNM frequencies will behave under a small change of the BH parameters.
In quantum two-level system, the energy eigenvalues such that the Hamiltonian matrix is diagonal are obtained by solving the eigenvalue equation.
Similarly, solving the eigenvalue equation, the resulting QNM frequencies are given by~\cite{Dias:2022oqm,Motohashi:2024fwt}
\begin{align} \label{wpm} \omega^2_\pm = \mathcal{E}_c \pm \sqrt{\mathcal{E}_d^2+\Delta^2} , \end{align}
where $\mathcal{E}_{c,d} \coloneqq (\mathcal{E}_{n_1}\pm \mathcal{E}_{n_2})/2$,  
$\mathcal{E}_{n_j} \coloneqq \omega_{n_j}^2 + \delta\omega_{n_j}^2$ with $j=1,2$, and $\Delta$ captures a small change in the BH parameters.
The present derivation of Eq.~\eqref{wpm} follows~\cite{Motohashi:2024fwt}. 
An analogous formula can be found in Eq.~(4.6) of~\cite{Dias:2022oqm}. 
However, in~\cite{Dias:2022oqm}, the squared frequencies $\omega^2$'s in~\eqref{wpm} are replaced with linear terms in $\omega$, predicting different values in general. 

From Eq.~\eqref{wpm}, we see that the frequencies $\omega_\pm$ are degenerate if $\mathcal{E}_d^2+\Delta^2=0$. 
Solutions to this equation correspond to exceptional points.
Since this condition involves complex variables, in general it cannot be satisfied by tuning a single real parameter.
This corresponds to the crossing between two QNM frequencies~\cite{Dias:2021yju,Dias:2022oqm,Davey:2023fin,Motohashi:2024fwt}. 
On the other hand, if one can tune two independent parameters, it is possible to satisfy the condition and reach the exceptional point.
The hysteresis-like effects in the QNM spectrum~\cite{Cavalcante:2024kmy,Cavalcante:2024swt} can also be understood as geometric phases associated with an exceptional point~\cite{Berry:2004ypy,PhysRevA.72.014104,PhysRevA.85.064103,Ryu:2023pqq}.

The anomalous enhancement of the excitation factors at the avoided crossing can be understood as follows.
The excitation factor $B_n$ for the QNM frequency $\omega_n$, defined in Eq.~\eqref{Bdef}, can be rewritten by using the norm squared $A_n^2$ defined by the biorthogonal product as~\cite{Motohashi:2024fwt}
\begin{align} 
\label{Bana} 
B_n = \f{i\mathcal{A}_{\rm out}^2}{2\omega_n A_n^2} ,
\end{align}
where $\mathcal{A}_{\rm out}$ is the asymptotic amplitude for the outgoing wave at infinity in Eq.~\eqref{Rin}.
In general, the norm vanishes at an exceptional point, and hence $A_n^2$ becomes small at the avoided crossing near the exceptional point.
From Eq.~\eqref{Bana}, this leads to the enhancement of the excitation factor observed above.
Further, arguments based on Eqs.~\eqref{wpm} and \eqref{Bana} can be used to
derive analytically some peculiar features of the phenomenon we reviewed in this
section, namely, the hyperbolic avoidance curve in the complex frequency plane
and the lemniscate-shaped resonance in the excitation factor plane,
characterized by a quarter-power Lorentzian peak (see~\cite{Motohashi:2024fwt}
for more details). Note that different features can appear in resonances of QNMs
in coupled systems~\cite{Takahashi:2025uwo}. Additional insights regarding the
structure of QNMs near exceptional points and avoided crossings, along with
observational implications for time domain (TD) waveforms, are provided in Refs.~\cite{Yang:2025dbn,Kubota:2025hjk,PanossoMacedo:2025xnf}.

In summary: numerical and analytical calculations in Ref.~\cite{Motohashi:2024fwt} have firmly established that the resonant excitation of QNMs at the avoided crossings is a universal phenomenon in BH QNM spectra.
In analogy with other areas of physics~\cite{Wiersig:2020dgv,Parto2021,Smirnov:2003da,Wurm:2017cmm,Giganti:2017fhf}, the recently discovered exceptional points, avoided crossings and resonances in BH spectra can be expected to lead to further theoretical developments and possibly to new observational discoveries, especially in view of upcoming accurate GW observations of ringdown waves over a wide range of spin parameters (see Section~\ref{subsec:future_tests}).

\subsection{Light-ring physics}
\label{subsec:LR}

\vspace{-.1cm}

\noindent \textit{Initial contributors: Fransen}

\vspace{.2cm}

The event horizon is the defining feature of a BH. Yet BHs are also the only known bodies (with the possible exception of exotic ultracompact objects~\cite{Cardoso:2019rvt}) supporting a ``light-ring,'' i.e., bound orbits of null geodesics~\cite{Dodelson:2023nnr}. 
The importance of these geodesics for the ringdown rests on two observations. First, congruences of null geodesics provide the leading order geometrical optics approximation to the linearized wave equation. Second,  congruences which come neither from the BH nor from asymptotic infinity in the past must instead asymptote to bound null geodesics. The light-ring thus provides the boundary conditions to start approximating QNMs in the high-frequency limit. As a result, essential features of these QNMs are fixed by spacetime properties around the light-ring, and the ringdown carries the imprint of light-ring physics.

\subsubsection{Geometrical optics}

The geometrical optics approximation constructs solutions for (say) trace-reversed gravitational perturbations of the following form~\cite{Isaacson:1968hbi,Isaacson:1968zza,Misner:1973prb,poisson2014gravity,Dolan:2017zgu}
\begin{equation}\label{eqn:geomoptics}
	\bar{h}_{\mu \nu}(x) = \epsilon_{\mu \nu}(x;\lambdabar)e^{i S(x)/\lambdabar} \, , \quad \epsilon_{\mu \nu}(x;\lambdabar) =  \sum_{i = 0}^{\infty} \epsilon^{(i)}_{\mu \nu}(x) \lambdabar^i \, .
\end{equation}
More properly, one should take the real part of Eq.~\eqref{eqn:geomoptics}.
Here $\lambdabar$ is a small parameter, while $\epsilon_{\mu \nu}(x;\lambdabar)$ and $S(x)$ are to be determined. To leading order in $\lambdabar$, the wave equation reduces to the ``eikonal'' equation~\cite{guillemin2013semi}. For both electromagnetic and gravitational perturbations in the Lorenz gauge, the eikonal equation and its first subleading correction are fixed by the d'Alembertian. Explicitly, in our gravitational example~\cite{Poisson:2011nh,Dolan:2017zgu}:
\begin{equation}\label{eqn:nullgeodesicHJ}
g^{\mu \nu}(\nabla_{\mu}S)(\nabla_{\nu}S) = 0 \, , \quad \left((\nabla^{\mu}S) \nabla_{\mu}  + \frac{1}{2}\left(\nabla^{\mu}\nabla_{\mu}S\right)\right) \epsilon^{(0)}_{\alpha \beta}=0  \, , \quad (\nabla^{\mu}S)\epsilon^{(0)}_{\mu \nu}(x) = 0  \, .
\end{equation} 
The geodesic equation follows from the first equation in \eqref{eqn:nullgeodesicHJ} by taking a covariant derivative and using $[\nabla_{\mu},\nabla_{\nu}]S = 0$. Higher order $\epsilon^{(i)}_{\alpha \beta}$ will satisfy the second equation in \eqref{eqn:nullgeodesicHJ}, but with source terms constructed from lower-order quantities $\epsilon^{(k < i)}_{\alpha \beta}$. Often, it is useful to write $\epsilon^{(0)}_{\mu \nu} = \cA^{(0)} \hat{\epsilon}^{(0)}_{\mu \nu}$ in terms of an amplitude and a normalized, parallel transported polarization tensor~\cite{Isaacson:1968hbi,Misner:1973prb,Dolan:2017zgu}.

As an example, the null geodesic congruence with critical impact parameter $b_c=L/E=3\sqrt{3}M$ and $L=L_{\phi}$ around the equatorial bound null geodesic, which we shall denote $\gamma$, at $r_c = 3M$ of a Schwarzschild BH is described to first subleading order by~\cite{Hadar:2022xag}
\begin{equation}\label{eqn:Scongruenceexpansion}
S(x) = E \left(-t+\frac{\phi}{\Omega_{\rm orb}} +\frac{\sqrt{3}}{2 M} \delta r^2 + i 3 \sqrt{3} M \delta \theta^2 + O(\delta r^3 , \delta \theta^3) \right) \, , \; \delta r = r-r_c \, , \; \delta \theta=\theta - \frac{\pi}{2} \, ,
\end{equation}
where it follows from \eqref{eqn:nullgeodesicHJ} that $\Omega_{\rm orb} = (\nabla_{t}S/\nabla_{\phi} S)|_{\gamma} = 1/3 \sqrt{3} M$ and
\begin{equation}\label{eqn:nullsubleadinggeodesicHJSchwarzschild}
\left((\nabla^{\mu}S) \nabla_{\mu}  + \frac{1+i}{2 \sqrt{3}} \frac{E}{M} \right)  \cA^{(0)}(t, \phi)=0  \, , \quad \cA^{(0)}(t, \phi) =  F(\Omega_{\rm orb} t-  \phi)e^{-\frac{1}{2}\left(1+i\right)\left(\Omega_{\rm orb} t+  \phi\right)}   \, .
\end{equation}
Here, $F(x)$ is an arbitrary function, and $\cA^{(0)}$ is evaluated at $r=r_c$ and  $\theta=\frac{\pi}{2}$. Demanding that the full azimuthal mode number is unchanged fixes $F(x)$ up to a constant and allows us to determine, by comparison with $\sim e^{-i \omega t + i \ell \phi}$, the ``eikonal'' $\omega_{\ell 0}$ mode. Instead, a slightly more general congruence around the bound null geodesic leads to the benchmark relation between QNMs and ``light-ring'' physics~\cite{Press:1971wr,Ferrari:1984zz,Iyer:1986nq,Cardoso:2008bp}
\begin{equation}
\label{eqn:subleadingeikonal}
\omega_{\ell n}  = \Omega_{\rm orb} \left(\ell + \frac{1}{2}\right) - i \Lambda_L \left( n + \frac{1}{2}\right) + O(\ell^{-1}) \, , \quad \Lambda_L = \frac{1}{3 \sqrt{3} M} \, .
\end{equation}
Here, $\Lambda_L$ highlights the relation to the instability timescale of the reference bound null geodesic, as could be deduced from \eqref{eqn:Scongruenceexpansion} by considering $(dS)^{\mu}$ at linear order in $\delta r$~\cite{Cardoso:2008bp}.

Similar eikonal QNM frequencies have been derived in higher-dimensional~\cite{Berti:2003si,Konoplya:2003ii}, Reissner-Nordstr\"om~\cite{Kokkotas:1988fm,Andersson1993,Mashhoon:1985cya}, Kerr~\cite{Dolan:2010wr,Yang:2012he}, Myers-Perry~\cite{Cardoso:2008bp}, Kerr-Newman~\cite{Mashhoon:1985cya,Berti:2005eb}, and various other BHs~\cite{Fernando:2012yw,Glampedakis:2019dqh}. Typically, an expansion similar to \eqref{eqn:geomoptics} is applied only after a symmetry reduction. When the reduction results in an ODE, Eq.~\eqref{eqn:geomoptics} essentially becomes the WKB method, which has been extensively applied to study BH perturbations~\cite{Schutz:1985km,Iyer:1986np,Kokkotas:1988fm,Seidel:1989bp, Froeman:1992gp,Konoplya:2003ii,Dolan:2009nk,Daghigh:2011ty}. Rather than reviewing these well-known methods, we highlight a more recent variation of the geometrical approach which, instead of approximately solving the wave equation, directly approximates the spacetime around the light ring, and is therefore well-suited to studying the interplay between QNMs and light-ring physics~\cite{Fransen:2023eqj,Kapec:2024lnr}.

\subsubsection{Near-ring spacetime}

As a starting point, observe that the eikonal equations \eqref{eqn:nullgeodesicHJ} can be simplified by going to adapted coordinates (see e.g.~\cite{Papadopoulos:2020qik} for a construction of adapted coordinates for Kerr BHs)
\begin{equation}\label{eqn:adapted}
ds^2 = 2 dU dS + D(x^{\mu}) dS^2 + B_i(x^{\mu}) dS dX^i + C_{ij}(x^{\mu})dX^i dX^j \, ,
\end{equation}
with $D(x^{\mu})$, $B_i(x^{\mu})$, and $C_{ij}(x^{\mu})$ functions of the coordinates $x^{\mu} = \left\lbrace U, S, X^i \right\rbrace$ and the indices $i$, $j$ running over the $d-2$  ``transverse'' directions ($i,j, \ldots \in \left\lbrace 1, \ldots, d-2 \right \rbrace$). Specifically, $S(x)$ itself has been made into a coordinate, and the null congruence is generated by the coordinate vector field $\partial_U$. Therefore, Eq.~\eqref{eqn:nullgeodesicHJ} is automatically satisfied, and the geometrical optics solution \eqref{eqn:geomoptics} to first subleading order can be written explicitly as
\begin{equation}\label{eqn:geomopticsadapted}
\bar{h}_{\mu \nu}(x) = \left( \frac{\cA^{(0)}_0(X^i,S) \hat{\epsilon}^{(0)}_{\mu \nu}(x^{\mu})}{\left(\text{det}\left(C_{ij}(x^{\mu})\right)\right)^{1/4}} + O(\lambdabar)\right)e^{i S/\lambdabar} \, .
\end{equation}
To obtain a high-frequency approximation to QNMs it is appropriate (as illustrated by the example \eqref{eqn:Scongruenceexpansion}) to simultaneously expand around a reference bound null geodesic, say $\gamma$, with conserved charges appropriate to the QNM of interest. The adapted coordinate system \eqref{eqn:adapted} can be chosen such that $\gamma$ is located at $S = X^i = 0$. Now, an expansion around $S = X^i = 0$ which keeps $U \sim \lambdabar^0$ but lets $S \sim \lambdabar$ and $X^i \sim \sqrt{\lambdabar}$ corresponds, to leading order, to a spacetime limit proposed by Penrose~\cite{Penrose1976}. In this Penrose limit, the metric \eqref{eqn:adapted} reduces to a plane wave spacetime in Rosen coordinates~\cite{Rosen1937}:
\begin{equation}\label{eqn:Rosen}
ds^2_{\gamma} = 2 dU dS + C_{ij}(U) dX^i dX^j \, , \quad  C_{ij}(U) =  C_{ij}(x^{\mu})|_{S = X^i = 0}\, .
\end{equation}
The approximate solution \eqref{eqn:geomopticsadapted}, with $\cA^{(0)}_0$ reduced to a constant by the limit, satisfies the wave equation on the Penrose limit spacetime exactly. In this sense, the Penrose limit acts on the spacetime analogously to a truncation after the first subleading order in the geometrical optics approximation, at least when simultaneously expanded close to a reference null geodesic as in \eqref{eqn:Scongruenceexpansion}.

Neither the adapted coordinates \eqref{eqn:adapted} nor the Rosen coordinates \eqref{eqn:Rosen} are globally well-defined. They (roughly speaking) break down at caustics of the congruence $\partial_U$. The related breakdown of the geometrical optics approximation, in the local form \eqref{eqn:geomoptics}, is a starting point for more sophisticated global mathematical treatments~\cite{guillemin2013semi}. However, for the plane wave \eqref{eqn:Rosen} it is well-known~\cite{Penrose:1965rx,Penrose1976} that the coordinate singularities of the Rosen coordinates can be removed by going to Brinkmann coordinates $\left\lbrace u, v, x^a \right\rbrace$ with $a, b, \ldots \in \left\lbrace 1, \ldots, d-2 \right \rbrace$, in which the metric takes the form~\cite{Brinkmann:1925fr}
\begin{equation}\label{eqn:Brinkmann}
ds_{\gamma}^2 = 2 du dv - H_{ab}(u)x^a x^b du^2 + \delta_{ab} dx^a dx^b  \, .
\end{equation}
The Penrose limit of an arbitrary null geodesic can be found in~\cite{Chawla:2024mse}, or more implicitly in~\cite{Penrose:1995cu,Shipley:2019kfq}. Closely related geodesic deviation equations for Kerr are also discussed in~\cite{Cariglia:2018erv}.
In terms of the original spacetime, the ``wave-profile'' $H_{ab}(u)$ is exactly related to the geodesic deviation along the reference null geodesic~\cite{Blau:2006ar}
\begin{equation}\label{eqn:penroseframe}
H_{ab}(u) = \left(R_{\mu \nu \alpha \beta} E^{\mu}_a u^{\nu} E^{\alpha}_b u^{\beta} \right)|_{\gamma} \, ,
\end{equation}
where $E^{\mu}_a$ is a transverse frame, parallel propagated along $\gamma$, while $u^{\mu}$ is its tangent; locally, $u^{\mu} = (dS)^{\mu}|_{\gamma}$.  The Penrose limit \eqref{eqn:Brinkmann} can be viewed as the leading order of a Fermi null coordinate expansion, closely related to keeping higher orders in adapted coordinates \eqref{eqn:adapted} beyond the Rosen coordinate plane wave \eqref{eqn:Rosen}~\cite{Blau:2006ar}.

\subsubsection{Eikonal spectrum}

Plane wave metrics of the form \eqref{eqn:Brinkmann} are particularly simple. They generally act precisely as a harmonic oscillator with square frequency matrix $H_{ab}$. Explicitly, for a massless free scalar $\Psi(x) = e^{i v p_v}\psi(u,x_1,x_2)$ we have
\begin{equation} \label{eqn:quantumoscillator}
\frac{i}{p_v} \partial_u \psi =\left(  -\frac{1}{2p_v^2} \delta^{ab}\partial_a \partial_b + \frac{1}{2} H_{ab}(u) x^a x^b \right) \psi  \, .
\end{equation}

Electromagnetic and gravitational perturbations can be constructed from massless free scalar solution to \eqref{eqn:quantumoscillator} by a spin-raising operator~\cite{Adamo:2017nia,Araneda:2022lgu}. This is the origin of the spin-independence of the eikonal QNMs at the order \eqref{eqn:subleadingeikonal} from the Penrose limit point of view. Note also that $v|_{x^a=0} = S$ and $p_v \sim 1/\lambdabar$ acts as the inverse Planck constant in \eqref{eqn:quantumoscillator}. Its value is fixed by the choice of $\gamma$ and a single-valuedness condition, just like $E$ was fixed in \eqref{eqn:Scongruenceexpansion} at $\delta r = \delta \theta = 0$. Instead, the $\lambdabar^0$ pieces in eikonal QNMs directly correspond to the eigenstates of \eqref{eqn:quantumoscillator} under the affine time evolution $u$, with two important subtleties. First, $H_{ab}(u)$ is explicitly time-dependent in important cases of interest, such as nonequatorial orbits of a Kerr BH. On the other hand, in that case, it is periodic, so there is a well-defined notion of Floquet states~\cite{Sambe:1973cnm}. Second, there is an unstable direction in the oscillator \eqref{eqn:quantumoscillator}, which is of course expected as the modes are quasinormal~\cite{Ferrari:1984zz,Parmentier:2023axg}.

For a Schwarzschild BH, $H_{ab}$ is diagonal, and using the same implicit affine parameter as in \eqref{eqn:Scongruenceexpansion} one finds $-H_{11} = H_{22} = 3 L_c^2 M/r_c^5 = E^2/9M^2$. Taking into account the transformation from the affine parameter $u$ to the Schwarzschild time $t$, Eq.~\eqref{eqn:quantumoscillator} represents a stable harmonic oscillator with frequency $\Omega_{\rm prec} = 1/3\sqrt{3}M$ and an unstable one with Lyapunov exponent $\Lambda_L = 1/3\sqrt{3}M$, consistent with Eq.~\eqref{eqn:subleadingeikonal}. From the perspective of the plane wave \eqref{eqn:Brinkmann}, $-H_{11} = H_{22}$ follows directly from the Einstein equations. In combination with spherical symmetry, this fixes \eqref{eqn:quantumoscillator} entirely up to $\Omega_{\rm orb}$. The eikonal QNMs of a $d$-dimensional (vacuum) spherically symmetric BH follow from the same logic, with $d-3$ stable oscillators and an unstable one~\cite{Berti:2003si,Konoplya:2003ii}. While more complicated for less symmetric cases, it is still striking how directly the Einstein equations are reflected in the $O(\lambdabar^0)$ eikonal QNM frequencies.

The gravitational wave functions near the light ring can be constructed by solving Eq.~\eqref{eqn:quantumoscillator} and using a spin raising operator~\cite{Adamo:2017nia,Araneda:2022lgu}. Alternatively, (mode) raising and lowering operators can be constructed from the isometry algebra of \eqref{eqn:Brinkmann}, which in that example includes $\partial_u$ in addition to the algebra of Killing vectors $\left\lbrace Q^{(a)},P^{(a)},Z \right \rbrace$
\begin{equation}\label{eqn:planewaveisometryalgebra}
\left[Q_{(a)}, P_{(b)}\right] = \delta_{ab} Z \, , \quad \left[Q_{(a)}, Q_{(b)}\right] = \left[P_{(a)}, P_{(b)}\right] = \left[Z, P_{(b)}\right] = \left[Z, P_{(b)}\right] = 0 \, ,
\end{equation}
present in any plane wave spacetime. In Eq.~\eqref{eqn:planewaveisometryalgebra}, the central element $Z = \partial_v$  while $\left\lbrace Q^{(a)},P^{(a)} \right \rbrace$ are more complicated (see e.g.~\cite{Blau:2002mw,Harte:2012uw}), but note that $d-2$ of these are manifest as coordinate vector fields in Rosen coordinates \eqref{eqn:Rosen}. After constructing raising and lowering operators using the additional symmetry $X = \partial_u$ from
\begin{equation}
	\left[X, Q_{(a)}\right] = Q_{(a)}  \, , \quad 	\left[X, P_{(a)}\right] = - H_{ab}Q_{(b)} \, ,
\end{equation}
one can act on the Gaussian solutions to \eqref{eqn:quantumoscillator},
associated to \eqref{eqn:Scongruenceexpansion} with resulting QNM frequency
$\omega_{\ell 0}$, to obtain the overtone wavefunctions corresponding to
$\omega_{\ell n}$. For an alternative approach to this ``emergent'' symmetry
structure in eikonal QNMs see~\cite{Hadar:2022xag}. The nature of the
simplifications involved perhaps necessitates that some effective emergent
symmetry governs the eikonal QNM spectrum. It is less obvious that this symmetry
should be realized in terms of local spacetime isometries, which reflects that,
while QNMs are ordinarily sensitive to the global spacetime structure, eikonal
QNMs become localized at the light-ring.

Another instance where a subset of QNMs organizes according to a ``spacetime-limit'' isometry are the ZDMs of near-extremal rotating BHs (see also Section \eqref{sec:near_extremal_branching}). These QNMs take the form~\cite{Detweiler:1980gk,Cardoso:2004hh,Yang:2012pj,Casals:2019vdb}
\begin{equation}\label{eqn:zerodamped}
	\omega_{\ell m n} \approx m \Omega_{\rm H} - i(n + h_{\ell m}) 2 \pi T_{\rm H} \, ,
\end{equation}
where $T_{\rm H} \to 0$ is the BH temperature, $\Omega_H \to 1/2M$ is the horizon angular velocity, and $\text{Re}(h_{\ell m}) = 1/2$ for modes with $m \sim \ell$~\cite{Yang:2013uba}. The modes  \eqref{eqn:zerodamped} are closely related to interesting phenomenology near-extremal rotating BHs, such as a ``polynomial ringdown'' phase~\cite{Yang:2013uba,Yang:2014tla,Gralla:2016sxp,Compere:2017hsi,Compere:2019wfw}. They organize themselves (approximately) into representations of the $\text{SL}(2,\mathbb{R})\times U(1)$ isometry of the near-horizon, near-extremal limit metric~\cite{Bardeen:1999px,Chen:2017ofv}, a warped version of the limiting $\text{AdS}_2 \times \text{S}^2$ ($\text{AdS}_{p+2} \times \text{S}^{d-p-2}$) near-horizon of extremal charged BHs (p-branes)~\cite{Gibbons:1993sv}.

The Penrose limit can capture the eikonal limit of Eq.~\eqref{eqn:zerodamped} as (near-equatorial) corotating bound null geodesics go into the near-horizon region~\cite{Bardeen:1972fi}: see~\cite{Kapec:2022dvc,Fransen:2023eqj} for the interplay between the near-horizon and the Penrose limit descriptions. Notably, the isometries of a spacetime are a ``hereditary property''; any limit spacetime will have at least as much isometries as its parent~\cite{Geroch:1969ca,Blau:2002mw,Blau:2002dy}. Nevertheless, the nature of the isometries changes through a group contraction~\cite{Inonu:1953sp}, which is clearly illustrated in the interplay between the near-horizon and Penrose limits~\cite{Hatsuda:2002xp}. For the interpretation of the near-horizon Penrose limit within the context of the AdS/CFT conjecture, see~\cite{Berenstein:2002jq,Sadri:2003pr}.

As discussed in detail in Section~\ref{sec:spectral_environmental}, even
``small'' changes can drastically alter the QNM spectrum. The connection between
the light ring and the eikonal QNMs can similarly be broken. For instance, if an
effective cavity forms -- either because of exotic physics at the horizon or
because the BH is immersed in some environment -- waves can go back and forth
between the light ring and the other features involved, such as AdS asymptotics
or the surface of an exotic compact star (see also
Section~\ref{subsec:echoes_theory}).
The argument that null congruences around the light ring are the only ones
consistent with the QNM boundary conditions no longer holds, and the direct
relation of the light ring with the eikonal QNMs is broken. Alternatively, the
leading eikonal equation can be modified such that null geodesic congruences no
longer provide the relevant geometrical optics limit. Given that the eikonal
equation is determined by the ``principal symbol'' of a PDE, the terms with the
maximal number of differential operators, this routinely happens in
higher-derivative theories~\cite{Cano:2024wzo}.

The remainder of the review indicates that observational opportunities are
pushing for an understanding of the ringdown in more quantitative detail. This
comes with computational challenges and results that are not always easy to
interpret. The eikonal limit can provide helpful clarifications. However, to do
so, it too must be developed further. The Penrose limit perspective has natural
(Penrose-G\"uven limit) extensions to include additional fields~\cite{Gueven:2000ru} which would for instance be applicable to the Kerr-Newman
case (see Section \ref{sec:KNqnm}). The limiting plane waves are almost
universal solutions to higher-derivative gravity theories in which they are
naturally incorporated~\cite{Gueven:1987ad}: see
Section~\ref{subsec:theory-spec} for potential applications. The Penrose limit
also may also lead to a better understanding of qualitative aspects of
nonlinearities (see Sections~\ref{sec:nonlinSch} and \ref{sec:nonlinKerr} and
Refs.~\cite{Kehagias:2024sgh,Bucciotti:2025rxa,Kehagias:2025ntm, Fransen:2025cgv} for preliminary
work in this direction). Higher-order Fermi null coordinate expansions could
help relate higher-order eikonal results directly to curvature properties around
the light-ring~\cite{Blau:2006ar}. Finally, it is precisely the nearly bound
null geodesics that are responsible for a distinctive ``photon ring'' in BH
images~\cite{Johnson:2019ljv,Gralla:2019drh}. Current Event Horizon Telescope
images are not yet able to resolve the universal, accretion-independent features
of this photon ring, but future observations have the potential to do
so~\cite{Gralla:2020pra,Johnson:2024ttr}. For this reason, the eikonal limit of
QNMs is a direct link between GW and very-long-baseline interferometry
observations of BHs.

\subsection{Mode scalar products: completeness, orthogonality, and other properties}  \label{sec:scalar_products} 

\vspace{-.1cm}

\noindent \textit{Initial contributors: Green, London, Panosso Macedo, Sberna}

\vspace{.2cm}

\subsubsection{Introduction}
Perturbed BHs are inherently nonconservative systems, as they generally emit gravitational radiation that can escape to infinity or fall into the BH horizon. This means that to understand the properties of their QNM spectrum -- in mathematical terms, of QNMs as a kind of vector space -- we cannot rely on the familiar theory of conservative systems.
In conservative systems (e.g., a particle in a box in quantum mechanics), the equations of motion are Hermitian, their spectrum consists of real eigenvalues, and the related eigenfunctions are referred to as normal modes~\cite{Morse:1955aqj,ARFKEN2013401,Courant1954}. Normal modes are orthogonal with respect to an inner product and constitute a complete set.
These properties can be exploited to build general solutions, to compute mode excitation coefficients, and to build perturbative solutions to more complex problems. They also ensure the stability of the spectrum~\cite{Kato1995}.

Nonconservative (or non-Hermitian) problems are not unique to gravity, but rather they are ubiquitous in physics, from electromagnetism to nuclear physics, in either classical or quantum systems~\cite{Ashida:2020dkc}.
It is therefore not surprising that the first scalar product for BH QNMs, appearing as far back as 1993, was inspired by QNMs in optical cavities~\cite{Ching:1993gt,Ching:1995rt} (see also~\cite{Sauvan:2021edw,alsheikh:tel-04116011} for reviews). 
The scalar product introduced there was a \emph{bilinear form} for Schwarzschild QNMs.
Here, a distinction needs to be made between inner products (commonly used in conservative systems) and bilinear forms. An inner product is a map from a vector space to a field that is (i) conjugate-symmetric, i.e., linear in one argument and conjugate linear in the other; and (ii) positive definite. An example of an inner product in quantum mechanics is the integral $\langle f, g\rangle = \int_a^b \overline{f(x)}g(x) \mathrm{d}x$ between two continuous, complex-valued functions on the interval $[a,\,b]$. A \emph{bilinear form} is also a map from a vector space to a field, but one that is linear in both arguments and is not guaranteed to be positive definite. %
Such a Schwarzschild QNM bilinear form was connected to the QNM excitation coefficients~\cite{Ching:1993gt}, and it was used to compute frequency shifts due to a perturbation in the potential of the mode equation~\cite{Leung:1997was,Leung:1999iq}. 
This approach was later extended to Kerr, and applied to compute perturbative corrections to the Kerr frequencies~\cite{Yang:2014tla,Zimmerman:2014aha,Mark:2014aja,Yang:2015jja}. 
Later, motivated by the practicalities of GW signal modeling, situations in which the Kerr QNM spheroidal harmonics are complete have been pointed out~\cite{London:2020uva}, and this motivates special functions which enable QNM orthogonality over the solid angle~\cite{Thompson:2023ase, Hamilton:2021pkf, Pratten:2020ceb, London:2022urb}. 
More recently, a product under which the Kerr QNMs display full spatial orthogonality was presented~\cite{Green:2022htq}.

\begin{figure*}[t]
    \centering
    \includegraphics[width=0.45\linewidth]{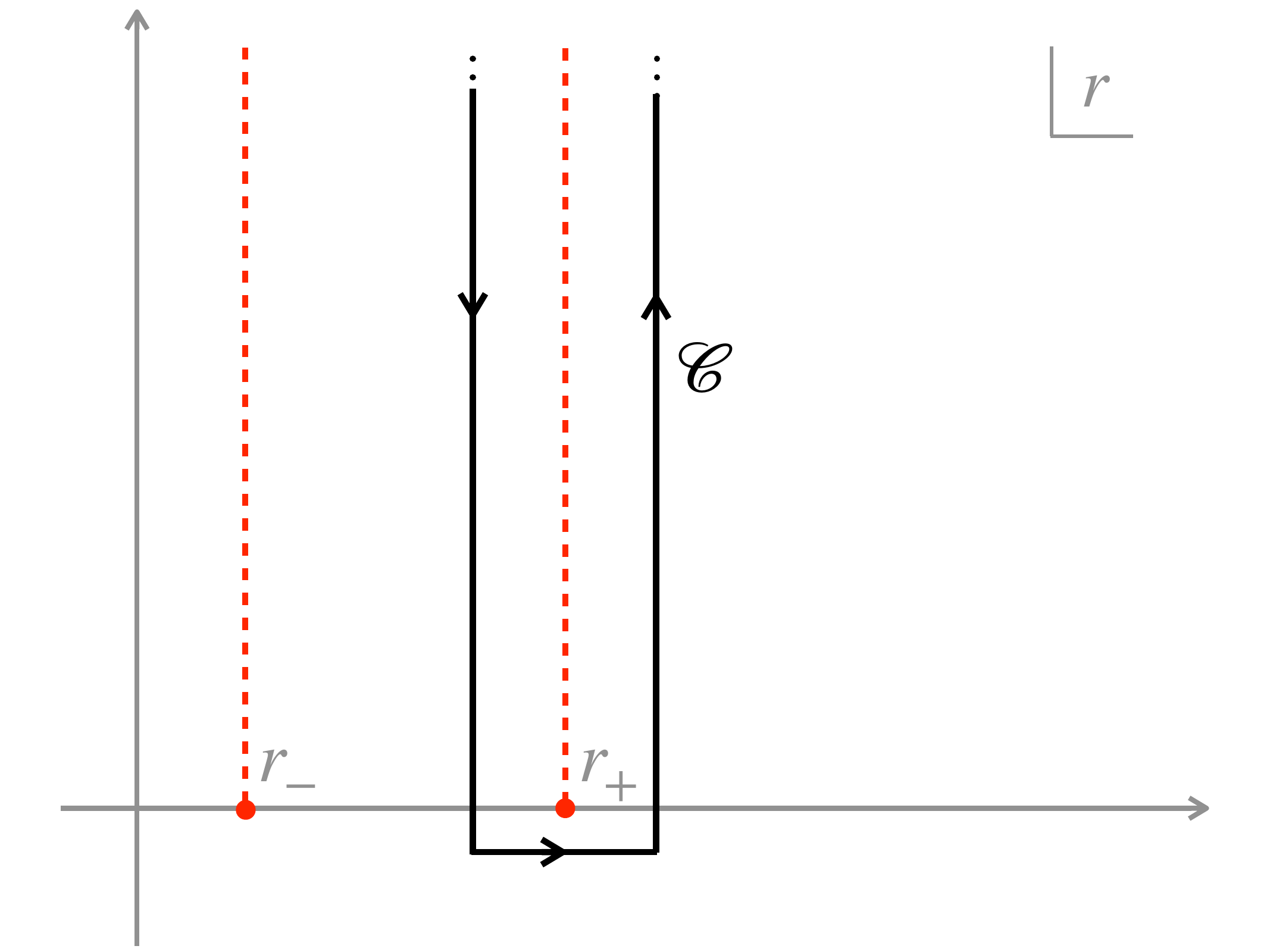}
    \caption{Integration contour for QNM products ($\mathcal{C}$, solid black line) in the Boyer-Lindquist coordinate $r$. We also show the branch cuts of the radial function, stemming from $r_{\pm}$ (dashed red).}
    \label{fig:contour_products}
\end{figure*}
A notable technical complication in the definition of Kerr QNM products 
arises from the asymptotic behavior of QNMs at the bifurcation surface and spatial infinity, located at $r= r_+$ and $r\to \infty$, respectively, in the Boyer-Lindquist coordinates. The QNM radial eigenfunctions diverge exponentially in these two limits, so inner products must be modified to give finite results (this same issue is encountered in other physical systems with QNMs, such as resonators in electromagnetism~\cite{Wu:2025sjt}). The simplest resolution is to compute the inner product by deforming the radial integration into regions of the complex plane where modes decay exponentially~\cite{Leaver:1986gd,Green:2022htq,Ma:2024qcv,Arnaudo:2025bnm}. An example of such a contour in the Boyer-Lindquist radial coordinate, for modes with ${\rm Re} (\omega_1+\omega_2)>0$, is given in Fig.~\ref{fig:contour_products}. 
In certain cases, convergence may be accomplished also by subtracting appropriate boundary terms: see e.g.~\cite{Ching:1993gt,Ching:1995rt,Sberna:2021eui,Cannizzaro:2023jle,Motohashi:2024fwt}.
Alternatively, it has long been suggested~\cite{gowdy_wave_1981,schmidt_relativistic_1993} that selecting suitable hypersurfaces, correctly foliating both the BH horizon and future null infinity, should yield QNM regular functions asymptotically. The hyperboloidal framework in BH perturbation theory, discussed in Section~\ref{subsubsec:hyperboloidal} below, exploits this geometrical strategy. By constructing time-translation invariant hyperboloidal coordinates in BH spacetimes~\cite{Zenginoglu:2007jw} and identifying their relation to regularization schemes in the FD~\cite{Zenginoglu:2011jz, Ansorg:2016ztf}, the framework has become a widespread approach in BH perturbation theory over the past decades (see~\cite{PanossoMacedo:2023qzp, PanossoMacedo:2024nkw} for recent reviews). Besides addressing the asymptotic divergence of QNMs, the hyperboloidal framework offers a powerful approach to go beyond the linear regime and to properly resolve late-time tails (see Sections~\ref{sec:nonlinKerr}, \ref{sec:nonlin_num_expe}, and \ref{sec:tails}). 

This body of work, reviewed in more detail in the following subsections, demonstrated that some properties of conservative systems can be recovered -- at least to some extent, and at the cost of additional mathematical complexity. There is indeed a sense in which Kerr QNMs are vectors within a topological space and orthogonal. Simple extensions of time-independent perturbation theory can be used to compute QNM frequency shifts. We will review a number of these product spaces for QNMs, as summarized in Table~\ref{tab:mode-products}. Scalar products of QNMs have been used in several applications, including the calculation of perturbative frequency shifts~\cite{Mark:2014aja,Hussain:2022ins,Cannizzaro:2023jle}, the study of avoided crossings and resonances in Kerr QNMs (see Section~\ref{sec:EP}), and the modeling of GWs from BBHs~\cite{London:2022urb,Garcia-Quiros:2020qpx}.
Other properties of conservative systems, such as spectral stability (see Section~\ref{sec:spectral_environmental}) and completeness (discussed briefly here and in Section~\ref{subsec:greensfunc}), are unfortunately lost for Kerr QNM spectra. For this reason we will also review some recent developments in the field achieved by importing tools from the theory of non-normal operators~\cite{trefethen2005spectra, Sjostrand2019} to gravity via the hyperboloidal framework, including the notion of pseudospectrum~\cite{trefethen2005spectra,Jaramillo:2020tuu,Jaramillo:2022kuv} and the so-called Keldysh projection scheme~\cite{Besson:2024adi}.

\begin{table}[]
    \centering
    \begin{tblr}{X[.85,l]X[1,l]X[1.6,l]}
    \hline\hline
    Product & Definition & Key properties \\
    \hline
    Angular product, $\brak{\cdot}{\cdot}_{\theta}$& $\int_0^{\pi}d\theta \sin\theta \tilde{S}_\ell(\theta)S_{\ell'}(\theta)$& QNMs with identical $n$ index are biorthogonal over the solid angle. The adjoint-spheroidal function, $\tilde{S}_\ell$, is defined in Eq.~\eqref{S-adj-fun}. \\
    $(r,\theta)$ product, $\brak{\cdot}{\cdot}_{r\theta}$& $\int_{r_+}^\infty \int_0^\pi dr d\theta \Delta^s \sin\theta \psi_1 \psi_2$ & $(r,\theta)$ Teukolsky operator is symmetric for fixed $s$, $m$ and $w$.\\
    QNM-orthogonality product, $\langle\langle\cdot,\cdot\rangle\rangle$  & $\int {\rm d}\Sigma_a \pi^a\left[\Psi_2^{2s/3}\mathcal{J}\psi_1,\psi_2\right]$, see Eq.~\eqref{eq:product_green_coord}  %
     & QNMs are orthogonal for $\omega_1 \ne \omega_2$. Connected to the $(r,\theta)$ product via Eq.~\eqref{orthog-r-theta}. \\ 
    Energy product, $\braket{\cdot|\cdot}_{\rm E}$ & $\int (q_1^{\mu \nu} \bar \phi^\ast_{,\bar x^\mu} \bar \psi_{,\bar x^\nu} + q_0 \bar \phi^\ast \bar \psi)$, see Eq.~\eqref{eq:Energy_Norm} & Positive definite. Evaluated on hyperboloidal hypersurfaces. Operator non-self-adjoint due to fluxes at ${\cal H}^+$ and $\scri^+$.\\
    High-derivative $H^{\rm p}$ product, $\braket{\cdot|\cdot}_{H^{\rm p}}$ & $\sum_{k=0}^{\rm p} \int q_k^{\mu \nu}(\bar x)\partial^k_{\bar \mu} \bar \phi^\ast \partial^k_{\bar \nu} \bar \psi$, see Eq.~\eqref{eq:Hp_norm} & Positive definite. Evaluated on hyperboloidal hypersurfaces. QNM definition in terms of Gevrey-2 class of functions.\\
    \hline\hline
    \end{tblr}
    \caption{
    Various scalar products on QNMs. The first two are based on the Sturm-Liouville form of the angular and radial Teukolsky equations. The third is designed such that the Geroch-Held-Penrose~\cite{Geroch:1973am} Lie-$t$ derivative is symmetric. For integrations extending to the bifurcation surface or spatial infinity, care must be taken to properly regularize divergences. Alternatively, the hyperboloidal framework offers a geometrical regularization for the QNM eigenfunctions. In this context, inner products make the non-self-adjoint character of wave operators explicitly due to energy fluxes, and one formally defines QNMs as eigenvectors in a proper Hilbert space.}
    \label{tab:mode-products}
\end{table}

In the subsections that follow we will pay considerable attention to the spatial properties of Kerr QNMs, in particular their spatial orthogonality~\cite{London:2020uva,Green:2022htq,Ma:2017bxq}, as well as ongoing efforts to understand whether QNMs may be spatially complete~\cite{London:2020uva,London:2023aeo,London:2023idh}.

\vspace{0.25cm}
\par
\noindent
\textit{Primer: QNMs as a vector space in the azimuthal variable $\phi$.}
We will begin by considering the azimuthal dependence $e^{i m \phi}$ in \eqn{eq:Teukolsky_separation_form}.
This provides an unambiguous yet restrictive avenue to better understand QNM orthogonality and completeness.
It is also a reference relative to which some nontrivial aspects of QNM vector spaces may be framed.

The requirement that QNM eigenfunctions be continuous and smooth as one rotates around the BH spin direction (i.e., along $\phi$) means that $m$ may only take on integer values.
Therefore, for fixed $s$, $\ell$ and $n$, the QNMs constitute a countably infinite set whose {natural label} or \textit{order} is $m \in \mathbb{Z}$. 
The related complex exponential functions, $e^{i m \phi}$, are known to be orthogonal and complete: 
\begin{subequations}
\begin{align}
    \label{phi-orthog}
    \brak{\psi^*_m}{\psi_{m'}}_\phi   &=  \int_{0}^{2 \pi} \, e^{-i m  \phi} \, e^{i m'  \phi} \, d\phi = {2 \pi} \, \delta_{m'm} \; ,
    \\
    \label{phi-complete}
    \mathbb{I}_\phi  \; &= \; \sum_{m=-\infty}^{\infty} \, \ketbra{\psi_m}{\psi^*_{m}}_\phi \; .
\end{align}
\label{phi-complete-both}    
\end{subequations}
In \eqn{phi-orthog}, we define $\psi_m = e^{i m \phi}$, and $\psi^*_m$ is the complex conjugate of $\psi_m$. 
In \eqn{phi-complete}, $\ketbra{\psi_m}{\psi^*_{m}}_\phi$ signals that all products must be performed according to $\brak{\cdot}{\cdot}_\phi$, as is done in the right-hand side of \eqn{phi-orthog}.
Note that the scalar product in \eqn{phi-orthog} differs from an inner product in that complex conjugation is not treated as an inherent property of the bra-ket operation $\brak{\cdot}{\cdot}_\phi$; rather, we treat $\psi^*_m$ and $\psi_m$ as members of different sets.
This is characteristic of the scalar products encountered throughout this section, and reflects the complex nature of the QNM frequencies.
In \eqn{phi-complete}, $\mathbb{I}_\phi$ is simply the identity operator, meaning that if $\Psi(\phi)$ is an arbitrary function of $\phi$ with domain $[0,2 \pi]$, then  
\begin{align}
    \label{phi-decomp}
    \mathbb{I}_\phi \ket{\Psi}  = \sum_{m=-\infty}^{\infty} \, \ket{\psi_m}\brak{\psi^*_m}{\Psi}_\phi \; .
\end{align} 
Together, Eqs.~\eqref{phi-complete-both} and \eqref{phi-decomp} define a topological vector space, specifically: a set of vectors, $\psi_m$, and a way to compute distances between them with $\brak{\cdot}{\cdot}_\phi$.
We will refer to this as the $\phi$ scalar product space, or $\mathcal{Q}_\phi$, where
    \begin{align}
        \label{Q-phi-a}
        \mathcal{Q}_{\phi}  &=  \{ \brak{\cdot}{\cdot}_\phi,\mathcal{U}_\phi \}
    \end{align}
In \eqn{Q-phi-a}, we have defined $\mathcal{U}_\phi=\{e^{i m \phi} \, |  \, m\in\mathbb{Z}\}$.
Since $\mathcal{U}_\phi$ is complete, and norms may be defined via $\brak{\cdot}{\cdot}_\phi$, $\mathcal{Q}_\phi$ is a Hilbert space. 
\par In summary, the azimuthal functions used in QNM solutions are orthogonal and complete, and can thereby serve as a basis for general decomposition. Indeed, these functions are widely used throughout GW theory~(see e.g.~\cite{Ruiz:2007yx,Blanchet:2013haa}).
Note that, since Eqs.~\eqref{phi-complete-both} and \eqref{phi-decomp} are independent of the QNM frequency, they apply to all QNM eigenfunctions equally.
In the following, we will discuss how QNMs may share these properties, either as full spatial functions, or in single spatial dimensions. 
Historically, these possibilities have been most commonly studied using \SLT{}.

\subsubsection{Sturm-Liouville theory in the Teukolsky formalism} 

Given Teukolsky's master equation, one can apply ideas from Sturm-Liouville theory to construct scalar-product spaces for the QNMs.
These spaces provide a foundation for understanding whether it is possible to use the QNMs for projection and decomposition, much as is done with special functions in electromagnetism and quantum mechanics.
Key to this program is the intersection between \SLT{} and complex operator theory~\cite{London:2023aeo,teschl2000jacobi,lax2002functional}.

According to Sturm-Liouville theory, a multi- or uni-variate ordinary differential operator may be used to (i) construct a scalar product under which the operator is self-adjoint, (ii) combine the operator's eigenfunctions with the scalar product to define a vector space, and (iii) study the properties of that scalar product space in the context of physical applications.
These applications include (but are not limited to) the estimation of separation constants, the approximation of QNM angular and radial functions, and potential ``spectral decomposition'' with QNMs, for which it is required that the QNMs be \textit{spatially complete} in some way. 
The points above apply to the full spatial dependence of the QNMs, as well as to their angular and radial functions, ${_{s}S_{\ell}(\theta;\tilde{\omega}_{\ell m n})}$ and ${_{s}R_{\ell}(\theta;\tilde{\omega}_{\ell m n})}$, respectively. 
Therefore, it is possible to construct a scalar product space over $(r,\theta,\phi)$, as well as spaces over $r$, $\theta$ and $\phi$ separately.
We have already encountered the scalar product space in $\phi$, $\mathcal{Q}_\phi$. 
Like $\mathcal{Q}_\phi$, the remaining spaces are defined by a scalar product, as well as a set of functions (abstract vectors) which span the space. 

In this subsection, we will briefly review the construction and properties of select \SL{} scalar product spaces, namely $\mathcal{Q}_{\theta}$ and $\mathcal{Q}_{r\theta}$~\cite{Zimmerman:2014aha,London:2020uva,Green:2022htq}.
The reader should note that, since Teukolsky's master equation is an operator with complex coefficients, it is possible to have degenerate eigenvalues, as well as related Jordan chains~\cite{Axler:2015,Finster:2015xma}. 
As it has been observed that this is not the case for Schwarzschild BHs and for Kerr BHs below the extremal limit, we will henceforth focus only on cases in which Teukolsky's radial and angular problems have unique eigenvalues~\cite{London:2020uva,London:2023aeo,London:2023idh}. 
Lastly, in what follows we will include within \SLT{} two types of situations: one in which we consider a single QNM frequency (``single-frequency'' case), and another in which we consider two QNM frequencies concurrently (``two-frequency'' case).
These two cases are built upon slightly different versions of \SL{} arguments~\cite{Morse:1955aqj,ARFKEN2013401}. 
\vspace{0.25cm}
\par
\noindent
\textit{\SLT{} and indefinite scalar products.}
Let $\mathcal{T}_{t r \theta \phi}$ be the linear differential operator corresponding to Teukolsky's equation in Boyer-Lindquist coordinates, as shown in \eqn{eq:Teukolsky_Master}. 
With a QNM frequency $\omega$ in mind, we wish to consider the action of $\mathcal{T}_{t r \theta \phi}$ on functions whose $t$ and $\phi$ dependence is defined by an overall proportionality to $e^{-i  (\omega  t  -  m  \phi )}$, as in Eq.~\eqref{eq:Teukolsky_separation_form}.
Since the result will still be proportional to $e^{-i  (\omega  t  -  m  \phi )}$, it is useful to consider a new operator, $\mathcal{T}_{r \theta}$, %
 \begin{align}
        \label{eq:Teukolsky_Master_Op2a}
        \mathcal{T}_{r \theta} \; &= \; e^{i  (\omega  t  -  m  \phi )} \; \mathcal{T}_{t r \theta \phi} \;e^{-i  (\omega  t  -  m  \phi )} \, \mathbb{I} \; .
    \end{align}
On the right-hand side of \eqn{eq:Teukolsky_Master_Op2a}, $\mathcal{T}_{t r \theta \phi}$ acts on the quantity $( e^{-i  (\omega  t  -  m  \phi )} \; \mathbb{I} )$, where $\mathbb{I}$ (the identity operator) stands in for an arbitrary function of $r$ and $\theta$.
\SLT{} provides a simple framework for the construction of scalar products.
Since $\mathcal{T}_{r \theta}$ involves derivatives scaled by complex-valued functions, related scalar products will typically be nonpositive definite or, equivalently, \textit{indefinite}~\cite{adkins2012algebra,gohberg2006indefinite}.
\par The key idea is that one may attempt to define a \textit{weight function}, $\mathrm{W}$, {such that} $\mathrm{W}  \mathcal{T}_{r \theta}$ is formally self-adjoint: 
\begin{align}
    \label{T-self-adj}
    \brak{f}{\mathrm{W} \mathcal{T}_{r \theta}\,g} = \brak{\mathrm{W}\mathcal{T}_{r \theta}\,f}{ g} \, ,
\end{align}
where $\brak{\cdot}{\cdot}$ refers to an integral over a domain volume. 
Note that formally self-adjoint operators may not satisfy \SL{} boundary conditions~\cite{ARFKEN2013401}. Any second order linear operator may be formatted by coordinate and/or similarity transformation to be formally self-adjoint; however, the required transformations may involve impractical or nonalgebraic functions~\cite{ARFKEN2013401,Courant1954}.
One then \textit{redefines} the scalar product to implicitly include the weight function, making the operator self-adjoint with respect to the newly defined product~\cite{ARFKEN2013401}. 
As demonstrated in many texts~\cite{pinchover_rubinstein_2005,ARFKEN2013401,Courant1954,Morse:1955aqj,Arnold1978,abramowitz+stegun,ByronFuller}, the property of self-adjointness imposes constraints on an operator's form, and has the simplifying benefit of requiring that an operator's matrix elements are symmetric in a suitably chosen vector space. 
All linear differential operators may be reformatted (via coordinate transformation and/or appropriate scaling of a weight function) such they are formally self-adjoint.
\vspace{0.25cm}
\par
\noindent
\textit{Single-frequency \SL{} weights.} For example, let us consider the polar space, $\mathcal{Q}_\theta$.
Using \SLT{}, we will construct a polar scalar product, $\brak{\cdot}{\cdot}_{\theta}$, and then briefly discuss the radial scalar product.
According to \SLT{}, $\brak{\cdot}{\cdot}_{\theta}$ is defined such that $\mathcal{T}_{r \theta}$ is self-adjoint by first rewriting it as
\begin{align}
    \label{eq:Teukolsky_Master_Op3}
    \mathcal{T}_{r \theta}  =  p_0(r,\theta,\partial_r) + p_1(\theta) \frac{\partial}{\partial\theta} + p_2(\theta)\frac{\partial^2}{\partial\theta^2} \; .
\end{align}
In \eqn{eq:Teukolsky_Master_Op3}, the functions $p_0$, $p_1$ and $p_2$ are simply the coefficients of derivatives with respect to $\theta$. 
Their explicit form may be determined by inspection.
Since there are no cross-derivatives between $r$ and $\theta$, we have grouped derivatives in $r$ (i.e., $\partial_r$) into $p_0$. 
Following standard \SL{} arguments~\cite{ARFKEN2013401,Courant1954}, a weight $\mathrm{W}_\theta$ for which $\mathrm{W}_\theta\mathcal{T}_{r \theta }$ is self-adjoint is
\begin{subequations}
    \label{W_theta_all}
    \begin{align}
        \label{W-phi-a}
        \mathrm{W}_\theta  &=  \frac{1}{p_2(\theta)} \, \int^{\theta} \frac{p_1(\theta')}{p_2(\theta')} \, d\theta' \;
        \\
        \label{W_theta}
        &=  \sin\theta\,.
    \end{align}
\end{subequations}
The resulting \SL{} scalar product is $\brak{f}{g}_{\theta}=\int_{0}^{\pi}fg \mathrm{W}_\theta d\theta$.
Unlike $t$ and $\phi$, $r$ and $\theta$ appear explicitly in $\mathcal{T}_{tr\theta \phi}$, making it possible to derive single variable weights from \eqn{eq:Teukolsky_Master_Op3}. 
Applying the algorithm encapsulated by Eqs.~\eqref{eq:Teukolsky_Master_Op3} and \eqref{W-phi-a} to $r$ yields 
\begin{align}
    \label{W_r}
    \mathrm{W}_{r}  &=  \Delta(r)^s  =  (r-r_+)^s \, (r-r_-)^s\,.
\end{align}
Recall that $r_+$ and $r_-$ are defined below \eqn{eq:Kerr_Metric}.
\Eqnsa{W_theta_all}{W_r} may also be derived by considering Teukolsky's radial and angular equations separately. 
\par \SL{} weights depend strongly on coordinate and scaling choices.
Coordinate choices affect the weight function via the chain rule applied to the scalar product's integrand~\cite{London:2023aeo}.
Choices about spacetime-slicing affect the spatial asymptotic behavior of the QNMs, and are therefore interchangeable with different similarity transformations, akin to \eqn{eq:Teukolsky_Master_Op2a}~\cite{Minucci:2024qrn,London:2023aeo,Teukolsky:1972my}.
These similarity transformations are sometimes referred to as scaling or ``s-homotopic'' transformations~\cite{London:2023aeo,ronveaux1995heun,leaver1985analytic}.
In the context of Frobenius-series solutions to the QNM problem, these transformations correspond to solutions with different spatial asymptotic behaviors~\cite{London:2023aeo,London:2023idh,Teukolsky:1974yv}.

The spatial behavior of QNM solutions is constrained by their phase velocity near the BH horizon and spatial infinity~\cite{Teukolsky:1973ha}.
Therefore, while one may consider arbitrary similarity transformations on the QNM problem, the underlying physics is only consistent with transformations that preserve the QNM ingoing-outgoing boundary conditions~\cite{London:2023aeo,Teukolsky:1973ha,Leaver:1985ax,Cook:2014cta}. 
\par For example, $\mathrm{W}_{r}$ has been used in the context of operator perturbation theory for QNMs~\cite{Zimmerman:2014aha,Ma:2024qcv}; however, as it fails to incorporate the QNM asymptotic boundary conditions, this weight is not the most appropriate for studying the potential spatial completeness of QNMs~\cite{London:2023aeo,London:2023idh}.
Upon imposing the QNM boundary conditions~\cite{London:2023aeo}, $\mathrm{W}_{r}$ 
transforms into a complex valued Pollaczek-Jacobi weight~\cite{Chen:2010,Chen:2019},
\begin{align}
    \label{W_xi}
    \W_\xi  =  \xi^{\mathrm{B}_0} \, (1-\xi)^{\mathrm{B}_1} \,  e^{\frac{\mathrm{B}_2}{1-\xi}} 
\end{align}
where $\xi=(r-r_+)/(r-r_-)$, and $\mathrm{B}_0$ through $\mathrm{B}_2$ are defined in terms of $a$, $s$, $M$ and $\omega$~\cite{London:2023aeo}.
Polynomials derived from $\W_\xi$ enable one to express Teukolsky's radial operator for QNMs as a symmetric tridiagonal matrix, and \eqn{eq:radialR:Diff_Eqn} as a simple matrix eigenvalue problem~\cite{London:2023idh}.
These observations point to intersections between QNMs, matrix representations of \SL{} operators, and the study of orthogonal polynomial systems~\cite{Fackerell:1977shn,London:2020uva,London:2023aeo,London:2023idh, chihara2011introduction,teschl2000jacobi}.
In the case of the spheroidal harmonics for QNMs, these intersections are one avenue through which QNM spatial completeness may be studied~\cite{London:2020uva,London:2022urb}.
\par The \SL{} weights defined in Eqs.~\eqref{W_theta_all} and \eqref{W_r} have a few important properties that are relevant to their potential use in defining a {unique} projection operator, similar to $\mathbb{I}_\phi$ in \eqn{phi-decomp}.
Both $\W_r$ and $\W_\theta$ are independent of the QNM frequency; as a result, the related scalar products could be applied equally to different QNMs that make up GW signals.
Concurrently, $\W_r$ and $\W_\theta$ were derived using $\mathcal{T}_{r \theta}$ which is, by construction, frequency-dependent~\cite{London:2023aeo}: see 
Eq.~\eqref{eq:Teukolsky_Master}.
This is illustrated by the fact that $\W_r$ can be similarity-transformed into the explicitly frequency-dependent weight $\W_\xi$, without affecting the underlying physics. 
Thus, similarity transformations enable the conversion of frequency-agnostic \SL{} weights into single-frequency weights, and vice-versa. 
\vspace{0.25cm}
\par
\noindent
\textit{Two-frequency weights, and the $2$D space $\mathcal{Q}_{r\theta}$.}
Using a variant of the \SL{} construction~\cite{Morse:1955aqj}, two-frequency weights may be constructed from the action of the $(r,\theta)$ part of Teukolsky's operator on two different QNM solutions~\cite{Motohashi:2024fwt,Ma:2024qcv,Green:2022htq}. 
The $(r,\theta)$ part of Teukolsky's operator, $\mathcal{T}_{r\theta}$, was shown in \eqn{eq:Teukolsky_Master_Op2a}.
Here, it is useful to refer to this operator as $\mathcal{T}_{r\theta}(w)$, to explicitly indicate its dependence on a single QNM frequency. 
Following the logic of \eqn{W-phi-a}, $\mathcal{T}_{r \theta}(\omega)$ is self-adjoint with respect to the $2$D product whose weight is $\W_r\,\W_\theta$, and whose domain of integration is the BH exterior. 
For two arbitrary functions, $f(r,\theta)$ and $g(r,\theta)$, this product is
\begin{align}
    \label{eq:brak_r_theta_phi}
    \brak{f}{g}_{r\theta}  =  \int_{r_+}^{\infty}\int_{0}^{\pi}\; f(r,\theta)\,g(r,\theta)\;\W^\mathrm{SL}_\theta \, \W^\mathrm{SL}_r \, d\theta \, dr \;.
\end{align}
\par We now consider two different QNM frequencies, $\omega_j$ and $\omega_k$. 
For compactness, let $\mathcal{T}_k=\mathcal{T}_{r \theta}(\omega_k)$, and let $\psi_k$ be the $(r,\theta)$ dependence of a single QNM.
Thus, for QNMs, $\mathcal{T}_k\,\psi_k=0$.
Using this fact, and the properties of $\brak{\cdot}{\cdot}$, it can be shown that~\cite{Ma:2024qcv} 
\begin{align}
    \label{Tjk-d}
    \brak{ \psi_j }{\mathcal{T}_j \, \psi_k}_{r\theta} - \brak{\psi_j}{\mathcal{T}_k \, \psi_k}_{r\theta}  =  \brak{ \psi_j }{(\mathcal{T}_j - \mathcal{T}_k) \,| \psi_k}_{r\theta}  =  0 \; ,
\end{align}
where $\mathcal{T}_j - \mathcal{T}_k$ can be written as
\begin{align}
    \label{Tjk-e}
    (\mathcal{T}_j - \mathcal{T}_k)  =  \left(\omega _j-\omega _k\right) \W_{r\theta} \, ,
\end{align}
and where $\W_{r\theta}$ is
\begin{align}
    \label{W-r-theta}
    \nonumber
    \W_{r\theta} = -\left(\omega _j+\omega _k\right) &\left(\frac{\left(a^2+r^2\right)^2}{\Delta (r)}-a^2 \sin ^2(\theta )\right) 
   \\ 
   &+2 i s
   \left(\frac{M(r^2-a^2)}{\Delta (r)}-i a \cos (\theta )-r\right)
   +\frac{4 M a m r}{\Delta (r)}  .
\end{align}
In effect, \eqn{Tjk-d} implies that $\W_{r\theta}$ is a frequency-dependent weight function such that two QNM functions of $r$ and $\theta$ are orthogonal,
\begin{align}
    \label{orthog-r-theta}
    \brak{ \psi_j }{\W_{r\theta} \,| \psi_k}_{r\theta}  \propto  \delta_{jk} \; .
\end{align}
Equations~\eqref{Tjk-d}--\eqref{orthog-r-theta} echo standard arguments for the orthogonality of linearly independent eigenfunctions of a self-adjoint operator~\cite{Courant1954, London:2020uva, ARFKEN2013401}, but differ in that one focuses on the operator's null space.
With \eqn{orthog-r-theta}, we may define the $r$-$\theta$ scalar product space as
\begin{align}
    \label{Q-r-theta}
    \mathcal{Q}_{r\theta}  &=  \{ \brak{\cdot}{\W_{r\theta}|\cdot}_\phi, \mathcal{U}_{r\theta} \} \, ,
\end{align}
where $ \mathcal{U}_{r\theta}  =  \{ {_{s}S_{\ell}(\theta;a{\omega}_{\ell m n})} \, {_{s}R_{n}(r;{\omega}_{\ell m n})} \; | \; \ell\ge \max(|s|,|m|)\text{, } n\ge 0  \}$.
It is presently unknown whether this space is complete.
It may be over-complete~\cite{Cook:2014cta,Cook:2016ngj, London:2023aeo,London:2023idh}, which would mean that an identity operator constructed from $\mathcal{Q}_{r\theta}$ would not be unique.
\par  Like $\W_\theta$ and $\W_r$, the weight $\W_{r\theta}$ is strongly coordinate-dependent. 
Although $\W_{r\theta}$ depends on two QNM frequencies, the sum of QNM frequencies in \eqn{W-r-theta} may be replaced with time derivatives, resulting in a weighting operator that is frequency-independent~\cite{Green:2022htq,Ma:2024qcv}.
The product \eqref{orthog-r-theta} is equivalent to the result of~\cite{Green:2022htq}, discussed in the next subsection (see~\cite{Ma:2024qcv}).
\par Lastly, the algorithm encapsulated in 
Eqs.~\eqref{Tjk-d}--\eqref{orthog-r-theta} also applies to Teukolsky's radial and angular equations.
For example, the resulting angular weight is 
\begin{align}
    \label{W-theta-2}
    \W_\theta^{(2)}  =  1 +  a  \cos(\theta)  \, \frac{ \omega_j-\omega_k }{A(\omega_j)-A(\omega_k)} \, \left[\;-2s  +  a\cos(\theta)\,\left(\omega _j+\omega _k\right) \;\right] \, .
\end{align}
Since the separation constants, $A(\omega)$, are generally nonlinear functions of $\omega$ (see e.g.~\cite{Cook:2004kt,Fackerell:1977shn,Seidel:1988ue}), it is unlikely that $\W_\theta^{(2)}$ can be rewritten in a frequency-independent manner. 
Consequently, it is unlikely that $\W_\theta^{(2)}$ can be used to construct a frequency-agnostic projection operator. %
Equivalently, it is unclear whether $\W_\theta^{(2)}$ is compatible with numerical decomposition.
\vspace{0.25cm}
\par
\noindent
{\textit{The polar space, $\mathcal{Q}_{\theta}$}.} 
An alternative perspective~\cite{London:2020uva,London:2022urb} concludes that decomposition is possible with the QNM spheroidal harmonics, but that it requires a generalization of orthogonality called \textit{biorthogonality} (i.e., orthogonality between members of two different sets), as well as the use of special functions called adjoint-spheroidal harmonics.
The resulting decomposition can be applied, for example, to extreme and comparable mass-ratio BBHs~\cite{London:2022urb}.
Below, we briefly review the treatment of the QNM spheroidal harmonics as a Hilbert space.
In what follows, we will drop the labels of $s$, $m$ and $n$, as they will all be considered fixed, and we will refer only to the $\theta$-dependence of the spherical and spheroidal harmonics; e.g., $S_\ell={_s}S_\ell(\theta;a\omega_{\ell m n})$.

For fixed values of $s$, $m$ and $n$, the following properties hold~\cite{London:2020uva}: (i) the QNM angular separation constants are unique, even for very large values of $\ell$, and (ii) there is a close relationship between the spherical and spheroidal harmonics, meaning that there exists a one-to-one mapping between them.
The related spherical-spheroidal mixing coefficients may be used to construct an invertible operator, $\mathbb{T}$, that converts spin-weighted spherical harmonics into the QNM spheroidal harmonics,
\begin{align}
    \label{S-Tjk-a}
    \mathbb{T} = \sum_{\ell,\ell'} \, \ket{Y_{\ell'}} \brak{S_{\ell'}}{Y_{\ell}}_\theta \bra{Y_{\ell}} \; . 
\end{align}
In \eqn{S-Tjk-a}, all scalar products are defined using the \SL{} weight, $\W_\theta=\sin(\theta)$, with $\theta\in[0,2\pi]$.
Note that we have represented $\mathbb{T}$ in the basis of spherical harmonics, where it is a matrix with elements $\brak{S_{\ell'}}{Y_{\ell}}_\theta$.
Using the orthogonality and completeness of the spherical harmonics, it is easy to show that $\mathbb{T}\, Y_{\ell } = S_{\ell }$.
The invertibility of $\mathbb{T}$ is underpinned by the bounded inverse theorem~\cite{London:2020uva}.
Given $\mathbb{T}^{-1}$, it follows that $ Y_{\ell } = \mathbb{T}^{-1}\,S_{\ell }$.
One can then recast the orthogonality relationship for the spherical harmonics in terms of the QNM spheroidal harmonics,
\begin{align}
    \label{S-bio}
    \brak{Y_{\ell}}{Y_{\ell'}}_\theta
    =\brak{Y_{\ell}} {\mathbb{T}^{-1}\mathbb{T}Y_{\ell'}} _\theta
    =\brak{\mathbb{T}^{-\dagger}Y_{\ell}} {\mathbb{T}Y_{\ell'}} _\theta
    = \brak{\tilde{S}_{\ell}} {S_{\ell'}}_\theta =   \delta_{\ell' \ell}  \; .
\end{align}
$\mathbb{T}^{-\dagger}Y_{\ell}$ can be identified~\cite{London:2020uva} as an ``adjoint-spheroidal'' harmonic
\begin{align}
    \label{S-adj-fun}
    \tilde{S}_\ell = \mathbb{T}^{-\dagger} \, Y_{\ell} \; .
\end{align}
In \eqn{S-adj-fun}, one should read that $\tilde{S}_\ell$ is represented in the basis of spherical harmonics, where the expansion coefficients are the matrix elements of $\mathbb{T}$ inverted and conjugate-transposed~\cite{London:2020uva}.
Using similar arguments, the QNM spheroidal harmonics may be shown to be complete, 
\begin{align}
    \label{S-complete}
    \mathbb{I}_\theta = \sum_{\ell} \, \ket{S_\ell} \bra{\tilde{S}_\ell}_\theta \; . 
\end{align}
With the weight $W_{\theta}$, we may define the scalar product space in $\theta$ as 
\begin{align}
    \label{Q-theta}
    \mathcal{Q}_\theta = \{ \brak{\cdot}{\cdot}_\theta, \mathcal{U}_\theta \} \; ,
\end{align}
where $\mathcal{U}_\theta = \{ \, {_{s}}S_{\ell}(\theta;a\omega_{\ell m n})\, |\, \ell \ge \max(|m|,|s|)\, \}$~\cite{London:2020uva,London:2023aeo}.
Te small spin expansion of $\tilde{S}_\ell$ was computed and $\mathcal{Q}_\theta$ was used to estimate ringdown amplitudes by projection for extreme mass-ratio, comparable mass-ratio, and precessing BBH simulations in~\cite{London:2022urb}.
An approximation of $\mathcal{Q}_\theta$ is used in \texttt{PhenomXHM}~\cite{Garcia-Quiros:2020qpx} and related GW signal models~\cite{Thompson:2023ase,Hamilton:2021pkf,Pratten:2020ceb}.
\par We have the freedom to define the norm of the spheroidal harmonics based on the scalar product~\cite{London:2020uva}.
This, along with completeness of the spheroidal harmonics~\cite{Finster:2015xma,London:2020uva}, means that $\mathcal{Q}_\theta$ is a Hilbert space. 
\subsubsection{QNM orthogonality product.}

A bilinear form with respect to which the time-translation operator is symmetric can be constructed to show that QNMs of the spin-$s$ Teukolsky equation are orthogonal~\cite{Green:2022htq}. 
This construction, inspired by the physics of leaky optical cavities~\cite{PhysRevA.49.3057} and other open systems, and already applied to scalar waves in Schwarzschild~\cite{Ching:1993gt,Ching:1995rt,Leung:1997was,Leung:1999iq}, provides a more general derivation of the orthogonality of QNMs.

The starting point is a conserved current %
defined by the Teukolsky operator. Here the Teukolsky operator acting on spin-$s$ solutions $\Psi$ [the left-hand side of Eq.~\eqref{eq:Teukolsky_separation_form}] is denoted by $\mathcal{O}$. The current also involves the spacetime adjoint of the Teukolsky equation, denoted by $\mathcal{O}^\dagger$ and acting on spin $-s$ solutions $\tilde{\Psi}$ (we denote this ``spacetime adjoint'' to stress the fact, often left implicit in the literature, that the product defining the adjoint is the spacetime integral). The current $\pi^a$ is a functional of a spin $s$ and a spin $-s$ solution, and is implicitly defined by
\begin{equation}
\label{las-eq1}
(\mathcal{O}^\dagger \tilde{\psi})\psi - \tilde{\psi} (\mathcal{O}\psi) = \nabla_a \pi^a[\tilde{\psi},\psi] .
\end{equation}
The current is evidently conserved on solutions, i.e., $\nabla_a \pi^a[\tilde{\psi},\psi]=0$ for $\psi \in \ker \mathcal O, \tilde\psi \in \ker \mathcal O^\dagger$. A ``base product'' can then be built as the associated conserved quantity on any spacelike hypersurface,
\begin{equation}
\Pi[\tilde{\psi},\psi] = \int {\rm d}\Sigma_a \pi^a[\tilde{\psi},\psi] .
\end{equation}

To build a product on two modes of the same spin $s$, one can map a mode of spin $s$ into a mode of spin $-s$ using the operator $\mathcal{J}$ associated to the discrete $t-\phi$ reflection symmetry of Kerr~\cite{Green:2022htq}. The resulting product is
\begin{equation}\label{eq:product_green}
\langle\langle\psi_1,\psi_2\rangle\rangle = \int {\rm d}\Sigma_a \pi^a[\Psi_2^{2s/3}\mathcal{J}\psi_1,\psi_2] ,
\end{equation}
where $\Psi_2$ is the only nonzero Weyl scalar of the Kerr metric. The product possesses the following key properties~\cite{Green:2022htq}: (i) it does not rely on the spatial separability of the Teukolsky equation; (ii) it is bilinear (a bilinear form);
(iii) it is symmetric, $\langle\langle \psi_1, \psi_2 \rangle \rangle = \langle\langle \psi_2, \psi_1 \rangle \rangle$. Moreover, the time-translation symmetry operator $L_t$ of Kerr is symmetric with respect to the product, $\langle\langle \psi_1, L_t \psi_2 \rangle \rangle = \langle\langle L_t \psi_1, \psi_2 \rangle \rangle$. Consequently, for two QNMs, $(\omega_1 - \omega_2)\langle\langle \psi_1, \psi_2 \rangle \rangle = 0 $, implying that modes with different frequencies are orthogonal. Equivalently, $\{\psi_n\}$ and $\{\tilde{\psi}_n \equiv \Psi_2^{2s/3}\mathcal{J}\psi_n\}$ form biorthogonal sets under the product $\Pi$. In other words, QNM orthogonality arises from the orthogonality between QNMs and ``anti-QNMs''~\cite{Arnaudo:2025bnm}. Thanks to this understanding, this construction has been extended to scalar fields on black holes in AdS, exploiting the CPT operator to map the two families of QNM solutions \cite{Arnaudo:2025bnm}.

In Boyer-Lindquist coordinates, %
the product takes the form
\begin{subequations}
\begin{align}\label{eq:product_green_coord}
  &\langle \langle \psi_{1}, \psi_{2} \rangle \rangle \nonumber\\
  &={} 4 M^{4/3} %
  \int_\Sigma \mathrm{d}  r \, \mathrm{d}\theta \mathrm{d}\phi\, \sin\theta \Delta^s \Bigg[
  \psi_1\Big|_{\substack{t\to-t \\ \phi\to-\phi}} \left( \frac{\Lambda}{\Delta}\partial_t + \frac{2Mra}{\Delta}\partial_\phi  -2 s \left[ -r - ia\cos\theta + \frac{M}{\Delta}(r^2 - a^2)\right] \right) \psi_2
    \nonumber\\
  & 
    \quad + \psi_2 \left( \frac{\Lambda}{\Delta}\partial_t + \frac{2Mra}{\Delta}\partial_\phi  -2 s \left[ -r - ia\cos\theta + \frac{M}{\Delta}(r^2 - a^2)\right] \right) \psi_1\Big|_{\substack{t\to-t \\ \phi\to-\phi}}
   \Bigg] \\
\label{eq:product_green_coord_qnm}
  &={}  8\pi M^{4/3} \delta_{m_1m_2} e^{-i(\omega_2-\omega_1)t}
  \int_{\mathcal{C}} {\rm d} r \int_0^\pi  {\rm d} \theta\, \bigg\{ \sin\theta \Delta^s S_1(\theta) S_2(\theta)  R_1(r)  R_2(r) \nonumber\\
  & \quad \left( - \frac{i\Lambda}{\Delta}(\omega_1+\omega_2) + \frac{2iMra}{\Delta}(m_1+m_2) -2 s \left[ -r - ia\cos\theta + \frac{M}{\Delta}(r^2 - a^2)\right] \right)\bigg\},
\end{align}
\end{subequations}
where $\Lambda\equiv (r^2+a^2)^2-\Delta a^2\sin^2\theta$, and in the second line we have explicitly applied it to QNMs.
An example of the regularization contour, $\mathcal{C}$, is shown in Fig.~\ref{fig:contour_products}.

The bilinear form defined in Eq.~\eqref{eq:product_green_coord} can be used, for example, to compute the excitation coefficients of QNMs given data $\psi$ on a space-like hypersurface~\cite{Green:2022htq},
\begin{equation}
    C_{n\ell m} = \frac{\langle\langle\psi_{n\ell m},\psi\rangle\rangle }{\langle\langle\psi_{n\ell m},\psi_{n\ell m}\rangle\rangle} \, ,
\end{equation}
providing an alternative to the Green's function formalism described in Sections~\ref{subsec:greensfunc} and~\ref{subsec:Kerr_amplitudes} below.
It also provides an alternative to the Sturm-Liouville %
product Eq.~\eqref{eq:brak_r_theta_phi} to compute perturbative frequency shifts~\cite{Cannizzaro:2023jle}. 
Finally, it enables to calculate corrections to linear ringdown predictions~\cite{Sberna:2021eui,Redondo-Yuste:2023ipg,Iuliano:2024ogr}.

\subsubsection{The hyperboloidal framework, the energy product, and the pseudospectrum\label{subsubsec:hyperboloidal}}

The study of scalar products was initially motivated by attempts to understand the completeness properties of the QNM eigenfunctions~\cite{Nollert:1996rf,Nollert:1998ys}. 
As previously mentioned, however, the radial integration used in all of the products introduced so far requires a regularization of the QNM eigenfunctions via a deformation of the relevant integration contour in the complex $r$ plane, as in Fig.~\ref{fig:contour_products}. Over the past decades, the so-called hyperboloidal framework~\cite{Zenginoglu:2007jw, Zenginoglu:2011jz, Ansorg:2016ztf,PanossoMacedo:2023qzp, PanossoMacedo:2024nkw} 
has become a widespread geometrical approach in BH perturbation theory to avoid the divergence of QNMs on spacelike hypersurfaces near the bifurcation sphere and spatial infinity.

In Riemannian geometry, spherical coordinates (or their spheroidal deformations) play a crucial role to capture symmetries of the manifold or to simplify the equations. Hyperboloids 
play a similar role in Lorentzian geometries~\cite{Zenginoglu:2024bzs}, as they capture key features of the spacetime's casual structure. In the particular case of BH spacetimes, the level sets of constant hyperboloidal time coordinate are regular
spacelike hypersurfaces, extending smoothly from the BH event horizon to future
null infinity. Most importantly, they reveal that QNM eigenfunctions are regular in the BH exterior~\cite{Horowitz:1999jd,Zenginoglu:2011zz,vasy2013microlocal,Warnick:2013hba,Ansorg:2016ztf,Gajic:2019oem,Gajic:2019qdd,Gajic:2024xrn}. 

In essence, a hyperboloidal foliation $\bar x^\mu = (\tau, \sigma, \theta, \varphi)$ follows from the usual Boyer-Lindquist coordinates $x^\mu = (t, r, \theta, \phi)$ via
\begin{equation}
\label{eq:HypTrasfo}
t = r_+ \bigg( \tau - H(\sigma) \bigg), \quad r = r_0 + \dfrac{r_+ - r_0}{\sigma},  \quad \phi = \varphi - \phi_*(r).
\end{equation}
Each element in Eq.~\eqref{eq:HypTrasfo} contributes to the regularity of the QNM eigenfunctions at the BH horizon and at future null infinity in different ways. For instance, the azimuthal angle transformation defined by the relation $d\phi_*/dr = a/\Delta$ is well-known for the ingoing Kerr coordinates, and it ensures the regularity of observables across the BH horizon. 

To properly obtain a hyperboloidal  surface, the height function $H(\sigma)$ in Eq.~\eqref{eq:HypTrasfo} must satisfy specific asymptotic conditions~\cite{Zenginoglu:2007jw,Zenginoglu:2011jz,PanossoMacedo:2023qzp}. The so-called minimal gauge retains just these minimal conditions when constructing a hyperboloidal foliation, and $H(\sigma)$ reads~\cite{PanossoMacedo:2019npm, PanossoMacedo:2023qzp}
  \begin{equation}
  \label{eq:H_MinGaug}
  H(\sigma) = -\dfrac{1 - r_0/r_+}{\sigma} + \dfrac{2M}{r_+} \ln \sigma + \dfrac{1}{2 r_+ \kappa_+ } \ln(1- \sigma) + \dfrac{1}{2 r_+ \kappa_- } \ln\left(1- \dfrac{\sigma}{\sigma_-}\right),
  \end{equation}
with $\kappa_\pm = (4M)^{-1} \left( 1 - a^2/r_\pm^2 \right)$ denoting the surface gravity of the horizons. 

The radial compactification in Eq.~\eqref{eq:HypTrasfo} includes only the minimal structure necessary to bring the infinitely far wave zone into a finite range of coordinate values. Indeed, $\sigma = 0$ represents future null infinity, whereas the choice in Eq.~\eqref{eq:HypTrasfo} conveniently fixes the BH horizon $r = r_+$ to $\sigma_+ = 1$. This minimal compactification is not unique, as it allows for an arbitrary initial offset $r_0$ in the radial coordinate. In terms of this parameter, the Cauchy horizon at $r=r_-$ is mapped into $\sigma_- = (r_+ - r_0)/(r_- - r_0)$.  

This arbitrary constant offers some freedom to understand and explore the Kerr spacetime extremal limit from this geometrical perspective~\cite{PanossoMacedo:2019npm}.
A straightforward option is to choose $r_0=0$, which implies $\sigma_- = r_+/r_-$, i.e., the coordinate value of the Cauchy horizon depends on the BH rotation parameter $a$. As a consequence, the hyperboloidal hypersurfaces $\tau = $ constant foliate the Cauchy horizon smoothly, and the extremal Kerr limit is directly achieved when $|a|\rightarrow M$ and $r_- \rightarrow r_+$~\cite{PanossoMacedo:2019npm,Minucci:2024qrn}. This option is equivalent to the hyperboloidal implementation used in~\cite{Ripley:2022ypi}.

A second option is to fix the Cauchy horizon at a coordinate value independent of the BH rotation parameter. The choice $r_0=r_-$, for instance, is such that $\sigma_-\rightarrow \infty$ for all values of $a$. In this way, the hypersurfaces $\tau = $ constant do not cross the Cauchy horizon~\cite{Minucci:2024qrn}, and the extremal limit provides a discontinuous transition into the Kerr near-horizon geometry~\cite{PanossoMacedo:2019npm}. This option is the spacetime counterpart of Leaver's strategy~\cite{leaver1985analytic, Leaver:1990zz,Ansorg:2016ztf, PanossoMacedo:2018hab, PanossoMacedo:2019npm}.

The construction of Eqs.~\eqref{eq:HypTrasfo},\eqref{eq:H_MinGaug} automatically yields a configuration where outgoing boundary conditions at both the BH horizon and the wave zone are satisfied. The physically relevant scenario is described by the regular solution to the resulting hyperboloidal wave equation~\cite{Zenginoglu:2011jz, PanossoMacedo:2023qzp, PanossoMacedo:2024nkw}, $\partial_{\tau} \vec u = {\hat L} \vec u$. The operator ${\hat L}$, acting only on the spatial dimensions, is  schematically represented as
\begin{equation}
\label{eq:operator_L}
\quad {\hat L} = \left( 
\begin{array}{cc}
0 & 1 \\
{\boldsymbol L}_1 & {\boldsymbol L}_2
\end{array}
\right).
\end{equation} 
The second-order differential operator ${\boldsymbol L}_1$ assumes a singular Sturm–Liouville form, and the first-order differential operator ${\boldsymbol L}_2$ incorporates the energy flow into the BH and towards $\scri^+$~\cite{Jaramillo:2020tuu, Gasperin:2021kfv}.

An early objective of this approach was to utilize the fact that QNMs defined on hyperboloidal surfaces ``are represented by proper eigenvalues and eigenfunctions''~\cite{schmidt_relativistic_1993}. A FD transformation $\vec u \sim e^{ \mathfrak{s} \tau} \vec u$ leads to the eigenvalue problem $\hat L \vec u = \mathfrak{s} \vec u$, with the dimensionless parameter $\mathfrak{s} = - i r_+ \omega$.  
 Consequently, the eigenvalues and eigenvectors of ${\hat L}$ correspond directly to the QNM eigenfrequencies and eigenfunctions~\cite{Jaramillo:2020tuu}. Advances in this framework for spherically symmetric spacetimes have been significant~\cite{Jaramillo:2020tuu, PanossoMacedo:2023qzp},  with successful applications to the Kerr spacetime~\cite{PanossoMacedo:2019npm, Ripley:2022ypi}.

The regularity of the QNM eigenfunctions allows one to revisit the completeness properties of the QNM eigenfunctions~\cite{Nollert:1996rf,Nollert:1998ys} via  the representation of solutions as a complete spectral decomposition in terms of a superposition of a discrete spectrum (the QNMs) and of a continuous spectrum from the branch cut contribution~\cite{Ansorg:2016ztf, PanossoMacedo:2018hab}:
\begin{equation}
\label{eq:spec_decomp}
\bar \phi_{\ell m}(\tau, \sigma) = \sum_{n=0}^{\infty} \eta_{n \ell m} \bar \phi_{n\ell m}(\sigma) e^{\mathfrak{s}_{n \ell m} \tau} + \underset{{\rm Re}(\mathfrak{s})\leq 0 }\int \eta_{\ell m}(\omega) e^{\mathfrak{s} \tau} \bar \phi_{n\ell m}(\mathfrak{s}; \sigma) d \omega.
\end{equation}
The notation $\bar \phi$ emphasizes that these fields are defined in terms of the hyperboloidal coordinate system, with $\bar \phi_{n\ell m}(\sigma)$ the regular QNM eigenfunctions. An extension of Leaver's algorithm~\cite{Ansorg:2016ztf, PanossoMacedo:2018hab} allows one to define QNMs as the poles of a discrete Green's function (see e.g. Eq.~(73) in~\cite{Ansorg:2016ztf}) and to compute the QNM and branch cut amplitudes, $\eta_n$ and $\eta(\mathfrak{s})$, from a given set of initial data via a discrete projection operator (see e.g. Eqs.~(106) and (109) in~\cite{Ansorg:2016ztf}). The orthogonality of QNMs is verified in terms of these discrete projection operators. The spectral decomposition of Eq.~\eqref{eq:spec_decomp} has been recently revisited in terms of the so-called Keldysh scheme~\cite{Gasperin:2021kfv,alsheikh:tel-04116011,Besson:2024adi}, that extends the analysis in~\cite{Ansorg:2016ztf} beyond one-dimensional problems. This scheme relies on defining the QNM projection solely in terms of quantities in the dual vector space~\cite{Besson:2024adi}, in particular not requiring a scalar product. Proof-of-principle studies indicate that the transpose operator $L^t$ provides the elements to express Eq.~\eqref{eq:spec_decomp} with QNM/tail amplitudes fixed by projections of the initial data into the dual vector space~\cite{Besson:2024adi}.

The early studies on the completeness properties of the QNM eigenfunctions~\cite{Nollert:1996rf,Nollert:1998ys} also noticed a peculiar feature: the QNM spectrum changed drastically when considering a discrete approximation to the Regge-Wheeler potential. In particular, the (continuum representation) QNMs were not recovered, even in the limit where the discrete representation was pushed closer to the continuous counterpart. Since this FD instability did not seem to directly affect the TD signal at moderately early times, the issue was regarded as a mostly technical curiosity.
However, spectral instability is now understood to be a broad property of BH spectra which occurs quite generically in the presence of environmental perturbations (see~\cite{Aguirregabiria:1996zy,Vishveshwara:1996jgz} for early work, and Section~\ref{sec:spectral_environmental} for a more extensive discussion). 
The issue of spectral instabilities -- possibly introduced by dilute astrophysical environments or even corrections to Einstein's theory -- is of paramount importance. We discuss it more thoroughly below in Section~\ref{sec:spectral_environmental}, but note that detection strategies and systematic effects, discussed in Section~\ref{subsec:RD_systematics}, are directly impacted by such possible instabilities.

The development of a hyperboloidal framework was a crucial tool in our understanding of these instabilities as it also enabled the application of non-self-adjoint operator theory~\cite{trefethen2005spectra, Sjostrand2019} to gravitational studies~\cite{Jaramillo:2020tuu}. In particular, the notion of pseudospectrum~\cite{trefethen2005spectra} provides a formal explanation for QNM spectral instability, and it facilitates a nonmodal analysis~\cite{Jaramillo:2022kuv}. The $\epsilon$-pseudospectrum $\varsigma_{\epsilon}(\omega)$ is defined as the set of points in the complex frequency plane such that
\begin{equation}
\varsigma_{\epsilon}(\mathfrak{s}) \equiv\left|\left| \left(\hat L - \mathfrak{s} \right)^{-1} \right| \right|_{\rm E} > 1/\epsilon  .  \label{eq:def_pseudospec}
\end{equation}
The limit $\epsilon \rightarrow 0$ recovers the spectrum of the operator $\hat L$. For $\epsilon\gtrsim 0$, the pseudospectrum reveals the spectral properties of the operator as a topographic map on the complex plane, with peaks located at the complex eigenvalues. Unlike the spectrum, the computation of the pseudospectrum requires a well-defined notion of norm to distinguish ``large'' from ``small'' $\epsilon$ values. A norm reflecting the field's energy on a constant-time hypersurface, based on the scalar product of the field and its derivatives, is given schematically by~\cite{Jaramillo:2020tuu, Gasperin:2021kfv}
\begin{equation}
\label{eq:Energy_Norm}
\braket{\bar \phi|\bar \psi}_{\rm E} \propto \underset{\tau = {\rm const.}}\int (q_1^{\mu \nu}(\bar x) \bar \phi^\ast_{,\bar x^\mu} \bar \psi_{,\bar x^\nu} + q_0(\bar x) \bar \phi^\ast \bar \psi) \, d\Sigma_{\tau}.
\end{equation}
The functions $q_i^{\mu \nu}(\bar x)$ ($i=0,1$) are directly related to the spacetime metric and to the structure of the wave equation. A common example of such an energy norm is derived from the energy-momentum tensor of a scalar field, where the integrand components reflect the field's kinetic and potential energy contributions. The product~\eqref{eq:Energy_Norm} is positive-definite and makes the non-self-adjoint character of the operator $\hat L$ evident~\cite{Jaramillo:2020tuu, Gasperin:2021kfv}. Indeed, calculating the formal adjoint with respect to the energy norm $\braket{\bar \phi | \hat L \bar \psi}_{\rm E} = \braket{ \hat L^\dagger \bar \phi |  \bar \psi}_{\rm E}$ yields $\hat L^\dagger = \hat L + \hat L_2^\partial$, with $L_2^\partial$ a boundary term resulting from $\boldsymbol{L}_2$ that carries information about the energy loss at the horizon and at future null infinity.

\begin{figure}[t]
    \centering
\includegraphics[width=0.7\linewidth]{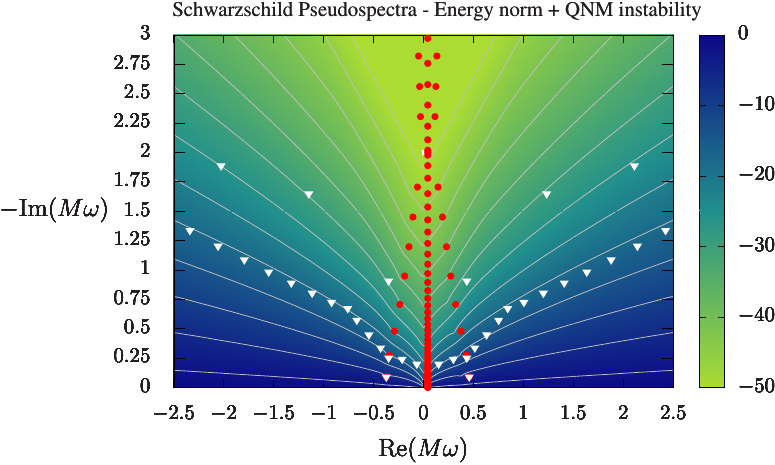}
    \caption{Schwarzschild pseudospectra based on the energy norm. Open contour lines extending across the complex plane signify QNM spectral instability. Small perturbations to the wave equation's potential cause significant destabilization of QNM overtones~\cite{Jaramillo:2020tuu, Gasperin:2021kfv}. The color code corresponds to $\log_{10} \varsigma_{\epsilon}(\mathfrak{s})$ with the $\epsilon$-pseudospectra defined in Eq.~\eqref{eq:def_pseudospec} with contour lines of $\epsilon=$constant.
    Figure adapted from~\cite{Jaramillo:2020tuu}.
    }
    \label{fig:pseudospectrum}
\end{figure}

As an example, Fig.~\ref{fig:pseudospectrum} shows the pseudospectra for the Schwarzschild spacetime (Regge-Wheeler operator)  constructed under the energy norm \eqref{eq:Energy_Norm}. Open contour lines in the complex plane indicate spectral instability: small perturbations of the potential completely destabilize the QNMs overtones. The astrophysical implications of this phenomenon are discussed in Section~\ref{sec:spectral_environmental} below.

Despite the solid mathematical foundation of this program,  there are some technical open issues. The numerical calculation of the pseudospectrum with the energy norm does not always converge to a fixed, nonvanishing value with increasing numerical resolution~\cite{Jaramillo:2021tmt,Boyanov:2023qqf,Cownden:2023dam, Boyanov:2024fgc,Cai:2025irl}. In such cases, the continuum limit indicates that the pseudospectrum goes to zero in the entire complex half-plane ${\rm Re}(\omega)<0$. This result would imply that {\em any point $\omega$} in this half-plane is an eigenvalue of the operator, and not only the discrete values corresponding to the well-known QNMs. Thus, the numerical calculation of pseudospectral contour lines under the energy norm offers a correct {\em qualitative} understanding and interpretation of the underlying structure of the operator, but quantitative assessments must be performed carefully~\cite{Jaramillo:2021tmt,Boyanov:2023qqf,Cownden:2023dam, Boyanov:2024fgc,Cai:2025irl}.

The origin of this feature is not directly related to the particular numerical
scheme used in the calculation, but it is rooted in the definition of QNMs. The
intuitive notion that QNMs are mode solutions to Eq.~\eqref{eq:swSF_DiffEqn}
that ``forbid waves to travel in from infinity or out of the event horizon of
the BH'' (see Section~\ref{sec_21}) is a necessary, but not sufficient
condition~\cite{Nollert:1999ji,Nollert:1992ifk,bachelot1993resonances}. In other
words, outgoing boundary conditions at the horizon and at infinity are not the
only defining structure of QNMs, as there exist further unphysical solutions
satisfying such boundary conditions~\cite{Nollert:1999ji, Warnick:2013hba,
  Ansorg:2016ztf, Gajic:2019oem, Gajic:2019qdd, PanossoMacedo:2024nkw,
  Gajic:2024xrn} precisely in the entire complex half-plane
${\rm Re}(\omega)<0$. The numerical nonconvergence of the pseudospectrum in the
energy norm captures this subtlety.

The traditional definition of QNMs as poles of the Green's function~\cite{Nollert:1999ji,Nollert:1992ifk,bachelot1993resonances} removes the ambiguities from the intuitive approach based on the boundary conditions. By exploiting the hyperboloidal framework, an alternative definition for QNMs arises as isolated eigenvalues of the infinitesimal generator of time translations acting on an appropriately defined Hilbert space~\cite{horowitz2000quasinormal, vasy2013microlocal,Warnick:2013hba,Ansorg:2016ztf,Gajic:2019oem,Gajic:2019qdd,Gajic:2024xrn}. 

This definition relies on an inner product controlling higher derivatives of the functions belonging to the functional space. Such product generalizes the energy norm, and it is often known in the mathematical literature as the $H^{\rm p}$ Sobolev norm
\begin{equation}
\label{eq:Hp_norm}
\braket{\bar \phi|\bar \psi}_{H^{\rm p}} \propto \sum_{k=0}^{\rm p} \underset{\tau = {\rm const.}}\int q_k^{\mu \nu}(\bar x)\partial^k_{\bar \mu} \bar \phi^\ast \partial^k_{\bar \nu} \bar \psi \, d\Sigma_{\tau}.
\end{equation}
Then, QNM eigenfunctions belong to the so-called \emph{Gevrey-2 class}: roughly speaking, they are more regular than a smooth function, but less than an analytic function~\cite{horowitz2000quasinormal, vasy2013microlocal, Warnick:2013hba, Gajic:2019oem, Gajic:2019qdd,Gajic:2024xrn}.
In this context, recent investigations show that the higher-order derivatives in the $H^{\rm p}$ scalar product permit to address certain technical issues in the calculation of the pseudospectrum, providing a better mathematical control of the convergence of the pseudospectrum and offering additional insight into the stability properties of the overtones~\cite{Besson:2024adi,Besson:2025}.

All in all, the discussion in this section summarizes the ongoing effort to tackle open mathematical questions underlying the theory of BH QNMs. By revisiting seminal works from the nineties~\cite{Nollert:1996rf,Nollert:1998ys} with novel approaches and strategies, the scalar products listed here have found application in the study of many physical problems, such as  mode avoidance
 (Section~\ref{sec:avoidance}), environmental perturbations (Section~\ref{sec:spectral_environmental}), nonlinearities (Sections~\ref{sec:nonlinKerr} and \ref{sec:nonlin_num_expe}) and tails (Section~\ref{sec:tails}).
 There is an ongoing effort to explore and synthesize these different strategies to establish a unified understanding of QNM theory that reconciles their orthogonality properties, (in)completeness, non-self-adjointness, and the geometric insights offered by the various methods. 
\subsection{Spectral instabilities and environmental effects}
\label{sec:spectral_environmental}

\vspace{-.1cm}

\noindent \textit{Initial contributors: Cardoso, Destounis, Duque, Panosso Macedo}

\vspace{.2cm}

So far, we have reviewed the structure of the QNM spectrum for \textit{electrovacuum} spacetimes. This assumption is justified for BHs in isolation, but the Universe is filled with various forms of matter, including interstellar clouds, gas, plasma and dust in galaxies, dark matter, populations of compact objects, and cosmic radiation. 
There are hints that some of the GW events already observed by ground-based detectors may originate in gas-rich environments, ranging from somewhat controversial claims of electromagnetic counterparts to BBH mergers~\cite{Graham:2020gwr} to statistical hints that at least a fraction of the observed BBH population may have originated in Active Galactic Nuclei~\cite{Santini:2023ukl}. In fact, there is an ongoing effort to place direct bounds on the density of the medium surrounding these BBH merger events~\cite{CanevaSantoro:2023aol}.
The effect of environments is definitely a concern for space-based instruments like LISA that will target the merger of supermassive BH binaries in galactic centers, where the two BHs are brought to coalesce by dynamical friction and migration driven either by gas or stars~\cite{Begelman:1980vb,LISA:2022yao}. 
It is therefore important to ask how astrophysical environments will affect the BH spectroscopy program~\cite{Barausse:2014tra,Destounis:2023ruj}. This is the topic of this section.

One would expect that the BH response should be only slightly perturbed in the presence of external matter with sufficiently small density, localized away from the BH light ring. 
However, numerical investigations revealed the surprising fact that even small perturbations of the BH potential, due to e.g. astrophysical environments, can lead to considerable changes in the BH vibrational spectra, potentially orders of magnitude away from the QNM frequencies computed in vacuum~\cite{Nollert:1996rf,Nollert:1998ys,Aguirregabiria:1996zy,Vishveshwara:1996jgz,Barausse:2014tra,Jaramillo:2020tuu,Daghigh:2020jyk,Shen:2025yiy,Jaramillo:2021tmt,Cheung:2021bol,Konoplya:2022hll,Cardoso:2024mrw,Siqueira:2025lww}. 
This phenomenon is known as \emph{spectral instability}. 

The BH spectroscopy program is based on perturbative calculations of the vacuum BH spectrum in GR~\cite{Dreyer:2003bv,Berti:2005ys,Baibhav:2023clw}
and it implicitly assumes a reasonable physical robustness against ``small'' changes in the system. Thus, understanding spectral instabilities is of paramount importance for the interpretation of GW data. 

\subsubsection{Spectral instabilities}

\begin{figure*}[t]
    \centering
    \includegraphics[width=0.75\linewidth]{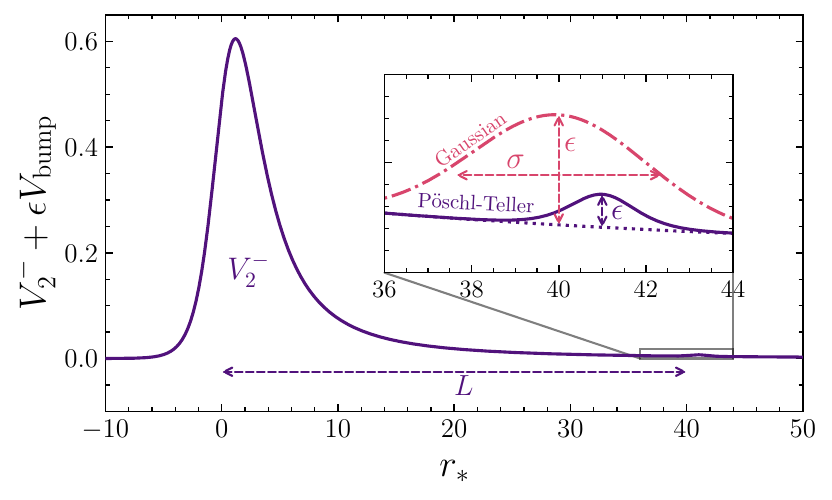}
    \caption{Schematic illustration of the effective Regge-Wheeler potential of Eq.~\eqref{eq:RW-potential} for $\ell=2$ perturbed by a small, {\em ad hoc} ``bump'' of amplitude $\epsilon$ (inset), with a P\"oschl-Teller or a Gaussian shape, centered around $r_*=L$. Figure adapted from~\cite{Cheung:2021bol}.}
    \label{fig:potential}
\end{figure*}

There are two major classes of spectral instabilities: \emph{fundamental mode} and \emph{high overtone} instabilities. A simple toy model illustrating fundamental mode instabilities in the context of massless perturbations of nonspinning BHs (see Section~\ref{sec_21}) is helpful to better understand the issue. To mimic possible environments, let us assume the master equation to be of the form of Eq.~\eqref{eq:RWZ master}, but with a perturbed potential $V_{\epsilon}$~\cite{Barausse:2014tra,Cheung:2021bol}:
\begin{equation}
\dfrac{d^2 \Psi_2^-}{d r_*^2} + \left[\omega^2 - V_{\epsilon}\right] \Psi_2^- = 0\,, \label{eq:RW}
\end{equation}
where the tortoise coordinate is defined as $dr/dr_*=1-2M/r$, and a small, localized ``bump'' is added to the effective potential $V_2^-$ in vacuum (see Fig.~\ref{fig:potential}), i.e.,
\begin{equation} 
V_{\epsilon}\equiv V_2^-+\epsilon \, V_{\rm bump}\,.\label{eq:replacement}
\end{equation} 
Note that when $\epsilon = 0$, Eq.~\eqref{eq:RW} is equivalent to the RWZ equation~\eqref{eq:RWZ master}, i.e., it governs gravitational perturbations around nonrotating BHs in GR. The functional form of the unperturbed potential is not crucial for the discussion,
and for simplicity we will consider the Regge-Wheeler potential $V_2^-$ defined in Eq.~\eqref{eq:RW-potential}.

In this \emph{ad hoc} modification, external environments conspire to produce the ``bump'' located at $r_* = L$, such that $V_{\rm bump}$ goes to zero as $r_*\to \pm \infty$. In Fig.~\ref{fig:potential} we consider two possible choices for $V_{\rm bump}$, i.e., a  P\"oschl-Teller~\cite{Poschl:1933zz} potential and a Gaussian:
\begin{equation}
V_{\rm bump}= {\rm sech}^2\left(r_* - L\right),\qquad V_{\rm bump} = \exp \left(-\frac{(r_* - L)^2}{2\sigma}\right)\,.\label{eq:bumps}
\end{equation}

\begin{figure*}[t]
    \centering
    \includegraphics[width=\linewidth]{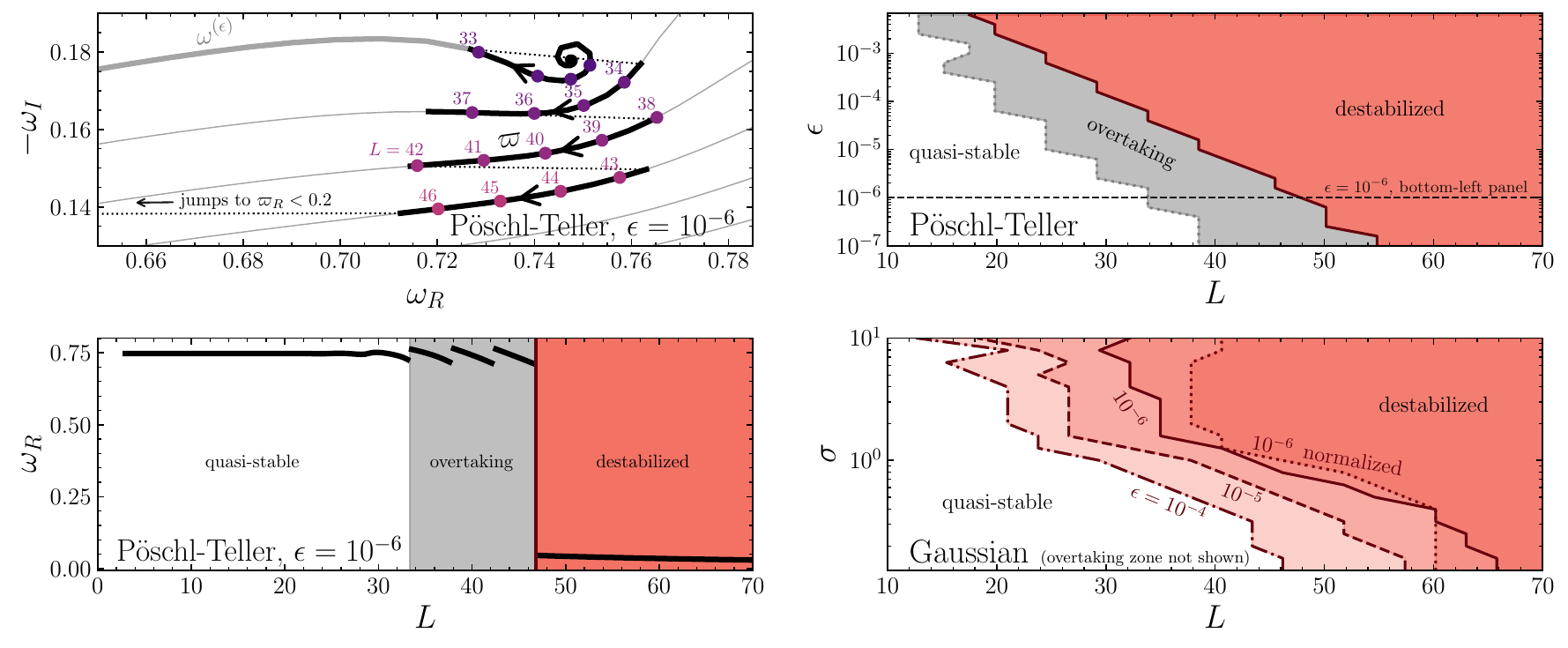}
    \caption{Top left: Migration of the fundamental mode $\varpi$ in the complex plane as the distance $L$ between the bump and the BH (marked next to the purple dots) is increased, for a P\"oschl-Teller bump and $\epsilon = 10^{-6}$.
	Bottom left: Variation of the real part of the fundamental mode $\varpi$ from the top left panel as a function of $L$. The gray area designates the $L$ range where overtaking instabilities take place, while the orange area designates the region of $L$ where the new fundamental mode $\varpi$ has been completely destabilized with respect to the scale of the perturbation $\epsilon$.
	Top right: ``Phase diagram'' illustrating the regime where the three kinds of instabilities occur in the  ($\epsilon,\,L$) plane for a P\"oschl-Teller bump.
	Bottom right: The analogous ``phase diagram'' in the ($\sigma,\,L$) plane for a Gaussian bump with different values of $\epsilon$. Figure adapted from~\cite{Cheung:2021bol}.}
    \label{fig:destabilization}
\end{figure*}

The QNM spectra of the perturbed potentials are shown in Fig.~\ref{fig:destabilization} (top left panel) for $\epsilon=10^{-6}$. 
Here, $\omega^{(\epsilon)}$ denotes the fundamental mode resulting from the original potential under a perturbation of order $\epsilon$, while $\varpi$ is the actual perturbed dominant/fundamental mode -- i.e., the mode with the smallest (in absolute value) imaginary part, which does not necessarily coincide with $\omega^{(\epsilon)}$.  
For small $L$, $\varpi$ is initially quasi-stable and it coincides with the perturbed fundamental mode of the original potential, $\omega^{(\epsilon)}$, which spirals outward from its unperturbed Schwarzschild value (thick gray line). Notice how the fundamental mode can change by a factor of two or more even for such a small value of the perturbation amplitude $\epsilon$, indicating spectral instability.

As $L$ increases, something even more dramatic happens: the trapped modes localized between the two potential barriers migrate downwards and to the left in the complex plane. Eventually (at $L\simeq 34$) one of these modes overtakes $\omega^{(\epsilon)}$. This causes the first of several discontinuous jumps in $\varpi$ (shown with dotted lines, and highlighted in the bottom-right panel). 
Figure~\ref{fig:destabilization} illustrates that, quite independently of the functional form of $V_{\rm bump}$, the fundamental mode $\varpi$ is unstable in large regions of the relevant phase space ($[L,\epsilon]$ for the P\"oschl-Teller bump, and $[L,\sigma]$ for the Gaussian bump). 
The nature of this FD instability, its relation with TD waveforms, and the location of the edges between the various regions were clarified analytically in various works~\cite{Ianniccari:2024ysv,Yang:2024vor}.

The crucial observation is that ``two bump'' systems have distinct scales. When the characteristic distance $L$ between the bump and the peak of the unperturbed BH potential (located near the light ring) is much larger than the other scale in the problem (i.e., the BH mass), a cavity is formed by the two potential peaks. This cavity supports ``trapped modes'' of frequency $\sim 1/L$ which are trapped as long as $1/L^2\ll
\epsilon V_{\rm bump}$. In these conditions, the fundamental BH  QNM is destabilized~\cite{Barausse:2014tra,Cheung:2021bol,Berti:2022xfj,Cardoso:2024mrw}: for any arbitrarily small $\epsilon V_{\rm bump}$ there is a scale $L$ where an extremely low-frequency mode appears ends up dominating over the fundamental BH QNM associated with the light ring~\cite{Cardoso:2016rao,Cardoso:2019rvt,Berti:2022xfj}, and this low-frequency mode is never ``close'' to the original fundamental mode (``close'' meaning that the mode perturbation is at most of order $\epsilon$~\cite{Berti:2022xfj}).
Similar phenomenology occurs if new scales are present {\it close} to the horizon, under the form of matter or new boundary conditions~\cite{Cardoso:2016rao,Cardoso:2019rvt};
see e.g.~\cite{Motohashi:2024fwt,Yang:2024vor,Ianniccari:2024ysv,DeLuca:2024uju,Boyanov:2024fgc} for more detailed analyses of the spectra of double-peak potentials and of the consequent spectral instabilities, and~\cite{Boyanov:2024fgc} for a discussion of the high energetic content of the apparent ``small'' bump. 
The mode avoidance and resonance phenomena discussed in Section~\ref{sec:avoidance} occur quite generally in double-peak potentials~\cite{Motohashi:2024fwt}. 

The above discussion shows that the fundamental mode is generally unstable when a second (large) length scale is present in the problem. What about the overtones? 

In an attempt to capture possible ultraviolet quantum effects and to identify spectral instabilities under small-scale changes to the effective potential, one can use the notion of the $\epsilon$-pseudospectrum defined in Section~\ref{subsubsec:hyperboloidal} to predict the occurrence of spectral instabilities without the need of introducing \emph{ad hoc} perturbations of order $\mathcal{O}(\epsilon)$ of the BH potential~\cite{Jaramillo:2020tuu}. 
The pseudospectrum is very useful to understand the phenomenon of QNM spectral instability in perturbation theory beyond the usual modal analysis~\cite{Jaramillo:2020tuu,Jaramillo:2022kuv}. Mathematical tools from the theory of non-normal operators~\cite{trefethen2005spectra, Sjostrand2019} were successfully adapted to gravitational physics through a careful treatment of the QNM boundary conditions, in which a geometrical strategy was used to adapt the time coordinates to the asymptotic spacetime structure around the BH horizon and in the wave zone, i.e., at future null infinity~\cite{Zenginoglu:2011jz, Jaramillo:2020tuu} (see Section~\ref{subsubsec:hyperboloidal}).
An appropriate norm is necessary to measure the ``strength'' or ``smallness'' of the perturbation: see the discussion around Eq.~\eqref{eq:Energy_Norm}. The energy norm is well-defined under certain conditions~\cite{Jaramillo:2020tuu,Gasperin:2021kfv,Cardoso:2024mrw,Boyanov:2024fgc}, and it is a useful discriminator to understand why the spectrum is destabilized under apparent ``small'' perturbations of the potential~\cite{Boyanov:2024fgc}.
By plotting pseudospectral contour lines, that identify the maximal migration of perturbed QNMs in the complex plane under a perturbation of order $\left|\epsilon \right|\ll 1$, one finds that asymptotically flat~\cite{Jaramillo:2020tuu,Destounis:2021lum,Jaramillo:2022kuv,Cao:2024oud}, de Sitter~\cite{Sarkar:2023rhp,Destounis:2023nmb,Luo:2024dxl}, and anti-de Sitter~\cite{Arean:2023ejh,Cownden:2023dam,Boyanov:2023qqf,Chen:2024mon} BHs all exhibit spectral instabilities. See discussion in Section~\ref{subsubsec:hyperboloidal} and fig.~\ref{fig:pseudospectrum}.

An example of spectral instability in the high-overtone regime concerns charged BHs~\cite{Motl:2003cd,Andersson:2003fh,Natario:2004jd,Berti:2009kk,Daghigh:2024wcl,Cardoso:2024mrw,Daghigh:2020jyk,Shen:2025yiy}. The asymptotic structure of the spectrum for any massless field perturbation of nonspinning, neutral BHs is
\begin{equation}
M\omega \sim \frac{\log 3-i(2n+1)\pi}{8\pi} + {\cal O}(n^{-1/2})\,,\quad Q=0\,,
\end{equation}
in the large-$n$ limit~\cite{Berti:2009kk}.
On the other hand, the asymptotic spectrum of slightly charged BHs is found to be
\begin{equation}
M\omega \sim \frac{\log 5-i(2n+1)\pi}{8\pi} + {\cal O}(n^{-1/2})\,,\quad Q\to 0\,,
\end{equation}
As a consequence, for any arbitrarily small charge $Q$ the BH spectrum differs by ${\cal O}(Q^0)$ from the spectrum of a Schwarzschild BH.
This is a clear example of spectral instability in the large-$n$ regime~\cite{Cardoso:2024mrw}. 

To conclude, a few important remarks are in order concerning the spectral instability of both the fundamental mode and overtones. 
Despite the large shift to the spectrum found above, in all realistic cases studied in the literature the corresponding early-time signal in the TD is smooth and receives only small corrections (as quantified e.g. by the unfaithfulness, see Section~\ref{sec:environments}).
In other words, the initial prompt response in the TD is very close to the vacuum BH signal, while the destabilized modes dominate only at late times.
The information of the dominant ringdown at early time signals is imprinted in the phase shift defined analogously to quantum scattering~\cite{Kyutoku:2022gbr}, which is only perturbatively modified even with the spectral instability.
This fact can also be easily captured by causality argument for signals propagating in a cavity~\cite{Barausse:2014tra, Berti:2022xfj}: 
at late times, the double-bump-cavity filters the high-frequency content of the outgoing GW, and all that remains is the new low-frequency fundamental mode, which is approached via a sequence of
``echoes''~\cite{Cardoso:2016rao,Cardoso:2019rvt,Cardoso:2016oxy,Maggio:2019zyv,Maggio:2020jml,Maggio:2021ans} (see Section~\ref{subsec:echoes_theory} below). 
The second observation concerns the existence of actual physical systems that produce ``double-bump'' potentials as solutions to the Einstein field equations. There are no known viable astrophysical setups whose perturbations to the gravitational field are described by a single decoupled master equation as in Eq.~\eqref{eq:RW}. Indeed, there are indications that spacetimes described by a double-peak effective potential have at least two light rings, and are therefore {\it not} a small perturbation of a vacuum BH spacetime~\cite{Cardoso:2024mrw}. 

Spectral instability has nevertheless been observed and studied in a consistent description of BH environments~\cite{Cardoso:2021wlq,Cardoso:2022whc,Spieksma:2024voy}, which requires coupling matter to gravitational perturbations.
No matter how small the strength of the coupling, the gravitational sector will always be ``contaminated'' (or destabilized) by the mode content in the matter sector. 
However, there is now widespread consensus~\cite{Barausse:2014tra,Berti:2022xfj,Kyutoku:2022gbr,Cardoso:2024mrw} that the prompt ringdown is only perturbatively affected,
as discussed in detail in Section~\ref{sec:environments} below.
\subsubsection{Environmental effects\label{sec:environments}}

The analysis outlined above leads to the intriguing conclusion that astrophysical environments can, in principle, produce considerable changes in BH QNM spectra. 
What is left to understand is \textit{how} the perturbed QNMs appear in the TD ringdown signal, as computed from self-consistent solutions of the Einstein field equations.
This question is of paramount importance because large deviations from the vacuum predictions, if they occur in nature, may either hamper our ability to infer the properties of BHs 
or (if properly understood) carry important information on BH environments.

There are several physical mechanisms that can affect the BH spectroscopy program in astrophysical environments (see~\cite{Barausse:2014tra} for a thorough list and simple order-of-magnitude estimates).
Here we will focus on the effect of galaxies, dark matter halos, and dark matter spikes. To understand why environmental corrections could be relevant in this context, recall for example that the BH at the center of the Milky Way, with a mass $M\sim 4\times 10^6 M_{\odot}$, is surrounded by stars and dark matter of total mass $M_{\rm H}\sim 5\times 10^{11}M_\odot$ within a radius of $a_\text{H} \sim 20 \, \text{kpc}$ (corresponding to $a_\text{H} \sim 10^{4}\, M_H$ in geometrical units)~\cite{Posti:2019tbp,Jiao:2023aci}.

In the simple case of spherical symmetry, a fully relativistic solution can be found in closed form~\cite{Cardoso:2021wlq}. Other matter distributions are, of course, possible, but the qualitative conclusions discussed below are expected to hold also in other cases~\cite{Speeney:2024mas}. 

\begin{figure*}[t]
    \centering
    \includegraphics[width=0.6\linewidth]{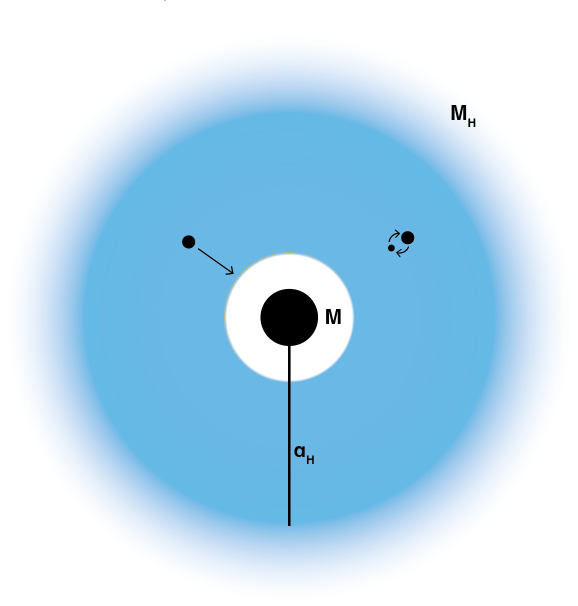}
    \caption{Illustration of a BH of mass $M$  sitting at the center of a galaxy, surrounded by stars and dark matter (in blue). The halo of matter extends over a length scale $a_{\rm H}$ and has mass $M_{\rm H}$. If a secondary object plunges into the massive BH at the center, or a light binary merges somewhere in the galaxy, a ringdown signal is triggered in this environment. The ringdown will be redshifted as GWs propagate out of the galactic potential, accompanied by low-frequency fluid modes imprinted in the GW signal.
    \label{fig:anatomy_galaxy}}
\end{figure*}
The qualitative setup
is shown in Fig.~\ref{fig:anatomy_galaxy}.
A BH of mass $M$ is surrounded by a halo of matter (stars, dark matter, etc.) of mass $M_{\rm H}$ with characteristic length scale $\text{a}_\text{H}$. The environment's compactness $\mathcal{C}$ and typical density $\rho$ are given by
\begin{equation}
    \mathcal{C} = \frac{M_\text{H}}{\text{a}_\text{H}}, \qquad \rho \sim \frac{M_\text{H}}{\text{a}_\text{H}^3} \, .
\end{equation}
For the Milky Way, an order of magnitude estimate yields
$\mathcal{C} \sim 10^{-6}$~\cite{Jiao:2023aci}. 
In general, the halo size and mass are orders of magnitude larger than the BH's. 

Stationary and spherically symmetric solutions to the Einstein equations for an anisotropic fluid surrounding a BH are known in closed form for specific density profiles~\cite{Cardoso:2021wlq}, but similar procedures can be used to model generic astrophysical environments around BHs either analytically or numerically~\cite{Jusufi:2022jxu,Feng:2022evy,Konoplya:2022hbl,Figueiredo:2023gas,Shen:2023erj,Datta:2023zmd,Stelea:2023yqo,Speeney:2024mas,Heydari-Fard:2024wgu,Mollicone:2024lxy,Maeda:2024tsg}. 
Given these fully relativistic background spacetimes, computing geodesic motion and  perturbations is a relatively simple matter~\cite{Cardoso:2021wlq,Cardoso:2022whc,Destounis:2022obl,Figueiredo:2023gas,Speeney:2024mas,Mollicone:2024lxy}.

The scales in the problem can be sufficiently distinct to make the computation of QNMs in the general case challenging. However the axial sector of gravitational perturbations does not couple to matter (under mild assumptions, such as being nondissipative~\cite{Redondo-Yuste:2023ipg,Boyanov:2024jge}),
and it is easier to handle~\cite{Cardoso:2021wlq,Pezzella:2024tkf}. An accurate calculation of the corresponding QNMs relies on casting the wave equation in terms of an eigenvalue problem of a (non-self-adjoint) operator, achieved by means of the hyperboloidal approach described in Section~\ref{subsubsec:hyperboloidal}~\cite{Ansorg:2016ztf,PanossoMacedo:2023qzp,PanossoMacedo:2024nkw}.

\begin{figure*}[t]
    \centering
    \includegraphics[width=\linewidth]{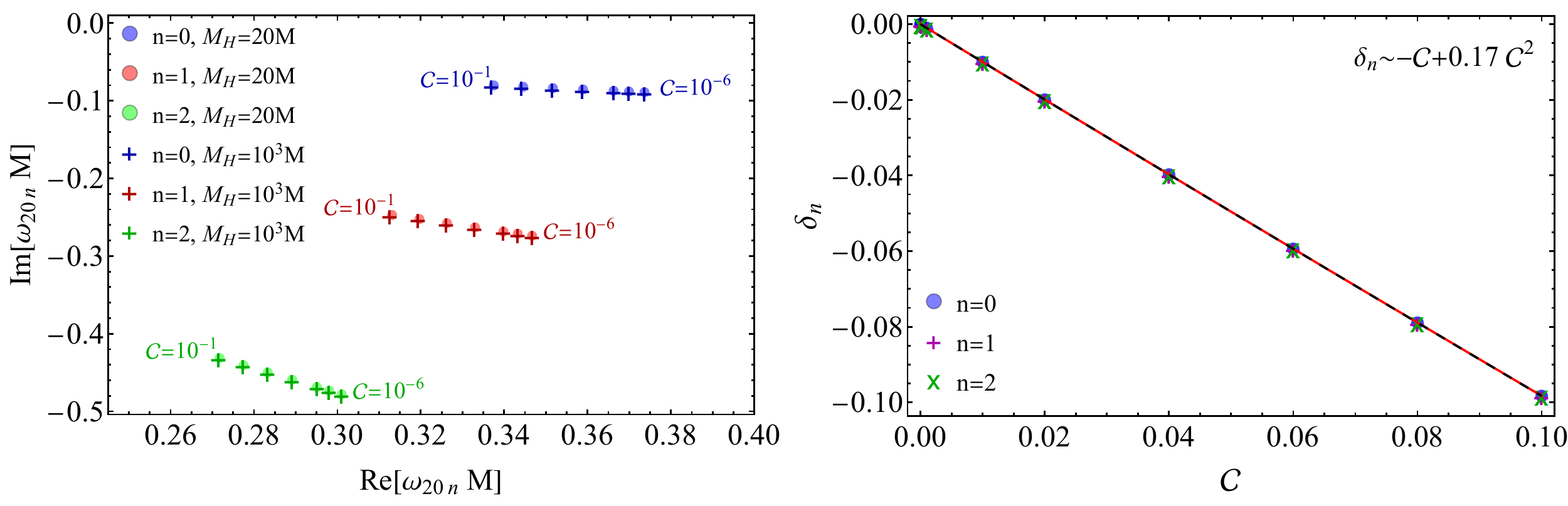}
    \caption{Left panel: Gravitational axial, quadrupolar QNMs $\omega_{20n}$ of a static BH with mass $M$ immersed in the center of a halo with mass $M_{\rm H}$. Circles represent QNMs for $M_{\rm H}=20M$ while crosses represent QNMs for $M_{\rm H}=10^3 M$. The compactness $\mathcal{C}$ is varied from $10^{-6}$ to $10^{-1}$. Right panel: Fit of $\delta_{n}\equiv(\omega_{20n}-\omega^\textrm{vacuum}_{20n})/\omega^\textrm{vacuum}_{20n}$, where $\omega_{20n}$ are the real (black line) and imaginary (red dashed line) parts of axial QNMs resulting from a halo of mass $M_{\rm H}=20 M,\,10^3 M$ and $\omega^\textrm{vacuum}_{20n}$ are the real and imaginary parts of axial QNMs in vacuum with respect to the compactness $\mathcal{C}$.}
    \label{fig:Axial_redshift}
\end{figure*}

The resulting (nonvacuum) axial QNM frequencies are shown in Fig.~\ref{fig:Axial_redshift}, and they clearly depend on the properties of the BH environment. As the compactness $\mathcal{C}$ of the halo increases from $10^{-6}$ to $10^{-1}$, the QNM frequencies $\omega_{20n}$ are redshifted towards smaller values with respect to the vacuum frequencies $\omega^\textrm{vacuum}_{20n}$. For $\ell=2$, a fit gives
\begin{equation}
\delta_n\equiv \frac{\omega_{20n}-\omega_{20n}^{\rm vacuum}}{\omega_{20n}^{\rm vacuum}}\approx -{\cal C}+0.17{\cal C}^2\,.
\end{equation}
This behavior, which is common to other multipoles and applies over a large range of halo masses, is in essence a redshift of axial waves as they cross the galactic gravitational potential, after being generated close to the light ring. A calculation in the geometric optics approximation predicts $\delta_n=-{\cal C}+{\cal C}^2/6$ in the small compactness limit, in very good agreement with the numerical results. Indeed, it can be shown that any sufficiently dilute environment produces, to leading order, a redshift $1+U$ of the axial frequencies and damping times, where
$U\sim -{\cal C}$ is the Newtonian potential of the environment~\cite{Pezzella:2024tkf}. 

\begin{figure*}[t]
    \centering
    \includegraphics[width=0.75\linewidth]{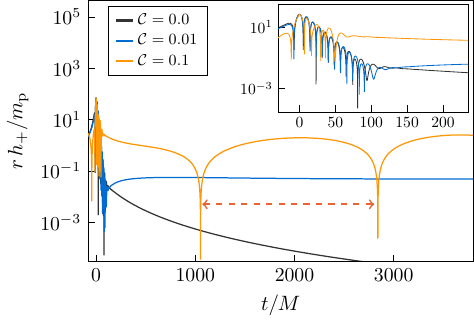}
    \caption{GW signal $h_+$ from a radial plunge of a point particle (of mass $m_p$) from rest at a distance of $100 M$, onto a BH which is immersed in a halo of varying compactness and mass $M_{\rm H} = 10M$. The signal is extracted at $3000M$. The vacuum signal is in black, while the inset shows a zoom-in of the prompt ringdown. Waveforms are aligned in time and amplitude, such that the peak strain is at $t = 0$. The red dashed line indicates the presence of a slowly-decaying fluid mode. Figure taken from~\cite{Spieksma:2024voy}. Inset shows zoom-in of the prompt ringdown.}
    \label{fig:PlungeEnvironment}
\end{figure*}

The polar sector of GWs couples to matter and triggers richer phenomenology~\cite{Cardoso:2022whc,Spieksma:2024voy}. A QNM analysis in the FD is still lacking, but this case can be studied through TD evolutions.
The TD signal for a BH excited by a particle plunging radially, computed in a fully relativistic setting, is shown in Fig.~\ref{fig:PlungeEnvironment}~\cite{Cardoso:2022whc,Spieksma:2024voy}.
The plunge excites a prompt ringdown signal (highlighted in the inset) corresponding to direct excitation of the light ring QNMs. 
As in the axial sector, these ringdown waves cross the gravitational potential and are redshift to lower frequencies as a consequence of the additional gravitational pull exerted by the halo mass.
However, the signal structure is substantially different from vacuum after this prompt ringdown. 
The intermediate-time strain in vacuum is dominated by a ``tail,'' i.e., a superposition of a large number of power laws~\cite{DeAmicis:2024not} due to waves backscattering off the nonzero curvature of the background~\cite{Price:1971fb,Leaver:1986gd,Gundlach:1993tp}.
These terms slowly decay, leaving place to a single power law $h_+ \sim t^{-p}$ (with $p \in \mathbb{N}$), as discussed in Section~\ref{sec:tails} below. 
In the presence of a halo of matter, one observes instead a slowly decaying \emph{fluid mode} with characteristic period $T\sim a_\text{H}/c_{s,r}$, where $c_{s,r}$ is the speed of sound in the radial direction, and with amplitude determined mostly by the halo's compactness. The excitation of this fluid mode is an interesting example of how spectral instabilities can arise quite generically via mode coupling~\cite{Cardoso:2024mrw}.
In conclusion, astrophysical environments can significantly alter the QNM spectrum of a BH, but these modifications in the spectrum do not necessarily imply that the signal will drastically change at early times, when it has the highest amplitude. 

\begin{figure*}[t]
    \centering
    \includegraphics[width=0.85\linewidth]{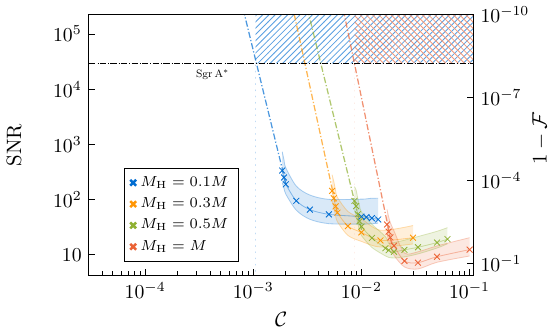}
    \caption{Signal-to-noise ratio (SNR) required to distinguish the ringdown signal from vacuum for different values of the halo compactness, ${\cal C}$, and halo mass, $M_{\rm H}$. For a chosen SNR (corresponding to some horizontal line) and halo mass, only values of ${\cal C}$ to the right of the intersection between the horizontal line and the colored curves are distinguishable from vacuum.
    The black horizontal line represents a putative signal from $\mathrm{Sgr}\,\mathrm{A}^{*}$ with mass ratio $10^{-5}$. 
    The right axis shows the corresponding mismatch value.
    Colored dash-dotted lines are power-law fits through the last few points of each curve. The shaded regions are estimates of extrapolation errors in the waveform. Figure adapted from~\cite{Spieksma:2024voy}.}
    \label{fig:FaithEnvironment}
\end{figure*}

A measure of the contribution of environmental effects in the ringdown signal is the so-called ``faithfulness'' between the numerical signal generated in the nonvacuum case and the numerical vacuum waveform~\cite{Spieksma:2024voy}.
In real observations the BH mass is \textit{a priori} unknown, so the faithfulness must be maximized over a ``stretching'' of the time component which is different at each spacetime location, to model redshift as the wave propagates.
The result of this calculation is shown in Fig.~\ref{fig:FaithEnvironment}. 
For example, for a halo configuration with $M_\textrm{H} = M$ and $a_\text{H}=58 M$, the faithfulness is maximized for a redshift factor $0.987$, whereas $1-M_\text{H}/a_\text{H} = 0.982$. 
For smaller values of $\mathcal{C}$, the mismatch follows a power-law decay. This allows extrapolation to even smaller values of $\mathcal{C}$, for which numerical calculations are computationally prohibitive. 

Are differences between the ringdown template and the real signal measurable at very large signal-to-noise ratios?
Current observations from the motion of the S2 star in our own Galactic center limit the additional mass enclosed within its orbit to be $\lesssim 10^3 \, M_\odot$~\cite{GRAVITY:2024tth}, but distributed over a semi-major axis $a_\text{S2} \sim  2\times 10^4 M_\text{Sgr~A*}$ (in units of Sgr~A* mass).
Figure~\ref{fig:FaithEnvironment} suggests that
for a typical galactic halo compactness ${\cal C} \lesssim 10^{-4}$ the SNR must to be larger than several thousands to detect any environmental effects other than the gravitational redshift. 
Therefore BH spectroscopy is not expected to be able to probe astrophysical environments, even in the optimistic case of a signal from a plunge of a $40 \, M_\odot$ compact object into Sgr~A* (unless the plunge happens to occur in some exotic, highly compact environment).

Recent work explored NR simulations of collapsing and merging neutron stars leading to BH formation, and used them to perform BH spectroscopy by extracting QNMs as a function of the starting time~\cite{Steppohn:2025kbh}. The resulting ringdown signals can be
classified into ``clean'', ``modified'', and ``distorted'' cases, depending on the amount of matter present in the system. Even in the presence of matter, possible
deviations of the QNM frequencies from the expected Kerr values seem to be dominated by ringdown modeling systematics. Incorporating multiple QNMs drastically reduces mismatches
and errors in estimating the final BH mass at early times. If not treated carefully, deviations in the fundamental QNM might artificially be overestimated and falsely attributed to the presence of matter or violations of GR.

In summary, the last decade has seen substantial progress in our understanding of how astrophysical environments affect BH spectroscopy.
The emerging consensus is that environments have a small impact, but further research is necessary to confirm this finding. 
First of all, the conclusions summarized above are mostly based on the assumption of a time-independent, nonrotating backgrounds.
As discussed in Section~\ref{sec:mtaft} below, mass accretion can change the QNM spectrum~\cite{Sberna:2021eui,Redondo-Yuste:2023ipg,May:2024rrg,Zhu:2024dyl,Capuano:2024qhv}, but in most situations of astrophysical interest the resulting modifications are expected to be negligibly small~\cite{Barausse:2014tra,Bamber:2021knr}. 
The ringdown radiation produced by spinning BHs within astrophysical environments is poorly understood. 
Tidal modifications to the spectrum of Schwarzschild and Kerr BHs also deserve further scrutiny~\cite{Cardoso:2021qqu}. 
Even if environmental modifications to the vacuum QNM spectrum turn out to be large, it may be possible to test the BH nature of the underlying spacetime in other ways (e.g. through perturbative quantities that may be observable and have been found to be largely unaffected by instabilities in the spectrum, such as the greybody factors~\cite{Kyutoku:2022gbr,Rosato:2024arw,Oshita:2024fzf}). %

\clearpage
\section{Ringdown beyond general relativity and the Standard Model}
\label{sec:beyondGR}

\noindent
{\em Ci sono soltanto due possibili conclusioni: se il risultato conferma l'ipotesi, allora hai appena fatto una misura; se il risultato \`e contrario all'ipotesi, allora hai fatto una scoperta. 
(There are only two possible conclusions: if the result confirms the hypothesis, then you have just made a measurement; if the result disproves the hypothesis, then you have made a discovery.)} 

\vspace{.2cm}

\noindent
\begin{flushright}
Enrico Fermi
\end{flushright}

\vspace{.2cm}

Part of the current experimental interest in BH spectroscopy comes from
the possibility of finding experimental smoking guns of new physics or
modifications of GR in the large and growing catalog of GW events. This
possibility has motivated an intense effort, reviewed in this chapter, to (i)
compute QNM spectra in specific classes of modified theories
(Section~\ref{subsec:theory-spec}); (ii) develop theory-agnostic ringdown
parameterizations, similar in spirit to the parameterized post-Newtonian
formalism~\cite{Will:2014kxa} (Section~\ref{subsec:theory-agnostic}); and (iii)
investigate the effects of physics beyond the Standard Model that could produce
characteristic signatures -- e.g., ``echoes'' -- in the ringdown waveform
(Section~\ref{subsec:echoes_theory}).

\subsection{Theory-specific spectra}
\label{subsec:theory-spec}

\vspace{-.1cm}

\noindent \textit{Initial contributors: Bl\'azquez-Salcedo, Cano, Chung, Khoo, Yunes}

\vspace{.2cm}

In recent years, there has been substantial progress in understanding the QNM spectrum of rotating BHs in theories beyond GR and the Standard Model of particle physics. Most of the work concerns higher-derivative theories that modify the Einstein equations in a continuous way in terms of free parameters (coupling constants). The regime where these couplings are small and the new theory is close to GR is currently the best understood scenario, but there are also results that apply to large couplings.

The space of possible theories is too large to enumerate~\cite{Clifton:2011jh,Berti:2015itd}.
The theories we will discuss here can be split into two main classes.

The first class assumes a pure metric theory, for which a general effective field theory (EFT) extension of GR~\cite{Endlich:2017tqa,Cano:2019ore}, up to eight derivatives, reads
\begin{equation}\label{eq:GREFT}
\begin{aligned}
S_{\rm EFT}=\frac{1}{16\pi G}\int& d^4x\sqrt{|g|}\bigg[R+L^4\left(\lambda_{\rm ev}\mathcal{R}^3+\lambda_{\rm odd}\tilde{\mathcal{R}}^3\right)
+L^6\left(\lambda_{1}\mathcal{C}^2+\lambda_{2}\tilde{\mathcal{C}}^2+\lambda_{3}\mathcal{C}\tilde{\mathcal{C}}\right) \bigg]\, ,
\end{aligned}
\end{equation}
with
\begin{equation}
\mathcal{R}^3=R_{\mu\nu }^{\,\,\,\,\,\,\rho\sigma}R_{\rho\sigma }^{\,\,\,\,\,\,\delta\gamma}R_{\delta\gamma }^{\,\,\,\,\,\,\mu\nu }\, ,\quad  \tilde{\mathcal{R}}^3=R_{\mu\nu }^{\,\,\,\,\,\,\rho\sigma}R_{\rho\sigma }^{\,\,\,\,\,\,\delta\gamma } \tilde{R}_{\delta\gamma }^{\,\,\,\,\,\,\mu\nu }\, ,\quad
\mathcal{C}=R_{\mu\nu\rho\sigma} R^{\mu\nu\rho\sigma}\, ,\quad
\tilde{\mathcal{C}}=R_{\mu\nu\rho\sigma} \tilde{R}^{\mu\nu\rho\sigma}\, .
\end{equation}
Here $\tilde{R}_{\mu\nu\rho\sigma}=\frac{1}{2}\epsilon_{\mu\nu\alpha\beta}R^{\alpha\beta}_{\,\,\,\,\,\,\rho\sigma}$ is the dual Riemann tensor, $\epsilon_{\mu\nu\alpha\beta}$ is the Levi-Civita tensor density, defined such that $\epsilon_{0123}=+\sqrt{|g|}$, 
$\lambda_{\rm q}$ (with $\rm q= ev, odd, 1,2,3$) are free dimensionless coupling constants, and $L$ is the characteristic length scale of new physics.
The EFT is applicable for BHs larger than $L$, i.e. $r_{+}\gg L$, and it describes the most general diffeomorphism-invariant metric modification of GR.  For a BH of mass $M$, it is useful to introduce the dimensionless coupling constants
\begin{equation}\label{eq:zetaqdef}
    \zeta_{\rm q}=\lambda_{\rm q}\left(\frac{L}{M}\right)^{n}\, ,
\end{equation}
where $n=4$ for ${\rm q = ev, odd}$ and $n=6$ for ${\rm q = 1, 2, 3}$. These control the relative deviations with respect to GR and are assumed to satisfy $|\zeta_{q}|\ll 1$. We will work at first order in these couplings. 

The second class of theories that we will consider involves additional fields (typically chosen to be scalar fields, for simplicity) coupled to curvature invariants.
Including up to quadratic curvature invariants, these scalar-tensor theories are described by the action 
\begin{equation}\label{eq:LagrangianGBdCS}
S_{\rm ST} =\frac{1}{16\pi G}\int d^{4}x\sqrt{|g|}\left[ R - \frac{1}{2} \nabla_{\mu} \Phi \nabla^{\mu} \Phi- V(\Phi) + \alpha f(\Phi) \mathcal{Q} \right]\,,
\end{equation}
where $\Phi$ is a scalar field that couples to the spacetime metric, $V(\Phi)$ is a potential governing the self-interactions of the scalar field, $\alpha$ is a dimensionful coupling constant that measures the strength of modifications to GR, $f(\Phi)$ is a coupling function, and $\mathcal{Q}$ is a scalar constructed from the curvature tensor. 
The quantities $V(\Phi)$, $\mathcal{Q}$, and $f(\Phi)$ determine the nature of the gravity theory. 
For example, for the class of Einstein-scalar-Gauss-Bonnet (EsGB) theories~\cite{Torii:1996yi, Kanti:1995vq}, %
$\mathcal{Q} = \mathcal{G} \equiv R^2 - 4 R_{\alpha \beta} R^{\alpha \beta} + R_{\alpha \beta \gamma \delta} R^{\alpha \beta \gamma \delta}$, where $\mathcal{G}$ is the Gauss-Bonnet invariant, while different choices of $f(\Phi)$ define different members in the class.
For dynamical Chern-Simons (dCS) gravity, $\mathcal{Q} = \tilde{\mathcal{C}} = R_{\nu \mu \rho \sigma} \tilde{R}^{\mu \nu \rho \sigma}$, where $\tilde{\mathcal{C}}$ is the parity-odd Pontryagin invariant, while $f(\Phi) = \Phi$ and $V(\Phi) = 0$, so that the theory is shift-symmetric~\cite{Alexander:2009tp,Wagle:2021tam}.
In both of these cases, $\mathcal{Q}$ is built from quadratic curvature invariants, so $\alpha$ has units of length squared. Analogously to \eqref{eq:zetaqdef}, we introduce 
\begin{equation}\label{eq:zetaGBdef}
    \zeta_{\text{EsGB/dCS}}=\frac{\alpha}{M^2}\, ,
\end{equation}
as the dimensionless coupling that determines the order of magnitude of the deviations with respect to GR.

A theory with two scalars that couple simultaneously to the Gauss-Bonnet and Pontryagin invariants can also be considered~\cite{Cano:2021rey}. 
Such a theory is a natural prediction of string theory~\cite{Bergshoeff:1989de}, where one of the scalars is the dilaton and the second one is an axion dual to the Kalb-Ramond two-form. However, QNM calculations in this theory are in their infancy, hence it will not be discussed below.

For other classes of theories beyond the higher-derivative corrections discussed above, the study of QNMs is less complete. 
One such example concern so-called DHOST theories, for which it is not possible (in general) to write down the linearized perturbations in a compact Schr\"odinger-like form~\cite{Langlois:2021aji,Langlois:2022ulw,Tattersall:2017erk}. Moreover, the presence of additional terms and couplings may change the implementation of the boundary conditions~\cite{Langlois:2021xzq}.
A possible way out for studying perturbations of BHs in alternative theories of
gravity is to use full-fledged TD numerical relativity (NR) evolutions. Although the required numerical resolution to extract QNMs is not always met, a few preliminary results in this direction have been achieved~\cite{Okounkova:2019zep,Okounkova:2018pql,Evstafyeva:2022rve}.
One of the biggest challenges in studying linear perturbations and computing QNMs in Lorentz-violating theories of gravity is that BH solutions in these theories possess multiple horizons, 
and there is ongoing research on how to impose boundary conditions~\cite{Cardoso:2024qie}.
There are exceptional cases (such as a subclass of the so-called Ho\v{r}ava gravity theories~\cite{Franchini:2021bpt}) in which all of the additional degrees of freedom have the same propagation speed as the gravitational ones, but the QNM spectrum in these cases is indistinguishable from GR. 

The remainder of this section is structured as follows. We first review rotating BH solutions in the theories defined by the actions~\eqref{eq:GREFT},\,\eqref{eq:LagrangianGBdCS}, which are no longer given by the Kerr metric. Then we discuss several methods to analyze perturbations and QNMs in these (and more general) theories. Finally, we summarize current state-of-the-art calculations of QNM frequencies in specific cases. 

\subsubsection{Modified Kerr black holes}\label{sec:mod_Kerr} 
The calculation of the modified Kerr metric describing rotating BHs in the theories under consideration is not a completely solved problem: there are no exact analytic solutions for BHs with arbitrary rotation and coupling. Most current efforts focus either on analytical approximations valid for small couplings or slow-rotation, or on numerical solutions. 

\vspace{0.5cm}
\noindent
\textit{Small-spin and small-coupling expansion} A common strategy to obtain corrections to the Kerr metric is to find approximate analytical solution through a double series expansion in small angular momentum and small coupling~\cite{Yunes:2009hc,Pani:2009wy,Pani:2011gy,Yagi:2012ya,Ayzenberg:2014aka,Maselli:2015tta,Cardoso:2018ptl}. While the small coupling approximation can be justified from the perspective of EFTs and current tests of GR~\cite{Will:2014kxa,Yunes:2013dva,Berti:2015itd,LIGOScientific:2021sio,Yunes:2024lzm}, restricting to small spin is not sufficient to describe the BH remnants resulting from binary mergers, which typically have dimensionless spin $\chi\simeq 0.7$. This problem can be addressed by extending the spin expansion to very high orders~\cite{Cano:2019ore}. 

The corrected Kerr metric is typically written in Boyer-Lindquist like coordinates $(t,r,x,\phi)$, where $x = \cos{\theta}$, as
\begin{align}\label{eq:beyondKerr}
    ds^2 =& -\left(1-\frac{2M r}{\Sigma}-H_1\right)dt^2
    -(1+ H_2)\frac{4a M r (1-x^2)}{\Sigma}dtd\phi
    \\\notag
    +&\left(1+H_3\right)\Sigma\left(\frac{dr^2}{\Delta}+\frac{dx^2}{1-x^2}\right)+(1+H_4)\left(r^2 + a^2+\frac{2a^2Mr(1-x^2)}{\Sigma}\right)(1-x^2) d\phi^2\, , %
\end{align}
where $(\Sigma,\Delta)$ are the known functions of $r$ and $\theta$ that define the Kerr metric, while the four functions $H_{i}(r,x)$ admit the expansion
\begin{equation}\label{eq:spin_expansion}
    H_{i} = \sum_{n=0}^{\infty}\chi^n  \sum_{p=0}^{n} \sum_{k=0}^{k_\text{max}(n)}H_{i}^{(n,p,k)}\left(\frac{M}{r}\right)^k x^p \,.
\end{equation}
Here $\chi = a/M$ is the dimensionless spin parameter for a BH with mass $M$ and angular momentum $M a$. The coefficients $H_{i}^{(n,p,k)}$ are functions of the various coupling constants that can be determined by solving the modified field equations. We will consider only the solution to linear order in the $\zeta$ couplings \eqref{eq:zetaqdef},  \eqref{eq:zetaGBdef}, but expressions of the form \eqref{eq:spin_expansion} in principle also work at higher order in the small coupling expansion. 
For theories defined by the action~\eqref{eq:LagrangianGBdCS} with $V(\Phi)=0$, the scalar field can also be written as 
\begin{equation}\label{eq:spin_expansion_theta}
    \Phi = \sum_{n=0}^{\infty}\chi^n  \sum_{p=0}^{n} \sum_{k=0}^{k_\text{max}(n)}\vartheta^{(n,p,k)}\left(\frac{M}{r}\right)^k x^p \, ,
\end{equation}
where the coefficients $\vartheta^{(n,p,k)}$ are functions of the coupling constants that can similarly be determined by solving the modified field equations. 
The spin expansion is convergent, and 10--20 terms are usually enough to accurately describe typical post-merger BHs with $\chi \in [0.5, 0.9]$~\cite{Cano:2023qqm}. 
The functions $H_{i}$ and $\Phi$ have been derived to the $14$-th order in $\chi$ in~\cite{Cano:2019ore}, but these calculations can in principle be extended to arbitrary orders. A \texttt{Mathematica} notebook allowing to construct such extensions is publicly available~\cite{Cano:2019ore,RotatingBHWolfram} (see Appendix~\ref{sec:public_codes}).
The small-spin expansion, however, becomes inefficient when approaching extremality 
(see e.g.~\cite{McNees:2015srl} for a discussion in dCS gravity).
\vspace{0.25cm}
\par
\noindent
{\textit{Numerical solutions} 
Full numerical approaches are typically required to search for solutions with arbitrary rotation and coupling constants, or  solutions with nontrivial coupling functions $f(\Phi)$.
For example, in Einstein-dilaton-Gauss-Bonnet (EdGB) gravity (a member of the EsGB class $\mathcal{Q} = \mathcal{G}$ with $V(\Phi) = 0$ and $f(\Phi)= e^{-\gamma\Phi}$, where $\gamma$ is the dilaton coupling constant), both static spherically symmetric~\cite{Kanti:1995vq,Sullivan:2019vyi} and rotating configuration~\cite{Kleihaus:2011tg,Kleihaus:2014lba, Kleihaus:2015aje,Sullivan:2020zpf} have been obtained.
Static spherically symmetric~\cite{Sotiriou:2014pfa,Sullivan:2019vyi} and rotating BH solutions~\cite{Sullivan:2020zpf,Delgado:2020rev} are also known in linear EsGB gravity (another member of the massless EsGB class, but with coupling function $f(\Phi)= \Phi$).
Interestingly, when the coupling function satisfies certain conditions (examples of which include $f(\Phi) \propto \Phi^2$~\cite{Silva:2017uqg,Silva:2018qhn,Macedo:2019sem,Herdeiro:2020wei} and $f(\Phi) \propto \exp(-\Phi^2)$~\cite{Doneva:2017bvd}), BHs can become ``spontaneously scalarized,'' with a stable branch of modified BH solutions arising nonperturbatively for large enough curvature: see~\cite{Doneva:2022ewd} for a recent review.
In dCS gravity, static and spherically-symmetric BH solutions are described by the Schwarzschild metric just as in GR, while rotating configurations were studied in~\cite{Stein:2014xba,Okounkova:2018abo,Delsate:2018ome,Sullivan:2020zpf,Richards:2023xsr}. 

To obtain stationary and axially symmetric BHs one typically uses ``quasi-isotropic'' spherical coordinates, with the generic metric ansatz
\begin{eqnarray}
\label{metric}
ds^2=- f dt^2 +  \frac{b}{f} \left( d r^2+ r^2d\theta^2 \right) 
           +  \frac{l}{f} r^2\sin^2\theta (d\phi-\frac{w}{r} dt)^2 ,
\end{eqnarray}
where $f,\,b,\,l,\,w$ are functions of $r$ and $\theta$.
With this ansatz for the metric, together with the background field $\Phi$ (also a function of $r$ and $\theta$), the field equations are reduced to a system of \textit{elliptic}, PDEs that are solved numerically.
The metric and scalar field functions satisfy a number of boundary conditions that are derived from requiring the standard conditions for astrophysical BHs: regularity of the solution on the rotation axis, asymptotic flatness of the metric at spatial infinity, and regularity of the Killing horizon. 

The domain of existence of the resulting solutions in some modified theories can sometimes have interesting properties, as shown e.g.~in Fig.~5 of~\cite{Kleihaus:2015aje} for EdGB gravity. 
For instance, in EdGB gravity with $\gamma=1$ and fixed $\alpha$, the scalar charge of the solution grows as the mass of the static BH decreases, and eventually a critical solution with a minimum mass is reached~\cite{Kanti:1995vq, Torii:1996yi, Guo:2008hf}. At this point, the BH mass increases again along a small unstable branch of solutions before a singular solution is reached. For an analysis of this branch structure with other values of $\gamma$, see~\cite{Blazquez-Salcedo:2017txk}.
The situation is similar when introducing angular momentum~\cite{Kleihaus:2011tg, Kleihaus:2014lba, Kleihaus:2015aje}: for fixed values of coupling constant $\alpha$ and $\chi$, there is a BH solution with minimum mass.
For large enough angular momentum there is no second branch, and the singular solutions also possess a minimum mass. 
For even larger values of the angular momentum, it is possible to reach a set of extremal BHs in EdGB gravity. 
Hence, the boundary of the domain of BH existence is given by the set of static EdGB BHs, the set of critical solutions, the set of extremal BHs, and the set of Kerr BHs.

\subsubsection{Black hole perturbations in modified gravity: methods}
Besides the lack of analytical BH solutions in modified gravity, other challenges plague the analysis of perturbations in these theories. 
First of all, these solutions are not of Petrov type D~\cite{Owen:2021eez}, and the Teukolsky equation is not valid. 
Separability is also generically lost, even in the case of the wave operator for a scalar field. In addition, the equations of motion for these theories are extremely intricate, containing higher-order derivatives and/or extra degrees of freedom. 
For these reasons, up until recently, QNM frequencies were known only for static and slowly rotating BHs~\cite{Cardoso:2009pk,Blazquez-Salcedo:2016enn,Cardoso:2018ptl,deRham:2020ejn,Moura:2021eln,Pierini:2021jxd,Wagle:2021tam,Srivastava:2021imr,Bryant:2021xdh,Cano:2021myl,Pierini:2022eim,Silva:2024ffz}, since these can be computed by conventional methods using metric perturbations. These results are useful to learn about some generic features of beyond-GR effects, like the breaking of isospectrality (we will say more about this below), but they are not applicable to the typical angular momenta of post-merger BHs. 
Two different classes of methods, reviewed below, have been devised to overcome these challenges and compute QNMs in the case of large rotation.
\vspace{0.25cm}
\par
\noindent
\textit{Modified Teukolsky equation} 
The first approach aims at extending the Teukolsky equation~\cite{Teukolsky:1973ha} to encompass theories of gravity that permit algebraically general BH spacetimes.
It was first constructed within the Newman-Penrose formalism~\cite{Li:2022pcy, Hussain:2022ins}, and reformulated using the Geroch-Held-Penrose formalism~\cite{Cano:2023tmv}. 
Practical implementations of this ``modified Teukolsky formalism'' have been developed for higher-derivative gravity~\cite{Cano:2023tmv,Cano:2023jbk,Cano:2024ezp} and dCS gravity~\cite{Wagle:2023fwl, Li:2025fci}.
The main idea behind these approaches is to find an equation, valid for arbitrary theories, that reduces to the usual Teukolsky equation when the theory reduces to GR and the background is of Petrov type D. 
\emph{Any equation} with such properties is enough to perform perturbation theory in a regime close to GR.

The key is the introduction of a two-parameter expansion scheme, which helps circumvent certain algebraic constraints due to the possible non-Petrov-type-D nature of the background BH spacetime. Using $\zeta$ (denoting one of the dimensionless couplings defined in \eqref{eq:zetaqdef}, \eqref{eq:zetaGBdef} above) to characterize the deviation from GR  and $\epsilon$ to characterize the strength of gravitational perturbations, one can expand all geometrical quantities in a bivariate series. For example, the Weyl scalars $\Psi_n$ with $n \in (0,4)$ can be expanded as
\begin{equation}\label{modWeylScalar}
\Psi_n=\Psi_n^{(0,0)}+\zeta\Psi_n^{(1,0)}+\epsilon\Psi_n^{(0,1)}+\zeta\epsilon\Psi_n^{(1,1)}+\cdots\,.
\end{equation}
The quantities $\Psi_n^{(0,0)}$ and $\Psi_n^{(1,0)}$ are Weyl scalars evaluated on the BH background and expanded to zeroth- and first-order in $\zeta$, respectively. Similarly, $\Psi_n^{(0,1)}$ and $\Psi_n^{(1,1)}$ represent the Weyl scalars induced by GWs in GR and their leading-order, beyond-GR corrections, respectively. 

To generalize the Teukolsky equation one can consider the same combination of Bianchi identities of the Riemann tensor that lead to the usual Teukolsky equation, but without making any assumptions on the form of the curvature tensor, the equations of motion, or the Petrov type of the background spacetime. 
Using these combinations and the two-parameter expansion, the perturbed Newman-Penrose equations can be decoupled to obtain the modified Teukolsky equation at first order in $(\zeta,\epsilon)$
\begin{align}
\label{eq:MT0}
\mathcal{O}_{0}^{(0,0)}\Psi_{0}^{(1,1)}=\mathcal{S}_{0}^{(1,0)}[\Psi_{n}^{(0,1)},e^{a(0,1)},\gamma_{abc}^{(0,1)}]+\mathcal{T}_{0}^{(0,0)}[\phi_s^{(1,1)}]\,,
    \\
\mathcal{O}_{4}^{(0,0)}\Psi_{4}^{(1,1)}=\mathcal{S}_{4}^{(1,0)}[\Psi_{n}^{(0,1)},e^{a(0,1)},\gamma_{abc}^{(0,1)}]+\mathcal{T}_{4}^{(0,0)}[\phi_s^{(1,1)}]\,,
\label{eq:MT4}
\end{align}
where $\mathcal{O}_{0,4}$ are the original Teukolsky operators as defined in Eq.~\eqref{eq:Teukolsky_Master}, acting on a spin 2 or spin $-2$ scalar, respectively,  while $\mathcal{S}_{0}^{(1,0)}$, $\mathcal{S}_{4}^{(1,0)}$, $\mathcal{T}_{0}^{(0,0)}$, and $\mathcal{T}_{4}^{(0,0)}$ are source operators. The latter depend on background Newman-Penrose quantities and act on the perturbed Weyl scalars $\Psi_{n}^{(0,1)}$, the perturbed tetrad $e^{a(0,1)}$, and the perturbed spin coefficients $\gamma_{abc}^{(0,1)}$. Similarly, the extra field $\phi_{s}$ satisfies
\begin{equation}
\label{eq:MT3}
    \mathcal{O}_{s}^{(0,0)}\phi_s^{(1,1)}=\mathcal{S}_{\phi}^{(1,0)}[\Psi_n^{(0,1)},e^{a(0,1)},\gamma_{abc}^{(0,1)}]\,,
\end{equation}
where $\mathcal{O}_{s}^{(0,0)}$ is the original Teukolsky operator acting on the perturbed spin-weight-s field $\phi_s^{(1,1)}$, while  $\mathcal{S}_{\phi}^{(1,0)}$ is another source operator. 
These equations are still coupled PDEs, complicating the analysis.

To proceed, one needs to compute the solution to the Teukolsky equation in GR for a Kerr background, and reconstruct the metric perturbation in vacuum $h_{\mu\nu}^{(0,1)}$ using standard methods~\cite{Chrzanowski:1975wv,Cohen_Kegeles_1975,Keidl:2006wk,Kegeles:1979an,Yunes:2005ve,Whiting:2005hr}.
In turn, from this metric perturbation one can calculate the perturbed Weyl scalars, tetrad and spin coefficients, obtaining explicit expressions for the right-hand sides of Eqs.~\eqref{eq:MT0}-\eqref{eq:MT3}. 
At this point, the shift in the QNM frequencies is obtained by applying eigenvalue perturbation theory~\cite{Hussain:2022ins}. 
The first-order in $\zeta$ correction to the frequency can then be found as an integral of $\mathcal{S}^{(1,0)}_{0,4,\phi}[\delta \Psi_{n},\delta e^{a (0,1)}, \delta \gamma_{abc}^{(0,1)}]$ and $\mathcal{T}^{(0,0)}_{0,4}[\phi_s^{(1,1)}]$ over the angular coordinates and over a complex contour for the radial coordinate. 
Alternatively, Eqs.~\eqref{eq:MT0}-\eqref{eq:MT3} can be reduced to a more manageable system of radial equations by assuming a decomposition of the Teukolsky variables as 
\begin{align}
\label{eq:deltapsiexpansion0}
\Psi_0^{(1,1)}=e^{-i\omega t+i m \phi} \sum_{\ell} R^{\ell m}_{0}(r)S^{\ell m}_{0}(x;a\omega)\, , \\
\rho^{-4} \Psi_4^{(1,1)}=e^{-i\omega t+i m \phi} \sum_{\ell} R^{\ell m}_{4}(r)S^{\ell m}_{4}(x;a\omega)\, ,\\
\phi_s^{(1,1)}=e^{-i\omega t+i m \phi} \sum_{\ell} {R}^{\ell m}_{s}(r)S^{\ell m}_{s}(x;a\omega)\, ,
\label{eq:deltapsiexpansion3}
\end{align}
where $m$ is the magnetic mode number, $\omega$ is the complex frequency of the QNM, $S^{\ell m}_{s}(x;a\omega)$ are spin-weighted spheroidal harmonics, $\rho=(r-i a x)^{-1}$, and $[{R}^{\ell m}_{0}(r), {R}^{\ell m}_{4}(r), {R}^{\ell m}_{s}(r)]$ are unknown radial functions.
One can then project Eqs.~\eqref{eq:MT0}-\eqref{eq:MT3} onto the spheroidal harmonics in order to get decoupled equations for the radial variables~\cite{Cano:2023tmv,Wagle:2023fwl} (see also~\cite{Cano:2020cao,Ghosh:2023etd}).

For the EFT of GR \eqref{eq:GREFT}, containing no extra fields $\phi_s$, this procedure yields a modified Teukolsky radial equation which, after redefinitions of the radial variable, can be written as~\cite{Cano:2023jbk,Cano:2024ezp}
\begin{equation}\label{eq:correctedradial} 
\Delta^{-s}\frac{d}{dr}\left[\Delta^{s+1}\frac{dR_{s}}{dr}\right]+\left(V_s+ \zeta_{\rm q} \delta V^{(\rm q)}_s\right) R_{s}=0\, ,
\end{equation}  
for the total Teukolsky variable $\Psi_{0,4}^{(0,1)}+\zeta_{\rm q}\Psi_{0,4}^{(1,1)}$, where $s=\pm 2$.
For nonextremal BHs, the correction to the potential can always be put in the form
\begin{equation}\label{eq:dVm2}
\delta V^{(\rm q)}_{-2}=\frac{1}{\Delta}\left[\frac{A^{\rm (q)}_{-2}}{r^{2}}+A^{\rm (q)}_0+A^{\rm (q)}_1 r+A^{\rm (q)}_2 r^2\right]\, . 
\end{equation}
The coefficients $A^{(\rm q)}_k$ can be found analytically as a power series in the angular momentum. They depend on the polarization of the perturbation, hence leading to a breaking of isospectrality.  
For EsGB and dCS gravity, a similar process can be implemented by decomposing the scalar fields in spheroidal harmonics. In this case, one expects that scalar and gravitational modes of different $\ell$ will be coupled, leading to a set of coupled radial PDEs that need to be solved numerically.
In the case of dCS gravity, these coupled Teukolsky radial equations have been analyzed  at first order in the angular momentum expansion~\cite{Wagle:2023fwl}, and the corresponding QNM frequencies have been computed~\cite{Li:2025fci}.

Further insights can be gained by employing the parity operator~\cite{Li:2023ulk}. 
This operator enables separation of the modified Teukolsky equation into two distinct components: an odd-parity equation coupled to the scalar field perturbations and an even-parity equation that remains entirely decoupled,  thereby exhibiting isospectrality breaking in dCS gravity. Similarly, in EsGB gravity, one obtains an even-parity equation coupled to the scalar field perturbations and an odd-parity equation that remains entirely decoupled, again exhibiting isospectrality breaking.

All these findings are consistent with those obtained through the metric perturbation framework discussed below, reinforcing the  validity of both approaches~\cite{Wagle:2021tam,Srivastava:2021imr,Molina:2010fb,Pierini:2021jxd,Pierini:2022eim,Blazquez-Salcedo:2016enn,Blazquez-Salcedo:2017txk,Chung:2024ira,Chung:2024vaf, Blazquez-Salcedo:2024oek,Khoo:2024agm,Blazquez-Salcedo:2024dur}. 

\vspace{0.25cm}
\par
\noindent
{\textit{The METRICS approach}} 
A different approach, known as Metric pErTuRbations wIth speCtral methodS (METRICS)~\cite{Chung:2023zdq,Chung:2023wkd}, models both gravitational and scalar perturbations by expressing them in a product decomposition (by peeling off a certain asymptotic factor) and then applying a  
spectral expansion in Legendre polynomials of (cosine of) the polar angle and Chebyshev polynomials of a (compactified) radial variable~~\cite{Chung:2023zdq,Chung:2023wkd}.
The linearized field equations then become a linear algebra problem for the spectral coefficients and the QNMs frequencies. 

The METRICS approach is as follows. First, small perturbations of both the background metric $(g_{\mu \nu}^{(0)})$ and scalar field background $(\Phi^{(0)})$ in Boyer-Lindquist-like coordinates are written as
\begin{equation}\label{eq:field_perts}
\begin{split}
g_{\mu \nu} & = g^{\rm (0)}_{\mu \nu} + \epsilon \; e^{i m \phi - i \omega t} \hat{h}_{\mu \nu}, \\
\Phi & = \Phi^{\rm (0)} + \epsilon \; e^{i m \phi - i \omega t} \hat{h}_{\vartheta}\,,
\end{split}
\end{equation}
where $\hat{h}_{\mu \nu}$ and $\hat{h}_{\vartheta}$ are functions of $r$ and $\chi$ only, and the quantity $\epsilon \ll 1$ is a perturbative bookkeeping parameter.
Enforcing the Regge-Wheeler gauge, the metric perturbation is completely specified by 6 unknown metric variable $h_{j=1, ..., 6}$~\cite{Regge:1957td}, plus $\hat{h}_{\vartheta}$.
The linearized field equations governing $h_{j=1, ..., 7}$ read
\begin{equation}\label{eq:PDEs}
\sum_{\beta = \gamma = 0} \sum_{j=1}^7 \mathscr{C}_{k, \beta, \gamma, j}(r, \cos \theta) \partial^{\beta}_{r} \partial^{\gamma}_{\cos \theta} h_j = 0\,, 
\end{equation}
where $\beta$ and $\gamma$ are nonnegative integers, $\mathscr{C}_{k, \beta, \gamma, j}$ are complex functions of $(r, \cos \theta)$ that also depend on $(M, a, m, \omega,\zeta)$, and $k \in [1, 11]$ labels the equations (10 metric field and 1 scalar field equation).
Instead, the upper limit of $\beta$ and $\gamma$ depends on the theory.
For example, in GR and EsGB gravity, the summation ranges from $\beta = \gamma = 0$ to $\beta + \gamma = 2$; in dCS gravity, $0 \leq \beta + \gamma \leq 3 $.
Assuming the expansion of Eq.~\eqref{eq:spin_expansion}, $\mathscr{C}_{k, \beta, \gamma, j}$ are rational functions.
Upon appropriate factorization and manipulation, $\mathscr{C}_{k, \beta, \gamma, j}$ can be converted into polynomials~\cite{Chung:2023wkd, Chung:2023zdq}. 

The divergent behavior of $h_{j}$ at the event horizon and spatial infinity due to coordinate singularities, governed by the linearized field equations, is accounted for by factorizing or ``peeling off'' an asymptotic factor $A_j (r)$, 
which in GR reads
\begin{equation}
A_j (r) = e^{i \omega r} r^{2 i M \omega + \rho_{\infty}^{(j)}} \left( \frac{r-r_+}{r}\right)^{-i M \frac{\omega - m \Omega_{H}}{2 \kappa_+}- \rho_H^{(j)}}, 
\end{equation}
where $\rho_{\infty}^{(j)}$ and $\rho_H^{(j)}$ are positive integers that depend on $j$, and $\kappa_+=\sqrt{M^2-a^2}/(2Mr_+)$ is the surface gravity of the outer horizon.
For general, axially symmetric BHs whose event horizon is a Killing horizon, $A_j (r) $ can be derived by imposing ingoing and outgoing wave boundary conditions at the event horizon and at spatial infinity, while using the beyond-GR $\Omega_{H}$ and $\kappa_+$ expressions for the modified Killing horizon~\cite{Chung:2024vaf}.
In terms of $A_j (r)$, one can express $h_{j}$ as 
\begin{equation}\label{eq:spectral_expansion}
h_j = A_j (r) \sum_{p=1}^{N_r} \sum_{l=1}^{N_{l}} v_{j, p, l} ~ \varphi_{p, l} (r, \cos \theta)\,.
\end{equation} 
Here $\varphi_{p, l} (r, \cos \theta)$ encodes all of the radial and polar angle dependence, which in the METRICS approach is represented through a spectral expansion. That is, $\varphi_{p, l} (r, \cos \theta)$ is modeled through a complete and orthogonal spectral basis of $r$ and $\cos \theta$ (typically, Legendre polynomials in $\cos \theta$ and Chebyshev polynomials in a compactified radial coordinate~\cite{Chung:2024vaf});
$N_r$ and $N_{l}$ are the spectral order of the spectral projection along the radial and angular coordinates; finally, $p$ and $l$ label the degree of the spectral basis, with $v_{j, p, l}$ constants. 
The approach has been validated in GR, showcasing its high accuracy, with relative fractional errors smaller than 
$10^{-5}$ for all dimensionless spins below up to 0.95.

For EsGB and dCS gravity, this method simultaneously solves all 11 partial differential equations for the 7 unknowns by transforming Eq.~\eqref{eq:spectral_expansion} into a generalized eigenvalue problem.
The solution has been found using the Newton-Raphson algorithm and the generalized Moore-Penrose inverse~\cite{Chung:2024vaf}.
\vspace{0.25cm}
\par
\noindent
{\textit{Generalized spectral method (collocation)}} 
An alternative approach is the generalized spectral collocation method~\cite{Blazquez-Salcedo:2024oek}.
The method can be used for a range of theories (e.g., linear EsGB gravity), but for concreteness is exemplified below for EdGB gravity.
Under linear perturbations of the metric tensor and dilaton fields, Eq.~\eqref{eq:field_perts}, the equations take the form
\begin{eqnarray}
\mathcal{G}_{\mu\nu} &=& \mathcal{G}_{\mu\nu}^{(bg)} + \epsilon \, \delta\mathcal{G}_{\mu\nu} (r,\theta) e^{i(m\phi-\omega t)}  =0 \, , \\
\mathcal{S} &=& \mathcal{S}^{(bg)} + \epsilon \, \delta\mathcal{S} (r,\theta) e^{i(m\phi-\omega t)}   =0 \,.
\end{eqnarray}
where the subscript ``$(bg)$'' denotes background quantities (see~\cite{Blazquez-Salcedo:2024oek} for the full expressions). The background can be generic and numerical if desired. The perturbations are also generic and do not require special coordinates, leading to results valid beyond the small-coupling  and small-spin approximations.
Rotating background solutions satisfy $\mathcal{G}_{\mu\nu}^{(bg)}=0$ and
$\mathcal{S}^{(bg)}=0$, leaving a system of coupled PDEs in $r$ and $\theta$ for the perturbation functions, $\delta\mathcal{G}_{\mu\nu}=0$ and $\delta\mathcal{S}=0$.
Imposing standard QNM boundary conditions and regularity on the rotation axis yields a set of constraints through a perturbative analysis of the asymptotic behavior at asymptotic infinity and near the horizon~\cite{Blazquez-Salcedo:2024oek}.

The perturbations are then expanded using Chebyshev polynomials and Legendre functions, employing a compactified radial coordinate $z = (r-r_+)/(r+1)$ and $x=\cos(\theta)$.
After the Chebyshev and Legendre expansions 
of the perturbation functions, 
the partial differential system can be expressed as a quadratic eigenvalue problem
\begin{equation}
    \left( \mathcal{M}_0 + \mathcal{M}_1 \omega + \mathcal{M}_2 \omega^2 \right) \vec{C} = 0 \, ,
    \label{matrix_eq}
\end{equation}
with square matrices $\mathcal{M}_0$, $\mathcal{M}_1$ and $\mathcal{M}_2$ that depend on the background solution. 

At variance with the METRICS approach, the quadratic eigenvalue is obtained directly, without factorization and simplification of the perturbed field equations, on a discretized grid of $N_x \times N_y$ points (typically of order 20 in both directions).
The unknown expansion coefficients that lie in the vector $\vec{C}$ can be accurately determined.
One can then find accurate numerical solutions~\cite{Blazquez-Salcedo:2024oek} for the eigenvalue $\omega$, for example through \texttt{Maple} and \texttt{MATLAB} with Multiprecision Computing Toolbox Advanpix~\cite{mct2015}.
The method has been validated in the GR rotating case, confirming its high accuracy: at least $10^{-6}$ accuracy up to about $80\%$ of the maximum angular momentum~\cite{Blazquez-Salcedo:2023hwg}.
Besides the modified gravity results covered below, this method was also applied to study the QNMs of rapidly rotating wormholes~\cite{Khoo:2024yeh}. 

\subsubsection{Black hole perturbations in modified gravity: results}\label{sec:modgrav_res}
We now summarize the state of the art of QNM frequency calculations that applied the methods described above to a variety of theories.  

\vspace{0.25cm}
\par
\noindent
{\textit{EFT extensions of GR}}
In these theories, defined by the action~\eqref{eq:GREFT}, the modified Teukolsky equation allows one to obtain the modifications to the Kerr QNM frequencies to linear order in the coupling constants $\zeta_{\rm q}$:
\begin{equation}\label{eq:lineardeltaomega}
\omega_{\ell mn\pm}=\omega^{\rm Kerr}_{\ell mn}+\frac{\zeta_{\rm q}}{M}\delta\omega_{\ell mn\pm}^{({\rm q})}(\chi)+\mathcal{O}\left(\zeta_{\rm q}^2\right)\, .
\end{equation}
Here $\pm$ denotes the mode polarization, and we have introduced the factor of $1/M$ so that the target of the computations, $\delta\omega_{\ell mn\pm}^{({\rm q})}(\chi)$, is dimensionless.

The frequency shifts can be found by solving Eq.~\eqref{eq:correctedradial} through, e.g., a generalized continued fraction method through a higher-step recursive relation~\cite{Cano:2024jkd} or eigenvalue perturbation techniques~\cite{Zimmerman:2014aha,Hussain:2022ins}. 
The frequency shifts can also be obtained by mapping Eq.~\eqref{eq:correctedradial} to the parameterized formalism reviewed in Section~\ref{sec_param_pot} below.
Achieving a reliable result for large angular momentum requires computing Eq.~\eqref{eq:correctedradial} at very high order in the spin expansion, which is computationally challenging. 

Shifts $\delta\omega_{\ell mn\pm}^{({\rm q})}(\chi)$ for all the EFT theories, for the modes $\ell=2,3,4$, $m\in [-\ell,\ell]$, $n=0,1,2$ and up to angular momentum $\chi\sim 0.7-0.9$ were computed up to expansion order $\chi^{18}$~\cite{Cano:2024ezp}. 
In Fig.~\ref{fig:cubiccorrections} we show the $(\ell,m,n)=(2,m,0), (2,m,1)$ modes in the even-parity cubic theory case. 
The results agree with previous computations for static and slowly rotating BHs obtained via modified RWZ equations~\cite{Cardoso:2018ptl,deRham:2020ejn,Cano:2021myl} and are publicly available~\cite{gitbeyondkerr} (see Appendix~\ref{sec:public_codes}).
\begin{figure*}[t]
    \centering
\includegraphics[width=0.95\linewidth]{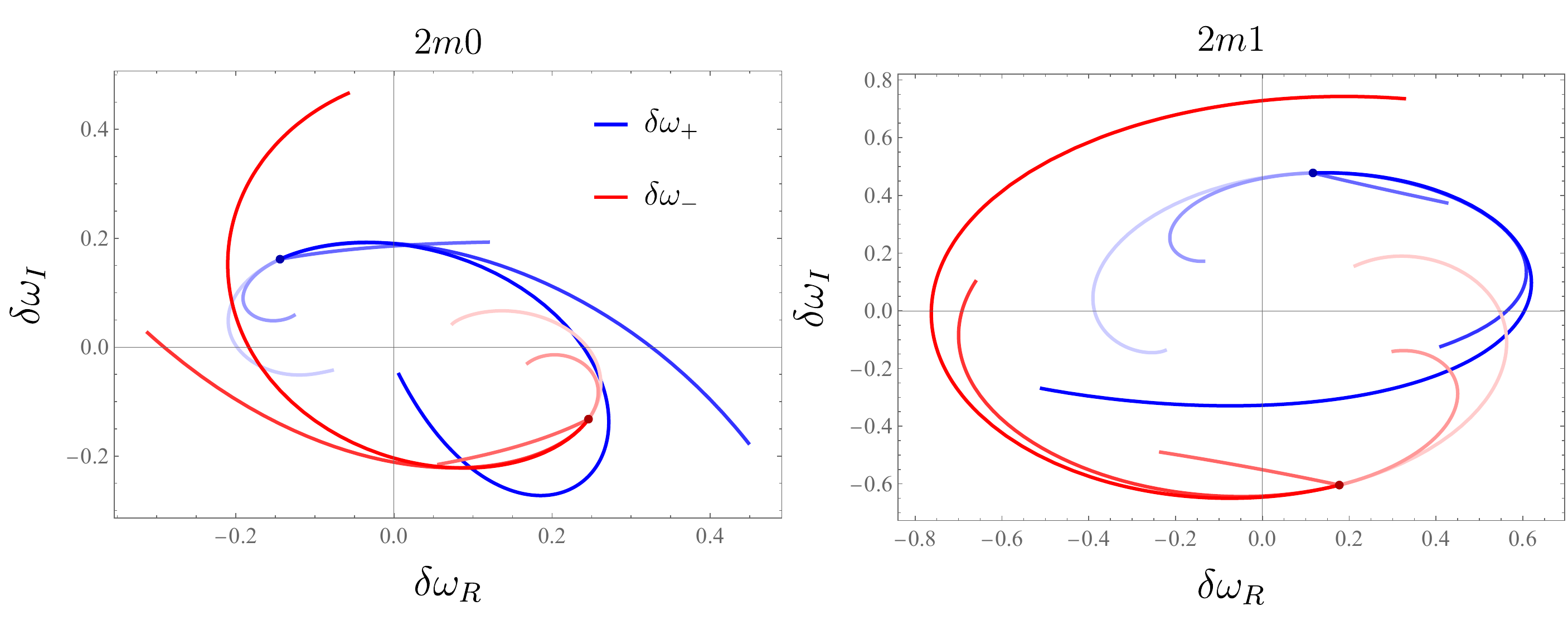}
    \caption{Linear shifts in the Kerr QNM frequencies due to cubic curvature corrections: see Eq.~\eqref{eq:lineardeltaomega}. Left panel: $2m0$ modes for BHs of spin $\chi\in [0,0.85]$. Right panel: $2m1$ modes for BHs of spin $\chi\in [0,0.75]$. Blue curves correspond to the ``+'' polarization modes, red curves to the ``$-$'' polarization, and the intensity of the color grows from $m=-2$ (fainter) to $m=+2$ (brighter). The dots mark the case of static BHs, where the different $m$ modes are degenerate.}
    \label{fig:cubiccorrections}
\end{figure*}

One of the most generic beyond-GR effects is the breaking of isospectrality, so that even and odd modes receive different corrections~\cite{Cardoso:2019mqo,McManus:2019ulj}. This works differently for parity-preserving and parity-breaking theories. In the former case, $\delta\omega_{+}$ and $\delta\omega_{-}$ are unrelated to each other, while in the latter $\delta\omega_{+}=-\delta\omega_{-}$. Interestingly, combining parity-preserving corrections (with shifts $\delta\omega_{\rm pres\pm}$) and parity-breaking ones (with shifts $\delta\omega_{\pm}=\pm\delta\omega_{\rm break}$) the total shift is not just the sum of each of them. Instead, the following combination rule holds~\cite{McManus:2019ulj,Cano:2021myl,Cano:2023jbk}
\begin{equation}
    \delta\omega_{\rm tot\pm}=\frac{1}{2}\left(\delta\omega_{\rm pres+}+\delta\omega_{\rm pres-}\right)\pm\sqrt{\frac{1}{4}\left(\delta\omega_{\rm pres+}-\delta\omega_{\rm pres-}\right)^2+\delta\omega_{\rm break}^2}\, .
\end{equation}
This relationship arises from the coupling between perturbations of different parity.  While all of the analyzed higher-derivative theories break isospectrality, it has been recently observed that the special quartic theory $\mathcal{C}^2+\tilde{\mathcal{C}}^2$ preserves isospectrality in the eikonal limit~\cite{Cano:2024wzo}. In this theory, isospectrality is still broken for low-$\ell$ modes, but the difference between even and odd modes is small.    
Additional results for other theories and more modes are available online~\cite{gitbeyondkerr} (see Appendix~\ref{sec:public_codes}).

These results have recently been applied to test EFT corrections against real
ringdown data from LIGO-Virgo-KAGRA (LVK) observations in~\cite{Maenaut:2024oci}, finding a bound on the scale of new physics $\ell\lesssim 35$ km.  Future ground-based detectors will substantially improve these bounds, as discussed in Section~\ref{Section~7.2.3} below.

\vspace{0.25cm}
\par
\noindent
{\textit{dCS gravity}}
In this theory, the QNM frequencies are also obtained as a linear expansion in the coupling, similar to Eq.~\eqref{eq:lineardeltaomega}. 
Results are known for the static and spherically symmetric (Schwarzschild) background~\cite{Cardoso:2009pk,Molina:2010fb, Kimura:2018nxk, McManus:2019ulj}, and for rotating (non-Kerr) BHs at linear order in the small-spin expansion~\cite{Wagle:2021tam,Srivastava:2021imr, Wagle:2023fwl} and for $\chi$ up to $\sim 0.75$ using METRICS~\cite{Chung:2025gyg}.
All of these studies relied on a metric perturbation approach, except for~\cite{Wagle:2023fwl}, which employed the modified Teukolsky approach, expanded to linear order in the spin.  

\begin{figure*}[t]
\centering  
\includegraphics[width=0.47\linewidth]{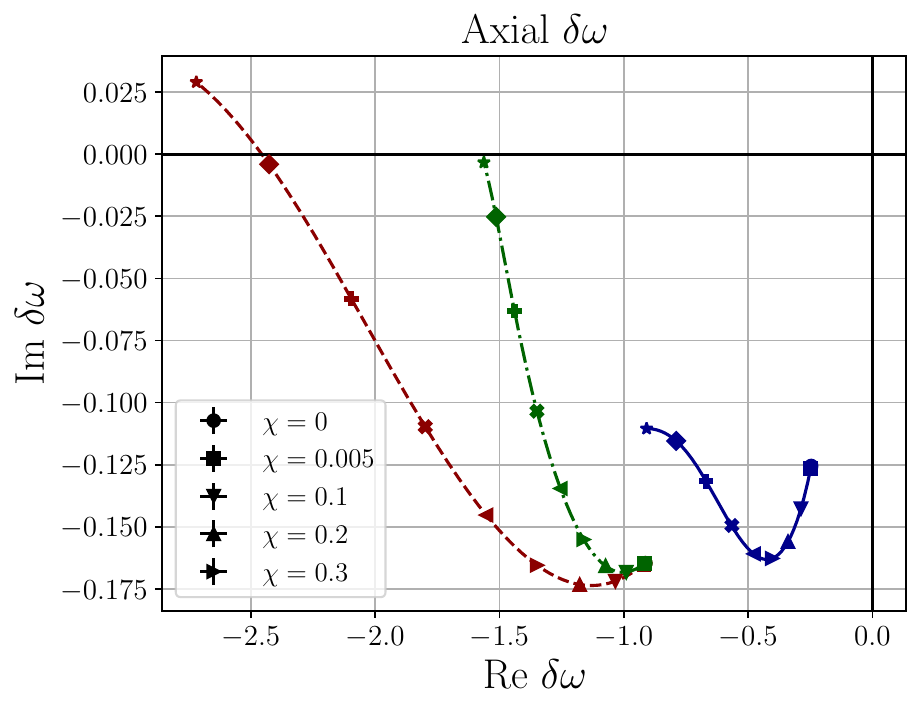}
\includegraphics[width=0.47\linewidth]{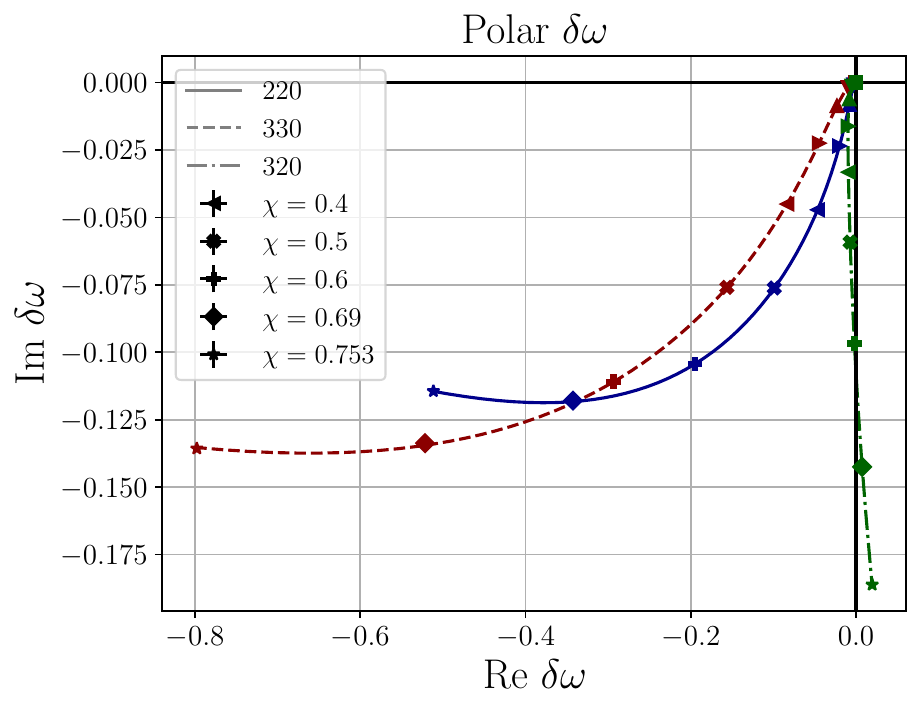}
\caption{Complex trajectories of $\delta \omega_{\ell m n}$ in dCS gravity
  computed using METRICS for the axial (middle) and polar (right) perturbations,
  and of the $220$ (solid), $330$ (dashed), and $320$ (dashed-dotted) modes as
  $\chi$ increases from $0$ (circle) to $0.753$ (star).  The real and imaginary
  axes are marked by solid black lines.  Figure adapted from~\cite{Chung:2025gyg}.  }
\label{fig:dCS}
\end{figure*}

All the existing calculations show that, in dCS gravity, the axial frequency is shifted more significantly than the polar frequency, due to the stronger coupling of the perturbations of the pseudoscalar to axial metric perturbations. 
In particular, when the background has zero spin ($a=0$), the perturbations of the pseudoscalar field completely decouple from the polar metric perturbations. 
Fig.~\ref{fig:dCS} shows the lead-order QNM-frequency shift in $\zeta$ (i.e. $\delta \omega$ defined by Eq.~(\ref{eq:lineardeltaomega})) of the axial metric perturbations for the $\ell m n = 220, 330$ and 320 modes of a rotating BH background with $0 \leq \chi \leq 0.753 $. 
Observe that $\delta \omega $ of the polar perturbations to nonrotating BHs is zero for all modes, because the polar metric perturbations are decoupled from the dCS terms when $\chi = 0 $.
Analytical fitting expressions of the QNM-frequency shift obtained by METRICS
that are valid for $\chi$ up to $0.75$ are available in~\cite{Chung:2025gyg},
and obtained by a separate analytical calculation valid up to linear order in
the spin are also available~\cite{Srivastava:2021imr}.

\vspace{0.25cm}
\par
\noindent
{\textit{Linear EsGB gravity}}
Initial results for this theory were obtained using 
either numerical solutions or analytic (slow-rotation) approximations for the background metric and scalar field~\cite{Blazquez-Salcedo:2016enn, Blazquez-Salcedo:2017txk, Pierini:2021jxd, Pierini:2022eim, Langlois:2022eta}. 
The extension to rapidly rotating BH backgrounds, with $a \lesssim 0.85M$, and small coupling were instead computed using the METRICS approach~\cite{Chung:2024ira, Chung:2024vaf}.
Results are shown in Fig.~\ref{fig:METRICS_EsGB} for $n \ell m = 022$, both axial and polar perturbations, and an illustrative value of the coupling $\zeta_{\text{EsGB}}=0.1$, where we recall that $\zeta_{\text{EsGB}}$ is defined in \eqref{eq:zetaGBdef}.  
The figures confirms the breaking of isospectrality, as in the case of cubic EFTs and dCS gravity.
Fitting formulas and calculations for other modes are also available~\cite{Chung:2024vaf}.

\begin{figure*}[t]
\centering  
\includegraphics[width=12cm]{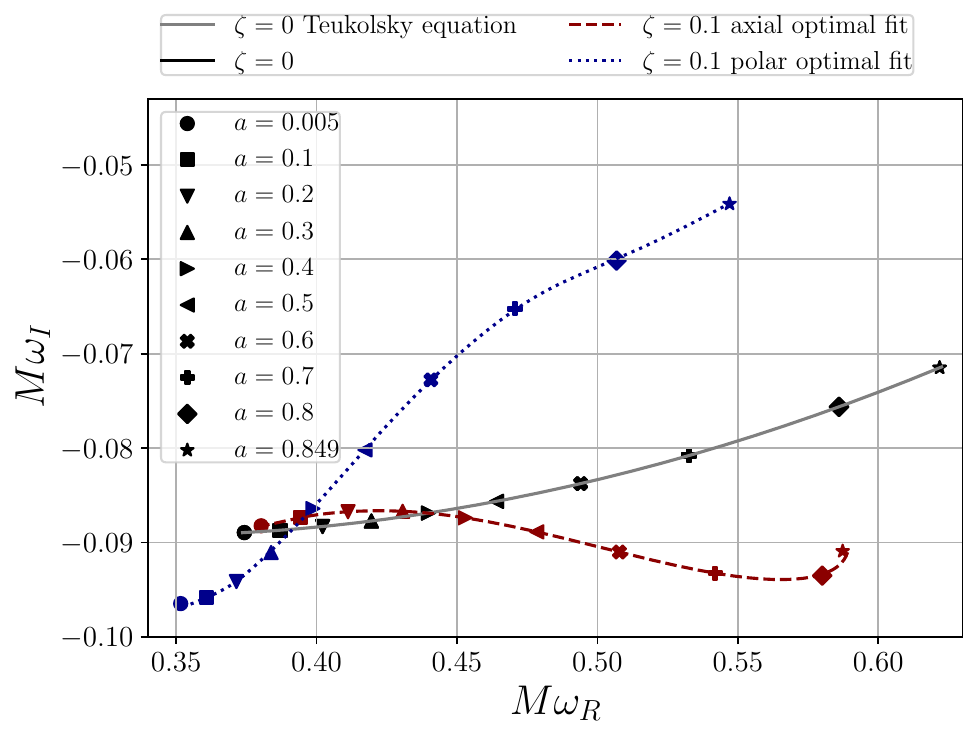}
\caption{Trajectory in the complex plane of the QNM frequencies of the $n \ell m = 022$ mode in linear EsGB gravity for increasing BH spin, computed using METRICS~\cite{Chung:2024ira, Chung:2024vaf} setting $\zeta_{\text{EsGB}} = 0.1$ (note that this value of $\zeta_{\text{EsGB}}$ was chosen for illustrative purposes, and it is outside the domain of existence of background solutions if the theory is treated as exact).
The symbols correspond to the numerical frequencies, while the lines are polynomial fits.
Figure taken from~\cite{Chung:2024vaf}.
}
\label{fig:METRICS_EsGB}
\end{figure*}

For small spin, the numerical results are consistent with second-order in spin perturbative predictions~\cite{Blazquez-Salcedo:2016enn, Blazquez-Salcedo:2017txk, Pierini:2021jxd, Pierini:2022eim}, confirming that polar frequency corrections are shifted by a larger extent compared to axial ones. 
The relative fractional uncertainty of the real (imaginary) frequency varies between $[10^{-8}, 10^{-2}]$ ($[10^{-6}, 10^{-1}]$), depending on the spin~\cite{Chung:2024ira}. This level of accuracy is good enough to analyze individual ringdown GW signals observed by the LVK ground-based detectors in the third observing run~\cite{LIGOScientific:2020tif, LIGOScientific:2021sio}.
The polar frequencies were used to construct an effective one body (EOB) waveform model~\cite{Julie:2024fwy} employed to search for modified gravity signatures in GW data (see Section~\ref{subsec:TGR_theory-specific}).

A calculation of the QNM frequencies in linear EsGB gravity
that is fully nonperturbative in the coupling can be found in~\cite{Khoo:2024agm}.

\vspace{0.25cm}
\par
\noindent
{\textit{EdGB gravity}}
In Fig.~\ref{fig_js_bkkk} we show fully nonperturbative results for the QNM frequencies of EdGB BHs for a dilaton coupling with $\gamma=1$ as a function of coupling constant $\zeta_{\text{EdGB}}$ for two representative values of the spin. The frequencies of the longest-lived modes with $\ell=2,3$ and $m=2$ were computed using the spectral collocation method described above~\cite{Blazquez-Salcedo:2024oek}, and the vertical lines mark the critical values of $\zeta_{\text{EdGB}}$ beyond which no regular BHs exist in EdGB theory for the given dimensionless spin $\chi=J/M^2$.

\begin{figure}[t]
\begin{center}
\includegraphics[width=7cm]{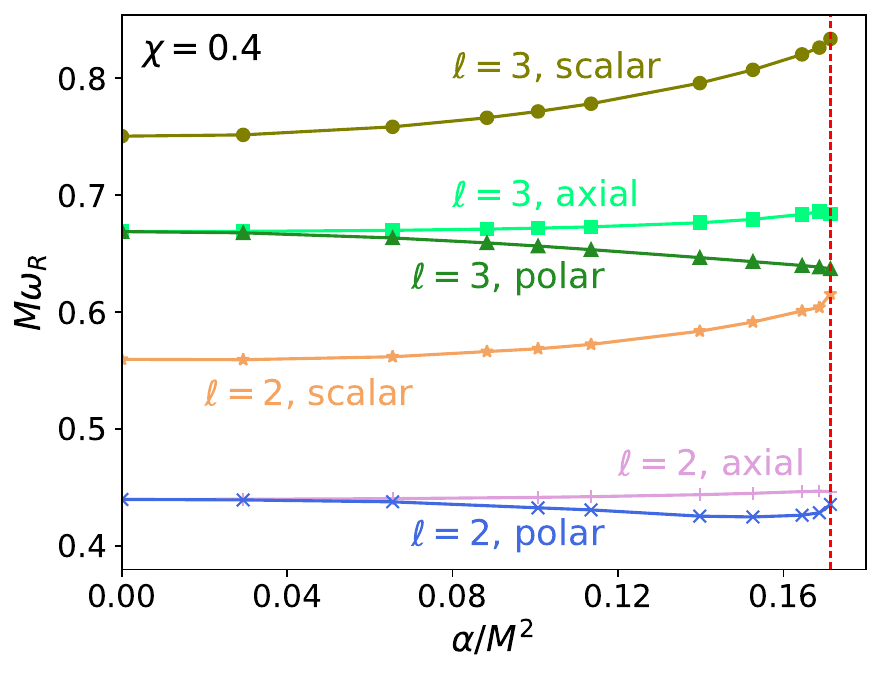}
\includegraphics[width=7.4cm]{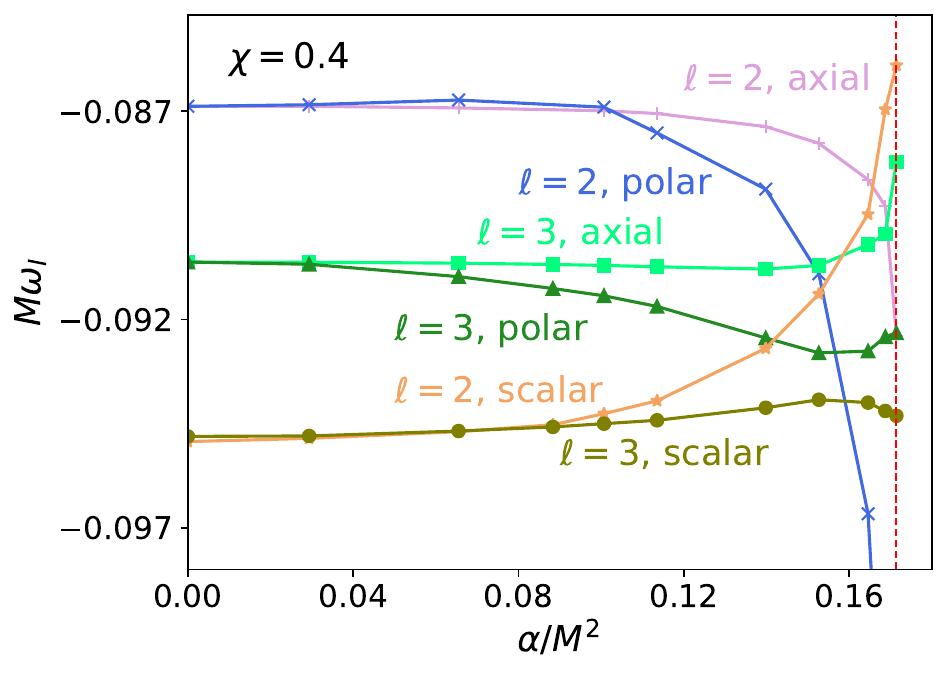}
\includegraphics[width=7cm]{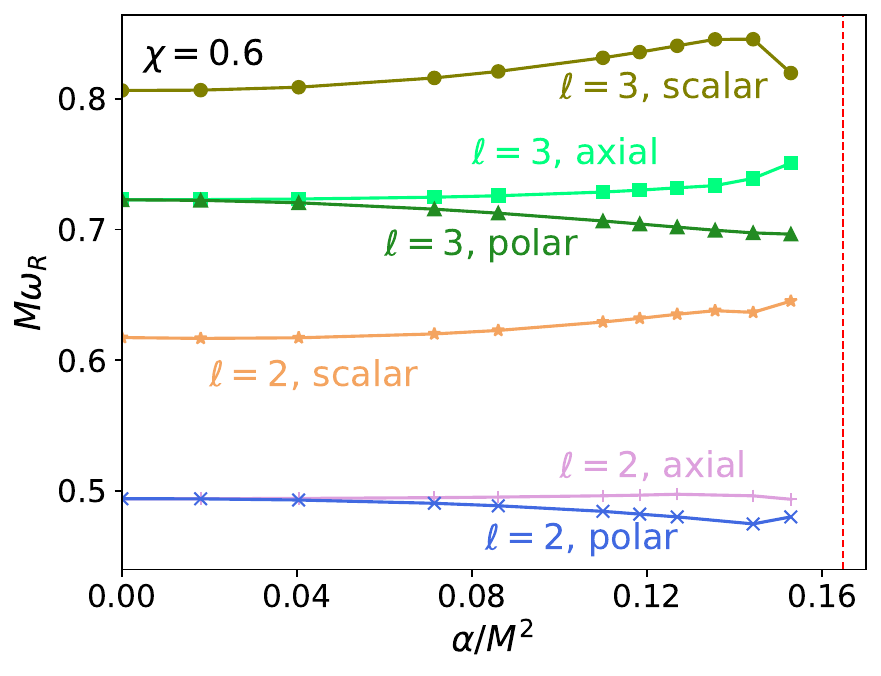}
\includegraphics[width=7.3cm]{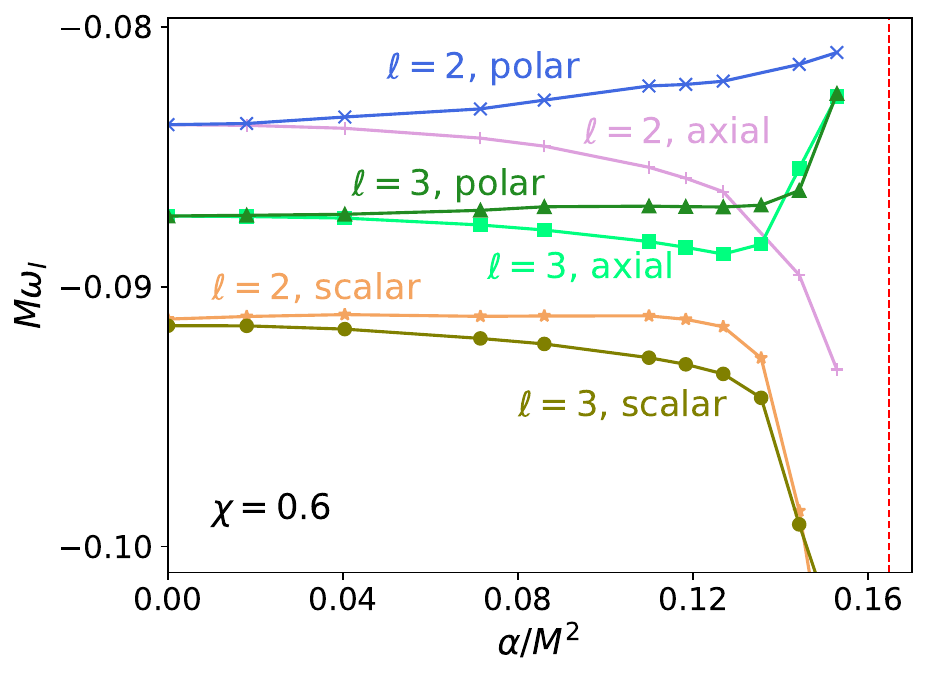}
\end{center}
\caption{
Fundamental modes of EdGB BHs with $\chi=0.4, 0.6$ (higher values are also known~\cite{Blazquez-Salcedo:2024oek}) and $\ell=2,3, m=2$. The left (right) panels show the scaled real part $M\omega_R$ (scaled imaginary part $M\omega_I$) as functions of the scaled coupling constant $\zeta_{\text{EdGB}}=\alpha/M^2$.
The red vertical dashed lines indicate the largest scaled coupling constant allowed for each $\chi$.
Figure adapted from~\cite{Blazquez-Salcedo:2024oek}.
}
\label{fig_js_bkkk}
\end{figure} 
As we see in Figure \ref{fig_js_bkkk}, the isospectrality of the Kerr axial-led and polar-led modes is broken in EdGB gravity.
This happens already in the static case~\cite{Blazquez-Salcedo:2016enn,Blazquez-Salcedo:2017txk}, becoming more prominent as the coupling strength and BH spin increase, approaching the boundary of the domain of existence of BH solutions.
The figure also shows the scalar-led modes of the dilaton field, smoothly connecting with the modes of a minimally coupled scalar field on the Kerr background as the coupling strength is decreased.
The lowest value of the frequency $M\omega_R$ always corresponds to the $\ell=2$ metric modes. 
Typically, polar modes have smaller values of $M \omega_R$, but the difference with respect to the axial modes is always small.

Another interesting property of the spectrum is the switch of mode dominance as the critical coupling strength is approached, and the scalar charge of the solutions increases. 
While for Kerr the gravity-led modes are always longer lived than the scalar-led modes, here, depending on the angular momentum and coupling strength, it is possible to find solutions for which (vice versa) the scalar-led modes are longer lived than the gravity-led modes. 
This dominance switch is displayed in Fig.~\ref{fig_js_bkkk} for $\chi=0.4$ and $\ell=2$, stressing the importance of scalar-led modes when modeling the ringdown phase in EdGB theory. 

These general results also reproduce with excellent accuracy the static limit~\cite{Blazquez-Salcedo:2016enn,Blazquez-Salcedo:2017txk} obtained with a different shooting method, and the analytical predictions in the slow-rotation approximation~\cite{Pierini:2021jxd,Pierini:2022eim}. 
The computed QNMs satisfy all the PDEs and boundary conditions of the perturbations with an accuracy of at least $10^{-4}$.
More details can be found in~\cite{Blazquez-Salcedo:2024oek} as well as~\cite{Blazquez-Salcedo:2024dur}, where the modes with a negative real part are also computed. Like the positive-frequency modes, the modes with negative frequency have damping times that depend strongly on the angular momentum and on the coupling constant, and they present dominance switch between gravity-led and scalar-led modes in some cases.

\subsubsection*{Future directions}
Despite the remarkable progress summarized above, several outstanding problems remain in order to complete our understanding of the linear ringdown in theories beyond GR. An important open question is the case of very highly rotating BHs, $\chi\gtrsim 0.95$. None of the methods reviewed above has achieved reliable results in this regime yet, as this is technically challenging and may require alternative approaches~\cite{Cano:2024bhh} (see also Section~\ref{sec:near_extremal_branching}). The computation of overtones has also been quite limited so far. 
In addition, some of the results we have discussed only consider corrections to the QNMs perturbative in the coupling. For the purpose of testing these theories, it is important to have reliable predictions for the QNM frequencies for arbitrary values of the coupling constants. 
Beyond the QNM frequencies, one should also analyze the effect of beyond-GR physics on the QNM amplitudes~\cite{Silva:2024ffz} and, more broadly, on the TD signal. In a context where extensions of GR are treated as EFTs, it is important to have reliable predictions for the QNMs for arbitrary values of the coupling constants or an estimate of the regime validity of QNM calculations. This validity regime may depend on the harmonic or overtone number~\cite{Hirano:2024fgp,Silva:2024ffz,Cano:2024ezp}.

Addressing these issues is important in order to perform accurate ringdown-based
tests of theories beyond GR with next-generation (XG) GW observatories (see Section~\ref{sec:nextgen}). 

\subsection{Theory-agnostic parameterization} \label{subsec:theory-agnostic}

\vspace{-.1cm}

\noindent \textit{Initial contributors: Franchini, Kimura, V\"olkel}

\vspace{.2cm}

Rather than computing QNM frequencies on a theory-by-theory basis, an alternative approach is to develop a theory-agnostic, parametric framework to compute QNMs when the underlying theory of gravity or the corresponding BH solutions are unknown. 
This approach is strongly motivated by the large amount of modified theories of gravity or new physics scenarios that could induce deviations from the Kerr spectrum in the data.
Note however that the terminology ``theory-agnostic'' can be somewhat misleading, as specific theoretical constraints are often imposed, either implicitly or explicitly.
For example, if we start from the ansatz that modifications to the QNM spectrum can be described by modifications of the perturbation potentials in GR we must necessarily assume separability, which selects only certain classes of beyond-GR theories (see e.g.~\cite{Roussille:2022vfa}). 
Similarly, the eikonal limit (see Section \ref{subsec:LR}), may not hold in some theories; approaches using parameterized BH metrics requires a choice of test fields (e.g., the Klein-Gordon equation in a background spacetime); and so on. 

Any parameterization must resolve a tension between generality and simplicity. The parameterization should be general enough to encompass the largest possible amount of new physics, yet limit the number of free parameters so that they will not be degenerate or unconstrained, thus making measurements uninformative.
If the aim is to understand fundamental properties of QNMs beyond GR without focusing on a given theory, one must assume some general structure for the equations of motion or even for the action of the theory, either starting from first principles or using known cases as a guide. 
Finally, assuming a metric theory, one might be interested in linking the modifications to the QNM spectrum (as induced, e.g., by environmental effects) to modifications in the underlying spacetime metric. 

Despite these limitations, theory-agnostic approaches allow us to study a rich phenomenology capturing several classes of modifications, and they provide a valuable, interpretable observational framework to search for new physics.
In this chapter we review the most commonly used parameterized models.
We first introduce the simplest parameterization of possible deviations from the
gravitational QNM spectrum in Section~\ref{sec_agnostic}, then we discuss the
``parameterized spectroscopy'' (or \texttt{ParSpec}) framework in
Section~\ref{sec_parspec}, and finally we discuss parameterizations of the
perturbation potentials in Section~\ref{sec_param_pot}.
In Section~\ref{sec_param_metrics_eikonal} we describe approaches based on parameterized BH metrics and ideas based on the eikonal limit (see Section~\ref{subsec:LR}).
Finally, in Section~\ref{sec_eft} we discuss parameterizations at the level of the action and the so-called ``EFT of QNMs.''

\subsubsection{Agnostic QNM deviations}\label{sec_agnostic}
Quite generally, if there are any deviations from vacuum GR, the gravitational QNMs can be parameterized in the form
\begin{align}\label{eq:agno_QNM}
f_{\ell{m}n} &= f_{\ell{m}n}^\mathrm{Kerr}\, (1 + \delta f_{\ell{m}n}) \,,\\
\tau_{\ell{m}n} &= \tau_{\ell{m}n}^\mathrm{Kerr}\, (1 +\delta\tau_{\ell{m}n}) \,,
\end{align}
where here $f$ refers to the real component of the complex QNM frequency $\omega$, and $\tau$ to the inverse of its imaginary component.
This parameterization has been proposed and used extensively in GW data analysis for tests of GR and searches of new physics~\cite{Gossan:2011ha,Meidam:2014jpa,Carullo:2018sfu,LIGOScientific:2020tif,LIGOScientific:2021sio} (see also Section~\ref{sec:DataAnalysis} below).
The parameterization is extremely flexible and general, as it can accommodate any shift to a given QNM with respect to the Kerr spectrum.
Its main limitation is that Eq.~\eqref{eq:agno_QNM} does not take into account new modes that are not given by continuous deformations of the Kerr gravitational spectrum which may appear in the signal (see~\cite{Crescimbeni:2024sam, Lestingi:2025jyb}). In general, additional modes and polarizations will appear in the spectrum whenever a specific modified theory of gravity involves additional degrees of freedom (such as dynamical scalar, vector or tensor fields).
While the parameterization will not capture the power contained in such new modes,
any additional degrees of freedom should only contain a small fraction of the signal power if the modified theory is close enough to GR. In this sense, the deviation parameters are still expected to capture the main modifications to the GR signal.
Any deviation from the Kerr spectrum would flag the detection of an anomalous feature, that can later be followed up with informed searches including additional modes.

From the data analysis point of view, the main disadvantage of this parameterized model is that it requires two new degrees of freedom \textit{for each mode} $(\ell, m, n)$ to describe potential deviations from the GR spectrum.
For this reason the parameterization can be very inefficient when the number of observed modes increases, resulting in uninformative measurements (see Section~\ref{sec:DataAnalysis}).
A second important limitation is that the parameterization is ``too general,''
in the sense that it does not fold in the dependence of the frequencies and of
their modifications on the BH parameters (mass and spin). This makes it
impossible to exploit the properties of the underlying source population:
individual sources with different masses and spins are better at probing
different modifications from GR, and by combining multiple sources it should be
possible to obtain better bounds on any specific modification.
More fundamentally, if a deviation is detected, one would like a physical understanding of its origin -- something which is not provided by Eq.~\eqref{eq:agno_QNM}.

In conclusion, the parameterization of Eq.~\eqref{eq:agno_QNM} is useful for {\em null tests}, but it cannot be used to extract the underlying physics without further assumptions.
Due to the generality of the parameterization, it is also very difficult to give a physical interpretation of nonzero deviation parameters in multi-modal scenarios if the underlying GR signal model is incomplete.
This can happen when the number of detectable modes is unknown (and especially for overtones), even in idealized situations~\cite{Baibhav:2023clw,Nee:2023osy,Cheung:2023vki,Takahashi:2023tkb}.

In the following we will show how including more assumptions based on the underlying physics affects QNM parameterizations.

\subsubsection{The \texttt{ParSpec} framework}\label{sec_parspec}
A simple and reasonable ansatz is to assume the deviations from GR to be small, i.e.,
\begin{equation}
	\delta f_{\ell{m}n} \ll 1\,, \qquad \delta\tau_{\ell{m}n} \ll 1 \,.
\end{equation}
This ansatz is based on the reasoning that even if modifications of GR are present in nature, the stringent experimental bounds imposed by existing tests of GR~\cite{Will:2014kxa,Berti:2015itd,Will:2018bme,LIGOScientific:2016lio,Yunes:2016jcc,LIGOScientific:2018dkp,LIGOScientific:2019fpa,LIGOScientific:2020tif,LIGOScientific:2021sio,Yunes:2024lzm} imply that the modifications must be hidden in the noise, and the relative size of any frequency modifications to the Kerr spectrum can be expected to be much smaller than unity. From a theoretical perspective, this assumption fits well with an EFT description of gravity, where GR is the low-energy limit of some as yet unknown theory
(see Section~\ref{subsec:theory-spec} for theory-specific examples). 

A phenomenological model for ``universal'' modifications of spinning BH spectra that incorporate the key features of the remnant mass and spin dependence was proposed in~\cite{Maselli:2019mjd}.
In this parameterized spectroscopy, or ``\texttt{ParSpec}'' model, the modifications are written in the form of parameterized ringdown spin expansion coefficients as follows:
\begin{align}\label{eq:parspec_par}
	f_{\ell m n} 	& = \frac{1}{M} \sum_{k=0}^{K} \, \chi^k \, f_{\ell m n}^{(k)} \left( 1 + \gamma \, \delta f_{\ell m n}^{(k)} \right)\,, \\
	\tau_{\ell m n} 	& = M \sum_{k=0}^{K} \, \chi^k \, \tau_{\ell m n}^{(k)} \left( 1 + \gamma \, \delta\tau_{\ell m n}^{(k)} \right)\,,
\end{align}
where $f^{(k)}_{\ell m n}$ and $\tau^{(k)}_{\ell m n}$ come from a Taylor expansion of the GR values around 
$\chi = 0$, 
$\gamma$ is a theory-dependent and (possibly) source-dependent parameter, and $\delta f^{(k)}_{\ell m n}$ and $\delta \tau^{(k)}_{\ell m n}$ are the source-independent beyond-GR deviations expanded around 
$\chi = 0$.

Depending on the underlying theory of modified gravity, we can consider three different cases:

\noindent
{\em (i) Scale-free corrections} If the corrections to GR are given by a single {\em dimensionless} coupling $\alpha$, we can set $\gamma = \alpha$ and reabsorb it within $\delta f^{(k)}_{\ell m n}$ and $\delta \tau^{(k)}_{\ell m n}$.
This case includes certain scalar-tensor and Lorentz-violating theories, such as Einstein-Aether and Ho\v{r}ava gravity (to leading order)~\cite{Barausse:2013nwa}.

\noindent
{\em (ii) Single dimensionful coupling} The corrections to GR are given by a dimensionful coupling $\alpha$ with dimensions $[\alpha]= \hat{M}^p$, with $p$ a (usually positive) integer, and $\hat{M}$ the typical length scale of the problem. We can set $\gamma = \alpha/M_\mathrm{s}^p = \alpha (1+z)^p/M^p$, where $M_\mathrm{s}$ is the BH mass in the source frame and $z$ is the source redshift. For example, for $p=4$ this captures the scaling of the curvature-dependent modifications in EsGB and dCS gravity, and for $p=6$ it captures the scaling of some of the EFTs discussed in Section~\ref{subsec:theory-spec}.

\noindent
{\em (iii) Individual charges} BHs in the modified theory possess what is sometimes called ``primary hair,'' i.e.,~a new charge $Q$ that is completely independent on the BH mass and spin, with dimensions $[Q] = \hat{M}^p$. The simplest examples are Kerr-Newman BHs (see~\cite{Herdeiro:2015waa} for a review of ``hairy'' BH solutions). In this case, we can set $\gamma  = Q/M_\mathrm{s}^p$.

This parameterization was used to derive constraints from current observational data in the case of scale-free and dimensionful couplings~\cite{Carullo:2021dui}, in the case of individual charges with $p=2$ (i.e., Kerr-Newman BHs)~\cite{Carullo:2021oxn,Gu:2023eaa}, and in some specific theories of gravity with selected values of $p$~\cite{Silva:2022srr}.
Projected bounds with XG detectors were computed in~\cite{Maselli:2023khq} (see Sections~\ref{subsec:deviations_current} and \ref{subsec:future_tests} for an extensive discussion).

\subsubsection{Parameterized QNM framework}\label{sec_param_pot}

This section deals with the \textit{parameterized QNM framework}, which is one
step closer to the theory-specific case.  Instead of assuming a mass and spin
dependence in the parameterization of the QNM spectrum, in this framework one
assumes a parameterization of the effective potentials. In this way we can make
direct contact with the numerous specific models in which the perturbation
equations can be cast in the form of a modified RWZ equation (in the
nonrotating case) or of a modified Teukolsky equation (in the rotating
case). The power of the method is that one can compute the QNM spectrum in all
of these beyond-GR theories in one sweep without assuming any specific
theory. The main (important) assumption is that, in the perturbative spirit
discussed above, the modifications of the RWZ or Teukolsky potentials must be
small, in a sense that can be made precise.  We will first focus on the
nonrotating case for simplicity, and then generalize the treatment to rotating
BHs.

The parameterized QNM framework was first introduced to compute linear QNM corrections due to small modifications of single-field (scalar, gravitational axial, gravitational polar) perturbation equations of the Schwarzschild BH~\cite{Cardoso:2019mqo}. Then it was extended to include quadratic corrections, systems with multiple coupled fields~\cite{McManus:2019ulj}, QNM overtones~\cite{Volkel:2022aca,Hirano:2024fgp} and the TD evolution of initial data~\cite{Thomopoulos:2025nuf}.

In the nonrotating or slowly rotating case, let us assume that single or multi-field perturbations $\mathbf{\Phi}$ are well approximated by the wave equation
\begin{equation}\label{eq:mastersystem}
	f \frac{\mathrm{d}}{\mathrm{d} r}\left( f \frac{\mathrm{d} \mathbf{\Phi}}{\mathrm{d} r} \right) + \left[ \omega^2 -f \mathbf{V} \right]\mathbf{\Phi} = 0\,,
\end{equation}
where $f = 1-r_+/r$, $r_+=2M$, and the effective potential is
\begin{align}
	V_{ij} &= V^\mathrm{GR}_{ij} + \delta V_{ij}, \\
	\delta V_{ij} &= \frac{1}{r_+^2} \sum_{k=0}^{\infty} \alpha^{(k)}_{ij} \left(\frac{r_+}{r} \right)^k\,.\label{deltaVij}
\end{align}
Here the diagonal elements of the matrix $V^\mathrm{GR}_{ij}$ stand for the scalar ($s=0$) potential, the gravitational axial (Regge-Wheeler) potential, or the polar gravitational (Zerilli) potential.
This way of writing the unperturbed wave equations differs slightly from standard conventions in the literature due to the extra factor of $f$.
Many models are included in this parameterized potential~\cite{Cardoso:2018ptl, Cardoso:2019mqo, McManus:2019ulj, Hatsuda:2020egs, Volkel:2022aca, Volkel:2022khh, Tattersall:2019nmh, deRham:2020ejn, Sirera:2023pbs, Mukohyama:2023xyf, Hirano:2024fgp}, which allows for the inclusion of coupling functions between multiple fields through the off-diagonal matrix elements.

The QNM frequencies $\omega$ (where we omit the $\ell,\,m,\,n$ indices for simplicity) can be expanded in powers of the small modifications $\alpha^{(k)}_{ij}$, and the resulting corrections up to quadratic order can be written as
\begin{equation}\label{eq:omega_mod}
	\omega \approx \omega^0 + \alpha^{(k)}_{ij} d_{(k)}^{ij} 
     + \alpha_{ij}^{(k)} \alpha{}_{pq}^{\prime (s)} d^{ij}_{(k)}d^{pq}_{(s)} 
    +  
	\frac{1}{2} \alpha^{(k)}_{ij} \alpha^{(s)}_{pq} e^{ijpq}_{(ks)} \,,
\end{equation}
where $\omega^0$ is the unperturbed QNM. 
In Eq.~\eqref{eq:omega_mod}, $\alpha_{ij}^{(k)}$ and its derivative with respect to $\omega$, denoted by a prime, are evaluated at $\omega^0$.
Note that the coefficients $d_{(k)}^{ij}$ and $e^{ijpq}_{(ks)}$ do not depend on $\alpha^{(k)}_{ij}$, and thus only need to be computed once. 
Numerical data for the coefficients are publicly available in~\cite{GRIT,JHU,CoG, Hirano:2024fgp} (see Appendix~\ref{sec:public_codes}).

It was shown that the number of deviation parameters can be conveniently reduced by suitable field redefinitions, thus limiting the total number of free parameters to just a few low-order coefficients~\cite{Kimura:2020mrh, Hatsuda:2023geo}.
This makes the computation of QNM corrections extremely fast and stable, which is particularly interesting for data analysis applications or theoretical explorations of the parameter space~\cite{Volkel:2022khh}. 
The parameterized QNM framework was applied to describe deviations at the level of the background metric~\cite{Franchini:2022axs}
and to reconstruct modifications of the potential in the case where a deviation is detected~\cite{Volkel:2022aca}.

In the large-$k$ limit, the amplitude of the coefficients $d_{(k)}^{ij}$ in the decoupled case is proportional to $k^{-1 - 2 r_+ \omega_I^0}$, where $\omega_I^0$ is the imaginary part of the unperturbed QNM~\cite{Cardoso:2019mqo,Hirano:2024fgp}.
Since $- r_+ \omega_I^0$ is large and positive for higher overtones, $k^{-1 - 2 r_+ \omega_I^0}$ also becomes large as $k$ increases.
This implies that the deviation from the GR spectra in the large-$k$ limit is particularly significant for higher overtones.
Since large-$k$ modifications of the effective potential are largest close to the horizon, at $r_* \simeq -r_+ \ln(k/e)$,
this is yet another illustration of the spectral QNM instability discussed in Section~\ref{sec:spectral_environmental}: overtones are easily destabilized by near-horizon modifications of the effective potential~\cite{Konoplya:2022pbc, Hirano:2024fgp}.

The parameterized QNM framework has been extended to encompass modifications of the Teukolsky equation, which describes curvature perturbations of the Kerr BH~\cite{Cano:2024jkd}.
There are several nontrivial differences with the nonrotating case, which are only briefly summarized below.
First, as in GR, the angular separation constant becomes $\omega$-dependent. 
The linear QNM shifts and separation constants are given by %
\begin{align}\label{eq:dOmega_dB}
    \omega_{n\ell m} & \simeq \omega_{n\ell m}^0 + \sum_{k} d_{\omega, n\ell m}^{(k)} \alpha^{(k)} \,, \\
    B_{\ell m}(a\omega) & \simeq B_{\ell m}^0(a\omega) + \sum_{k} d_{B, \ell m}^{(k)} \alpha^{(k)} \,,
\end{align}
where the $\alpha^{(k)}$ describe deviations from the generalized Teukolsky equation. 
In Fig.~\ref{pqnms_kerr_coeffs} we plot the thresholds at which the linear QNM shifts differ by more than $1\,\%$ from fully nonlinear results for different $\alpha^{(k)}$'s, focusing on the fundamental mode.
The regime in which the beyond-Teukolsky framework is valid clearly shrinks as the BH spin increases, underlining the limitations of the nonrotating parameterized QNM framework to model the typical BBH ringdown signals observed in GWs (which have $\chi \simeq 0.7)$.
The beyond-Teukolsky framework has been applied to specific higher-derivative gravity theories in~\cite{Cano:2024ezp}. 

The coefficients of the parameterized QNM framework beyond Schwarzschild/Kerr can be found in a public \texttt{github} repository~\cite{sebastian_volkel_2024_14001739}, presented in~\cite{Cano:2024jkd} (see Appendix~\ref{sec:public_codes}).

\begin{figure}
\centering
\includegraphics[width=1.0\linewidth]{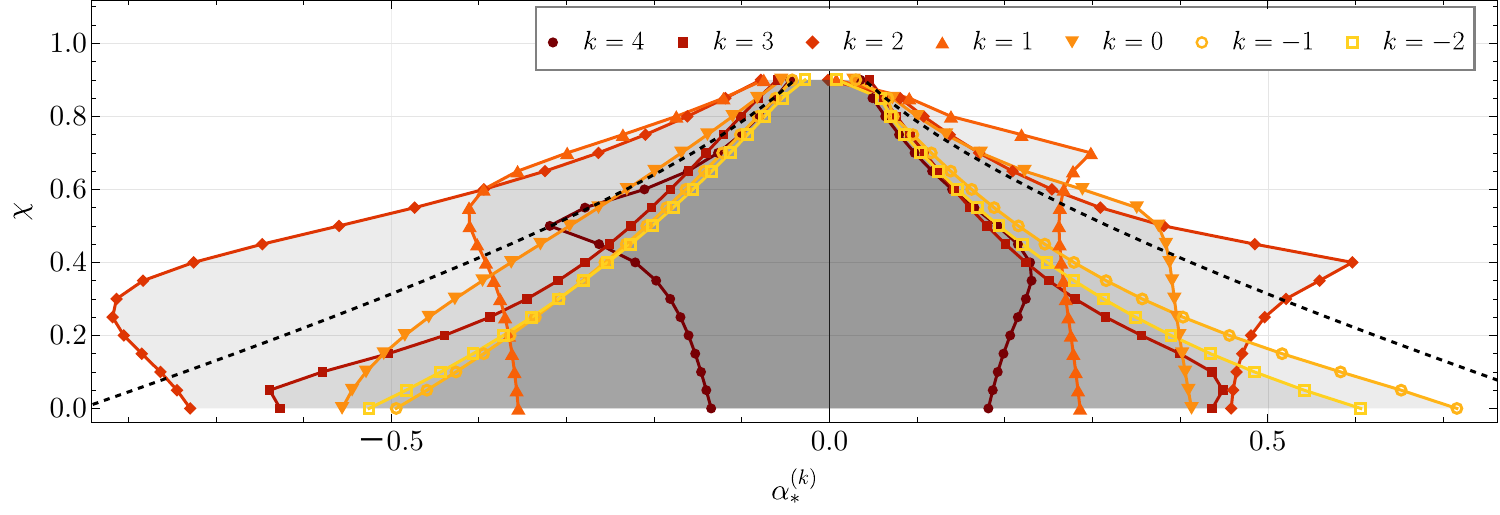}
\caption{
Here we show threshold values $\alpha^{(k)}_*$ of the beyond-Teukolsky %
framework against the spin for different values of $k$ for the $n=0$, $\ell=m=2$ mode. 
The threshold describes the value at which the linear approximation and accurate result differ by $1\%$. 
An estimate for this error is provided in~\cite{Cano:2024jkd} and here shown as black dashed lines. 
Figure adapted from~\cite{Cano:2024jkd}.
}
\label{pqnms_kerr_coeffs}
\end{figure}

\subsubsection{Parameterized black hole metrics with test fields and the eikonal approximation}\label{sec_param_metrics_eikonal}

In the following, we discuss two closely related approaches for exploring QNMs beyond GR. 
We start by studying test fields on a parameterized BH background, and then we outline how the eikonal limit can be used in this context. 

While the parameterized QNM framework from Section~\ref{sec_param_pot} connects QNM deviations with modifications of the effective potential, it cannot directly be used to infer the background metric unless the equations of motions beyond GR are being provided (e.g., in a theory-specific application).
However, deviations from the GR QNMs in terms of the background spacetime deviations can be studied under certain assumptions -- for instance considering test-field perturbations (as discussed below), or assuming the validity of the eikonal limit (discussed in Section~\ref{subsec:LR}).
Since perturbations of test fields, e.g. the massless Klein-Gordon equation on the background spacetime, share qualitatively similar properties with gravitational perturbations in GR, one might still identify certain smoking gun features of new physics. 
A comprehensive overview of the literature on parameterized BH metrics is beyond the scope of this review~\cite{Vigeland:2011ji,Johannsen:2011dh,Johannsen:2013szh,Rezzolla:2014mua,Konoplya:2016jvv,Papadopoulos:2018nvd,Cardoso:2014rha,Yagi:2023eap}; here we discuss QNM studies based on some of the most commonly used parameterizations.

The simplest cases are those where the background metric is static and spherically symmetric. 
Here, the master equation for a massless scalar field is simply given by 
\begin{align}
	\frac{\text{d}^2 \Phi}{\text{d}r_*^2} + \left[ \omega^2 - V_0(r) \right]\Phi = 0\,,
\end{align}
where
\begin{align}
    V_0(r) = -g_{tt}(r) \left[\frac{\ell(\ell+1)}{r^2} + \frac{1}{2 r g_{tt}(r)} \frac{\text{d}}{\text{d}r} \frac{g_{tt}(r)}{g_{rr}(r)} \right] 
\end{align}
and $\text{d} r = \sqrt{-g_{rr}(r)/g_{tt}(r)} \text{d} r_* $. 
Scalar~\cite{Rezzolla:2014mua,Volkel:2019muj} and axial gravitational perturbations (assuming the GR background equations)~\cite{Volkel:2020daa,Cardoso:2024mrw,
Siqueira:2025lww} have been studied for some of these parameterizations.
It is also possible to construct scalar-tensor theories of gravity  such that these static, spherically symmetric parameterized metrics can be ``reverse engineered'' to be exact solutions of the theory, so that one can consistently compute the QNMs in the full theory~\cite{Suvorov:2021amy}. 
The QNM overtones are very sensitive to the properties of the parameterized metrics close to the horizon~\cite{Konoplya:2022pbc}. 

The problem becomes significantly more complicated for rotating BHs because even test fields do not, in general, yield separable perturbation equations. 
Some progress is possible for special classes of axisymmetric BHs for which separability holds~\cite{Konoplya:2018arm}. 
In particular, scalar QNMs of rotating 
spacetimes that describe deviations from Kerr BHs through a single parameter have been analyzed using Leaver's method~\cite{Siqueira:2022tbc}. 
There has also been some progress on computing the QNMs of nonseparable scalar perturbations in non-Kerr backgrounds, with analytical or numerical tools~\cite{Cano:2020cao,Ghosh:2023etd,Khoo:2025qjc}.

Next we outline how the eikonal approximation, reviewed in Section~\ref{subsec:LR}, has been used in the present context. The key aspect of the ``eikonal limit'' (the limit of large angular numbers $\ell$) 
is that the QNMs of perturbations induced by test fields can be directly related to properties of the unstable photon orbit at the light ring. This connection implies that, in GR, the QNM frequencies can be expressed solely in terms of the properties of the background metric. This is a very convenient simplification  that opens a variety of theory-agnostic tests based on beyond-Kerr spacetimes. 

The correspondence was first established by connecting the QNM frequency to the peak of the effective potential and to circular photon orbits~\cite{Press:1971wr,1972ApJ...172L..95G}. 
It has then been shown that the orbital period $\Omega$ and Lyapunov exponent $\lambda$ can be used to approximate both the real and imaginary part of the fundamental QNM~\cite{Ferrari:1984ozr,Ferrari:1984zz,Mashhoon:1985cya,Cardoso:2008bp}. This is also in agreement with a WKB analysis~\cite{Schutz:1985km,Cardoso:2008bp}, yielding
\begin{align}
\omega  = \Omega \ell - \mathrm{i} \left(n+\frac{1}{2} \right) |\lambda|\,.
\end{align}
Further GR-related studies can be found in~\cite{Cardoso:2008bp,Dolan:2010wr,Yang:2012he,Yang:2012pj,Yang:2013uba,Li:2021zct}.

In a series of works~\cite{Cardoso:2008bp,Glampedakis:2017dvb,Glampedakis:2019dqh,Silva:2019scu,Chen:2019dip,Bryant:2021xdh}, the eikonal limit was also studied in theories beyond GR. 
In a theory-agnostic approach, one could attempt to use the eikonal limit in modified theories of gravity in order to compute QNMs from the modified background metric, provided there are no additional fields that lead to coupled perturbation equations~\cite{Glampedakis:2017dvb}. 
However, the correspondence does not always hold~\cite{Khanna:2016yow,Konoplya:2017wot}, in particular when considering theories with additional scalar or vector field that lead to coupled perturbation equations.
Moreover, a practical limitation of the method is that small-$\ell$ multipoles are dominant in GW data, whereas the eikonal approximation (while still accurate at the percent level even for $\ell=2$) is more accurate for large $\ell$. 
Despite complications due to its low accuracy for the dominant quadrupolar mode $\ell=2$, the eikonal approximation has been applied to obtain bounds for different types of modifications to the Kerr spacetime using both current (observational) and future (simulated) GW data~\cite{Carson:2020iik,Uchikata:2020wsp,Dey:2022pmv,Ahmed:2024ykc}.

\subsubsection{Effective field theory of QNMs}\label{sec_eft}
Yet another possibility for parameterizing the QNMs of a gravity theory is to do so directly at the level of the action. 
In the Schwarzschild case this can be done by listing all the possible covariant terms at first order in perturbations, including possible contributions from scalar and vector fields~\cite{Tattersall:2017erk}, following a covariant version of the formalism developed for cosmological perturbations in~\cite{Lagos:2016wyv,Lagos:2016gep}.
A more general result is the EFT of QNMs of any spherically symmetric metric plus a scalar field coupled to gravity~\cite{Franciolini:2018uyq}.
By assuming a spin-2, unitary, Lorentz invariant theory of gravity coupled to a spin-0 field, perturbations of a generic nonrotating background metric of the form
\begin{equation}
	\mathrm{d} s^2 = -a(r)^2 \mathrm{d} t^2 + \frac{\mathrm{d} r^2}{b(r)^2} + c(r)^2 \left( \mathrm{d}\theta^2 + \sin^2\!\theta \, \mathrm{d}\phi^2 \right)
\end{equation}
can be studied for a generic background scalar field $\Phi(r)$. 
By choosing the unitary gauge, the scalar field perturbations can be set to vanish (i.e., $\delta\Phi=0$), and the most general action for linear perturbations 
at leading order in derivative terms for both axial and polar perturbations reads~\cite{Franciolini:2018uyq}
\begin{align}
	S & = \int \mathrm{d}^4x \sqrt{-g} \left[ R - \Lambda(r) - f(r) g^{rr} + M_2(r) \left(\delta g^{rr} \right)^2\right] \,.
\end{align}
The only free functions of this action are $\Lambda(r)$, $f(r)$ and $M_2(r)$.
However, a condition from the constraint equation can be imposed to relate $\Lambda(r)$ and $f(r)$ with the background metric. 
These two functions are the only ones responsible for the perturbations in the odd sector, meaning that the modified Regge-Wheeler equation only depends on the background metric, and it reads
\begin{equation}
	\frac{\mathrm{d}^2\Psi}{\mathrm{d}r_*^2} + \left[\omega^2 + a^2 b^2 \left( \frac{c''}{c} - 2\frac{c'^2}{c^2} + \frac{a'c'}{a c}  + \frac{b'c'}{bc} - \frac{\ell^2 + \ell -2}{b^2c^2} \right) \right] \Psi = 0 \,,
\end{equation}
where the tortoise coordinate is $\mathrm{d}r_* = \mathrm{d} r/ab$. 
The equation for polar perturbations is more involved~\cite{Franciolini:2018uyq}.
This calculation was also extended to the slowly rotating case~\cite{Hui:2021cpm}. 
These EFTs of QNMs have been constructed for the case of a scalar field coupled to gravity, but one could in principle follow the same strategy to add relevant operators for other cases, including different couplings.
At present, the EFT of QNMs for the fully rotating case is still lacking.

\subsection{New physics at the horizon \label{subsec:echoes_theory}}

\vspace{-.1cm}

\noindent \textit{Initial contributors: Afshordi, Maggio, Pani}

\vspace{.2cm}

\subsubsection{Motivation} 
Einstein's theory of GR is in agreement with all of our experimental evidence for BHs so far, but it leaves unsolved some of the deepest theoretical puzzles unveiled by these mysterious objects. In particular, the semi-classical picture of Hawking's BH evaporation is incompatible with quantum unitarity, the microstates underlying the enormous BH entropy are unknown, and the singularities hidden by the BH horizon signal a breakdown of classical physics inside BHs.
These far-reaching problems are dramatic manifestations of the incompleteness of our fundamental laws of nature.

Motivated by Penrose's weak cosmic censorship conjecture~\cite{Penrose:1969pc}, the incompleteness of GR has traditionally been overlooked in observations of astrophysical BHs, as it was assumed to manifest only near singularities, safely concealed behind event horizons. However, a drastic violation of the classical causal structure of event horizons, beyond small corrections~\cite{Mathur:2009hf,Giddings:2017mym}, appears necessary to evaporate away BH information. In particular, the so-called ``firewall paradox''~\cite{Almheiri:2012rt} has been invoked to suggest that it is impossible to reconcile the three (apparently modest) assumptions of locality, causality, and unitarity outside regular event horizons if the entropy $S_{\rm BH}$ of an evaporating BH follows its horizon area $A_{\rm H}=4\pi(r_+^2+a^2)$.
The relation between the two,  according to Bekenstein and Hawking, reads:
\begin{equation}
    S_{\rm BH} = \frac{A_{\rm H}}{4G\hbar}\,. \label{eq:BH_entropy}
\end{equation}
While the paradox only manifests itself about halfway through the BH evaporation (after the so-called Page time $t_{\rm Page} \sim G^2 M^3/\hbar$), it has been argued~\cite{Almheiri:2012rt} that due to the fast scrambling of quantum states within BHs~\cite{Sekino:2008he}, a significant quantum modification of the classical horizon, or ``firewall,'' would happen much earlier, i.e., within a ``scrambling time'':
\begin{equation}
    t_{\rm scr} \simeq \frac{\hbar}{2\pi k_B T_{\rm H}} \ln S_{\rm BH} = \kappa_+^{-1} \ln\left(A_{\rm H} \over 4\hbar G\right) \sim GM \ln \left(G M^2\over \hbar \right) \ll t_{\rm Page}, \label{eq:scrambling}
\end{equation}
where $T_{\rm H}=2(r_+-M)/A_{\rm H}$ is the horizon (or Hawking) temperature, and $\kappa_+$ is the surface gravity of the event horizon. This coincides with the time scale to reflect a signal off a mirror, within a Planck proper distance from the BH horizon~\cite{Hayden:2007cs}.  

An alternative derivation comes from studying a BH in equilibrium with a thermal bath of radiation at the Hawking temperature. Due to the blueshift effect near the BH horizon, the total entropy of the radiation bath (even within a finite box) is infinite. However, if we consider the microstates of radiation near the horizon as contributing to the BH entropy, then we would require $S_{\rm rad} < S_{\rm BH}$, which suggests that the free radiation needs to be cut off by a ``firewall'' near the horizon. The inequality $S_{\rm rad} < S_{\rm BH}$ provides an upper limit on the time for null signals to be reflected off the firewall, which coincides with the scrambling time in Eq.~\eqref{eq:scrambling}~\cite{Oshita:2023tlm}. If the bath temperature deviates from the Hawking temperature, then the dynamical backreaction of the quantum vacuum stress on the geometry would truncate the classical spacetime even farther out, resulting in shorter time scales, as small as half of Eq.~\eqref{eq:scrambling}~\cite{Mathur:2024mvo}.     

The executive summary of the state of affairs is that the assertion of a Bekenstein-Hawking area law for thermodynamic entropy (\ref{eq:BH_entropy}) suggests significant deviations from classical GR dynamics near BH event horizons (e.g.,~\cite{Mathur:2009hf,Hayden:2020vyo}). This challenges the classical BH paradigm, and thus cosmic censorship, which posits that the fate of the gravitational collapse of sufficiently massive and compact stars is a classical BH, surrounded by regular event horizons. %
Therefore, horizonless compact objects can emerge due to quantum modifications at the horizon scale (e.g., gravastars~\cite{Mazur:2004fk} and wormholes~\cite{Morris:1988cz,Visser:1995cc,Damour:2007ap}) or due to a modified theory of gravity describing the interior of compact objects (e.g., fuzzballs~\cite{Mathur:2005zp,Bena:2007kg,Balasubramanian:2008da,Bena:2022rna} and nonlocal stars~\cite{Buoninfante:2018xif,Buoninfante:2019swn}).

\subsubsection{General signatures of horizon-scale structures}

If the remnant of a compact binary coalescence is a horizonless compact object,
the QNM spectrum is predicted to deviate from the BH
one~\cite{Cardoso:2016rao,Cardoso:2016oxy}, also showing a breaking of
isospectrality between axial and polar
modes~\cite{Maggio:2020jml,Lenzi:2021wpc,Lenzi:2021njy,Saketh:2024ojw}. When the
horizonless remnant is ultracompact (i.e., featuring a photon
sphere~\cite{Cardoso:2019rvt}), a characteristic cavity in the effective
potential of the perturbation exists between the light ring and the object. In
the TD, the prompt ringdown signal would be nearly indistinguishable
from the signal from a BH, since it is excited at the light ring. After a travel
time in the cavity of the effective potential, an additional signal could be
emitted in the form of GW
echoes~\cite{Cardoso:2016rao,Cardoso:2016oxy,Price:2017cjr}.  Therefore, GW
echoes in the postmerger phase of a compact binary coalescence are possible
smoking guns for the formation of a horizonless ultracompact
object~\cite{Kokkotas:1999bd,Kokkotas:1995av,Ferrari:2000sr,Cardoso:2016oxy}
(see
also~\cite{Tominaga:1999iy,Andrade:1999mj,Tominaga:2000cs,Andrade:2001hk}). Assuming
that these horizonless compact objects are stabilized outside a proper Planck
length of the would-be classical horizon results in echo time delays that
precisely match the quantum scrambling time $\tau_{\rm echo} = t_{\rm scr}$ of
Eq.~\eqref{eq:scrambling} in generic circumstances~\cite{Saraswat:2019npa}.

Since their conception, GW echoes have been studied in a variety of contexts, including near-horizon quantum structures~\cite{Cardoso:2016rao,Cardoso:2016oxy,Wang:2019rcf}, ultracompact stars~\cite{Pavlidou:2000cs,Ferrari:2000sr,Raposo:2018rjn,Pani:2018flj,Mannarelli:2018pjb}, 
BHs in modified theories of gravity in which the graviton reflects effectively on a hard wall~\cite{Zhang:2017jze,Oshita:2018fqu}, or any other scenarios producing a BH effective potential with multiple peaks.
Echoes have also been studied in the fuzzball scenario.
The cavity reflections of low-energy scalars can be expressed
as a sum over the number of bounces at the throat of these geometries~\cite{Lunin:2001jy,Giusto:2004ip}. 
More recently, GW echoes have been studied for specific BH (or fuzzball) microstates in string theory~\cite{Ikeda:2021uvc,Heidmann:2023ojf,Dima:2024cok}.

The cavity picture of GW echoes we have presented so far is semi-classical, and
limited to linearized perturbations. Therefore, it is fair to ask whether such a
naive description would survive a fully quantum and/or nonlinear treatment, a
point that we shall come back to later in this subsection. However, it is
interesting to note that a fully quantum language for the very same dynamics was
already proposed by Bekenstein back in 1972~\cite{Bekenstein:1972tm}, in terms
of quantized BH horizon area. Roughly speaking, the QNMs of the cavity
(significantly different and much more long-lived than those of a GR BH) have
frequencies $\omega_{n} \simeq 2\pi n \tau^{-1}_{\rm echo}$ + const., where $n$
is an integer.  If we conceive the ringdown of a horizonless compact object as
the spontaneous emission of gravitons, the gravitons appear to be emitted at
discrete energy quanta of $\Delta E = 2\pi \hbar \tau^{-1}_{\rm echo}$. Using
the first law of BH thermodynamics,
$\kappa_+ \Delta A_{\rm H}/(8\pi) = \Delta E$ (and ignoring the BH spin for
now), the BH area/entropy can only vary by a discrete amount:
\begin{equation}
    \Delta A_{\rm H} = \frac{16\pi^2\hbar G}{\kappa_+\tau_{\rm echo}}.
\end{equation}
This suggests that the discretization of the horizon area proposed by Bekenstein~\cite{Bekenstein:1972tm} is directly related to the presence of echoes in the semi-classical cavity description~\cite{Cardoso:2019apo}.
This connection was also extended to spinning BHs~\cite{Agullo:2020hxe}, where the picture is more complex~\cite{Coates:2019bun}.
The connection was initially proposed to apply to the prompt ringdown QNM emission~\cite{Foit:2016uxn}, but this would require a more drastic departure from current understanding~\cite{Coates:2021dlg}.

A similar quantum description of the emission mechanism for GW echoes can be provided. Considering Hawking radiation as quantum {\it spontaneous emission} of gravitons in vacuum, incident GWs onto a quantum BH should lead to quantum {\it stimulated emission}~\cite{1916DPhyG..18..318E}, amplifying the amplitude of Hawking radiation. This stimulation is caused by the classical infalling GWs during ringdown, which in turn can lead to the emergence of GW echoes~\cite{Oshita:2019sat,Wang:2019rcf}.

Given the uncertainty on the nature of quantum physics that replaces the classical description of event horizons, it is reasonable to seek a model-agnostic EFT approach allowing for generic deviations, as a systematic expansion~\cite{Burgess:2018pmm}. An ambiguity in this approach is to decide on the correct small parameter (e.g., distance, frequency, energy, curvature, or something else) that controls the expansion. For example, strictly local EFT approaches to modifications of GR (e.g.,~\cite{Cardoso:2018ptl,Kehagias:2024yyp}) are unlikely to accommodate the nonlocal or highly unusual physics necessary to replace horizons by ultracompact objects, or to address the BH information paradox.

\subsubsection{QNM spectrum beyond Kerr}
If the remnant is not compact enough to feature a photon sphere, cavity effects are absent, and so are echoes in the post-merger phase.
Nevertheless, the ringdown would still show deviations from the ordinary BH QNMs, different QNM
excitation~\cite{Forteza:2022tgq}, or excitation of new nongravitational modes~\cite{Crescimbeni:2024sam} (e.g. extra fluid modes in objects containing matter~\cite{Yunes:2016jcc,Brustein:2023gea}, or scalar excitations in boson star models~\cite{Macedo:2016wgh,Palenzuela:2017kcg,Bezares:2022obu,Siemonsen:2024snb}).
These deviations are mostly degenerate with those predicted by beyond-GR theories (see Sections~\ref{subsec:theory-spec} and \ref{subsec:theory-agnostic}). They may also resemble the post-merger phenomenology of neutron star remnants, featuring long-lived modes, fluid-mode excitations, and in general being qualitatively different from the ordinary BH ringdown.
The QNMs of exotic compact objects have been computed for boson stars~\cite{Yoshida:1994xi,Macedo:2013jja}, gravastars~\cite{Chirenti:2007mk,Pani:2009ss,Chirenti:2016hzd}, BH microstates within the fuzzball paradigm~\cite{Ikeda:2021uvc,Heidmann:2023ojf,Bena:2024hoh,Dima:2024cok}, and generic compact objects described within the membrane paradigm~\cite{Maggio:2020jml,Saketh:2024ojw}.

\subsubsection{Echo morphology}
For a classical BH, the ringdown phase can be understood as the spacetime's response to perturbations occurring near the photon sphere~\cite{Price:2015gia}. Part of the signal produced near the light ring propagates outward, reaching distant observers, while the remainder travels inward towards the horizon, without influencing the observables at large distances.
This occurs because classical BHs are by definition perfect absorbers, with their horizon serving as a one-way, null hypersurface.

However, this property relies on the existence of a classical horizon and is dramatically different for horizonless ultracompact objects or in any model predicting new physics at the horizon scale (see~\cite{Cardoso:2019rvt} for a detailed overview). 
In this case, perturbations can probe the inner boundary of compact objects (such as the regular core of an ultracompact object, or a modified horizon) and reach an observer at large distance, carrying crucial information about putative effects occurring at the horizon scale.

Simply by causality, the effects of the horizon-scale structure cannot show up in the signal before a time of the order of the (round-trip) light crossing time from the photon ring to the boundary. Assuming that the background spacetime is sufficiently close to the Kerr metric, this time scale is no shorter than~\cite{Cardoso:2016rao,Cardoso:2016oxy,Abedi:2016hgu} 
\begin{equation}
\tau_{\rm echo}\sim 2M \left[1+(1-\chi^2)^{-1/2}\right]|\ln\epsilon|\,, \label{tauechospin}
\end{equation}
where $\epsilon\sim d^2/M^2\ll1$, and $d$ is the proper distance between the would-be horizon and any putative effective surface. 
If such distance is Planckian, $d\sim\hbar^{1/2}\approx 10^{-35}{\rm m}$, then $\epsilon \sim S^{-1}_{\rm BH}$, and this time scale is comparable to the scrambling time~\eqref{eq:scrambling}.
Depending on the model, the radiation can take significant time to probe the object's interior, in which case the time scale above should be augmented by the corresponding time delay.
Therefore, if the spacetime geometry is similar to Kerr, the prompt ringdown is essentially indistinguishable from that of an ordinary BH for times shorter than $\tau_{\rm echo}$.

On longer time scales, 
the reflected pulse is semi-trapped between the object and the light ring. With each interaction at the light ring, a portion escapes to distant observers, producing a series of echoes with progressively diminishing amplitudes. These reflections occur at intervals defined by a characteristic echo delay time, which is no shorter than $\tau_{\rm echo}$.
Remarkably, owing to the logarithmic dependence of Eq.~\eqref{tauechospin}, even if the proper distance of the effective surface is Planckian, the delay time is only a factor $|\ln \epsilon| = |\ln(\hbar/M^2)|$ ($\approx 175$ for a solar-mass object) longer than the BH light crossing time $4M$, and therefore potentially observable in the post-merger phase following the prompt ringdown.

Consideration of possible exotic classical interiors for some ultracompact horizonless objects have led to proposals for much longer (geometric) echo time delays~\cite{Brustein:2018ixz,Zimmerman:2023hua,Arrechea:2024nlp}. However, as noted above, the quantum vacuum backreaction~\cite{Mathur:2024mvo} appears to truncate these geometries, and/or their entropy is likely to significantly exceed the Bekenstein-Hawking entropy (\ref{eq:BH_entropy})~\cite{Oshita:2023tlm}. This suggests that, even if such long echo time delays with $\tau_{\rm echo} \gtrsim t_{\rm scr}$ exist, they cannot be understood within a semi-classical geometric description.

The photon sphere barrier acts as a frequency dependent high-pass filter. The characteristic frequencies governing the prompt ringdown are very similar to the BH QNM frequencies, being essentially governed by the excitation of the photon sphere~\cite{Cardoso:2016rao}. The frequencies governing each subsequent GW echo become progressively smaller and, at very late times, the GW signal is dominated by low-frequency QNMs associated with the partially reflecting boundary conditions at the inner boundary~\cite{Mark:2017dnq,Wang:2018gin,Maggio:2019zyv}.

\begin{figure}[t]
\centering
\includegraphics[width=0.475\textwidth]{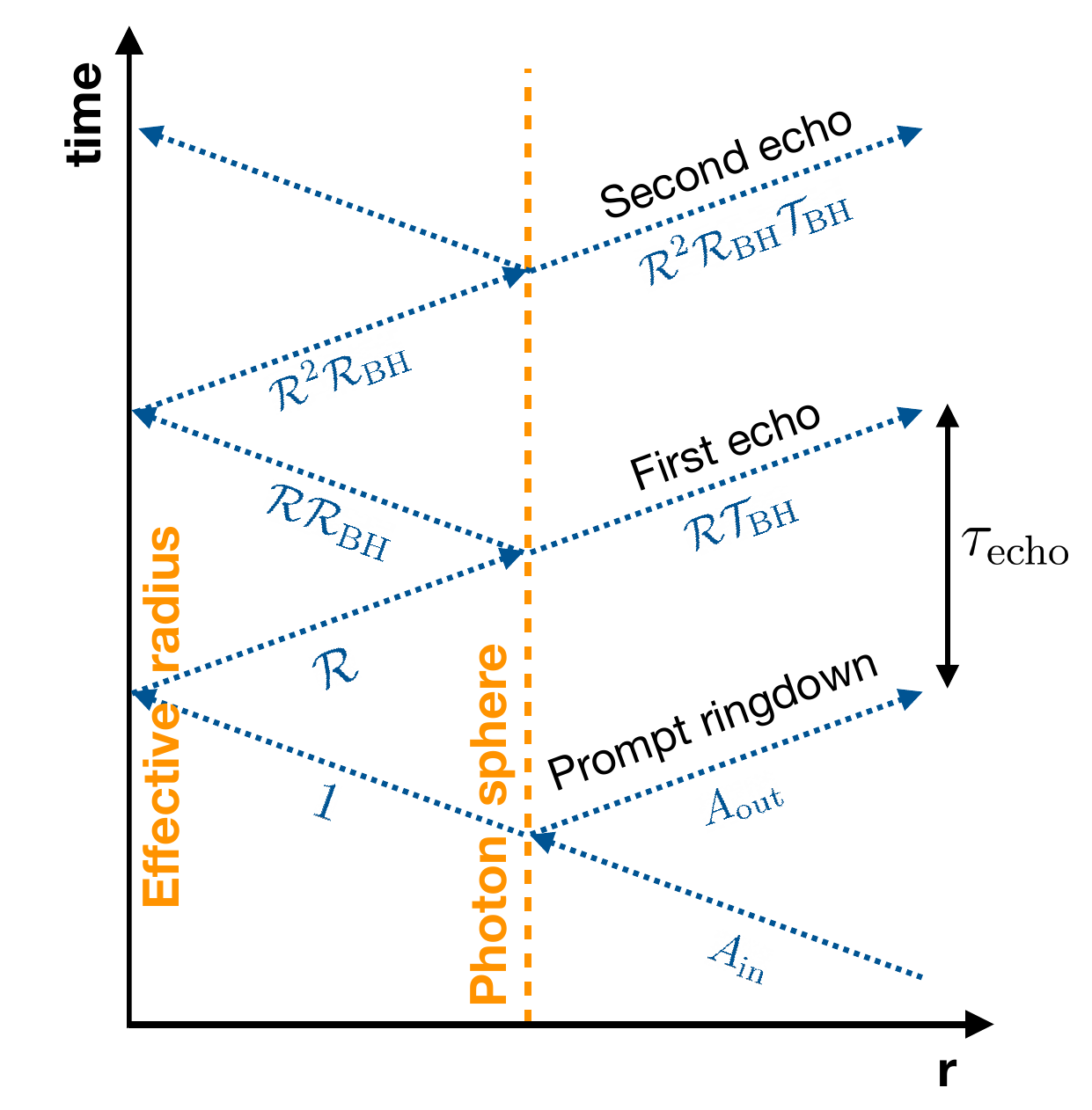}
\includegraphics[width=0.37\textwidth]{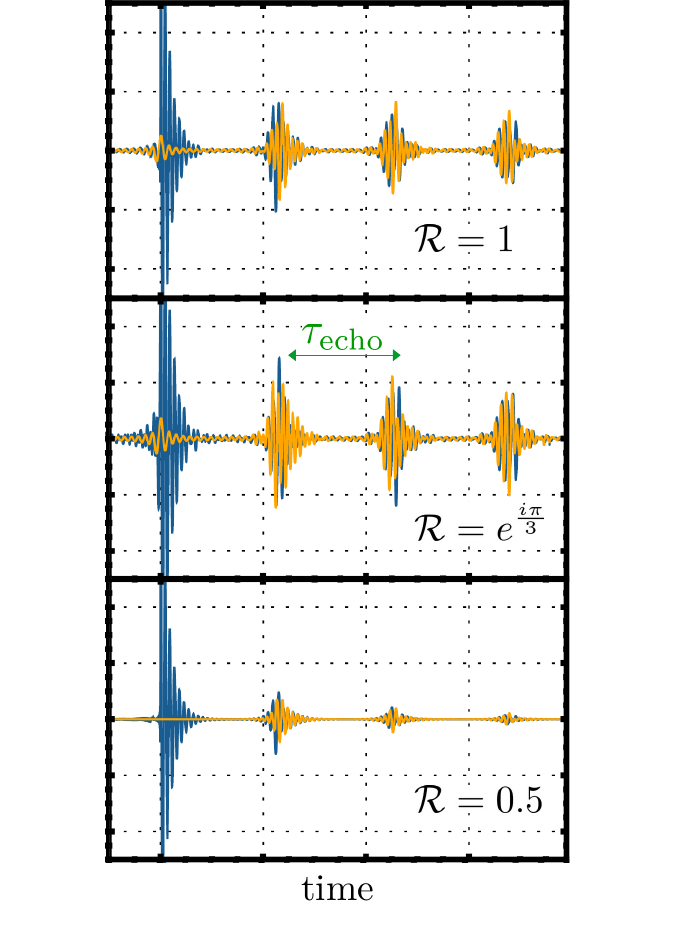}
\caption{Left: Schematic diagram for the wave propagation in a spacetime with new physics at the horizon scale parameterized by an effective reflectivity ${\cal R}$~\cite{Vilenkin:1978uc,Abedi:2016hgu,Cardoso:2019rvt,Maggio:2021ans}.
Here, ${\cal R}_{\rm BH}$ and ${\cal T}_{\rm BH}$ are respectively the reflection and transmission coefficient of the BH effective potential. 
Figure adapted from~\cite{Maggio:2021ans}.
Right: Schematic representation of the two polarizations (blue and yellow waveforms) of a GW in the presence of echoes for different effective reflectivities. The first two upper panels have both $|{\cal R}|=1$, while in the lower panel the reflectivity is halved, and the amplitude of each subsequent echo is suppressed by a factor of $2$.
Figure adapted from~\cite{Maggio:2019zyv}.} 
\label{fig:cavityechoes}
\end{figure}

Echoes appear in the \emph{transient regime} following the universal prompt ringdown, and developing while the remnant reaches stationarity on long time scales.
While such echoes are generic smoking guns of nonconventional horizon-scale physics, their actual detectability crucially depends on the object's effective reflectivity, ${\cal R}$, since the amplitude of each subsequent echo is suppressed by a factor of ${\cal R}$ relative to the previous one (see Fig.~\ref{fig:cavityechoes} for an illustration).
The reflectivity is a model-dependent, complex function of the frequency which can be conveniently defined from the asymptotic behavior of the perturbations near the effective surface~\cite{Maggio:2020jml}
\begin{equation}
  \Psi\sim e^{-i\omega (x-x_0)}+ {\cal R}(\omega) e^{i\omega (x-x_0)} \,,\label{defReflectivity}
\end{equation}
where $\Psi$ is a suitably-defined wave function describing the radial component of the perturbation~\cite{Maggio:2018ivz,Chen:2020htz}, and $x_0$ is the location of the effective surface in the tortoise coordinate. The ordinary boundary condition for a classical BH is ${\cal R}=0$ for any frequency (perfect absorption). On the contrary, an object whose interaction with GWs is negligible would have $|{\cal R}|^2=1$ (perfect ``reflection,'' which can actually correspond to radiation crossing the object without being absorbed). Overall, the frequency-dependent phase factor in ${\cal R}$ accounts for any time delay within the object.
As discussed in the next section, the effective reflectivity modulates the signal and governs the amplitude of subsequent echoes, which depends also on the initial conditions for the perturbation~\cite{LongoMicchi:2020cwm,Annulli:2021ccn,Xin:2021zir,Srivastava:2021uku,Ma:2022xmp}. 
A list of models for which the reflectivity has been explicitly computed is given in Table~\ref{tab:reflectivity}.
Recently, the effective reflectivity has been linked to the shear and bulk viscosity parameters within the membrane paradigm~\cite{Silvestrini:2025lbe}. 

Finally, using WKB theory, the effective perturbation potential of certain types of exotic compact objects can be reconstructed from a knowledge of their QNM spectrum~\cite{Volkel:2017kfj,Volkel:2018hwb}, which is imprinted in the ringdown at very late times.

\begin{table}
\centering
   \begin{tabular}{c|c|c}
  \hline
  \hline
  Model  & ${\cal R}(\omega)$ & References  \\
   \hline
  BH   & $0$ & \\
  Perfect-fluid star  & $e^{i\phi(\omega)}$ &~\cite{Silvestrini:2025lbe} \\
  Gravastar           & $e^{i\phi(\omega)}$ &~\cite{Silvestrini:2025lbe} \\
  Wormhole            & ${\cal R}_{\rm BH}(\omega)e^{-2i\omega x_0}$ &~\cite{Mark:2017dnq,Bueno:2017hyj} \\
  Boltzmann           & $e^\frac{-|\omega-m\Omega_{\rm H}|}{2 T_{\rm H}}e^{i\phi(\omega)}$  &~\cite{Oshita:2019sat,Wang:2019rcf}\\
  Area quantization   & $\Theta(\omega_1-\omega)+\left|\sin\frac{\pi\omega}{\omega_N(\alpha)}\right|^\delta \Theta(\omega-\omega_1)$  &~\cite{Deppe:2024qrk} \\
  Membrane paradigm & function of viscosity parameters &~\cite{Silvestrini:2025lbe}
  \\
  \hline
  \hline
\end{tabular} 
\caption{Effective reflectivities for several models of horizonless compact objects. For a nondissipative system (e.g., perfect-fluid star, gravastar) $|{\cal R}(\omega)|^2=1$ and the phase, $\phi(\omega)$, is model dependent.
For perfect-fluid neutron stars, $\phi(\omega)$ depends on the equation of state and the stellar compactness, while for gravastars it can be computed analytically in terms of the compactness.
The unspecified parameters in the table are defined in the corresponding references.
} \label{tab:reflectivity}
\end{table}

\subsubsection{Echo templates}
\label{subsubsec:echotemplates}

Several phenomenological templates for GW echoes have been developed based on standard GR ringdown templates, augmented with additional parameters representing the echoes emission~\cite{Cardoso:2019rvt}.

A first proposal is a TD template based on a GR inspiral-merger-ringdown (IMR) template $\mathcal{M}(t)$ with five additional parameters~\cite{Abedi:2016hgu},
\begin{equation}
    h(t) = A \sum_{n=0}^{\infty} (-1)^{n+1} \gamma^n \mathcal{M}(t+t_{\rm{merger}} - t_{\rm{echo}}-n \Delta t_{\rm{echo}},t_0) \,,
\end{equation}
where $A$ is the overall amplitude of the echo template, 
$\gamma$ is the damping factor of successive echoes, 
$t_{\rm echo}$ is the time of arrival of the first echo, $\tau_{\rm echo}$ is the time-interval in between 
successive echoes, $(-1)^{n+1}$ is the term due to the phase inversion in each reflection, $\mathcal{M}(t,t_0)=\Theta(t,t_0)\mathcal{M}(t)$, and $t_0$ quantifies which part of the GR 
merger template is truncated to produce the subsequent echoes. The smooth cut-off function is defined as
\begin{equation}\label{cutoff}
    \Theta(t,t_0) = \frac{1}{2} \left\{ 1 + \tanh \left[ \frac{1}{2} \omega(t) \left(t-t_{\rm merger}-t_0\right)\right] \right\} \,,
\end{equation}
where $\omega(t)$ is the frequency of the model as a function of time, and $t_{\rm merger}$ is the time at which the GR template peaks. 
This template (publicly available at~\cite{echowfm}; see also
Appendix~\ref{sec:public_codes}), was used to search for GW echoes using
Bayesian model selection~\cite{Lo:2018sep}, and to analyze the events in the
second Gravitational Wave Transient Catalog (GWTC) of the LVK Collaboration~\cite{LIGOScientific:2020tif}.
Phenomenological TD templates for the echoes in terms of superposition of individual modes were also studied~\cite{Maselli:2017tfq}, together with their detectability via a Fisher-information matrix approach.

A different FD template is based on a standard GR ringdown signal with two additional parameters, related to the object's compactness and reflectivity~\cite{Maggio:2019zyv}. At infinity, the perturbation is a purely outgoing wave
\begin{equation}
    \Psi \sim  \tilde{Z}_{\rm{HCO}}^\infty(\omega) \ e^{i \omega r_*} \,, \qquad r_* \to \infty \,,
\end{equation}
where the response of a horizonless compact object at infinity, $\tilde{Z}_{\rm{HCO}}^\infty(\omega)$, can be written in terms of the BH response at infinity, $\tilde{Z}_{\rm{BH}}^\infty(\omega)$, and the BH reflection coefficient, $\mathcal{R}_{\rm{BH}}(\omega)$, through a transfer function~\cite{Mark:2017dnq,Testa:2018bzd}
\begin{equation}
        \tilde{Z}_{\rm{ECO}}^\infty(\omega) = \tilde{Z}_{\rm{BH}}^\infty(\omega) \left[ 1+\frac{\mathcal{R}_{\rm{BH}}(\omega) \mathcal{R}(\omega) e^{-2 i \tilde{\omega} r_*^0}}{1-\mathcal{R}_{\rm{BH}}(\omega) \mathcal{R}(\omega)e^{-2 i \tilde{\omega} r_*^0} }\right] \,.
        \label{template}
\end{equation}
The BH response at infinity for the fundamental $\ell=m=2$ %
QNM is the Fourier transform of 
\begin{equation}
        Z_{\rm{BH}}^\infty(t) = \theta(t-t_0) \left[ \mathcal{A}_+ \cos(\omega_R t + \phi_+) + i \mathcal{A}_\times \sin(\omega_R t + \phi_\times)\right] e^{-t/\tau} \,,
\end{equation}
where $t_0$ is the starting time of the ringdown, $\mathcal{A}_{+,\times}\in \Re$ and $\phi_{+,\times} \in \Re$ are the amplitude and phase of the ringdown polarizations $h_{+,\times}$, and $\omega_R$ and $\omega_I=-1/\tau$ are the real and the imaginary part of the QNM, respectively. The BH reflection coefficient is defined for waves coming from the left of the photon sphere potential barrier as
\begin{eqnarray}
        \Psi &\sim& e^{i \tilde{\omega} r_*} + \mathcal{R}_{\rm{BH}} e^{-i \tilde{\omega} r_*}\,, \qquad r_* \to -\infty \,, \\
        \Psi &\sim& \mathcal{T}_{\rm{BH}} e^{i \omega r_*} \,, \qquad \qquad \qquad r_* \to +\infty \,,
\end{eqnarray}
where $\tilde{\omega}=\omega-m\Omega_H$ and $\Omega_H=a/(2Mr_+)$ is the angular velocity at the BH horizon. The template depends on the BH ringdown parameters ($M, \chi, \mathcal{A}_{+,\times}, \phi_{+, \times}, t_0$) and on the parameters of the horizonless compact object, i.e., the reflectivity $\mathcal{R}(\omega)$ and the location of the object's radius  $r_0$.
This template was used to assess the detectability of GW echoes with XG detectors with a Fisher matrix analysis~\cite{Branchesi:2023mws}.

A FD template with three search parameters was also proposed~\cite{Nakano:2017fvh} (publicly available at~\cite{echowfm}; see also Appendix~\ref{sec:public_codes}):
\begin{equation}
        \tilde{h}(\omega) = \sqrt{1-\mathcal{R}_{\rm{BH}}^2(\omega)} \ \tilde{h}_0(\omega) \sum_{n=1}^{N_{\rm echo}} \mathcal{R}_{\rm{BH}}(\omega)^{n-1} e^{-i (\omega \tau_{\rm echo}+\phi) (n-1)} \,,
\end{equation}
where $\tilde h_0(\omega)$ is obtained by cutting off the inspiral part of the best fit IMR waveform with a smooth function as in Eq.~\eqref{cutoff} with $t_0 = -0.1 \tau_{\rm echo}$, and $\phi$ is a phase shift due to reflections between the surface and the potential barrier.
The search parameters of the template are therefore $(M, \chi, \phi)$, where $\tau_{\rm echo}$ is determined by $(M, \chi)$. This template was used to search for GW echoes from BBH events in the third observing run of the LVK Collaboration~\cite{Uchikata:2023zcu}.
For additional details on the data analysis methodology and results, see Section~\ref{subsec:echoes_DS}.

\subsubsection{Nonlinear effects and echoes in numerical relativity}

In closing, let us address the question of whether our understanding of GW echoes (or any signature of new horizon physics) may survive beyond linear perturbation theory. 
Energy loss due to emitted echoes inevitably changes the background spacetime at third order in perturbation theory, decreasing $\tau_{\rm echo}$ for subsequent echoes~\cite{Vellucci:2022hpl,Dailey:2023mvn}. More drastically, one may imagine that the expansion of an apparent horizon due to the backreaction of an infalling wave packet may trap it before any potential reflection is possible~\cite{Guo:2022umn} (but see~\cite{Dailey:2023mvn} for a counterpoint).     

Arguably, the holy grail for modeling any nonlinear novel horizon physics in the ringdown phase is its implementation into NR codes, which currently provide state-of-the-art full gravitational waveforms from compact binary mergers. This is a hard problem, especially given the uncertain nature of the quantum gravity physics under consideration. Studies of the post-merger signal from coalescing boson stars in the large mass-ratio regime~\cite{Siemonsen:2024snb} showcase the subsequent emission of bursts at high frequencies, sourced by perturbations of the remnant's internal degrees of freedom.
However, in this case the maximum compactness is far from the values discussed above, which makes a systematic separation of ringdown and echoes impossible. 
Another approach~\cite{Ma:2022xmp} stitches a numerical GR simulated volume into an outer region that obeys linearized perturbations, with a dissipative dispersion relation that models Boltzmann echoes~\cite{Oshita:2019sat,Wang:2019rcf}. 
Recently, a numerical implementation of novel near-horizon physics has been carried out in the form of a numerical solver for the GR initial-boundary value problem~\cite{Dailey:2023mvn,Dailey:2024kjg}. While these studies have provided examples of fully nonlinear evolution of GR within an effective numerical ``membrane paradigm,'' the correct scheme to evolve this membrane depends on the new physics responsible for its formation and thus remains a challenging open problem, both theoretically and numerically. 

\subsubsection{Other proposals for new physics at the horizon scale}

In addition to predicting deviations in the standard BH QNMs, new modes in the ringdown, or echoes in the post-merger phase, specific models can introduce other subtle effects.

For example, while models of horizonless ultracompact objects are expected to have the same ringdown phenomenology at least qualitatively, peculiar effects departing from this picture might arise.
The ringdown of BH microstates emerging in the fuzzball scenario is much richer and more complex than in other models due to the absence of spatial isometries~\cite{Ikeda:2021uvc}. This introduces mode mixing and effectively quenches the echoes, even in the absence of effective absorption.

In the frozen-star model~\cite{Brustein:2018web,Brustein:2019bou}, the pressure is maximally negative in the radial direction and vanishes in the tangential directions. The interior is not described by a classical geometry but, at the microscopic level, is described by a fluid of highly excited, interacting, closed, fundamental strings. Due to the peculiar equation of state, perturbations in the interior vanish identically at the classical level, but display low-frequency, long-lived modes whose lifetime depends on the string length, which is parametrically longer than the Planck length~\cite{Brustein:2017koc}.

Finally, if massless modes propagate at different speeds (as predicted, for instance, in Lorentz-violating theories), each would possess its own distinct horizon. One mode could potentially probe the interior of the other's horizon, leading to energy extraction from the BH and a distinctive late-time relaxation behavior~\cite{Cardoso:2024qie}.

Compact objects with an ergoregion but without a horizon might be affected by an ergoregion instability when spinning sufficiently fast~\cite{Penrose:1969pc,Brito:2015oca}.
The instability time scale is much longer than the decay time of the BH ringdown and the GW echo delay time~\cite{Maggio:2018ivz}.
The ergoregion instability was studied in uniform-density stars~\cite{Kokkotas:2002sf}, gravastars~\cite{Chirenti:2008pf}, and boson stars~\cite{Cardoso:2007az}.
The only way to prevent the instability is by absorbing the negative-energy states in the ergoregion with dissipation mechanisms~\cite{Maggio:2018ivz}.

\clearpage
\section{Quasinormal mode amplitudes}
\label{sec:amplitudes}

\noindent
{\em One thought spectra are marvelous, but it is not possible to make progress there. Just as if you have the wings of a butterfly, then certainly it is very regular with the colors and so on, but nobody thought one could get the basis of biology from the coloring of the wing of a butterfly. So that was the way to look at it.}

\vspace{.2cm}

\noindent
\begin{flushright}
Niels Bohr, Interview by T.S.~Kuhn, L.~Rosenfeld, E.~Rudinger, and A.~Petersen,\\October 31, 1962, AHQP.
\end{flushright}

\vspace{.2cm}

After decades of work, we now have a deep understanding of the ornate structure of BH QNM spectra in GR and beyond GR. However, to make progress in the experimental implementation of BH spectroscopy we need a similar understanding of QNM {\it excitation}; more specifically, we must be able to predict the amplitudes of the dominant modes resulting from a binary merger, and their dynamical excitation in the full nonlinear theory. The mode amplitudes are usually computed by solving the radial perturbation equation using Green's functions techniques (Section~\ref{subsec:greensfunc}), and then computing the so-called ``excitation factors'' and ``excitation coefficients'' (Section~\ref{subsec:Kerr_amplitudes}).
These basic techniques have recently been extended to compute the QNM amplitudes at quadratic order, both in Schwarzschild (Section~\ref{sec:nonlinSch}) and in Kerr (Section~\ref{sec:nonlinKerr}). The analytical results obtained in this way are complementary to more empirical approaches, where the amplitudes of linear and nonlinear QNMs are extracted by direct fits of TD evolutions of the perturbation equations for single perturbed BHs, or from NR simulations of BBH mergers. These efforts are reviewed in Section~\ref{sec:nonlin_num_expe}.

\subsection{Green's functions}
\label{subsec:greensfunc}

\vspace{-.1cm}

\noindent \textit{Initial contributor: Lagos}

\vspace{.2cm}

Standard spectroscopic tests of gravity are concerned with testing the exponential temporal evolution of the signal by measuring the QNM frequencies of different angular harmonics. However, the complete ringdown solution is characterized by additional nontrivial temporal dependence, as well as a spatial evolution. Both of these bring information about the causal structure of the signal, when QNMs are expected to dominate the emission, and the process setting the initial conditions of ringdown relaxation (e.g., the merger of a BBH system).

At linear order in perturbation theory, the ringdown signal satisfies the RWZ equations~\eqref{eq:RWZ master} for Schwarzschild BHs and the Teukolsky equation~\eqref{eq:radialR:Diff_Eqn} for Kerr BHs. These are linear differential equations that can be formally solved with the Green's function approach.

The properties of the Green's function of ringdown signals have been thoroughly analyzed~\cite{Leaver:1986gd, Nollert:1992ifk, Andersson:1996cm, Nollert:1998ys, Szpak:2004sf, Dolan:2009nk, Yang:2013shb,Casals:2016soq, Hui:2019aox, Lagos:2022otp, Casals:2024ynr}.
Since the equations of interest have a separable angular dependence in the FD, let us consider the following representative Schr\"odinger-like 1D equation in space:
\begin{equation}\label{Radial_eq}
    \left[ \partial_{r_*}^2 +\omega^2- V_2(\omega,r_*) \right] \Psi_{2}(\omega, r_*)=0\,,
\end{equation}
where $r_*$ is the tortoise coordinate, defined by $dr_*/dr=(r^2+a^2)/\Delta$ with $\Delta$ given in Eq.~\eqref{def:delta}, and ranging from $r_* \rightarrow -\infty$ towards the horizon ($r\rightarrow r_+$) and $r_* \rightarrow +\infty$ towards spatial infinity ($r\rightarrow +\infty$).
Here, contrary to the ansatz made in Eq.~\eqref{eq:Teukolsky_separation_form}, a Fourier transform of the solution has been performed
\begin{equation}
\label{Psi_decompose}
\Psi(r_*,t,\theta,\omega) = \sum_{\ell,m} \int_{-\infty}^{+\infty} {d\omega \over 2\pi} e^{-i \omega t}
  \Psi_{2}(\omega,r_*){}_{s}S_{\ell m}(\theta,\phi;a\omega).
\end{equation}
Avoiding the exact ansatz in Eq.~\eqref{eq:Teukolsky_separation_form} is important for the solution to include not only QNM contributions, but to allow for additional time dependence. 

The exact form of the effective potential $V_{2}(\omega,r_*)$ depends on the type of BH and perturbation variable. 
For Schwarzschild BHs, $V_{2}(\omega,r_*)$ matches the effective potentials $V_2^{\pm}$ defined in Eqs.\ (\ref{eq:RW-potential})-(\ref{eq:Zerilli-potential}), which are purely real, and do not depend on $m$ nor $\omega$. 
For Kerr BHs, the original radial Teukolsky equation~\eqref{eq:radialR:Diff_Eqn} in terms of the Boyer-Lindquist coordinates does not take the form of Eq.~\eqref{Radial_eq}. 
Nonetheless, a change of variables for the field will bring Eq.~\eqref{eq:radialR:Diff_Eqn} into the form of Eq.~\eqref{Radial_eq} for the tortoise coordinate~\cite{Detweiler:1976zz, Chandrasekhar:1976zz, Detweiler:1977gy}. 
In this case, the potential $V_2(\omega,r_*)$ of Kerr will generally depend on $m$ and $\omega$, and it can be made real when $\omega$ is real~\cite{Detweiler:1976zz, Detweiler:1977gy}. 
For both types of BHs, Schwarzschild and Kerr, the effective potentials are short range.
Specifically, the Kerr potential decays to zero towards spatial infinity, and to $V_2\rightarrow \omega^2-(\omega-am/(2Mr_+))^2$ towards the horizon $r_+$~\cite{Detweiler:1977gy}.
For Schwarzschild, both potentials decay to zero in both limits.

A different radial field redefinition
transforms the radial Teukolsky equation into the Sasaki-Nakamura equation~\cite{Sasaki:1981sx}, which has the form of Eq.~\eqref{Radial_eq} with an additional first-order derivative on the left-hand side, and a short-ranged potential that reduces to the Regge-Wheeler potential when $a=0$. 
The Green's function for Kerr can also be studied in the Sasaki-Nakamura formalism, but here our starting point will be that fields are chosen such that no first derivatives are present, and hence Eq.~\eqref{Radial_eq} holds.

The radial Green's function, $G_{\omega \ell m}$, satisfies
\begin{equation}
\label{Gdelta}
\left[ \partial_{r_*}^2 +\omega^2 - V_{2}(\omega, r_*) \right] 
G_{\omega \ell m} ( r_* | \bar r_*) = \delta (r_* -
  \bar r_*) \, ,
\end{equation}
and it is related to the Green's function in the TD by a Laplace transform:
\begin{equation}
\label{G_in_t}
G(t,r_*, \theta,\phi| \bar t, \bar r_*, \bar\theta, \bar\phi)   = \sum_{\ell m}\int_{-\infty+i\epsilon}^{
+\infty +i\epsilon
} {d\omega \over 2\pi} e^{-i \omega (t-\bar t)}
  G_{\omega \ell m}(r_* | \bar r_*){}_sS_{\ell m}(\theta,\phi; a\omega){}_sS_{\ell m}^*(\bar \theta,\bar \phi; a\omega).
\end{equation}
The integral is performed slightly above the real axis, such that $G = 0$ if $t-\bar t < 0$ -- i.e., this is a retarded Green's function. 

Once the Green's function is known, one can use it to evolve any initial conditions on the perturbation field $\Psi$ and obtain its general solution. 
Consider the initial conditions
\begin{eqnarray}
\psi_0 (r_*,\theta,\phi) \equiv \Psi (t=0) \,, \quad
\dot\psi_0  (r_*,\theta,\phi) \equiv \partial_t \Psi (t)
  |_{t=0} \,,
\end{eqnarray}
where we set $t=0$ as the initial time. The Green's function can be used to evolve the initial linear perturbation forward in time as follows: 
\begin{align}\label{Phi1_Gral} 
    &\Psi(t,r_*,\theta,\phi)=  \int \frac{d\omega}{2\pi} d\bar{r}_* d\bar{\Omega} 
     e^{-i\omega t}G_{\omega \ell m}(r_*|\bar{r}_*)  \mathcal{I}_0(\bar r_*,\bar \theta, \bar \phi) {}_sS_{\ell m}(\theta,\phi; a\omega){}_sS_{\ell m}^*(\bar \theta,\bar \phi; a\omega), \\
     & \mathcal{I}_0(\bar r_*,\bar \theta, \bar \phi)\equiv i\omega \psi_{0}(\bar{r}_*, \bar \theta,\bar \phi) - \dot{\psi}_{0}(\bar{r}_*, \bar \theta,\bar \phi) .\nonumber
\end{align}

Therefore, the Green's function allows us to understand the time evolution and causality properties of ringdown signals. 
The Green's function solution can be constructed from two homogeneous solutions
$g_{\rm out}(r_*)$ and $g_{\rm in}(r_*)$ satisfying Eq.~\eqref{Radial_eq} with the appropriate asymptotic boundary conditions:
\begin{eqnarray}
&& g_{\rm in} \rightarrow e^{-ik_H r_*} \,\,\,\,\quad\quad\quad \quad\quad\quad \quad \quad {\rm for} \quad r_* \rightarrow
  -\infty, \nonumber \\
&& g_{\rm in} \rightarrow {\cal A}_{\rm in} e^{-i\omega r_*}  + {\cal B}_{\rm in}
  e^{i\omega r_*} \,\,\quad \quad\quad {\rm for} \quad r_* \rightarrow +\infty, \label{g_in}\\
&& g_{\rm out} \rightarrow {\cal A}_{\rm out} e^{i k_H r_*}  + {\cal B}_{\rm out}
  e^{-ik_H r_*} \quad {\rm for} \quad r_* \rightarrow
  -\infty, \nonumber \\
&& g_{\rm out} \rightarrow e^{i\omega r_*}  \quad \quad\quad \quad\quad \quad\quad \quad\,\,\quad {\rm for} \quad r_* \rightarrow
  +\infty \, ,\label{g_out}
\end{eqnarray}
where $k_H= \omega - m a/(2Mr_+)$, we have used the fact that the potential $V$ vanishes towards spatial infinity and asymptotes a constant towards the horizon (which vanishes when $a=0$), 
and we have suppressed the dependence of the coefficients ${\cal A}_{\rm in} , {\cal B}_{\rm in} , {\cal A}_{\rm out}, {\cal B}_{\rm out}$ on $(\omega,m, \ell)$ to avoid clutter. 
Combining this asymptotic radial dependence with the decomposition in Eq.~\eqref{Psi_decompose}, $g_{\rm out}(r_*)$ describes outgoing plane waves toward spatial infinity, while $g_{\rm in}(r_*)$ describes ingoing plane waves toward the horizon.

The radial Green's function $G_{\omega\ell m}$ can then be constructed as:
\begin{eqnarray}
\label{radialG}
G_{\omega\ell m} (r_* | \bar r_*) = {1\over W} g_{\rm out} (r_{*>})g_{\rm in} (r_{*<}),
\end{eqnarray}
where $r_{*>}=\max(r_*,\bar{r}_*)$, $r_{*<}=\min(r_*,\bar{r}_*)$, and $W$ is the Wronskian:
\begin{eqnarray}
W(\omega) \equiv g_{\rm in} (r_*) \partial_{r_*}
  g_{\rm out} (r_*) - g_{\rm out} (r_*) \partial_{r_*} g_ {\rm
  in}(r_*) \, .
\end{eqnarray}
Note that the Wronskian depends on $\omega$, but it is independent of $r_*$, since both $g_{\rm out}(r_*)$  and $g_{\rm in}(r_*)$ satisfy Eq.~\eqref{Radial_eq}. 
In particular, one obtains: 
\begin{equation}\label{Wronskian}
 W = 2i\omega {\cal A}_{\rm
  in} = 2i\omega {\cal A}_{\rm out}.
\end{equation}
Replacing into Eq.~\eqref{Phi1_Gral}, the general ringdown solution is
\begin{align}
  \Psi(r_*,t,\theta,\omega) = &\int \frac{d\omega}{2\pi}  d\bar{\Omega} 
     e^{-i\omega t}  {}_sS_{\ell m}(\theta,\phi; a\omega){}_sS_{\ell m}^*(\bar \theta,\bar \phi; a\omega)  \nonumber\\
     & \times \left[\frac{g_{\rm out}(r_*)}{W}\int_{-\infty}^{r_*}  d\bar{r}_* g_{\rm in}(\bar r_*)\mathcal{I}_0(\bar r_*,\bar \theta, \bar \phi)  + \frac{g_{\rm in}(r_*)}{W}\int^{+\infty}_{r_*}  d\bar{r}_* g_{\rm out}(\bar r_*)\mathcal{I}_0(\bar r_*,\bar \theta, \bar \phi)  \right].
\end{align}
Hence, the overall radial evolution of ringdown signals is fully determined by $g_{\rm out}(r_*)$ and $g_{\rm in}(r_*)$, whereas the temporal evolution is fully determined by the $\omega$ dependence of the Green's function $G_{\omega \ell m}$ and the spheroidal harmonics. The initial conditions determine the value of the constant factors in the ringdown signal.

The exact form of $G_{\omega \ell m}$ depends on the potential $V_2$, and to date an
analytical closed-form expression for $G_{\omega \ell m}$ in the full parameter space $(\omega,r_*)$ is not known, although expressions as infinite expansion series~\cite{Dolan:2009nk,Dolan:2011fh,Casals:2013mpa}, qualitative, and asymptotic features are well known, at least for Schwarzschild BHs~\cite{Leaver:1986gd,Andersson:1996cm,Szpak:2004sf}. 
As discussed below, approximate expansions have also been calculated for Kerr BHs, and they have analogous features. Since the derivations for Kerr are more complicated, from now on we will focus on Schwarzschild BHs and discuss general features of the Green's function.

In particular, the functions $g_{\rm in}$ and $g_{\rm out}$ for Schwarzschild BHs can be expressed as infinite series~\cite{Leaver:1986gd} 
\begin{align}
    g_{\rm in}(r) &= r^{2i\omega}(r-1)^{-i\omega}e^{i\omega(r-2)}\sum_{n=0}^{\infty}a_n(1-1/r)^n, \\
    g_{\rm out}(r) & =(2\omega)^{-i\omega}e^{i\phi_+} (1-1/r)^{-i\omega}\sum_{\ell=-\infty}^{\infty}b_{\ell} \left[  G_{\ell}(\eta,\rho)+iF_{\ell}(\eta,\rho) \right]  ,
\end{align}
in terms of the Schwarzschild radial coordinate $r$, instead of the tortoise coordinate $r_*$. Here, $a_n$ and $b_\ell$ are coefficients satisfying specific recurrence relations, the functions $G_{\ell}(\eta,\rho)$ and $F_{\ell}(\eta,\rho)$ are the Coulomb wave functions with $\eta=-\omega$ and $\rho=\omega r$, and $\phi_+$ is a normalization phase. Using Eq.~(\ref{radialG}), these expressions for $g_{\rm in}$ and $g_{\rm out}$ can be used to obtain an expansion series for the Green's function.

Since GW signals are observed as time series, the temporal/frequency dependence of the Green's function deserves particular attention. 
For Schwarzschild BHs, the spheroidal harmonics do not depend on $\omega$, and thus only the $\omega$ dependence of $G_{\omega\ell m}$ in Eq.~\eqref{radialG} determines the TD Green's function:
\begin{align}
& G(t,r_*, \theta,\phi| \bar t, \bar r_*, \bar\theta, \bar\phi)   = \sum_{\ell m}
  G_{\ell m}(t,r_*|\bar t, \bar r_*){}_sY_{\ell m}(\theta,\phi){}_sY_{\ell m}^*(\bar \theta,\bar \phi), \nonumber \\
    & G_{\ell m}(t,r_*|\bar t, \bar r_*)=\int_{-\infty +i\epsilon}^{+\infty +i\epsilon} {d\omega \over 2\pi} e^{-i \omega (t-\bar t)}
  G_{\omega \ell m}(r_* | \bar r_*).
\end{align}\label{eq:GF_TD}

For a general value of $\omega$, the boundary conditions for $g_{\rm out}$ and $g_{\rm in}$ cannot be satisfied at the same time; therefore these are two independent solutions of the homogeneous equation, and $W\not=0$. However, for $\omega$ values that coincide with
the linear QNM spectrum, $g_{\rm out}$ and $g_{\rm in}$ are given by
the same single solution. In that case, ${\cal A}_{\rm in}={\cal A}_{\rm out}=0$ and thus  $W=0$: see Eq.~\eqref{Wronskian}. As a consequence of the fact that $G\propto 1/W$, $G_{\omega \ell m}$ has poles at the (linear)
QNM frequencies $\omega=\omega_{\ell m n}$. 

The integral on the displaced real axis involved in the computation of the TD Green's function $G_{\ell m}$, Eq.~\eqref{eq:GF_TD}, is obtained by deforming the integration contour in the complex-frequency plane according to the features of $G_{\omega\ell m}$~\cite{Leaver:1986gd}.
The main features are: 
(i) it contains first-order poles at the linear QNM frequencies $\omega=\omega_{\ell m n}$~\cite{Detweiler:1977gy};
(ii) it contains a pole at $\omega=0$, due not to the Wronskian alone, but its combination with $g_{\rm out}$ and $g_{\rm in}$; 
(iii) it contains a branch cut running along the negative imaginary $\omega$ values, connected to the pole at $\omega=0$. 
According to these features, the Green's function in the TD may be obtained using the residue theorem, following a closed contour integral as shown in Fig.~\ref{fig:G_contour}.

\begin{figure}[t]
\centering
\includegraphics[width = 0.60\textwidth]{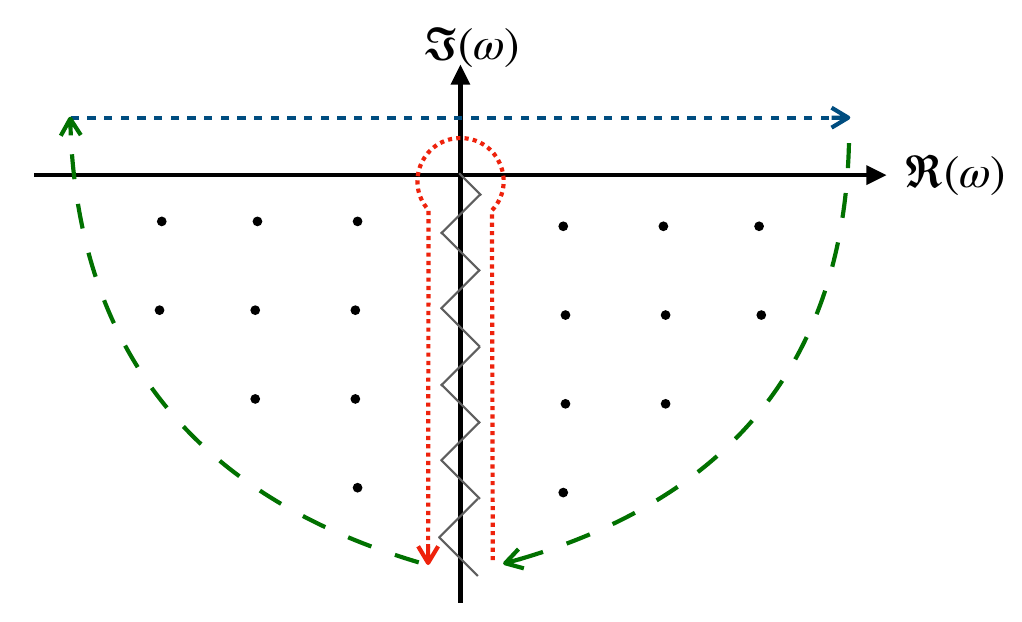}
 \caption{Contour integration in the complex frequency plane for the Green's function. Black dots schematically show the simple poles corresponding to the linear QNM frequencies with negative imaginary frequency. The contour contains a piece above the real axis (blue short-dashed line), a large semi-circle (green long-dashed line) and a piece surrounding the branch cut and pole at $\omega=0$ (red dotted line).}
 \label{fig:G_contour}
\end{figure}

As a result of the frequency integration, the TD Green's function can be expressed in terms of three contributions: 
\begin{equation}
 G_{\ell m}(t,r_*|\bar t, \bar r_*)=G_{Q}+ G_F+ G_B,
\end{equation}
with $G_Q$, $G_F$ and $G_{B}$ known as the QNM, flat, and branch cut contributions, respectively.
\begin{itemize}
    \item[(1)] $G_Q$ comes from a sum over the residues at the linear QNM frequency poles (black dots in Fig.~\ref{fig:G_contour}), and leads to exponentials with frequencies $\omega=\omega_{\ell m n}$, thus producing the usual QNM solutions. 
    This implies that the linear QNM frequencies will be contained in the ringdown signal, unless the initial conditions are fine-tuned~\cite{Chavda:2024awq}.
    \item[(2)] $G_F$ comes from the arcs of the semi-infinite circle of the integration contour (green long-dashed lines in  Fig.~\ref{fig:G_contour}), and its contribution includes information about high-magnitude frequency contributions and asymptotically far signals, that propagate effectively in free space since they are insensitive to the potential.
    For this reason, $G_F$ is expected to be responsible for the prompt response in ringdown solutions, which will mostly depend on the details of the initial conditions.
    \item[(3)] $G_B$ comes from the branch cut contribution (red dotted line in Fig.~\ref{fig:G_contour}) and is interpreted to arise from the long-range decay of the potential, corresponding to low-frequencies. 
    It generally leads to polynomial tails in ringdown solutions, which can dominate the late-time signal, and are understood as the result of waves back-scattering off the potential~\cite{Ching:1994bd} (see Section~\ref{sec:tails}). 
\end{itemize}

While their complete expressions are not known, details on different asymptotic limits for $G_Q$ and $G_B$ can be found in~\cite{Leaver:1986gd}. 
An intuition on their effect can be gained by looking at the limit of large $|r_*|$ and $|\bar r_*|$, using the asymptotic forms of $g_{\rm in }$ and $g_{\rm out}$ in Eqs.~\eqref{g_in}-\eqref{g_out}, leading to
\begin{align}
\label{GF}
& G_F \sim -{1\over 2} \left[ \Theta(t - \bar t - |r_* - \bar r_*|) - 
\Theta(t - \bar t - |r_*| - |\bar r_*|) \right] \, , \\
& G_Q  \sim \sum_n {-i f_{\ell m n}\over W'_{\ell m n}}
e^{-i \omega_{\ell m n}^{(1)} (t - \bar t - |r_*| - |\bar r_*|)} 
\Theta(t - \bar t - |r_*| - |\bar r_*|),\label{GQ}
\end{align}
where $\Theta(x)$ is the step function (unity if $x > 0$, zero
otherwise), and $W'_{\ell m n}$ is the derivative of the Wronskian with respect to the frequency, evaluated at the QNM frequencies $\omega_{\ell m n}$.
The term $f_{\ell m n}$ is a constant factor depending on the QNM frequencies $\omega_{\ell m n}$, with $f_{\ell mn}=1$ when $r_*$ and $\bar r_*$ have opposite signs, $f_{\ell m n}=\mathcal{B}_{\rm in}(\omega_{\ell m n})$ when  $r_*, {\bar r}_*>0$, and $f_{\ell m n}=\mathcal{B}_{\rm out}(\omega_{\ell m n})$ when  $r_*, {\bar r}_*<0$.
The step functions in the above expressions represent
nontrivial causality constraints coming from how the contour
in the $\omega$ integral is closed.

\begin{figure}[t]
\centering
\includegraphics[width = 0.60\textwidth]{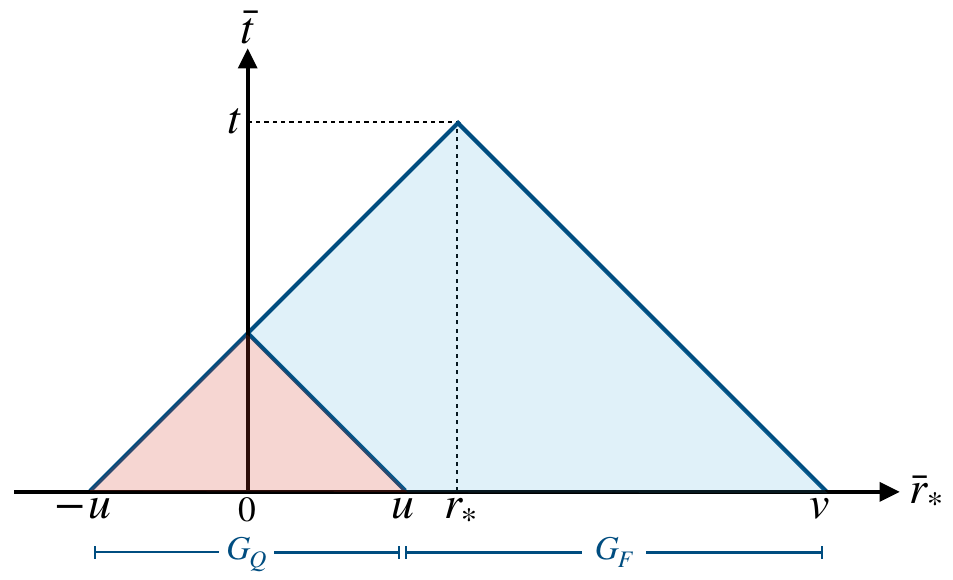}
 \caption{Support of asymptotic $G_{F}$ (shaded blue) and $G_{Q}$ (shaded red) for a given point $(t,r_*)$.  Here, $u=t-r_*$ and $v=t+r_*$, and the potential peak is around $\bar{r}_*\approx 0$. In the general case, the boundaries of these regions are expected to be fuzzy, but this figure schematically illustrates the role of causality constraints. Horizontal blue lines indicate the maximum size of the spatial region causally connected to $(t,r_*)$ through $G_{Q}$ and $G_{F}$. Figure taken from~\cite{Lagos:2022otp}.}
 \label{fig:Gsupport}
\end{figure}

The causality constraints that ringdown solutions have to satisfy in the asymptotic limit considered here are shown in Fig.~\ref{fig:Gsupport}. In this figure $(\bar{t}, \bar{r}_*)$ is interpreted as an event with a nonzero initial condition, and $(t, r_*)$
corresponds to the observer. In particular, the step function for $G_Q$ implies that the QNM piece of the Green's function
does not vanish only if the point $(\bar t , \bar r_*)$ is
causally connected to $(t, r_*)$ via the potential (approximated to be at $r_*=0$ in these asymptotic expressions). This means that QNM solutions will build in time, as a given initial condition gets causally connected to the potential and then has enough time to travel towards the observer.
For initial conditions with compact support, the QNM solutions will eventually reach a stationary regime, when the entire initial condition has been connected to the potential and the observer, and the solution will be well approximated by a superposition of QNM frequencies with constant amplitudes.
This causality constraint will thus lead to a time variation of the linear QNM amplitudes beyond the complex exponential dependence in the initial ringdown regime (e.g., near the merger of two BHs)~\cite{Andersson:1996cm,Lagos:2022otp, Baibhav:2023clw}. 
The time taken to stabilize will depend on the initial condition, as exemplified by calculations with toy models~\cite{Chavda:2024awq}. 
Since these results were obtained in the asymptotic limit of the Green's function, the actual causality conditions at smaller radii are expected to be analogous but with fuzzier boundaries.

In addition, the causality constraints from $G_F$ imply that part of the initial conditions will travel directly towards the observer. Compared to $G_Q$, a signal produced by $G_F$ is thus expected to arrive earlier than those produced by $G_Q$. For this reason, $G_F$ is said to generate a ``prompt response.''
The effect of $G_Q$ and $G_F$ has been studied with toy model initial conditions~\cite{Leaver:1986gd, Andersson:1996cm}. An example is given in Fig.~\ref{fig:Gaussian_Response}.

\begin{figure}[t]
\centering
\includegraphics[width = 0.50\textwidth]{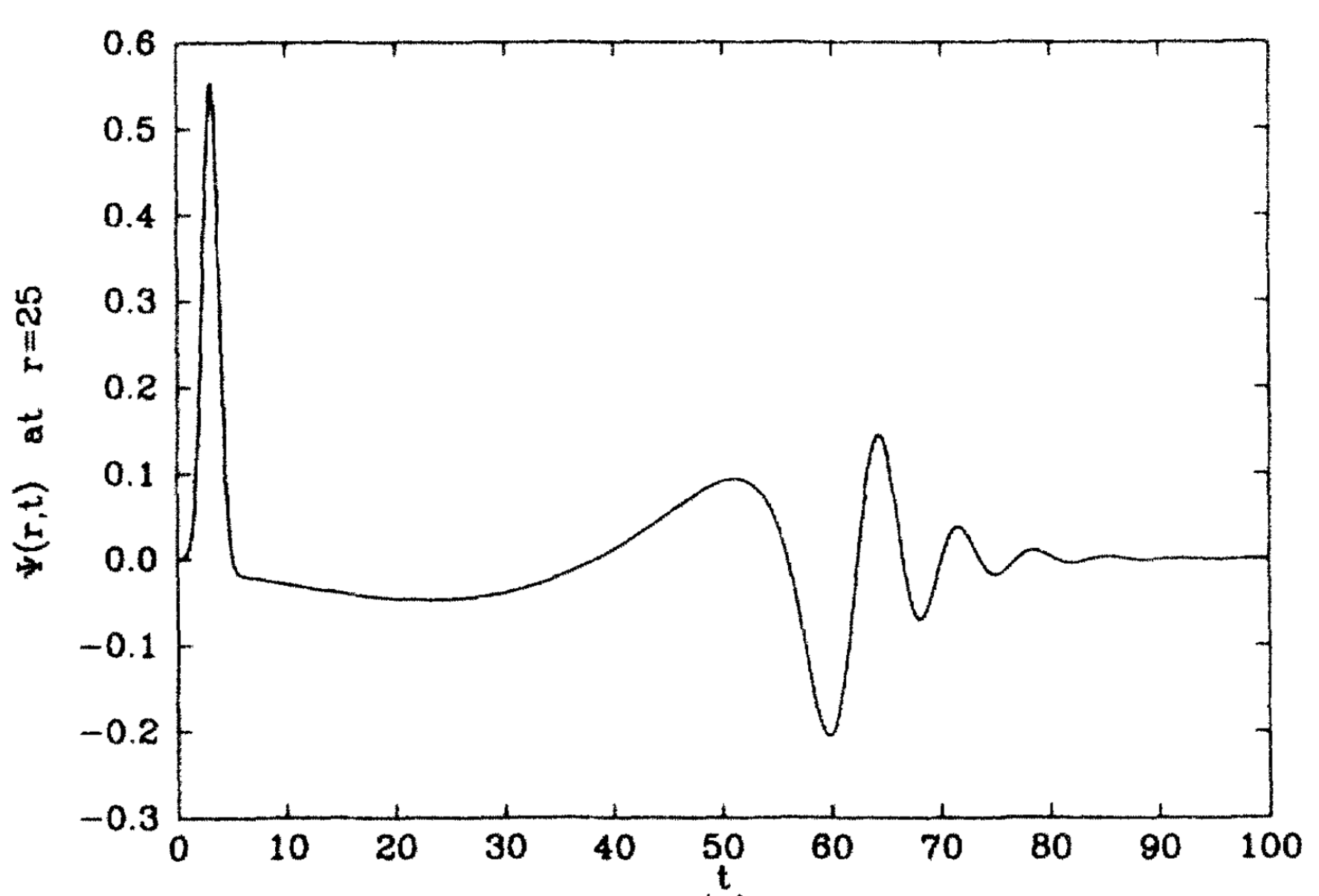}
 \caption{Solution to the Regge-Wheeler equation with $\ell=2$ and spin 1, as a function of time, for an observer at $r=25$ (in units $2M=1$). The initial condition is a Gaussian at $r=28$ traveling toward the horizon. The first peak is the prompt response, and the following oscillations correspond to the QNM solutions. Figure taken from~\cite{Leaver:1986gd}.}
 \label{fig:Gaussian_Response}
\end{figure}

The $G_B$ component does not appear in the asymptotic study of the Green's function, since the potential vanishes in that limit. 
However, a number of studies of its polynomial form have been performed in other limits~\cite{Price:1972pw, Leaver:1986gd, Ching:1994bd, Ching:1995tj, Nakano:2000ne, Casals:2015nja, Casals:2016soq}. 
For instance, at late times such that $r_*, {\bar r_*}\ll (t-{\bar t})$ and for static initial conditions one finds
\begin{equation}
    G_B\sim (r_* \bar{r}_*)^{\ell +1}(t-\bar{t})^{-2\ell -3} \,,
\end{equation}
which describes the propagation of low-frequency waves. 
From this expression, the dominant multipole will have $\ell=2$, with higher multipoles decaying faster.
This Green's function component also leads to polynomial tails in ringdown solutions. 
Since QNM solutions decay exponentially fast, polynomial tails dominate ringdown signals at late times. 
For a detailed discussion of their physical interpretation and impact on signal modeling, see Section~\ref{sec:tails}.

Analogous features for the Green's function are expected for Kerr BHs. 
In particular, as previously mentioned, the Schwarzschild and Kerr cases can both be cast in the form of Eq.~\eqref{Radial_eq}, with the same asymptotic behavior towards the horizon and infinity. 
The main difference between the two cases lies in the fact that the angular separation in Kerr uses spheroidal harmonics, which depend on $\omega$ and may thus introduce additional nontrivial time dependence on the TD Green's function.
As discussed in~\cite{Casals:2016soq}, which analyzes the Kerr Green's function to obtain late-time tail solutions, the $\omega$ dependence of the spheroidal harmonics leads to apparent branch points that are not present for Schwarzschild BHs, yet such angular branch cuts are spurious artifacts of the spheroidal decomposition, which will vanish when performing the infinite sum over $\ell$ to obtain the full Green's function. Analytical series approximations of the Kerr Green's function have been obtained in~\cite{Shibata:1994jx,Tagoshi:1996gh, Mano:1996vt, Mano:1996gn} (see also the review~\cite{Sasaki:2003xr}). The Green's function for a scalar field around a Kerr BH has also been studied in~\cite{Yang:2013shb}. As opposed to the Teukolsky formalism discussed here, a different approach to the Kerr Green's function  useful for reconstructing the full metric perturbation $h_{\mu\nu}$ can be found in~\cite{Casals:2024ynr}.

An important observation is that the Green's function approach can also be used to analyze the behavior of higher-order BH perturbations, iteratively.
In fact, at any perturbative order, perturbations will satisfy the same linear differential equation, with the only difference being the presence of an additional source term fixed by lower order perturbations. For example, second-order ringdown solutions $\Psi^{(2)}_2$ satisfy the equation
\begin{equation}
    \left[ \partial_{r_*}^2 +\omega^2- V_{2}(\omega,r_*) \right] \Psi^{(2)}_2(\omega, r_*)=S^{(2)}[\Psi_2],
\end{equation}
where $S^{(2)}$ is a second-order source, a function of quadratic terms in the linear perturbation $\Psi_2$ and its derivatives. 
While a homogeneous solution to this equation will simply renormalize the linear solution, its particular solution can be obtained with the Green's function in a way analogous to Eq.~\eqref{Phi1_Gral}:
\begin{equation}
    \Psi^{(2)}_2(t,r_*,\theta,\phi)=  \int  d\bar{r}_* d\bar{t} d\bar{\Omega} \;   G(t,r_*,\theta,\phi|\bar{t},\bar{r}_*,\bar{\theta},\bar{\phi})  S^{(2)}(\bar{t}, \bar r_*,\bar \theta, \bar \phi).
\end{equation}
Applications of the Green's function approach to second-order ringdown perturbations for Schwarzschild BHs predict the presence of nonlinear polynomial tails in toy models~\cite{Okuzumi:2008ej} and shed light on the causal formation and evolution of nonlinearities~\cite{Lagos:2022otp}. Analogous results are expected for Kerr BHs.

\subsection{Kerr amplitudes in the time domain and excitation factors} \label{subsec:Kerr_amplitudes}

\vspace{-.1cm}

\noindent \textit{Initial contributors: Oshita}

\vspace{.2cm}

\subsubsection{Kerr perturbations}
Gravitational waveforms in the Kerr background can be computed by solving the Teukolsky equation (see Section~\ref{sec_21}).
A useful feature of the Teukolsky equation is that it can be separated into two ODEs for the angular and radial part of the perturbations, Eqs.~\eqref{eq:swSF_DiffEqn} and~\eqref{eq:radialR:Diff_Eqn}.
The full waveform can be obtained by solving the Teukolsky equation using the Green's function approach, possibly including a matter source term. In this ``Teukolsky formalism,''  an apparent obstacle is that the required integrals can be divergent. 
An appropriate regularization, equivalent to introducing a Green's function that is well-behaved at infinity, can resolve this issue~\cite{Poisson:1996ya}.
Alternatively, the effective potential in the wave equation can be made short-range by transforming the Teukolsky equation using the ``generalized Darboux'' or ``Chandrasekhar'' transformation~\cite{Chandrasekhar:1976zz,1976RSPSA.348...39C,Detweiler:1977gy,Detweiler:1978ge}, which reads
\begin{equation}\label{4_2eq:Chandrasekhar_transformation}
    X_{\ell m \omega} \equiv \alpha (r) R_{\ell m \omega} + \frac{\beta (r)}{\Delta} \frac{d R_{\ell m \omega}}{dr}\,,
\end{equation}
in terms of certain functions $\alpha (r)$ and $\beta (r)$. 
By properly choosing the functions $\alpha$ and $\beta$ in \eqref{4_2eq:Chandrasekhar_transformation}, the radial Teukolsky equation is transformed into another ODE for $X_{\ell m \omega}$ with a short-range potential (see e.g.~\cite{Glampedakis:2017rar} for a detailed review of the generalized Darboux transformation).
With a specific choice of  $\alpha (r)$ and $\beta (r)$, %
the radial perturbation equation has a short-range {\em real} potential, and it is known as the Chandrasekhar-Detweiler equation~\cite{Chandrasekhar:1976zz}.
This equation features a potential with four distinct branches. 
For certain combinations of the Kerr parameter and of the mode frequency, one of the potential branches may become singular, necessitating a switch between branches.
A different choice of $\alpha(r)$ and $\beta(r)$ leads to a second radial perturbation equation with a short-range complex potential, known as the Sasaki-Nakamura equation~\cite{Sasaki:1981sx}, 
which (for perturbations with $s=-2$) has the form
\begin{equation}\label{eq:SasakiNakamura}
\left[ \frac{d^2}{dr_*^2} - {\cal F}(r) \frac{d}{dr_*} - {\cal U}(r)\right] X_{\ell m \omega}(r) = I_{\ell m \omega} (r).
\end{equation}
The explicit expressions of ${\cal F} (r)$, ${\cal U} (r)$ and $I_{\ell m \omega}(r)$ are given in~\cite{Sasaki:1981sx}.
The Sasaki-Nakamura equation is convenient for numerical work due to its short-range complex potential, and it has also been extended to other values of the spin weight~\cite{Hughes:2000pf}.
The equation reduces to the Regge-Wheeler equation in the limit $a \to 0$.
In the Sasaki-Nakamura formalism, the source term $I_{\ell m \omega}$ is also short-range, hence the full waveform can be obtained by performing a convergent integral. 
The calculation of the source term requires a double integral of the Teukolsky source term, with two corresponding integration constants determined so that the inhomogeneous solution of the Teukolsky equation is purely outgoing at infinity~\cite{Sasaki:1981sx, Sasaki:1981}.
The source term of the Sasaki-Nakamura equation for a point particle in geodesic motion on the Kerr background was computed in~\cite{Sasaki:1981sx,Kojima:1984cj,Saijo:1996iz}.  An important property of the source term is that it redshifts to zero as the particle approaches the Kerr event horizon.  Since solutions of the source-free Teukolsky equation are QNMs, this ensures that the waveform for a plunging particle terminates in QNMs which are phase-coherent with the inspiral preceding the plunge. 

Another possibility to study the excitation of Kerr amplitudes is to directly solve the Teukolsky equation~\eqref{eq:Teukolsky_Master} with a point-particle source, without switching to an equation which uses a short-range potential~\cite{Sundararajan:2007jg, Sundararajan:2008zm}.
Key to this method is that the source is approximated as a variant of an impulse response function which preserves integral identities of the Dirac delta function and its derivatives. 
By combining adiabatic backreaction data (see e.g.~\cite{Hughes:2021exa} for an overview) with a prescription to describe how the adiabatic evolution ends as the particle begins plunging into the Kerr BH, one can construct a worldline describing a small body's inspiral and plunge for a wide range of initial conditions. 
This method, initially developed in~\cite{Ori:2000zn}, was extended to a larger class of inspirals and improved in~\cite{Apte:2019txp}.
Given the worldline, the Teukolsky equation source term (e.g.~\cite{Zenginoglu:2011zz,Field:2020rjr}) and the gravitational waveforms can be computed by numerically solving the evolution equation.  
The final cycles of these waveforms consist of QNMs, so these time evolutions can be used to numerically study the relative excitation of the modes as a function of the geometry of the point-particle orbit.

\subsubsection{Kerr waveforms in the time domain} \label{subsec:kerr_waveform_TD}
The Sasaki-Nakamura equation with source describing a propagating mode with frequency $\omega$ takes the canonical form (see Appendix~B of~\cite{Glampedakis:2017rar})
\begin{equation}
\left[ \partial_{x}^2 + V(\omega, x) \right] \tilde{\xi} (\omega, x) = S(\omega, x)\,,
\label{4_2_eq_example}
\end{equation}
where $x$ is the tortoise coordinate and $V(x)$ has the asymptotic limits
\begin{align}
V(\omega, x) \sim
\begin{cases}
    k_{\rm H}^2 & x \to -\infty\,,\\
    \omega^2 - 1/x^2 & x \to +\infty\,,
\end{cases}
\end{align}
with $k_{\rm H} \equiv \omega - m \Omega_{\rm H}$. 
Here we omit the $(\ell,\,m)$ subscripts for brevity.
The TD signal $\xi(t-x)$ 
is given by the inverse Laplace transform of $\tilde{\xi} (\omega, x)$. 
Employing the Green's function technique (see Section~\ref{subsec:greensfunc}), the inhomogeneous solution is
\begin{equation}\label{4_2eq:inhomo_solution}
\tilde{\xi} (\omega, x) = \frac{\tilde{\xi}_{\infty} (\omega, x)}{W(\omega)} \int_{- \infty}^{x} dx' \tilde{\xi}_+ (\omega, x') S (\omega, x') + \frac{\tilde{\xi}_+ (\omega, x)}{W(\omega)} \int_{x}^{\infty} dx' \tilde{\xi}_{\infty} (\omega, x') S (\omega, x')\,,
\end{equation}
where $W(\omega)$ is the Wronskian of $\tilde{\xi}_{\infty}$ and $\tilde{\xi}_+$, and the two homogeneous solutions, $\tilde{\xi}_+(\omega, x)$ and $\tilde{\xi}_{\infty}(\omega, x)$, have the asymptotic form
\begin{align}
\tilde{\xi}_+ \sim
\begin{cases}
    e^{-i k_{\rm H} x} & x\to - \infty\,,\\
    {\cal A}_{\rm out}(\omega) e^{i \omega x} + {\cal A}_{\rm in}(\omega) e^{-i \omega x} & x\to  + \infty\,,
\end{cases}\\
\tilde{\xi}_{\infty} \sim
\begin{cases}
    {\cal B}_{\rm out}(\omega) e^{i k_{\rm H} x} + {\cal B}_{\rm in}(\omega) e^{-i k_{\rm H} x} & x\to - \infty\,,\\
    e^{i \omega x}  & x\to + \infty\,.
\end{cases}
\end{align}

For signals observed in the far zone ($x \to \infty$) we have
\begin{equation}
\tilde{\xi} (\omega, x) \simeq \frac{\tilde{\xi}_{\infty} (\omega, x)}{2 i \omega {\cal A}_{\rm in}}  \int_{- \infty}^{\infty} dx' \tilde{\xi}_+ (\omega, x') S (\omega, x') 
= \frac{e^{i\omega x} {\cal A}_{\rm out}}{2 i \omega {\cal A}_{\rm in}}  \int_{- \infty}^{\infty} dx' \frac{\tilde{\xi}_+ (\omega, x')}{{\cal A}_{\rm out}} S (\omega, x') \,,
\end{equation}
which can be rewritten as
\begin{equation}
\tilde{\xi} (\omega, x) = [\tilde{h}_{\rm G} (\omega) \times \tilde{T} (\omega)] e^{i \omega x}\,,
\end{equation}
with 
\begin{equation}\label{4_2eq:Gringdown_source}
\tilde{h}_{\rm G} (\omega) \equiv \frac{{\cal A}_{\rm out}}{2i\omega {\cal A}_{\rm in}}\,, \ \ \tilde{T} (\omega) \equiv \int_{- \infty}^{\infty} dx' \frac{\tilde{\xi}_+ (\omega, x')}{{\cal A}_{\rm out}} S (\omega, x')\,.
\end{equation}
Note that the two functions $\tilde{h}_{\rm G}$ and $\tilde{T}$ %
are independent of the overall scale of the homogeneous solutions $\tilde{\xi}_+$ and $\tilde{\xi}_{\infty}$, since ${\cal A_{\rm out}}/{\cal A_{\rm in}}$ and ${\tilde{\xi}_+}/{\cal A_{\rm out}}$ are invariant with respect to the overall scale transformation of $\tilde{\xi}_{+/\infty}$.
While $\tilde{h}_{\rm G}$ is determined only by the Green's function, $\tilde{T}$ contains all the initial data and source information.
Neglecting the time dependence of QNM amplitudes, the TD waveform is then given by
\begin{align}
\begin{split}
h (t-x) = \frac{1}{2 \pi}\int_{-\infty+i \epsilon}^{\infty+i \epsilon} d\omega \tilde{\xi} (\omega, x) e^{-i \omega t}
&= \sum_{n} B_n^{\rm (SN)} T_n e^{-i \omega_n (t-x)} \\
&+ \text{(prompt response term)} + \text{(branch cut term)}\,,
\end{split}
\label{eq:lrdown}
\end{align}
where for simplicity $\tilde{T} (\omega)$ is assumed to not have any poles in the complex-$\omega$ plane, and $T_n \equiv \tilde{T}(\omega_n)$. 
The relevant integration contour is shown in Fig.~\ref{fig:G_contour}. The prompt response and branch cut terms correspond to the green-dashed and red-dotted lines in Fig.~\ref{fig:G_contour}, respectively.
The first term in Eq.~\eqref{eq:lrdown} above represents the QNM contribution to the waveform.
The total QNM amplitude is $B_n^{\rm (SN)} \times T_n$, where the complex ``excitation factors'' $B_n^{\rm (SN)}$, defined as in Eq.~\eqref{Bdef} in terms of the asymptotic amplitudes ${\cal A}_{\rm out/in}$ of the Sasaki-Nakamura equation, are determined solely by the Green's function
(here we omit the $\ell,m$ subscripts for brevity). 
The ``source factors'' $T_n$ depend on the initial data and on the specific source exciting the perturbations. The overall amplitude of each QNM, i.e., the so-called ``excitation coefficient'' $C_n$, is given by the product $C_n = B_n^{\rm (SN)} T_n$.

The ringdown component of $\xi$, i.e., the first term on the right-hand side of Eq.~\eqref{eq:lrdown}, can be derived using a QNM eigenfunction decomposition~\cite{Leaver:1986gd}. Substituting Eq.~\eqref{4_2eq:inhomo_solution} into Eq.~\eqref{eq:lrdown}, one finds that the QNM part has the form
\begin{align}\label{4_2eq:QNMeigenfunction}
\begin{split}
\xi &\sim 2 \pi i \sum_n \text{Res}_{\omega= \omega_n} \left[ \tilde{\xi}(\omega) e^{-i \omega t}\right]\,,\\
&=  \sum_n \left[ \frac{{\cal A}_{\rm out}(\omega_n)}{2 \omega_n (d{\cal A}_{\rm in}/d\omega)_{\omega = \omega_n}} \tilde{\xi}_{\rm q} (\omega_n,x) \int_{-\infty}^{\infty} dx' \tilde{\xi}_{\rm q} (\omega_n,x) S(\omega_n, x') \right]e^{-i \omega_n t}\,,
\end{split}
\end{align}
where 
\begin{equation}\label{4_2eq:QNMfunction}
\tilde{\xi}_{\rm q} (\omega_n, x) \equiv \tilde{\xi}_{\infty}(\omega_n,x) = \tilde{\xi}_+ (\omega_n,x)/{\cal A}_{\rm out}
\end{equation}
is the QNM eigenfunction.
It can be checked that Eq.~\eqref{4_2eq:QNMeigenfunction} reduces to the ringdown component of Eq.~\eqref{eq:lrdown} in the limit $x \to \infty$.
Note that the QNM eigenfunction $\tilde{\xi}_{\rm q}$ is divergent both at the horizon and at infinity. To obtain a finite value of the source factor $T_n$ in Eq.~\eqref{4_2eq:Gringdown_source}, it is useful to perform an analytic continuation of the areal coordinate $r'(x')$ and to change the contour of the spatial integral so that the integral has no divergence at the horizon and at infinity~\cite{Leaver:1986gd,Sun:1988tz}.

For simplicity of exposition, in the above expression the time dependence of the QNM amplitudes has been neglected, as taking it into account gives rise to more complicated expressions~\cite{Leaver:1986gd}.
The full waveform $\xi(t-x)$ including these contributions, as well as the prompt response and branch cut portions, is instead given by the convolution of the two TD waveforms
\begin{equation}
    \xi_{\ell m}(t-x) = h_{{\rm G},\ell m}(t-x) * T_{\ell m}(t)\,,
\end{equation}
where the convolution is defined as 
\begin{equation}
(f * g)(t) \equiv \int_{- \infty}^{\infty} d\tau f(\tau) g (t-\tau)
\end{equation}
in terms of the functions
\begin{align}
h_{{\rm G},\ell m}(t-x) &\equiv \frac{1}{2 \pi} \int_{-\infty +i \epsilon}^{\infty + i \epsilon} d \omega \tilde{h}_{{\rm G},\ell m}(\omega) e^{i\omega x} e^{-i \omega t}\,,
\label{4_2_eq_greenwaveform}
\\
T_{\ell m}(t) &\equiv \frac{1}{2 \pi} \int_{-\infty}^{\infty} d \omega \tilde{T}_{\ell m}(\omega) e^{-i \omega t}\,,
\end{align}
where we have restored the subscript $(\ell,m)$, that had been omitted for simplicity earlier. 
\begin{figure*}[t]
    \centering    \includegraphics[width=0.70\linewidth]{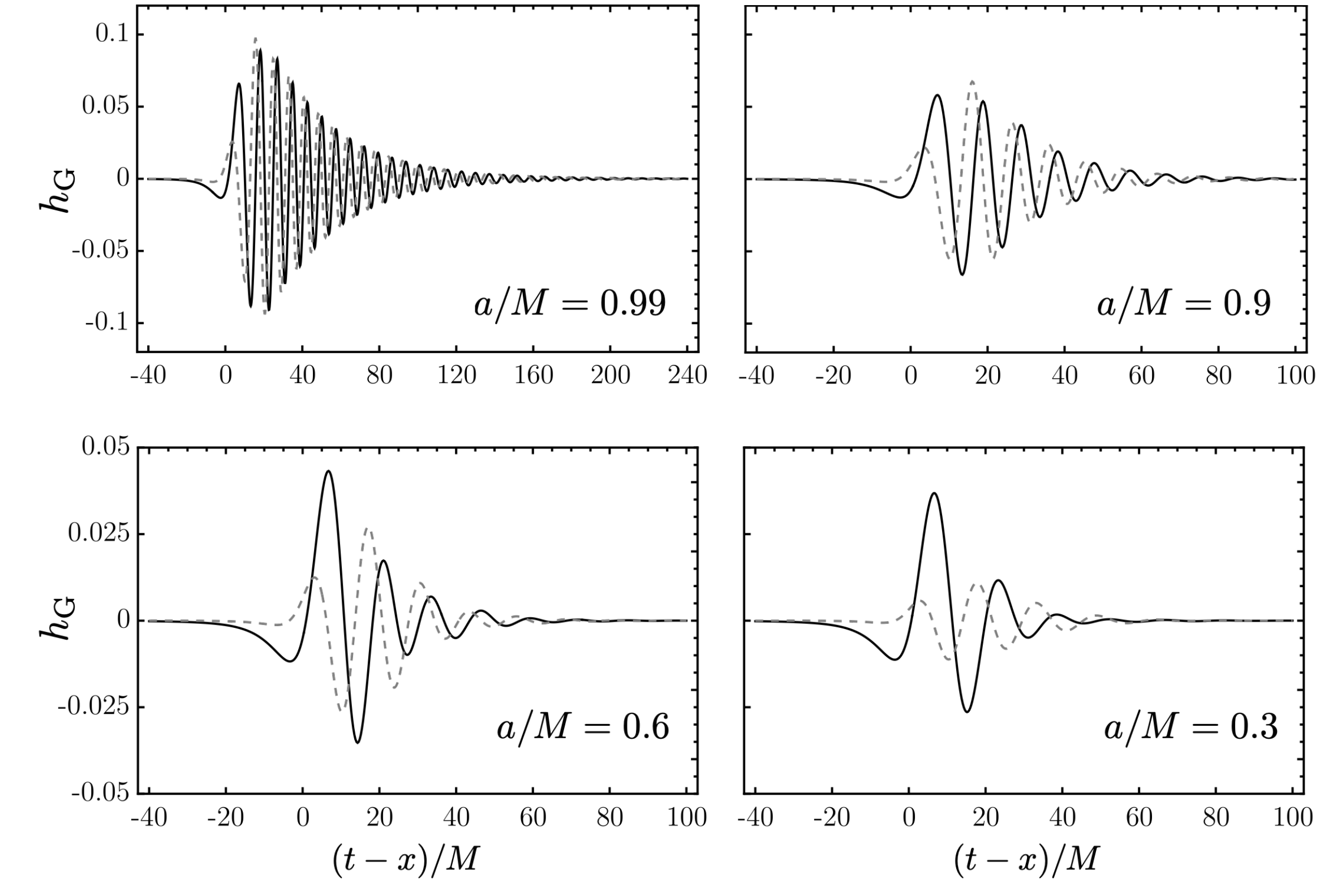}
    \caption{
    The Green's-function waveform, $h_{{\rm G},22} (t-x)$, for various spin parameters ($a/M = 0.99, 0.9, 0.6, 0.3$). 
    The solid and dashed lines indicate the real and imaginary parts of $h_{{\rm G},22}$.
    }
    \label{fig:4_2_hg}
\end{figure*}

The Kerr Green's function contribution to the TD waveform for the dominant quadrupolar mode, $h_{{\rm G},22}(t-x)$, 
is shown in Fig.~\ref{fig:4_2_hg}, and it exhibits a ringdown phase at late times. The waveform is computed by numerically integrating the Sasaki-Nakamura equation and by directly reading off ${\cal A}_{\rm in/out}$ from the homogeneous solution. The TD waveform is found by using a contour on the real-$\omega$ axis and an integration interval in the range $[-4 M,\,4M]$.
The full waveform $\xi(t-x)$ shown in Fig.~\ref{fig:4_2_particle} 
was found by taking the convolution with the numerically computed source term of~\cite{Kojima:1984cj}, and it corresponds to a particle plunging along an equatorial geodesic into a Kerr BH with varying spin, neglecting radiation reaction on the particle trajectory.
\begin{figure*}[t]
    \centering    \includegraphics[width=1\linewidth]{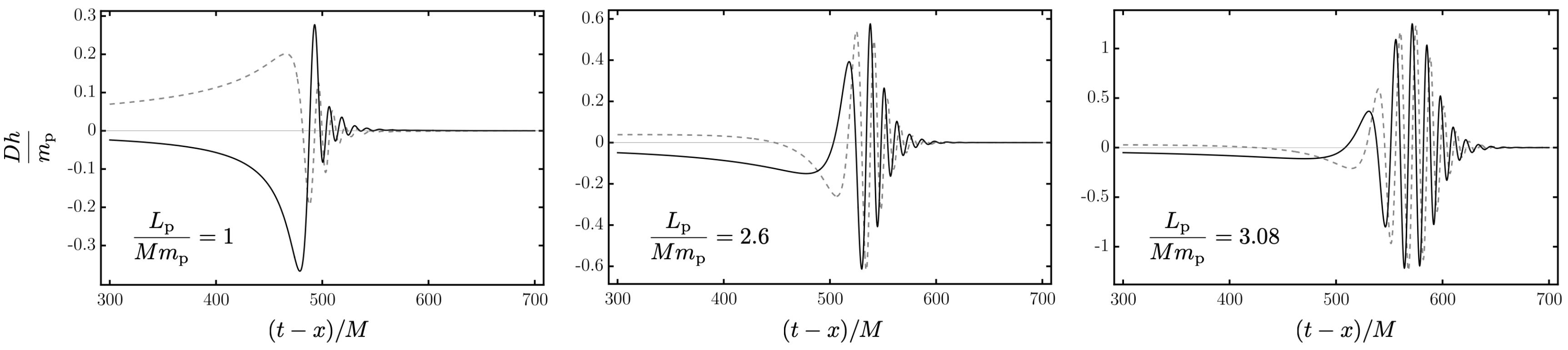}
    \caption{Real (solid lines) and imaginary (dashed lines) parts of the TD waveform sourced by a particle plunging into a Kerr BH with spin $a/M = 0.7$ at (luminosity) distance $D$. 
    The trajectory is restricted to the equatorial plane with energy $E= m_{\rm p}$ and orbital angular momentum $L_{\rm p}/ (M m_{\rm p}) = 1$ (left), $2.6$ (center), and $3.08$ (right), where $m_{\rm p}$ is the mass of the particle. 
    }
    \label{fig:4_2_particle}
\end{figure*}

The excitation of QNMs in the extreme mass ratio limit has been studied using different methods to construct the source term.
Early papers computed complete gravitational waveforms in the extreme mass-ratio limit within the EOB framework~\cite{Nagar:2006xv, Barausse:2009xi, Bernuzzi:2010ty, Barausse:2011kb}, considering a PN-driven trajectory with a generic test-particle source term~\cite{Nagar:2005ea}.
In other works, the loss of energy and angular momentum of the small body was taken into account by generalizing the Ori-Thorne procedure~\cite{Ori:2000zn} to describe the transition from inspiral to plunge.
For example, the dependence of QNM excitation on the inspiral-to-plunge transition in the Schwarzschild background can be studied using an analytical form of the trajectory~\cite{Hadar:2009ip,Hadar:2011vj}.
Several works have explored trajectories describing the transition from inspiral to plunge~\cite{Mummery:2022ana,Mummery:2023hlo,Dyson:2023fws}
and the worldline of small bodies following initially circular but misaligned orbits~\cite{Apte:2019txp} in the Kerr background. 
This work was the basis for subsequent studies of the excitation of Kerr QNMs~\cite{Lim:2019xrb,Hughes:2019zmt} where the Teukolsky equation was solved with high precision in the TD, following methods developed in~\cite{Zenginoglu:2011zz,Field:2020rjr}.

As in this approach it is not possible to split the QNM, prompt and tail components of the Green's function, the QNM amplitudes need to be extracted {\em a posteriori} from the full solution by fitting a superposition of damped sinusoids with frequencies corresponding to the QNM spectrum.
This methodology is described in detail in Section~\ref{sec:waveforms}.

Rather than finding the QNM amplitudes from fits of the gravitational waveforms, an alternative is to analytically compute the excitation factor $E_{\ell m n}$ and the source factor $T_{\ell m n}$ from first principles. 
The source factor calculation in the Schwarzschild case was pioneered by Leaver~\cite{Leaver:1986gd}. 
It involves the integration of the QNM eigenfunctions $\xi_{\rm q} (\omega_n, x)$ in Eq.~\eqref{4_2eq:QNMfunction} in standard Schwarzschild coordinates, which yields divergent results at both the horizon and infinity. 
The source factor integral can be regularized, either by subtracting divergent terms (as in~\cite{Detweiler:1979xr}) or by analytical continuation of the radial coordinate into the complex plane and contour deformation (as in~\cite{Leaver:1986gd}, see also~\cite{Andersson:1995zk,Glampedakis:2003dn,Silva:2024ffz,Lo:2025njp}).
A similar source factor calculation for a particle plunging along the spin axis of a Kerr BH, based on the regularization scheme developed in~\cite{Detweiler:1979xr}, can be found in~\cite{Zhang:2013ksa} (see also recent work in~\cite{Watarai:2024huy,DellaRocca:2025zbe}).
A generalization of these calculations to generic orbits is required to understand QNM excitation and to improve waveform models for extreme mass ratio binaries, but it is technically complicated. 
A promising avenue in this direction is the extension of self-force techniques~\cite{Pound:2021qin} beyond the inspiral phase~\cite{Kuchler:2024esj,Honet:2025dho}.

\subsubsection{Kerr excitation factors}
The excitation factors $B_{\ell m n}$,
defined as the residues of the Green's function at the QNM poles, quantify the ``intrinsic excitability'' of QNMs. It is sometimes useful to introduce also the quantity $E_{\ell m n} \equiv B_{\ell m n}/\omega_{\ell m n}^2$~\cite{Oshita:2021iyn}, which is useful to take into account the fact that the strain amplitude $h$ and the Newman-Penrose scalar are related by two time derivatives ($\Psi_4 \sim \ddot{h}$).
The excitation factors were first computed in the Schwarzschild case~\cite{Leaver:1986gd,Sun:1988tz,Andersson:1995zk, Andersson:1996cm,Berti:2006wq}. 
The Kerr case was initially investigated in~\cite{Glampedakis:2003dn,Berti:2006wq}, and later extended to $n=3$ and up to $\ell = 7$~\cite{Zhang:2013ksa}, 
to $n=20$ for $\ell=2$~\cite{Oshita:2021iyn}, and to  multipoles with $\ell>2$~\cite{Oshita:2022pkc}.
The excitation factors are now known in the scalar, electromagnetic and gravitational case up to $\ell=7$ and to $n=7$, thanks to technical improvements to achieve high accuracy~\cite{motohashi_2024_12696858,Motohashi:2024fwt, Lo:2025njp}.
Excitation factor calculations require the homogeneous solution of the perturbation equation for general complex frequencies, which can be achieved using the Mano-Suzuki-Takasugi (MST) method~\cite{Mano:1996vt}. This technique was applied in~\cite{Zhang:2013ksa,Motohashi:2024fwt}. 
In~\cite{Lo:2025njp}, the Sasaki-Nakamura equation with a general complex frequency was solved instead.
An alternative is to use Heun functions to obtain the relevant expressions for the Kerr-de Sitter spacetime, and then extrapolate to the Kerr case by setting the cosmological constant to zero~\cite{Oshita:2021iyn,Oshita:2025ibu}.
The high-$\ell$ expansion method was also applied to the computation of Schwarzschild excitation factors~\cite{Dolan:2009nk,Dolan:2011fh}. The accuracy of this method for low $\ell$'s can be improved by including higher-order terms~\cite{Chen:2024hum}.

\begin{figure*}[t]
    \centering    \includegraphics[width=1\linewidth]{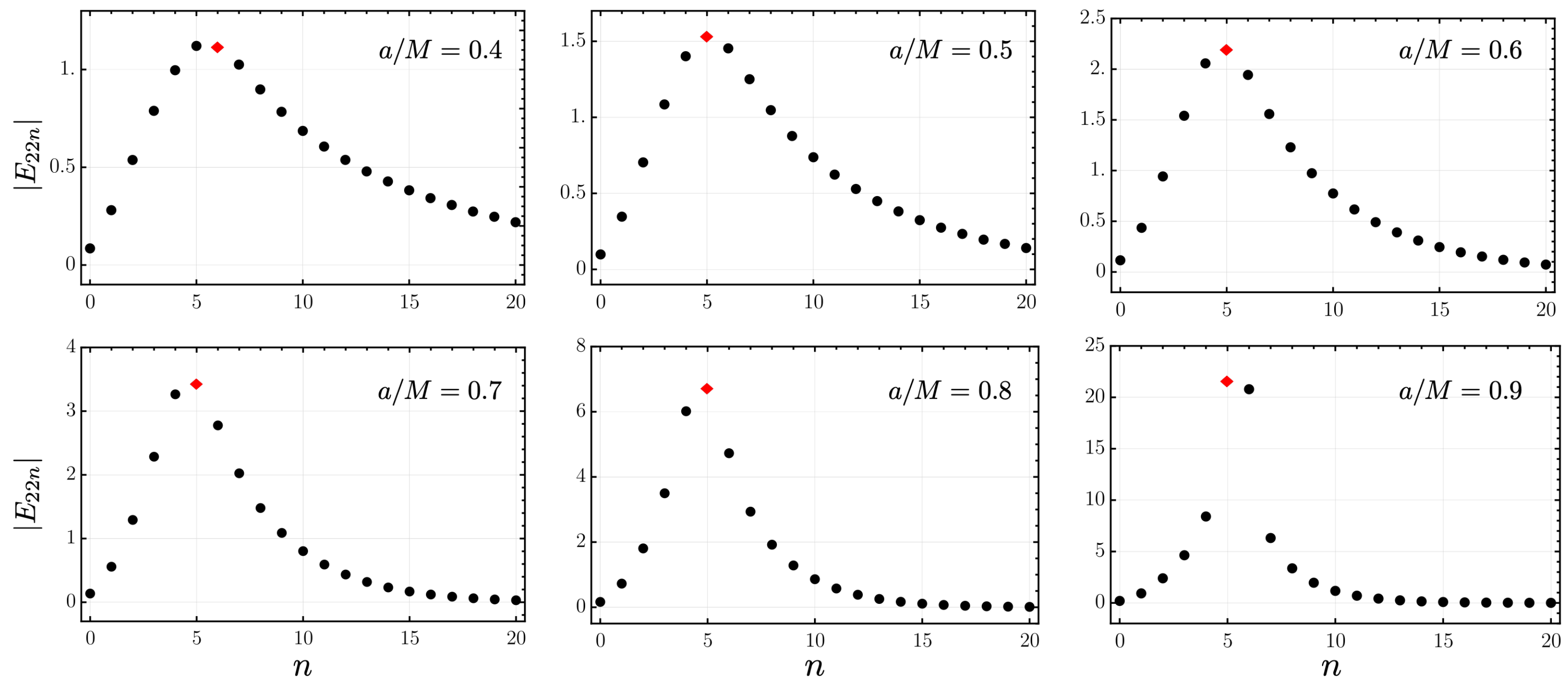}
    \caption{The absolute value of the excitation factor, $|E_{22n}|$, for overtones up to $n=20$. The red marker indicates the peak of $|E_{22n}|$. Figure adapted from~\cite{Oshita:2021iyn}.}
    \label{fig:4_2_e22_n}
\end{figure*}
\begin{figure*}[t]
    \centering    \includegraphics[width=0.80\linewidth]{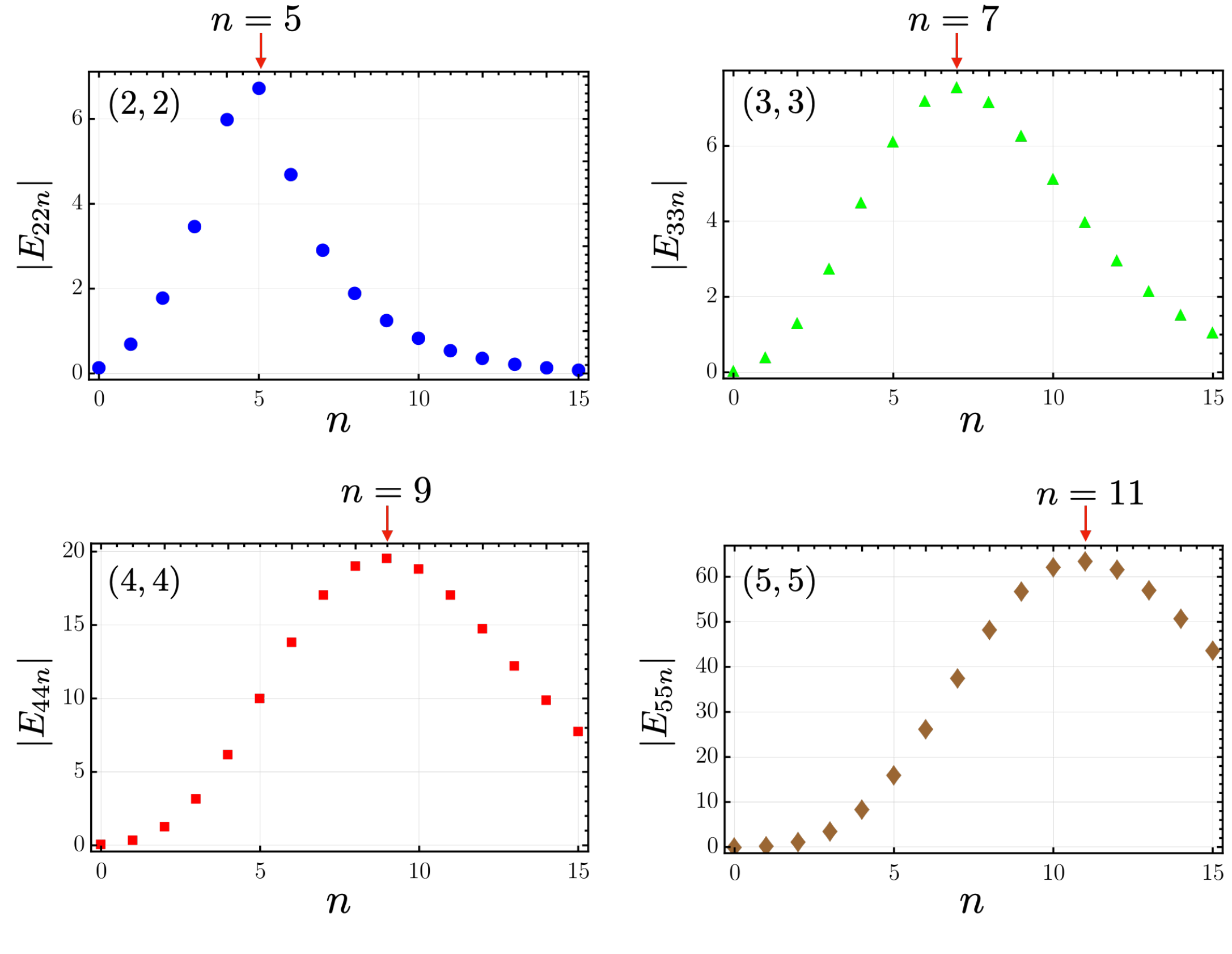}
    \caption{Absolute values of the excitation factors $|E_{\ell m n}|$ for $(\ell, m) = (2,2)$, $(3,3)$, $(4,4)$, and $(5,5)$, and for $a=0.8M$. Figure adapted from~\cite{Oshita:2022pkc}.}
    \label{fig:4_2_eellm_n}
\end{figure*}

\begin{figure*}[t]
    \centering    \includegraphics[width=1\linewidth]{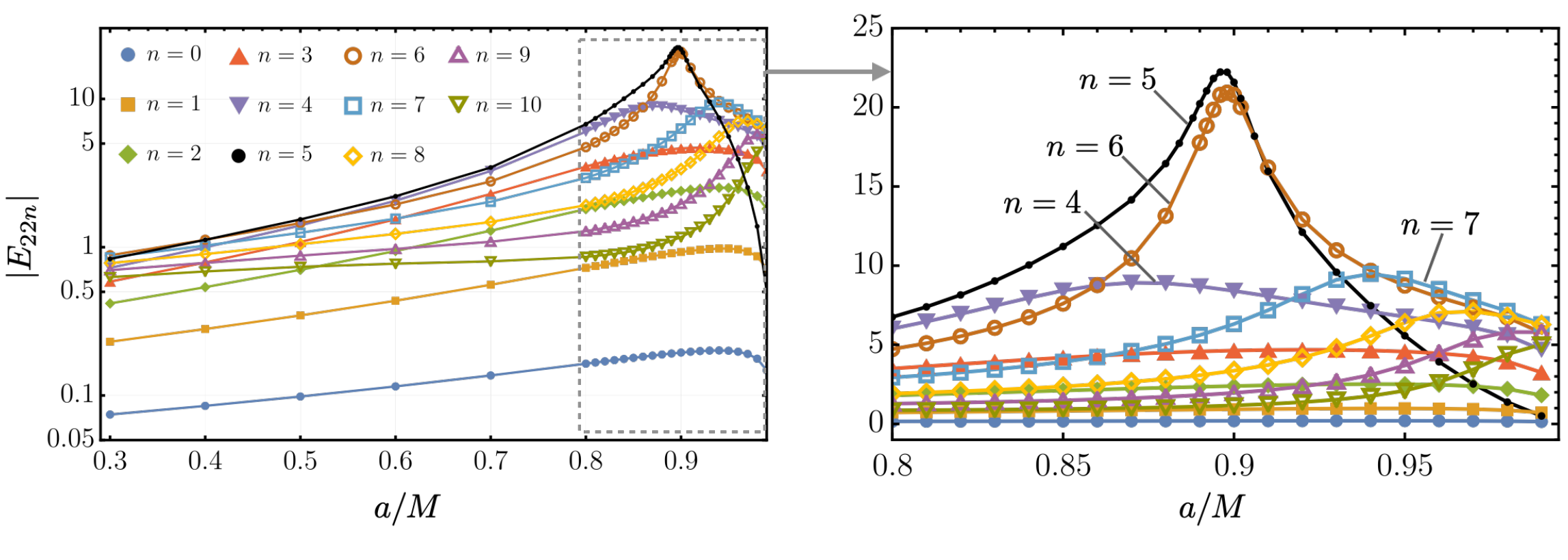}
    \caption{Spin dependence of the absolute value of the excitation factors, $|E_{22n}|$, for overtones up to $n=10$. Figure adapted from~\cite{Oshita:2021iyn}.}
    \label{fig:4_2_e22_spin}
\end{figure*}

The Kerr excitation factors are shown in Fig.~\ref{fig:4_2_e22_n} for $\ell = m = 2$ and varying overtone number.
The absolute value of the excitation factor peaks around $n=5,6$.
However, the overtone excitation coefficients are ultimately weighted by the source factor, so it is not possible to draw definitive implications on the  excitations of overtones in actual ringdown signals just by looking at the excitation factors.
The peak of the excitation factor shifts towards higher overtone numbers for higher multipoles (see Fig.~\ref{fig:4_2_eellm_n}).
As shown in Fig.~\ref{fig:4_2_e22_spin}, $E_{22n}$ grows with spin for all overtones up to $a \simeq 0.9M$, at which point the $n=5,\,6$ overtones display a peak, with the $n=5$ overtone being strongly suppressed in the near-extremal case. This behavior is caused by mode repulsion (see Section~\ref{sec:avoidance}).

\begin{figure*}[t]
    \centering    \includegraphics[width=0.70\linewidth]{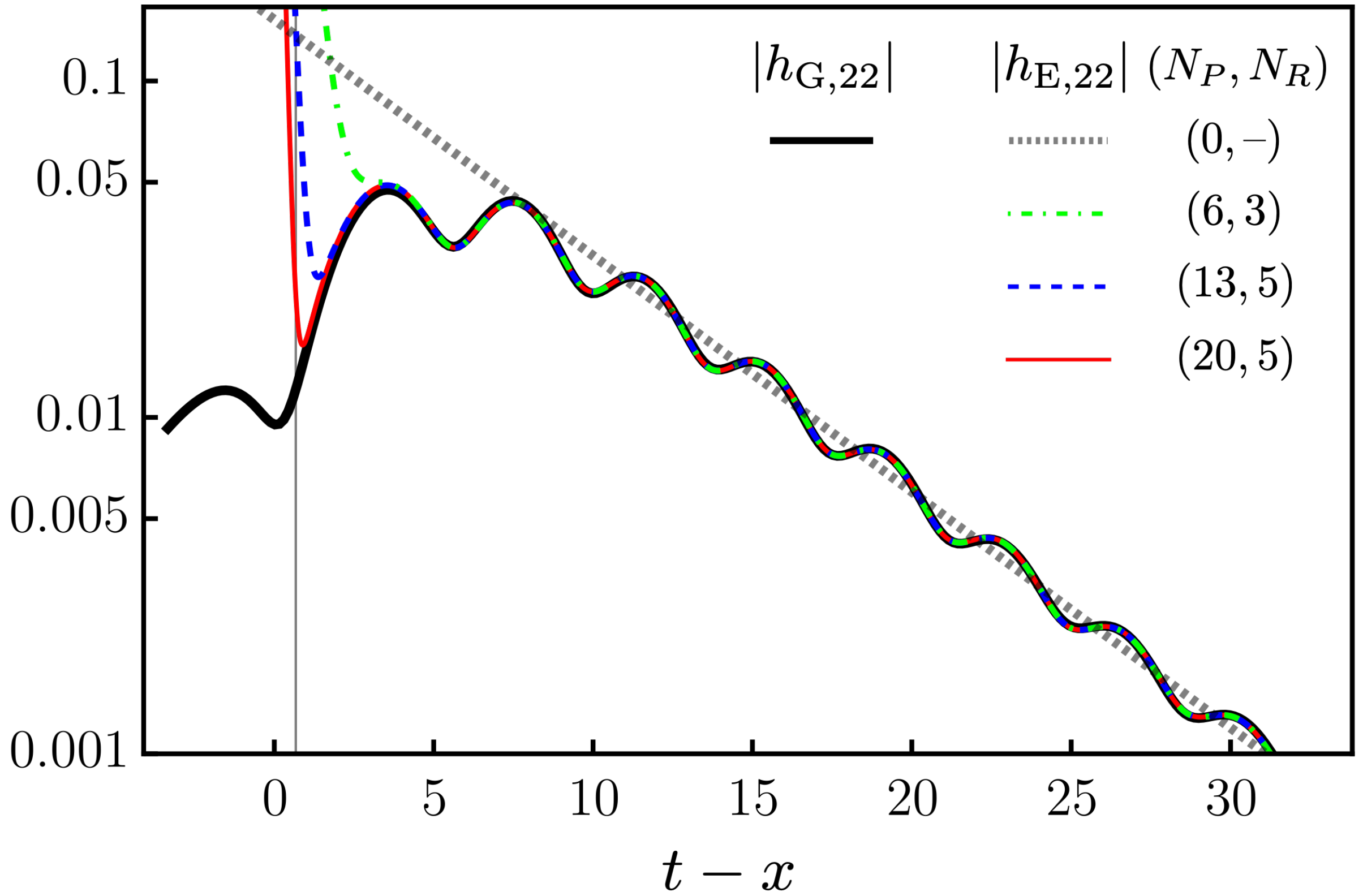}
    \caption{Comparison of $h_{\rm G}$ and $h_{\rm E}$ for $\ell=m=2$ and $a=0.7M$~\cite{Oshita:2024wgt}. Time is measured in units $2M=1$, and $D_* = -2M \log[1-(a/M)^2]$~\cite{Oshita:2024wgt}. Only the prograde fundamental mode, i.e., no retrograde mode, is included in the gray dotted line. Figure adapted from~\cite{Oshita:2024wgt}.
    }
    \label{fig:4_2_hg_tx}
\end{figure*}

By definition, the excitation factors $E_{\ell m n}$ correspond to the QNM amplitudes in the Green's function waveform $h_{{\rm G}, \ell m}$ of Eq.~\eqref{4_2_eq_greenwaveform}.
Since constant-valued QNM amplitudes depend on the time at which the exponential decay starts, some care is required in specifying the time at which the excitation factor is defined. 
In other words, $E_{\ell m n}$ is ambiguous up to a factor of $e^{i \omega_{\ell m n} t_*}$, where the parameter $t_*$ encodes the ambiguity in the starting time of ringdown, and it should be defined at a time such that the QNM expansion is convergent~\cite{Oshita:2021iyn, Oshita:2022pkc, Oshita:2024wgt}, as carefully investigated in~\cite{Oshita:2024wgt}. The ringdown phase in $h_{{\rm G}, \ell m}$ 
can be reconstructed using a superposition of QNMs weighted by the excitation factors, $h_{\rm E}$:
\begin{equation}
h_{{\rm G}, \ell m} \simeq h_{{\rm E}, \ell m} \equiv 
 \sum^{N_{\rm P}}_{n_{\rm P}=0}  E_{\ell m n_{\rm P}} e^{-i \omega_{\ell m n_{\rm P}} (t-x-D_*)}
+ \sum^{N_{\rm R}}_{n_{\rm R}=0}  E_{\ell m n_{\rm R}} e^{-i \omega_{\ell m n_{\rm R}} (t-x-D_*)}\,.
\end{equation}
Here $t_*$ is fixed to $t_* = D_* \equiv -2M \log[1-(a/M)^2]$~\cite{Oshita:2024wgt}, i.e. the ``starting time'' of the ringdown, in the sense that for $t-x < D_*$ the QNM expansion does not converge (see Fig.~\ref{fig:4_2_hg_tx}). 
The two indices $n_{\rm P}$ and $n_{\rm R}$ label prograde and retrograde overtones, respectively. The number of prograde and retrograde overtones in the sum, $N_{\rm P}$ and $N_{\rm R}$, must be large enough to precisely reconstruct the ringdown.
With $N_{\rm P} = 20$ and $N_{\rm R} = 5$, one can reconstruct the ringdown part of $h_{{\rm G}, \ell m}$ within a mismatch of 
${\cal M} \lesssim 10^{-3}$ (see Fig.~\ref{fig:4_2_hg_tx}), where the mismatch is defined as
\begin{equation}
{\cal M} \equiv 1 - \left| \frac{\braket{h_{{\rm G},\ell m}|h_{{\rm E},\ell m}}}{\sqrt{\braket{h_{{\rm G},\ell m}|h_{{\rm G},\ell m}} \braket{h_{{\rm E},\ell m}|h_{{\rm E},\ell m}}}} \right|\,,
\end{equation}
in terms of the scalar product 
\begin{equation}
\braket{a(u)|b(u)} \equiv \int_{D_*}^{\infty} du' a(u') b^*(u')\,,
\end{equation}
where $u\equiv t-x$. Note that this analysis does not rely on any fits, %
as the excitation factors are computed from first principles.
It turns out that the excitation factors computed in~\cite{Oshita:2021iyn} are defined at the earliest time at which the QNM expansion is convergent~\cite{Oshita:2024wgt}.
Even more accurate reconstructions of the TD signal (including early times and even late-time tails) are possible by considering expansions in terms of the QNMs of destabilized potentials~\cite{Oshita:2025ibu}.

\subsection{Quadratic modes in Schwarzschild} \label{sec:nonlinSch}

\vspace{-.1cm}

\noindent \textit{Initial contributors: Bucciotti, Kuntz, Riotto}

\vspace{.2cm}

Until recently, most spectroscopy efforts have been dedicated to understanding the physical excitation mechanisms and detectability of the linear ringdown spectrum. However, a debate on the domain of validity of linear perturbation theory in describing BH binary dynamics~\cite{Giesler:2019uxc, Baibhav:2023clw} has sparked a community-wide initiative to systematically investigate nonlinear effects, such as quadratic mode excitation, absorption-induced changes in BH mass and spin, and transient effects.
Among these, significant attention has been given to the new modes appearing in the spectrum at higher orders in perturbation theory~\cite{Gleiser:1995gx,Brizuela:2009qd,Ioka:2007ak,Nakano:2007cj,Loutrel:2020wbw,Ripley:2020xby}.
The \textit{quadratic modes} or quadratic QNMs (QQNMs) appearing at second order are the topic of this section. 
These additional families of QNM frequencies naturally arise from the nonlinear nature of Einstein's equations, and were recently confidently extracted from nonlinear binary merger simulations~\cite{Cheung:2022rbm,Mitman:2022qdl,Ma:2022wpv,Baibhav:2023clw,Cheung:2023vki}.

Following these breakthroughs, various analytical and numerical techniques were swiftly employed to characterize the amplitudes and the dependence on initial data of these modes for Schwarzschild and Kerr BHs~\cite{Bucciotti:2023ets, Perrone:2023jzq, Kehagias:2023ctr, Redondo-Yuste:2023seq,Ma:2024qcv,Bucciotti:2024zyp,Bucciotti:2024jrv,Bourg:2024jme,Zhu:2024rej,Ma:2024qcv,Khera:2024yrk,Kehagias:2024sgh,Kehagias:2025ntm}. These predictions are currently being incorporated as additional components in the construction of BBH waveform templates.
Remarkably, not only the QQNM frequencies are calculable from perturbation theory, but also their relative \textit{amplitudes} with respect to the linear modes can be fully obtained within perturbation theory only. This makes quadratic modes very clean probes of BH physics, as all of their properties are determined from perturbation theory alone. This section deals with the \sch case, while Kerr QQNMs are described in Section~\ref{sec:nonlinKerr} below.

\subsubsection{Quadratic modes in equatorial symmetry}\label{sec:quadmodes_eq}
Let us expand around the Schwarzschild metric $\bar g_{\mu \nu}$ to second order in a bookkeeping parameter $\varepsilon$ representing the small amplitude of the perturbations: 
\begin{equation}
g_{\mu\nu} = \bar g_{\mu\nu} + \varepsilon h^{(1)}_{\mu \nu} + \varepsilon^2 h^{(2)}_{\mu \nu} + \mathcal{O}(\varepsilon^3)\,.
\end{equation}
Linear perturbations $h_{\mu \nu}^{(1)}$ obey the standard RWZ equations~\eqref{eq:RWZ master}, so that their propagating modes $h_+^{(1)}$ and $h_\times^{(1)}$ can be written far from the BH as
\begin{equation}\label{eq:linearStrain}
    h_+^{(1)} - i h_\times^{(1)} = \frac{M}{r} \sum_{\ell m n \mathfrak{m}}  \mathcal{A}_{\ell m n}^{(1) \mathfrak{m}} e^{-i \omega_{\ell m n}^{\mathfrak{m}} (t-r)} \hphantom{a}_{-2}Y^{\ell m}(\theta, \phi) \, .
\end{equation}
In this equation, the frequencies $\omega_{\ell m n}^{\mathfrak{m}}$ belong to the linear QNM frequency spectrum, and $ \mathcal{A}_{\ell m n}^{(1) \mathfrak{m}}$ are the linear QNM amplitudes.
Both depend on the mirror mode number $\mathfrak m = \pm$, which
distinguishes ordinary and mirror modes (see Eq.~\eqref{eq:mirror mode symmetry}).
Note that for a \sch BH, the QNM frequencies $\omega_{\ell m n}^{\mathfrak{m}} = \omega_{\ell 0 n}^{\mathfrak{m}}$ do not depend on $m$.
In the following discussion, the ${}_{\ell m n}^{\mathfrak{m}}$ indices will often be suppressed to avoid clutter. 

We will start by making the simplifying assumption that the perturbations satisfy equatorial (or reflection, or planar) symmetry.
In the general case where the progenitors' intrinsic angular momenta (``spin'') are not aligned with the orbital angular momentum, the spins precess around the total angular momentum and this assumption is violated (see Section~\ref{sec:waveforms}), but for the moment let us neglect spin precession and focus on nonprecessing binaries; this assumption will be dropped in Section~\ref{sec:nonlinSch_dropequatorialsym}.
In this case the parity of a mode is $(-1)^{\ell +m}$, i.e., all metric perturbations satisfy the Zerilli equation for $\ell+m$ even and the Regge-Wheeler equation for $\ell+m$ odd (in a more generic setting, metric perturbations would contain both even and odd components for a fixed $\ell+m$). Furthermore, equatorial symmetry imposes a relation between the amplitudes of mirror and ordinary modes:
\begin{equation} \label{eq:equatorial_sym_relation_amplitudes}
     \mathcal{A}_{\ell (-m) n}^{(1) (-\mathfrak{m})} = (-1)^{\ell} {\mathcal{A}^*}_{\ell m n}^{(1) \mathfrak{m}}
\end{equation}
The second-order perturbations $h_{\mu \nu}^{(2)}$ can be split into parity-even (Zerilli) and parity-odd (Regge-Wheeler) sectors, each characterized by a Regge-Wheeler or Zerilli variable $\Psi^
{\pm(2)}$, with frequency $\omega$ and angular momentum numbers $(\ell,m)$, using the same definitions as at first order. By plugging the metric decomposition into Einstein's equations, at order $\varepsilon^2$ one obtains the equations of motion of $\Psi^{\pm (2)}$,
\begin{align}
    \label{eq:RWZ_equation_2ndorder}
    \frac{\mathrm{d}^2 \Psi^{\pm (2)}}{\mathrm{d} r_*^2}+\omega^2 \Psi^{\pm (2)}-V_2^\pm(r) \Psi^{\pm (2)} = S \; ,
\end{align}
where $V_2^\pm$ is the Regge-Wheeler or Zerilli potential (see Eqs.~\eqref{eq:RW-potential} and~\eqref{eq:Zerilli-potential}), we have suppressed the subscript denoting the spin of the perturbation from $\Psi^{\pm (2)}_2=\Psi^{\pm (2)}$ to avoid confusion, and $S$ is a source term quadratic in the linear perturbation $h_{\mu \nu}^{(1)}$. This simple quadratic dependence allows us to focus on a source given by a single pair of linear QNMs, the general case being obtained by superposition. Three essential properties of the quadratic source term $S$ can be derived by squaring the linear QNMs in Eq.~\eqref{eq:linearStrain}:
\begin{itemize}
    \item[(1)] It contains a sum of terms oscillating at frequencies $\omega_1+\omega_2$, where $\omega_i$ are the frequencies of the two linear modes $1$ and $2$ entering the square. In particular, since regular modes can couple to mirror modes, the real parts of the QNM frequencies can be added or subtracted, while the imaginary parts always add.
    \item[(2)] The product of two spin-weighted spherical harmonics ${}_{s}Y_{\ell_1 m_1}$ and ${}_{s}Y_{\ell_2 m_2}$ needs to be expressed as a sum of spin-weighted spherical harmonics, which means the source will be proportional to a $3j$ symbol
    $\begin{pmatrix}
        \ell_1&\ell_2&\ell\\
        m_1&m_2&-m
    \end{pmatrix}$ 
    ensuring the usual selection rules: 
    \begin{equation}
        \ell = |\ell_1-\ell_2|, \dots, \ell_1+\ell_2\,,\qquad
        m = m_1+m_2 \, .
    \end{equation}
    In fact, the only dependence of the source term on $m$, $m_1$ and $m_2$ is fully contained in the $3j$ symbol because of angular momentum conservation.
    \item[(3)] The amplitude of the source is quadratic in the linear mode amplitudes $ \mathcal{A}^{(1)}$.
\end{itemize}
This nonlinear source term was computed in early works~\cite{Gleiser:1995gx,Gleiser:1998rw,Nicasio:1998aj,Nicasio:2000ge} for the case $\ell=\ell_1=\ell_2=2$, and in~\cite{Nakano:2007cj,Ioka:2007ak} for $\ell=4$, $\ell_1=\ell_2=2$. 
The generic $\ell, \ell_1, \ell_2$ case was tackled in~\cite{Brizuela:2006ne,Brizuela:2007zza,Brizuela:2009qd}, although only in recent work~\cite{Bucciotti:2024jrv,Bucciotti:2024zyp} expressions for the source term became publicly available (see Appendix~\ref{sec:public_codes}).
In the most general case, the source for a single product of two linear modes has the following form:
\begin{align}
    &S = \mathcal F_1(r) \Psi_1^{(1)} \Psi_2^{(1)} + \mathcal F_2(r)  (\Psi_1^{(1)})' \Psi_2^{(1)} 
    + \mathcal F_3(r)   \Psi_1^{(1)} (\Psi_2^{(1)})' + \mathcal F_4(r) (\Psi_1^{(1)})' (\Psi_2^{(1)})' \, , \label{eq:sourceSchwarzschild}
\end{align}
where $\Psi_1^{(1)}$ and $\Psi_2^{(1)}$ are linear mode wavefunctions (Regge-Wheeler or Zerilli variables, depending on the parity of $\ell+m$) and the four functions $\mathcal F_i$ depend on $(\ell, m)$ as well as $(\ell_1, m_1, n_1, \mathfrak m_1) \times (\ell_2, m_2, n_2, \mathfrak m_2)$. 
A priori, the quadratic frequency $\omega$ appearing in $S$ could belong to the linear QNM spectrum. 
If this happens, the resulting QQNM amplitude would only renormalize the linear amplitude. 
This does not occur for Schwarzschild QNMs in vacuum, because the sum of two QNM frequencies does not belong to the discrete spectrum.
For nonvacuum spacetimes, where models for the source term $S$ matter, this case is instead realized, because $S$ contains in general a continuum of frequencies.
While the QQNM frequencies are trivial to find ($\omega = \omega_1+\omega_2$), the calculation of their amplitudes requires solving Eq.~\eqref{eq:RWZ_equation_2ndorder}.

By definition, QQNMs are solutions to the second-order equations~\eqref{eq:RWZ_equation_2ndorder} with QNM boundary conditions, $\psi^{(2)} \underrel{{r_* \rightarrow \infty}}{\propto} e^{i \omega r_*}$ and $\psi^{(2)} \underrel{{r_* \rightarrow -\infty}}{\propto} e^{-i \omega r_*}$. 
To impose these boundary conditions, observe that Eq.~\eqref{eq:RWZ_equation_2ndorder} is a linear inhomogeneous equation, whose solutions are the sum of a particular solution and a combination of homogeneous solutions.
Imposing the QNM boundary conditions completely fixes the coefficients of the homogeneous solutions, thus removing any ambiguity. 
Instead, the inhomogeneity of Eq.~\eqref{eq:RWZ_equation_2ndorder} is what completely determines the amplitude of QQNMs, in contrast with linear QNMs, where the equation is homogeneous.

The second-order equation~\eqref{eq:RWZ_equation_2ndorder} can be solved with standard methods already used to solve for the linear spectrum, e.g., WKB~\cite{Perrone:2023jzq}, uniform approximation~\cite{Bucciotti:2023ets}, or the well-known Leaver algorithm~\cite{Bucciotti:2024jrv,Bucciotti:2024zyp} (see Appendix~\ref{sec:public_codes}). 
However, as noted in~\cite{Gleiser:1995gx,Gleiser:1998rw,Nicasio:1998aj,Nicasio:2000ge,Brizuela:2006ne,Brizuela:2007zza,Brizuela:2009qd}, a complication arises due to the source term diverging at the boundaries. 
In the particular case of the QNM problem, imposing QNM boundary conditions e.g. at infinity in Eq.~\eqref{eq:RWZ_equation_2ndorder} implies that $S \underrel{{r_* \rightarrow \infty}}{\propto} e^{i \omega r_*}/r^2 $, but one finds from expression~\eqref{eq:sourceSchwarzschild} that $S \underrel{{r_* \rightarrow \infty}}{\propto} e^{i \omega r_*} $ instead~\cite{Nakano:2007cj,Bucciotti:2024jrv}. 
Thus, it seems at first that QNM boundary conditions are incompatible with second-order perturbations. The solution to this conundrum is that these divergences are spurious, and they should cancel out when expressing QNM amplitudes in the GW strain~\eqref{eq:linearStrain} in a regular gauge at infinity, while the Regge-Wheeler gauge employed to derive Eq.~\eqref{eq:RWZ_equation_2ndorder} entails a poor asymptotic scaling~\cite{Brizuela:2009qd}. In the particular case of the QNM setting, a workaround was proposed in~\cite{Nakano:2007cj,Bucciotti:2024jrv}: one can redefine a new, regular master scalar
\begin{equation}
    \label{eq:regulated_master_scalar}
    \tilde \Psi^{(2)}(r) = \Psi^{(2)}(r) + \Delta(r) \Psi_1^{(1)}\Psi_2^{(1)} \, ,
\end{equation}
where $\Delta(r) = a_2 r^2 + a_1 r$ is chosen to cancel the divergences. The regular variable obeys a differential equation at second order similar to Eq.~\eqref{eq:RWZ_equation_2ndorder}, but with a modified source term obeying the correct asymptotics. One can also check that $\Delta(r)$ encodes the divergences in the change of gauge needed at second order to go from the Regge-Wheeler gauge to the transverse-traceless gauge, where the GW strain~\eqref{eq:linearStrain} is most easily extracted~\cite{Bucciotti:2024jrv}. 

Employing this gauge, the amplitude prediction for all relevant QQNMs in the strain~\eqref{eq:linearStrain} can be computed using Leaver's algorithm~\cite{Bucciotti:2024zyp,Bucciotti:2024jrv}. 
Since the source is proportional to a product of linear amplitudes, the ratio of quadratic to linear amplitudes
\begin{equation}
    \mathcal{R} = \mathcal{A}^{(2)}/(\mathcal{A}^{(1)}_1 \mathcal{A}^{(1)}_2)
\end{equation}
can be predicted~\cite{Bucciotti:2024zyp}, and it is shown for several combinations of linear modes (including mirror modes) in Fig.~\ref{fig:qqnm_ampl}. 
\begin{figure}[t]
    \centering
    \includegraphics[width=1.1\linewidth]{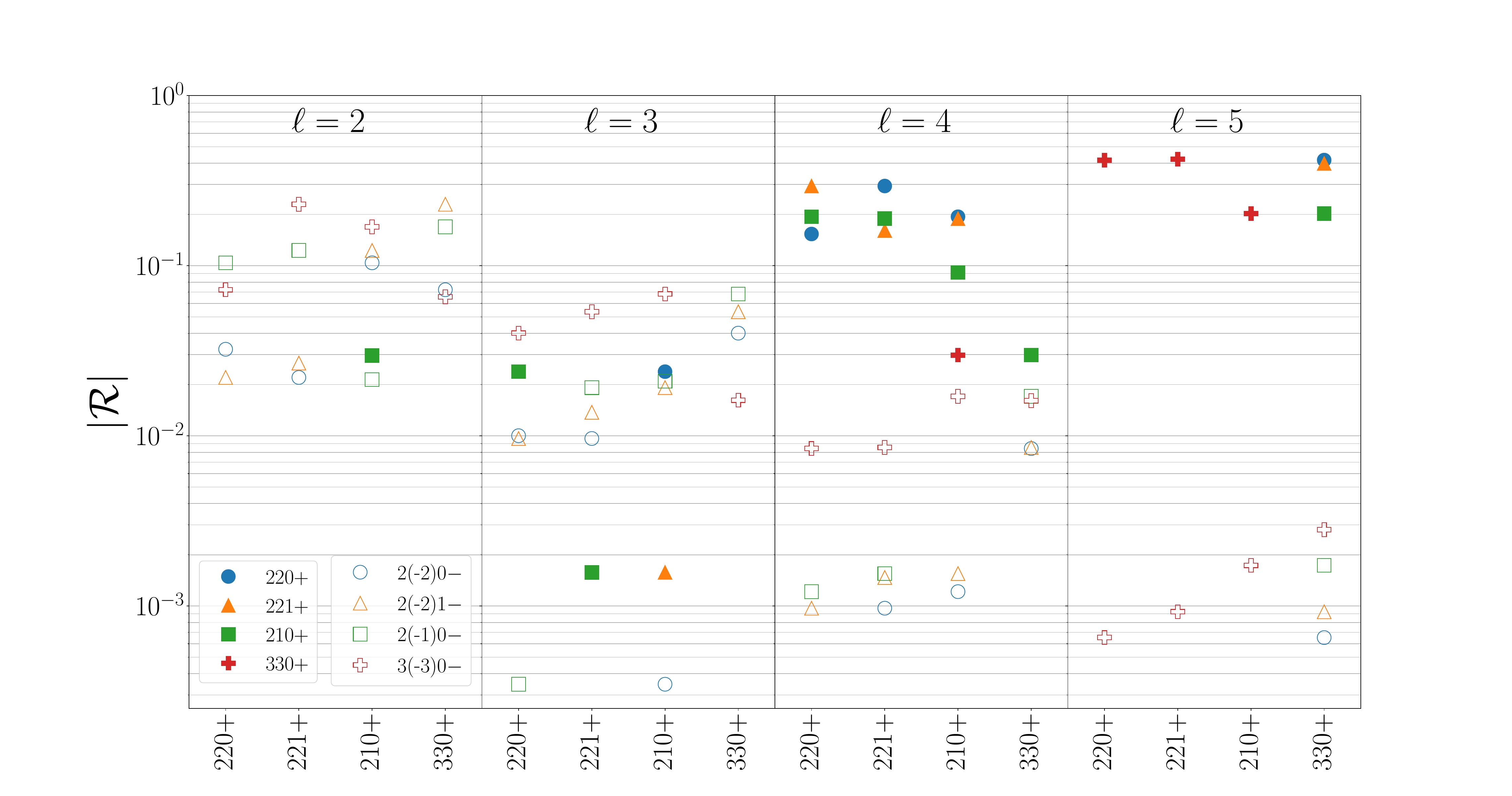}
    \caption{Nonlinear ratio of amplitudes  $|\mathcal{R}|$ for different mode numbers. The $x$ axis represents the mode numbers $\ell m n \mathfrak m$ of the first linear parent mode, while the second linear parent mode is represented with different markers. Filled symbols represents the combination of two ordinary modes with $m>0$ ($\omega = \omega_{\ell_1 m_1 n_1}^+ + \omega_{\ell_2 m_2 n_2}^+$), while empty symbols represents an ordinary mode with $m>0$ combined with a mirror mode with $m<0$ ($\omega = \omega_{\ell_1 m_1 n_1}^+ - \big(\omega_{\ell_2 (-m_2) n_2}^+ \big)^*$). A missing marker means that $|\mathcal{R}|<2.5 \times 10^{-4}$. Notice that most points in this figure correspond to contributions to different spherical harmonics, as each point corresponds to its own $m = m_1+m_2$ mode, and $m_1$, $m_2$ vary. Figure taken from~\cite{Bucciotti:2024zyp}.}
    \label{fig:qqnm_ampl}
\end{figure}
Note in particular that $\mathcal{R}=0.154e^{-0.068i}$ for the QQNM $(220+)\times(220+)\rightarrow(44)$, which is typically the most excited nonlinear mode in quasi-circular binary mergers (see Section~\ref{sec:nonlin_num_expe}). 
This is a small number (even if its smallness was not required for perturbation theory to be valid, as this only requires $\mathcal{A}^{(2)} \ll \mathcal{A}^{(1)}$), but could be within reach of XG detectors~\cite{Yi:2024elj,Khera:2024yrk,Lagos:2024ekd} (see Section~\ref{sec:quad_nextgen} for a detailed discussion).
Combinations of regular modes with mirror modes or with overtones can also have non-negligible $\mathcal{R}$, e.g. in the dominant $\ell=2$ multipole~\cite{Bucciotti:2024zyp,Bucciotti:2024jrv}, as also observed in numerical simulations~\cite{Giesler:2024hcr}.
In particular, combining a mode with its exact mirror is interesting because it produces quadratic modes which are purely imaginary and still have a nonzero amplitude (see also Section~\ref{sec:bms-frames-memory}).

\subsubsection{Beyond equatorial symmetry}
\label{sec:nonlinSch_dropequatorialsym}
The previous discussion was based on assuming equatorial symmetry for the QNM amplitudes. Going beyond that assumption, each single $(\ell, m)$ linear metric perturbation will contain both even ($P=+$, Zerilli) and odd ($P=-$, Regge-Wheeler) parities.
It was thus first conjectured in~\cite{Ma:2024qcv}, and then explicitly shown~\cite{Bourg:2024jme}, that in full generality quadratic modes are not completely characterized by a single number (the ratio $\mathcal{R}$), but they also depend on the linear modes' parity. 
This dependence could explain some of the discrepancies found in the literature concerning the value of the ratio $\mathcal{R}$~\cite{Ma:2024qcv,Redondo-Yuste:2023seq,Mitman:2022qdl,Zhu:2024rej,Cheung:2022rbm,Cheung:2023vki}.
A single quadratic mode thus depends on four numbers, which represent the four different ratios $\mathcal{R}_{P_1 \times P_2 \rightarrow P}$ combining different parities among each other~\cite{Bucciotti:2024jrv}. 
Indeed, although there seem to be eight different ways to combine $\pm \times \pm \rightarrow \pm $, they are in fact restricted by parity invariance, which imposes
\begin{equation}
    \label{eq:selection_rule_parity}
    (-1)^{\ell_1} P_1 \times (-1)^{\ell_2} P_2 = (-1)^{\ell} P\, ,
\end{equation}
so that the parity $P$ of the QQNM is fixed by the individual parities of linear modes. 
To find the value of the quadratic amplitude in this generic case, a \textit{polarization parameter} 
can be introduced, encoding the deviation from equatorial symmetry:
\begin{equation}
    \mathfrak{p}_{\ell m n}^\mathfrak{m} = (-1)^m \frac{{\mathcal{A}^*}_{\ell (-m) n}^{(-\mathfrak{m})}}{\mathcal{A}_{\ell m n}^\mathfrak{m}} \,.
\end{equation}
As before, the ${}_{\ell m n}^\mathfrak{m}$ indices will be suppressed in the following ($\mathfrak{p}_{\ell m n}^\mathfrak{m}\to \mathfrak{p}$) to avoid clutter. 

\begin{figure}
    \centering
    \includegraphics[width=0.75\linewidth]{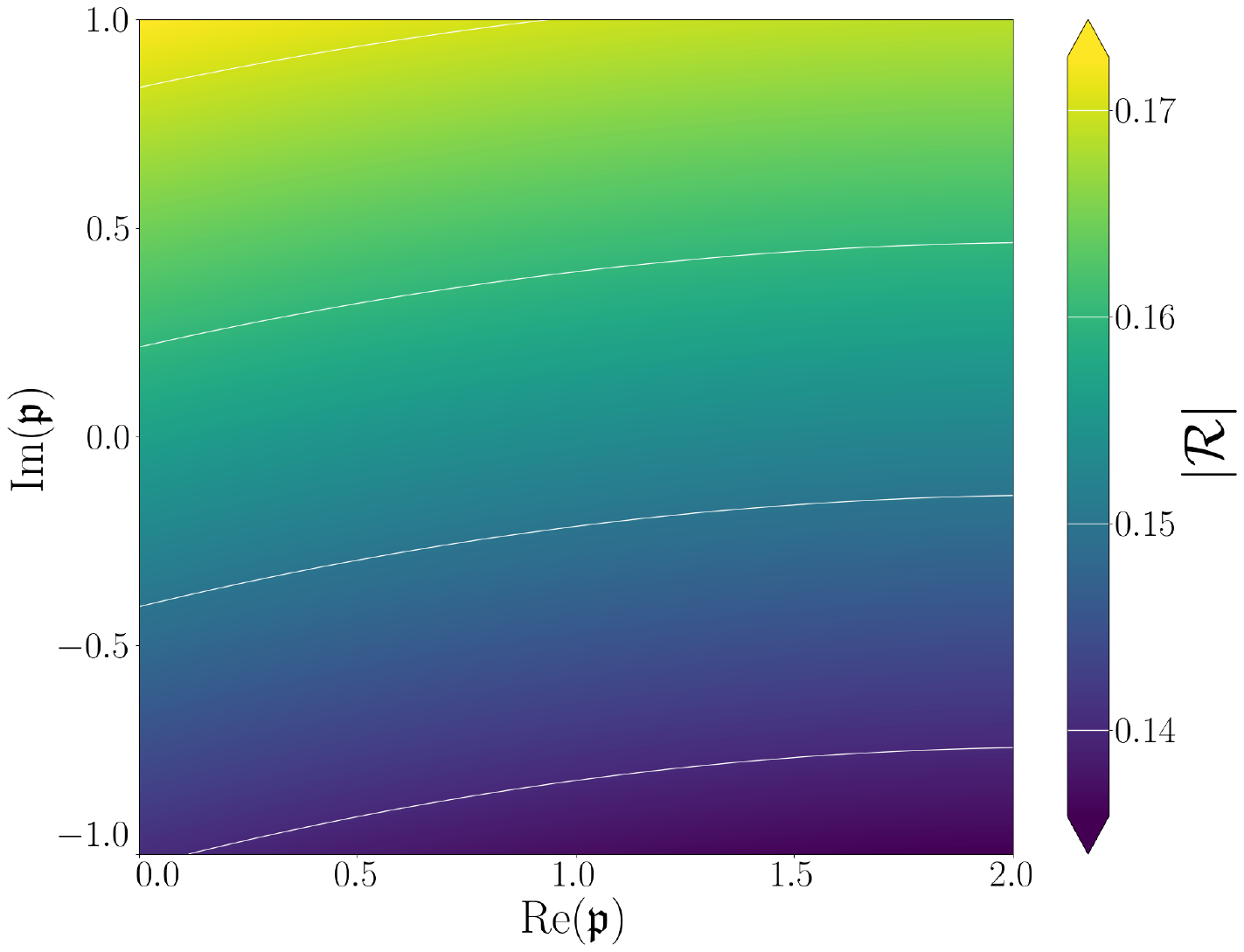}
    \caption{Dependence of the amplitude ratio $\mathcal{R}$ for the fundamental quadratic mode $(220+)\times(220+)\rightarrow(44)$ on the polarization parameter $\mathfrak{p}$ of the linear mode $(220+)$. The center of the figure represents the equatorial symmetry case, $\mathfrak{p}=1$.}
    \label{fig:qqnm_ratio}
\end{figure}

The parameter $\mathfrak{p}$ can also be thought of as a measure of the parity content of a given mode, i.e.,
$\mathfrak{p} =  (\mathcal{A}_{+} + i \mathcal{A}_{-})/(\mathcal{A}_{+} - i \mathcal{A}_{-})$,
where $\mathcal{A}_{\pm}$ is the amplitude of the even (odd) component of the GW strain in Eq.~\eqref{eq:linearStrain}~\cite{Bucciotti:2024jrv}. 
Equatorial symmetry imposes $\mathfrak{p} = (-1)^{\ell + m}$ (consistent with either $\mathcal{A}_{+}$ or $\mathcal{A}_{-}$ being zero), but in the general case $\mathfrak{p}$ is a complex number. 
At second order, the quadratic ratio $\mathcal{R}$ and the polarization parameter $\mathfrak{p}^{(2)}$ of the quadratic mode are given by
\begin{align}
    \mathcal R &= \frac{1}{4} \sum_{P_1, P_2} \mathcal{R}_{P_1 \times P_2 \rightarrow P} \big(1 + P_1 \mathfrak{p}_1^{(1)} \big) \big(1+ P_2 \mathfrak{p}_2^{(1)} \big)\,, \label{eq:quadraticRatio_schwarschild} \\
    \mathfrak{p}^{(2)} &= \frac{\sum_{P_1, P_2} P \, \mathcal{R}_{P_1 \times P_2 \rightarrow P} \big(1 + P_1 \mathfrak{p}_1^{(1)} \big) \big(1+ P_2 \mathfrak{p}_2^{(1)} \big)}{\sum_{P_1, P_2} \mathcal{R}_{P_1 \times P_2 \rightarrow P} \big(1 + P_1 \mathfrak{p}_1^{(1)} \big) \big(1+ P_2 \mathfrak{p}_2^{(1)} \big)}\,,
\end{align}
where the sum runs over all parities of the linear modes $P_1$, $P_2$, $P$ being given by Eq.~\eqref{eq:selection_rule_parity}, and $\mathcal{R}_{P_1 \times P_2 \rightarrow P}$ are the quadratic ratios for individual parities. 
In equatorial symmetry $\mathfrak{p}_i = (-1)^{\ell_i + m_i}$ ($i=1,2$), recovering previous results with parities given by $P_i =  (-1)^{\ell_i + m_i}$. 
It is interesting to consider $\mathcal{R}$ as a function of the linear polarization parameters $\mathcal{R}(\mathfrak{p}_1^{(1)},\,\mathfrak{p}_2^{(1)})$, fixing the remaining quantum numbers. The variations of $\mathcal{R}$ coming from its dependence on $\mathfrak{p}$ are plotted in Fig.~\ref{fig:qqnm_ratio} (see also~\cite{Bourg:2024jme}) for the self-coupling case $(220+)\times(220+)\rightarrow(44)$.
In addition, it was numerically noticed that the the following combination vanishes identically:
\begin{equation}
    \mathcal{R}(+1,+1)+\mathcal{R}(-1,-1)-2\mathcal{R}(0,0)\equiv 0\,.
\end{equation}
As a consequence, there is a linear relation between the four definite parity ratios $\mathcal{R}_{P_1\times P_2\to P}$, only three of them being independent. This property generalizes to Kerr, as will be explained in the next section.

\subsection{Quadratic modes in Kerr} \label{sec:nonlinKerr}

\vspace{-.1cm}

\noindent \textit{Initial contributors: Ma, Yang}

\vspace{.2cm}

An essential tool to investigate quadratic effects in Kerr BHs is the second-order Teukolsky formalism~\cite{Campanelli:1998jv,Loutrel:2020wbw,Ripley:2020xby,Green:2019nam,soton469806}, that has been applied to compute QQNMs and mode couplings~\cite{Ma:2024qcv,Khera:2024yrk}. 
Here we summarize the main results of this body of work.

\subsubsection{Classification of mode couplings.}
As in Section~\ref{sec:Teukolsky_symms}, a QQNM can be represented in terms of the linear modes that drive it with the notation $(\ell_1, m_1, n_1, \mathfrak{m}_1; \ell_2, m_2, n_2, \mathfrak{m}_2)$, and its time and azimuthal dependence are given by $e^{-i(\omega_{\ell_1,m_1,n_1}^{\mathfrak{m}_1}+\omega_{\ell_2,m_2,n_2}^{\mathfrak{m}_2} ) t} e^{i(m_1+m_2)\phi}$. By the usual angular momentum selection rules, this QQNM appears in the harmonics with azimuthal number $m_Q=(m_1+m_2)$. In the following, to reduce clutter in the notation, we denote $(l,m,n,\mathfrak{m})$ by a single multi-index $L$, and $(l,-m,n,-\mathfrak{m})$ by $\bar{L}$.

A QQNM can be excited by interactions between four distinct linear modes: $L_1, \bar{L}_1,L_2$, and $\bar{L}_2$, with respective amplitudes $A_1$, $A_{-1}$, $A_2$, and $A_{-2}$. These interactions can be categorized into four channels:
\begin{itemize}
    \item $L_1\otimes L_2$ (hereafter $++$): The two linear modes, $L_1$ and $L_2$, generate frequency components $\omega_{\ell_1,m_1,n_1}^{\mathfrak{m}_1}$ and $\omega_{\ell_2,m_2,n_2}^{\mathfrak{m}_2}$ in the linearized metric. The interaction between these components results in the QQNM frequency.
    \item $\bar{L}_1\otimes L_2$ (hereafter $-+$): The linear mode $\bar{L}_1$ generates a frequency component $-\bar{\omega}_{\ell_1-m_1n_1}^{-\mathfrak{m}_1}$ in the linearized metric, which is numerically equal to $\omega_{\ell_1m_1n_1}^{\mathfrak{m}_1}$ due to Eq.~\eqref{eq:mirror mode symmetry}. 
    At the same time, the second linear mode $L_2$ produces a frequency component $\omega_{\ell_2,m_2,n_2}^{\mathfrak{m}_2}$. The product of these components leads to the QQNM frequency.
    \item $L_1\otimes \bar{L}_2$ (hereafter $+-$): This case is analogous to $-+$, but with the roles of $L_1$ and $L_2$ exchanged $(1\leftrightarrow2)$.
    \item $\bar{L}_1\otimes \bar{L}_2$ (hereafter $--$): The two linear modes, $\bar{L}_1$ and $\bar{L}_2$, generate frequency components $-\bar{\omega}_{\ell_1,-m_1,n_1}^{-\mathfrak{m}_1}$ and $-\bar{\omega}_{\ell_2,-m_2,n_2}^{-\mathfrak{m}_2}$ in the linearized metric, which are numerically equal to $\omega_{\ell_1m_1n_1}^{\mathfrak{m}_1}$ and $\omega_{\ell_2m_2n_2}^{\mathfrak{m}_2}$. Their coupling leads to the QQNM frequency. Intriguingly, a careful calculation~\cite{Khera:2024yrk,Ma:2024qcv} shows that this coupling channel always vanishes identically, regardless of the types of parent modes and the spin of the Kerr BH.
\end{itemize}
The representation of $L$ and $\bar{L}$ is a more natural basis than the parity basis for analyzing QQNMs for a generic Kerr background, as the time and azimuthal dependence are conveniently factorized to calculate individual channels using the Teukolsky formalism. In addition, the vanishing coupling between two conjugate modes $\bar{L}_1$ and $\bar{L}_2$ is nontrivial, and is most conveniently expressed in the $L$ basis.
As a result, the total amplitude of the QQNM, $A_Q$, can be written as 
\begin{align} \label{eq:quadraticAmpl_Kerr}
    A_Q=&\mathcal{R}_{++}A_1A_2+\mathcal{R}_{+-}A_1\bar{A}_{-2}+\mathcal{R}_{-+}\bar{A}_{-1}A_2\,,
\end{align}
where $\mathcal{R}_{++},\mathcal{R}_{-+}$, and $\mathcal{R}_{+-}$ depend only on the dimensionless spin of the Kerr BH. 
This equation is the Kerr counterpart of Eq.~\eqref{eq:quadraticRatio_schwarschild} in the Schwarzschild case.
In general, the four linear modes $L_{1,2}$ and $\bar{L}_{1,2}$ are excited independently, and so are their amplitudes $A_{\pm1,\pm2}$. 
Therefore, the connection between the quadratic amplitude and the linear amplitudes is initial-data dependent.
In the special case of parity symmetry (as is the case for, e.g., nonprecessing binary mergers), mirror and ordinary mode amplitudes satisfy Eq.~\eqref{eq:equatorial_sym_relation_amplitudes}.
Consequently we have
\begin{align}
    \mathcal{R}^{\rm total} \equiv \frac{A_Q}{A_1A_2}=&\mathcal{R}_{++}+(-1)^{\ell_2}\mathcal{R}_{+-}+(-1)^{\ell_1}\mathcal{R}_{-+}. \label{eq:QQNM_kerr_R_total}
\end{align}
In this case, the amplitude ratio becomes an intrinsic property of the Kerr BH, and reflects the total excitability of the QQNM.

\subsubsection{Excitation of a mirror QQNM}
Similarly to linear QNMs, where $L$ and $\bar{L}$ are mirrored counterparts, a QQNM $(L_1;L_2)$ also has its mirror, represented by $(\bar{L}_1;\bar{L}_2)$. Here we show that their respective amplitude ratios are directly related via the parity transformation $P_-$, defined by $(\theta,\phi)\to(\pi-\theta,\phi+\pi)$.

Under the parity transformation, a strain $h(\theta,\phi)$ is mapped to $\bar{h}(\pi-\theta,\phi+\pi)$~\cite{Boyle:2014ioa}. Consequently, a linear QNM transforms as follows:
\begin{align}
    A_{L} e^{-i\omega_L u} \swSH{-2}{\ell{m}}{\theta,\phi}{c_L} &\to \bar{A}_{L} e^{i\bar{\omega}_L u} {}_{-2}\bar{S}_{\ell m}(\pi-\theta,\phi+\pi;c_L)\notag \\
    &=(-1)^{\ell+m}\bar{A}_{L} e^{-i\omega_{\bar{L}} u} \swSH{-2}{\ell{-m}}{\theta,\phi}{c_{\bar{L}}}, \label{eq:Kerr_QQNM_parity_transformation}
\end{align}
where we have used Eqs.~\eqref{eq:swSF_all_ident} and \eqref{eq:mirror mode symmetry}. Therefore, the parity transformation takes the mode $L$ with amplitude $A_L$ to the mode $\bar{L}$ with amplitude $(-1)^{\ell+m}\bar{A}_{L}$.

Now consider the mirror QQNM $(\bar{L}_1;\bar{L}_2)$ and its $++$ channel, where $\bar{L}_1$ with amplitude $A_{\bar{L}_1}$ interacts with $\bar{L}_2$ with amplitude $A_{\bar{L}_2}$. The quadratic amplitude excited through this channel can be written as
\begin{align}
    A^{++}_{(\bar{L}_1;\bar{L}_2)}= \mathcal{R}_{++}^{(\bar{L}_1;\bar{L}_2)} A_{\bar{L}_1}A_{\bar{L}_2}.
\end{align}
Since the Kerr BH spin is an axial vector and is unaffected by parity transformations, the parity-transformed $++$ channel remains a real process for the same Kerr BH. This leads to a QQNM $(L_1;L_2)$ with amplitude $(-1)^{\ell_Q+m_Q}\bar{A}^{++}_{(\bar{L}_1;\bar{L}_2)}$, excited by two linear modes $L_1$ and $L_2$ with amplitudes $(-1)^{\ell_1+m_1}\bar{A}_{\bar{L}_1}$ and $(-1)^{\ell_2+m_2}\bar{A}_{\bar{L}_2}$, respectively. By definition, we have 
\begin{align}
    \mathcal{R}_{++}^{(L_1;L_2)}\equiv \frac{(-1)^{\ell_Q+m_Q}\bar{A}^{++}_{(\bar{L}_1;\bar{L}_2)}}{(-1)^{\ell_1+\ell_2+m_1+m_2}\bar{A}_{\bar{L}_1}\bar{A}_{\bar{L}_2}}=(-1)^{\ell_Q+\ell_1+\ell_2} \bar{\mathcal{R}}_{++}^{(\bar{L}_1;\bar{L}_2)},
\end{align}
where we have used the relation $m_Q=m_1+m_2$. In fact, this argument extends to any channel, leading to
\begin{align}
    \mathcal{R}_{c_1c_2}^{(L_1;L_2)} =(-1)^{\ell_Q+\ell_1+\ell_2} \bar{\mathcal{R}}_{c_1c_2}^{(\bar{L}_1;\bar{L}_2)}, && {\rm with}\quad c_1,c_2=(\pm,\pm). \label{eq:QQNM_mirror_amp_ratio}
\end{align}
The physical interpretation of Eq.~\eqref{eq:QQNM_mirror_amp_ratio} is clear: the amplitude ratio for a given channel of a QQNM is directly related to the amplitude ratio of the corresponding channel of the mirrored QQNM.

\subsubsection{Quadratic effect of parity-definite QNMs.} By leveraging Eq.~\eqref{eq:QQNM_mirror_amp_ratio}, we now show that the QQNM excited by parity-definite QNMs also has a definite parity.

Let us first consider two linear parity-definite modes 
\begin{align}
    A_{L_i} e^{-i \omega_{L_i}u} +(-1)^{\ell_i+m_i}P_i\bar{A}_{L_i} e^{-i\omega_{\bar{L}_i}u}, && {\rm with}\quad i=1,2,
\end{align}
where we have omitted the angular dependence for brevity. It is straightforward to show, by Eq.~\eqref{eq:Kerr_QQNM_parity_transformation}, that $P_i=\pm 1$ corresponds to even (odd) parity. The linear QNMs excite a QQNM $(L_1;L_2)$ and its mirror $(\bar{L}_1;\bar{L}_2)$, which read
\begin{align}
    &A_{L_1}A_{L_2} e^{-i(\omega_{L_1}+\omega_{L_2})u}\left[\mathcal{R}_{++}^{(L_1;L_2)}+(-1)^{\ell_2+m_2}P_2\mathcal{R}_{+-}^{(L_1;L_2)}+(-1)^{\ell_1+m_1}P_1\mathcal{R}_{-+}^{(L_1;L_2)}\right] + e^{-i(\omega_{\bar{L}_1}+\omega_{\bar{L}_2})u}  \notag \\
    &\times P_1P_2\bar{A}_{L_1}\bar{A}_{L_2} \left[(-1)^{\ell_1+\ell_2+m_1+m_2}\mathcal{R}_{++}^{(\bar{L}_1;\bar{L}_2)}+(-1)^{\ell_1+m_1}P_2\mathcal{R}_{+-}^{(\bar{L}_1;\bar{L}_2)}+(-1)^{\ell_2+m_2}P_1\mathcal{R}_{-+}^{(\bar{L}_1;\bar{L}_2)}\right]. \label{eq:Kerr_QQNM_parity_definite}
\end{align}
Using Eq.~\eqref{eq:QQNM_mirror_amp_ratio}, the second term (namely, that involving $\omega_{\bar{L}_1}+\omega_{\bar{L}_2}$) can be converted to 
\begin{align}
    &(\ldots) + e^{-i(\omega_{\bar{L}_1}+\omega_{\bar{L}_2})u}  \notag \\
    &\times P_1P_2(-1)^{\ell_Q+m_Q}\bar{A}_{L_1}\bar{A}_{L_2} \left[\bar{\mathcal{R}}_{++}^{(L_1;L_2)}+(-1)^{\ell_2+m_2}P_2\bar{\mathcal{R}}_{+-}^{(L_1;L_2)}+(-1)^{\ell_1+m_1}P_1\bar{\mathcal{R}}_{-+}^{(L_1;L_2)}\right].
\end{align}
Notice that the bracket of the mirror QQNM is now the complex conjugate of the first bracket in Eq.~\eqref{eq:Kerr_QQNM_parity_definite}, which immediately reveals that the parity of the QQNM $(L_1;L_2)$ is $P_1P_2$. This can be summarized into the rule:
\begin{align}
    ({\rm even})\otimes({\rm even})\to ({\rm even}), \quad 
    ({\rm odd})\otimes({\rm odd})\to ({\rm even}), \quad
    ({\rm  odd})\otimes({\rm even})&\to ({\rm odd}), \notag
\end{align}
which extends the rule in Eq.~\eqref{eq:selection_rule_parity} to the Kerr case.
Similar to nonprecessing binaries, we can define a total amplitude ratio $R^{\rm total}_{\rm parity}$ for the parity-definite QQNM
\begin{align}
    \label{eq:quadraticRatio_kerr_parity_definite}
    \mathcal{R}^{\rm total}_{\rm parity}=\mathcal{R}_{++}^{(L_1;L_2)}+(-1)^{\ell_2+m_2}P_2\mathcal{R}_{+-}^{(L_1;L_2)}+(-1)^{\ell_1+m_1}P_1\mathcal{R}_{-+}^{(L_1;L_2)}\,,
\end{align}
which is also initial-data independent. 

\subsubsection{Computational methods.} 
The discussion of quadratic mode couplings in the Kerr BH spacetime presented so far has been somewhat qualitative. We now turn to two quantitative methods to compute the values of the amplitude ratios $\mathcal{R}_{c_1c_2}$.

The first method adopts an analytical approach~\cite{Ma:2024qcv}. Consider the second-order perturbation of $\Psi_4$,
\begin{align}
    \Psi_4=\epsilon \Psi_4^{(1)} + \epsilon^2 \Psi_4^{(2)} + \mathcal{O}(\epsilon^3)\,,
\end{align}
with $\epsilon$ being a book-keeping parameter. The dynamics of $\Psi_4^{(2)}$ is governed by~\cite{Campanelli:1998jv} 
\begin{align}
    \mathcal{T} \left[\bar{\Gamma}^4\Psi_4^{(2)}\right] = S_4^{(2)},  \label{eq:QQNM_second_order_Teukolsky_eqn}
\end{align}
where the Teukolsky operator $\mathcal{T}$ is defined in Eq.~\eqref{eq:Teukolsky_Master}, $\Gamma=r+ia\cos\theta$, and the source term $S_4^{(2)}$ is a complicated function of the linearized metric $h_{ab}^{(1)}$. The perturbative metric $h_{ab}^{(1)}$ is defined by the expansion
\begin{equation}
    g_{ab} = g^{(0)}_{ab} + \epsilon h_{ab}^{(1)} + \epsilon^2 h_{ab}^{(2)} + \mathcal{O}\left(\epsilon^3\right)\,. \label{eq:QQNM_kerr_metric_second_order_pert}
\end{equation}
For a Kerr BH, a linear QNM is described in terms of $\Psi_4^{(1)}$. Therefore, to derive the source term $S_4^{(2)}$ in Eq.~\eqref{eq:QQNM_second_order_Teukolsky_eqn}, it is necessary to solve for $h_{ab}^{(1)}$ from $\Psi_4^{(1)}$ through a process known as \emph{metric reconstruction}. This method was pioneered by Chrzanowski~\cite{Chrzanowski:1975wv} and Cohen and Kegeles~\cite{COHEN19755,Kegeles:1979an} for source-free scenarios. To be specific, it introduces an auxiliary scalar called the ``Hertz potential,'' which satisfies the $s=2 \,(s=-2)$ Teukolsky equation in the outgoing (ingoing) radiation gauge. Once determined, the linearized metric $h_{ab}^{(1)}$ and all Newman-Penrose spinors can be derived by differentiating the Hertz potential. Wald elegantly showed that this is feasible through an operator identity linking the Teukolsky operator and the linearized Einstein operator, which governs the dynamics of the linearized metric~\cite{Wald:1978vm}.

Metric reconstruction is performed analytically in the outgoing radiation gauge in~\cite{Ma:2024qcv}, as this gauge directly relates the solution to physical observables at infinity.
The procedure is made analytically tractable by using commutation relations between three derivative operators and other quantities in the Newman-Penrose formalism~\cite{Ma:2024qcv}. 
It was later realized that this approach is very similar to the Held formalism~\cite{1974CMaPh..37..311H,Spiers:2024src}.

The reconstruction yields an analytic expression for the source $S_4^{(2)}$, which depends on both $r$ and $\theta$:
\begin{align}
     S_4^{(2)}=&A_1(r,\theta) \left[\mathcal{L}_2^\dagger S(\theta)\right]^2 + A_2(r,\theta) \left[S(\theta)\,\mathcal{L}_1^\dagger\mathcal{L}_2^\dagger S(\theta)\right]+ A_3(r,\theta) \left[3ia\sin\theta\,S \, \mathcal{L}_2^\dagger S(\theta) \right] \notag \\
     &+ A_4(r,\theta) \left[a\sin\theta S(\theta)\right]^2,
\end{align}
where $S(\theta)$ is the spin-weighted $-2$ spheroidal harmonic. 
The operator $\mathcal{L}_n^\dagger$ is given by
\begin{align}
    \mathcal{L}_n^\dagger=\partial_\theta+ i\csc\theta\partial_\phi+ia \sin\theta\partial_t+n\cot\theta.
\end{align}
The angular dependence of the coefficients $A_{1..4}$ is solely determined by $\Gamma$: $A_{1..4}\sim \Gamma^j/\bar{\Gamma}^k$,
with $j=0,1$ and $k=1..7$. 
Asymptotically, the source follows $S_4^{(2)}\sim\mathcal{O}(r^2)$.
Since the source $S_4^{(2)}$ cannot be separated into a product of purely angular and radial components, an angular projection of the equation is necessary. This can be performed by using the orthogonality of spin-weighted spheroidal harmonics with different $\ell$, but same $\omega$ and $m$~\cite{Ma:2024qcv}.

Finally, the second-order Teukolsky equation in Eq.~\eqref{eq:QQNM_second_order_Teukolsky_eqn} can be solved using the shooting method along a complex contour, as shown in Fig.~\ref{fig:contour_products}. Two solutions satisfying the QNM boundary condition are shot separately on the left and right vertical paths. The value of $\mathcal{R}_{c_1c_2}$ can be obtained by matching the two solutions in the middle~\cite{Ma:2024qcv}.

The second method~\cite{Khera:2024yrk} uses hyperboloidal slicing~\cite{Ripley:2022ypi,Zenginoglu:2011jz,Zenginoglu:2007jw,PanossoMacedo:2024nkw} to solve the reduced second-order Teukolsky equation~\cite{Green:2019nam,soton469806}.
The coordinates $(T,\Phi,R)$ are defined by the transformations
\begin{align}
    \mathrm{d} T = \mathrm{d} t + \frac{r^2 + a^2}{\Delta} \mathrm{d} r - 2\left(1+\frac{2M}{r}\right) \mathrm{d} r\,,\quad \mathrm{d}\Phi = \mathrm{d}\phi + \frac{a \mathrm{d} r}{\Delta}\,, \quad R = \frac{1}{r}\,.
\end{align}
A pseudospectral method, where all quantities are expanded in Chebyshev polynomials in the radial direction and spin-weighted spherical harmonics in the angular directions, is adopted to treat the system fully numerically.

\begin{figure}[t]
    \centering
    \includegraphics[width=\textwidth]{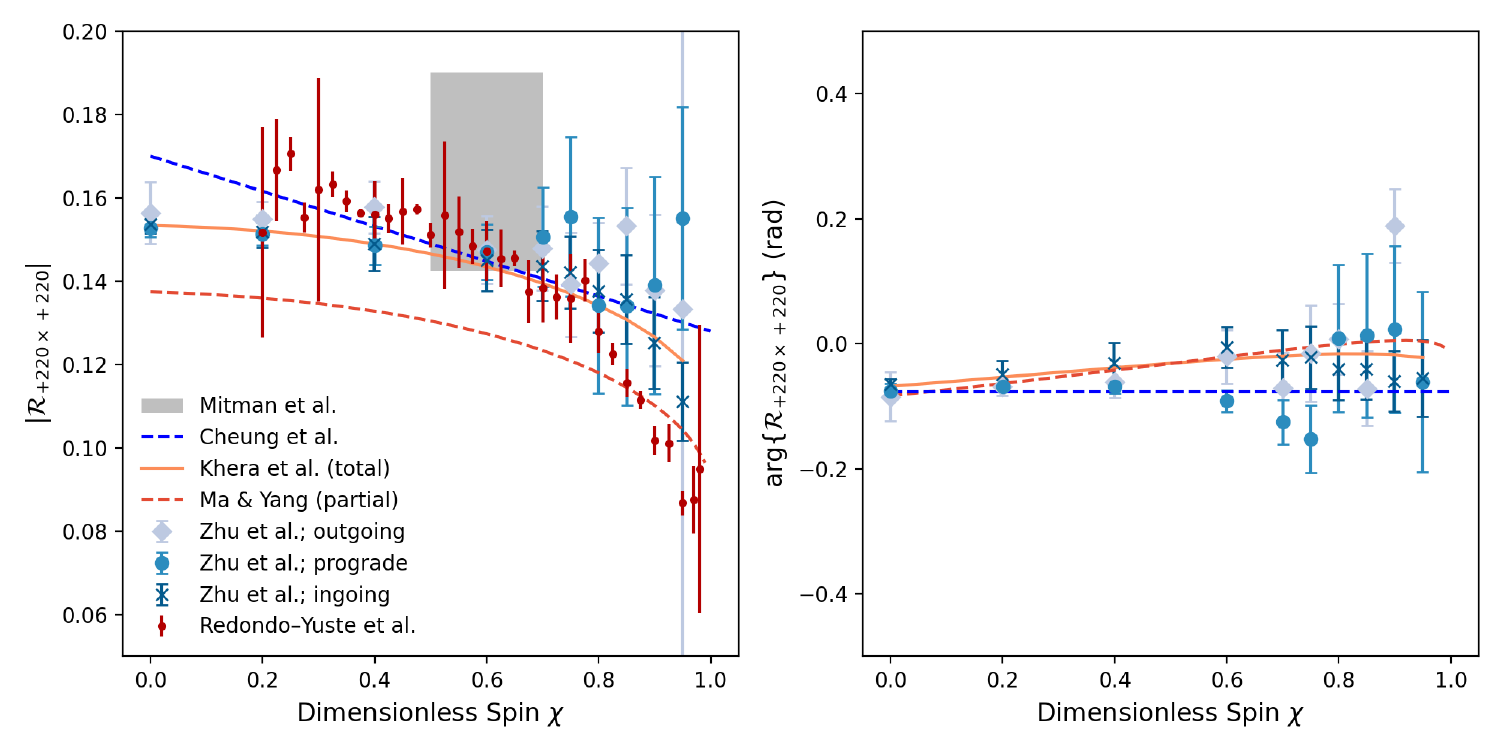}
    \caption{Absolute value (left) and phase (right) of the quadratic ratio $\mathcal{R}_{220\times220}$, assuming equatorial symmetry, as computed using different approaches (see legend). The results of Ma \& Yang~\cite{Ma:2024qcv} (referred to as partial, since they only include the $(220+)\times(220+)$ channel), Khera et al.~\cite{Khera:2024yrk} (referred to as total, since it includes all possible contributions in equatorial symmetry), and Redondo-Yuste et al.~\cite{Redondo-Yuste:2023seq} are based on second-order perturbation theory, while the results of Zhu et al.~\cite{Zhu:2024rej} use nonlinear scattering experiments, for different configurations of the initial Gaussian pulse (ingoing, outgoing, and prograde). The amplitudes from Mitman et al.~\cite{Mitman:2022qdl} and Cheung et al.~\cite{Cheung:2023vki} are extracted from fully nonlinear, quasi-circular BBH merger simulations.
    Figure adapted from~\cite{Zhu:2024rej}.
    }
    \label{fig:Quadratic_Ratio}
\end{figure}

The source of the reduced second-order Teukolsky equation is computed numerically from the Einstein tensor. Specifically, considering the second-order perturbation of the metric in Eq.~\eqref{eq:QQNM_kerr_metric_second_order_pert},
the Einstein tensor can be written as
\begin{align}
    G_{ab}[g_{ab}] = \epsilon G_{ab}^{(1)}[h_{ab}^{(1)}] + \epsilon^2 G_{ab}^{(1)}[h_{ab}^{(2)}] + \epsilon^2G_{ab}^{(2)}[h_{ab}^{(1)}, h_{ab}^{(1)}] +\mathcal{O}\left(\epsilon^3\right)\,. \label{eq:QQNM_Einstein_Tensor}
\end{align}
Einstein's equations at second order lead to
\begin{equation}
    G_{ab}^{(1)}[h_{ab}^{(2)}] = - G_{ab}^{(2)}[h_{ab}^{(1)}, h_{ab}^{(1)}]\,.
\end{equation}
Therefore, $- G_{ab}^{(2)}[h_{ab}^{(1)}, h_{ab}^{(1)}]/8\pi$ acts as an effective stress-energy tensor driving the metric at second order, which can then be converted to the source term for the reduced Teukolsky equation~\cite{Khera:2024yrk}. 
Using the outgoing radiation gauge, the reduced second-order Teukolsky equation~\cite{Green:2019nam,soton469806} is equivalent to the Campanelli-Lousto approach~\cite{Campanelli:1998jv}: $\Psi_4^{(2)}$ arises solely from $h_{ab}^{(2)}$, and the quadratic contribution from $h_{ab}^{(1)}$ vanishes. Finally, the equation is solved with the QNM boundary conditions to compute $\mathcal{R}_{c_1c_2}$. 
As an example, in Fig.~\ref{fig:Quadratic_Ratio} we show the quadratic ratios $\mathcal{R}_{++}$ and $\mathcal{R}^{\rm total}$, as defined in Eq.~\eqref{eq:QQNM_kerr_R_total}, for the QQNM $(220+;220+)$, across BH spins $\chi\in[0,0.95]$. They are in remarkable agreement with the values obtained from numerical simulations.

One limitation of both methods is their inability to handle purely decaying QQNMs, which have vanishing real frequencies. 
For an alternative treatment of these modes, see Section~\ref{sec:bms-frames-memory}.

\subsection{Nonlinear numerical experiments: perturbation theory vs. full numerical relativity} \label{sec:nonlin_num_expe}

\vspace{-.1cm}

\noindent \textit{Initial contributors: Cheung, Mitman, Redondo-Yuste, Zhu}

\vspace{.2cm}

The physics driving the relaxation of a BH to equilibrium, for instance, after a BBH merger, is mostly understood through the lens of linear perturbation theory. 
After a transient phase, the newly formed BH is assumed to be perturbatively close to its final state, and therefore the relaxation is expected to be well described as a superposition of QNMs with amplitudes that depend on the initial perturbation, followed by a late-time tail. 
However, GR is a nonlinear theory, and especially in the strong-field regime, nonlinearities are expected to have a significant impact. The goal of this section is to review the description of BH ringdown by comparing nonlinear and perturbative results. 

Because of their nonlinear character, Einstein's equations are very hard to solve analytically. This is particularly true in the dynamical, strong-field regime, where most known approximation schemes fail. 
However, continuous progress in the field of NR allows for numerical solutions
with increasingly high accuracy.

Some problems in NR can also be addressed within BH perturbation theory. Paradigmatic examples are wave scattering off a BH~\cite{Vishveshwara:1970zz, Press:1971wr, Press:1973zz, Abrahams:1992wm, Zlochower:2003yh} and the head-on collision of two BHs~\cite{Anninos:1993zj,Anninos:1994gp,Sperhake:2011ik,Cheung:2022rbm,DeAmicis:2024eoy}. 
It is by virtue of these numerical experiments, where perturbation theory can be compared directly with fully nonlinear solutions, that we can understand more precisely the nonlinear content of the ringdown.

We first review numerical methods to deal with gravitational perturbations of BHs at the linear level and at higher orders, focusing on the comparison between the amplitude ratio between QQNMs and QNMs found from perturbative calculations and from scattering experiments. 
Then we consider fully nonlinear numerical experiments, including wave scattering and BBH mergers. We close by discussing how a time-varying background affects the relaxation of BHs at both the perturbative and nonlinear level.

\subsubsection{Perturbation theory}
Consider the problem of gravitational radiation scattering off a BH. In perturbation theory, the metric can be written as $g_{ab} = g^{(0)}_{ab} + \sum_{n=1} \epsilon^n h^{(n)}_{ab}$, where $g^{(0)}_{ab}$ is the background Kerr metric, 
and $\epsilon \ll 1$ is a small book-keeping parameter. The stress-energy tensor is assumed to be zero at all perturbative orders, i.e., we are not concerned with how the gravitational perturbation is generated, but rather with its evolution and propagation. Thus, we consider the time evolution of some initial data, 
\begin{equation}
    h_{ab}^{(n)}(t=0) = \mathring{h}_{ab}^{(n)} \, , \qquad \partial_t h_{ab}^{(n)}(t=0) = \mathring{k}_{ab}^{(n)} \, , 
\end{equation}
where $t$ is a causal coordinate.
For example, one can choose $t$ to be the Boyer--Lindquist coordinate time $t$ or, alternatively, the hyperboloidal time parameter $\tau$ (see Section~\ref{sec:scalar_products}).
The quantities $\mathring{h}_{ab}^{(n)}$ and  $\mathring{k}_{ab}^{(n)}$ are functions of the spatial coordinates alone. 
The evolution of the initial data is determined uniquely by \emph{linear} equations at each order that can be written in the form 
\begin{equation}\label{eq:Perturbative_EFE}
    (\mathcal{E} h^{(n)})_{ab} = \mathcal{S}^{(n)}(h^{(1)},\dots,h^{(n-1)}) \, , 
\end{equation}
where $(\mathcal{E} h)_{ab}$ is the linearized Einstein tensor (given by the Lichnerowicz operator in the Lorenz gauge), and $\mathcal{S}^{(n)}$ are source terms that involve second- and higher-order perturbations of the Einstein equations~\cite{Wardell:2024yoi}. As discussed previously, lower-order metric perturbations act as an effective stress-energy tensor for the higher-order perturbations. 

Perhaps the most relevant question in scattering experiments is how to specify the initial data $\{\mathring{h}_{ab}^{(n)},\mathring{k}_{ab}^{(n)}\}$. Once this is done, solving Eqs.~\eqref{eq:Perturbative_EFE} order by order is relatively straightforward from the mathematical point of view. Most early works on scattering in linear perturbation theory considered Gaussian-like profiles~\cite{Vishveshwara:1970zz, Press:1971wr, Krivan:1996da, Krivan:1997hc, Dorband:2006gg}. These have the advantage of being simple to implement, and of illustrating interesting physical content. For instance, one can directly characterize the excitation of different QNMs as a function of the width or of the spatial spread of the initial data~\cite{Dorband:2006gg,Bernuzzi:2008rq}. The disadvantage of using simple initial data like Gaussian wave packets is that it is harder to directly compare the results with nonlinear evolutions of the Einstein equations. 

There are three ways to overcome this obstacle. 

The simplest approach is to prescribe initial data for nonlinear simulations that mimic those chosen in perturbation theory~\cite{Abrahams:1992wm, Zhu:2024rej, Zhu:2024dyl}. Brill wave initial data~\cite{Abrahams:1992wm} provide a natural way of doing this when the perturbed BH does not rotate. Another alternative for rotating BHs is to specify an initial guess for $\{\mathring{h}_{ab}^{(n)},\mathring{k}_{ab}^{(n)}\}$, and rescale the metric by a conformal factor. The shift and conformal factor that solve the Hamiltonian and momentum constraints can be found numerically using, e.g., the conformal thin sandwich method~\cite{Zhu:2024rej,Zhu:2024dyl}. Then, one can read off the actual metric perturbations and their time derivatives. 

An alternative approach, the so-called ``close limit approximation''
(CLA)~\cite{Price:1994pm}, consists of starting from exact initial data that
describe two BHs. Originally, the initial data were based on the Misner wormhole
solution that describes a wormhole with a single asymptotic
region~\cite{Misner:1960zz}, and parameterized by the distance between the two
mouths. If this distance is small enough, both mouths are enclosed by a common
apparent horizon, and the Misner wormhole can be written explicitly as a
perturbation around Schwarzschild, order by order in the distance between the
mouths of the wormhole. This problem was studied at second order in perturbation
theory in a series of papers~\cite{Price:1994pm, Gleiser:1996yc, Gleiser:1997ng}
that found very good agreement with nonlinear simulations, even for cases where
the perturbative solution was not expected to hold up. The CLA was extended to
Brill--Lindquist initial data in~\cite{Abrahams:1994qu, Abrahams:1995wd}, and to
Bowen--York initial data in~\cite{Khanna:1999mh}. An extension to rotating
backgrounds is more challenging since the Kerr metric cannot be written as a
conformally flat metric, although there are ways to overcome this
difficulty~\cite{Krivan:1998er, Krivan:1998td, Dain:2001iw}.

A third approach is the Lazarus program~\cite{Campanelli:1997un, Campanelli:1998jv, Campanelli:1998uh, Campanelli:1998yt, Baker:1999sj, Baker:2001sf, Campanelli:2005ia}. The idea here is to read out the initial data for the perturbations around a Kerr BH directly from a snapshot of a numerical simulation, once the two merging BHs are very close to each other. This allowed to obtain hybrid waveforms (fully nonlinear during the late stages of the inspiral, and based on perturbation theory to describe the merger-ringdown) which, nonetheless, showed excellent agreement with NR simulations that were evolved through the merger, once those became available~\cite{Campanelli:2005dd}.

Once the initial data are prescribed, one has to solve the perturbation equations. We now describe the main practical methods employed for the solution, with a focus on perturbations around rotating BHs. We refer the interested reader to~\cite{Pazos:2010xf} for an equivalent calculation for second-order perturbations in Schwarzschild. The method presented here is based on~\cite{Loutrel:2020wbw, Ripley:2020xby}; alternative methods designed to describe perturbations sourced by the motion of point masses~\cite{Spiers:2023cip,Wardell:2024yoi} are not covered here.

Let $\{l^a,n^a,m^a,\bar{m}^a\}$ be a null tetrad aligned along the principal null directions. We work in the Newman-Penrose formalism (see~\cite{Pound:2021qin} for a review), and denote by $D=l^a\nabla_a$, $\Delta = n^a\nabla_a$ the contractions of the covariant derivatives along the principal null directions. We will use the Geroch-Held-Penrose operators $\eth,\eth^\prime$ to compute the covariant derivatives in the remaining directions. As discussed in Section~\ref{sec_21}, the vacuum first-order metric perturbation in the radiation gauge $h_{ab}^{(1)}$ is governed only by one homogeneous equation for the linear perturbation of the Weyl scalar with maximum spin weight, $\Psi_4^{(1)}$. 
This is the Teukolsky equation, which we can write in terms of Newman-Penrose quantities (see e.g.~\cite{Chandrasekhar:1985kt} for definitions) as  
\begin{equation}\label{eq:teuk_NP}
    \mathcal{O}[\Psi_4^{(1)}] \equiv \Bigl[(\Delta+4\mu+\bar{\mu})(D+4\epsilon-\rho)-(\eth^\prime+4\pi-\bar{\tau})(\eth-\tau)-3\Psi_2\Bigr]\Psi_4^{(1)}=0 \, .
\end{equation}
The second-order metric perturbations can also be encoded in the perturbation of the maximal spin weight Weyl scalars, e.g., $\Psi_4^{(2)}$, and they also satisfy the Teukolsky equation, but in this case, with a source~\cite{Campanelli:1998jv}:
\begin{equation}
    \mathcal{O}[\Psi_4^{(2)}] = \mathcal{T}[h^{(1)}_{ab}] \, .
\end{equation}
The source term cannot be written directly in terms of $\Psi_4^{(1)}$, but instead it involves different metric components of $h^{(1)}_{ab}$, perturbations of spin coefficients, as well as perturbations of other Weyl scalars. The issue of finding these quantities from a solution $\Psi_4^{(1)}$ of the homogeneous Teukolsky equation is known as the metric reconstruction problem, and many different approaches have been proposed to solve it~\cite{Chrzanowski:1975wv,Kegeles:1979an, Wald:1978vm, Green:2019nam, Toomani:2021jlo, Hollands:2024iqp, Aksteiner:2016pjt}. The approach put forward in~\cite{Loutrel:2020wbw} is easily implemented in order to reconstruct $h^{(1)}_{ab}$ in the absence of matter fields, and it casts the metric in the outgoing radiation gauge, which is well behaved numerically. In simple words, it consists in solving a set of seven nested transport equations (i.e., first order differential equations which do not involve time derivatives). The procedure is lengthy, but the resulting equations can easily be solved numerically. However, this framework is not well suited to push the formalism to higher orders in perturbation theory~\cite{Loutrel:2020wbw, Ripley:2020xby}. 

Once the source term is known, the inhomogeneous Teukolsky equation can be solved. A particularly efficient solution technique uses pseudospectral methods in hyperboloidal slicing. By choosing hyperboloidal coordinates $\{T,R,\theta,\phi\}$, the boundary conditions that select QNMs are satisfied automatically, since the constant-$T$ slices become null at the horizon and at future null infinity (see Section~\ref{subsubsec:hyperboloidal}). A generic unknown $f$ (e.g., $\Psi^{(1)}$) is then decomposed in a spectral basis, 
\begin{equation}
    f = \sum_{\ell,n} f_{\ell,n}(T) T_n(x){}_sP_{\ell m}(-\cos\theta)e^{-im\phi} \, , 
\end{equation}
where $x=2R/r_+-1$ is the compact radial coordinate, $T_n(x)$ are Chebsyhev polynomials, and ${}_sP_{\ell m}$ are spin-weighted associated Legendre polynomials. The Teukolsky equation can be written in a first-order form as $\partial_T \mathbf{v} = \mathbf{A}\partial_R \mathbf{v}+\mathbf{B}\mathcal{D}^2\mathbf{v}+\mathbf{C}\mathbf{v}$, where $\mathcal{D}^2$ is the Laplace--Beltrami operator on the sphere, and the matrices $\mathbf{A},\mathbf{B},\mathbf{C}$ are read from the equation. In this form, the variables are evolved in time using, e.g., an explicit Runge--Kutta method. 

We can choose initial data corresponding to a Gaussian profile for $\Psi_4^{(1)}$, setting $\Psi_4^{(2)}=0$. Initially, the rest of the metric fluctuations are assumed to be zero (e.g., $h^{(1)}_{ll} = 0$). This is not consistent with $\Psi_4^{(1)}\neq 0$. However, if the initial data have compact support, one can show~\cite{Ripley:2020xby} that constraint violating modes propagate towards the horizon, and therefore they will exit the computational domain after a short amount of time $T=T_R$. Then, in a certain sense, the ``real'' initial data are only specified at $T=T_R$, the time after which the evolution of $\Psi_4^{(2)}$ is started.

This method was used in~\cite{Redondo-Yuste:2023seq, Zhu:2024rej} to extract the amplitude ratio between the QQNM $(220)\times(220)$ and its parent mode $(220)$. In~\cite{Redondo-Yuste:2023seq} the amplitude of the modes was inferred with an agnostic algorithm based on Bayesian methods; in~\cite{Zhu:2024rej}, it was found by minimizing the mismatch between the signal and a template containing a superposition of QNMs. In both cases the recovered amplitudes were checked to be stable and independent of the starting time of the fit (within a certain appropriate range), and the two methods are in very good agreement. More interestingly, these studies indicate that the ratio between these modes as computed from perturbation theory is in very good agreement with the value extracted with nonlinear, NR simulations of merging BHs. In addition, both works show that the ratio between QQNMs and QNMs is only mildly dependent on the properties of the Gaussian pulse used as initial data. A summary of the different results for this nonlinear ratio is shown in Fig.~\ref{fig:Quadratic_Ratio}.
Note the good agreement between second-order perturbation theory calculations (including both TD scattering experiments and FD calculations) and the fully nonlinear evolutions.

\subsubsection{Fully nonlinear numerical experiments}
The perturbative regime can also be studied through numerical experiments which solve the full Einstein equations. Fully nonlinear simulations of BH spacetimes have a long history~\cite{Choptuik:2015mma}. In fact, the first attempts at simulating BBH collisions by Hahn and Lindquist in 1964~\cite{1964AnPhy..29..304H} predate the usage of the term ``black hole'' to describe these objects. 
Here we discuss simulations of waves scattering off isolated BHs, as well as BBH mergers.

\paragraph{Wave packet scattering}
One of the first ``modern'' fully nonlinear evolutions of a single perturbed BH~\cite{Abrahams:1992wm} studies a time-symmetric GW falling into a single BH in full NR and compares the results to the corresponding first-order evolution. The outgoing GW in the fully nonlinear simulation is well-modeled by the expected fundamental QNM and the result agrees remarkably well with the perturbative solution, as long as  the amplitude of the infalling perturbation is small. For larger perturbations, however, the mass of the BH can change by a large fraction, and the outgoing radiation can be significantly different from the predictions of first-order perturbation theory.

More recent work considers infalling radiation on an isolated BH to investigate whether certain modes of the outgoing radiation exhibit a quadratic scaling with respect to the initial amplitude of the perturbation~\cite{Zlochower:2003yh}. This work simulates infalling radiation of varying amplitude with angular structure corresponding to $(\ell,m)=(2,0)$ or $(2,2)$. In the $(2,0)$ case, excitations in the $(4,0)$ and $(6,0)$ modes scale as the square and cube of the initial perturbation's amplitude. By Fourier transforming the resulting waveform, the dominant frequency in the $(4,0)$ mode is found to be twice that of the dominant frequency in the $(2,0)$ mode, while two frequencies are found in the $(6,0)$ mode: one is three times the $(2,0)$ mode frequency, and the other is very close to the frequency of the $(2,0)$ mode. For the $(2,2)$ case, the $(2,0)$, $(4,0)$, and $(4,4)$ modes exhibit a very similar quadratic behavior, while the $(4,2)$, $(6,2)$, and $(6,6)$ modes exhibit a cubic behavior. 

In conclusion, these pioneering investigations already provide strong numerical evidence that a clear nonlinear structure in the ringdown should generically be present (and predictable) given the initial amplitude and angular structure of the perturbation.

\paragraph{Head-on mergers}

To date, all of the observed GW events, including those with a clear ringdown signal, come from binary compact object mergers. If nonlinear QNMs are within reach of current observatories, the first detections will most likely come from the ringdown of a remnant BH produced by a binary merger.
To understand the experimental data,
it is essential to study the excitation of nonlinear QNMs in NR simulations of binary mergers.

Astrophysically realistic simulations should target BBHs in inspiraling (and typically approximately circular) orbits, that eventually merge and form a single, spinning remnant.
The advantage of studying head-on collisions is that these systems are more symmetric (hence easier to simulate) and that the amplitude of the emitted GWs, including linear and nonlinear QNMs, can be controlled by tuning the initial speed (or equivalently, the center-of-mass energy) of the colliding BHs~\cite{Sperhake:2008ga,Cardoso:2014uka,Sperhake:2015siy,Healy:2015mla,Bozzola:2022uqu,Cheung:2022rbm}.
This is particularly useful to measure the scaling relationship between the amplitudes of nonlinear QNMs and their linear parent modes, because a set of simulations with a wide range of initial speeds can probe the relationship over a large range of amplitudes.
Quasi-circular simulations, while more realistic, typically span a smaller range of collision energies, 
so the scaling relationship can only be probed in a more restricted range of amplitudes.

In~\cite{Cheung:2022rbm}, head-on collisions of nonspinning, equal mass BHs are simulated with the \texttt{GRChombo} code~\cite{Andrade:2021rbd} over a relatively broad range of center-of-mass Lorentz factor boosts $\gamma \in [1.2, 3.2]$ (i.e., $\beta = v/c\in [0.553, 0.95]$) with the specific goal of measuring the excitation of nonlinear QNMs.
Under these symmetry assumptions, the simulation domain can be reduced to two dimensions. 
The remnant is a nonrotating Schwarzschild BH, so the frequencies of QNMs with the same azimuthal number $m$ are degenerate, meaning that each mode can be labeled by $(\ell,n)$.
By extracting the gravitational perturbations in terms of $\Psi_4$, this work finds QQNMs sourced by the $(2, 0)\times(2,0)$ and $(2,0)\times(4,0)$ linear QNMs  in the $\ell = 4$ and $\ell = 6$ harmonics, respectively, with amplitudes spanning an order of magnitude over the simulations.
The second-order mode amplitudes scale quadratically with  the amplitudes of their parent linear modes and the phases are the sum of the corresponding linear phases, as expected, implying that nonlinear QNMs are definitely present in these ultrarelativistic head-on simulations.

\paragraph{Inspiralling mergers}
The first evidence of QQNMs in BBH ringdown waveforms was presented more than a decade ago in~\cite{London:2014cma}. 
The work is mostly focused on the construction of fits of the linear QNM amplitudes as a function of the parameters of the progenitor through the analysis of 68 nonspinning BBH mergers with mass ratios between [1, 15]. Interestingly, the QNM superpositions that seem to best describe the $(4,4)$ and $(5,5)$ multipoles of the $\Psi_{4}$ Weyl scalar are found to contain mode frequencies close to the $(2,2,0)\times(2,2,0)$ and $(2,2,0)\times(3,3,0)$ QQNMs. 

More recent studies focusing on QQNMs in BBH simulations~\cite{Cheung:2022rbm,Mitman:2022qdl}, independently confirm the presence of QQNMs in quasi-circular, nonprecessing NR simulations using the \texttt{SpEC} code~\cite{SpECwebsite}. The amplitudes of the QQNMs scale quadratically with the amplitudes of their parent modes, providing clear evidence that QQNMs with the expected nonlinear behavior are present in quasi-circular BBH merger simulations.
This result has since received even more support through the analysis of a larger set of quasi-circular, nonprecessing BBH simulations presented in~\cite{Mitman:2025hgy}: see, e.g., Fig. 9 of.~\cite{Mitman:2025hgy}.
Another recent analysis focused on the stability of overtone fitting in NR waveforms finds evidence of at least four QQNMs in the dominant $(2,2)$ mode of quasi-circular, nonprecessing BH mergers~\cite{Giesler:2024hcr}. For highly-spinning remnants, there is evidence for QQNMs sourced by the linear QNM pairs $(2,-2,0,-)\times(4,4,0,+)$, $(3,-2,0,-)\times(4,4,0,+)$, $(2,2,0,+)\times(2,0,0,+)$, and $(2,2,0,+)\times(2,0,0,-)$. A second high-accuracy simulation producing a remnant with smaller spin shows evidence of the same four modes, as well as weaker evidence of the presence of QQNMs sourced by the $(2,2,0,+)\times(3,0,0,+)$, $(2,2,0,+)\times(3,0,0,-)$, $(2,2,0,+)\times(4,0,0,+)$, and $(2,2,0,+)\times(4,0,0,-)$ pairs.

While the QQNM frequencies can be simply predicted from BH perturbation theory, their angular structure is nontrivial. In particular, even though the angular structure of the parent modes is represented by spheroidal harmonics with an oblateness parameter equal to the product of the spin and the QNM frequency, the source at second order in perturbation theory is not separable, and its angular dependence is a complicated combination of spin-weighted spheroidal harmonics~\cite{Ma:2024qcv}.
This complexity can be illustrated numerically by leaving the spherical-spheroidal QQNM mixing coefficients as free parameters when fitting the ringdown phase of NR waveforms over the two-sphere. This analysis finds that the angular structure of the QQNMs does not match simple (and naive) expectations~\cite{Dyer:2024jfz}, as suggested in~\cite{Ma:2024qcv}.
This complexity offers interesting opportunities: spatial information can be extracted from NR by fitting a feature with known time dependence to all of the spherical harmonic modes, allowing the shape of the feature to be reconstructed. This program was initiated in~\cite{Dyer:2024jfz} and called ``BH cartography.'' BH cartography allows us, at least in principle, to determine the viewing angles that maximize the amplitude of the quadratic QNMs. This can be an important guide for future searches.

\paragraph{Horizons}
The QNMs observed in binary merger signals represent perturbations propagating out to infinity.
However, GWs also propagate into the BH horizon.
It has been shown that the shear of the out-going light rays at the common horizon of a BBH merger can be accurately described by a superposition of QNMs that match those observed at infinity~\cite{Mourier:2020mwa}.
It is then natural to expect that nonlinear QNMs should also be observable within the evolution of the shear, as the dynamics of the horizon should be nonlinear.
Indeed, the $(2,0)\times(2,0), (2,0)\times(4,0)$ and possibly the $(2,0)\times(6,0)$ or $(4,0)\times(4,0)$ QQNM are found, respectively, in the $\ell = 2,4,6$ horizon shear modes in head-on BBH collisions~\cite{Khera:2023oyf}.
The amplitudes of the QQNMs follow the expected quadratic scaling relationship with respect to the parent linear QNMs.
However, the scaling coefficient is found to be different when studying a set of simulations with two BHs of fixed mass ratios but varying initial momentum, versus simulations with varying mass ratios but fixed initial momentum.
This is unexpected, given that quadratic amplitude ratios should be independent of initial conditions, beside parity.
Further work is needed to understand the source of the discrepancy.

\subsubsection{Changing background}\label{sec:mtaft}
During a BBH coalescence, a significant portion of the system's energy resides outside the newly formed horizon as perturbations, some of which fall back into the BH while others propagate to infinity, producing the ringdown waveform.
Infalling of perturbations increase the remnant BH's mass (as computed from the apparent horizon properties)
by as much as 5\% in equal-mass mergers~\cite{Pretorius:2005gq, Zhu:2024dyl}.
This substantial change in the BH background raises questions about the validity of traditional perturbation theory at the peak of the waveform.
Specifically, can the GWs be modeled as a superposition of QNMs when the underlying spacetime itself is evolving? If so, to what extent? 

This issue is not new, as debates around nonlinearity in ringdown go back to early efforts like the Lazarus project~\cite{Campanelli:1997un, Campanelli:1998jv, Campanelli:1998uh, Campanelli:1998yt, Baker:1999sj, Baker:2001sf, Campanelli:2005ia}, prior to the first nonlinear simulation of BBH merger~\cite{Pretorius:2005gq}, and continued when these simulations first became available~\cite{Buonanno:2006ui, Berti:2007fi}. 
Early comparisons between time evolutions of identical initial data describing a single perturbed BH in NR and in linear perturbation theory show that nonlinear waveforms can significantly deviate from their linear counterparts when the amplitude of the initial perturbation is sufficiently large~\cite{Baker:1999sj}.

Another way to quantify this nonlinear effect is to perform perturbation theory on a time-dependent background. A good toy model is provided by the Vaidya spacetime, where the change in the background is driven by accretion of null dust (instead of the in-falling perturbations characteristic of BBH mergers) and a nonlinearly changing mass function can be arbitrarily specified. 
The scalar QNMs of the Vaidya spacetime are known~\cite{Abdalla:2006vb} and its nonadiabatic behavior was quantified in a study of time-dependent Reissner-Nordstr\"om BHs~\cite{Chirenti:2011rc}.
The imprint on changing QNM frequencies in a Vaidya spacetime with a mass function resembling what is observed in BBH mergers was studied for gravitational perturbations in~\cite{Redondo-Yuste:2023ipg}. 
If the mass function changes slowly, the frequencies follow an adiabatic evolution such that the product $\omega M$ remains approximately constant. 
This simple model can be improved by studying the problem perturbatively in the derivatives of the mass function~\cite{Capuano:2024qhv}.
Including this information in the ringdown model can significantly improve the fit of the waveform in a changing-mass background, even when the change due to the mass function is far from adiabatic~\cite{Redondo-Yuste:2023ipg}.

Alternatively, one can try to capture the time-evolution of the background through an adiabatic approximation~\cite{Sberna:2021eui}. This work compares the nonlinear and linear evolution of a spherically symmetric scalar field in an asymptotically AdS Schwarzschild BH background, with initial data given by an exact QNM profile. Because of spherical symmetry and the ``compact'' nature of the spacetime provided by the AdS boundary conditions, the numerical evolutions are very accurate. 
For the linear evolution, the mass flux of the scalar field is tracked through the horizon, and the background mass is updated as a function of time during the course of the evolution. Reference~\cite{Sberna:2021eui} shows that an infalling linear mode can nonlinearly excite other scalar modes due to the changing mass of the background. 
A similar study of linear gravitational perturbations of (asymptotically flat) Kerr BHs yields similar results~\cite{May:2024rrg}.

In general, these {\em ad hoc} adiabatic approximations are in agreement with the qualitative predictions based on Vaidya spacetimes, yielding credibility to both. 
However these approximations do not fully capture the nonlinear features present in the BH ringdown. In particular, the key assumptions of spherical symmetry (in the Vaidya spacetime) and adiabaticity (in postulating the mass-varying Teukolsky equation in~\cite{May:2024rrg}) are only rough approximations.
A more self-consistent numerical assessment of the imprints of a changing mass and spin~\cite{Zhu:2024dyl} compares the ringdown waveform of a linearly and nonlinearly perturbed Kerr BH with spin $a=0.7$  using a setup similar to~\cite{Baker:1999sj}. 
The perturbation is a Gaussian pulse. Constraint-satisfying initial data are constructed using the conformal thin sandwich method~\cite{Pfeiffer:2004qz} and evolved both linearly and nonlinearly with consistent gauge conditions to first order. 
A comparison of the waveforms and of their decomposition into QNMs shows that the the fully nonlinear QNM amplitudes deviate from the linear predictions at third order in the initial perturbation amplitude, as expected from  perturbation theory. 
It is perhaps more surprising that perturbations leading to $\sim5\%$ changes in BH mass and spin (typical in astrophysical mergers) lead to nonlinear distortions in the mode amplitudes that can reach $\sim 50\%$ near the peak of the waveform. 
In addition, the phase difference between the linear and nonlinear waveforms displays a nonlinear drift in the frequency of the waveform, causing rapid dephasing between nonlinear and linear waveforms for large perturbations. 

Despite these significant nonlinear effects, a linear QNM model using fixed QNM frequencies based on the remnant BH mass and spin remains surprisingly accurate from the peak amplitude onward, in the sense that the residual and stability of the fit for linear and nonlinear waveforms are comparable~\cite{Zhu:2024dyl}. 
This remarkable success of the linear QNM model can be  attributed to the overlap between the nonlinear phase (where the BH mass and spin are rapidly evolving) and the transient, which is already present at the linear level. 
This suggests that the linear transient is the bottleneck for measuring and constraining nonlinear effects, such as the time dependence of the QNM frequency.
However, the simple time-symmetric Gaussian pulse initial data used in~\cite{Zhu:2024dyl} are unlikely to be a good approximation of post-merger perturbations for generic binaries. 
More realistic initial data are required to better understand nonlinear effects in realistic BBH mergers. 

Some promising results in this direction emerge 
from a comparison of fully nonlinear head-on BH collisions with perturbative evolutions of test-particle infalls with compatible initial data. The resulting ringdown waveforms are in remarkable agreement (see~\cite{DeAmicis:2024eoy} and Fig.~\ref{fig:tails_fullNR_news}), suggesting that nonlinear effects are suppressed in this case.
More work is needed to understand if this suppression occurs for more general, astrophysically realistic mergers, and how the evolution of the BH spin (which is identically zero in head-on collisions) affects these results.

\clearpage
\section{Ringdown modeling in gravitational waveforms}\label{sec:waveforms}

\noindent
{\em Misner proposed the following problem for my Ph.D. thesis. Take two of these entities that are now called black holes. Revolving around each other, they come close as energy is radiated away in the form of gravitational waves. They coalesce into an ellipsoidal ``Schwarzschild surface'' still rotating and radiating. Study the whole process, computing all the characteristics of the emitted gravitational radiation. Fine, I said, thy will be done! At the time, I did not realize the magnitude of this problem. Had I pursued it, I might have entered the Guinness book of records as the oldest graduate student alive, that too without financial support.}

\vspace{.2cm}

\noindent
C.V.~Vishveshwara, {\em On the black hole trail...: a personal journey}, text of the lecture delivered on February 17, 1996 on the occasion of the XVII Meeting of the Indian Association for General Relativity and Gravitation held at the Institute of Mathematical Sciences, Madras.

\vspace{.2cm}

The ultimate goal of the techniques described in the previous chapter is to build a model of the ringdown waveform that can be used in GW data analysis. This is the formidable problem that Misner assigned to Vishveshwara for his Ph.D. thesis more than 50 years ago. In this chapter we consider various aspects of this problem. We start by reviewing a large body of work to extract stationary QNM amplitudes from numerical simulations of BBH mergers, and to understand their dependence on the binary parameters (Section~\ref{subsec:WF_modeling_dependence}). The extraction of the QNM amplitudes is conceptually challenging because they generally become time-dependent as we approach the nonlinear merger. This has been addressed through several phenomenological models, reviewed in Section~\ref{sec:effective-one-body}.
In Section~\ref{sec:tails} we discuss our current understanding of the power-law tails following the ringdown regime, both in the test-particle limit and for comparable mass mergers. To ensure that the
ringdown physics extracted from numerical simulations is frame-invariant, one must transform from the initial binary frame used in the simulations to an appropriate remnant frame through a ``BMS frame fixing.'' This procedure, and the permanent net change in the GW strain between early and late times (``GW memory''), are discussed in
Section~\ref{sec:bms-frames-memory}.
Finally, we review recent progress in our understanding of the correspondence between QNMs and horizon dynamics (Section~\ref{sec:horizons}).

\subsection{Stationary amplitudes dependence on binary parameters: numerical extraction} \label{subsec:WF_modeling_dependence}

\vspace{-.1cm}

\noindent \textit{Initial contributors: Baibhav, Carullo, Forteza, Hughes, London}

\vspace{.2cm}

The QNM frequencies and damping times are solely dictated by the mass and spin of the remnant BH. 
Instead, the initial conditions arising during the perturbation phase, prior to the ringdown, influence the excitation of the modes, i.e., their amplitudes and phases.
In the context of BBH mergers, the ringdown follows immediately after the merger of the two progenitor BHs, which thus provides the initial conditions for the ringdown phase, and sets the QNM amplitudes and phases.
Thus there is an intrinsic link between the parameters of the initial binary system (determining the merger configuration) and the values of the ringdown complex amplitudes $C_{n'}$ capturing the spacetime geometry right after merger.
All GW ringdown models based on QNM superpositions embed this information in the values of $C_{n'}$, either implicitly or explicitly. 
Since the amplitudes $C_{n'}$ depend on the contribution of the initial perturbation, $T_{n'}$, a complete characterization of the ringdown amplitudes would require a knowledge of $T_{n'}$, but these coefficients are currently unknown with the exception of a few simplified scenarios (see Section~\ref{subsec:kerr_waveform_TD}).
Lacking a direct analytical computation, the coefficients $C_{n'}$ can be estimated using NR simulations of coalescing BBH systems with varying masses and spins. This is the main topic of this section.

For brevity we will use a multi-index $n'\rightarrow ( \ell mn,\pm)$, where the $(\ell m n)$ indices refer to the decomposition of the signal in spheroidal harmonics $S_{\ell mn}(\theta,\varphi, a\omega)$, and the index $\pm$ labels either prograde or retrograde modes (also referred to as co-/counter-rotating modes).
For most of the parameter space of interest, $C_{+} \gg C_{-}$.
Notable exceptions are binaries with large spins counter-aligned with respect to the orbital angular momentum, highly precessing binaries, or Schwarzschild BHs, in which the two branches of the solution converge to a single one with $C_{+}=C_{-}^*$.
We will usually drop the $\pm$ indices to simplify the notation, and restore them only in cases where counter-rotating modes play a significant role.
Typically, these complex amplitudes are split into an amplitude and phase as follows:
\begin{equation}
    C_{\ell mn} = A_{\ell mn} e^{i  \phi_{\ell mn}}\,.
\end{equation}

To understand how the $C_{\ell mn}$'s enter the gravitational waveform detected by far-away observers, it should first be recalled that ringdown is a \textit{dynamical process}, as extensively discussed in Section~\ref{subsec:greensfunc} (see also Chapter 4.B of~\cite{Buoninfante:2024oxl}).
Hence, the early ringdown phase will include the \textit{dynamical} excitation of QNMs, with corresponding \textit{time-dependent} amplitudes $C_{\ell mn}(t)$.
Such a dynamical ``activation'' regime exists even in the linear theory.
The merger waveform from BBH numerical simulations will include this dynamical regime, driven by the dynamical multipolar structure of the background, together with other nonlinear contributions, such as those due to a time-varying mass $M(t)$ and spin $\chi(t)$.
Following the terminology introduced in Chapter 4 of~\cite{Buoninfante:2024oxl}, it is only when the transient effects have decayed that the ringdown enters into a ``stationary'' regime, where amplitudes can be represented as constants $C_{\ell mn}(t) = C_{\ell mn}$, as most commonly assumed.
Given the current lack of first-principles models for the dynamical excitation of QNMs, in the following we will also assume to be in the ``stationary'' regime.
It is important to select the appropriate time window where this approximation is valid in order to construct a physically meaningful QNM amplitudes model and to avoid overfitting.
The lower limit of this time window is often (somewhat improperly) called the ``linear ringdown starting time,'' but it would be more appropriate to refer to it as the ``linear \textit{stationary} ringdown starting time.''
In Section~\ref{sec:effective-one-body} we present a phenomenological extension to the dynamical regime (compare and contrast with Sections~\ref{sec:nonlinSch} and \ref{sec:nonlinKerr}, where we discussed nonlinear contributions).

In the stationary regime, QNM amplitudes contribute to the linear GW BBH ringdown strain as 
\begin{equation}
   h_+ - i h_\times= \frac{M}{D} \sum_{\ell\ge 2,m,n,\pm}A_{\ell mn}^{\pm} e^{i  \phi_{\ell mn}^{\pm} } e^{i  \omega_{\ell mn}^{\pm} (t-t_{\rm ref})} S_{\ell mn}(\theta,\varphi, a\omega) + \text{(tails)}\,,
   \label{eq:rdown_complete}
\end{equation}
where $\omega_{\ell mn}^{\pm}$ represent the two branches of prograde/retrograde modes (also known as co/counter-rotating or ordinary/mirror modes) introduced in Eq.~\eqref{eq:mirror mode symmetry}.
The amplitude parameters of the merger remnant can be modeled as a function of the parameters of the binary progenitors, namely the mass ratio $q$ (the total mass scales out), the progenitor BH spins $\vec{\chi}_{i=1,2}$, and the orbital parameters related to eccentricity, so that
\begin{align}
    A_{\ell mn} &\rightarrow  A_{\ell mn}(q,\vec{\chi}_{i||},\vec{\chi}_{i\perp}, E, J) \,,\nonumber\\
    \phi_{\ell mn} &\rightarrow \phi_{\ell mn}(q,\vec{\chi}_{i||},\vec{\chi}_{i\perp}, E, J) \,,
    \label{eq:ins_to_ring_rels}
\end{align}
where $E,J$ are any two independent gauge-invariant quantities representing eccentric degrees of freedom.
We have also decomposed each spin vector into components that are parallel and perpendicular to $\vec{L}$, the orbital angular momentum vector, as $\vec{\chi}_{i} = (\vec{\chi}_{i||},\vec{\chi}_{i\perp})$.
Other useful combinations in the case where the progenitor spins are aligned with the orbital angular momentum (``spin-aligned'' case) are $\chi_{\pm} = (q \chi_1 {\pm} \chi_2) / (1 + q)$, with $\chi_{1,2}$ the components of the spins along the direction of $\vec{L}$.
In the broader GW literature, $\chi_+$ (the spin combination entering at the lowest PN order in the phasing~\cite{Blanchet:2013haa}) is also referred to as the ``effective spin,'' $\chi_{\rm eff}$.

The relevance of modes with different ${\ell mn}$'s varies in different regions of the parameter space of the progenitor parameters.
Greater asymmetry in the system enhances the contribution of higher harmonics with $\ell > 2$. 
For instance, for nonprecessing systems with $\vec{\chi}_{i\perp}=0$, asymmetric ringdown modes with $\ell > 2$ are typically excited when the progenitors have unequal masses and misaligned spins.
Similarly, these modes are excited for systems with significant orbital precession.
Therefore, the ringdown phase can also reveal information about the past dynamics of BBH systems~\cite{Kamaretsos:2011um}, provided that the relations in Eq.~\eqref{eq:ins_to_ring_rels} are known. 

The simplest way to determine these relations is through phenomenological fits of NR waveforms. 
In general, the ansatzes used for these fits are informed by:
(i) numerical waveforms from the extreme-mass ratio case, a simplified limiting scenario for which analytical intuition is more easily developed; and
(ii) PN predictions for the amplitude and phase in the inspiral. 
Albeit not trivial in principle, soon after NR simulations of coalescing binaries became available, it was realized that most modes exhibit a dependence on the progenitors parameters that is similar to those derived from PN theory~\cite{Buonanno:2006ui,Berti:2007fi,Berti:2007nw,Kamaretsos:2011um,Kamaretsos:2012bs,Borhanian:2019kxt}. 
This has led to the widespread usage of PN-inspired fitting formulas for predicting the QNM amplitudes.
To date, there is no analytical computation extending these scalings to the merger regime for comparable-mass binaries.

\subsubsection{Fitting methods and ringdown starting times}\label{subsubsec:fitting_methods}

There are various methods to extract all ``detectable'' QNMs from NR and perturbative waveforms while avoiding overfitting.
Standard nonlinear least squares techniques, like the Levenberg-Marquardt algorithm, are suboptimal and ill-conditioned (see~\cite{Giesler:2024hcr} for a discussion). Variants of Prony's algorithm (namely, the Kumaresan-Tufts and matrix pencil methods) should be better suited to extracting QNM parameters from GW signals~\cite{Berti:2007dg}.
Early studies of the QNM content of binary merger waveforms used either nonlinear least squares~\cite{Buonanno:2006ui} or Prony methods~\cite{Berti:2007fi}.

A later, comprehensive study used ordinary linear least-squares (OLS) fitting method to estimate the amplitudes of spheroidal QNMs for each spherical multipole, and a greedy algorithm  to select the optimal set of QNMs that best fit NR waveforms $10 M$ after the peak of the Penrose scalar $\Psi_4$~\cite{London:2014cma}. 
This greedy algorithm starts with a single QNM and iteratively adds more QNMs, ensuring that each addition significantly reduces the fitting error. The process ends when adding more than two QNMs fails to improve, or worsens, the fit.
To ensure the extraction of physical QNM components, this method also checks for the stability of the extracted QNM amplitudes.
Namely, given the assumption of constant $C_{\ell mn}$ within the stationary regime in which Eq.~\eqref{eq:rdown_complete} is valid, the algorithm checks that extracting them at different times should yield the same value within a given tolerance (``amplitudes time stability  criterion'').
Similarly, physical QNMs are expected to show smooth changes under variation of binary parameters  (``amplitudes parametric stability criterion''), something that can be imposed by constructing ``global fit'' models across the parameter space of the progenitor binary. 

Another approach to find the QNM content of the ringdown is to decompose the waveform into two components: one ``parallel'' and one ``perpendicular'' to the QNM.
This projection is achieved through a normalized time-integral overlap between a QNM superposition and the numerical waveform: see e.g.~\cite{Baibhav:2017jhs}.
The ringdown start time is defined as the moment when the energy parallel to the QNM reaches its maximum. 
The ringdown starting time corresponds to the lower limit of integration at which the energy along the QNM is maximized. 
This concept, introduced by Nollert~\cite{nollertthesis}, is referred to as the ``energy maximized orthogonal projection'' (EMOP) and it has been used to calculate the mode excitation for both nonspinning~\cite{Berti:2007fi} and nonprecessing waveforms~\cite{Baibhav:2017jhs}.
Other studies find the peak of the strain of each individual mode, and fit at a fixed time after the peak~\cite{JimenezForteza:2020cve, Carullo:2024smg}.
Yet another method chooses the ringdown starting time in the observed waveform by assessing how closely the strong-field spacetime resembles that of a Kerr BH. The idea is to generate a map that associates each time in the gravitational waveform with a value of a certain ``Kerrness'' measure, by tracking outgoing null characteristics from the strong and near-field regions to the wave zone~\cite{Bhagwat:2017tkm}.

A systematic TD fitting algorithm can be specifically designed to enforce time and parametric stability while avoiding overfitting~\cite{Cheung:2023vki}.
First, the excited modes in NR data are identified agnostically, i.e., without assuming which specific QNMs are present in the data~\cite{Baibhav:2023clw}.
The algorithm then iteratively discards time-unstable modes, i.e., those whose amplitudes and phases strongly deviate from a constant under changes in the fitting window. 
The starting time for each mode is determined by imposing a stability criterion on its amplitude. 

In alternative, one can use a FD filter~\cite{Ma:2022wpv}.
The idea is to remove a chosen QNM component from the TD ringdown signal. After filtering out the primary QNMs, the method allows for the extraction of subdominant effects, such as retrograde modes and second-order contributions, that might otherwise be masked by the stronger dominant modes.
Filters yield results that are equivalent to conventional fitting methods employing amplitude maximization. They offer a significant speed increase, at the cost of preventing an agnostic identification of the mode content of the waveform.

While most studies fit the QNM content of individual spherical harmonic components of the waveform, one can also fit various dominant modes across all angles on the two-sphere, accounting for mode-mixing and mapping the NR waveforms to the super-rest frame (see Section~\ref{sec:bms-frames-memory}), as required by the Teukolsky formalism~\cite{MaganaZertuche:2021syq}. This has been incorporated in many recent works, including those exploring the robustness of QNM fits, such as Refs.~\cite{Giesler:2024hcr,Gao:2025zvl,MaganaZertuche:2025bua,Mitman:2025hgy}. 

Most of the studies mentioned above use PN-inspired formulas to fit the amplitudes and phases as functions of binary properties. 
This approach is useful to understand the physical content of  amplitude models, but it can miss complex patterns that are not captured by the analytical predictions (e.g. for precessing and eccentric binaries), or lead to fits of limited accuracy.
Higher accuracy can be achieved with Gaussian Process Regression (GPR). Ringdown waveform models based on GPR (rather than closed-form analytic expressions) starting at $20 M$ after the peak typically have higher accuracy~\cite{MaganaZertuche:2024ajz,Pacilio:2024tdl,Nobili:2025ydt}.
The GPR implementation of~\cite{Pacilio:2024tdl} is available in the \texttt{postmerger} package~\cite{pacilio_2024_13220424}.
Another advantage of GPR is that it provides an estimate of the fitting uncertainty that can be used in data analysis~\cite{Moore:2014pda,Breschi:2022ens,Pompili:2024yec}.

As we pointed out at the beginning of this section, least squares fitting can be ill-conditioned when applied to damped sinusoids with unknown frequencies, making it highly sensitive to numerical noise and unmodeled physics. 
The Variable Projection (\texttt{VarPro}) algorithm used in~\cite{Giesler:2024hcr} helps mitigate the ill-conditioning associated with least squares fitting, allowing for a more stable extraction of the amplitudes of highly damped modes, such as overtones.

\subsubsection{Fundamental mode harmonics in spin-aligned quasi-circular systems}

Initial systematic fits of the amplitudes of nonspinning~\cite{Buonanno:2006ui,Berti:2007fi} and spin-aligned~\cite{Kamaretsos:2011um, Kamaretsos:2012bs} binaries as a function of the parameters of the binary progenitors
were later extended to include both amplitudes and phases for the dominant multipoles of nonspinning~\cite{London:2014cma} and spin-aligned~\cite{London:2018gaq} binaries. These fits, that rely on the MAYA~\cite{gatechcatalog} and Cardiff~\cite{Hamilton:2023qkv} NR simulation catalogs and take into account the effect of spherical-spheroidal mode-mixing,
were later revisited and extended by several authors~\cite{Baibhav:2017jhs, Cheung:2023vki, Pacilio:2024tdl, MaganaZertuche:2024ajz}, exploiting improved fitting algorithms and higher-accuracy simulations from the SXS catalog~\cite{Boyle:2019kee}.

\begin{figure}
\centering
\includegraphics[width=\linewidth]{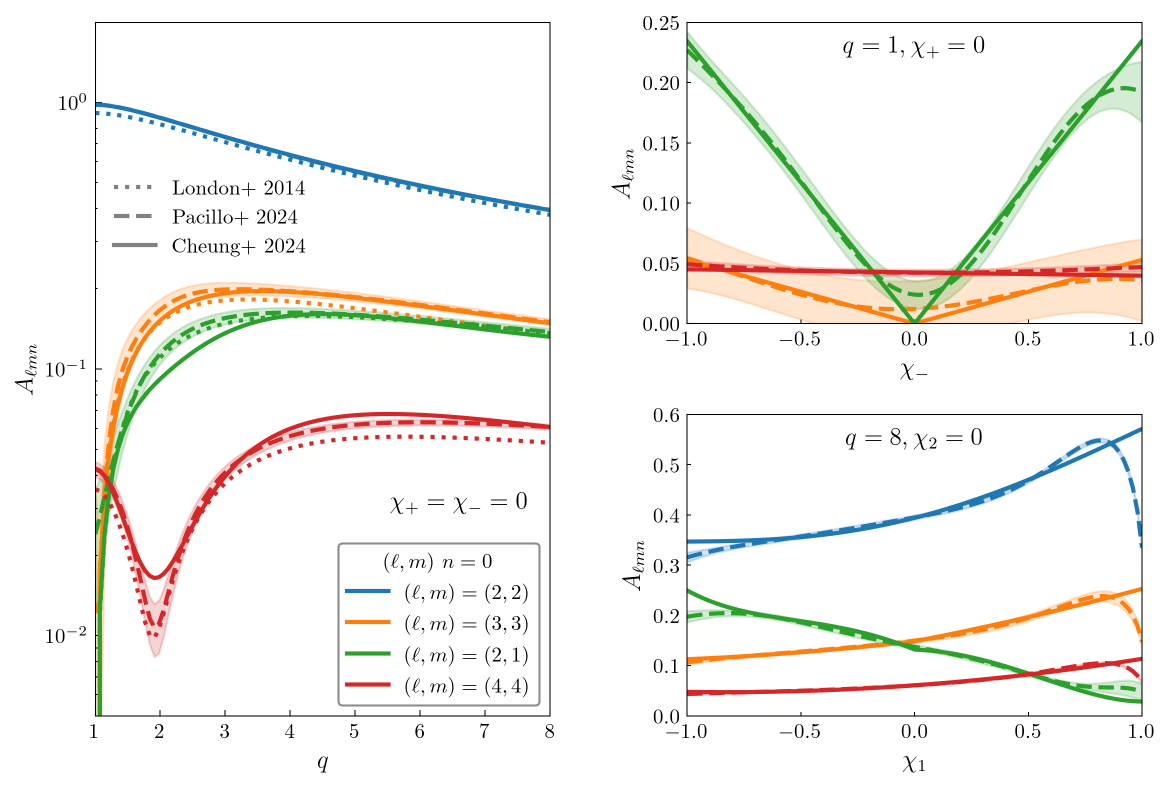}
\caption{
Amplitude of dominant $(\ell, m)$ modes at the peak for nonspinning (left) and aligned-spin (right) systems, calculated from the fits by London et al.~\cite{London:2014cma} (dotted), Cheung et al.~\cite{Cheung:2023vki} (solid), and Pacilio et al.~\cite{Pacilio:2024tdl} (dashed; in this case, colored bands mark the 2$\sigma$ error bars from GPR).
Left panel: Amplitudes as a function of mass ratio for nonspinning waveforms: $(2,2)$, $(3,3)$, $(2,1)$, and $(4,4)$.
Right panel: Amplitudes for spin-aligned binaries. 
Top right: amplitudes for equal-mass binaries ($q = 1$) as a function of $\chi_-$, assuming $\chi_+ = 0$ (i.e., the spins are anti-aligned with each other: $\chi_- = \chi_1=-\chi_2 $). The $\ell = m$ multipoles, e.g. $(2,2)$ and $(4,4)$, depend only mildly on $\chi_-$. We focus on subdominant modes and we do not even show the $(2,2)$ mode, because its amplitude has values larger than the plot range and it is nearly independent of $\chi_-$.
Bottom right: amplitude for a binary with $q = 8$ as a function of $\chi_1$, assuming $\chi_2=0$.
}
\label{fig:A_n0}
\end{figure}

In Fig.~\ref{fig:A_n0} we compare the amplitude at the peak of the fundamental QNMs for some of the dominant multipolar indices $(\ell,m)$ found with some of the fitting algorithms described above~\cite{London:2014cma, Cheung:2023vki, Pacilio:2024tdl}. 
In the left panel we plot the amplitude as a function of the mass ratio for nonspinning binary waveforms.
The hierarchy of the modes follows the sequence $(\ell, m) = (2,2), (3,3), (2,1), (4,4)$, except as the system approaches the equal-mass (and equal-spin) limit.
As the mass ratio approaches unity, odd-$m$ modes are suppressed~\cite{Berti:2007fi,Kidder:2007rt} and the $(4,4)$ mode becomes the most important subdominant mode. 
This transition is driven by the symmetry of the system: for equal-mass, equal-spin binaries ($q = 1$ and $\chi_1 = \chi_2$), the modes with odd $m$ vanish due to the requirement that the system is invariant under a rotation by an angle $\pi$ around the axis of the binary~\cite{Blanchet:2013haa}.
However, for asymmetric mass ratios ($q \gtrsim 2$), the relative amplitudes of higher modes increase very rapidly.

In the right panel of Fig.~\ref{fig:A_n0} we show how the multipolar QNM amplitude hierarchy changes when the BH spins are included. 
The amplitudes of the $(2,2)$, $(3,3)$, and $(4,4)$ fundamental QNMs depend only weakly on the spin parameters, while the $(2,1)$ mode shows a more significant sensitivity to spin~\cite{Berti:2007nw}, as expected for $m = \ell-1$ ``magnetic'' multipoles~\cite{Blanchet:2013haa}.
In the top right panel we plot the amplitudes of equal-mass systems ($q = 1$) as a function of $\chi_-$ for $\chi_+ = 0$ and $q = 1$. In this case the BH spins are anti-aligned with each other, i.e., $\chi_1 = \chi_- = - \chi_2$. 
Again, the odd-$m$ modes vanish from symmetry arguments at $q=1$ and $\chi_-=0$, with the amplitude growing proportionally to $|\chi_-|$. 
In particular, the $(2,1)$ mode amplitude increases significantly with spin, as predicted by PN theory, where the amplitude of this mode scales with $\chi_-$~\cite{Berti:2007nw,Blanchet:2013haa}. 
This is because the $(2,1)$ mode contains spin-dependent terms at a lower PN order~\cite{Berti:2007nw,Pan:2010hz}, making it more sensitive to spin effects compared to other modes. 
In contrast, multipoles with even $m$, such as the $(2, 2)$ and $(4, 4)$ modes, exhibit a much weaker dependence on the spins. 
For moderately large asymmetric spins ($|\chi_-| \gtrsim 0.25$), the $(2,1)$ multipole dominates over the $(3,3)$ and $(4,4)$ multipoles.

In the bottom right panel of Fig.~\ref{fig:A_n0},  we show the amplitude of the modes as a function of the spin of the more massive BH for $q=8$. 
For large positive spins ($\chi_1 \gtrsim 0.5$), the $(2,1)$ mode becomes subdominant compared to the $(4,4)$ mode~\cite{Cotesta:2018fcv}; conversely, for negative spins ($\chi_1 \lesssim -0.25$), the $(2,1)$ mode grows larger than the $(3,3)$ mode~\cite{Cotesta:2018fcv}. 

In~\cite{Forteza:2022tgq}, similar fits have been used to point out that the amplitude-phase correlation predicted  by NR simulations spans only a limited portion of the allowed parameter space for spin-aligned quasi-circular systems.
This suggests the possibility to carry out an ``amplitude-phase ringdown consistency test'': anomalous amplitude-phase measurements could be flagged to reveal unmodeled physics, or even effects beyond GR~\cite{Forteza:2022tgq}.
  
\subsubsection{Overtones in spin-aligned quasi-circular systems}\label{subsec:overtones}

One of the earliest comparisons between analytical predictions and  BBH waveforms computed within NR~\cite{Buonanno:2006ui} shows that the inclusion of three overtones (with complex frequencies fixed to the asymptotic values of the BH mass and spin) improves the mismatch between QNM superpositions and the NR signal at early times.
However, these early comparisons already note that a simple superposition of large numbers of overtones with fixed mass and spin close to the waveform peak does not take into account near-peak transient or nonlinear effects~\cite{Buonanno:2006ui, Berti:2007fi}, and so their physical interpretation is unclear.
Later work finds robust fits of the $n=1$ overtone at late times (in the stationary regime) that display both time stability and parametric stability~\cite{London:2014cma, London:2018gaq}.

In the attempt to match the entire post-peak waveform using QNM superpositions, EOB models often employ the superposition of a large number ($N=7$) of overtones up to the waveform peak.
In fact, additional phenomenological components with frequencies that do {\em not} belong to the Kerr QNM spectrum, called ``pseudo-QNMs,'' are needed to bridge the gap between the lowest QNM frequency and the merger frequency~\cite{Pan:2013rra,Taracchini:2013rva,Babak:2016tgq}.
Over time, QNM superpositions of overtones and pseudo-QNMs have become a crucial component of early EOB models~\cite{Pan:2013rra,Taracchini:2013rva,Babak:2016tgq}. 
Because of issues with the parametric stability of these phenomenological fits, together with the improved accuracy of alternative phenomenological templates~\cite{Damour:2014yha} (see Section~\ref{sec:effective-one-body}), these QNM superpositions are no longer used in most of the latest EOB models.

The inclusion of one or two overtones close to the waveform peak  significantly improves the extraction of the fundamental mode frequency and damping time (and hence of the remnant's mass and spin), typically leading to $\sim 1\%$ accuracy~\cite{Baibhav:2017jhs}. 
However, this study did not address the issue of the parametric stability of global fits across the parameter space (that is achieved instead for the fundamental mode).

A later study considered a superposition of $N=7$ overtones with frequencies fixed to their asymptotic Kerr values, showing that this combination of QNMs can fit a GW150914-like NR simulation from the SXS catalog with very high accuracy~\cite{Giesler:2019uxc}.
This study stressed the validity of linear and stationary QNMs superpositions in accurately describing the post-merger waveform, and sparked a lively debate on the problem of fitting a large number of modes to the entire post-peak waveform, on the quantification of the nonlinear and transient content of these signals, and on the implications of these modeling assumptions for observational follow-up studies on real data (see Section~\ref{sec:DataAnalysis}).
Given the short lifetime of the overtones, these follow-up studies highlighted the importance of accounting for time and parametric stability to discern between physical QNMs and other contributions to the waveform.
Refs.~\cite{Bhagwat:2019dtm,Forteza:2021wfq} challenged the parametric stability of overtone amplitudes with $n>1$, and showed that stable fits across the parameter space can not be achieved from the waveform peak, but only at later times (during what we called the ``stationary regime'').

As previously mentioned, the fact that multiple overtones can be extracted by fitting does not necessarily guarantee their physical significance. 
For example, adding more modes can help by overfitting the nonlinear and transient structure of the signal, particularly near the peak. 
Comprehensive linear and nonlinear studies show that these additional modes (especially the highest overtones) can enable a more accurate extraction of the fundamental mode and of the first overtone, which at late times carry the majority of information about the remnant properties, by ``fitting-away'' the poorly understood nonlinear and transient portion of the signal~\cite{Baibhav:2023clw,Nee:2023osy, Takahashi:2023tkb}.
The risk of overfitting can be illustrated by e.g. considering ``unphysical hybrids'' (Appendix E of~\cite{Baibhav:2023clw}) constructed by stitching together two different signals $h_1$ and $h_2$: in this case, a superposition of overtones ($h_1$) can accurately -- but unphysically -- fit the post-merger of $h_2$.

An agnostic extraction of QNMs is necessary to identify the mode content of the ringdown and to mitigate the risk of overfitting~\cite{Baibhav:2023clw, Redondo-Yuste:2023seq, Cheung:2023vki}.
Rather than assuming the presence of known overtones in a given $(\ell, m)$ multipole, agnostic fits treat the complex frequencies as free parameters. 
The agnostic approach leads to stronger tests of mode excitation and it is conceptually more similar to the spirit of the original BH spectroscopy proposal (in fact, it can be thought of as the infinite-SNR limit of the observational BH spectroscopy program). 
Without considering nonlinear contributions, the agnostic fits of  the $\ell=m=2$ multipole in~\cite{Baibhav:2023clw} can only reliably identify the fundamental mode, the first overtone, and some retrograde modes (in addition to modes related to spherical-spheroidal mode mixing).
Higher overtones ($n > 1$) are not as easy to identify using free-frequency fittings and standard least squares methods on extrapolated SXS waveforms~\cite{Baibhav:2023clw}. 
This conclusion remains valid even when fitting with as many as 10 damped sinusoids~\cite{Cheung:2023vki}.

While agnostic fitting using least squares methods allows for unbiased identification of excited modes, it may struggle to detect highly damped modes in the presence of even small numerical noise. 
This is because agnostic QNM fitting is numerically ill-conditioned -- i.e., it is extremely sensitive to numerical noise and other underlying unmodeled physics.
Being very short-lived, overtones are particularly sensitive to unmodeled contributions and hard to extract: even subdominant tail effects can obscure the identification of highly damped overtones~\cite{Baibhav:2023clw,Giesler:2024hcr}.
This presents a challenge for performing BH spectroscopy with high overtones.

These issues can be partially addressed by reducing the impact of numerical noise and improving the robustness of the fitting process.
By adopting a superior fitting method known as Variable Projection (\texttt{VarPro}) to help mitigate the issue related to ill-conditioning in least square fitting, correctly normalizing the time unit by the Christodoulou mass, and using simulations with higher accuracy augmented by Cauchy Characteristic Evolution (CCE) rather than polynomial extrapolation at future null infinity -- which avoids spurious contributions related to recoil that could spoil the fit (see also Section~\ref{sec:tails}) -- it is possible to identify as many as nine overtones with relatively time-stable amplitudes~\cite{Giesler:2024hcr}.
This is achievable for simulations that produce remnants spinning faster than the GW150914 remnant, when accounting for quadratic contributions, and by ``dropping'' unstable modes outside of the initial fitting window, which results in partial fits discontinuities for highly damped modes.
A stable fit could be achieved starting at $4-8M$ after the peak strain (depending on the binary parameters), after transient contributions have sufficiently decayed away.

While at intermediate times most QNMs were found to follow the expected decay, sizable departures from constant amplitudes are present for certain modes. 
The $n=8,9$ QNMs deviate from a constant scaling from the start of the time-grid, and the jumps due to mode dropping further contribute to constant-scaling deviation. 
As previously highlighted, the employed model assumes that \textit{all} modes are constant in the same time-window. 
Extracting constant amplitudes of different modes in different windows leaves open the possibility that higher overtones can ``fit away'' other physical contributions~\cite{Baibhav:2023clw, Nee:2023osy, Zhu:2023mzv}, such as the QNMs or mass-spin dynamical growth, prompt response and nonperturbative terms.
Future studies will need to address these remaining issues related to time-stability.

The parametric stability of overtone amplitudes across the binary parameter space was not addressed in~\cite{Giesler:2024hcr}, but it has been achieved for the $n=1$ overtone in the stationary regime (typically $\sim 10M$ after the peak strain, depending on the binary parameters) by several authors: see e.g.~\cite{London:2014cma, Forteza:2021wfq, Cheung:2023vki, MaganaZertuche:2024ajz} for global fits of the complex amplitudes of spin-aligned BBH constructed. 
The amplitude ratio between the first overtone and the fundamental mode is mainly determined by $\chi_f$, and this holds true for both the $(2,2)$ and $(3,3)$ modes. 
In the right panel of Fig.~\ref{fig:A_OvertoneRetro}, 
based on~\cite{Cheung:2023vki}, 
we show the amplitude ratios between the first overtone and the fundamental mode, $A_{221} / A_{220}$ (purple) and $A_{331} / A_{330}$ (orange), as functions of the remnant spin $\chi$, along with the ratio of excitation factors $E_{221} / E_{220}$ (purple dashed) and $E_{331} / E_{330}$ (orange dashed). 
Recall that the amplitudes $A_{\ell mn}$ are the product of the excitation factor $E_{\ell mn}$ and the integral term $T_{\ell mn}$, with $E_{\ell mn}$ solely determined by the remnant spin and independent of the initial conditions of the binary system, while $T_{\ell m n}$ depends on both the initial configuration of the system and the source of perturbations.
For low values of the remnant spin $\chi$, the amplitude ratios and excitation factor ratios have similar values, but as the remnant spin increases, these two ratios diverge significantly.
This interesting numerical result has not yet been confirmed by analytical arguments.

\begin{figure}
\centering
\includegraphics[width=\linewidth]{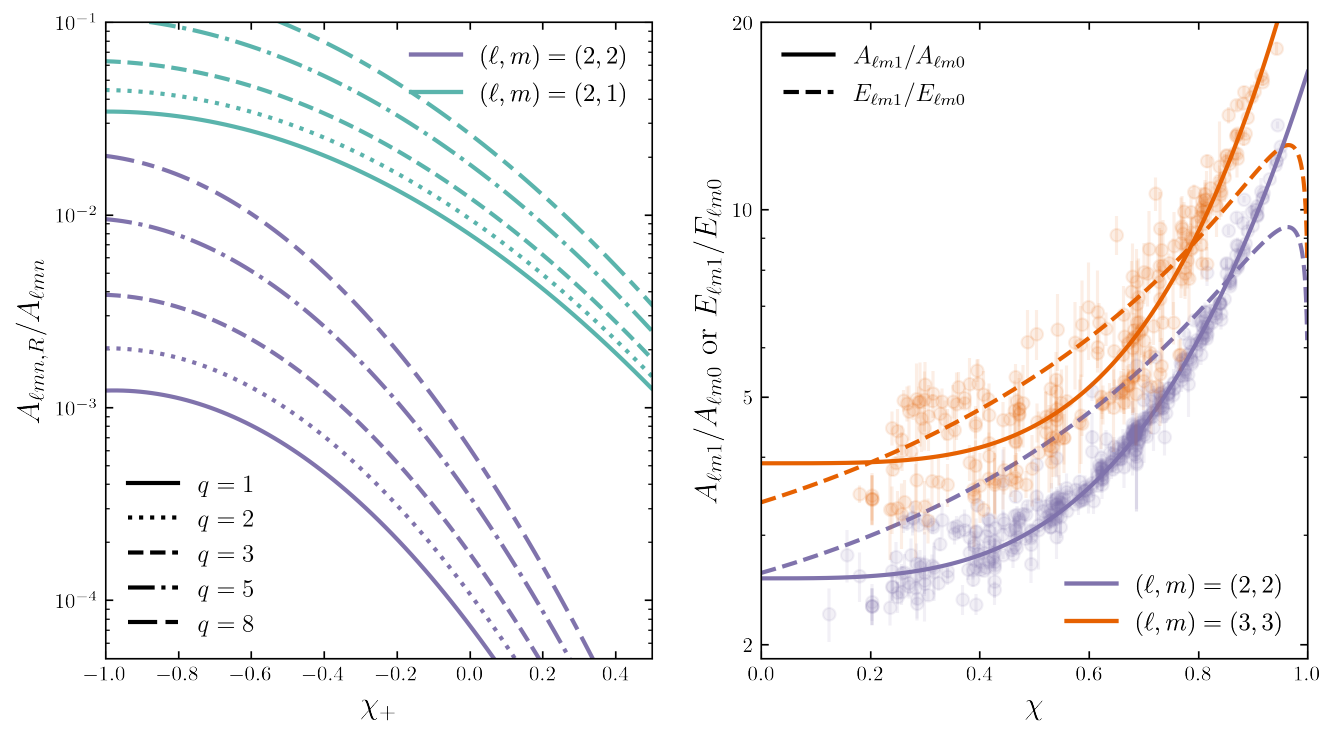}
\caption{
Left: Ratios ${A_{\ell m n, R}}/{A_{\ell m n}}$ for $(\ell, m) = (2, 2)$ and $(2, 1)$ (purple and green) as functions of $\chi_+$, for discrete values of $q = [1, 2, 3, 5, 8]$ (different line styles).
{Right:} Solid lines represent the amplitude ratio between first overtone and the fundamental mode  [${A_{221}}/{A_{220}}$  (purple) and ${A_{331}}/{A_{330}}$ (orange)] as functions of the remnant spin $\chi$. 
The dashed lines indicate the ratio of excitation factors [${E_{221}}/{E_{220}}$ (purple) and ${E_{331}}/{E_{330}}$ (orange)]. 
Based on fits from~\cite{Cheung:2023vki}.
}
\label{fig:A_OvertoneRetro}
\end{figure}
\begin{figure}
\centering
\includegraphics[width=\textwidth]{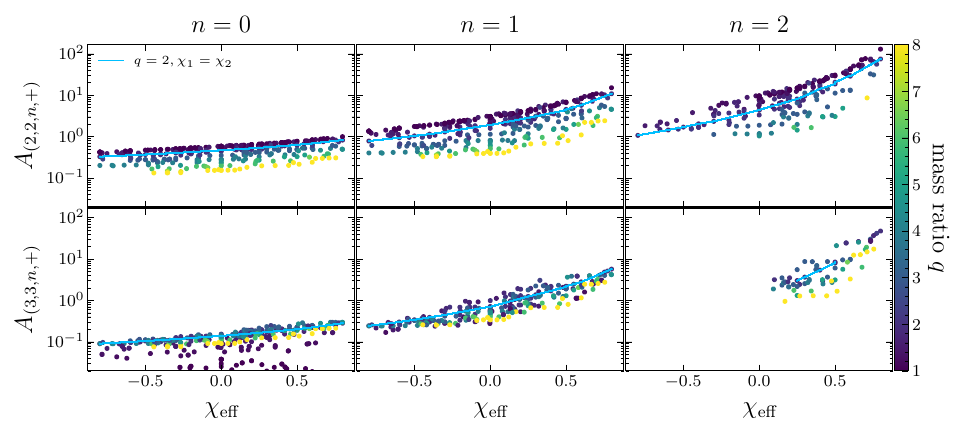}
\caption{Amplitude of the $(2,2,n\leq2,+)$ and $(3,3,n\leq2,+)$ overtones extracted from quasi-circular, nonprecessing BBH merger simulations as a function of the effective spin $\chi_{\mathrm{eff}}$ (horizontal axis) and the mass ratio $q$ (color) of the binary.
For each of the six panels, we show only the QNMs found using the algorithm presented in~\cite{Mitman:2025hgy}: for example, for the $(3,3,2,+)$ QNM in the bottom right panel there are fewer points because fewer QNMs were found across the simulations. Amplitudes are extrapolated to the time of peak strain over the two-sphere.
We show the equal-spin $q=2$ case in light blue as an example of a case where a 1-D parametric relationship is expected to hold.
Data from~\cite{Mitman:2025hgy}.}
\label{fig:overtone_chi_eff}
\end{figure}

The fits of~\cite{Mitman:2025hgy} display parametric stability across quasi-circular, nonprecessing BBH mergers for the $(2,\pm2,n\leq2,\pm)$ QNMs, as well as for various other QNMs, such as the $(3,\pm3,n\leq2,\pm)$ and $(2,\pm1,n\leq1,\pm)$.
This is possible through an automated QNM extraction routine that performs a reverse search in time for QNM frequencies using \texttt{VarPro}, and only adds QNMs to the model if their complex amplitudes can be shown to be stable (within some tolerance) as a function of the starting time of the fit over some time interval of fitting times whose length depends on the QNM's damping time. 
On average, the overtones are found to be stable at times $\gtrsim8M$ after the peak of the strain, with some slight variation across parameter space and overtone number~\cite{Mitman:2025hgy}. 
Stability was enforced through tolerance thresholds derived from test cases.
In the future it will be important to extend these fits to higher overtones by adopting ``mode-dropping'' techniques or GPR methods to incorporate the noise level in the fit through a Bayesian formulation (rather than tolerance thresholds).

While the $(2,2,n\leq1,+)$ QNMs are found across every simulation, the $(2,2,2,+)$ QNM is found in $\sim 60\%$ of the simulations and the $(2,2,3,+)$ QNM is found only in $\sim 3\%$ (primarily equal-mass and high-spinning) binaries.
Overtones in other $(\ell, m)$ modes are found slightly less often than in the $(2,2)$ mode. 
The dependence of the $n\leq2$ prograde overtones for the $(2,2)$ and $(3,3)$ multipoles on the binary's effective spin and mass ratio is shown in Fig.~\ref{fig:overtone_chi_eff}.

\begin{figure}
\centering
\includegraphics[width=\textwidth]{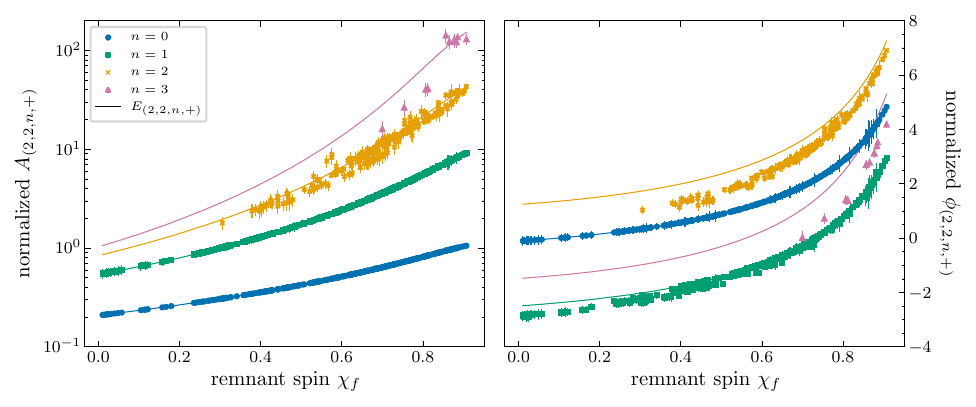}
\caption{
Amplitudes (left) and phases (right) of the $(2,2,n,+)$ QNMs, normalized such that the $(2,2,0,+)$ QNM's complex amplitude matches its excitation factor and the $(2,2,1,+)$ QNM's amplitude matches that of its excitation factor.
Data points correspond to fits to NR simulations, with error bars representing the variation of the complex QNM amplitude over a series of fits with varying start times (see~\cite{Mitman:2025hgy} for more details).
Solid lines are the Teukolsky excitation factors computed in~\cite{Zhang:2013ksa}, scaled by a factor of $1/(-i\omega_{(2,2,n,+)}^{2})$ so they correspond to the strain rather than $\Psi_{4}$.
Figure taken from~\cite{Mitman:2025hgy}.}
\label{fig:overtone_vs_EF}
\end{figure}

When several overtones can be found across many NR simulations, it is possible to perform a more robust analysis of the relationship between the $(2,\pm2,n,\pm)$ amplitudes and their excitation factors~\cite{Giesler:2019uxc,Oshita:2021iyn,Cheung:2023vki,Mitman:2025hgy}.
By fixing three degrees of freedom (two for the overall scale of the source function and one for the time at which the QNMs are extracted), it can be shown that the $(2,\pm2,n\leq2,\pm)$ QNMs track almost exactly their excitation factors, with very good agreement as a function of $n$ for the amplitudes, and slightly larger disagreement for the phases~\cite{Mitman:2025hgy}. This is shown in Fig.~\ref{fig:overtone_vs_EF}, which appears to differ with the right panel of Fig.~\ref{fig:A_OvertoneRetro} because the amplitude ratios in~\cite{Cheung:2023vki} are always taken at the peak of the strain. 
This suggests that there may be something inherent to the BH scattering potential or the remnant perturbation in BBH mergers that naturally excites the overtones at the level predicted by their excitation factors. Consequently, there may be a physical reason to expect the excitation of a large number of overtones, even if demonstrating their stability in numerical simulations is challenging~\cite{Baibhav:2023clw,Nee:2023osy,Takahashi:2023tkb,Giesler:2024hcr}. More work on overtone excitation using either scattering experiments or NR simulations of BBH mergers is needed to confidently confirm this hypothesis.

In summary, despite many recent advancements in the characterization of low ($n\leq2$) overtones at intermediate times, many of the complications related to the excitation of higher-order overtones close to the waveform peak remain.
A definitive solution to the debate on the physical meaning of the overtone contributions close to the peak may only be possible through analytical models of overtone excitation on a dynamical background, including transient effects.
The extent to which these effects are at play in comparable mass systems, together with the robust determination of the time-window in which a stationary description is valid are still subject of active investigation.

\subsubsection{Amplitudes in precessing quasi-circular systems}
\label{subsec:precession}

Precession effects in the ringdown resulting from a BBH coalescence appear due to a misalignment between the progenitors spins $\boldsymbol{{\chi}}_{i}$ and the orbital angular momentum $\textbf{L}$.
In the inspiral, this is most conveniently described in two frames: one with the $z$-axis along the total angular momentum $\textbf{J}$, the other along $\textbf{L}$.
Wigner D-matrices $D(R)^{\ell}_{\mu m}$ can be used~\cite{Gualtieri:2008ux,Campanelli:2008nk} to transform between the two frames:
\begin{equation}
    h'_{\ell m} =\sum_{\mu=-l}^l D(R)^\ell_{\mu m} h_{\ell \mu}\,,
    \label{eq:prec_mode_mixing}
\end{equation}
where $R$ is the quaternion representing the time-dependent rotation.
This rotation changes the strain modes detected by far-away inertial observers, and can be interpreted as a ``mode-mixing'' yielding a ``spread'' of dominant harmonics into sub-dominant harmonics.
This implies a corresponding re-ordering of their amplitudes compared to the spin-aligned case.
As an example, binaries with strong in-plane spin components induce larger excitation amplitudes for modes with $\ell = m \neq 2$.
This effect has been well-studied in the context of IMR modeling~\cite{Buonanno:2002fy,Schmidt:2012rh, OShaughnessy:2012iol,Boyle:2011gg,Pan:2013rra, Hamilton:2023znn, Babak:2016tgq, Ossokine:2020kjp, Ramos-Buades:2021adz}.

In the ringdown, these two frames become those anchored to either the spin of the final BH $\boldsymbol{{\chi}}_f$ (the ``preferred direction'' of the background) or to $\textbf{L}$ (the direction of the perturbation). 
Knowledge of the rotation matrix allows to un-mix  individual modes from the higher harmonic modes~\cite{Finch:2021iip}.
Such approach shows how QNM superpositions with free amplitudes achieve small mismatches and remnant errors if higher harmonics are added and unmixed through Eq.~\eqref{eq:prec_mode_mixing}.

\begin{figure}[t]
\centering
\includegraphics[width=\textwidth]{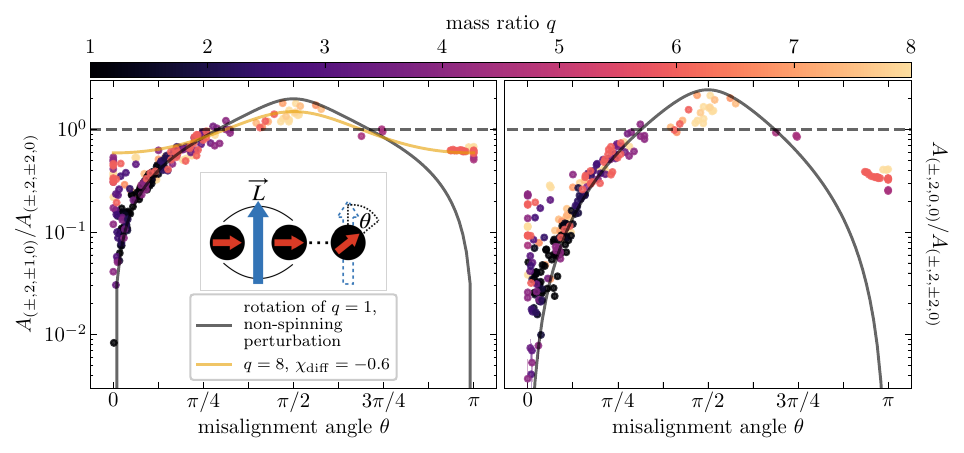}
\caption{Amplitude ratios of the $(2,1,0)$ vs. $(2,2,0)$ QNM amplitudes (left) and $(2,0,0)$ vs. $(2,2,0)$ QNM amplitudes (right) as a function of the misalignment angle $\theta$: the angle between the angular momentum flux at the time of peak luminosity and the remnant's final spin (for more details, see~\cite{Zhu:2023fnf}).
Color shows the binary mass ratio.
The black line represents the QNM amplitude ratio expected when a purely $(2,2,0)$ perturbation is rotated by $\theta$ to excite other $(2,m,0)$ QNMs.
The yellow curve is the same as the black curve, but also includes the $(2,1,0)$ QNM expected for a $q=8$, $\chi_{\mathrm{diff}}=-0.8$ system in the initial perturbation.
Figure taken from~\cite{Zhu:2023fnf}.}
\label{fig:QNM_prec}
\end{figure}
Moreover, the QNM amplitudes are also affected by the intrinsically distinct nature of the precessing perturbation. 
In particular, NR fits of QNM superpositions with free amplitudes~\cite{Zhu:2023fnf} can be used to identify systems where strong precession leads to, e.g., a louder excitation of the $(2,1,0,\pm)$ or $(2,0,0,\pm)$ QNMs than the $(2,2,0,\pm)$ QNMs, as shown in Fig.~\ref{fig:QNM_prec}.
It is particularly interesting that subleading modes can become dominant: as shown in Fig.~\ref{fig:QNM_prec}, it is possible for the amplitude $|A_{(2,1,0)}|$ to be larger than $|A_{(2,2,0)}|$ for large mass-ratio systems~\cite{Zhu:2023fnf}.
Specializing the above picture to a single time-independent rotation between the remnant's spin and the total angular momentum flux $d \textbf{J}/dt$, this enhancement can be well-explained by ``rotating'' the spin-aligned amplitudes predicted in, e.g.,~\cite{Kamaretsos:2012bs}.
This approach is motivated by perturbative studies in the extreme-mass-ratio case~\cite{Apte:2019txp, Lim:2019xrb, Hughes:2019zmt, Ghosh:2023mhc}, detailed below in Section~\ref{sec:QNM_EMRIs}.
Each individual mode in Fig.~\ref{fig:QNM_prec} is fit in the canonical super-rest frame of the remnant, the one used in BH perturbation theory, mitigating the mode-mixing effects due to the remnant kick velocity~\cite{Mitman:2021xkq,MaganaZertuche:2021syq,Mitman:2024uss}.
Apart from this effect, precessing systems can also lead to distinct excitation of $\pm m$ modes due to the breaking of reflection symmetry, as shown in Fig. 3 of~\cite{Zhu:2023fnf}.
This reflection symmetry breaking is also important in the modeling of the entire waveform~\cite{Ossokine:2020kjp, Ghosh:2023mhc}.

\begin{figure}[t]
    \centering
    \includegraphics[width=0.6\columnwidth]{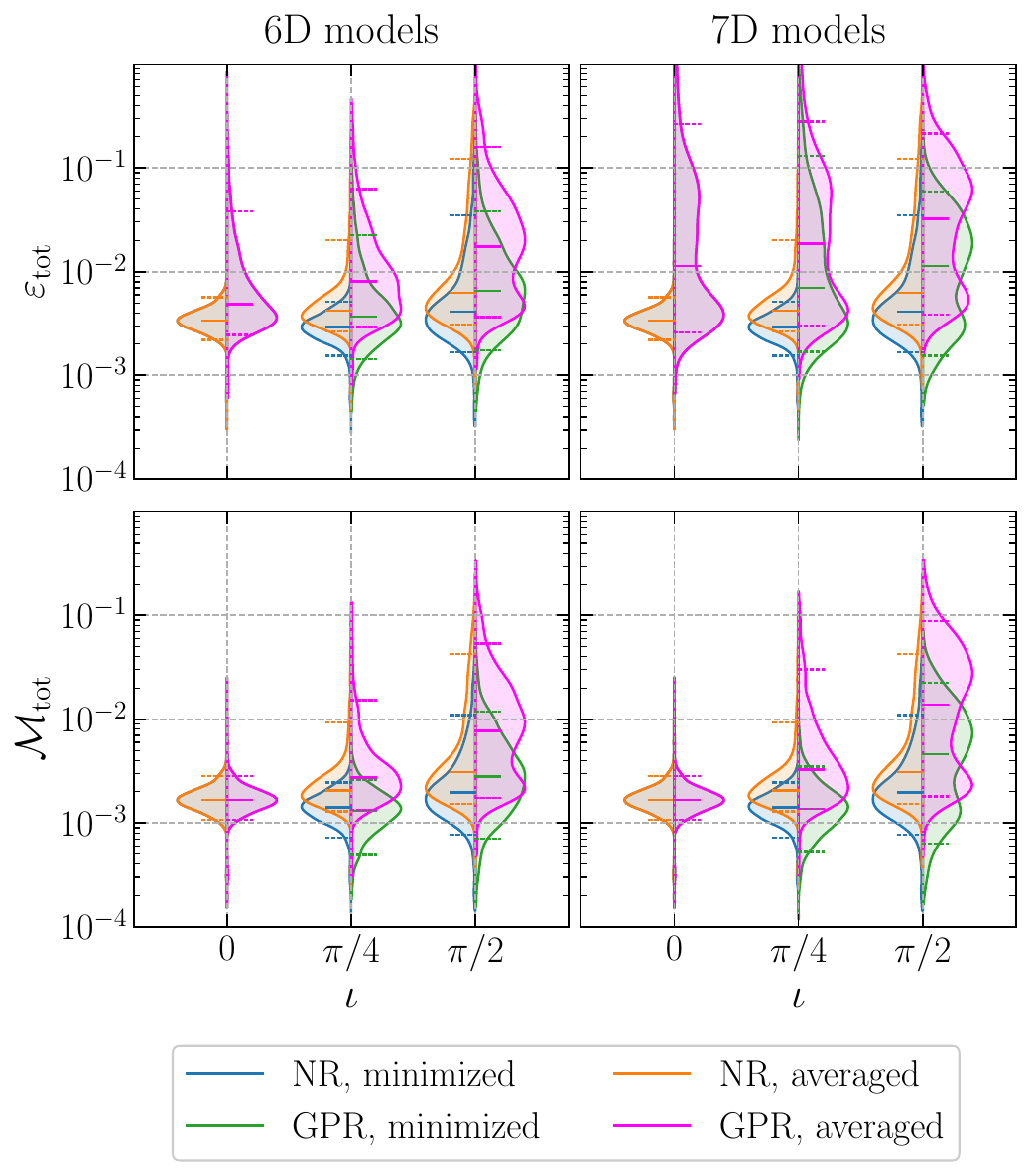}
    \caption{Top panel: distribution of relative amplitude strain errors $\varepsilon^{\rm NR}$. Bottom panel: standard mismatches computed between the GPR fit and the NR waveform. Solid ticks refer to the median values, and dotted ticks mark the 5\% and 95\% percentiles. The left and right panels correspond to GPR fits in the 6D and 7D parameter spaces, respectively. Figure taken from~\cite{Nobili:2025ydt}.
}
\label{fig:Nobili-Bhagwat-precessing-fit-score}
\end{figure}

An alternative approach involves the development of surrogate models for the ringdown mode amplitudes of quasi-circular precessing BBHs~\cite{Nobili:2025ydt}. The fits are provided in a coordinate frame where the $z$-axis is aligned with the spin direction of the remnant BH. 
The amplitudes are fitted as a function of six parameters (6D  model): the asymmetric mass ratio $\delta = (q-1)/(q+1)$, with $q = m_1/m_2 \geq 1$; the symmetric and antisymmetric combinations of the spin components parallel to the orbital angular momentum, $\chi_{\rm s} = (q\chi_{1z} + \chi_{2z})/(1+q)$ and $\chi_{\rm a} = (q\chi_{1z} - \chi_{2z})/(1+q)$; the angle $\theta_f$ between the orbital angular momentum at the ISCO and the spin of the remnant $\chi_f$; the angle $\phi_k$ between $\chi_f$ and the direction of the recoil; and the magnitude of the recoil velocity $v_{\rm k}$. Additionally, slightly less accurate fits are also provided for a 7D model involving $\delta$ and the individual BHs' Cartesian spin components. Note that these two models include only the fundamental modes ($n = 0$) and do not take into account spherical-spheroidal mode mixing. The waveforms are aligned at a reference time defined by the point of maximum energy projection ($t_{\text{EMOP}}$), as introduced in~\cite{Berti:2007fi}. In particular, the amplitudes are modeled starting at $20M$ after $t_{\text{EMOP}}$. The surrogates are constructed for all $\ell = 2$ modes, as well as for the $\ell = 3, m = 3$ mode, using a Gaussian process regression method and the SXS NR catalog.

In Fig.~\ref{fig:Nobili-Bhagwat-precessing-fit-score} we show the mismatch between the calibrated amplitude ringdown templates and the waveforms of the SXS NR catalog. A metric $\varepsilon^{\rm NR}$ that quantifies the goodness of fit is defined as
\begin{equation}
    \varepsilon^{\rm NR} = \frac{\int_{t_0}^{100M}\left| h_{lm}^{\rm NR} - h_{lm}^{\rm fit} \right|^2 dt}{\int_{t_0}^{100M}\left| h_{lm}^{\rm NR} \right|^2 dt},
\end{equation}
where $h_{lm}^{\rm fit}$ is the QNM model with amplitude determined by the fitting procedure, and $h_{lm}^{\rm NR}$ is the NR waveform. The median percentage errors for these fits are 3.1\% and 6.4\% for the $(2,2)$ and $(2,1)$ modes, respectively, and approximately 13\% for the $(3,3)$ and the real components of the $(2,0)$  modes. While the model's accuracy is sufficient for current detector sensitivities, it will need to be improved for use with future and more sensitive detectors. The current limitations arise dominantly from the sparsity and uneven distribution of available NR simulations across the parameter space.

\subsubsection{Eccentricity}

Similar to other binary properties, orbital eccentricity leaves an imprint on the ringdown amplitudes and phases. 
However, since eccentricity tends to be radiated away during the binary inspiral~\cite{Peters:1963ux}, its effects become apparent in the ringdown only if it has not been fully radiated away.
This implies that eccentric effects can observable in the ringdown only if the system formed shortly before merger, or if eccentricity has been enhanced by other mechanisms.
This is possible in dense environments (such as e.g. globular or nuclear star clusters~\cite{Mapelli:2021taw}), due e.g. to the presence of a third body or matter distributions.
These considerations lead to the expectation that only a fraction of systems will be affected by eccentric corrections.
Nevertheless, in order to achieve searches of new physics using BH ringdown (one of the major goals of the BH spectroscopy program), it is imperative to construct \textit{complete} models of the binary dynamics, including sub-dominant effects.
Failure to do so would result in false violations of the predictions of GR~\cite{Gupta:2024gun}.

\begin{figure}
\centering
\includegraphics[width=0.7\textwidth]{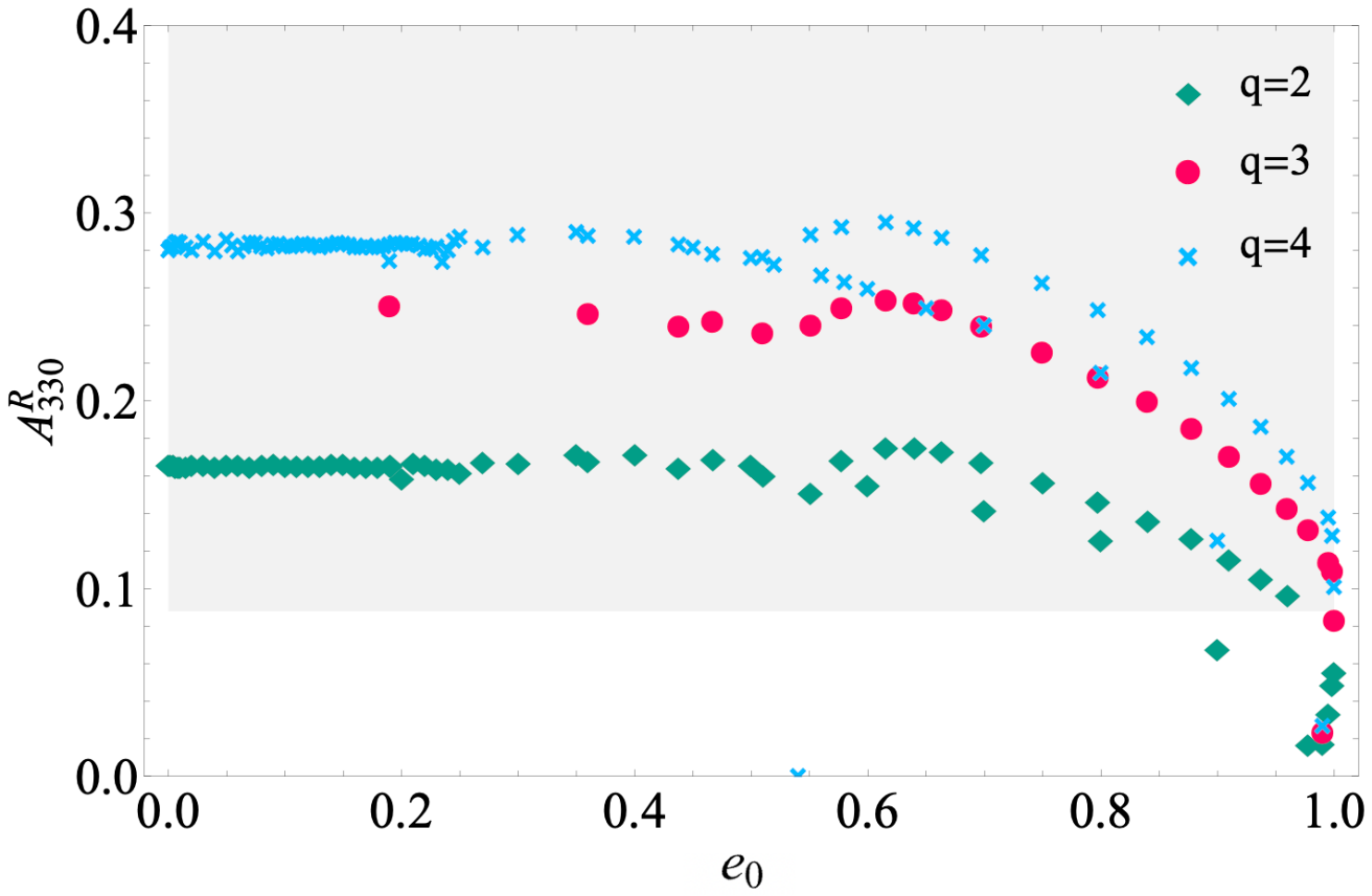}
\caption{Impact of the initial eccentricity $e_0$ at 10Hz on the estimates of the amplitude ratio $A_{330}/A_{220}$ for a set of mass ratios $q =[2,3,4]$ and nonspinning binaries, using the RIT NR catalog~\cite{ritcatalog}. The shaded area represents the 95\% credible intervals for the posteriors on the same quantities obtained from the analysis on GW190521. Figure adapted from~\cite{Forteza:2022tgq}.
\label{fig:eccentricity}}
\end{figure}

In Fig.~\ref{fig:eccentricity} we report the values of the $A_{330}/A_{220}$ amplitude ratio as a function of the residual orbital eccentricity $e_0$ at 10 Hz (i.e., when the binary enters the LVK band)~\cite{Forteza:2022tgq}.
Significant departures from the quasi-circular case start around $e_0\simeq0.6$,  while an initial eccentricity $e_0>0.3$ would be enough to have significant departures in the phase difference $\delta \phi_{330} = m/2\, \phi_{330}- \phi_{220}$~\cite{Forteza:2022tgq}. 
The amplitudes shown in 
Fig.~\ref{fig:eccentricity} are multi-valued functions, 
because two parameters are required to fully specify a noncircular orbit, and these amplitudes do not include the required dependence on a second parameter.
In addition the amplitudes are plotted as functions of a gauge-dependent parameter, so it is not immediately obvious how to compare amplitudes across different simulations or how to use them data-analysis.
Moreover, to successfully model noncircular orbits we must incorporate the time-dependence of the noncircular parameter(s).

\begin{figure}
\begin{center}
\hspace{0.5cm}
\includegraphics[width=0.5\textwidth]{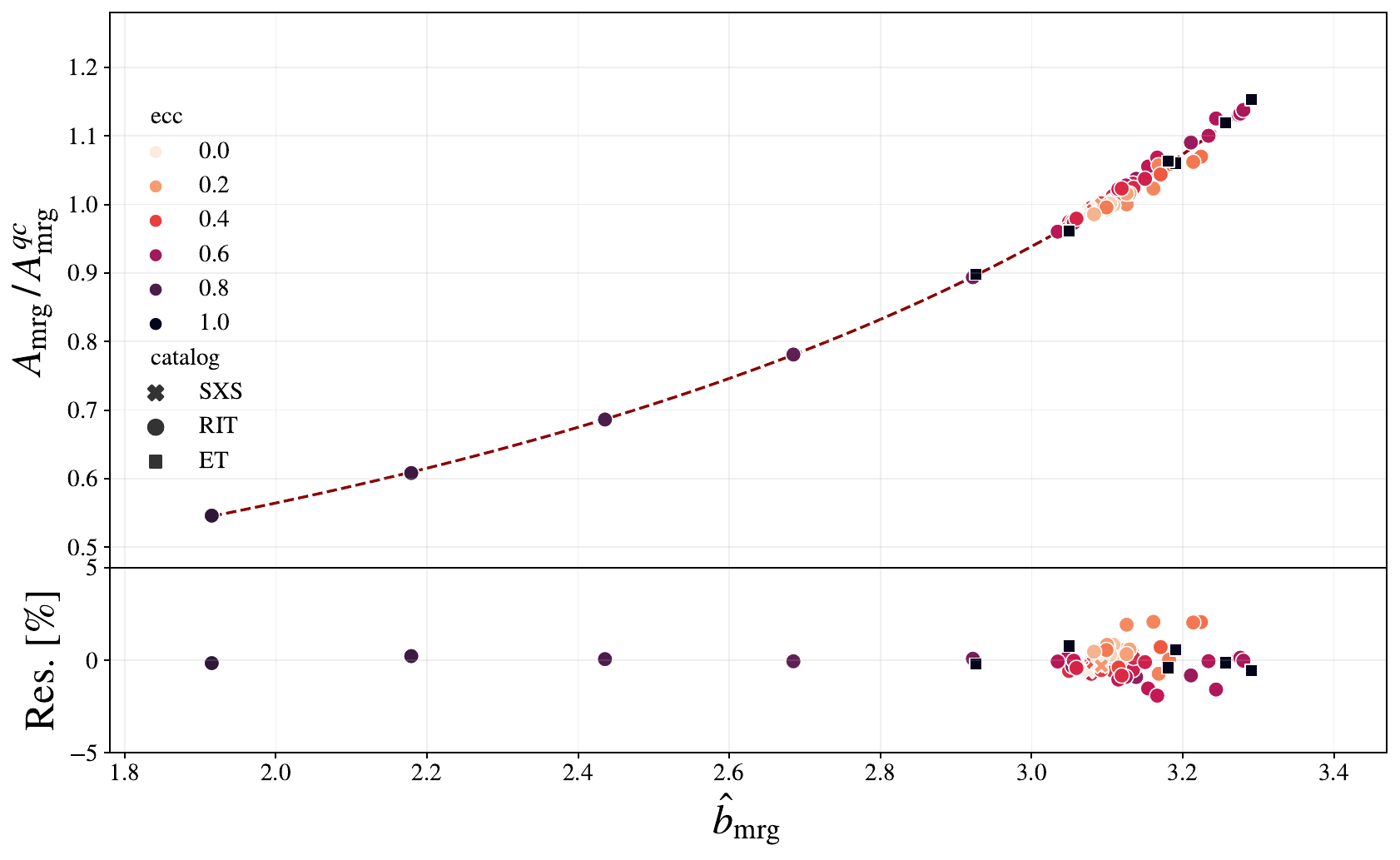}
\hspace{0.3cm}
\includegraphics[width=0.42\textwidth]{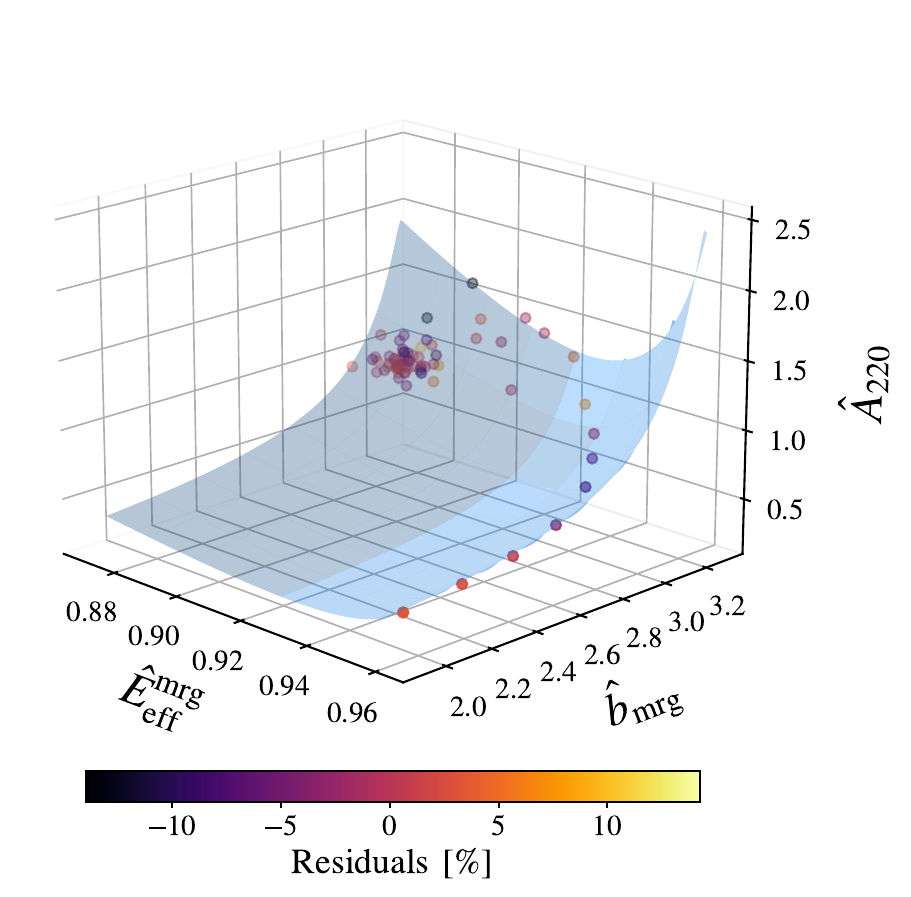}
\hspace{-2cm}
\caption{
Left: The $h_{22}$ merger amplitude (normalized by the quasi-circular value) as a function of a suitably defined dynamical impact parameter evaluated
at merger.
Markers correspond to NR simulations of different (bounded and dynamically bounded) systems, while the line is a fit to the data in terms of rational functions.
The amplitude has a simple monotonic
dependence, even for hyperbolic-like systems (square markers), hinting at a ``quasi-universality'' of the merger process. 
Right:
Ratio of the QNM amplitude with respect to the quasi-circular predictions of~\cite{Cheung:2023vki} for the fundamental quadrupolar mode as a function of time-evolved energy and angular momentum combinations.
Dots correspond to numerical simulations from the RIT catalog~\cite{ritcatalog}, while the surface corresponds to a closed-form fit across the parameter space in terms of rational functions.
Both figures refer to equal-mass, nonspinning progenitors.
Figures taken from~\cite{Carullo:2023kvj} and~\cite{Carullo:2024smg}.
\label{fig:eccentricity_EJ}}
\end{center}
\end{figure}

These issues can be addressed by using a parameterization based on gauge-invariant combinations of the binary's energy and angular momentum was~\cite{Carullo:2023kvj}.
By shifting the focus from orbit-based to dynamics-based quantities (i.e., ADM energy and angular momentum), this parameterization can account for generic noncircular planar orbits, including bounded eccentric orbits and dynamical captures.
The time dependence is incorporated by considering combinations of the ADM energy and angular momentum evolved up to the merger time by using the GW fluxes.
This approach differs from previous approaches that define time-dependent eccentric parameters in the inspiral by interpolating the waveform~\cite{Ramos-Buades:2022lgf, Shaikh:2023ypz, Boschini:2024scu}, and it can be applied to arbitrarily short waveforms.

A specific combination of these parameters -- an effective ``impact parameter'' at merger $ b_{\text{mrg}}$, inspired by the test-mass limit in an EOB description~\cite{Albanesi:2023bgi} (see Section~\ref{sec:effective-one-body}) -- captures the main dependence of the quantities characterizing the merger remnant.  
In Fig.~\ref{fig:eccentricity_EJ} we observe that the merger amplitude is a monotonic function of $\hat{b}_{\text{mrg}}$ in the equal-mass, nonspinning case.
Similar results hold for the merger frequency, the remnant BH mass and the remnant BH spin for spin-aligned binaries with different mass ratios.
This simple dependence makes it possible to construct global fits  of these quantities (that enter as inputs of IMR models when attached to inspiral predictions) across the parameter space of spin-aligned binaries.
The same parameterization can be used to fit the ringdown amplitudes of the dominant fundamental modes~\cite{Carullo:2024smg}. In the right panel of Fig.~\ref{fig:eccentricity_EJ} for $(\ell, m,n)=(2,2,0)$.
we see that noncircular orbits can have a large impact on the merger and ringdown amplitudes, with relative changes larger than $50\%$ compared to the quasi-circular case in the domain of NR calibration, and even higher when extrapolating beyond this domain.
Models of the post-merger amplitudes of eccentric systems in the extreme mass-ratio regime exist within the EOB framework~\cite{Albanesi:2023bgi} (see Section~\ref{sec:effective-one-body} below for more details).

A noncircular binary is characterized, in general, by two additional parameters (e.g. the eccentricity and radial phase/anomaly in the bounded case, or the energy and angular momentum $(E,J)$ in the generic case), but the left panel of Fig.~\ref{fig:eccentricity_EJ} shows a monotonic dependence in terms of a \textit{single} parameter. This suggests some sort of ``quasi-universality'' of the merger regime, which to a good approximation seems to be entirely captured by $\hat{b}_{\text{mrg}}$.
An interpretation of this result in terms of the orbital features can be found in~\cite{Carullo:2024smg}, but an analytical explanation would be desirable.

\begin{figure*}[t]
    \centering    \includegraphics[width=0.6\linewidth]{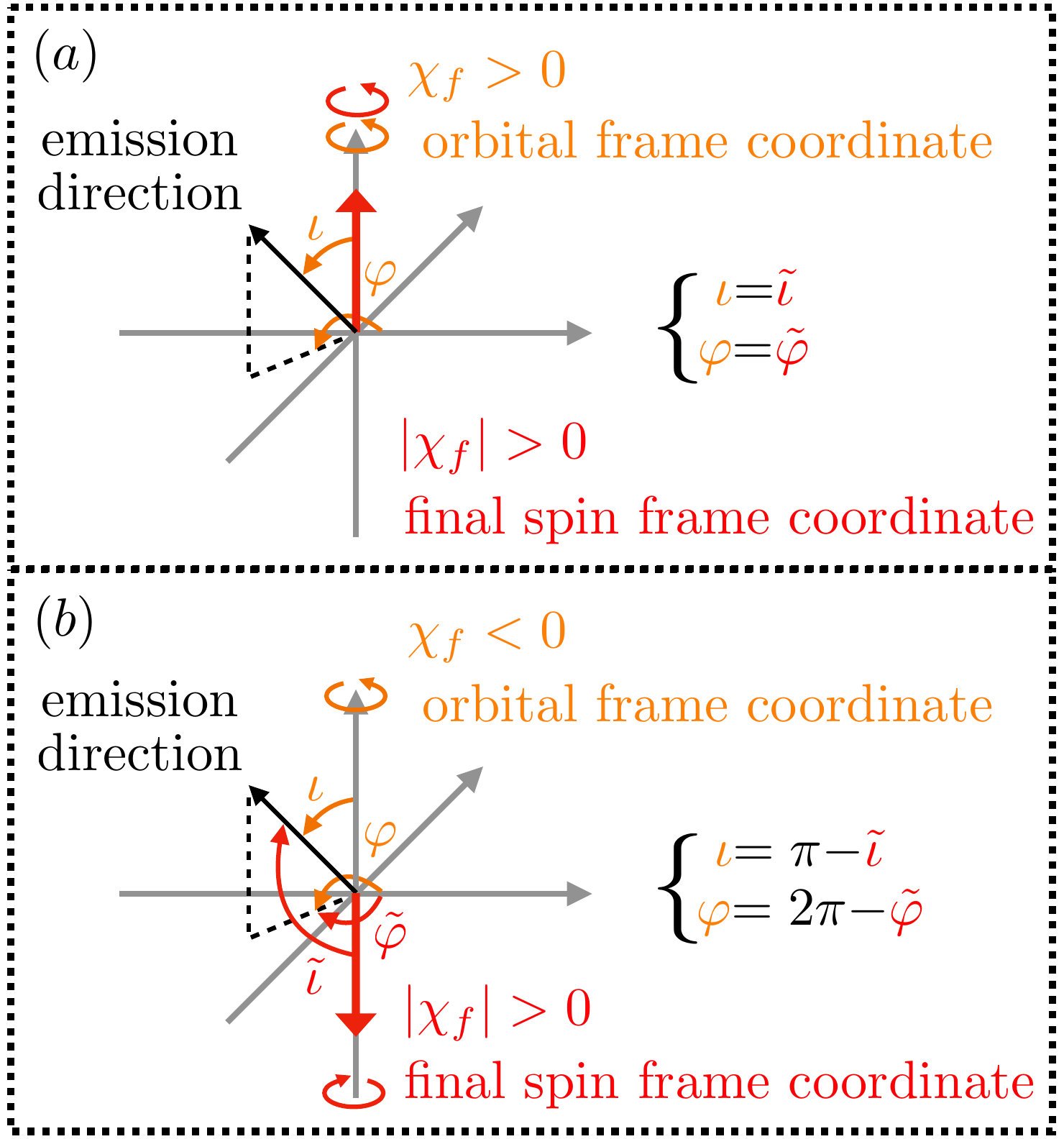}
    \caption{Depiction of coordinate conventions when the remnant spin is not aligned with the orbital angular momentum.
    Counter-rotating modes are most excited in the situation depicted in panel b).
    Figure taken from~\cite{Li:2021wgz}.}
    \label{fig:mirror_modes}
\end{figure*}

\subsubsection{Retrograde modes}\label{subsubsec:retrograde_modes_excitations}
The excitation of retrograde modes is related to the presence of a perturbation which rotates in the opposite direction relative to the spin of the background spacetime (see e.g.~\cite{Barausse:2011kb}).
These modes matter most when the progenitor spins are aligned antiparallel to the binary's orbital angular momentum.
This is pedagogically depicted in Fig.~\ref{fig:mirror_modes}, depicting two situations in which: 
(a) the remnant spin of a BH resulting from a binary merger is parallel to the progenitors orbital angular momentum;
(b) the remnant spin is opposite to the orbital angular momentum.
This latter case arises depending on the binary mass ratio, spin magnitudes and spin directions (see e.g.~\cite{Jimenez-Forteza:2016oae, Varma:2018aht}).
In situation (a) the prograde modes dominate the signal, while retrograde modes are maximally excited in situation (b).
This intrinsic excitation is observed-independent, as it is set by the binary properties.

The excitation of retrograde modes should instead not be confused with the magnitude of the $m = \pm |m|$ signal components.
The $m$ index accounts for the spherical decomposition of the signal, and the relevance of the $+ |m|$ vs $-|m|$ component depends on the observer location on the sky, set by a reference frame $(\iota, \varphi)$.
Focusing for concreteness on the $m=\pm2$ mode, an observer located on the positive side of the z-axis will only receive contributions from the $m=+ 2$ modes, and will see the binary rotate with a positive orbital frequency.
Instead, an observer located on the negative z-axis will only receive contributions from the $m=- 2$ modes, observing a binary rotating with an orbital frequency equal in magnitude and opposite in sign to observers located on the positive z-axis.
For $m=0$, both descriptions lose meaning, and the two modes are degenerate.

The intuitive excitation picture presented above is confirmed by NR fits~\cite{London:2018gaq, JimenezForteza:2020cve, Cheung:2023vki}.
The retrograde mode is found more strongly excited in the $ (2, 1)$ multipole than in the $ (2, 2)$ multipole~\cite{Cheung:2023vki}.

For binary modeling purposes, it is instead more convenient to define ``prograde'' and ``retrograde'' modes as those co/counter-rotating with the BBH orbit, rather than with the remnant BH spin. This definition coincides with the remnant-spin based one (used in Section~\ref{sec_21}) when the remnant spin is aligned with the BBH orbital direction, but it becomes opposite when the remnant spin is flipped (see also Fig.~12 of~\cite{Cheung:2023vki}).
This alternative definition offers practical advantages. In BBH simulations, the orbital direction is easy to control, while the remnant spin direction depends nontrivially on the initial conditions. By defining modes relative to the binary's orbital motion, one ensures that as the initial parameters (e.g., $q, \chi_{+}, \chi_{-}$) are varied smoothly, the amplitudes and phases of the dominant component also varies smoothly, even when the remnant spin direction flips. In contrast, adopting the (perturbative) remnant-spin based definition would lead to discontinuous transitions in QNM frequency, amplitudes, and phases when the remnant spin flips.
Comparing to the remnant-spin based definition used in Section~\ref{sec_21}, IMR models always include the dominant component co- and counter-rotating modes, switching among the two depending on the final spin direction.
Under the orbit-based definition, IMR models typically include only prograde modes, as these capture the dominant late-time frequency trend and allow for smooth fits even when the final spin direction changes, even when including a single component. In the future, both co-rotating and counter-rotating modes should be included. 
Their combination gives rise to mode-mixing, as observed in several studies conducted in the test-mass limit~\cite{Damour:2007xr,Bernuzzi:2010ty,Bernuzzi:2010xj,Barausse:2011kb,Taracchini:2014zpa,Albanesi:2023bgi} (see also Section~\ref{sec:effective-one-body}).
This effect has been modeled for test particles falling into a Kerr BH using QNM superpositions with constant amplitudes~\cite{Taracchini:2014zpa}, and more recently, a procedure to incorporate this effect in QNM-factorized phenomenological ringdown models was applied to test particles falling into a Schwarzschild BH~\cite{Albanesi:2023bgi}. However, counter-rotating modes have yet to be included in IMR models for binaries with generic mass ratios.

\subsubsection{Inclined extreme mass ratio inspirals}\label{sec:QNM_EMRIs}
As we discussed in Section~\ref{subsec:Kerr_amplitudes}, test-particle studies are valuable laboratories to develop an intuitive picture in a rigorous perturbative framework, which can then be applied to the modeling of comparable-mass systems.
A detailed QNM analysis of extreme-mass-ratio inspirals can be found  in~\cite{Lim:2019xrb} (which builds on previous work in~\cite{Hughes:2019zmt}). 
This study uses the procedure described in~\cite{Apte:2019txp} to construct worldlines describing the inspiral and plunge of a small body on initially circular but misaligned orbits of Kerr BHs.  
These worldlines in turn are used to build the Teukolsky source term for point particles following such trajectories and to
compute the resulting gravitational waveforms following~\cite{Zenginoglu:2011zz}. 
The final cycles of the resulting waveforms are Kerr QNMs, with different modes excited in a predictable way depending on the larger BH's spin parameter $a$, an angle $I$ describing the misalignment of the smaller body's orbital plane relative to the BH spin, and an angle $\theta_{\rm fin}$ which describes where on the event horizon the smaller body plunges into the BH~\cite{Lim:2019xrb, Hughes:2019zmt}.
Similarly to what is done in the comparable-mass case, by fitting to the final cycles of the waveforms to the functional form~\cite{Berti:2005ys}
\begin{eqnarray}
h(t) &=& \frac{\mu}{D}\sum_{\ell mn} \bigg[\mathcal{A}_{\ell mn}e^{-i[\omega_{\ell mn}(t-t_0) - \varphi_{\ell mn}]} S^{a\omega_{\ell mn}}_{\ell mn}(\theta,\phi)
\nonumber\\
& &+ \mathcal{A}'_{\ell mn} e^{i[\omega^*_{\ell mn}(t-t_0)+\varphi'_{\ell mn}]}S^{a\omega_{\ell mn}}_{\ell mn}(\pi-\theta,\phi)^*\bigg]\;.
\label{eq:spheroidaldecomp}
\end{eqnarray}
(where $\mu$ is the mass of the secondary body, $D$ is distance from the system to point where $h(t)$ is measured and $^*$ denotes complex conjugation), 
one can determine the amplitude $\mathcal{A}_{\ell mn}$ and phase $\varphi_{\ell mn}$ of mode $(\ell, m, n)$ in the waveform.  
Figure~\ref{fig:misalign_qnm_I} illustrates how the geometry of the system influences the resulting GWs, and the mixture of modes which enter the TD waveform.

\begin{figure*}[t]
    \centering    \includegraphics[width=0.85\linewidth]{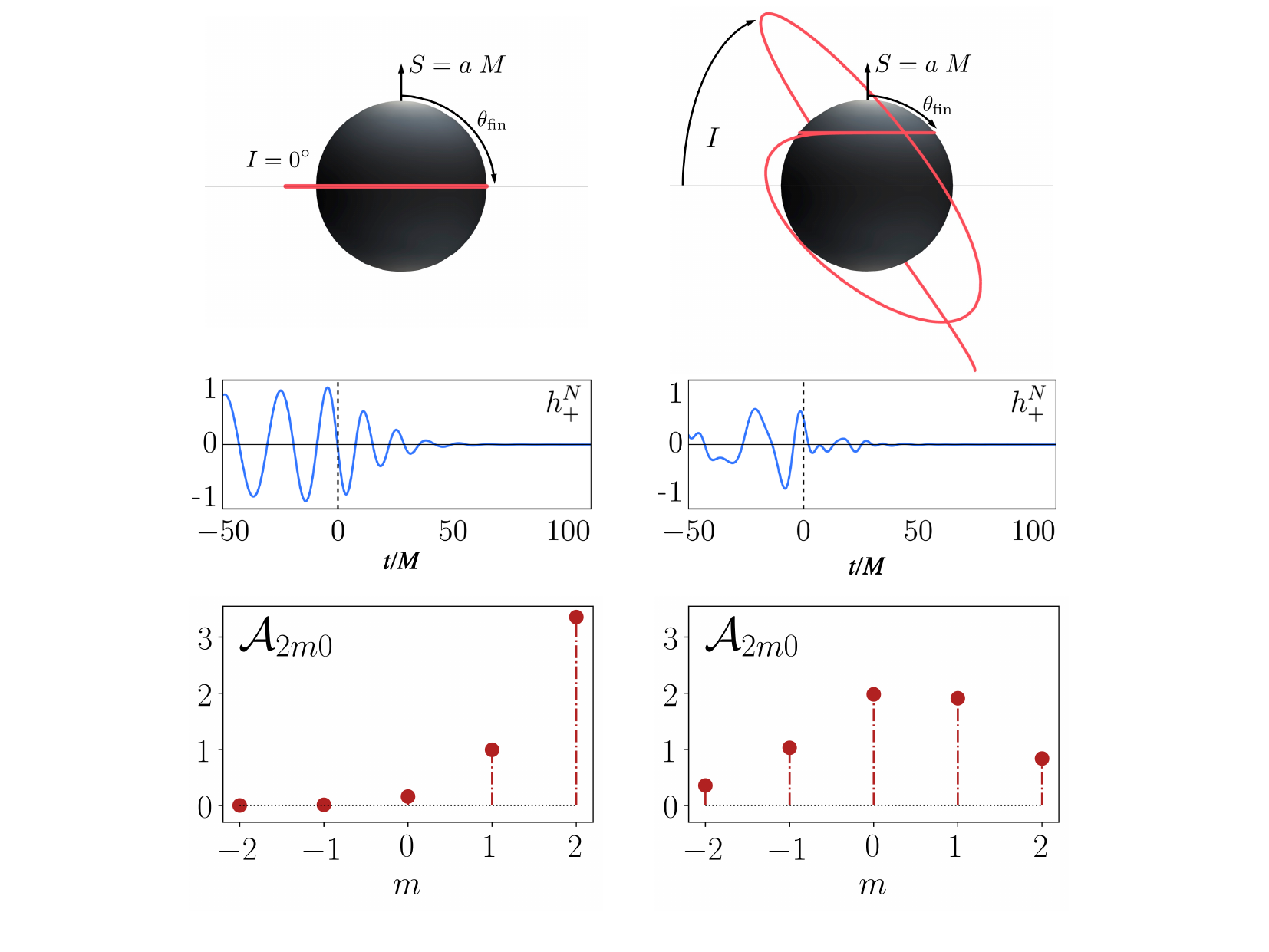}
    \caption{Two inspirals and plunges into a Kerr BH with spin $a = 0.5M$, the final several waveform cycles that result, and some of the waveforms' mode content.  
    Panels on the left show results for an equatorial configuration, $I = 0^\circ$; those on the right are for a system with inclination $I = 60^\circ$.  
    The top panels show segments of the inspiral and plunge trajectories.
    On the right, the small body's motion oscillates in the range $30^\circ \le \theta \le 150^\circ$ before plunging at $\theta_{\rm fin} = 52^\circ$.  
    The middle panels show the final several cycles of the TD waveform. The bottom panels show the contribution of the $(\ell, m, n) = (2,m,0)$ QNM to these waves.
    Notice how the equatorial case is dominated by the $m  = 2$ and $m = 1$ modes; the misaligned case has strong contributions from many different values of $m$.  
    Figure adapted from~\cite{Hughes:2019zmt}.}
    \label{fig:misalign_qnm_I}
\end{figure*}

It is worth noting that $\theta_{\rm fin}$ is essentially an accidental phase parameter: a system with inclination angle $I$ will have $90^\circ - I \le \theta_{\rm fin} \le 90^\circ + I$ (for $I \le 90^\circ$) or $I - 90^\circ \le \theta_{\rm fin} \le 270^\circ - I$ (for $I > 90^\circ$).  
Two systems which are identical in all ways except for their initial position along their orbit will plunge with different values of $\theta_{\rm fin}$, and as a consequence excite very different mixtures of QNMs.  
This is illustrated in Fig.~\ref{fig:mode_vs_thfin}, which shows how the mode amplitude $\mathcal{A}_{220}$ and phase $\varphi_{220}$ vary with $\theta_{\rm fin}$ for a set of systems which all have $a = 0.5M$, $I = 60^\circ$.  
Despite the fact that the systems are identical in many of their most important properties, the amplitude of this important QNM can vary significantly depending on how the plunging body crosses the horizon. 
A rapidly spinning remnant BH with extreme mass ratio may be advantageous for detecting QNMs, as it results in a long-lived ringdown and enhanced excitation of higher multipole modes~\cite{Oshita:2022yry,Watarai:2024vni}. 
In an extension of these studies, significant excitation of higher multipole modes was also observed in intermediate mass ratio mergers~\cite{Watarai:2024huy}, where the energy and angular momentum losses are taken into account by utilizing the Ori-Thorne procedure~\cite{Ori:2000zn}.

\begin{figure*}[t]
    \centering    \includegraphics[width=0.75\linewidth]{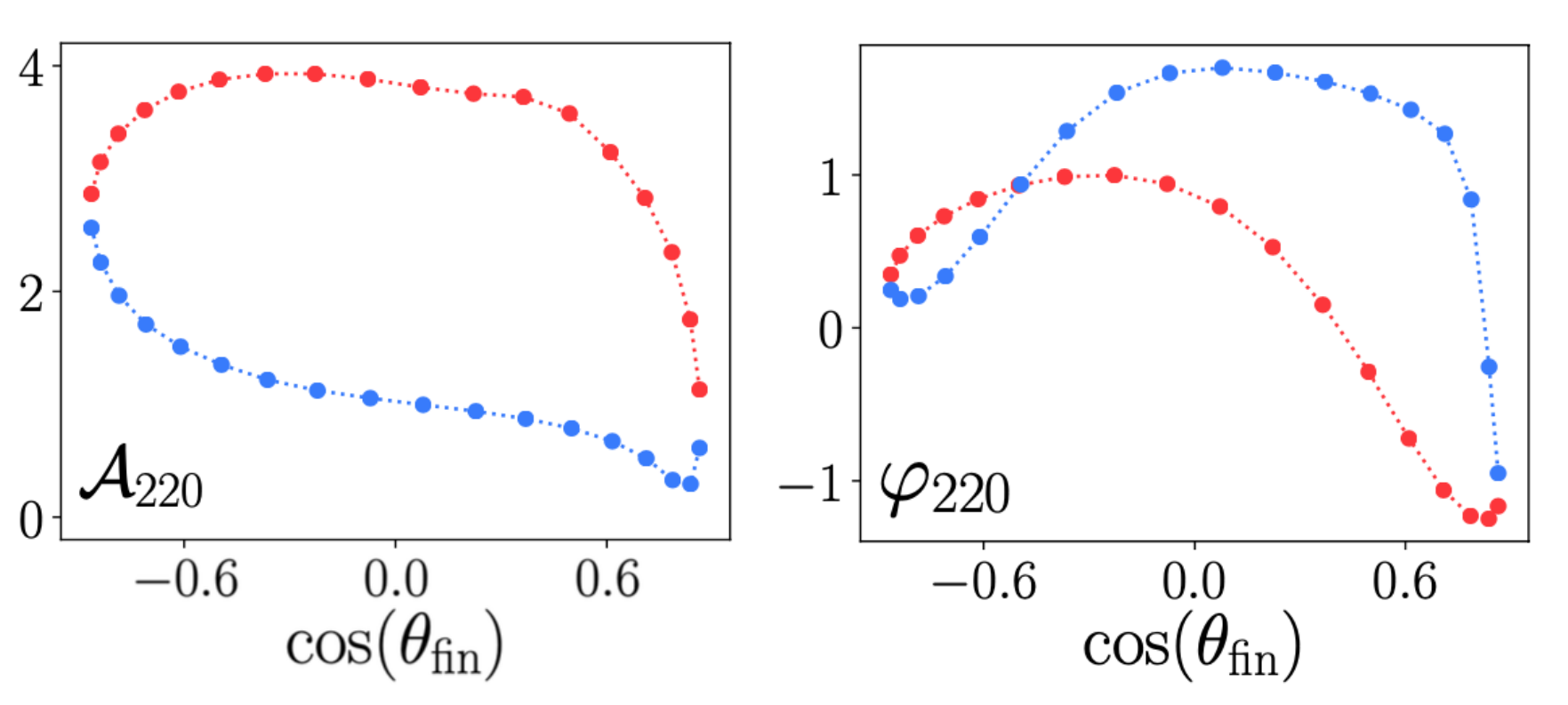}
    \caption{Example showing how the $(\ell, m, n) = (2,2,0)$ amplitude and phase vary for a inspirals and plunges differing only by the final plunge angle $\theta_{\rm fin}$.  The spin of the larger BH is $a = 0.5$, and the system's orbital inclination angle $I = 60^\circ$.  Notice that the amplitude and phase of this mode varies significantly as a function of $\theta_{\rm fin}$.  Notice also that these quantities are double valued: for each value of $\theta_{\rm fin}$, the mode varies depending on whether the plunging object has $\dot\theta > 0$ before crossing the horizon (red dots), or has $\dot\theta < 0$ (blue dots).  Figure adapted from~\cite{Hughes:2019zmt}.}
    \label{fig:mode_vs_thfin}
\end{figure*}

\subsection{Time-dependent amplitudes: phenomenological models}
\label{sec:effective-one-body}

\vspace{-.1cm}

\noindent \textit{Initial contributors: Albanesi, Pompili}

\vspace{.2cm}

\subsubsection{Basic concepts}
The EOB approach, originally introduced in~\cite{Buonanno:1998gg,Buonanno:2000ef,Damour:2000we,Damour:2001tu,Buonanno:2005xu}, provides a framework to construct semi-analytic waveform models covering the entire coalescence of compact-object binaries (i.e., the inspiral, plunge, merger and ringdown). 
The basic idea of EOB models is to map the motion of the two bodies, typically described in terms of PN equations of motion, into the motion of a particle in an effective metric. 
The latter is a deformation of a BH solution, where the parameter that tunes this deformation is the symmetric mass ratio $\eta \equiv q/(1+q)^2$. 
Through this mapping, EOB models naturally incorporate the test-mass limit and, as a consequence, information from perturbation theory can be used to improve their reliability. 
Moreover, since the advent of numerical simulations~\cite{Pretorius:2005gq,Campanelli:2005dd,Baker:2005vv}, the EOB framework has integrated information from NR simulations, offering highly accurate waveform models that are routinely used in GW data analysis of IMR signals. 
Numerical simulations are used both to calibrate analytically unknown coefficients in the EOB Hamiltonian, adding phenomenological contributions to the poorly modeled near-merger waveform, to select the unknown optimal polynomial orders to be chosen in resummation procedures, and to build phenomenological post-merger models that are matched to the inspiral-plunge waveform. 

The complete EOB waveform is usually constructed by attaching a merger-ringdown waveform, $h_{\ell m}^{\rm merger-RD}(t)$, to an inspiral-plunge one, $h_{\ell m}^{\rm insp-plunge}(t)$, at a suitable matching time $t=t_{\rm match}$, around the peak of the EOB orbital frequency (which approximately corresponds to merger time), that is~\cite{Buonanno:2000ef}:
\begin{equation}
    h_{\ell m}(t) = h_{\ell m}^{\rm insp-plunge}(t)\hat{h}_{\ell m}^{\rm NQC}(t) \, \Theta \left( t_{\rm match}^{\ell m}-t \right)
    + h_{\ell m}^{\rm merger-RD}(t) \, \Theta \left( t-t_{\rm match}^{\ell m} \right) \,,
\end{equation}
where $\Theta(t)$ is the Heaviside step function.
The inspiral-plunge waveform modes are based on a resummation of the PN GW modes~\cite{Damour:2008gu, Damour:2007xr, Damour:2007yf,Pan:2010hz,Messina:2018ghh,vandeMeent:2023ols,Nagar:2024oyk}, evaluated on the dynamics obtained from the EOB equations of motion~\cite{Buonanno:1998gg,Buonanno:2000ef,Damour:2001tu}.
The smooth connection of the inspiral-plunge waveform with the merger-ringdown waveform is achieved by applying the numerically tuned next-to-quasi-circular corrections~\cite{Damour:2007xr} $\hat{h}_{\ell m}^{\rm NQC}$, which are particularly relevant during the plunge, when the radial motion dominates the dynamics.
In the rest of this section, we focus on the modeling of the merger-ringdown waveform modes.

\subsubsection{Ringdown as a QNM superposition}
As previously mentioned, early EOB models represented the merger-ringdown waveform as a linear superposition of QNMs~\cite{Buonanno:2000ef,Buonanno:2006ui,Buonanno:2007pf,Damour:2007xr,Damour:2007vq,Damour:2008te,Damour:2009kr,Buonanno:2009qa,Bernuzzi:2010ty,Bernuzzi:2010xj,Pan:2011gk,Barausse:2011kb,Taracchini:2012ig,Damour:2012ky,Pan:2013rra,Taracchini:2013rva,Damour:2013tla,Babak:2016tgq}.
This approach, inspired by results for the infall of a test mass into a BH~\cite{Davis:1972dm} and the close-limit approximation~\cite{Price:1994pm}, was initiated in~\cite{Buonanno:2000ef}, where the inspiral waveform was matched to the least-damped QNM of the remnant BH at the light-ring crossing. 
Notably, the final mass and spin were estimated from the EOB Hamiltonian and angular momentum at the end of the plunge, thus providing the first semi-analytical estimate of such quantities for a remnant generated by a BBH merger.
Later works confirmed the goodness of the EOB estimates for the remnant properties~\cite{Buonanno:2006ui,Damour:2007cb,Damour:2013tla}, but recent models compute these quantities from more accurate NR fits~\cite{Jimenez-Forteza:2016oae, Hofmann:2016yih}.
To ensure continuity between the EOB inspiral-plunge waveform and the QNM ansatz, the introduction of two free coefficients in the amplitude and phase of the merger-ringdown waveform was needed, and determined by imposing $C_1$ continuity conditions at the matching time.
With the 2005 NR breakthrough~\cite{Pretorius:2005gq,Campanelli:2005dd,Baker:2005vv}, the picture delineated in~\cite{Buonanno:2000ef}, which put forward the first model for the entire IMR waveform in the comparable-mass regime, was revealed to be qualitatively, and to some extent quantitatively, correct~\cite{Buonanno:2006ui}. 
There, the underlying picture was inspired by the study of the test-mass case, a useful laboratory for many EOB developments in the comparable-mass regime.

The inclusion of three overtones up to the waveform peak was first considered in~\cite{Buonanno:2006ui} (see Section~\ref{subsec:overtones}). The relevance of higher overtones in the comparable-mass regime was confirmed and used to complete a pre-merger EOB model in~\cite{Buonanno:2007pf}, with free parameters fixed by requiring continuity. 
A study in the test-mass limit also found that higher QNM overtones are required in order to improve the frequency of the inspiral-plunge waveform, and thus achieve a smoother transition~\cite{Damour:2007xr}.
Further, the latter work introduced the idea of performing the match of the QNM ansatz with the inspiral-plunge EOB waveform on a time-extended interval around the peak of the orbital frequency, a procedure also known as {\it matching comb}.
While these refined methods led to reliable results for nonspinning binaries~\cite{Damour:2007xr,Damour:2007vq,Damour:2009kr,Taracchini:2012ig,Damour:2013tla}, the ringdown frequency obtained from the QNM superposition was found to rise too quickly after merger for spin-aligned binaries and higher harmonics.
To address this problem, {\it pseudo}-QNMs, i.e., fictitious modes, were introduced~\cite{Buonanno:2009qa}.
These complex frequencies were phenomenologically determined to improve the transition from the inspiral-plunge EOB frequency, to the one of QNM superpositions assuming the asymptotic Kerr mass and spin values.

\subsubsection{QNM-factorized phenomenological ringdown models}
The necessity to include {\it pseudo}-QNMs highlighted the need to include other physical effects beyond the pure QNM description, such as transient contributions. 
While analytical computations from first principles of such effects are still ongoing, a practical solution is to build physical intuition, and then provide phenomenological ansatzes.
The seminal work of~\cite{Baker:2008mj} described the full waveform in terms of a rotating source and produced closed-form fits of the waveform frequency. 
Later, a phenomenological description of the ringdown which is still employed in state-of-the-art EOB and TD Phenom models was proposed~\cite{Damour:2014yha}.
The basic idea is to consider a {\it multiplicative} decomposition of the waveform by factorizing the contribution of the co-rotating fundamental QNM. 
Focusing on the dominant $(\ell,m)=(2,2)$ mode $h_{22}/\eta = A_{22}e^{-i \phi_{22}}$, the QNM-rescaled signal reads 
\begin{equation}
\label{eq:hbar_eob}
\hat{h}_{22}(\tau) = e^{i\omega_{220}\tau + i\phi_{22}^0} h_{22}(\tau),
\end{equation}
where $\omega_{220}$ is the $M$-rescaled complex fundamental frequency, $\tau=(t-t_0)/M$, and $t_0$ corresponds to the peak of the (2,2) amplitude.
Under the assumption of a pure QNM superposition with constant amplitudes $c_i$, the rescaled signal can be written as
\begin{equation}
    \label{eq:hbar_eob_qnm}
    \hat{h}_{22}^{\rm QNM}(\tau)= c_0 + c_1 e^{-i (\omega_{221}-\omega_{220})\tau}+c_2 e^{-i (\omega_{222}-\omega_{220})\tau}+ \cdots .
\end{equation}
Note that while the damping coefficients of the QNMs increase with $n$, the real parts remain relatively unchanged, especially if the spin of the remnant is large. 
As a consequence, the differences $\omega_{221}-\omega_{220}$, $\omega_{222}-\omega_{220}$, $\dots$ in Eq.~\eqref{eq:hbar_eob_qnm} are approximately constant and imaginary. 
Therefore, Eq.~\eqref{eq:hbar_eob_qnm} describes approximate straight lines in the complex plane. 
However, when looking at the QNM-rescaled waveform obtained from equal-mass NR simulations, only the ones which correspond to nonspinning or anti-aligned configurations are approximately straight, as shown in the left panel Fig.~\ref{fig:snail_qnm}. On the contrary, spin-aligned configurations show a {\it snail-shaped} behavior, thus indicating that a pure QNM ansatz is not suitable for their description.
From this perspective, the role of pseudo-QNMs $\omega_{221}'$~\cite{Buonanno:2009qa,Pan:2011gk,Taracchini:2012ig,Taracchini:2013rva} for positive-spin configurations was needed to account for the rotation of the curve in the ${\bar h}_{22}$ plane by having $\omega_{221}'-\omega_{220}$ with significant nonzero real frequency difference.
However, a different strategy was employed in~\cite{Damour:2014yha}. 
In the following, we drop the $(\ell,m)$ indices to simplify the notation.
By decomposing the QNM-rescaled waveform $\hat{h}$ in amplitude and phase as
\begin{equation}
\hat{h} = A_{\hat{h}} e^{i \phi_{\hat{h}}},
\end{equation}
one can see that the shapes of the amplitude $A_{\hat{h}}$ and phase $\phi_{\hat{h}}$ are reminiscent of activation-like functions, as shown in the right panels of Fig.~\ref{fig:snail_qnm} for the equal-mass case with individual BH spins $\chi\equiv\chi_1=\chi_2=0.97$ (dashed blue). 
To model such activation-like behavior, the following ansatz was proposed

\begin{align}
\label{eq:Atemp}
\tilde A(\tau)    & = c_1^A \tanh\left(c_2^A \tau + c_3^A\right) + c_4^A,\\
\label{eq:phitemp}
\tilde \phi(\tau) & = -c_1^\phi\ln\left(\dfrac{1+c_3^\phi e^{-c_2^\phi\tau}+c_4^\phi e^{-2c_2^\phi\tau}}{1+c_3^\phi+c_4^\phi}\right).
\end{align}
Some of these coefficients were constrained by requiring continuity conditions at $t_0$ and fixing the asymptotic behavior for large $\tau$ (see~\cite{Damour:2014yha} for the explicit expressions).
The remaining coefficients were instead determined by a least-square fit performed on NR waveforms. 

\begin{figure*}[t]
    \centering
    \includegraphics[width=0.48\linewidth]{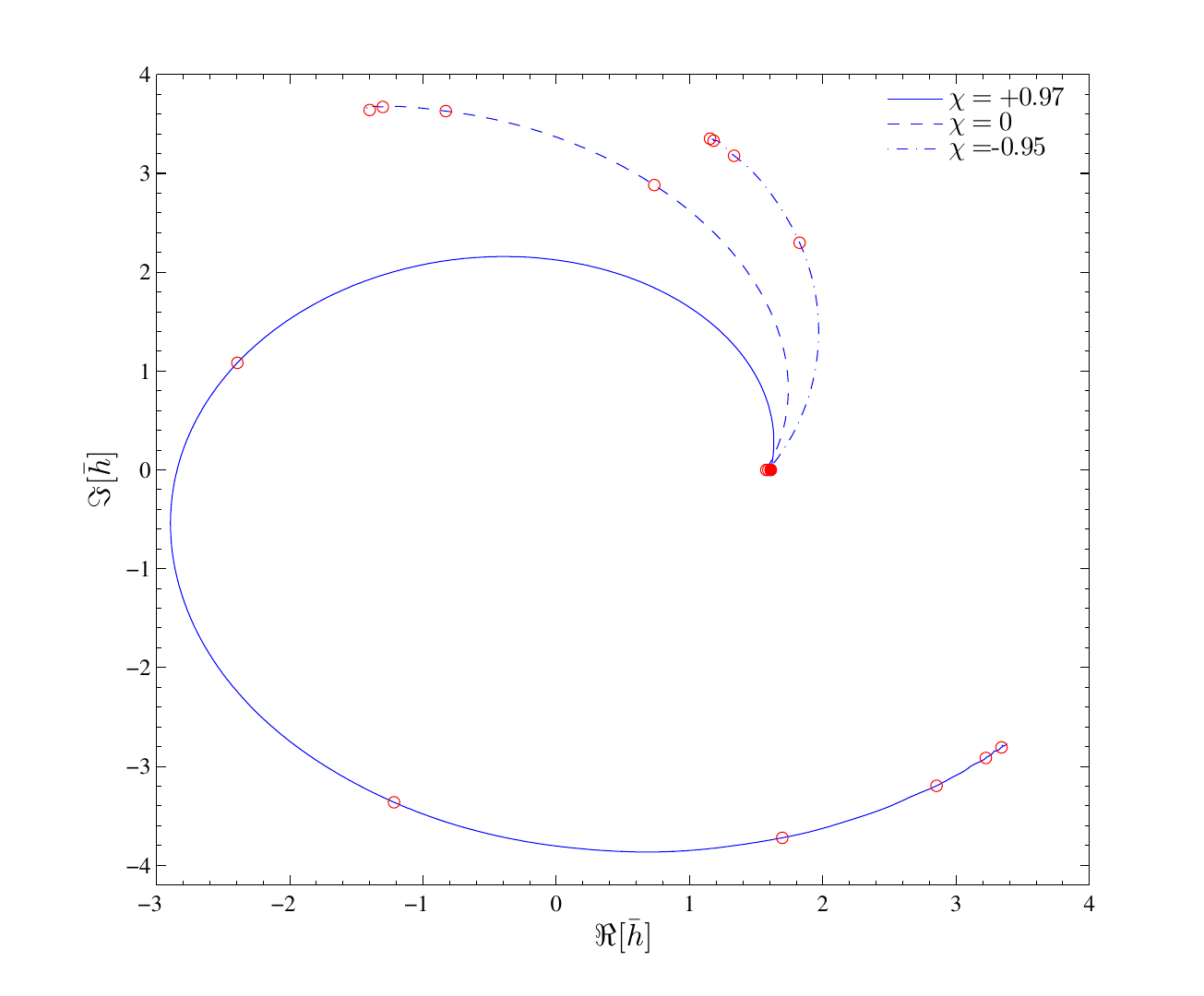}
    \includegraphics[width=0.50\linewidth]{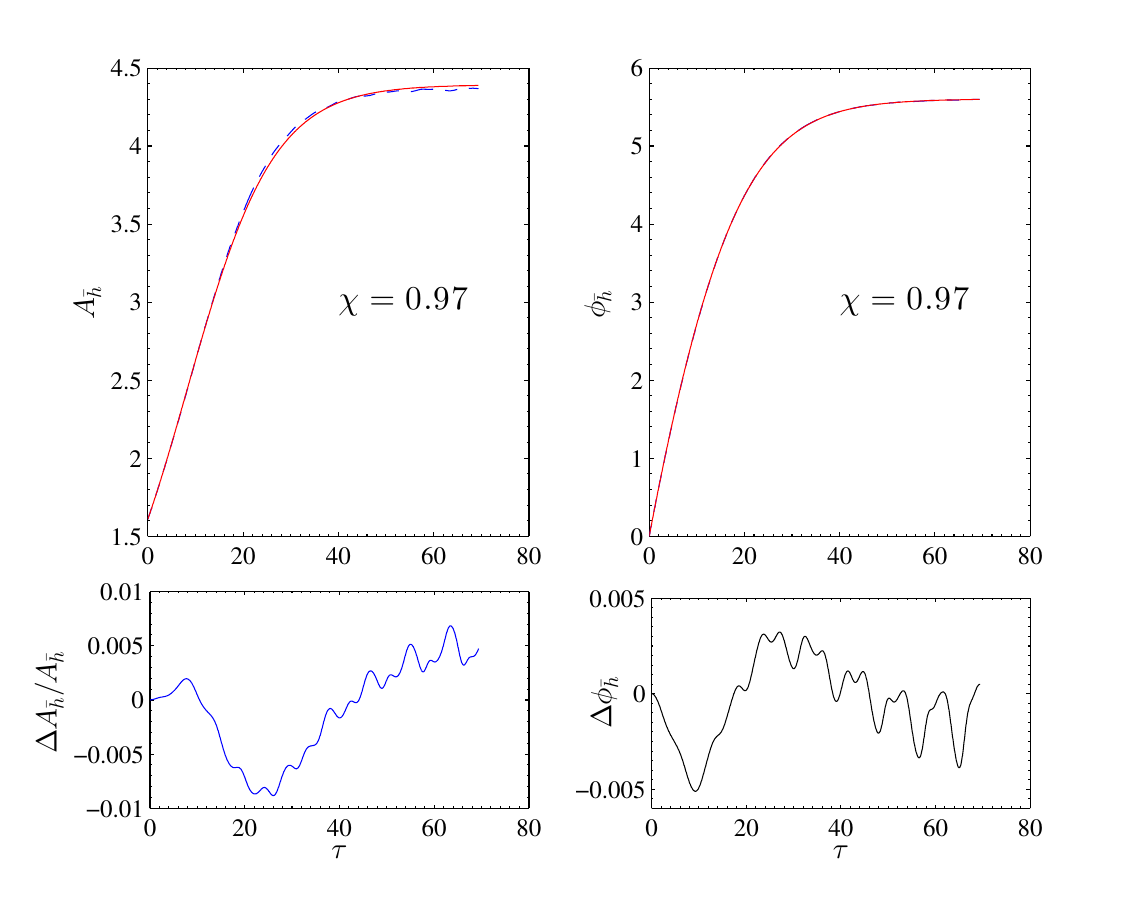}
    \caption{Left: QNM-rescaled (2,2) waveform $\hat{h}$ for three different equal-mass NR simulations of the SXS catalog. Each line corresponds to a different value for the BH spins $\chi\equiv \chi_1=\chi_2$. The filled circle corresponds to $\tau = 0$, while the empty ones mark intervals of $\Delta \tau = 10$. Right: amplitude and phase of $\hat{h}$ for the $\chi=0.97$ case (dashed blue) and fits performed using Eqs.~\eqref{eq:Atemp} and~\eqref{eq:phitemp} (red). Residuals are shown in the bottom subpanels. Figure taken from~\cite{Damour:2014yha}.}
    \label{fig:snail_qnm}
\end{figure*}

The results of this fit are shown for the $\chi=0.97$ case in the right panels of Fig.~\ref{fig:snail_qnm} (red). 
Even though transient effects, overtones, and nonlinearities are not explicitly encapsulated in Eq.~\eqref{eq:Atemp} and~\eqref{eq:phitemp}, these effects are effectively included in the merger-ringdown model, since these phenomenological ansatzes are calibrated on NR simulations.
Note that at late times $\hat{h}$ is approximately constant, indicating that in the late ringdown, before the transition to the tail regime (see Section~\ref{sec:tails}), only the contribution of the fundamental QNM is still relevant.

The fitted coefficients, together with the numerical quantities needed for the continuity conditions at $t_0$, depend smoothly on $\chi=\chi_1=\chi_2$,  thus allowing a simple polynomial fit approximation over the whole spin range (see, e.g., Fig.~6 of~\cite{Damour:2014yha}). 
Later works~\cite{DelPozzo:2016kmd, Bohe:2016gbl, Nagar:2020pcj, Pompili:2023tna,Nagar:2023zxh} considered also unequal-mass binaries and progenitors with different individual spins, thus providing closed-form descriptions of the post-merger waveform over the whole parameter space of quasi-circular spin-aligned binaries, which are currently used in state-of-the-art EOB and phenomenological TD models~\cite{Estelles:2020osj, Estelles:2020twz, Estelles:2021gvs}.

While initial work focused on the $(2,2)$ mode, the same methodology can be readily applied to higher-modes.
For each complex $(\ell,m)$ mode, the fit can be started from the peak of the corresponding amplitude~\cite{Nagar:2019wds}. However, fits for the higher harmonics can robustly be performed even starting from the peak of the $(2,2)$ amplitude~\cite{Cotesta:2018fcv}.
This choice facilitates the matching of the post-merger model with the inspiral-plunge EOB waveform, especially for the multipoles whose peaks are strongly delayed with respect to the $(2,2)$ amplitude, such as the $(2,1)$ mode. 

The templates in Eq.~\eqref{eq:Atemp} and~\eqref{eq:phitemp} describe a monotonic amplitude and frequency evolution, so they are not directly applicable to modes with post-merger oscillations due to mode mixing~\cite{Buonanno:2006ui, Kelly:2012nd, Berti:2014fga}. 
For comparable-mass binaries, this occurs primarily in the $(3,2)$ and $(4,3)$ modes, and is related to the mismatch between the {\it spherical harmonic} basis used for extraction in NR simulations, and the {\it spheroidal harmonic} basis employed in perturbation theory~\cite{Kelly:2012nd}. 
The previous template was recently extended~\cite{Pompili:2023tna} to such contributions in the $(3,2)$ and $(4,3)$ modes by applying Eqs.~\eqref{eq:Atemp},~\eqref{eq:phitemp} to the spheroidal harmonics~\cite{Berti:2005gp} $(3,2,0)$ and $(4,3,0)$, which maintain a monotonic amplitude and frequency evolution. 
The $(3,2)$ and $(4,3)$ spherical harmonics can be reconstructed by combining the $(3,2,0)$ and $(4,3,0)$ spheroidal harmonics with the $(2,2)$ and $(3,3)$ spherical harmonics, using suitable mode mixing coefficients~\cite{Berti:2014fga}.
Another source of mode mixing is the beating between co-rotating and counter-rotating modes, as noticed in several studies conducted in the test-mass limit~\cite{Damour:2007xr,Bernuzzi:2010ty,Bernuzzi:2010xj,Barausse:2011kb,Taracchini:2014zpa,Albanesi:2023bgi}. 
While this effect is also visible in the (2,2) ringdown waveform generated by the infall of a test particle in a Schwarzschild BH, it is even more relevant in some higher harmonics, such as the (2,1) and (3,2) ones, or for orbital angular momenta anti-aligned with the Kerr spin~\cite{Taracchini:2014zpa}. 
A procedure to include this effect in QNM-factorized phenomenological ringdown models at late times has been recently proposed and applied to test particles falling in a Schwarzschild BH~\cite{Albanesi:2023bgi}. While this methodology seems to retain its validity in the comparable-mass case, as shown in the example reported in Fig.~32 of~\cite{Nagar:2023zxh}, it has yet to be included in EOB models for generic mass ratios.

The methodology of~\cite{Damour:2014yha} has also been recently extended to the oscillatory contribution in real-valued $m=0$ modes by applying a complexification of the physical signal with a Hilbert transform~\cite{Albanesi:2024fts}. 
In these cases, the description has to be completed by including null memory contributions, discussed in more detail in Section~\ref{sec:bms-frames-memory}.

The closed-form nature of the merger-ringdown waveform also makes it well-suited for BH spectroscopy analyses, either by introducing parameterized deviations to the QNM complex frequencies within the complete IMR EOB model~\cite{Brito:2018rfr, Ghosh:2021mrv, Maggio:2022hre, Toubiana:2023cwr,Pompili:2025cdc}, or by using the merger-ringdown waveform as a standalone post-merger model~\cite{Gennari:2023gmx}. 
The same model has also been applied to theories of gravity beyond GR~\cite{Silva:2022srr, Julie:2024fwy}, provided the QNM frequencies in the alternative theory are known.
These applications are discussed in more detail in Section~\ref{sec:DataAnalysis}.

\subsubsection{Eccentric and precessing orbits}
In recent years, EOB models have been generalized to describe generic orbits~\cite{Hinderer:2017jcs,Chiaramello:2020ehz,Nagar:2020xsk,Khalil:2021txt,Placidi:2021rkh,Ramos-Buades:2021adz,Gamba:2024cvy,Nagar:2024dzj,Nagar:2024oyk,Gamboa:2024hli}.
However, the post-merger signal of these models is still described using the quasi-circular ringdown waveform described above.
For moderate eccentricity, such modeling still yields accurate results due to the circularization of the dynamics during the inspiral, and works reasonably well even in the dynamical capture scenario~\cite{Gamba:2021gap,Andrade:2023trh,Albanesi:2024xus}.
However, the accuracy of these models could be improved by including noncircular information from NR data in the ringdown waveform, as shown in~\cite{Andrade:2023trh}.
Motivated by these considerations, gauge-invariant combinations of evolved energy and angular momentum, well suited to model post-merger effects were introduced in~\cite{Carullo:2023kvj}.
Closed-form descriptions of the remnant properties and waveform quantities at merger were also provided, which could be used to improve the post-merger waveforms of EOB models for generic orbits. 
Fully eccentric post-merger models have instead been explored in the test-mass limit~\cite{Albanesi:2021rby,Albanesi:2023bgi}, where an alternative to Eq.~\eqref{eq:Atemp} has also been proposed~\cite{Albanesi:2021rby}. 

For binaries with generic spins, the same construction is applied to the waveform in the co-precessing frame~\cite{Ossokine:2020kjp, Estelles:2021gvs, Gamba:2021ydi, Ramos-Buades:2023ehm}. 
The co-precessing frame is a noninertial frame that tracks the instantaneous motion of the orbital plane, in which the gravitational radiation resembles the one of an aligned-spin binary~\cite{Buonanno:2002fy, Schmidt:2010it, Boyle:2011gg, OShaughnessy:2011pmr, Schmidt:2012rh} (up to asymmetries over the orbital plane~\cite{Boyle:2014ioa, Ramos-Buades:2020noq}).
However, spin precession introduces specific effects that require additional consideration.
To convert the waveform from the co-precessing frame to the inertial observer's frame, one must describe the rotation between the two, often parameterized by a set of time-dependent Euler angles $(\alpha(t), \beta(t), \gamma(t))$, which extends beyond merger.
A phenomenological prescription, informed by NR simulations~\cite{OShaughnessy:2012iol}, was then employed~\cite{Ossokine:2020kjp, Estelles:2021gvs, Gamba:2021ydi, Ramos-Buades:2023ehm} to show that the co-precessing frame continues to approximately precess around the direction of the final angular momentum, with a precession frequency $\omega_{\rm{prec}}=\omega_{R, 220} - \omega_{R, 210}$, where $\omega_{R, \ell m n}$ denotes the real part of the complex $\omega_{\ell m n}$ QNM frequency.
As noted in Section~\ref{subsec:precession}, such a rotation induces a re-ordering of the amplitudes of different modes as compared to the nonprecessing scenario~\cite{Schmidt:2010it, Boyle:2013nka, Finch:2021iip}, leading to particularly large amplitudes for modes with $\ell = m \neq 2$~\cite{Hughes:2019zmt, Lim:2022veo, Siegel:2023lxl, Zhu:2023fnf}.
The rotation prescription used in EOB and Phenom models provides an accurate approximation if the $(\ell,m)=(2,0)$ QNM amplitude in the remnant frame is negligible. 
However, for highly precessing binaries in which the $(2,0)$ QNM can be strongly excited, this assumption may be a source of systematics~\cite{Zhu:2023fnf}.

Finally, the QNM frequencies determined from BH perturbation theory are applicable in a frame in which the final spin of the remnant BH is aligned with the $z$-direction (the so-called $\mathbf{J}$-frame). 
A relationship between the QNM frequencies in the $\mathbf{J}$-frame and an effective ringdown frequency in the co-precessing frame was derived in~\cite{Hamilton:2023znn}:
    \begin{equation}
    \omega_{R, \ell m 0}^{\mathrm{CP}}=\omega_{R, \ell m 0}^{\mathrm{J}}-m\left(1-\left|\cos \beta \right|\right) \omega_{\text {prec }} ,
\end{equation}
This relation was employed in the construction of recent phenomenological IMR models~\cite{Ramos-Buades:2023ehm, Thompson:2023ase}.

\subsection{Tails}
\label{sec:tails}

\vspace{-.1cm}

\noindent \textit{Initial contributors: De Amicis, Khanna}

\vspace{.2cm}

A signal characterized by a slow decay in time, persisting after the main body of radiation has passed, is usually called a ``tail effect.'' 
In this section we review past and recent progress on modeling the tail signal that characterizes the late-time response of a BH spacetime to a perturbation.
We will refer to effects originated by initial data perturbations of a BH in vacuum as ``initial-data driven'' tails, and to spacetime effects due to moving perturbing bodies (arising e.g. in the relativistic two-body problem) as ``source-driven'' tails.

\subsubsection{Initial data-driven tails: spherically symmetric case}
Late-time tails were first predicted by Price in the context of linear relaxation of the gravitational field following nonspherical collapse~\cite{Price:1971fb,Price:1972pw}.
He showed that for an observer at finite distances, the signal
follows an inverse power-law falloff; in particular, that $\sim t^{-(2\ell+2+x)}$, where $x=0$ if a static mode is initially present, and $x=1$ otherwise. 

This result was later generalized in~\cite{Leaver:1986gd,Ching:1994bd, Andersson:1996cm,Poisson:2002jz,Hod:2009my} by means of a Green's function approach. 
These studies found that tails are generated by the back-scattering of low frequency signals with the long-range curvature of the background.
In particular, Leaver~\cite{Leaver:1986gd} computed the late-time signal observed at null infinity as $u^{-\ell-2}$ for all retarded times $u$, where $\ell$ the waveform's multipole index, at leading order in a small-frequency expansion. 
At finite distances, the late-time response is first dominated by a transient radiative tail $u^{-\ell-2}$, while Price's prediction is recovered only after some time.
The duration of this radiative transient increases with the distance of the observer's location~\cite{Zenginoglu:2008wc}: this work showed that the radiative transient is the most relevant for real observations of astrophysical sources.

Corrections to the leading-order expansion of~\cite{Leaver:1986gd} were computed in Refs.~\cite{Andersson:1996cm,Hod:2009my}, for tails observed at finite distances.
For example, a superposition of faster decaying power laws $\sim t^{-(2\ell+3+2m)}$, originating from sub-leading orders in the small frequency expansion $\omega M\ll 1$, was found in~\cite{Andersson:1996cm}. 
It is important to note that these faster decaying power laws come from corrections to the Green's function that propagate the \textit{late-time} signal.
Corrections to Price's law including more terms in the large distance ($r\gg M$) expansion for the background potential were computed in~\cite{Hod:2009my}.
Such corrections are logarithmic and belong to the class of potentials analyzed also in Refs.~\cite{Ching:1994bd,Ching:1995tj}, leading to logarithmic corrections in the coordinate time $t$ for the late-time signal.

Different approaches to compute late-time tails, likewise exploiting the key role of back-scattering off the large-scale background curvature, can be found in~\cite{Barack:1998bv,Barack:1998bw,Price:2004mm, Poisson:2002jz}.

Analytical predictions for late-time tails as observed at finite distances~\cite{Price:1971fb,Price:1972pw}, and radiative tails at null infinity $\mathcal{I}^+$~\cite{Leaver:1986gd}, were later investigated numerically: see in particular~\cite{Gundlach:1993tn,Burko:1997tb} for numerical evolutions of the spherical collapse of a scalar field, and~\cite{Price:1994pm,Krivan:1997hc,Bernuzzi:2008rq,Zenginoglu:2008wc,Zenginoglu:2009ey} for perturbative numerical evolutions. 

\subsubsection{Initial data-driven tails: Kerr case}

The asymptotic late-time, power-law ($t^n$) decay rates of matter fields in Kerr BH spacetime have been a matter of significant debate for several years. 
Initial difficulties were related to the complex mode coupling which arises in the spinning case, and required the development of sophisticated mathematical and computational techniques~\cite{Tiglio:2007jp,Gleiser:2007ti,Zenginoglu:2009hd,Burko:2007ju,Zenginoglu:2012us,Racz:2011qu,Burko:2013bra,Angelopoulos:2021cpg}.
The following late-time decay rate expressions for scalar fields in sub-extremal BHs were found in~\cite{Burko:2013bra,Angelopoulos:2021cpg}:  
\begin{equation}  \label{eq:rates}\quad
n = \left\{ \begin{array}{ll}
-(\ell'+\ell + 3) & \mathrm{for}\quad \ell'=0, 1, \\
-(\ell'+\ell + 1) & \mathrm{otherwise}.\quad \end{array}\right.\\
\end{equation}
along $r={\rm const}$ and $\mathscr{H}^+$, and
\begin{equation} \label{eq:rates2}\quad
n^{\scri^+} = \left\{ \begin{array}{ll}
-\ell' & \mathrm{for}\quad \ell\leq \ell'-2,  \\
-(\ell+2) & \mathrm{for} \quad \ell\geq \ell', \end{array}\right.
\end{equation}
along null infinity $\mathscr{I}^+$,
where $\ell'$ refers to the initial field multipole, and $\ell$ is the multipole of interest. 
These rates refer to axisymmetric multipoles, and were obtained by carefully studying the inter-mode coupling effects that are present in the Kerr spacetime due to frame-dragging. 
The nonaxisymmetric results do not change, except for fact that both $\ell'$ and $\ell$ are bounded from below by the value of $m$~\cite{Burko:2010zj}.

\begin{figure}
\centering
\includegraphics[width=3in]{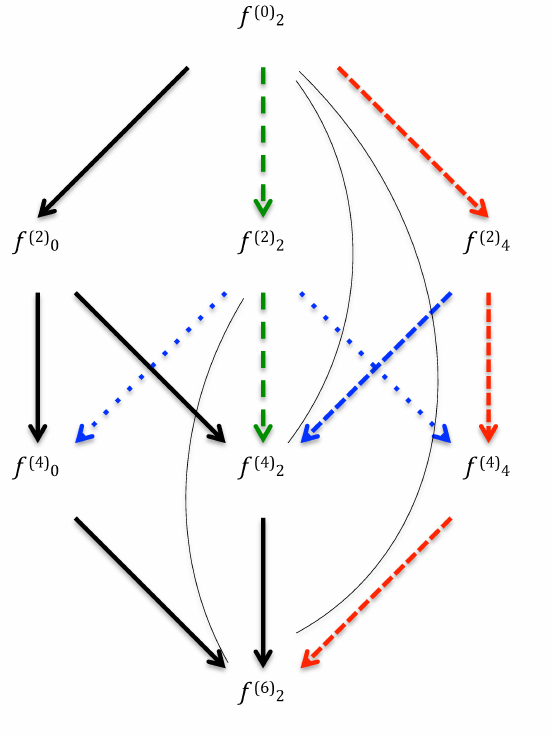}
\caption{The ten possible channels of excitations leading from the initial $\ell'=2$ ($f^{(0)}_2$) multipole to the final $\ell=2$ ($f^{(6)}_2$). 
The dominant channels (shown with solid arrows) have one leg through $f^{(4)}_0$ that is excited through channel~(i), and a second leg through $f^{(4)}_2$ that is excited through channel~(ii). 
The thin curves are excitations that couple directly modes separated by more than one order. 
The dominant channel is the one that predicts the final decay rate of the $\ell=2$ multipole. 
Figure taken from~\cite{Burko:2013bra}.
}
\label{fig:all_channels}
\end{figure}

To illustrate in some detail the role of mode coupling, let us consider for example an initial multipole $\ell'=2$, and study the multipoles $\ell=0,2,4,6$. 
The evolution equations for each of these multipoles may be found in~\cite{Burko:2010zj}. They are expanded in powers of the BH spin $\chi = a/M$ and denoted as $f^{(2n)}_{\ell}$, where $2n$ refers to the power of $\chi$ in the expansion, and $\ell$ to the multipole moment. 
Each equation has the form
\begin{equation}
\Box f^{(2n)}_{\ell}+V_{\ell}(\rho)\, f^{(2n)}_{\ell}=S^{(2n)}_{\ell}\, ,
\end{equation}
where the effective potential is given by 
\begin{equation}
V_{\ell}(\rho)=\frac{1}{\rho^2}\,\left(1-\frac{2M}{\rho}\right)\,\left[ \ell(\ell+1)+\frac{2M}{\rho}\right]\, ,
\end{equation}
and $S^{(2n)}_{\ell}$ is a sum of terms involving differential operators operating on lower--order iteration modes with multipoles $\ell-2$, $\ell$, and $\ell+2$. 
The radial variable $\rho$ is defined via $r=M+\sqrt{\rho^2-2M\rho+M^2(1-\chi^2)}$, so that the horizon is always at $2M$. 

Each of the solutions of these inhomogeneous equations is the sum of a general solution of the homogeneous equation, plus a particular solution of the inhomogeneous equation. 
The nontrivial, nonzero homogeneous solution is the one corresponding to the mode present in the initial data.
There are ten possible channels of excitation from the initial $f^{(0)}_2$ mode to the final $f^{(6)}_2$ mode, schematically represented in Fig.~\ref{fig:all_channels}.
Notice that three channels include direct coupling between modes that are separated by more than one order. 
A detailed study of such excitation channels established the dominant channels that predict the final decay rates of the $\ell=2$ multipole~\cite{Burko:2010zj}. 
The same rates were recently confirmed by a rigorous mathematical analysis of mode-coupling effects in Kerr spacetime~\cite{Angelopoulos:2021cpg}. 

\begin{table}[t]
\centering
\begin{tabular}{cll}
\hline
 & Horizon Data & No Horizon Data \\
\hline\hline
$r={\rm const}$
 & $-n=\left\{
   \begin{matrix}
     \ell^{\prime}+\ell+2\,\,\, , & \ell'=0,1 \\
     \ell^{\prime}+\ell \,\,\, , & {\rm else}
   \end{matrix}\right.$
 & $-n=\left\{
   \begin{matrix}
     \ell^{\prime}+\ell+3\,\,\, , & \ell'=0,1 \\
     \ell^{\prime}+\ell+ 1 \,\,\, , & {\rm else}
   \end{matrix}\right.$ \\
\hline
$\mathscr{I}^+$
 & $-n=\left\{
   \begin{matrix}
     \ell^{\prime}\,\,\,\,\,\,\,\,\,\,\;\;\; , & \ell\le\ell'-2 \\
     \ell+2 \,\,\,\,\,\, , & \ell\ge\ell^{\prime}
   \end{matrix}\right.$
 & $-n=\left\{
   \begin{matrix}
     \ell^{\prime}\,\,\,\,\,\,\,\,\,\,\;\;\; , & \ell\le\ell'-2 \\
     \ell+2 \,\,\,\,\,\, , & \ell\ge\ell^{\prime}
   \end{matrix}\right.$ \\
\hline
$\mathscr{H}^+$
 & $-n=\left\{
   \begin{matrix}
     \ell^{\prime}-1\,\;\;\; , & \ell\le\ell'-2 \\
     \ell+1 \,\,\,\,\,\, , & \ell\ge\ell^{\prime}
   \end{matrix}\right.$
 & $-n=\left\{
   \begin{matrix}
     \ell^{\prime}\,\,\,\,\,\,\,\,\,\,\;\;\; , & \ell\le\ell'-2 \\
     \ell+2 \,\,\,\,\,\, , & \ell\ge\ell^{\prime}
   \end{matrix}\right.$ \\
\hline
\end{tabular}
\caption{\label{table1} Power-law indices $n$ for late-time decays: along $r={\rm const}$, $t^{n}$; along $\mathscr{I}^+$,  $u^{n}$; and along $\mathscr{H}^+$, $v^{n}$.
Table taken from~\cite{Burko:2023uag}.} 
\end{table} 

In Table~\ref{table1} we summarize late-time power-law decay rates for extremal Kerr. The table, built using extensive numerical solutions~\cite{Burko:2023uag}, lists the fall-off rates for scalar perturbations along $r={\rm const}$, along $\mathscr{I}^+$, and along $\mathscr{H}^+$, for the case of no initial data supported on $\mathscr{H}^+$.
It also reports the corresponding decay rates when the initial data are supported on $\mathscr{H}^+$.
It is interesting to note that along $\mathscr{I}^+$, horizon data do not change the decay rate. 

There are relatively few generalizations of these results beyond the scalar case (e.g.~\cite{Csukas:2019kcb}).
However, the spin-weight of the matter field does not impact the asymptotic rates~\cite{Gundlach:1993tp}. 
Therefore, it is expected that the dominant GW mode ($h_{\ell m}$ with $\ell=m=2$) at future null infinity should have a late-time tail decay of $1/u^4$ which has been confirmed in studies such as Ref.~\cite{Csukas:2019kcb}.

\subsubsection{Source-driven tails: perturbative test-particle results}

The initial data studies reviewed in the previous section suggest that tail effects are typically strongly suppressed compared to the QNM emission.
Hence, while the study of late-time tails has attracted considerable attention from the mathematical point of view, their contributions to GW templates has typically been considered negligible for any near-future observations.
Here we discuss work showing that the tail can be considerably enhanced when BH perturbations are driven by an orbiting body along noncircular orbits. This indicates that the relevance of tails in GW source modeling should be revisited.

Perturbative numerical evolutions of a test particle falling along generic orbits into a Schwarzschild BH found an enhancement of late-time tails as the progenitor's binary eccentricity increases~\cite{Albanesi:2023bgi}.
This observation was highlighted and exploited in~\cite{Carullo:2023tff} to attempt the first extraction of late-time tails in nonlinear simulations (see Section~\ref{subsubsec:tailsNR}), and later investigated in depth in~\cite{DeAmicis:2024not,Islam:2024vro}. 

A perturbative analytical model for late-time tails emitted by eccentric particle trajectories developed in~\cite{DeAmicis:2024not} following a strategy similar to~\cite{Leaver:1986gd} predicts late-time, post-merger tails observed at $\mathcal{I}^+$ as an integral over the system past history of the form
\begin{equation}
        \Psi^{\rm tail}_{\ell m}(u)=\int_{-\infty}^{u}dt'\frac{\mathcal{S}_{\ell m}\left[r(t'),\varphi(t')\right]}{\left(u-t'\right)^{\ell+2}} \, ,
    \label{eq:tail_model}
\end{equation}
where $\mathcal{S}_{\ell m}$ a function of the test particle's trajectory $\left(r(t'),\varphi(t')\right)$, and $\Psi_{\ell m}$ is a master function linearly related to the GW strain at future null infinity~\cite{DeAmicis:2024not}.
This expression illustrates the \textit{hereditary} nature of late-times tails: the post-ringdown strain is inherited from the inspiral in a nonlocal fashion, similarly to what is observed in PN theory~\cite{Blanchet:1992br}.

\begin{figure}
\includegraphics[width=0.9\textwidth]{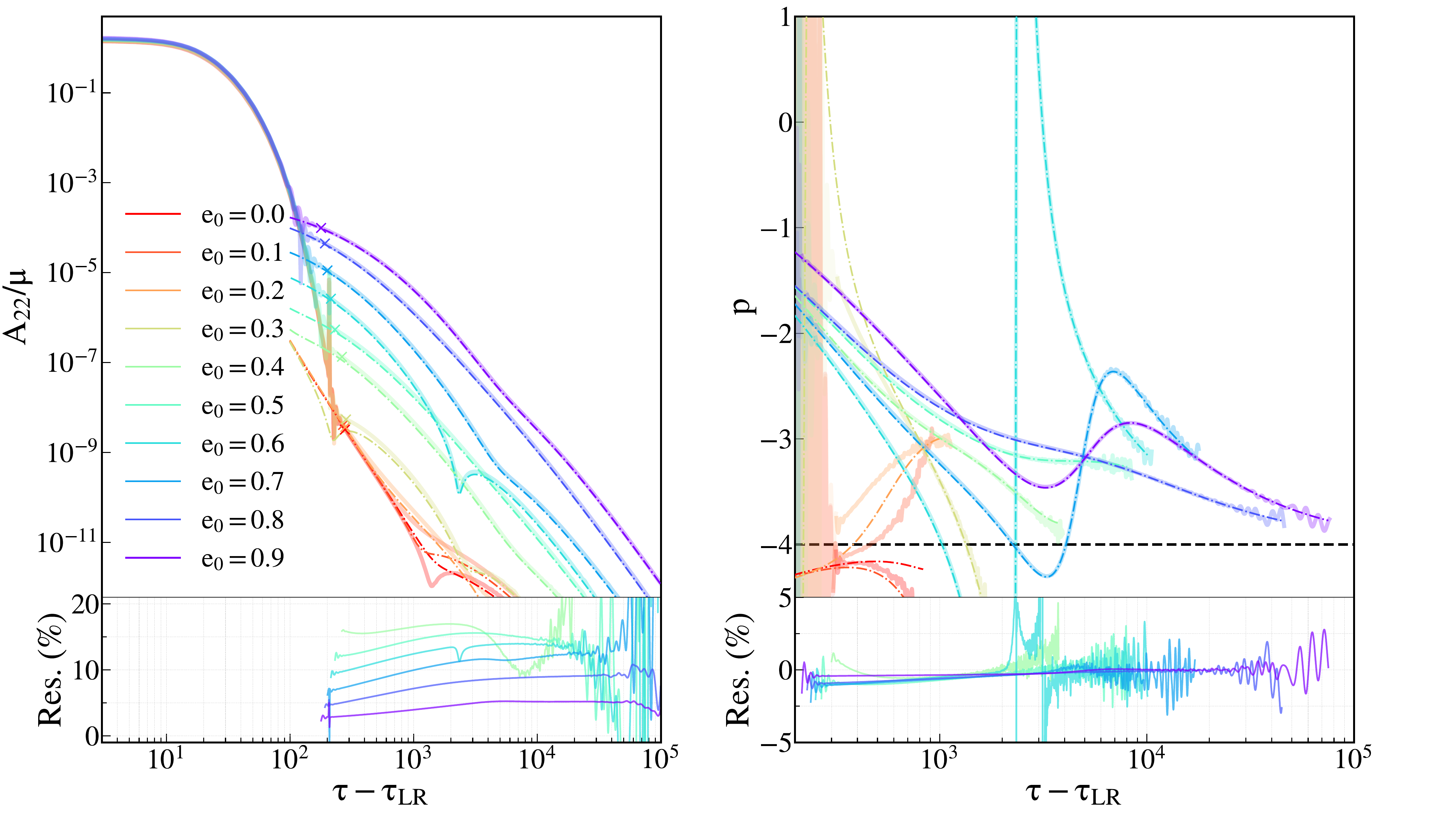}
\caption{Amplitude of the $(22)$ strain multipole (left) and tail exponent $p= d(\ln A_{22})/d (\ln \left(\tau-\tau_{\rm LR}\right)$ (right) vs time, rescaled with respect to the light-ring crossing time. 
Each color is relative to a bounded orbit with different initial eccentricity~$e_0$. 
Thick opaque lines are perturbative numerical evolutions, while dot-dashed lines are obtained through the model of Eq.~\eqref{eq:tail_model}. 
The bottom panels display the residuals between numerical evolutions and Eq.~\eqref{eq:tail_model}.
Figure taken from~\cite{DeAmicis:2024not}.}
\label{fig:Tail_vs_ecc_Schw}
\end{figure}

From Eq.~\eqref{eq:tail_model} one can accurately predict the scaling of late-time tails with the eccentricity, as shown by the comparison with perturbative numerical waveforms in Fig.~\ref{fig:Tail_vs_ecc_Schw}~\cite{DeAmicis:2024not}.
The perturbative numerical waveforms are computed with the \texttt{RWZHyp} code~\cite{Bernuzzi:2010ty,Bernuzzi:2011aj}, by simultaneously solving the RWZ equations (initially setting the $\Psi$ field and its derivative to zero) and the Hamiltonian equation of motion for the test particle.
The signal is computed at $\mathcal{I}^+$ and the trajectories are driven by a (fully analytical) PN-expanded radiation-reaction, which allows for the evolution of generic orbits up to merger, instead of focusing on geodesic plunges.
The semi-analytical predictions were obtained by feeding the numerical trajectory into the model of Eq.~\eqref{eq:tail_model}.
A similar agreement holds for generic planar bounded orbits, including dynamical captures and radial infalls~\cite{DeAmicis:2024not}.

\begin{figure}[t]
\centering
\includegraphics[width=0.5\linewidth]{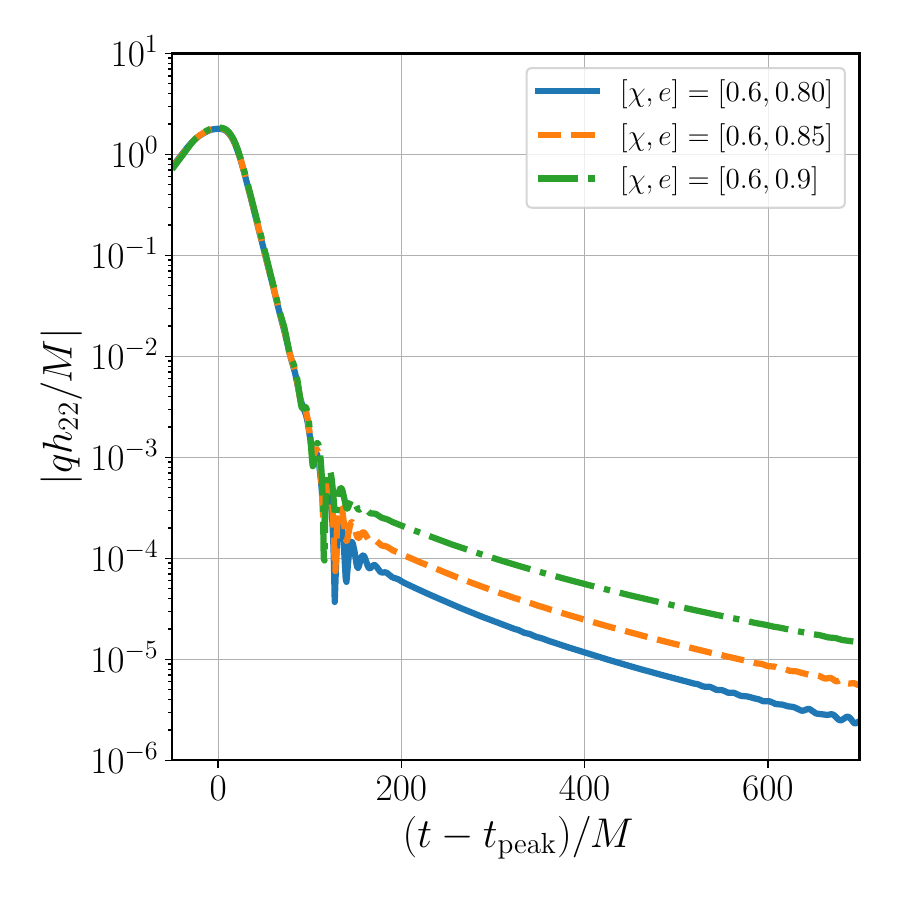}
\caption{The $(2,2)$ mode ringdown amplitude for a test particle  plunging along geodesics with eccentricities $e=[0.8, 0.85, 0.9]$ onto a Kerr BH with spin $\chi=0.6$. 
The simulation starts at the last apocenter passage, before the last stable orbit. 
Figure taken from~\cite{Islam:2024vro}.}
\label{fig:tail_long_geodesic}
\end{figure}
An enhancement of the tails with increasing eccentricity was found also in the Kerr background~\cite{Islam:2024vro}, as shown in Fig.~\ref{fig:tail_long_geodesic}.
While the tail amplitude is clearly enhanced with increasing eccentricity, its dependence on the BH spin $\chi$ is less clear, and analytical predictions are not available yet. 
The numerical results in the range $\chi =[-0.6, 0.6]$ show clear evidence that the amplitude increases for large negative spins (anti-aligned with the orbital angular momentum), but they are inconclusive for co-aligned spins. 
Through a series of numerical experiments on generic orbits, the last apastron before merger was isolated as the portion of the trajectory contributing the most to tail enhancement, both in the Schwarzschild and in the Kerr cases~\cite{DeAmicis:2024not,Islam:2024vro}. 

When the trajectories include the last apastron passage, a strong tail excitation is observed. On the contrary, no tails are observed when that part of the trajectory is ignored.
An expansion of the late-time tail model of Eq.~\eqref{eq:tail_model} valid in the limits $r\gg 1$ and $L\ll 1$ (where $L$ is the angular momentum of the test particle) yields, for the motion near apastron~\cite{DeAmicis:2024not}:
\begin{equation}
\begin{gathered}
        \Psi^{\rm Tail}_{\ell m}(u)=\int_{T_{in}}^{T_{f}}dt'\frac{r^{\ell}(t')e^{-im\varphi(t')}P_{\ell m}\left(\cos\theta_0\right)}{\left(u-t'\right)^{\ell+2}}
        \cdot\left[a_1-\frac{a_1}{2}\dot{r}^2+a_2\dot{r}\frac{L}{r}+\left(a_3+\frac{a_1}{2}\right)\frac{L^2}{r^2}\right] \, ,
\end{gathered}
\label{eq:tail_expanded_large_R_small_L}
\end{equation}
where
\begin{equation}
    \begin{split}
&a_1=a_0\left(\ell+1\right)\left(\ell+2\right) \, ,\ \ \ \ \ \ \ \ \ \ \ \ a_2=a_04i m \, ,\\
&a_3=a_0\left[\ell(\ell+1)-2m^2-2\right] \, , \ \ \ a_0=c_{\ell}\frac{8\pi\mu}{\ell(\ell+1)\left[\ell(\ell+1)-2\right]} \, .
\end{split}
\end{equation}
Here $\mu$ is the test particle mass, and $\theta_0$ is a constant angle determining the planar motion.
This equation implies that the integral (hence the tail effect) is enhanced for motion at large $r$ and with small angular velocity, due to the oscillating phase $e^{-im\phi(t')}$. 
Therefore the late-time tail is maximized for radial infalls from large distances, as corroborated by a series of numerical evolutions~\cite{DeAmicis:2024not}.
For generic binaries, the larger is the last orbit's eccentricity, the further is the last apastron location, and the smaller is the angular velocity near this point (due to Kepler's second law). 
These conditions imply an enhanced emission of tails at the last apastron before merger, an effect that is later observed in the post-ringdown strain.
The oscillatory contribution and alternating sign of the expansion terms also imply an ``interference'' pattern that suppresses tail contributions for more ``circular'' motions: tail effects are maximized for systems with small phase variations, such as radial infalls.
Since the oscillatory factor in the integral always reduces to unity for $m=0$, the tail will be enhanced in the $m=0$ modes of generic (eccentric and quasi-circular) orbits, as shown in~\cite{Albanesi:2024fts}.

Long-lived, perturbative numerical waveforms~\cite{DeAmicis:2024not} show that late-time tails converge slowly and (in general) nonmonotonically towards Price's law prediction $u^{-\ell-2}$~\cite{Leaver:1986gd}.
Expanding Eq.~\eqref{eq:tail_model} for $u\gg T_f$, with $T_f$ the time of merger, it was found that source-driven late-time tails, computed through Price's law propagator~\cite{Leaver:1986gd}, are not an inverse power law, but rather a {\em superposition} of an infinite number of inverse power laws:
\begin{equation}
\begin{gathered}
        \Psi^{\rm Tail}_{\ell m}(u)=\frac{1}{u^{\ell+2}}\cdot \int_{-\infty}^{T_{\rm f}}dt'\mathcal{S}_{\ell m}\left[r(t'),\varphi(t')\right]\left[1+\sum_{n=1}^{\infty}\frac{\left(\ell+1+n\right)!}{n!\left(\ell+1\right)!}
\left(\frac{t'}{u}\right)^n\right] \,.
\end{gathered}
\label{eq:tail_analytic_expansion_power_laws}
\end{equation}
A large number of these power-law contributions (at least $O(100)$) is required to obtain the transient emission to a reasonable degree of accuracy~\cite{DeAmicis:2024eoy}.

\begin{figure}[t]
\centering
\includegraphics[width=0.75\linewidth]{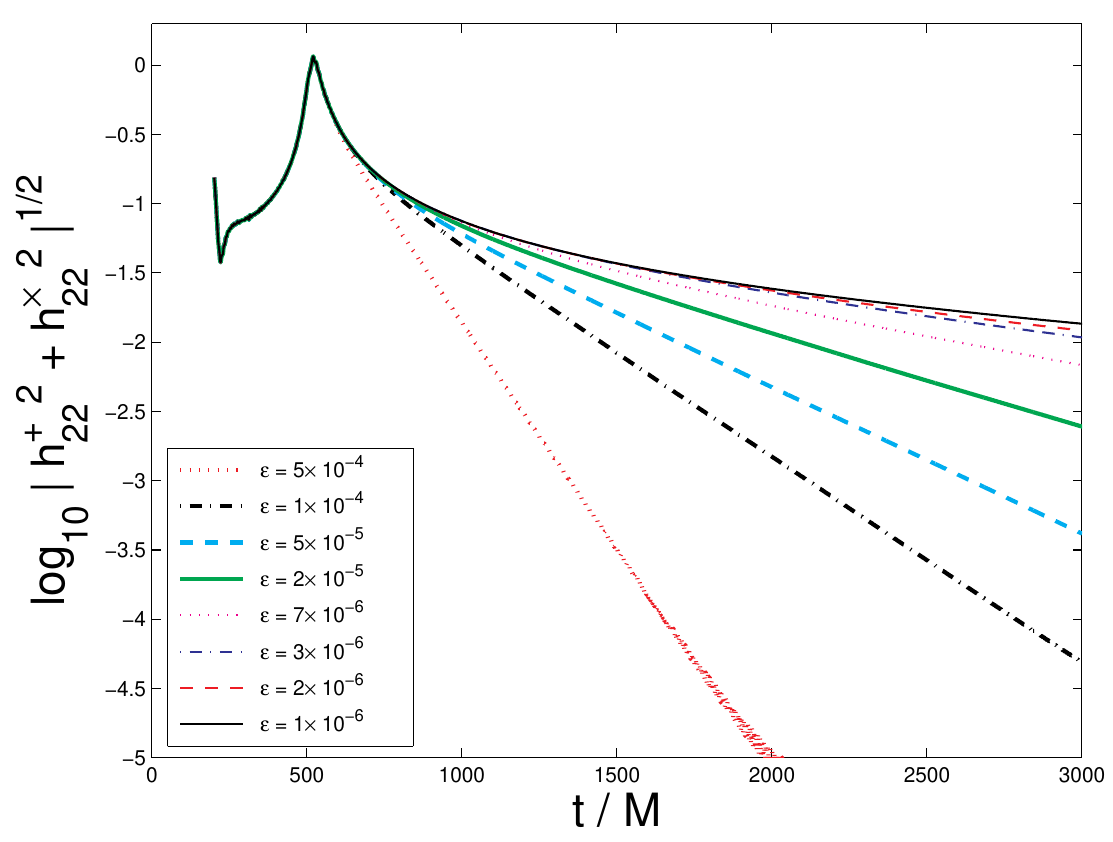}
\caption{The amplitude of the field $h_{22}$ as a function of the time $u$ at $\scri^+$ for several values of $\epsilon = 1-\chi$. As the BH spin approaches extremality, the behavior of the decay transitions from an exponential to a power law. 
Figure taken from~\cite{Burko:2016sfi}.}
\label{amp}
\end{figure}

The excitation coefficient of each fast-decaying contribution depends on the history of the system and on the details of the specific orbit it describes, enclosed in the trajectory function $\mathcal{S}_{\ell m}$.
For generic orbits, corrections to Price's law due to the source give rise to a long-lived transient, while homogeneous perturbation theory results can only be recovered at asymptotically late times. 
Note how faster-decaying corrections in Eq.~\eqref{eq:tail_analytic_expansion_power_laws} are source-driven, hence conceptually different from higher-order corrections to the tail propagator computed (e.g.) in~\cite{Andersson:1996cm}.

The above discussion pertains to the sub-extremal Kerr case.
In the near-extremal regime, the late-time behavior is different and rather peculiar.
As the BH approaches extremality, a remarkable superposition of overtones due to a ``superradiance resonance cavity'' implies that the ringdown transitions from an exponential decay to a $1/u$ behavior~\cite{Andersson:1999wj,Yang:2013uba}.
Numerical investigations~\cite{Burko:2016sfi} confirm this prediction, as shown in Fig.~\ref{amp}.

Moreover, this overtone superposition imparts a peculiar evolution to QNM frequencies during ringdown~\cite{Rifat:2019fkt}. 
In the near-extremal regime, the $u^{-1}$ decay is due to the collective excitation of zero-damped modes~\cite{Yang:2013uba}.
For extremal BHs, the $u^{-1}$ decay is eternal instead of transient.
\subsubsection{Tails in nonlinear evolutions}\label{subsubsec:tailsNR}

The above discussion was limited to perturbation theory, as until recently the identification of late-time tails in nonlinear binary mergers seemed out of reach because of the small amplitude of tails in quasi-circular binaries.
Motivated by tails enhancement in highly eccentric binaries, and by the understanding of the large-radius origin of this effect discussed above, two independent studies extracted late-time tails in fully nonlinear 3+1 evolutions of head-on BBH mergers~\cite{DeAmicis:2024eoy,Ma:2024hzq}.

Tails identification in previous simulations had been challenging because standard evolutions employed imperfect boundary conditions at the outer simulation boundary, imposing no incoming radiation towards the source. 
In curved spacetime this condition does not hold due to the presence of tails. 
Hence, if the boundary is too close to the source, enforcing no incoming radiation implies cutting a large portion of spacetime inside which tails are generated, and thus quenching the tail signal.
Moreover, due to the nonexact nature of the boundary conditions,  when the evolved radiation reaches the boundary it generates a nonphysical signal, that later travels back to the observer's location and spoils the extraction of the tail~\cite{Allen:2004js,Dafermos:2004wt}.
This second subtlety depends on the observer's location: if the extraction radius is far from the source, the extrapolated late-time signal is characterized by an initial transient following the proper $\mathcal{I}^+$ behavior (i.e., what would be observed by real detectors~\cite{Zenginoglu:2008wc}); if instead the extraction radius is close to the source, the tail signal is suppressed.

By analyzing several Cauchy evolutions, in which the signal was computed at different finite-radii locations and then extrapolated towards null infinity~\cite{Iozzo:2020jcu,scri,scrirepo,Boyle:2013nka,Boyle:2014ioa,Boyle:2015nqa}, it is possible to robustly extract the late-time tail  by fixing the observer's radii and the outer boundary to be very far from the GW source~\cite{DeAmicis:2024eoy}. 
 
A different strategy based on Cauchy-Characteristic Matching~\cite{PhysRevLett.76.4303,Bishop:1998uk,Szilagyi:2000xu,Winicour:2012znc,Ma:2023qjn,Ma:2024bed}  
was adopted in~\cite{Ma:2024hzq}.
This method combines (i) a Cauchy integration performed up to an outer boundary, used to evolve a characteristic system from this location; and (ii) characteristic evolution to impose correct boundary conditions at each time step of the Cauchy evolution. 
In this way it is possible to include the back-scattered signal (i.e., tail effects) in the full numerical integration.

\begin{figure}[t]
\includegraphics[width=0.8\linewidth]{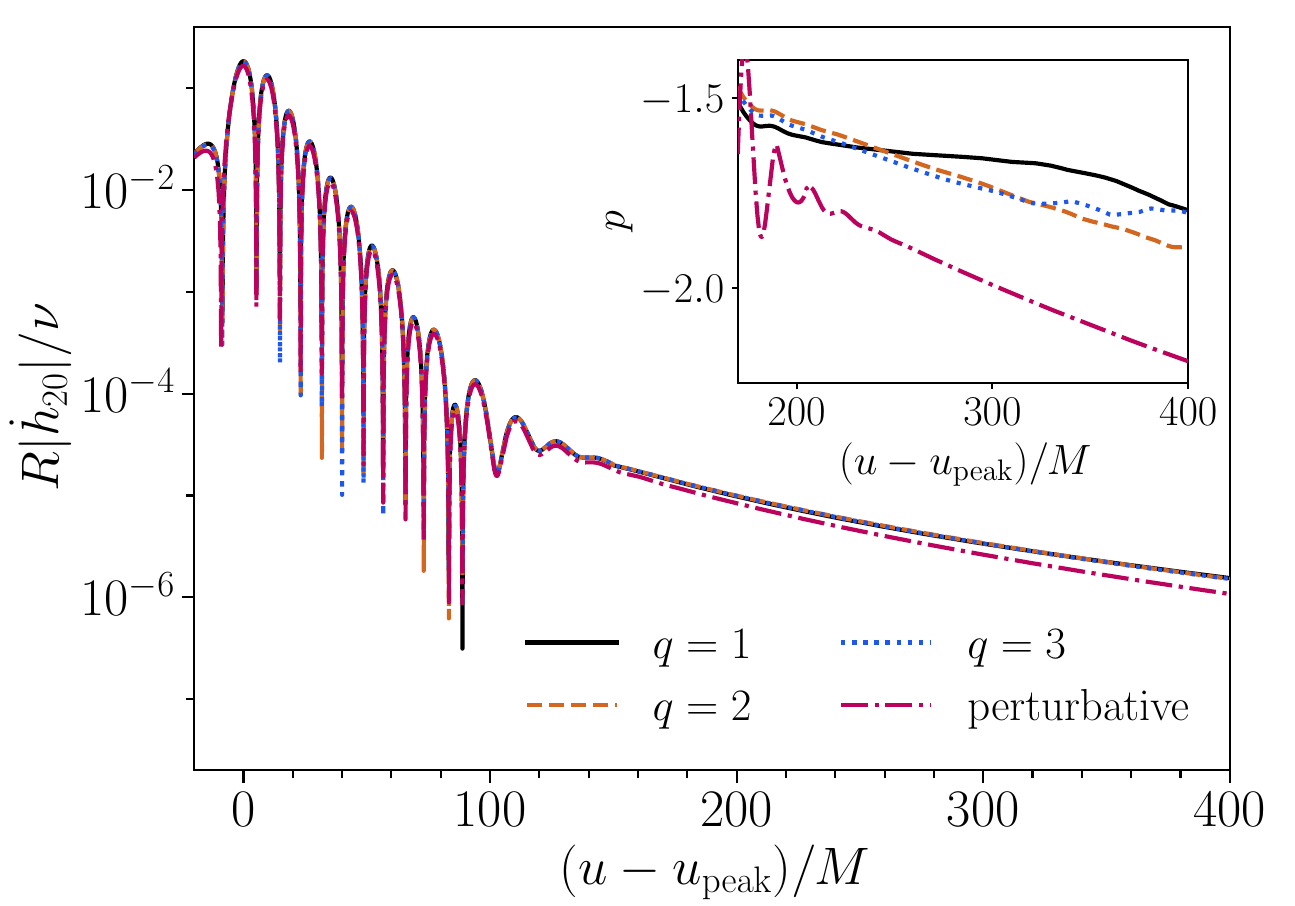}
\caption{Amplitude of the news $(20)$ multipole as a function of time, rescaled with respect to the light-ring crossing time.
Colored thick lines are nonlinear evolutions of head-on collisions for different mass-ratios (as indicated in the legend). The dot-dashed line refers to the perturbative evolution of a radial infall with compatible initial data.
Figure taken from~\cite{DeAmicis:2024eoy}.}
\label{fig:tails_fullNR_news}
\end{figure}

These results are summarized in Figs.~\ref{fig:tails_fullNR_news} and~\ref{fig:tails_ccm_psi4}, respectively. 

\begin{figure}[t]
\centering
\includegraphics[width=0.8\linewidth]{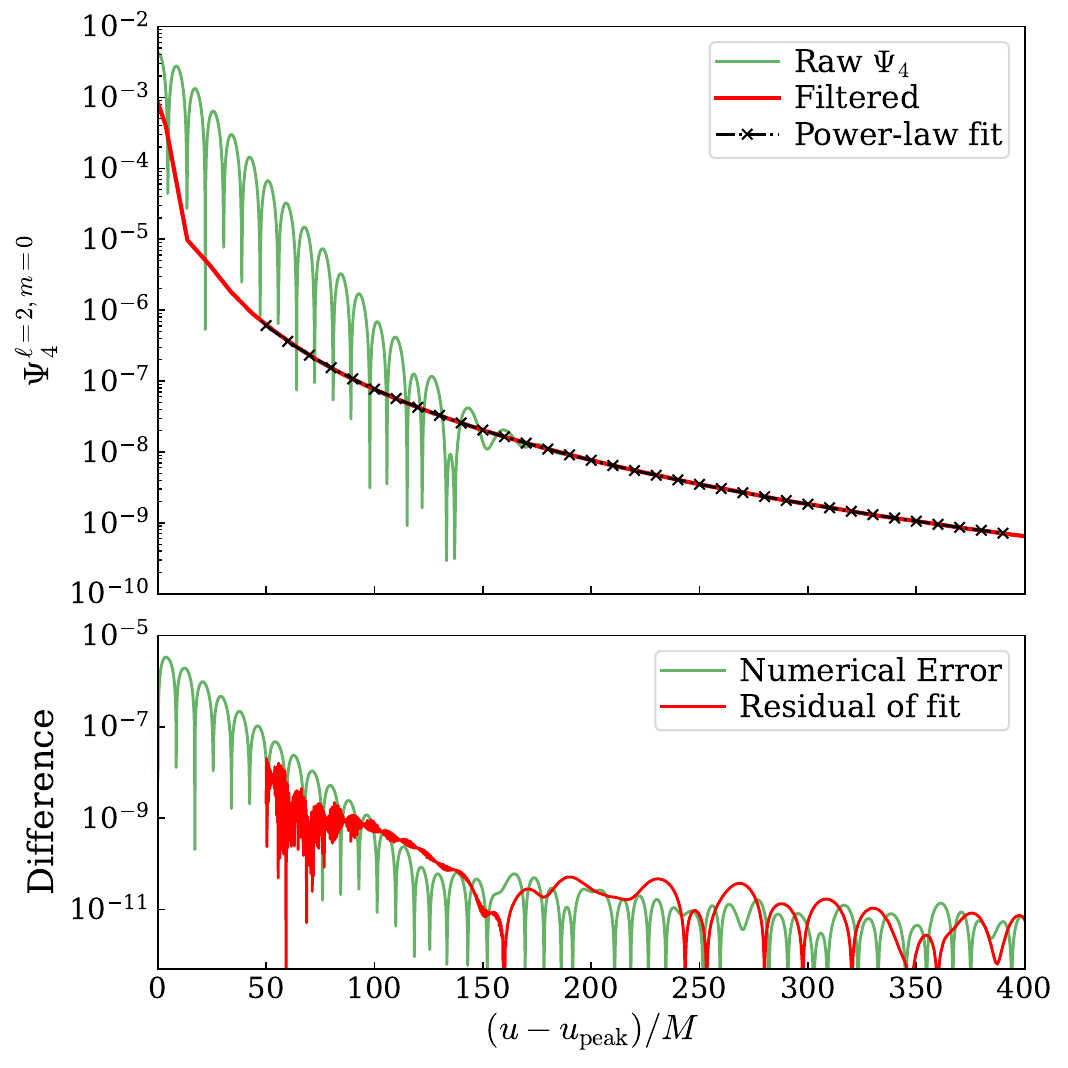}
\caption{Top panel: amplitude of the $(20)$ harmonic of $\Psi_4$ as a function of time. The green curve comes from the nonlinear evolution of a head-on BBH system, simulated with Cauchy-Characteristic Matching. The red curve corresponds to the reconstructed tail signal obtained via QNM rational filters and it is compared to the best power-law fit (denoted by black crosses) over the interval $[50M,400M]$. Bottom panel: the residual of the power-law fit (red), compared to the numerical error of the Cauchy-Characteristic Matching simulation (green).
Figure taken from~\cite{Ma:2024hzq}.}
\label{fig:tails_ccm_psi4}
\end{figure}

In Fig.~\ref{fig:tails_fullNR_news} we show fully nonlinear numerical evolutions of head-on collisions with mass ratios $q=[1,2,3]$ from an initial separation $D_0=100 M$, with initial ADM energy equal to the total rest-mass of the system $M$. 
The NR results are compared with perturbative waveforms computed with the \texttt{RWZHyp} code~\cite{Bernuzzi:2010xj,Bernuzzi:2011aj} for a test particle of mass $\mu$ falling from the same distance and with initial energy $E_0=\mu$.
Beyond clear signatures of a slow tail decay, Fig.~\ref{fig:tails_fullNR_news} shows that the binary mass ratio has a small effect on the excitation of tails,
and that there is remarkable agreement between perturbative and fully nonlinear waveforms from the peak of the strain of the $(20)$ multipole up to late-times.
This observation adds to a large body of literature indicating the surprising predictive power of BH perturbation theory in the nonlinear comparable-mass case~\cite{Nagar:2022icd, Wardell:2021fyy, Islam:2023aec}.
Radial infalls are admittedly ``simpler'' than quasi-circular inspirals as radiation reaction is suppressed, but it would be interesting to better understand this striking agreement by analytical methods.

The results in Fig.~\ref{fig:tails_ccm_psi4} refer instead to a head-on collision with mass-ratio $q=1$ and initial separation $D_0=110M$. 
Here the tail signal is reconstructed starting from early times, and the QNM contributions to the strain are removed using the filters developed in~\cite{Ma:2022wpv,Ma:2023cwe, Ma:2023vvr}. 
The tail signal is always present (although initially subdominant) during the BH ringdown. The impact of tails on waveform modeling has yet to be investigated.
A fit of the late-time behavior of the Weyl scalar $\Psi_4$ is compatible with an inverse power-law $(u+19.1)^{-3.79}$. This decay is slower than Price's prediction, and in agreement with the perturbative predictions of~\cite{DeAmicis:2024eoy}.

The nonlinear signal shown in Fig.~\ref{fig:tails_fullNR_news} displays a slower late-time decay with respect to perturbative source-driven results, in agreement with~\cite{Ma:2024hzq}.
This can be tentatively explained by the presence of nonlinearities: hereditary tails depend, at any time, on the entire history of the system, hence they can amplify nonlinearities and make them relevant even at very late times. 

Recent work focused on a new late-time nonlinear effect that does not originate from long-range back-scattering and is not hereditary~\cite{Okuzumi:2008ej,Cardoso:2024jme,Ling:2025wfv,Kehagias:2025xzm}. This signal, referred to as a ``nonlinear tail'' because of its inverse power-law behavior arising at second-order in perturbation theory, decays slower than Price's law and it presumably originates from coupled linear QNMs in the second-order source, mediated through the prompt response propagator~\cite{Okuzumi:2008ej,Cardoso:2024jme,Ling:2025wfv,Kehagias:2025xzm}.
\subsection{BMS frames and memory}
\label{sec:bms-frames-memory}

\vspace{-.1cm}

\noindent \textit{Initial contributors: Magaña Zertuche, Mitman, Stein}

\vspace{.2cm}

In GR, the perturbed remnant formed at the end of a BBH coalescence asymptotes to a Kerr BH, but the metric is not expressed in the coordinates normally used in BH perturbation theory. Instead, the coordinates are transformed
within the symmetry group of future null infinity $\mathcal{I}^{+}$ --- the final
destination of outgoing gravitational radiation. For example, in general the remnant BH receives a ``kick'' (i.e., a boost)
relative to center-of-momentum frame of its progenitor
binary. Consequently, the radiation that one observes from the boosted
remnant BH will not be the usual ringdown expected from
perturbation theory: because of the boost, the frequencies will be Doppler shifted and the angular structure will be beamed by relativistic aberration. To ensure that the extracted ringdown physics is frame-invariant, one must transform from the initial binary frame to an appropriate remnant frame.
In this section we review the characterization of this frame and its implications.

\subsubsection{Basic formalism}

In the 1960s, Bondi, van der Burg, Metzner, and Sachs
(BMS)~\cite{Bondi:1960jsa,Sachs:1961zz,Bondi:1962px,Sachs:1962wk,Sachs:1962zza} sought to better understand the asymptotics of gravitational radiation by fixing as much coordinate gauge freedom as possible. They began by considering an arbitrary metric written in Bondi-Sachs coordinates with gauge conditions $g_{rr}=0$, $g_{rA}=0$, and $\partial_{r}\mathrm{det}\left(\gamma_{AB}\right)=0$:
\begin{align}
    \label{eq:BSmetric}
    ds^{2}&=-Ue^{2\beta}du^{2}-2e^{2\beta}dudr
    +r^{2}\gamma_{AB}\left(dx^{A}-\mathcal{U}^{A}du\right)\left(dx^{B}-\mathcal{U}^{B}du\right)
    \,,
\end{align}
where capital Latin indices range over $\{\theta,\phi\}$, $u\equiv
t-r$, and $U$, $\beta$, $\mathcal{U}^{A}$, and $\gamma_{AB}$ are
arbitrary functions of the coordinates $(u,r,\theta,\phi)$. Since
radiation decays with distance, they specialized to metrics that approach the standard Minkowski metric in the large radius limit. Consequently, as $r\rightarrow\infty$ they imposed the conditions
\begin{align}
    \label{eq:falloffconditions}
    U\rightarrow1,\quad\beta\rightarrow0,\quad\mathcal{U}^{A}\rightarrow 1,\quad\gamma_{AB}\rightarrow\begin{pmatrix}1&0\\0&\sin^{2}\theta\end{pmatrix}.
\end{align}
This asymptotic restriction is often called being in ``Bondi gauge''
or some ``Bondi frame'' and provided an early notion of ``asymptotic
flatness.'' Gravitational radiation is now entirely encoded in the $1/r $ part of
$\gamma_{AB}$. However, there is a remaining ambiguity regarding the coordinate freedom of radiation on the boundary of the spacetime, i.e., what is the set of transformations of the asymptotic coordinates $(u,\theta,\phi)$ that preserve the form of the metric described in Eqs.~\eqref{eq:BSmetric} and~\eqref{eq:falloffconditions}.

One may expect that since the conditions in
Eq.~\eqref{eq:falloffconditions} make the metric asymptote to the
Minkowski metric as $r\to\infty$, then these coordinate freedoms
should simply be the ten elements of the Poincaré group. A bit
surprisingly, this intuition turns out to be only nearly
correct. Instead, the symmetry group of future null infinity is an
extension of the more familiar Poincaré group, in which the usual
spacetime translations are replaced by a larger set of
\emph{direction-dependent} spacetime translations called
``supertranslations.'' Fortunately, one can intuit why
supertranslations are symmetries of future null infinity through the
following realization. Take a family of inertial observers placed
about a 2-sphere of radius $r$, and consider the limit as $r\to\infty$
along outgoing null rays, thus approaching $\mathcal{I}^{+}$.  Their
world lines will asymptote to null geodesics tangent to
$\mathcal{I}^{+}$.  These generators are causally disconnected from
each other, as all future-directed null rays starting on
$\mathcal{I}^{+}$ stay tangent to $\mathcal{I}^{+}$.
Therefore, one can freely translate each generator, i.e., an arbitrary
point on the two-sphere, without changing this asymptotic boundary.

Thus the symmetry group of future null infinity is not the usual Poincaré group, but an infinite-dimensional extension called the BMS group~\cite{Bondi:1962px,Sachs:1962wk}. As a result, whenever one works with radiation from the perturbed BH formed by a BBH coalescence, one must always control the BMS freedoms of future null infinity to ensure that their analysis is in the frame of the remnant~\cite{Mitman:2021xkq,MaganaZertuche:2021syq,Mitman:2022qdl,Mitman:2024uss}.

To do this, one needs to understand how a BMS transformation, i.e., an element of the BMS group, transforms asymptotic data. As shown by the original works~\cite{Bondi:1960jsa,Sachs:1961zz,Bondi:1962px,Sachs:1962wk,Sachs:1962zza}, a BMS transformation changes the asymptotic coordinates via
\begin{align}
    (u,\zeta)\mapsto(u',\zeta')\equiv\left(k(\zeta,\bar{\zeta})\left(u-\alpha(\zeta,\bar{\zeta})\right),\frac{a\zeta+b}{c\zeta+d}\right),
\end{align}
where $\zeta\equiv e^{i\phi}\cot(\theta/2)$ is the complex stereographic projective coordinate,
\begin{align}
    k(\zeta,\bar{\zeta})\equiv\frac{1+|\zeta|^{2}}{|a\zeta+b|^{2}+|c\zeta+d|^{2}}
\end{align}
is the conformal factor, $(a,b,c,d)$ are complex coefficients with $ad-bc=1$ that encode the Lorentz rotation and boost, and $\alpha(\zeta,\bar{\zeta})$ is a real, smooth function that encodes the supertranslation. From this coordinate transformation, by examining how the tetrad on $\mathcal{I}^{+}$ transforms~\cite{Boyle:2015nqa}, one can then ascertain that the usual GW strain $h$ and the five Weyl scalars transform as
\begin{align}
  \label{eq:h-BMS-transformed}
  h'(u',\zeta')&=\frac{1}{k}e^{-2i\bar{\lambda}}\left[h(u,\zeta)-2\eth^{2}\alpha\right],\\
    \Psi_{A}'(u',\zeta')&=\frac{1}{k^{3}}e^{(2-A)i\lambda}\sum\limits_{a=A}^{4}\begin{pmatrix}4-A\\a-A\end{pmatrix}\left(-\frac{1}{k}\eth u'\right)^{a-A}\Psi_{a}(u,\zeta),
\end{align}
where $A\in\{0,1,2,3,4\}$, $\lambda$ is the spin phase with
\begin{align}
    e^{i\lambda}\equiv\left[\frac{\partial\bar{\zeta}'}{\partial\bar{\zeta}}\left(\frac{\partial\zeta'}{\partial\zeta}\right)^{-1}\right]^{\frac{1}{2}}=\frac{c\zeta+d}{\bar{c}\bar{\zeta}+\bar{d}}\,,
\end{align}
and
\begin{align}
    \eth f=-\frac{1}{\sqrt{2}}(\sin\theta)^{+s}(\partial_{\theta}+i\csc\theta\partial_{\phi})\left[(\sin\theta)^{-s}f\right]
\end{align}
is the usual Geroch-Held-Penrose spin-weight operator~\cite{Geroch:1973am} (see Appendix~\ref{sec:NP_GHP}).

Apart from this, the remaining information one needs is what BMS transformation must actually be applied to the system to map it to the canonical frame of the remnant. Extracting this transformation is typically referred to as ``BMS frame fixing''~\cite{Mitman:2021xkq,Mitman:2022qdl,Mitman:2024uss}. It consists of constructing functions of asymptotic data which measure the failure of the system to be in some canonical frame. For example, to measure how much a system is boosted out of its center-of-mass frame, one can study the center-of-mass charge and extract the boost from said function. For fixing the entire BMS freedom, one requires three charges: a rotation charge to fix the rotation, a center-of-mass charge to fix the boost and translation, and a supermomentum charge, which extends the usual notion of a four-momentum charge, to fix the supertranslation. Definitions of these charges and how one can use them to fix the BMS frame are provided in~\cite{Mitman:2022qdl,Mitman:2024uss}.

\subsubsection{Conserved charges and memory effect}

When a generalization of Noether's
theorem~\cite{Dray:1984rfa,Wald:1999wa} is applied to the BMS group,
we find new (non)conservation laws that extend the usual Poincaré conservation laws. In particular, when considering the conservation law stemming from supertranslations, one finds
\begin{align}
\label{eq:supertranslationconservationlaw}
    \mathrm{Re}\left[\eth^{2}h\right]=\frac{1}{2}\left(m+\frac{1}{4}\int_{-\infty}^{u}|\dot{h}|^{2}du\right)
  \,,
\end{align}
where
\begin{align}
    m\equiv\mathrm{Re}\left[\Psi_{2}+\frac{1}{4}\dot{h}\bar{h}\right]
\end{align}
is the Bondi mass aspect, a direction-dependent generalization of the mass on future null infinity $\mathcal{I}^{+}$. The second term in Eq.~\eqref{eq:supertranslationconservationlaw} is a directional energy flux.
When these equations are restricted to the $\ell=0$ and $\ell=1$ spherical harmonics, this conservation law yields the usual mass and momentum Poincaré conservation laws. However, for the $\ell\geq2$ components, a brand new conservation law is encoded, yielding additional information on gravitational radiation. In particular, the first term is simply the usual mass multipole moment source, e.g., the mass quadrupole, while the second term is a source that corresponds to the flux of gravitational radiation sourcing even more radiation. Furthermore, because the flux of radiation can never be negative, the second term implies that the strain will experience a permanent net change between early and late times.

\begin{figure}[t]
\label{fig:memory}
\centering
\includegraphics{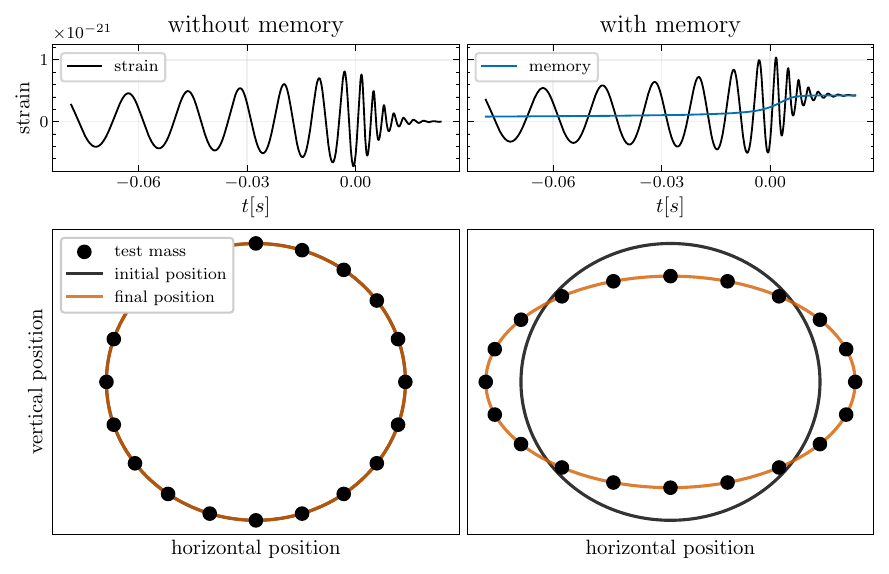}
\caption{Illustration of the GW memory effect according to a NR waveform from an equal-mass, aligned spin system with dimensionless spins of magnitude $0.6$ in the direction of the orbital angular momentum, with total mass $60 M_{\odot}$, luminosity distance $400$ Mpc, and an edge-on orientation (maximizing the impact of memory effects). Top: comparison of a GW without memory, i.e., an incorrect solution to Einstein's equations, to a GW with memory, i.e., the correct solution to Einstein's equations. Bottom: the initial (black) and final (orange) positions of test particles before and after the passage of the GW shown in the top panel. Figure taken from~\cite{Mitman:2024uss}.}
\end{figure}

This phenomenon is known as the ``GW memory effect''~\cite{Zeldovich:1974gvh, Braginsky:1985vlg, Braginsky:1987kwo,Christodoulou:1991cr, Thorne:1992sdb}. It corresponds to the permanent net displacement that two inertial observers will experience due to a flux of gravitational radiation passing by them. Figure~\ref{fig:memory} illustrates what this effect looks like for a realistic BBH merger. The strain exhibits a clear net change in its average value between early and late times. The blue curve in the top right panel is exactly the contribution from the second term in Eq.~\eqref{eq:supertranslationconservationlaw}.

Furthermore, Figure~\ref{fig:memory} also highlights the importance of fixing the BMS frame to match that of perturbation theory. The strain in the top-right panel of Fig.~\ref{fig:memory} does not decay to zero at late times, which means it is in complete disagreement with the exponentially decaying QNMs that perturbation theory predicts. This, however, is simply because the remnant BH formed in this binary merger is supertranslated relative to what perturbation theory expects. Consequently, to analyze this system with BH perturbation theory, one must first map the BMS charges of this system to match those of the usual Kerr metric, which will then make the strain decay to zero at late times, as well as have the expected temporal and angular structure.
This is because a supertranslation not only changes the modes of the strain by a constant, but also mixes modes due to the change in the time coordinate: see, e.g., Eq.~(19) of~\cite{Boyle:2015nqa}.

Finally, it is worth noting that memory can also be interpreted as a second-order perturbative effect in the context of BH ringdown.  In particular, since the strain enters into the second term in Eq.~\eqref{eq:supertranslationconservationlaw} nonlinearly, one can imagine that if the strain was replaced with a purely first-order perturbation in Eq.~\eqref{eq:supertranslationconservationlaw}, a quadratic QNM would appear through the memory contribution. For example, if the perturbation is primarily the $(2,2,0,+)$ QNM, then the quadratic QNM that arises would be $(2,2,0,+)\times(2,-2,0,-)$, and it would predominantly appear in the $(2,0)$ spherical harmonic.

\begin{figure}[t]
\label{fig:superrest_qnm_fitting}
\centering
\includegraphics[width=\linewidth]{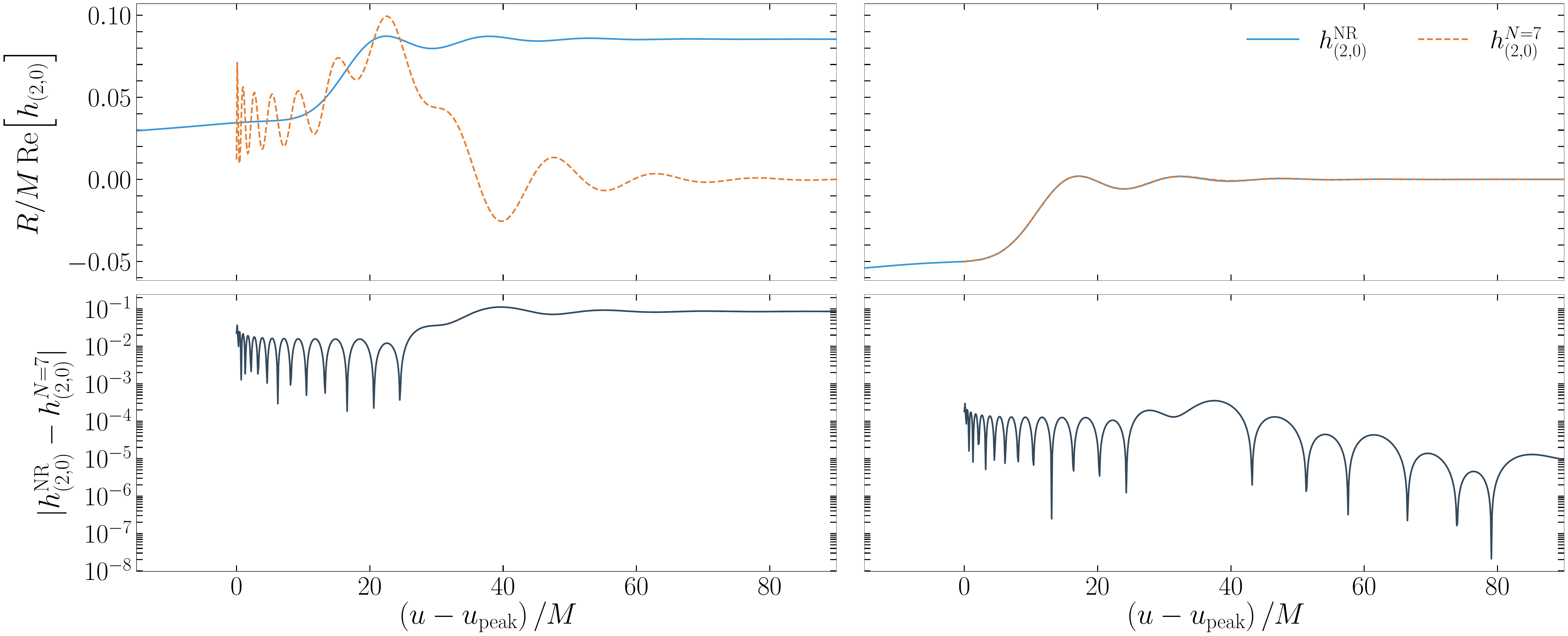}
\caption{%
		The impact of QNM modeling when using a waveform in the PN BMS frame (left) versus one in the superrest frame (right) of the remnant BH.
		Top: The real component of the $(2,0)$ mode of a NR waveform and its corresponding QNM model built from the $(2,0)$ mode with 7 overtones at the peak of the strain, i.e., $u_0 - u_{\mathrm{peak}} = 0 M$ .
		Bottom: Residuals between the NR waveform and the analytical QNM model. Figure taken from~\cite{MaganaZertuche:2021syq}.}
\end{figure}

\subsubsection{Effects on waveform modeling}

As mentioned above, to have a correct description of ringdown in the
presence of GW memory one must apply a BMS transformation so that the
remnant BH is in its superrest frame, i.e., the strain decays to zero
for late times. Being in the wrong frame, like the waveform in the top right
panel of Fig.~\ref{fig:memory}, produces two effects: a
time-independent shift in the waveform (last term of
Eq.~\eqref{eq:h-BMS-transformed}), and mode mixing distinct from
the usual spherical-spheroidal mixing.  The supertranslation mixing
can be easily seen from Eq.~\eqref{eq:h-BMS-transformed},
specializing to a pure supertranslation by setting
$(a,b,c,d)=(1,0,0,1)$ and any nonconstant
$\alpha=\alpha(\theta,\phi)$.  Then the transformation of the retarded
time $u'=u-\alpha$ results in
\begin{align}
  h(u,\theta,\phi) = h(u'+\alpha,\theta,\phi) =
  \sum_{n=0}^{\infty} \frac{1}{n!} \left[ \alpha(\theta,\phi) \frac{\partial}{\partial u'} \right]^{n} h(u',\theta,\phi)
  \,.
\end{align}
This is simply a Taylor expansion in the time argument, but
the resulting product between (powers of) $\alpha$ and (time
derivatives of) the strain $h$ generates mode mixing, controlled by
the mode content in $\alpha = \sum \alpha_{\ell m} Y_{\ell m}$.

Fitting a ringdown waveform in the wrong BMS frame introduces errors illustrated in Fig.~\ref{fig:superrest_qnm_fitting}.
The blue curves in the top panels show the $(2,0)$ ringdown strain of NR simulation \texttt{SXS:BBH:0305} with memory in the PN BMS frame (left) and the remnant's superrest frame (right). If one were to extract the complex QNM amplitudes from the simulation in the PN BMS frame and build an analytical model with those amplitudes, the model would fail and the QNMs would decay to zero at late times. However, when one is in the correct BMS frame (right), the amplitudes and, therefore, the model can faithfully represent the ringdown. The effect of using the correct BMS frame can also be seen in the residuals between the NR waveform and the QNM model, shown as black curves in the lower panels. Although this effect is most pronounced in the $(2,0)$ mode, it is present in all modes (see~\cite{MaganaZertuche:2021syq} for the effect on the $(2,2)$ and $(3,2)$ modes). Fortunately, the effect of fixing the BMS frame on the dominant $(2,2)$ mode is minimal and does not strongly affect models that have relied on this mode in the past. Nevertheless, it is important to take the BMS frame into account for high-precision ringdown modeling, especially as the demand for improved gravitational waveform models increases for next-generation detectors.

As of now, two different waveform surrogate models have been built using BMS
frame fixing: \texttt{NRHybSur3dq8\_CCE} and \texttt{NRSur3dq8\_RD}. The first
is an IMR surrogate built by mapping the NR simulations to the PN BMS frame for
the inspiral portion, and then hybridized with PN and EOB waveforms (see
Section~\ref{sec:effective-one-body})~\cite{Yoo:2023spi}. The second model,
\texttt{NRSur3dq8\_RD}, is a purely ringdown surrogate trained on waveforms that
are in the BMS superrest frame of the remnant BH and predicts the complex QNM
amplitudes for a wide range of spin-aligned BH
configurations~\cite{MaganaZertuche:2024ajz}. The QNM amplitudes used in the
training are extracted using the \texttt{python} package
\texttt{qnmfits}~\cite{qnmfitscode,MaganaZertuche:2025bua}. Additionally, there
are two recent models that do not fix the BMS frame but add the memory effect as
a post-processing step through analytical calculations using the BMS balance
laws~\cite{Rossello-Sastre:2024zlr,Albanesi:2024fts}. These models are built as
extensions of \texttt{IMRPhenomTHM} and \texttt{TEOBResumS-GIOTTO} to include
the $(2,0)$ mode and memory contributions~\cite{Estelles:2020twz,Nagar:2023zxh}.
For an extended review dedicated to BMS frame fixing and memory effects in NR, see~\cite{Mitman:2024uss}.

\subsection{Horizon physics}
\label{sec:horizons}

\vspace{-.1cm}

\noindent \textit{Initial contributors: Bonga, Krishnan}

\vspace{.2cm}

There is a close correspondence between QNMs and horizon dynamics. %
As we have discussed at length, the GW signal from a compact binary merger is conventionally divided into three regimes: inspiral, merger and ringdown. 
This division can be thought of in two different ways based on: (1) the morphology of the GW signal, and (2) the dynamics of the two compact objects. 
In particular, during the inspiral regime, the GW signal is quasi-periodic with slowly varying amplitude. This corresponds to the two objects being on a slowly decaying binary orbit. 
The merger regime is a burst of radiation corresponding to the merger of the two objects. 
The final, ringdown stage characterized by damped sinusoids corresponds to the dynamics of the post-merger remnant. 
This correspondence between the morphology of the GW and dynamics of the objects generating the wave is so commonplace that it often escapes a deeper analysis. 
It is nonetheless nontrivial, in that GW observations can be used to infer properties of dynamics in the strong field dynamical spacetime that is otherwise inaccessible to distant observers.

While it is not practically feasible to travel to a nearby neutron star and measure its properties, this is at least possible as a thought experiment. For BHs, on the other hand, as a matter of principle within classical GR, there could be no signal from the horizon which could ever reach outside observers.  Nevertheless, GW observations can still be used to infer some of their properties, just as for any star.

The details of the correspondence between the different coalescence regimes often contain numerous assumptions worth closer attention.  For example, it is common to associate the peak of the GW signal with the merger. What does this definition of ``merger'' correspond to in terms of the compact objects is not simple to determine.
One option is the formation of the post-merger remnant, e.g. the formation of the common horizon in the case of a BBH merger.
Alternatively, it could be taken to indicate when the two individual horizons (or compact objects) get sufficiently close to each other.
In either case, making a precise correlation between the two phenomena which occur in such very different spacetime regions is a complex task.

To conceptually clarify the above picture, let us consider a different heuristic way of looking at this problem, originally presented in~\cite{Rezzolla:2010df,Jaramillo:2012rr}. 
The observed GW signals are emitted by the dynamical spacetime region in the vicinity of the two BHs (this region lies, of course, outside the event horizon).  The emitted signal consists of an outgoing part which reaches GW detectors, but it also has an ingoing part which is absorbed by the BHs. Given that both of these signals have the same origin, it is not surprising that the infalling down-horizon radiation bears similar features as the observed GW signal. It does however lead to a different interpretation of the ringdown process: it is incorrect to say that the remnant BH horizon loses its distortions by emitting GW signals. Rather, the remnant BH loses its distortions by \emph{absorbing} just the right amount of radiation. Furthermore, following the heuristic picture above, this radiation might plausibly also be a superposition of damped sinusoids.
In the remainder of this section, this description will be supported with evidence from numerical simulations and perturbative calculations. 

\subsubsection{Black hole horizons}

The heuristic picture formulated above applies to the event horizon in a BBH merger. Thus, one can consider inferring properties of the radiation crossing the event horizon from GW observations, and use them in numerical studies of BH mergers. However, event horizons are not
practical in numerical simulations since they are global objects
which require knowledge of the entire spacetime in order to be located.
Moreover, event horizons have teleological properties (i.e, they require knowledge of the entire future history), which make them
unsuitable \textit{in principle} for such studies. There are well-known examples in gravitational collapse where event horizons are formed in
flat spacetime regions and thus grow without any infalling radiation, in anticipation of a future gravitational collapse~\cite{Ashtekar:2004cn,Ashtekar:2025wnu}.  
Hence, the heuristic picture above cannot strictly apply to the true event horizon. This problem
can be avoided in perturbative situations or in cases when we specify the final state, but not generally.

In numerical simulations one instead uses ``quasi-local'' notions of
horizons, often referred to as ``apparent horizons.''  These objects were
originally introduced by Penrose in the context of the singularity
theorems~\cite{Penrose:1964wq,Hawking:1969sw}, and have proven to
be extremely useful in numerous applications. Apparent horizons are closed
2-surfaces of spherical topology $S^2$, with the nontrivial
property that outgoing light rays emanating from the surfaces are
\emph{marginally trapped}, i.e., have vanishing expansion. 
Such surfaces are often referred to as marginally-outer-trapped-surfaces
(MOTSs).  There is an extensive literature on the geometry and
dynamics of such surfaces (see~\cite{Booth:2005qc,Gourgoulhon:2005ng,Ashtekar:2004cn}). 
For the purposes of this review, the following will suffice: in a numerical simulation of a BH spacetime, at each instant of time one can locate MOTSs. The MOTSs can be used to evaluate the BH parameters such as mass, spin, and higher multipole mass and spin moments. As the MOTS evolves, it traces out a 3-dimensional surface $\mathcal{H}$ with topology $S^2\times \mathbb{R}$, referred to as a ``dynamical horizon'' (DH)~\cite{Ashtekar:2002ag,Ashtekar:2003hk,Andersson:2005gq,Andersson:2007fh}.  
On $\mathcal{H}$, it is then possible to trace out in an unambiguous manner the evolution of the multipole moments. Moreover, on $\mathcal{H}$ it is also possible to evaluate certain fields (to be defined below), which represent in a precise, well-defined sense the infalling radiation. 
The time evolution of the multipole moments and the infalling radiation turn out to be described by damped sinusoids, with the same frequency and damping times as the well-known QNMs associated with the remnant BH.

Let us introduce some basic definitions which will play an important role in the discussion.  
Start with a MOTS $\mathcal{S}_t$ at a given time $t$ (corresponding to a certain time in the numerical evolution).  The union of all the $\mathcal{S}_t$ is the DH $\mathcal{H}$.  
In the specific case of the remnant BH formed in a merger, $\mathcal{H}$ is spacelike, and it becomes ``closer'' to a null surface as the BH approaches equilibrium.  On $\mathcal{H}$, there is a well-defined ``time evolution'' vector field, so that one can meaningfully speak about the time evolution of various quantities as functions of $t$~\cite{Ashtekar:2013qta}. This allows us to ``connect'' quantities at different times, despite complications that can arise when the horizon has angular momentum. The quantities of interest include the horizon mass and spin multipole moments, originally defined for axisymmetric horizons in~\cite{Ashtekar:2004gp}. 
For the axisymmetric case, the mass multipole
moments will be denoted $M_n$, with $M_0$ being the mass, and where $M_1$ vanishes
identically.  The spin multipole moments are denoted $J_n$, with $J_0$
vanishing identically, and $J_1$ being the spin.  Extensions beyond
axisymmetry are also possible (see e.g.~\cite{Owen:2009sb,Ashtekar:2013qta,Ashtekar:2021wld}).  

To define radiation,  the outgoing null normal $l^a$ is needed: this is the null, future-directed and outward pointing vector orthogonal to $\mathcal{S}_t$ (defined up to a positive rescaling).  The infalling radiation field is given by the shear $\sigma_{ab}$ of this outgoing null-normal $l^a$. Being a symmetric-tracefree tensor, $\sigma_{ab}$ can be decomposed into angular modes $\sigma_{\ell m}$ ($\ell\geq 2$, $-\ell\leq m\leq \ell$) using suitable spin-weighted spherical harmonics on $\mathcal{S}_t$, that can be evaluated as functions of time: $\sigma_{\ell m}(t)$.  Thus, the basic objects of interest below will be $(M_n(t), J_n(t), \sigma_{\ell m}(t))$.

\subsubsection{Binary black hole mergers} 

We can now present numerical results for the head-on collision of two unequal mass BHs (see~\cite{Ashtekar:2025wnu} for more details). The initial data will consist of two nonspinning BHs at a moment of time-symmetry. During the evolution, as the BHs approach each other, a common horizon is formed, which settles down to a Schwarzschild BH. This situation is axially symmetric, so we can limit attention to shear modes $\sigma_{\ell m}(t)$ with $m=0$, labeled $\sigma_\ell(t)$ for ease of notation. Furthermore, all the spin multipoles $J_n(t)$ vanish identically, and only the mass moments $M_n(t)$ need to be considered.

Since the final BH is Schwarzschild, all multipole moments beyond $\ell\geq 2$ will decay to zero. Fig.~\ref{fig:l2-overtonefits-multipole} shows the mass quadrupole moment for the common horizon as a function of time, fitted to two models based on QNMs associated with the asymptotic remnant BH.  A close look shows that the dynamics of the quadrupole moment shows distinct regimes with different damping rates: a rapid decay (prompt response) followed by damped oscillations (QNM regime). Under the hypothesis that this decay is related to the QNMs of the asymptotic remnant, two fits are shown. The first includes just the fundamental $\ell=2$ mode, while the second uses a combination of four modes: the fundamental and first three overtones. Much better agreement is obtained when the overtones are considered. A detailed study considering moments up to $\ell=12$ can be found in~\cite{Mourier:2020mwa}.

\begin{figure}
  \centering    
  \includegraphics[width=0.4\columnwidth]{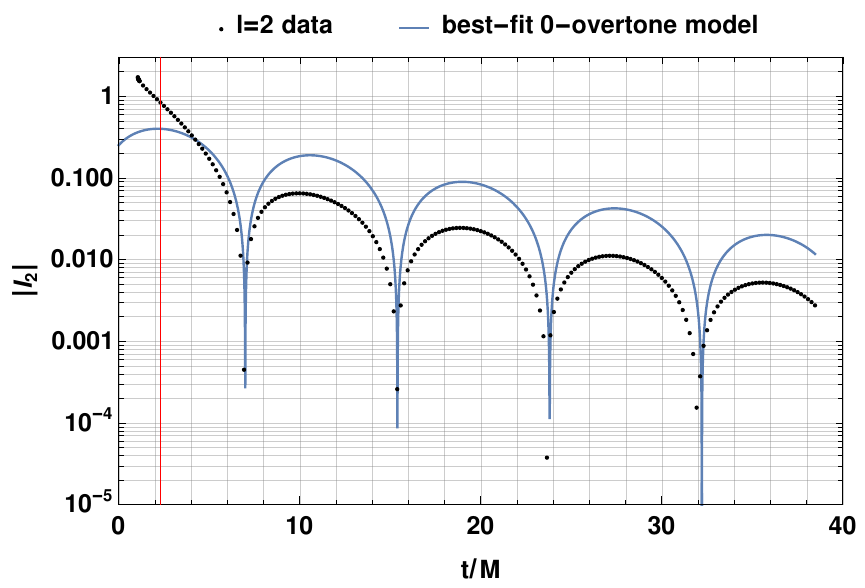}
  \includegraphics[width=0.4\columnwidth]{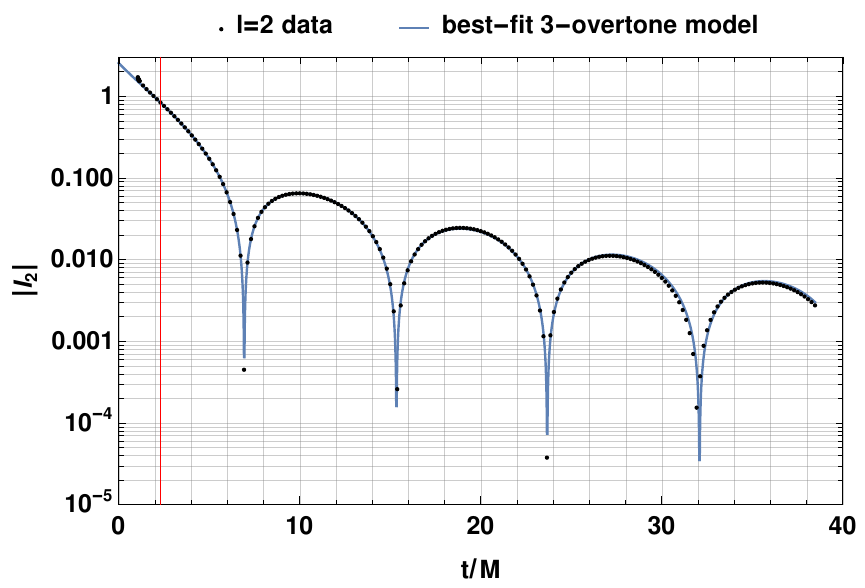}\\
  \caption{Time evolution of the $\ell=2$ multipole moment $M_2$. The plots shows the multipole moment
    as a function of time (black dots) and also two fits modeled using
    combinations of QNMs.  The fit on the left only
    includes the fundamental tone, while the fit shown on the right
    includes modes up to the third overtone.  Only data to the right
    of the vertical red line is used for the fitting procedure. The
    time stamps refer to the simulation time scaled with the total
    mass, while the data series (the black dots) begin from the moment
    when the common horizon forms. 
    Figure taken from~\cite{Mourier:2020mwa}.}
  \label{fig:l2-overtonefits-multipole}
\end{figure}

Turning now to the shear, more directly related to the infalling gravitational radiation, $\sigma_\ell$ can be expanded as a sum of damped sinusoidal signals closely related to the QNMs of the asymptotic remnant BH, including also the overtones. In fact, it was empirically found that for a given mode $\sigma_\ell$, an adequate fit is provided by a QNM expansion including $\ell+1$ overtones. An example is shown in Fig.~\ref{fig:shear_overtone} for the dominant $\ell=2$ mode, which is well described by a fit including three overtones.  
Using the overtones allows one to start closer to the merger and to also capture the rapid decay at the beginning.

\begin{figure*}
  \centering    
  \includegraphics[width=0.4\textwidth]{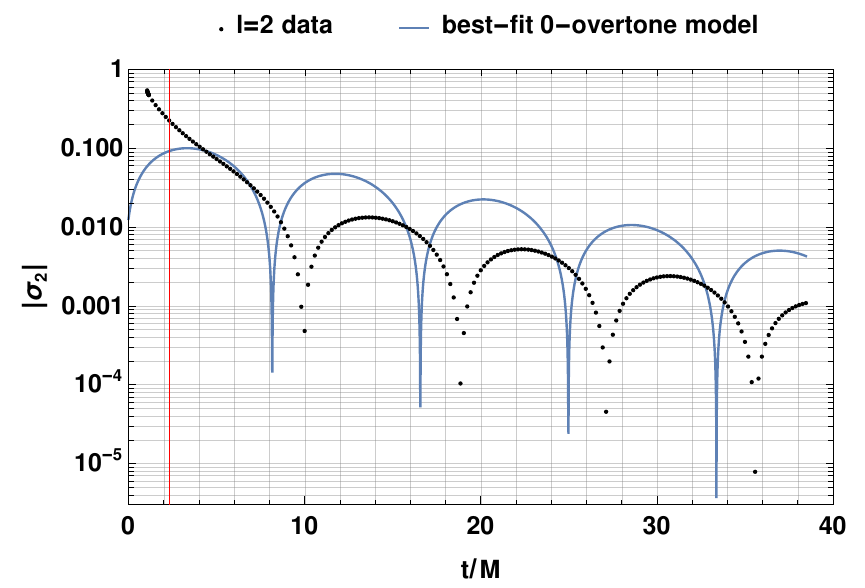}
  \includegraphics[width=0.4\textwidth]{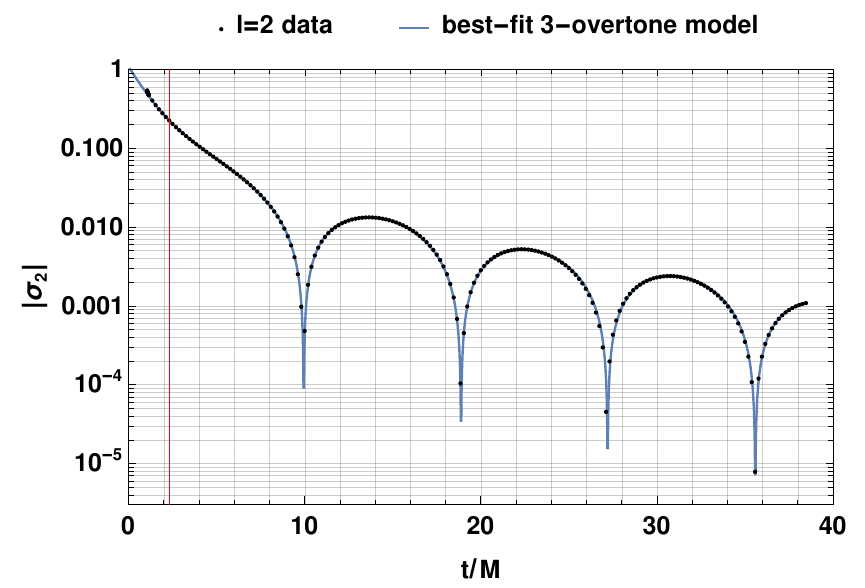}
  \caption{%
    The shear mode $\sigma_{20}$ and its fits with a QNM expansion
    including three overtones.  The vertical red line and the description
    of the data are the same as in
    Fig.~\ref{fig:l2-overtonefits-multipole}.
    Figure taken from~\cite{Mourier:2020mwa}.}
    \label{fig:shear_overtone}
\end{figure*}

The above results concern head-on collision of two comparable mass BHs, i.e. in the presence of axisymmetry.  The case of two equal-mass, nonspinning BHs in a quasi-circular orbit was studied in~\cite{Chen:2022dxt}, exploiting the formalism of~\cite{Ashtekar:2013qta} to consider time evolution of quantities on a DH. In this case, it becomes necessary to go beyond the axisymmetric multipoles $(M_n, J_n)$ and to instead consider $(M_{\ell m}, J_{\ell m})$.  As shown in~\cite{Chen:2022dxt}, $M_{22}(t)$ on the common horizon admits an excellent description in terms of QNMs associated with the remnant BH. An important aspect here is the phenomenon of mode-mixing: in order to describe $M_{22}(t)$, one needs not only the $\ell=m=2$ QNMs including overtones, but additional modes as well (in particular, the $\ell=3, m=2, n=0$ mode). Similar results are found for other moments. For example, the $(\ell=3,m=2)$ spin-multipole moment requires contributions from the $(\ell=2,m=2)$ QNM and its overtones, as well as the $(\ell=3,m=2,n=0)$ and $(\ell=4,m=2,n=0)$ modes.

For all of the above results, including the head-on case, a detailed analytical understanding is still lacking.
For example, despite an improved agreement of the QNM template with the numerical simulation when including overtones, all concerns regarding overfitting reported in Section~\ref{subsec:overtones} apply to these results as well, and therefore it is unclear if these features can be unambiguously identified as being due to the physical excitation of overtones.
Resolving the question of the physical description of the horizon signal requires analytically computing the excitation of QNMs for a given multipole moment and relating the mode amplitudes to the initial configuration parameters, rather than fitting arbitrary combinations of modes.
The same considerations apply to the waveform in the wavezone.  

Going beyond linear theory, a closer investigation of the same
$\ell=2$ shear data discussed above reveals the presence of yet more
interesting features~\cite{Khera:2023oyf}.  
As shown in Fig.~\ref{fig:shear_quadratic}, the $\ell=2$ shear mode $\sigma_2$
contains also a \textit{quadratic} combination of the $\ell=2$ QNMs. This is
analogous to the results in Sections~\ref{sec:nonlinSch}
and~\ref{sec:nonlinKerr}, where quadratic ringdown modes are found in the
outgoing gravitational radiation in the ringdown regime. There is also evidence
for quadratic modes in $\sigma_4(t)$ and $\sigma_6(t)$~\cite{Khera:2023oyf}, but
now including a boost for the initial BHs. A puzzling finding is that the
amplitude ratio of the $\ell=2$ quadratic modes divided by its linear parent
modes is different for the sets of simulations with unequal masses relative to
those with equal masses, but boosted initial data. This difference might be due
to the difference in parity-mode content at the linear level (as highlighted in
Sections~\ref{sec:nonlinSch} and~\ref{sec:nonlinKerr}), but it should be
investigated in more detail. Overall, the results are in agreement with those
reported in~\cite{Cheung:2022rbm, Mitman:2022qdl, Cheung:2023vki}, and they
provide further evidence for the correspondence between GW signals and horizon
dynamics.

\begin{figure*}[t!]
    \centering
    \includegraphics[width=0.6\textwidth]{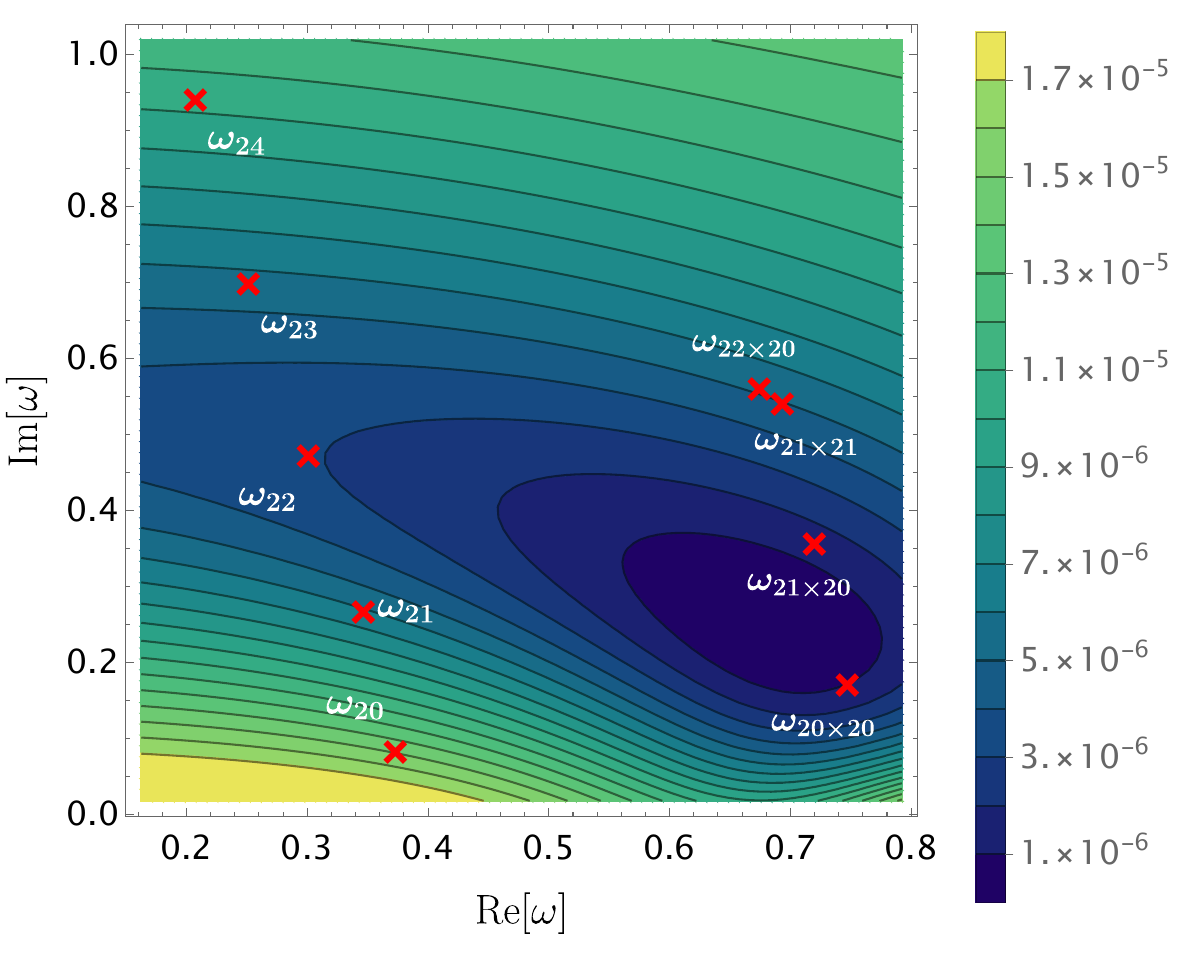}
    \caption{%
      Quadratic modes in the shear data $\sigma_2$. The figure shows the mismatch of the $\sigma_2(t)$ data from models containing various QNM combinations. The lowest mismatch, i.e. the best fit, is obtained by including the $\omega_{20}\times\omega_{20}$ quadratic mode. 
      Figure taken from~\cite{Khera:2023oyf}.
      }
    \label{fig:shear_quadratic}
\end{figure*}

\subsubsection{A Bondi-Sachs framework perspective}

As discussed above, the QNM problem is traditionally formulated within the framework of BH perturbation theory using a specific slice of the Schwarzschild spacetime.  On the other hand, the Bondi-Sachs framework~\cite{Bondi:1962px,Sachs:1962wk} of Section~\ref{sec:bms-frames-memory} is commonly used to deal with gravitational radiation at null infinity.  Here one expands the metric in a neighborhood of $\mathcal{I}^+$, and uses the Einstein equations to identify the appropriate GW fluxes and other physical effects.  As shown in~\cite{Mongwane:2024vao}, the Bondi-Sachs framework can also be used successfully to reformulate the QNM problem for a Schwarzschild black hole.  More generally, BH perturbation theory can be reformulated within a Bondi-Sachs framework, including also data on a perturbed Schwarzschild horizon~\cite{RibesMetidieri:2024tpk,RibesMetidieri:2025lxr}.  In the usual framework, the definition of the QNM requires a resonance condition, i.e., purely outgoing boundary conditions at infinity and at the horizon. The analog of this condition turns out to be the requirement that there is no incoming radiation from $\mathcal{I}^-$.  With this requirement, assuming also analyticity of $\Psi_0$ and $\Psi_4$ along the radial direction, the usual QNM frequencies are recovered~\cite{RibesMetidieri:2025lxr}.

\clearpage
\section{Data analysis}
\label{sec:DataAnalysis}

\noindent
{\em In every branch of knowledge the progress is proportional to the amount of facts on which to build, and therefore to the facility of obtaining data.}

\vspace{.2cm}

\noindent
James Clerk Maxwell, {\em The Scientific Letters and Papers of James Clerk Maxwell: 1846-1862}, p. 209, Cambridge University Press Archive (1990).

\vspace{.2cm}

The primary goal of the BH spectroscopy program is, in principle, simple: measure at least two frequencies belonging to the QNM spectrum to infer the Kerr BH mass and spin, and then test the Kerr nature of the object by performing any additional (agnostic) frequency measurements.
If the predictions disagree with those of vacuum GR, this may lead to the conclusion that the BH spacetime must be described by additional parameters beyond the mass and spin, or even to the discovery of new gravitational degrees of freedom and exotic states of matter.
Within GR, spectroscopic measurements can also inform astrophysical inference, for example because different orbital configurations lead to different predictions on which QNMs must be excited and on their relative amplitudes.

In practice, realizing these goals in GW data analysis involves several subtleties, that we discuss in this section.
The initial proposal of agnostically inferring a few damped sinusoids is hard to implement when one deals with low-SNR signals with complicated morphologies and large uncertainties.
As we argue in Section~\ref{subsec:bayesian_ringdown}, these measurements are affected by technical and conceptual issues, that must be addressed by comparing different data analysis strategies and by carefully handling systematic uncertainties.
In Section~\ref{subsec:current_observations} we summarize the latest experimental results on QNM measurements from post-merger signals (now routinely performed within the LVK network) and the underlying assumptions.
In Section~\ref{subsec:deviations_current} we review current bounds on new physics in the context of theory-agnostic and theory-specific proposals.
Finally, in Section~\ref{subsec:echoes_DS} we discuss searches for post-ringdown ``echo'' signals that may be present if BH horizons are not perfect one-way membranes.

\subsection{The formulation of ringdown data analysis}\label{subsec:bayesian_ringdown}

\vspace{-.1cm}

\noindent \textit{Initial contributors: Capano, Carullo, Del Pozzo, Finch, Gennari, Isi}

\vspace{.2cm}

The problem of measuring ringdown signals in noisy GW data is surprisingly subtle, and comes with its own unique set of challenges relative to other standard GW analyses~\cite{Isi:2021iql,Carullo:2019flw,Isi:2019aib,Cabero:2017avf,Prix:T1500618,Finch:2021qph,Capano:2021etf}.
Given a segment of data containing a GW signal, the goal of a ringdown analysis is to fit some template based on perturbation theory (typically, a combination of damped sinusoids: see Section~\ref{subsec:ringdown_models}) to some subsegment of the data containing the ringdown relaxation process.
In the context of LVK observations, this segment is the last portion of a BBH coalescence signal.
From a data analysis point of view, the challenge lies in properly fitting this ringdown model ``alone'' in noisy detector data, independently of the rest of the signal that preceded it.

When analyzing datasets in which the level of noise is negligible, isolating
ringdown data for a fit is operationally straightforward: one may simply discard
data points before a certain time of interest (say, $t_0$), and then fit or
otherwise manipulate the remaining data in the TD or FD as desired to obtain the best fitting parameters.
This is a common procedure in, e.g., the analysis of NR simulations, for which the ``noise level'' is typically small compared to the targeted effect, and assumed uncorrelated across time steps.
Real GW signals, however, are buried in detector noise. This requires a method for separating the signal and the noise in order to characterize the former. In GW astronomy, Bayesian inference is typically used to perform this separation and to produce a probability distribution (known as the posterior distribution) that quantifies our best estimate of the signal's parameters. Computationally efficient and rigorous methods have been devised to perform inference on binary merger signals that are detected by GW detectors~\cite{Veitch:2014wba,Biwer:2018osg,Ashton:2018jfp,Romero-Shaw:2020owr,Dax:2021tsq,Wong:2023lgb}.
Inferring the parameters of the ringdown signal alone complicates the standard Bayesian analysis. The key problem is that the noise in real GW detectors is best characterized as a stationary \emph{colored} Gaussian process~\cite{LIGOScientific:2019hgc}.
This applies, to a good approximation, to individual segments of LVK data lasting $\mathcal{O}(1\,\mathrm{s} {-} 100\,\mathrm{s})$, as typically considered for BH data analysis. 
The noise is nonstationary over longer periods of time: the baseline level changes slowly over time, and there are transient outbursts of power (``glitches'') that are individually cleaned from the data stream if they happen to overlap with a signal~\cite{LIGO:2021ppb,Davis:2022ird}.
For the purpose of this discussion, we assume that glitches have been removed whenever required.
In colored Gaussian noise, points in time are correlated, so it is challenging to isolate a specific sub-portion of the signal to properly fit a ringdown model. Below we review GW Bayesian inference and the different methods that have been devised to isolate the ringdown signal.

\subsubsection{Bayesian inference in gravitational-wave astronomy}\label{subsec:bayesian_inference}

Given some global assumptions $I$ (that in this context we will call a ``model'' -- not to be confused with a ``waveform model'')
subsuming our understanding of the data-generation process (signal and noise), we wish to measure parameters $\params$ conditional on the observation of some noisy data $\vdata$ by obtaining the posterior probability density function $p(\params|\vdata, I)$. By Bayes' theorem, this can be factorized as
\begin{equation} \label{eq:bayes}
    p(\params | \vdata, I) = \frac{1}{\mathcal{Z}}\,\, p(\vdata| \params, I)\, p(\params | I),
\end{equation}
where $p(\vdata | \params, I)$ is the \emph{likelihood function}, $p(\params|I)$ is the \emph{prior}, and $\mathcal{Z}  = p(\vdata | I) = \int p(\vdata | \params, I)\, p(\params | I)\, {\rm d}\params$ is a normalization constant known as the model \emph{evidence}, or marginal likelihood. Estimates on a single parameter are obtained by marginalizing (i.e., integrating) the posterior over all other parameters; marginalizing over all parameters yields the evidence.
For different hypotheses, the ratio of evidences gives the odds that one hypothesis is favored over the other, providing a statistical metric for model selection.
In addition, the parameters $\params$ encode the information on the waveform model used in the analysis (see Section~\ref{subsec:ringdown_models}).
For simplicity of discussion, here we will not consider the inclusion of additional parameters characterizing the noise (e.g. calibration uncertainties~\cite{Vitale:2020gvb} and power spectral density (PSD) parameters~\cite{Littenberg:2014oda}), since their inclusion is conceptually identical to standard full-signal analyses.

Per Eq.~\eqref{eq:bayes}, observations inform our measurement only through the likelihood function $p(\vdata | \params, I)$, which encapsulates our assumptions about the data generation process. Consider a GW detector, which we sample every $\Delta t$ seconds over a time $T$ to obtain $N = \lceil T/\Delta t \rceil$ uniformly time-ordered samples $\vdata = \{\data_0, ..., \data_{N-1}\}$. A network of $Q$ detectors sampled in this way will produce a set of samples $\vdata_{\net} = \{\vdata_0, ..., \vdata_{Q-1}\}$.
Assuming the noise $\vnoise$ is additive at each detector (namely that $\vdata = \vnoise + \vsignal$ for some signal $\vsignal$, if present), then we can obtain the likelihood function by noting that the residuals for the true parameters should be drawn from the detector noise distribution $p(\vnoise_{\net} | I)$.
In other words, for residuals $\vresidual(\params) = \vdata - \vsignal(\params)$, the network likelihood is $p(\vdata_{\net} | \params, I) = p(\vnoise_{\net} = \vresidual_{\net}(\params) | I)$.

In GW astronomy it is common to assume that, in the absence of a signal, the detectors output stochastic Gaussian noise that has zero mean and is independent across detectors~\cite{LIGOScientific:2019hgc}. Under this assumption, the probability density function describing the network of time-ordered noise samples $\vnoise_{\net}$ is a product of %
$N$-dimensional multivariate normal distributions~\cite{LIGOScientific:2019hgc}, %
\begin{equation}
\label{eq:pnoise}
p(\vnoise_{\net} | I) = \frac{1}{\sqrt{(2\pi)^{NQ} \prod_{q=1}^{Q}\det \vcovmat_{q}}} \cdot \exp\left[
    -\frac{1}{2}\sum_{q=0}^{Q-1} \vnoise_{q}^{\top}\vcovmat_{q}^{-1} \vnoise_{q}\right].
\end{equation}
Here, $\vcovmat_{q}$ is the covariance matrix of the noise in detector $q$, which is most often prescribed a priori based on knowledge of the instrument (so that $I$ contains the set of all $\vcovmat_q$'s, besides the assumption of Gaussianity itself).
Without considering the cost of obtaining $\vresidual(\boldsymbol{\theta})$, a single evaluation of Eq.~\eqref{eq:pnoise} in general requires $O(N^2)$ computations if the covariance matrices have been preprocessed appropriately (e.g., through a Cholesky decomposition).
However, the cost can be reduced if the problem has further structure.

If a detector's noise is wide-sense stationary (a good approximation for current GW detectors, as discussed above), then the covariance matrix will only be a function of time differences. 
Assuming a segment of data without gaps, we can introduce an autocovariance function (ACF) that is a function of sample lags, i.e., $\rho(k) = \rho_k$, for some integer $0 \leq k < N$, such that
\begin{equation}
    \covmat_{ij} = \rho(|i - j|)\, .
\end{equation}
The covariance matrix is therefore symmetric with a \emph{Toeplitz} form, meaning that $\covmat_{ij} = \covmat_{i+1,j+1}$ for all $i,j < N-1$.
There exist tailored algorithms to compute the inverse of Toeplitz matrices (like Levinson recursion~\cite{Levinson1946,Durbin1960}) entering in Eq.~\eqref{eq:pnoise}, but they are also $O(N^2)$ in complexity and they do not offer a significant speed-up.

Further efficiency, however, can be achieved if the covariance matrix has more structure.
If we impose periodic boundary conditions on the Gaussian process, as would be the case for noise defined on a circular domain (e.g., the celestial sphere), then the covariance matrix is said to be \emph{circulant}.
Although this is exactly the case in some applications (e.g., studies of the cosmic microwave background), it is \emph{not} the case for (finite) segments of GW strain data~\cite{Talbot:2021igi}.
Nevertheless, if $T$ is large compared to the characteristic length of the ACF, then \vcovmat{} will primarily have support along the central diagonal and mostly vanish in the upper and lower triangles; consequently, to excellent approximation~\cite{gray:2006},
\begin{equation}
\vcovmat \approx \vcircmat,
\label{eq:circapprox}
\end{equation}
where \vcircmat{} is a circulant matrix with elements
\begin{equation}
\circmat_{i,j} =
\begin{cases}
      \covmat_{N-i+2j, j} & i-j > \lceil N/2 \rceil,\\
      \covmat_{i, N-j+2i}, & i-j < -\lceil N/2 \rceil, \\
      \covmat_{i,j}, & \text{otherwise}.
    \end{cases}
\end{equation}
The approximation in Eq.~\eqref{eq:circapprox} is equivalent to assuming that the noise has periodic boundary conditions.

The imposition of periodicity allows us to leverage discrete Fourier transforms (DFTs), which can be efficiently computed through the fast Fourier transform (FFT) algorithm.
The DFT of any TD $N$-vector $\mathbf{x}$ is denoted $\mathbf{\tilde{x}}$ and defined by
\begin{equation}
\tilde{x}_{j} = \sum_{k=0}^{N-1} \exp\left(-2\pi i \frac{j k}{N}\right) x_k = \sum_{k=0}^{N-1} \dftmat_{jk}\, x_k\, ,
\end{equation}
where $i$ is now the imaginary unit, and the second equality defines the DFT matrix $\vdftmat$.
As it turns out, all circulant matrices have the same eigenvectors, $e^{-2\pi i j k /N}/{\sqrt{N}}$~\cite{gray:2006,UNSER1984231}, meaning that $\vdftmat$ diagonalizes $\vcircmat$.
Thus, there exists a function of frequency $S(f)$ such that
\begin{equation} \label{eq:fourier-covariance}
\sum_{p,q} \dftmat_{jp}\, \circmat_{pq}\, \dftmat_{qk}^* = \frac{1}{2} T\, S_j\, \delta_{jk} = \tilde{\circmat}_{jk} ,
\end{equation}
where $S_j = S(f_j)$ for a frequency bin $f_j = j/\Delta t$, and the prefactor in the first equality is chosen such that $S(f)$ corresponds to the one-sided PSD of the instrumental noise, which is itself related to the (circular) ACF by $S_j = 2 \Delta t\, \tilde{\rho}_j$; the second equality defines the FD covariance matrix $\tilde{\vcircmat}$.
In fact, a consequence of Eq.~\eqref{eq:fourier-covariance} is that $\tilde{\vnoise}$ must be drawn from a multivariate Gaussian process with diagonal covariance matrix given by $\tilde{\vcircmat}$, such that for a given detector $q$~\cite{Allen:2001ay}
\begin{equation} \label{eq:fdlikelihood}
    p( \tilde{\vnoise}_q | I) = \prod_{j=0}^{\lfloor N / 2 \rfloor} \frac{1}{2\pi} \left(T S_{q,j} \right)^{-2} \exp \left( - \frac{2\left|\tilde{\noise}_{q,j}\right|^2}{T S_{q,j}} \right) .
\end{equation}
Assuming additive noise under this hypothesis, the residual $\vresidual_q = \vdata_q - \vsignal_q$ is a realization of Gaussian noise; the probability of observing it is given by Eq.~\eqref{eq:pnoise}. \emph{As long as the circulant approximation Eq.~\eqref{eq:circapprox} holds}, we can instead use Eq.~\eqref{eq:fdlikelihood} with the $\vnoise_q$ replaced by the residuals $\vresidual_q$ for the likelihood function of the signal hypothesis, using the FFT to efficiently obtain FD quantities and then evaluating Eq.~\eqref{eq:fdlikelihood} with complexity $O(N)$ -- a vast improvement over the original $O(N^2)$ complexity of Eq.~\eqref{eq:pnoise}.
Clearly, any covariance matrix can be diagonalized by \emph{some} coordinate transformation matrix; the special feature of the circulant Toeplitz case is that its diagonalization matrix is the DFT matrix, simplifying the complexity of the transformation and guaranteeing it is independent of the specific form of $S(f)$.

\subsubsection{The ringdown problem}
Equation \eqref{eq:fdlikelihood} is the basis for the vast majority of inferences in GW astronomy;
however, to obtain it we had to assume not only Gaussianity and stationarity, but also circularity.
In most GW applications, this is not an issue: one may window the data to enforce circularity (at the expense of stationarity)~\cite{Talbot:2021igi}, or otherwise truncate the response of the TD whitening filter (proportional to the inverse Fourier transform of $1/S^{-1/2}$) and then discard data at the edges of the whitened segment~\cite{Allen:2005fk}.
All these procedures destroy or corrupt information localized in time close to the beginning or end of the analysis data; this is unimportant if no relevant signal content is present at the edges, or if the segment can be extended until this is satisfied (either by directly taking a longer data segment, or by applying a rolling whitening filter with inverse spectrum truncation).

Ringdown analyses are not easily amenable to any of those standard strategies that enable the direct application of Eq.~\eqref{eq:fdlikelihood}.
Since we are trying to censor data before a certain time, it is not possible to expand the analysis segment without encompassing the region of data we want to exclude.
Windowing the analysis segment is also generally not an optimal way forward, as that would either decrease the amount of ringdown signal that can be analyzed~\cite{Carullo:2018sfu} or alter data near the edges where the ringdown signal is located (see, e.g., Fig.~7 in Ref.~\cite{Isi:2021iql}).
Moreover, windowing parameters would need to be optimized given the signal morphology, rendering the entire procedure rather cumbersome.
Instead, padding the segment (e.g., with zeros) breaks stationarity and causes ringdown data near the start of the segment to be corrupted by the prepended timestamps through the correlations imposed by the covariance matrix.
Finally, without any treatment Eq.~\eqref{eq:fdlikelihood} will cause data points at the beginning of the segment to be correlated with those at the end.
In all of these situations, the derived posterior distribution will be biased~\cite{Isi:2021iql}.

There are two classes of approaches to addressing this problem: (1) give up on fitting a pure-ringdown model alone, prescribing a (potentially very flexible) model also for the pre-ringdown data~\cite{Finch:2021qph,Brito:2018rfr}, or (2) abandon the assumption of circular boundary conditions that enabled the simplification of Eq.~\eqref{eq:circapprox}~\cite{Carullo:2019flw,Isi:2019aib,Isi:2021iql,Capano:2021etf}.
The former comes with the computational advantage of Eq.~\eqref{eq:fdlikelihood} and the well-tested infrastructure to implement it that has been developed for traditional GW analyses; however, it comes at the risk of undesirable interactions between the two pieces of the model (ringdown and pre-ringdown).
The latter has the advantage of allowing one to define a model for the ringdown data alone, while remaining fully agnostic about the preceding times; yet, it has the disadvantage of higher computational complexity, plus an inability to directly vary the start time of the ringdown fit (see the next section for a discussion of this point).

The first strategy for computing the ringdown likelihood is already covered in standard treatments of GW inference (see~\cite{LIGOScientific:2019hgc} and references therein); interesting notions specific to such ringdown analyses have to do with the prescription of the waveform model and are discussed in Section~\ref{subsec:ringdown_models} below.
On the other hand, there are subtleties regarding the second strategy that are worth discussing here, including the two main approaches for implementing it: (1) direct truncation in the TD with a TD likelihood (Section~\ref{subsub:td}), and (2) effective censoring of certain timestamps with a modified FD likelihood (Section~\ref{subsub:gating}).
The two approaches are formally equivalent, and we discuss them below.

\subsubsection{Time-domain likelihood} \label{subsub:td}
The first strategy consists of isolating the target data in the TD~\cite{Carullo:2019flw} and computing Eq.~\eqref{eq:pnoise} directly~\cite{Isi:2019aib,Isi:2021iql}.
Once the fitting time is specified, the only remaining subtleties have to do with the construction of the noncirculant covariance matrix $\vcovmat$, and with any preconditioning applied to the data (see Section~\ref{subsub:data_conditioning} below).
Assuming stationarity, all that is needed to construct $\vcovmat$ is an estimate of the ACF $\rho$, which can be obtained directly from the empirical autocorrelation of noise-only samples of data (subject to some boundary conditions prescription) or by inverse-Fourier transforming a pre-existing estimate of the PSD (obtained, e.g., through Welch averaging, or by Bayesian modeling~\cite{Chatziioannou:2019zvs,Littenberg:2014oda}).
It is crucial to truncate the ACF estimate to destroy the circularity that will necessarily be imposed by the Fourier transformation~\cite{Isi:2019aib}; this requires that the analysis segment $T$ be much shorter than the segment length used in either estimating the PSD (see, e.g., Fig.~9 in Ref.~\cite{Isi:2021iql}) or in computing the empirical autocorrelation (to avoid boundary effects). 
If this is not satisfied, the resulting covariance matrix will be circulant, and the analysis equivalent to that of Eq.~\eqref{eq:fdlikelihood}.

The resulting $\vcovmat$ matrix will be of Toeplitz form, as mentioned above, allowing us to solve the likelihood efficiently through dedicated algorithms, like Levinson recursion.
In practice, a Cholesky decomposition (which only leverages the positive definiteness of the matrix, not its Toeplitz structure) often does just as well~\cite{Isi:2021iql}.
The Cholesky decomposition ($\vcovmat = \mathbf{L} \mathbf{L}^\transpose$ where $\mathbf{L}$ is lower triangular) allows for natural definitions of whitened (decorrelated) quantities in the TD: for any TD vector $\mathbf{x}$, the corresponding TD whitened quantity is $\bar{\mathbf{x}} = \mathbf{L}^{-1} \mathbf{x}$.
Since both this operation as well as the subsequent
$\vcovmat$ inversion using $\mathbf{L}^{-1}$
involve a matrix inversion, care must must be taken in ensuring the numerical stability of such operations. 
The stability is quantified by the condition number of the matrix $\vcovmat$. 
The condition number, defined as 
$\kappa (\vcovmat) = ||\vcovmat|| \cdot ||\vcovmat^{-1} ||$,
characterizes the sensitivity of the solution to numerical perturbations (see e.g.~\cite{golub13}). 
If  $\vcovmat$ is ill-conditioned ($\kappa \gg 1$), even small floating-point errors can lead to large errors in the likelihood computation.
If the ACF was correctly estimated, $\bar{\vnoise}$ will be drawn from a standard normal distribution, as can be deduced from Eq.~\eqref{eq:pnoise}; the same should be true of $\bar{\vresidual}(\params)$ if the parameters are a good fit.
This allows us to construct statistical tests of the hypotheses entering the likelihood construction.

Based on Eq.~\eqref{eq:pnoise}, we can use the covariance matrix to define a time-limited noise-weighted inner product,
\begin{equation}
   \left. \left\langle \mathbf{x} | \mathbf{y} \right\rangle \right|_{t_0}^{t_0+T} = \mathbf{x}\, \vcovmat^{-1}\, \mathbf{y} = \bar{\mathbf{x}}^\transpose \bar{\mathbf{y}} \, ,
\end{equation}
for any TD vectors $\mathbf{x}$ and $\mathbf{y}$ extending from time $t_0$ to $t_0 + T$ (equivalently, from index $0$ to $N-1$).
With this definition, we can introduce a notion of SNR that is also limited to the times of interest, as the normalized time-limited inner product
\begin{equation}
    \left. \mathrm{SNR}(\mathbf{x}, \mathbf{y})\right|_{t_0}^{t_0+T} = \frac{\left. \left\langle \mathbf{x} | \mathbf{y} \right\rangle \right|_{t_0}^{t_0+T} }{\sqrt{ \left. \left\langle \mathbf{y} | \mathbf{y} \right\rangle \right|_{t_0}^{t_0+T}}} \, .
\end{equation}
With $\mathbf{x} = \vdata$ and $\mathbf{y} = \vsignal$, this is the matched-filter SNR at a given detector; with $\mathbf{x}=\mathbf{y} = \vsignal$, this is known as the optimal SNR.
This quantity represents the expectation for the SNR when marginalized over noise realizations, which is not ``optimal'' in the sense of being the highest: any given instantiation of noise may increase or decrease the SNR relative to this value.
Such a time-limited notion of SNR is the most appropriate for ringdown analyses~\cite{Isi:2021iql}.

\subsubsection{Gating and in-painting}\label{subsub:gating}

Gating and in-painting is a method that is mathematically equivalent to the TD likelihood, but uses a FD formulation, Eq.~\eqref{eq:fdlikelihood}. Suppose that a signal becomes observable at time $t = a \Delta t$, with the portion of data to be isolated starting at a later time $t = (a+M)\Delta t$. We wish to only evaluate the truncated set $\vdata_{tr} = \{\data_0, ..., \data_a, \data_{a+M}, ..., \data_{N-1}\}$. The data between the time steps $[a,a+M)$ is said to be ``gated.''

Define $\vnoise' = \vnoise_g + \mathbf{x}$, where $\vnoise_g$ is the noise with the gated times $t\in [a, a+M)\Delta t$ zeroed out, and $\mathbf{x}$ is a vector that is zero everywhere except in the gated times. If the nonzero elements of $\mathbf{x}$ are such that $(\vcovmat^{-1} \vnoise')[k] = 0$ for all $k \in [a, a+M)$, then $\vnoise'^\transpose \vcovmat^{-1} \vnoise'$ will  be the same as the truncated version $\vnoise_{tr}^\transpose \vcovmat_{tr}^{-1} \vnoise_{tr}$. Our aim is to solve the equation $\vcovmat^{-1}(\vnoise_g + \mathbf{x}) = \mathbf{0}$ in the gated region. Since $\mathbf{x}$ is zero outside of the gated region, $\vcovmat^{-1}\mathbf{x}$ only involves  the $[a, a+M)$ rows and columns of $\vcovmat^{-1}$, which form an $M \times M$ Toeplitz matrix [cf.~Eq.~\eqref{eq:fourier-covariance}]. We therefore solve for $\mathbf{x}$ such that
\begin{equation}
    \overline{\vcovmat^{-1}}\,\overline{\mathbf{x}} = -\overline{\vcovmat^{-1}\vnoise_g},
\label{eq:gatecondition}
\end{equation}
where the overbar indicates the $[a, a+M)$ rows (and columns) of the given vector (matrix). This can be solved numerically using a Toeplitz solver~\cite{Virtanen:2019joe,Harris:2020xlr}. Adding $\mathbf{x}$ to the gated noise (``in-painting'') will then yield the same result as if we had truncated the noise and the covariance matrix. To evaluate the likelihood for a signal, we use $\vnoise_g = \vdata_g - \vsignal_g$ (i.e., the residual with the gated region zeroed out) in Eq.~\eqref{eq:gatecondition} and solve for $\mathbf{x}$. We can then use $\mathbf{x} + \vdata_g - \vsignal_g$ in the standard likelihood, Eq.~\eqref{eq:fdlikelihood}.

For $a = 0$, this method is mathematically equivalent to the TD method of Section~\ref{subsub:td}, as is shown in Section~IIIB of~\cite{Isi:2021iql}.
The main difference between the two methods lies in their computational costs. A GW Bayesian inference analysis is usually dominated by the cost of the likelihood function, as it can take millions of likelihood evaluations to obtain an accurate estimate of the posterior distribution. Although inverting $\vcovmat$ costs $O((N-M)^{3})$ operations in the TD method of Section~\ref{subsub:td}, this only needs to be done once at start-up. From then on, each likelihood evaluation costs $N'^2$ operations to do the matrix multiplication in the exponent of Eq.~\eqref{eq:pnoise}, where $N'=N-M$. In gating and in-painting the dominant cost is solving the Toeplitz equation \eqref{eq:gatecondition}, which needs to be done for each likelihood evaluation. This requires $O(M^2)$ operations. In addition, $O(N\log N)$ operations are needed to FFT in-painted data to the FD. Gating and in-painting is thus more computationally efficient for $M \ll N$; for $M \sim N$, the straight TD method of Section~\ref{subsub:td} is more efficient.
A summary of the different analyses techniques outlined, and of the ones used below is available in Table~\ref{tab:methods}.

\begin{table}[h!]
\centering
\begin{tabular}{p{4.5cm}p{10cm}}
\toprule
\textbf{Method} & \textbf{Description} \\
\midrule
TD & Formulate the GW likelihood in the TD~\cite{Carullo:2019flw,Isi:2019aib} and isolate the ringdown portion of the signal by cutting the timeseries~\cite{Isi:2021iql}. \newline Used in, for example, Refs.~\cite{Carullo:2019flw, Isi:2019aib, Cotesta:2022pci, Isi:2022mhy, Carullo:2023gtf}. \\
\arrayrulecolor{gray}\cmidrule(lr){1-2}
Gating and in-painting & Isolate the ringdown portion of the signal whilst remaining in the FD via a gating and in-painting procedure~\cite{Capano:2021etf}. Also referred to as a ``modified FD likelihood'' in Ref.~\cite{Isi:2021iql}. \newline Used in Refs.~\cite{Capano:2021etf, Wang:2023xsy}. \newline See also Refs.~\cite{Correia:2023bfn, Correia:2023ipz}, where the gate time can be varied by doing a simultaneous analysis of the pre-ringdown and ringdown signals. \\
\arrayrulecolor{gray}\cmidrule(lr){1-2}
pSEOB & Model the full IMR signal with an SEOB GR waveform baseline, allowing for deviations in the QNM spectrum of the ringdown~\cite{Brito:2018rfr,Ghosh:2021mrv,Maggio:2022hre,Pompili:2025cdc} (see also Section~\ref{sec:effective-one-body}). \\
\arrayrulecolor{gray}\cmidrule(lr){1-2}
Wavelet+ringdown & ``Marginalize'' over the pre-ringdown signal via a flexible sum of wavelets~\cite{Finch:2021qph}. \newline Used in Ref.~\cite{Finch:2022ynt}. \\
\arrayrulecolor{gray}\cmidrule(lr){1-2}
SBI & Simulation-based (i.e., likelihood-free) inference whereby the posterior is learned from simulated data realizations~\cite{Crisostomi:2023tle, Pacilio:2024qcq}. \\
\arrayrulecolor{gray}\cmidrule(lr){1-2}
Rational filter & Remove power from specific QNMs in the data~\cite{Ma:2022wpv}. \newline Used in Refs.~\cite{Ma:2023vvr, Ma:2023cwe}. \\
\arrayrulecolor{gray}\cmidrule(lr){1-2}
$\mathcal{F}$-statistic & Analytically maximize over QNM amplitudes and phases in a TD analysis~\cite{Wang:2024jlz}. \newline Used in Ref.~\cite{Wang:2024yhb}. \\
\arrayrulecolor{black}\bottomrule
\end{tabular}
\caption{Summary of methods used in ringdown data analysis.}
\label{tab:methods}
\end{table}

\subsubsection{Data conditioning}\label{subsub:data_conditioning}

Analyses of GW observations most often require some kind of data conditioning before the inference step.
Besides identifying a segment of data to analyze, this includes operations like downsampling, filtering and cleaning designed to reduce the amount of data that needs to be processed, thus speeding up calculations.
Even if no such manipulations are explicitly applied, it is rarely the case that analysts have access to truly raw data, so it is important to understand the effects of these operations on inference in general, and ringdown analyses in particular.
For example, LIGO calibrated data are produced at 16,384 Hz with filters applied by the data acquisition and calibration procedures, and typically also some subtraction of known spectral lines and other well-understood sources of noise; public LVK data has traditionally been released at 4,096~Hz and 16,384~Hz, both with a high pass filter at 8 Hz and 4,096~Hz with a slow-rolling Butterworth antialiasing filter.

There is a vast literature studying the effects of data cleaning and PSD estimation methods in the context of the standard FD GW likelihood of Eq.~\eqref{eq:fdlikelihood}, see e.g.~\cite{Talbot:2020auc} and references therein, but this is less well studied in the context of the TD ringdown likelihood of Eq.~\eqref{eq:pnoise}.
In the context of ringdown analysis, the authors of Ref.~\cite{Siegel:2024jqd} quantify the impact of downsampling, filtering, and line-cleaning (or lack thereof), whereas Refs.~\cite{Wang:2023mst,Wang:2023xsy} explore different noise (ACF and PSD) estimation methods.

After selecting a target start time, two basic decisions need to be made regarding data for a ringdown analysis by picking (1) the duration of the analysis segment, and (2) the sampling rate of the data. Naively, one might expect the duration of the required segment to be prescribed by the intrinsic decay rate of the targeted signal. However, that is not the only factor to be considered: since real detector noise follows (to good approximation) a \emph{colored} Gaussian distribution that correlates points in time, interactions between the signal and the whitening filter can necessitate an analysis segment that is up to an order of magnitude longer than the decay rate of the signal alone~\cite{Siegel:2024jqd}.
In particular, this can be caused by narrow lines in the noise PSD, and can thus be mitigated by cleaning such features as part of the conditioning step in preprocessing~\cite{Siegel:2024jqd}.
An analysis segment that is too short will typically cause spurious broadening of the posterior, thus leading to overestimates of the statistical uncertainty~\cite{Isi:2023nif}.

Downsampling the data can also impact the resulting posterior.
To reduce computational costs, it is desirable to analyze as few data points as possible; this leads one to analyze data with the lowest sampling rate possible that does not compromise the integrity of the posterior.
It is useful to consider the FD representation of damped sinusoid templates as a useful, albeit imperfect~\cite{Siegel:2024jqd}, guide for deciding on a safe sampling rate: it is generally safe to downsample to a given Nyquist frequency if the Lorentzian representing the fastest decaying mode in the template has fallen significantly below the PSD by that point.
The details of the downsampling method are also important. Many traditional decimation schemes use a slow-roll antialiasing filter that corrupts frequencies significantly below Nyquist, which is undesirable for TD analyses. If the signal has considerable power near Nyquist, templates cannot be filtered in the same way, and depending on the template, this mismatch can measurably alter posteriors~\cite{Siegel:2024jqd,Wang:2023mst}.
These considerations explain why standard LVK ringdown analyses~\cite{LIGOScientific:2020iuh,LIGOScientific:2020ufj,LIGOScientific:2021sio, LIGOScientific:2020tif} are typically performed with conservatively high sampling rates (e.g. $4096 \, \rm Hz$ during the third observing run O3).
The problem can be avoided by doing away with such filters, which are not necessary for the TD analyses, or by producing an ACF from a PSD that has been manipulated to censor the affected frequencies~\cite{Siegel:2024jqd}.

\subsubsection{Ringdown parameterizations} \label{subsec:ringdown_models}
The complex GW polarizations in the ringdown regime can be expressed as~\cite{LIGOScientific:2020tif, LIGOScientific:2021sio}
\begin{equation} \label{eq:RD_waveform}
    h_{+} -i h_{\times} = \sum_{\ell=2}^{+\infty}\sum_{m=-\ell}^{\ell}\sum_{n=0}^{+\infty}\, h_{\ell m n}\:{}_{-2}S_{\ell m n}, \quad h_{\ell m n} \equiv A_{\ell m n}(t) \,e^{-(t-t_0)/\tau_{\ell m n}}\,e^{-2\pi if_{\ell m n }(t-t_0) + i\phi_{\ell m n}},
\end{equation}
where for the moment we neglect the contribution of retrograde modes (defined in Section~\ref{sec:Teukolsky_symms}), which are hardly excited for quasi-circular binaries at current sensitivities~\cite{Li:2021wgz, MaganaZertuche:2021syq, Dhani:2021vac, Isi:2022mbx}.
For the same reason, we also neglect quadratic modes and tails (Sections~\ref{sec:nonlin_num_expe} and~\ref{sec:tails}).
The spin-weighted spheroidal harmonics ${}_{-2}S_{\ell m n}$ encode the angular dependence of the GW signal (see Section~\ref{sec_21}), so the only part of the waveform that depends only on the intrinsic parameters of the system is $h_{\ell m n}$.

A parameterization of $h_{\ell m n}$ in terms of a set of parameters $\params$, measured in the inference process through Eq.~\eqref{eq:bayes}, is referred to as a \textit{ringdown model} (see Sections~\ref{sec:amplitudes}, \ref{sec:waveforms}).
This parameterization is one of the main ingredients of ringdown data analysis, because in a Bayesian framework the information that we are able to extract from the data depends on the model 
(see Sections~\ref{subsec:bayesian_ringdown}).
For this reason, the use of different parameterizations is a valuable tool for ringdown analysis, as it allows us to compare different assumptions and characterize the robustness of the  results.
In this section we discuss the ``hierarchy'' of different models routinely used in current ringdown analyses~\cite{LIGOScientific:2021sio, Carullo:2019flw, Isi:2019aib, Finch:2021qph, Capano:2021etf, Gennari:2023gmx, Maggio:2022hre}.
In fact, more assumptions come with less flexibility, possibly favoring biases if the data violate the model hypothesis, and preventing us from capturing hypothetical new physics that may induce significant deviations from the assumed model.
On the other hand, if the data do not contain any additional physics beyond that included in the model, more general ringdown parameterizations will allow for more information to be retrieved, in turn providing more sensitive measurements~\cite{LIGOScientific:2020tif, LIGOScientific:2021sio}.
This is qualitatively represented in Fig.~\ref{fig:corner_models_hierarchy}, where we compare posterior distributions obtained with the different ringdown models listed below.
A summary of the different parameterizations outlined below is in Table~\ref{tab:models}.

\begin{figure*}[t!]
    \begin{center}
    \includegraphics[width=0.8\textwidth]{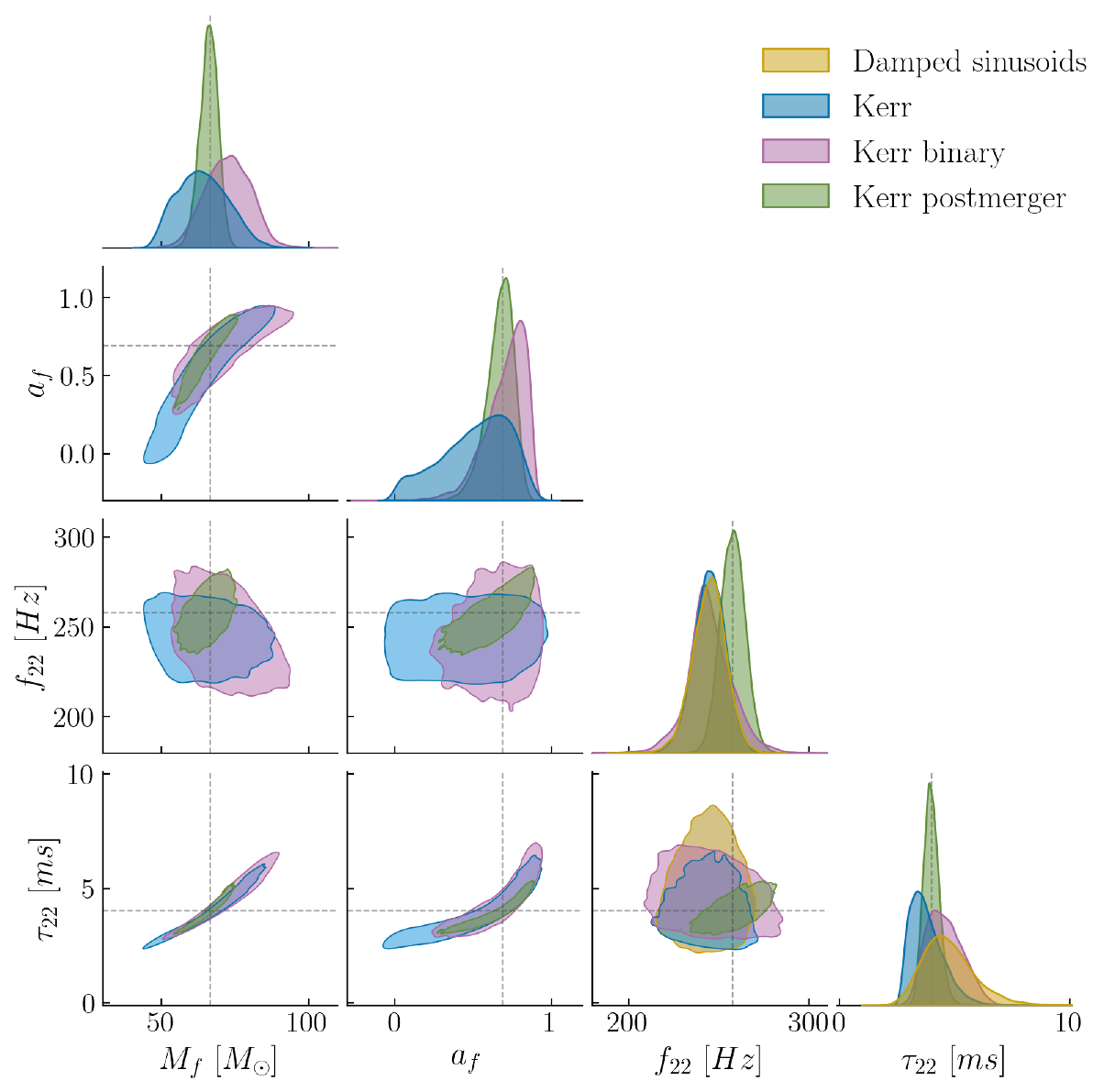}
    \caption{Model hierarchy comparison on the measurement of the remnant parameters $M_f, a_f$ and the $(2,2,0)$ mode frequency and damping time, $f_{22},\tau_{22}$.
    Different colors denote the posterior distributions for the \texttt{Damped sinusoids} (\textit{yellow}), \texttt{Kerr} (\textit{blue}), \texttt{KerrBinary} (\textit{pink}), and \texttt{KerrPostmerger} (\textit{green}) models obtained using \texttt{pyRing}~\cite{pyRing}.
    The data consist of the $(\ell,m)=(2,2)$ mode of the numerical simulation $\text{SXS:BBH:}0305$ from the SXS catalog~\cite{Lindblom:2005qh, Boyle:2019kee}, and the dashed lines correspond to the true injected values.
    Consistent with each template's validity regime, the analysis starting times are $10M_f$ after the IMR waveform peak time for the \texttt{Damped sinusoids} and \texttt{Kerr} templates, $15M_f$ for \texttt{KerrBinary}, and $0M_f$ for \texttt{KerrPostmerger}.
    The recovered SNR with the \texttt{Kerr} (\texttt{KerrPostmerger}) model is $\sim 10$ ($\sim 14$, respectively).
    }
    \label{fig:corner_models_hierarchy} 
    \end{center}
\end{figure*}

\begin{table}[h!]
\centering
    \begin{tabular}{p{3.8cm}p{2.8cm}p{3.0cm}p{4.4cm}}
    \toprule
    \textbf{Models} & \textbf{\shortstack[l]{Free intrinsic \\ parameters}} & \textbf{Assumptions} & \textbf{Fits} \\
    \midrule
    \texttt{Damped sinusoids} & \footnotesize{$f_{j}, \tau_{j}, A_{j}, \phi_{j}$} & & \small{Used in Ref.~\cite{LIGOScientific:2016lio, Carullo:2019flw}}. \\
    \arrayrulecolor{gray}\cmidrule(lr){1-4}
    \texttt{Kerr} \newline (Kerr-informed $f, \tau$) & \footnotesize{$M_f, a_f, A_{\ell m n}, \phi_{\ell m n}$} & \footnotesize{$f_{\ell m n}(M_f,a_f)$, $\tau_{\ell m n}(M_f,a_f)$} & \small{Generic Kerr perturbations~\cite{Berti:2005ys}.} \newline \small{Used, for example, in Refs.~\cite{LIGOScientific:2020tif, LIGOScientific:2021sio, Carullo:2019flw, Isi:2019aib, Capano:2021etf, Ma:2023vvr, Siegel:2023lxl, Crisostomi:2023tle, Pacilio:2024qcq, Wang:2024jlz}.} \\
    \arrayrulecolor{gray}\cmidrule(lr){1-4}
    \texttt{KerrBinary} \newline (NR-informed $f, \tau, A, \phi$) & \footnotesize{$m_{1,2}, \mathbf{\chi}_{1,2}$} & \footnotesize{$f_{\ell m n}(M_f, a_f)$, $\tau_{\ell m n}(M_f, a_f)$, $A_{\ell m n}(m_{1,2}, \mathbf{\chi}_{1,2})$, $\phi_{\ell m n}(m_{1,2}, \mathbf{\chi}_{1,2})$} & \small{BH mergers aligned-spins and quasi-circular~\cite{London:2018gaq, JimenezForteza:2020cve, Cheung:2023vki} or nonspinning and eccentric~\cite{Carullo:2024smg}}. \newline \small{Used in Refs.~\cite{LIGOScientific:2020tif, LIGOScientific:2021sio, Carullo:2018sfu}}. \\
    \arrayrulecolor{gray}\cmidrule(lr){1-4}
    \texttt{KerrPostmerger} \newline (NR-informed $f, \tau, A(t), \phi(t)$) & \footnotesize{$m_{1,2}, \mathbf{\chi}_{1,2}$} & \footnotesize{$f_{\ell m n}(M_f, a_f)$, $\tau_{\ell m n}(M_f, a_f)$, $A_{\ell m n}(t; m_{1,2}, \mathbf{\chi}_{1,2})$, $\phi_{\ell m n}(t; m_{1,2}, \mathbf{\chi}_{1,2})$} & \small{BH mergers precessing and quasi-circular~\cite{Pratten:2020ceb, Gamba:2021ydi, Pompili:2023tna}. \newline \small{Used in Ref.~\cite{Gennari:2023gmx} for aligned-spins binaries.}} \\
    \arrayrulecolor{black}\bottomrule
    \end{tabular}
\caption{Summary of ringdown parameterizations.}
\label{tab:models}
\end{table}

\noindent
\textit{Damped sinusoids template }
The most agnostic model, labeled \texttt{Damped sinusoids}, consists of a superposition of exponentially damped harmonic oscillations
\begin{equation} \label{eq:DS_waveform}
    h_{+} -i h_{\times} = \sum_{j}^{+\infty}\, h_{j}, \quad h_{j} \equiv A_{j}(t) \,e^{-(t-t_0)/\tau_{j}}\,e^{-2\pi if_{j}(t-t_0) + i\phi_{j}}.
\end{equation}
This choice can be motivated either theoretically~\cite{Vishveshwara:1968ksg, Cunningham:1978zfa}, being related to the ``open'' boundary conditions of BH perturbations at both the horizon and infinity (see Sections~\ref{sec_21} and \ref{subsec:echoes_DS}), or observationally, from the exponentially decaying behavior of measured GW signals~\cite{LIGOScientific:2016lio, LIGOScientific:2020tif, LIGOScientific:2021sio}.
If we assume the analysis takes place during the ``stationary'' relaxation regime~\cite{Buoninfante:2024oxl}, i.e., late enough that the QNM amplitudes and the background mass and spin have stabilized, but early enough that tail effects do not play a significant role, we can further assume constant QNM amplitudes ($A_{j}(t) = A_{j}$).
In this parameterization, the frequencies, damping times, initial amplitudes and phases $\{f_{j}, \tau_{j}, A_{j}, \phi_{j}\}$ of the damped sinusoids are the only variables needed to describe the signal, and represent the free parameters measured in the data analysis inference.
Here, and below when discussing constant-amplitude QNM superpositions, we will be assuming that we are in a regime such that all modes have already been excited, so that we can ignore time delays between modes (see Section~\ref{sec:waveforms}); this is not the case for phenomenological models with time-dependent amplitudes, which include such delays.

This model captures the behavior of linear Kerr perturbations in Eq.~\eqref{eq:RD_waveform}, but does not contain any information on the relationships between different modes.
For the same reason, note that the sum in Eq.~\eqref{eq:DS_waveform} is on generic modes $j$, which are not projected onto spin-weighted spheroidal harmonics as in Eq.~\eqref{eq:RD_waveform}.
Thus, the $\{f_{j}, \tau_{j}, A_{j}, \phi_{j}\}$ parameterization allows us to fit ringdown signals that are not necessarily Kerr perturbations, although it still includes the Kerr spectrum as a special case.
Such parameterizations can capture scenarios where the set of excited modes is not easily predictable, e.g. because of environmental effects (see Section~\ref{sec:environments}).

\noindent
\textit{Kerr template }
As described in Section~\ref{sec_21}, the Kerr QNM spectrum in GR is determined uniquely by the BH mass and spin, $\{M,a\}$, and therefore the modes' complex frequencies are not independent under the Kerr assumption.
This allows us to construct an upgraded class of models, labeled \texttt{Kerr}, by assuming a functional form between the QNMs and the BH parameters, $\{f_{\ell m n}(M,a),\tau_{\ell m n}(M,a)\}$, leading to a different set of model parameters $\{M,a,A_{\ell m n},\phi_{\ell m n}\}$.
Note that, compared to the \texttt{Damped sinusoids} parameterization, now the modes are labeled by the angular indices $(\ell,m)$ and by the overtone index $n$, as in Eq.~\eqref{eq:RD_waveform}.
The \texttt{Kerr} model assumes that the data correspond to GW linear perturbations radiated by a Kerr background spacetime.

This parameterization is sufficiently agnostic while including the basic features expected from BH perturbations and allowing us to extract a considerable amount of information on BH perturbations, hence it is the most commonly adopted ringdown parameterization~\cite{LIGOScientific:2020tif, LIGOScientific:2021sio, Carullo:2019flw, Isi:2019aib, Capano:2021etf, Ma:2023vvr, Siegel:2023lxl, Crisostomi:2023tle, Pacilio:2024qcq, Wang:2024jlz}.
As astrophysical BHs are expected to rapidly shed off any electromagnetic charge, the Kerr solution is often assumed to be the only astrophysically relevant one, however the same parameterization has been applied to Kerr-Newman BHs~\cite{Carullo:2021oxn,Gu:2023eaa} or beyond-GR theories (see Section~\ref{subsec:TGR_theory-specific} and references therein).

\noindent
\textit{Numerical relativity informed late-ringdown models }
The $\{M,a,A_{\ell m n},\phi_{\ell m n}\}$ parameterization contains GR information only in the QNM spectrum, leaving the mode amplitudes and phases as free parameters.
In fact, if the complex frequencies are intimately tied to the Kerr spacetime by the uniqueness theorems~\cite{Loutrel:2020wbw, Penrose1982SOMEUP, Ginzburg, Zeldovich, Israel:1967wq, Carter:1971zc, Hawking:1971vc, Robinson:1975bv, Bunting, Mazur:1982db}, the excitation amplitudes and phases of the different modes are a manifestation of the initial conditions of the perturbations (see Section~\ref{subsec:Kerr_amplitudes}), meaning they depend on the specific physical process that leads to a perturbed BH.
Hence, any dynamics leading to the formation of a perturbed BH is imprinted in the BH spectrum, providing initial conditions for the Kerr perturbations.
Under the assumption that the observed ringdown signals stem from the merger of stellar-mass BH binaries~\cite{LIGOScientific:2016aoc, LIGOScientific:2020ibl, LIGOScientific:2018mvr, KAGRA:2021vkt}, the complex amplitudes can be modeled through NR simulations~\cite{Kamaretsos:2011um}, leading to a class of models referred to as \texttt{KerrBinary}.
Incorporating this mapping in the ringdown waveforms is a nontrivial task (see Section~\ref{sec:waveforms}), and several works have developed such fits using a wealth of numerical techniques~\cite{Kamaretsos:2011um, Kamaretsos:2012bs, London:2014cma, London:2018gaq, Bhagwat:2019bwv, Bhagwat:2019dtm, JimenezForteza:2020cve, Ota:2021ypb, Cheung:2022rbm, Hughes:2019zmt, Lim:2019xrb, JimenezForteza:2020cve, Lim:2022veo, Zhu:2023fnf, Capano:2021etf, Carullo:2024smg}.
The outcome of these investigations are parameterizations for the mode amplitudes and phases as a function of progenitor parameters, such as the masses and spins $\{A_{\ell m n}(m_{1,2}, \mathbf{\chi}_{1,2}), \phi_{\ell m n}(m_{1,2}, \mathbf{\chi}_{1,2})\}$ or orbital parameters $\{A_{\ell m n}(E_0, J_0), \phi_{\ell m n}(E_0, J_0)\}$.

At the expense of assuming the validity of GR, imposing these functional relationship significantly decreases the number of free parameters and the related prior volume (increasingly so as more modes are added), improving constraints on the BH parameters.
Compared to a simpler $\{M,a,A_{\ell m n}\}$ parameterization, this can have a strong impact on the search for GR deviations.
For example, an NR-informed amplitude parameterization breaks the degeneracy between $\{M,a\}$ and the spectral deviation parameters $\{\delta f, \delta\tau\}$ that may be present for a single QNM, opening the possibility of measuring deviations even with a single resolvable mode~\cite{LIGOScientific:2021sio, Gennari:2023gmx}.
Instead of completely fixing amplitudes and phases, an ``intermediate'' constraint can be achieved by imposing an upper bound on the amplitude ratios (typically constraining $A_{\ell m n}/A_{220}$ to be smaller than a certain value), reflecting the relative excitation of different modes over regions of the progenitor parameter space extracted from NR~\cite{Capano:2021etf, Forteza:2022tgq}.

\noindent
\textit{NR-informed postmerger models }
As discussed in Sections~\ref{subsec:WF_modeling_dependence} and~\ref{subsec:RD_systematics}, the identification of the regime in which the signal can be described by a superposition of constant-amplitude QNMs is one of the main systematics in current ringdown data analysis.
In practice, any ringdown model made up of a sum of constant-amplitude damped sinusoids is a meaningful description of the data only at sufficiently late times after the merger, depending on the system parameters and on the SNR of the signal~\cite{Buonanno:2006ui, Berti:2007fi, London:2014cma, Bhagwat:2017tkm, Carullo:2018sfu, London:2018gaq}.
At the root of this problem is our limited understanding of the merger regime in the two-body problem, and of the related dynamical excitation of QNMs (see Section~\ref{sec:waveforms}).
Hence, phenomenological descriptions are currently used to effectively capture nonlinearities and transient effects, that produce time-dependent amplitudes and phases $\{A_{\ell m n}(t),\phi_{\ell m n}(t)\}$ in this part of the signal (see Section~\ref{sec:effective-one-body}).
This class of models is referred to as \texttt{KerrPostmerger} .

Such parameterizations share the same set of free parameters as stationary QNM models with NR-informed amplitudes $\{m_{1,2}, \mathbf{\chi}_{1,2}\}$, but can be applied up to the peak of the signal.
They can thus capture larger SNR, obtaining more information from the data~\cite{Gennari:2023gmx} (see Fig.~\ref{fig:corner_models_hierarchy}).
While a complete understanding of amplitude excitations for precessing systems is still missing~\cite{Hamilton:2023znn, Zhu:2023fnf}, current state-of-the-art IMR waveforms account for spin precession using phenomenological time-dependent rotations (e.g.~\cite{Pratten:2020ceb, Gamba:2021ydi, Pompili:2023tna}).
Instead of relying on direct NR calibration, other intermediate approaches have been developed to phenomenologically capture the early postmerger phase, exploiting the use of overtones~\cite{LIGOScientific:2020tif, LIGOScientific:2021sio, Isi:2019aib, Giesler:2019uxc} or wavelets~\cite{Finch:2021qph, Finch:2022ynt}.
This gain comes at the cost of losing physical interpretability, often required when performing searches for new physics.

Taking this idea further, full IMR templates can be used to test ringdown properties and provide accurate models for sensitive tests of GR~\cite{Brito:2018rfr, LIGOScientific:2020tif, LIGOScientific:2021sio, Ghosh:2021mrv, Maggio:2022hre,Pompili:2025cdc}, at the cost of being more susceptible to waveform systematics~\cite{Maggio:2022hre, Gupta:2024gun}.

\noindent
\textit{Beyond-GR parameterizations }
All the models and parameterizations discussed above are used to characterize the resolvable modes in ringdown signals.
Much of the interest in this analysis is due to the possibility of testing the nature of the relaxing object and our understanding of gravity in the regime where gravitational fields are strong and dynamical.
The parameterizations employed in searches of new physics are the subject of Section~\ref{subsec:deviations_current} below.

\subsubsection{Systematics in ringdown data analysis}\label{subsec:RD_systematics}

In BH spectroscopy we are interested in isolating a specific portion (the ringdown) of a larger signal (typically the full IMR signal). This requires dedicated data analysis techniques, and relies on models describing only portions of the entire signal.
Given a data analysis framework and a model parameterization in terms of a sum of QNM, one has to make careful choices for the starting time of the analysis and data conditioning in order to ensure that the model is applicable and to avoid significant systematic errors.
In this section, we review these systematic effects and some of the strategies developed to mitigate them.

\noindent
\textit{Starting time: regime of validity of the model}
Selecting a ringdown analysis starting time ($t_{\rm start}$) is a compromise between maximizing the SNR (which requires early values of $t_{\rm start}$) and ensuring that the QNM model is applied within the domain of validity of BH perturbation theory (which requires late values of $t_{\rm start}$).

In the case of noisy data, it might seem that this amounts to ensuring any systematic error from applying the model too early is less than the statistical error in the analysis.
This criterion is correct when considering phenomenological postmerger models, such as \texttt{KerrPostmerger} or \texttt{pSEOB}.
In these cases, the situation is similar to standard parameter estimation for IMR signals: a waveform model may be imperfect or incomplete, but ``good enough'' for the analysis given the available SNR~\cite{Lindblom:2008cm}.
However, when considering more standard ``BH spectroscopy'' templates made of sums of damped sinusoids (as in the \texttt{Damped sinusoids} or \texttt{Kerr} models), this criterion is no longer sufficient.
The flexibility of damped-sinusoid sums implies that such models may still provide a good fit to the data even outside of the perturbative regime, making them prone to overfitting (see Section~\ref{subsec:overtones}).
For example, a fit of GW150914 with multiple damped sinusoids before the peak of the strain would favor a two-mode model over a one-mode model.
This should not be interpreted as evidence of two modes being present in the data, because pre-peak data should not be physically understood as a simple superposition of multiple QNMs. 
These considerations are related to the choice of the QNM content in the model.
As discussed in Section~\ref{subsec:overtones}, when one includes multiple QNMs (most notably overtones),  ringdown models can effectively fit the data at earlier times, and yield remnant parameter estimates compatible with IMR values.
However, the interpretation of these results in terms of the physical QNMs being excited is questionable because of overfitting.
For these reasons, overfitting complicates the usage of consistency between IMR and ringdown estimates of remnant parameters to determine the regime of validity of a ringdown model~\cite{Baibhav:2023clw}.

This issue is an inherent feature of the assumed signal model, and it cannot be resolved by data analysis studies.
The regime of validity of QNM superpositions is something that must be predetermined based on a theoretical understanding of the system, which can then inform observational analyses.
Based on NR studies and criteria of parametric stability (see Section~\ref{sec:waveforms}), the validity regime of a constant-amplitude $(2,2)$ model including overtones with $n\leq 2$ was found to start around $\sim 8 \, M_f$ past the signal peak~\cite{Mitman:2025hgy}, although this is highly dependent on the binary parameter space and mode content.
The expected level of systematic error obtained from NR studies can then be used to decide whether any given model is sufficient at the observed SNR (e.g.,~\cite{Giesler:2019uxc, Gennari:2023gmx, Cheung:2023vki}).
A systematic and accurate determination of the regime of validity of ringdown models as a function of the parameter space and of the  SNR is among the most important open problems in BH spectroscopy.

\noindent
\textit{Starting time: inspiral-merger-ringdown inputs}
Even assuming a theoretical prediction for an appropriate $t_{\rm start} = t_{\rm peak} + X \cdot M_f$ at which a model should be applied, when analyzing real data both $t_{\rm peak}$ and $M_f$ carry some measurement uncertainty.
While $M_f$ can be determined by the ringdown analysis itself, the uncertainty in $t_{\rm peak}$ has nontrivial implications.
In the spirit of an agnostic methodology, one might hope to estimate a suitably defined ``ringdown start time'' \textit{directly} from the data, marginalizing over its uncertainty.
This task has proved to be challenging.

As discussed in Section~\ref{subsec:bayesian_ringdown}, the main motivation for switching to the TD is to remove any pre-ringdown data from the analysis.
This means that, without assuming a model for the pre-peak data, the choice of the ringdown starting time determines by construction the exact data stretch to be analyzed. 
Varying the data used within a given analysis would introduce additional complications, as discussed below.
To avoid this, $t_{\rm start}$ needs to be fixed to a predetermined value compatible with a probability distribution $p(t_{\rm peak})$ obtained from a previous analysis of the full signal~\cite{Isi:2021iql}.
Given data from multiple detectors, the sky location of the signal  (which determines the time shift between the data from each detector) also needs to be predetermined.

This uncertainty can have a sizable impact on QNM detection (see Section~\ref{subsec:current_observations}).
Consequently, it is standard practice to perform a ``grid-coverage'' of $p(t_{\rm peak})$, performing multiple analyses over a series of fixed starting times within the support of $p(t_{\rm peak})$.
Once this set of grid analyses has concluded, one may attempt to combine the result into a sort of ``marginalized'' final estimate of the quantities of interest, now incorporating the timing uncertainty. 
The precise method to accomplish this marginalization is again not trivial, and discussed below (see the discussion of \textit{Detection criteria} below).

Typically one uses probability distributions from an IMR analysis to estimate these uncertainties.
In these cases, $p(t_{\rm peak}) = p(t_{\rm peak} | D_{\rm IMR}, \mathcal{I}_{\rm IMR})$ (with $\mathcal{I}_{\rm IMR}$ incorporating the IMR model assumptions, and $\vdata = D_{\rm IMR}$ the full-signal data).
This can represent an additional source of systematics.
If IMR models are not accurate enough, inaccuracies will propagate into the ringdown analysis, possibly spoiling tests of gravity. 
One way to account for such uncertainty could be to extend the grid to cover the range of all $p(t_{\rm peak} | D_{\rm IMR}, \mathcal{I}^i_{\rm IMR})$, with $i$ ranging over different IMR models.
Alternatively, one can ``marginalize'' over the $p(t_{\rm peak})$ distribution obtained from unmodeled full-signal reconstructions~\cite{Cornish:2014kda, Drago:2020kic}.
Currently, this full-signal systematic uncertainty is typically neglected, and it should be accounted for in future high-precision measurements.

\noindent
\textit{Starting time: inference }
Alternative data-analysis methods have been proposed to avoid fixing the starting time and sky position.
Despite initial attempts~\cite{Carullo:2019flw}, it has now become clear that marginalizing over the start time requires modeling the data before ringdown, and the resulting posterior on the starting time will be a result of the interplay between the ringdown and pre-ringdown models.
Some hybrid approaches attempt to model the entirety of the signal.
In Refs.~\cite{Finch:2021qph, Finch:2022ynt}, the pre-ringdown signal is modeled as a flexible sum of sine-Gaussian wavelets, allowing the analysis to return to the FD and introducing the ringdown starting time as a free parameter in the model.
Within this framework one could in principle infer the earliest time at which a ringdown model fits the data, although in practice this has proven to yield uninformative results (see Fig.~10 in Ref.~\cite{Finch:2022ynt}). The authors instead opt to use more informative priors on the ringdown starting time (either a set of narrow Gaussians centered on a discrete set of time samples, or the posterior on $t_\mathrm{peak}$ from an IMR analysis).
In the same spirit, Ref.~\cite{Correia:2023ipz} proposes a different approach based on gating and in-painting (see Section~\ref{subsub:gating}), by which it is possible to sharply divide the signal in two independent parts separated by the starting time and marginalize over the starting time, as well as the location of the source in the sky~\cite{Correia:2023bfn}.
Even if the gate time varies as a free parameter in the analysis (i.e., their ringdown starting time), the authors find that the pre-ringdown signal must be modeled, or the ringdown starting time would correspond to very late times (essentially removing all of the signal from the likelihood).
To obtain a ringdown inference as independent as possible from the pre-ringdown data, the two portions only share the sky position and transition time.
When the ringdown starting time is varied, one must carefully disentangle the ringdown inference results from the reference time at which QNM amplitudes are defined in the signal model $t_{\rm ref}$ (see Eq.~\eqref{eq:rdown_complete}). This complicates the comparison of amplitude posteriors across different analyses.

\noindent
\textit{Detection criteria }
A key goal of BH spectroscopy is to detect multiple QNMs.
A robust detection criterion could use Bayes' theorem. Define the hypotheses $\mathcal{H}_n$ and $\mathcal{H}_m$ as the statements ``$n$ (or $m$) detectable modes are present in the ringdown signal.'' Then the odds ratio
\begin{equation} \label{eq:odds_multi}
    \mathcal{O}_{n,m} \equiv \frac{p(\mathcal{H}_n|\mathbf{d},I)}{p(\mathcal{H}_m|\mathbf{d},I)} = \frac{p(\mathcal{H}_n|I)}{p(\mathcal{H}_m|I)} \frac{p(\mathbf{d}|\mathcal{H}_n,I)}{p(\mathbf{d}|\mathcal{H}_m,I)} \equiv \frac{p(\mathcal{H}_n|I)}{p(\mathcal{H}_m|I)} \frac{\mathcal{Z}_n}{\mathcal{Z}_m} \equiv \frac{p(\mathcal{H}_n|I)}{p(\mathcal{H}_m|I)}\: \mathcal{B}_{n,m}
\end{equation}
quantifies the relative probability that $\mathcal{H}_n$ or $\mathcal{H}_m$ are true, given their respective prior and background assumptions, as long as the truth is encompassed by one of the models under consideration~\cite{10.1214/13-BA826}.
If $\mathcal{O}_{n,m} >1$, then $\mathcal{H}_n$ is more probable, and vice versa.
If the two competing hypotheses are considered equally likely a priori, then
$p(\mathcal{H}_n|I) = p(\mathcal{H}_m|I)$, and the odds ratio coincides with the
\textit{Bayes factor} (BF), $\mathcal{B}_{n,m}$.
Whenever new parameters are added to a ringdown model, the dimensionality of the parameter space increases. Even if the ability to fit the data improves, the prior probability density for the model diminishes, naturally penalizing the model by lowering its evidence (an Occam penalty)~\cite{Jefferys:1992, Jaynes:2003}.
Hence, unless a mode has contributions to the signal that are substantial relative to the dominant $(\ell, m, n) = (2,2,0)$ mode, the increase in prior volume always tends to disfavor multi-modal templates.
This effect naturally favors more conservative results and reduces our ability to resolve multiple modes, especially at low SNR.

One way to mitigate the problem is by assuming NR-informed complex amplitudes parameterized as a function of the progenitor parameters (see Section~\ref{subsec:ringdown_models}), so multiple modes can be added without increasing the dimensionality of the problem.
Such parameterizations, however, amount to assuming in the background information $I$ that GR is correctly predicting the mode amplitudes~\cite{LIGOScientific:2021sio, Gennari:2023gmx, Maggio:2022hre}.
This still enables ``weaker'' spectroscopic tests by allowing the spectrum to deviate from GR (see Section~\ref{subsec:deviations_current}), but such tests are less agnostic than more traditional proposals in which the mode amplitudes are left free to vary.

A complication with using BFs alone to assess the presence of a mode arises whenever one is trying to decide between \emph{nested models}, i.e., models that encompass one another for a given choice of the parameters.
This is the case when we consider two ringdown models that differ because of the addition of an extra mode in the template: the more generic model reduces to the narrower one if the additional mode's amplitude is equal to zero.
In this context, whether the BF favors the addition of a mode typically depends on the chosen prior for all parameters, and in particular for the amplitudes. Crucially, this dependence extends to regions of parameter space not constrained by the likelihood~\cite{Isi:2022cii,Isi:2022mhy}.
For ringdown signals, for example, this means that the amount of support for an additional mode can be decreased arbitrarily by enlarging the maximum allowed value of its amplitude.
The lack of physical guidance in setting prior expectations for the mode amplitudes exacerbates this issue.

Given these difficulties in calculating the evidence, some authors reject the idea of using BFs as a detection statistic, and instead prefer to rely on other strategies~\cite{Clarke:2024lwi, Isi:2019aib, Isi:2022mhy, Isi:2023nif, Capano:2021etf, Siegel:2023lxl}.

One approach is to assess detection significance based on the degree to which a nonzero mode amplitude is supported by the posterior without direct reference to the prior, e.g., by computing the ratio of the posterior mean or median to its standard deviation or some other measure of the spread, or by computing the credible level of zero (valid also for asymmetric posteriors).
In fact, the BF discussed above is intimately related to the posterior support at zero, as it can be shown to be exactly the ratio of the posterior to the prior at that point in parameter space~\cite{Dickey1971TheWL}.
Additionally, the evolution of the amplitude posterior recovered for different fitting times can be used as a safeguard against mismodeling. For example, if the data being fitted are not well described by any of the QNM models under consideration, the posterior may nonetheless favor a nonzero amplitude at a given starting time, but it is less likely to also evolve consistently with the expected QNM decay rate (since the amplitude posterior can be pushed away from zero due to mismodeling systematics, so can the BF spuriously favor the addition of a mode for the same reason).

Another approach focuses on establishing whether the model is a good description of the data without explicitly referencing an alternative model.
This can be achieved by studying the statistics of residuals left after subtracting templates drawn from the posterior.
That is the intent of leave-one-out cross-validation (LOOCV), which, being largely independent of priors, could provide more reliable statistical prescriptions when dealing with noninformative priors~\cite{Siegel:2023lxl,2015arXiv150704544V}.

Because of these subtleties, the question of mode detectability is still partly open, and deeply intertwined with the systematics that affect ringdown data analysis.
Future ringdown signals with higher SNR will mitigate these difficulties.

Other authors have argued that while the above concerns are in principle valid, in practice the dependence of the BFs on the priors is rather weak, and can be easily overcome by running an analysis with wide priors and subsequently ``reweighting'' the BF to exclude the region without likelihood support~\cite{Cotesta:2022pci, Carullo:2023gtf}.
BFs have been employed by the LVK collaboration in the spectroscopic search of the O1, O2, O3a and O3b datasets~\cite{LIGOScientific:2020tif, LIGOScientific:2021sio}.

This would normally conclude the discussion of mode detectability in a standard scenario in which the entire GW signal could be well-described by a first-principles QNM model (either with constant or nonconstant amplitudes).
Unfortunately, there are additional complications related to ringdown data analysis formulation and to our incomplete understanding of how QNMs are dynamically generated.
These two elements imply that some care is required when comparing BFs across different models.

The major conceptual complication in this regard is inherited from the TD truncation.
As discussed in Section~\ref{subsec:RD_systematics}, given multiple runs performed over a series of starting times $t_{\rm start}$, the evidence for each run is itself a function of $t_{\rm start}$, $\mathcal{Z} = \mathcal{Z}(t_{\rm start})$, since the chunk of data analyzed varies for different choices of $t_{\rm start}$.
For this reason, the comparison of analyses at different starting times is statistically ill-defined, and models can be meaningfully compared only for the same choice of $t_{\rm start}$.
Unfortunately $t_{\rm start}$ itself is uncertain, with an associated probability distribution $p(t_{\rm start})$.
How can one produce a combined evidence that takes into account this uncertainty?
Assume that $p(t_{\rm start})$ is bounded within $[t_{\rm min}, t_{\rm max}]$, and that we have repeated the analysis $\mathcal{N}$ times for certain values of $t_i$ in this range, with $i=0,...,\mathcal{N}$.
A simple guess would be to define an average evidence over the starting times, $\mathcal{Z} \equiv \frac{p(t_i) \cdot Z(t_i)}{\mathcal{N}}$, and impose a detection threshold on this quantity.
However, $\mathcal{Z}$ inherits the ill-defined properties discussed before, since it would combine evidences computed on different data portions.
Moreover, since the evidence decays exponentially with time, earlier times will bear significant weight in this criterion.
Ringdown models are known to be subject to \textit{increasing} systematics for earlier times, hence one should conversely try to penalize them.

This problem does not seem to admit a simple solution, and there is no completely satisfactory proposal to deal with time-truncated ringdown signals.
Assuming that a mode is considered detected for $\mathcal{Z} > X_{\rm det}$, a conservative criterion often adopted is simply to require that this condition holds over the entire support of $p(t_{\rm start})$, so that $\mathcal{Z}(t_i) > X_{\rm det}$, with $i=0, ..., \mathcal{N}$.
Here, we have discussed evidence for simplicity, but the same argument can be repeated for any other quantity of interest (e.g., the mode amplitude).
New proposals are needed to improve on this over-conservative criterion.

\noindent
\textit{Noise systematics}
The noise in real detectors, which is often difficult to model, can significantly affect the calculation of the evidence and criteria for mode detection, or even induce false detections for tests of gravity~\cite{Gupta:2024gun, Kwok:2021zny}. 
This is true of any signal, but even more so for ringdowns, which typically have low SNR.

Currently, the likelihood applied to ringdown data analysis is derived assuming stationary and Gaussian noise (see Section~\ref{subsec:bayesian_ringdown}).
These approximations are often violated in real data~\cite{LIGOScientific:2019hgc}.
Given the flexibility of damped sinusoid sums, additional modes in the ringdown template can, in principle, improve the fit by capturing nonstationarities, increasing the probability of false detections when the model contains multiple modes.
The inclusion of multiple detectors characterized by independent noise processes helps to alleviate the impact of specific noise realizations that necessarily affect each signal.

We can assess the impact of noise on a given event in two steps, by adding a simulated signal (``injection'') with parameters close to those expected (e.g. estimated from some preferred analysis) into: (i) simulated Gaussian noise; and (ii) real noise around the GW event.
Repeating step (ii) by injecting at different times, we can ``sample'' the true detector noise process and quantify the impact of real noise on the results (see e.g.  Refs.~\cite{LIGOScientific:2021sio,Cotesta:2022pci, Maggio:2022hre}).
These studies can rapidly become computationally intensive because of high noise variability, uncertainty in the parameters of the detected signal, and the necessity to include different QNM combinations and starting times~\cite{LIGOScientific:2021sio}.

\subsection{Current status of observations}
\label{subsec:current_observations}

\vspace{-.1cm}

\noindent \textit{Initial contributors: Caneva, Carullo, Farr, Finch, Westerweck}

\begin{figure*}[t!]
    \begin{center}
    \includegraphics[width=0.8\textwidth]{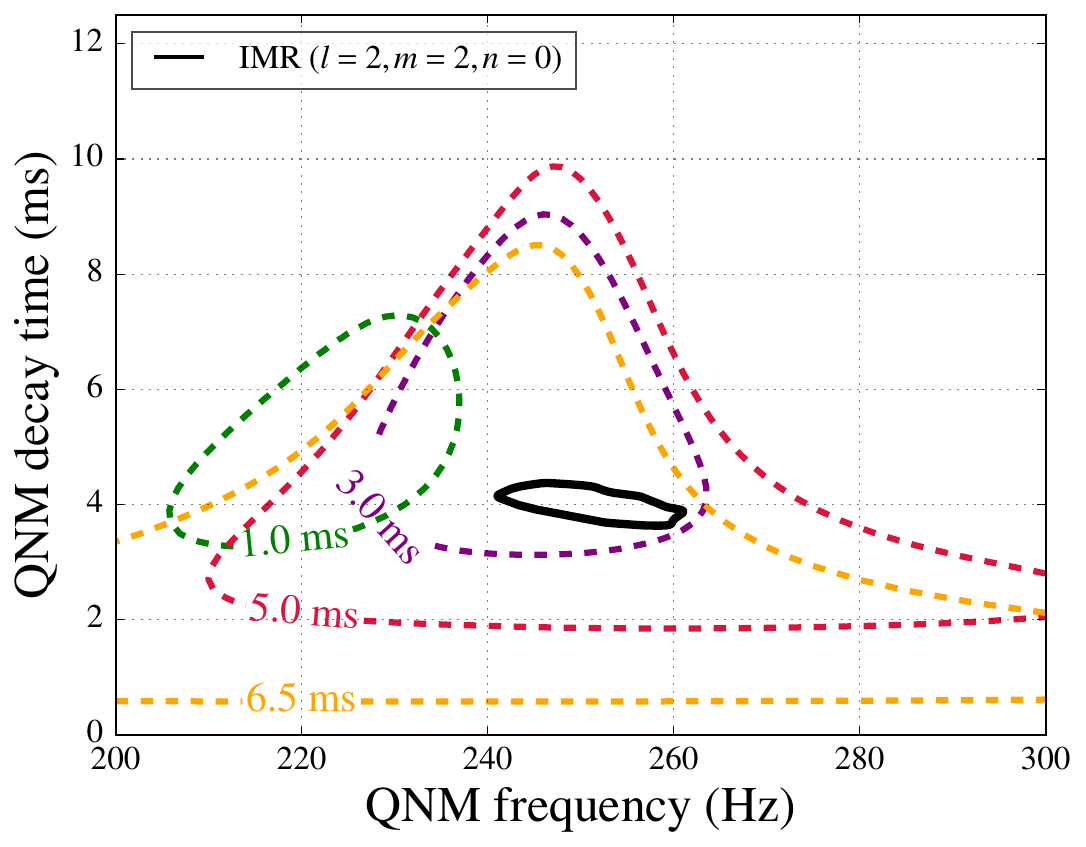}
    \caption{Two-dimensional posterior distribution of the frequency and damping time of a single damped sinusoids compared against GW150914 data, starting at different times (dashed lines). 
    As expected, starting at late enough times (roughly corresponding to $t_{\rm peak} + 10 \, M_f$), the result are in agreement with the predictions from the full IMR analysis (solid line).
    Figure taken from~\cite{LIGOScientific:2016lio}.}
    \label{fig:LVK_GW150914} 
    \end{center}
\end{figure*}

\vspace{.2cm}

\subsubsection{Fundamental quadrupolar mode observations}
\label{sec:fundametal_l2_obs}

The observation of a coincident transient signal in the two LIGO interferometers on September 14, 2015~\cite{LIGOScientific:2016aoc} marked not only the historical discovery of GWs and the first observation of a binary BH merger, but also the first detection of the dominant QNM frequency of a BH. 
The signal power within the detector band was sufficiently large to isolate the post-peak signal and extract the dominant frequency and damping time in the ringdown waveform.
Given the remnant BH parameters extracted under the assumption that GR correctly describes the entire signal, the expected frequency and damping time of the fundamental $(\ell,m,n)=(2,2,0)$ mode correspond to $f_{220} \simeq 250\,\mathrm{Hz}, \, \tau_{220} \simeq 4\,\mathrm{ms}$.
The LVK collaboration compared this expectation against a measurement obtained under the simplest possible assumption: a single damped sinusoid post-peak~\cite{LIGOScientific:2016lio,Prix:T1500618}.
Given the statistical uncertainty on the signal peak location and the uncertainty on the starting time of a single-QNM ringdown, the analysis was repeated at multiple starting times past the waveform peak.
For the system parameters under consideration, an analysis assuming a single QNM was predicted to yield an estimate in agreement with the fundamental mode around a timescale of $t_{\rm start} \simeq 3 \,\mathrm{ms} \simeq 10 M$, with $M$ the total mass of the binary system expressed in seconds units.
The analysis result is in agreement with this expectation, as shown in Fig.~\ref{fig:LVK_GW150914}. 

Despite its consistency with the predictions of GR, the LVK ringdown-only
analysis of GW150914 was based on a FD technique that did not apply a window or gating and in-painting filters to the data~\cite{Prix:T1500618}.
As discussed in the previous subsection, this approach can lead to biased results.
This realization is what sparked the development of TD and nonstandard FD methodologies capable of isolating a specific signal portion.
These were applied to a re-analysis of the presence of the fundamental QNM in GW150914 at late times~\cite{LIGOScientific:2020tif}.
This confirmed the robustness of the detection of the dominant mode against differences in details of the data analysis (such as PSD estimation, data conditioning, etc.) with high statistical confidence, as quantified e.g. by the BF comparing the hypothesis for the presence of a damped sinusoid signal hypothesis vs. pure noise, $\mathcal{B}^{s}_{n} \simeq e^{14.6}$~\cite{LIGOScientific:2020tif}.
Additional searches found no evidence for multiple fundamental modes in the signal beyond the quadrupolar harmonic~\cite{Carullo:2019flw,LIGOScientific:2020tif}.\\

\noindent
\textit{GWTC-3 } GW150914 was the first GW event shown to be consistent with a BBH signal as predicted by GR. Since then, the LVK collaboration has validated the predictions of GR in the strong-field regime by performing a variety of tests on an expanding collection of GW events~\cite{LIGOScientific:2021sio, LIGOScientific:2019fpa, LIGOScientific:2020tif}. 
Over the years, this list of events has been extended with additional binary merger detections during the first, second, and third observing runs (O1–O3).
The third GWTC~\cite{KAGRA:2021vkt} includes 90 compact binary coalescence events; however, not all events are used for testing GR (TGR). 
The TGR analyses apply stringent event selection criteria to ensure the astrophysical origin of the candidate event.
Only events detected by at least two interferometers that would be generated by a random noise fluctuations in less than a thousand years are considered. 
Applying these conditions to GWTC-3 leads to a subset of 48 events passing the FAR threshold and the two-detector detection criterion. Once this subset is established, additional criteria are imposed depending on the specific strategy of each GR test.
Probing the nature of the merger remnant is a crucial part of the TGR studies, which the LVK achieves using two complementary approaches: the TD analysis \texttt{pyRing} and the parameterized analysis \texttt{pSEOB}.\\

\noindent
\textit{pyRing }
The \texttt{pyRing} analysis uses the truncated TD likelihood formulation described in Section~\ref{subsub:td}.
A hierarchical analysis strategy using templates with increasing levels of complexity (described in Section~\ref{subsec:ringdown_models}) is used to extract information from the observed signal with varying degrees of agnosticism. 
The templates employed in the analysis are, in order of encoded information: \texttt{Damped sinusoids} (here abbreviated to \texttt{DS}), \texttt{Kerr} and \texttt{KerrBinary}. 
The analysis is repeated for each template over a grid of fixed starting times to ensure that the estimated quantities converge toward their GR values, as expected. 
For simplicity, results for each model are reported only at a nominal start time \(t_{\text{nom}}\), defined as the time after the peak \(t_0\) when the QNM superposition model is expected to become valid. 
For analyses up to GWTC-3, the \texttt{Kerr} template, including a combination of the ($\ell = m = 2$, $n = 0$) and ($\ell = m = 2$, $n = 1$) QNMs, was the most sensitive. 
The inclusion of an overtone phenomenologically models the early time portion of the signal, thereby capturing more power.
Additional selection criteria are imposed on the list of events passing the TGR selection process to ensure that only events with detectable ringdown signals are thoroughly analyzed, reducing computational cost.
This is determined by a first analysis of all events using the most sensitive template: if the analysis favors the presence of a signal over Gaussian noise, as well as yielding estimates of the remnant's mass and spin which are informative relative to the prior distributions, the event is included in the ``ringdown catalog.''
Out of the 48 events, only 22 make the cut.
The analysis starting time for each model is defined relative to the best estimate of the peak strain $h_+^2 + h_\times^2$ time $t_0$, as determined by the IMR analysis under the assumption of GR. 
The sky position is also fixed to values pre-determined by IMR analyses.

The simplest model, \texttt{DS}, is an agnostic framework for testing the consistency of ringdown emission with GR in which the signal is a simple superposition of QNMs. The frequency, damping time, logarithmic amplitude and phase of each mode are free parameters, each with uniform priors.
Spectroscopic measurements of QNMs are achieved by comparing the observed frequencies and damping times with the theoretical predictions from BH perturbation theory.  The comparison between the frequency and damping time obtained for the 22 analyzed events agrees with the IMR predictions for the fundamental $\ell = |m| = 2, n = 0$ mode at $t_{start}=10M$.
Searches based on other templates are discussed in detailed below.

\subsubsection{Searches for additional modes}
\label{sec:Observations_Additional_Modes}

\begin{figure}
\label{fig:overtone_results}
\centering
\includegraphics[width=\linewidth]{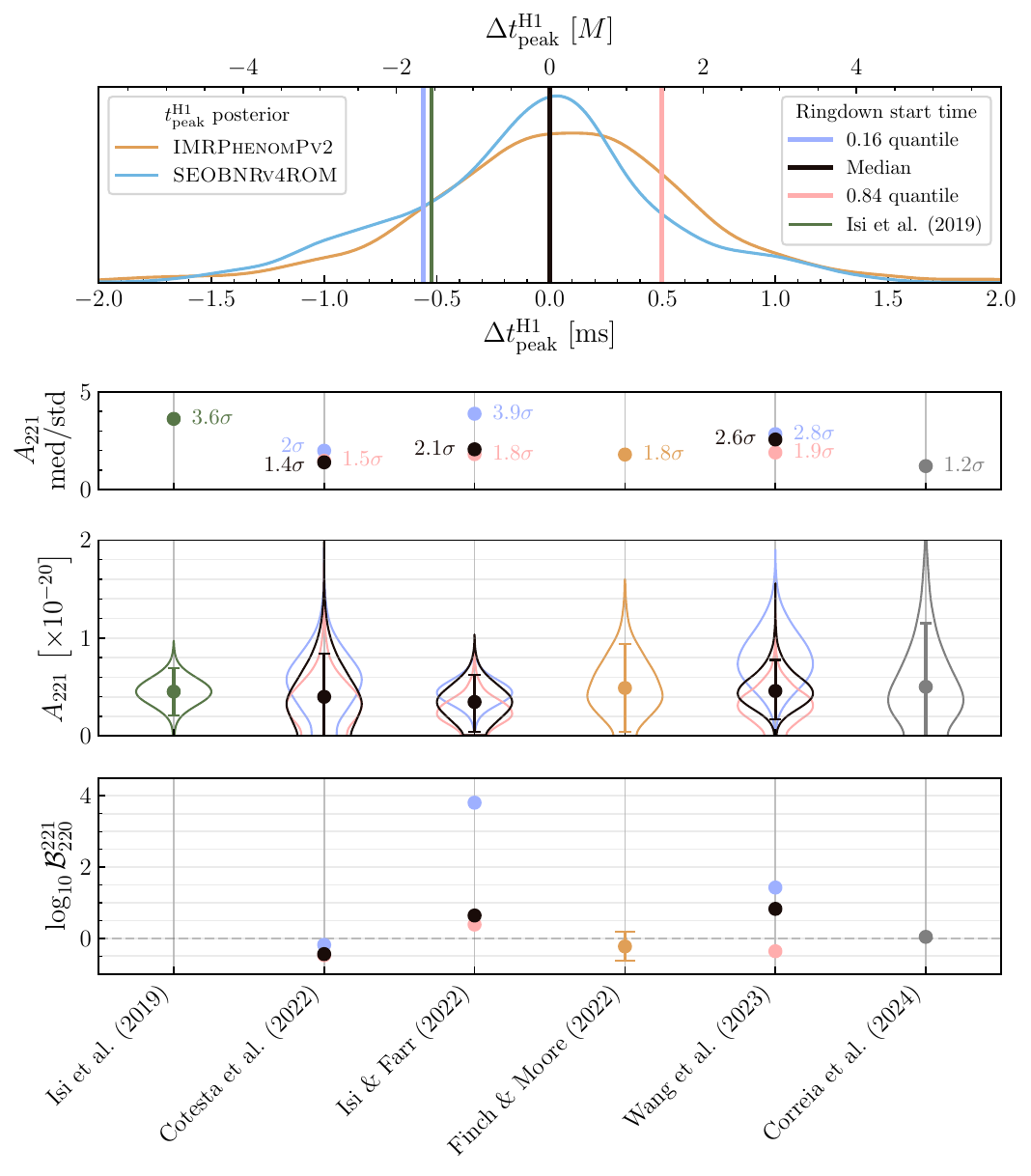}
\caption{
Overtone amplitude posteriors and log base-10 BFs in favor of an overtone for select analyses.
Top panel: posteriors on the GW150914 time of peak strain in Hanford from the $\textsc{IMRPhenomPv2}$~\cite{Husa:2015iqa,Khan:2015jqa,Schmidt:2010it} and $\textsc{SEOBNRv4ROM}$~\cite{Taracchini:2013rva,Purrer:2014fza,Purrer:2015tud} IMR analyses (data obtained from Ref.~\cite{maximiliano_isi_2022_5965773}).
Times are given relative to the (joint) $t_\mathrm{peak}^\mathrm{H1}$ median $1126259462.42352\,\mathrm{s}$.
We also give times in units of the remnant BH mass, with $M = 69M_\odot$.
Data from:
Isi et al. (2019)~\cite{Isi:2019aib} (updated in Isi \& Farr (2022)~\cite{Isi:2022mhy});
Cotesta et al. (2022)~\cite{Cotesta:2022pci} (reporting the updated version of Carullo et al. (2023)~\cite{Carullo:2023gtf});
Finch \& Moore (2022)~\cite{Finch:2022ynt};
Yi-Fan Wang et al. (2023)~\cite{Wang:2023xsy};
and Correia et al. (2024)~\cite{Correia:2023bfn}.
}
\end{figure}

\begin{figure}
\label{fig:overtone_results_time_variability}
\centering
\includegraphics[width=\linewidth]{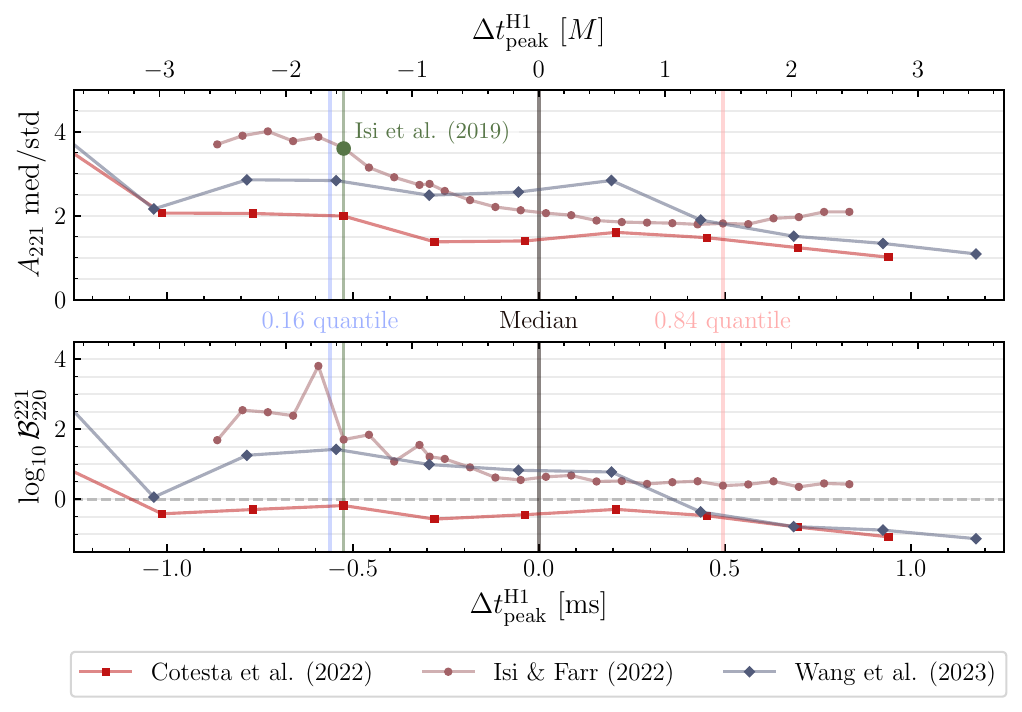}
\caption{
Cotesta et al. (2022)~\cite{Cotesta:2022pci} (reporting the updated version of Carullo et al. (2023)~\cite{Carullo:2023gtf}), Isi \& Farr (2022)~\cite{Isi:2022mhy} and Wang et al. (2023)~\cite{Wang:2023xsy} perform their analysis at multiple (fixed) ringdown start times.
In Fig.~\ref{fig:overtone_results} we show results from three ringdown starting times for these analyses, chosen based on the 0.16, 0.5, and 0.84 quantiles of an IMR $t_\mathrm{peak}$ posterior (top panel).
Here, in contrast, we show the overtone amplitude posterior median over standard deviation (top panel) and log base-10 BFs in favor of an overtone (bottom panel) at {\em every} ringdown starting time considered in these works.
As in the top panel of Fig.~\ref{fig:overtone_results},  times are measured relative to the (joint \textsc{IMRPhenomPv2}--\textsc{SEOBNRv4ROM}) $t_\mathrm{peak}^\mathrm{H1}$ median $1126259462.42352\,\mathrm{s}$ in either milliseconds (bottom axis) or in units of the remnant BH mass $M = 69M_\odot$ (top axis).
}
\end{figure}

Spectroscopic tests of the Kerr nature of the merger remnant require the detection of more than one mode. Here we review the current status of searches for multiple modes.\\

\noindent 
\textit{Overtones in GW150914 }
As discussed in Section~\ref{subsec:overtones}, the authors of~\cite{Giesler:2019uxc}
suggested that the post-peak GW signal from BBH
mergers like GW150914 could be described to high accuracy by a sum of damped
sinusoids corresponding to the final-state linear ringdown modes of the remnant
BH, provided that enough overtones are included. Motivated by this suggestion,
Isi et al.~\cite{Isi:2019aib} fit a \texttt{Kerr} model to post-peak GW150914 data using the TD methods discussed in Section
\ref{subsub:td}. They consider a fixed time offset
between the Hanford and Livingston detectors, corresponding to a signal
originating at right ascension $\alpha = 1.95 \, \mathrm{rad}$ and declination
$\delta = -1.27 \, \mathrm{rad}$; additionally, they constrain the signal to
be left-circularly polarized (consistent with a face-off viewing angle, $\iota =
180^\circ$). These values are consistent with the sky location and orientation
for the GW150914 event determined by an analysis of the full signal~\cite{LIGOScientific:2016vlm}. 
Isi et al.~\cite{Isi:2019aib} show that a
single-tone Kerr ringdown model, with only the $(2,2,0)$ fundamental mode, fitted
to the Hanford and Livingston detector data sampled at a 2 kHz sample rate
starting at $t_H = 1126259462.423$ (the peak time employed in Ref.~\cite{LIGOScientific:2016lio}), recovers a final-state mass and spin that are
inconsistent with the values inferred from the full signal analysis.
The analysis duration is $T=0.5 \,$s, and a direct correlation method is used to estimate the ACF.

In contrast, using a two-tone (fundamental $(2,2,0)$ plus overtone $(2,2,1)$) Kerr model at
the estimated peak time, Ref.~\cite{Isi:2019aib} recovers a mass and spin that
are instead consistent with the analysis of the full signal. %
In the two-tone model, the posterior mean amplitude of
the overtone is 3.6 standard deviations from zero; this is the leftmost result shown in Fig.~\ref{fig:overtone_results}. 
Relaxing
the Kerr assumption and allowing the frequency and damping rate of the overtone
to vary from the Kerr value consistently with the fundamental mode's frequency and
damping rate, as described in Section~\ref{subsec:theory-agnostic}, recovers
posteriors on the deviations $\delta f_{221}$ and $\delta \tau_{221}$ compatible with $\delta f_{221} = \delta \tau_{221} = 0$; the measurement of
$\delta f_{221} = -0.05 \pm 0.2$ establishes that the data are consistent with the Kerr
overtone frequency at the $20 \%$ level~\cite{Isi:2019aib}. 
In the
three-tone model, neither the first nor second overtone amplitudes are
individually bounded away from zero, but their joint posterior excludes the
origin at similarly high credibility.  
Subsequently, Isi et al.~\cite{Isi:2021iql} demonstrated similar results for two-tone
models applied to GW150914-like waveforms from NR calculations
in realistic detector noise at SNRs comparable to GW150914, at
least partially validating the applicability of the two-tone Kerr model in these
contexts as an effective description of the post-peak data.  
Their inference framework is implemented in the publicly available \texttt{ringdown} package (see Appendix~\ref{sec:public_codes}).

Cotesta et al.~\cite{Cotesta:2022pci} re-assess these results. They note that (i) the peak-time distribution of GW150914
spans a width of $2\,$ms (at $2\sigma$ credibility), or several $t_M$, and this uncertainty can not be neglected; (ii) the fixed reference time used in Isi et al.~\cite{Isi:2019aib} underestimates the average peak time of $h_+^2 + h_{\times}^2$ (see the top panel of Fig.~\ref{fig:overtone_results}).
The analysis of Cotesta et al. is similar to Isi et al.~\cite{Isi:2019aib}, but it uses the independent and publicly available \texttt{pyRing} package (see Appendix~\ref{sec:public_codes}).
They report a fluctuating posterior on the
overtone amplitude across a range of analysis times consistent with the
estimates from full-signal models, finding a much smaller significance for the overtone presence and negative Bayesian evidence (see Figs.~\ref{fig:overtone_results} and \ref{fig:overtone_results_time_variability}).
They repeat a similar analysis for GW150914-like signals added to LIGO detector noise, which suggests that noise fluctuations could induce spurious evidence for the presence of an overtone.
Such a significantly different posterior on the overtone amplitude is not only due to the inclusion of peak-time uncertainty -- it is also found at a reference time compatible with Isi et al.~\cite{Isi:2019aib}.
Compared to Isi et al., Cotesta et al. use a higher sampling rate ($\sim 16 \, \rm kHz$ instead of $\sim 2 \, \rm kHz$), an analysis duration of $T=0.1s$, and an identical method to estimate the ACF through the \texttt{ringdown} package.

Isi \& Farr~\cite{Isi:2022mhy} performed a subsequent re-analysis at a sampling rate of $\sim 2 \, \mathrm{kHz}$, with a Fourier-transform based ACF estimation and with a duration $T=0.2 \, \mathrm{s}$ (arguing that this is sufficiently long to capture virtually the entire signal power).
The re-analysis notes how the estimated overtone amplitude decays consistently with
the expectation of exponential damping~\cite{Isi:2022mhy}, even in a range of times consistent with estimates of the peak of the strain.
A subsequent {\em Comment}~\cite{Isi:2023nif} later identified two issues with the analysis in Cotesta et al.~\cite{Cotesta:2022pci}: an insufficient analysis duration ($T=0.1\, \mathrm{s}$, instead of the original $T=0.2\, \mathrm{s}$), and a shift of one timestamp (or $0.06\, \mathrm{ms}$) in the likelihood time axis with respect to the detector data axis.
A re-analysis accounting for these issues, presented in Carullo et al.~\cite{Carullo:2023gtf}, finds negligible differences with respect to previous work by the same authors in Cotesta et al.~\cite{Cotesta:2022pci}.
The remaining differences with the analyses of Isi et al.~\cite{Isi:2019aib, Isi:2022mhy} are attributed to different sampling rates and ACF estimation methods: recall that Ref.~\cite{Isi:2022mhy} switched to an FFT-based estimate, similar to the one employed within the standard LVK spectroscopy searches~\cite{Carullo:2019flw, LIGOScientific:2021sio}, but different from the direct correlation originally employed in Isi et al.~\cite{Isi:2019aib} and (in the attempt to reproduce their results) by Cotesta et al.~\cite{Cotesta:2022pci}.

Finch \& Moore~\cite{Finch:2022ynt} also analyze GW150914 with the method
discussed in Section~\ref{subsec:RD_systematics}, which allows for a variable
damped-sinusoid starting time in the analysis at the cost of introducing a
wavelet-based model for the early time signal.
Marginalizing over the starting time estimates from~\cite{LIGOScientific:2016vlm}, they find that the Bayesian evidence for the
two-tone model is lower than for the single-tone model by a small factor
($\log_{10} B_{1\mathrm{QNM}}^{2\mathrm{QNM}} \simeq -0.2$). The
time-marginalized posterior on the overtone amplitude has a mean $\sim 1.8$
times its standard deviation away from zero. 
This marginalized result is shown in Fig.~\ref{fig:overtone_results}.
At the same starting
time used by Isi et al.~\cite{Isi:2019aib}, the posterior on the overtone amplitude agrees
with the results of Ref.~\cite{Isi:2019aib}, and the evidence for the overtone
is positive, $\log_{10} B_{1\mathrm{QNM}}^{2\mathrm{QNM}} \simeq 0.9$. 
However, when assuming a fixed starting time corresponding to the peak-time median (as in the top panel of Fig.~\ref{fig:overtone_results}) the evidence drops to $\log_{10} B_{1\mathrm{QNM}}^{2\mathrm{QNM}} \simeq 0$.

Ma et al.~\cite{Ma:2023cwe,Ma:2023vvr} use a rational filter to target and remove specific QNM frequencies from the data.
They report a log base-10 BF as high as $\sim 2.8$ in favor of a model with an overtone, but note that this is sensitive to the ringdown starting time.
We chose not to include their results in Figs.~\ref{fig:overtone_results} and \ref{fig:overtone_results_time_variability}, because the interpretation of their reported BF is affected by the filtering technique they use.

Crisostomi et al.~\cite{Crisostomi:2023tle} report a mild ($\sim 2\sigma$) evidence for the presence of an overtone, making use of a calibrated simulation-based inference pipeline. 
This number is based on the credible region at which the (calibrated) two-dimensional $(A_{220},\,A_{221})$ posterior excludes $A_{221} = 0$, for some preferred choice of ringdown start time and sky position.

Yi-Fan Wang et al.~\cite{Wang:2023xsy} employ gating and in-painting (see Section~\ref{subsub:gating}) to perform the analysis at multiple (fixed) ringdown starting times, focusing on data analysis systematics. Their ``fiducial'' analysis with a sampling rate of 8192 Hz (shown in Figs.~\ref{fig:overtone_results} and \ref{fig:overtone_results_time_variability}) reports log base-10 BFs ranging from $\sim -0.4$ to $1.4$ and an overtone amplitude median over standard deviation of $\sim 1.9$ to $2.8$ (over the 1-sigma uncertainty on $t_\mathrm{peak}$).

Correia et al.~\cite{Correia:2023bfn} also use gating and in-painting. However, by allowing the switch-on time of the gate ($t_a$) to vary as a free parameter within the analysis, they are able to marginalize over the ringdown starting time and sky position.
They find that simply allowing the switch-on time of the gate to vary would result in the gate being pushed as late as possible, effectively excising all signal from the data (see also~\cite{Correia:2023ipz} for details).
To address this issue, the pre-ringdown signal is simultaneously modeled with the $\texttt{IMRPhenomXPHM}$ approximant, where now the data after $t_a$ is gated out.
The pre-ringdown and ringdown analyses only share as parameters the gate time and sky position.
As in Finch \& Moore, their result consists of a single marginalized analysis with a log base-10 BF of $\sim 0.04$ in favor of the overtone, and an overtone amplitude median over standard deviation of $\sim 1.2$.

Finally, Hai-Tian Wang et al.~\cite{Wang:2024yhb} employ the $\mathcal{F}$-statistic method.
This consists of analytically maximizing over the ringdown amplitudes and phases. The overtone significance is quantified in terms of an ``$\mathcal{F}$-statistic information criterion'' (analogous to, but different from, a BF) with reported values ranging between $\sim -1.3$ and $0.3$.

In Figs.~\ref{fig:overtone_results} and
\ref{fig:overtone_results_time_variability} we summarize the amplitude estimates
for the overtone in a two-tone Kerr model applied to GW150914 at various starting times, from a selection of the works summarized above. 
It is now broadly accepted that in order to obtain
agreement for the final mass and spin between the ringdown-only and full-signal analyses, it is necessary to include at least one overtone when the ringdown analysis starts close to the peak strain time. 
This conclusion, however, is SNR-dependent: louder signals would require an increasing number of overtones.  
The level of statistical significance for a nonzero overtone amplitude, quantified via the posterior median over standard deviation, varies between $1\sigma$ and $4\sigma$ for a range of times within the estimated peak time of the signal.
Values at the median time quoted in Fig.~\ref{fig:overtone_results} range within $[1.4, 2.6] \sigma$, while time-marginalized estimates range instead in the $[1.2, 1.8] \sigma$ interval.
The Bayesian evidence in favor of the overtone model presents wider variations, with log base-10 BFs ranging between $-0.5$ and $4$.
Time-marginalized evidences are uninformative or slightly disfavor the inclusion of an overtone ($[-0.2, 0.04]$), while estimates at the median time range within $[-0.4, 0.8]$.

Significant disagreement remains on the interpretation of the results. Several authors have questioned the physical significance of a constant-amplitude QNM model starting at the peak, which ignores the dynamical nature of the BH relaxation process (including the time-dependent nature of the background and of the QNM amplitudes, as well as possible nonlinear couplings): see the discussion in Section~\ref{subsec:overtones}.\\

\noindent
\textit{Higher harmonic searches in GWTC-3}
The multimodal search employed by \texttt{pyRing} involves performing model selection between different hypotheses, represented by models with varying mode content in the templates used to fit the data. These models include \texttt{Kerr} and \texttt{KerrBinary}~\cite{London:2018gaq} templates. The start times of the analysis are chosen to be $10M$ and $15M$, respectively.
Two baseline models represent the fundamental mode hypothesis: the \texttt{Kerr} template and the \texttt{KerrBinary} template with only the \((2,2)\) mode. The higher mode hypotheses are constructed by augmenting the fundamental mode with additional modes:  
\begin{itemize}
    \item \texttt{Kerr} templates include \((2,2,0) + (\ell, m, n)\) and \((2,2,0) + (2,2,1) + (\ell, m, n)\), where \((\ell, m, n) = (3,3,0), (3,2,0), (2,1,0), (2,0,0)\), added individually and independently.
    \item \texttt{KerrBinary} templates include all \((3,3), (2,1), (3,2), (4,4), (4,3)\) modes simultaneously (overtones are only ``implicitly'' included in this parameterization). 
\end{itemize}  
Additional hypotheses are introduced to refine the sensitivity of multi-modal analyses. 
First, the search assumes reflection symmetry in the Kerr amplitudes, such that modes with opposite \(m\) signs are related by the nonprecessing symmetry
\begin{equation}
A_{\ell,-|m|,n} = (-1)^{\ell} A^{\ast}_{\ell,|m|,n}\,.
\end{equation}
This symmetry is valid for remnants of nonprecessing binary mergers.
To enhance sensitivity to higher modes, the analysis is performed under two conditions. First, no restriction is imposed on the relative magnitude of the modes.
Second, an amplitude hierarchy constraint is imposed between the fundamental and higher order modes: $A_{\ell,m,n}/A_{2,2,0} < 0.9$, consistent with expectations for quasi-circular spin-aligned binaries with moderate mass ratios~\cite{Forteza:2022tgq, Capano:2021etf}. 
This suppresses bi-modalities arising in the previous unconstrained search, that degrade sensitivity to multi-modal detections.
Since the above templates do not account for precession, the reflection symmetry assumption is relaxed in an additional analysis to accommodate scenarios where the progenitor binary's spins are misaligned with the orbital angular momentum. 
This approach includes precessing binaries, which break the above symmetry between the mode amplitudes~\cite{Blanchet:2013haa, Pekowsky:2013ska, Kolitsidou:2024vub}.
A log BF is used to quantify the evidence for the presence of higher modes relative to the fundamental mode, providing a statistical basis for hypothesis selection.

\begin{figure}[t]
\centering
\includegraphics[width=0.95\textwidth]{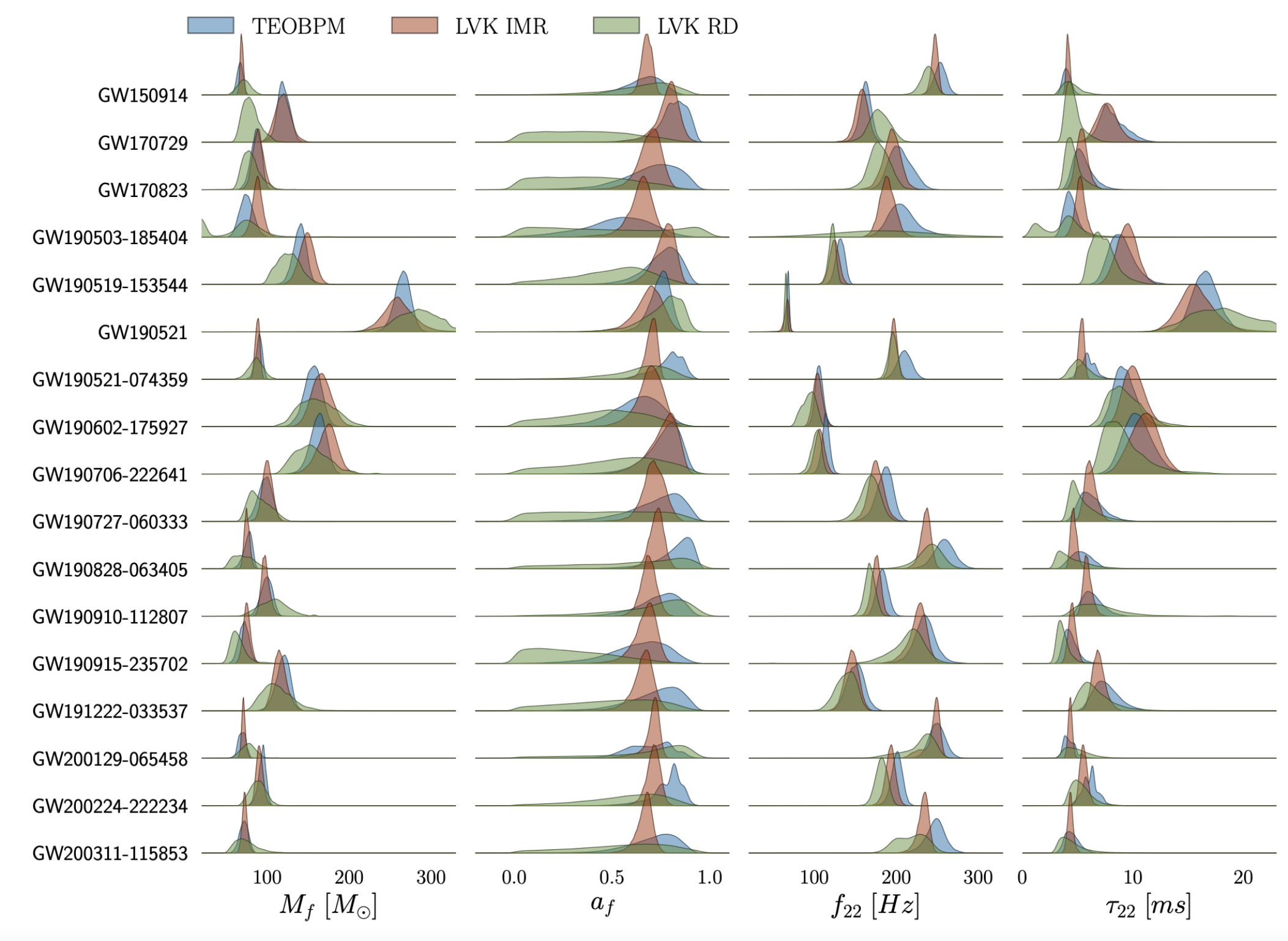}
\caption{Posterior distributions of the remnant parameters \(M_f\) and \(a_f\) for various events, listed along the y-axis, assuming the LVK (combined) IMR results (``LVK IMR''), the \texttt{Kerr} template including the (2,2,0) mode (``LVK RD'') \texttt{KerrPostmerger} template based on the \texttt{TEOBPM} version.
The only event showing significant discrepancy is GW170729, not included in the LVK ringdown catalog because below the noise contamination selection criteria.
Figure taken from~\cite{Gennari:2023gmx}.}
\label{fig:LVK_comparison}
\end{figure}

No significant evidence for the presence of higher modes was found in the LVK events using any of the search strategies we discussed so far~\cite{LIGOScientific:2021sio}, with the exception of one event: GW191109\_010717. 
However, the apparent detection in this case was attributed to nonstationarities in the data, so GW191109\_010717 was excluded from the combined results.
This particular event shows weak evidence for the \((3,2,0)\) and \((2,1,0)\) modes at \(t < 5M\), but with overestimated values of the remnant mass \(M_f\) and spin \(a_f\), consistently with a starting time outside of the model validity regime.
Another event (GW200224\_222234) shows weak, but not fully convincing, statistical evidence for the presence of the \((2,2,1)\) overtone.

In Fig.~\ref{fig:LVK_comparison} we compare the GR remnant parameters predicted by the full IMR waveforms and by the \texttt{Kerr} ringdown template including the $(2,2,0)$ mode.
For all events that pass the noise contamination selection criteria, the IMR predictions agree with the properties inferred from the ringdown alone.
The one exception is GW170729, which however does not pass the noise contamination selection criteria. This is a good example of how noise contamination can lead to false violations of GR.

The \texttt{KerrPostmerger} template, based on the \texttt{TEOBPM}~\cite{Damour:2014yha} ansatz, has been implemented in \texttt{pyRing}~\cite{Gennari:2023gmx} in preparation for the fourth LVK observing run (O4).
This template is the most sensitive ringdown-only data analysis model to date. 
While similar to the \texttt{KerrBinary} template, \texttt{KerrPostmerger} is NR-calibrated and designed to model the time-dependent amplitudes and phase, so it can effectively capture nonlinearities, overtone excitation, and transient effects in the early post-merger signal.
A total of 18 events from GWTC-3 pass the threshold of having a \(\ln B > 3\) in favor of a signal over noise with the \texttt{KerrPostmerger} template. These events were re-analyzed using the \texttt{KerrPostmerger} template, considering the modes \((\ell, m) = [(2,2), (3,3), (2,1)]\).
More specifically, the multi-modal search uses four different \texttt{KerrPostmerger} templates at the peak of the IMR waveform, $t_{\text{nom}}$, which include the following combinations:
\begin{itemize}
    \item only the \((2,2)\) mode;
    \item \((2,2) + (\ell,m)\), with \((\ell,m) = [(2,1)], [(3,3)], [(2,1)+(3,3)]\).
\end{itemize}
These higher-order modes are chosen because they are more likely to be detectable. 
Their detectability 
can be estimated semi-analytically using the relation
\begin{equation}
\ln B^{\ell m}_{22} \propto \frac{1}{2}(1 - \text{FF}^2) \, \text{SNR}^2_{\text{opt}}\,,
\end{equation}
valid in the high-SNR limit. Here, \(\text{FF}\) denotes the fitting factor, defined as the normalized best-match between a \texttt{KerrPostmerger} template containing higher modes and one with only the \((2,2)\) mode. The quantity $\text{SNR}_{\text{opt}}$ is the optimal SNR, calculated as the signal power divided by the noise power, integrated over the detector's bandwidth.

\begin{figure}[t]
\centering
\includegraphics[width=0.95\textwidth]{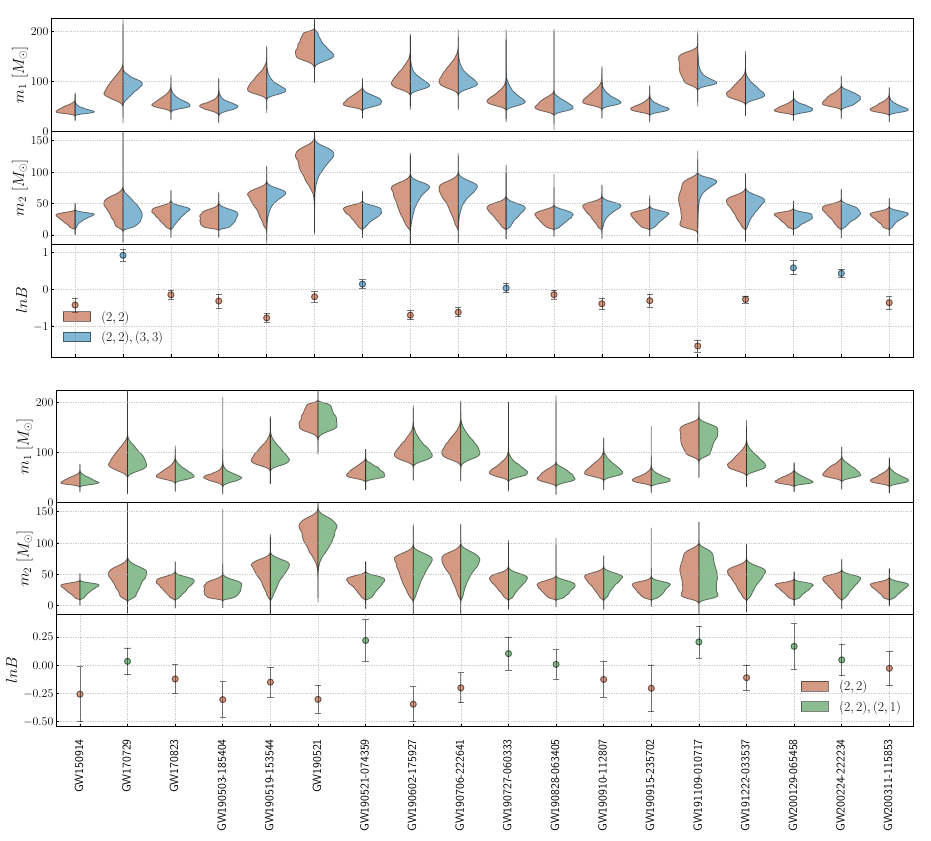}
\caption{Posterior distributions of the initial masses \(m_1\) and \(m_2\) for various events, listed along the x-axis, assuming the \texttt{KerrPostmerger} template based on the \texttt{TEOBPM} version. 
The results of the analysis using only the fundamental mode \((2,2)\) are represented in red, while those incorporating both the fundamental mode and the \((3,3)\) mode are shown in blue. Additionally, the posteriors derived from the analysis including the fundamental mode and the \((2,1)\) mode are shown in green. The two subplots in the bottom row present the logarithmic BF, \(\ln B_{\ell m,22}\), which compares the higher modes hypothesis under consideration with the fundamental mode alone. The bars are colored to indicate which hypothesis is preferred in each case.
Figure taken from~\cite{Gennari:2023gmx}.}
\label{fig:HM_teob}
\end{figure}

For nearly equal mass binaries (\(q \sim 1\)), which are prevalent among current GW observations, the excitation of higher modes during the ringdown phase is minimal, and their detection is challenging. 
For example, detecting the \((3,3)\) mode with a detection threshold of $\ln B^{\ell m}_{22} = 3$ would require an $\text{SNR}_{\text{opt}}\gtrsim 50$, while typical events in the current catalogs have ringdown $\text{SNR}_{\text{opt}} \lesssim 15$.

The higher modes are more excited for large mass ratio binaries (\(q \in [2,3]\)), reducing the required \(\text{SNR}_{\text{opt}}\) for detection. For these systems, the \((3,3)\) mode becomes detectable in loud events with \(\text{SNR}_{\text{opt}} \sim [10,15]\), particularly at larger inclinations (\(\iota > \pi/4\)).
This highlights the importance of mass ratio and inclination for the observability of higher modes. The \((3,3)\) mode in particular stands out as the easiest to observe, often requiring a lower \(\text{SNR}_{\text{opt}}\) compared to modes like \((2,1)\) for systems with moderate mass ratios. Higher modes such as \((3,2)\) and \((4,4)\), on the other hand, require substantially larger \(\text{SNR}_{\text{opt}}\) for detection -- typically above 25 for mass ratios \(q \lesssim 2\) and inclinations \(\iota > \pi/4\). 

These observations reinforce the need to carefully account for the system's parameters when predicting the detectability of higher modes. 
Furthermore, the detection threshold itself plays a significant role. Reducing the threshold increases the likelihood of observing higher modes, even at lower \(\text{SNR}\), but it also necessitates careful control of false positives due to noise. Future studies should refine these estimates by incorporating realistic noise simulations and accounting for systematic uncertainties, such as modeling inaccuracies and the determination of the ringdown starting time.

Model selection between the \((2,2)\) and \((2,2) + \ell m\) templates yields
BFs that overall do not show a preference for higher modes in
the analyzed data, as shown in Fig.~\ref{fig:HM_teob}. Four events (GW170729,
GW190521\_074359, GW200129\_065458, GW200224\_222234) exhibit a low but positive
BF, and for these events, adding the \((3,3)\) mode yields posteriors that favor
higher mass ratios.
Once can probe the sensitivity of the results to the starting time of the analysis by varying it within the 95\% credible interval of the IMR peak time. 
Within this range, some events show BF peaks, with corresponding posterior distributions shifting towards higher mass ratios. Even when the \((3,3)\) mode is not confidently detected, its inclusion in the model improves the recovered parameter estimates.
For GW200129\_065458 and GW200224\_222234, a BF increase is observed at later times (outside the IMR peak time support region). However, since the \texttt{KerrPostmerger} template assumes given values for the amplitudes defined at the peak time, starting the fit outside the domain of definition for the amplitudes could introduce biases. Once again, a careful selection of the starting time is important to avoid systematic errors due to modeling assumptions.\\

\noindent
\textit{Higher harmonics: GW190521}
On May 21, 2019, 03:02:29 UTC, the two LIGO detectors and the Virgo interferometers detected the GW190521 signal~\cite{LIGOScientific:2020iuh}.
Assuming the quasi-circular waveform families \texttt{NRSur}, \texttt{SEOB} and \texttt{Phenom} (see Section~\ref{sec:waveforms}), all including both higher harmonics and spin precession effects, the signal is found to be consistent with a merger of two nearly equal mass BHs, with a high total mass, and component spins preferentially large and oriented along the equatorial plane.
This is the heaviest BH merger detected with high confidence to date.
The high mass leads to the merger occurring at comparatively low frequency, placing the ringdown signal at the lower end of the detector spectrum -- around $60 \, \rm Hz$, a region with degraded detector sensitivity and larger impact from detector noise~\cite{LIGO:2021ppb}.

A later study found support for large unequal mass ratio, highlighting the significant impact of prior distributions on the inferred parameters~\cite{Nitz:2020mga}.
The study employed an early version of the \texttt{IMRPhenomXPHM} model, at the time still under development. 
Subsequent follow-ups~\cite{Estelles:2021jnz, Olsen:2021qin}, using an improved version of \texttt{IMRPhenomXPHM}, confirmed a multi-modal likelihood.
The support for the unequal mass configuration was reduced, with the bulk of the posterior distribution falling close to the equal mass regime~\cite{Estelles:2021jnz}.
However, the unequal mass peak yielded the highest likelihood, with a peak larger by a factor $\sim 7$~\cite{Olsen:2021qin} compared to the equal mass case.
The significant impact of the prior distribution on the mass ratio was also confirmed.
These findings enhanced prospects for the excitation of modes beyond the fundamental quadrupolar one.
Due to the prominence of the merger-ringdown signal component, several studies investigated in detail the post-merger emission of GW190521, searching for the presence of multiple modes.

\begin{figure}[ht]
\centering
\includegraphics[width=0.6\textwidth]{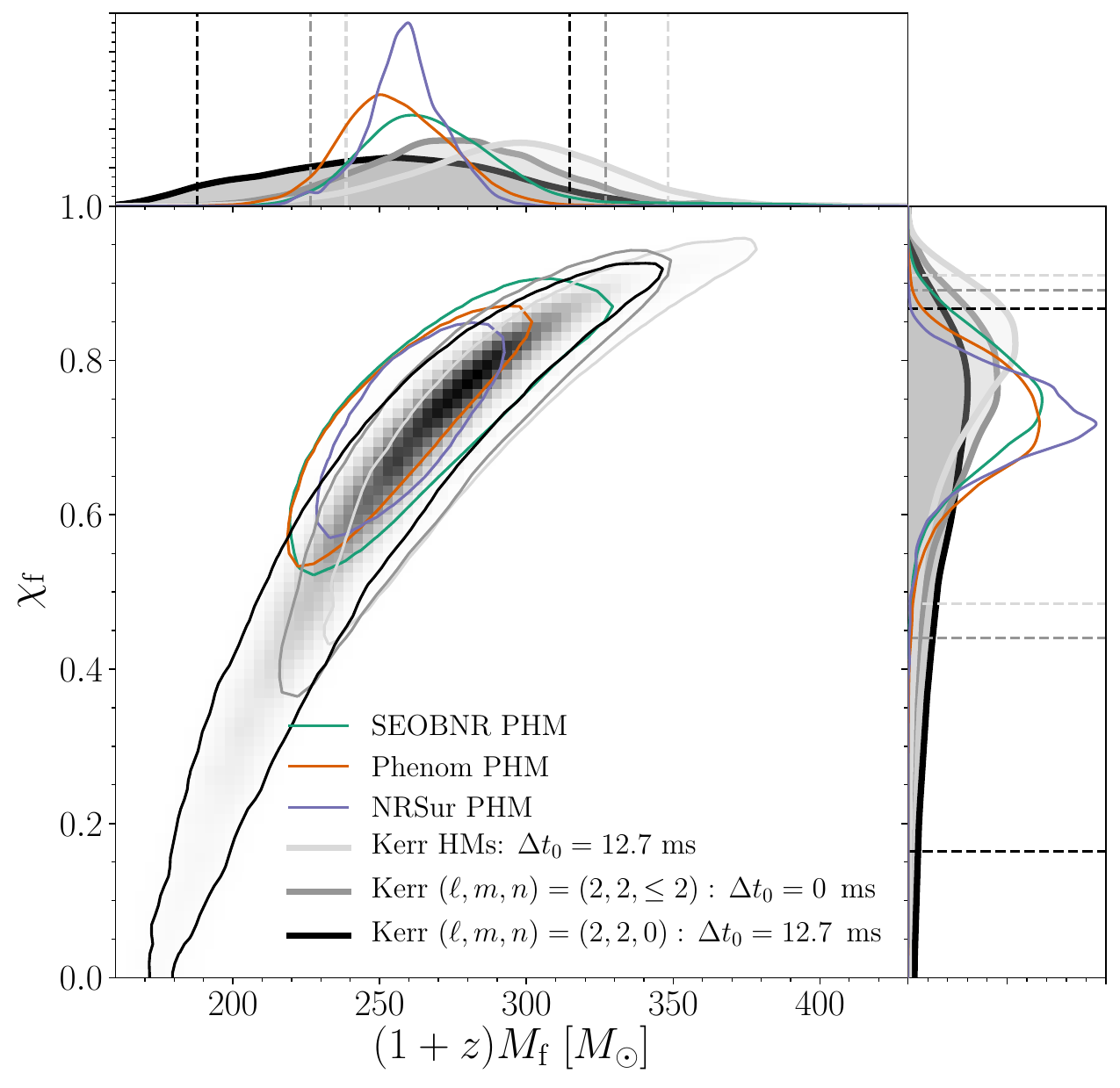}
\caption{Posteriors on final mass and spin for several IMR and ringdown-only models for GW190521.
\texttt{KerrHM} corresponds to the \texttt{KerrBinary} model~\cite{London:2018gaq}.
Figure taken from~\cite{LIGOScientific:2020ufj}.
}
\label{fig:LVC_GW190521_HM_posterior}
\end{figure}

A first search for multiple modes by the LVK collaboration~\cite{LIGOScientific:2020ufj} employed the \texttt{Damped sinusoids}, \texttt{Kerr} (assuming unconstrained amplitudes parameterizations compatible with spin precession) and \texttt{KerrBinary} (spin aligned) models, with a $t_{\rm peak}$ estimate based on the \texttt{NRSur} IMR model.
The \texttt{Kerr} model contained the modes with $\ell = m = 2$, and up to $n = 0,1,2$, targeting overtones of the fundamental quadrupolar mode.
The \texttt{KerrBinary} model contained instead all modes $\ell = 2, 3, 4$ with $m = \ell, \ell - 1$ and $n = 0$, targeting multiple fundamental modes.
The ringdown analysis used the \texttt{pyRing} software package (see Appendix~\ref{sec:public_codes}) and a TD truncated likelihood function~\cite{Isi:2021iql}.
Multi-mode models were found to yield posteriors on final mass and spin consistent with those of the full IMR signal, and also with those from the single-mode model with $\ell = m = 2$, $n = 0$, as shown in Figure~\ref{fig:LVC_GW190521_HM_posterior}.
There was no strong evidence in favor of modes beyond the fundamental quadrupolar one, as quantified by BFs.

\begin{figure}[ht]
\centering
\includegraphics[width=\textwidth]{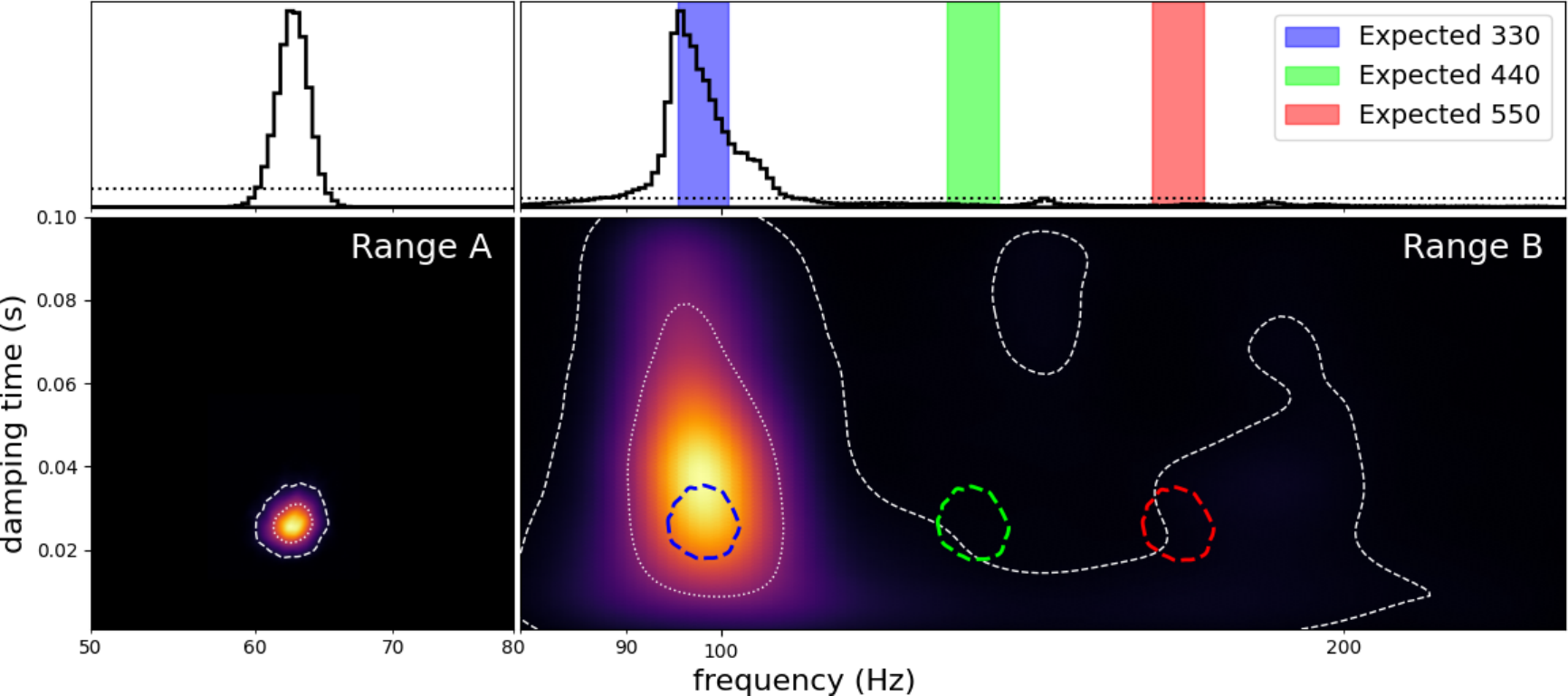}
\caption{Marginal posteriors on frequency and damping time from a \texttt{Damped Sinusoids} analysis of GW190521 at  $6\, \rm ms$ after the reference time used (see Table~\ref{tab:GW190521_studies_setup} for details about the reference time). A single QNM was searched for in each preselected frequency range shown. The mode found in frequency range B is consistent with the expected frequency and damping time of the $(3,3,0)$ mode from a Kerr BH, assuming the mode found in range A is the $(2,2,0)$ mode.
Figure taken from~\cite{Capano:2021etf}.}
\label{fig:Capano_GW190521_agnostic_posterior}
\end{figure}

\begin{table}[t]
\centering
\begin{tabular}[t]{c c c c c}
\hline \hline
Study & Reference & Time range & Sky location & Amplitude priors \\ 
\hline
LVK & $t^{\rm pol}_{\rm peak} =+0.4306$ s & $[+6.4, +19.1]$ ms & $(0.1, -1.14)$ &  ``Generic'' \\
Capano et al. & $t^{\rm harm}_{\rm peak}=+0.4259$ s & $[-9, +24]$ ms & $(3.5,0.73)$ & ``Aligned-circular''\\
Siegel et al. & $t^{\rm harm}_{\rm peak}=+0.4278$ s & $[-12.7, +25.4]$ ms & $(5.75,-0.42)$ & ``Generic''\\
\hline \hline
\end{tabular}
\caption{Reference times relative to a GPS time of $1242442967.0$ seconds, each representing the estimate of the time of peak strain (see the main text for the definitions) in the H1 detector.
The time ranges considered are relative to this reference time. Each analysis fixed the sky location (ra, dec) to the value listed in the table.
Amplitude priors referred to as ``Generic'' do not assume any constraints on the relative $(\ell,m,n)$ modes excitations, while ``Aligned-circular'' priors assume equatorial symmetry (i.e., spins aligned with the orbital angular momentum) and constraints on amplitude ratios dictated by NR simulations of spin-aligned quasi-circular binaries (see Section~\ref{sec:waveforms}).
}
\label{tab:GW190521_studies_setup}
\end{table}

A subsequent study employed \texttt{Damped sinusoids} and \texttt{Kerr} models to search for multiple modes~\cite{Capano:2021etf}.
The gating and inpainting method implemented in \texttt{PyCBC} (see Appendix~\ref{sec:public_codes}) was used to restrict the analysis to a data segment following the chosen signal starting time.

In the \texttt{Damped sinusoids} search, additional constraints were imposed to extract multiple modes from the data.
Each sinusoid was restricted to an individual frequency band around its central frequency.
The resulting posterior distribution in frequency and damping time, shown in Fig.~\ref{fig:Capano_GW190521_agnostic_posterior}, displays a narrow peak near the values expected for the $(2,2,0)$ mode from the IMR analysis.
A secondary peak was found at a higher frequency.
Assuming the first peak to match $(\ell,m,n) = (2,2,0)$ indicates that the secondary peak is consistent with the expected $(\ell,m,n) = (3,3,0)$ mode.
This feature was most clearly present at $t_{\rm ref} + 6\, \rm ms$ after the reference time used in this analysis (see Table~\ref{tab:GW190521_studies_setup}). It suggests the presence of a secondary feature compatible with a damped sinusoid, but this does not on its own provide statistical evidence, since filtering bands have to be selected \textit{a priori}.

Equatorial reflection symmetry of the initial perturbation and priors on the amplitudes of the subdominant modes restricted to $A_{330} / A_{220} < 0.5$, $A_{221} / A_{220} < 5$, valid only for binaries with aligned spins~\cite{Forteza:2022tgq}, were assumed.
The authors looked for various mode combinations, including $(\ell, m, n) = [(2,2,0),(2,2,1),(2,1,0),(3,3,0)]$. Statistical evidence for these mode combinations in terms of BFs within the \texttt{Kerr} model is quantified in Fig.~\ref{fig:Capano_GW190521_BF}.
Evidence favoring the presence of the $(2,2,0)+(3,3,0)$ and $(2,2,0)+(3,3,0)+(2,1,0)$ mode combinations was found, compared to models containing only the $(2,2,0) + (2,1,0)$ or the $(2,2,0) + (2,2,1)$ combinations.
The excitation of the $(3,3,0)$ mode is indicative of a high mass ratio binary, in disagreement with the \texttt{NRSur} posterior measurement, which is points instead to a more equal-mass system.
The largest BF ($\sim 56$) was found at $t^{\rm harm}_{\rm peak} + 6\, \rm ms$, where $t^{\rm harm}_{\rm peak} = \underset{t}{\max} \sum_{\ell m} h_{\ell m}(t) = 1242442967.445$ GPS time is estimated from the \texttt{NRSur} IMR model assuming a uniform prior on the mass ratio.
Using the reported $M_f^{\rm det} \simeq 330 M_{\odot}$~\cite{Capano:2021etf}, this translates to $t^{\rm harm}_{\rm peak} + 3.7 \, M_f^{\rm det}$.
The analysis was repeated for the model with no equatorial symmetry, while maintaining the priors on amplitude ratios. In this case, the BF for the $(2,2,0)+(3,3,0)$ model peaked at 42, also at $t^{\rm harm}_{\rm peak} + 6\, \rm ms$.
Later studies on populations of simulated signals, intended to mimic the full signal of GW190521 under the same assumptions of this initial analysis, found the BFs to be robust against noise fluctuations.
They also assigned a consistent statistical quantification to the \texttt{Damped sinusoids} results~\cite{Capano:2022zqm}.
There, support for a mode compatible with larger mass ratios was found when assuming the \texttt{IMRPhenomTPHM} model, unlike the \texttt{IMRPhenomXPHM} and \texttt{NRSur} models.
Assuming the \texttt{IMRPhenomTPHM} model, they also validated the correlation between unequal mass-ratios and Bayes Factors, consistent with the GW190521 analysis~\cite{Abedi:2023kot}.

\begin{figure}[t]
\centering
\includegraphics[width=0.6\textwidth]{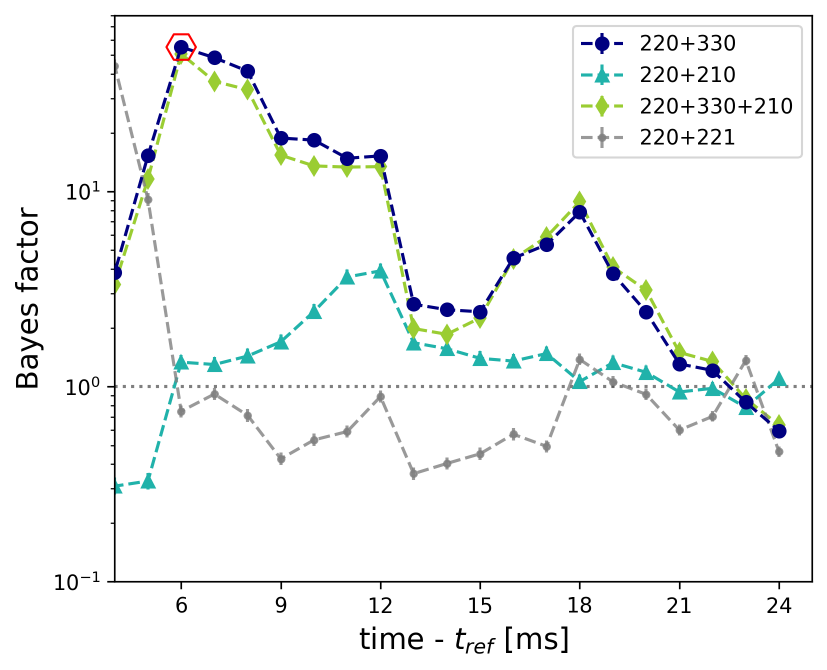}
\caption{BFs for the presence of multiple modes using a \texttt{Kerr} model, compared to a single $(2,2,0)$ mode, assuming spin-aligned quasi-circular amplitude constraints and $t_{\rm peak}$ estimates from the \texttt{IMRPhenom} model.
Figure taken from~\cite{Capano:2021etf}.}
\label{fig:Capano_GW190521_BF}
\end{figure}

There is some uncertainty related to the choice of the reference $t_{\rm peak}$: the difference of the first two entries in Table~\ref{tab:GW190521_studies_setup} results in a shift of $\sim 4.7 \, {\rm ms}$ compared to the estimate of $t^{\rm pol}_{\rm peak} = \underset{t}{\max} (h_+^2 + h_{\times}^2)$ used in~\cite{LIGOScientific:2020iuh}, which is based on the \texttt{NRSur} model with a uniform prior on the source-frame component masses.
Thus, depending on the waveform model and on the definition of $t_{\rm peak}$, the largest BF is found in the range $[1,\,3.7] \, M_f$ after the signal peak.
If the $t_{\rm peak}$ estimate from the \texttt{NRSur} is unbiased, this is well outside the regime where one would expect the fundamental mode to be excited.
Compared to $t^{\rm harm}_{\rm peak}$, the $t^{\rm pol}_{\rm peak}$ reference time makes use of the polarizations, yielding a larger statistical uncertainty $O(7-10 \, {\rm ms})$~\cite{LIGOScientific:2020ufj}.
Such uncertainty should also be accounted for in computing the ``time-marginalized'' evidence (see Section~\ref{subsec:RD_systematics}).
Moreover, higher harmonics in spin-aligned systems can peak at significantly later times compared to the fundamental mode: from Table VI of~\cite{Nagar:2019wds} or Appendix E of~\cite{Cotesta:2018fcv}, the $(3,3)$ multipole can peak up to $5 M$ after the $(2,2)$ mode, while the $(2,1)$ multipole should can peak up to $12 M$ after the $(2,2)$ mode, for nonspinning systems with moderate mass ratios.
Such time delays are not taken into account by QNM superpositions, which assume that all modes are already excited at the starting time of the analysis.
In addition, the prior assumptions on the spin-aligned quasi-circular amplitudes do not take into account the effect of spin precession and eccentricity, while both effects may be relevant for this particular event~\cite{LIGOScientific:2020ibl, Gayathri:2020coq, Gamba:2021gap}.

If one relaxes either the assumption of equatorial symmetry or the amplitude ratio constraints, and assumes the \texttt{NRSur} $t^{\rm pol}_{\rm peak}$ estimate, there is no evidence for a secondary mode in the regime of validity of QNM superpositions~\cite{LIGOScientific:2021sio}.

A subsequent study~\cite{Siegel:2023lxl}, based on the \texttt{ringdown} software package (see Appendix~\ref{sec:public_codes}), assumes a \texttt{Kerr} model and implements a TD likelihood function. This study uses relaxed assumptions on the amplitudes, computes $t^{\rm harm}_{\rm peak}$ through the \texttt{NRSur} model and uses improved data conditioning, looking for evidence of multiple modes and analyzing the consistency of the corresponding model parameters with those from the full IMR waveform.
It notes that assuming a single $(\ell, m, n) = (2,2,0)$ mode, the remnant parameter posteriors do not fully overlap with the \texttt{NRSur} IMR analysis.
Instead, a fit assuming $(2,2,0) + (2,1,0)$ or $(2,2,0) + (2,1,0)+(3,2,0)$ yields complete overlap with the IMR analysis, suggesting that these multi-modal templates recover signal content that was not captured in the previous model.
Nonvanishing amplitudes for these three modes are preferred up until the signal peak, with the $(2,1,0)$ amplitude being comparable in magnitude to the $(2,2,0)$ amplitude.
This surprising feature is conjectured to arise from strong precession. However it is unclear if such large amplitudes are compatible with the near-equal mass system and inclination angle inferred with \texttt{NRSur}: in fact, studies of ringdown modeling for precessing binaries do not predict such large intrinsic amplitudes for comparable-mass systems, but only for larger mass ratios~\cite{Zhu:2023fnf, Nobili:2025ydt};
nevertheless, viewing angle effects could conceivably amplify the observed $(2,1,0)$ or $(2,0,0)$ amplitudes over the $(2,2,0)$ mode in nearly equal-mass binaries~\cite{Zhu:2023fnf}.

\begin{figure}[t]
\centering
    {%
  \includegraphics[width=0.7\columnwidth]{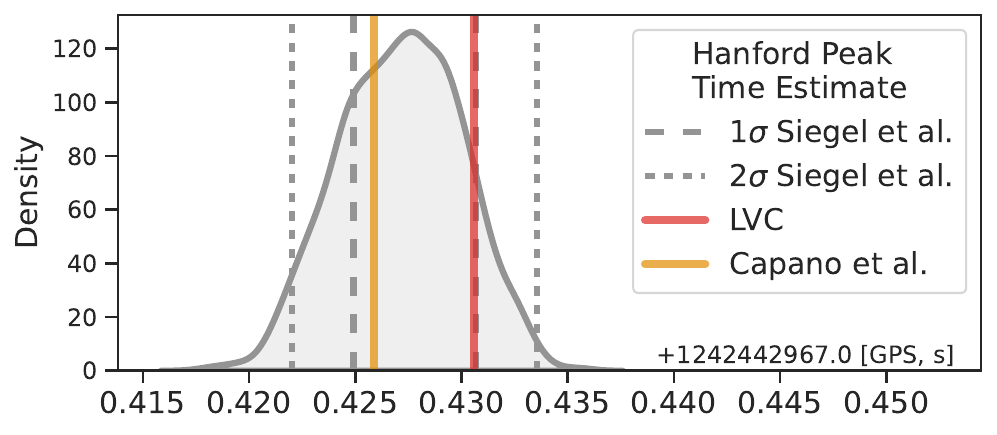}%
    }
    {%
  \includegraphics[width=0.7\columnwidth]{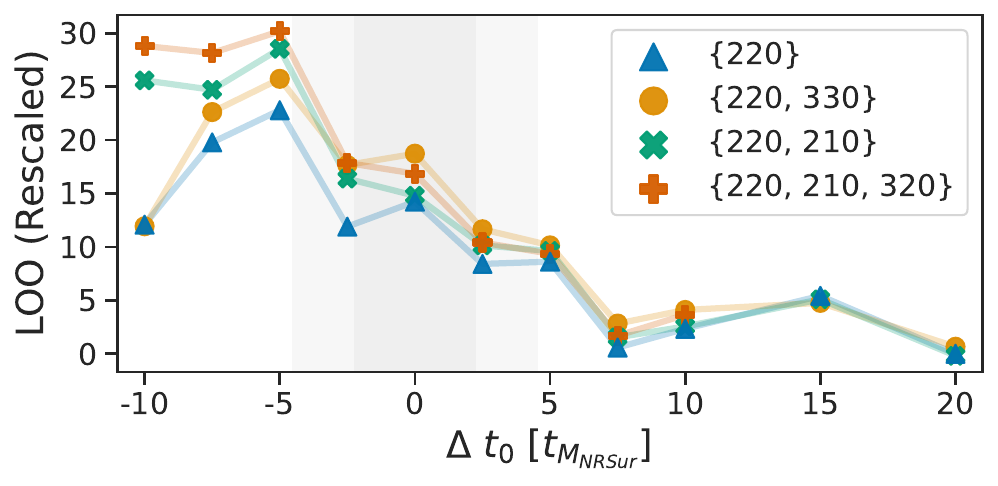}%
    }
\caption{Leave-one-out (LOO) cross-validation  (bottom) for the presence of multiple modes using a \texttt{Kerr} model, assuming generic amplitudes and $t^{\rm harm}_{\rm peak}$ estimates from the \texttt{NRSur} model (top, where the estimate is compared to the estimate of~\cite{Capano:2021etf} used in Fig.~\ref{fig:Capano_GW190521_agnostic_posterior}).
The LOO metric does not indicate statistically significant preferences for any model after the peak of the signal.
Figure taken from~\cite{Siegel:2023lxl}.}
\label{fig:Siegel_GW190521_3mode_posterior}
\end{figure}

The authors of Ref.~\cite{Siegel:2023lxl} use the leave-one-out cross-validation to quantify the statistical evidence of mode detection and allow for model comparison.
As shown in Fig.~\ref{fig:Siegel_GW190521_3mode_posterior} (which shows also the statistical uncertainty on $t^{\rm harm}_{\rm peak}$ computed from \texttt{NRSur}), they find no significant statistical preference for additional modes after the peak.

The assumptions in these different studies are compared in Table~\ref{tab:GW190521_studies_setup}.
In summary, there are three primary hypotheses to explain the ringdown of GW190521: 
(i) there are no observable sub-dominant ringdown modes, consistent with the equal-mass posterior from the LVK \texttt{NRSur} analysis~\cite{LIGOScientific:2020iuh};
(ii) the (3,3,0) mode is observable, but the remnant mass, coalescence time, and mass ratio are inconsistent with the \texttt{NRSur} analysis~\cite{Capano:2021etf};
(iii) other mode combinations -- including the (2,1,0) and possibly the (3,2,0) modes -- are preferred, as they agree with the remnant parameter posteriors inferred from the \texttt{NRSur} analysis if restricted to the highly precessing part of the posterior. The (2,1,0) and (2,2,0) mode amplitudes are comparable.
In both cases (ii) and (iii), statistical preference for additional modes is observed close to the peak, where the validity of a QNM model is questionable.

It is hard to find conclusive evidence for any of these competing hypotheses for various reasons: there are no waveforms modeling precession with an accuracy comparable to \texttt{NRSur} beyond $q=6$; it is hard to marginalize over the starting time of models that use only the fundamental QNMs; and we do not have compelling analytical models for the early ringdown regime, where constant-amplitude QNM superpositions are generally not valid. Even if we were to overcome these difficulties, the short duration of the GW190521 event would likely still make its interpretation problematic.

\begin{figure}[t]
\centering
    {%
  \includegraphics[width=0.48\columnwidth]{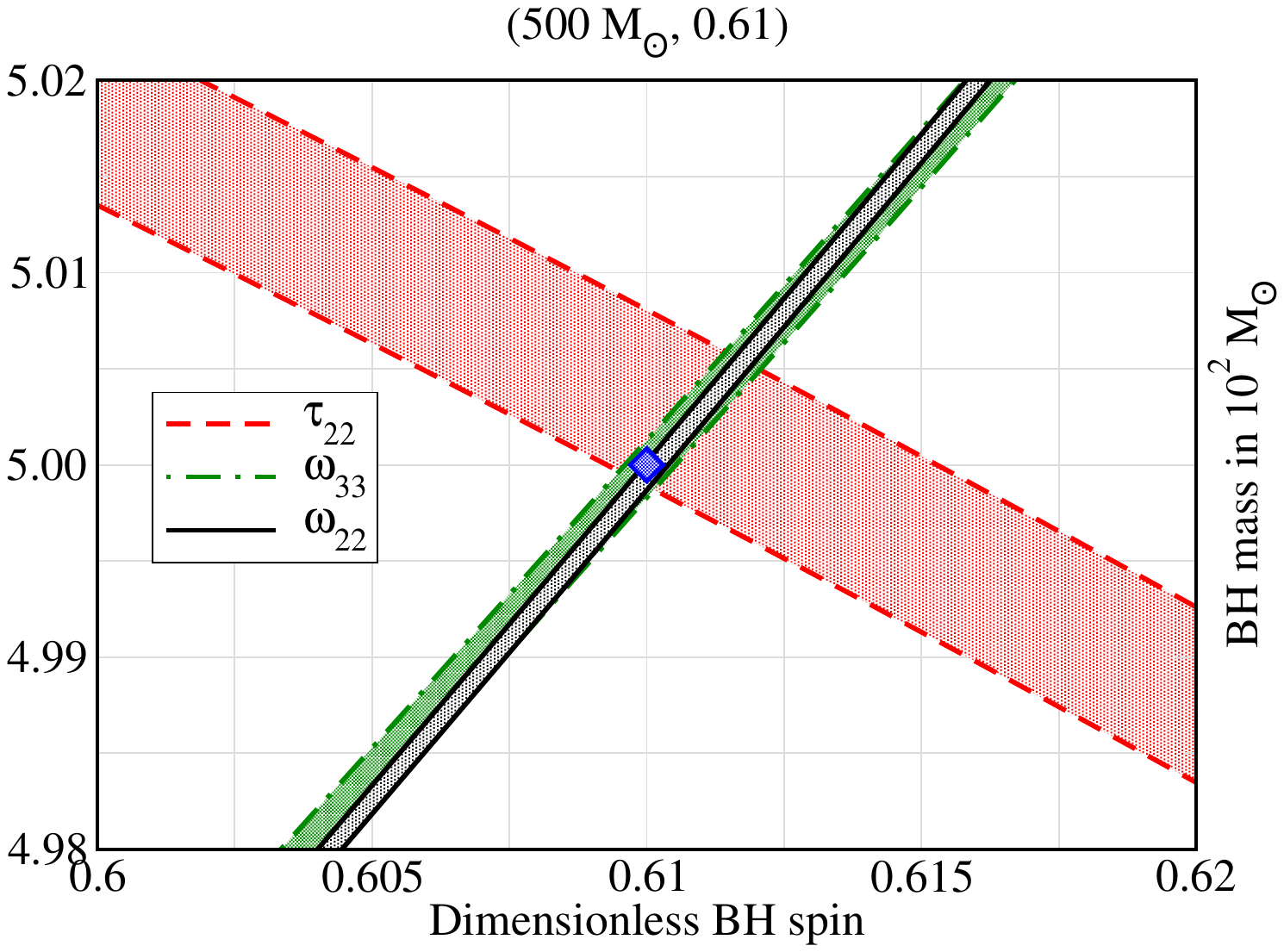}%
    }
    {%
  \includegraphics[width=0.48\columnwidth]{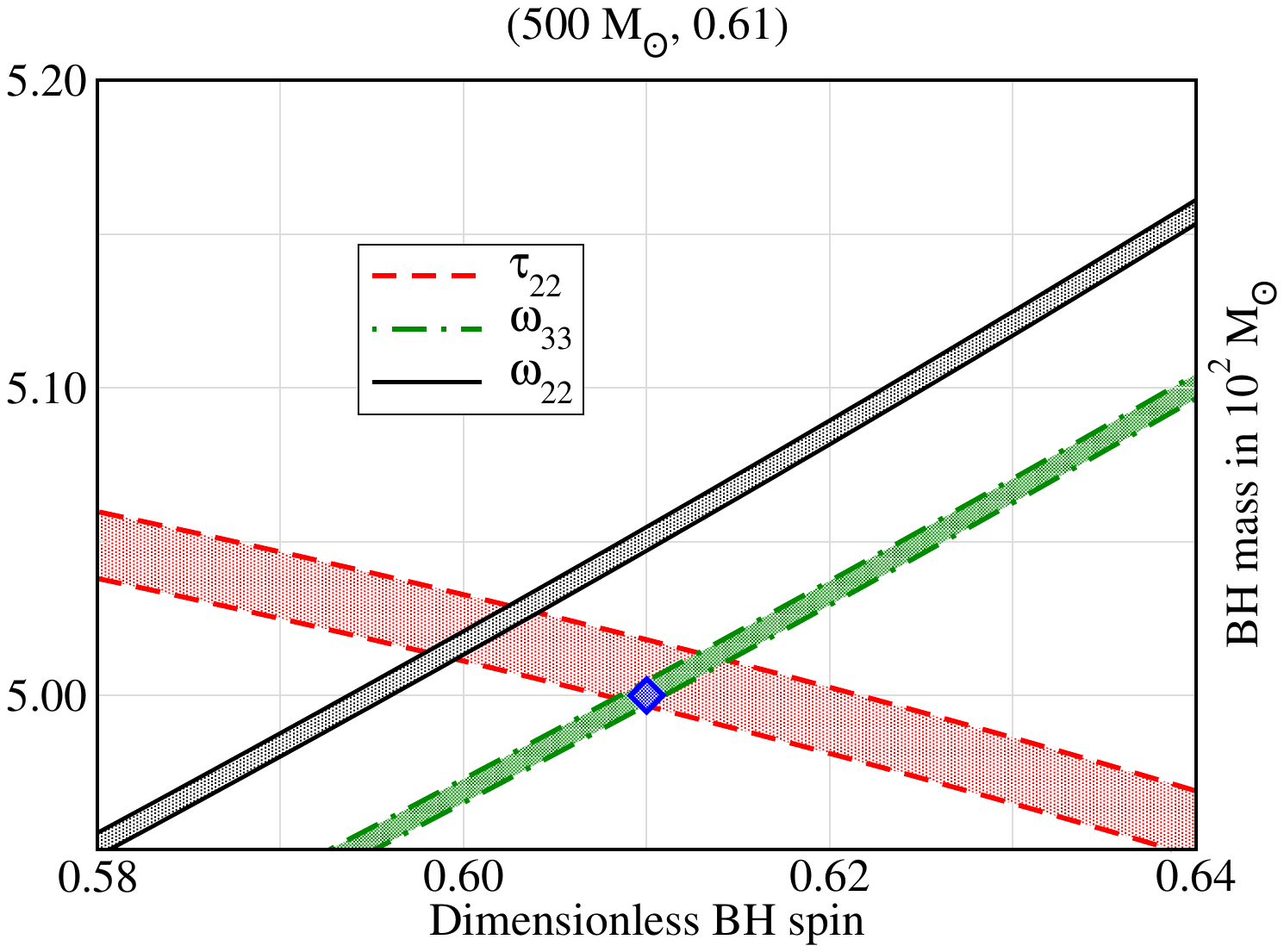}%
    }
\caption{Remnant BH parameters as inferred by inverting two frequencies and one damping time agnostic measurements, under the assumption that the $(2,2,0), (3,3,0)$, modes are present in the simulated data.
In the case of a GR simulation (left) the three inversion bands cross each other, signaling consistency with GR.
In the case of a non-GR simulation (here, $\delta\omega_{220} = 0.1$), there is no overlap region between all three bands, signaling a violation from GR.
Figure taken from~\cite{Gossan:2011ha}.}
\label{fig:No_hair_sim}
\end{figure}

\subsection{Tests of general relativity: observational results}\label{subsec:deviations_current}

\vspace{-.1cm}

\noindent \textit{Initial contributors: Carullo, Silva}

\vspace{.2cm}

\subsubsection{Parameterized (theory-agnostic) tests}
\label{sec:theory-independent}

\noindent
In the data analysis language of this chapter, the original, agnostic BH spectroscopy program consists of (i) assuming a \texttt{Damped sinusoids} template, (ii) measuring at least three frequencies or damping times, and (iii) inverting these measurements to check that they yield estimates of the remnant mass and spin consistent with the predictions of GR. 
The measurements of two parameters are, in principle, sufficient to obtain the Kerr BH mass and spin.
The third measurement (and any further measurements) can then be used to yield null tests of GR~\cite{Dreyer:2003bv,Berti:2005ys}, similar to those routinely performed for binary pulsars (see e.g. Fig.~8 of~\cite{Will:2014kxa}).
The idea is illustrated in Fig.~\ref{fig:No_hair_sim}.

This simple, robust approach allows for model-independent tests of gravity and investigations of the nature of the compact binary merger remnant.
However, even in this simplest version of the test, we can only map the frequency measurements to the remnant parameters if we know which modes are excited.
As we discussed earlier, in reality the starting time is unknown, and effects such as spin precession and eccentricity drastically complicate the mode excitation pattern, so identifying which modes are excited can be challenging.
This issue is particularly relevant for the faint signals observed at current sensitivities, but it should be less pressing in the loud observations that will be possible with XG detectors (see Section~\ref{sec:nextgen}).

These simple, agnostic BH spectroscopy tests were initially proposed and explored on simulated LISA data assuming loud, pure ringdown signals and using the Fisher matrix approximation~\cite{Berti:2005ys,Berti:2007zu}.
A first Bayesian analysis~\cite{Gossan:2011ha} proposed to use a set of agnostic parameterized deviations from the GR spectrum of the form
\begin{subequations}\label{eq:tgr_par}
\begin{align}
    f_{\ell m n }(M, a) &= f^{\rm Kerr}_{\ell m n }(M, a)  \cdot [1+\delta f_{\ell m n }],\\
    \tau_{\ell m n }(M, a)   &= \tau^{\rm Kerr}_{\ell m n }(M, a) \cdot [1+\delta\tau_{\ell m n }],
\end{align}
\end{subequations}
where $\delta\omega_{\ell m n }$ and $\delta\tau_{\ell m n }$ control the deviations from the Kerr QNM oscillation frequencies and damping times, respectively (see Section~\ref{sec_agnostic}).
This parameterization was leveraged to construct posterior probability distributions for the set of independent parameters ($M, a, \delta f_{\ell m n }, \delta\tau_{\ell m n }$), obtained using Monte Carlo or Nested Sampling~\cite{Skilling:2006gxv} techniques for simulated pure ringdown signals in XG detectors.
In the spirit of the agnostic approach described earlier, the measurement of nonzero deviation parameters in the presence of more than a single damped sinusoid would indicate a deviation from GR.
Compared to an agnostic superposition of damped sinusoids, this parameterization reduces the number of free parameters from $4N$ to $2+2N$, where $N$ is the number of modes. This  increases the sensitivity to possible GR deviations and reduces correlations in the presence of a large number of modes.
It is also possible to establish a direct connection to the mass and spin parameters of the remnant, instead of identifying them \emph{a posteriori} (as in the agnostic approach).
The evidence for deviations from GR can be quantified through an overall BF between all the possible combinations of parameterized GR deviations and the GR case, where all the deviation parameters are equal to zero~\cite{Li:2011cg, Agathos:2013upa}.

These initial studies made the simplifying assumption that the simulated data consisted of a pure superposition of damped sinusoids, with no preceding signal.
In Ref.~\cite{Carullo:2018sfu} this assumption was overcome through a windowing function, so that the inference can be formulated in the FD with small SNR loss.
The parameterized tests described above were applied to forecast the capabilities of the LVK network to constrain $\delta f_{220}$ and $\delta \tau_{220}$ at design sensitivity.
A similar windowing technique, now applied to the whitened signal, was introduced in~\cite{Cabero:2017avf}. This work highlighted the key role of sky position and starting times and proposed some strategies to take into account the uncertainty on these parameters.
A summary of these early forecasts and more recent developments can be found in Section~\ref{subsec:future_rates} below.

The parameterization of Eq.~\eqref{eq:tgr_par} can in principle capture any deviations from the GR spectrum.
However, at finite SNR, including a large number of deviation parameters will increase the prior volume and broaden the posterior distribution, and make the result uninformative.
Besides, due to the sensitivity of GW interferometers to phase measurements, frequency deviations are typically better measured than the corresponding damping times~\cite{Meidam:2014jpa}.
These considerations imply that certain combinations of deviation parameters are better suited to capturing GR deviations without unnecessarily increasing the prior volume.
The optimal parameter combinations also depend on the functional dependence of the Kerr QNM spectrum on the BH spin.
For example, in a search relying on the $(\ell,m,n) = (2,2,0)+(2,2,1)$ mode combination, it is more convenient to add parameterized modifications to the $(\ell,m,n)=(2,2,1)$ mode, rather than the $(\ell,m,n)=(2,2,0)$ mode~\cite{Isi:2021iql}.
For an arbitrary number of modes, to avoid useless repetitions of posterior samples representing the same physical state, the sampling algorithm must distinguish the different modes by attaching a ``label'' to each mode~\cite{Buscicchio:2019rir}.
For an agnostic superposition of damped sinusoids, this mode switching issue is typically addressed by ordering the modes by increasing frequency, since frequencies are better measured.
This choice is not unique: we can order in terms of any other mode parameter, at the expenses of sampling efficiency if the posterior distributions for the same parameter of two different modes overlap.
There is no loss of generality by labeling in terms of a single parameter (but there would be if we imposed more than one ordering condition).
For Kerr BHs, the ordering is intrinsically given by the GR spectrum. However, when we consider tests based on Eq.~\eqref{eq:tgr_par} mode switching can (in principle) happen, especially for overtones (which are in close in frequency and have poorly measured damping times), and the (required) ordering choice can affect one-dimensional marginalized constraints~\cite{LIGOScientific:2020tif}.
Similar ideas are discussed in Refs.~\cite{Biscoveanu:2020are, Gerosa:2024ojv} in the context of the identification of binary merger progenitors.

So far, complex amplitudes and other extrinsic parameters were considered free, in the sense that they are directly constrained from the data.
When the complex amplitudes are not free to vary, but predicted by GR as well (as in the \texttt{KerrBinary}, \texttt{KerrPostmerger} or templates), consistency tests become possible even with a single mode~\cite{Gennari:2023gmx, Forteza:2022tgq}.
The result is a model-dependent test of GR that can be performed using weaker signals, but now under more stringent assumptions on the emission process (for example, at present, most amplitude models are limited to quasi-circular orbits).

Pushing this idea even further, it is possible to measure frequency deviations using ringdown signals by considering the entire IMR waveform. 
In the \texttt{pSEOB} framework~\cite{Brito:2018rfr, Ghosh:2021mrv, Maggio:2022hre, Toubiana:2023cwr,Pompili:2025cdc}, this idea is implemented with a parameterized EOB waveform model (see Section~\ref{sec:effective-one-body} for more details).
The analysis includes the entire IMR waveform rather than isolating the post-merger ringdown phase. Its assumes that the GR predictions apply up to the merger phase and describes the merger-ringdown with phenomenological time-dependent amplitudes, similarly to the \texttt{KerrPostmerger} search based on the \texttt{TEOB} template~\cite{Gennari:2023gmx} (cf.~Section~\ref{sec:effective-one-body}). Each QNM present in the baseline GR waveform model is modified using the parameterization of Eq.~\eqref{eq:tgr_par}.
For example, focusing on the dominant (2,2,0) mode, the \texttt{pSEOB} analysis would yield a posterior distribution on $\delta f_{220}$ and $\delta\tau_{220}$.
This approach has two main advantages: it leverages the SNR of the full IMR signal, and it does not require a separate estimation of the ringdown starting time parameters. 
A drawback of the approach is that it is inevitably subject to a larger number of GR constraints. The mass and the spin that determine the QNM frequencies are largely constrained from the inspiral contributions, the inspiral-plunge and merger-ringdown ``stitching'' time. In addition, the mode amplitudes and phases are calibrated against NR simulations in GR, and the modifications to the QNM frequencies assume continuity with the GR spectrum.
For these reasons, the \texttt{pSEOB} analysis will in general work best if the deviations from the baseline GR model -- either from modifications to the QNM frequencies or from unmodeled physics -- are perturbatively small. 
If instead the deviations lay outside the \texttt{pSEOB}'s ``functional space,'' the test is expected to be biased towards GR, and it should not be as effective at characterizing possible sources of new physics.
In fact, when considering the performance of parameterized tests of GR, it is important to distinguish between the ``detection'' of a beyond-GR signature and its ``characterization''~\cite{Carullo:2021dui}.
Highly informed templates like \texttt{pSEOB} should be able to flag an inconsistency with GR even if deviations are ``large,'' but they are expected to under-perform more agnostic templates when it comes to identifying physical shifts in the system parameters.\\

\begin{figure*}[t]
    \includegraphics[width=0.5\textwidth]{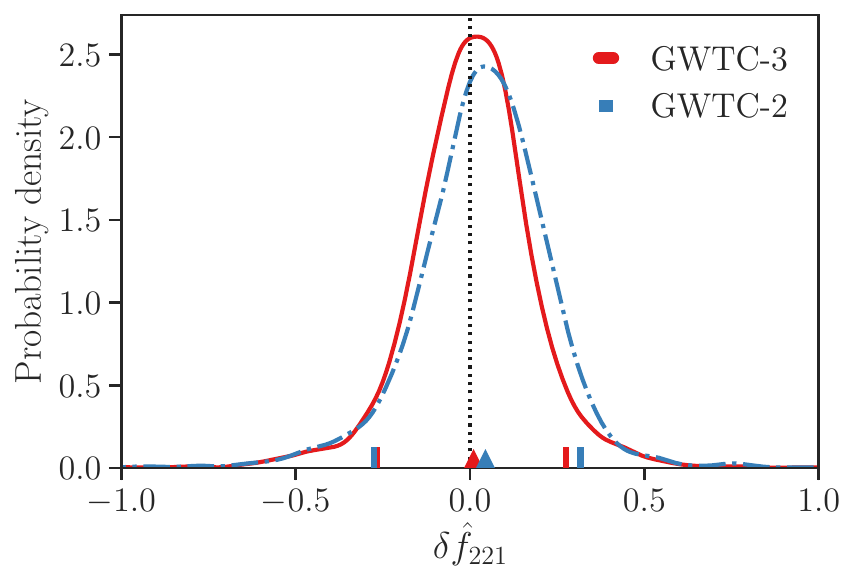}
    \includegraphics[width=0.45\textwidth]{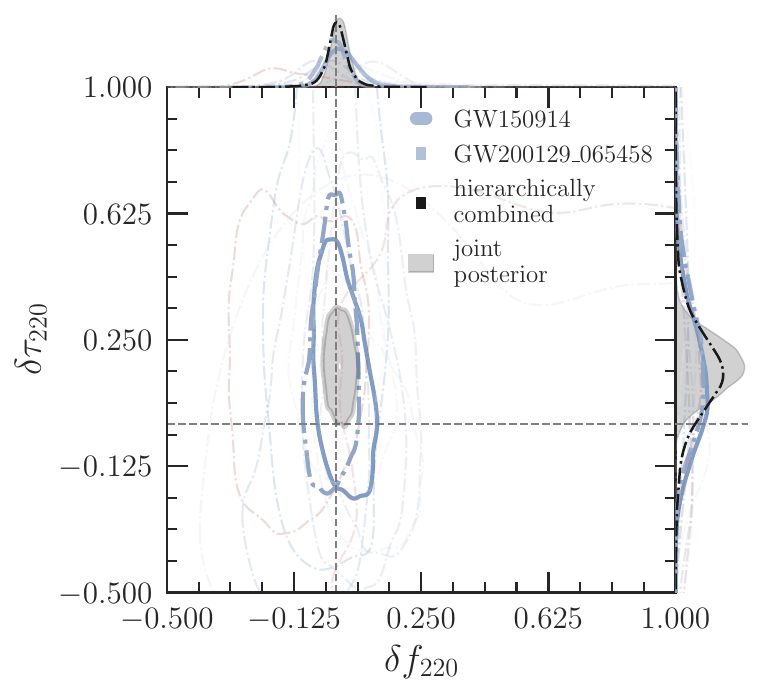}
    \caption{Combined posterior distribution on the frequency and damping time deviation parameters defined in Eq.~\eqref{eq:tgr_par} as inferred from LVK observations, including all GWTC-3 events passing the selection criteria of the \texttt{pyRing} (left) and \texttt{pSEOB} (right) pipelines.
    In both cases the results are consistent with GR.
    Figures taken from~\cite{LIGOScientific:2021sio, Pompili:2025cdc}.
    }
    \label{fig:LVK_GWTC3_pyRing_pSEOB} 
\end{figure*}

\noindent
\textit{Results} 
Observational constraints on parameterized deviations from the fundamental mode of the form~\eqref{eq:tgr_par} using GW150914 can be found in~\cite{Brito:2018rfr, Carullo:2019flw}, while other papers consider catalogs of LVK events~\cite{Ghosh:2021mrv, LIGOScientific:2020tif, LIGOScientific:2021sio, Gennari:2023gmx,Pompili:2025cdc}.
The first ringdown-only constraints based on $(2,2,0)+(2,2,1)$ QNM superpositions starting at the signal peak found $| \delta f_{221} | < 0.4$, $| \delta \tau_{221} | < 0.8 $ at $90\%$ credibility~\cite{Isi:2019aib}.
A re-analysis using a wavelet approach found similar results~\cite{Finch:2022ynt}.
These constraints were later refined by combining multiple events using the GWTC-2~\cite{LIGOScientific:2020tif} and GWTC-3~\cite{LIGOScientific:2021sio} LVK catalogs. The most up-to-date analysis, shown in the left panel of Fig.~\ref{fig:LVK_GWTC3_pyRing_pSEOB}, finds $| \delta f_{221}| < 0.28$ at $90 \%$ credibility.
These constraints are based on a hierarchical method to combine single-event inferences which allows for the possibility of having different deviations in different GW events~\cite{Zimmerman:2019wzo, Isi:2019asy}.
The idea is to impose that the parent distribution of each realization is a Gaussian with parameters $\mu$, $\sigma$, and quantify the agreement with GR (in which case the distribution would be a delta function).
The latest LVK data imply $\mu = 0.01 \pm 0.18$, $\sigma < 0.22$, indicating no evidence for GR violations at the population level~\cite{LIGOScientific:2021sio}.
The combined BF is $\mathcal{O}_{GR}^{modGR} = -0.90 \pm 0.45$ in favor of GR, while the damping time deviations were found to be uninformative~\cite{LIGOScientific:2021sio}.

Similar tests on GW190521 can be found in~\cite{Capano:2021etf}. Their analysis using the dominant $(2,2,0) + (3,3,0)$ \texttt{Kerr} combination and an amplitude model compatible with spin-aligned binaries starting at $\sim t_{\rm peak}^{\rm harm}+ 3.7 \, M_f$ finds $| \delta f_{330} | < 0.09$ at $90 \%$ credibility.
A generic \texttt{Kerr} amplitude model assuming the $(2,2,0)+(2,1,0)+(3,2,0)$ combinations starting at $\sim t_{\rm peak}^{\rm harm}$ finds that measurements of the deviation parameters on the fundamental mode, $\delta f_{220}$ and $\delta \tau_{220}$, are uninformative; however $\delta f_{320} < 0.2$ at $90 \%$ credibility, while the lower bound on this deviation parameter is prior driven.
A \texttt{KerrPostmerger} combination of the $(2,2) + (3,3)$ modes at $t_{\rm peak}^{\rm pol}$ yields $| \delta f_{22} | < 0.4$, $\delta \tau_{22} < 0.5$ for the loudest signals, with uninformative bounds on the $(3,3)$ deviation parameters.

These results are valid phenomenological consistency tests of GR. They are somewhat analogous to similar results obtained by varying the phenomenological $(\alpha, \beta)$ merger-ringdown calibration parameters of \texttt{Phenom} waveform families~\cite{LIGOScientific:2016lio,LIGOScientific:2019fpa, LIGOScientific:2020tif}, which yield constraints at the $10 \%$ level ($90 \%$ credibility).
However these bounds are not necessarily genuine ``no-hair'' tests in the spirit of the original spectroscopy proposal, because constant-amplitude superpositions at the peak need not represent the physical QNM content of the system (see~\cite{Baibhav:2023clw, Khera:2023oyf, Nee:2023osy, Carullo:2023gtf, Zhu:2023mzv} and Section~\ref{subsec:overtones}).
Similar considerations apply to GW190521 constraints involving the $(3,3,0)$ mode~\cite{Capano:2021etf} and other mode combinations~\cite{Siegel:2023lxl}, because the starting time of the analysis is close to the peak and model systematic uncertainties are important. 
For other results based on IMR consistency see~\cite{Isi:2020tac} and Section~IVB in~\cite{LIGOScientific:2021sio}.

Less agnostic (and more informative) constraints can be found using the \texttt{ParSpec} parameterization~\eqref{eq:parspec_par} following~\cite{Maselli:2019mjd}. Here, the idea is to perform a series expansion of the frequencies in terms of the remnant spin, in which the deviations from GR are introduced as a perturbative correction in the (small) coupling constant that characterizes beyond-GR modifications. These perturbative corrections can be either scale-free or dimensionful. This parameterization was implemented in \texttt{pyRing} and applied to all events with detectable ringdown signals in the GWTC-2 catalog~\cite{Carullo:2021dui}.
For dimensionless couplings ($p=0$ in the \texttt{ParSpec} notation), the bounds is $\delta f^{(0)}_{220} = -0.05 \pm 0.05$, while for dimensionful corrections of the form $\ell_p = \sqrt[p]{\alpha}$ one finds $\ell_2 < 23 \mathrm{km}$, $\ell_4 < 35 \mathrm{km}$, $\ell_6 < 42 \mathrm{km}$.
The case of an individual charge (corresponding to Kerr-Newman BHs in GR for $p=2$) was studied in~\cite{Carullo:2021oxn,Gu:2023eaa}. At current sensitivities, the LVK spin measurements  yield upper bounds on the normalized BH charge parameter $q <  0.33$ ($90\%$ credibility) for the best measured events.
Ringdown measurements with future detectors would be able to directly {\it measure} hypothetical charges $q \simeq 0.5$, and even lower values of the charge should be measurable through IMR consistency tests~\cite{Carullo:2021oxn}.
Even better constraints may be possible by developing full IMR templates of Kerr-Newman mergers, based e.g. on the numerical simulations in Refs.~\cite{Bozzola:2020mjx, Bozzola:2021elc}.

The larger SNR available in the \texttt{pSEOB} pipeline -- initially introduced in Ref.~\cite{Brito:2018rfr} building upon the \texttt{EOBNRv2HM} waveform model for quasi-circular, nonspinning BBHs~\cite{Pan:2011gk} --
leads to more optimistic predictions: it would take approximately 30 GW150914-like
BBHs detected by Advanced LIGO and Virgo at design sensitivity to carry out no-hair tests by measuring the frequencies of both the $(2,2,0)$ and $(3,3,0)$ modes, and the decay time of the $(2,2,0)$ mode with an
accuracy of $\lesssim 5\%$ at the $2\sigma$ level~\cite{Brito:2018rfr}.
Follow-up work~\cite{Ghosh:2021mrv} incorporated the parameterization~\eqref{eq:tgr_par} into the quasi-circular spin-aligned waveform model \texttt{SEOBNRv4HM}~\cite{Cotesta:2018fcv}.
By hierarchically combining the results from individual events up to the O3a observing run, they find the constraints $\delta f_{220} = 0.03^{+0.10}_{-0.09}$ and 
$\delta \tau_{220} = 0.10^{+0.44}_{-0.39}$ at 90\% credibility.

The \texttt{pSEOB} analysis was augmented in~\cite{Maggio:2022hre} (using \texttt{SEOBNRv4HM\_PA}~\cite{Mihaylov:2021bpf} as the baseline GR model) to incorporate additional fractional changes to (i) the instant at which the GW mode amplitude peaks, (ii) the instantaneous frequency at this time, and (iii) the value of the peak amplitude. 
By modifying (a subset) of the quantities that describe the merger, the resulting \texttt{pSEOB} model becomes more flexible in capturing potential strong-field deviations from GR, and expands the functional space covered by the model. 
As an example, Ref.~\cite{Maggio:2022hre} analyzed GW150914
allowing for fractional changes in the GW mode amplitude $\delta A$ (assumed to be the same for all GW modes present in \texttt{SEOBNRv4HM\_PA}), in addition
to $\delta f_{220}$ and $\delta \tau_{220}$. These three parameters were constrained to be $\delta A = 0.03^{+0.29}_{-0.20}$, $\delta f_{220} = {0.041}^{+0.151}_{-0.084}$, and $\delta \tau_{220} = {0.04}^{+0.27}_{-0.29}$ at 90\% credibility. The frequency and damping times constraints are compatible with those found for the same event in Ref.~\cite{Ghosh:2021mrv} (where amplitude variations were not considered) -- namely, $\delta f_{220} =  0.05^{+0.11}_{-0.07}$ and $\delta \tau_{220} = 0.07^{+0.26}_{-0.23}$.

A particularly interesting result of Ref.~\cite{Maggio:2022hre} is that the GW200129 event
displays strong {\it violations} of GR on the peak-amplitude parameter $\delta A$ (the QNM deviation parameters remain compatible with GR). This result was interpreted as a false violation of GR originating either from waveform systematics (mismodeling of spin precession that is not present in the baseline GR waveform model) or from data-quality issues, depending on the interpretation of this particular event. This case study is an important cautionary tale: waveform systematics, noise artifacts or astrophysics can easily lead to false violations of GR even at current sensitivities~\cite{Gupta:2024gun}.

The \texttt{pSEOB} framework was recently extended to include spin precession effects~\cite{Pompili:2025cdc}, using the quasi-circular, spin-precessing \texttt{SEOBNRv5PHM} model~\cite{Ramos-Buades:2023ehm} as the GR baseline.The analysis of a hierarchical combination of events finds $\delta f_{220} = 0.00^{+0.06}_{-0.06}$ and $\delta \tau_{220} = 0.15^{+0.26}_{-0.24}$ at 90\% credibility.  These results are broadly consistent with those reported by the LVK Collaboration (using an earlier version of the \texttt{pSEOB} model for aligned-spin binaries) for the same events.

The \texttt{pSEOB} pipeline is now among the standard battery of GR tests performed by the LVK Collaboration. The latest bounds~\cite{Pompili:2025cdc} from a hierarchical combination of events are
\begin{equation}
\delta {f}_{220} = 0.01 \pm 0.04, 
\quad \textrm{and} \quad
\delta {\tau}_{220} = 0.17^{+0.14}_{-0.13}.
\end{equation}
Because a sufficiently high inspiral SNR is typically required to break the degeneracy between $\delta f_{220}$ and the remnant's mass $M_f$, events are normally analyzed if $\rm SNR \geq 8$ in both the pre- and post-merger regimes~\cite{Ghosh:2021mrv}. 
The corresponding two-dimensional posterior distributions are shown in Fig.~\ref{fig:LVK_GWTC3_pyRing_pSEOB}. There, we report both the hierarchically-combined results mentioned above and a joint deviation obtained by multiplying the single-event likelihoods, which includes the GR value only in the tails of the posterior distribution. This result may still be compatible with the statistical uncertainty of the analysis, given the finite number of events~\cite{Pacilio:2023uef}.

\subsubsection{Theory-specific tests}
\label{subsec:TGR_theory-specific}

Recent progress in calculations of QNMs emitted by rotating BHs in modified theories of gravity (cf.~Section~\ref{subsec:theory-spec}) allows us to place constraints on these theories using ringdown data.
These searches for new physics have improved sensitivity relative to previous agnostic tests, at the expense of assuming a specific theory of gravity.
Reference~\cite{Silva:2022srr} used the $(\ell, m, n) = (2,2,0)$ QNM calculations to linear order in spin in EdGB, dCS, and cubic and quartic EFTs of GR from Refs.~\cite{Pierini:2021jxd,Wagle:2021tam,Cano:2021myl}.
These computations fix the theory-agnostic 
parameters up to linear order in the \texttt{ParSpec} spin expansion, while all other theory-agnostic parameters (at quadratic and higher orders in spin) were set to zero. 
In other words, the two lowest powers in the spin expansion include beyond-GR correction, whereas higher-order terms in the spin expansion are the same as in GR.
The resulting theory-informed \texttt{ParSpec} parameterization was then used to modify the $h_{22}$ mode of the pSEOB waveform model (cf.~Section~\ref{sec:fundametal_l2_obs}). 
and to carry out a full Bayesian parameter estimation analysis of two events: GW150914 and GW200129. 
The analysis does not constrain EdGB gravity, but it places upper bounds on the fundamental length scale of dCS gravity ($\ell_{\rm dCS} \leqslant 38.7$~km) as well as cubic- ($\ell_{\rm cEFT} \leqslant 38.2$~km) and quartic-order ($\ell_{\rm qEFT} \leqslant 51.3$~km) EFTs of GR at 90\% credible level. 
Note that dCS gravity is a good example of a theory which is presently unconstrained by inspiral-only analyses, while it can be constrained using merger-ringdown radiation. 

This ``\texttt{pSEOB} plus \texttt{ParSpec}'' approach has some limitations. For example, it does not accommodate the two branches of gravitational QNMs that generally arise in modified gravity theories because of isospectrality breaking.
In particular, Ref.~\cite{Silva:2022srr} made an educated guess by adopting the branch that has the slowest decaying $f_{220}$ QNM. 
Moreover, linear-in-spin beyond-GR predictions are not accurate enough to describe typical BBH merger remnants, which have Kerr parameters $a_f \simeq 0.7$. 
A third important limitation (and a typical assumption of the \texttt{pSEOB} approach) is that the inspiral part of the waveform was assumed to be the same as in GR.

\begin{figure*}[t]
\centering
\includegraphics[width=0.48\textwidth]{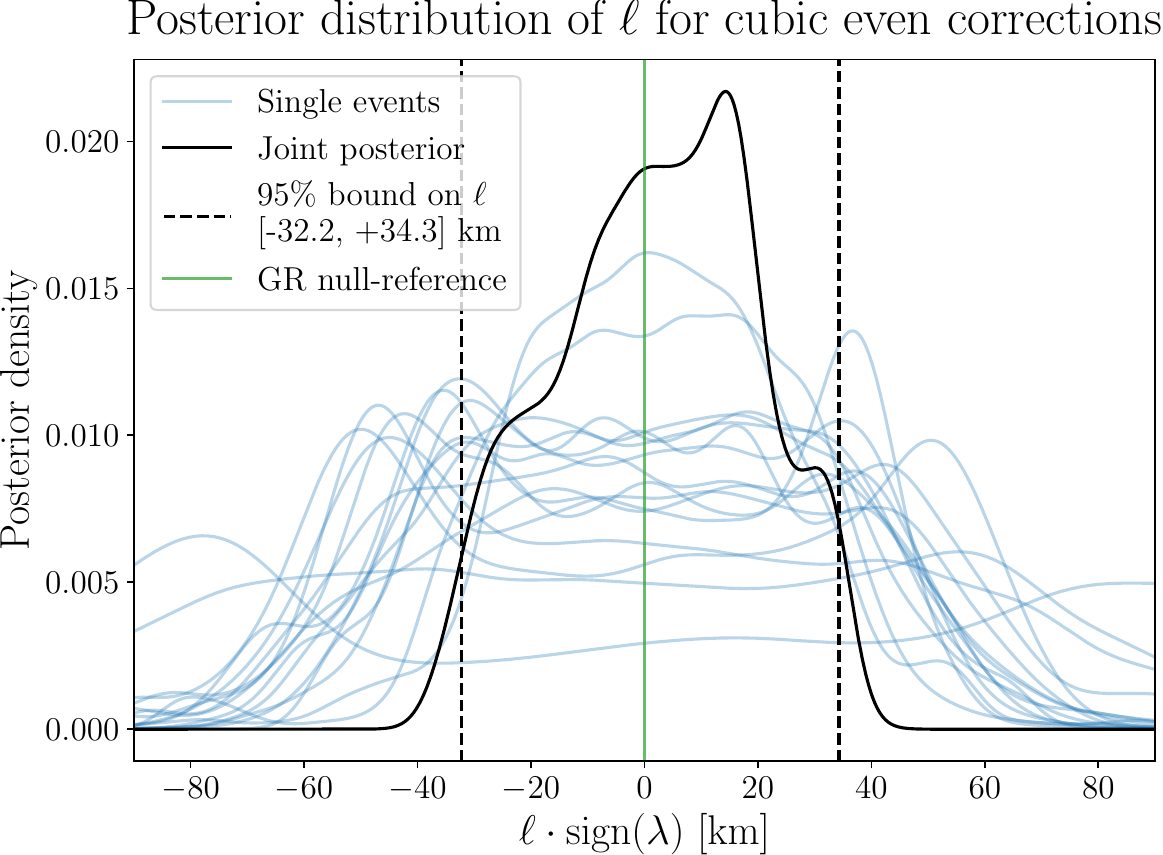}
\includegraphics[width=0.46\textwidth]{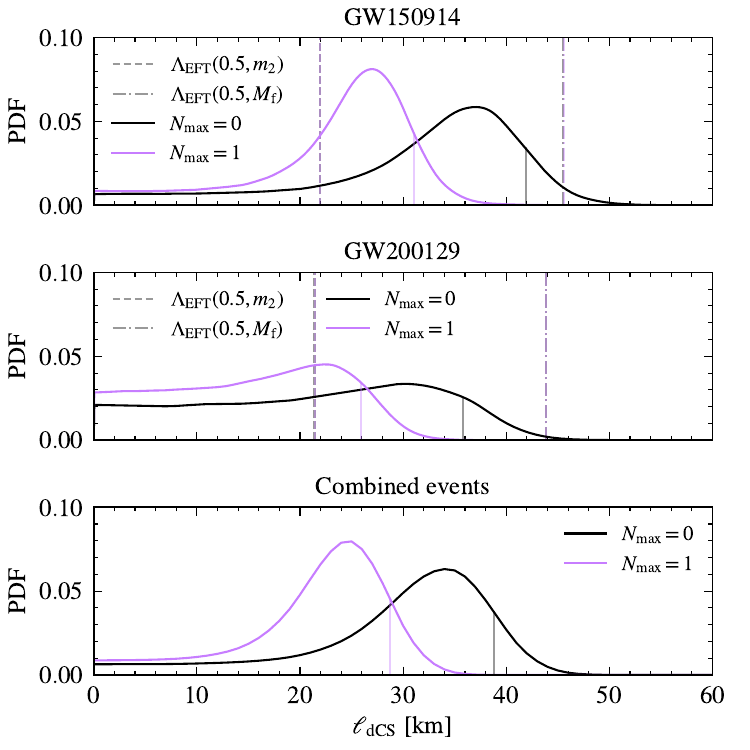}
\caption{Constraints on the length scale of new physics $\ell$.
Left: joint constraints on cubic even corrections (which are valid up to large spins) including all LVK GWTC-3 signals with detectable ringdown signatures.
Right: dCS couplings on selected informative events found by using linear-in-spin corrections.
Figures taken from~\cite{Maenaut:2024oci,Silva:2022srr}.}
\label{fig:eft_dcs_bounds}
\end{figure*}

These shortcomings were addressed to some extent in Refs.~\cite{Maenaut:2024oci,Liu:2024atc,Chung:2025wbg}.
Specifically, Ref.~\cite{Maenaut:2024oci} used the results from Ref.~\cite{Cano:2023jbk,Cano:2024ezp} to construct a prograde $(\ell,m,n) =(2,2,0),(2,2,1)$ ringdown model, including corrections to the spectrum for the cubic and quartic EFTs of GR up to twelfth order in a small spin expansion, and allowing for isospectrality breaking.
This model was used to perform a complete Bayesian analysis of all events in the GWTC-3 catalog with clear and confident ringdown signatures within the \texttt{pyRing} infrastructure (cf.~Section~\ref{sec:fundametal_l2_obs}).
By combining posterior distributions of single events from GWTC-3, the length scale associated to these higher-curvature terms is bound to be $\ell_{\rm EFT} \lesssim 35$~km.
In~\cite{Chung:2025wbg}, the results from~\cite{Chung:2024ira, Chung:2024vaf, Chung:2025gyg} were used to construct a similar ringdown model, accurate up to $a \lesssim 0.75$, incorporating isospectrality breaking for quadratic gravity theories.
This model was applied to analyze the most informative ringdown signals in the GWTC-3 catalogs using \texttt{pyRing}.
By combining the posterior distributions of all analyzed events, the authors find a constraint on the quadratic-gravity coupling length scale of $\ell_{\rm EFT} \lesssim 34 - 47$~km, as well as the first constraint on axi-dilaton gravity, $\ell_{\rm AD} \lesssim 34$~km. 
Results from Refs.~\cite{Silva:2022srr, Maenaut:2024oci} are shown in Fig.~\ref{fig:eft_dcs_bounds}.

Full IMR analyses including both inspiral and ringdown deviations induced by an EsGB term and EFT higher-curvature corrections~\cite{Julie:2024fwy, Liu:2024atc} analyzed only low-mass signals, for which the ringdown contributions are negligible.

\subsection{Echoes: data analysis methods and observational status}\label{subsec:echoes_DS}

\vspace{-.1cm}

\noindent \textit{Initial contributors: Abedi, Lo}

\vspace{.2cm}

Here we summarize observational efforts to search for GW echoes (see Section~\ref{subsec:echoes_theory}). 
Some initial studies suggested positive evidence for echoes in the data following BBH merger events in LVK data~\cite{Abedi:2016hgu, Conklin:2017lwb}. 
However, subsequent efforts have reported both positive~\cite{Abedi:2018npz, Holdom:2019bdv, Conklin:2021cbc, Abedi:2021tti} and negative or inconclusive findings~\cite{Westerweck:2017hus,Nielsen:2018lkf,Salemi:2019uea,Abedi:2022bph,Lo:2018sep,Uchikata:2019frs,Tsang:2019zra,LIGOScientific:2020tif,Wang:2020ayy, Westerweck:2021nue,Ren:2021xbe, LIGOScientific:2021sio,Miani:2023mgl,Uchikata:2023zcu}. 
The disagreement stem from different choices of statistical methodologies and/or models, but there is full agreement on the fact that current data show no conclusive detection of echoes.
The detection debate has stimulated improvements in both models and statistical techniques, highlighting that extensive detection campaigns and higher SNRs are required to robustly detect and characterize echoes.

\subsubsection{Classification of echo analysis approaches}

Efforts to detect echoes in GW signals have predominantly relied on three different methodologies:

\begin{itemize}
\item \textbf{Waveform-dependent approaches}: These studies use specific assumptions about the expected form of the echo signal, often based on physical parameters of the post-merger remnant, such as its mass and spin. Examples include studies that assume echo templates from scenarios with potential quantum gravitational structures near the horizon~\cite{Abedi:2016hgu,Westerweck:2017hus,Nielsen:2018lkf,Lo:2018sep,Uchikata:2019frs,LIGOScientific:2020tif,Wang:2020ayy,Westerweck:2021nue,LIGOScientific:2021sio,Abedi:2021tti,Uchikata:2023zcu,Abedi:2022bph}. 
These approaches are limited by the degree of agreement between template models and actual echo waveforms, but would deliver the most sensitive constraints if the data were described by the echo models under consideration.
Highly faithful models are hard to achieve, given the vast landscape of theoretical possibilities.
An imperfect representation of the echo signal (catching most but not all of the signal power) is expected to be effective at \textit{detecting} echoes~\cite{Abedi:2021tti}, but the unbiased characterization of the underlying physics depends heavily on accurate waveform templates.
Some of the waveform models for GW echoes are implemented in a publicly available \texttt{python} package, \texttt{echoes\_waveform\_model}~\cite{echowfm} (see Appendix~\ref{sec:public_codes}).

\item \textbf{Weakly modeled approaches}: These methods try to avoid specific waveform assumptions by focusing on generic signatures or patterns indicative of echoes, allowing for a broader search for potential deviations from the GR waveform. Model-agnostic analyses, such as those by~\cite{Abedi:2018npz,Conklin:2017lwb,Salemi:2019uea,Holdom:2019bdv,Ren:2021xbe,Wu:2023wfv,Abedi:2021tti,Conklin:2021cbc,Miani:2023mgl}, but may be more sensitive to noise and statistical interpretation issues.
For well-modeled scenarios, they will also naturally be less sensitive.

\item \textbf{Electromagnetic confirmation}: 
A complementary line of investigation explores potential electromagnetic counterparts to GW echoes. 
For example, astrophysical considerations suggest that the remnant of GW170817 collapsed into a BH  within $t_{\rm{coll}} \lesssim 2 \, \mathrm{s}$~\cite{Granot:2017tbr, Gottlieb:2017pju, Nakar:2018cbe, Metzger:2018qfl, Xie:2018vya, Gill:2019bvq, vanPutten:2019kca, Hamidani:2019qyx, Murguia-Berthier:2020tfs}.
If the collapse were delayed, an echo signal for this event at around $\simeq 1$s after the merger could independently confirm or refute the existence of echoes~\cite{Abedi:2018npz, Abedi:2020sgg,Abedi:2020ujo}.
In this scenario, the echoes are generated by GWs with a resonant harmonic frequency, arising from the cavity between the angular momentum barrier and the near-horizon membrane formed following the collapse of a binary neutron star merger into a BH.

\end{itemize}

\subsubsection{Early claims and up-to-date understanding}

We will now discuss two different, early echo searches in GW data and their current understanding.

\noindent
\textit{Binary black hole mergers in O1}
A simple phenomenological template for echoes, the ``ADA model,'' was introduced in~\cite{Abedi:2016hgu}.
Searches using this template indicated tentative evidence for echoes at a significance level of approximately \(2.5\sigma\) in data collected during the first LIGO observing run (``O1'')~\cite{LIGOScientific:2018mvr}.
Significance levels were interpreted using a two-tailed Gaussian probability, where p-values of 68\% and 95\% correspond to \(1\sigma\) and \(2\sigma\), respectively. 

These data were subsequently re-analyzed by different groups using different data analysis methodologies, finding no evidence for echoes~\cite{Westerweck:2017hus,Nielsen:2018lkf,Salemi:2019uea,Abedi:2022bph,Lo:2018sep,Uchikata:2019frs,Tsang:2019zra,LIGOScientific:2020tif,Wang:2020ayy, Westerweck:2021nue,Ren:2021xbe, LIGOScientific:2021sio,Miani:2023mgl,Uchikata:2023zcu}.
Another independent study~\cite{Salemi:2019uea} used a weakly-modeled approach, based on the \texttt{coherent WaveBurst (cWB}) software, not originally designed to search for echoes.
They initially identified post-merger features that could be interpreted as echoes, but subsequent rigorous re-analysis by the same group found no evidence for these features~\cite{Miani:2023mgl}.

In Table~\ref{table_echo} we summarize the results obtained by various groups for the O1 events. Here, the smaller the p-value, the stronger the evidence against the null hypothesis. A p-value of $=2.87 \times 10^{-7}$ corresponds to the probability of observing a result $\geq5\sigma$ (in one tail) under a Gaussian distribution if the null hypothesis is true. Jeffreys' scale for interpreting BFs~\cite{10.2307/2291091} classifies a BF in the range of 10 to 100 as strong evidence, and a BF greater than 100 as decisive evidence. 
It is important to note that p-values and BFs are distinct statistical quantities with no known mathematical relationship. Since both quantities are used for hypothesis testing, we report them both in this table and elsewhere in this section.\\

\begin{table}
\begin{center}
\resizebox{\textwidth}{!}{%
\begin{tabular}{ |c|c|c|c|c|c|c| }
\hline
 & \multicolumn{5}{c|}{p-value} & Bayes factor \\
\hline
Event & ADA~\cite{Abedi:2016hgu} & Westerweck et al.~\cite{Westerweck:2017hus} & Uchikata et al.~\cite{Uchikata:2019frs} & Lo et al.~\cite{Lo:2018sep} & Salemi et al.~\cite{Salemi:2019uea} & Nielsen et al.~\cite{Nielsen:2018lkf} \\
\hline
GW150914 & 0.11 & $0.238 \pm 0.043$ & $0.157 \pm 0.035$ & 0.806 & $0.94 \pm 0.02$ & 0.16\\
\hline
GW151012 & - & $0.063 \pm 0.022$ & $0.047 \pm 0.019$ & 0.0873 & $0.0037 \pm 0.0014$ & 3.5 \\
\hline
GW151226 & - & $0.476 \pm 0.061$ & $0.598 \pm 0.069$ & 0.254 & $0.025 \pm 0.005$ & 1.5\\
\hline
Total & 0.011 & $0.032 \pm 0.016$ & $0.055 \pm 0.021$ & - & - & - \\
\hline
\end{tabular}%
}
\caption{Results of searches for echoes in the events detected during the first LVK observing run (O1).}\label{table_echo}
\end{center}
\begin{center}
\end{center}
\end{table}

\noindent
\textit{Binary neutron star merger (GW170817)}
A search for echoes following the binary neutron star merger GW170817 used the weakly-modeled approach proposed in~\cite{Abedi:2018npz}, and focused on scenarios where the merger results in either prompt or rapid collapse into a BH. Echoes are hypothesized to arise due to Planck-scale modifications near the event horizon, repeating with a characteristic time delay $(\Delta t_{\rm echo})$ estimated to be $\sim 4.7 ^{\rm{ms}}\ (1+1/\sqrt{1-\chi^2})$ for a $2.7 M_\odot$ BH.

The search targeted fundamental echo frequencies $f_{\rm echo} = \Delta t_{\rm echo}^{-1}$ in the range $[63,\,92]\,$Hz, and identified potential signals within the first second post-merger. By cross-correlating spectrograms from LIGO detectors, periodic peaks corresponding to echo harmonics were claimed with \(4.2\sigma\) significance. 
This signal, observed approximately $1\,$s after the BNS merger GW170817~\cite{Abedi:2018npz}, was interpreted as evidence that the remnant collapsed into a BH at that time.
The claim was disputed by a subsequent reanalysis~\cite{Tsang:2019zra} and contested on theoretical grounds~\cite{Pani:2018flj}.

\begin{figure*}[t]
\centering
\includegraphics[width=1\textwidth]{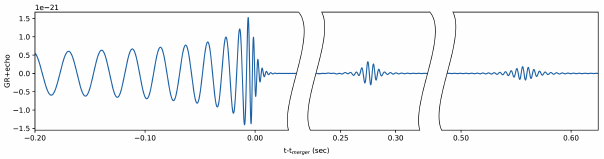}
\caption{Boltzmann GW echoes template for a GW150914-like signal with amplitude $A=1$. A Boltzmann echo waveform code can be found in~\cite{GW190521echo}. Figure taken from~\cite{Abedi:2022bph}.}
\label{echo_pic_3}
\end{figure*}

\subsubsection{Combined event search}

Bayesian inference and advanced statistical inference pipelines have been used to quantify the likelihood of echo signals across multiple events. In particular, a recent effort~\cite{Abedi:2022bph} applied a waveform inspired by stimulated Hawking radiation (the Boltzmann echo model~\cite{Abedi:2021tti,Abedi:2022bph} shown in Fig.~\ref{echo_pic_3}) to a set of 47 highly-confident LVK BBH merger events in the GWTC-3 catalog~\cite{KAGRA:2021vkt}, each with a false alarm rate of less than $10^{-3} \ \text{yr}^{-1}$~\cite{LIGOScientific:2020tif}.
The study used a fixed echo amplitude $A$ across events, allowing for the calculation of combined BFs as a measure of signal strength across multiple events.
It also assumed that the echo model is identical across all events; while this assumption is not expected to be physically well-motivated, it is a reasonable minimal framework to facilitate the combination of multiple events.
The analysis found no statistically significant evidence for this signal in GWTC-3 data. 
The Bayesian evidence for most events falls within the range of 0.3–1.6, with the hypothesis of a common (nonvanishing) echo amplitude for all mergers being weakly disfavored at 2:5 odds.
The one exception was GW190521, the most massive event  observed so far, which shows a positive evidence of 9.2, but the significant waveform systematic uncertainties affecting this event complicate its interpretation~\cite{Gamba:2021gap, Gayathri:2020coq, Gupte:2024jfe}.
An optimal combination of posteriors yields an upper limit on the universal echo amplitude of  $A < 0.4$ at 90\% confidence level, whereas typical models predict $A \sim 1$.

Later work used Bayesian analysis techniques suitable to combine multiple events, such as hierarchical methods~\cite{Mandel:2018mve, Isi:2019asy}, to further improve detection reliability~\cite{Abedi:2022bph}. 
By assuming that individual events follow a parent distribution in echo amplitude, with parameters fitted across the dataset, the method can analyze weak or inconclusive signals by integrating information across a population of potential echo candidates.
The result is a posterior function that estimates the parameters of interest while accounting for uncertainties and selection biases in individual detections, improving over previous work that assumed equal echo properties across events.
The analysis assumes a target Gaussian distribution $\bar{\mathcal{P}}(A,\bar{A},\sigma) = e^{-(A-\bar{A})^{2}/2\sigma^{2}}/\sqrt{2\pi\sigma^{2}}$ characterized by a mean $\bar{A}$ and a standard deviation $\sigma$. These parameters are then combined hierarchically~\cite{Abedi:2022bph,Isi:2019asy,Mandel:2018mve}, as described by the equation:
\begin{equation}
\mathcal{P}(\bar{A},\sigma)=\int\prod_{i=\rm{Events}}dA_{i} B_{i}(A_{i}) \bar{\mathcal{P}}(A_{i},\bar{A},\sigma)\,.
\label{eq:6}
\end{equation}
Here, $p(\bar{A},\sigma)$ represents the posterior function for the hierarchical combination of events, yielding the parameters $\bar{A}$ and $\sigma$, and $\mathcal{B}_{i}(A_{i})$ is the BF as a function of echo amplitude $A_{i}$ for event $i$. The values of $\bar{A}$ and $\sigma$ inferred from this analysis are shown in Fig.~\ref{fig:echoes_hierarchical}. The mean amplitude $\bar{A}$ peaks around $0.35$, while $\sigma$ peaks close to zero. This indicates minimal spread among the events, suggesting consistency across the dataset.

\begin{figure}[t]
\centering
\includegraphics[width=0.7\textwidth]{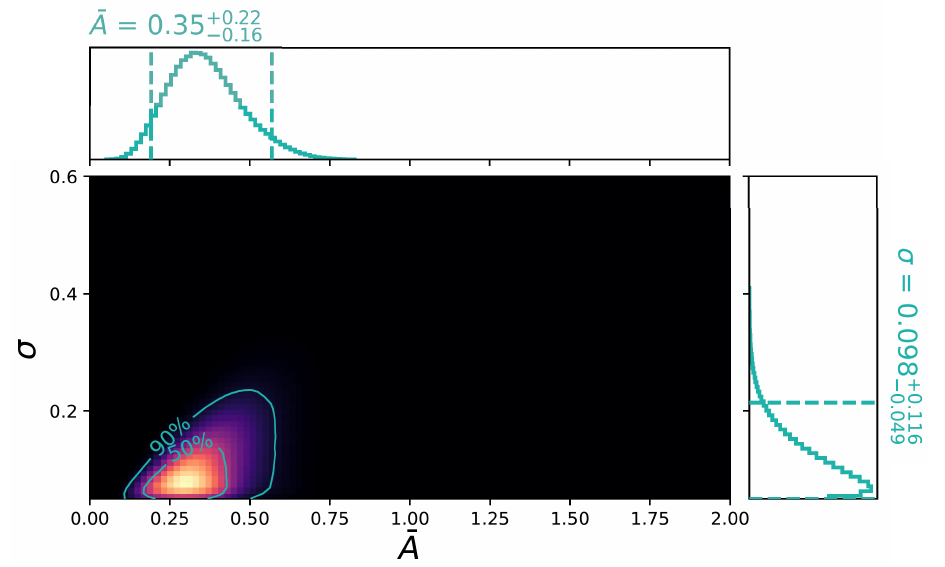}
\caption{
Posterior distribution for the extracted parameters $\bar{A}$ and $\sigma$ based on a target Gaussian distribution, $\bar{\mathcal{P}}(\rm{A},\bar{A},\sigma) = e^{-(A-\bar{A})^{2}/2\sigma^{2}}/\sqrt{2\pi\sigma^{2}}$, with mean $\bar{A}$ and standard deviation $\sigma$ for the measured echo amplitude $A$ across 47 events analyzed in~\cite{Abedi:2022bph} and hierarchically combined~\cite{Mandel:2018mve,Isi:2019asy}. 
Figure taken from~\cite{Abedi:2022bph}.}
\label{fig:echoes_hierarchical}
\end{figure}

Subsequent analyses from the LVK collaboration and affiliated groups also searched for echoes in multiple signals,  using both waveform-dependent~\cite{LIGOScientific:2021sio,Uchikata:2023zcu} and weakly-modeled~\cite{Tsang:2018uie,Tsang:2019zra, Miani:2023mgl} approaches.
They obtained no evidence for echoes, as shown in Fig.~\ref{fig:echoes_pp}.

\begin{figure}[h!]
\centering
\includegraphics[width=0.65\textwidth]{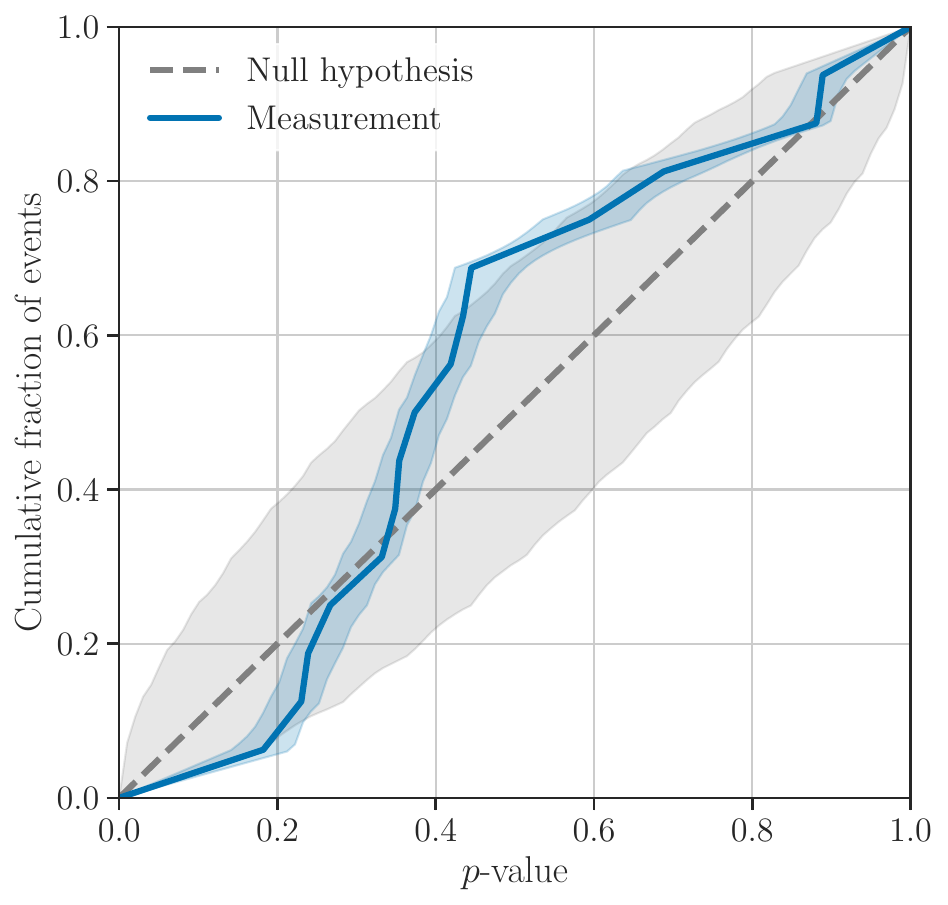}
\caption{
Probability-probability plot for the unmodeled echoes search carried out by the LVK on the GWTC-3 catalog~\cite{Tsang:2018uie, Tsang:2019zra, LIGOScientific:2021sio}. 
The diagonal line is the null hypothesis. Gray bands are $90\%$ uncertainties, while the light-blue band corresponds to the $90\%$ credible interval of the measured p-values.
Figure taken from~\cite{LIGOScientific:2021sio}.}
\label{fig:echoes_pp}
\end{figure}

\subsubsection{Summary of current status and future prospects}
Despite extensive efforts, a definitive detection of echoes remains elusive. 
Finding a consistent and statistically significant signal across multiple events is challenging due to the sensitivity of the echo templates to waveform modeling, noise, and the inherent complexities of GW signals. 
Continued developments in waveform modeling and more extensive datasets from future observing will improve the situation. 

Echo searches are among the science goals outlined in the LISA White Papers on waveform modeling~\cite{LISAConsortiumWaveformWorkingGroup:2023arg} and fundamental physics~\cite{LISA:2022kgy}. 
One of the challenges lies in accurately modeling the near-horizon reflection effects due to quantum phenomena. 
The vast range of possible models and large prior uncertainties make it less likely to detect weak signals. 
Current models often rely on simplifying assumptions for both the near-horizon regime and the angular momentum barrier. 
Additionally, the effects of nonlinearities are typically excluded from these models, which can significantly reduce their accuracy.
On the data analysis side, the practical challenges are equally significant. 
Echoes are expected to be faint signals, often buried in noise, making their detection difficult. 
The sensitivity of echo templates to small variations in the waveform, noise characteristics, and even the underlying BH properties demands highly precise and computationally expensive data analysis techniques. 
For example, injection studies have shown that the true positive detection probability for large BFs (greater than 8) is only $35 \pm 7$\%~\cite{Abedi:2021tti}. 
This probability is expected to be even lower if the models used are not sufficiently accurate, which is currently the case.
These challenges require further methodological refinements, independent confirmation (e.g. from potential electromagnetic counterparts), advancements in waveform modeling (e.g. by incorporating nonlinear effects and more complex boundary conditions), and more extensive observational datasets.

Future ground-based XG detectors are expected to probe the reflectivity of
compact objects up to values close to the BH one due to the large SNR of compact
binary coalescences~\cite{Testa:2018bzd,Maggio:2019zyv,Ma:2022xmp}. If measured
with the Einstein Telescope (ET), an event like GW150914 would have a ringdown
SNR of $\mathcal{O}(200)$~\cite{Bhagwat:2023jwv}, improving the measurability of
the reflectivity by one order of magnitude with respect to the LVK detector
network~\cite{Branchesi:2023mws}. Similar improvements are predicted also for
QNM frequency measurements and ``ordinary'' BH spectroscopy
tests~\cite{Berti:2016lat,Perkins:2020tra,Bhagwat:2021kwv,Bhagwat:2023jwv}.

A population of horizonless compact objects would give rise to a stochastic GW background of GW echoes. Reference~\cite{Du:2018cmp} estimated the detectability of this background with Voyager and ET as a function of the amplitude of the first echo relative to the ringdown.
Furthermore, if the remnant of a compact binary coalescence is a highly spinning horizonless compact object, the ergoregion instability would spin down the remnant by GW emission, until the condition for stability is satisfied. The incoherent superposition of the GW signals from all sources in the population would produce a stochastic GW background due to spin loss~\cite{Barausse:2018vdb}. The LVK network at design sensitivity, ET and LISA can place stringent constraints on the percentage of perfectly reflecting horizonless compact objects from the astrophysical population~\cite{Barausse:2018vdb}.

Finally, assuming the quantization of the BH horizon area dictates the QNMs of the prompt ringdown~\cite{Foit:2016uxn}. Current LVK observations already place interesting constraints on the quantization parameter~\cite{Laghi:2020rgl}.
Thanks to the discontinuous limit of this proposal against the spectrum of QNMs predicted in classical GR, this scenario can be entirely ruled out from a population of $\mathcal{O}(20)$ GW150914-like events at design sensitivity~\cite{Laghi:2020rgl}.

\clearpage
\section{Future prospects}
\label{sec:nextgen}

\noindent
{\em It is evident that if we adopt this point of view toward concepts, namely that the proper definition of a concept is not in terms of its properties but in terms of actual operations, we need run no danger of having to revise our attitude toward nature. [...]
Furthermore, if we remember that the operations to which a physical concept are equivalent are actual physical operations, the concepts can be defined only in the range of actual experiment, and are undefined and meaningless in regions as yet untouched by experiment. It follows that strictly speaking we cannot make statements at all about regions as yet untouched, and that when we do make such statements, as we inevitably shall, we are making a conventionalized extrapolation, of the looseness of which we must be fully conscious, and the justification of which is in the experiment of the future.}

\noindent
\begin{flushright}
Percy W.~Bridgman,\\{\em The logic of modern physics},\\The Macmillan Company (1927)\\
\end{flushright}

\vspace{.2cm}

Future detectors such as ET~\cite{Punturo:2010zz}, Cosmic Explorer~(CE)~\cite{Evans:2023euw} and LISA~\cite{LISA:2024hlh} will attain at least an order of magnitude increase in sensitivity with respect to current interferometers, open up entirely new observational windows, and pave the way to new explorations of gravity and physics beyond the Standard Model~\cite{Abac:2025saz,Gupta:2023lga,Barausse:2020rsu}.
Among these, a long-sought goal is the robust measurement of multiple QNMs in ringdown waveforms.
However, the large SNR of future detectors means that systematic errors will become dominant over statistical errors, and demands improvements in both GW source modeling and data analysis~\cite{Gupta:2024gun}.

In this section we review the prospects for multi-mode ringdown detections that will allow for straightforward, linear tests of gravity (Section~\ref{subsec:future_rates}); we expand the discussion to nonlinear tests of gravity and parameterized searches for new physics that will be possible with XG detectors (Section~\ref{subsec:future_tests}); and we outline the data analysis challenges that we must overcome to take full advantage of these observations (Section~\ref{subsec:future_challenges}).

\subsection{Multi-mode observations and rates}
\label{subsec:future_rates}

\vspace{-.1cm}

\noindent \textit{Initial contributors: Bhagwat, Chirenti, Pacilio}

\vspace{.2cm}

The detection of multiple ringdown modes is the first step in realizing the promise of BH spectroscopy as a tool for testing GR~\cite{Detweiler:1980gk,Dreyer:2003bv,Berti:2005ys}. At its heart the idea is simple: each mode detected in the signal provides a pair $(\omega_R, \omega_I)$, which can be related to the parameters $M$ and $a$ of the resulting Kerr BH~\cite{Echeverria:1989hg,Finn:1992wt}. Detecting multiple modes then allows to perform consistency tests that the same parameters $M$ and $a$ are recovered from each mode -- a condition that can be formulated in rigorous statistical terms~\cite{Gossan:2011ha,Meidam:2014jpa,Brito:2018rfr}. Thus, BH spectroscopy is a probe of the background geometry and of the dynamical equations governing the ringdown perturbations~\cite{Barausse:2008xv,Berti:2018vdi}. Moreover, using NR fits for the amplitudes and phases of each mode provides additional constraints~\cite{Carullo:2018sfu,Bhagwat:2021kfa, Forteza:2022tgq}.

Current detectors have provided, in a few cases, initial hints for the presence of a secondary mode in the ringdown (see Section~\ref{sec:Observations_Additional_Modes}). Theoretically, the ringdown is composed of an infinite sum of modes, starting at the fundamental quadrupolar mode with $(\ell, m, n) = (2,2,0)$ and adding fundamental higher harmonics with $(\ell,m) \ne (2,2)$, as well as overtones with $n > 0$. 
Fundamental higher harmonics are expected to be more easily excited in BBH mergers with unequal binary component masses $m_1$ and $m_2$~\cite{Berti:2007fi,Berti:2007zu,Kamaretsos:2011um,Kamaretsos:2012bs,Ota:2019bzl}, whereas the first overtone of the quadrupolar mode should be more relevant for binaries with $m_1 \approx m_2$~\cite{Ota:2019bzl,JimenezForteza:2020cve}.
In the first case, fundamental higher harmonics decay at a similar rate to the fundamental quadrupolar mode. 
In the second case, the overtone oscillates with similar frequency, but decays faster than the fundamental quadrupolar mode. 
Detecting these subdominant contributions poses different challenges in each case (see Section~\ref{sec:DataAnalysis}).

The modeling of the secondary mode is relevant also for its identification, which is needed for tests of GR (see, e.g., Section~\ref{subsec:future_tests}). One possible modeling approach is to assume that the labeling $(\ell,m,n)$ of the secondary mode is known in advance, e.g. from the binary masses inferred from the full IMR waveform. In this case, adding the secondary mode to the ringdown model description requires only {\it two} additional parameters: the relative amplitude and phase of the mode relative to the fundamental quadrupolar mode. Alternatively, the secondary mode may be assumed not to be known in advance, either because the mode ranking can not be determined, or because we want to perform agnostic tests of GR~\cite{Baibhav:2023clw}. Adding an unspecified secondary mode to the ringdown model requires {\it four} additional parameters: the relative amplitude and phase (as above), plus the frequency and damping time of the mode. The additional parameters raise the bar for detectability of the secondary mode by lowering the number of degrees of freedom in the evaluation of the reduced $\chi^2$ in a frequentist approach, or by penalizing the evidence in favor of this model with a larger volume in parameter space.

After the first detection of GW150914, it became natural to ask: when are we going to have a first incontrovertible observation of at least two QNM frequencies in order to do BH spectroscopy?

In order to confidently detect a secondary mode in the stationary regime (see Section~\ref{sec:waveforms}), conservative estimates require at least a typical ringdown SNR $50 \lesssim \rho \lesssim 100$~\cite{Berti:2007zu,Ota:2019bzl,Pacilio:2023mvk}. 
The SNR requirements for multi-mode BH spectroscopy can be translated to an expected rate of less than one event per decade for heavy stellar mass BBH mergers with the advanced LVK network at design sensitivity~\cite{Ota:2021ypb}.
These rates are expected to be much higher with planned XG ground-based detectors, that will probe a volume approximately $10^3 - 10^4$ times larger. As the rate of events is proportional to the observable volume, we expect hundreds to thousands of events per year that would be within the BH spectroscopy horizon, and therefore enable multi-mode observations~\cite{Bhagwat:2023jwv,Ota:2021ypb}. 

Future detectors will probe different frequency ranges, with XG ground-based
detectors such as CE~\cite{Evans:2023euw} and
ET~\cite{Punturo:2010zz,Abac:2025saz} being sensitive to stellar-mass BH
remnants, while space-based detectors such as LISA~\cite{LISA:2017pwj} or
the TianQin and Taiji observatories~\cite{TianQin:2020hid,Luo:2021qji} will
probe massive BHs (MBHs). We discuss below the multi-mode detection prospects for both stellar-mass and MBH binaries with XG detectors (Section~\ref{subsubsec:multi_mode_hist}). We also briefly comment on the
less explored case of binaries containing at least one neutron star
(Section~\ref{sec:multimode_NS}). %

\subsubsection{Prospects for multi-mode detections over the years}\label{subsubsec:multi_mode_hist}

Early estimates of the capabilities of XG detectors focused on the $(2,2,0)$ and $(3,3,0)$ modes~\cite{Berti:2007zu, Gossan:2011ha, Meidam:2014jpa, Berti:2016lat,Maselli:2017kvl, Cabero:2019zyt}. 
For example, some early work demonstrated that deviations from the GR predictions for the dominant mode (i.e., $\delta\omega_{220}$) can be constrained at the 10\% level  for 
a remnant of mass $500 M_{\odot}$ at luminosity distance $\sim 6\,$Gpc detected by ET, or for a $10^6 M_{\odot}$ source at approximately the same distance $\sim 6\,$Gpc detected by LISA~\cite{Gossan:2011ha}.
These order-of-magnitude estimates were revised in~\cite{Meidam:2014jpa} using improved QNM fits, updated astrophysical models, and a Bayesian framework to combine multiple sources~\cite{Li:2011cg, Agathos:2013upa}, with roughly consistent conclusions: $\sim 10\%$ constraints can be achieved by combining $\mathcal{O}(10)$ sources out to $50 \,$Gpc.

After the detection of GW150914, the authors of Ref.~\cite{Berti:2016lat}
estimated that Voyager-class detectors (i.e., the best detectors that can be
hosted within current LIGO facilities) would be necessary to have at least one
BH spectroscopy test per year, and that high-sensitivity instruments like ET and
a 40-km CE would be necessary to perform BH spectroscopy at cosmological
distances ($z\gtrsim3$).
On the other hand, they pointed out that the high-SNR observations possible with LISA can yield precise QNM measurements out to $z\approx5$, $10$, or even higher, depending on the underlying MBH population model~\cite{Berti:2016lat}. 
The pessimistic outlook for ground-based detectors was due in part to conservative assumptions: the authors of~\cite{Berti:2016lat} used stellar-mass BH formation models that did not allow for binaries with large total mass and/or significantly unequal masses (while we have now observed BBH mergers with these properties), they did not consider the possibility of coherently combining multiple observations~\cite{Yang:2017zxs}, and they imposed rather stringent criteria on the detectability of the second mode. 
Later on, more optimistic forecasts~\cite{Cabero:2017avf,Carullo:2018sfu} found that by leveraging information from NR and combining multiple events, semi-agnostic tests of GR at the $\mathcal{O}(1\%)$ level could be achieved with the LVK detector network at design sensitivity.

To further understand BH spectroscopy capabilities with XG detectors, the authors of Ref.~\cite{Baibhav:2017jhs} showed that the addition of one or two overtones to the $\ell=m=2$ fundamental QNM at late times significantly improves the estimate of the QNM frequencies inferred from either SXS waveforms, or waveforms computed in the extreme mass ratio limit using BH perturbation theory. The inclusion of overtones also improves the estimate of the remnant BH mass and spin, in the spirit of BH spectroscopy. They also estimated the energy emitted by the first few dominant ringdown modes for nonprecessing binaries, as well as the starting time of these modes. As detector sensitivities increase, the energy radiated in each mode will be crucial for understanding which QNMs are expected to be observed in a given binary BH merger event, and the starting time is needed to understand when the signal can be accurately modeled by a superposition of QNMs. These energy estimates were used to confirm that, while the median redshift at which multiple ringdown modes can be observed with ET is around $z\sim3$, LISA can detect four or more modes out to $z>20$ within certain mass ranges~\cite{Baibhav:2018rfk}. In fact, LISA ringdown signals can be so loud that more QNMs will be visible than those that can accurately be computed in most NR codes~\cite{Baibhav:2018rfk} -- a stark reminder that our ability to do BH spectroscopy is currently limited by waveform modeling and systematics.

Continuing the investigation of which subdominant modes (overtones or higher harmonics) are most likely to be detected in different scenarios, the authors of Ref.~\cite{Ota:2021ypb} found that for
nearly equal-mass binaries, it can be beneficial to measure the $(2,2,1)$ and
$(3,3,0)$ subdominant modes in addition to the $(2,2,0)$ mode, whereas the
$(3,3,0)$ and $(4,4,0)$ modes are generally easier to detect and distinguish
than the overtone for more asymmetric binaries. In this study, the authors also
computed BH spectroscopy horizons for LISA, ET, and CE, i.e., the maximum
distances at which an event with a given mass and mass ratio has more than one
QNM that is resolvable by the detector. For CE, they found up to
$\mathcal{O}(1000)$ events per year with multiple detectable QNMs that can be
used to perform BH spectroscopy.

As BBH catalogs and data analysis tools have continued to expand and improve, more detailed studies of XG capabilities to achieve multi-mode observations have been performed for both ground-based and space-based detectors. 
Prospects for multi-mode ringdown observations on the ground can be investigated 
using a synthetic population of stellar-mass BH binaries, as described in
e.g.~\cite{Mapelli:2021gyv}. This population model contains a mixture of
dynamical and isolated formation channels, and is consistent with LVK
observations~\cite{LIGOScientific:2020ibl}. The future network of XG
ground-based detectors, although still uncertain, can be assumed to consist of
(e.g.) a triangular-shaped ET with power spectral density following the ET-D
model~\cite{Hild:2010id} and an L-shaped CE with 20km
arms~\cite{Gupta:2023lga}. The mass ratio distribution for the assumed
population peaks at equal masses (favoring the detection of the quadrupolar
overtone mode), but there is large support for higher mass ratio binaries ---
with $\sim 10^3$ events/yr with $q\geq3$ and $\sim 10^2$ events/yr with $q\geq6$
--- that are more promising for resolving secondary angular
modes. In~\cite{Bhagwat:2023jwv}, for simplicity, all binaries are modeled as
quasi-circular and nonprecessing, including the $(2,2,0)$ dominant mode
alongside the $(3,3,0)$, $(4,4,0)$ and $(2,1,0)$ subdominant higher modes. Given
the large number of events in a synthetic population analysis, the SNR of the
events, and the large astrophysical uncertainties, a Fisher matrix approximation
is typically adequate for these estimates~\cite{Berti:2005ys,Bhagwat:2023jwv}.

\begin{figure*}[t]
    \centering
    \includegraphics[width=0.48\linewidth]{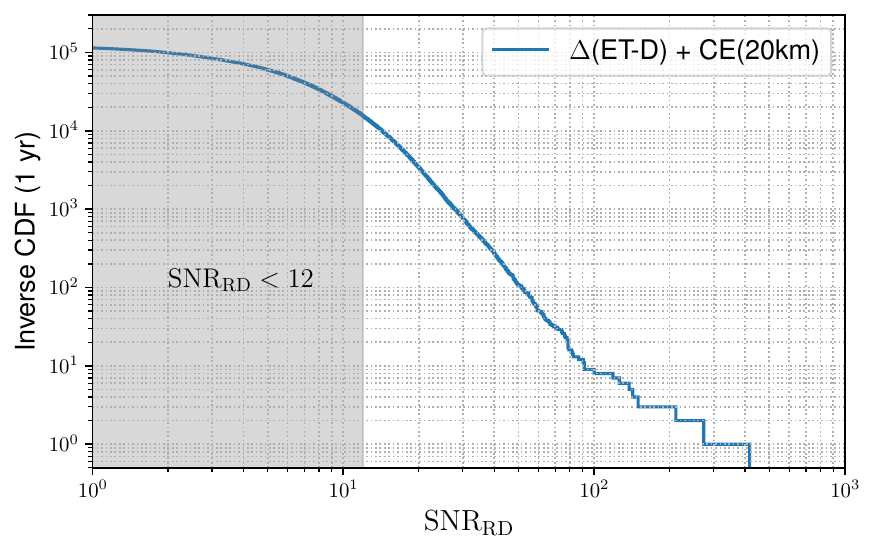}
    \includegraphics[width=0.48\linewidth]{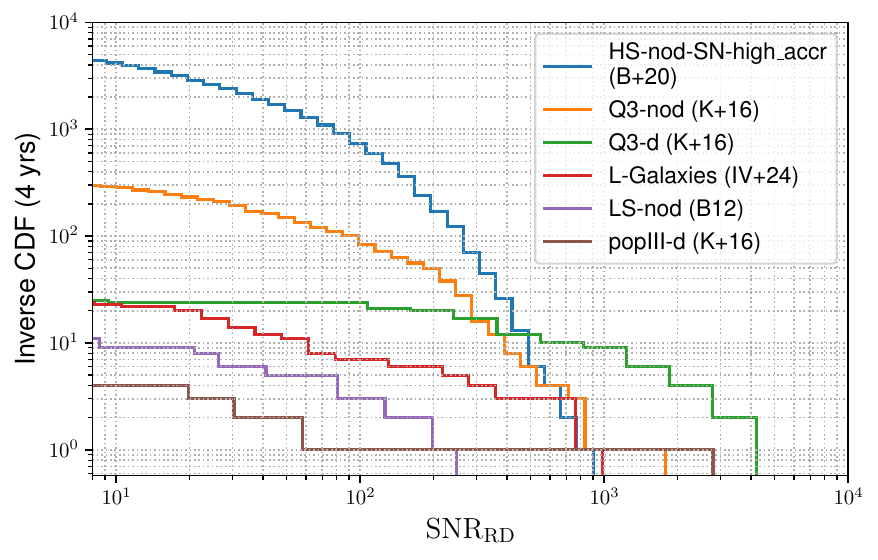}
    
\caption{Left: Inverse cumulative distribution function (CDF) of the ringdown signal-to-noise ratio ${\rm SNR}_{\rm RD}$ for a population of stellar-mass BH binaries compatible with the most recent LVK constraints~\cite{Mapelli:2021gyv}. Figure adapted from~\cite{Bhagwat:2023jwv}. Right: Same, but for semi-analytic models of MBH binary populations agreeing with the latest PTA data releases, as measured by LISA. Figure adapted from~\cite{Bhagwat:2021kwv}.}
    \label{fig:snr:cdf}
\end{figure*}

In Fig.~\ref{fig:snr:cdf} (left) we show the inverse cumulative distribution of the ringdown SNR, ${\rm SNR}_{\rm RD}$, for a population of stellar-mass BBHs~\cite{Bhagwat:2023jwv}. Most of the events will be close to the threshold for detectability, but a non-negligible fraction of $\sim10^2$ events/yr will have ${\rm SNR}_{\rm RD}\gtrsim50$, and $\mathcal{O}(10)$ events/yr will be observed with ${\rm SNR}_{\rm RD}\gtrsim100$. Assuming that the noise treatment and other complications arising in realistic detections are under control (see Section~\ref{sec:DataAnalysis}), these numbers translate directly into corresponding prospects for constraining deviations away from GR. In Fig.~\ref{fig:sigma:omega:cdf} (left) we show the cumulative distribution for the 1-$\sigma$ uncertainty on the posterior of the relative deviations $\delta\omega_{{\rm R},\ell m n}$ in the QNM frequencies~\cite{Gossan:2011ha,Isi:2021iql,Pacilio:2023mvk}), as predicted by Fisher matrix estimates: the $(3,3,0)$ and the $(4,4,0)$ modes, which have the best measurement prospects, will be constrained at the $\sim10\%$ level in $\sim10^3$ events/yr, and at the $\sim1\%$ level in $\lesssim10$ events/yr. These numbers imply that a coherent analysis of a large number of events could yield even better constraints~\cite{Meidam:2014jpa,Yang:2017zxs}.

\begin{figure*}[t]
    \centering
    \includegraphics[width=0.49\linewidth]{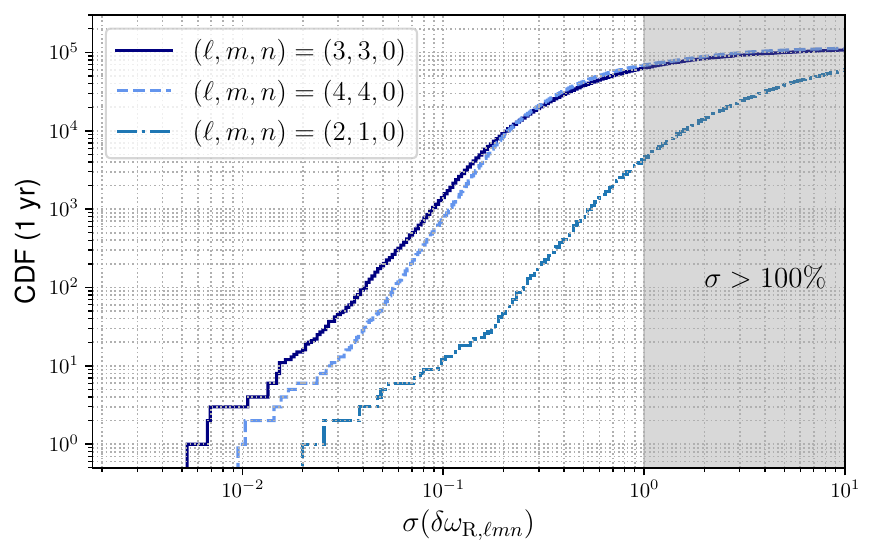}
    \includegraphics[width=0.49\linewidth]{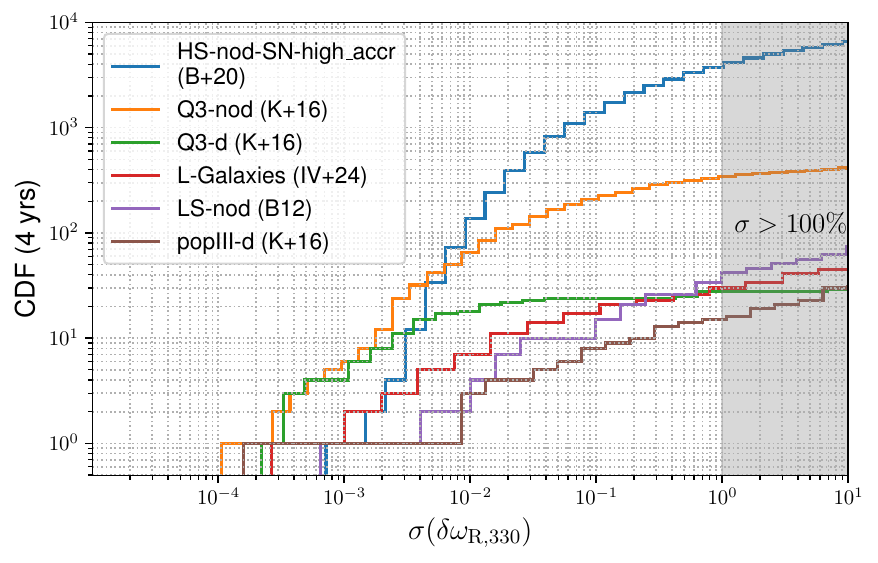}
    \caption{Left: Cumulative distribution function (CDF) of the 1-$\sigma$
      uncertainty in the deviation parameters $\delta\omega_{{\rm R},\ell m n}$,
      as predicted by Fisher matrix estimates, assuming a population of
      stellar-mass BH binaries~\cite{Mapelli:2021gyv} observed by an ET+CE
      detector network. Figure adapted from~\cite{Bhagwat:2023jwv}. Right: Same
      for $(\ell,m,n)=(3,3,0)$ and for semi-analytic models of MBH binary
      populations agreeing with the latest PTA data releases, as measured by
      LISA. Figure adapted from~\cite{Bhagwat:2021kwv}.}
    \label{fig:sigma:omega:cdf}
\end{figure*}

The prospects improve further for MBH binaries detected by space-based observatories. Decades of astronomical observations show that MBHs with masses ranging from approximately $\sim 10^5 M_\odot$ (or even lower) to $\sim 10^9 M_\odot$ are prevalent in massive galaxies~\cite{1984ApJ...278...11G,Kormendy:1995er} and in dwarf galaxies~\cite{Reines:2011na,Reines:2013pia,Baldassare:2019yua}. Their mass spectrum and redshift distribution are intricately linked to their dynamical evolution within their host galaxies. However, these processes remain poorly understood, resulting in significant uncertainty on the mass and redshift distribution of MBH binaries. Current models predict MBH merger rates ranging from a few events to tens of thousands per year. To account for these uncertainties, several  distinct synthetic models of MBH binaries have been proposed. State-of-the-art models align with the most recent data releases from pulsar timing arrays (PTAs)~\cite{EPTA:2023fyk,Tarafdar:2022toa,NANOGrav:2023gor,Reardon:2023gzh,Xu:2023wog}. For example, the semi-analytic model of~\cite{Barausse:2012fy,Barausse:2023yrx} has been used to generate models that comprise binaries originating from a population of high-mass seeds~\cite{Klein:2015hvg,Barausse:2020mdt,Barausse:2023yrx}, low-mass seeds~\cite{Barausse:2012fy,Klein:2015hvg} and a mix of low-mass and high-mass seeds~\cite{Izquierdo-Villalba:2024bhc}.

For each selected population, it is possible to forecast the ability to test GR using ringdowns as detected by LISA, a joint ESA-NASA space mission designed to target MBH binaries with masses between $10^4 M_\odot$ and $10^7 M_\odot$ up to very high redshifts. Here, the FD response of the detector assumes a four-year observational period and LISA's sensitivity as described in~\cite{Babak:2021mhe}, while accounting for confusion noise from the population of Galactic white dwarfs.

In Fig.~\ref{fig:snr:cdf} (right) we update the analogous Fig.~2 of Ref.~\cite{Bhagwat:2021kwv} by focusing on MBH binary population models compatible with the most recent PTA results~\cite{Barausse:2023yrx,Izquierdo-Villalba:2024bhc},
and present the inverse cumulative distribution function of the ringdown SNR. While predictions vary significantly depending on the assumed intrinsic population, all models anticipate the detection of at least a few events with high ringdown SNR (${\rm SNR}_{\rm RD}\gtrsim 100$) over a four-year observation period. The detection rate increases to approximately $\sim 10^2$ and $\sim 10^3$ for the most optimistic models (with high-mass seeds).
Some individual detections are expected to exhibit exceptional SNRs, on the order of ${\rm SNR}_{\rm RD}\sim(10^3)$. In Fig.~\ref{fig:sigma:omega:cdf} (right) we focus on Fisher matrix predictions for the 1-$\sigma$ uncertainties $\delta\omega_{{\rm R},330}$. Remarkably, ringdown observations with LISA will have the potential to produce constraints at the sub-percent level from individual sources. 
However, high-SNR ringdown observations with small statistical errors of QNM parameters are also more susceptible to biases from systematic effects, such as unmodeled QNMs (including overtones or quadratic modes) and tails, which can be estimated using linear signal analysis~\cite{Capuano:2025kkl,Volkel:2025jdx}.

In addition to consistency tests between the QNM frequencies, complementary tests of GR can be performed using measurements of the {\it amplitudes and phases} of different QNMs, which are restricted to a fairly small region of phase space within GR~\cite{Forteza:2022tgq}. For the GW190521 event (if we assume a multi-mode detection in this event, which is debatable: see Section~\ref{sec:Observations_Additional_Modes}), the results are compatible with GR. Constraints on GR with the amplitude-phase consistency test will be much stronger with the increased sensitivity of XG detectors. 

Finally, we note that besides allowing for consistency tests of GR, measuring multiple modes in the ringdown can also enhance LISA's measurements of the sky localization and luminosity distance without relying on the modulations of the signal due to LISA's orbital motion~\cite{Baibhav:2020tma}. This is particularly important in the context of multi-messenger astronomy with MBH sources. Measurements of multiple harmonics in the ringdown also allow for improved measurements of the remnant mass and spin, the mass ratio, and the inclination of the binary~\cite{Baibhav:2020tma}.

\subsubsection{Binaries containing at least one neutron star}
\label{sec:multimode_NS}
Binary neutron stars can result in the formation of a BH with mass in the
approximate range $2 - 5\, M_{\odot}$, that is, in the so-called ``lower mass
gap''. The fundamental quadrupolar mode of this BH will have an oscillation
frequency in the approximate range $2.5 - 6$ kHz. Such high frequencies are
virtually inaccessible to current detectors, but could be detectable by XG
ground-based detectors such as CE or ET. Additionally, a detector concept called
Neutron Star Extreme Matter Observatory (NEMO)~\cite{Ackley:2020atn} has been
proposed to achieve sensitivity comparable to CE and ET at frequencies above 1
kHz, bringing the expected event rates for post-merger remnants from binary
neutron star mergers to a few per year. Multi-mode detection for the low-mass
BHs formed in binary neutron star mergers may require even more sensitive
detectors, but it would open exciting new avenues of investigation. As some
flavors of alternative theories of gravity propose higher-order curvature
corrections~\cite{Berti:2015itd,Yunes:2013dva}, such low-mass BHs would be the
most promising for constraining beyond GR corrections. Similar arguments can be
made for mixed binaries containing one neutron star and one BH, which also
benefit from a larger detection range and lower ringdown frequencies due to the
higher total mass of the binary, when compared with the binary neutron star
case.

\subsection{Nonlinear and parameterized tests with next-generation detectors}
\label{subsec:future_tests}

\vspace{-.1cm}

\noindent \textit{Initial contributors: Berti, Maselli, Sathyaprakash, Yi, Yunes}

\vspace{.2cm}

Given the promising possibility of a large number of high-SNR multi-modal observations with future detectors, we now review how these observations can be used in nonlinear (Section~\ref{sec:quad_nextgen}) and parameterized (Section~\ref{Section~7.2.3}) tests of GR.

\subsubsection{Nonlinear black hole spectroscopy}\label{sec:quad_nextgen}

As discussed in Section~\ref{sec:nonlin_num_expe},
QQNMs are present in binary BH waveforms~\cite{London:2014cma,Cheung:2022rbm,Mitman:2022qdl} and potentially detectable. In addition to improving our models of the ringdown signal, confident detections of these nonlinear modes would allow us to perform more stringent tests of GR as a nonlinear theory of gravity.

The detectability of nonlinear modes through ringdown observations with both ground- and space-based XG detectors was explored in~\cite{Yi:2024elj} (see~\cite{Qiu:2023lwo} for ground-based prospects).
The study of~\cite{Yi:2024elj} focused on the dominant QQNM arising from the self-interaction of the $(2,2,0)$ linear mode. 
The authors computed the SNR of the QQNM for different catalogs, including: (i)
stellar-mass BH populations expected to emit in the frequency band of CE and ET
-- this class comprised intermediate-mass binaries with total masses
$\mathcal{O}(100)M_\odot$, identified as promising candidates for detecting
nonlinear modes with CE; and (ii) various populations of MBHs, key
observational targets for LISA.  They found that ET and CE could observe up to a
few tens of events per year with a quadratic mode SNR larger than 8. Within a
4-year mission lifetime, LISA has the potential to detect anywhere from a few
events -- or even none -- to up to $\mathcal{O}(1000)$ events with a QQNM SNR
above 8, depending on the MBH population model.

A Fisher matrix analysis of all events with SNR $>8$ showed that all parameters
of both the linear and the nonlinear $(4,4,0)$ QNMs can be determined with a
relative accuracy smaller than 100\%. In a follow-up study~\cite{Khera:2024yrk},
several other quadratic modes were found to have SNR $>8$ in some regions of
parameter space, as observed by CE.  With LISA,
several of the QQNMs have the potential to be observed with SNR
$\mathcal{O}(100)$. In another separate study of the QNM detectability and
resolvability for nonspinning binaries, it was similarly found that quadratic
modes have the potential of being detected with an SNR $>8$ in up to 51 and 24
events for the $(2,2,0)\times(2,2,0)$ and $(3,3,0)\times(2,2,0)$ quadratic
modes, respectively, as seen with LISA and using three different MBH
catalogs~\cite{Shi:2024ttu}. The authors further use the Rayleigh criterion to
determine that out of these, up to 24 and 17 events will have quadratic mode
complex frequencies that are fully resolvable from adjacent QNM frequencies. In
addition, the authors of~\cite{Shi:2024ttu} examined detectability prospects for
the space-based detector TianQin~\cite{TianQin:2015yph}, finding somewhat lower
numbers detectability and resolvability rates (generally by a factor of a few).

The prospects for detecting and resolving the dominant quadratic mode with CE
and ET were further investigated in~\cite{Lagos:2024ekd} using two different
data analysis strategies. The parameters of nonlinear modes can be considered
independent from those of linear QNMs, allowing their measurements to be used as
consistency checks, similar to standard practice with linear QNMs.  This is the
approach previously considered in Ref.~\cite{Yi:2024elj} and revisited in
Ref.~\cite{Lagos:2024ekd}.  The second strategy explored in
Ref.~\cite{Lagos:2024ekd} relies on the fact that, within GR, the parameters of
nonlinear QNMs are determined by those of linear modes, with additional
excitation factors (see Section~\ref{sec:amplitudes}).
This consideration reduces the number of parameters that need to be constrained
by data, as we can focus only on those of the linear QNMs.  By using this
strategy, the inclusion of the quadratic $(4,4,0)$ component can improve the
measurement of the linear $(4,4,0)$ QNM by a factor of 2. According to
Ref.~\cite{Lagos:2024ekd}, this allows a confident detection and resolvability
between the nonlinear and linear QNMs up to redshift $z\sim20$ and $z\sim35$
for CE and ET, respectively.

Since the ratios between nonlinear and linear mode amplitudes are uniquely
determined in GR, as discussed in Section~\ref{sec:amplitudes}, measurements of
these ratios with XG detectors could be powerful null tests of the
nonlinear nature of Einstein's theory.

\subsubsection{Ringdown beyond general relativity: parameterized spectroscopy}\label{Section~7.2.3}

The sensitivity of XG detectors will allow us to use ringdown observations for searches of GR deviations in the QNM spectrum of detected events. Tests of GR can be done either by comparing ringdown models computed in alternative theories of gravity, or by performing agnostic analyses.

As extensively discussed in Section~\ref{sec:beyondGR}, calculating QNMs beyond GR is nontrivial because of the coupling with new degrees of freedom and because the perturbation equations are, in general, nonseparable.
A common approach to address these challenges is to use perturbative expansions in the BH spin, starting from a spherically symmetric solution and then adding rotational corrections~\cite{Hartle:1968si}. 

Recent studies have successfully applied this method to either calculate QNMs at a given order in spin for specific theories, or employed agnostic approaches, augmenting Kerr frequencies and damping times with shifts that are expected to vanish in the GR limit.
Although immediately applicable, the constraints on these QNM shifts must be mapped to specific theories to go beyond consistency tests of GR. 
Moreover, these deviations are often assumed to be constant, but calculations in specific beyond-GR models show that they depend on the source parameters, such as the remnant BH mass and spin.  
Historically, such tests have directly modified QNM frequencies, but recent studies emphasize the importance of also considering changes in amplitude due to coupling with non-GR degrees of freedom~\cite{DAddario:2023erc,Crescimbeni:2024sam}.

Bridging the gap between these two strategies, the parameterized spectroscopy (\texttt{ParSpec}) approach~\cite{Maselli:2019mjd}, introduced in Section~\ref{sec:beyondGR}, employs a two-parameter expansion to describe shifts in mode frequencies and damping times as perturbative corrections to the Kerr values.
These shifts are expressed as functions of the BH spin and of a coupling function that can depend on the source properties. 
Within this framework, the QNM frequencies are written in the form
\begin{align}
    M_i \omega^{\ell m}_i &= \sum^{n_1}_{k_1=0} \chi_i^{k_1} \Bar{\omega}_{\ell m}^{(k_1)}+\sum^{n_2}_{k_2=0} \chi_i^{k_2} \Bar{\omega}_{\ell m}^{(k_2)} \gamma_i \delta \omega^{(k_2)}_{\ell m}\;,\label{eqn:parspec1}\\
    \tau^{\ell m}_i/M_i  &= \sum^{n_1}_{k_1=0} \chi_i^{k_1} \Bar{\tau}_{\ell m}^{(k_1)}+\sum^{n_2}_{k_2=0} \chi_i^{k_2} \Bar{\tau}_{\ell m}^{(k_2)} \gamma_i \delta \tau^{(k_2)}_{\ell m}\;,\label{eqn:parspec2}
\end{align}
where we have explicitly allowed for different orders in the GR and beyond-GR spin-expansions, and since we are going to focus on the fundamental mode, we dropped the overtone index to simplify the notation.
Here $M$ and $\chi$ are the BH mass and the dimensionless BH spin, respectively. The index $i$ in Eqs.~\eqref{eqn:parspec1}-\eqref{eqn:parspec2} runs over the number of sources used in the analysis.  The first term in each equation gives the small-spin expansion of the frequency and damping time within GR, while the second terms encode deviations from the Kerr values, given at order $k_2$ in $\chi$ by $(\delta \omega^{(k_2)}_{\ell m},\delta \tau^{(k_2)}_{\ell m})$.  The parameter $\gamma_i$, which governs the amplitude of these GR corrections, generally depends on the specific theory of gravity. In Ref.~\cite{Maselli:2023khq}, the authors assumed the following form for $\gamma_i$:
\begin{equation}
  \gamma_i = \frac{\alpha}{M^p_i}(1+z_i)^p \ ,
  \label{eqn:parspec_coupling}
\end{equation}
where $\alpha$ is the dimensionful coupling constant of the theory, with $[\alpha]=(\mathrm{mass})^p$.  Equation~\eqref{eqn:parspec_coupling} reproduces the scaling of GR deviations for EsGB and dCS gravity ($p=4$) and for different classes of EFTs ($p=4$ and $p=6$).

Ringdown tests with \texttt{ParSpec} can be performed in an agnostic manner, leaving $\delta \omega^{(k_2)}_{\ell m}$ and $\delta \tau^{(k_2)}_{\ell m}$ as free parameters to be constrained by data.
Alternatively, the formalism can be used as a theory-specific tool, by mapping the shifts (and the coupling function) to actual QNM calculations, e.g. using the modified Teukolsky approach: see e.g.~\cite{Carullo:2021dui} for an application of the formalism to GW data, and~\cite{Carullo:2021oxn} for constraints on the Kerr-Newman QNMs computed via spectral methods.

The \texttt{ParSpec} parameterization allows the combination of data from multiple ringdown events and/or multiple modes within a single detection.
In Ref.~\cite{Maselli:2023khq}, constraints on GR deviations were computed by combining simulated XG detector observations based on astrophysical binary BH population models, for both a model-agnostic approach and for specific theories.

In the model-agnostic case, their analysis implies that only constraints on nonrotating corrections in Eqs.~\eqref{eqn:parspec1}-\eqref{eqn:parspec2} are generally informative, even when prior information from the inspiral-merger phase is used to fix the remnant's mass and spin.
When spin terms are introduced, the correlations among parameters widen the posterior distributions, resulting in measurable shifts only for quadratic frequency modifications. 
In Fig.~\ref{fig:parspec-agnostic} we show, for illustration, the posterior
distributions for $\delta \omega^{(0)}_{22}$ and $\delta \tau^{(0)}_{22}$
inferred by combining datasets with different ringdown observations made by ET
and CE.
The analysis is performed assuming detection of the fundamental mode only, and for two binary BH catalogs that assume different prescriptions for the progenitor spins.

\begin{figure*}[t]
    \centering
\includegraphics[width=0.49\linewidth]{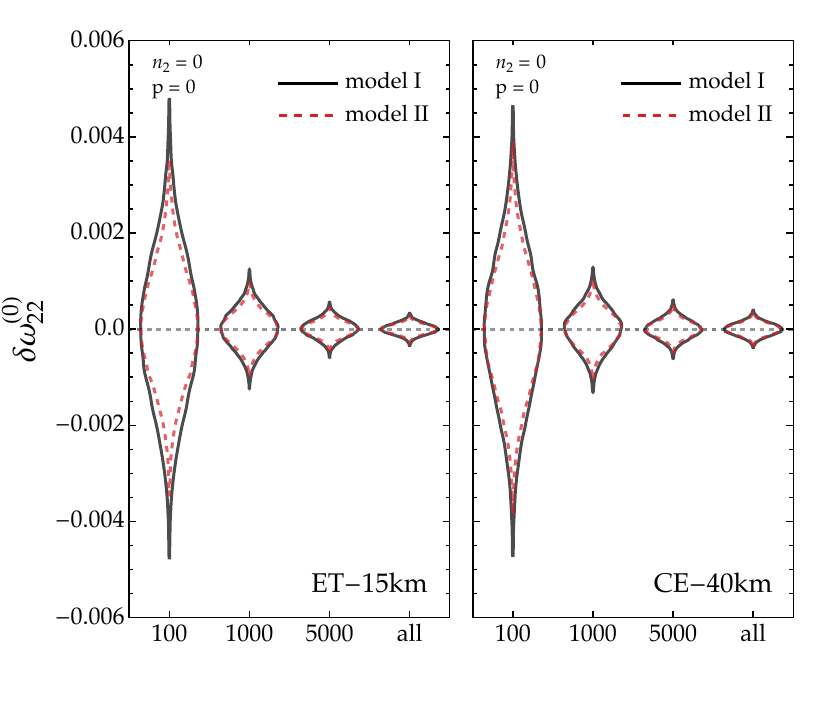}
\includegraphics[width=0.49\linewidth]{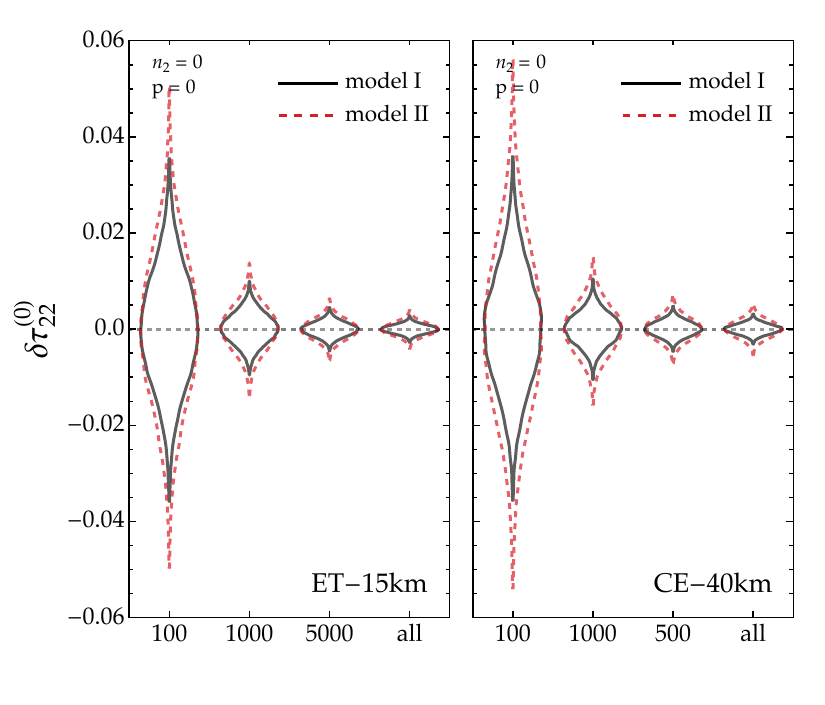}
\caption{Posterior probability densities for nonrotating corrections in
  Eqs.~\eqref{eqn:parspec1}-\eqref{eqn:parspec2} by stacking different numbers
  of ringdown events (x-axis) detected by either ET or CE.  These posteriors are
  inferred using only detections of the $(2,2,0)$ mode, and for an underlying
  theory with dimensionless coupling ($p=0$).  Solid and dashed contours refer
  to the two population models adopted in the analysis, which assume different
  prescriptions for the BH progenitor spins.
  Figure taken from~\cite{Maselli:2023khq}.}
    \label{fig:parspec-agnostic}
\end{figure*}

The authors of~\cite{Maselli:2023khq} also assessed the ability of ET and CE to
measure the coupling constants of specific theories (EsGB, dCS, and EFT gravity)
through multiple QNM observations. The inferred constraints are weak for all of
these theories: the posteriors on the coupling constants are always consistent
with the GR hypothesis.  This is mainly due to the nature of the Kerr shifts,
which are strongly suppressed by the function $\gamma_i$ in
Eqs.~\eqref{eqn:parspec1}-\eqref{eqn:parspec2} -- specifically,
$\gamma_{\rm GB,CS} \propto M^{-4}$, $\gamma_{\rm EFT} \propto M^{-4}$ or
$\gamma_{\rm EFT} \propto M^{-6}$, depending on the model. For this reason the
most massive events, which generally yield higher SNRs, also tend to produce
small GR deviations, which are difficult to observe.

The constraints found on EsGB and dCS gravity were limited by the fact that at the time of that study, the QNM frequencies were only known at low order in the spin expansion.
Incorporating higher-order terms, such as those computed for EsGB gravity~\cite{Chung:2024ira,Chung:2024vaf,Blazquez-Salcedo:2024oek,Khoo:2024agm,Blazquez-Salcedo:2024dur} or dCS~\cite{Chung:2025gyg}, could improve these bounds. 
However, LVK observations of the full inspiral-merger waveform for the ``mass gap'' event GW230529~\cite{Gao:2024rel,Sanger:2024axs} already provided stringent limits on this theory. This suggests that ringdown observations alone have limited constraining power for higher-order curvature scenarios, even at the increased sensitivity of XG detectors.
The latter constraints could be overcome for hypothetical theories that would triggers deviations from GR only around the merger.

For EFT gravity, deviations in the QNM frequencies are known at higher order in the spin expansion. This means that we can better investigate the effect of truncating the spin expansion on the resulting constraints. In Fig.~\ref{fig:parspec-EFTs} we show the posterior distribution of the coupling constant for one of the EFT models considered in~\cite{Maselli:2023khq}. The ringdown recovery template is truncated at different orders in the spin up to $\mathcal{O}(\chi^{12})$, corresponding to the value used for the injected signal. Clearly, a zero-order spin template introduces a significant bias in the posterior distribution.  However, a sixth-order expansion in $\chi$ is already accurate enough to recover the injected value of the coupling (but as usual the posterior is very broad and compatible with GR, for the reasons explained earlier).

\begin{figure*}[t]
    \centering
\includegraphics[width=0.6\linewidth]{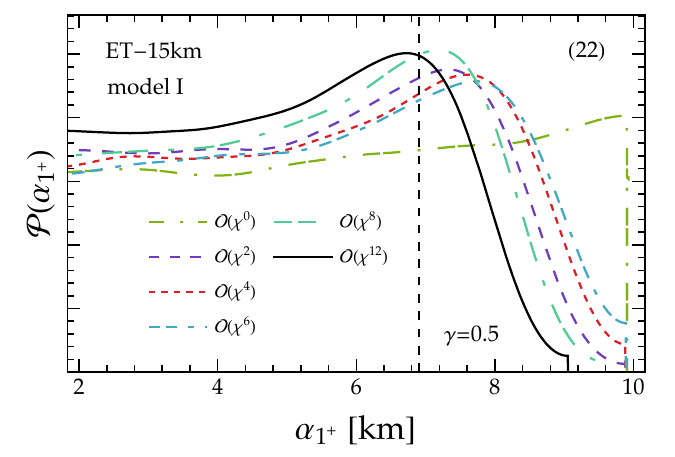}
\caption{Posterior distribution for the coupling constant of an EFT gravity theory with quartic curvature corrections~\cite{Cano:2023jbk}, as inferred by combining ringdown BH observations by ET. The analysis is performed by using only the fundamental (2,2) mode, and assuming a coupling constant of $\gamma=0.5$.  Different colors refer to recovery templates that include spin corrections at different orders, as shown in the legend. The vertical dashed line marks the injected value of the coupling. Figure taken from~\cite{Maselli:2023khq}.}
    \label{fig:parspec-EFTs}
\end{figure*}

The potential for detecting non-GR effects through LISA observations was also studied in~\cite{Pitte:2024zbi}. Working with a toy model of the ringdown consisting of a superposition of QNMs, the most stringent constraints can be achieved by conducting a targeted search for fractional deviations in the $(3,3,0)$ and $(4,4,0)$ QNMs. Then the injected frequency deviations in these modes can be measured with uncertainties of 5\% and 3\%, respectively, while deviations in the damping times can be inferred with an accuracy of 10\% and 17\%, respectively. Events with source masses in the range $[10^6-10^7]M_\odot$ and redshift $z \leq 5$ could enable constraints on deviations in the real part of the $(3,3,0)$ mode with an uncertainty of $\leq 0.05\%$~\cite{Pitte:2024zbi}.

Constraints on QNMs from LISA observations were also studied within the \texttt{pSEOB} framework~\cite{Toubiana:2023cwr}. For astrophysically realistic systems, the study finds that it will be possible to measure relative deviations of $\mathcal{O}(10^{-3})$ and $\mathcal{O}(10^{-2})$ simultaneously in several of the frequencies and damping times, respectively. This work demonstrates that systematic effects due to waveform inaccuracy could spoil such tests already in the near future, as discussed in the next section.

\subsection{Data analysis for future detectors}
\label{subsec:future_challenges}

\vspace{-.1cm}

\noindent \textit{Initial contributors: Maenaut, Pitte, Toubiana}

\vspace{.2cm}

The high SNR of ``golden'' merger events in XG detectors can lead to the detection of as many as five or more QNMs.
This leads to challenges both on the model and on the data analysis front, that we will discuss in this section.

\subsubsection{Ringdown modeling challenges}
Extending spectroscopy to earlier times and around merger, where the SNR is larger, remains the most significant (and rewarding) challenge on the modeling side.
Compared to numerical approaches, analytical, first-principles templates may allow us to extend fundamental physics explorations to larger classes of modified theories or to include environmental effects.
It is not possible to construct large catalogs of NR simulations (and hence
phenomenological EOB-like templates, see Section~\ref{sec:effective-one-body})
for each of these cases.
As discussed in Section~\ref{sec:waveforms}, significant advancements in analytical techniques are necessary to handle the dynamical ringdown regime near the peak. 

In addition, effects beyond the linear ringdown approximation at late times, that are likely irrelevant for current observations, will become important for XG detectors. 
As discussed in Section~\ref{sec:nonlinSch}, quadratic modes could be as important as higher angular modes and overtones, and will need to be included in the analysis. Their presence can complicate the identification of modes found in agnostic ringdown analyses and affect the performance of ``Kerr-like'' templates. 
Moreover, the time dependence of the final mass and spin
generally leads to cubic corrections that accumulate secularly and might affect the signal (see Section~\ref{sec:mtaft}). 
In the late portion of the signal, backscattering gives rise to a tail that cannot be described by a superposition of damped sinusoids (see Section~\ref{sec:tails}). 
For highly eccentric systems, this tail can become dominant already at $30M$ after the peak. Its measurability is an interesting topic for future work. 

In some tests of GR, the goal is to verify the consistency between the inspiral and the ringdown portion of the signal. These tests must include the effect of relativistic recoils (or ``kicks'') imparted to the remnant~\cite{Gerosa:2018qay, Varma:2020nbm}.
If the time variation of the mass is sufficiently small, the kick can be reabsorbed into the redshifted (constant) mass, and as such, it should have no significant impact on ringdown-only tests of GR, because all QNM frequencies are uniformly shifted  under such a transformation. 
If instead the remnant receives a large kick, estimates of the final mass and spin from the inspiral and from the ringdown could be inconsistent with each other, erroneously suggesting a violation of GR.
The magnitude of the kick strongly depends on the mass ratio, the magnitude and orientation of the binary spins, and the eccentricity. We can already infer that large kicks should be present in some of the systems observed so far~\cite{Varma:2022pld}.
The MBH binaries observed with LISA can have large spin  misalignment~\cite{Barausse:2020mdt,Toubiana:2021iuw} and large eccentricities even at the time of merger~\cite{Bonetti:2018tpf}, possibly leading to large kicks. Surrogate waveforms~\cite{Varma:2018mmi} obtained by interpolation of NR waveforms can accurately describe the merger signal, including the effect of kicks, and may be well suited to address the issues described above, but they have not been extensively used to test GR.

Finally, NR catalogs have expanded significantly since the development of the first ringdown amplitude and phase models~\cite{Berti:2007dg,Berti:2007zu,London:2014cma}. 
Currently, there are approximately 4500 publicly available NR waveforms from catalogs such as RIT, MAYA, Cardiff, and SXS~\cite{ritcatalog,gatechcatalog,sxscatalog,Hamilton:2023qkv}, see Appendix~\ref{sec:public_codes}.
These catalogs cover a broad range of physical parameter space in the mass ratio $q \in [1,15]$, component spin magnitudes $\chi_{1,2}\in [0,1]$ and eccentricity $e \in [0,1]$, 
and they are well adapted to the low mass-ratio and presumably low-spin BBH population currently observed in LVK data (although spin measurements are still uncertain). 
The coverage of the parameter space is rather unequal, with most simulations at $q\lesssim 4$ and moderate-low spins and bigger uncertainties in the large mass ratio, high-spin regime. 
The different NR catalogs rely on different initial data, evolution formalism and gauge/coordinate choices: for instance, the RIT, MAYA, Cardiff are evolved using the BSSNOK conformal formulation of the 3+1 Einstein field equations~\cite{Nakamura:1987zz,Evans:1989fnr,Shibata:1995we,Baumgarte:1998te}, while the SXS catalog is based on the generalized harmonic gauge formulation~\cite{Firedrich:1985,Garfinkle:2001ni,Pretorius:2005gq}.
They also use different wave extraction techniques and resolution schemes.
These factors affect the accuracy of the waveforms also in the ringdown regime~\cite{hinder:2013oqa}, but currently there is limited understanding of how these errors propagate across the parameter space, or how they accumulate when combining different NR catalogs.
This is particularly important for the calibration of subdominant or less explored effects (such as eccentricity, precession, quadratic modes, and tails), as these are some of the most challenging aspects of NR simulations. 

Understanding the relative importance of these physical effects is important to control ringdown systematics.
Waveform systematics affect also ringdown tests that make use of the full IMR signal, such as the \texttt{pSEOB} pipeline discussed in Section~\ref{sec:DataAnalysis}.
In fact, waveform mismodeling can lead to false violations of GR already for SNRs of $\sim 100$, which may be achievable in the near future~\cite{Toubiana:2023cwr}.
For both ringdown-only and IMR-based tests, one possibility is to acknowledge the incompleteness of our templates directly in the likelihood formulation, and marginalize over physical effects that are not included by accounting for them with an additional ``noise component.''
A first step in this direction for direct fits to NR simulations is discussed in~\cite{Redondo-Yuste:2023seq,Carullo:2023tff,Clarke:2024lwi}, and some data analysis applications have been presented in~\cite{Breschi:2022ens,Pompili:2024yec,Mezzasoma:2025moh}.

\subsubsection{Ringdown data analysis challenges}

On the data analysis side, the number of detectable modes increases with large SNR. This will cause difficulties for both the agnostic and \texttt{Kerr} approaches for measuring multiple modes (using the terminology introduced in Section~\ref{sec:DataAnalysis}). 
For agnostic searches, Bayesian inference for a high number of modes is a
challenging task for Markov chain Monte Carlo (MCMC) samplers.
There is a high number of parameters (4 per mode), and if no ordering is imposed, these modes can be interchanged when exploring the parameter space. This can affect the convergence of the samplers and the interpretation of the results. 
On the other hand, in the \texttt{Kerr} approach we must decide in advance which modes are included in the analysis. 
This is nontrivial, because the relative importance of higher angular modes and overtones is still debated.
This difficulty already affects current observations: for example, different analyses of GW190521 disagree on the mode content of the signal~\cite{LIGOScientific:2021sio,Capano:2021etf,Siegel:2023lxl}. 
To avoid performing Bayesian studies for all possible combinations of modes, one possibility is to use reversible-jump MCMC techniques~\cite{Green:1995mxx} and sample over the different mode combinations at once.
This would yield the BF between different mode combinations and the posterior distribution of the parameters of the model for each combination, providing a good fit to data.  

The complexity of upcoming interferometers will also complicate the treatment of the likelihood. 
The post-processing technique of choice for LISA is time-delay interferometry (TDI)~\cite{Tinto:2004wu}. The idea is to suppress laser fluctuation noise (that would otherwise be orders of magnitude stronger than other noise sources and the expected GW signals) by considering time-shifted linear combinations of the Doppler measurements between pairs of spacecraft. 
In contrast to ground-based detectors, where this noise is canceled when the laser light from the two
arms recombines at the beam splitter, LISA will measure changes in the length of each arm, which varies due to the motion of the spacecraft on their respective orbits. 
The output of this process are TDI variables called $X$, $Y$, and $Z$, that can be constructed to replicate Michelson interferometers between each pair of arms.
These variables can be further combined (under idealized conditions) to yield
two independent variables $A$ and $E$ in signal space, and one variable $T$ in null space.
Other sets of variables -- such as the Sagnac variables $\alpha, \beta, \gamma$, new variables named $\mathcal{A}, \mathcal{E}$ and $\mathcal{T}$ constructed upon the Sagnac variables, and a null variable called $\zeta$ -- have been proposed as combinations of different links in order to suppress instrumental noises, even in the case of a malfunctioning link~\cite{Vallisneri:2005ji, Tinto:2020fcc,Muratore:2020mdf,Hartwig:2021mzw, Muratore:2022nbh}. 
The use of $A$, $E$ and $\zeta$ is the optimal recommended combination to perform noise characterization and parameter estimation in the case of unequal arm lengths and unequal noise levels~\cite{Hartwig:2023pft}.

In the context of ringdown, TDI leads to combining parts of the signal from before and after the selected ringdown starting time $t_{\rm start}$, eventually polluting the ringdown signal in the data stream with parts of the signal that are not well-described by a ringdown template. One proposed solution is to start the analysis later than $t_{\rm start}$ by accounting for the delays between the spacecraft~\cite{Pitte:2024zbi}, which could add up to $\sim 70\,{\rm s}$ for the current TDI implementation. 
The downside is that this cropping procedure means that we would lose the loudest part of the signal, with a considerable loss of SNR. 

\begin{figure}[t]
    \centering
    \includegraphics[width=0.49\textwidth]{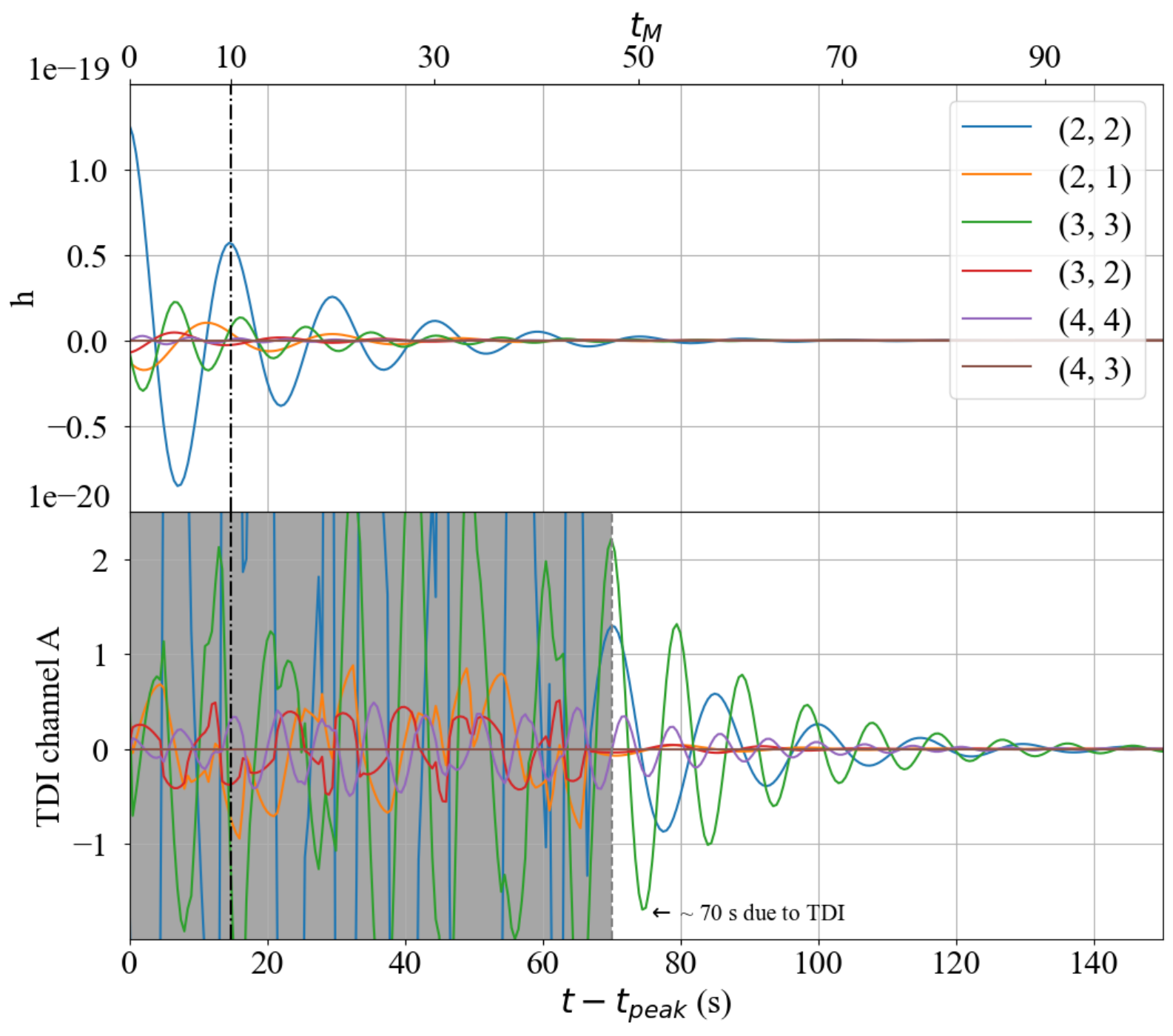}
    \includegraphics[width=0.49\textwidth]{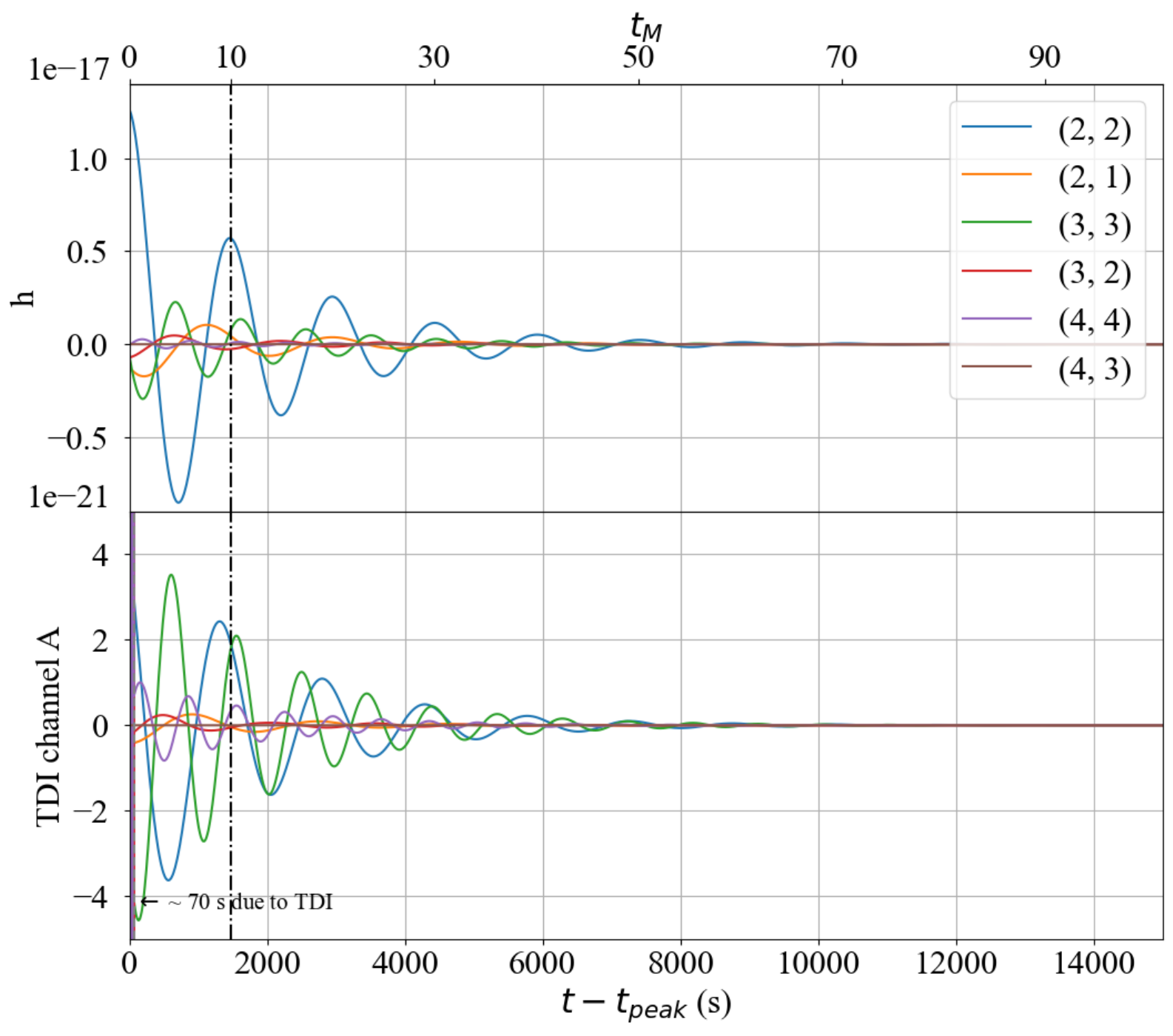}
    \caption{\label{fig:sec_7_ringdown}Ringdown signal for the strain $h$ and the TDI variable $A$, for systems with total mass $3 \times 10^{5} M_\odot$ (left) and $3 \times 10^{7} M_\odot$ (right), aligned spins ($\chi_1=\chi_2=0,5$), and mass ratio $q=2$ at redshift $z=4$. The upper $x$-axis shows time in units of $M$ and the lower one in physical units (note the different scales between the left and right panels). The initial $70\,{\rm s}$ are not included in the analysis to avoid polluting the ringdown signal in the TDI variables with the corrupted portion of the signal.}
\end{figure}

As an example, in Fig.~\ref{fig:sec_7_ringdown} we show the transformation of a ringdown waveform as it would appear in the final data set, i.e., including the LISA instrument response and TDI (the figure refers to the TDI variable $A$).
For $t<0$, the signal is padded with zeros. 
The gray shaded area shows the $70\,{\rm s}$ that are lost due to the delays. 
Since this time window is not mass-dependent, for light sources of $\mathcal{O} (10^{5}M_\odot)$, whose ringdown signal are shorter, this discarded time could almost encompass the entirety of the signal.
For instance, the ringdown SNR of a source with mass $M=3\times 10^{5} M_\odot$ starting at $10M$, without the response of LISA nor the TDI post-processing, would be around 260. 
Now, applying the response to the data means that the first $\sim 70$ s will be corrupted, and that chunk of data must be cropped. The SNR of the remaining data decreases to a value of 8. 
For this low-mass source, the observed ringdown represents only $3\%$ of the total SNR achievable without truncation, which translates into a loss of $\sim 97 \%$ of the signal.
In contrast, for a heavier source with mass $M=3\times 10^{7} M_\odot$, the corrupted data merely induce a $\sim 4.5\%$ SNR loss: when including the response and starting at $10M$, we would have a total SNR of $475$.
It is important to mention that the TDI generally does not corrupt the data. This issue only arises for ringdown models (or any other models) where the signal has been padded before the starting time.

Figure~\ref{fig:sec_7_ringdown} highlights another peculiarity of the LISA instrument: its response is frequency-dependent. 
Therefore, when converted into TDI variables, sub-dominant modes might be more excited than the dominant one~\cite{Pitte:2023ltw}. 
For instance, in Fig.~\ref{fig:sec_7_ringdown} we can see that the $(3,3)$ mode in the $A$ channel has amplitude comparable to the dominant $(2,2)$ mode for both systems, precisely because of the frequency dependence of the response.

Moreover, performing ringdown data analysis in the TD involves dealing with Toeplitz matrices. This is trickier than the standard FD analysis, for reasons discussed in Section~\ref{sec:DataAnalysis}. Since LISA's PSD is nonstationary~\cite{Tinto:2020fcc, Hartwig:2021mzw, Hartwig:2023pft,Rosati:2024lcs,Buscicchio:2024wwm,Piarulli:2024yhj} and not well conditioned, the inversion of the Toeplitz matrix to compute the likelihood requires special care.
State-of-the-art LISA data analysis uses the TDI channels $A$, $E$ and $T$, assuming them to be noise-independent. However, this is only true under certain conditions (namely, equal arm lengths and stationary instrumental noise~\cite{Prince:2002hp}).

In a more realistic scenario, one must account for the unequal arm lengths and the presence of correlated noise between the channels. In the TD, this requires the inversion of a matrix of size $(3N \times 3N)$, where $N$ is the number of time points (there are ways to deal with these kinds of matrices as block matrices). Alternatively, one could use the noise spectral densities written as a $(3\times 3)$ matrix of frequency functions, and perform the inversion in the FD. However, novel techniques are needed to robustly transform this to the TD in order to compute the likelihood. These issues make the TD ringdown analysis even more complicated for LISA.

An alternative solution of the problems mentioned above could be to extend the analysis to the pre-ringdown signal using either IMR waveforms~\cite{Correia:2023ipz} or agnostic approaches to describe that portion of the signal (such as wavelets~\cite{Finch:2021qph} or in-painting~\cite{Wang:2023xsy}). These techniques, currently being tested in LVK data analysis, have two advantages: (i) they include a larger portion of the signal in the analysis, and (ii) they allow us to perform the analysis in the FD rather than the TD, since one can more safely window the signal and avoid Gibbs noise when taking discrete Fourier transforms. In the case of LISA, the application of these techniques will require ensuring smoothness between the pre-ringdown and ringdown portions, in order to ensure continuity in the TDI channels.
The LISA response acts effectively as a second derivative of the waveform, so the overall waveform must be at least $C^{2}$.

It has also been suggested that using wavelet transforms and working in the time-frequency domain might be better suited for LISA data analysis~\cite{Cornish:2020odn}, one of the reasons being that the noise covariance matrix in the time-frequency domain remains diagonal even in presence of small nonstationarities. Since ringdown data analysis depends on both time and frequency, these ideas deserve further research.

In general, LISA data analysis will be challenging due to the presence of a large number of overlapping signals within the data stream at any given time. In order to disentangle the sources, it is necessary to perform a ``global fit''~\cite{Littenberg:2023xpl,Katz:2024oqg}, where all sources are fitted at the same time. In particular, the detector noise must be characterized together with the properties of different sources, and its characterization will improve with the mission lifetime.
Global fitting techniques are rapidly being expanded to include additional signal classes and more realistic assumptions~\cite{Edwards:2020tlp, Hartwig:2023pft, Castelli:2024sdb, Piarulli:2024yhj, Houba:2025ckw}, but there is no robust demonstration of unbiased tests of GR under realistic assumptions.
In this sense, our ability to perform robust tests of GR with LISA signals has not been demonstrated. 
For example, is still unclear if we can first perform a global fit assuming GR to be correct and perform tests of GR \textit{a posteriori} on the detected signals, because in principle the omission of deviations from GR could bias the global fit. 

For XG ground-based detectors, such as ET and CE, some of the challenges for
ringdown data analysis are similar to those of LISA. The very large number of
anticipated signals each year
($\sim 10^5-10^7$~\cite{Relton:2022whr,Samajdar:2021egv}) increases the chance
of overlapping signals. In the presence of so many signals, noise
characterization becomes challenging, since no detector data stretch will be
free of signals. If the detector has a null stream (as would be the case with a
triangular configuration for ET), this could be leveraged to characterize the
noise~\cite{Wong:2021eun,Goncharov:2022dgl}. A possible alternative would be to
simultaneously fit the signal model and an agnostic noise
model~\cite{Littenberg:2014oda}. The presence of overlapping signals can be a
concern for parameter estimation.  When the merger times of the overlapping
signals are close to each other (a relatively rare occurrence), this can
introduce biases in the
estimates~\cite{Relton:2022whr,Pizzati:2021apa,Samajdar:2021egv}. Furthermore,
systematics due to waveform modeling inaccuracies and overlapping signals can
lead to false GR deviations when multiple signal overlaps occur and are not
taken into account~\cite{Hu:2022bji}. Correctly modeling GW signals from other
compact binary mergers will be important to avoid an extra residual
background. The techniques for ``source subtraction'' that are being developed
for LISA will also be relevant and could be adopted to tackle these
challenges~\cite{Littenberg:2023xpl, Katz:2024oqg, Strub:2024kbe, Deng:2025wgk}.

Depending on the timeline of future detectors, multi-band observations by different instruments across different frequency bands could be revolutionary. 
Binaries as massive as GW150914 or intermediate-mass BH binaries (if they exist in the Universe) may be observed by a space-based detector such as LISA years or months before merging in the ground-based frequency band~\cite{Amaro-Seoane:2009vjl,Sesana:2017vsj}. 
This opens up the possibility of ``tuning'' ground-based detectors to maximize the science output from these signals.
If experimentally viable, the detuning of the signal-recycling cavity in coordination with LISA early-warning observations can drastically increase the sensitivity of a ground-based detector to the ringdown phase of a GW150914-like signal~\cite{Tso:2018pdv}. 
A similar concept would be applicable in the context of binary neutron star post-merger signals. 
The prospect of coordinated observational campaigns fine-tuned to extract specific source properties highlights the new and exciting possibilities opened by multi-band GW observations~\cite{Carson:2019rda, Carson:2019kkh, Datta:2020vcj, Gupta:2020lxa}.

\clearpage
\section{Conclusions}
\label{sec:conclusions}

The remarkable uniqueness theorems of GR imply that the resonant frequencies of Kerr BHs  depend only on their mass and spin. As we highlighted at the beginning of Chapter~\ref{chap2}, Chandrasekhar and Detweiler soon realized that {\em ``after the advent of gravitational wave astronomy, the observation of these resonant frequencies might finally provide direct evidence of BHs with the same certainty as, say, the 21cm line identifies interstellar hydrogen''}~\cite{Detweiler:1980gk} -- i.e., that BHs can be treated as ``gravitational atoms''.

What was initially wishful thinking has become reality. While this manuscript was under review, on September 10 2025 -- almost exactly ten year after the first GW detection -- the LVK Collaboration published the landmark observation of GW250114: an event with total SNR of 80, so loud that its SNR {\em in the ringdown alone} was comparable to the SNR of the first detection (GW150914)~\cite{LIGOScientific:2025rid,LIGOScientific:2025obp}. This remarkable event allowed the LVK Collaboration to infer that postmerger data (excluding the near-peak region, and thus bypassing many of the controversial issues discussed in this review) are consistent with the dominant $\ell=|m|=2$ mode of a Kerr black hole and with its first overtone; to constrain the frequency of the first overtone to $\pm 30$\% of the Kerr spectrum, providing the first spectroscopic test of the Kerr nature of the merger remnant; and to verify Hawking’s area law (also known as the second law of black hole mechanics), which states that the total area of the BH event horizons cannot decrease with time.

This review was a collective effort to summarize our current understanding of BH QNM spectra in GR (Chapter~\ref{chap2}) and beyond GR (Chapter~\ref{sec:beyondGR}); attempts to better understand QNM excitation in BH binary mergers, either by pushing perturbation theory to higher orders, or by using numerical relativity simulations (Chapter~\ref{sec:amplitudes}); state-of-the-art analytical models of post-merger GWs (Chapter~\ref{sec:waveforms}); ringdown data analysis methods and results (Chapter~\ref{sec:DataAnalysis}); and the new possibilities that will be enabled by future detectors on the ground and in space (Chapter~\ref{sec:nextgen}).

In hindsight, GW250114 was not only the culmination of more than five decades of theoretical and experimental studies of BH spectroscopy, but also the beginning of a new era in precision measurements of BH dynamics. The last decade of LVK observations has truly triggered a second golden age of BH perturbation theory. The hope is that in the near future, exquisitely accurate measurements of ringdown radiation will shed light on some of the deepest mysteries in gravitational physics (such as the information loss paradox, strong cosmic censorship, the nature of dark matter and dark energy), very much like the first decades of atomic spectroscopy shaped our understanding of the laws of quantum mechanics.

\clearpage
\section*{Appendices}
\begin{appendices}

\section{Useful primers, public codes, and public data}
\label{sec:public_codes}

Besides Chandrasekhar's monograph~\cite{Chandrasekhar:1985kt} and the reviews mentioned in the introduction, pedagogical introductions to BH ringdown, perturbation theory and QNM physics can be found in:

\begin{itemize}

    \item Lecture notes of the ``\textit{Summer School on Gravitational-Wave Astronomy}'' held at ICTS-Bangalore~\cite{icts_school};
    
    \item Lecture notes in ``\textit{NRHEP Spring School}'' held at IST-Lisbon~\cite{Pani:2013pma};

    \item Chapter~12 of the book~\cite{Maggiore:2018sht};

    \item Chapter~17 of the book~\cite{Andersson:2019yve}.

\end{itemize}

\noindent Below we list publicly available software and data mentioned in different sections of this review.

\begin{itemize}
\item From Section~\ref{sec:Solving for modes}: Extensive sets of gravitational
  Kerr QNM and TTM solutions are freely
  available~\cite{GRIT,CoG,JHU,cook_2024_14024959, motohashi_2024_12696858,
    Lo:2025njp}, along with the \texttt{Mathematica} routines used to compute
  them~\cite{GRIT,CoG,JHU,KerrModes_2024_cook}.
  An open-source \texttt{python} package for computing QNMs is also
  available~\cite{Stein:2019mop}.

\item From Section~\ref{sec:KNqnm}: Kerr-Newman QNM data for the equal
  charge-spin case can be downloaded from the arXiv source
  of~\cite{Dias:2015wqa}, and closed-form fits for the dominant modes are
  available in~\cite{Carullo:2021oxn}.
    
\item From Section~\ref{sec:avoidance_KerrGWs}: The full dataset of Kerr QNM
  frequencies and excitation factors is publicly
  available~\cite{motohashi_2024_12696858,GRIT}.

\item From Section~\ref{sec:hysteresis}: Datasets and numerical codes associated
  with massive scalar QNMs around Kerr BHs, based on Leaver's
  method~\cite{Leaver:1985ax,Nollert:1993zz,Konoplya:2006br,Dolan:2007mj,Konoplya:2013rxa}
  and on the isomonodromic
  method~\cite{CarneirodaCunha:2015hzd,daCunha:2021jkm}, are publicly
  available~\cite{zenodo13961216}.

\item From Section~\ref{sec:mod_Kerr}: The functions $H_{i}$ and $\Phi$ have
  been derived up to $14$-th order in a dimensionless spin
  expansion~\cite{Cano:2019ore}, but these calculations can in principle be
  extended to arbitrary orders. A \texttt{Mathematica} notebook to construct
  such extensions is publicly available~\cite{Cano:2019ore,RotatingBHWolfram}.

\item From Section~\ref{sec:modgrav_res}: The results agree with previous
  computations for static and slowly rotating BHs obtained via modified RWZ
  equations~\cite{Cardoso:2018ptl,deRham:2020ejn,Cano:2021myl}, and are publicly
  available~\cite{gitbeyondkerr}.

\item From Section~\ref{sec_param_pot}: Numerical data of the beyond-GR
  coefficients in Eq.~\eqref{eq:omega_mod} are publicly available
  in~\cite{GRIT,CoG,JHU, Hirano:2024fgp}.
    
\item From Section~\ref{sec_param_pot}: The expansion coefficients of the
  parameterized QNM framework beyond Schwarzschild/Kerr presented
  in~\cite{Cano:2024jkd} can be found in a public \texttt{github}
  repository~\cite{sebastian_volkel_2024_14001739,GRIT,CoG}.

\item From Section~\ref{subsubsec:echotemplates} and~\ref{subsec:echoes_DS}:
  Waveform models for GW echoes, based on~\cite{Abedi:2016hgu, Nakano:2017fvh},
  are publicly available~\cite{echowfm} as a \texttt{python} package.

\item From Section~\ref{sec:quadmodes_eq}: Numerical routines to obtain
  solutions to the Regge-Wheeler and Teukolsky equations, spin weighted
  spheroidal harmonics, and the second-order source terms of the linearized
  Einstein and Teukolsky equations (only for Schwarzschild) are publicly
  available in the \texttt{Black Hole Perturbation Toolkit}~\cite{BHPToolkit}.

\item From Section~\ref{sec:quadmodes_eq}: The generic $\ell, \ell_1, \ell_2$
  case was tackled in~\cite{Brizuela:2006ne,Brizuela:2007zza,Brizuela:2009qd},
  although only the recent work in~\cite{Bucciotti:2024jrv,Bucciotti:2024zyp}
  provided an expression for the source term publicly available to the
  community.

\item From Section~\ref{sec:waveforms}: Accurate fitting routines of NR data
  employing Bayesian techniques and stochastic samplers, allowing to include
  intrinsic NR errors, are available through the \texttt{bayRing}
  package~\cite{carullo_gregorio_2023_8284026}, introduced in
  Ref.~\cite{Redondo-Yuste:2023seq}.

\item From Section~\ref{subsubsec:fitting_methods}: Tools to fit ringdown
  amplitudes with Gaussian Process Regression are available in the
  \texttt{postmerger} package~\cite{pacilio_2024_13220424} introduced
  in~\cite{Pacilio:2024tdl}.

\item From Section~\ref{subsubsec:fitting_methods}:
  \texttt{qnmfits}~\cite{qnmfitscode} and
  \texttt{KerrRingdown}~\cite{KerrRingdownCode} are \texttt{python} and
  \texttt{Mathematica} packages, respectively, built for high-precision QNM
  amplitude fits. While \texttt{qnmfits} includes a greedy algorithm to fit for
  the most important QNMs of a given BBH system, \texttt{KerrRingdown} can use
  different methods, such as the limited-mode eigenvalue
  method~\cite{Cook:2014cta}, to perform the fits. Both codes have the
  capability to fit for the remnant mass and spin of a BBH system using any
  given waveform, and at any given starting time. Handling of CCE waveforms and
  supertranslations is included for easy use.
     
\item From Section~\ref{subsubsec:fitting_methods}:
  \texttt{NRSur3dq8\_RD}~\cite{MaganaZertuche:2024ajz} is a ringdown surrogate
  model that gives high-precision QNM amplitude fits as well as remnant mass and
  spin values using Gaussian Process Regression. This surrogate model was
  trained using CCE waveforms in the superrest frame of the remnant, with
  amplitudes extracted from the \texttt{qnmfits} package. \texttt{NRSur3dq8\_RD}
  is available as part of the larger \texttt{python} package
  \texttt{surfinBH}~\cite{vijay_varma_2018_1435832}.

\item From Section~\ref{subsec:current_observations}:
  \texttt{pyRing}~\cite{pyRing} is a \texttt{python} package for ringdown
  analysis, model comparison and parameter estimation. Tailored to the
  post-merger phase of CBC GW signals, it uses a native TD likelihood
  formulation. Integrated with standard LVK software, natively \texttt{pyRing}
  incorporates a variety of ringdown waveform models, including \texttt{Damped
    sinusoids} and \texttt{Kerr}~\cite{Berti:2009kk} (including variants
  enforcing amplitude symmetries and ratios consistent with nonprecessing
  binaries), \texttt{KerrBinary}~\cite{Kamaretsos:2011um, Kamaretsos:2012bs,
    London:2014cma, London:2018gaq, Cheung:2023vki},
  \texttt{KerrPostmerger}~\cite{Damour:2014yha}, parameterized tests of GR, and
  QNM spectrum predictions for some alternatives to the Kerr
  solution~\cite{Dias:2015wqa, Cano:2023jbk, Pierini:2022eim}. The majority of
  its features have been internally reviewed for scientific usage by the LVK
  collaboration. The code is routinely used to produce catalogs of BH ringdown
  properties and tests of GR in the merger-ringdown regime by the LVK
  collaboration~\cite{LIGOScientific:2020iuh, LIGOScientific:2020ufj,
    LIGOScientific:2019fpa, LIGOScientific:2021sio}.
  Configuration files for all available models and configurations are available
  in the main repository~\cite{pyRing}.
    
\item From Section~\ref{subsec:current_observations}: \texttt{ringdown}~\cite{ringdown} is a \texttt{python} package for Bayesian inference of
  ringdown signals in TD GW data. It leverages auto-differentiation, hardware
  acceleration (CPU and GPU) and coordinate transformations to efficiently
  compute posterior samples for the frequencies, damping rates, amplitudes and
  phases of QNMs that may be present in a stretch of data. The semi-analytic
  treatment of linear parameters and the use of a gradient-based sampler
  (no-U-turn Hamiltonian Markov chain Monte Carlo) facilitate scaling to large
  number of modes, and reduce computation time -- e.g., currently achieving
  ${\sim}1$ minute sampling for a two-mode analysis on a consumer grade
  laptop. \texttt{ringdown} currently supports \texttt{Damped sinusoids} and
  \texttt{Kerr} templates, including arbitrary QNMs polarization structure or
  enforcing $z$-parity symmetry (expected for nonprecessing systems). The
  package also includes tools to facilitate handling noise power spectra and
  autocovariance functions, as well as simulating data with damped-sinusoid or
  full-CBC signals.
    
\item From Section~\ref{subsec:current_observations}: \texttt{PyCBC Inference}~\cite{Biwer:2018osg} provides Bayesian parameter inference and model selection
  methods within the \texttt{PyCBC} \texttt{python} package for general GW data
  analysis. It performs ringdown analyses when supplied with the appropriate
  configuration file.  The implemented ringdown models include \texttt{Damped
    sinusoids} and \texttt{Kerr}, allowing for parameterized deviations from the
  GR mode parameters, using symmetries expected for nonprecessing systems or
  generic polarizations.  A waveform plug-in interface enables inclusion of
  external models.  Data treatment enabling ringdown analysis is given by the
  gating and in-painting method, which replaces data to be excluded from the
  analysis such that they do not contribute to the likelihood. The likelihood
  uses the FD formulation common in GW analysis.  This method also allows to
  compare models using different data such as ringdown models with different
  start times, and thus to sample and marginalize over uncertainties in
  coalescence time and sky location.  \texttt{PyCBC} is available at~\cite{nitz_pycbc_zenodo}, and a ringdown-specific tutorial is provided at~\cite{nitz_2020_pycbc_tutorial}.  Example configuration files are available
  for a multi-mode analysis of GW190521 at~\cite{capano_2021_10493247}, and for
  a study marginalizing over sky location and coalescence time for GW150914
  at~\cite{correia_2024_10493247}.

\item From Section~\ref{subsec:deviations_current}: The parameterized EOB
  waveform models employed in the \texttt{pSEOB} analysis are implemented in
  \texttt{LALSuite}~\cite{lalsuite} and
  \texttt{pySEOBNR}~\cite{Mihaylov:2023bkc}. Wrappers for running the analysis
  using the \texttt{Bilby} parameter estimation package are provided in
  \texttt{Bilby TGR}~\cite{ashton_2024_10940210}.
    
\item From Section~\ref{subsec:future_challenges}: Currently, there are
  approximately 4500 publicly available NR waveforms from catalogs such as RIT,
  MAYA, Cardiff, and
  SXS~\cite{ritcatalog,gatechcatalog,sxscatalog,Hamilton:2023qkv}.

\end{itemize}

\clearpage

\noindent 
\section{The Newman-Penrose and Geroch-Held-Penrose formalism}\label{sec:NP_GHP}

\vspace{-.1cm}

\noindent \textit{Initial contributor: Redondo-Yuste}

\vspace{.2cm}

The simplicity of the equations describing the perturbations of a Kerr BH relies on the algebraic and geometric properties of the Kerr spacetime. In this appendix we review some of these properties, which allow for the decoupling of the equations describing gravitational perturbations in Kerr. We also introduce the formalisms based on this intuition developed first by Newman and Penrose (NP)~\cite{Newman:1961qr}, and later by Geroch, Held and Penrose (GHP)~\cite{Geroch:1973am}. 

As a starting point, let us revisit an important property of any Lorentzian geometry. Consider the Bianchi identity
\begin{equation}\label{eq:Derivative_Bianchi}
    \mathcal{B}_{abcdf} \equiv\nabla_f R_{abcd}+\nabla_b R_{facd}+\nabla_a R_{bfcd}=0 \,,
\end{equation}
and note that by taking another covariant derivative, i.e., $\nabla^f \mathcal{B}_{abcdf}=0$, we find an equation involving the wave operator $\Box = \nabla_a \nabla^a$ acting on the Riemann tensor: 
\begin{equation}\label{eq:Intermediate_one}
    \begin{aligned}
        \Box R_{abcd} &+ \nabla_a\nabla^f R_{bfcd}-\nabla_b\nabla^fR_{afcd}+2\Bigl(R_{a\,d}^{\,f\,h}R_{bfch}-R_{a\,c}^{\,f\,h}R_{bfdh}-R_{a\,b}^{\,f\,h}R_{cdfh}\Bigr) \\ 
        &+R_{a\,cd}^{\,f} R_{bf}-R_{b\,cd}^{\,f}R_{af} = 0 \, .
    \end{aligned}
\end{equation}
Now, by using the contractions $g^{hf}\nabla_a\mathcal{B}_{cdhbf}=0$ and $g^{hf}\nabla_b\mathcal{B}_{cdahf}=0$, the second-order derivatives of the Riemann tensor can be traded for second-order derivatives of the Ricci tensor, that vanishes for vacuum solutions of the Einstein equations. Thus, the Riemann tensor itself satisfies a nonlinear wave equation, as first observed by Penrose~\cite{Penrose:1960eq}. In fact, we can write the Riemann tensor in terms of the Weyl tensor $C_{abcd}$ and find, instead, a nonlinear wave equation for the Weyl tensor:
\begin{equation}\label{eq:Penrose_Wave_Eq}
    \Box C_{abcd} + 2\Bigl(C_{a\,d}^{\,f\,h}C_{bfch}-C_{a\,c}^{\,f\,h}C_{bfdh}-C_{a\,b}^{\,f\,h}C_{cdfh}\Bigr) = f(\mathrm{Ric}) \overset{\rm Vacuum}{\equiv} 0 \, , 
\end{equation}
where the right-hand side contains terms proportional to the Ricci tensor and its derivatives or contractions, so that it vanishes in vacuum. Remarkably, this equation shows clearly how gravity sources itself. Even in vacuum, the equation describing the propagation of curvature fluctuations is nonlinear. It is only for small-curvature fluctuations near flat space that $\Box C_{abcd}\simeq 0$.

Motivated by the role that the Weyl tensor plays in the dynamics of gravitational fields, Penrose examined its algebraic structure from a spinorial point of view. Ref.~\cite{Penrose:1985bww} provides a pedagogical introduction to the topic. In four dimensions, the Weyl tensor can be written uniquely in terms of a single spinor $\psi_{ABCD}$ which is completely symmetric in the spinor indices $A,B,C,D=0,1$. A completely symmetric tensor with four indices taking values in two (complex) dimensions is uniquely described by $5$ complex scalars. These are the well known Weyl scalars $\Psi_{0,\dots,4}$. Alternatively, it can be shown that the Weyl spinor can be decomposed as $\psi_{ABCD}= \kappa^{(1)}_{(A}\kappa^{(2)}_{B}\kappa^{(3)}_{C}\kappa^{(4)}_{D)}$ in terms of four spinors $\kappa^{(i)}_A$. The directions formed by these spinors, defined as $X^{(i)}_a = \kappa^{(i)}_A\bar{\kappa}^{(i)}_{A'}$ are the principal null directions (PNDs) of the spacetime. Petrov's~\cite{1969eisp.book.....P} algebraic classification now simply states how many of these directions are equal to each other. Type II spacetimes are those where two of the directions are equal, say, $\kappa^{(1)}_A=\kappa^{(2)}_A$, while the other two PNDs are different. The Kerr metric corresponds to an even more special class of spacetimes, labeled type D, where there are two repeated PNDs, i.e., $\kappa^{(1)}_A=\kappa^{(2)}_A$ and $\kappa^{(3)}_A=\kappa^{(4)}_A$. These repeated PNDs can each be associated to ingoing and outgoing null geodesics. 

The NP formalism~\cite{Newman:1961qr} was introduced shortly after Penrose's seminal paper~\cite{Penrose:1960eq} connecting the tetrad and spinor languages. A null tetrad is a set of four null vectors $E_a^{\bf a} = (l_a, n_a, m_a, \bar{m}_a)$, where all of them are null, satisfying $l^an_a=-1, m^a\bar{m}_a=1$, where all other contractions of tetrad vectors vanish. (It is also common to study the NP formalism using the mostly minus metric signature, in which case the normalization is flipped, $l^a n_a=1, m^a\bar{m}_a=-1$; however in this Appendix, as in the rest of the review, we use the ``mostly plus'' convention.) Thus, the tetrad indices can be raised and lowered with the tetrad metric $\eta_{\bf ab}$, 
\begin{equation}\label{eq:metric_to_tetrad}
    g_{ab} = \eta_{\bf ab}E_a^{\bf a}E_b^{\bf b}=-2l_{(a}n_{b)}+2m_{(a}\bar{m}_{b)} \, , \qquad \eta_{\bf ab} = \begin{pmatrix}
        0 & -1 & 0 & 0 \\
        -1 & 0 & 0 & 0 \\
        0 & 0 & 0 & 1 \\
        0 & 0 & 1 & 0
    \end{pmatrix}  \, .
\end{equation}
Based on this tetrad, we can define directional derivatives, given by the projection along the tetrad vectors of the usual covariant derivative:
\begin{equation}\label{eq:NP_derivatives}
    D = l^a\nabla_a \, , \qquad \Delta = n^a\nabla_a \, , \qquad \delta = m^a\nabla_a \,  ,\qquad \bar{\delta} = \bar{m}^a\nabla_a \, .
\end{equation}
Similarly we can introduce the Ricci rotation coefficients $\gamma_{\bf abc} = E_{\bf a}^aE_{\bf b}^b \nabla_a (E_{\bf c})_b$. With the NP normalization, the Ricci rotation coefficients are described uniquely by twelve independent components, which are given specific variable names. Their definitions in terms of the tetrad components of the Ricci rotation coefficients are summarized below (for concreteness, we use the notation $\kappa = -\gamma_{\bf 131} = -l^a m^b \nabla_a l_b = -m^b Dl_b$):
\begin{equation}\label{eq:Spin_coefs}
    \begin{aligned}
        \kappa =& -\gamma_{\bf 131} \, , \quad \tau = -\gamma_{\bf 231} \, , \quad \sigma=-\gamma_{\bf 331} \, , \quad \rho = -\gamma_{\bf 431} \, , \\
        \pi =& \gamma_{\bf 142} \, , \qquad  \, \nu = \gamma_{\bf 242} \, ,\qquad  \mu = \gamma_{\bf 342} \,  \qquad \lambda=\gamma_{\bf 442} \, , \\
        \varepsilon =& -\frac{1}{2}\Bigl(\gamma_{\bf121}-\gamma_{\bf 143}\Bigr) \, , \qquad  \gamma = -\frac{1}{2}\Bigl(\gamma_{\bf221}-\gamma_{\bf 243}\Bigr) \, , \\
        \beta =& -\frac{1}{2}\Bigl(\gamma_{\bf321}-\gamma_{\bf 343}\Bigr) \, , \qquad  \alpha = -\frac{1}{2}\Bigl(\gamma_{\bf421}-\gamma_{\bf 443}\Bigr) \, ,
    \end{aligned}
\end{equation}
Now, the tetrad components of the Riemann tensor are directly related to first-order derivatives of the Ricci rotation coefficients. Indeed, by the Ricci identity, $2\nabla_{[b}\nabla_{c]}E^{\bf a}_a=-E^{\bf a}_d R^{\,\,\,d}_{abc}$. Rewriting this relation, we find 
\begin{equation}\label{eq:Riemann_tetrad}
    R_{\bf abcd} = E_{\bf a}^a\nabla_a\gamma_{\bf bcd}-E_{\bf b}^a\nabla_a \gamma_{\bf acd} - \gamma_{\bf acf}\gamma_{\bf bd}^{\,\,\,\,\bf f}+\gamma_{\bf bcf}\gamma_{\bf ad}^{\,\,\,\,\bf f} - \gamma_{\, \, \bf cd}^{\bf f}\Bigl(\gamma_{\bf afb}-\gamma_{\bf bfa}\Bigr) \, .
\end{equation}
Each of the tetrad components of Eq.~\eqref{eq:Riemann_tetrad} leads to one of the $18$ NP equations. These will relate the first-order derivatives of the spin coefficients~\eqref{eq:Spin_coefs} to terms which are quadratic in the spin coefficients, and linear terms in the Riemann tensor. In order to complete the NP equations, we need to expand the Riemann tensor appropriately. First, recall that the Riemann can be written in terms of the Weyl and Ricci tensors, through
\begin{equation}\label{eq:Riemann_to_Weyl}
    R_{\bf abcd} = C_{\bf abcd} +\frac{1}{2}\Bigl(\eta_{\bf ac}R_{\bf bd}-\eta_{\bf ad}R_{\bf bc}+\eta_{\bf bd}R_{\bf ac}-\eta_{\bf bc}R_{\bf ad}\Bigr)+\frac{R}{6}\Bigl(\eta_{\bf ad}\eta_{\bf bc}-\eta_{\bf ac}\eta_{\bf bd}\Bigr) \, .
\end{equation}
As discussed previously, the Weyl tensor can be decomposed uniquely into $5$ complex scalars. These scalars are denoted by $\Psi_i$ ($i=0,\dots,4$) and are given, in terms of its tetrad components, by
\begin{equation}\label{eq:Weyl_scalars}
    \Psi_0 = C_{\bf 1313} \, , \quad \Psi_1 = C_{\bf 1213} \, , \quad \Psi_2 = C_{\bf 1342} \, , \quad \Psi_3 = C_{\bf 1242} \, , \quad \Psi_4 = C_{\bf 2424} \, .
\end{equation}
Recall that, in terms of the spacetime components, this just means, e.g., $\Psi_4 = n^a\bar{m}^bn^c\bar{m}^dC_{abcd}$. The Ricci scalar is captured by a single real variable, which we label $\Lambda = R$. Finally, the traceless components of the Ricci tensor are given by $3$ real scalars $\Phi_{ii}$  ($i=0,1,2$) and $3$ complex scalars, $\Phi_{10},\Phi_{20},\Phi_{21}$, given by 
\begin{equation}\label{eq:Ricci_scalars}
    \begin{aligned}
        \Phi_{00} =& \frac{1}{2}R_{\bf 11} \, , \qquad \Phi_{11} = \frac{1}{4}\Bigl(R_{\bf 12}+R_{\bf 34}\Bigr) \, , \qquad \Phi_{22} = \frac{1}{2}R_{\bf 22} \, , \\
        \Phi_{10} =& \frac{1}{2}R_{\bf 14} \, , \qquad \Phi_{20} = \frac{1}{2}R_{\bf 44} \, , \qquad \qquad \qquad \Phi_{12} = \frac{1}{2}R_{\bf 24} \, .
    \end{aligned}
\end{equation}
Therefore, each of the components of Eq.~\eqref{eq:Riemann_tetrad} can be written in terms of NP derivatives of spin coefficients, products of spin coefficients, and terms linear in $\{\Lambda, \Phi_{ij}, \Psi_i\}$. These are known as the NP equations. Their exact expressions, together with their spinorial derivation, can be found, e.g., in~\cite{Newman:1961qr}.

This construction is generic for any spacetime; in particular, we have not made any assumptions about the spacetime's algebraic type. However, as discussed previously, BH geometries are a remarkable example of the so-called type D spacetimes, where there are two repeated PNDs, which can be associated with ingoing and outgoing null geodesics. 

About a decade after the introduction of the NP formalism, Geroch, Held, and Penrose realized that the formalism can be made even more clear~\cite{Geroch:1973am}. The key observation is that one can choose an NP tetrad where two of the basis vectors are aligned with the ingoing and outgoing PNDs. Let $l_a$ (respectively $n_a$) be aligned with the outgoing (respectively ingoing) PND. Then, there is a residual two-dimensional freedom, which can be summarized through the following transformation:
\begin{equation}\label{eq:Ctimes_Gauge_Action}
    (l_a,n_a,m_a) \mapsto (\lambda \bar{\lambda} l_a, \lambda^{-1}\bar{\lambda}^{-1}n_a, \lambda \bar{\lambda}^{-1} m_a) \, , \qquad \lambda \in \mathbb{C}_\times \, , 
\end{equation}
where $\lambda$ is a nonzero complex number. These transformations correspond to Lorentz boosts and spatial rotations. This defines a principal $\mathbb{C}_\times$ frame bundle, which we label $P$. A representation of this gauge group with weight $(p,q)$ can be defined through the left group action $\eta \mapsto \lambda^p \bar{\lambda}^q \eta$ under the corresponding tetrad Lorentz transformation~\eqref{eq:Ctimes_Gauge_Action}. This defines associated $(p,q)$ line bundles $L_{p,q} = P\times_{(p,q)}\mathbf{C}$, and sections of this bundle are called GHP scalars, with GHP weight $\eta \overset{\circ}= (p,q)$.

The NP derivatives~\eqref{eq:NP_derivatives} do not map GHP scalars to GHP scalars with definite weight. However, a connection on the line bundle $L_{p,q}$ can be defined through the $1$-form 
\begin{equation}
    \omega_a = -\varepsilon^\prime l_a + \varepsilon n_a +\beta^\prime m_a - \beta \bar{m}_a \, ,
\end{equation}
where the prime operation corresponds to permuting simultaneously $l_a \leftrightarrow n_a$ and $m_a \leftrightarrow \bar{m}_a$. Thus, the covariant derivative on $L_{p,q}$ is 
\begin{equation}
    \Theta_a = \nabla_a -p\omega_a - q\bar{\omega}_a \, .
\end{equation}
Now, the tetrad projections of this covariant derivative have the right transformation properties, and map quantities with a well-defined GHP weight to quantities which also have a well-defined, albeit different, GHP weight. They are denoted by
\begin{equation}\label{eq:GHP_derivatives}
    \tho = l^a \Theta_a \,  , \quad \tho^\prime = n^a \Theta_a \, , \quad \edt = m^a\Theta_a \, , \quad  \edt^\prime = \bar{m}^a \Theta_a \, .
\end{equation}
These operators have GHP weight $\tho \overset{\circ}{=}(1,1),\tho^\prime \overset{\circ}{=}(-1,-1),\edt \overset{\circ}{=}(1,-1),\edt^\prime \overset{\circ}{=}(-1,1)$. Similarly, we can examine the NP spin coefficients and notice that $\varepsilon,\gamma,\beta,\alpha$ do not have definite GHP weight. Moreover, four of the spin coefficients with definite GHP weight are related to the other four through the prime operation defined above. These are
\begin{equation}
    \lambda = -\sigma^\prime \, , \quad \mu = -\rho^\prime \, ,\quad \nu = -\kappa^\prime \, , \quad \pi = -\tau^\prime \, .
\end{equation}
Thus, the fundamental ingredients of the GHP formalism are the four spin coefficients $\{\kappa,\tau,\rho,\sigma\}$, with GHP weights $\kappa \overset{\circ}{=}(3,1)$, $\tau \overset{\circ}{=}(1,-1)$, $\rho \overset{\circ}=(1,1)$, and $\sigma \overset{\circ}{=}(3,-1)$; their four primed counterparts; and the $4$ GHP derivatives~\eqref{eq:GHP_derivatives}. One of the main advantages of introducing all of this structure is that now we can always deal with quantities with a definite GHP weight. For example, all the Weyl and Ricci scalars~\eqref{eq:Weyl_scalars}--\eqref{eq:Ricci_scalars} have definite GHP weights. Thus, one can easily check the correctness of many equations through the sometimes lengthy derivations by making sure that the GHP weights on each side of an equality agree. This also simplifies remarkably the NP equations, given by (2.21)--(2.26) in~\cite{Geroch:1973am}, together with (2.30)--(2.32), as well as their primed versions, and it shows explicitly the frame independence of some equations.

Focusing on the case of the Kerr spacetime, we now show how the Teukolsky equation, of major importance in this review, can be derived by applying this formalism. The Kerr BH is a particular case of a type D geometry, so we can define an NP tetrad with $l_a,n_a$ aligned with the outgoing and ingoing repeated PNDs, respectively. A particularly common tetrad choice is the Kinnersley tetrad, which in Boyer--Lindquist coordinates is given by  
\begin{equation}\label{eq:Kinnersley_tetrad}
    \begin{aligned}
        l^a_{\rm Kin}\partial_a =& \frac{r^2+a^2}{\Delta} \partial_t + \partial_r + \frac{a}{\Delta}\partial_\phi \, , \quad n^a_{\rm Kin}\partial_a = \frac{1}{2\Sigma}\Bigl[(r^2+a^2)\partial_t-\Delta \partial_r + a\partial_\phi\Bigr] \, , \\
        m^a_{\rm Kin}\partial_a =& \frac{1}{\sqrt{2}(r+ia\cos\theta)}\Bigl(i a \sin\theta\partial_t+\partial_\theta+\frac{i}{\sin\theta}\partial_\phi\Bigr) \, .
    \end{aligned}
\end{equation}
In this case one can show that $\kappa = \kappa^\prime = \sigma = \sigma^\prime= 0$, and $\Psi_i = 0$ for $i=0,1,3,4$. Since the Kerr metric is a vacuum solution, it is Ricci flat, and thus $\Phi_{ij}=\Lambda = 0$ as well. Thus, the only nonzero spin coefficients (out of those with definite GHP weight) are
\begin{equation}
    \begin{aligned}
        \rho =& -(r-ia\cos\theta)^{-1} \, , \quad \tau = -\frac{ia}{\sqrt{2}}|\rho|^2 \sin\theta \, ,  \\
        \rho^\prime =& -\frac{|\rho|^2 \Delta}{2} \, , \quad \tau^\prime = -\frac{ia}{2}\rho^2\sin\theta \, , \quad \Psi_2 = M\rho^3 \, .
    \end{aligned}
\end{equation}
Let us now start from the Penrose wave equation~\eqref{eq:Penrose_Wave_Eq}. We will further assume a vacuum spacetime, i.e., that we deal with purely gravitational perturbations (note that the Teukolsky equation was first derived from the Penrose wave equation in~\cite{Ryan:1974nt}). Then, the right-hand side of Eq.~\eqref{eq:Penrose_Wave_Eq} is always zero. The left-hand side can be written solely in terms of GHP quantities as 
\begin{equation}
    \begin{aligned}
        24\bigl(&- \rho \sigma + \kappa \tau \bigr)\Psi_{2} + 4\Bigl[\sigma(\tau'+\bar{\tau}-2\edt')-\kappa(\rho'+\bar{\rho}'-2\tho')-\tau(\bar{\rho}-2\tho)+\rho(\bar{\tau}'-2\edt)\\
        &+3(\Psi_1 + \tho\tau+\tho'\kappa-\edt\rho-\edt'\sigma)\Bigr]\Psi_1 + \Bigl[\tho\tho'+\tho'\tho-\edt\edt'-\edt'\edt-(\rho'+\bar{\rho}')\tho-(\rho+\bar{\rho})\tho'\\
        &+(\bar{\tau}+\tau')\edt+(\tau+\bar{\tau}')\edt'+4(\tau\tau'+\kappa\kappa'-\rho\rho'-\sigma\sigma')-12\Psi_2\Bigr]\Psi_0 = 0 \, .
    \end{aligned}
\end{equation}
This equation is fully nonlinear, and valid for any vacuum solution. Let us now consider perturbations around the Kerr background, in the Kinnersley tetrad. We denote by $\delta X$ the first-order perturbation of any quantity $X$, so that $X = X^{\rm Kerr} + \delta X$. Notice that we can write the previous equation as
\begin{equation}\label{eq:General_Teukolsky_NL}
    \mathcal{O}_0 \Psi_0 + \mathcal{O}_1 \Psi_1 + \mathcal{O}_2 \Psi_2 = 0 \, .
\end{equation}
The Teukolsky equation is obtained by decoupling this equation and reducing it to a single equation for $\delta\Psi_0$. Since in the background, $\Psi_1 \overset{\rm Kerr}=0$, it will be sufficient to reduce the equation to a form where $\mathcal{O}_1 = 0$. Additionally, $\Psi_2\overset{\rm Kerr}{\neq}0$, so we need to make $\mathcal{O}_2 = 0$. First, we write commutators between the GHP derivatives, and use the commutator rules to reduce the number of second-order derivatives acting on $\Psi_0$. Then we can use the Bianchi identities to replace some of the derivatives acting on $\Psi_1$, namely
\begin{equation}
    \begin{aligned}
        \tho \Psi_{1} =& -3 \Psi_{2} \kappa + 4 \Psi_{1} \rho -  \Psi_{0} \tau' + \edt' \Psi_{0} \, , \\
        \edt \Psi_{1} =& - \Psi_{0} \rho' - 3 \Psi_{2} \sigma + 4 \Psi_{1} \tau + \tho' \Psi_{0} \, .
    \end{aligned}
\end{equation}
Finally, we use the NP equations to add to the equation the following vanishing quantity:
\begin{equation}
    \edt \tau' -\Bigl( \Psi_{2} + \kappa \kappa' -  \bar{\rho} \rho' -  \sigma \sigma' + \tau' \bar{\tau}' + \tho \rho'\Bigr)=0 \, ,
\end{equation}
and replace
\begin{equation}
    \edt \rho = -2 \Psi_{1} -  \kappa \rho ' + \kappa \bar{\rho}' -  \bar{\rho} \tau -  \sigma \bar{\tau} + \sigma \tau ' + \rho \bar{\tau}' + \tho \tau -  \tho' \kappa + \edt' \sigma \, , 
\end{equation}
once again by virtue of the NP equations. As a result, we obtain an equation of the form~\eqref{eq:General_Teukolsky_NL}, with 
\begin{equation}\label{eq:Teukolsky_Operators}
    \begin{aligned}
        \mathcal{O}_0 =& 2\Bigl[(\tho-4\rho-\bar{\rho})\tho'-(\edt-4\tau-\bar{\tau}')\edt'-\rho'\tho+\tau'\edt \Bigr] \\
        &+8(\rho\rho'-\tau\tau')+7(\kappa\kappa'-\sigma\sigma')+\bar{\rho}\rho'-\tau'\bar{\tau}'-\tho\rho'+\edt\tau'-5\Psi_2\, , \\
        \mathcal{O}_1 =& 8\Bigl(\kappa\tho'-\sigma\edt'-\kappa\bar{\rho}'+\sigma\bar{\tau}-\edt'\sigma+\tho'\kappa\Bigr) +20\Psi_1\, , \\ 
        \mathcal{O}_2 =& 0 \, .
    \end{aligned}
\end{equation}
It is now clear that $\mathcal{O}_1=0$, since it only depends on $\kappa,\sigma,\Psi_1$. Thus, the linearization of Eq.~\eqref{eq:General_Teukolsky_NL} with the operators defined in Eq.~\eqref{eq:Teukolsky_Operators} is just given by 
\begin{equation}
    \mathcal{O}_0 \delta\Psi_0 = 0 \, ,
\end{equation}
which agrees with Eq.~(3.26) in~\cite{Stewart:1974uz} after commuting the order of the GHP derivatives. Evaluating this equation in coordinates results in the usual Teukolsky equation. An alternative derivation following similar steps, but using only the NP formalism, can be found in~\cite{Bini:2002jx}. 

\clearpage

\section{Black hole spectra in (Anti-)de Sitter and higher dimensions}
\label{sec:GRcosmoHighD}

\vspace{-.1cm}

\noindent \textit{Initial contributors: Dias, Santos}

\vspace{.2cm}

Within GR, it is important to explore BHs and their perturbations in spacetimes with a nonvanishing cosmological constant and/or in $d \geq 4$ spacetime dimensions. In this section, we briefly review key developments that have occurred since the 2009 review~\cite{Berti:2009kk}. We refer the reader to~\cite{Berti:2009kk} for earlier foundational studies.

\subsection{QNMs and instabilities in AdS spacetimes}
\label{sec:GR-AdS}

Anti-de Sitter (AdS) spacetime is a maximally symmetric solution of GR endowed with a negative cosmological constant. The study of linear
perturbations of asymptotically AdS BHs is particularly interesting due to two
important aspects. First, there are regions in the parameter space where the QNM
spectrum becomes unstable, resulting in the emergence of superradiant
instabilities. These instabilities arise because the asymptotic AdS gravitational
potential acts as a reflecting barrier, confining superradiant modes. Second,
within the framework of the AdS/CFT
correspondence~\cite{Maldacena:1997re, Aharony:1999ti}, BHs in AdS spacetime are
dual to states in certain quantum field theories. In this context, the QNM
frequencies correspond to the thermalization frequencies of the dual quantum
field theory~\cite{Horowitz:1999jd, Danielsson:1999fa, Birmingham:2001pj,
  Son:2002sd, Kovtun:2005ev, Policastro:2002se, Friess:2006kw,
  Michalogiorgakis:2006jc}. Many of the key studies of AdS QNM frequencies and
their relevance for the AdS/CFT correspondence were already covered in the 2009
review~\cite{Berti:2009kk} (and also in the 2015 lecture notes~\cite{Brito:2015oca}). Therefore,
here, we will only focus our attention on key developments that have occurred
since 2009 and that have contributed to our current understanding of
instabilities in AdS, their endpoint, and novel hairy or resonator BHs
associated with them. To clarify the terminology, recall that static
Schwarzschild-like BHs in AdS can have spatial horizon cross-sections with
planar, spherical, or hyperboloidal topologies. Each of these spacetimes has a
distinct conformal boundary metric: $\mathbb{R}_t \times \mathbb{R}^{d-2}$,
$\mathbb{R}_t \times \mathbb{S}^{d-2}$, or
$\mathbb{R}_t \times \mathbb{H}^{d-2}$ for the planar, spherical, and hyperbolic
BHs, respectively. When BHs are absent, all these different geometries reduce to
pure AdS, with the different boundary metrics corresponding to different
coordinate patches of AdS. The patch asymptotic to
$\mathbb{R}_t \times \mathbb{R}^{d-2}$ is often referred to as the Poincar\'e
patch of AdS and covers only part of AdS, while the global coordinates (which
asymptote to the Einstein Static Universe
$\mathbb{R}_t \times \mathbb{S}^{d-2}$) cover the entire AdS spacetime. Rotating
AdS BHs are necessarily asymptotically global AdS, as a rotational shift is a
pure gauge transformation on a plane. Unlike rotation, all of the above solutions admit
charged versions in any dimension $d$, becoming AdS Reissner-Nordstr\"om BHs.

When perturbed by a charged scalar field, planar AdS
Reissner-Nordstr\"om BHs exhibit a linear scalar condensation instability, which
can be traced back to the violation of the near-horizon AdS$_2$
Breitenlohner-Freedman bound~\cite{Gubser:2008px} (see also
\cite{Horowitz:2009ij,Dias:2010ma}). 
(Incidentally, even neutral scalar fields with sufficiently negative square mass, but above the $d$-dimensional Breitenlohner-Freedman, can condense if the BH is cold enough.)
Interestingly, in the phase diagram of
static solutions, the onset of this scalar condensation instability signals a
merger between planar Reissner-Nordstr\"om BHs (the {\it normal phase}) and
novel planar ``hairy'' BHs~\cite{Hartnoll:2008vx,Hartnoll:2008kx} (the {\it
  superconducting phase} with a nonvanishing scalar field). For energies and
charges where both solutions coexist, the planar AdS hairy BH has higher
entropy, meaning it dominates the microcanonical ensemble. Accordingly, the
endpoint of the time evolution of an unstable planar Reissner-Nordstr\"om BH is
a planar hairy BH~\cite{Murata:2010dx}. This discovery initiated a new research
avenue known as the {\it holographic superconductor program} or {\it
  gravity/condensed matter correspondence}, with far-reaching consequences that
are too vast to cover in the present review. Reviews on the subject can be found
in~\cite{Herzog:2009xv,Horowitz:2010gk,Hartnoll:2011fn,Iqbal:2011ae,Musso:2013vtg,Cai:2015cya}. On
the other hand, global AdS Reissner-Nordstr\"om BHs are also unstable when
perturbed by a charged scalar field
\cite{Basu:2010uz,Dias:2011tj,Gentle:2011kv,Dias:2016pma}. In this case, the
origin of the instability can also be traced to the violation of the AdS$_2$
Breitenlohner-Freedman bound and/or to charged superradiance. Finally,
hyperboloidal BHs are also afflicted by similar instabilities
\cite{Dias:2010ma}.

Let us now turn our attention to rotating AdS BHs. Kerr-AdS BHs can be unstable to superradiance if their angular velocity (in units of the AdS radius $L$) exceeds the speed of light, $\Omega_H L>1$. This was first conjectured in~\cite{Hawking:1999dp}, explicitly established in~\cite{Cardoso:2004hs} (for scalar field modes) and in~\cite{Dias:2013sdc,Cardoso:2013pza} (for gravitational modes), and a formal mathematical proof was provided  in~\cite{Green:2015kur}. The $\ell=m=2$ QNM/instability spectrum of Kerr-AdS can be found in Figs.~4 and~7 of~\cite{Cardoso:2013pza}.   

A puzzling question has endured along the years and is now partially
understood: what is the time evolution and endpoint of superradiantly unstable
Kerr-AdS BHs? This question started being addressed in~\cite{Dias:2011at,Stotyn:2011ns}, where it was found that (as partially
conjectured in~\cite{Reall:2002bh,Kunduri:2006qa}), in a phase diagram of
stationary solutions, the onset of scalar field superradiance in Kerr-AdS
describes a bifurcation to a new family of rotating {\it hairy} BHs that are
also known as {\it black resonators} (the zero horizon radius limit of these
solutions is a regular hairy soliton, a.k.a. a boson star). Quite remarkably,
black resonators also exist in Einstein-AdS theory without matter
\cite{Dias:2015rxy,Ishii:2018oms,Ishii:2020muv} (the zero horizon radius limit
of these solutions is a regular geon~\cite{Dias:2011ss, Horowitz:2014hja,
  Dias:2015rxy, Martinon:2017uyo, Ishii:2018oms, Ishii:2020muv}), i.e. Kerr-AdS
is not the unique BH solution of Einstein-AdS gravity. Kerr-AdS and black
resonators merge (in a phase diagram of solutions) at the onset of gravitational
superradiance~\cite{Dias:2013sdc,Cardoso:2013pza}.

These black resonators (whether with scalar hair or in pure AdS gravity) have a remarkable property: they are neither time-independent nor axisymmetric (i.e., $\partial_t$ and $\partial_\phi$ are not Killing vector fields), but they are time-periodic, since $\partial_t + \Omega_H \partial_\phi$ is a helical Killing vector field of the solution, which turns out to generate the horizon.
(Asymptotically flat black resonators with a helical Killing field also exist if the theory has a massive scalar field~\cite{Herdeiro:2014goa}; in this case, the mass of the scalar field provides the gravitational wall that confines superradiant modes.)
The existence of such time-periodic BHs was a significant development, as these BHs were expected to be ruled out by Hawking's rigidity theorem~\cite{Hawking:1971vc}. However, black resonators evade one of the assumptions of this theorem: their only Killing field is precisely the horizon generator.

For a given energy and angular momentum, when the two families coexist, black resonators have more entropy than Kerr-AdS BHs. This suggests that they can be, at least,  metastable configurations in the time evolution of the superradiant instability in the microcanonical ensemble. However, this cannot be the whole story, as there is a family of black resonators for each superradiant $m$-mode, and $m$-mode black resonators remain superradiantly unstable to higher $m$-modes (since all of these black resonators have $\Omega_H L > 1$).
This led the authors of~\cite{Dias:2011at} (see also~\cite{Cardoso:2013pza,Dias:2015rxy,Niehoff:2015oga}) to conjecture that the most plausible scenario for the time evolution of the superradiant instability of Kerr-AdS is one where the system evolves through a cascade of metastable black resonator configurations with increasingly higher $m$. In this process, the angular velocity of the system would decrease as time evolves. As the small length scale limit $m \to \infty$ is approached (i.e., as one reaches the Planck scale), GR should break down, and quantum gravity effects should become relevant. Hence, weak cosmic censorship~\cite{Penrose:1969pc} would be violated, in the sense that GR would no longer be able to describe the endpoint of the time evolution. This was the first proposed counterexample to weak cosmic censorship in a $d=4$ system that can be formed by gravitational collapse, as the authors of~\cite{Choptuik:2017cyd} showed that scalar hairy black resonators can also form through the gravitational collapse of initial data in global AdS. 

Available time evolution simulations~\cite{Chesler:2018txn,Chesler:2021ehz} provide evidence that the conjecture of~\cite{Dias:2011at} appears to be correct (in the sense that it observes the system going through a series of metastable states that are well approximated by the black resonators of~\cite{Dias:2015rxy}), but the numerical code breaks down well before the endpoint can be inferred. However, two further developments support the conjecture of~\cite{Dias:2011at}. First, the existence of a classical regular black resonator with $\Omega_H L = 1$ is excluded~\cite{Niehoff:2015oga}; thus, it seems inevitable that the superradiant time evolution must lead to a breakdown of GR. Second, once GR breaks down and quantum effects are considered, a candidate for the endpoint of the superradiant instability was recently proposed~\cite{Kim:2023sig}. It is a configuration consisting of a Kerr-AdS BH with $\Omega_H L = 1$ at the central core and a quantum thermal gas of gravitons orbiting it (also with $\Omega_H L = 1$) at a very large distance (proportional to the inverse of Newton's constant). This was called a {\it grey galaxy}~\cite{Kim:2023sig}. 

The existence of black resonators also solves a long-standing puzzle of AdS/CFT. Indeed, from the conformal field theory perspective, one expects the existence of thermal states with energies and angular momenta in the region between extremal Kerr-AdS (i.e., with zero temperature) and the so-called BPS line $E = J/L$ (this curve also describes the geons). Although no Kerr-AdS BHs exist in this region of the phase diagram, black resonators completely fill it. The phase diagram of asymptotically global AdS stationary BHs can be found in Fig.~1 of~\cite{Dias:2015rxy}.

There is one aspect that deserves further discussion. As described above, even within pure Einstein gravity, Kerr-AdS BHs are unstable to gravitational superradiance~\cite{Dias:2013sdc,Cardoso:2013pza}. However, studying the properties of this system was not trivial due to the thorny issue of boundary conditions. When studying gravitational perturbations of Kerr-AdS, one must impose boundary conditions that preserve the AdS asymptotics, i.e., those that do not alter the Einstein Static Universe conformal boundary (such boundary conditions are often called ``reflective'' because they do not permit the flux of energy or angular momentum across the conformal boundary). Similar to the Kerr BH, gravitational perturbations of Kerr-AdS are best studied using the Teukolsky equation (the reader can find the Teukolsky equations in Kerr-AdS for any spin, e.g., in Eq.~(2.8) of~\cite{Dias:2012pp}). The nontrivial question in Kerr-AdS is determining the appropriate boundary conditions to impose on the Teukolsky master (gauge-invariant) variable to ensure that the linear perturbations preserve the conformal Einstein Static Universe. There is no reason to expect that imposing homogeneous Dirichlet boundary conditions on the Teukolsky master variable will suffice. Indeed,~\cite{Dias:2013sdc} found that the desired conditions are intricate mixed or Robin boundary conditions; see Eqs.~(2.18)-(2.22) of~\cite{Dias:2013sdc}. Once these physical boundary conditions were identified, the gravitational QNM spectra of Kerr-AdS and the associated superradiant modes were computed in~\cite{Cardoso:2013pza}. Despite the intricate form of the boundary conditions for the Teukolsky master variable in Kerr-AdS (and the elaborate pathway to derive them~\cite{Dias:2012pp}), their correctness has been confirmed in several ways. To begin with, the sub-sector of long-wavelength (hydrodynamic) gravitational perturbations, within the fluid/gravity limit~\cite{Baier:2007ix,Bhattacharyya:2007vjd,Hubeny:2011hd} of the AdS/CFT correspondence, have a fluid description where they are captured by solving the perturbed Navier-Stokes equations emerging from the conservation of the holographic stress tensor in the Einstein Static Universe. The frequencies obtained from this exercise indeed match the hydrodynamic gravitational QNMs~\cite{Cardoso:2013pza}. Moreover, formal mathematical studies have confirmed these boundary conditions~\cite{Graf:2022fve,Graf:2024yug}.

Finally, it is important to highlight a significant development in the study of
perturbations of AdS spacetimes that occurred in the last decade. Any discussion
of BH stability should begin by considering the stability of its asymptotic
background. For example, when focusing on asymptotically flat or de Sitter BHs,
it is reassuring to know that Minkowski and de Sitter spacetimes have been
proven to be nonlinearly stable~\cite{friedrich86, Christodoulou:1993uv}.  This
is in sharp contrast to the global AdS solution (to which Kerr-AdS and black
resonators asymptote), which is nonlinearly unstable. This has been an active
topic of research in recent years~\cite{Dafermos2006, DafermosHolzegel2006,
  Bizon:2011gg, Dias:2011ss, Dias:2012tq, Buchel:2012uh, Buchel:2013uba,
  Maliborski:2013jca, Bizon:2013xha, Maliborski:2012gx, Maliborski:2013ula,
  Baier:2013gsa, Jalmuzna:2013rwa, Basu:2012gg, Gannot:2012pb,
  Fodor:2013lza, Friedrich:2014raa, Maliborski:2014rma, Abajo-Arrastia:2014fma,
  Balasubramanian:2014cja, Bizon:2014bya, Balasubramanian:2015uua,
  daSilva:2014zva, Craps:2014vaa, Basu:2014sia, Okawa:2014nea, Deppe:2014oua,
  Dimitrakopoulos:2014ada, Horowitz:2014hja, Buchel:2014xwa, Craps:2014jwa,
  Basu:2015efa, Yang:2015jha, Fodor:2015eia, Okawa:2015xma, Bizon:2015pfa,
  Dimitrakopoulos:2015pwa, Green:2015dsa, Deppe:2015qsa, Craps:2015iia,
  Craps:2015xya, Evnin:2015gma, Menon:2015oda, Jalmuzna:2015hoa, Evnin:2015wyi,
  Freivogel:2015wib, Dias:2016ewl, Evnin:2016mjx, Deppe:2016gur,
  Dimitrakopoulos:2016tss, Dimitrakopoulos:2016euh, Rostworowski:2016isb,
  Rostworowski:2017tcx, Martinon:2017uyo, Moschidis:2017lcr, Moschidis:2017llu}.

The fact that linear perturbations of global AdS have a normal spectrum (i.e., its frequencies are purely real) led~\cite{DafermosHolzegel2006,Dafermos2006} to conjecture that global AdS should be nonlinearly unstable. The backreaction of initial data involving a combination of stable linear modes would likely give rise to nonlinear instabilities. Indeed, this conjecture proved correct, as a turbulent mechanism known as {\it weak turbulence} occurs in global AdS~\cite{Bizon:2011gg,Dias:2011ss,Dias:2012tq}, driving the nonlinear instability of AdS. Reflecting boundary conditions in AdS (which preserve the conformal metric and rule out energy flux) allow arbitrarily small finite-energy perturbations to grow large and collapse into a BH. Reference~\cite{Bizon:2011gg} provided the first numerical evidence that AdS is indeed unstable: the time evolution of spherical scalar field initial data in global AdS shows that no matter how small the amplitude, its energy cascades to higher frequencies (similar to turbulent flows) and forms a BH. Moreover, this instability was formally proven for the spherically symmetric and pressureless Einstein-massless Vlasov system~\cite{Moschidis:2017lcr,Moschidis:2017llu}.

In a complementary analysis using perturbation theory, it was found that the interaction of two (or more) modes introduces irremovable secular resonances at third order in the amplitude of the seed perturbation (in addition to resonances that are removed by a Poincaré-Lindstedt resummation). This phenomenon is the {\it weak turbulent mechanism}~\cite{Bizon:2011gg,Dias:2011ss}. The fact that weak turbulence predicts a breakdown of perturbation theory at third order suggests that it accurately describes the mechanism responsible for the onset of the instability, as it is consistent with the observation that numerical simulations show the timescale for BH formation scales with the inverse square of the amplitude of the initial data (i.e., it scales with the inverse of the energy).

Reference~\cite{Dias:2012tq} further argued that a necessary condition (though not sufficient) for these resonances to occur is that the linear spectrum of frequencies of the system is commensurable (i.e., the sum or difference of two frequencies remains in the spectrum). The AdS nonlinear instability appears to be present for a wide variety of initial data. However, there are also ``islands of stability''~\cite{Dias:2012tq}, i.e., perturbations that do not necessarily lead to an instability~\cite{Dias:2012tq,Buchel:2012uh,Maliborski:2013jca,Dimitrakopoulos:2015pwa,Green:2015dsa,Craps:2015xya}. An open question in the field is whether we can understand the size of these islands as a function of the properties of the initial data, particularly what happens when the energy of the system becomes arbitrarily small.
Research in this direction often involves numerical simulations with different initial data and/or the use of improved perturbation theory schemes. These methods are essentially equivalent and include the two-timescale formalism~\cite{Balasubramanian:2014cja}, renormalization group perturbation techniques~\cite{Craps:2014vaa,Craps:2014jwa}, and resonant approximation~\cite{Bizon:2015pfa}.

So far, much of the above discussion has focused on the spherically symmetric collapse of a scalar field in global AdS. A pertinent question is whether the instability persists when moving away from these fine-tuned spherical conditions. {\it A priori}, there is a strong argument suggesting that the nonlinear instability and associated gravitational collapse into a BH should cease once spherical symmetry is broken.
Indeed, once we break spherical symmetry, we also break axial symmetry. In other words, we can consider initial data or perturbation modes with nonvanishing azimuthal number, $m\neq 0$, which introduces angular momentum into the system. In this case, it is natural to expect that the associated centrifugal effects could counteract gravitational collapse and might prevent BH formation. Surprisingly, perturbative analyses carried out in~\cite{Dias:2011ss,Dias:2016ewl,Rostworowski:2016isb,Rostworowski:2017tcx,Martinon:2017uyo,Dias:2017tjg,Choptuik:2017cyd} show that this intuition is overly simplistic and, in fact, incorrect. The only time evolution numerical simulation performed so far in the non-spherically-symmetric sector confirms these predictions. Specifically, the authors of~\cite{Choptuik:2017cyd} showed that the endpoint of the gravitational collapse of initial data consisting of a scalar field doublet in global AdS$_5$ is either a Kerr-AdS$_5$ BH at high energies (where black resonators do not exist and Kerr-AdS$_5$ is superradiantly stable) or a hairy black resonator from~\cite{Dias:2011at} at low energies. Moreover, although the inclusion of angular momentum delays the collapse time, it still retains a timescale that scales with the inverse of the energy (similar to the collapse of spherically symmetric matter). Furthermore,~\cite{Choptuik:2017cyd} also studied the perturbation and evolution of rotating boson stars and found that boson stars near AdS are stable, while those farther from AdS are unstable. This aligns with the existence of the islands of stability~\cite{Dias:2012tq} observed in spherically symmetric numerical simulations.

To conclude this brief review, it is important to note that black resonators and boson stars/geons establish a connection between the nonlinear instability of AdS and the superradiant instability of AdS BHs. On the one hand, the zero-size limit of a black resonator is a geon or boson star, which plays a key role in the weak turbulent nonlinear instability of AdS. On the other hand, black resonators merge with the Kerr or Reissner-Nordstr\"om BHs of the theory at the onset of the superradiant instability~\cite{Basu:2010uz,Dias:2011tj,Dias:2011ss,Dias:2011at,Dias:2015rxy}. For a more detailed review of the nonlinear instability of AdS, the reader is invited to consult the introduction of~\cite{Dias:2017tjg}.

AdS spacetime is special in the sense that, unlike for nonnegative cosmological constant, one can have AdS black holes already in three spacetime dimensions, the most notable one being the Ba\~nados-Teitelboim-Zanelli (BTZ) black hole~\cite{Banados:1992wn}. Naturally, some of the physics described above for four-dimensional AdS spacetimes has been investigated also in asymptotically AdS$_3$ backgrounds (as well as in higher dimensions, as referenced above without specifically commenting on the dimension). For example, QNMs of BTZ have been analyzed in detail, starting with the studies of~\cite{Cardoso:2001hn,Birmingham:2001hc,Birmingham:2001pj} (see~\cite{Berti:2009kk} for a review).
AdS$_3$ instabilities and associated boson stars and resonators have also been studied (see e.g.~\cite{Ishibashi:2004wx,Astefanesei:2003rw,Stotyn:2012ap,Stotyn:2013spa,Iizuka:2015vsa,Ferreira:2017cta,Dappiaggi:2017pbe,Gao:2023rqc}).
``Quantum induced superradiance'' and ``quantum black resonators'' (via braneworld models~\cite{Emparan:1999wa,Emparan:1999fd}) have also been addressed~\cite{Cartwright:2024iwc,Cartwright:2025fay}.
For studies of the nonlinear instability of AdS$_3$, see e.g.~\cite{Bizon:2013xha,Jalmuzna:2015hoa} and references therein.
\subsection{QNMs in de Sitter spacetimes and strong cosmic censorship}
\label{sec:GR-dS}

In recent years, the study of QNMs of asymptotically de Sitter (dS) BHs has received much attention in the context of potential violations of strong cosmic censorship (SCC).

Christodoulou's formulation of SCC posits that, for smooth generic initial data prescribed on a Cauchy hypersurface $\Sigma$, the maximal Cauchy development of this data cannot be extended beyond the Cauchy horizon as a \textit{weak solution} of the Einstein equations (eventually coupled to matter fields)~\cite{Christodoulou:2008nj}.
This reformulates Penrose's original $C^0$ version of SCC, which conjectured that the solution should be inextendible as a {\it continuous} solution of the Einstein equations~\cite{penrose1978}, a conjecture that is now known to be false~\cite{McNamara121,Ori:1991zz,Dafermos:2003wr,Franzen:2014sqa,Dafermos:2017dbw}. A failure of Christodoulou's SCC would imply a loss of predictability in GR (unless a new reformulation of SCC is found), since events occurring beyond the Cauchy horizon would no longer depend solely on the initial data on $\Sigma$, but would also be influenced by the dynamics beyond the horizon.

Asymptotically flat BHs do respect Christodoulou's formulation of SCC~\cite{Luk:2015qja,Dafermos:2012np,Dafermos:2015bzz}. Essentially, a blueshift effect caused by the late-time tails of perturbations in the exterior of the event horizon (known as ``mass inflation'' when its backreaction is included~\cite{Poisson:1990eh,Ori:1991zz}) prevents the crossing of the Cauchy horizon. This effect generates infinite tidal forces on observers attempting to cross the Cauchy horizon, causing the solution to break down as a weak solution of the system (although the solution can remain continuous at the Cauchy horizon)~\cite{Price:1971fb,Simpson:1973ua,penrose1978,mcnamara1978instability,chandrasekhar1982crossing,Poisson:1990eh}.

The situation is considerably different for asymptotically dS spacetimes. In this case, the blueshift effect competes with a cosmological redshift effect. As a result, perturbations {\it may} decay too quickly to render the Cauchy horizon unstable, ultimately preventing the formation of a singularity that would otherwise enforce SCC. More concretely, the behavior of perturbations at the Cauchy horizon depends on the decay rate of perturbations along the event horizon~\cite{Hintz:2015jkj,Hintz:2015koq}. For de Sitter BHs, perturbations decay exponentially along the event horizon~\cite{MR1432814,MR2426141,Dyatlov:2011jd,Dyatlov:2013hba,Hintz:2016gwb,Hintz:2016jak}. This means that for linearized matter or gravitational perturbations in de Sitter BHs, the outcome of the blueshift/redshift competition depends on whether the ratio $\beta$ between the spectral gap (i.e., the magnitude of the imaginary part of the least damped QNM of the BH) and the surface gravity of the Cauchy horizon is above or below a critical value~\cite{Hintz:2015jkj}. This is why QNMs play such a pivotal role in the discussion of SCC in de Sitter BHs, and why this problem has prompted a detailed study of de Sitter QNMs.

Performing the actual computation to determine the outcome of this competition (i.e., the value of $\beta$), it has been established that Christodoulou's formulation of SCC is -- unlike in the asymptotically flat case -- {\it violated} in (near-extremal) Reissner-Nordstr\"om-de Sitter BHs for linear scalar field perturbations~\cite{Cardoso:2017soq}. Charged scalar fields (which are required in a theory that forms Reissner-Nordstr\"om-de Sitter BHs through gravitational collapse) can not prevent this violation~\cite{Cardoso:2018nvb, Mo:2018nnu, Dias:2018ufh}. Christodoulou's SCC is even more severely violated in Reissner-Nordstr\"om-de Sitter BHs when we focus on the gravito-electromagnetic sector of perturbations~\cite{Dias:2018etb}. Thus, it is somewhat surprising that Christodoulou's formulation of SCC does hold for Kerr-de Sitter BHs, both for the scalar field and gravitational sectors of linear perturbations~\cite{Dias:2018ynt}. This also means that for Kerr-Newman-de Sitter BHs, there exists a boundary in the parameter space that marks the transition between solutions that respect and violate SCC~\cite{Casals:2020uxa, Davey:2024xvd}.

In this context, a natural question that arises is whether there are reformulations of the problem that could rescue SCC and, thus, predictability for Reissner-Nordstr\"om and weakly rotating Kerr-Newman de Sitter BHs. One possibility is that nonlinear corrections could resolve the issue. However, the few existing nonlinear studies that have attempted to address this possibility remain inconclusive~\cite{Luna:2019olw, Zhang:2019nye, Luna:2019olwaddendum}. Alternatively, one might consider dropping the assumption of smooth initial data and instead adopt a ``rough'' formulation of SCC that allows for nonsmooth initial data. In this context, it has been demonstrated that the degree of regularity of the solutions at the Cauchy horizon is generally lower than that of the nonsmooth initial data. In this sense, it can be argued that the nonsmooth version of SCC still holds~\cite{Dafermos:2018tha}. The connection between the formulation in~\cite{Dafermos:2018tha} and previous studies~\cite{Mellor:1989ac, Mellor1992, Brady:1998au} is discussed in~\cite{Dias:2018etb}.

Finally, moving beyond the realm of classical discussions of SCC, one might question whether a reformulation of SCC that includes quantum perturbations could enforce predictability. This was conjectured in~\cite{Dias:2018etb, Dias:2019ery}, and indeed such a quantum formulation of SCC was proven to hold for Reissner-Nordstr\"om-de Sitter BHs~\cite{Hollands:2019whz, Hollands:2020qpe, Klein:2024sdd, Emparan:2020rnp, Emparan:2020znc}. These studies have shown that for any nonsingular quantum state within the domain of dependence of the initial data, the (quantum) expectation value of the energy-stress tensor diverges quickly enough to make the Cauchy horizon unstable. When both effects coexist, the divergence caused by the quantum fields turns out to be even stronger than the one caused by the classical late-time blueshift instability. Therefore, predictability in charged de Sitter spacetimes is restored once quantum effects are included in the late-time behavior of BH perturbations.

\subsection{QNMs and instabilities in higher dimensional spacetimes}
\label{sec:GR-higherD}
In the last three decades, the Schwarzschild black string~\cite{Horowitz:1991cd}
-- i.e., the product spacetime of a Schwarzschild BH and a circle -- has
served as a prototype solution for exploring black objects in higher-dimensional
Einstein gravity (and supergravity). In particular, these solutions are unstable
due to the so-called Gregory-Laflamme instability~\cite{Gregory:1993vy,
  Gregory:1994bj}, which is similar to the Rayleigh-Plateau instability in fluid
jets~\cite{Cardoso:2006ks}. From a phase diagram perspective, the
Gregory-Laflamme instability is ultimately responsible for the existence of
novel static solutions that describe nonuniform black strings and localized BHs on a circle~\cite{Gubser:2001ac, Harmark:2002tr, Kol:2002xz,
  Wiseman:2002zc, Kol:2003ja, Harmark:2003dg, Harmark:2003yz, Kudoh:2003ki,
  Sorkin:2004qq, Gorbonos:2004uc, Kudoh:2004hs, Dias:2007hg, Harmark:2007md,
  Wiseman:2011by, Figueras:2012xj, Dias:2017coo}. On the other hand, the time
evolution of the Gregory-Laflamme instability~\cite{Lehner:2010pn,
  Emparan:2015gva, Figueras:2022zkg} appears to provide the first example of a
violation of weak cosmic censorship~\cite{Penrose:1969pc}. This solution is a
prototype because other solutions and their associated instabilities, which have
been found subsequently, can be motivated by it and share similar underlying
physics. For example, the gedankenexperiment of bending a black string
while adding momentum along the circle to balance the system against
gravitational collapse foreshadows the existence of black rings
\cite{Emparan:2001wn} (and, consequently, black Saturns~\cite{Elvang:2007rd} and
concentric black rings~\cite{Evslin:2007fv, Iguchi:2007is}), which themselves
are unstable to the Gregory-Laflamme instability~\cite{Santos:2015iua}, with
time evolution leading to a violation of weak cosmic censorship~\cite{Figueras:2015hkb}. A review of higher-dimensional Einstein gravity that
covers developments up until 2008 can be found in~\cite{Emparan:2008eg}.

Rotating black strings -- i.e., the product spacetime of a Kerr or Myers-Perry
BH and a circle -- also exist. These black strings exhibit the Gregory-Laflamme
instability, which is further enhanced by rotational effects~\cite{Dias:2010eu,
  Dias:2022mde}; see, for example, Fig.~1 of~\cite{Dias:2022mde}. Additionally,
these strings carry momentum along the string direction, which produces an
effective mass term. Together with the presence of an ergoregion, this provides
a form of confinement that can (but does not necessarily) trigger a superradiant
instability~\cite{Marolf:2004fya, Cardoso:2004zz, Cardoso:2005vk, Dias:2006zv,
  Dias:2022mde}; see Fig.~4 of~\cite{Dias:2022mde}. This represents the first
system in which both Gregory-Laflamme and superradiant instabilities are
present, with regimes where they compete~\cite{Dias:2022mde}. Interestingly, the
onset of these superradiant instabilities signals the existence of novel {\it
  black resonator strings}~\cite{Dias:2022str} and {\it helical black
  strings}~\cite{Dias:2023nbj}. Under Kaluza-Klein reduction, these solutions
have 4-dimensional counterparts~\cite{Dias:2022str, Dias:2023nbj}.

Of course, besides black strings, higher-dimensional GR also includes BHs with spherical horizon topology, such as the Schwarzschild-Tangherlini, Reissner-Nordstr\"om, and Myers-Perry BHs~\cite{Tangherlini:1963bw, Myers:1986un}. The Kodama-Ishibashi master variable formalism~\cite{Kodama:2003jz, Kodama:2003kk} generalizes the RWZ formalism~\cite{Regge:1957td, Zerilli:1970se} to address perturbations of higher-dimensional Schwarzschild-Tangherlini and Reissner-Nordstr\"om BHs. Despite numerous efforts over the years, no decoupled equation (similar to the Teukolsky equation) describing how gravitational perturbations propagate on Myers-Perry BHs has been found. For a discussion of the main attempts, challenges, and obstacles, see~\cite{Durkee:2010qu, Godazgar:2011sn}. In practice, one must solve the coupled system of linearized Einstein equations to study perturbations of higher-dimensional Kerr BHs. A broad (but necessarily incomplete) study of QNMs for Schwarzschild-Tangherlini and Myers-Perry BHs has been conducted in~\cite{Kunduri:2006qa, Murata:2008yx, Kodama:2009bf, Dias:2009iu, Dias:2010maa, Dias:2010eu, Hartnett:2013fba, Dias:2014eua, Emparan:2014cia}.

An interesting property of Myers-Perry BHs is that their rotation can reach ultraspinning regimes (up to the point where singly spinning Myers-Perry BHs have no upper bound on their angular momentum). This suggests that such BHs should develop certain instabilities. Indeed, the authors of~\cite{Emparan:2003sy} argued that for very large spins, Myers-Perry BHs should exhibit two classes of instabilities: an axisymmetric {\it ultraspinning instability} and a nonaxisymmetric {\it bar-mode instability}. The existence of ultraspinning instabilities was established in~\cite{Dias:2009iu, Dias:2010maa, Dias:2010eu, Dias:2010gk, Dias:2011jg, Dias:2014eua, Emparan:2014jca}. This represents the first example of a classical gravitational instability of a vacuum asymptotically flat BH. It has a Gregory-Laflamme nature, as for large spins the solution becomes ``pancaked,'' leading to the appearance of a hierarchy of scales~\cite{Emparan:2003sy}. On the other hand, bar-mode instabilities were found in~\cite{Shibata:2009ad, Shibata:2010wz, Dias:2014eua, Bantilan:2019bvf}.

For sufficiently small spins, the endpoint of the bar-mode instability is a stable Myers-Perry BH with a smaller spin, after it radiates excess angular momentum through GW emission. However, for large enough spins, the ``fragmentation'' process conjectured in~\cite{Emparan:2003sy} is observed and leads to a violation of weak cosmic censorship~\cite{Shibata:2009ad, Shibata:2010wz, Bantilan:2019bvf}. On the other hand, ultraspinning instabilities have the added interest of being triggered by a time-independent zero-mode~\cite{Dias:2009iu, Dias:2010maa}. As a result, the onset of the instability signals (one for each harmonic $\ell$) a bifurcation to a novel family of stationary BHs, known as {\it lumpy} BHs~\cite{Dias:2014cia, Emparan:2014pra}, in the phase diagram of stationary solutions (see Fig.~1 of~\cite{Dias:2009iu}). Lumpy BHs have an axisymmetric spherical horizon, but with deformations along the polar direction (hence the name). In the phase diagram of stationary solutions~\cite{Dias:2014cia, Emparan:2014pra}, lumpy BHs connect Myers-Perry BHs (with horizon topology $S^{d-2}$) to black rings (with horizon topology $S^{d-3} \times S^1$) or black Saturns~\cite{Elvang:2007rd} or black di-rings~\cite{Evslin:2007fv, Iguchi:2007is}, depending on the ultraspinning harmonic $\ell$ being considered: see Figs.~1-4 of~\cite{Dias:2014cia} or Fig.~10 of~\cite{Emparan:2014pra} for more details. For energies and angular momenta where lumpy and Myers-Perry BHs coexist, lumpy BHs always have higher entropy. Depending on the initial data, the time evolution of an unstable ultraspinning Myers-Perry BH ends with the formation of (concentric) black rings or black Saturns (after passing through metastable states resembling lumpy BHs)~\cite{Figueras:2017zwa}. In the transition to the final configuration, the system violates weak cosmic censorship.

To conclude, we mention that there are two analytical formalisms that can be used to identify BH solutions and study key properties of their perturbations and instabilities. These are the {\it blackfold approach} (when the system has horizons with a separation of scales; see e.g.~\cite{Emparan:2009cs,Emparan:2009at,Armas:2010hz,Emparan:2011br}), and the {\it large-$D$ limit of gravity}  reviewed in~\cite{Emparan:2020inr}.

\clearpage

\section{Superradiant instabilities of massive bosonic perturbations\label{sec:superradiant_instabilities}}

\vspace{-.1cm}

\noindent \textit{Initial contributor: Pani}

\vspace{.2cm}

The confining mechanism responsible for the superradiant instabilities of spinning BHs in AdS (see previous section) is provided by the AdS boundary. Another confining mechanism that triggers the superradiant instability and occurs also in asymptotically flat spacetime is provided by \emph{massive} bosonic perturbations, since massive bosonic fields naturally confine low-frequency radiation due to a Yukawa-like suppression.

The existence of this instability was originally suggested in~\cite{Damour:1976kh} for massive scalar perturbations around a Kerr BH, and has been thoroughly investigated in a variety of cases since then. For a detailed review, we refer the reader to~\cite{Brito:2015oca}; here we summarize the main results.

Massive perturbations with frequency $\omega<\mu$ (where $\mu\hbar$ is the mass of the field, and $\hbar$ is Planck's constant) support quasibound states which have the same boundary conditions of QNMs near the horizon, but decay exponentially at infinity as $\sim e^{\sqrt{\mu^2-\omega^2}}$. Bosonic modes are  amplified if their frequency satisfies the superradiant condition, $0<\omega_R<m\Omega_{\rm H}$. When confined coherently, such amplification triggers a superradiant instability~\cite{Brito:2015oca}.

While the qualitative aspects of this phenomenon are valid for any bosonic field, regardless of its spin, a crucial ingredient is the instability timescale of the dominant unstable mode, which strongly depends on the gravitational coupling $M\mu$ and on the spin of the massive boson~\cite{Brito:2015oca}.  
Scalar (spin-0)~\cite{Damour:1976kh,Ternov:1978gq,Zouros:1979iw,Detweiler:1980uk,Dolan:2007mj,Arvanitaki:2014wva,Arvanitaki:2016qwi,Brito:2017wnc,Brito:2017zvb} and, more recently, vector (spin-1)~\cite{Pani:2012vp,Pani:2012bp,Witek:2012tr,Endlich:2016jgc,East:2017mrj,East:2017ovw,Baryakhtar:2017ngi, 
East:2018glu,Frolov:2018ezx,Dolan:2018dqv,Siemonsen:2019ebd} and tensor (spin-2)~\cite{Brito:2013wya,Brito:2020lup,Dias:2023ynv,East:2023nsk} fields have been studied in great detail.
In general, 
these modes are localized at a distance from the BH of the order of the Compton wavelength $1/\mu$. In the small-coupling limit, $M\mu\ll1$, the spectrum of these modes is hydrogenic:
\begin{equation}
\omega_R \sim \mu +{\cal O}(M\mu^2)\,, \qquad 
\omega_I\sim-\alpha^{N}(\omega_R-m\Omega_{\rm H})\,,
\label{hydrogenic}
\end{equation}
where the ${\cal O}(M\mu^2)$ corrections are negative and, along with the positive prefactor and exponent $N$ of the imaginary part $\omega_I$, they depend on the total angular momentum and spin projection of the specific mode.
Beyond this analytical small-coupling regime and a small-spin expansion~\cite{Pani:2013pma}, the frequency and instability timescales have been computed numerically for any coupling and spin for scalar~\cite{Dolan:2007mj}, vector~\cite{Dolan:2018dqv}, and tensor~\cite{Dias:2023ynv} modes.

As an estimate, the superradiant instability is most effective when
its Compton wavelength is comparable to the BH gravitational radius, i.e., when $M\mu={\cal O}(0.1)$. This translates into the optimal condition for the instability $\mu \hbar\sim 10^{-11}(M_\odot/M)\,{\rm eV}$, but in certain cases the range of relevant masses for which the instability is efficient can encompass several orders of magnitude.
This makes BHs ideal systems to study the effects of ultralight bosons~\cite{Arvanitaki:2009fg,Arvanitaki:2014wva,Brito:2015oca}, which are common in theories beyond the Standard Model and as dark matter candidates.

In the superradiant regime, the BH loses energy and angular momentum to a predominantly dipolar ($m=1$) boson condensate, spinning down until the condition $\omega_R \sim \Omega_{\rm H}$ is met, at which point the instability saturates. The condensate then primarily dissipates by emitting nearly monochromatic quadrupolar GWs~\cite{Arvanitaki:2014wva,Brito:2014wla}, with a frequency determined by the boson mass. For higher modes ($m > 1$), this process proceeds on longer timescales~\cite{Ficarra:2018rfu}. The observational signatures of these condensates both in the GW and in the electromagnetic band are manifold: see~\cite{Brito:2015oca} for a review.

\clearpage

\section{Quasinormal modes and analog black holes}\label{sec:analog}

\vspace{-.1cm}

\noindent \textit{Initial contributors: Destounis, Richartz}

\vspace{.2cm}

The study of BH dynamics has so far been proven to be a feasible, with some caveats,  
by observing GW emission from BH binaries.  Nevertheless, the phenomenology of
BH mechanics is extremely vast; current detectors cannot observe, yet, an
abundance of phenomena that are theoretically well established and expected to
occur. In such cases, we usually rely upon the phenomenology of \emph{analog
  systems} that can successfully reproduce known results and further emulate
other phenomena that have not been detected yet in strong field gravity. In
fact, strong field effects are sometimes subtle and challenging to reproduce
even with terrestrial experiments. The seminal work by Moncrief, Unruh, and
Visser~\cite{Moncrief:1980ApJ,Unruh:1980cg,Visser:1997ux} demonstrated that key
features of BHs can be effectively reproduced in laboratories using analog BH
settings (see~\cite{Barcelo:2005fc,Almeida:2022otk} for reviews
and~\cite{Jacquet:2020bar} for recent developments in analog gravity).

Although analog BHs do not provide a direct insight into Einstein's field equations, they are a valuable platform for exploring the BH dynamics at the classical and quantum level. Phenomena such as Hawking radiation~\cite{Rousseaux:2007is,Weinfurtner:2010nu,Euve:2015vml,Steinhauer:2015saa,MunozdeNova:2018fxv,Drori:2018ivu,Kolobov:2019qfs,Shi:2021nkx}, superradiance~\cite{Torres:2016iee,Cromb:2020ldn,Braidotti:2021nhw}, and quasinormal ringing~\cite{Torres:2020tzs,Smaniotto:2025hqm}, among others~\cite{Philbin:2007ji,Jaskula:2012ab,Eckel:2017uqx,Patrick:2019kis,Steinhauer:2021fhb,Viermann:2022wgw,Tajik:2022lyt,Gregory:2024ogi}, have been investigated in controlled laboratory settings through the lens of analog gravity.

Unruh's pioneering idea~\cite{Unruh:1980cg}
was based on the propagation of sound waves in classical fluids. The analogy with BHs has since been extended to a vast range of physical systems~\cite{Barcelo:2005fc,Jacquet:2020bar}. A widely employed laboratory configuration to study analog BHs is based on the mathematical correlation between the propagation of fields in general-relativistic spacetimes and long wavelength (shallow water) surface waves on moving water~\cite{Schutzhold:2002rf}. Other experiments that mimic BHs rely, for instance, on superfluids~\cite{Volovik:1995ja,Volovik:1999fc,Svancara:2023yrf,Smaniotto:2025hqm},  ultra-cold atomic gases~\cite{Garay:2000jj}, optical systems~\cite{Leonhardt:2000fd,Marino:2008kk}, and superconducting circuits~\cite{Schutzhold:2004tv,Nation:2009xb}.

The plurality of potential experimental settings has paved the way for the realization of a vast diversity of analog BH effects in laboratories. In particular, analog BHs possess analog/acoustic horizons that act as one-way membranes for perturbations. When they are driven out of equilibrium, they exhibit a ringdown phase characterized by the emission of a series of QNMs~\cite{Berti:2004ju,Cardoso:2004fi}, much like general-relativistic BHs. The associated QNM spectra encode information about the underlying system. Thus, analog BH spectroscopy~\cite{Torres:2019sbr,Smaniotto:2025hqm} has been proposed as a tool for directly measuring (super)fluid flows, as well as indirectly extrapolating results obtained with analog spectroscopy for general-relativistic systems~\cite{Barcelo:2018ynq}.

Theoretical investigations of QNMs within analog gravity began in the 2000s~\cite{Berti:2004ju,Cardoso:2004fi,Nakano:2004ha,Lepe:2004kv,Saavedra:2005ug,Xi:2007yb,Okuzumi:2007hf,Barcelo:2007ru,Abdalla:2007dz} and continue to be a vibrant area of research~\cite{Dolan:2010zza,Dolan:2011ti,Daghigh:2014mwa,Chaverra:2015aya,Patrick:2018orp,Assumpcao:2018bka,Herdeiro:2019fps,Guo:2020blq,Ge:2019our,Patrick:2020yyy,Geelmuyden:2021sdh,Jacquet:2021scv,Vieira:2021ozg,Vieira:2021xqw,deOliveira:2023qxe,Burgess:2023pny,Vieira:2023ylz,Liu:2024vde,Matyjasek:2024uwo,Liu:2024wch,Keshet:2024hlm,Albuquerque:2025eny,Vieira:2025ljl,dePaula:2025fqt}. Experimental progress is gradual and recently gained significant momentum, especially towards the direction of QNMs and analog BH spectroscopy. The first observation of QNM oscillations in analog gravity~\cite{Torres:2020tzs} has marked a major milestone by experimentally detecting the oscillation frequencies of the first 25 counter-rotating modes.
These were simultaneously excited during a
nonequilibrium analog BH experiment in a draining bathtub flow that emulates a rotating BH spacetime~\cite{Visser:1997ux,Schutzhold:2002rf}. Such detection further enhanced the relationship between the angular frequency of null particles in light-rings and the corresponding QNMs~\cite{Cardoso:2008bp,Torres:2017vaz}. Additional QNM experimental research, particularly in ultra-cold atomic gases~\cite{Keshet:2024hlm}, fluids of light~\cite{Jacquet:2021scv,Falque:2023ctx}, fiber optical solitons~\cite{Burgess:2023pny}, and superfluids~\cite{Svancara:2023yrf,Smaniotto:2025hqm}, has the potential to yield novel and groundbreaking discoveries in the near future.

In summary, the study of BH phenomenology in controlled laboratory environments enhances our understanding of the underlying, nongravitational dual systems while providing valuable insights that can be implemented to the gravitational sector~\cite{Barcelo:2018ynq}. When attempting to replicate BH phenomena in the laboratory, including analog ringdown signals and their respective QNM frequencies, it is crucial to assess the robustness of the gravity duality. Depending on the specific analog system, factors such as dispersion~\cite{Barcelo:2007ru,Patrick:2020yyy},  vorticity~\cite{Patrick:2018orp,Liberati:2018uev}, asymmetries~\cite{Assumpcao:2018bka}, and finite-size effects~\cite{Smaniotto:2025hqm,Solidoro:2024yxi} must be carefully considered. The interplay between analog systems and gravitational physics fosters mutual advancements in both fields, offering a promising pathway for further progress in ringdown physics.

\end{appendices}

\clearpage
\section*{Acknowledgments}
\markboth{{\scshape Acknowledgments}}{}
\addcontentsline{toc}{section}{Acknowledgments}

We thank Sebastiano Bernuzzi, J\'er\'emy Besson, Javier Carballo, Bruno Carneiro da Cunha, João Paulo Cavalcante, Roberto Cotesta, Ramin Daghigh, Alex Gaina, Luis Manuel González-Romero, Jan Harms, David Izquierdo-Villalba, José Luis Jaramillo, Alex Kehagias, Burkhard Kleihaus, Jutta Kunz, Philippe Lalanne, Jacopo Lestingi, Dongjun Li, Francisco Navarro-Lérida, Alessandro Nagar, Juan Pedraza, David Pere\~n\'iguez, Davide Perrone, Jing Ren, Thomas Spieksma, Saul Teukolsky, Bobur Turimov, Pratik Wagle and Benjamin Withers for useful discussions, comments and suggestions.

N.~Afshordi is supported by the Natural Sciences and Engineering Research Council of Canada, the University of Waterloo, and the Perimeter Institute for Theoretical Physics. Research at Perimeter Institute is supported in part by the Government of Canada through the Department of Innovation, Science, and Economic Development and by the Province of Ontario through the Ministry of Colleges and Universities.
S.~Albanesi acknowledges support from the Deutsche Forschungsgemeinschaft (DFG) project ``GROOVHY'' (BE 6301/5-1 Projektnummer: 523180871).
V.~Baibhav acknowledges support from the NASA Hubble Fellowship grant HST-HF2-51548.001-A awarded by the Space Telescope Science Institute, which is operated by the Association of Universities for Research in Astronomy, Inc., for NASA, under contract NAS5-26555.
E.~Berti, M.~H.-Y.~Cheung and S.~Yi are supported by NSF Grants No.~AST-2307146, No.~PHY-2513337, No.~PHY-090003, and No.~PHY-20043, by NASA Grant No.~21-ATP21-0010, by the John Templeton Foundation Grant No.~62840, by the Simons Foundation [MPS-SIP-00001698, E.B.], by the Simons Foundation International [SFI-MPS-BH-00012593-02], and by the Italian Ministry of Foreign Affairs and International Cooperation Grant No.~PGR01167.
J.~L.~Blázquez-Salcedo acknowledges support from MICINN project CNS2023-144089 ``Quasinormal modes,'' MICINN project PID2021-125617NB-I00 ``QuasiMode,'' and FCT project PTDC/FIS-AST/3041/2020.
S.~Bhagwat would like to acknowledge the UKRI Stephen Hawking Fellowship funded by the Engineering and Physical Sciences Research Council (EPSRC) with grant reference number EP/W005727 for support during this project.
B.~Bucciotti is partly supported by the Italian MIUR under contract 20223ANFHR (PRIN2022).
The work of P.A.~Cano received the support of a fellowship from ``la Caixa'' Foundation (ID 100010434) with code LCF/BQ/PI23/11970032.
C.~Capano acknowledges support from NSF Grants No.~PHY-2412341 and AST-2407454.
We acknowledge the support by VILLUM Foundation (grant No.~VIL37766) and the DNRF Chair program (grant No.~DNRF162) by the Danish National Research Foundation.
V.~Cardoso\ is a Villum Investigator and a DNRF Chair.  
V.~Cardoso acknowledges financial support provided under the European Union's H2020 ERC Advanced Grant ``Black holes: gravitational engines of discovery'' grant agreement No.~Gravitas–101052587. 
Views and opinions expressed are however those of the author only and do not necessarily reflect those of the European Union or the European Research Council. Neither the European Union nor the granting authority can be held responsible for them.
This project has received funding from the European Union's Horizon 2020 research and innovation programme under the Marie Sklodowska-Curie grant agreement No 101007855 and No 101131233.
The Center of Gravity is a Center of Excellence funded by the Danish National Research Foundation under grant No.~DNRF184.
K.~Destounis acknowledges financial support provided by FCT – Fundação para a Ciência e a Tecnologia, I.P., under the Scientific Employment Stimulus – Individual Call – Grant No.~2023.07417.CEECIND/CP2830/CT0008.
E.~Finch acknowledges support from the Department of Energy under award number DE-SC0023101.
N.~Franchini acknowledges financial support provided by FCT – Fundação para a Ciência e a Tecnologia, I.P., under the Scientific Employment Stimulus – Individual Call – Grant No.~2023.06263.CEECIND/CP2830/CT0004 and support from the Center for Astrophysics and Gravitation (CENTRA/IST/ULisboa) through FCT grant No.~UID/PRR/00099/2025 and grant No.~UID/00099/2025.
K.~Fransen is supported by the Heising-Simons Foundation ``Observational Signatures of Quantum Gravity'' collaboration grant 2021-2817, the U.S. Department of Energy, Office of Science, Office of High Energy Physics, under Award No.~DE-SC0011632, and the Walter Burke Institute for Theoretical Physics.
V.~Gennari acknowledges financial support form the French space agency CNES in the framework of LISA, from the CNRS through the AMORCE funding framework and from the Agence Nationale de la Recherche (ANR) through the MRSEI project ANR-24-MRS1-0009-01.
S.~R.~Green is supported by a UKRI Future Leaders Fellowship (grant no.~MR/Y018060/1). 
S.~A.~Hughes gratefully acknowledges support from National Science Foundation grants PHY-1403261, PHY-1707549, PHY-2110384, and PHY-240964.
X.~Jimenez Forteza is supported by the
Spanish Ministerio de Ciencia, Innovaci\'on y Universidades (Beatriz Galindo, BG22-00034) and cofinanced
by UIB; the Spanish Agencia Estatal de Investigaci\'on
Grants No.~PID2022-138626NB-I00, No.~RED2022-
134204-E, and No.~RED2022-134411-T, funded by
MCIN/AEI/10.13039/501100011033/FEDER, UE; the
MCIN with funding from the European Union NextGenerationEU/PRTR (No.~PRTR-C17.I1); the Comunitat
Auton\'oma de les Illes Balears through the Direcci\'o General de Recerca, Innovaci\'o I Transformaci\'o Digital with
funds from the Tourist Stay Tax Law (No.~PDR2020/11 -
ITS2017-006), and the Conselleria d'Economia, Hisenda i
Innovaci\'o Grant No.~SINCO2022/6719.
G.~Khanna acknowledges support from the US National Science Foundation Grants No.~PHY-2307236 and DMS-2309609. All computations were performed on the UMass-URI UNITY HPC/AI cluster at the Massachusetts Green High-Performance Computing Center (MGHPCC).
F.~S.~Khoo acknowledges support from ``Atracci\'on de Talento Investigador Cesar Nombela'' of the Comunidad de Madrid, grant No.~2024-T1/COM-31385, DFG project Ku612/18-1, MICINN project PID2021-125617NB-I00 ``QuasiMode,'' and FCT project PTDC/FIS-AST/3041/2020.
M.~Kimura was supported by Japan Society for the Promotion of Science (JSPS) Grants-in-Aid for Scientific Research (KAKENHI) Grant No.~JP22K03626.
A.~Kuntz acknowledges funding from the FCT project ``Gravitational waves as a new probe of fundamental physics and astrophysics'' grant agreement 2023.07357.CEECIND/CP2830/CT0003.
L.~London acknowledges funding from UK Royal Society grant ({URF{\textbackslash}R1{\textbackslash}211451}).
S.~Maenaut acknowledges support from the Flemish inter-university project IBOF/21/084.
E.~Maggio, H.~O.~Silva, and S.\,H.~V\"olkel acknowledge funding from the Deutsche Forschungsgemeinschaft (DFG)~-~Project No.:~386119226.
E.~Maggio is supported by the European Union's Horizon Europe research and innovation programme under the Marie
Skłodowska-Curie grant agreement No.~101107586.
A.~Maselli acknowledges financial support from MUR PRIN Grants No.~2022-Z9X4XS and No.~2020KB33TP.
K.~Mitman is supported by NASA through the NASA Hubble
Fellowship grant No.~HST-HF2-51562.001-A awarded by
the Space Telescope Science Institute, which is operated
by the Association of Universities for Research in Astronomy, Incorporated, under NASA contract NAS5-26555, the National Science Foundation under Grants No.~PHY-2407742, No.~PHY-2207342,
and No.~OAC-2209655, and the Sherman Fairchild Foundation. 
H.~Motohashi was supported by Japan Society for the Promotion of Science (JSPS) Grants-in-Aid for Scientific Research (KAKENHI) Grant No.~JP22K03639.
N.~Oshita was supported by Japan Society for the Promotion of Science (JSPS) KAKENHI Grant No.~JP23K13111 and by the Hakubi project at Kyoto University.
C.~Pacilio and A.~Toubiana are supported by ERC Starting Grant No.~945155--GWmining, Cariplo Foundation Grant No.~2021-0555, MUR PRIN Grant No.~2022-Z9X4XS, MUR Grant ``Progetto Dipartimenti di Eccellenza 2023-2027'' (BiCoQ), and the ICSC National Research Centre funded by NextGenerationEU. 
P.~Pani is partially supported by the MUR FIS2 Advanced Grant ET-NOW (CUP:~B53C25001080001), by the PRIN Grant 2020KR4KN2 ``String Theory as a bridge between Gauge Theories and Quantum Gravity,'' by the FARE programme (GW-NEXT, CUP:~B84I20000100001), and by the INFN TEONGRAV initiative.
R.~Panosso Macedo acknowledges support from the Villum Investigator program supported by the VILLUM Foundation (grant No.~VIL37766) and the Danish National Research Foundation (grants No.~DNRF162 and 184)and the European Union’s Horizon 2020 research and innovation programme under the Marie Sklodowska-Curie grant agreement No.~101131233.
C.~Pitte acknowledges support from CNES Grant No.~51/20082, CEA Grant No.~2022-039, and the Agenzia Spaziale Italiana (ASI), Project No.~2024-36-HH.0, ``Attività per la fase B2/C della missione LISA.''
M.~Richartz acknowledges partial support from the Conselho
Nacional de Desenvolvimento Científico e Tecnológico
(CNPq, Brazil), Grant 315991/2023-2, and from the São
Paulo Research Foundation (FAPESP, Brazil), Grants
2022/08335-0 and 2024/00923-6.
A.~Riotto acknowledges support from the Swiss National Science Foundation
(project number CRSII5 213497) and from the Boninchi Foundation for the project ``PBHs in
the Era of GW Astronomy.''
B.S.~Sathyaprakash was supported in part by NSF Grants No.~PHY-2207638, AST-2307147, PHY-2308886, and PHY-2309064.
L.~Sberna~acknowledges support from the UKRI guarantee funding (project
No.~EP/Y023706/1) and by the University of Nottingham Anne McLaren Fellowship.
L.~C.~Stein acknowledges support from CAREER Award No.~PHY-2047382 and a Sloan
Foundation Research Fellowship.
A.~Toubiana is supported by MUR Young Researchers Grant No.~SOE2024-0000125
S.~Yi is supported by the NSF Graduate Research Fellowship Program under Grant No.~DGE2139757. 
N.~Yunes acknowledges support from the Simons Foundation through Award No.~896696, the National Science Foundation (NSF) through Grant No.~PHY-2207650, and NASA through Grant No.~80NSSC22K0806.

\clearpage

\section*{Affiliation list}
\markboth{{\scshape Affiliation list}}{}
\addcontentsline{toc}{section}{Affiliation list}
\begingroup
\let\clearpage\relax
\affilInfo{} %
\endgroup

\clearpage

\section*{References}
\markboth{{\scshape References}}{}
\addcontentsline{toc}{section}{References}

\bibliographystyle{iopart-num}
\bibliography{references}

\providecommand{\newblock}{}
\begin{thebibliography}{1000}
\expandafter\ifx\csname url\endcsname\relax
  \def\url#1{{\tt #1}}\fi
\expandafter\ifx\csname urlprefix\endcsname\relax\def\urlprefix{URL }\fi
\providecommand{\eprint}[2][]{\href{http://arxiv.org/abs/#2}{arXiv:#2}}

\bibitem{Chandrasekhar:1985kt}
Chandrasekhar S 1985 {\em {The mathematical theory of black holes}\/} (Oxford
  University Press) ISBN 978-0-19-850370-5

\bibitem{Kokkotas:1999bd}
Kokkotas K~D and Schmidt B~G 1999 \href{http://dx.doi.org/10.12942/lrr-1999-2}{
  {\em Living Rev. Rel.\/} {\bf 2} 2 } [\eprint{gr-qc/9909058}]

\bibitem{Nollert:1999ji}
Nollert H~P 1999 \href{http://dx.doi.org/10.1088/0264-9381/16/12/201}{ {\em
  Class. Quant. Grav.\/} {\bf 16} R159--R216 }

\bibitem{Ferrari:2007dd}
Ferrari V and Gualtieri L 2008
  \href{http://dx.doi.org/10.1007/s10714-007-0585-1}{ {\em Gen. Rel. Grav.\/}
  {\bf 40} 945--970 } [\eprint{0709.0657}]

\bibitem{Berti:2009kk}
Berti E, Cardoso V and Starinets A~O 2009
  \href{http://dx.doi.org/10.1088/0264-9381/26/16/163001}{ {\em Class. Quant.
  Grav.\/} {\bf 26} 163001 } [\eprint{0905.2975}]

\bibitem{Konoplya:2011qq}
Konoplya R~A and Zhidenko A 2011
  \href{http://dx.doi.org/10.1103/RevModPhys.83.793}{ {\em Rev. Mod. Phys.\/}
  {\bf 83} 793--836 } [\eprint{1102.4014}]

\bibitem{Berti:2018vdi}
Berti E, Yagi K, Yang H and Yunes N 2018
  \href{http://dx.doi.org/10.1007/s10714-018-2372-6}{ {\em Gen. Rel. Grav.\/}
  {\bf 50} 49 } [\eprint{1801.03587}]

\bibitem{Cardoso:2019rvt}
Cardoso V and Pani P 2019 \href{http://dx.doi.org/10.1007/s41114-019-0020-4}{
  {\em Living Rev. Rel.\/} {\bf 22} 4 } [\eprint{1904.05363}]

\bibitem{2025PhT..2025d4178B}
{Berti} E, {Cheung} M~H~Y and {Yi} S 2025
  \href{http://dx.doi.org/10.1063/pt.fvtp.lpxx}{ {\em Physics Today\/} {\bf
  2025} 44178 }

\bibitem{ringdownIO}
\url{https://strong-gr.com/ringdown-inside-and-out/}

\bibitem{Chandrasekhar:1984mgh}
Chandrasekhar S 1984 \href{http://dx.doi.org/10.1098/rspa.1984.0021}{ {\em
  Proceedings of the Royal Society of London. A. Mathematical and Physical
  Sciences\/} {\bf 392} 1--13 }

\bibitem{Detweiler:1980gk}
Detweiler S~L 1980 \href{http://dx.doi.org/10.1086/158109}{ {\em Astrophys.
  J.\/} {\bf 239} 292--295 }

\bibitem{Regge:1957td}
Regge T and Wheeler J~A 1957 \href{http://dx.doi.org/10.1103/PhysRev.108.1063}{
  {\em Phys. Rev.\/} {\bf 108} 1063--1069 }

\bibitem{Zerilli:1970wzz}
Zerilli F~J 1970 \href{http://dx.doi.org/10.1103/PhysRevD.2.2141}{ {\em Phys.
  Rev. D\/} {\bf 2} 2141--2160 }

\bibitem{Teukolsky:1973ha}
Teukolsky S~A 1973 \href{http://dx.doi.org/10.1086/152444}{ {\em Astrophys.
  J.\/} {\bf 185} 635--647 }

\bibitem{Sasaki:1981}
Sasaki M and Nakamura T 1981
  \href{http://dx.doi.org/10.1016/0375-9601(81)90568-5}{ {\em Phys. Lett.\/}
  {\bf A87} 85 }

\bibitem{Sasaki:1981kj}
Sasaki M and Nakamura T 1982
  \href{http://dx.doi.org/10.1016/0375-9601(82)90507-2}{ {\em Phys. Lett. A\/}
  {\bf 89} 68--70 }

\bibitem{Sasaki:1981sx}
Sasaki M and Nakamura T 1982 \href{http://dx.doi.org/10.1143/PTP.67.1788}{ {\em
  Prog. Theor. Phys.\/} {\bf 67} 1788 }

\bibitem{Newman:1961qr}
Newman E and Penrose R 1962 \href{http://dx.doi.org/10.1063/1.1724257}{ {\em
  J.Math.Phys.\/} {\bf 3} 566--578 }

\bibitem{Geroch:1973am}
Geroch R~P, Held A and Penrose R 1973
  \href{http://dx.doi.org/10.1063/1.1666410}{ {\em J.Math.Phys.\/} {\bf 14}
  874--881 }

\bibitem{Kerr:1963ud}
Kerr R~P 1963 \href{http://dx.doi.org/10.1103/PhysRevLett.11.237}{ {\em Phys.
  Rev. Lett.\/} {\bf 11} 237--238 }

\bibitem{Newman:1965my}
Newman E~T, Couch R, Chinnapared K, Exton A, Prakash A and Torrence R 1965
  \href{http://dx.doi.org/10.1063/1.1704351}{ {\em J. Math. Phys.\/} {\bf 6}
  918--919 }

\bibitem{Israel:1967wq}
Israel W 1967 \href{http://dx.doi.org/10.1103/PhysRev.164.1776}{ {\em Phys.
  Rev.\/} {\bf 164} 1776--1779 }

\bibitem{Hawking:1971vc}
Hawking S~W 1972 \href{http://dx.doi.org/10.1007/BF01877517}{ {\em Commun.
  Math. Phys.\/} {\bf 25} 152--166 }

\bibitem{Robinson:1975bv}
Robinson D~C 1975 \href{http://dx.doi.org/10.1103/PhysRevLett.34.905}{ {\em
  Phys. Rev. Lett.\/} {\bf 34} 905--906 }

\bibitem{Mazur:1982db}
Mazur P~O 1982 \href{http://dx.doi.org/10.1088/0305-4470/15/10/021}{ {\em J.
  Phys. A\/} {\bf 15} 3173--3180 }

\bibitem{Bekenstein:1996pn}
Bekenstein J~D 1996 {Black hole hair: 25 - years after} {\em {2nd International
  Sakharov Conference on Physics}\/} pp 216--219 [\eprint{gr-qc/9605059}]

\bibitem{Carter:1997im}
Carter B 1997 {Has the black hole equilibrium problem been solved?} {\em {8th
  Marcel Grossmann Meeting on Recent Developments in Theoretical and
  Experimental General Relativity, Gravitation and Relativistic Field Theories
  (MG 8)}\/} pp 136--155 [\eprint{gr-qc/9712038}]

\bibitem{Mazur:2000pn}
Mazur P~O 2000  [\eprint{hep-th/0101012}]

\bibitem{Robinson:2004zz}
Robinson D 2004 {Four decades of black holes uniqueness theorems} {\em {Kerr
  Fest: Black Holes in Astrophysics, General Relativity and Quantum Gravity}\/}

\bibitem{Chrusciel:2012jk}
Chrusciel P~T, Lopes~Costa J and Heusler M 2012
  \href{http://dx.doi.org/10.12942/lrr-2012-7}{ {\em Living Rev. Rel.\/} {\bf
  15} 7 } [\eprint{1205.6112}]

\bibitem{Ginzburg}
Ginzburg V and Ozernoy L 1964 {\em Zh. Eksp. Teor. Fiz.\/} {\bf 147} 1030--1040

\bibitem{Zeldovich}
Doroshkevic A, Zeldovich Y~B and Novikov I 1966 {\em Sov. Phys. JETP 22, 122\/}

\bibitem{Bunting}
Bunting G 1983 {\em Ph. D. Thesis (unpublished) University of New England,
  Armidale, N. S. W.\/}

\bibitem{Teukolsky:1972my}
Teukolsky S~A 1972 \href{http://dx.doi.org/10.1103/PhysRevLett.29.1114}{ {\em
  Phys. Rev. Lett.\/} {\bf 29} 1114--1118 }

\bibitem{Zerilli:1970se}
Zerilli F~J 1970 \href{http://dx.doi.org/10.1103/PhysRevLett.24.737}{ {\em
  Phys. Rev. Lett.\/} {\bf 24} 737--738 }

\bibitem{Nagar:2005ea}
Nagar A and Rezzolla L 2005
  \href{http://dx.doi.org/10.1088/0264-9381/22/16/R01}{ {\em Class. Quant.
  Grav.\/} {\bf 22} R167 } [Erratum: Class.Quant.Grav. 23, 4297 (2006)]
  [\eprint{gr-qc/0502064}]

\bibitem{Berti:2014bla}
Berti E 2014 {A Black-Hole Primer: Particles, Waves, Critical Phenomena and
  Superradiant Instabilities} [\eprint{1410.4481}]

\bibitem{Berti:2005gp}
Berti E, Cardoso V and Casals M 2006
  \href{http://dx.doi.org/10.1103/PhysRevD.73.109902}{ {\em Phys. Rev. D\/}
  {\bf 73} 024013 } [Erratum: Phys.Rev.D 73, 109902 (2006)]
  [\eprint{gr-qc/0511111}]

\bibitem{London:2020uva}
London L~T 2023 \href{http://dx.doi.org/10.1103/PhysRevD.107.044056}{ {\em
  Phys. Rev. D\/} {\bf 107} 044056 } [\eprint{2006.11449}]

\bibitem{ronveaux1995heun}
Ronveaux A (ed) 1995 {\em Heun's Differential Equations\/} (Oxford University,
  New York)

\bibitem{London:2023aeo}
London L~T 2023  [\eprint{2312.17678}]

\bibitem{London:2023idh}
London L and Gurevich M 2023  [\eprint{2312.17680}]

\bibitem{Cardoso:2004hh}
Cardoso V 2004 \href{http://dx.doi.org/10.1103/PhysRevD.70.127502}{ {\em Phys.
  Rev. D\/} {\bf 70} 127502 } [\eprint{gr-qc/0411048}]

\bibitem{Hod:2008zz}
Hod S 2008 \href{http://dx.doi.org/10.1103/PhysRevD.78.084035}{ {\em Phys. Rev.
  D\/} {\bf 78} 084035 } [\eprint{0811.3806}]

\bibitem{Yang:2012pj}
Yang H, Zhang F, Zimmerman A, Nichols D~A, Berti E and Chen Y 2013
  \href{http://dx.doi.org/10.1103/PhysRevD.87.041502}{ {\em Phys. Rev. D\/}
  {\bf 87} 041502 } [\eprint{1212.3271}]

\bibitem{Yang:2013uba}
Yang H, Zimmerman A, Zengino\u{g}lu A, Zhang F, Berti E and Chen Y 2013
  \href{http://dx.doi.org/10.1103/PhysRevD.88.044047}{ {\em Phys. Rev. D\/}
  {\bf 88} 044047 } [\eprint{1307.8086}]

\bibitem{Cook:2014cta}
Cook G~B and Zalutskiy M 2014
  \href{http://dx.doi.org/10.1103/PhysRevD.90.124021}{ {\em Phys. Rev. D\/}
  {\bf 90} 124021 } [\eprint{1410.7698}]

\bibitem{Li:2021wgz}
Li X, Sun L, Lo R~K~L, Payne E and Chen Y 2022
  \href{http://dx.doi.org/10.1103/PhysRevD.105.024016}{ {\em Phys. Rev. D\/}
  {\bf 105} 024016 } [\eprint{2110.03116}]

\bibitem{MaganaZertuche:2021syq}
Maga\~na Zertuche L {\em et~al.\/} 2022
  \href{http://dx.doi.org/10.1103/PhysRevD.105.104015}{ {\em Phys. Rev. D\/}
  {\bf 105} 104015 } [\eprint{2110.15922}]

\bibitem{Cook:2016fge}
Cook G~B and Zalutskiy M 2016
  \href{http://dx.doi.org/10.1088/0264-9381/33/24/245008}{ {\em Class. Quant.
  Grav.\/} {\bf 33} 245008 } [\eprint{1603.09710}]

\bibitem{Leaver:1985ax}
Leaver E 1985 \href{http://dx.doi.org/10.1098/rspa.1985.0119}{ {\em
  Proc.Roy.Soc.Lond.\/} {\bf A402} 285--298 }

\bibitem{Onozawa:1996ux}
Onozawa H 1997 \href{http://dx.doi.org/10.1103/PhysRevD.55.3593}{ {\em Phys.
  Rev. D\/} {\bf 55} 3593--3602 } [\eprint{gr-qc/9610048}]

\bibitem{Berti:2003jh}
Berti E, Cardoso V, Kokkotas K~D and Onozawa H 2003
  \href{http://dx.doi.org/10.1103/PhysRevD.68.124018}{ {\em Phys.Rev.\/} {\bf
  D68} 124018 } [\eprint{hep-th/0307013}]

\bibitem{wald1973perturbations}
Wald R~M 1973 {\em Journal of Mathematical Physics\/} {\bf 14} 1453--1461

\bibitem{Cook:2022kbb}
Cook G~B and Lu S 2023 \href{http://dx.doi.org/10.1103/PhysRevD.107.044043}{
  {\em Phys. Rev. D\/} {\bf 107} 044043 } [\eprint{2211.14955}]

\bibitem{Cook:2018ses}
Cook G~B, Annichiarico L~S and Vickers D~J 2019
  \href{http://dx.doi.org/10.1103/PhysRevD.99.024008}{ {\em Phys. Rev. D\/}
  {\bf 99} 024008 } [\eprint{1808.00987}]

\bibitem{blandin1983general}
Blandin J, Pons R and Marcilhacy G 1983 {\em Nuovo Cimento Lettere\/} {\bf 38}
  561--567

\bibitem{Batic:2007it}
Batic D and Schmid H 2007 \href{http://dx.doi.org/10.1063/1.2720277}{ {\em J.
  Math. Phys.\/} {\bf 48} 042502 } [\eprint{gr-qc/0701064}]

\bibitem{Fiziev:2009wn}
Fiziev P~P 2010 \href{http://dx.doi.org/10.1088/0264-9381/27/13/135001}{ {\em
  Class. Quant. Grav.\/} {\bf 27} 135001 } [\eprint{0908.4234}]

\bibitem{Gautschi:1967cat}
Gautschi W 1967 {\em SIAM Rev.\/} {\bf 9} 24--82

\bibitem{Nollert:1993zz}
Nollert H~P 1993 \href{http://dx.doi.org/10.1103/PhysRevD.47.5253}{ {\em Phys.
  Rev. D\/} {\bf 47} 5253--5258 }

\bibitem{Onozawa:1995vu}
Onozawa H, Mishima T, Okamura T and Ishihara H 1996
  \href{http://dx.doi.org/10.1103/PhysRevD.53.7033}{ {\em Phys. Rev. D\/} {\bf
  53} 7033--7040 } [\eprint{gr-qc/9603021}]

\bibitem{Richartz:2015saa}
Richartz M 2016 \href{http://dx.doi.org/10.1103/PhysRevD.93.064062}{ {\em Phys.
  Rev. D\/} {\bf 93} 064062 } [\eprint{1509.04260}]

\bibitem{Cook:2016ngj}
Cook G~B and Zalutskiy M 2016
  \href{http://dx.doi.org/10.1103/PhysRevD.94.104074}{ {\em Phys. Rev. D\/}
  {\bf 94} 104074 } [\eprint{1607.07406}]

\bibitem{Ansorg:2016ztf}
Ansorg M and Panosso~Macedo R 2016
  \href{http://dx.doi.org/10.1103/PhysRevD.93.124016}{ {\em Phys. Rev. D\/}
  {\bf 93} 124016 } [\eprint{1604.02261}]

\bibitem{Ferrari:1984zz}
Ferrari V and Mashhoon B 1984 \href{http://dx.doi.org/10.1103/PhysRevD.30.295}{
  {\em Phys. Rev. D\/} {\bf 30} 295--304 }

\bibitem{Cardoso:2008bp}
Cardoso V, Miranda A~S, Berti E, Witek H and Zanchin V~T 2009
  \href{http://dx.doi.org/10.1103/PhysRevD.79.064016}{ {\em Phys. Rev. D\/}
  {\bf 79} 064016 } [\eprint{0812.1806}]

\bibitem{Dolan:2010wr}
Dolan S~R 2010 \href{http://dx.doi.org/10.1103/PhysRevD.82.104003}{ {\em Phys.
  Rev. D\/} {\bf 82} 104003 } [\eprint{1007.5097}]

\bibitem{Yang:2012he}
Yang H, Nichols D~A, Zhang F, Zimmerman A, Zhang Z and Chen Y 2012
  \href{http://dx.doi.org/10.1103/PhysRevD.86.104006}{ {\em Phys. Rev. D\/}
  {\bf 86} 104006 } [\eprint{1207.4253}]

\bibitem{Keshet:2012uq}
Keshet U and Ben-Meir A 2014
  \href{http://dx.doi.org/10.1016/j.physletb.2014.09.011}{ {\em Phys. Lett.
  B\/} {\bf 737} 379--382 } [\eprint{1207.2460}]

\bibitem{Hod:2012dtv}
Hod S 2012 \href{http://dx.doi.org/10.1016/j.physletb.2012.09.057}{ {\em Phys.
  Lett. B\/} {\bf 717} 462--464 } [\eprint{1304.0529}]

\bibitem{Hod:2013sna}
Hod S 2013 \href{http://dx.doi.org/10.1103/PhysRevD.87.064017}{ {\em Phys. Rev.
  D\/} {\bf 87} 064017 } [\eprint{1304.4683}]

\bibitem{Casals:2018cgx}
Casals M, Ottewill A~C and Warburton N 2019
  \href{http://dx.doi.org/10.1098/rspa.2018.0701}{ {\em Proc. Roy. Soc. Lond.
  A\/} {\bf 475} 20180701 } [\eprint{1810.00432}]

\bibitem{Vickers:2022ivk}
Vickers D~J and Cook G~B 2022
  \href{http://dx.doi.org/10.1103/PhysRevD.106.104037}{ {\em Phys. Rev. D\/}
  {\bf 106} 104037 } [\eprint{2208.06259}]

\bibitem{MaassenvandenBrink:2000iwh}
Maassen van~den Brink A 2000
  \href{http://dx.doi.org/10.1103/PhysRevD.62.064009}{ {\em Phys. Rev. D\/}
  {\bf 62} 064009 } [\eprint{gr-qc/0001032}]

\bibitem{Yang:2014tla}
Yang H, Zimmerman A and Lehner L 2015
  \href{http://dx.doi.org/10.1103/PhysRevLett.114.081101}{ {\em Phys. Rev.
  Lett.\/} {\bf 114} 081101 } [\eprint{1402.4859}]

\bibitem{Gralla:2016sxp}
Gralla S~E, Zimmerman A and Zimmerman P 2016
  \href{http://dx.doi.org/10.1103/PhysRevD.94.084017}{ {\em Phys. Rev. D\/}
  {\bf 94} 084017 } [\eprint{1608.04739}]

\bibitem{Casals:2016mel}
Casals M, Gralla S~E and Zimmerman P 2016
  \href{http://dx.doi.org/10.1103/PhysRevD.94.064003}{ {\em Phys. Rev. D\/}
  {\bf 94} 064003 } [\eprint{1606.08505}]

\bibitem{Richartz:2017qep}
Richartz M, Herdeiro C~A~R and Berti E 2017
  \href{http://dx.doi.org/10.1103/PhysRevD.96.044034}{ {\em Phys. Rev. D\/}
  {\bf 96} 044034 } [\eprint{1706.01112}]

\bibitem{Casals:2019vdb}
Casals M and Longo~Micchi L~F 2019
  \href{http://dx.doi.org/10.1103/PhysRevD.99.084047}{ {\em Phys. Rev. D\/}
  {\bf 99} 084047 } [\eprint{1901.04586}]

\bibitem{Whiting:1988vc}
Whiting B~F 1989 \href{http://dx.doi.org/10.1063/1.528308}{ {\em
  J.Math.Phys.\/} {\bf 30} 1301 }

\bibitem{TeixeiradaCosta:2019skg}
Teixeira~da Costa R 2020 \href{http://dx.doi.org/10.1007/s00220-020-03796-z}{
  {\em Commun. Math. Phys.\/} {\bf 378} 705--781 } [\eprint{1910.02854}]

\bibitem{Aretakis:2011gz}
Aretakis S 2012 \href{http://dx.doi.org/10.1016/j.jfa.2012.08.015}{ {\em J.
  Funct. Anal.\/} {\bf 263} 2770--2831 } [\eprint{1110.2006}]

\bibitem{Lucietti:2012sf}
Lucietti J and Reall H~S 2012
  \href{http://dx.doi.org/10.1103/PhysRevD.86.104030}{ {\em Phys. Rev. D\/}
  {\bf 86} 104030 } [\eprint{1208.1437}]

\bibitem{Aretakis:2012ei}
Aretakis S 2015 \href{http://dx.doi.org/10.4310/ATMP.2015.v19.n3.a1}{ {\em Adv.
  Theor. Math. Phys.\/} {\bf 19} 507--530 } [\eprint{1206.6598}]

\bibitem{Klainerman:2022ric}
Klainerman S and Szeftel J 2024
  \href{http://dx.doi.org/10.4310/pamq.2024.v20.n4.a8}{ {\em Pure Appl. Math.
  Quart.\/} {\bf 20} 1721--1761 } [\eprint{2210.14400}]

\bibitem{Shlapentokh-Rothman:2023bwo}
Shlapentokh-Rothman Y and da~Costa R~T 2023  [\eprint{2302.08916}]

\bibitem{GRIT}
\url{http://blackholes.ist.utl.pt/?page=Files}

\bibitem{CoG}
\url{https://the-center-of-gravity.com/}

\bibitem{JHU}
\url{https://pages.jh.edu/~eberti2/ringdown/}

\bibitem{PaniRome}
\url{https://paolopani.weebly.com/notebooks.html}

\bibitem{cook_2024_14024959}
Cook G~B 2024 Kerr {M}odes: Phase fixed gravitational {QNM}s and {TTM}s
  \href{https://zenodo.org/records/14024959}{Zenodo}

\bibitem{motohashi_2024_12696858}
Motohashi H 2024 Kerr quasinormal mode frequencies and excitation factors
  \href{https://doi.org/10.5281/zenodo.12696857}{Zenodo}

\bibitem{Lo:2025njp}
Lo R~K~L, Sabani L and Cardoso V 2025
  \href{http://dx.doi.org/10.1103/PhysRevD.111.124002}{ {\em Phys. Rev. D\/}
  {\bf 111} 124002 } [\eprint{2504.00084}]

\bibitem{KerrModes_2024_cook}
Cook G~B 2024 Kerr{M}odes
  \href{https://github.com/cookgb/KerrModes/releases}{GitHub}

\bibitem{Stein:2019mop}
Stein L~C 2019 \href{http://dx.doi.org/10.21105/joss.01683}{ {\em J. Open
  Source Softw.\/} {\bf 4} 1683 } [\eprint{1908.10377}]

\bibitem{Gibbons:1975kk}
Gibbons G~W 1975 \href{http://dx.doi.org/10.1007/BF01609829}{ {\em Commun.
  Math. Phys.\/} {\bf 44} 245--264 }

\bibitem{Blandford:1977ds}
Blandford R~D and Znajek R~L 1977
  \href{http://dx.doi.org/10.1093/mnras/179.3.433}{ {\em Mon. Not. Roy. Astron.
  Soc.\/} {\bf 179} 433--456 }

\bibitem{Palenzuela:2011es}
Palenzuela C, Bona C, Lehner L and Reula O 2011
  \href{http://dx.doi.org/10.1088/0264-9381/28/13/134007}{ {\em Class. Quant.
  Grav.\/} {\bf 28} 134007 } [\eprint{1102.3663}]

\bibitem{Cardoso:2016olt}
Cardoso V, Macedo C~F~B, Pani P and Ferrari V 2016
  \href{http://dx.doi.org/10.1088/1475-7516/2016/05/054}{ {\em JCAP\/} {\bf 05}
  054 } [Erratum: JCAP 04, E01 (2020)] [\eprint{1604.07845}]

\bibitem{Mark:2014aja}
Mark Z, Yang H, Zimmerman A and Chen Y 2015
  \href{http://dx.doi.org/10.1103/PhysRevD.91.044025}{ {\em Phys. Rev. D\/}
  {\bf 91} 044025 } [\eprint{1409.5800}]

\bibitem{Dias:2015wqa}
Dias O~J~C, Godazgar M and Santos J~E 2015
  \href{http://dx.doi.org/10.1103/PhysRevLett.114.151101}{ {\em Phys. Rev.
  Lett.\/} {\bf 114} 151101 } [\eprint{1501.04625}]

\bibitem{Dias:2022oqm}
Dias O~J~C, Godazgar M and Santos J~E 2022
  \href{http://dx.doi.org/10.1007/JHEP07(2022)076}{ {\em JHEP\/} {\bf 07} 076 }
  [\eprint{2205.13072}]

\bibitem{Zerilli:1974ai}
Zerilli F 1974 \href{http://dx.doi.org/10.1103/PhysRevD.9.860}{ {\em
  Phys.Rev.\/} {\bf D9} 860--868 }

\bibitem{Chandrasekhar:1975nkd}
Chandrasekhar S 1975 \href{http://dx.doi.org/10.1098/rspa.1975.0066}{ {\em
  Proc. Roy. Soc. Lond. A\/} {\bf 343} 289--298 }

\bibitem{Dias:2013sdc}
Dias O~J and Santos J~E 2013 \href{http://dx.doi.org/10.1007/JHEP10(2013)156}{
  {\em JHEP\/} {\bf 1310} 156 } [\eprint{1302.1580}]

\bibitem{Dias:2021yju}
Dias O~J~C, Godazgar M, Santos J~E, Carullo G, Del~Pozzo W and Laghi D 2022
  \href{http://dx.doi.org/10.1103/PhysRevD.105.084044}{ {\em Phys. Rev. D\/}
  {\bf 105} 084044 } [\eprint{2109.13949}]

\bibitem{Dias:2015nua}
Dias O~J~C, Santos J~E and Way B 2016
  \href{http://dx.doi.org/10.1088/0264-9381/33/13/133001}{ {\em Class. Quant.
  Grav.\/} {\bf 33} 133001 } [\eprint{1510.02804}]

\bibitem{Trefethen}
Trefethen L~N 2000 {\em Spectral Methods in MATLAB\/} (SIAM, Philadelphia)

\bibitem{Boyd}
Boyd J~P 2001 {\em Chebyshev and Fourier Spectral Methods\/} (Dover Books on
  Mathematics)

\bibitem{CanutoBook}
{Canuto} C, {Hussaini} M~Y, {Quarteroni} A and {Zang} T~A 2006 {\em {Spectral
  Methods}\/} (Springer-Verlag)

\bibitem{Dias:2009iu}
Dias O~J, Figueras P, Monteiro R, Santos J~E and Emparan R 2009
  \href{http://dx.doi.org/10.1103/PhysRevD.80.111701}{ {\em Phys.Rev.\/} {\bf
  D80} 111701 } [\eprint{0907.2248}]

\bibitem{Dias:2010eu}
Dias O~J, Figueras P, Monteiro R, Reall H~S and Santos J~E 2010
  \href{http://dx.doi.org/10.1007/JHEP05(2010)076}{ {\em JHEP\/} {\bf 1005} 076
  } [\eprint{1001.4527}]

\bibitem{Dias:2010maa}
Dias O~J, Figueras P, Monteiro R and Santos J~E 2010
  \href{http://dx.doi.org/10.1103/PhysRevD.82.104025}{ {\em Phys.Rev.\/} {\bf
  D82} 104025 } [\eprint{1006.1904}]

\bibitem{Dias:2010gk}
Dias O~J, Figueras P, Monteiro R and Santos J~E 2010
  \href{http://dx.doi.org/10.1007/JHEP12(2010)067}{ {\em JHEP\/} {\bf 1012} 067
  } [\eprint{1011.0996}]

\bibitem{Dias:2011jg}
Dias O~J, Monteiro R and Santos J~E 2011
  \href{http://dx.doi.org/10.1007/JHEP08(2011)139}{ {\em JHEP\/} {\bf 1108} 139
  } [\eprint{1106.4554}]

\bibitem{Dias:2010ma}
Dias O~J~C, Monteiro R, Reall H~S and Santos J~E 2010
  \href{http://dx.doi.org/10.1007/JHEP11(2010)036}{ {\em JHEP\/} {\bf 11} 036 }
  [\eprint{1007.3745}]

\bibitem{Dias:2011tj}
Dias O~J, Figueras P, Minwalla S, Mitra P, Monteiro R {\em et~al.\/} 2012
  \href{http://dx.doi.org/10.1007/JHEP08(2012)117}{ {\em JHEP\/} {\bf 1208} 117
  } [\eprint{1112.4447}]

\bibitem{Cardoso:2013pza}
Cardoso V, Dias O~J, Hartnett G~S, Lehner L and Santos J~E 2014
  \href{http://dx.doi.org/10.1007/JHEP04(2014)183}{ {\em JHEP\/} {\bf 1404} 183
  } [\eprint{1312.5323}]

\bibitem{Dias:2014eua}
Dias O~J~C, Hartnett G~S and Santos J~E 2014
  \href{http://dx.doi.org/10.1088/0264-9381/31/24/245011}{ {\em Class. Quant.
  Grav.\/} {\bf 31} 245011 } [\eprint{1402.7047}]

\bibitem{Dias:2018etb}
Dias O~J~C, Reall H~S and Santos J~E 2018
  \href{http://dx.doi.org/10.1007/JHEP10(2018)001}{ {\em JHEP\/} {\bf 10} 001 }
  [\eprint{1808.02895}]

\bibitem{Zilhao:2014wqa}
Zilh\~ao M, Cardoso V, Herdeiro C, Lehner L and Sperhake U 2014
  \href{http://dx.doi.org/10.1103/PhysRevD.90.124088}{ {\em Phys.Rev.\/} {\bf
  D90} 124088 } [\eprint{1410.0694}]

\bibitem{Bozzola:2020mjx}
Bozzola G and Paschalidis V 2021
  \href{http://dx.doi.org/10.1103/PhysRevLett.126.041103}{ {\em Phys. Rev.
  Lett.\/} {\bf 126} 041103 } [\eprint{2006.15764}]

\bibitem{Bozzola:2021elc}
Bozzola G and Paschalidis V 2021
  \href{http://dx.doi.org/10.1103/PhysRevD.104.044004}{ {\em Phys. Rev. D\/}
  {\bf 104} 044004 } [\eprint{2104.06978}]

\bibitem{Bozzola:2022uqu}
Bozzola G 2022 \href{http://dx.doi.org/10.1103/PhysRevLett.128.071101}{ {\em
  Phys. Rev. Lett.\/} {\bf 128} 071101 } [\eprint{2202.05310}]

\bibitem{Bozzola:2023nzo}
Bozzola G and Paschalidis V 2023
  \href{http://dx.doi.org/10.1103/PhysRevD.108.064010}{ {\em Phys. Rev. D\/}
  {\bf 108} 064010 } [\eprint{2309.04368}]

\bibitem{Berti:2005eb}
Berti E and Kokkotas K~D 2005
  \href{http://dx.doi.org/10.1103/PhysRevD.71.124008}{ {\em Phys.Rev.\/} {\bf
  D71} 124008 } [\eprint{gr-qc/0502065}]

\bibitem{Pani:2013ija}
Pani P, Berti E and Gualtieri L 2013
  \href{http://dx.doi.org/10.1103/PhysRevLett.110.241103}{ {\em
  Phys.Rev.Lett.\/} {\bf 110} 241103 } [\eprint{1304.1160}]

\bibitem{Pani:2013wsa}
Pani P, Berti E and Gualtieri L 2013
  \href{http://dx.doi.org/10.1103/PhysRevD.88.064048}{ {\em Phys.Rev.\/} {\bf
  D88} 064048 } [\eprint{1307.7315}]

\bibitem{Blazquez-Salcedo:2022eik}
Bl\'azquez-Salcedo J~L and Khoo F~S 2023
  \href{http://dx.doi.org/10.1103/PhysRevD.107.084031}{ {\em Phys. Rev. D\/}
  {\bf 107} 084031 } [\eprint{2212.00054}]

\bibitem{Zimmerman:2015trm}
Zimmerman A and Mark Z 2016
  \href{http://dx.doi.org/10.1103/PhysRevD.93.044033}{ {\em Phys. Rev. D\/}
  {\bf 93} 044033 } [Erratum: Phys.Rev.D 93, 089905 (2016)]
  [\eprint{1512.02247}]

\bibitem{Davey:2023fin}
Davey A, Dias O~J~C and Santos J~E 2023
  \href{http://dx.doi.org/10.1007/JHEP12(2023)101}{ {\em JHEP\/} {\bf 12} 101 }
  [\eprint{2305.11216}]

\bibitem{Landau1981Quantum}
Landau L~D and Lifshitz L~M 1981 {\em Quantum Mechanics Non-Relativistic
  Theory, Third Edition: Volume 3\/} 3rd ed (Butterworth-Heinemann) ISBN
  0750635398 \urlprefix\url{http://www.worldcat.org/isbn/0750635398}

\bibitem{Cohen-Tannoudji:1977}
Cohen-Tannoudji C, Diu B and Laloë F 1977 {\em {Quantum mechanics; 1st ed.}\/}
  (New York, NY: Wiley) trans. of : Mécanique quantique. Paris : Hermann, 1973
  \urlprefix\url{https://cds.cern.ch/record/101367}

\bibitem{Hund1927}
Hund F 1927 \href{http://dx.doi.org/10.1007/BF01400234}{ {\em Z. Physik\/} {\bf
  40} 742--764 }

\bibitem{1929PhyZ...30..467V}
{von Neuman} J and {Wigner} E 1929
  \href{http://dx.doi.org/10.1007/978-3-662-02781-3_19}{ {\em Physikalische
  Zeitschrift\/} {\bf 30} 467--470 }

\bibitem{LandauQM}
Landau L and Lifshitz E 1977 {\em Quantum Mechanics: Non-Relativistic Theory\/}
  3rd ed ({\em Course of theoretical physics\/} vol~3) (Pergamon Press)

\bibitem{Arnold1978}
Arnold V~I 1978 {\em Mathematical Methods of Classical Mechanics\/} Graduate
  Texts in Mathematics (Springer New York)

\bibitem{ashcroft1976solid}
Ashcroft N and Mermin N 1976 {\em Solid State Physics\/} HRW international
  editions (Holt, Rinehart and Winston) ISBN 9780030839931

\bibitem{1932PhyZS...2...46L}
{Landau} L~D 1932 {\em Phys. Zs. Sowjet\/} {\bf 2} 46

\bibitem{Zener:1932ws}
Zener C 1932 \href{http://dx.doi.org/10.1098/rspa.1932.0165}{ {\em Proc. Roy.
  Soc. Lond. A\/} {\bf 137} 696--702 }

\bibitem{Majorana:1932ga}
Majorana E 1932 \href{http://dx.doi.org/10.1007/BF02960953}{ {\em Nuovo Cim.\/}
  {\bf 9} 43--50 }

\bibitem{Stu1932}
St{\"u}ckelberg E~C~G 1932 {\em Atomen, Helv. Phys. Acta\/} {\bf 5} 369

\bibitem{Ivakhnenko:2022sfl}
Ivakhnenko O~V, Shevchenko S~N and Nori F 2023
  \href{http://dx.doi.org/10.1016/j.physrep.2022.10.002}{ {\em Phys. Rept.\/}
  {\bf 995} 1--89 } [\eprint{2203.16348}]

\bibitem{PhysRev.69.674}
Purcell E~M 1946 Spontaneous emission probabilities at ratio frequencies {\em
  Proceedings of the American Physical Society\/} vol~69 (American Physical
  Society) p 681

\bibitem{Herzberg1991}
Herzberg G 1991 {\em Molecular spectra and molecular structure. Volume 3,
  Electronic spectra and electronic structure of polyatomic molecules\/}
  (Krieger Publ., Malabar, FL) ISBN 9780894642708

\bibitem{Smirnov:2003da}
Smirnov A~Y 2003 {The MSW effect and solar neutrinos} {\em {10th International
  Workshop on Neutrino Telescopes}\/} pp 23--43 [\eprint{hep-ph/0305106}]

\bibitem{Wurm:2017cmm}
Wurm M 2017 \href{http://dx.doi.org/10.1016/j.physrep.2017.04.002}{ {\em Phys.
  Rept.\/} {\bf 685} 1--52 } [\eprint{1704.06331}]

\bibitem{Giganti:2017fhf}
Giganti C, Lavignac S and Zito M 2018
  \href{http://dx.doi.org/10.1016/j.ppnp.2017.10.001}{ {\em Prog. Part. Nucl.
  Phys.\/} {\bf 98} 1--54 } [\eprint{1710.00715}]

\bibitem{Dias:2020ncd}
Dias O~J~C and Santos J~E 2020
  \href{http://dx.doi.org/10.1103/PhysRevD.102.124039}{ {\em Phys. Rev. D\/}
  {\bf 102} 124039 } [\eprint{2005.03673}]

\bibitem{Davey:2022vyx}
Davey A, Dias O~J~C, Rodgers P and Santos J~E 2022
  \href{http://dx.doi.org/10.1007/JHEP07(2022)086}{ {\em JHEP\/} {\bf 07} 086 }
  [\eprint{2203.13830}]

\bibitem{Davey:2024xvd}
Davey A, Dias O~J~C and Gil D~S 2024
  \href{http://dx.doi.org/10.1007/JHEP07(2024)113}{ {\em JHEP\/} {\bf 07} 113 }
  [\eprint{2404.03724}]

\bibitem{Motohashi:2024fwt}
Motohashi H 2025 \href{http://dx.doi.org/10.1103/PhysRevLett.134.141401}{ {\em
  Phys. Rev. Lett.\/} {\bf 134} 141401 } [\eprint{2407.15191}]

\bibitem{Leaver:1986gd}
Leaver E~W 1986 \href{http://dx.doi.org/10.1103/PhysRevD.34.384}{ {\em Phys.
  Rev. D\/} {\bf 34} 384--408 }

\bibitem{Mano:1996gn}
Mano S and Takasugi E 1997 \href{http://dx.doi.org/10.1143/PTP.97.213}{ {\em
  Prog. Theor. Phys.\/} {\bf 97} 213--232 } [\eprint{gr-qc/9611014}]

\bibitem{Mano:1996vt}
Mano S, Suzuki H and Takasugi E 1996
  \href{http://dx.doi.org/10.1143/PTP.95.1079}{ {\em Prog. Theor. Phys.\/} {\bf
  95} 1079--1096 } [\eprint{gr-qc/9603020}]

\bibitem{Sasaki:2003xr}
Sasaki M and Tagoshi H 2003 \href{http://dx.doi.org/10.12942/lrr-2003-6}{ {\em
  Living Rev. Rel.\/} {\bf 6} 6 } [\eprint{gr-qc/0306120}]

\bibitem{Glampedakis:2001js}
Glampedakis K and Andersson N 2001
  \href{http://dx.doi.org/10.1103/PhysRevD.64.104021}{ {\em Phys. Rev. D\/}
  {\bf 64} 104021 } [\eprint{gr-qc/0103054}]

\bibitem{Berti:2004md}
Berti E 2004 {\em Conf. Proc. C\/} {\bf 0405132} 145--186
  [\eprint{gr-qc/0411025}]

\bibitem{Giesler:2019uxc}
Giesler M, Isi M, Scheel M~A and Teukolsky S 2019
  \href{http://dx.doi.org/10.1103/PhysRevX.9.041060}{ {\em Phys. Rev. X\/} {\bf
  9} 041060 } [\eprint{1903.08284}]

\bibitem{Oshita:2021iyn}
Oshita N 2021 \href{http://dx.doi.org/10.1103/PhysRevD.104.124032}{ {\em Phys.
  Rev. D\/} {\bf 104} 124032 } [\eprint{2109.09757}]

\bibitem{Berti:2006wq}
Berti E and Cardoso V 2006 \href{http://dx.doi.org/10.1103/PhysRevD.74.104020}{
  {\em Phys. Rev. D\/} {\bf 74} 104020 } [\eprint{gr-qc/0605118}]

\bibitem{Zhang:2013ksa}
Zhang Z, Berti E and Cardoso V 2013
  \href{http://dx.doi.org/10.1103/PhysRevD.88.044018}{ {\em Phys. Rev. D\/}
  {\bf 88} 044018 } [\eprint{1305.4306}]

\bibitem{Andersson:1996cm}
Andersson N 1997 \href{http://dx.doi.org/10.1103/PhysRevD.55.468}{ {\em Phys.
  Rev. D\/} {\bf 55} 468--479 } [\eprint{gr-qc/9607064}]

\bibitem{Cavalcante:2024swt}
Cavalcante J~a~P, Richartz M and da~Cunha B~C 2024
  \href{http://dx.doi.org/10.1103/PhysRevLett.133.261401}{ {\em Phys. Rev.
  Lett.\/} {\bf 133} 261401 } [\eprint{2407.20850}]

\bibitem{Cavalcante:2024kmy}
Cavalcante J~a~P, Richartz M and da~Cunha B~C 2024
  \href{http://dx.doi.org/10.1103/PhysRevD.110.124064}{ {\em Phys. Rev. D\/}
  {\bf 110} 124064 } [\eprint{2408.13964}]

\bibitem{Konoplya:2006br}
Konoplya R~A and Zhidenko A 2006
  \href{http://dx.doi.org/10.1103/PhysRevD.73.124040}{ {\em Phys. Rev. D\/}
  {\bf 73} 124040 } [\eprint{gr-qc/0605013}]

\bibitem{Dolan:2007mj}
Dolan S~R 2007 \href{http://dx.doi.org/10.1103/PhysRevD.76.084001}{ {\em Phys.
  Rev. D\/} {\bf 76} 084001 } [\eprint{0705.2880}]

\bibitem{Konoplya:2013rxa}
Konoplya R~A and Zhidenko A 2013
  \href{http://dx.doi.org/10.1103/PhysRevD.88.024054}{ {\em Phys. Rev. D\/}
  {\bf 88} 024054 } [\eprint{1307.1812}]

\bibitem{CarneirodaCunha:2015hzd}
Carneiro~da Cunha B and Novaes F 2015
  \href{http://dx.doi.org/10.1007/JHEP11(2015)144}{ {\em JHEP\/} {\bf 11} 144 }
  [\eprint{1506.06588}]

\bibitem{daCunha:2021jkm}
da~Cunha B~C and Cavalcante J~a~P 2021
  \href{http://dx.doi.org/10.1103/PhysRevD.104.084051}{ {\em Phys. Rev. D\/}
  {\bf 104} 084051 } [\eprint{2105.08790}]

\bibitem{zenodo13961216}
Cavalcante J~P, Richartz M and Carneiro~da Cunha B 2024
  \href{http://dx.doi.org/10.5281/zenodo.13961216}{ }
  \urlprefix\url{https://doi.org/10.5281/zenodo.13961216}

\bibitem{Andersson:1999wj}
Andersson N and Glampedakis K 2000
  \href{http://dx.doi.org/10.1103/PhysRevLett.84.4537}{ {\em Phys. Rev.
  Lett.\/} {\bf 84} 4537--4540 } [\eprint{gr-qc/9909050}]

\bibitem{Hod:2011zzd}
Hod S 2011 \href{http://dx.doi.org/10.1103/PhysRevD.84.044046}{ {\em Phys. Rev.
  D\/} {\bf 84} 044046 } [\eprint{1109.4080}]

\bibitem{Cavalcante:2025abr}
Cavalcante J~P, Richartz M and da~Cunha B~C 2025  [\eprint{2511.16640}]

\bibitem{Hod:2012px}
Hod S 2012 \href{http://dx.doi.org/10.1103/PhysRevD.86.129902}{ {\em Phys. Rev.
  D\/} {\bf 86} 104026 } [Erratum: Phys.Rev.D 86, 129902 (2012)]
  [\eprint{1211.3202}]

\bibitem{Herdeiro:2014goa}
Herdeiro C~A~R and Radu E 2014
  \href{http://dx.doi.org/10.1103/PhysRevLett.112.221101}{ {\em Phys. Rev.
  Lett.\/} {\bf 112} 221101 } [\eprint{1403.2757}]

\bibitem{Damour:1976kh}
Damour T, Deruelle N and Ruffini R 1976
  \href{http://dx.doi.org/10.1007/BF02725534}{ {\em Lett. Nuovo Cim.\/} {\bf
  15} 257--262 }

\bibitem{Ternov:1978gq}
Ternov I~M, Khalilov V~R, Chizhov G~A and Gaina A~B 1978
  \href{http://dx.doi.org/10.1007/BF00894575}{ {\em Sov. Phys. J.\/} {\bf 21}
  1200--1204 }

\bibitem{Zouros:1979iw}
Zouros T~J~M and Eardley D~M 1979
  \href{http://dx.doi.org/10.1016/0003-4916(79)90237-9}{ {\em Annals Phys.\/}
  {\bf 118} 139--155 }

\bibitem{Detweiler:1980uk}
Detweiler S~L 1980 \href{http://dx.doi.org/10.1103/PhysRevD.22.2323}{ {\em
  Phys. Rev. D\/} {\bf 22} 2323--2326 }

\bibitem{Dolan:2012yt}
Dolan S~R 2013 \href{http://dx.doi.org/10.1103/PhysRevD.87.124026}{ {\em Phys.
  Rev. D\/} {\bf 87} 124026 } [\eprint{1212.1477}]

\bibitem{Arvanitaki:2009fg}
Arvanitaki A, Dimopoulos S, Dubovsky S, Kaloper N and March-Russell J 2010
  \href{http://dx.doi.org/10.1103/PhysRevD.81.123530}{ {\em Phys. Rev. D\/}
  {\bf 81} 123530 } [\eprint{0905.4720}]

\bibitem{Arvanitaki:2010sy}
Arvanitaki A and Dubovsky S 2011
  \href{http://dx.doi.org/10.1103/PhysRevD.83.044026}{ {\em Phys. Rev. D\/}
  {\bf 83} 044026 } [\eprint{1004.3558}]

\bibitem{Brito:2015oca}
Brito R, Cardoso V and Pani P 2015
  \href{http://dx.doi.org/10.1007/978-3-319-19000-6}{ {\em Lect. Notes Phys.\/}
  {\bf 906} pp.1--237 } [\eprint{1501.06570}]

\bibitem{Siegert:1939zz}
Siegert A~J~F 1939 \href{http://dx.doi.org/10.1103/PhysRev.56.750}{ {\em Phys.
  Rev.\/} {\bf 56} 750--752 }

\bibitem{Gamow:1928zz}
Gamow G 1928 \href{http://dx.doi.org/10.1007/BF01343196}{ {\em Z. Phys.\/} {\bf
  51} 204--212 }

\bibitem{Kukulin1989}
Kukulin V~I, Krasnopol'sky V~M and Hor\'{a}\v{c}ek J 1989 {\em Theory of
  Resonances: Principles and Applications\/} Reidel Texts in the Mathematical
  Sciences (Springer Dordrecht) ISBN 978-90-481-8432-3

\bibitem{moiseyev_2011}
Moiseyev N 2011 {\em Non-Hermitian Quantum Mechanics\/} (Cambridge University
  Press)

\bibitem{El-Ganainy:2018ksn}
El-Ganainy R, Makris K~G, Khajavikhan M, Musslimani Z~H, Rotter S and
  Christodoulides D~N 2018 \href{http://dx.doi.org/10.1038/nphys4323}{ {\em
  Nature Phys.\/} {\bf 14} 11--19 }

\bibitem{Bergholtz:2019deh}
Bergholtz E~J, Budich J~C and Kunst F~K 2021
  \href{http://dx.doi.org/10.1103/revmodphys.93.015005}{ {\em Rev. Mod.
  Phys.\/} {\bf 93} 015005 } [\eprint{1912.10048}]

\bibitem{Ashida:2020dkc}
Ashida Y, Gong Z and Ueda M 2021
  \href{http://dx.doi.org/10.1080/00018732.2021.1876991}{ {\em Adv. Phys.\/}
  {\bf 69} 249--435 } [\eprint{2006.01837}]

\bibitem{Kato1995}
Kato T 1995 {\em Perturbation Theory for Linear Operators\/} (Springer-Verlag,
  Berlin) ISBN 978-3-540-58661-6

\bibitem{Ozdemir:2019iqe}
\"Ozdemir c~K, Rotter S, Nori F and Yang L 2019
  \href{http://dx.doi.org/10.1038/s41563-019-0304-9}{ {\em Nature Materials\/}
  {\bf 18} 783--798 }

\bibitem{Wiersig:2020dgv}
Wiersig J 2020 \href{http://dx.doi.org/10.1364/PRJ.396115}{ {\em Photonics
  Res.\/} {\bf 8} 1457--1467 }

\bibitem{Parto2021}
Parto M, Liu Y~G~N, Bahari B, Khajavikhan M and Christodoulides D~N 2021
  \href{http://dx.doi.org/doi:10.1515/nanoph-2020-0434}{ {\em Nanophotonics\/}
  {\bf 10} 403--423 }

\bibitem{Ding:2022juv}
Ding K, Fang C and Ma G 2022
  \href{http://dx.doi.org/10.1038/s42254-022-00516-5}{ {\em Nature Rev.
  Phys.\/} {\bf 4} 745--760 } [\eprint{2204.11601}]

\bibitem{PhysRevE.61.929}
Heiss W~D 2000 \href{http://dx.doi.org/10.1103/PhysRevE.61.929}{ {\em Phys.
  Rev. E\/} {\bf 61}(1) 929--932 }

\bibitem{Heiss:2012dx}
Heiss W~D 2012 \href{http://dx.doi.org/10.1088/1751-8113/45/44/444016}{ {\em J.
  Phys. A\/} {\bf 45} 444016 } [\eprint{1210.7536}]

\bibitem{Zeldovich:1961a}
Zel’dovich Y~B 1961 {\em JETP\/} {\bf 12} 542
  \urlprefix\url{http://jetp.ras.ru/cgi-bin/e/index/e/12/3/p542?a=list}

\bibitem{Kapur1938}
Kapur P~L, Peierls R and Fowler R~H 1938
  \href{http://dx.doi.org/10.1098/rspa.1938.0093}{ {\em Proceedings of the
  Royal Society of London. Series A. Mathematical and Physical Sciences\/} {\bf
  166} 277--295 }

\bibitem{PeZe1998}
Perelomov A~M and Zel’dovich Y~B 1998 {\em Quantum Mechanics: Selected
  Topics\/} Selected Topics Series (World Scientific Publishing Company) ISBN
  978-9810235505

\bibitem{Ching:1993gt}
Ching E~S~C, Leung P~T, Suen W~M and Young K 1995
  \href{http://dx.doi.org/10.1103/PhysRevLett.74.4588}{ {\em Phys. Rev.
  Lett.\/} {\bf 74} 4588--4591 } [\eprint{gr-qc/9408043}]

\bibitem{Leung:1997was}
Leung P~T, Liu Y~T, Suen W~M, Tam C~Y and Young K 1997
  \href{http://dx.doi.org/10.1103/PhysRevLett.78.2894}{ {\em Phys. Rev.
  Lett.\/} {\bf 78} 2894--2897 } [\eprint{gr-qc/9903031}]

\bibitem{Leung_1998}
Leung P~T, Liu Y~T, Suen W~M, Tam C~Y and Young K 1998
  \href{http://dx.doi.org/10.1088/0305-4470/31/14/013}{ {\em Journal of Physics
  A: Mathematical and General\/} {\bf 31} 3271 } [\eprint{math-ph/9712037}]

\bibitem{Leung:1999iq}
Leung P~T, Liu Y~T, Suen W~M, Tam C~Y and Young K 1999
  \href{http://dx.doi.org/10.1103/PhysRevD.59.044034}{ {\em Phys. Rev. D\/}
  {\bf 59} 044034 } [\eprint{gr-qc/9903032}]

\bibitem{Zimmerman:2014aha}
Zimmerman A, Yang H, Mark Z, Chen Y and Lehner L 2015
  \href{http://dx.doi.org/10.1007/978-3-319-10488-1_19}{ {\em Astrophys. Space
  Sci. Proc.\/} {\bf 40} 217--223 } [\eprint{1406.4206}]

\bibitem{Yang:2015jja}
Yang H, Zhang F, Green S~R and Lehner L 2015
  \href{http://dx.doi.org/10.1103/PhysRevD.91.084007}{ {\em Phys. Rev. D\/}
  {\bf 91} 084007 } [\eprint{1502.08051}]

\bibitem{Cannizzaro:2023jle}
Cannizzaro E, Sberna L, Green S~R and Hollands S 2024
  \href{http://dx.doi.org/10.1103/PhysRevLett.132.051401}{ {\em Phys. Rev.
  Lett.\/} {\bf 132} 051401 } [\eprint{2309.10021}]

\bibitem{Berry:2004ypy}
Berry M~V 2004 \href{http://dx.doi.org/10.1023/B:CJOP.0000044002.05657.04}{
  {\em Czech. J. Phys.\/} {\bf 54} 1039--1047 }

\bibitem{PhysRevA.72.014104}
Mailybaev A~A, Kirillov O~N and Seyranian A~P 2005
  \href{http://dx.doi.org/10.1103/PhysRevA.72.014104}{ {\em Phys. Rev. A\/}
  {\bf 72}(1) 014104 }

\bibitem{PhysRevA.85.064103}
Lee S~Y, Ryu J~W, Kim S~W and Chung Y 2012
  \href{http://dx.doi.org/10.1103/PhysRevA.85.064103}{ {\em Phys. Rev. A\/}
  {\bf 85}(6) 064103 }

\bibitem{Ryu:2023pqq}
Ryu J~W, Han J~H, Yi C~H, Park M~J and Park H~C 2024
  \href{http://dx.doi.org/10.1038/s42005-024-01822-3}{ {\em Commun. Phys.\/}
  {\bf 7} 340 } [\eprint{2306.06967}]

\bibitem{Takahashi:2025uwo}
Takahashi T, Motohashi H and Takahashi K 2025
  \href{http://dx.doi.org/10.1103/59tb-2x9r}{ {\em Phys. Rev. D\/} {\bf 112}
  064006 } [\eprint{2505.03883}]

\bibitem{Yang:2025dbn}
Yang Y, Berti E and Franchini N 2025  [\eprint{2504.06072}]

\bibitem{Kubota:2025hjk}
Kubota K~i and Motohashi H 2026 \href{http://dx.doi.org/10.1103/pwyd-nv6v}{
  {\em Phys. Rev. D\/} {\bf 113} 043053 } [\eprint{2509.06411}]

\bibitem{PanossoMacedo:2025xnf}
Panosso~Macedo R, Katagiri T, Kubota K~i and Motohashi H 2025
  [\eprint{2512.02110}]

\bibitem{Dodelson:2023nnr}
Dodelson M, Iossa C, Karlsson R, Lupsasca A and Zhiboedov A 2024
  \href{http://dx.doi.org/10.1007/JHEP07(2024)046}{ {\em JHEP\/} {\bf 07} 046 }
  [\eprint{2310.15236}]

\bibitem{Isaacson:1968hbi}
Isaacson R~A 1968 \href{http://dx.doi.org/10.1103/PhysRev.166.1263}{ {\em Phys.
  Rev.\/} {\bf 166} 1263--1271 }

\bibitem{Isaacson:1968zza}
Isaacson R~A 1968 \href{http://dx.doi.org/10.1103/PhysRev.166.1272}{ {\em Phys.
  Rev.\/} {\bf 166} 1272--1279 }

\bibitem{Misner:1973prb}
Misner C~W, Thorne K~S and Wheeler J~A 1973 {\em {Gravitation}\/} (San
  Francisco: W. H. Freeman) ISBN 978-0-7167-0344-0, 978-0-691-17779-3

\bibitem{poisson2014gravity}
Poisson E and Will C~M 2014 {\em Gravity: Newtonian, post-newtonian,
  relativistic\/} (Cambridge University Press)

\bibitem{Dolan:2017zgu}
Dolan S~R 2017 \href{http://dx.doi.org/10.1142/S0218271818430101}{ {\em Int. J.
  Mod. Phys. D\/} {\bf 27} 1843010 } [\eprint{1806.08617}]

\bibitem{guillemin2013semi}
Guillemin V and Sternberg S 2013 {\em Semi-classical analysis\/} (International
  Press Boston, MA)

\bibitem{Poisson:2011nh}
Poisson E, Pound A and Vega I 2011
  \href{http://dx.doi.org/10.12942/lrr-2011-7}{ {\em Living Rev. Rel.\/} {\bf
  14} 7 } [\eprint{1102.0529}]

\bibitem{Hadar:2022xag}
Hadar S, Kapec D, Lupsasca A and Strominger A 2022
  \href{http://dx.doi.org/10.1088/1361-6382/ac8d43}{ {\em Class. Quant.
  Grav.\/} {\bf 39} 215001 } [\eprint{2205.05064}]

\bibitem{Press:1971wr}
Press W~H 1971 \href{http://dx.doi.org/10.1086/180849}{ {\em Astrophys. J.
  Lett.\/} {\bf 170} L105--L108 }

\bibitem{Iyer:1986nq}
Iyer S 1987 \href{http://dx.doi.org/10.1103/PhysRevD.35.3632}{ {\em Phys. Rev.
  D\/} {\bf 35} 3632 }

\bibitem{Berti:2003si}
Berti E, Cavaglia M and Gualtieri L 2004
  \href{http://dx.doi.org/10.1103/PhysRevD.69.124011}{ {\em Phys. Rev. D\/}
  {\bf 69} 124011 } [\eprint{hep-th/0309203}]

\bibitem{Konoplya:2003ii}
Konoplya R~A 2003 \href{http://dx.doi.org/10.1103/PhysRevD.68.024018}{ {\em
  Phys. Rev. D\/} {\bf 68} 024018 } [\eprint{gr-qc/0303052}]

\bibitem{Kokkotas:1988fm}
Kokkotas K~D and Schutz B~F 1988
  \href{http://dx.doi.org/10.1103/PhysRevD.37.3378}{ {\em Phys. Rev. D\/} {\bf
  37} 3378--3387 }

\bibitem{Andersson1993}
Andersson N 1993 {\em Proceedings of the Royal Society of London. Series A:
  Mathematical and Physical Sciences\/} {\bf 442} 427--436

\bibitem{Mashhoon:1985cya}
Mashhoon B 1985 \href{http://dx.doi.org/10.1103/PhysRevD.31.290}{ {\em Phys.
  Rev. D\/} {\bf 31} 290--293 }

\bibitem{Fernando:2012yw}
Fernando S and Correa J 2012
  \href{http://dx.doi.org/10.1103/PhysRevD.86.064039}{ {\em Phys. Rev. D\/}
  {\bf 86} 064039 } [\eprint{1208.5442}]

\bibitem{Glampedakis:2019dqh}
Glampedakis K and Silva H~O 2019
  \href{http://dx.doi.org/10.1103/PhysRevD.100.044040}{ {\em Phys. Rev. D\/}
  {\bf 100} 044040 } [\eprint{1906.05455}]

\bibitem{Schutz:1985km}
Schutz B~F and Will C~M 1985 \href{http://dx.doi.org/10.1086/184453}{ {\em
  Astrophys. J. Lett.\/} {\bf 291} L33--L36 }

\bibitem{Iyer:1986np}
Iyer S and Will C~M 1987 \href{http://dx.doi.org/10.1103/PhysRevD.35.3621}{
  {\em Phys. Rev. D\/} {\bf 35} 3621 }

\bibitem{Seidel:1989bp}
Seidel E and Iyer S 1990 \href{http://dx.doi.org/10.1103/PhysRevD.41.374}{ {\em
  Phys. Rev. D\/} {\bf 41} 374--382 }

\bibitem{Froeman:1992gp}
Froeman N, Froeman P~O, Andersson N and Hoekback A 1992
  \href{http://dx.doi.org/10.1103/PhysRevD.45.2609}{ {\em Phys. Rev. D\/} {\bf
  45} 2609--2616 }

\bibitem{Dolan:2009nk}
Dolan S~R and Ottewill A~C 2009
  \href{http://dx.doi.org/10.1088/0264-9381/26/22/225003}{ {\em Class. Quant.
  Grav.\/} {\bf 26} 225003 } [\eprint{0908.0329}]

\bibitem{Daghigh:2011ty}
Daghigh R~G and Green M~D 2012
  \href{http://dx.doi.org/10.1103/PhysRevD.85.127501}{ {\em Phys. Rev. D\/}
  {\bf 85} 127501 } [\eprint{1112.5397}]

\bibitem{Fransen:2023eqj}
Fransen K 2023 \href{http://dx.doi.org/10.1088/1361-6382/acf26d}{ {\em Class.
  Quant. Grav.\/} {\bf 40} 205004 } [\eprint{2301.06999}]

\bibitem{Kapec:2024lnr}
Kapec D and Sheta A 2025 \href{http://dx.doi.org/10.1088/1361-6382/adecda}{
  {\em Class. Quant. Grav.\/} {\bf 42} 155002 } [\eprint{2412.08551}]

\bibitem{Papadopoulos:2020qik}
Papadopoulos G 2021 \href{http://dx.doi.org/10.1088/1361-6382/ac1cf8}{ {\em
  Class. Quant. Grav.\/} {\bf 38} 195018 } [\eprint{2007.04702}]

\bibitem{Penrose1976}
Penrose R 1976 Any space-time has a plane wave as a limit {\em Differential
  geometry and relativity\/} (Springer) pp 271--275

\bibitem{Rosen1937}
Rosen N 1937 {\em Phys. Z. Sowjetunion\/} {\bf 12} 366--372

\bibitem{Penrose:1965rx}
Penrose R 1965 \href{http://dx.doi.org/10.1103/RevModPhys.37.215}{ {\em Rev.
  Mod. Phys.\/} {\bf 37} 215--220 }

\bibitem{Brinkmann:1925fr}
Brinkmann H~W 1925 \href{http://dx.doi.org/10.1007/BF01208647}{ {\em Math.
  Ann.\/} {\bf 94} 119--145 }

\bibitem{Chawla:2024mse}
Chawla S, Fransen K and Keeler C 2024
  \href{http://dx.doi.org/10.1088/1361-6382/ad8f8c}{ {\em Class. Quant.
  Grav.\/} {\bf 41} 245015 } [\eprint{2406.14601}]

\bibitem{Penrose:1995cu}
Penrose R 1995 \href{http://dx.doi.org/10.1007/BF02153320}{ {\em Gen. Rel.
  Grav.\/} {\bf 27} 1323--1325 }

\bibitem{Shipley:2019kfq}
Shipley J~O 2019 {\em {Strong-field gravitational lensing by black holes}\/}
  Ph.D. thesis Sheffield U. [\eprint{1909.04691}]

\bibitem{Cariglia:2018erv}
Cariglia M, Houri T, Krtous P and Kubiznak D 2018
  \href{http://dx.doi.org/10.1140/epjc/s10052-018-6133-1}{ {\em Eur. Phys. J.
  C\/} {\bf 78} 661 } [\eprint{1805.07677}]

\bibitem{Blau:2006ar}
Blau M, Frank D and Weiss S 2006
  \href{http://dx.doi.org/10.1088/0264-9381/23/11/020}{ {\em Class. Quant.
  Grav.\/} {\bf 23} 3993--4010 } [\eprint{hep-th/0603109}]

\bibitem{Adamo:2017nia}
Adamo T, Casali E, Mason L and Nekovar S 2018
  \href{http://dx.doi.org/10.1088/1361-6382/aa9961}{ {\em Class. Quant.
  Grav.\/} {\bf 35} 015004 } [\eprint{1706.08925}]

\bibitem{Araneda:2022lgu}
Araneda B 2023 \href{http://dx.doi.org/10.1088/1361-6382/acaa83}{ {\em Class.
  Quant. Grav.\/} {\bf 40} 025006 } [\eprint{2204.13673}]

\bibitem{Sambe:1973cnm}
Sambe H 1973 \href{http://dx.doi.org/10.1103/PhysRevA.7.2203}{ {\em Phys. Rev.
  A\/} {\bf 7} 2203 }

\bibitem{Parmentier:2023axg}
Parmentier K 2024 \href{http://dx.doi.org/10.1007/JHEP06(2024)109}{ {\em
  JHEP\/} {\bf 06} 109 } [\eprint{2312.08430}]

\bibitem{Blau:2002mw}
Blau M, Figueroa-O'Farrill J~M and Papadopoulos G 2002
  \href{http://dx.doi.org/10.1088/0264-9381/19/18/310}{ {\em Class. Quant.
  Grav.\/} {\bf 19} 4753 } [\eprint{hep-th/0202111}]

\bibitem{Harte:2012uw}
Harte A~I and Drivas T~D 2012
  \href{http://dx.doi.org/10.1103/PhysRevD.85.124039}{ {\em Phys. Rev. D\/}
  {\bf 85} 124039 } [\eprint{1202.0540}]

\bibitem{Compere:2017hsi}
Comp\`ere G, Fransen K, Hertog T and Long J 2018
  \href{http://dx.doi.org/10.1088/1361-6382/aab99e}{ {\em Class. Quant.
  Grav.\/} {\bf 35} 104002 } [\eprint{1712.07130}]

\bibitem{Compere:2019wfw}
Comp\`ere G, Fransen K, Hertog T and Liu Y 2020
  \href{http://dx.doi.org/10.1103/PhysRevD.101.064006}{ {\em Phys. Rev. D\/}
  {\bf 101} 064006 } [\eprint{1910.02081}]

\bibitem{Bardeen:1999px}
Bardeen J~M and Horowitz G~T 1999
  \href{http://dx.doi.org/10.1103/PhysRevD.60.104030}{ {\em Phys. Rev. D\/}
  {\bf 60} 104030 } [\eprint{hep-th/9905099}]

\bibitem{Chen:2017ofv}
Chen B and Stein L~C 2017 \href{http://dx.doi.org/10.1103/PhysRevD.96.064017}{
  {\em Phys. Rev. D\/} {\bf 96} 064017 } [\eprint{1707.05319}]

\bibitem{Gibbons:1993sv}
Gibbons G~W and Townsend P~K 1993
  \href{http://dx.doi.org/10.1103/PhysRevLett.71.3754}{ {\em Phys. Rev.
  Lett.\/} {\bf 71} 3754--3757 } [\eprint{hep-th/9307049}]

\bibitem{Bardeen:1972fi}
Bardeen J~M, Press W~H and Teukolsky S~A 1972
  \href{http://dx.doi.org/10.1086/151796}{ {\em Astrophys. J.\/} {\bf 178} 347
  }

\bibitem{Kapec:2022dvc}
Kapec D, Lupsasca A and Strominger A 2023
  \href{http://dx.doi.org/10.1088/1361-6382/acc164}{ {\em Class. Quant.
  Grav.\/} {\bf 40} 095006 } [\eprint{2211.01674}]

\bibitem{Geroch:1969ca}
Geroch R~P 1969 \href{http://dx.doi.org/10.1007/BF01645486}{ {\em Commun. Math.
  Phys.\/} {\bf 13} 180--193 }

\bibitem{Blau:2002dy}
Blau M, Figueroa-O'Farrill J~M, Hull C and Papadopoulos G 2002
  \href{http://dx.doi.org/10.1088/0264-9381/19/10/101}{ {\em Class. Quant.
  Grav.\/} {\bf 19} L87--L95 } [\eprint{hep-th/0201081}]

\bibitem{Inonu:1953sp}
Inonu E and Wigner E~P 1953 \href{http://dx.doi.org/10.1073/pnas.39.6.510}{
  {\em Proc. Nat. Acad. Sci.\/} {\bf 39} 510--524 }

\bibitem{Hatsuda:2002xp}
Hatsuda M, Kamimura K and Sakaguchi M 2002
  \href{http://dx.doi.org/10.1016/S0550-3213(02)00258-4}{ {\em Nucl. Phys. B\/}
  {\bf 632} 114--120 } [\eprint{hep-th/0202190}]

\bibitem{Berenstein:2002jq}
Berenstein D~E, Maldacena J~M and Nastase H~S 2002
  \href{http://dx.doi.org/10.1088/1126-6708/2002/04/013}{ {\em JHEP\/} {\bf 04}
  013 } [\eprint{hep-th/0202021}]

\bibitem{Sadri:2003pr}
Sadri D and Sheikh-Jabbari M~M 2004
  \href{http://dx.doi.org/10.1103/RevModPhys.76.853}{ {\em Rev. Mod. Phys.\/}
  {\bf 76} 853 } [\eprint{hep-th/0310119}]

\bibitem{Cano:2024wzo}
Cano P~A and David M 2025
  \href{http://dx.doi.org/10.1103/PhysRevLett.134.191401}{ {\em Phys. Rev.
  Lett.\/} {\bf 134} 191401 } [\eprint{2407.12080}]

\bibitem{Gueven:2000ru}
Gueven R 2000 \href{http://dx.doi.org/10.1016/S0370-2693(00)00517-7}{ {\em
  Phys. Lett. B\/} {\bf 482} 255--263 } [\eprint{hep-th/0005061}]

\bibitem{Gueven:1987ad}
Gueven R 1987 \href{http://dx.doi.org/10.1016/0370-2693(87)90254-1}{ {\em Phys.
  Lett. B\/} {\bf 191} 275--281 }

\bibitem{Kehagias:2024sgh}
Kehagias A and Riotto A 2025
  \href{http://dx.doi.org/10.1103/PhysRevD.111.L041506}{ {\em Phys. Rev. D\/}
  {\bf 111} L041506 } [\eprint{2411.07980}]

\bibitem{Bucciotti:2025rxa}
Bucciotti B, Cardoso V, Kuntz A, Pere\~niguez D and Redondo-Yuste J 2025
  \href{http://dx.doi.org/10.1103/PhysRevD.111.L081502}{ {\em Phys. Rev. D\/}
  {\bf 111} L081502 } [\eprint{2501.17950}]

\bibitem{Kehagias:2025ntm}
Kehagias A, Perrone D and Riotto A 2025  [\eprint{2503.09350}]

\bibitem{Fransen:2025cgv}
Fransen K, Pere{\~n}iguez D and Redondo-Yuste J 2025  [\eprint{2509.03598}]

\bibitem{Johnson:2019ljv}
Johnson M~D {\em et~al.\/} 2020
  \href{http://dx.doi.org/10.1126/sciadv.aaz1310}{ {\em Sci. Adv.\/} {\bf 6}
  eaaz1310 } [\eprint{1907.04329}]

\bibitem{Gralla:2019drh}
Gralla S~E and Lupsasca A 2020
  \href{http://dx.doi.org/10.1103/PhysRevD.101.044031}{ {\em Phys. Rev. D\/}
  {\bf 101} 044031 } [\eprint{1910.12873}]

\bibitem{Gralla:2020pra}
Gralla S~E 2021 \href{http://dx.doi.org/10.1103/PhysRevD.103.024023}{ {\em
  Phys. Rev. D\/} {\bf 103} 024023 } [\eprint{2010.08557}]

\bibitem{Johnson:2024ttr}
Johnson M~D {\em et~al.\/} 2024 \href{http://dx.doi.org/10.1117/12.3019835}{
  {\em Proc. SPIE Int. Soc. Opt. Eng.\/} {\bf 13092} 130922D }
  [\eprint{2406.12917}]

\bibitem{Morse:1955aqj}
Harvey R 1955 \href{http://dx.doi.org/10.2307/3611130}{ {\em The Mathematical
  Gazette\/} {\bf 39} 719–722 }

\bibitem{ARFKEN2013401}
Arfken G~B, Weber H~J and Harris F~E 2013 Chapter 9 - partial differential
  equations {\em Mathematical Methods for Physicists (Seventh Edition)\/} ed
  Arfken G~B, Weber H~J and Harris F~E (Boston: Academic Press) pp 401--445
  seventh edition ed ISBN 978-0-12-384654-9
  \urlprefix\url{https://www.sciencedirect.com/science/article/pii/B9780123846549000098}

\bibitem{Courant1954}
Courant R and Hilbert D 1954 \href{http://dx.doi.org/10.1002/qj.49708034534}{
  {\em Quarterly Journal of the Royal Meteorological Society\/} {\bf 80}
  485--485 }

\bibitem{Ching:1995rt}
Ching E~S~C, Leung P~T, Suen W~M and Young K 1996
  \href{http://dx.doi.org/10.1103/PhysRevD.54.3778}{ {\em Phys. Rev. D\/} {\bf
  54} 3778--3791 } [\eprint{gr-qc/9507034}]

\bibitem{Sauvan:2021edw}
Sauvan C, Wu T, Zarouf R, Muljarov E~A and Lalanne P 2022
  \href{http://dx.doi.org/10.1364/OE.443656}{ {\em Opt. Express\/} {\bf 30}
  6846--6885 } [\eprint{2112.08103}]

\bibitem{alsheikh:tel-04116011}
Al~Sheikh L 2022 {\em {Scattering resonances and Pseudospectrum : stability and
  completeness aspects in optical and gravitational systems}\/} Theses
  {Universit{\'e} Bourgogne Franche-Comt{\'e}}
  \urlprefix\url{https://theses.hal.science/tel-04116011}

\bibitem{Thompson:2023ase}
Thompson J~E, Hamilton E, London L, Ghosh S, Kolitsidou P, Hoy C and Hannam M
  2024 \href{http://dx.doi.org/10.1103/PhysRevD.109.063012}{ {\em Phys. Rev.
  D\/} {\bf 109} 063012 } [\eprint{2312.10025}]

\bibitem{Hamilton:2021pkf}
Hamilton E, London L, Thompson J~E, Fauchon-Jones E, Hannam M, Kalaghatgi C,
  Khan S, Pannarale F and Vano-Vinuales A 2021
  \href{http://dx.doi.org/10.1103/PhysRevD.104.124027}{ {\em Phys. Rev. D\/}
  {\bf 104} 124027 } [\eprint{2107.08876}]

\bibitem{Pratten:2020ceb}
Pratten G {\em et~al.\/} 2021
  \href{http://dx.doi.org/10.1103/PhysRevD.103.104056}{ {\em Phys. Rev. D\/}
  {\bf 103} 104056 } [\eprint{2004.06503}]

\bibitem{London:2022urb}
London L and Hughes S~A 2022  [\eprint{2206.15246}]

\bibitem{Green:2022htq}
Green S~R, Hollands S, Sberna L, Toomani V and Zimmerman P 2023
  \href{http://dx.doi.org/10.1103/PhysRevD.107.064030}{ {\em Phys. Rev. D\/}
  {\bf 107} 064030 } [\eprint{2210.15935}]

\bibitem{Wu:2025sjt}
Wu T, Jaramillo J~L and Lalanne P 2025
  \href{http://dx.doi.org/10.1002/lpor.202402133}{ {\em Laser Photonics Rev.\/}
  {\bf 19} 2402133 }

\bibitem{Ma:2024qcv}
Ma S and Yang H 2024 \href{http://dx.doi.org/10.1103/PhysRevD.109.104070}{ {\em
  Phys. Rev. D\/} {\bf 109} 104070 } [\eprint{2401.15516}]

\bibitem{Arnaudo:2025bnm}
Arnaudo P, Carballo J and Withers B 2025  [\eprint{2505.04696}]

\bibitem{Sberna:2021eui}
Sberna L, Bosch P, East W~E, Green S~R and Lehner L 2022
  \href{http://dx.doi.org/10.1103/PhysRevD.105.064046}{ {\em Phys. Rev. D\/}
  {\bf 105} 064046 } [\eprint{2112.11168}]

\bibitem{gowdy_wave_1981}
Gowdy R~H 1981 \href{http://dx.doi.org/10.1063/1.524975}{ {\em Journal of
  Mathematical Physics\/} {\bf 22} 675--678 } ISSN 0022-2488

\bibitem{schmidt_relativistic_1993}
Schmidt B~G 1993 {\em Gravity Research Foundation essay\/}

\bibitem{Zenginoglu:2007jw}
Zenginoglu A 2008 \href{http://dx.doi.org/10.1088/0264-9381/25/14/145002}{ {\em
  Class. Quant. Grav.\/} {\bf 25} 145002 } [\eprint{0712.4333}]

\bibitem{Zenginoglu:2011jz}
Zenginoglu A 2011 \href{http://dx.doi.org/10.1103/PhysRevD.83.127502}{ {\em
  Phys. Rev.\/} {\bf D83} 127502 } [\eprint{1102.2451}]

\bibitem{PanossoMacedo:2023qzp}
Panosso~Macedo R 2024 \href{http://dx.doi.org/10.1098/rsta.2023.0046}{ {\em
  Phil. Trans. Roy. Soc. Lond. A\/} {\bf 382} 20230046 } [\eprint{2307.15735}]

\bibitem{PanossoMacedo:2024nkw}
Panosso~Macedo R and Zenginoglu A 2024 {Hyperboloidal Approach to Quasinormal
  Modes} [\eprint{2409.11478}]

\bibitem{Hussain:2022ins}
Hussain A and Zimmerman A 2022
  \href{http://dx.doi.org/10.1103/PhysRevD.106.104018}{ {\em Phys. Rev. D\/}
  {\bf 106} 104018 } [\eprint{2206.10653}]

\bibitem{Garcia-Quiros:2020qpx}
Garc\'\i{}a-Quir\'os C, Colleoni M, Husa S, Estell\'es H, Pratten G,
  Ramos-Buades A, Mateu-Lucena M and Jaume R 2020
  \href{http://dx.doi.org/10.1103/PhysRevD.102.064002}{ {\em Phys. Rev. D\/}
  {\bf 102} 064002 } [\eprint{2001.10914}]

\bibitem{trefethen2005spectra}
Trefethen L and Embree M 2005 {\em Spectra and Pseudospectra: The Behavior of
  Nonnormal Matrices and Operators\/} (Princeton University Press) ISBN
  9780691119465 \urlprefix\url{https://books.google.es/books?id=7gIbT-Y7-AIC}

\bibitem{Sjostrand2019}
Sjostrand J 2019 {\em Non-self-adjoint differential operators, spectral
  asymptotics and random perturbations\/} ISBN 978-3030108182

\bibitem{Jaramillo:2020tuu}
Jaramillo J~L, Panosso~Macedo R and Al~Sheikh L 2021
  \href{http://dx.doi.org/10.1103/PhysRevX.11.031003}{ {\em Phys. Rev. X\/}
  {\bf 11} 031003 } [\eprint{2004.06434}]

\bibitem{Jaramillo:2022kuv}
Jaramillo J~L 2022 \href{http://dx.doi.org/10.1088/1361-6382/ac8ddc}{ {\em
  Class. Quant. Grav.\/} {\bf 39} 217002 } [\eprint{2206.08025}]

\bibitem{Besson:2024adi}
Besson J and Jaramillo J~L 2025
  \href{http://dx.doi.org/10.1007/s10714-025-03438-6}{ {\em Gen. Rel. Grav.\/}
  {\bf 57} 110 } [\eprint{2412.02793}]

\bibitem{Ma:2017bxq}
Ma S 2020 \href{http://dx.doi.org/10.1007/s00220-020-03777-2}{ {\em Commun.
  Math. Phys.\/} {\bf 377} 2489--2551 } [\eprint{1708.07385}]

\bibitem{Ruiz:2007yx}
Ruiz M, Takahashi R, Alcubierre M and Nunez D 2008
  \href{http://dx.doi.org/10.1007/s10714-007-0570-8}{ {\em Gen. Rel. Grav.\/}
  {\bf 40} 2467 } [\eprint{0707.4654}]

\bibitem{Blanchet:2013haa}
Blanchet L 2014 \href{http://dx.doi.org/10.12942/lrr-2014-2}{ {\em Living Rev.
  Rel.\/} {\bf 17} 2 } [\eprint{1310.1528}]

\bibitem{teschl2000jacobi}
Teschl G 2000 {\em Jacobi Operators and Completely Integrable Nonlinear
  Lattices\/} Mathematical surveys and monographs (American Mathematical
  Society) ISBN 9780821819401
  \urlprefix\url{https://books.google.co.uk/books?id=8mE1_zzi94QC}

\bibitem{lax2002functional}
Lax P 2002 {\em Functional analysis\/} (New York: Wiley) ISBN 978-0-471-55604-6

\bibitem{Axler:2015}
Axler S 2015 {\em Linear algebra done right (eBook)\/} 3rd ed (Cham: Springer)
  \urlprefix\url{https://doi.org/10.1007/978-3-319-11080-6}

\bibitem{Finster:2015xma}
Finster F and Smoller J 2016
  \href{http://dx.doi.org/10.4310/MAA.2016.v23.n1.a2}{ {\em Methods Appl.
  Anal.\/} {\bf 23} 35--118 } [\eprint{1507.05756}]

\bibitem{adkins2012algebra}
Adkins W and Weintraub S 2012 {\em Algebra: An Approach via Module Theory\/}
  Graduate Texts in Mathematics (Springer New York) ISBN 9781461209232
  \urlprefix\url{https://books.google.co.uk/books?id=RFzdBwAAQBAJ}

\bibitem{gohberg2006indefinite}
Gohberg I, Lancaster P and Rodman L 2006 {\em Indefinite Linear Algebra and
  Applications\/} (Birkh{\"a}user Basel) ISBN 9783764373504
  \urlprefix\url{https://books.google.co.uk/books?id=Q1-ZJjY8Q9UC}

\bibitem{pinchover_rubinstein_2005}
Pinchover Y and Rubinstein J 2005 {\em Equations in high dimensions\/}
  (Cambridge University Press) p 226–281

\bibitem{abramowitz+stegun}
Abramowitz M and Stegun I~A {\em Handbook of Mathematical Functions with
  Formulas, Graphs, and Mathematical Tables\/} (New York: Dover)

\bibitem{ByronFuller}
Byron F~W and Fuller R~W 1992 {\em Mathematics of classical and quantum
  physics\/} (New York: Dover)

\bibitem{Minucci:2024qrn}
Minucci M and Panosso~Macedo R 2025
  \href{http://dx.doi.org/10.1007/s10714-025-03364-7}{ {\em Gen. Rel. Grav.\/}
  {\bf 57} 33 } [\eprint{2411.19740}]

\bibitem{leaver1985analytic}
Leaver E~W 1985 {\em Proceedings of the Royal Society of London. A.
  Mathematical and Physical Sciences\/} {\bf 402} 285--298

\bibitem{Teukolsky:1974yv}
Teukolsky S~A and Press W~H 1974 \href{http://dx.doi.org/10.1086/153180}{ {\em
  Astrophys. J.\/} {\bf 193} 443--461 }

\bibitem{Chen:2010}
Chen Y and Dai D 2010 \href{http://dx.doi.org/10.1016/j.jat.2010.07.005}{ {\em
  Journal of Approximation Theory\/} {\bf 162} 2149--2167 }

\bibitem{Chen:2019}
Chen Y, Filipuk G and Zhan L 2019 \href{http://dx.doi.org/10.1063/1.5102102}{
  {\em Journal of Mathematical Physics\/} {\bf 60} } ISSN 1089-7658
  [\eprint{1905.04869}]

\bibitem{Fackerell:1977shn}
Fackerell E~D and Crossman R~G 1977 \href{http://dx.doi.org/10.1063/1.523499}{
  {\em J. Math. Phys.\/} {\bf 18} 1849--1854 }

\bibitem{chihara2011introduction}
Chihara T 2011 {\em An Introduction to Orthogonal Polynomials\/} Dover Books on
  Mathematics (Dover Publications) ISBN 9780486479293
  \urlprefix\url{https://books.google.co.uk/books?id=71CVAwAAQBAJ}

\bibitem{Cook:2004kt}
Cook G~B and Pfeiffer H~P 2004
  \href{http://dx.doi.org/10.1103/PhysRevD.70.104016}{ {\em Phys. Rev.\/} {\bf
  D70} 104016 } [\eprint{gr-qc/0407078}]

\bibitem{Seidel:1988ue}
Seidel E 1989 \href{http://dx.doi.org/10.1088/0264-9381/6/7/012}{ {\em Class.
  Quant. Grav.\/} {\bf 6} 1057 }

\bibitem{PhysRevA.49.3057}
Leung P~T, Liu S~Y and Young K 1994
  \href{http://dx.doi.org/10.1103/PhysRevA.49.3057}{ {\em Phys. Rev. A\/} {\bf
  49}(4) 3057--3067 }

\bibitem{Redondo-Yuste:2023ipg}
Redondo-Yuste J, Pere\~niguez D and Cardoso V 2024
  \href{http://dx.doi.org/10.1103/PhysRevD.109.044048}{ {\em Phys. Rev. D\/}
  {\bf 109} 044048 } [\eprint{2312.04633}]

\bibitem{Iuliano:2024ogr}
Iuliano C, Hollands S, Green S~R and Zimmerman P 2025
  \href{http://dx.doi.org/10.1103/PhysRevD.111.124038}{ {\em Phys. Rev. D\/}
  {\bf 111} 124038 } [\eprint{2412.02821}]

\bibitem{Nollert:1996rf}
Nollert H~P 1996 \href{http://dx.doi.org/10.1103/PhysRevD.53.4397}{ {\em Phys.
  Rev. D\/} {\bf 53} 4397--4402 } [\eprint{gr-qc/9602032}]

\bibitem{Nollert:1998ys}
Nollert H~P and Price R~H 1999 \href{http://dx.doi.org/10.1063/1.532698}{ {\em
  J. Math. Phys.\/} {\bf 40} 980--1010 } [\eprint{gr-qc/9810074}]

\bibitem{Zenginoglu:2024bzs}
Zengino\u{g}lu A 2024 \href{http://dx.doi.org/10.1119/5.0214271}{ {\em Am. J.
  Phys.\/} {\bf 92} 965--974 } [\eprint{2404.01528}]

\bibitem{Horowitz:1999jd}
Horowitz G~T and Hubeny V~E 2000
  \href{http://dx.doi.org/10.1103/PhysRevD.62.024027}{ {\em Phys. Rev. D\/}
  {\bf 62} 024027 } [\eprint{hep-th/9909056}]

\bibitem{Zenginoglu:2011zz}
Zenginoglu A and Khanna G 2011
  \href{http://dx.doi.org/10.1103/PhysRevX.1.021017}{ {\em Phys. Rev. X\/} {\bf
  1} 021017 } [\eprint{1108.1816}]

\bibitem{vasy2013microlocal}
Vasy A 2013 {\em Inventiones mathematicae\/} {\bf 194} 381--513

\bibitem{Warnick:2013hba}
Warnick C~M 2015 \href{http://dx.doi.org/10.1007/s00220-014-2171-1}{ {\em
  Commun. Math. Phys.\/} {\bf 333} 959--1035 } [\eprint{1306.5760}]

\bibitem{Gajic:2019oem}
Gajic D and Warnick C 2020 \href{http://dx.doi.org/10.1063/5.0024699}{ {\em J.
  Math. Phys.\/} {\bf 61} 102501 } [\eprint{1910.08481}]

\bibitem{Gajic:2019qdd}
Gajic D and Warnick C 2021 \href{http://dx.doi.org/10.1007/s00220-021-04137-4}{
  {\em Commun. Math. Phys.\/} {\bf 385} 1395--1498 } [\eprint{1910.08479}]

\bibitem{Gajic:2024xrn}
Gajic D and Warnick C~M 2024 {Quasinormal modes on Kerr spacetimes}
  [\eprint{2407.04098}]

\bibitem{PanossoMacedo:2019npm}
Panosso~Macedo R 2020 \href{http://dx.doi.org/10.1088/1361-6382/ab6e3e}{ {\em
  Class. Quant. Grav.\/} {\bf 37} 065019 } [\eprint{1910.13452}]

\bibitem{Ripley:2022ypi}
Ripley J~L 2022 \href{http://dx.doi.org/10.1088/1361-6382/ac776d}{ {\em Class.
  Quant. Grav.\/} {\bf 39} 145009 } [\eprint{2202.03837}]

\bibitem{Leaver:1990zz}
Leaver E~W 1990 \href{http://dx.doi.org/10.1103/PhysRevD.41.2986}{ {\em Phys.
  Rev. D\/} {\bf 41} 2986--2997 }

\bibitem{PanossoMacedo:2018hab}
Panosso~Macedo R, Jaramillo J~L and Ansorg M 2018
  \href{http://dx.doi.org/10.1103/PhysRevD.98.124005}{ {\em Phys. Rev. D\/}
  {\bf 98} 124005 } [\eprint{1809.02837}]

\bibitem{Gasperin:2021kfv}
Gasperin E and Jaramillo J~L 2022
  \href{http://dx.doi.org/10.1088/1361-6382/ac5054}{ {\em Class. Quant.
  Grav.\/} {\bf 39} 115010 } [\eprint{2107.12865}]

\bibitem{Aguirregabiria:1996zy}
Aguirregabiria J~M and Vishveshwara C~V 1996
  \href{http://dx.doi.org/10.1016/0375-9601(95)00937-X}{ {\em Phys. Lett. A\/}
  {\bf 210} 251--254 }

\bibitem{Vishveshwara:1996jgz}
Vishveshwara C~V 1996 {\em Curr. Sci.\/} {\bf 71} 824--830

\bibitem{Jaramillo:2021tmt}
Jaramillo J~L, Panosso~Macedo R and Sheikh L~A 2022
  \href{http://dx.doi.org/10.1103/PhysRevLett.128.211102}{ {\em Phys. Rev.
  Lett.\/} {\bf 128} 211102 } [\eprint{2105.03451}]

\bibitem{Boyanov:2023qqf}
Boyanov V, Cardoso V, Destounis K, Jaramillo J~L and Panosso~Macedo R 2024
  \href{http://dx.doi.org/10.1103/PhysRevD.109.064068}{ {\em Phys. Rev. D\/}
  {\bf 109} 064068 } [\eprint{2312.11998}]

\bibitem{Cownden:2023dam}
Cownden B, Pantelidou C and Zilh\~ao M 2024
  \href{http://dx.doi.org/10.1007/JHEP05(2024)202}{ {\em JHEP\/} {\bf 05} 202 }
  [\eprint{2312.08352}]

\bibitem{Boyanov:2024fgc}
Boyanov V 2024 {On destabilising quasi-normal modes with a radially
  concentrated perturbation} [\eprint{2410.11547}]

\bibitem{Cai:2025irl}
Cai R~G, Cao L~M, Chen J~N, Guo Z~K, Wu L~B and Zhou Y~S 2025
  \href{http://dx.doi.org/10.1103/PhysRevD.111.084011}{ {\em Phys. Rev. D\/}
  {\bf 111} 084011 } [\eprint{2501.02522}]

\bibitem{Nollert:1992ifk}
Nollert H~P and Schmidt B~G 1992
  \href{http://dx.doi.org/10.1103/PhysRevD.45.2617}{ {\em Phys. Rev. D\/} {\bf
  45} 2617 }

\bibitem{bachelot1993resonances}
Bachelot A and Motet-Bachelot A 1993 Les r{\'e}sonances d'un trou noir de
  schwarzschild {\em Annales de l'IHP Physique th{\'e}orique\/} vol~59 pp 3--68

\bibitem{horowitz2000quasinormal}
Horowitz G~T and Hubeny V~E 2000 {\em Physical Review D\/} {\bf 62} 024027

\bibitem{Besson:2025}
Besson J, Boyanov V and Jaramillo J~L  In preparation

\bibitem{Graham:2020gwr}
Graham M~J {\em et~al.\/} 2020
  \href{http://dx.doi.org/10.1103/PhysRevLett.124.251102}{ {\em Phys. Rev.
  Lett.\/} {\bf 124} 251102 } [\eprint{2006.14122}]

\bibitem{Santini:2023ukl}
Santini A, Gerosa D, Cotesta R and Berti E 2023
  \href{http://dx.doi.org/10.1103/PhysRevD.108.083033}{ {\em Phys. Rev. D\/}
  {\bf 108} 083033 } [\eprint{2308.12998}]

\bibitem{CanevaSantoro:2023aol}
Caneva~Santoro G, Roy S, Vicente R, Haney M, Piccinni O~J, Del~Pozzo W and
  Martinez M 2024 \href{http://dx.doi.org/10.1103/PhysRevLett.132.251401}{ {\em
  Phys. Rev. Lett.\/} {\bf 132} 251401 } [\eprint{2309.05061}]

\bibitem{Begelman:1980vb}
Begelman M~C, Blandford R~D and Rees M~J 1980
  \href{http://dx.doi.org/10.1038/287307a0}{ {\em Nature\/} {\bf 287} 307--309
  }

\bibitem{LISA:2022yao}
Seoane P~A {\em et~al.\/} (LISA) 2023
  \href{http://dx.doi.org/10.1007/s41114-022-00041-y}{ {\em Living Rev. Rel.\/}
  {\bf 26} 2 } [\eprint{2203.06016}]

\bibitem{Barausse:2014tra}
Barausse E, Cardoso V and Pani P 2014
  \href{http://dx.doi.org/10.1103/PhysRevD.89.104059}{ {\em Phys. Rev. D\/}
  {\bf 89} 104059 } [\eprint{1404.7149}]

\bibitem{Destounis:2023ruj}
Destounis K and Duque F 2023 {Black-hole spectroscopy: quasinormal modes,
  ringdown stability and the pseudospectrum} [\eprint{2308.16227}]

\bibitem{Daghigh:2020jyk}
Daghigh R~G, Green M~D and Morey J~C 2020
  \href{http://dx.doi.org/10.1103/PhysRevD.101.104009}{ {\em Phys. Rev. D\/}
  {\bf 101} 104009 } [\eprint{2002.07251}]

\bibitem{Shen:2025yiy}
Shen S~F, Li G~R, Daghigh R~G, Morey J~C, Green M~D, Qian W~L and Yue R~H 2025
  [\eprint{2507.11663}]

\bibitem{Cheung:2021bol}
Cheung M~H~Y, Destounis K, Macedo R~P, Berti E and Cardoso V 2022
  \href{http://dx.doi.org/10.1103/PhysRevLett.128.111103}{ {\em Phys. Rev.
  Lett.\/} {\bf 128} 111103 } [\eprint{2111.05415}]

\bibitem{Konoplya:2022hll}
Konoplya R~A, Zinhailo A~F, Kunz J, Stuchlik Z and Zhidenko A 2022
  \href{http://dx.doi.org/10.1088/1475-7516/2022/10/091}{ {\em JCAP\/} {\bf 10}
  091 } [\eprint{2206.14714}]

\bibitem{Cardoso:2024mrw}
Cardoso V, Kastha S and Panosso~Macedo R 2024
  \href{http://dx.doi.org/10.1103/PhysRevD.110.024016}{ {\em Phys. Rev. D\/}
  {\bf 110} 024016 } [\eprint{2404.01374}]

\bibitem{Siqueira:2025lww}
Siqueira P~H~C, de~Paula L~T, Panosso~Macedo R and Richartz M 2025
  \href{http://dx.doi.org/10.1103/PhysRevD.111.104039}{ {\em Phys. Rev. D\/}
  {\bf 111} 104039 } [\eprint{2501.13815}]

\bibitem{Dreyer:2003bv}
Dreyer O, Kelly B~J, Krishnan B, Finn L~S, Garrison D and Lopez-Aleman R 2004
  \href{http://dx.doi.org/10.1088/0264-9381/21/4/003}{ {\em Class. Quant.
  Grav.\/} {\bf 21} 787--804 } [\eprint{gr-qc/0309007}]

\bibitem{Berti:2005ys}
Berti E, Cardoso V and Will C~M 2006
  \href{http://dx.doi.org/10.1103/PhysRevD.73.064030}{ {\em Phys. Rev. D\/}
  {\bf 73} 064030 } [\eprint{gr-qc/0512160}]

\bibitem{Baibhav:2023clw}
Baibhav V, Cheung M~H~Y, Berti E, Cardoso V, Carullo G, Cotesta R, Del~Pozzo W
  and Duque F 2023 \href{http://dx.doi.org/10.1103/PhysRevD.108.104020}{ {\em
  Phys. Rev. D\/} {\bf 108} 104020 } [\eprint{2302.03050}]

\bibitem{Poschl:1933zz}
Poschl G and Teller E 1933 \href{http://dx.doi.org/10.1007/BF01331132}{ {\em Z.
  Phys.\/} {\bf 83} 143--151 }

\bibitem{Ianniccari:2024ysv}
Ianniccari A, Iovino A~J, Kehagias A, Pani P, Perna G, Perrone D and Riotto A
  2024 \href{http://dx.doi.org/10.1103/PhysRevLett.133.211401}{ {\em Phys. Rev.
  Lett.\/} {\bf 133} 211401 } [\eprint{2407.20144}]

\bibitem{Yang:2024vor}
Yang Y, Mai Z~F, Yang R~Q, Shao L and Berti E 2024
  \href{http://dx.doi.org/10.1103/PhysRevD.110.084018}{ {\em Phys. Rev. D\/}
  {\bf 110} 084018 } [\eprint{2407.20131}]

\bibitem{Berti:2022xfj}
Berti E, Cardoso V, Cheung M~H~Y, Di~Filippo F, Duque F, Martens P and
  Mukohyama S 2022 \href{http://dx.doi.org/10.1103/PhysRevD.106.084011}{ {\em
  Phys. Rev. D\/} {\bf 106} 084011 } [\eprint{2205.08547}]

\bibitem{Cardoso:2016rao}
Cardoso V, Franzin E and Pani P 2016
  \href{http://dx.doi.org/10.1103/PhysRevLett.116.171101}{ {\em Phys. Rev.
  Lett.\/} {\bf 116} 171101 } [Erratum: Phys.Rev.Lett. 117, 089902 (2016)]
  [\eprint{1602.07309}]

\bibitem{DeLuca:2024uju}
De~Luca V, Franciolini G and Riotto A 2024
  \href{http://dx.doi.org/10.1103/PhysRevD.110.104041}{ {\em Phys. Rev. D\/}
  {\bf 110} 104041 } [\eprint{2408.14207}]

\bibitem{Destounis:2021lum}
Destounis K, Macedo R~P, Berti E, Cardoso V and Jaramillo J~L 2021
  \href{http://dx.doi.org/10.1103/PhysRevD.104.084091}{ {\em Phys. Rev. D\/}
  {\bf 104} 084091 } [\eprint{2107.09673}]

\bibitem{Cao:2024oud}
Cao L~M, Chen J~N, Wu L~B, Xie L and Zhou Y~S 2024
  \href{http://dx.doi.org/10.1007/s11433-024-2435-5}{ {\em Sci. China Phys.
  Mech. Astron.\/} {\bf 67} 100412 } [\eprint{2401.09907}]

\bibitem{Sarkar:2023rhp}
Sarkar S, Rahman M and Chakraborty S 2023
  \href{http://dx.doi.org/10.1103/PhysRevD.108.104002}{ {\em Phys. Rev. D\/}
  {\bf 108} 104002 } [\eprint{2304.06829}]

\bibitem{Destounis:2023nmb}
Destounis K, Boyanov V and Panosso~Macedo R 2024
  \href{http://dx.doi.org/10.1103/PhysRevD.109.044023}{ {\em Phys. Rev. D\/}
  {\bf 109} 044023 } [\eprint{2312.11630}]

\bibitem{Luo:2024dxl}
Luo S 2024 \href{http://dx.doi.org/10.1103/PhysRevD.110.084071}{ {\em Phys.
  Rev. D\/} {\bf 110} 084071 } [\eprint{2408.08139}]

\bibitem{Arean:2023ejh}
Are\'an D, Fari\~na D~G and Landsteiner K 2023
  \href{http://dx.doi.org/10.1007/JHEP12(2023)187}{ {\em JHEP\/} {\bf 12} 187 }
  [\eprint{2307.08751}]

\bibitem{Chen:2024mon}
Chen J~N, Wu L~B and Guo Z~K 2024
  \href{http://dx.doi.org/10.1088/1361-6382/ad89a1}{ {\em Class. Quant.
  Grav.\/} {\bf 41} 235015 } [\eprint{2407.03907}]

\bibitem{Motl:2003cd}
Motl L and Neitzke A 2003 \href{http://dx.doi.org/10.4310/ATMP.2003.v7.n2.a4}{
  {\em Adv. Theor. Math. Phys.\/} {\bf 7} 307--330 } [\eprint{hep-th/0301173}]

\bibitem{Andersson:2003fh}
Andersson N and Howls C~J 2004
  \href{http://dx.doi.org/10.1088/0264-9381/21/6/021}{ {\em Class. Quant.
  Grav.\/} {\bf 21} 1623--1642 } [\eprint{gr-qc/0307020}]

\bibitem{Natario:2004jd}
Natario J and Schiappa R 2004
  \href{http://dx.doi.org/10.4310/ATMP.2004.v8.n6.a4}{ {\em Adv. Theor. Math.
  Phys.\/} {\bf 8} 1001--1131 } [\eprint{hep-th/0411267}]

\bibitem{Daghigh:2024wcl}
Daghigh R~G, Green M~D and Morey J~C 2024
  \href{http://dx.doi.org/10.1103/PhysRevD.109.104076}{ {\em Phys. Rev. D\/}
  {\bf 109} 104076 } [\eprint{2403.13074}]

\bibitem{Kyutoku:2022gbr}
Kyutoku K, Motohashi H and Tanaka T 2023
  \href{http://dx.doi.org/10.1103/PhysRevD.107.044012}{ {\em Phys. Rev. D\/}
  {\bf 107} 044012 } [\eprint{2206.00671}]

\bibitem{Cardoso:2016oxy}
Cardoso V, Hopper S, Macedo C~F~B, Palenzuela C and Pani P 2016
  \href{http://dx.doi.org/10.1103/PhysRevD.94.084031}{ {\em Phys. Rev. D\/}
  {\bf 94} 084031 } [\eprint{1608.08637}]

\bibitem{Maggio:2019zyv}
Maggio E, Testa A, Bhagwat S and Pani P 2019
  \href{http://dx.doi.org/10.1103/PhysRevD.100.064056}{ {\em Phys. Rev. D\/}
  {\bf 100} 064056 } [\eprint{1907.03091}]

\bibitem{Maggio:2020jml}
Maggio E, Buoninfante L, Mazumdar A and Pani P 2020
  \href{http://dx.doi.org/10.1103/PhysRevD.102.064053}{ {\em Phys. Rev. D\/}
  {\bf 102} 064053 } [\eprint{2006.14628}]

\bibitem{Maggio:2021ans}
Maggio E, Pani P and Raposo G 2020 {\em Testing the Nature of Dark Compact
  Objects with Gravitational Waves\/} (Singapore: Springer Singapore) pp 1--37
  ISBN 978-981-15-4702-7 [\eprint{2105.06410}]

\bibitem{Cardoso:2021wlq}
Cardoso V, Destounis K, Duque F, Macedo R~P and Maselli A 2022
  \href{http://dx.doi.org/10.1103/PhysRevD.105.L061501}{ {\em Phys. Rev. D\/}
  {\bf 105} L061501 } [\eprint{2109.00005}]

\bibitem{Cardoso:2022whc}
Cardoso V, Destounis K, Duque F, Panosso~Macedo R and Maselli A 2022
  \href{http://dx.doi.org/10.1103/PhysRevLett.129.241103}{ {\em Phys. Rev.
  Lett.\/} {\bf 129} 241103 } [\eprint{2210.01133}]

\bibitem{Spieksma:2024voy}
Spieksma T~F~M, Cardoso V, Carullo G, Della~Rocca M and Duque F 2025
  \href{http://dx.doi.org/10.1103/PhysRevLett.134.081402}{ {\em Phys. Rev.
  Lett.\/} {\bf 134} 081402 } [\eprint{2409.05950}]

\bibitem{Posti:2019tbp}
Posti L and Helmi A 2019 \href{http://dx.doi.org/10.1051/0004-6361/201833355}{
  {\em Astron. Astrophys.\/} {\bf 621} A56 }

\bibitem{Jiao:2023aci}
Jiao Y, Hammer F, Wang H, Wang J, Amram P, Chemin L and Yang Y 2023
  \href{http://dx.doi.org/10.1051/0004-6361/202347513}{ {\em Astron.
  Astrophys.\/} {\bf 678} A208 } [\eprint{2309.00048}]

\bibitem{Speeney:2024mas}
Speeney N, Berti E, Cardoso V and Maselli A 2024
  \href{http://dx.doi.org/10.1103/PhysRevD.109.084068}{ {\em Phys. Rev. D\/}
  {\bf 109} 084068 } [\eprint{2401.00932}]

\bibitem{Jusufi:2022jxu}
Jusufi K 2023 \href{http://dx.doi.org/10.1140/epjc/s10052-023-11264-w}{ {\em
  Eur. Phys. J. C\/} {\bf 83} 103 } [\eprint{2202.00010}]

\bibitem{Feng:2022evy}
Feng J~C, Chakraborty S and Cardoso V 2023
  \href{http://dx.doi.org/10.1103/PhysRevD.107.044050}{ {\em Phys. Rev. D\/}
  {\bf 107} 044050 } [\eprint{2211.05261}]

\bibitem{Konoplya:2022hbl}
Konoplya R~A and Zhidenko A 2022
  \href{http://dx.doi.org/10.3847/1538-4357/ac76bc}{ {\em Astrophys. J.\/} {\bf
  933} 166 } [\eprint{2202.02205}]

\bibitem{Figueiredo:2023gas}
Figueiredo E, Maselli A and Cardoso V 2023
  \href{http://dx.doi.org/10.1103/PhysRevD.107.104033}{ {\em Phys. Rev. D\/}
  {\bf 107} 104033 } [\eprint{2303.08183}]

\bibitem{Shen:2023erj}
Shen Z, Wang A, Gong Y and Yin S 2024
  \href{http://dx.doi.org/10.1016/j.physletb.2024.138797}{ {\em Phys. Lett.
  B\/} {\bf 855} 138797 } [\eprint{2311.12259}]

\bibitem{Datta:2023zmd}
Datta S 2024 \href{http://dx.doi.org/10.1103/PhysRevD.109.104042}{ {\em Phys.
  Rev. D\/} {\bf 109} 104042 } [\eprint{2312.01277}]

\bibitem{Stelea:2023yqo}
Stelea C, Dariescu M~A and Dariescu C 2023
  \href{http://dx.doi.org/10.1016/j.physletb.2023.138275}{ {\em Phys. Lett.
  B\/} {\bf 847} 138275 } [\eprint{2309.13651}]

\bibitem{Heydari-Fard:2024wgu}
Heydari-Fard M, Heydari-Fard M and Riazi N 2025
  \href{http://dx.doi.org/10.1007/s10714-025-03382-5}{ {\em Gen. Rel. Grav.\/}
  {\bf 57} 49 } [\eprint{2408.16020}]

\bibitem{Mollicone:2024lxy}
Mollicone A and Destounis K 2025
  \href{http://dx.doi.org/10.1103/PhysRevD.111.024017}{ {\em Phys. Rev. D\/}
  {\bf 111} 024017 } [\eprint{2410.11952}]

\bibitem{Maeda:2024tsg}
Maeda K~i, Cardoso V and Wang A 2025
  \href{http://dx.doi.org/10.1103/PhysRevD.111.044060}{ {\em Phys. Rev. D\/}
  {\bf 111} 044060 } [\eprint{2410.04175}]

\bibitem{Destounis:2022obl}
Destounis K, Kulathingal A, Kokkotas K~D and Papadopoulos G~O 2023
  \href{http://dx.doi.org/10.1103/PhysRevD.107.084027}{ {\em Phys. Rev. D\/}
  {\bf 107} 084027 } [\eprint{2210.09357}]

\bibitem{Boyanov:2024jge}
Boyanov V, Cardoso V, Kokkotas K~D and Redondo-Yuste J 2024
  [\eprint{2411.16861}]

\bibitem{Pezzella:2024tkf}
Pezzella L, Destounis K, Maselli A and Cardoso V 2025
  \href{http://dx.doi.org/10.1103/PhysRevD.111.064026}{ {\em Phys. Rev. D\/}
  {\bf 111} 064026 } [\eprint{2412.18651}]

\bibitem{DeAmicis:2024not}
De~Amicis M, Albanesi S and Carullo G 2024
  \href{http://dx.doi.org/10.1103/PhysRevD.110.104005}{ {\em Phys. Rev. D\/}
  {\bf 110} 104005 } [\eprint{2406.17018}]

\bibitem{Price:1971fb}
Price R~H 1972 \href{http://dx.doi.org/10.1103/PhysRevD.5.2419}{ {\em Phys.
  Rev. D\/} {\bf 5} 2419--2438 }

\bibitem{Gundlach:1993tp}
Gundlach C, Price R~H and Pullin J 1994
  \href{http://dx.doi.org/10.1103/PhysRevD.49.883}{ {\em Phys. Rev. D\/} {\bf
  49} 883--889 } [\eprint{gr-qc/9307009}]

\bibitem{GRAVITY:2024tth}
Abd El~Dayem K {\em et~al.\/} (GRAVITY) 2024
  \href{http://dx.doi.org/10.1051/0004-6361/202452274}{ {\em Astron.
  Astrophys.\/} {\bf 692} A242 } [\eprint{2409.12261}]

\bibitem{Steppohn:2025kbh}
Steppohn O, V{\"o}lkel S~H and Dietrich T 2026
  \href{http://dx.doi.org/10.1103/hnhc-m4lw}{ {\em Phys. Rev. D\/} {\bf 113}
  024011 } [\eprint{2508.15534}]

\bibitem{May:2024rrg}
May T, Ma S, Ripley J~L and East W~E 2024
  \href{http://dx.doi.org/10.1103/PhysRevD.110.084034}{ {\em Phys. Rev. D\/}
  {\bf 110} 084034 } [\eprint{2405.18303}]

\bibitem{Zhu:2024dyl}
Zhu H {\em et~al.\/} 2024 \href{http://dx.doi.org/10.1103/PhysRevD.110.124028}{
  {\em Phys. Rev. D\/} {\bf 110} 124028 } [\eprint{2404.12424}]

\bibitem{Capuano:2024qhv}
Capuano L, Santoni L and Barausse E 2024
  \href{http://dx.doi.org/10.1103/PhysRevD.110.084081}{ {\em Phys. Rev. D\/}
  {\bf 110} 084081 } [\eprint{2407.06009}]

\bibitem{Bamber:2021knr}
Bamber J, Tattersall O~J, Clough K and Ferreira P~G 2021
  \href{http://dx.doi.org/10.1103/PhysRevD.103.124013}{ {\em Phys. Rev. D\/}
  {\bf 103} 124013 } [\eprint{2103.00026}]

\bibitem{Cardoso:2021qqu}
Cardoso V and Foschi A 2021
  \href{http://dx.doi.org/10.1103/PhysRevD.104.024004}{ {\em Phys. Rev. D\/}
  {\bf 104} 024004 } [\eprint{2106.06551}]

\bibitem{Rosato:2024arw}
Rosato R~F, Destounis K and Pani P 2024
  \href{http://dx.doi.org/10.1103/PhysRevD.110.L121501}{ {\em Phys. Rev. D\/}
  {\bf 110} L121501 } [\eprint{2406.01692}]

\bibitem{Oshita:2024fzf}
Oshita N, Takahashi K and Mukohyama S 2024
  \href{http://dx.doi.org/10.1103/PhysRevD.110.084070}{ {\em Phys. Rev. D\/}
  {\bf 110} 084070 } [\eprint{2406.04525}]

\bibitem{Will:2014kxa}
Will C~M 2014 \href{http://dx.doi.org/10.12942/lrr-2014-4}{ {\em Living Rev.
  Rel.\/} {\bf 17} 4 } [\eprint{1403.7377}]

\bibitem{Clifton:2011jh}
Clifton T, Ferreira P~G, Padilla A and Skordis C 2012
  \href{http://dx.doi.org/10.1016/j.physrep.2012.01.001}{ {\em Phys. Rept.\/}
  {\bf 513} 1--189 } [\eprint{1106.2476}]

\bibitem{Berti:2015itd}
Berti E {\em et~al.\/} 2015
  \href{http://dx.doi.org/10.1088/0264-9381/32/24/243001}{ {\em Class. Quant.
  Grav.\/} {\bf 32} 243001 } [\eprint{1501.07274}]

\bibitem{Endlich:2017tqa}
Endlich S, Gorbenko V, Huang J and Senatore L 2017
  \href{http://dx.doi.org/10.1007/JHEP09(2017)122}{ {\em JHEP\/} {\bf 09} 122 }
  [\eprint{1704.01590}]

\bibitem{Cano:2019ore}
Cano P~A and Ruip\'erez A 2019
  \href{http://dx.doi.org/10.1007/JHEP05(2019)189}{ {\em JHEP\/} {\bf 05} 189 }
  [Erratum: JHEP 03, 187 (2020)] [\eprint{1901.01315}]

\bibitem{Torii:1996yi}
Torii T, Yajima H and Maeda K~i 1997
  \href{http://dx.doi.org/10.1103/PhysRevD.55.739}{ {\em Phys. Rev. D\/} {\bf
  55} 739--753 } [\eprint{gr-qc/9606034}]

\bibitem{Kanti:1995vq}
Kanti P, Mavromatos N~E, Rizos J, Tamvakis K and Winstanley E 1996
  \href{http://dx.doi.org/10.1103/PhysRevD.54.5049}{ {\em Phys. Rev. D\/} {\bf
  54} 5049--5058 } [\eprint{hep-th/9511071}]

\bibitem{Alexander:2009tp}
Alexander S and Yunes N 2009
  \href{http://dx.doi.org/10.1016/j.physrep.2009.07.002}{ {\em Phys. Rept.\/}
  {\bf 480} 1--55 } [\eprint{0907.2562}]

\bibitem{Wagle:2021tam}
Wagle P, Yunes N and Silva H~O 2022
  \href{http://dx.doi.org/10.1103/PhysRevD.105.124003}{ {\em Phys. Rev. D\/}
  {\bf 105} 124003 } [\eprint{2103.09913}]

\bibitem{Cano:2021rey}
Cano P~A and Ruip\'erez A 2022
  \href{http://dx.doi.org/10.1103/PhysRevD.105.044022}{ {\em Phys. Rev. D\/}
  {\bf 105} 044022 } [\eprint{2111.04750}]

\bibitem{Bergshoeff:1989de}
Bergshoeff E~A and de~Roo M 1989
  \href{http://dx.doi.org/10.1016/0550-3213(89)90336-2}{ {\em Nucl. Phys. B\/}
  {\bf 328} 439--468 }

\bibitem{Langlois:2021aji}
Langlois D, Noui K and Roussille H 2021
  \href{http://dx.doi.org/10.1103/PhysRevD.104.124044}{ {\em Phys. Rev. D\/}
  {\bf 104} 124044 } [\eprint{2103.14750}]

\bibitem{Langlois:2022ulw}
Langlois D, Noui K and Roussille H 2022
  \href{http://dx.doi.org/10.1088/1475-7516/2022/08/040}{ {\em JCAP\/} {\bf 08}
  040 } [\eprint{2205.07746}]

\bibitem{Tattersall:2017erk}
Tattersall O~J, Ferreira P~G and Lagos M 2018
  \href{http://dx.doi.org/10.1103/PhysRevD.97.044021}{ {\em Phys. Rev. D\/}
  {\bf 97} 044021 } [\eprint{1711.01992}]

\bibitem{Langlois:2021xzq}
Langlois D, Noui K and Roussille H 2021
  \href{http://dx.doi.org/10.1103/PhysRevD.104.124043}{ {\em Phys. Rev. D\/}
  {\bf 104} 124043 } [\eprint{2103.14744}]

\bibitem{Okounkova:2019zep}
Okounkova M 2019 \href{http://dx.doi.org/10.1103/PhysRevD.100.124054}{ {\em
  Phys. Rev. D\/} {\bf 100} 124054 } [\eprint{1909.12251}]

\bibitem{Okounkova:2018pql}
Okounkova M, Scheel M~A and Teukolsky S~A 2019
  \href{http://dx.doi.org/10.1103/PhysRevD.99.044019}{ {\em Phys. Rev. D\/}
  {\bf 99} 044019 } [\eprint{1811.10713}]

\bibitem{Evstafyeva:2022rve}
Evstafyeva T, Agathos M and Ripley J~L 2023
  \href{http://dx.doi.org/10.1103/PhysRevD.107.124010}{ {\em Phys. Rev. D\/}
  {\bf 107} 124010 } [\eprint{2212.11359}]

\bibitem{Cardoso:2024qie}
Cardoso V, Mukohyama S, Oshita N and Takahashi K 2024
  \href{http://dx.doi.org/10.1103/PhysRevD.109.124036}{ {\em Phys. Rev. D\/}
  {\bf 109} 124036 } [\eprint{2404.05790}]

\bibitem{Franchini:2021bpt}
Franchini N, Herrero-Valea M and Barausse E 2021
  \href{http://dx.doi.org/10.1103/PhysRevD.103.084012}{ {\em Phys. Rev. D\/}
  {\bf 103} 084012 } [\eprint{2103.00929}]

\bibitem{Yunes:2009hc}
Yunes N and Pretorius F 2009
  \href{http://dx.doi.org/10.1103/PhysRevD.79.084043}{ {\em Phys. Rev. D\/}
  {\bf 79} 084043 } [\eprint{0902.4669}]

\bibitem{Pani:2009wy}
Pani P and Cardoso V 2009 \href{http://dx.doi.org/10.1103/PhysRevD.79.084031}{
  {\em Phys. Rev. D\/} {\bf 79} 084031 } [\eprint{0902.1569}]

\bibitem{Pani:2011gy}
Pani P, Macedo C~F~B, Crispino L~C~B and Cardoso V 2011
  \href{http://dx.doi.org/10.1103/PhysRevD.84.087501}{ {\em Phys. Rev. D\/}
  {\bf 84} 087501 } [\eprint{1109.3996}]

\bibitem{Yagi:2012ya}
Yagi K, Yunes N and Tanaka T 2012
  \href{http://dx.doi.org/10.1103/PhysRevD.86.044037}{ {\em Phys. Rev. D\/}
  {\bf 86} 044037 } [Erratum: Phys.Rev.D 89, 049902 (2014)]
  [\eprint{1206.6130}]

\bibitem{Ayzenberg:2014aka}
Ayzenberg D and Yunes N 2014
  \href{http://dx.doi.org/10.1103/PhysRevD.90.044066}{ {\em Phys. Rev. D\/}
  {\bf 90} 044066 } [Erratum: Phys.Rev.D 91, 069905 (2015)]
  [\eprint{1405.2133}]

\bibitem{Maselli:2015tta}
Maselli A, Pani P, Gualtieri L and Ferrari V 2015
  \href{http://dx.doi.org/10.1103/PhysRevD.92.083014}{ {\em Phys. Rev. D\/}
  {\bf 92} 083014 } [\eprint{1507.00680}]

\bibitem{Cardoso:2018ptl}
Cardoso V, Kimura M, Maselli A and Senatore L 2018
  \href{http://dx.doi.org/10.1103/PhysRevLett.121.251105}{ {\em Phys. Rev.
  Lett.\/} {\bf 121} 251105 } [Erratum: Phys.Rev.Lett. 131, 109903 (2023)]
  [\eprint{1808.08962}]

\bibitem{Yunes:2013dva}
Yunes N and Siemens X 2013 \href{http://dx.doi.org/10.12942/lrr-2013-9}{ {\em
  Living Rev. Rel.\/} {\bf 16} 9 } [\eprint{1304.3473}]

\bibitem{LIGOScientific:2021sio}
Abbott R {\em et~al.\/} (LIGO Scientific, VIRGO, KAGRA) 2021
  [\eprint{2112.06861}]

\bibitem{Yunes:2024lzm}
Yunes N, Siemens X and Yagi K 2024  [\eprint{2408.05240}]

\bibitem{Cano:2023qqm}
Cano P~A, Deich A and Yunes N 2024
  \href{http://dx.doi.org/10.1103/PhysRevD.109.024048}{ {\em Phys. Rev. D\/}
  {\bf 109} 024048 } [\eprint{2305.15341}]

\bibitem{RotatingBHWolfram}
Cano P~A 2021 {Rotating black holes: corrections to the Kerr metric package}
  \urlprefix\url{https://community.wolfram.com/groups/-/m/t/2249688}

\bibitem{McNees:2015srl}
McNees R, Stein L~C and Yunes N 2016
  \href{http://dx.doi.org/10.1088/0264-9381/33/23/235013}{ {\em Class. Quant.
  Grav.\/} {\bf 33} 235013 } [\eprint{1512.05453}]

\bibitem{Sullivan:2019vyi}
Sullivan A, Yunes N and Sotiriou T~P 2020
  \href{http://dx.doi.org/10.1103/PhysRevD.101.044024}{ {\em Phys. Rev. D\/}
  {\bf 101} 044024 } [\eprint{1903.02624}]

\bibitem{Kleihaus:2011tg}
Kleihaus B, Kunz J and Radu E 2011
  \href{http://dx.doi.org/10.1103/PhysRevLett.106.151104}{ {\em Phys. Rev.
  Lett.\/} {\bf 106} 151104 } [\eprint{1101.2868}]

\bibitem{Kleihaus:2014lba}
Kleihaus B, Kunz J and Mojica S 2014
  \href{http://dx.doi.org/10.1103/PhysRevD.90.061501}{ {\em Phys. Rev. D\/}
  {\bf 90} 061501 } [\eprint{1407.6884}]

\bibitem{Kleihaus:2015aje}
Kleihaus B, Kunz J, Mojica S and Radu E 2016
  \href{http://dx.doi.org/10.1103/PhysRevD.93.044047}{ {\em Phys. Rev. D\/}
  {\bf 93} 044047 } [\eprint{1511.05513}]

\bibitem{Sullivan:2020zpf}
Sullivan A, Yunes N and Sotiriou T~P 2021
  \href{http://dx.doi.org/10.1103/PhysRevD.103.124058}{ {\em Phys. Rev. D\/}
  {\bf 103} 124058 } [\eprint{2009.10614}]

\bibitem{Sotiriou:2014pfa}
Sotiriou T~P and Zhou S~Y 2014
  \href{http://dx.doi.org/10.1103/PhysRevD.90.124063}{ {\em Phys. Rev. D\/}
  {\bf 90} 124063 } [\eprint{1408.1698}]

\bibitem{Delgado:2020rev}
Delgado J~F~M, Herdeiro C~A~R and Radu E 2020
  \href{http://dx.doi.org/10.1007/JHEP04(2020)180}{ {\em JHEP\/} {\bf 04} 180 }
  [\eprint{2002.05012}]

\bibitem{Silva:2017uqg}
Silva H~O, Sakstein J, Gualtieri L, Sotiriou T~P and Berti E 2018
  \href{http://dx.doi.org/10.1103/PhysRevLett.120.131104}{ {\em Phys. Rev.
  Lett.\/} {\bf 120} 131104 } [\eprint{1711.02080}]

\bibitem{Silva:2018qhn}
Silva H~O, Macedo C~F~B, Sotiriou T~P, Gualtieri L, Sakstein J and Berti E 2019
  \href{http://dx.doi.org/10.1103/PhysRevD.99.064011}{ {\em Phys. Rev. D\/}
  {\bf 99} 064011 } [\eprint{1812.05590}]

\bibitem{Macedo:2019sem}
Macedo C~F~B, Sakstein J, Berti E, Gualtieri L, Silva H~O and Sotiriou T~P 2019
  \href{http://dx.doi.org/10.1103/PhysRevD.99.104041}{ {\em Phys. Rev. D\/}
  {\bf 99} 104041 } [\eprint{1903.06784}]

\bibitem{Herdeiro:2020wei}
Herdeiro C~A~R, Radu E, Silva H~O, Sotiriou T~P and Yunes N 2021
  \href{http://dx.doi.org/10.1103/PhysRevLett.126.011103}{ {\em Phys. Rev.
  Lett.\/} {\bf 126} 011103 } [\eprint{2009.03904}]

\bibitem{Doneva:2017bvd}
Doneva D~D and Yazadjiev S~S 2018
  \href{http://dx.doi.org/10.1103/PhysRevLett.120.131103}{ {\em Phys. Rev.
  Lett.\/} {\bf 120} 131103 } [\eprint{1711.01187}]

\bibitem{Doneva:2022ewd}
Doneva D~D, Ramazano\u{g}lu F~M, Silva H~O, Sotiriou T~P and Yazadjiev S~S 2024
  \href{http://dx.doi.org/10.1103/RevModPhys.96.015004}{ {\em Rev. Mod.
  Phys.\/} {\bf 96} 015004 } [\eprint{2211.01766}]

\bibitem{Stein:2014xba}
Stein L~C 2014 \href{http://dx.doi.org/10.1103/PhysRevD.90.044061}{ {\em Phys.
  Rev. D\/} {\bf 90} 044061 } [\eprint{1407.2350}]

\bibitem{Okounkova:2018abo}
Okounkova M, Scheel M~A and Teukolsky S~A 2019
  \href{http://dx.doi.org/10.1088/1361-6382/aafcdf}{ {\em Class. Quant.
  Grav.\/} {\bf 36} 054001 } [\eprint{1810.05306}]

\bibitem{Delsate:2018ome}
Delsate T, Herdeiro C and Radu E 2018
  \href{http://dx.doi.org/10.1016/j.physletb.2018.09.060}{ {\em Phys. Lett.
  B\/} {\bf 787} 8--15 } [\eprint{1806.06700}]

\bibitem{Richards:2023xsr}
Richards C, Dima A and Witek H 2023
  \href{http://dx.doi.org/10.1103/PhysRevD.108.044078}{ {\em Phys. Rev. D\/}
  {\bf 108} 044078 } [\eprint{2305.07704}]

\bibitem{Guo:2008hf}
Guo Z~K, Ohta N and Torii T 2008 \href{http://dx.doi.org/10.1143/PTP.120.581}{
  {\em Prog. Theor. Phys.\/} {\bf 120} 581--607 } [\eprint{0806.2481}]

\bibitem{Blazquez-Salcedo:2017txk}
Bl\'azquez-Salcedo J~L, Khoo F~S and Kunz J 2017
  \href{http://dx.doi.org/10.1103/PhysRevD.96.064008}{ {\em Phys. Rev. D\/}
  {\bf 96} 064008 } [\eprint{1706.03262}]

\bibitem{Owen:2021eez}
Owen C~B, Yunes N and Witek H 2021
  \href{http://dx.doi.org/10.1103/PhysRevD.103.124057}{ {\em Phys. Rev. D\/}
  {\bf 103} 124057 } [\eprint{2103.15891}]

\bibitem{Cardoso:2009pk}
Cardoso V and Gualtieri L 2009
  \href{http://dx.doi.org/10.1103/PhysRevD.81.089903}{ {\em Phys. Rev. D\/}
  {\bf 80} 064008 } [Erratum: Phys.Rev.D 81, 089903 (2010)]
  [\eprint{0907.5008}]

\bibitem{Blazquez-Salcedo:2016enn}
Bl\'azquez-Salcedo J~L, Macedo C~F~B, Cardoso V, Ferrari V, Gualtieri L, Khoo
  F~S, Kunz J and Pani P 2016
  \href{http://dx.doi.org/10.1103/PhysRevD.94.104024}{ {\em Phys. Rev. D\/}
  {\bf 94} 104024 } [\eprint{1609.01286}]

\bibitem{deRham:2020ejn}
de~Rham C, Francfort J and Zhang J 2020
  \href{http://dx.doi.org/10.1103/PhysRevD.102.024079}{ {\em Phys. Rev. D\/}
  {\bf 102} 024079 } [\eprint{2005.13923}]

\bibitem{Moura:2021eln}
Moura F and Rodrigues J~a 2021
  \href{http://dx.doi.org/10.1016/j.physletb.2021.136407}{ {\em Phys. Lett.
  B\/} {\bf 819} 136407 } [\eprint{2103.09302}]

\bibitem{Pierini:2021jxd}
Pierini L and Gualtieri L 2021
  \href{http://dx.doi.org/10.1103/PhysRevD.103.124017}{ {\em Phys. Rev. D\/}
  {\bf 103} 124017 } [\eprint{2103.09870}]

\bibitem{Srivastava:2021imr}
Srivastava M, Chen Y and Shankaranarayanan S 2021
  \href{http://dx.doi.org/10.1103/PhysRevD.104.064034}{ {\em Phys. Rev. D\/}
  {\bf 104} 064034 } [\eprint{2106.06209}]

\bibitem{Bryant:2021xdh}
Bryant A, Silva H~O, Yagi K and Glampedakis K 2021
  \href{http://dx.doi.org/10.1103/PhysRevD.104.044051}{ {\em Phys. Rev. D\/}
  {\bf 104} 044051 } [\eprint{2106.09657}]

\bibitem{Cano:2021myl}
Cano P~A, Fransen K, Hertog T and Maenaut S 2022
  \href{http://dx.doi.org/10.1103/PhysRevD.105.024064}{ {\em Phys. Rev. D\/}
  {\bf 105} 024064 } [\eprint{2110.11378}]

\bibitem{Pierini:2022eim}
Pierini L and Gualtieri L 2022
  \href{http://dx.doi.org/10.1103/PhysRevD.106.104009}{ {\em Phys. Rev. D\/}
  {\bf 106} 104009 } [\eprint{2207.11267}]

\bibitem{Silva:2024ffz}
Silva H~O, Tambalo G, Glampedakis K, Yagi K and Steinhoff J 2024
  \href{http://dx.doi.org/10.1103/PhysRevD.110.024042}{ {\em Phys. Rev. D\/}
  {\bf 110} 024042 } [\eprint{2404.11110}]

\bibitem{Li:2022pcy}
Li D, Wagle P, Chen Y and Yunes N 2023
  \href{http://dx.doi.org/10.1103/PhysRevX.13.021029}{ {\em Phys. Rev. X\/}
  {\bf 13} 021029 } [\eprint{2206.10652}]

\bibitem{Cano:2023tmv}
Cano P~A, Fransen K, Hertog T and Maenaut S 2023
  \href{http://dx.doi.org/10.1103/PhysRevD.108.024040}{ {\em Phys. Rev. D\/}
  {\bf 108} 024040 } [\eprint{2304.02663}]

\bibitem{Cano:2023jbk}
Cano P~A, Fransen K, Hertog T and Maenaut S 2023
  \href{http://dx.doi.org/10.1103/PhysRevD.108.124032}{ {\em Phys. Rev. D\/}
  {\bf 108} 124032 } [\eprint{2307.07431}]

\bibitem{Cano:2024ezp}
Cano P~A, Capuano L, Franchini N, Maenaut S and V\"olkel S~H 2024
  \href{http://dx.doi.org/10.1103/PhysRevD.110.124057}{ {\em Phys. Rev. D\/}
  {\bf 110} 124057 } [\eprint{2409.04517}]

\bibitem{Wagle:2023fwl}
Wagle P, Li D, Chen Y and Yunes N 2024
  \href{http://dx.doi.org/10.1103/PhysRevD.109.104029}{ {\em Phys. Rev. D\/}
  {\bf 109} 104029 } [\eprint{2311.07706}]

\bibitem{Li:2025fci}
Li D, Wagle P, Chen Y and Yunes N 2025
  \href{http://dx.doi.org/10.1103/yfw6-32yl}{ {\em Phys. Rev. D\/} {\bf 112}
  044005 } [\eprint{2503.15606}]

\bibitem{Chrzanowski:1975wv}
Chrzanowski P~L 1975 \href{http://dx.doi.org/10.1103/PhysRevD.11.2042}{ {\em
  Phys. Rev. D\/} {\bf 11} 2042--2062 }

\bibitem{Cohen_Kegeles_1975}
Cohen J~M and Kegeles L~S 1975
  \href{http://dx.doi.org/10.1016/0375-9601(75)90583-6}{ {\em Physics Letters
  A\/} {\bf 54} 5--7 } ISSN 0375-9601

\bibitem{Keidl:2006wk}
Keidl T~S, Friedman J~L and Wiseman A~G 2007
  \href{http://dx.doi.org/10.1103/PhysRevD.75.124009}{ {\em Phys. Rev. D\/}
  {\bf 75} 124009 } [\eprint{gr-qc/0611072}]

\bibitem{Kegeles:1979an}
Kegeles L~S and Cohen J~M 1979
  \href{http://dx.doi.org/10.1103/PhysRevD.19.1641}{ {\em Phys. Rev. D\/} {\bf
  19} 1641--1664 }

\bibitem{Yunes:2005ve}
Yunes N and Gonzalez J 2006
  \href{http://dx.doi.org/10.1103/PhysRevD.89.089902}{ {\em Phys. Rev. D\/}
  {\bf 73} 024010 } [Erratum: Phys.Rev.D 89, 089902 (2014)]
  [\eprint{gr-qc/0510076}]

\bibitem{Whiting:2005hr}
Whiting B~F and Price L~R 2005
  \href{http://dx.doi.org/10.1088/0264-9381/22/15/003}{ {\em Class. Quant.
  Grav.\/} {\bf 22} S589--S604 }

\bibitem{Cano:2020cao}
Cano P~A, Fransen K and Hertog T 2020
  \href{http://dx.doi.org/10.1103/PhysRevD.102.044047}{ {\em Phys. Rev. D\/}
  {\bf 102} 044047 } [\eprint{2005.03671}]

\bibitem{Ghosh:2023etd}
Ghosh R, Franchini N, V\"olkel S~H and Barausse E 2023
  \href{http://dx.doi.org/10.1103/PhysRevD.108.024038}{ {\em Phys. Rev. D\/}
  {\bf 108} 024038 } [\eprint{2303.00088}]

\bibitem{Li:2023ulk}
Li D, Hussain A, Wagle P, Chen Y, Yunes N and Zimmerman A 2024
  \href{http://dx.doi.org/10.1103/PhysRevD.109.104026}{ {\em Phys. Rev. D\/}
  {\bf 109} 104026 } [\eprint{2310.06033}]

\bibitem{Molina:2010fb}
Molina C, Pani P, Cardoso V and Gualtieri L 2010
  \href{http://dx.doi.org/10.1103/PhysRevD.81.124021}{ {\em Phys. Rev. D\/}
  {\bf 81} 124021 } [\eprint{1004.4007}]

\bibitem{Chung:2024ira}
Chung A~K~W and Yunes N 2024
  \href{http://dx.doi.org/10.1103/PhysRevLett.133.181401}{ {\em Phys. Rev.
  Lett.\/} {\bf 133} 181401 } [\eprint{2405.12280}]

\bibitem{Chung:2024vaf}
Chung A~K~W and Yunes N 2024
  \href{http://dx.doi.org/10.1103/PhysRevD.110.064019}{ {\em Phys. Rev. D\/}
  {\bf 110} 064019 } [\eprint{2406.11986}]

\bibitem{Blazquez-Salcedo:2024oek}
Bl\'azquez-Salcedo J~L, Khoo F~S, Kleihaus B and Kunz J 2025
  \href{http://dx.doi.org/10.1103/PhysRevD.111.L021505}{ {\em Phys. Rev. D\/}
  {\bf 111} L021505 } [\eprint{2407.20760}]

\bibitem{Khoo:2024agm}
Khoo F~S, Bl\'azquez-Salcedo J~L, Kleihaus B and Kunz J 2024
  [\eprint{2412.09377}]

\bibitem{Blazquez-Salcedo:2024dur}
Blazquez-Salcedo J~L, Khoo F~S, Kleihaus B and Kunz J 2025
  \href{http://dx.doi.org/10.1103/PhysRevD.111.064015}{ {\em Phys. Rev. D\/}
  {\bf 111} 064015 } [\eprint{2412.17073}]

\bibitem{Chung:2023zdq}
Chung A~K~W, Wagle P and Yunes N 2023
  \href{http://dx.doi.org/10.1103/PhysRevD.107.124032}{ {\em Phys. Rev. D\/}
  {\bf 107} 124032 } [\eprint{2302.11624}]

\bibitem{Chung:2023wkd}
Chung A~K~W, Wagle P and Yunes N 2024
  \href{http://dx.doi.org/10.1103/PhysRevD.109.044072}{ {\em Phys. Rev. D\/}
  {\bf 109} 044072 } [\eprint{2312.08435}]

\bibitem{mct2015}
LLC A Multiprecision computing toolbox for matlab
  \urlprefix\url{http://www.advanpix.com/}

\bibitem{Blazquez-Salcedo:2023hwg}
Bl\'azquez-Salcedo J~L, Khoo F~S, Kunz J and Gonz\'alez-Romero L~M 2024
  \href{http://dx.doi.org/10.1103/PhysRevD.109.064028}{ {\em Phys. Rev. D\/}
  {\bf 109} 064028 } [\eprint{2312.10754}]

\bibitem{Khoo:2024yeh}
Khoo F~S, Azad B, Bl\'azquez-Salcedo J~L, Gonz\'alez-Romero L~M, Kleihaus B,
  Kunz J and Navarro-L\'erida F 2024
  \href{http://dx.doi.org/10.1103/PhysRevD.109.084013}{ {\em Phys. Rev. D\/}
  {\bf 109} 084013 } [\eprint{2401.02898}]

\bibitem{Cano:2024jkd}
Cano P~A, Capuano L, Franchini N, Maenaut S and V\"olkel S~H 2024
  \href{http://dx.doi.org/10.1103/PhysRevD.110.104007}{ {\em Phys. Rev. D\/}
  {\bf 110} 104007 } [\eprint{2407.15947}]

\bibitem{gitbeyondkerr}
Cano P~A 2024 {BeyondKerrQNM}
  \urlprefix\url{https://github.com/pacmn91/BeyondKerrQNM}

\bibitem{Cardoso:2019mqo}
Cardoso V, Kimura M, Maselli A, Berti E, Macedo C~F~B and McManus R 2019
  \href{http://dx.doi.org/10.1103/PhysRevD.99.104077}{ {\em Phys. Rev. D\/}
  {\bf 99} 104077 } [\eprint{1901.01265}]

\bibitem{McManus:2019ulj}
McManus R, Berti E, Macedo C~F~B, Kimura M, Maselli A and Cardoso V 2019
  \href{http://dx.doi.org/10.1103/PhysRevD.100.044061}{ {\em Phys. Rev. D\/}
  {\bf 100} 044061 } [\eprint{1906.05155}]

\bibitem{Maenaut:2024oci}
Maenaut S, Carullo G, Cano P~A, Liu A, Cardoso V, Hertog T and Li T~G~F 2026
  \href{http://dx.doi.org/10.1103/k3hv-cqfn}{ {\em Phys. Rev. D\/} {\bf 113}
  044039 } [\eprint{2411.17893}]

\bibitem{Kimura:2018nxk}
Kimura M 2018 \href{http://dx.doi.org/10.1103/PhysRevD.98.024048}{ {\em Phys.
  Rev. D\/} {\bf 98} 024048 } [\eprint{1807.05029}]

\bibitem{Chung:2025gyg}
Chung A~K~W, Lam K~K~H and Yunes N 2025
  \href{http://dx.doi.org/10.1103/g83n-rrlj}{ {\em Phys. Rev. D\/} {\bf 111}
  124052 } [\eprint{2503.11759}]

\bibitem{Langlois:2022eta}
Langlois D, Noui K and Roussille H 2022
  \href{http://dx.doi.org/10.1088/1475-7516/2022/09/019}{ {\em JCAP\/} {\bf 09}
  019 } [\eprint{2204.04107}]

\bibitem{LIGOScientific:2020tif}
Abbott R {\em et~al.\/} (LIGO Scientific, Virgo) 2021
  \href{http://dx.doi.org/10.1103/PhysRevD.103.122002}{ {\em Phys. Rev. D\/}
  {\bf 103} 122002 } [\eprint{2010.14529}]

\bibitem{Julie:2024fwy}
Juli\'e F~L, Pompili L and Buonanno A 2025
  \href{http://dx.doi.org/10.1103/PhysRevD.111.024016}{ {\em Phys. Rev. D\/}
  {\bf 111} 024016 } [\eprint{2406.13654}]

\bibitem{Cano:2024bhh}
Cano P~A and David M 2024 \href{http://dx.doi.org/10.1103/PhysRevD.110.064067}{
  {\em Phys. Rev. D\/} {\bf 110} 064067 } [\eprint{2407.02017}]

\bibitem{Hirano:2024fgp}
Hirano S, Kimura M, Yamaguchi M and Zhang J 2024
  \href{http://dx.doi.org/10.1103/PhysRevD.110.024015}{ {\em Phys. Rev. D\/}
  {\bf 110} 024015 } [\eprint{2404.09672}]

\bibitem{Roussille:2022vfa}
Roussille H 2022 {\em {Black hole perturbations in modified gravity
  theories}\/} Ph.D. thesis Diderot U., Paris [\eprint{2211.01103}]

\bibitem{Gossan:2011ha}
Gossan S, Veitch J and Sathyaprakash B~S 2012
  \href{http://dx.doi.org/10.1103/PhysRevD.85.124056}{ {\em Phys. Rev. D\/}
  {\bf 85} 124056 } [\eprint{1111.5819}]

\bibitem{Meidam:2014jpa}
Meidam J, Agathos M, Van Den~Broeck C, Veitch J and Sathyaprakash B~S 2014
  \href{http://dx.doi.org/10.1103/PhysRevD.90.064009}{ {\em Phys. Rev. D\/}
  {\bf 90} 064009 } [\eprint{1406.3201}]

\bibitem{Carullo:2018sfu}
Carullo G {\em et~al.\/} 2018
  \href{http://dx.doi.org/10.1103/PhysRevD.98.104020}{ {\em Phys. Rev. D\/}
  {\bf 98} 104020 } [\eprint{1805.04760}]

\bibitem{Crescimbeni:2024sam}
Crescimbeni F, Forteza X~J, Bhagwat S, Westerweck J and Pani P 2024
  [\eprint{2408.08956}]

\bibitem{Lestingi:2025jyb}
Lestingi J, D'Addario G and Sotiriou T~P 2025  [\eprint{2505.18261}]

\bibitem{Nee:2023osy}
Nee P~J, V\"olkel S~H and Pfeiffer H~P 2023
  \href{http://dx.doi.org/10.1103/PhysRevD.108.044032}{ {\em Phys. Rev. D\/}
  {\bf 108} 044032 } [\eprint{2302.06634}]

\bibitem{Cheung:2023vki}
Cheung M~H~Y, Berti E, Baibhav V and Cotesta R 2024
  \href{http://dx.doi.org/10.1103/PhysRevD.109.044069}{ {\em Phys. Rev. D\/}
  {\bf 109} 044069 } [Erratum: Phys.Rev.D 110, 049902 (2024)]
  [\eprint{2310.04489}]

\bibitem{Takahashi:2023tkb}
Takahashi K and Motohashi H 2024
  \href{http://dx.doi.org/10.1088/1361-6382/ad72c9}{ {\em Class. Quant.
  Grav.\/} {\bf 41} 195023 } [\eprint{2311.12762}]

\bibitem{Will:2018bme}
Will C~M 2018 {\em {Theory and Experiment in Gravitational Physics}\/}
  (Cambridge University Press) ISBN 978-1-108-67982-4, 978-1-107-11744-0

\bibitem{LIGOScientific:2016lio}
Abbott B~P {\em et~al.\/} (LIGO Scientific, Virgo) 2016
  \href{http://dx.doi.org/10.1103/PhysRevLett.116.221101}{ {\em Phys. Rev.
  Lett.\/} {\bf 116} 221101 } [Erratum: Phys.Rev.Lett. 121, 129902 (2018)]
  [\eprint{1602.03841}]

\bibitem{Yunes:2016jcc}
Yunes N, Yagi K and Pretorius F 2016
  \href{http://dx.doi.org/10.1103/PhysRevD.94.084002}{ {\em Phys. Rev. D\/}
  {\bf 94} 084002 } [\eprint{1603.08955}]

\bibitem{LIGOScientific:2018dkp}
Abbott B~P {\em et~al.\/} (LIGO Scientific, Virgo) 2019
  \href{http://dx.doi.org/10.1103/PhysRevLett.123.011102}{ {\em Phys. Rev.
  Lett.\/} {\bf 123} 011102 } [\eprint{1811.00364}]

\bibitem{LIGOScientific:2019fpa}
Abbott B~P {\em et~al.\/} (LIGO Scientific, Virgo) 2019
  \href{http://dx.doi.org/10.1103/PhysRevD.100.104036}{ {\em Phys. Rev. D\/}
  {\bf 100} 104036 } [\eprint{1903.04467}]

\bibitem{Maselli:2019mjd}
Maselli A, Pani P, Gualtieri L and Berti E 2020
  \href{http://dx.doi.org/10.1103/PhysRevD.101.024043}{ {\em Phys. Rev. D\/}
  {\bf 101} 024043 } [\eprint{1910.12893}]

\bibitem{Barausse:2013nwa}
Barausse E and Sotiriou T~P 2013
  \href{http://dx.doi.org/10.1088/0264-9381/30/24/244010}{ {\em Class. Quant.
  Grav.\/} {\bf 30} 244010 } [\eprint{1307.3359}]

\bibitem{Herdeiro:2015waa}
Herdeiro C~A~R and Radu E 2015
  \href{http://dx.doi.org/10.1142/S0218271815420146}{ {\em Int. J. Mod. Phys.
  D\/} {\bf 24} 1542014 } [\eprint{1504.08209}]

\bibitem{Carullo:2021dui}
Carullo G 2021 \href{http://dx.doi.org/10.1103/PhysRevD.103.124043}{ {\em Phys.
  Rev. D\/} {\bf 103} 124043 } [\eprint{2102.05939}]

\bibitem{Carullo:2021oxn}
Carullo G, Laghi D, Johnson-McDaniel N~K, Del~Pozzo W, Dias O~J~C, Godazgar M
  and Santos J~E 2022 \href{http://dx.doi.org/10.1103/PhysRevD.105.062009}{
  {\em Phys. Rev. D\/} {\bf 105} 062009 } [\eprint{2109.13961}]

\bibitem{Gu:2023eaa}
Gu H~P, Wang H~T and Shao L 2024
  \href{http://dx.doi.org/10.1103/PhysRevD.109.024058}{ {\em Phys. Rev. D\/}
  {\bf 109} 024058 } [\eprint{2310.10447}]

\bibitem{Silva:2022srr}
Silva H~O, Ghosh A and Buonanno A 2023
  \href{http://dx.doi.org/10.1103/PhysRevD.107.044030}{ {\em Phys. Rev. D\/}
  {\bf 107} 044030 } [\eprint{2205.05132}]

\bibitem{Maselli:2023khq}
Maselli A, Yi S, Pierini L, Vellucci V, Reali L, Gualtieri L and Berti E 2024
  \href{http://dx.doi.org/10.1103/PhysRevD.109.064060}{ {\em Phys. Rev. D\/}
  {\bf 109} 064060 } [\eprint{2311.14803}]

\bibitem{Volkel:2022aca}
V\"olkel S~H, Franchini N and Barausse E 2022
  \href{http://dx.doi.org/10.1103/PhysRevD.105.084046}{ {\em Phys. Rev. D\/}
  {\bf 105} 084046 } [\eprint{2202.08655}]

\bibitem{Thomopoulos:2025nuf}
Thomopoulos S, V{\"o}lkel S~H and Pfeiffer H~P 2025
  \href{http://dx.doi.org/10.1103/xtzl-lyn6}{ {\em Phys. Rev. D\/} {\bf 112}
  064054 } [\eprint{2504.17848}]

\bibitem{Hatsuda:2020egs}
Hatsuda Y and Kimura M 2020
  \href{http://dx.doi.org/10.1103/PhysRevD.102.044032}{ {\em Phys. Rev. D\/}
  {\bf 102} 044032 } [\eprint{2006.15496}]

\bibitem{Volkel:2022khh}
V\"olkel S~H, Franchini N, Barausse E and Berti E 2022
  \href{http://dx.doi.org/10.1103/PhysRevD.106.124036}{ {\em Phys. Rev. D\/}
  {\bf 106} 124036 } [\eprint{2209.10564}]

\bibitem{Tattersall:2019nmh}
Tattersall O~J 2020 \href{http://dx.doi.org/10.1088/1361-6382/ab839b}{ {\em
  Class. Quant. Grav.\/} {\bf 37} 115007 } [\eprint{1911.07593}]

\bibitem{Sirera:2023pbs}
Sirera S and Noller J 2023
  \href{http://dx.doi.org/10.1103/PhysRevD.107.124054}{ {\em Phys. Rev. D\/}
  {\bf 107} 124054 } [\eprint{2301.10272}]

\bibitem{Mukohyama:2023xyf}
Mukohyama S, Takahashi K, Tomikawa K and Yingcharoenrat V 2023
  \href{http://dx.doi.org/10.1088/1475-7516/2023/07/050}{ {\em JCAP\/} {\bf 07}
  050 } [\eprint{2304.14304}]

\bibitem{Kimura:2020mrh}
Kimura M 2020 \href{http://dx.doi.org/10.1103/PhysRevD.101.064031}{ {\em Phys.
  Rev. D\/} {\bf 101} 064031 } [\eprint{2001.09613}]

\bibitem{Hatsuda:2023geo}
Hatsuda Y and Kimura M 2024
  \href{http://dx.doi.org/10.1103/PhysRevD.109.044026}{ {\em Phys. Rev. D\/}
  {\bf 109} 044026 } [\eprint{2307.16626}]

\bibitem{Franchini:2022axs}
Franchini N and V\"olkel S~H 2023
  \href{http://dx.doi.org/10.1103/PhysRevD.107.124063}{ {\em Phys. Rev. D\/}
  {\bf 107} 124063 } [\eprint{2210.14020}]

\bibitem{Konoplya:2022pbc}
Konoplya R~A and Zhidenko A 2024
  \href{http://dx.doi.org/10.1016/j.jheap.2024.10.015}{ {\em JHEAp\/} {\bf 44}
  419--426 } [\eprint{2209.00679}]

\bibitem{sebastian_volkel_2024_14001739}
Völkel S and Franchini N 2024 {sebastianvoelkel/parametrized\_qnm\_framework:
  Parametrized QNM Framework}
  \urlprefix\url{https://doi.org/10.5281/zenodo.14001739}

\bibitem{Vigeland:2011ji}
Vigeland S, Yunes N and Stein L 2011
  \href{http://dx.doi.org/10.1103/PhysRevD.83.104027}{ {\em Phys. Rev. D\/}
  {\bf 83} 104027 } [\eprint{1102.3706}]

\bibitem{Johannsen:2011dh}
Johannsen T and Psaltis D 2011
  \href{http://dx.doi.org/10.1103/PhysRevD.83.124015}{ {\em Phys. Rev. D\/}
  {\bf 83} 124015 } [\eprint{1105.3191}]

\bibitem{Johannsen:2013szh}
Johannsen T 2013 \href{http://dx.doi.org/10.1103/PhysRevD.88.044002}{ {\em
  Phys. Rev. D\/} {\bf 88} 044002 } [\eprint{1501.02809}]

\bibitem{Rezzolla:2014mua}
Rezzolla L and Zhidenko A 2014
  \href{http://dx.doi.org/10.1103/PhysRevD.90.084009}{ {\em Phys. Rev. D\/}
  {\bf 90} 084009 } [\eprint{1407.3086}]

\bibitem{Konoplya:2016jvv}
Konoplya R, Rezzolla L and Zhidenko A 2016
  \href{http://dx.doi.org/10.1103/PhysRevD.93.064015}{ {\em Phys. Rev. D\/}
  {\bf 93} 064015 } [\eprint{1602.02378}]

\bibitem{Papadopoulos:2018nvd}
Papadopoulos G~O and Kokkotas K~D 2018
  \href{http://dx.doi.org/10.1088/1361-6382/aad7f4}{ {\em Class. Quant.
  Grav.\/} {\bf 35} 185014 } [\eprint{1807.08594}]

\bibitem{Cardoso:2014rha}
Cardoso V, Pani P and Rico J 2014
  \href{http://dx.doi.org/10.1103/PhysRevD.89.064007}{ {\em Phys. Rev. D\/}
  {\bf 89} 064007 } [\eprint{1401.0528}]

\bibitem{Yagi:2023eap}
Yagi K, Lomuscio S, Lowrey T and Carson Z 2024
  \href{http://dx.doi.org/10.1103/PhysRevD.109.044017}{ {\em Phys. Rev. D\/}
  {\bf 109} 044017 } [\eprint{2311.08659}]

\bibitem{Volkel:2019muj}
V\"olkel S~H and Kokkotas K~D 2019
  \href{http://dx.doi.org/10.1103/PhysRevD.100.044026}{ {\em Phys. Rev. D\/}
  {\bf 100} 044026 } [\eprint{1908.00252}]

\bibitem{Volkel:2020daa}
V\"olkel S~H and Barausse E 2020
  \href{http://dx.doi.org/10.1103/PhysRevD.102.084025}{ {\em Phys. Rev. D\/}
  {\bf 102} 084025 } [\eprint{2007.02986}]

\bibitem{Suvorov:2021amy}
Suvorov A~G and V\"olkel S~H 2021
  \href{http://dx.doi.org/10.1103/PhysRevD.103.044027}{ {\em Phys. Rev. D\/}
  {\bf 103} 044027 } [\eprint{2101.09697}]

\bibitem{Konoplya:2018arm}
Konoplya R~A, Stuchl\'\i{}k Z and Zhidenko A 2018
  \href{http://dx.doi.org/10.1103/PhysRevD.97.084044}{ {\em Phys. Rev. D\/}
  {\bf 97} 084044 } [\eprint{1801.07195}]

\bibitem{Siqueira:2022tbc}
Siqueira P~H~C and Richartz M 2022
  \href{http://dx.doi.org/10.1103/PhysRevD.106.024046}{ {\em Phys. Rev. D\/}
  {\bf 106} 024046 } [\eprint{2205.00556}]

\bibitem{Khoo:2025qjc}
Khoo F~S 2025 \href{http://dx.doi.org/10.1103/f35l-m8n5}{ {\em Phys. Rev. D\/}
  {\bf 111}(12) 124025 }
  \urlprefix\url{https://link.aps.org/doi/10.1103/f35l-m8n5}

\bibitem{1972ApJ...172L..95G}
{Goebel} C~J 1972 \href{http://dx.doi.org/10.1086/180898}{ {\em Astrophys. J.
  Lett.\/} {\bf 172} L95 }

\bibitem{Ferrari:1984ozr}
Ferrari V and Mashhoon B 1984
  \href{http://dx.doi.org/10.1103/PhysRevLett.52.1361}{ {\em Phys. Rev.
  Lett.\/} {\bf 52} 1361 }

\bibitem{Li:2021zct}
Li P~C, Lee T~C, Guo M and Chen B 2021
  \href{http://dx.doi.org/10.1103/PhysRevD.104.084044}{ {\em Phys. Rev. D\/}
  {\bf 104} 084044 } [\eprint{2105.14268}]

\bibitem{Glampedakis:2017dvb}
Glampedakis K, Pappas G, Silva H~O and Berti E 2017
  \href{http://dx.doi.org/10.1103/PhysRevD.96.064054}{ {\em Phys. Rev. D\/}
  {\bf 96} 064054 } [\eprint{1706.07658}]

\bibitem{Silva:2019scu}
Silva H~O and Glampedakis K 2020
  \href{http://dx.doi.org/10.1103/PhysRevD.101.044051}{ {\em Phys. Rev. D\/}
  {\bf 101} 044051 } [\eprint{1912.09286}]

\bibitem{Chen:2019dip}
Chen C~Y and Chen P 2020 \href{http://dx.doi.org/10.1103/PhysRevD.101.064021}{
  {\em Phys. Rev. D\/} {\bf 101} 064021 } [\eprint{1910.12262}]

\bibitem{Khanna:2016yow}
Khanna G and Price R~H 2017
  \href{http://dx.doi.org/10.1103/PhysRevD.95.081501}{ {\em Phys. Rev. D\/}
  {\bf 95} 081501 } [\eprint{1609.00083}]

\bibitem{Konoplya:2017wot}
Konoplya R~A and Stuchl\'\i{}k Z 2017
  \href{http://dx.doi.org/10.1016/j.physletb.2017.06.015}{ {\em Phys. Lett.
  B\/} {\bf 771} 597--602 } [\eprint{1705.05928}]

\bibitem{Carson:2020iik}
Carson Z and Yagi K 2020 \href{http://dx.doi.org/10.1103/PhysRevD.101.084050}{
  {\em Phys. Rev. D\/} {\bf 101} 084050 } [\eprint{2003.02374}]

\bibitem{Uchikata:2020wsp}
Uchikata N, Narikawa T, Sakai K, Takahashi H and Nakano H 2020
  \href{http://dx.doi.org/10.1103/PhysRevD.102.024007}{ {\em Phys. Rev. D\/}
  {\bf 102} 024007 } [\eprint{2003.06791}]

\bibitem{Dey:2022pmv}
Dey K, Barausse E and Basak S 2023
  \href{http://dx.doi.org/10.1103/PhysRevD.108.024064}{ {\em Phys. Rev. D\/}
  {\bf 108} 024064 } [\eprint{2212.10725}]

\bibitem{Ahmed:2024ykc}
Ahmed Z, Kastha S and Nielsen A~B 2024  [\eprint{2401.06049}]

\bibitem{Lagos:2016wyv}
Lagos M, Baker T, Ferreira P~G and Noller J 2016
  \href{http://dx.doi.org/10.1088/1475-7516/2016/08/007}{ {\em JCAP\/} {\bf 08}
  007 } [\eprint{1604.01396}]

\bibitem{Lagos:2016gep}
Lagos M and Ferreira P~G 2017
  \href{http://dx.doi.org/10.1088/1475-7516/2017/01/047}{ {\em JCAP\/} {\bf 01}
  047 } [\eprint{1610.00553}]

\bibitem{Franciolini:2018uyq}
Franciolini G, Hui L, Penco R, Santoni L and Trincherini E 2019
  \href{http://dx.doi.org/10.1007/JHEP02(2019)127}{ {\em JHEP\/} {\bf 02} 127 }
  [\eprint{1810.07706}]

\bibitem{Hui:2021cpm}
Hui L, Podo A, Santoni L and Trincherini E 2021
  \href{http://dx.doi.org/10.1007/JHEP12(2021)183}{ {\em JHEP\/} {\bf 12} 183 }
  [\eprint{2111.02072}]

\bibitem{Penrose:1969pc}
Penrose R 1969 \href{http://dx.doi.org/10.1023/A:1016578408204}{ {\em Riv.
  Nuovo Cim.\/} {\bf 1} 252--276 }

\bibitem{Mathur:2009hf}
Mathur S~D 2009 \href{http://dx.doi.org/10.1088/0264-9381/26/22/224001}{ {\em
  Class. Quant. Grav.\/} {\bf 26} 224001 } [\eprint{0909.1038}]

\bibitem{Giddings:2017mym}
Giddings S~B 2017 \href{http://dx.doi.org/10.1007/JHEP12(2017)047}{ {\em
  JHEP\/} {\bf 12} 047 } [\eprint{1701.08765}]

\bibitem{Almheiri:2012rt}
Almheiri A, Marolf D, Polchinski J and Sully J 2013
  \href{http://dx.doi.org/10.1007/JHEP02(2013)062}{ {\em JHEP\/} {\bf 02} 062 }
  [\eprint{1207.3123}]

\bibitem{Sekino:2008he}
Sekino Y and Susskind L 2008
  \href{http://dx.doi.org/10.1088/1126-6708/2008/10/065}{ {\em JHEP\/} {\bf 10}
  065 } [\eprint{0808.2096}]

\bibitem{Hayden:2007cs}
Hayden P and Preskill J 2007
  \href{http://dx.doi.org/10.1088/1126-6708/2007/09/120}{ {\em JHEP\/} {\bf 09}
  120 } [\eprint{0708.4025}]

\bibitem{Oshita:2023tlm}
Oshita N and Afshordi N 2023
  \href{http://dx.doi.org/10.1016/j.physletb.2023.137901}{ {\em Phys. Lett.
  B\/} {\bf 841} 137901 } [\eprint{2302.08964}]

\bibitem{Mathur:2024mvo}
Mathur S~D and Mehta M 2024 \href{http://dx.doi.org/10.1088/1361-6382/ad869e}{
  {\em Class. Quant. Grav.\/} {\bf 41} 235011 } [\eprint{2402.13166}]

\bibitem{Hayden:2020vyo}
Hayden P and Penington G 2020  [\eprint{2012.07861}]

\bibitem{Mazur:2004fk}
Mazur P~O and Mottola E 2004 \href{http://dx.doi.org/10.1073/pnas.0402717101}{
  {\em Proc. Nat. Acad. Sci.\/} {\bf 101} 9545--9550 } [\eprint{gr-qc/0407075}]

\bibitem{Morris:1988cz}
Morris M~S and Thorne K~S 1988 \href{http://dx.doi.org/10.1119/1.15620}{ {\em
  Am. J. Phys.\/} {\bf 56} 395--412 }

\bibitem{Visser:1995cc}
Visser M 1995 {\em {Lorentzian wormholes: From Einstein to Hawking}\/} ISBN
  978-1-56396-653-8

\bibitem{Damour:2007ap}
Damour T and Solodukhin S~N 2007
  \href{http://dx.doi.org/10.1103/PhysRevD.76.024016}{ {\em Phys. Rev. D\/}
  {\bf 76} 024016 } [\eprint{0704.2667}]

\bibitem{Mathur:2005zp}
Mathur S~D 2005 \href{http://dx.doi.org/10.1002/prop.200410203}{ {\em Fortsch.
  Phys.\/} {\bf 53} 793--827 } [\eprint{hep-th/0502050}]

\bibitem{Bena:2007kg}
Bena I and Warner N~P 2008
  \href{http://dx.doi.org/10.1007/978-3-540-79523-0_1}{ {\em Lect. Notes
  Phys.\/} {\bf 755} 1--92 } [\eprint{hep-th/0701216}]

\bibitem{Balasubramanian:2008da}
Balasubramanian V, de~Boer J, El-Showk S and Messamah I 2008
  \href{http://dx.doi.org/10.1088/0264-9381/25/21/214004}{ {\em Class. Quant.
  Grav.\/} {\bf 25} 214004 } [\eprint{0811.0263}]

\bibitem{Bena:2022rna}
Bena I, Martinec E~J, Mathur S~D and Warner N~P 2022  [\eprint{2204.13113}]

\bibitem{Buoninfante:2018xif}
Buoninfante L, Cornell A~S, Harmsen G, Koshelev A~S, Lambiase G, Marto J~a and
  Mazumdar A 2018 \href{http://dx.doi.org/10.1103/PhysRevD.98.084041}{ {\em
  Phys. Rev. D\/} {\bf 98} 084041 } [\eprint{1807.08896}]

\bibitem{Buoninfante:2019swn}
Buoninfante L and Mazumdar A 2019
  \href{http://dx.doi.org/10.1103/PhysRevD.100.024031}{ {\em Phys. Rev. D\/}
  {\bf 100} 024031 } [\eprint{1903.01542}]

\bibitem{Lenzi:2021wpc}
Lenzi M and Sopuerta C~F 2021
  \href{http://dx.doi.org/10.1103/PhysRevD.104.084053}{ {\em Phys. Rev. D\/}
  {\bf 104} 084053 } [\eprint{2108.08668}]

\bibitem{Lenzi:2021njy}
Lenzi M and Sopuerta C~F 2021
  \href{http://dx.doi.org/10.1103/PhysRevD.104.124068}{ {\em Phys. Rev. D\/}
  {\bf 104} 124068 } [\eprint{2109.00503}]

\bibitem{Saketh:2024ojw}
Saketh M~V~S and Maggio E 2024
  \href{http://dx.doi.org/10.1103/PhysRevD.110.084038}{ {\em Phys. Rev. D\/}
  {\bf 110} 084038 } [\eprint{2406.10070}]

\bibitem{Price:2017cjr}
Price R~H and Khanna G 2017 \href{http://dx.doi.org/10.1088/1361-6382/aa8f29}{
  {\em Class. Quant. Grav.\/} {\bf 34} 225005 } [\eprint{1702.04833}]

\bibitem{Kokkotas:1995av}
Kokkotas K~D 1995 {Pulsating relativistic stars} {\em {Les Houches School of
  Physics: Astrophysical Sources of Gravitational Radiation}\/} pp 89--102
  [\eprint{gr-qc/9603024}]

\bibitem{Ferrari:2000sr}
Ferrari V and Kokkotas K~D 2000
  \href{http://dx.doi.org/10.1103/PhysRevD.62.107504}{ {\em Phys. Rev. D\/}
  {\bf 62} 107504 } [\eprint{gr-qc/0008057}]

\bibitem{Tominaga:1999iy}
Tominaga K, Saijo M and Maeda K~i 1999
  \href{http://dx.doi.org/10.1103/PhysRevD.60.024004}{ {\em Phys. Rev. D\/}
  {\bf 60} 024004 } [\eprint{gr-qc/9901040}]

\bibitem{Andrade:1999mj}
Andrade Z and Price R~H 1999
  \href{http://dx.doi.org/10.1103/PhysRevD.60.104037}{ {\em Phys. Rev. D\/}
  {\bf 60} 104037 } [\eprint{gr-qc/9902062}]

\bibitem{Tominaga:2000cs}
Tominaga K, Saijo M and Maeda K~i 2001
  \href{http://dx.doi.org/10.1103/PhysRevD.63.124012}{ {\em Phys. Rev. D\/}
  {\bf 63} 124012 } [\eprint{gr-qc/0009055}]

\bibitem{Andrade:2001hk}
Andrade Z 2001 \href{http://dx.doi.org/10.1103/PhysRevD.63.124002}{ {\em Phys.
  Rev. D\/} {\bf 63} 124002 } [\eprint{gr-qc/0103062}]

\bibitem{Saraswat:2019npa}
Saraswat K and Afshordi N 2020
  \href{http://dx.doi.org/10.1007/JHEP04(2020)136}{ {\em JHEP\/} {\bf 04} 136 }
  [\eprint{1906.02653}]

\bibitem{Wang:2019rcf}
Wang Q, Oshita N and Afshordi N 2020
  \href{http://dx.doi.org/10.1103/PhysRevD.101.024031}{ {\em Phys. Rev. D\/}
  {\bf 101} 024031 } [\eprint{1905.00446}]

\bibitem{Pavlidou:2000cs}
Pavlidou V, Tassis K, Baumgarte T~W and Shapiro S~L 2000
  \href{http://dx.doi.org/10.1103/PhysRevD.62.084020}{ {\em Phys. Rev. D\/}
  {\bf 62} 084020 } [\eprint{gr-qc/0007019}]

\bibitem{Raposo:2018rjn}
Raposo G, Pani P, Bezares M, Palenzuela C and Cardoso V 2019
  \href{http://dx.doi.org/10.1103/PhysRevD.99.104072}{ {\em Phys. Rev. D\/}
  {\bf 99} 104072 } [\eprint{1811.07917}]

\bibitem{Pani:2018flj}
Pani P and Ferrari V 2018 \href{http://dx.doi.org/10.1088/1361-6382/aacb8f}{
  {\em Class. Quant. Grav.\/} {\bf 35} 15LT01 } [\eprint{1804.01444}]

\bibitem{Mannarelli:2018pjb}
Mannarelli M and Tonelli F 2018
  \href{http://dx.doi.org/10.1103/PhysRevD.97.123010}{ {\em Phys. Rev. D\/}
  {\bf 97} 123010 } [\eprint{1805.02278}]

\bibitem{Zhang:2017jze}
Zhang J and Zhou S~Y 2018 \href{http://dx.doi.org/10.1103/PhysRevD.97.081501}{
  {\em Phys. Rev. D\/} {\bf 97} 081501 } [\eprint{1709.07503}]

\bibitem{Oshita:2018fqu}
Oshita N and Afshordi N 2019
  \href{http://dx.doi.org/10.1103/PhysRevD.99.044002}{ {\em Phys. Rev. D\/}
  {\bf 99} 044002 } [\eprint{1807.10287}]

\bibitem{Lunin:2001jy}
Lunin O and Mathur S~D 2002
  \href{http://dx.doi.org/10.1016/S0550-3213(01)00620-4}{ {\em Nucl. Phys. B\/}
  {\bf 623} 342--394 } [\eprint{hep-th/0109154}]

\bibitem{Giusto:2004ip}
Giusto S, Mathur S~D and Saxena A 2005
  \href{http://dx.doi.org/10.1016/j.nuclphysb.2005.01.009}{ {\em Nucl. Phys.
  B\/} {\bf 710} 425--463 } [\eprint{hep-th/0406103}]

\bibitem{Ikeda:2021uvc}
Ikeda T, Bianchi M, Consoli D, Grillo A, Morales J~F, Pani P and Raposo G 2021
  \href{http://dx.doi.org/10.1103/PhysRevD.104.066021}{ {\em Phys. Rev. D\/}
  {\bf 104} 066021 } [\eprint{2103.10960}]

\bibitem{Heidmann:2023ojf}
Heidmann P, Speeney N, Berti E and Bah I 2023
  \href{http://dx.doi.org/10.1103/PhysRevD.108.024021}{ {\em Phys. Rev. D\/}
  {\bf 108} 024021 } [\eprint{2305.14412}]

\bibitem{Dima:2024cok}
Dima A, Melis M and Pani P 2024
  \href{http://dx.doi.org/10.1103/PhysRevD.110.084067}{ {\em Phys. Rev. D\/}
  {\bf 110} 084067 } [\eprint{2406.19327}]

\bibitem{Bekenstein:1972tm}
Bekenstein J~D 1972 \href{http://dx.doi.org/10.1007/BF02757029}{ {\em Lett.
  Nuovo Cim.\/} {\bf 4} 737--740 }

\bibitem{Cardoso:2019apo}
Cardoso V, Foit V~F and Kleban M 2019
  \href{http://dx.doi.org/10.1088/1475-7516/2019/08/006}{ {\em JCAP\/} {\bf 08}
  006 } [\eprint{1902.10164}]

\bibitem{Agullo:2020hxe}
Agullo I, Cardoso V, Rio A~D, Maggiore M and Pullin J 2021
  \href{http://dx.doi.org/10.1103/PhysRevLett.126.041302}{ {\em Phys. Rev.
  Lett.\/} {\bf 126} 041302 } [\eprint{2007.13761}]

\bibitem{Coates:2019bun}
Coates A, V\"olkel S~H and Kokkotas K~D 2019
  \href{http://dx.doi.org/10.1103/PhysRevLett.123.171104}{ {\em Phys. Rev.
  Lett.\/} {\bf 123} 171104 } [\eprint{1909.01254}]

\bibitem{Foit:2016uxn}
Foit V~F and Kleban M 2019 \href{http://dx.doi.org/10.1088/1361-6382/aafcba}{
  {\em Class. Quant. Grav.\/} {\bf 36} 035006 } [\eprint{1611.07009}]

\bibitem{Coates:2021dlg}
Coates A, V\"olkel S~H and Kokkotas K~D 2022
  \href{http://dx.doi.org/10.1088/1361-6382/ac4618}{ {\em Class. Quant.
  Grav.\/} {\bf 39} 045007 } [\eprint{2201.03245}]

\bibitem{1916DPhyG..18..318E}
{Einstein} A 1916 {\em Deutsche Physikalische Gesellschaft\/} {\bf 18} 318--323

\bibitem{Oshita:2019sat}
Oshita N, Wang Q and Afshordi N 2020
  \href{http://dx.doi.org/10.1088/1475-7516/2020/04/016}{ {\em JCAP\/} {\bf 04}
  016 } [\eprint{1905.00464}]

\bibitem{Burgess:2018pmm}
Burgess C~P, Plestid R and Rummel M 2018
  \href{http://dx.doi.org/10.1007/JHEP09(2018)113}{ {\em JHEP\/} {\bf 09} 113 }
  [\eprint{1808.00847}]

\bibitem{Kehagias:2024yyp}
Kehagias A and Riotto A 2024  [\eprint{2411.12428}]

\bibitem{Forteza:2022tgq}
Forteza X~J, Bhagwat S, Kumar S and Pani P 2023
  \href{http://dx.doi.org/10.1103/PhysRevLett.130.021001}{ {\em Phys. Rev.
  Lett.\/} {\bf 130} 021001 } [\eprint{2205.14910}]

\bibitem{Brustein:2023gea}
Brustein R, Medved A~J~M and Shindelman T 2023
  \href{http://dx.doi.org/10.1103/PhysRevD.108.044058}{ {\em Phys. Rev. D\/}
  {\bf 108} 044058 } [\eprint{2304.04984}]

\bibitem{Macedo:2016wgh}
Macedo C~F~B, Cardoso V, Crispino L~C~B and Pani P 2016
  \href{http://dx.doi.org/10.1103/PhysRevD.93.064053}{ {\em Phys. Rev. D\/}
  {\bf 93} 064053 } [\eprint{1603.02095}]

\bibitem{Palenzuela:2017kcg}
Palenzuela C, Pani P, Bezares M, Cardoso V, Lehner L and Liebling S 2017
  \href{http://dx.doi.org/10.1103/PhysRevD.96.104058}{ {\em Phys. Rev. D\/}
  {\bf 96} 104058 } [\eprint{1710.09432}]

\bibitem{Bezares:2022obu}
Bezares M, Bo\v{s}kovi\'c M, Liebling S, Palenzuela C, Pani P and Barausse E
  2022 \href{http://dx.doi.org/10.1103/PhysRevD.105.064067}{ {\em Phys. Rev.
  D\/} {\bf 105} 064067 } [\eprint{2201.06113}]

\bibitem{Siemonsen:2024snb}
Siemonsen N 2024 \href{http://dx.doi.org/10.1103/PhysRevLett.133.031401}{ {\em
  Phys. Rev. Lett.\/} {\bf 133} 031401 } [\eprint{2404.14536}]

\bibitem{Yoshida:1994xi}
Yoshida S, Eriguchi Y and Futamase T 1994
  \href{http://dx.doi.org/10.1103/PhysRevD.50.6235}{ {\em Phys. Rev. D\/} {\bf
  50} 6235--6246 }

\bibitem{Macedo:2013jja}
Macedo C~F~B, Pani P, Cardoso V and Crispino L~C~B 2013
  \href{http://dx.doi.org/10.1103/PhysRevD.88.064046}{ {\em Phys. Rev. D\/}
  {\bf 88} 064046 } [\eprint{1307.4812}]

\bibitem{Chirenti:2007mk}
Chirenti C~B~M~H and Rezzolla L 2007
  \href{http://dx.doi.org/10.1088/0264-9381/24/16/013}{ {\em Class. Quant.
  Grav.\/} {\bf 24} 4191--4206 } [\eprint{0706.1513}]

\bibitem{Pani:2009ss}
Pani P, Berti E, Cardoso V, Chen Y and Norte R 2009
  \href{http://dx.doi.org/10.1103/PhysRevD.80.124047}{ {\em Phys. Rev. D\/}
  {\bf 80} 124047 } [\eprint{0909.0287}]

\bibitem{Chirenti:2016hzd}
Chirenti C and Rezzolla L 2016
  \href{http://dx.doi.org/10.1103/PhysRevD.94.084016}{ {\em Phys. Rev. D\/}
  {\bf 94} 084016 } [\eprint{1602.08759}]

\bibitem{Bena:2024hoh}
Bena I, Di~Russo G, Morales J~F and Ruip\'erez A 2024
  \href{http://dx.doi.org/10.1007/JHEP10(2024)071}{ {\em JHEP\/} {\bf 10} 071 }
  [\eprint{2406.19330}]

\bibitem{Price:2015gia}
Price R~H, Nampalliwar S and Khanna G 2016
  \href{http://dx.doi.org/10.1103/PhysRevD.93.044060}{ {\em Phys. Rev. D\/}
  {\bf 93} 044060 } [\eprint{1508.04797}]

\bibitem{Abedi:2016hgu}
Abedi J, Dykaar H and Afshordi N 2017
  \href{http://dx.doi.org/10.1103/PhysRevD.96.082004}{ {\em Phys. Rev. D\/}
  {\bf 96} 082004 } [\eprint{1612.00266}]

\bibitem{Brustein:2018ixz}
Brustein R, Medved A~J~M and Yagi K 2019
  \href{http://dx.doi.org/10.1103/PhysRevD.100.104009}{ {\em Phys. Rev. D\/}
  {\bf 100} 104009 } [\eprint{1811.12283}]

\bibitem{Zimmerman:2023hua}
Zimmerman A, George R~N and Chen Y 2023  [\eprint{2306.11166}]

\bibitem{Arrechea:2024nlp}
Arrechea J, Liberati S and Vellucci V 2024
  \href{http://dx.doi.org/10.1088/1475-7516/2024/12/004}{ {\em JCAP\/} {\bf 12}
  004 } [\eprint{2407.08807}]

\bibitem{Mark:2017dnq}
Mark Z, Zimmerman A, Du S~M and Chen Y 2017
  \href{http://dx.doi.org/10.1103/PhysRevD.96.084002}{ {\em Phys. Rev. D\/}
  {\bf 96} 084002 } [\eprint{1706.06155}]

\bibitem{Wang:2018gin}
Wang Q and Afshordi N 2018 \href{http://dx.doi.org/10.1103/PhysRevD.97.124044}{
  {\em Phys. Rev. D\/} {\bf 97} 124044 } [\eprint{1803.02845}]

\bibitem{Vilenkin:1978uc}
Vilenkin A 1978 \href{http://dx.doi.org/10.1016/0370-2693(78)90027-8}{ {\em
  Phys. Lett. B\/} {\bf 78} 301--303 }

\bibitem{Maggio:2018ivz}
Maggio E, Cardoso V, Dolan S~R and Pani P 2019
  \href{http://dx.doi.org/10.1103/PhysRevD.99.064007}{ {\em Phys. Rev. D\/}
  {\bf 99} 064007 } [\eprint{1807.08840}]

\bibitem{Chen:2020htz}
Chen B, Wang Q and Chen Y 2021
  \href{http://dx.doi.org/10.1103/PhysRevD.103.104054}{ {\em Phys. Rev. D\/}
  {\bf 103} 104054 } [\eprint{2012.10842}]

\bibitem{LongoMicchi:2020cwm}
Longo~Micchi L~F, Afshordi N and Chirenti C 2021
  \href{http://dx.doi.org/10.1103/PhysRevD.103.044028}{ {\em Phys. Rev. D\/}
  {\bf 103} 044028 } [\eprint{2010.14578}]

\bibitem{Annulli:2021ccn}
Annulli L, Cardoso V and Gualtieri L 2022
  \href{http://dx.doi.org/10.1088/1361-6382/ac6410}{ {\em Class. Quant.
  Grav.\/} {\bf 39} 105005 } [\eprint{2104.11236}]

\bibitem{Xin:2021zir}
Xin S, Chen B, Lo R~K~L, Sun L, Han W~B, Zhong X, Srivastava M, Ma S, Wang Q
  and Chen Y 2021 \href{http://dx.doi.org/10.1103/PhysRevD.104.104005}{ {\em
  Phys. Rev. D\/} {\bf 104} 104005 } [\eprint{2105.12313}]

\bibitem{Srivastava:2021uku}
Srivastava M and Chen Y 2021
  \href{http://dx.doi.org/10.1103/PhysRevD.104.104006}{ {\em Phys. Rev. D\/}
  {\bf 104} 104006 } [\eprint{2108.01329}]

\bibitem{Ma:2022xmp}
Ma S, Wang Q, Deppe N, H\'ebert F, Kidder L~E, Moxon J, Throwe W, Vu N~L,
  Scheel M~A and Chen Y 2022
  \href{http://dx.doi.org/10.1103/PhysRevD.105.104007}{ {\em Phys. Rev. D\/}
  {\bf 105} 104007 } [\eprint{2203.03174}]

\bibitem{Silvestrini:2025lbe}
Silvestrini M, Maggio E, Chakraborty S and Pani P 2025
  \href{http://dx.doi.org/10.1103/pky4-23jb}{ {\em Phys. Rev. D\/} {\bf 112}
  124021 } [\eprint{2506.16516}]

\bibitem{Volkel:2017kfj}
V\"olkel S~H and Kokkotas K~D 2017
  \href{http://dx.doi.org/10.1088/1361-6382/aa82de}{ {\em Class. Quant.
  Grav.\/} {\bf 34} 175015 } [\eprint{1704.07517}]

\bibitem{Volkel:2018hwb}
V\"olkel S~H and Kokkotas K~D 2018
  \href{http://dx.doi.org/10.1088/1361-6382/aabce6}{ {\em Class. Quant.
  Grav.\/} {\bf 35} 105018 } [\eprint{1802.08525}]

\bibitem{Bueno:2017hyj}
Bueno P, Cano P~A, Goelen F, Hertog T and Vercnocke B 2018
  \href{http://dx.doi.org/10.1103/PhysRevD.97.024040}{ {\em Phys. Rev. D\/}
  {\bf 97} 024040 } [\eprint{1711.00391}]

\bibitem{Deppe:2024qrk}
Deppe N, Heisenberg L, Inchausp{\'e} H, Kidder L~E, Maibach D, Ma S, Moxon J,
  Nelli K~C, Throwe W and Vu N~L 2025
  \href{http://dx.doi.org/10.1103/k7jh-rhgw}{ {\em Phys. Rev. D\/} {\bf 111}
  124035 } [\eprint{2411.05645}]

\bibitem{echowfm}
Lo R~K~L and Uchikata N 2025 \texttt{echoes\_waveform\_models}
  \urlprefix\url{https://git.ligo.org/echoes_template_search/echoes_waveform_models}

\bibitem{Lo:2018sep}
Lo R~K~L, Li T~G~F and Weinstein A~J 2019
  \href{http://dx.doi.org/10.1103/PhysRevD.99.084052}{ {\em Phys. Rev. D\/}
  {\bf 99} 084052 } [\eprint{1811.07431}]

\bibitem{Maselli:2017tfq}
Maselli A, V\"olkel S~H and Kokkotas K~D 2017
  \href{http://dx.doi.org/10.1103/PhysRevD.96.064045}{ {\em Phys. Rev. D\/}
  {\bf 96} 064045 } [\eprint{1708.02217}]

\bibitem{Testa:2018bzd}
Testa A and Pani P 2018 \href{http://dx.doi.org/10.1103/PhysRevD.98.044018}{
  {\em Phys. Rev. D\/} {\bf 98} 044018 } [\eprint{1806.04253}]

\bibitem{Branchesi:2023mws}
Branchesi M {\em et~al.\/} 2023
  \href{http://dx.doi.org/10.1088/1475-7516/2023/07/068}{ {\em JCAP\/} {\bf 07}
  068 } [\eprint{2303.15923}]

\bibitem{Nakano:2017fvh}
Nakano H, Sago N, Tagoshi H and Tanaka T 2017
  \href{http://dx.doi.org/10.1093/ptep/ptx093}{ {\em PTEP\/} {\bf 2017} 071E01
  } [\eprint{1704.07175}]

\bibitem{Uchikata:2023zcu}
Uchikata N, Narikawa T, Nakano H, Sago N, Tagoshi H and Tanaka T 2023
  \href{http://dx.doi.org/10.1103/PhysRevD.108.104040}{ {\em Phys. Rev. D\/}
  {\bf 108} 104040 } [\eprint{2309.01894}]

\bibitem{Vellucci:2022hpl}
Vellucci V, Franzin E and Liberati S 2023
  \href{http://dx.doi.org/10.1103/PhysRevD.107.044027}{ {\em Phys. Rev. D\/}
  {\bf 107} 044027 } [\eprint{2205.14170}]

\bibitem{Dailey:2023mvn}
Dailey C, Afshordi N and Schnetter E 2023
  \href{http://dx.doi.org/10.1088/1361-6382/acde2f}{ {\em Class. Quant.
  Grav.\/} {\bf 40} 195007 } [\eprint{2301.05778}]

\bibitem{Guo:2022umn}
Guo B and Mathur S~D 2022 \href{http://dx.doi.org/10.1142/S0218271822420093}{
  {\em Int. J. Mod. Phys. D\/} {\bf 31} 2242009 } [\eprint{2205.10921}]

\bibitem{Dailey:2024kjg}
Dailey C, Schnetter E and Afshordi N 2025
  \href{http://dx.doi.org/10.1088/1361-6382/ad9701}{ {\em Class. Quant.
  Grav.\/} {\bf 42} 025002 } [\eprint{2409.17970}]

\bibitem{Brustein:2018web}
Brustein R and Medved A~J~M 2019
  \href{http://dx.doi.org/10.1103/PhysRevD.99.064019}{ {\em Phys. Rev. D\/}
  {\bf 99} 064019 } [\eprint{1805.11667}]

\bibitem{Brustein:2019bou}
Brustein R and Medved A~J~M 2019
  \href{http://dx.doi.org/10.1002/prop.201900058}{ {\em Fortsch. Phys.\/} {\bf
  67} 1900058 } [\eprint{1902.07990}]

\bibitem{Brustein:2017koc}
Brustein R, Medved A~J~M and Yagi K 2017
  \href{http://dx.doi.org/10.1103/PhysRevD.96.064033}{ {\em Phys. Rev. D\/}
  {\bf 96} 064033 } [\eprint{1704.05789}]

\bibitem{Kokkotas:2002sf}
Kokkotas K~D, Ruoff J and Andersson N 2004
  \href{http://dx.doi.org/10.1103/PhysRevD.70.043003}{ {\em Phys. Rev. D\/}
  {\bf 70} 043003 } [\eprint{astro-ph/0212429}]

\bibitem{Chirenti:2008pf}
Chirenti C~B~M~H and Rezzolla L 2008
  \href{http://dx.doi.org/10.1103/PhysRevD.78.084011}{ {\em Phys. Rev. D\/}
  {\bf 78} 084011 } [\eprint{0808.4080}]

\bibitem{Cardoso:2007az}
Cardoso V, Pani P, Cadoni M and Cavaglia M 2008
  \href{http://dx.doi.org/10.1103/PhysRevD.77.124044}{ {\em Phys. Rev. D\/}
  {\bf 77} 124044 } [\eprint{0709.0532}]

\bibitem{Szpak:2004sf}
Szpak N 2004  [\eprint{gr-qc/0411050}]

\bibitem{Yang:2013shb}
Yang H, Zhang F, Zimmerman A and Chen Y 2014
  \href{http://dx.doi.org/10.1103/PhysRevD.89.064014}{ {\em Phys. Rev. D\/}
  {\bf 89} 064014 } [\eprint{1311.3380}]

\bibitem{Casals:2016soq}
Casals M, Kavanagh C and Ottewill A~C 2016
  \href{http://dx.doi.org/10.1103/PhysRevD.94.124053}{ {\em Phys. Rev. D\/}
  {\bf 94} 124053 } [\eprint{1608.05392}]

\bibitem{Hui:2019aox}
Hui L, Kabat D and Wong S~S~C 2019
  \href{http://dx.doi.org/10.1088/1475-7516/2019/12/020}{ {\em JCAP\/} {\bf 12}
  020 } [\eprint{1909.10382}]

\bibitem{Lagos:2022otp}
Lagos M and Hui L 2023 \href{http://dx.doi.org/10.1103/PhysRevD.107.044040}{
  {\em Phys. Rev. D\/} {\bf 107} 044040 } [\eprint{2208.07379}]

\bibitem{Casals:2024ynr}
Casals M, Hollands S, Pound A and Toomani V 2024
  \href{http://dx.doi.org/10.1088/1361-6382/ad7cbc}{ {\em Class. Quant.
  Grav.\/} {\bf 41} 215010 } [\eprint{2402.15468}]

\bibitem{Detweiler:1976zz}
Detweiler S~L 1976 \href{http://dx.doi.org/10.1098/rspa.1976.0069}{ {\em Proc.
  Roy. Soc. Lond. A\/} {\bf 349} 217--230 }

\bibitem{Chandrasekhar:1976zz}
Chandrasekhar S and Detweiler S~L 1976
  \href{http://dx.doi.org/10.1098/rspa.1976.0101}{ {\em Proc. Roy. Soc. Lond.
  A\/} {\bf 350} 165--174 }

\bibitem{Detweiler:1977gy}
Detweiler S~L 1977 \href{http://dx.doi.org/10.1098/rspa.1977.0005}{ {\em Proc.
  Roy. Soc. Lond. A\/} {\bf 352} 381--395 }

\bibitem{Dolan:2011fh}
Dolan S~R and Ottewill A~C 2011
  \href{http://dx.doi.org/10.1103/PhysRevD.84.104002}{ {\em Phys. Rev. D\/}
  {\bf 84} 104002 } [\eprint{1106.4318}]

\bibitem{Casals:2013mpa}
Casals M, Dolan S, Ottewill A~C and Wardell B 2013
  \href{http://dx.doi.org/10.1103/PhysRevD.88.044022}{ {\em Phys. Rev. D\/}
  {\bf 88} 044022 } [\eprint{1306.0884}]

\bibitem{Chavda:2024awq}
Chavda A, Lagos M and Hui L 2025
  \href{http://dx.doi.org/10.1088/1475-7516/2025/07/084}{ {\em JCAP\/} {\bf 07}
  084 } [\eprint{2412.03435}]

\bibitem{Ching:1994bd}
Ching E~S~C, Leung P~T, Suen W~M and Young K 1995
  \href{http://dx.doi.org/10.1103/PhysRevLett.74.2414}{ {\em Phys. Rev.
  Lett.\/} {\bf 74} 2414--2417 } [\eprint{gr-qc/9410044}]

\bibitem{Price:1972pw}
Price R~H 1972 \href{http://dx.doi.org/10.1103/PhysRevD.5.2439}{ {\em Phys.
  Rev. D\/} {\bf 5} 2439--2454 }

\bibitem{Ching:1995tj}
Ching E~S~C, Leung P~T, Suen W~M and Young K 1995
  \href{http://dx.doi.org/10.1103/PhysRevD.52.2118}{ {\em Phys. Rev. D\/} {\bf
  52} 2118--2132 } [\eprint{gr-qc/9507035}]

\bibitem{Nakano:2000ne}
Nakano H and Sasaki M 2001 \href{http://dx.doi.org/10.1143/PTP.105.197}{ {\em
  Prog. Theor. Phys.\/} {\bf 105} 197--218 } [\eprint{gr-qc/0010036}]

\bibitem{Casals:2015nja}
Casals M and Ottewill A~C 2015
  \href{http://dx.doi.org/10.1103/PhysRevD.92.124055}{ {\em Phys. Rev. D\/}
  {\bf 92} 124055 } [\eprint{1509.04702}]

\bibitem{Shibata:1994jx}
Shibata M, Sasaki M, Tagoshi H and Tanaka T 1995
  \href{http://dx.doi.org/10.1103/PhysRevD.51.1646}{ {\em Phys. Rev. D\/} {\bf
  51} 1646--1663 } [\eprint{gr-qc/9409054}]

\bibitem{Tagoshi:1996gh}
Tagoshi H, Shibata M, Tanaka T and Sasaki M 1996
  \href{http://dx.doi.org/10.1103/PhysRevD.54.1439}{ {\em Phys. Rev. D\/} {\bf
  54} 1439--1459 } [\eprint{gr-qc/9603028}]

\bibitem{Okuzumi:2008ej}
Okuzumi S, Ioka K and Sakagami M~a 2008
  \href{http://dx.doi.org/10.1103/PhysRevD.77.124018}{ {\em Phys. Rev. D\/}
  {\bf 77} 124018 } [\eprint{0803.0501}]

\bibitem{Poisson:1996ya}
Poisson E 1997 \href{http://dx.doi.org/10.1103/PhysRevD.55.639}{ {\em Phys.
  Rev. D\/} {\bf 55} 639--649 } [\eprint{gr-qc/9606078}]

\bibitem{1976RSPSA.348...39C}
{Chandrasekhar} S 1976 \href{http://dx.doi.org/10.1098/rspa.1976.0022}{ {\em P.
  R. Soc. Lond. A\/} {\bf 348} 39--55 }

\bibitem{Detweiler:1978ge}
Detweiler S~L 1978 \href{http://dx.doi.org/10.1086/156529}{ {\em Astrophys.
  J.\/} {\bf 225} 687--693 }

\bibitem{Glampedakis:2017rar}
Glampedakis K, Johnson A~D and Kennefick D 2017
  \href{http://dx.doi.org/10.1103/PhysRevD.96.024036}{ {\em Phys. Rev. D\/}
  {\bf 96} 024036 } [\eprint{1702.06459}]

\bibitem{Hughes:2000pf}
Hughes S~A 2000 \href{http://dx.doi.org/10.1103/PhysRevD.62.044029}{ {\em Phys.
  Rev. D\/} {\bf 62} 044029 } [Erratum: Phys.Rev.D 67, 089902 (2003)]
  [\eprint{gr-qc/0002043}]

\bibitem{Kojima:1984cj}
Kojima Y and Nakamura T 1984 \href{http://dx.doi.org/10.1143/PTP.71.79}{ {\em
  Prog. Theor. Phys.\/} {\bf 71} 79--90 }

\bibitem{Saijo:1996iz}
Saijo M, Shinkai H~a and Maeda K~i 1997
  \href{http://dx.doi.org/10.1103/PhysRevD.56.785}{ {\em Phys. Rev. D\/} {\bf
  56} 785--797 } [\eprint{gr-qc/9701001}]

\bibitem{Sundararajan:2007jg}
Sundararajan P~A, Khanna G and Hughes S~A 2007
  \href{http://dx.doi.org/10.1103/PhysRevD.76.104005}{ {\em Phys. Rev. D\/}
  {\bf 76} 104005 } [\eprint{gr-qc/0703028}]

\bibitem{Sundararajan:2008zm}
Sundararajan P~A, Khanna G, Hughes S~A and Drasco S 2008
  \href{http://dx.doi.org/10.1103/PhysRevD.78.024022}{ {\em Phys. Rev. D\/}
  {\bf 78} 024022 } [\eprint{0803.0317}]

\bibitem{Hughes:2021exa}
Hughes S~A, Warburton N, Khanna G, Chua A~J~K and Katz M~L 2021
  \href{http://dx.doi.org/10.1103/PhysRevD.103.104014}{ {\em Phys. Rev. D\/}
  {\bf 103} 104014 } [Erratum: Phys.Rev.D 107, 089901 (2023)]
  [\eprint{2102.02713}]

\bibitem{Ori:2000zn}
Ori A and Thorne K~S 2000 \href{http://dx.doi.org/10.1103/PhysRevD.62.124022}{
  {\em Phys. Rev. D\/} {\bf 62} 124022 } [\eprint{gr-qc/0003032}]

\bibitem{Apte:2019txp}
Apte A and Hughes S~A 2019
  \href{http://dx.doi.org/10.1103/PhysRevD.100.084031}{ {\em Phys. Rev. D\/}
  {\bf 100} 084031 } [\eprint{1901.05901}]

\bibitem{Field:2020rjr}
Field S~E, Gottlieb S, Grant Z~J, Isherwood L~F and Khanna G 2023
  \href{http://dx.doi.org/10.1007/s42967-021-00129-2}{ {\em Appl. Math.
  Comput.\/} {\bf 5} 97--115 } [\eprint{2010.04760}]

\bibitem{Sun:1988tz}
Sun Y and Price R~H 1988 \href{http://dx.doi.org/10.1103/PhysRevD.38.1040}{
  {\em Phys. Rev. D\/} {\bf 38} 1040--1052 }

\bibitem{Nagar:2006xv}
Nagar A, Damour T and Tartaglia A 2007
  \href{http://dx.doi.org/10.1088/0264-9381/24/12/S08}{ {\em Class. Quant.
  Grav.\/} {\bf 24} S109--S124 } [\eprint{gr-qc/0612096}]

\bibitem{Barausse:2009xi}
Barausse E and Buonanno A 2010
  \href{http://dx.doi.org/10.1103/PhysRevD.81.084024}{ {\em Phys. Rev. D\/}
  {\bf 81} 084024 } [\eprint{0912.3517}]

\bibitem{Bernuzzi:2010ty}
Bernuzzi S and Nagar A 2010
  \href{http://dx.doi.org/10.1103/PhysRevD.81.084056}{ {\em Phys. Rev. D\/}
  {\bf 81} 084056 } [\eprint{1003.0597}]

\bibitem{Barausse:2011kb}
Barausse E, Buonanno A, Hughes S~A, Khanna G, O'Sullivan S and Pan Y 2012
  \href{http://dx.doi.org/10.1103/PhysRevD.85.024046}{ {\em Phys. Rev. D\/}
  {\bf 85} 024046 } [\eprint{1110.3081}]

\bibitem{Hadar:2009ip}
Hadar S and Kol B 2011 \href{http://dx.doi.org/10.1103/PhysRevD.84.044019}{
  {\em Phys. Rev. D\/} {\bf 84} 044019 } [\eprint{0911.3899}]

\bibitem{Hadar:2011vj}
Hadar S, Kol B, Berti E and Cardoso V 2011
  \href{http://dx.doi.org/10.1103/PhysRevD.84.047501}{ {\em Phys. Rev. D\/}
  {\bf 84} 047501 } [\eprint{1105.3861}]

\bibitem{Mummery:2022ana}
Mummery A and Balbus S 2022
  \href{http://dx.doi.org/10.1103/PhysRevLett.129.161101}{ {\em Phys. Rev.
  Lett.\/} {\bf 129} 161101 } [\eprint{2209.03579}]

\bibitem{Mummery:2023hlo}
Mummery A and Balbus S 2023
  \href{http://dx.doi.org/10.1103/PhysRevD.107.124058}{ {\em Phys. Rev. D\/}
  {\bf 107} 124058 } [\eprint{2302.01159}]

\bibitem{Dyson:2023fws}
Dyson C and van~de Meent M 2023
  \href{http://dx.doi.org/10.1088/1361-6382/acf552}{ {\em Class. Quant.
  Grav.\/} {\bf 40} 195026 } [\eprint{2302.03704}]

\bibitem{Lim:2019xrb}
Lim H, Khanna G, Apte A and Hughes S~A 2019
  \href{http://dx.doi.org/10.1103/PhysRevD.100.084032}{ {\em Phys. Rev. D\/}
  {\bf 100} 084032 } [\eprint{1901.05902}]

\bibitem{Hughes:2019zmt}
Hughes S~A, Apte A, Khanna G and Lim H 2019
  \href{http://dx.doi.org/10.1103/PhysRevLett.123.161101}{ {\em Phys. Rev.
  Lett.\/} {\bf 123} 161101 } [\eprint{1901.05900}]

\bibitem{Detweiler:1979xr}
Detweiler S~L and Szedenits E 1979 \href{http://dx.doi.org/10.1086/157182}{
  {\em Astrophys. J.\/} {\bf 231} 211--218 }

\bibitem{Andersson:1995zk}
Andersson N 1995 \href{http://dx.doi.org/10.1103/PhysRevD.51.353}{ {\em Phys.
  Rev. D\/} {\bf 51} 353--363 }

\bibitem{Glampedakis:2003dn}
Glampedakis K and Andersson N 2003
  \href{http://dx.doi.org/10.1088/0264-9381/20/15/312}{ {\em Class. Quant.
  Grav.\/} {\bf 20} 3441--3464 } [\eprint{gr-qc/0304030}]

\bibitem{Watarai:2024huy}
Watarai D 2024 \href{http://dx.doi.org/10.1103/PhysRevD.110.124029}{ {\em Phys.
  Rev. D\/} {\bf 110} 124029 } [\eprint{2408.16747}]

\bibitem{DellaRocca:2025zbe}
Della~Rocca M, Pezzella L, Berti E, Gualtieri L and Maselli A 2025
  [\eprint{2512.07959}]

\bibitem{Pound:2021qin}
Pound A and Wardell B 2020 {\em Black Hole Perturbation Theory and
  Gravitational Self-Force\/} (Singapore: Springer Singapore) pp 1--119 ISBN
  978-981-15-4702-7 [\eprint{2101.04592}]

\bibitem{Kuchler:2024esj}
K\"uchler L, Comp\`ere G, Durkan L and Pound A 2024
  \href{http://dx.doi.org/10.21468/SciPostPhys.17.2.056}{ {\em SciPost Phys.\/}
  {\bf 17} 056 } [\eprint{2405.00170}]

\bibitem{Honet:2025dho}
Honet L, K{\"u}chler L, Pound A and Comp{\`e}re G 2025  [\eprint{2510.13958}]

\bibitem{Oshita:2022pkc}
Oshita N 2023 \href{http://dx.doi.org/10.1088/1475-7516/2023/04/013}{ {\em
  JCAP\/} {\bf 04} 013 } [\eprint{2208.02923}]

\bibitem{Oshita:2025ibu}
Oshita N, Berti E and Cardoso V 2025
  \href{http://dx.doi.org/10.1103/ht2n-vvvh}{ {\em Phys. Rev. Lett.\/} {\bf
  135} 031401 } [\eprint{2503.21276}]

\bibitem{Chen:2024hum}
Chen C~H, Cho H~T, Chrysostomou A and Cornell A~S 2025
  \href{http://dx.doi.org/10.1103/PhysRevD.111.064030}{ {\em Phys. Rev. D\/}
  {\bf 111} 064030 } [\eprint{2407.18644}]

\bibitem{Oshita:2024wgt}
Oshita N and Cardoso V 2025
  \href{http://dx.doi.org/10.1103/PhysRevD.111.104043}{ {\em Phys. Rev. D\/}
  {\bf 111} 104043 } [\eprint{2407.02563}]

\bibitem{Gleiser:1995gx}
Gleiser R~J, Nicasio C~O, Price R~H and Pullin J 1996
  \href{http://dx.doi.org/10.1088/0264-9381/13/10/001}{ {\em Class. Quant.
  Grav.\/} {\bf 13} L117--L124 } [\eprint{gr-qc/9510049}]

\bibitem{Brizuela:2009qd}
Brizuela D, Martin-Garcia J~M and Tiglio M 2009
  \href{http://dx.doi.org/10.1103/PhysRevD.80.024021}{ {\em Phys. Rev. D\/}
  {\bf 80} 024021 } [\eprint{0903.1134}]

\bibitem{Ioka:2007ak}
Ioka K and Nakano H 2007 \href{http://dx.doi.org/10.1103/PhysRevD.76.061503}{
  {\em Phys. Rev. D\/} {\bf 76} 061503 } [\eprint{0704.3467}]

\bibitem{Nakano:2007cj}
Nakano H and Ioka K 2007 \href{http://dx.doi.org/10.1103/PhysRevD.76.084007}{
  {\em Phys. Rev. D\/} {\bf 76} 084007 } [\eprint{0708.0450}]

\bibitem{Loutrel:2020wbw}
Loutrel N, Ripley J~L, Giorgi E and Pretorius F 2021
  \href{http://dx.doi.org/10.1103/PhysRevD.103.104017}{ {\em Phys. Rev. D\/}
  {\bf 103} 104017 } [\eprint{2008.11770}]

\bibitem{Ripley:2020xby}
Ripley J~L, Loutrel N, Giorgi E and Pretorius F 2021
  \href{http://dx.doi.org/10.1103/PhysRevD.103.104018}{ {\em Phys. Rev. D\/}
  {\bf 103} 104018 } [\eprint{2010.00162}]

\bibitem{Cheung:2022rbm}
Cheung M~H~Y {\em et~al.\/} 2023
  \href{http://dx.doi.org/10.1103/PhysRevLett.130.081401}{ {\em Phys. Rev.
  Lett.\/} {\bf 130} 081401 } [\eprint{2208.07374}]

\bibitem{Mitman:2022qdl}
Mitman K {\em et~al.\/} 2023
  \href{http://dx.doi.org/10.1103/PhysRevLett.130.081402}{ {\em Phys. Rev.
  Lett.\/} {\bf 130} 081402 } [\eprint{2208.07380}]

\bibitem{Ma:2022wpv}
Ma S, Mitman K, Sun L, Deppe N, H\'ebert F, Kidder L~E, Moxon J, Throwe W, Vu
  N~L and Chen Y 2022 \href{http://dx.doi.org/10.1103/PhysRevD.106.084036}{
  {\em Phys. Rev. D\/} {\bf 106} 084036 } [\eprint{2207.10870}]

\bibitem{Bucciotti:2023ets}
Bucciotti B, Kuntz A, Serra F and Trincherini E 2023
  \href{http://dx.doi.org/10.1007/JHEP12(2023)048}{ {\em JHEP\/} {\bf 12} 048 }
  [\eprint{2309.08501}]

\bibitem{Perrone:2023jzq}
Perrone D, Barreira T, Kehagias A and Riotto A 2024
  \href{http://dx.doi.org/10.1016/j.nuclphysb.2023.116432}{ {\em Nucl. Phys.
  B\/} {\bf 999} 116432 } [\eprint{2308.15886}]

\bibitem{Kehagias:2023ctr}
Kehagias A, Perrone D, Riotto A and Riva F 2023
  \href{http://dx.doi.org/10.1103/PhysRevD.108.L021501}{ {\em Phys. Rev. D\/}
  {\bf 108} L021501 } [\eprint{2301.09345}]

\bibitem{Redondo-Yuste:2023seq}
Redondo-Yuste J, Carullo G, Ripley J~L, Berti E and Cardoso V 2024
  \href{http://dx.doi.org/10.1103/PhysRevD.109.L101503}{ {\em Phys. Rev. D\/}
  {\bf 109} L101503 } [\eprint{2308.14796}]

\bibitem{Bucciotti:2024zyp}
Bucciotti B, Juliano L, Kuntz A and Trincherini E 2024
  \href{http://dx.doi.org/10.1103/PhysRevD.110.104048}{ {\em Phys. Rev. D\/}
  {\bf 110} 104048 } [\eprint{2405.06012}]

\bibitem{Bucciotti:2024jrv}
Bucciotti B, Juliano L, Kuntz A and Trincherini E 2024
  \href{http://dx.doi.org/10.1007/JHEP09(2024)119}{ {\em JHEP\/} {\bf 09} 119 }
  [\eprint{2406.14611}]

\bibitem{Bourg:2024jme}
Bourg P, Panosso~Macedo R, Spiers A, Leather B, Bonga B and Pound A 2025
  \href{http://dx.doi.org/10.1103/PhysRevLett.134.061401}{ {\em Phys. Rev.
  Lett.\/} {\bf 134} 061401 } [\eprint{2405.10270}]

\bibitem{Zhu:2024rej}
Zhu H {\em et~al.\/} 2024 \href{http://dx.doi.org/10.1103/PhysRevD.109.104050}{
  {\em Phys. Rev. D\/} {\bf 109} 104050 } [\eprint{2401.00805}]

\bibitem{Khera:2024yrk}
Khera N, Ma S and Yang H 2025
  \href{http://dx.doi.org/10.1103/PhysRevLett.134.211404}{ {\em Phys. Rev.
  Lett.\/} {\bf 134} 211404 } [\eprint{2410.14529}]

\bibitem{Gleiser:1998rw}
Gleiser R~J, Nicasio C~O, Price R~H and Pullin J 2000
  \href{http://dx.doi.org/10.1016/S0370-1573(99)00048-4}{ {\em Phys. Rept.\/}
  {\bf 325} 41--81 } [\eprint{gr-qc/9807077}]

\bibitem{Nicasio:1998aj}
Nicasio C~O, Gleiser R~J, Price R~H and Pullin J 1999
  \href{http://dx.doi.org/10.1103/PhysRevD.59.044024}{ {\em Phys. Rev. D\/}
  {\bf 59} 044024 } [\eprint{gr-qc/9802063}]

\bibitem{Nicasio:2000ge}
Nicasio C~O, Gleiser R and Pullin J 2000
  \href{http://dx.doi.org/10.1023/A:1001994318436}{ {\em Gen. Rel. Grav.\/}
  {\bf 32} 2021 } [\eprint{gr-qc/0001021}]

\bibitem{Brizuela:2006ne}
Brizuela D, Martin-Garcia J~M and Mena~Marugan G~A 2006
  \href{http://dx.doi.org/10.1103/PhysRevD.74.044039}{ {\em Phys. Rev. D\/}
  {\bf 74} 044039 } [\eprint{gr-qc/0607025}]

\bibitem{Brizuela:2007zza}
Brizuela D, Martin-Garcia J~M and Marugan G~A~M 2007
  \href{http://dx.doi.org/10.1103/PhysRevD.76.024004}{ {\em Phys. Rev. D\/}
  {\bf 76} 024004 } [\eprint{gr-qc/0703069}]

\bibitem{Yi:2024elj}
Yi S, Kuntz A, Barausse E, Berti E, Cheung M~H~Y, Kritos K and Maselli A 2024
  \href{http://dx.doi.org/10.1103/PhysRevD.109.124029}{ {\em Phys. Rev. D\/}
  {\bf 109} 124029 } [\eprint{2403.09767}]

\bibitem{Lagos:2024ekd}
Lagos M, Andrade T, Rafecas-Ventosa J and Hui L 2025
  \href{http://dx.doi.org/10.1103/PhysRevD.111.024018}{ {\em Phys. Rev. D\/}
  {\bf 111} 024018 } [\eprint{2411.02264}]

\bibitem{Giesler:2024hcr}
Giesler M {\em et~al.\/} 2025
  \href{http://dx.doi.org/10.1103/PhysRevD.111.084041}{ {\em Phys. Rev. D\/}
  {\bf 111} 084041 } [\eprint{2411.11269}]

\bibitem{Campanelli:1998jv}
Campanelli M and Lousto C~O 1999
  \href{http://dx.doi.org/10.1103/PhysRevD.59.124022}{ {\em Phys. Rev. D\/}
  {\bf 59} 124022 } [\eprint{gr-qc/9811019}]

\bibitem{Green:2019nam}
Green S~R, Hollands S and Zimmerman P 2020
  \href{http://dx.doi.org/10.1088/1361-6382/ab7075}{ {\em Class. Quant.
  Grav.\/} {\bf 37} 075001 } [\eprint{1908.09095}]

\bibitem{soton469806}
Spiers A~R~C 2022 {\em Second-order gravitational self-force in Kerr
  spacetime\/} Ph.D. thesis University of Southampton
  \urlprefix\url{https://eprints.soton.ac.uk/469806/}

\bibitem{Boyle:2014ioa}
Boyle M, Kidder L~E, Ossokine S and Pfeiffer H~P 2014  [\eprint{1409.4431}]

\bibitem{COHEN19755}
Cohen J and Kegeles L 1975
  \href{http://dx.doi.org/https://doi.org/10.1016/0375-9601(75)90583-6}{ {\em
  Physics Letters A\/} {\bf 54} 5--7 } ISSN 0375-9601
  \urlprefix\url{https://www.sciencedirect.com/science/article/pii/0375960175905836}

\bibitem{Wald:1978vm}
Wald R~M 1978 \href{http://dx.doi.org/10.1103/PhysRevLett.41.203}{ {\em Phys.
  Rev. Lett.\/} {\bf 41} 203--206 }

\bibitem{1974CMaPh..37..311H}
{Held} A 1974 \href{http://dx.doi.org/10.1007/BF01645944}{ {\em Commun. Math.
  Phys.\/} {\bf 37} 311--326 }

\bibitem{Spiers:2024src}
Spiers A 2024 \href{http://dx.doi.org/10.1103/PhysRevD.109.104059}{ {\em Phys.
  Rev. D\/} {\bf 109} 104059 } [\eprint{2402.00604}]

\bibitem{Vishveshwara:1970zz}
Vishveshwara C~V 1970 \href{http://dx.doi.org/10.1038/227936a0}{ {\em Nature\/}
  {\bf 227} 936--938 }

\bibitem{Press:1973zz}
Press W~H and Teukolsky S~A 1973 \href{http://dx.doi.org/10.1086/152445}{ {\em
  Astrophys. J.\/} {\bf 185} 649--674 }

\bibitem{Abrahams:1992wm}
Abrahams A, Bernstein D, Hobill D, Seidel E and Smarr L 1992
  \href{http://dx.doi.org/10.1103/PhysRevD.45.3544}{ {\em Phys. Rev. D\/} {\bf
  45} 3544--3558 }

\bibitem{Zlochower:2003yh}
Zlochower Y, Gomez R, Husa S, Lehner L and Winicour J 2003
  \href{http://dx.doi.org/10.1103/PhysRevD.68.084014}{ {\em Phys. Rev. D\/}
  {\bf 68} 084014 } [\eprint{gr-qc/0306098}]

\bibitem{Anninos:1993zj}
Anninos P, Hobill D, Seidel E, Smarr L and Suen W~M 1993
  \href{http://dx.doi.org/10.1103/PhysRevLett.71.2851}{ {\em Phys. Rev.
  Lett.\/} {\bf 71} 2851--2854 } [\eprint{gr-qc/9309016}]

\bibitem{Anninos:1994gp}
Anninos P, Hobill D, Seidel E, Smarr L and Suen W~M 1995
  \href{http://dx.doi.org/10.1103/PhysRevD.52.2044}{ {\em Phys. Rev. D\/} {\bf
  52} 2044--2058 } [\eprint{gr-qc/9408041}]

\bibitem{Sperhake:2011ik}
Sperhake U, Cardoso V, Ott C~D, Schnetter E and Witek H 2011
  \href{http://dx.doi.org/10.1103/PhysRevD.84.084038}{ {\em Phys. Rev. D\/}
  {\bf 84} 084038 } [\eprint{1105.5391}]

\bibitem{DeAmicis:2024eoy}
De~Amicis M {\em et~al.\/} 2024  [\eprint{2412.06887}]

\bibitem{Wardell:2024yoi}
Wardell B, Kavanagh C and Dolan S~R 2024  [\eprint{2406.12510}]

\bibitem{Krivan:1996da}
Krivan W, Laguna P and Papadopoulos P 1996
  \href{http://dx.doi.org/10.1103/PhysRevD.54.4728}{ {\em Phys. Rev. D\/} {\bf
  54} 4728--4734 } [\eprint{gr-qc/9606003}]

\bibitem{Krivan:1997hc}
Krivan W, Laguna P, Papadopoulos P and Andersson N 1997
  \href{http://dx.doi.org/10.1103/PhysRevD.56.3395}{ {\em Phys. Rev. D\/} {\bf
  56} 3395--3404 } [\eprint{gr-qc/9702048}]

\bibitem{Dorband:2006gg}
Dorband E~N, Berti E, Diener P, Schnetter E and Tiglio M 2006
  \href{http://dx.doi.org/10.1103/PhysRevD.74.084028}{ {\em Phys. Rev. D\/}
  {\bf 74} 084028 } [\eprint{gr-qc/0608091}]

\bibitem{Bernuzzi:2008rq}
Bernuzzi S, Nagar A and De~Pietri R 2008
  \href{http://dx.doi.org/10.1103/PhysRevD.77.044042}{ {\em Phys. Rev. D\/}
  {\bf 77} 044042 } [\eprint{0801.2090}]

\bibitem{Price:1994pm}
Price R~H and Pullin J 1994
  \href{http://dx.doi.org/10.1103/PhysRevLett.72.3297}{ {\em Phys. Rev.
  Lett.\/} {\bf 72} 3297--3300 } [\eprint{gr-qc/9402039}]

\bibitem{Misner:1960zz}
Misner C~W 1960 \href{http://dx.doi.org/10.1103/PhysRev.118.1110}{ {\em Phys.
  Rev.\/} {\bf 118} 1110--1111 }

\bibitem{Gleiser:1996yc}
Gleiser R~J, Nicasio C~O, Price R~H and Pullin J 1996
  \href{http://dx.doi.org/10.1103/PhysRevLett.77.4483}{ {\em Phys. Rev.
  Lett.\/} {\bf 77} 4483--4486 } [\eprint{gr-qc/9609022}]

\bibitem{Gleiser:1997ng}
Gleiser R~J, Nicasio C~O, Price R~H and Pullin J 1998
  \href{http://dx.doi.org/10.1103/PhysRevD.57.3401}{ {\em Phys. Rev. D\/} {\bf
  57} 3401--3407 } [\eprint{gr-qc/9710096}]

\bibitem{Abrahams:1994qu}
Abrahams A~M and Cook G~B 1994
  \href{http://dx.doi.org/10.1103/PhysRevD.50.R2364}{ {\em Phys. Rev. D\/} {\bf
  50} R2364--R2367 } [\eprint{gr-qc/9405051}]

\bibitem{Abrahams:1995wd}
Abrahams A~M and Price R~H 1996
  \href{http://dx.doi.org/10.1103/PhysRevD.53.1972}{ {\em Phys. Rev. D\/} {\bf
  53} 1972--1976 } [\eprint{gr-qc/9509020}]

\bibitem{Khanna:1999mh}
Khanna G, Baker J~G, Gleiser R~J, Laguna P, Nicasio C~O, Nollert H~P, Price R
  and Pullin J 1999 \href{http://dx.doi.org/10.1103/PhysRevLett.83.3581}{ {\em
  Phys. Rev. Lett.\/} {\bf 83} 3581--3584 } [\eprint{gr-qc/9905081}]

\bibitem{Krivan:1998er}
Krivan W and Price R~H 1999
  \href{http://dx.doi.org/10.1103/PhysRevLett.82.1358}{ {\em Phys. Rev.
  Lett.\/} {\bf 82} 1358--1361 } [\eprint{gr-qc/9810080}]

\bibitem{Krivan:1998td}
Krivan W and Price R~H 1998
  \href{http://dx.doi.org/10.1103/PhysRevD.58.104003}{ {\em Phys. Rev. D\/}
  {\bf 58} 104003 } [\eprint{gr-qc/9806017}]

\bibitem{Dain:2001iw}
Dain S 2001 \href{http://dx.doi.org/10.1103/PhysRevD.64.124002}{ {\em Phys.
  Rev. D\/} {\bf 64} 124002 } [\eprint{gr-qc/0103030}]

\bibitem{Campanelli:1997un}
Campanelli M and Lousto C~O 1998
  \href{http://dx.doi.org/10.1103/PhysRevD.58.024015}{ {\em Phys. Rev. D\/}
  {\bf 58} 024015 } [\eprint{gr-qc/9711008}]

\bibitem{Campanelli:1998uh}
Campanelli M, Krivan W and Lousto C~O 1998
  \href{http://dx.doi.org/10.1103/PhysRevD.58.024016}{ {\em Phys. Rev. D\/}
  {\bf 58} 024016 } [\eprint{gr-qc/9801067}]

\bibitem{Campanelli:1998yt}
Campanelli M, Lousto C~O, Baker J~G, Khanna G and Pullin J 1998
  \href{http://dx.doi.org/10.1103/PhysRevD.58.084019}{ {\em Phys. Rev. D\/}
  {\bf 58} 084019 } [Erratum: Phys.Rev.D 62, 069901 (2000)]
  [\eprint{gr-qc/9803058}]

\bibitem{Baker:1999sj}
Baker J~G, Brandt S, Campanelli M, Lousto C~O, Seidel E and Takahashi R 2000
  \href{http://dx.doi.org/10.1103/PhysRevD.62.127701}{ {\em Phys. Rev. D\/}
  {\bf 62} 127701 } [\eprint{gr-qc/9911017}]

\bibitem{Baker:2001sf}
Baker J~G, Campanelli M and Lousto C~O 2002
  \href{http://dx.doi.org/10.1103/PhysRevD.65.044001}{ {\em Phys. Rev. D\/}
  {\bf 65} 044001 } [\eprint{gr-qc/0104063}]

\bibitem{Campanelli:2005ia}
Campanelli M, Kelly B~J and Lousto C~O 2006
  \href{http://dx.doi.org/10.1103/PhysRevD.73.064005}{ {\em Phys. Rev. D\/}
  {\bf 73} 064005 } [\eprint{gr-qc/0510122}]

\bibitem{Campanelli:2005dd}
Campanelli M, Lousto C~O, Marronetti P and Zlochower Y 2006
  \href{http://dx.doi.org/10.1103/PhysRevLett.96.111101}{ {\em Phys. Rev.
  Lett.\/} {\bf 96} 111101 } [\eprint{gr-qc/0511048}]

\bibitem{Pazos:2010xf}
Pazos E, Brizuela D, Martin-Garcia J~M and Tiglio M 2010
  \href{http://dx.doi.org/10.1103/PhysRevD.82.104028}{ {\em Phys. Rev. D\/}
  {\bf 82} 104028 } [\eprint{1009.4665}]

\bibitem{Spiers:2023cip}
Spiers A, Pound A and Moxon J 2023
  \href{http://dx.doi.org/10.1103/PhysRevD.108.064002}{ {\em Phys. Rev. D\/}
  {\bf 108} 064002 } [\eprint{2305.19332}]

\bibitem{Toomani:2021jlo}
Toomani V, Zimmerman P, Spiers A, Hollands S, Pound A and Green S~R 2022
  \href{http://dx.doi.org/10.1088/1361-6382/ac37a5}{ {\em Class. Quant.
  Grav.\/} {\bf 39} 015019 } [\eprint{2108.04273}]

\bibitem{Hollands:2024iqp}
Hollands S and Toomani V 2024  [\eprint{2405.18604}]

\bibitem{Aksteiner:2016pjt}
Aksteiner S, Andersson L and B\"ackdahl T 2019
  \href{http://dx.doi.org/10.1103/PhysRevD.99.044043}{ {\em Phys. Rev. D\/}
  {\bf 99} 044043 } [\eprint{1601.06084}]

\bibitem{Choptuik:2015mma}
Choptuik M~W, Lehner L and Pretorius F 2015  [\eprint{1502.06853}]

\bibitem{1964AnPhy..29..304H}
{Hahn} S~G and {Lindquist} R~W 1964
  \href{http://dx.doi.org/10.1016/0003-4916(64)90223-4}{ {\em Annals of
  Physics\/} {\bf 29} 304--331 }

\bibitem{Sperhake:2008ga}
Sperhake U, Cardoso V, Pretorius F, Berti E and Gonzalez J~A 2008
  \href{http://dx.doi.org/10.1103/PhysRevLett.101.161101}{ {\em Phys. Rev.
  Lett.\/} {\bf 101} 161101 } [\eprint{0806.1738}]

\bibitem{Cardoso:2014uka}
Cardoso V, Gualtieri L, Herdeiro C and Sperhake U 2015
  \href{http://dx.doi.org/10.1007/lrr-2015-1}{ {\em Living Rev. Relativity\/}
  {\bf 18} 1 } [\eprint{1409.0014}]

\bibitem{Sperhake:2015siy}
Sperhake U, Berti E, Cardoso V and Pretorius F 2016
  \href{http://dx.doi.org/10.1103/PhysRevD.93.044012}{ {\em Phys. Rev. D\/}
  {\bf 93} 044012 } [\eprint{1511.08209}]

\bibitem{Healy:2015mla}
Healy J, Ruchlin I, Lousto C~O and Zlochower Y 2016
  \href{http://dx.doi.org/10.1103/PhysRevD.94.104020}{ {\em Phys. Rev. D\/}
  {\bf 94} 104020 } [\eprint{1506.06153}]

\bibitem{Andrade:2021rbd}
Andrade T {\em et~al.\/} 2021 \href{http://dx.doi.org/10.21105/joss.03703}{
  {\em J. Open Source Softw.\/} {\bf 6} 3703 } [\eprint{2201.03458}]

\bibitem{London:2014cma}
London L, Shoemaker D and Healy J 2014
  \href{http://dx.doi.org/10.1103/PhysRevD.90.124032}{ {\em Phys. Rev. D\/}
  {\bf 90} 124032 } [Erratum: Phys.Rev.D 94, 069902 (2016)]
  [\eprint{1404.3197}]

\bibitem{SpECwebsite}
\url{https://www.black-holes.org/for-researchers/spec}

\bibitem{Mitman:2025hgy}
Mitman K {\em et~al.\/} 2025 \href{http://dx.doi.org/10.1103/qq1g-jlnw}{ {\em
  Phys. Rev. D\/} {\bf 112} 064016 } [\eprint{2503.09678}]

\bibitem{Dyer:2024jfz}
Dyer R and Moore C~J 2025 \href{http://dx.doi.org/10.1103/PhysRevD.111.024002}{
  {\em Phys. Rev. D\/} {\bf 111} 024002 } [\eprint{2410.13935}]

\bibitem{Mourier:2020mwa}
Mourier P, Jim\'enez~Forteza X, Pook-Kolb D, Krishnan B and Schnetter E 2021
  \href{http://dx.doi.org/10.1103/PhysRevD.103.044054}{ {\em Phys. Rev. D\/}
  {\bf 103} 044054 } [\eprint{2010.15186}]

\bibitem{Khera:2023oyf}
Khera N, Ribes~Metidieri A, Bonga B, Jim\'enez~Forteza X, Krishnan B, Poisson
  E, Pook-Kolb D, Schnetter E and Yang H 2023
  \href{http://dx.doi.org/10.1103/PhysRevLett.131.231401}{ {\em Phys. Rev.
  Lett.\/} {\bf 131} 231401 } [\eprint{2306.11142}]

\bibitem{Pretorius:2005gq}
Pretorius F 2005 \href{http://dx.doi.org/10.1103/PhysRevLett.95.121101}{ {\em
  Phys. Rev. Lett.\/} {\bf 95} 121101 } [\eprint{gr-qc/0507014}]

\bibitem{Buonanno:2006ui}
Buonanno A, Cook G~B and Pretorius F 2007
  \href{http://dx.doi.org/10.1103/PhysRevD.75.124018}{ {\em Phys. Rev. D\/}
  {\bf 75} 124018 } [\eprint{gr-qc/0610122}]

\bibitem{Berti:2007fi}
Berti E, Cardoso V, Gonzalez J~A, Sperhake U, Hannam M, Husa S and Bruegmann B
  2007 \href{http://dx.doi.org/10.1103/PhysRevD.76.064034}{ {\em Phys. Rev.
  D\/} {\bf 76} 064034 } [\eprint{gr-qc/0703053}]

\bibitem{Abdalla:2006vb}
Abdalla E, Chirenti C~B~M~H and Saa A 2006
  \href{http://dx.doi.org/10.1103/PhysRevD.74.084029}{ {\em Phys. Rev. D\/}
  {\bf 74} 084029 } [\eprint{gr-qc/0609036}]

\bibitem{Chirenti:2011rc}
Chirenti C and Saa A 2011 \href{http://dx.doi.org/10.1103/PhysRevD.84.064006}{
  {\em Phys. Rev. D\/} {\bf 84} 064006 } [\eprint{1105.1681}]

\bibitem{Pfeiffer:2004qz}
Pfeiffer H~P, Kidder L~E, Scheel M~A and Shoemaker D 2005
  \href{http://dx.doi.org/10.1103/PhysRevD.71.024020}{ {\em Phys. Rev. D\/}
  {\bf 71} 024020 } [\eprint{gr-qc/0410016}]

\bibitem{Buoninfante:2024oxl}
Afshordi N {\em et~al.\/} 2024 {Black Holes Inside and Out 2024: visions for
  the future of black hole physics} ed Buoninfante L, Carballo-Rubio R, Cardoso
  V, Di~Filippo F and Eichhorn A [\eprint{2410.14414}]

\bibitem{Kamaretsos:2011um}
Kamaretsos I, Hannam M, Husa S and Sathyaprakash B~S 2012
  \href{http://dx.doi.org/10.1103/PhysRevD.85.024018}{ {\em Phys. Rev. D\/}
  {\bf 85} 024018 } [\eprint{1107.0854}]

\bibitem{Berti:2007nw}
Berti E, Cardoso V, Gonzalez J~A, Sperhake U and Bruegmann B 2008
  \href{http://dx.doi.org/10.1088/0264-9381/25/11/114035}{ {\em Class. Quant.
  Grav.\/} {\bf 25} 114035 } [\eprint{0711.1097}]

\bibitem{Kamaretsos:2012bs}
Kamaretsos I, Hannam M and Sathyaprakash B 2012
  \href{http://dx.doi.org/10.1103/PhysRevLett.109.141102}{ {\em Phys. Rev.
  Lett.\/} {\bf 109} 141102 } [\eprint{1207.0399}]

\bibitem{Borhanian:2019kxt}
Borhanian S, Arun K~G, Pfeiffer H~P and Sathyaprakash B~S 2020
  \href{http://dx.doi.org/10.1088/1361-6382/ab6a21}{ {\em Class. Quant.
  Grav.\/} {\bf 37} 065006 } [\eprint{1901.08516}]

\bibitem{Berti:2007dg}
Berti E, Cardoso V, Gonzalez J~A and Sperhake U 2007
  \href{http://dx.doi.org/10.1103/PhysRevD.75.124017}{ {\em Phys. Rev. D\/}
  {\bf 75} 124017 } [\eprint{gr-qc/0701086}]

\bibitem{Baibhav:2017jhs}
Baibhav V, Berti E, Cardoso V and Khanna G 2018
  \href{http://dx.doi.org/10.1103/PhysRevD.97.044048}{ {\em Phys. Rev. D\/}
  {\bf 97} 044048 } [\eprint{1710.02156}]

\bibitem{nollertthesis}
Nollert H~P 2000 {\em Characteristic Oscillations of Black Holes and Neutron
  Stars: From Mathematical Background to Astrophysical Applications\/}
  (Tubingen: Habilitationsschrift Der Fakultat fur Physik der
  Eberhard-Karls-Universitat, Tubingen)

\bibitem{JimenezForteza:2020cve}
Jim\'enez~Forteza X, Bhagwat S, Pani P and Ferrari V 2020
  \href{http://dx.doi.org/10.1103/PhysRevD.102.044053}{ {\em Phys. Rev. D\/}
  {\bf 102} 044053 } [\eprint{2005.03260}]

\bibitem{Carullo:2024smg}
Carullo G 2024 \href{http://dx.doi.org/10.1088/1475-7516/2024/10/061}{ {\em
  JCAP\/} {\bf 10} 061 } [\eprint{2406.19442}]

\bibitem{Bhagwat:2017tkm}
Bhagwat S, Okounkova M, Ballmer S~W, Brown D~A, Giesler M, Scheel M~A and
  Teukolsky S~A 2018 \href{http://dx.doi.org/10.1103/PhysRevD.97.104065}{ {\em
  Phys. Rev. D\/} {\bf 97} 104065 } [\eprint{1711.00926}]

\bibitem{Gao:2025zvl}
Gao L {\em et~al.\/} 2025 \href{http://dx.doi.org/10.1103/3jj6-jc8q}{ {\em
  Phys. Rev. D\/} {\bf 112} 024025 } [\eprint{2502.15921}]

\bibitem{MaganaZertuche:2025bua}
Maga\~na Zertuche L, Gao L, Finch E and Cook G~B 2025  [\eprint{2502.03155}]

\bibitem{MaganaZertuche:2024ajz}
Maga{\~n}a~Zertuche L {\em et~al.\/} 2025
  \href{http://dx.doi.org/10.1103/q7sy-g3kl}{ {\em Phys. Rev. D\/} {\bf 112}
  024077 } [\eprint{2408.05300}]

\bibitem{Pacilio:2024tdl}
Pacilio C, Bhagwat S, Nobili F and Gerosa D 2024
  \href{http://dx.doi.org/10.1103/PhysRevD.110.103037}{ {\em Phys. Rev. D\/}
  {\bf 110} 103037 } [\eprint{2408.05276}]

\bibitem{Nobili:2025ydt}
Nobili F, Bhagwat S, Pacilio C and Gerosa D 2025
  \href{http://dx.doi.org/10.1103/cl3k-3xt2}{ {\em Phys. Rev. D\/} {\bf 112}
  044058 } [\eprint{2504.17021}]

\bibitem{pacilio_2024_13220424}
Pacilio C, Swetha B, Francesco N and Gerosa D 2024 postmerger
  \urlprefix\url{https://doi.org/10.5281/zenodo.13220424}

\bibitem{Moore:2014pda}
Moore C~J and Gair J~R 2014
  \href{http://dx.doi.org/10.1103/PhysRevLett.113.251101}{ {\em Phys. Rev.
  Lett.\/} {\bf 113} 251101 } [\eprint{1412.3657}]

\bibitem{Breschi:2022ens}
Breschi M, Gamba R, Borhanian S, Carullo G and Bernuzzi S 2022
  [\eprint{2205.09979}]

\bibitem{Pompili:2024yec}
Pompili L, Buonanno A and P\"urrer M 2024  [\eprint{2410.16859}]

\bibitem{London:2018gaq}
London L~T 2020 \href{http://dx.doi.org/10.1103/PhysRevD.102.084052}{ {\em
  Phys. Rev. D\/} {\bf 102} 084052 } [\eprint{1801.08208}]

\bibitem{gatechcatalog}
Jani K, Healy J, Clark J~A, London L, Laguna P and Shoemaker D 2016 {Georgia
  Tech} catalog of binary black hole simulations
  \urlprefix\url{http://www.einstein.gatech.edu/catalog/}

\bibitem{Hamilton:2023qkv}
Hamilton E {\em et~al.\/} 2024
  \href{http://dx.doi.org/10.1103/PhysRevD.109.044032}{ {\em Phys. Rev. D\/}
  {\bf 109} 044032 } [\eprint{2303.05419}]

\bibitem{Boyle:2019kee}
Boyle M {\em et~al.\/} 2019 \href{http://dx.doi.org/10.1088/1361-6382/ab34e2}{
  {\em Class. Quant. Grav.\/} {\bf 36} 195006 } [\eprint{1904.04831}]

\bibitem{Kidder:2007rt}
Kidder L~E 2008 \href{http://dx.doi.org/10.1103/PhysRevD.77.044016}{ {\em Phys.
  Rev. D\/} {\bf 77} 044016 } [\eprint{0710.0614}]

\bibitem{Pan:2010hz}
Pan Y, Buonanno A, Fujita R, Racine E and Tagoshi H 2011
  \href{http://dx.doi.org/10.1103/PhysRevD.83.064003}{ {\em Phys. Rev. D\/}
  {\bf 83} 064003 } [Erratum: Phys.Rev.D 87, 109901 (2013)]
  [\eprint{1006.0431}]

\bibitem{Cotesta:2018fcv}
Cotesta R, Buonanno A, Boh\'e A, Taracchini A, Hinder I and Ossokine S 2018
  \href{http://dx.doi.org/10.1103/PhysRevD.98.084028}{ {\em Phys. Rev. D\/}
  {\bf 98} 084028 } [\eprint{1803.10701}]

\bibitem{Pan:2013rra}
Pan Y, Buonanno A, Taracchini A, Kidder L~E, Mrou\'e A~H, Pfeiffer H~P, Scheel
  M~A and Szil\'agyi B 2014
  \href{http://dx.doi.org/10.1103/PhysRevD.89.084006}{ {\em Phys. Rev. D\/}
  {\bf 89} 084006 } [\eprint{1307.6232}]

\bibitem{Taracchini:2013rva}
Taracchini A {\em et~al.\/} 2014
  \href{http://dx.doi.org/10.1103/PhysRevD.89.061502}{ {\em Phys. Rev. D\/}
  {\bf 89} 061502 } [\eprint{1311.2544}]

\bibitem{Babak:2016tgq}
Babak S, Taracchini A and Buonanno A 2017
  \href{http://dx.doi.org/10.1103/PhysRevD.95.024010}{ {\em Phys. Rev. D\/}
  {\bf 95} 024010 } [\eprint{1607.05661}]

\bibitem{Damour:2014yha}
Damour T and Nagar A 2014 \href{http://dx.doi.org/10.1103/PhysRevD.90.024054}{
  {\em Phys. Rev. D\/} {\bf 90} 024054 } [\eprint{1406.0401}]

\bibitem{Bhagwat:2019dtm}
Bhagwat S, Forteza X~J, Pani P and Ferrari V 2020
  \href{http://dx.doi.org/10.1103/PhysRevD.101.044033}{ {\em Phys. Rev. D\/}
  {\bf 101} 044033 } [\eprint{1910.08708}]

\bibitem{Forteza:2021wfq}
Forteza X~J and Mourier P 2021
  \href{http://dx.doi.org/10.1103/PhysRevD.104.124072}{ {\em Phys. Rev. D\/}
  {\bf 104} 124072 } [\eprint{2107.11829}]

\bibitem{Zhu:2023mzv}
Zhu H, Ripley J~L, C\'ardenas-Avenda\~no A and Pretorius F 2024
  \href{http://dx.doi.org/10.1103/PhysRevD.109.044010}{ {\em Phys. Rev. D\/}
  {\bf 109} 044010 } [\eprint{2309.13204}]

\bibitem{Gualtieri:2008ux}
Gualtieri L, Berti E, Cardoso V and Sperhake U 2008
  \href{http://dx.doi.org/10.1103/PhysRevD.78.044024}{ {\em Phys. Rev. D\/}
  {\bf 78} 044024 } [\eprint{0805.1017}]

\bibitem{Campanelli:2008nk}
Campanelli M, Lousto C~O, Nakano H and Zlochower Y 2009
  \href{http://dx.doi.org/10.1103/PhysRevD.79.084010}{ {\em Phys. Rev. D\/}
  {\bf 79} 084010 } [\eprint{0808.0713}]

\bibitem{Buonanno:2002fy}
Buonanno A, Chen Y~b and Vallisneri M 2003
  \href{http://dx.doi.org/10.1103/PhysRevD.67.104025}{ {\em Phys. Rev. D\/}
  {\bf 67} 104025 } [Erratum: Phys.Rev.D 74, 029904 (2006)]
  [\eprint{gr-qc/0211087}]

\bibitem{Schmidt:2012rh}
Schmidt P, Hannam M and Husa S 2012
  \href{http://dx.doi.org/10.1103/PhysRevD.86.104063}{ {\em Phys. Rev. D\/}
  {\bf 86} 104063 } [\eprint{1207.3088}]

\bibitem{OShaughnessy:2012iol}
O'Shaughnessy R, London L, Healy J and Shoemaker D 2013
  \href{http://dx.doi.org/10.1103/PhysRevD.87.044038}{ {\em Phys. Rev. D\/}
  {\bf 87} 044038 } [\eprint{1209.3712}]

\bibitem{Boyle:2011gg}
Boyle M, Owen R and Pfeiffer H~P 2011
  \href{http://dx.doi.org/10.1103/PhysRevD.84.124011}{ {\em Phys. Rev. D\/}
  {\bf 84} 124011 } [\eprint{1110.2965}]

\bibitem{Hamilton:2023znn}
Hamilton E, London L and Hannam M 2023
  \href{http://dx.doi.org/10.1103/PhysRevD.107.104035}{ {\em Phys. Rev. D\/}
  {\bf 107} 104035 } [\eprint{2301.06558}]

\bibitem{Ossokine:2020kjp}
Ossokine S {\em et~al.\/} 2020
  \href{http://dx.doi.org/10.1103/PhysRevD.102.044055}{ {\em Phys. Rev. D\/}
  {\bf 102} 044055 } [\eprint{2004.09442}]

\bibitem{Ramos-Buades:2021adz}
Ramos-Buades A, Buonanno A, Khalil M and Ossokine S 2022
  \href{http://dx.doi.org/10.1103/PhysRevD.105.044035}{ {\em Phys. Rev. D\/}
  {\bf 105} 044035 } [\eprint{2112.06952}]

\bibitem{Finch:2021iip}
Finch E and Moore C~J 2021
  \href{http://dx.doi.org/10.1103/PhysRevD.103.084048}{ {\em Phys. Rev. D\/}
  {\bf 103} 084048 } [\eprint{2102.07794}]

\bibitem{Zhu:2023fnf}
Zhu H {\em et~al.\/} 2025 \href{http://dx.doi.org/10.1103/PhysRevD.111.064052}{
  {\em Phys. Rev. D\/} {\bf 111} 064052 } [\eprint{2312.08588}]

\bibitem{Ghosh:2023mhc}
Ghosh S, Kolitsidou P and Hannam M 2024
  \href{http://dx.doi.org/10.1103/PhysRevD.109.024061}{ {\em Phys. Rev. D\/}
  {\bf 109} 024061 } [\eprint{2310.16980}]

\bibitem{Mitman:2021xkq}
Mitman K {\em et~al.\/} 2021
  \href{http://dx.doi.org/10.1103/PhysRevD.104.024051}{ {\em Phys. Rev. D\/}
  {\bf 104} 024051 } [\eprint{2105.02300}]

\bibitem{Mitman:2024uss}
Mitman K {\em et~al.\/} 2024 \href{http://dx.doi.org/10.1088/1361-6382/ad83c2}{
  {\em Class. Quant. Grav.\/} {\bf 41} 223001 } [\eprint{2405.08868}]

\bibitem{Peters:1963ux}
Peters P~C and Mathews J 1963 \href{http://dx.doi.org/10.1103/PhysRev.131.435}{
  {\em Phys. Rev.\/} {\bf 131} 435--439 }

\bibitem{Mapelli:2021taw}
Mapelli M 2021 {\em {Formation Channels of Single and Binary Stellar-Mass Black
  Holes}\/} [\eprint{2106.00699}]

\bibitem{Gupta:2024gun}
Gupta A {\em et~al.\/} 2025
  \href{http://dx.doi.org/10.21468/SciPostPhysCommRep.5}{ {\em SciPost Phys.
  Comm. Rep.\/}  5 } [\eprint{2405.02197}]

\bibitem{ritcatalog}
Campanelli M, Healy J, Lousto C and Zlochower Y 2022 {CCRG@RIT Catalog of
  Numerical Simulations} \urlprefix\url{http://ccrg.rit.edu/~RITCatalog/}

\bibitem{Carullo:2023kvj}
Carullo G, Albanesi S, Nagar A, Gamba R, Bernuzzi S, Andrade T and Trenado J
  2024 \href{http://dx.doi.org/10.1103/PhysRevLett.132.101401}{ {\em Phys. Rev.
  Lett.\/} {\bf 132} 101401 } [\eprint{2309.07228}]

\bibitem{Ramos-Buades:2022lgf}
Ramos-Buades A, van~de Meent M, Pfeiffer H~P, R\"uter H~R, Scheel M~A, Boyle M
  and Kidder L~E 2022 \href{http://dx.doi.org/10.1103/PhysRevD.106.124040}{
  {\em Phys. Rev. D\/} {\bf 106} 124040 } [\eprint{2209.03390}]

\bibitem{Shaikh:2023ypz}
Shaikh M~A, Varma V, Pfeiffer H~P, Ramos-Buades A and van~de Meent M 2023
  \href{http://dx.doi.org/10.1103/PhysRevD.108.104007}{ {\em Phys. Rev. D\/}
  {\bf 108} 104007 } [\eprint{2302.11257}]

\bibitem{Boschini:2024scu}
Boschini M, Loutrel N, Gerosa D and Fumagalli G 2025
  \href{http://dx.doi.org/10.1103/PhysRevD.111.024008}{ {\em Phys. Rev. D\/}
  {\bf 111} 024008 } [\eprint{2411.00098}]

\bibitem{Albanesi:2023bgi}
Albanesi S, Bernuzzi S, Damour T, Nagar A and Placidi A 2023
  \href{http://dx.doi.org/10.1103/PhysRevD.108.084037}{ {\em Phys. Rev. D\/}
  {\bf 108} 084037 } [\eprint{2305.19336}]

\bibitem{Jimenez-Forteza:2016oae}
Jim\'enez-Forteza X, Keitel D, Husa S, Hannam M, Khan S and P\"urrer M 2017
  \href{http://dx.doi.org/10.1103/PhysRevD.95.064024}{ {\em Phys. Rev. D\/}
  {\bf 95} 064024 } [\eprint{1611.00332}]

\bibitem{Varma:2018aht}
Varma V, Gerosa D, Stein L~C, H\'ebert F and Zhang H 2019
  \href{http://dx.doi.org/10.1103/PhysRevLett.122.011101}{ {\em Phys. Rev.
  Lett.\/} {\bf 122} 011101 } [\eprint{1809.09125}]

\bibitem{Damour:2007xr}
Damour T and Nagar A 2007 \href{http://dx.doi.org/10.1103/PhysRevD.76.064028}{
  {\em Phys. Rev. D\/} {\bf 76} 064028 } [\eprint{0705.2519}]

\bibitem{Bernuzzi:2010xj}
Bernuzzi S, Nagar A and Zenginoglu A 2011
  \href{http://dx.doi.org/10.1103/PhysRevD.83.064010}{ {\em Phys. Rev. D\/}
  {\bf 83} 064010 } [\eprint{1012.2456}]

\bibitem{Taracchini:2014zpa}
Taracchini A, Buonanno A, Khanna G and Hughes S~A 2014
  \href{http://dx.doi.org/10.1103/PhysRevD.90.084025}{ {\em Phys. Rev. D\/}
  {\bf 90} 084025 } [\eprint{1404.1819}]

\bibitem{Oshita:2022yry}
Oshita N and Tsuna D 2023 \href{http://dx.doi.org/10.1103/PhysRevD.108.104031}{
  {\em Phys. Rev. D\/} {\bf 108} 104031 } [\eprint{2210.14049}]

\bibitem{Watarai:2024vni}
Watarai D, Oshita N and Tsuna D 2024  [\eprint{2403.12380}]

\bibitem{Buonanno:1998gg}
Buonanno A and Damour T 1999
  \href{http://dx.doi.org/10.1103/PhysRevD.59.084006}{ {\em Phys. Rev. D\/}
  {\bf 59} 084006 } [\eprint{gr-qc/9811091}]

\bibitem{Buonanno:2000ef}
Buonanno A and Damour T 2000
  \href{http://dx.doi.org/10.1103/PhysRevD.62.064015}{ {\em Phys. Rev. D\/}
  {\bf 62} 064015 } [\eprint{gr-qc/0001013}]

\bibitem{Damour:2000we}
Damour T, Jaranowski P and Schaefer G 2000
  \href{http://dx.doi.org/10.1103/PhysRevD.62.084011}{ {\em Phys. Rev. D\/}
  {\bf 62} 084011 } [\eprint{gr-qc/0005034}]

\bibitem{Damour:2001tu}
Damour T 2001 \href{http://dx.doi.org/10.1103/PhysRevD.64.124013}{ {\em Phys.
  Rev. D\/} {\bf 64} 124013 } [\eprint{gr-qc/0103018}]

\bibitem{Buonanno:2005xu}
Buonanno A, Chen Y and Damour T 2006
  \href{http://dx.doi.org/10.1103/PhysRevD.74.104005}{ {\em Phys. Rev. D\/}
  {\bf 74} 104005 } [\eprint{gr-qc/0508067}]

\bibitem{Baker:2005vv}
Baker J~G, Centrella J, Choi D~I, Koppitz M and van Meter J 2006
  \href{http://dx.doi.org/10.1103/PhysRevLett.96.111102}{ {\em Phys. Rev.
  Lett.\/} {\bf 96} 111102 } [\eprint{gr-qc/0511103}]

\bibitem{Damour:2008gu}
Damour T, Iyer B~R and Nagar A 2009
  \href{http://dx.doi.org/10.1103/PhysRevD.79.064004}{ {\em Phys. Rev. D\/}
  {\bf 79} 064004 } [\eprint{0811.2069}]

\bibitem{Damour:2007yf}
Damour T and Nagar A 2008 \href{http://dx.doi.org/10.1103/PhysRevD.77.024043}{
  {\em Phys. Rev. D\/} {\bf 77} 024043 } [\eprint{0711.2628}]

\bibitem{Messina:2018ghh}
Messina F, Maldarella A and Nagar A 2018
  \href{http://dx.doi.org/10.1103/PhysRevD.97.084016}{ {\em Phys. Rev. D\/}
  {\bf 97} 084016 } [\eprint{1801.02366}]

\bibitem{vandeMeent:2023ols}
van~de Meent M, Buonanno A, Mihaylov D~P, Ossokine S, Pompili L, Warburton N,
  Pound A, Wardell B, Durkan L and Miller J 2023
  \href{http://dx.doi.org/10.1103/PhysRevD.108.124038}{ {\em Phys. Rev. D\/}
  {\bf 108} 124038 } [\eprint{2303.18026}]

\bibitem{Nagar:2024oyk}
Nagar A, Chiaramello D, Gamba R, Albanesi S, Bernuzzi S, Fantini V, Panzeri M
  and Rettegno P 2025 \href{http://dx.doi.org/10.1103/PhysRevD.111.064050}{
  {\em Phys. Rev. D\/} {\bf 111} 064050 } [\eprint{2407.04762}]

\bibitem{Buonanno:2007pf}
Buonanno A, Pan Y, Baker J~G, Centrella J, Kelly B~J, McWilliams S~T and van
  Meter J~R 2007 \href{http://dx.doi.org/10.1103/PhysRevD.76.104049}{ {\em
  Phys. Rev. D\/} {\bf 76} 104049 } [\eprint{0706.3732}]

\bibitem{Damour:2007vq}
Damour T, Nagar A, Dorband E~N, Pollney D and Rezzolla L 2008
  \href{http://dx.doi.org/10.1103/PhysRevD.77.084017}{ {\em Phys. Rev. D\/}
  {\bf 77} 084017 } [\eprint{0712.3003}]

\bibitem{Damour:2008te}
Damour T, Nagar A, Hannam M, Husa S and Bruegmann B 2008
  \href{http://dx.doi.org/10.1103/PhysRevD.78.044039}{ {\em Phys. Rev. D\/}
  {\bf 78} 044039 } [\eprint{0803.3162}]

\bibitem{Damour:2009kr}
Damour T and Nagar A 2009 \href{http://dx.doi.org/10.1103/PhysRevD.79.081503}{
  {\em Phys. Rev. D\/} {\bf 79} 081503 } [\eprint{0902.0136}]

\bibitem{Buonanno:2009qa}
Buonanno A, Pan Y, Pfeiffer H~P, Scheel M~A, Buchman L~T and Kidder L~E 2009
  \href{http://dx.doi.org/10.1103/PhysRevD.79.124028}{ {\em Phys. Rev. D\/}
  {\bf 79} 124028 } [\eprint{0902.0790}]

\bibitem{Pan:2011gk}
Pan Y, Buonanno A, Boyle M, Buchman L~T, Kidder L~E, Pfeiffer H~P and Scheel
  M~A 2011 \href{http://dx.doi.org/10.1103/PhysRevD.84.124052}{ {\em Phys. Rev.
  D\/} {\bf 84} 124052 } [\eprint{1106.1021}]

\bibitem{Taracchini:2012ig}
Taracchini A, Pan Y, Buonanno A, Barausse E, Boyle M, Chu T, Lovelace G,
  Pfeiffer H~P and Scheel M~A 2012
  \href{http://dx.doi.org/10.1103/PhysRevD.86.024011}{ {\em Phys. Rev. D\/}
  {\bf 86} 024011 } [\eprint{1202.0790}]

\bibitem{Damour:2012ky}
Damour T, Nagar A and Bernuzzi S 2013
  \href{http://dx.doi.org/10.1103/PhysRevD.87.084035}{ {\em Phys. Rev. D\/}
  {\bf 87} 084035 } [\eprint{1212.4357}]

\bibitem{Damour:2013tla}
Damour T, Nagar A and Villain L 2014
  \href{http://dx.doi.org/10.1103/PhysRevD.89.024031}{ {\em Phys. Rev. D\/}
  {\bf 89} 024031 } [\eprint{1307.2868}]

\bibitem{Davis:1972dm}
Davis M, Ruffini R, Tiomno J and Zerilli F 1972
  \href{http://dx.doi.org/10.1103/PhysRevLett.28.1352}{ {\em Phys. Rev.
  Lett.\/} {\bf 28} 1352--1355 }

\bibitem{Damour:2007cb}
Damour T and Nagar A 2007 \href{http://dx.doi.org/10.1103/PhysRevD.76.044003}{
  {\em Phys. Rev. D\/} {\bf 76} 044003 } [\eprint{0704.3550}]

\bibitem{Hofmann:2016yih}
Hofmann F, Barausse E and Rezzolla L 2016
  \href{http://dx.doi.org/10.3847/2041-8205/825/2/L19}{ {\em Astrophys. J.
  Lett.\/} {\bf 825} L19 } [\eprint{1605.01938}]

\bibitem{Baker:2008mj}
Baker J~G, Boggs W~D, Centrella J, Kelly B~J, McWilliams S~T and van Meter J~R
  2008 \href{http://dx.doi.org/10.1103/PhysRevD.78.044046}{ {\em Phys. Rev.
  D\/} {\bf 78} 044046 } [\eprint{0805.1428}]

\bibitem{DelPozzo:2016kmd}
Del~Pozzo W and Nagar A 2017
  \href{http://dx.doi.org/10.1103/PhysRevD.95.124034}{ {\em Phys. Rev. D\/}
  {\bf 95} 124034 } [\eprint{1606.03952}]

\bibitem{Bohe:2016gbl}
Boh\'e A {\em et~al.\/} 2017
  \href{http://dx.doi.org/10.1103/PhysRevD.95.044028}{ {\em Phys. Rev. D\/}
  {\bf 95} 044028 } [\eprint{1611.03703}]

\bibitem{Nagar:2020pcj}
Nagar A, Riemenschneider G, Pratten G, Rettegno P and Messina F 2020
  \href{http://dx.doi.org/10.1103/PhysRevD.102.024077}{ {\em Phys. Rev. D\/}
  {\bf 102} 024077 } [\eprint{2001.09082}]

\bibitem{Pompili:2023tna}
Pompili L {\em et~al.\/} 2023
  \href{http://dx.doi.org/10.1103/PhysRevD.108.124035}{ {\em Phys. Rev. D\/}
  {\bf 108} 124035 } [\eprint{2303.18039}]

\bibitem{Nagar:2023zxh}
Nagar A, Rettegno P, Gamba R, Albanesi S, Albertini A and Bernuzzi S 2023
  \href{http://dx.doi.org/10.1103/PhysRevD.108.124018}{ {\em Phys. Rev. D\/}
  {\bf 108} 124018 } [\eprint{2304.09662}]

\bibitem{Estelles:2020osj}
Estell\'es H, Ramos-Buades A, Husa S, Garc\'\i{}a-Quir\'os C, Colleoni M,
  Haegel L and Jaume R 2021
  \href{http://dx.doi.org/10.1103/PhysRevD.103.124060}{ {\em Phys. Rev. D\/}
  {\bf 103} 124060 } [\eprint{2004.08302}]

\bibitem{Estelles:2020twz}
Estell\'es H, Husa S, Colleoni M, Keitel D, Mateu-Lucena M,
  Garc\'\i{}a-Quir\'os C, Ramos-Buades A and Borchers A 2022
  \href{http://dx.doi.org/10.1103/PhysRevD.105.084039}{ {\em Phys. Rev. D\/}
  {\bf 105} 084039 } [\eprint{2012.11923}]

\bibitem{Estelles:2021gvs}
Estell\'es H, Colleoni M, Garc\'\i{}a-Quir\'os C, Husa S, Keitel D,
  Mateu-Lucena M, Planas M~d~L and Ramos-Buades A 2022
  \href{http://dx.doi.org/10.1103/PhysRevD.105.084040}{ {\em Phys. Rev. D\/}
  {\bf 105} 084040 } [\eprint{2105.05872}]

\bibitem{Nagar:2019wds}
Nagar A, Pratten G, Riemenschneider G and Gamba R 2020
  \href{http://dx.doi.org/10.1103/PhysRevD.101.024041}{ {\em Phys. Rev. D\/}
  {\bf 101} 024041 } [\eprint{1904.09550}]

\bibitem{Kelly:2012nd}
Kelly B~J and Baker J~G 2013
  \href{http://dx.doi.org/10.1103/PhysRevD.87.084004}{ {\em Phys. Rev. D\/}
  {\bf 87} 084004 } [\eprint{1212.5553}]

\bibitem{Berti:2014fga}
Berti E and Klein A 2014 \href{http://dx.doi.org/10.1103/PhysRevD.90.064012}{
  {\em Phys. Rev. D\/} {\bf 90} 064012 } [\eprint{1408.1860}]

\bibitem{Albanesi:2024fts}
Albanesi S 2025 \href{http://dx.doi.org/10.1103/PhysRevD.111.L121501}{ {\em
  Phys. Rev. D\/} {\bf 111} L121501 } [\eprint{2411.04024}]

\bibitem{Brito:2018rfr}
Brito R, Buonanno A and Raymond V 2018
  \href{http://dx.doi.org/10.1103/PhysRevD.98.084038}{ {\em Phys. Rev. D\/}
  {\bf 98} 084038 } [\eprint{1805.00293}]

\bibitem{Ghosh:2021mrv}
Ghosh A, Brito R and Buonanno A 2021
  \href{http://dx.doi.org/10.1103/PhysRevD.103.124041}{ {\em Phys. Rev. D\/}
  {\bf 103} 124041 } [\eprint{2104.01906}]

\bibitem{Maggio:2022hre}
Maggio E, Silva H~O, Buonanno A and Ghosh A 2023
  \href{http://dx.doi.org/10.1103/PhysRevD.108.024043}{ {\em Phys. Rev. D\/}
  {\bf 108} 024043 } [\eprint{2212.09655}]

\bibitem{Toubiana:2023cwr}
Toubiana A, Pompili L, Buonanno A, Gair J~R and Katz M~L 2024
  \href{http://dx.doi.org/10.1103/PhysRevD.109.104019}{ {\em Phys. Rev. D\/}
  {\bf 109} 104019 } [\eprint{2307.15086}]

\bibitem{Pompili:2025cdc}
Pompili L, Maggio E, Silva H~O and Buonanno A 2025
  \href{http://dx.doi.org/10.1103/ng8w-98sz}{ {\em Phys. Rev. D\/} {\bf 111}
  124040 } [\eprint{2504.10130}]

\bibitem{Gennari:2023gmx}
Gennari V, Carullo G and Del~Pozzo W 2024
  \href{http://dx.doi.org/10.1140/epjc/s10052-024-12550-x}{ {\em Eur. Phys. J.
  C\/} {\bf 84} 233 } [\eprint{2312.12515}]

\bibitem{Hinderer:2017jcs}
Hinderer T and Babak S 2017
  \href{http://dx.doi.org/10.1103/PhysRevD.96.104048}{ {\em Phys. Rev. D\/}
  {\bf 96} 104048 } [\eprint{1707.08426}]

\bibitem{Chiaramello:2020ehz}
Chiaramello D and Nagar A 2020
  \href{http://dx.doi.org/10.1103/PhysRevD.101.101501}{ {\em Phys. Rev. D\/}
  {\bf 101} 101501 } [\eprint{2001.11736}]

\bibitem{Nagar:2020xsk}
Nagar A, Rettegno P, Gamba R and Bernuzzi S 2021
  \href{http://dx.doi.org/10.1103/PhysRevD.103.064013}{ {\em Phys. Rev. D\/}
  {\bf 103} 064013 } [\eprint{2009.12857}]

\bibitem{Khalil:2021txt}
Khalil M, Buonanno A, Steinhoff J and Vines J 2021
  \href{http://dx.doi.org/10.1103/PhysRevD.104.024046}{ {\em Phys. Rev. D\/}
  {\bf 104} 024046 } [\eprint{2104.11705}]

\bibitem{Placidi:2021rkh}
Placidi A, Albanesi S, Nagar A, Orselli M, Bernuzzi S and Grignani G 2022
  \href{http://dx.doi.org/10.1103/PhysRevD.105.104030}{ {\em Phys. Rev. D\/}
  {\bf 105} 104030 } [\eprint{2112.05448}]

\bibitem{Gamba:2024cvy}
Gamba R, Chiaramello D and Neogi S 2024
  \href{http://dx.doi.org/10.1103/PhysRevD.110.024031}{ {\em Phys. Rev. D\/}
  {\bf 110} 024031 } [\eprint{2404.15408}]

\bibitem{Nagar:2024dzj}
Nagar A, Gamba R, Rettegno P, Fantini V and Bernuzzi S 2024
  \href{http://dx.doi.org/10.1103/PhysRevD.110.084001}{ {\em Phys. Rev. D\/}
  {\bf 110} 084001 } [\eprint{2404.05288}]

\bibitem{Gamboa:2024hli}
Gamboa A {\em et~al.\/} 2024  [\eprint{2412.12823}]

\bibitem{Gamba:2021gap}
Gamba R, Breschi M, Carullo G, Albanesi S, Rettegno P, Bernuzzi S and Nagar A
  2023 \href{http://dx.doi.org/10.1038/s41550-022-01813-w}{ {\em Nature
  Astron.\/} {\bf 7} 11--17 } [\eprint{2106.05575}]

\bibitem{Andrade:2023trh}
Andrade T {\em et~al.\/} 2024
  \href{http://dx.doi.org/10.1103/PhysRevD.109.084025}{ {\em Phys. Rev. D\/}
  {\bf 109} 084025 } [\eprint{2307.08697}]

\bibitem{Albanesi:2024xus}
Albanesi S, Rashti A, Zappa F, Gamba R, Cook W, Daszuta B, Bernuzzi S, Nagar A
  and Radice D 2025 \href{http://dx.doi.org/10.1103/PhysRevD.111.024069}{ {\em
  Phys. Rev. D\/} {\bf 111} 024069 } [\eprint{2405.20398}]

\bibitem{Albanesi:2021rby}
Albanesi S, Nagar A and Bernuzzi S 2021
  \href{http://dx.doi.org/10.1103/PhysRevD.104.024067}{ {\em Phys. Rev. D\/}
  {\bf 104} 024067 } [\eprint{2104.10559}]

\bibitem{Gamba:2021ydi}
Gamba R, Ak\c{c}ay S, Bernuzzi S and Williams J 2022
  \href{http://dx.doi.org/10.1103/PhysRevD.106.024020}{ {\em Phys. Rev. D\/}
  {\bf 106} 024020 } [\eprint{2111.03675}]

\bibitem{Ramos-Buades:2023ehm}
Ramos-Buades A, Buonanno A, Estell\'es H, Khalil M, Mihaylov D~P, Ossokine S,
  Pompili L and Shiferaw M 2023
  \href{http://dx.doi.org/10.1103/PhysRevD.108.124037}{ {\em Phys. Rev. D\/}
  {\bf 108} 124037 } [\eprint{2303.18046}]

\bibitem{Schmidt:2010it}
Schmidt P, Hannam M, Husa S and Ajith P 2011
  \href{http://dx.doi.org/10.1103/PhysRevD.84.024046}{ {\em Phys. Rev. D\/}
  {\bf 84} 024046 } [\eprint{1012.2879}]

\bibitem{OShaughnessy:2011pmr}
O'Shaughnessy R, Vaishnav B, Healy J, Meeks Z and Shoemaker D 2011
  \href{http://dx.doi.org/10.1103/PhysRevD.84.124002}{ {\em Phys. Rev. D\/}
  {\bf 84} 124002 } [\eprint{1109.5224}]

\bibitem{Ramos-Buades:2020noq}
Ramos-Buades A, Schmidt P, Pratten G and Husa S 2020
  \href{http://dx.doi.org/10.1103/PhysRevD.101.103014}{ {\em Phys. Rev. D\/}
  {\bf 101} 103014 } [\eprint{2001.10936}]

\bibitem{Boyle:2013nka}
Boyle M 2013 \href{http://dx.doi.org/10.1103/PhysRevD.87.104006}{ {\em Phys.
  Rev. D\/} {\bf 87} 104006 } [\eprint{1302.2919}]

\bibitem{Lim:2022veo}
Lim H, Hughes S~A and Khanna G 2022
  \href{http://dx.doi.org/10.1103/PhysRevD.105.124030}{ {\em Phys. Rev. D\/}
  {\bf 105} 124030 } [\eprint{2204.06007}]

\bibitem{Siegel:2023lxl}
Siegel H, Isi M and Farr W~M 2023
  \href{http://dx.doi.org/10.1103/PhysRevD.108.064008}{ {\em Phys. Rev. D\/}
  {\bf 108} 064008 } [\eprint{2307.11975}]

\bibitem{Poisson:2002jz}
Poisson E 2002 \href{http://dx.doi.org/10.1103/PhysRevD.66.044008}{ {\em Phys.
  Rev. D\/} {\bf 66} 044008 } [\eprint{gr-qc/0205018}]

\bibitem{Hod:2009my}
Hod S 2009 \href{http://dx.doi.org/10.1088/0264-9381/26/2/028001}{ {\em Class.
  Quant. Grav.\/} {\bf 26} 028001 } [\eprint{0902.0237}]

\bibitem{Zenginoglu:2008wc}
Zenginoglu A 2008 \href{http://dx.doi.org/10.1088/0264-9381/25/17/175013}{ {\em
  Class. Quant. Grav.\/} {\bf 25} 175013 } [\eprint{0803.2018}]

\bibitem{Barack:1998bv}
Barack L 1999 \href{http://dx.doi.org/10.1103/PhysRevD.59.044016}{ {\em Phys.
  Rev. D\/} {\bf 59} 044016 } [\eprint{gr-qc/9811027}]

\bibitem{Barack:1998bw}
Barack L 1999 \href{http://dx.doi.org/10.1103/PhysRevD.59.044017}{ {\em Phys.
  Rev. D\/} {\bf 59} 044017 } [\eprint{gr-qc/9811028}]

\bibitem{Price:2004mm}
Price R~H and Burko L~M 2004
  \href{http://dx.doi.org/10.1103/PhysRevD.70.084039}{ {\em Phys. Rev. D\/}
  {\bf 70} 084039 } [\eprint{gr-qc/0408077}]

\bibitem{Gundlach:1993tn}
Gundlach C, Price R~H and Pullin J 1994
  \href{http://dx.doi.org/10.1103/PhysRevD.49.890}{ {\em Phys. Rev. D\/} {\bf
  49} 890--899 } [\eprint{gr-qc/9307010}]

\bibitem{Burko:1997tb}
Burko L~M and Ori A 1997 \href{http://dx.doi.org/10.1103/PhysRevD.56.7820}{
  {\em Phys. Rev. D\/} {\bf 56} 7820--7832 } [\eprint{gr-qc/9703067}]

\bibitem{Zenginoglu:2009ey}
Zenginoglu A 2010 \href{http://dx.doi.org/10.1088/0264-9381/27/4/045015}{ {\em
  Class. Quant. Grav.\/} {\bf 27} 045015 } [\eprint{0911.2450}]

\bibitem{Tiglio:2007jp}
Tiglio M, Kidder L~E and Teukolsky S~A 2008
  \href{http://dx.doi.org/10.1088/0264-9381/25/10/105022}{ {\em Class. Quant.
  Grav.\/} {\bf 25} 105022 } [\eprint{0712.2472}]

\bibitem{Gleiser:2007ti}
Gleiser R~J, Price R~H and Pullin J 2008
  \href{http://dx.doi.org/10.1088/0264-9381/25/7/072001}{ {\em Class. Quant.
  Grav.\/} {\bf 25} 072001 } [\eprint{0710.4183}]

\bibitem{Zenginoglu:2009hd}
Zenginoglu A and Tiglio M 2009
  \href{http://dx.doi.org/10.1103/PhysRevD.80.024044}{ {\em Phys. Rev. D\/}
  {\bf 80} 024044 } [\eprint{0906.3342}]

\bibitem{Burko:2007ju}
Burko L~M and Khanna G 2009
  \href{http://dx.doi.org/10.1088/0264-9381/26/1/015014}{ {\em Class. Quant.
  Grav.\/} {\bf 26} 015014 } [\eprint{0711.0960}]

\bibitem{Zenginoglu:2012us}
Zengino\u{g}lu A, Khanna G and Burko L~M 2014
  \href{http://dx.doi.org/10.1007/s10714-014-1672-8}{ {\em Gen. Rel. Grav.\/}
  {\bf 46} 1672 } [\eprint{1208.5839}]

\bibitem{Racz:2011qu}
Racz I and Toth G~Z 2011
  \href{http://dx.doi.org/10.1088/0264-9381/28/19/195003}{ {\em Class. Quant.
  Grav.\/} {\bf 28} 195003 } [\eprint{1104.4199}]

\bibitem{Burko:2013bra}
Burko L~M and Khanna G 2014
  \href{http://dx.doi.org/10.1103/PhysRevD.89.044037}{ {\em Phys. Rev. D\/}
  {\bf 89} 044037 } [\eprint{1312.5247}]

\bibitem{Angelopoulos:2021cpg}
Angelopoulos Y, Aretakis S and Gajic D 2023
  \href{http://dx.doi.org/10.1016/j.aim.2023.108939}{ {\em Adv. Math.\/} {\bf
  417} 108939 } [\eprint{2102.11884}]

\bibitem{Burko:2010zj}
Burko L~M and Khanna G 2011
  \href{http://dx.doi.org/10.1088/0264-9381/28/2/025012}{ {\em Class. Quant.
  Grav.\/} {\bf 28} 025012 } [\eprint{1001.0541}]

\bibitem{Burko:2023uag}
Burko L~M, Khanna G and Sabharwal S 2023
  \href{http://dx.doi.org/10.1103/PhysRevD.107.124023}{ {\em Phys. Rev. D\/}
  {\bf 107} 124023 } [\eprint{2304.06210}]

\bibitem{Csukas:2019kcb}
Csuk\'as K, R\'acz I and T\'oth G~Z 2019
  \href{http://dx.doi.org/10.1103/PhysRevD.100.104025}{ {\em Phys. Rev. D\/}
  {\bf 100} 104025 } [\eprint{1905.09082}]

\bibitem{Carullo:2023tff}
Carullo G and De~Amicis M 2023  [\eprint{2310.12968}]

\bibitem{Islam:2024vro}
Islam T, Faggioli G, Khanna G, Field S~E, van~de Meent M and Buonanno A 2025
  \href{http://dx.doi.org/10.1103/191t-5svc}{ {\em Phys. Rev. D\/} {\bf 112}
  024061 } [\eprint{2407.04682}]

\bibitem{Blanchet:1992br}
Blanchet L and Damour T 1992 \href{http://dx.doi.org/10.1103/PhysRevD.46.4304}{
  {\em Phys. Rev. D\/} {\bf 46} 4304--4319 }

\bibitem{Bernuzzi:2011aj}
Bernuzzi S, Nagar A and Zenginoglu A 2011
  \href{http://dx.doi.org/10.1103/PhysRevD.84.084026}{ {\em Phys. Rev. D\/}
  {\bf 84} 084026 } [\eprint{1107.5402}]

\bibitem{Burko:2016sfi}
Burko L~M and Khanna G 2016
  \href{http://dx.doi.org/10.1103/PhysRevD.94.084049}{ {\em Phys. Rev. D\/}
  {\bf 94} 084049 } [\eprint{1608.02244}]

\bibitem{Rifat:2019fkt}
Rifat N~E~M, Khanna G and Burko L~M 2019
  \href{http://dx.doi.org/10.1103/PhysRevResearch.1.033150}{ {\em Phys. Rev.
  Research.\/} {\bf 1} 033150 } [\eprint{1910.03462}]

\bibitem{Ma:2024hzq}
Ma S, Scheel M~A, Moxon J, Nelli K~C, Deppe N, Kidder L~E, Throwe W and Vu N~L
  2025 \href{http://dx.doi.org/10.1103/jd26-8q5w}{ {\em Phys. Rev. D\/} {\bf
  112} 024003 } [\eprint{2412.06906}]

\bibitem{Allen:2004js}
Allen E~W, Buckmiller E, Burko L~M and Price R~H 2004
  \href{http://dx.doi.org/10.1103/PhysRevD.70.044038}{ {\em Phys. Rev. D\/}
  {\bf 70} 044038 } [\eprint{gr-qc/0401092}]

\bibitem{Dafermos:2004wt}
Dafermos M and Rodnianski I 2004  [\eprint{gr-qc/0403034}]

\bibitem{Iozzo:2020jcu}
Iozzo D~A~B, Boyle M, Deppe N, Moxon J, Scheel M~A, Kidder L~E, Pfeiffer H~P
  and Teukolsky S~A 2021 \href{http://dx.doi.org/10.1103/PhysRevD.103.024039}{
  {\em Phys. Rev. D\/} {\bf 103} 024039 } [\eprint{2010.15200}]

\bibitem{scri}
Boyle M, Iozzo D and Stein L~C 2020 moble/scri: v1.2
  \href{https://doi.org/10.5281/zenodo.4041972}{10.5281/zenodo.4041972}
  \urlprefix\url{https://doi.org/10.5281/zenodo.4041972}

\bibitem{scrirepo}
The scri package \url{https://github.com/sxs-collaboration/scri}

\bibitem{Boyle:2015nqa}
Boyle M 2016 \href{http://dx.doi.org/10.1103/PhysRevD.93.084031}{ {\em Phys.
  Rev. D\/} {\bf 93} 084031 } [\eprint{1509.00862}]

\bibitem{PhysRevLett.76.4303}
Bishop N~T, Gomez R, Holvorcem P~R, Matzner R~A, Papadopoulos P and Winicour J
  1996 \href{http://dx.doi.org/10.1103/PhysRevLett.76.4303}{ {\em Phys. Rev.
  Lett.\/} {\bf 76}(23) 4303--4306 }

\bibitem{Bishop:1998uk}
Bishop N~T, Gomez R, Lehner L, Szilagyi B, Winicour J and Isaacson R~A 1998
  {\em {Cauchy characteristic matching}\/} pp 383--408 [\eprint{gr-qc/9801070}]

\bibitem{Szilagyi:2000xu}
Szilagyi B 2000 {\em {Cauchy characteristic matching in general relativity}\/}
  Other thesis [\eprint{gr-qc/0006091}]

\bibitem{Winicour:2012znc}
Winicour J 2012 \href{http://dx.doi.org/10.12942/lrr-2012-2}{ {\em Living Rev.
  Rel.\/} {\bf 15} 2 }

\bibitem{Ma:2023qjn}
Ma S {\em et~al.\/} 2024 \href{http://dx.doi.org/10.1103/PhysRevD.109.124027}{
  {\em Phys. Rev. D\/} {\bf 109} 124027 } [\eprint{2308.10361}]

\bibitem{Ma:2024bed}
Ma S, Nelli K~C, Moxon J, Scheel M~A, Deppe N, Kidder L~E, Throwe W and Vu N~L
  2025 \href{http://dx.doi.org/10.1088/1361-6382/adaf6f}{ {\em Class. Quant.
  Grav.\/} {\bf 42} 055006 } [\eprint{2409.06141}]

\bibitem{Nagar:2022icd}
Nagar A, Healy J, Lousto C~O, Bernuzzi S and Albertini A 2022
  \href{http://dx.doi.org/10.1103/PhysRevD.105.124061}{ {\em Phys. Rev. D\/}
  {\bf 105} 124061 } [\eprint{2202.05643}]

\bibitem{Wardell:2021fyy}
Wardell B, Pound A, Warburton N, Miller J, Durkan L and Le~Tiec A 2023
  \href{http://dx.doi.org/10.1103/PhysRevLett.130.241402}{ {\em Phys. Rev.
  Lett.\/} {\bf 130} 241402 } [\eprint{2112.12265}]

\bibitem{Islam:2023aec}
Islam T and Khanna G 2023 \href{http://dx.doi.org/10.1103/PhysRevD.108.044012}{
  {\em Phys. Rev. D\/} {\bf 108} 044012 } [\eprint{2306.08767}]

\bibitem{Ma:2023cwe}
Ma S, Sun L and Chen Y 2023
  \href{http://dx.doi.org/10.1103/PhysRevLett.130.141401}{ {\em Phys. Rev.
  Lett.\/} {\bf 130} 141401 } [\eprint{2301.06705}]

\bibitem{Ma:2023vvr}
Ma S, Sun L and Chen Y 2023
  \href{http://dx.doi.org/10.1103/PhysRevD.107.084010}{ {\em Phys. Rev. D\/}
  {\bf 107} 084010 } [\eprint{2301.06639}]

\bibitem{Cardoso:2024jme}
Cardoso V, Carullo G, De~Amicis M, Duque F, Katagiri T, Pereniguez D,
  Redondo-Yuste J, Spieksma T~F~M and Zhong Z 2024
  \href{http://dx.doi.org/10.1103/PhysRevD.109.L121502}{ {\em Phys. Rev. D\/}
  {\bf 109} L121502 } [\eprint{2405.12290}]

\bibitem{Ling:2025wfv}
Ling S, Shah S and Wong S~S~C 2025 \href{http://dx.doi.org/10.1103/22lc-62gj}{
  {\em Phys. Rev. D\/} {\bf 112} 024008 } [\eprint{2503.19967}]

\bibitem{Kehagias:2025xzm}
Kehagias A and Riotto A 2025 \href{http://dx.doi.org/10.1103/j5fw-w3xc}{ {\em
  Phys. Rev. D\/} {\bf 112} 024023 } [\eprint{2504.06224}]

\bibitem{Bondi:1960jsa}
Bondi H 1960 \href{http://dx.doi.org/10.1038/186535a0}{ {\em Nature\/} {\bf
  186} 535--535 }

\bibitem{Sachs:1961zz}
Sachs R~K 1961 \href{http://dx.doi.org/10.1098/rspa.1961.0202}{ {\em Proc. Roy.
  Soc. Lond. A\/} {\bf 264} 309--338 }

\bibitem{Bondi:1962px}
Bondi H, van~der Burg M~G~J and Metzner A~W~K 1962
  \href{http://dx.doi.org/10.1098/rspa.1962.0161}{ {\em Proc. Roy. Soc. Lond.
  A\/} {\bf 269} 21--52 }

\bibitem{Sachs:1962wk}
Sachs R~K 1962 \href{http://dx.doi.org/10.1098/rspa.1962.0206}{ {\em Proc. Roy.
  Soc. Lond. A\/} {\bf 270} 103--126 }

\bibitem{Sachs:1962zza}
Sachs R 1962 \href{http://dx.doi.org/10.1103/PhysRev.128.2851}{ {\em Phys.
  Rev.\/} {\bf 128} 2851--2864 }

\bibitem{Dray:1984rfa}
Dray T and Streubel M 1984 \href{http://dx.doi.org/10.1088/0264-9381/1/1/005}{
  {\em Class. Quant. Grav.\/} {\bf 1} 15--26 }

\bibitem{Wald:1999wa}
Wald R~M and Zoupas A 2000 \href{http://dx.doi.org/10.1103/PhysRevD.61.084027}{
  {\em Phys. Rev. D\/} {\bf 61} 084027 } [\eprint{gr-qc/9911095}]

\bibitem{Zeldovich:1974gvh}
Zel'dovich Y~B and Polnarev A~G 1974 {\em Sov. Astron.\/} {\bf 18} 17

\bibitem{Braginsky:1985vlg}
Braginsky V~B and Grishchuk L~P 1985 {\em Sov. Phys. JETP\/} {\bf 62} 427--430

\bibitem{Braginsky:1987kwo}
Braginsky V~B and Thorne K~S 1987 \href{http://dx.doi.org/10.1038/327123a0}{
  {\em Nature\/} {\bf 327} 123--125 }

\bibitem{Christodoulou:1991cr}
Christodoulou D 1991 \href{http://dx.doi.org/10.1103/PhysRevLett.67.1486}{ {\em
  Phys. Rev. Lett.\/} {\bf 67} 1486--1489 }

\bibitem{Thorne:1992sdb}
Thorne K~S 1992 \href{http://dx.doi.org/10.1103/PhysRevD.45.520}{ {\em Phys.
  Rev. D\/} {\bf 45} 520--524 }

\bibitem{Yoo:2023spi}
Yoo J {\em et~al.\/} 2023 \href{http://dx.doi.org/10.1103/PhysRevD.108.064027}{
  {\em Phys. Rev. D\/} {\bf 108} 064027 } [\eprint{2306.03148}]

\bibitem{qnmfitscode}
Magaña~Zertuche L and Finch E 2025 qnmfits
  \urlprefix\url{https://doi.org/10.5281/zenodo.14806974}

\bibitem{Rossello-Sastre:2024zlr}
Rossell\'o-Sastre M, Husa S and Bera S 2024
  \href{http://dx.doi.org/10.1103/PhysRevD.110.084074}{ {\em Phys. Rev. D\/}
  {\bf 110} 084074 } [\eprint{2405.17302}]

\bibitem{Rezzolla:2010df}
Rezzolla L, Macedo R~P and Jaramillo J~L 2010
  \href{http://dx.doi.org/10.1103/PhysRevLett.104.221101}{ {\em Phys. Rev.
  Lett.\/} {\bf 104} 221101 } [\eprint{1003.0873}]

\bibitem{Jaramillo:2012rr}
Jaramillo J~L, Macedo R~P, Moesta P and Rezzolla L 2012
  \href{http://dx.doi.org/10.1063/1.4734411}{ {\em AIP Conf. Proc.\/} {\bf
  1458} 158--173 } [\eprint{1205.3902}]

\bibitem{Ashtekar:2004cn}
Ashtekar A and Krishnan B 2004 {\em Living Rev. Rel.\/} {\bf 7} 10
  [\eprint{gr-qc/0407042}]

\bibitem{Ashtekar:2025wnu}
Ashtekar A and Krishnan B 2025  [\eprint{2502.11825}]

\bibitem{Penrose:1964wq}
Penrose R 1965 \href{http://dx.doi.org/10.1103/PhysRevLett.14.57}{ {\em Phys.
  Rev. Lett.\/} {\bf 14} 57--59 }

\bibitem{Hawking:1969sw}
Hawking S~W and Penrose R 1970 \href{http://dx.doi.org/10.1098/rspa.1970.0021}{
  {\em Proc. Roy. Soc. Lond.\/} {\bf A314} 529--548 }

\bibitem{Booth:2005qc}
Booth I 2005 \href{http://dx.doi.org/10.1139/p05-063}{ {\em Can. J. Phys.\/}
  {\bf 83} 1073--1099 } [\eprint{gr-qc/0508107}]

\bibitem{Gourgoulhon:2005ng}
Gourgoulhon E and Jaramillo J~L 2006
  \href{http://dx.doi.org/10.1016/j.physrep.2005.10.005}{ {\em Phys. Rept.\/}
  {\bf 423} 159--294 } [\eprint{gr-qc/0503113}]

\bibitem{Ashtekar:2002ag}
Ashtekar A and Krishnan B 2002
  \href{http://dx.doi.org/10.1103/PhysRevLett.89.261101}{ {\em Phys. Rev.
  Lett.\/} {\bf 89} 261101 } [\eprint{gr-qc/0207080}]

\bibitem{Ashtekar:2003hk}
Ashtekar A and Krishnan B 2003
  \href{http://dx.doi.org/10.1103/PhysRevD.68.104030}{ {\em Phys. Rev.\/} {\bf
  D68} 104030 } [\eprint{gr-qc/0308033}]

\bibitem{Andersson:2005gq}
Andersson L, Mars M and Simon W 2005
  \href{http://dx.doi.org/10.1103/PhysRevLett.95.111102}{ {\em
  Phys.Rev.Lett.\/} {\bf 95} 111102 } [\eprint{gr-qc/0506013}]

\bibitem{Andersson:2007fh}
Andersson L, Mars M and Simon W 2008 {\em Adv.Theor.Math.Phys.\/} {\bf 12}
  [\eprint{0704.2889}]

\bibitem{Ashtekar:2013qta}
Ashtekar A, Campiglia M and Shah S 2013
  \href{http://dx.doi.org/10.1103/PhysRevD.88.064045}{ {\em Phys. Rev.\/} {\bf
  D88} 064045 } [\eprint{1306.5697}]

\bibitem{Ashtekar:2004gp}
Ashtekar A, Engle J, Pawlowski T and Van Den~Broeck C 2004
  \href{http://dx.doi.org/10.1088/0264-9381/21/11/003}{ {\em Class. Quant.
  Grav.\/} {\bf 21} 2549--2570 } [\eprint{gr-qc/0401114}]

\bibitem{Owen:2009sb}
Owen R 2009 \href{http://dx.doi.org/10.1103/PhysRevD.80.084012}{ {\em Phys.
  Rev. D\/} {\bf 80} 084012 } [\eprint{0907.0280}]

\bibitem{Ashtekar:2021wld}
Ashtekar A, Khera N, Kolanowski M and Lewandowski J 2022
  \href{http://dx.doi.org/10.1007/JHEP01(2022)028}{ {\em JHEP\/} {\bf 01} 028 }
  [\eprint{2111.07873}]

\bibitem{Chen:2022dxt}
Chen Y {\em et~al.\/} 2022
  \href{http://dx.doi.org/10.1103/PhysRevD.106.124045}{ {\em Phys. Rev. D\/}
  {\bf 106} 124045 } [\eprint{2208.02965}]

\bibitem{Mongwane:2024vao}
Mongwane B, Nkele S, Duniya D~G~A and Bishop N~T 2025
  \href{http://dx.doi.org/10.1103/PhysRevD.111.044002}{ {\em Phys. Rev. D\/}
  {\bf 111} 044002 } [\eprint{2407.06636}]

\bibitem{RibesMetidieri:2024tpk}
Ribes~Metidieri A, Bonga B and Krishnan B 2024
  \href{http://dx.doi.org/10.1103/PhysRevD.110.024069}{ {\em Phys. Rev. D\/}
  {\bf 110} 024069 } [\eprint{2403.17114}]

\bibitem{RibesMetidieri:2025lxr}
Ribes~Metidieri A, Bonga B and Krishnan B 2025 {Black Hole Tomography:
  Unveiling Black Hole Ringdown via Gravitational Wave Observations}
  [\eprint{2501.08964}]

\bibitem{Isi:2021iql}
Isi M and Farr W~M 2021  [\eprint{2107.05609}]

\bibitem{Carullo:2019flw}
Carullo G, Del~Pozzo W and Veitch J 2019
  \href{http://dx.doi.org/10.1103/PhysRevD.99.123029}{ {\em Phys. Rev. D\/}
  {\bf 99} 123029 } [Erratum: Phys.Rev.D 100, 089903 (2019)]
  [\eprint{1902.07527}]

\bibitem{Isi:2019aib}
Isi M, Giesler M, Farr W~M, Scheel M~A and Teukolsky S~A 2019
  \href{http://dx.doi.org/10.1103/PhysRevLett.123.111102}{ {\em Phys. Rev.
  Lett.\/} {\bf 123} 111102 } [\eprint{1905.00869}]

\bibitem{Cabero:2017avf}
Cabero M, Capano C~D, Fischer-Birnholtz O, Krishnan B, Nielsen A~B, Nitz A~H
  and Biwer C~M 2018 \href{http://dx.doi.org/10.1103/PhysRevD.97.124069}{ {\em
  Phys. Rev. D\/} {\bf 97} 124069 } [\eprint{1711.09073}]

\bibitem{Prix:T1500618}
Prix R 2016 {Bayesian QNM search on GW150914} Tech. Rep. LIGO-T1500618 LIGO
  Scientific Collaboration
  \urlprefix\url{https://dcc.ligo.org/LIGO-T1500618/public}

\bibitem{Finch:2021qph}
Finch E and Moore C~J 2021
  \href{http://dx.doi.org/10.1103/PhysRevD.104.123034}{ {\em Phys. Rev. D\/}
  {\bf 104} 123034 } [\eprint{2108.09344}]

\bibitem{Capano:2021etf}
Capano C~D, Cabero M, Westerweck J, Abedi J, Kastha S, Nitz A~H, Wang Y~F,
  Nielsen A~B and Krishnan B 2023
  \href{http://dx.doi.org/10.1103/PhysRevLett.131.221402}{ {\em Phys. Rev.
  Lett.\/} {\bf 131} 221402 } [\eprint{2105.05238}]

\bibitem{Veitch:2014wba}
Veitch J {\em et~al.\/} 2015
  \href{http://dx.doi.org/10.1103/PhysRevD.91.042003}{ {\em Phys. Rev. D\/}
  {\bf 91} 042003 } [\eprint{1409.7215}]

\bibitem{Biwer:2018osg}
Biwer C~M, Capano C~D, De S, Cabero M, Brown D~A, Nitz A~H and Raymond V 2019
  \href{http://dx.doi.org/10.1088/1538-3873/aaef0b}{ {\em Publ. Astron. Soc.
  Pac.\/} {\bf 131} 024503 } [\eprint{1807.10312}]

\bibitem{Ashton:2018jfp}
Ashton G {\em et~al.\/} 2019 \href{http://dx.doi.org/10.3847/1538-4365/ab06fc}{
  {\em Astrophys. J. Suppl.\/} {\bf 241} 27 } [\eprint{1811.02042}]

\bibitem{Romero-Shaw:2020owr}
Romero-Shaw I~M {\em et~al.\/} 2020
  \href{http://dx.doi.org/10.1093/mnras/staa2850}{ {\em Mon. Not. Roy. Astron.
  Soc.\/} {\bf 499} 3295--3319 } [\eprint{2006.00714}]

\bibitem{Dax:2021tsq}
Dax M, Green S~R, Gair J, Macke J~H, Buonanno A and Sch\"olkopf B 2021
  \href{http://dx.doi.org/10.1103/PhysRevLett.127.241103}{ {\em Phys. Rev.
  Lett.\/} {\bf 127} 241103 } [\eprint{2106.12594}]

\bibitem{Wong:2023lgb}
Wong K~W~K, Isi M and Edwards T~D~P 2023
  \href{http://dx.doi.org/10.3847/1538-4357/acf5cd}{ {\em Astrophys. J.\/} {\bf
  958} 129 } [\eprint{2302.05333}]

\bibitem{LIGOScientific:2019hgc}
Abbott B~P {\em et~al.\/} (LIGO Scientific, Virgo) 2020
  \href{http://dx.doi.org/10.1088/1361-6382/ab685e}{ {\em Class. Quant.
  Grav.\/} {\bf 37} 055002 } [\eprint{1908.11170}]

\bibitem{LIGO:2021ppb}
Davis D {\em et~al.\/} (LIGO) 2021
  \href{http://dx.doi.org/10.1088/1361-6382/abfd85}{ {\em Class. Quant.
  Grav.\/} {\bf 38} 135014 } [\eprint{2101.11673}]

\bibitem{Davis:2022ird}
Davis D, Littenberg T~B, Romero-Shaw I~M, Millhouse M, McIver J, Di~Renzo F and
  Ashton G 2022 \href{http://dx.doi.org/10.1088/1361-6382/aca238}{ {\em Class.
  Quant. Grav.\/} {\bf 39} 245013 } [\eprint{2207.03429}]

\bibitem{Vitale:2020gvb}
Vitale S, Haster C~J, Sun L, Farr B, Goetz E, Kissel J and Cahillane C 2021
  \href{http://dx.doi.org/10.1103/PhysRevD.103.063016}{ {\em Phys. Rev. D\/}
  {\bf 103} 063016 } [\eprint{2009.10192}]

\bibitem{Littenberg:2014oda}
Littenberg T~B and Cornish N~J 2015
  \href{http://dx.doi.org/10.1103/PhysRevD.91.084034}{ {\em Phys. Rev. D\/}
  {\bf 91} 084034 } [\eprint{1410.3852}]

\bibitem{Levinson1946}
Levinson N 1946 \href{http://dx.doi.org/10.1002/sapm1946251261}{ {\em Journal
  of Mathematics and Physics\/} {\bf 25} 261--278 }

\bibitem{Durbin1960}
Durbin J 1960 \href{http://dx.doi.org/10.2307/1401322}{ {\em Revue de
  l'Institut International de Statistique / Review of the International
  Statistical Institute\/} {\bf 28} 233--244 } ISSN 03731138
  \urlprefix\url{http://www.jstor.org/stable/1401322}

\bibitem{Talbot:2021igi}
Talbot C, Thrane E, Biscoveanu S and Smith R 2021
  \href{http://dx.doi.org/10.1103/PhysRevResearch.3.043049}{ {\em Phys. Rev.
  Res.\/} {\bf 3} 043049 } [\eprint{2106.13785}]

\bibitem{gray:2006}
Gray R~M 2006 \href{http://dx.doi.org/10.1561/0100000006}{ {\em Foundations and
  Trends® in Communications and Information Theory\/} {\bf 2} 155--239 } ISSN
  1567-2190

\bibitem{UNSER1984231}
Unser M 1984
  \href{http://dx.doi.org/https://doi.org/10.1016/0165-1684(84)90002-1}{ {\em
  Signal Processing\/} {\bf 7} 231 -- 249 } ISSN 0165-1684
  \urlprefix\url{http://www.sciencedirect.com/science/article/pii/0165168484900021}

\bibitem{Allen:2001ay}
Allen B, Creighton J~D~E, Flanagan E~E and Romano J~D 2002
  \href{http://dx.doi.org/10.1103/PhysRevD.65.122002}{ {\em Phys. Rev. D\/}
  {\bf 65} 122002 } [\eprint{gr-qc/0105100}]

\bibitem{Allen:2005fk}
Allen B, Anderson W~G, Brady P~R, Brown D~A and Creighton J~D~E 2012
  \href{http://dx.doi.org/10.1103/PhysRevD.85.122006}{ {\em Phys. Rev. D\/}
  {\bf 85} 122006 } [\eprint{gr-qc/0509116}]

\bibitem{Chatziioannou:2019zvs}
Chatziioannou K, Haster C~J, Littenberg T~B, Farr W~M, Ghonge S, Millhouse M,
  Clark J~A and Cornish N 2019
  \href{http://dx.doi.org/10.1103/PhysRevD.100.104004}{ {\em Phys. Rev. D\/}
  {\bf 100} 104004 } [\eprint{1907.06540}]

\bibitem{golub13}
Golub G~H and van Loan C~F 2013 {\em Matrix Computations\/} 4th ed (JHU Press)
  ISBN 1421407949 9781421407944
  \urlprefix\url{http://www.cs.cornell.edu/cv/GVL4/golubandvanloan.htm}

\bibitem{Virtanen:2019joe}
Virtanen P {\em et~al.\/} 2020
  \href{http://dx.doi.org/10.1038/s41592-019-0686-2}{ {\em Nature Meth.\/} {\bf
  17} 261 } [\eprint{1907.10121}]

\bibitem{Harris:2020xlr}
Harris C~R {\em et~al.\/} 2020
  \href{http://dx.doi.org/10.1038/s41586-020-2649-2}{ {\em Nature\/} {\bf 585}
  357--362 } [\eprint{2006.10256}]

\bibitem{Cotesta:2022pci}
Cotesta R, Carullo G, Berti E and Cardoso V 2022
  \href{http://dx.doi.org/10.1103/PhysRevLett.129.111102}{ {\em Phys. Rev.
  Lett.\/} {\bf 129} 111102 } [\eprint{2201.00822}]

\bibitem{Isi:2022mhy}
Isi M and Farr W~M 2022  [\eprint{2202.02941}]

\bibitem{Carullo:2023gtf}
Carullo G, Cotesta R, Berti E and Cardoso V 2023
  \href{http://dx.doi.org/10.1103/PhysRevLett.131.169002}{ {\em Phys. Rev.
  Lett.\/} {\bf 131} 169002 } [\eprint{2310.20625}]

\bibitem{Wang:2023xsy}
Wang Y~F, Capano C~D, Abedi J, Kastha S, Krishnan B, Nielsen A~B, Nitz A~H and
  Westerweck J 2023  [\eprint{2310.19645}]

\bibitem{Correia:2023bfn}
Correia A, Wang Y~F, Westerweck J and Capano C~D 2024
  \href{http://dx.doi.org/10.1103/PhysRevD.110.L041501}{ {\em Phys. Rev. D\/}
  {\bf 110} L041501 } [\eprint{2312.14118}]

\bibitem{Correia:2023ipz}
Correia A and Capano C~D 2024
  \href{http://dx.doi.org/10.1103/PhysRevD.110.044018}{ {\em Phys. Rev. D\/}
  {\bf 110} 044018 } [\eprint{2312.15146}]

\bibitem{Finch:2022ynt}
Finch E and Moore C~J 2022
  \href{http://dx.doi.org/10.1103/PhysRevD.106.043005}{ {\em Phys. Rev. D\/}
  {\bf 106} 043005 } [\eprint{2205.07809}]

\bibitem{Crisostomi:2023tle}
Crisostomi M, Dey K, Barausse E and Trotta R 2023
  \href{http://dx.doi.org/10.1103/PhysRevD.108.044029}{ {\em Phys. Rev. D\/}
  {\bf 108} 044029 } [\eprint{2305.18528}]

\bibitem{Pacilio:2024qcq}
Pacilio C, Bhagwat S and Cotesta R 2024
  \href{http://dx.doi.org/10.1103/PhysRevD.110.083010}{ {\em Phys. Rev. D\/}
  {\bf 110} 083010 } [\eprint{2404.11373}]

\bibitem{Wang:2024jlz}
Wang H~T, Yim G, Chen X and Shao L 2024
  \href{http://dx.doi.org/10.3847/1538-4357/ad7096}{ {\em Astrophys. J.\/} {\bf
  974} 230 } [\eprint{2409.00970}]

\bibitem{Wang:2024yhb}
Wang H~T, Wang Z, Dong Y, Yim G and Shao L 2024 {\em arXiv eprints\/}
  [\eprint{2411.13333}]

\bibitem{Talbot:2020auc}
Talbot C and Thrane E 2020
  \href{http://dx.doi.org/10.1103/PhysRevResearch.2.043298}{ {\em Phys. Rev.
  Res.\/} {\bf 2} 043298 } [\eprint{2006.05292}]

\bibitem{Siegel:2024jqd}
Siegel H, Isi M and Farr W~M 2025
  \href{http://dx.doi.org/10.1103/PhysRevD.111.044070}{ {\em Phys. Rev. D\/}
  {\bf 111} 044070 } [\eprint{2410.02704}]

\bibitem{Wang:2023mst}
Wang H~T and Shao L 2023 \href{http://dx.doi.org/10.1103/PhysRevD.108.123018}{
  {\em Phys. Rev. D\/} {\bf 108} 123018 } [\eprint{2311.13300}]

\bibitem{Isi:2023nif}
Isi M and Farr W~M 2023
  \href{http://dx.doi.org/10.1103/PhysRevLett.131.169001}{ {\em Phys. Rev.
  Lett.\/} {\bf 131} 169001 } [\eprint{2310.13869}]

\bibitem{LIGOScientific:2020iuh}
Abbott R {\em et~al.\/} (LIGO Scientific, Virgo) 2020
  \href{http://dx.doi.org/10.1103/PhysRevLett.125.101102}{ {\em Phys. Rev.
  Lett.\/} {\bf 125} 101102 } [\eprint{2009.01075}]

\bibitem{LIGOScientific:2020ufj}
Abbott R {\em et~al.\/} (LIGO Scientific, Virgo) 2020
  \href{http://dx.doi.org/10.3847/2041-8213/aba493}{ {\em Astrophys. J.
  Lett.\/} {\bf 900} L13 } [\eprint{2009.01190}]

\bibitem{Dhani:2021vac}
Dhani A and Sathyaprakash B~S 2021  [\eprint{2107.14195}]

\bibitem{Isi:2022mbx}
Isi M 2023 \href{http://dx.doi.org/10.1088/1361-6382/acf28c}{ {\em Class.
  Quant. Grav.\/} {\bf 40} 203001 } [\eprint{2208.03372}]

\bibitem{pyRing}
Carullo G, Del~Pozzo W and Veitch J 2023 \texttt{pyRing}: a time-domain
  ringdown analysis python package
  \href{https://git.ligo.org/lscsoft/pyring}{git.ligo.org/lscsoft/pyring}
  \urlprefix\url{https://doi.org/10.5281/zenodo.8165508}

\bibitem{Lindblom:2005qh}
Lindblom L, Scheel M~A, Kidder L~E, Owen R and Rinne O 2006
  \href{http://dx.doi.org/10.1088/0264-9381/23/16/S09}{ {\em
  Class.Quant.Grav.\/} {\bf 23} S447--S462 } [\eprint{gr-qc/0512093}]

\bibitem{Vishveshwara:1968ksg}
Vishveshwara C~V 1968 {\em {The Stability of the Schwarzschild Metric}\/} Ph.D.
  thesis Maryland U.

\bibitem{Cunningham:1978zfa}
Cunningham C~T, Price R~H and Moncrief V 1978
  \href{http://dx.doi.org/10.1086/156413}{ {\em Astrophys. J.\/} {\bf 224} 643
  }

\bibitem{Penrose1982SOMEUP}
Penrose R 1982 Some unsolved problems in classical general relativity
  \urlprefix\url{https://api.semanticscholar.org/CorpusID:116195610}

\bibitem{Carter:1971zc}
Carter B 1971 \href{http://dx.doi.org/10.1103/PhysRevLett.26.331}{ {\em Phys.
  Rev. Lett.\/} {\bf 26} 331--333 }

\bibitem{LIGOScientific:2016aoc}
Abbott B~P {\em et~al.\/} (LIGO Scientific, Virgo) 2016
  \href{http://dx.doi.org/10.1103/PhysRevLett.116.061102}{ {\em Phys. Rev.
  Lett.\/} {\bf 116} 061102 } [\eprint{1602.03837}]

\bibitem{LIGOScientific:2020ibl}
Abbott R {\em et~al.\/} (LIGO Scientific, Virgo) 2021
  \href{http://dx.doi.org/10.1103/PhysRevX.11.021053}{ {\em Phys. Rev. X\/}
  {\bf 11} 021053 } [\eprint{2010.14527}]

\bibitem{LIGOScientific:2018mvr}
Abbott B~P {\em et~al.\/} (LIGO Scientific, Virgo) 2019
  \href{http://dx.doi.org/10.1103/PhysRevX.9.031040}{ {\em Phys. Rev. X\/} {\bf
  9} 031040 } [\eprint{1811.12907}]

\bibitem{KAGRA:2021vkt}
Abbott R {\em et~al.\/} (KAGRA, VIRGO, LIGO Scientific) 2023
  \href{http://dx.doi.org/10.1103/PhysRevX.13.041039}{ {\em Phys. Rev. X\/}
  {\bf 13} 041039 } [\eprint{2111.03606}]

\bibitem{Bhagwat:2019bwv}
Bhagwat S, Cabero M, Capano C~D, Krishnan B and Brown D~A 2020
  \href{http://dx.doi.org/10.1103/PhysRevD.102.024023}{ {\em Phys. Rev. D\/}
  {\bf 102} 024023 } [\eprint{1910.13203}]

\bibitem{Ota:2021ypb}
Ota I and Chirenti C 2022 \href{http://dx.doi.org/10.1103/PhysRevD.105.044015}{
  {\em Phys. Rev. D\/} {\bf 105} 044015 } [\eprint{2108.01774}]

\bibitem{Lindblom:2008cm}
Lindblom L, Owen B~J and Brown D~A 2008
  \href{http://dx.doi.org/10.1103/PhysRevD.78.124020}{ {\em Phys. Rev. D\/}
  {\bf 78} 124020 } [\eprint{0809.3844}]

\bibitem{Cornish:2014kda}
Cornish N~J and Littenberg T~B 2015
  \href{http://dx.doi.org/10.1088/0264-9381/32/13/135012}{ {\em Class. Quant.
  Grav.\/} {\bf 32} 135012 } [\eprint{1410.3835}]

\bibitem{Drago:2020kic}
Drago M {\em et~al.\/} 2021
  \href{http://dx.doi.org/10.1016/j.softx.2021.100678}{ {\em SoftwareX\/} {\bf
  14} 100678 } ISSN 2352-7110 [\eprint{2006.12604}]

\bibitem{10.1214/13-BA826}
Clarke J~L, Clarke B and Yu C~W 2013 \href{http://dx.doi.org/10.1214/13-BA826}{
  {\em Bayesian Analysis\/} {\bf 8} 647 -- 690 }

\bibitem{Jefferys:1992}
{Jefferys} W~H and {Berger} J~O 1992 \href{http://dx.doi.org/stable/29774559}{
  {\em American Scientist\/} {\bf 80} }

\bibitem{Jaynes:2003}
{Jaynes} E~T 2003 {\em Probability Theory: The Logic of Science\/} (Cambridge
  University Press) ISBN 9780521592710

\bibitem{Isi:2022cii}
Isi M, Farr W~M and Chatziioannou K 2022
  \href{http://dx.doi.org/10.1103/PhysRevD.106.024048}{ {\em Phys. Rev. D\/}
  {\bf 106} 024048 } [\eprint{2204.10742}]

\bibitem{Clarke:2024lwi}
Clarke T~A {\em et~al.\/} 2024
  \href{http://dx.doi.org/10.1103/PhysRevD.109.124030}{ {\em Phys. Rev. D\/}
  {\bf 109} 124030 } [\eprint{2402.02819}]

\bibitem{Dickey1971TheWL}
Dickey J~M 1971 {\em Annals of Mathematical Statistics\/} {\bf 42} 204--223
  \urlprefix\url{https://api.semanticscholar.org/CorpusID:123029751}

\bibitem{2015arXiv150704544V}
{Vehtari} A, {Gelman} A and {Gabry} J 2015 {\em {}\/} [\eprint{1507.04544}]

\bibitem{Kwok:2021zny}
Kwok J~Y~L, Lo R~K~L, Weinstein A~J and Li T~G~F 2022
  \href{http://dx.doi.org/10.1103/PhysRevD.105.024066}{ {\em Phys. Rev. D\/}
  {\bf 105} 024066 } [\eprint{2109.07642}]

\bibitem{Husa:2015iqa}
Husa S, Khan S, Hannam M, P\"urrer M, Ohme F, Jim\'enez~Forteza X and Boh\'e A
  2016 \href{http://dx.doi.org/10.1103/PhysRevD.93.044006}{ {\em Phys. Rev.
  D\/} {\bf 93} 044006 } [\eprint{1508.07250}]

\bibitem{Khan:2015jqa}
Khan S, Husa S, Hannam M, Ohme F, P\"urrer M, Jim\'enez~Forteza X and Boh\'e A
  2016 \href{http://dx.doi.org/10.1103/PhysRevD.93.044007}{ {\em Phys. Rev.
  D\/} {\bf 93} 044007 } [\eprint{1508.07253}]

\bibitem{Purrer:2014fza}
P\"urrer M 2014 \href{http://dx.doi.org/10.1088/0264-9381/31/19/195010}{ {\em
  Class. Quant. Grav.\/} {\bf 31} 195010 } [\eprint{1402.4146}]

\bibitem{Purrer:2015tud}
P\"urrer M 2016 \href{http://dx.doi.org/10.1103/PhysRevD.93.064041}{ {\em Phys.
  Rev. D\/} {\bf 93} 064041 } [\eprint{1512.02248}]

\bibitem{maximiliano_isi_2022_5965773}
Isi M and Farr W~M 2022 Revisiting the ringdown of gw150914
  \urlprefix\url{https://doi.org/10.5281/zenodo.5965773}

\bibitem{LIGOScientific:2016vlm}
Abbott B~P {\em et~al.\/} (LIGO Scientific, Virgo) 2016
  \href{http://dx.doi.org/10.1103/PhysRevLett.116.241102}{ {\em Phys. Rev.
  Lett.\/} {\bf 116} 241102 } [\eprint{1602.03840}]

\bibitem{Pekowsky:2013ska}
Pekowsky L, O'Shaughnessy R, Healy J and Shoemaker D 2013
  \href{http://dx.doi.org/10.1103/PhysRevD.88.024040}{ {\em Phys. Rev. D\/}
  {\bf 88} 024040 } [\eprint{1304.3176}]

\bibitem{Kolitsidou:2024vub}
Kolitsidou P, Thompson J~E and Hannam M 2025
  \href{http://dx.doi.org/10.1103/PhysRevD.111.024050}{ {\em Phys. Rev. D\/}
  {\bf 111} 024050 } [\eprint{2402.00813}]

\bibitem{Nitz:2020mga}
Nitz A~H and Capano C~D 2021 \href{http://dx.doi.org/10.3847/2041-8213/abccc5}{
  {\em Astrophys. J. Lett.\/} {\bf 907} L9 } [\eprint{2010.12558}]

\bibitem{Estelles:2021jnz}
Estell\'es H {\em et~al.\/} 2022
  \href{http://dx.doi.org/10.3847/1538-4357/ac33a0}{ {\em Astrophys. J.\/} {\bf
  924} 79 } [\eprint{2105.06360}]

\bibitem{Olsen:2021qin}
Olsen S, Roulet J, Chia H~S, Dai L, Venumadhav T, Zackay B and Zaldarriaga M
  2021 \href{http://dx.doi.org/10.1103/PhysRevD.104.083036}{ {\em Phys. Rev.
  D\/} {\bf 104} 083036 } [\eprint{2106.13821}]

\bibitem{Capano:2022zqm}
Capano C~D, Abedi J, Kastha S, Nitz A~H, Westerweck J, Wang Y~F, Cabero M,
  Nielsen A~B and Krishnan B 2024
  \href{http://dx.doi.org/10.1088/1361-6382/ad84ae}{ {\em Class. Quant.
  Grav.\/} {\bf 41} 245009 } [\eprint{2209.00640}]

\bibitem{Abedi:2023kot}
Abedi J, Capano C~D, Kastha S, Nitz A~H, Wang Y~F, Westerweck J, Nielsen A~B
  and Krishnan B 2023 \href{http://dx.doi.org/10.1103/PhysRevD.108.104009}{
  {\em Phys. Rev. D\/} {\bf 108} 104009 } [\eprint{2309.03121}]

\bibitem{Gayathri:2020coq}
Gayathri V, Healy J, Lange J, O'Brien B, Szczepanczyk M, Bartos I, Campanelli
  M, Klimenko S, Lousto C~O and O'Shaughnessy R 2022
  \href{http://dx.doi.org/10.1038/s41550-021-01568-w}{ {\em Nature Astron.\/}
  {\bf 6} 344--349 } [\eprint{2009.05461}]

\bibitem{Berti:2007zu}
Berti E, Cardoso J, Cardoso V and Cavaglia M 2007
  \href{http://dx.doi.org/10.1103/PhysRevD.76.104044}{ {\em Phys. Rev. D\/}
  {\bf 76} 104044 } [\eprint{0707.1202}]

\bibitem{Skilling:2006gxv}
Skilling J 2006 \href{http://dx.doi.org/10.1214/06-BA127}{ {\em Bayesian
  Analysis\/} {\bf 1} 833--859 }

\bibitem{Li:2011cg}
Li T~G~F, Del~Pozzo W, Vitale S, Van Den~Broeck C, Agathos M, Veitch J, Grover
  K, Sidery T, Sturani R and Vecchio A 2012
  \href{http://dx.doi.org/10.1103/PhysRevD.85.082003}{ {\em Phys. Rev. D\/}
  {\bf 85} 082003 } [\eprint{1110.0530}]

\bibitem{Agathos:2013upa}
Agathos M, Del~Pozzo W, Li T~G~F, Van Den~Broeck C, Veitch J and Vitale S 2014
  \href{http://dx.doi.org/10.1103/PhysRevD.89.082001}{ {\em Phys. Rev. D\/}
  {\bf 89} 082001 } [\eprint{1311.0420}]

\bibitem{Buscicchio:2019rir}
Buscicchio R, Roebber E, Goldstein J~M and Moore C~J 2019
  \href{http://dx.doi.org/10.1103/PhysRevD.100.084041}{ {\em Phys. Rev. D\/}
  {\bf 100} 084041 } [\eprint{1907.11631}]

\bibitem{Biscoveanu:2020are}
Biscoveanu S, Isi M, Vitale S and Varma V 2021
  \href{http://dx.doi.org/10.1103/PhysRevLett.126.171103}{ {\em Phys. Rev.
  Lett.\/} {\bf 126} 171103 } [\eprint{2007.09156}]

\bibitem{Gerosa:2024ojv}
Gerosa D, De~Renzis V, Tettoni F, Mould M, Vecchio A and Pacilio C 2025
  \href{http://dx.doi.org/10.1103/PhysRevLett.134.121402}{ {\em Phys. Rev.
  Lett.\/} {\bf 134} 121402 } [\eprint{2409.07519}]

\bibitem{Zimmerman:2019wzo}
Zimmerman A, Haster C~J and Chatziioannou K 2019
  \href{http://dx.doi.org/10.1103/PhysRevD.99.124044}{ {\em Phys. Rev. D\/}
  {\bf 99} 124044 } [\eprint{1903.11008}]

\bibitem{Isi:2019asy}
Isi M, Chatziioannou K and Farr W~M 2019
  \href{http://dx.doi.org/10.1103/PhysRevLett.123.121101}{ {\em Phys. Rev.
  Lett.\/} {\bf 123} 121101 } [\eprint{1904.08011}]

\bibitem{Isi:2020tac}
Isi M, Farr W~M, Giesler M, Scheel M~A and Teukolsky S~A 2021
  \href{http://dx.doi.org/10.1103/PhysRevLett.127.011103}{ {\em Phys. Rev.
  Lett.\/} {\bf 127} 011103 } [\eprint{2012.04486}]

\bibitem{Mihaylov:2021bpf}
Mihaylov D~P, Ossokine S, Buonanno A and Ghosh A 2021
  \href{http://dx.doi.org/10.1103/PhysRevD.104.124087}{ {\em Phys. Rev. D\/}
  {\bf 104} 124087 } [\eprint{2105.06983}]

\bibitem{Pacilio:2023uef}
Pacilio C, Gerosa D and Bhagwat S 2024
  \href{http://dx.doi.org/10.1103/PhysRevD.109.L081302}{ {\em Phys. Rev. D\/}
  {\bf 109} L081302 } [\eprint{2310.03811}]

\bibitem{Liu:2024atc}
Liu H and Yunes N 2024 {Robust and improved constraints on higher-curvature
  gravitational effective-field-theory with the GW170608 event}
  [\eprint{2407.08929}]

\bibitem{Chung:2025wbg}
Chung A~K~W and Yunes N 2025  [\eprint{2506.14695}]

\bibitem{Conklin:2017lwb}
Conklin R~S, Holdom B and Ren J 2018
  \href{http://dx.doi.org/10.1103/PhysRevD.98.044021}{ {\em Phys. Rev. D\/}
  {\bf 98} 044021 } [\eprint{1712.06517}]

\bibitem{Abedi:2018npz}
Abedi J and Afshordi N 2019
  \href{http://dx.doi.org/10.1088/1475-7516/2019/11/010}{ {\em JCAP\/} {\bf 11}
  010 } [\eprint{1803.10454}]

\bibitem{Holdom:2019bdv}
Holdom B 2020 \href{http://dx.doi.org/10.1103/PhysRevD.101.064063}{ {\em Phys.
  Rev. D\/} {\bf 101} 064063 } [\eprint{1909.11801}]

\bibitem{Conklin:2021cbc}
Conklin R~S and Afshordi N 2021  [\eprint{2201.00027}]

\bibitem{Abedi:2021tti}
Abedi J, Longo~Micchi L~F and Afshordi N 2023
  \href{http://dx.doi.org/10.1103/PhysRevD.108.044047}{ {\em Phys. Rev. D\/}
  {\bf 108} 044047 } [\eprint{2201.00047}]

\bibitem{Westerweck:2017hus}
Westerweck J, Nielsen A, Fischer-Birnholtz O, Cabero M, Capano C, Dent T,
  Krishnan B, Meadors G and Nitz A~H 2018
  \href{http://dx.doi.org/10.1103/PhysRevD.97.124037}{ {\em Phys. Rev. D\/}
  {\bf 97} 124037 } [\eprint{1712.09966}]

\bibitem{Nielsen:2018lkf}
Nielsen A~B, Capano C~D, Birnholtz O and Westerweck J 2019
  \href{http://dx.doi.org/10.1103/PhysRevD.99.104012}{ {\em Phys. Rev. D\/}
  {\bf 99} 104012 } [\eprint{1811.04904}]

\bibitem{Salemi:2019uea}
Salemi F, Milotti E, Prodi G~A, Vedovato G, Lazzaro C, Tiwari S, Vinciguerra S,
  Drago M and Klimenko S 2019
  \href{http://dx.doi.org/10.1103/PhysRevD.100.042003}{ {\em Phys. Rev. D\/}
  {\bf 100} 042003 } [\eprint{1905.09260}]

\bibitem{Abedi:2022bph}
Abedi J 2022  [\eprint{2301.00025}]

\bibitem{Uchikata:2019frs}
Uchikata N, Nakano H, Narikawa T, Sago N, Tagoshi H and Tanaka T 2019
  \href{http://dx.doi.org/10.1103/PhysRevD.100.062006}{ {\em Phys. Rev. D\/}
  {\bf 100} 062006 } [\eprint{1906.00838}]

\bibitem{Tsang:2019zra}
Tsang K~W, Ghosh A, Samajdar A, Chatziioannou K, Mastrogiovanni S, Agathos M
  and Van Den~Broeck C 2020
  \href{http://dx.doi.org/10.1103/PhysRevD.101.064012}{ {\em Phys. Rev. D\/}
  {\bf 101} 064012 } [\eprint{1906.11168}]

\bibitem{Wang:2020ayy}
Wang Y~T and Piao Y~S 2020  [\eprint{2010.07663}]

\bibitem{Westerweck:2021nue}
Westerweck J, Sherf Y, Capano C~D and Brustein R 2021  [\eprint{2108.08823}]

\bibitem{Ren:2021xbe}
Ren J and Wu D 2021 \href{http://dx.doi.org/10.1103/PhysRevD.104.124023}{ {\em
  Phys. Rev. D\/} {\bf 104} 124023 } [\eprint{2108.01820}]

\bibitem{Miani:2023mgl}
Miani A, Lazzaro C, Prodi G~A, Tiwari S, Drago M, Milotti E and Vedovato G 2023
  \href{http://dx.doi.org/10.1103/PhysRevD.108.064018}{ {\em Phys. Rev. D\/}
  {\bf 108} 064018 } [\eprint{2302.12158}]

\bibitem{Wu:2023wfv}
Wu D, Gao P, Ren J and Afshordi N 2023
  \href{http://dx.doi.org/10.1103/PhysRevD.108.124006}{ {\em Phys. Rev. D\/}
  {\bf 108} 124006 } [\eprint{2308.01017}]

\bibitem{Granot:2017tbr}
Granot J, Guetta D and Gill R 2017
  \href{http://dx.doi.org/10.3847/2041-8213/aa991d}{ {\em Astrophys. J.
  Lett.\/} {\bf 850} L24 } [\eprint{1710.06407}]

\bibitem{Gottlieb:2017pju}
Gottlieb O, Nakar E, Piran T and Hotokezaka K 2018
  \href{http://dx.doi.org/10.1093/mnras/sty1462}{ {\em Mon. Not. Roy. Astron.
  Soc.\/} {\bf 479} 588--600 } [\eprint{1710.05896}]

\bibitem{Nakar:2018cbe}
Nakar E, Gottlieb O, Piran T, Kasliwal M~M and Hallinan G 2018
  \href{http://dx.doi.org/10.3847/1538-4357/aae205}{ {\em Astrophys. J.\/} {\bf
  867} 18 } [\eprint{1803.07595}]

\bibitem{Metzger:2018qfl}
Metzger B~D, Thompson T~A and Quataert E 2018
  \href{http://dx.doi.org/10.3847/1538-4357/aab095}{ {\em Astrophys. J.\/} {\bf
  856} 101 } [\eprint{1801.04286}]

\bibitem{Xie:2018vya}
Xie X, Zrake J and MacFadyen A 2018
  \href{http://dx.doi.org/10.3847/1538-4357/aacf9c}{ {\em Astrophys. J.\/} {\bf
  863} 58 } [\eprint{1804.09345}]

\bibitem{Gill:2019bvq}
Gill R, Nathanail A and Rezzolla L 2019
  \href{http://dx.doi.org/10.3847/1538-4357/ab16da}{ {\em Astrophys. J.\/} {\bf
  876} 139 } [\eprint{1901.04138}]

\bibitem{vanPutten:2019kca}
van Putten M~H~P~M, Della~Valle M and Levinson A 2019
  \href{http://dx.doi.org/10.3847/2041-8213/ab18a2}{ {\em Astrophys. J.
  Lett.\/} {\bf 876} L2 } [\eprint{1910.12730}]

\bibitem{Hamidani:2019qyx}
Hamidani H, Kiuchi K and Ioka K 2020
  \href{http://dx.doi.org/10.1093/mnras/stz3231}{ {\em Mon. Not. Roy. Astron.
  Soc.\/} {\bf 491} 3192--3216 } [\eprint{1909.05867}]

\bibitem{Murguia-Berthier:2020tfs}
Murguia-Berthier A, Ramirez-Ruiz E, De~Colle F, Janiuk A, Rosswog S and Lee W~H
  2021 \href{http://dx.doi.org/10.3847/1538-4357/abd08e}{ {\em Astrophys. J.\/}
  {\bf 908} 152 } [\eprint{2007.12245}]

\bibitem{Abedi:2020sgg}
Abedi J and Afshordi N 2020  [\eprint{2001.00821}]

\bibitem{Abedi:2020ujo}
Abedi J, Afshordi N, Oshita N and Wang Q 2020
  \href{http://dx.doi.org/10.3390/universe6030043}{ {\em Universe\/} {\bf 6} 43
  } [\eprint{2001.09553}]

\bibitem{10.2307/2291091}
Kass R~E and Raftery A~E 1995 {\em Journal of the American Statistical
  Association\/} {\bf 90} 773--795 ISSN 01621459
  \urlprefix\url{http://www.jstor.org/stable/2291091}

\bibitem{GW190521echo}
{PyCBC, cWB and cross-correlation data}
  \url{https://github.com/No-GAP/GW190521echo}

\bibitem{Gupte:2024jfe}
Gupte N {\em et~al.\/} 2024  [\eprint{2404.14286}]

\bibitem{Mandel:2018mve}
Mandel I, Farr W~M and Gair J~R 2019
  \href{http://dx.doi.org/10.1093/mnras/stz896}{ {\em Mon. Not. Roy. Astron.
  Soc.\/} {\bf 486} 1086--1093 } [\eprint{1809.02063}]

\bibitem{Tsang:2018uie}
Tsang K~W, Rollier M, Ghosh A, Samajdar A, Agathos M, Chatziioannou K, Cardoso
  V, Khanna G and Van Den~Broeck C 2018
  \href{http://dx.doi.org/10.1103/PhysRevD.98.024023}{ {\em Phys. Rev. D\/}
  {\bf 98} 024023 } [\eprint{1804.04877}]

\bibitem{LISAConsortiumWaveformWorkingGroup:2023arg}
Afshordi N {\em et~al.\/} (LISA Consortium Waveform Working Group) 2023
  [\eprint{2311.01300}]

\bibitem{LISA:2022kgy}
Arun K~G {\em et~al.\/} (LISA) 2022
  \href{http://dx.doi.org/10.1007/s41114-022-00036-9}{ {\em Living Rev. Rel.\/}
  {\bf 25} 4 } [\eprint{2205.01597}]

\bibitem{Bhagwat:2023jwv}
Bhagwat S, Pacilio C, Pani P and Mapelli M 2023
  \href{http://dx.doi.org/10.1103/PhysRevD.108.043019}{ {\em Phys. Rev. D\/}
  {\bf 108} 043019 } [\eprint{2304.02283}]

\bibitem{Berti:2016lat}
Berti E, Sesana A, Barausse E, Cardoso V and Belczynski K 2016
  \href{http://dx.doi.org/10.1103/PhysRevLett.117.101102}{ {\em Phys. Rev.
  Lett.\/} {\bf 117} 101102 } [\eprint{1605.09286}]

\bibitem{Perkins:2020tra}
Perkins S~E, Yunes N and Berti E 2021
  \href{http://dx.doi.org/10.1103/PhysRevD.103.044024}{ {\em Phys. Rev. D\/}
  {\bf 103} 044024 } [\eprint{2010.09010}]

\bibitem{Bhagwat:2021kwv}
Bhagwat S, Pacilio C, Barausse E and Pani P 2022
  \href{http://dx.doi.org/10.1103/PhysRevD.105.124063}{ {\em Phys. Rev. D\/}
  {\bf 105} 124063 } [\eprint{2201.00023}]

\bibitem{Du:2018cmp}
Du S~M and Chen Y 2018 \href{http://dx.doi.org/10.1103/PhysRevLett.121.051105}{
  {\em Phys. Rev. Lett.\/} {\bf 121} 051105 } [\eprint{1803.10947}]

\bibitem{Barausse:2018vdb}
Barausse E, Brito R, Cardoso V, Dvorkin I and Pani P 2018
  \href{http://dx.doi.org/10.1088/1361-6382/aae1de}{ {\em Class. Quant.
  Grav.\/} {\bf 35} 20LT01 } [\eprint{1805.08229}]

\bibitem{Laghi:2020rgl}
Laghi D, Carullo G, Veitch J and Del~Pozzo W 2021
  \href{http://dx.doi.org/10.1088/1361-6382/abde19}{ {\em Class. Quant.
  Grav.\/} {\bf 38} 095005 } [\eprint{2011.03816}]

\bibitem{Punturo:2010zz}
Punturo M {\em et~al.\/} 2010
  \href{http://dx.doi.org/10.1088/0264-9381/27/19/194002}{ {\em Class. Quant.
  Grav.\/} {\bf 27} 194002 }

\bibitem{Evans:2023euw}
Evans M {\em et~al.\/} 2023  [\eprint{2306.13745}]

\bibitem{LISA:2024hlh}
Colpi M {\em et~al.\/} (LISA) 2024  [\eprint{2402.07571}]

\bibitem{Abac:2025saz}
Abac A {\em et~al.\/} 2025  [\eprint{2503.12263}]

\bibitem{Gupta:2023lga}
Gupta I {\em et~al.\/} 2024 \href{http://dx.doi.org/10.1088/1361-6382/ad7b99}{
  {\em Class. Quant. Grav.\/} {\bf 41} 245001 } [\eprint{2307.10421}]

\bibitem{Barausse:2020rsu}
Barausse E {\em et~al.\/} 2020
  \href{http://dx.doi.org/10.1007/s10714-020-02691-1}{ {\em Gen. Rel. Grav.\/}
  {\bf 52} 81 } [\eprint{2001.09793}]

\bibitem{Echeverria:1989hg}
Echeverria F 1989 \href{http://dx.doi.org/10.1103/PhysRevD.40.3194}{ {\em Phys.
  Rev. D\/} {\bf 40} 3194--3203 }

\bibitem{Finn:1992wt}
Finn L~S 1992 \href{http://dx.doi.org/10.1103/PhysRevD.46.5236}{ {\em Phys.
  Rev. D\/} {\bf 46} 5236--5249 } [\eprint{gr-qc/9209010}]

\bibitem{Barausse:2008xv}
Barausse E and Sotiriou T~P 2008
  \href{http://dx.doi.org/10.1103/PhysRevLett.101.099001}{ {\em Phys. Rev.
  Lett.\/} {\bf 101} 099001 } [\eprint{0803.3433}]

\bibitem{Bhagwat:2021kfa}
Bhagwat S and Pacilio C 2021
  \href{http://dx.doi.org/10.1103/PhysRevD.104.024030}{ {\em Phys. Rev. D\/}
  {\bf 104} 024030 } [\eprint{2101.07817}]

\bibitem{Ota:2019bzl}
Ota I and Chirenti C 2020 \href{http://dx.doi.org/10.1103/PhysRevD.101.104005}{
  {\em Phys. Rev. D\/} {\bf 101} 104005 } [\eprint{1911.00440}]

\bibitem{Pacilio:2023mvk}
Pacilio C and Bhagwat S 2023
  \href{http://dx.doi.org/10.1103/PhysRevD.107.083021}{ {\em Phys. Rev. D\/}
  {\bf 107} 083021 } [\eprint{2301.02267}]

\bibitem{LISA:2017pwj}
Amaro-Seoane P {\em et~al.\/} (LISA) 2017  [\eprint{1702.00786}]

\bibitem{TianQin:2020hid}
Mei J {\em et~al.\/} (TianQin) 2021
  \href{http://dx.doi.org/10.1093/ptep/ptaa114}{ {\em PTEP\/} {\bf 2021} 05A107
  } [\eprint{2008.10332}]

\bibitem{Luo:2021qji}
Luo Z, Wang Y, Wu Y, Hu W and Jin G 2021
  \href{http://dx.doi.org/10.1093/ptep/ptaa083}{ {\em PTEP\/} {\bf 2021} 05A108
  }

\bibitem{Maselli:2017kvl}
Maselli A, Kokkotas K and Laguna P 2017
  \href{http://dx.doi.org/10.1103/PhysRevD.95.104026}{ {\em Phys. Rev. D\/}
  {\bf 95} 104026 } [\eprint{1702.01110}]

\bibitem{Cabero:2019zyt}
Cabero M, Westerweck J, Capano C~D, Kumar S, Nielsen A~B and Krishnan B 2020
  \href{http://dx.doi.org/10.1103/PhysRevD.101.064044}{ {\em Phys. Rev. D\/}
  {\bf 101} 064044 } [\eprint{1911.01361}]

\bibitem{Yang:2017zxs}
Yang H, Yagi K, Blackman J, Lehner L, Paschalidis V, Pretorius F and Yunes N
  2017 \href{http://dx.doi.org/10.1103/PhysRevLett.118.161101}{ {\em Phys. Rev.
  Lett.\/} {\bf 118} 161101 } [\eprint{1701.05808}]

\bibitem{Baibhav:2018rfk}
Baibhav V and Berti E 2019 \href{http://dx.doi.org/10.1103/PhysRevD.99.024005}{
  {\em Phys. Rev. D\/} {\bf 99} 024005 } [\eprint{1809.03500}]

\bibitem{Mapelli:2021gyv}
Mapelli M, Bouffanais Y, Santoliquido F, Sedda M~A and Artale M~C 2022
  \href{http://dx.doi.org/10.1093/mnras/stac422}{ {\em Mon. Not. Roy. Astron.
  Soc.\/} {\bf 511} 5797--5816 } [\eprint{2109.06222}]

\bibitem{Hild:2010id}
Hild S {\em et~al.\/} 2011
  \href{http://dx.doi.org/10.1088/0264-9381/28/9/094013}{ {\em Class. Quant.
  Grav.\/} {\bf 28} 094013 } [\eprint{1012.0908}]

\bibitem{1984ApJ...278...11G}
{Gehren} T, {Fried} J, {Wehinger} P~A and {Wyckoff} S 1984
  \href{http://dx.doi.org/10.1086/161763}{ {\em Astrophys. J.\/} {\bf 278}
  11--27 }

\bibitem{Kormendy:1995er}
Kormendy J and Richstone D 1995
  \href{http://dx.doi.org/10.1146/annurev.aa.33.090195.003053}{ {\em Ann. Rev.
  Astron. Astrophys.\/} {\bf 33} 581 }

\bibitem{Reines:2011na}
Reines A~E, Sivakoff G~R, Johnson K~E and Brogan C~L 2011
  \href{http://dx.doi.org/10.1038/nature09724}{ {\em Nature\/} {\bf 470} 66--68
  } [\eprint{1101.1309}]

\bibitem{Reines:2013pia}
Reines A~E, Greene J~E and Geha M 2013
  \href{http://dx.doi.org/10.1088/0004-637X/775/2/116}{ {\em Astrophys. J.\/}
  {\bf 775} 116 } [\eprint{1308.0328}]

\bibitem{Baldassare:2019yua}
Baldassare V~F, Geha M and Greene J 2020
  \href{http://dx.doi.org/10.3847/1538-4357/ab8936}{ {\em Astrophys. J.\/} {\bf
  896} 10 } [\eprint{1910.06342}]

\bibitem{EPTA:2023fyk}
Antoniadis J {\em et~al.\/} (EPTA, InPTA:) 2023
  \href{http://dx.doi.org/10.1051/0004-6361/202346844}{ {\em Astron.
  Astrophys.\/} {\bf 678} A50 } [\eprint{2306.16214}]

\bibitem{Tarafdar:2022toa}
Tarafdar P {\em et~al.\/} 2022 \href{http://dx.doi.org/10.1017/pasa.2022.46}{
  {\em Publ. Astron. Soc. Austral.\/} {\bf 39} e053 } [\eprint{2206.09289}]

\bibitem{NANOGrav:2023gor}
Agazie G {\em et~al.\/} (NANOGrav) 2023
  \href{http://dx.doi.org/10.3847/2041-8213/acdac6}{ {\em Astrophys. J.
  Lett.\/} {\bf 951} L8 } [\eprint{2306.16213}]

\bibitem{Reardon:2023gzh}
Reardon D~J {\em et~al.\/} 2023
  \href{http://dx.doi.org/10.3847/2041-8213/acdd02}{ {\em Astrophys. J.
  Lett.\/} {\bf 951} L6 } [\eprint{2306.16215}]

\bibitem{Xu:2023wog}
Xu H {\em et~al.\/} 2023 \href{http://dx.doi.org/10.1088/1674-4527/acdfa5}{
  {\em Res. Astron. Astrophys.\/} {\bf 23} 075024 } [\eprint{2306.16216}]

\bibitem{Barausse:2012fy}
Barausse E 2012 \href{http://dx.doi.org/10.1111/j.1365-2966.2012.21057.x}{ {\em
  Mon. Not. Roy. Astron. Soc.\/} {\bf 423} 2533--2557 } [\eprint{1201.5888}]

\bibitem{Barausse:2023yrx}
Barausse E, Dey K, Crisostomi M, Panayada A, Marsat S and Basak S 2023
  \href{http://dx.doi.org/10.1103/PhysRevD.108.103034}{ {\em Phys. Rev. D\/}
  {\bf 108} 103034 } [\eprint{2307.12245}]

\bibitem{Klein:2015hvg}
Klein A {\em et~al.\/} 2016
  \href{http://dx.doi.org/10.1103/PhysRevD.93.024003}{ {\em Phys. Rev. D\/}
  {\bf 93} 024003 } [\eprint{1511.05581}]

\bibitem{Barausse:2020mdt}
Barausse E, Dvorkin I, Tremmel M, Volonteri M and Bonetti M 2020
  \href{http://dx.doi.org/10.3847/1538-4357/abba7f}{ {\em Astrophys. J.\/} {\bf
  904} 16 } [\eprint{2006.03065}]

\bibitem{Izquierdo-Villalba:2024bhc}
Izquierdo-Villalba D, Sesana A, Colpi M, Spinoso D, Bonetti M, Bonoli S and
  Valiante R 2024 \href{http://dx.doi.org/10.1051/0004-6361/202449293}{ {\em
  Astron. Astrophys.\/} {\bf 686} A183 } [\eprint{2401.10983}]

\bibitem{Babak:2021mhe}
Babak S, Petiteau A and Hewitson M 2021  [\eprint{2108.01167}]

\bibitem{Capuano:2025kkl}
Capuano L, Vaglio M, Chandramouli R~S, Pitte C~L, Kuntz A and Barausse E 2025
  \href{http://dx.doi.org/10.1103/86yd-x1sl}{ {\em Phys. Rev. D\/} {\bf 112}
  104031 } [\eprint{2506.21181}]

\bibitem{Volkel:2025jdx}
V{\"o}lkel S~H and Dhani A 2025 \href{http://dx.doi.org/10.1103/g6sz-dw28}{
  {\em Phys. Rev. D\/} {\bf 112} 084076 } [\eprint{2507.22122}]

\bibitem{Baibhav:2020tma}
Baibhav V, Berti E and Cardoso V 2020
  \href{http://dx.doi.org/10.1103/PhysRevD.101.084053}{ {\em Phys. Rev. D\/}
  {\bf 101} 084053 } [\eprint{2001.10011}]

\bibitem{Ackley:2020atn}
Ackley K {\em et~al.\/} 2020 \href{http://dx.doi.org/10.1017/pasa.2020.39}{
  {\em Publ. Astron. Soc. Austral.\/} {\bf 37} e047 } [\eprint{2007.03128}]

\bibitem{Qiu:2023lwo}
Qiu Y, Forteza X~J and Mourier P 2024
  \href{http://dx.doi.org/10.1103/PhysRevD.109.064075}{ {\em Phys. Rev. D\/}
  {\bf 109} 064075 } [\eprint{2312.15904}]

\bibitem{Shi:2024ttu}
Shi C, Zhang Q and Mei J 2024
  \href{http://dx.doi.org/10.1103/PhysRevD.110.124007}{ {\em Phys. Rev. D\/}
  {\bf 110} 124007 } [\eprint{2407.13110}]

\bibitem{TianQin:2015yph}
Luo J {\em et~al.\/} (TianQin) 2016
  \href{http://dx.doi.org/10.1088/0264-9381/33/3/035010}{ {\em Class. Quant.
  Grav.\/} {\bf 33} 035010 } [\eprint{1512.02076}]

\bibitem{Hartle:1968si}
Hartle J~B and Thorne K~S 1968 \href{http://dx.doi.org/10.1086/149707}{ {\em
  Astrophys. J.\/} {\bf 153} 807 }

\bibitem{DAddario:2023erc}
D'Addario G, Padilla A, Saffin P~M, Sotiriou T~P and Spiers A 2024
  \href{http://dx.doi.org/10.1103/PhysRevD.109.084046}{ {\em Phys. Rev. D\/}
  {\bf 109} 084046 } [\eprint{2311.17666}]

\bibitem{Gao:2024rel}
Gao B, Tang S~P, Wang H~T, Yan J and Fan Y~Z 2024
  \href{http://dx.doi.org/10.1103/PhysRevD.110.044022}{ {\em Phys. Rev. D\/}
  {\bf 110} 044022 } [\eprint{2405.13279}]

\bibitem{Sanger:2024axs}
S\"anger E~M {\em et~al.\/} 2024  [\eprint{2406.03568}]

\bibitem{Pitte:2024zbi}
Pitte C, Baghi Q, Besan\c{c}on M and Petiteau A 2024
  \href{http://dx.doi.org/10.1103/PhysRevD.110.104003}{ {\em Phys. Rev. D\/}
  {\bf 110} 104003 } [\eprint{2406.14552}]

\bibitem{Gerosa:2018qay}
Gerosa D, H\'ebert F and Stein L~C 2018
  \href{http://dx.doi.org/10.1103/PhysRevD.97.104049}{ {\em Phys. Rev. D\/}
  {\bf 97} 104049 } [\eprint{1802.04276}]

\bibitem{Varma:2020nbm}
Varma V, Isi M and Biscoveanu S 2020
  \href{http://dx.doi.org/10.1103/PhysRevLett.124.101104}{ {\em Phys. Rev.
  Lett.\/} {\bf 124} 101104 } [\eprint{2002.00296}]

\bibitem{Varma:2022pld}
Varma V, Biscoveanu S, Islam T, Shaik F~H, Haster C~J, Isi M, Farr W~M, Field
  S~E and Vitale S 2022
  \href{http://dx.doi.org/10.1103/PhysRevLett.128.191102}{ {\em Phys. Rev.
  Lett.\/} {\bf 128} 191102 } [\eprint{2201.01302}]

\bibitem{Toubiana:2021iuw}
Toubiana A, Wong K~W~K, Babak S, Barausse E, Berti E, Gair J~R, Marsat S and
  Taylor S~R 2021 \href{http://dx.doi.org/10.1103/PhysRevD.104.083027}{ {\em
  Phys. Rev. D\/} {\bf 104} 083027 } [\eprint{2106.13819}]

\bibitem{Bonetti:2018tpf}
Bonetti M, Sesana A, Haardt F, Barausse E and Colpi M 2019
  \href{http://dx.doi.org/10.1093/mnras/stz903}{ {\em Mon. Not. Roy. Astron.
  Soc.\/} {\bf 486} 4044--4060 } [\eprint{1812.01011}]

\bibitem{Varma:2018mmi}
Varma V, Field S~E, Scheel M~A, Blackman J, Kidder L~E and Pfeiffer H~P 2019
  \href{http://dx.doi.org/10.1103/PhysRevD.99.064045}{ {\em Phys. Rev. D\/}
  {\bf 99} 064045 } [\eprint{1812.07865}]

\bibitem{sxscatalog}
{The SXS Collaboration} 2025 {SXS Gravitational Waveform Database}
  \urlprefix\url{http://www.black-holes.org/waveforms/}

\bibitem{Nakamura:1987zz}
Nakamura T, Oohara K and Kojima Y 1987
  \href{http://dx.doi.org/10.1143/PTPS.90.1}{ {\em Prog. Theor. Phys. Suppl.\/}
  {\bf 90} 1--218 }

\bibitem{Evans:1989fnr}
{Evans} C~R, {Finn} L~S and {Hobill} D~W 1989 {\em {Frontiers in numerical
  relativity}\/}

\bibitem{Shibata:1995we}
Shibata M and Nakamura T 1995
  \href{http://dx.doi.org/10.1103/PhysRevD.52.5428}{ {\em Phys. Rev.\/} {\bf
  D52} 5428--5444 }

\bibitem{Baumgarte:1998te}
Baumgarte T~W and Shapiro S~L 1998
  \href{http://dx.doi.org/10.1103/PhysRevD.59.024007}{ {\em Phys. Rev.\/} {\bf
  D59} 024007 } [\eprint{gr-qc/9810065}]

\bibitem{Firedrich:1985}
Friedrich H 1985 {\em On the hyperbolicity of Einstein's and other gauge field
  equations\/} (Springer)

\bibitem{Garfinkle:2001ni}
Garfinkle D 2002 \href{http://dx.doi.org/10.1103/PhysRevD.65.044029}{ {\em
  Phys. Rev.\/} {\bf D65} 044029 } [\eprint{gr-qc/0110013}]

\bibitem{hinder:2013oqa}
Hinder I {\em et~al.\/} 2014
  \href{http://dx.doi.org/10.1088/0264-9381/31/2/025012}{ {\em Class. Quant.
  Grav.\/} {\bf 31} 025012 } [\eprint{1307.5307}]

\bibitem{Mezzasoma:2025moh}
Mezzasoma S, Haster C~J, Owen C~B, Cornish N~J and Yunes N 2025
  [\eprint{2503.23304}]

\bibitem{Green:1995mxx}
Green P~J 1995 \href{http://dx.doi.org/10.1093/biomet/82.4.711}{ {\em
  Biometrika\/} {\bf 82} 711--732 }

\bibitem{Tinto:2004wu}
Tinto M and Dhurandhar S~V 2005 \href{http://dx.doi.org/10.12942/lrr-2005-4}{
  {\em Living Rev. Rel.\/} {\bf 8} 4 } [\eprint{gr-qc/0409034}]

\bibitem{Vallisneri:2005ji}
Vallisneri M 2005 \href{http://dx.doi.org/10.1103/PhysRevD.76.109903}{ {\em
  Phys. Rev. D\/} {\bf 72} 042003 } [Erratum: Phys.Rev.D 76, 109903 (2007)]
  [\eprint{gr-qc/0504145}]

\bibitem{Tinto:2020fcc}
Tinto M and Dhurandhar S~V 2021
  \href{http://dx.doi.org/10.1007/s41114-020-00029-6}{ {\em Living Rev. Rel.\/}
  {\bf 24} 1 }

\bibitem{Muratore:2020mdf}
Muratore M, Vetrugno D and Vitale S 2020
  \href{http://dx.doi.org/10.1088/1361-6382/ab9d5b}{ {\em Class. Quant.
  Grav.\/} {\bf 37} 185019 } [\eprint{2001.11221}]

\bibitem{Hartwig:2021mzw}
Hartwig O and Muratore M 2022
  \href{http://dx.doi.org/10.1103/PhysRevD.105.062006}{ {\em Phys. Rev. D\/}
  {\bf 105} 062006 } [\eprint{2111.00975}]

\bibitem{Muratore:2022nbh}
Muratore M, Hartwig O, Vetrugno D, Vitale S and Weber W~J 2023
  \href{http://dx.doi.org/10.1103/PhysRevD.107.082004}{ {\em Phys. Rev. D\/}
  {\bf 107} 082004 } [\eprint{2207.02138}]

\bibitem{Hartwig:2023pft}
Hartwig O, Lilley M, Muratore M and Pieroni M 2023
  \href{http://dx.doi.org/10.1103/PhysRevD.107.123531}{ {\em Phys. Rev. D\/}
  {\bf 107} 123531 } [\eprint{2303.15929}]

\bibitem{Pitte:2023ltw}
Pitte C, Baghi Q, Marsat S, Besan\c{c}on M and Petiteau A 2023
  \href{http://dx.doi.org/10.1103/PhysRevD.108.044053}{ {\em Phys. Rev. D\/}
  {\bf 108} 044053 } [\eprint{2304.03142}]

\bibitem{Rosati:2024lcs}
Rosati R and Littenberg T~B 2024  [\eprint{2410.17180}]

\bibitem{Buscicchio:2024wwm}
Buscicchio R, Klein A, Korol V, Di~Renzo F, Moore C~J, Gerosa D and Carzaniga A
  2025 \href{http://dx.doi.org/10.1140/epjc/s10052-025-14616-w}{ {\em Eur.
  Phys. J. C\/} {\bf 85} 887 } [\eprint{2410.08263}]

\bibitem{Piarulli:2024yhj}
Piarulli M, Buscicchio R, Pozzoli F, Burke O, Bonetti M and Sesana A 2025
  \href{http://dx.doi.org/10.1103/nfn4-pgr5}{ {\em Phys. Rev. D\/} {\bf 111}
  103047 } [\eprint{2410.08862}]

\bibitem{Prince:2002hp}
Prince T~A, Tinto M, Larson S~L and Armstrong J~W 2002
  \href{http://dx.doi.org/10.1103/PhysRevD.66.122002}{ {\em Phys. Rev. D\/}
  {\bf 66} 122002 } [\eprint{gr-qc/0209039}]

\bibitem{Cornish:2020odn}
Cornish N~J 2020 \href{http://dx.doi.org/10.1103/PhysRevD.102.124038}{ {\em
  Phys. Rev. D\/} {\bf 102}(12) 124038 } [\eprint{2009.00043}]

\bibitem{Littenberg:2023xpl}
Littenberg T~B and Cornish N~J 2023
  \href{http://dx.doi.org/10.1103/PhysRevD.107.063004}{ {\em Phys. Rev. D\/}
  {\bf 107} 063004 } [\eprint{2301.03673}]

\bibitem{Katz:2024oqg}
Katz M~L, Karnesis N, Korsakova N, Gair J~R and Stergioulas N 2025
  \href{http://dx.doi.org/10.1103/PhysRevD.111.024060}{ {\em Phys. Rev. D\/}
  {\bf 111} 024060 } [\eprint{2405.04690}]

\bibitem{Edwards:2020tlp}
Edwards M~C, Maturana-Russel P, Meyer R, Gair J, Korsakova N and Christensen N
  2020 \href{http://dx.doi.org/10.1103/PhysRevD.102.084062}{ {\em Phys. Rev.
  D\/} {\bf 102} 084062 } [\eprint{2004.07515}]

\bibitem{Castelli:2024sdb}
Castelli E, Baghi Q, Baker J~G, Slutsky J, Bobin J, Karnesis N, Petiteau A,
  Sauter O, Wass P and Weber W~J 2025
  \href{http://dx.doi.org/10.1088/1361-6382/adb931}{ {\em Class. Quant.
  Grav.\/} {\bf 42} 065018 } [\eprint{2411.13402}]

\bibitem{Houba:2025ckw}
Houba N 2025 \href{http://dx.doi.org/10.1103/6bjw-xjj2}{ {\em Phys. Rev.
  Res.\/} {\bf 7} 033115 } [\eprint{2503.10398}]

\bibitem{Relton:2022whr}
Relton P, Virtuoso A, Bini S, Raymond V, Harry I, Drago M, Lazzaro C, Miani A
  and Tiwari S 2022 \href{http://dx.doi.org/10.1103/PhysRevD.106.104045}{ {\em
  Phys. Rev. D\/} {\bf 106} 104045 } [\eprint{2208.00261}]

\bibitem{Samajdar:2021egv}
Samajdar A, Janquart J, Van Den~Broeck C and Dietrich T 2021
  \href{http://dx.doi.org/10.1103/PhysRevD.104.044003}{ {\em Phys. Rev. D\/}
  {\bf 104} 044003 } [\eprint{2102.07544}]

\bibitem{Wong:2021eun}
Wong I~C~F and Li T~G~F 2022
  \href{http://dx.doi.org/10.1103/PhysRevD.105.084002}{ {\em Phys. Rev. D\/}
  {\bf 105} 084002 } [\eprint{2108.05108}]

\bibitem{Goncharov:2022dgl}
Goncharov B, Nitz A~H and Harms J 2022
  \href{http://dx.doi.org/10.1103/PhysRevD.105.122007}{ {\em Phys. Rev. D\/}
  {\bf 105} 122007 } [\eprint{2204.08533}]

\bibitem{Pizzati:2021apa}
Pizzati E, Sachdev S, Gupta A and Sathyaprakash B 2022
  \href{http://dx.doi.org/10.1103/PhysRevD.105.104016}{ {\em Phys. Rev. D\/}
  {\bf 105} 104016 } [\eprint{2102.07692}]

\bibitem{Hu:2022bji}
Hu Q and Veitch J 2023 \href{http://dx.doi.org/10.3847/1538-4357/acbc18}{ {\em
  Astrophys. J.\/} {\bf 945} 103 } [\eprint{2210.04769}]

\bibitem{Strub:2024kbe}
Strub S~H, Ferraioli L, Schmelzbach C, St\"ahler S~C and Giardini D 2024
  \href{http://dx.doi.org/10.1103/PhysRevD.110.024005}{ {\em Phys. Rev. D\/}
  {\bf 110} 024005 } [\eprint{2403.15318}]

\bibitem{Deng:2025wgk}
Deng S, Babak S, Le~Jeune M, Marsat S, Plagnol E and Sartirana A 2025
  \href{http://dx.doi.org/10.1103/PhysRevD.111.103014}{ {\em Phys. Rev. D\/}
  {\bf 111} 103014 } [\eprint{2501.10277}]

\bibitem{Amaro-Seoane:2009vjl}
Amaro-Seoane P and Santamaria L 2010
  \href{http://dx.doi.org/10.1088/0004-637X/722/2/1197}{ {\em Astrophys. J.\/}
  {\bf 722} 1197--1206 } [\eprint{0910.0254}]

\bibitem{Sesana:2017vsj}
Sesana A 2017 \href{http://dx.doi.org/10.1088/1742-6596/840/1/012018}{ {\em J.
  Phys. Conf. Ser.\/} {\bf 840} 012018 } [\eprint{1702.04356}]

\bibitem{Tso:2018pdv}
Tso R, Gerosa D and Chen Y 2019
  \href{http://dx.doi.org/10.1103/PhysRevD.99.124043}{ {\em Phys. Rev. D\/}
  {\bf 99} 124043 } [\eprint{1807.00075}]

\bibitem{Carson:2019rda}
Carson Z and Yagi K 2020 \href{http://dx.doi.org/10.1088/1361-6382/ab5c9a}{
  {\em Class. Quant. Grav.\/} {\bf 37} 02LT01 } [\eprint{1905.13155}]

\bibitem{Carson:2019kkh}
Carson Z and Yagi K 2020 \href{http://dx.doi.org/10.1103/PhysRevD.101.044047}{
  {\em Phys. Rev. D\/} {\bf 101} 044047 } [\eprint{1911.05258}]

\bibitem{Datta:2020vcj}
Datta S, Gupta A, Kastha S, Arun K~G and Sathyaprakash B~S 2021
  \href{http://dx.doi.org/10.1103/PhysRevD.103.024036}{ {\em Phys. Rev. D\/}
  {\bf 103} 024036 } [\eprint{2006.12137}]

\bibitem{Gupta:2020lxa}
Gupta A, Datta S, Kastha S, Borhanian S, Arun K~G and Sathyaprakash B~S 2020
  \href{http://dx.doi.org/10.1103/PhysRevLett.125.201101}{ {\em Phys. Rev.
  Lett.\/} {\bf 125} 201101 } [\eprint{2005.09607}]

\bibitem{LIGOScientific:2025rid}
Abac A~G {\em et~al.\/} (LIGO Scientific, Virgo, KAGRA) 2025
  \href{http://dx.doi.org/10.1103/kw5g-d732}{ {\em Phys. Rev. Lett.\/} {\bf
  135} 111403 } [\eprint{2509.08054}]

\bibitem{LIGOScientific:2025obp}
Abac A~G {\em et~al.\/} (LIGO Scientific, VIRGO, KAGRA) 2025
  [\eprint{2509.08099}]

\bibitem{icts_school}
Bangalore I 2016 {Summer School on Gravitational-Wave Astronomy}
  \url{https://www.icts.res.in/event/page/3071}

\bibitem{Pani:2013pma}
Pani P 2013 \href{http://dx.doi.org/10.1142/S0217751X13400186}{ {\em Int. J.
  Mod. Phys. A\/} {\bf 28} 1340018 } [\eprint{1305.6759}]

\bibitem{Maggiore:2018sht}
Maggiore M 2018 {\em {Gravitational Waves. Vol. 2: Astrophysics and
  Cosmology}\/} (Oxford University Press) ISBN 978-0-19-857089-9

\bibitem{Andersson:2019yve}
Andersson N 2019 {\em {Gravitational-Wave Astronomy}\/} Oxford Graduate Texts
  (Oxford University Press) ISBN 978-0-19-856803-2

\bibitem{BHPToolkit}
{Black Hole Perturbation Toolkit}
  (\href{http://bhptoolkit.org/}{bhptoolkit.org})

\bibitem{carullo_gregorio_2023_8284026}
Carullo G, De~Amicis M and Redondo-Yuste J 2023 bayring
  \href{https://github.com/GCArullo/bayRing}{github.com/GCArullo/bayRing}
  \urlprefix\url{https://doi.org/10.5281/zenodo.8284026}

\bibitem{KerrRingdownCode}
Cook G~B and Gao L 2025 Kerr{R}ingdown
  \urlprefix\url{https://doi.org/10.5281/zenodo.14804284}

\bibitem{vijay_varma_2018_1435832}
Varma V, Stein L~C and Gerosa D 2018 {vijayvarma392/surfinBH: Surrogate Final
  BH properties}

\bibitem{ringdown}
Isi M and Farr W~M 2024 \textsc{ringdown} package
  \href{https://ringdown.readthedocs.io/en/latest/}{ringdown.readthedocs.io}
  \urlprefix\url{https://doi.org/10.5281/zenodo.5094067}

\bibitem{nitz_pycbc_zenodo}
Nitz A~H, Harry I, Brown D, Biwer C~M, Willis J, Dal~Canton T, Capano C~D, Dent
  T, Pekowsky L, Cabourn~Davies G~S, De S, Cabero M, Wu S, Williamson A~R,
  Machenschalk B, Macleod D, Pannarale F, Kumar P, Reyes S, Finstad D, Kumar S,
  Tápai M, Singer L, Kumar P, Villa~Ortega V, Trevor M, Gadre B, Khan S,
  Fairhurst S and Tolley A Pycbc
  \urlprefix\url{https://zenodo.org/records/10473621}

\bibitem{nitz_2020_pycbc_tutorial}
Nitz A~H, Harry I and Kanner J~B 2020 Pycbc tutorials
  \href{https://github.com/gwastro/PyCBC-Tutorials}{GitHub}
  \urlprefix\url{https://github.com/gwastro/PyCBC-Tutorials}

\bibitem{capano_2021_10493247}
Capano C~D, Westerweck J, Kastha S and Wang Y~F 2021 Bh spectroscopy with
  gw190521 \href{https://github.com/gwastro/BH-Spectroscopy-GW190521}{GitHub}
  \urlprefix\url{https://github.com/gwastro/BH-Spectroscopy-GW190521}

\bibitem{correia_2024_10493247}
Correia A and Capano C 2024 Sky marginalization in black hole spectroscopy
  \href{https://zenodo.org/records/10493247}{Zenodo}
  \urlprefix\url{https://zenodo.org/records/10493247}

\bibitem{lalsuite}
{LIGO Scientific Collaboration}, {Virgo Collaboration} and {KAGRA
  Collaboration} 2018 {LVK} {A}lgorithm {L}ibrary - {LALS}uite Free software
  (GPL)

\bibitem{Mihaylov:2023bkc}
Mihaylov D~P, Ossokine S, Buonanno A, Estelles H, Pompili L, P\"urrer M and
  Ramos-Buades A 2023  [\eprint{2303.18203}]

\bibitem{ashton_2024_10940210}
Ashton G, Talbot C, Roy S, Pratten G, Pang T~H, Agathos Michalis an~Baka T,
  Sänger E, Mehta A, Steinhoff J, Maggio E, Ghosh A and Vijaykumar A 2024
  Bilby tgr \urlprefix\url{https://doi.org/10.5281/zenodo.10940210}

\bibitem{Penrose:1960eq}
Penrose R 1960 \href{http://dx.doi.org/10.1016/0003-4916(60)90021-X}{ {\em
  Annals Phys.\/} {\bf 10} 171--201 }

\bibitem{Penrose:1985bww}
Penrose R and Rindler W 2011 {\em {Spinors and Space-Time}\/} Cambridge
  Monographs on Mathematical Physics (Cambridge, UK: Cambridge Univ. Press)
  ISBN 978-0-521-33707-6, 978-0-511-86766-8, 978-0-521-33707-6

\bibitem{1969eisp.book.....P}
{Petrov} A~Z 1969 {\em {Einstein spaces}\/}

\bibitem{Ryan:1974nt}
Ryan M~P 1974 \href{http://dx.doi.org/10.1103/PhysRevD.10.1736}{ {\em Phys.
  Rev. D\/} {\bf 10} 1736--1740 }

\bibitem{Stewart:1974uz}
Stewart J~M and Walker M 1974 \href{http://dx.doi.org/10.1098/rspa.1974.0172}{
  {\em Proc. Roy. Soc. Lond. A\/} {\bf 341} 49--74 }

\bibitem{Bini:2002jx}
Bini D, Cherubini C, Jantzen R~T and Ruffini R~J 2002
  \href{http://dx.doi.org/10.1143/PTP.107.967}{ {\em Prog. Theor. Phys.\/} {\bf
  107} 967--992 } [\eprint{gr-qc/0203069}]

\bibitem{Maldacena:1997re}
Maldacena J~M 1998 \href{http://dx.doi.org/10.4310/ATMP.1998.v2.n2.a1}{ {\em
  Adv. Theor. Math. Phys.\/} {\bf 2} 231--252 } [\eprint{hep-th/9711200}]

\bibitem{Aharony:1999ti}
Aharony O, Gubser S~S, Maldacena J~M, Ooguri H and Oz Y 2000
  \href{http://dx.doi.org/10.1016/S0370-1573(99)00083-6}{ {\em Phys. Rept.\/}
  {\bf 323} 183--386 } [\eprint{hep-th/9905111}]

\bibitem{Danielsson:1999fa}
Danielsson U~H, Keski-Vakkuri E and Kruczenski M 2000
  \href{http://dx.doi.org/10.1088/1126-6708/2000/02/039}{ {\em JHEP\/} {\bf 02}
  039 } [\eprint{hep-th/9912209}]

\bibitem{Birmingham:2001pj}
Birmingham D, Sachs I and Solodukhin S~N 2002
  \href{http://dx.doi.org/10.1103/PhysRevLett.88.151301}{ {\em Phys. Rev.
  Lett.\/} {\bf 88} 151301 } [\eprint{hep-th/0112055}]

\bibitem{Son:2002sd}
Son D~T and Starinets A~O 2002
  \href{http://dx.doi.org/10.1088/1126-6708/2002/09/042}{ {\em JHEP\/} {\bf 09}
  042 } [\eprint{hep-th/0205051}]

\bibitem{Kovtun:2005ev}
Kovtun P~K and Starinets A~O 2005
  \href{http://dx.doi.org/10.1103/PhysRevD.72.086009}{ {\em Phys. Rev. D\/}
  {\bf 72} 086009 } [\eprint{hep-th/0506184}]

\bibitem{Policastro:2002se}
Policastro G, Son D~T and Starinets A~O 2002
  \href{http://dx.doi.org/10.1088/1126-6708/2002/09/043}{ {\em JHEP\/} {\bf 09}
  043 } [\eprint{hep-th/0205052}]

\bibitem{Friess:2006kw}
Friess J~J, Gubser S~S, Michalogiorgakis G and Pufu S~S 2007
  \href{http://dx.doi.org/10.1088/1126-6708/2007/04/080}{ {\em JHEP\/} {\bf 04}
  080 } [\eprint{hep-th/0611005}]

\bibitem{Michalogiorgakis:2006jc}
Michalogiorgakis G and Pufu S~S 2007
  \href{http://dx.doi.org/10.1088/1126-6708/2007/02/023}{ {\em JHEP\/} {\bf 02}
  023 } [\eprint{hep-th/0612065}]

\bibitem{Gubser:2008px}
Gubser S~S 2008 \href{http://dx.doi.org/10.1103/PhysRevD.78.065034}{ {\em Phys.
  Rev. D\/} {\bf 78} 065034 } [\eprint{0801.2977}]

\bibitem{Horowitz:2009ij}
Horowitz G~T and Roberts M~M 2009
  \href{http://dx.doi.org/10.1088/1126-6708/2009/11/015}{ {\em JHEP\/} {\bf 11}
  015 } [\eprint{0908.3677}]

\bibitem{Hartnoll:2008vx}
Hartnoll S~A, Herzog C~P and Horowitz G~T 2008
  \href{http://dx.doi.org/10.1103/PhysRevLett.101.031601}{ {\em Phys. Rev.
  Lett.\/} {\bf 101} 031601 } [\eprint{0803.3295}]

\bibitem{Hartnoll:2008kx}
Hartnoll S~A, Herzog C~P and Horowitz G~T 2008
  \href{http://dx.doi.org/10.1088/1126-6708/2008/12/015}{ {\em JHEP\/} {\bf 12}
  015 } [\eprint{0810.1563}]

\bibitem{Murata:2010dx}
Murata K, Kinoshita S and Tanahashi N 2010
  \href{http://dx.doi.org/10.1007/JHEP07(2010)050}{ {\em JHEP\/} {\bf 07} 050 }
  [\eprint{1005.0633}]

\bibitem{Herzog:2009xv}
Herzog C~P 2009 \href{http://dx.doi.org/10.1088/1751-8113/42/34/343001}{ {\em
  J. Phys. A\/} {\bf 42} 343001 } [\eprint{0904.1975}]

\bibitem{Horowitz:2010gk}
Horowitz G~T 2011 \href{http://dx.doi.org/10.1007/978-3-642-04864-7_10}{ {\em
  Lect. Notes Phys.\/} {\bf 828} 313--347 } [\eprint{1002.1722}]

\bibitem{Hartnoll:2011fn}
Hartnoll S~A 2012 {\em {Horizons, holography and condensed matter}\/}
  (Cambridge University Press) pp 387--419 [\eprint{1106.4324}]

\bibitem{Iqbal:2011ae}
Iqbal N, Liu H and Mezei M 2011 {Lectures on holographic non-Fermi liquids and
  quantum phase transitions} {\em {Theoretical Advanced Study Institute in
  Elementary Particle Physics}: {String theory and its Applications: From meV
  to the Planck Scale}\/} pp 707--816 [\eprint{1110.3814}]

\bibitem{Musso:2013vtg}
Musso D 2013 \href{http://dx.doi.org/10.22323/1.201.0004}{ {\em PoS\/} {\bf
  Modave2013} 004 } [\eprint{1401.1504}]

\bibitem{Cai:2015cya}
Cai R~G, Li L, Li L~F and Yang R~Q 2015
  \href{http://dx.doi.org/10.1007/s11433-015-5676-5}{ {\em Sci. China Phys.
  Mech. Astron.\/} {\bf 58} 060401 } [\eprint{1502.00437}]

\bibitem{Basu:2010uz}
Basu P, Bhattacharya J, Bhattacharyya S, Loganayagam R, Minwalla S and Umesh V
  2010 \href{http://dx.doi.org/10.1007/JHEP10(2010)045}{ {\em JHEP\/} {\bf 10}
  045 } [\eprint{1003.3232}]

\bibitem{Gentle:2011kv}
Gentle S~A, Rangamani M and Withers B 2012
  \href{http://dx.doi.org/10.1007/JHEP05(2012)106}{ {\em JHEP\/} {\bf 05} 106 }
  [\eprint{1112.3979}]

\bibitem{Dias:2016pma}
Dias O~J~C and Masachs R 2017 \href{http://dx.doi.org/10.1007/JHEP02(2017)128}{
  {\em JHEP\/} {\bf 02} 128 } [\eprint{1610.03496}]

\bibitem{Hawking:1999dp}
Hawking S~W and Reall H~S 2000
  \href{http://dx.doi.org/10.1103/PhysRevD.61.024014}{ {\em Phys. Rev. D\/}
  {\bf 61} 024014 } [\eprint{hep-th/9908109}]

\bibitem{Cardoso:2004hs}
Cardoso V and Dias O~J~C 2004
  \href{http://dx.doi.org/10.1103/PhysRevD.70.084011}{ {\em Phys. Rev. D\/}
  {\bf 70} 084011 } [\eprint{hep-th/0405006}]

\bibitem{Green:2015kur}
Green S~R, Hollands S, Ishibashi A and Wald R~M 2016
  \href{http://dx.doi.org/10.1088/0264-9381/33/12/125022}{ {\em Class. Quant.
  Grav.\/} {\bf 33} 125022 } [\eprint{1512.02644}]

\bibitem{Dias:2011at}
Dias O~J~C, Horowitz G~T and Santos J~E 2011
  \href{http://dx.doi.org/10.1007/JHEP07(2011)115}{ {\em JHEP\/} {\bf 07} 115 }
  [\eprint{1105.4167}]

\bibitem{Stotyn:2011ns}
Stotyn S, Park M, McGrath P and Mann R~B 2012
  \href{http://dx.doi.org/10.1103/PhysRevD.85.044036}{ {\em Phys. Rev. D\/}
  {\bf 85} 044036 } [\eprint{1110.2223}]

\bibitem{Reall:2002bh}
Reall H~S 2003 \href{http://dx.doi.org/10.1103/PhysRevD.70.089902}{ {\em Phys.
  Rev. D\/} {\bf 68} 024024 } [Erratum: Phys.Rev.D 70, 089902 (2004)]
  [\eprint{hep-th/0211290}]

\bibitem{Kunduri:2006qa}
Kunduri H~K, Lucietti J and Reall H~S 2006
  \href{http://dx.doi.org/10.1103/PhysRevD.74.084021,
  10.1142/9789812834300_0151}{ {\em Phys.Rev.\/} {\bf D74} 084021 }
  [\eprint{hep-th/0606076}]

\bibitem{Dias:2015rxy}
Dias O~J~C, Santos J~E and Way B 2015
  \href{http://dx.doi.org/10.1007/JHEP12(2015)171}{ {\em JHEP\/} {\bf 12} 171 }
  [\eprint{1505.04793}]

\bibitem{Ishii:2018oms}
Ishii T and Murata K 2019 \href{http://dx.doi.org/10.1088/1361-6382/ab1d76}{
  {\em Class. Quant. Grav.\/} {\bf 36} 125011 } [\eprint{1810.11089}]

\bibitem{Ishii:2020muv}
Ishii T, Murata K, Santos J~E and Way B 2020
  \href{http://dx.doi.org/10.1007/JHEP07(2020)206}{ {\em JHEP\/} {\bf 07} 206 }
  [\eprint{2005.01201}]

\bibitem{Dias:2011ss}
Dias O~J~C, Horowitz G~T and Santos J~E 2012
  \href{http://dx.doi.org/10.1088/0264-9381/29/19/194002}{ {\em Class. Quant.
  Grav.\/} {\bf 29} 194002 } [\eprint{1109.1825}]

\bibitem{Horowitz:2014hja}
Horowitz G~T and Santos J~E 2015
  \href{http://dx.doi.org/10.4310/sdg.2015.v20.n1.a13}{ {\em Surveys Diff.
  Geom.\/} {\bf 20} 321--335 } [\eprint{1408.5906}]

\bibitem{Martinon:2017uyo}
Martinon G, Fodor G, Grandcl\'ement P and Forg\`acs P 2017
  \href{http://dx.doi.org/10.1088/1361-6382/aa6f48}{ {\em Class. Quant.
  Grav.\/} {\bf 34} 125012 } [\eprint{1701.09100}]

\bibitem{Niehoff:2015oga}
Niehoff B~E, Santos J~E and Way B 2016
  \href{http://dx.doi.org/10.1088/0264-9381/33/18/185012}{ {\em Class. Quant.
  Grav.\/} {\bf 33} 185012 } [\eprint{1510.00709}]

\bibitem{Choptuik:2017cyd}
Choptuik M~W, Dias O~J~C, Santos J~E and Way B 2017
  \href{http://dx.doi.org/10.1103/PhysRevLett.119.191104}{ {\em Phys. Rev.
  Lett.\/} {\bf 119} 191104 } [\eprint{1706.06101}]

\bibitem{Chesler:2018txn}
Chesler P~M and Lowe D~A 2019
  \href{http://dx.doi.org/10.1103/PhysRevLett.122.181101}{ {\em Phys. Rev.
  Lett.\/} {\bf 122} 181101 } [\eprint{1801.09711}]

\bibitem{Chesler:2021ehz}
Chesler P~M 2022 \href{http://dx.doi.org/10.1103/PhysRevD.105.024026}{ {\em
  Phys. Rev. D\/} {\bf 105} 024026 } [\eprint{2109.06901}]

\bibitem{Kim:2023sig}
Kim S, Kundu S, Lee E, Lee J, Minwalla S and Patel C 2023
  \href{http://dx.doi.org/10.1007/JHEP11(2023)024}{ {\em JHEP\/} {\bf 11} 024 }
  [\eprint{2305.08922}]

\bibitem{Dias:2012pp}
Dias O~J~C, Santos J~E and Stein M 2012
  \href{http://dx.doi.org/10.1007/JHEP10(2012)182}{ {\em JHEP\/} {\bf 10} 182 }
  [\eprint{1208.3322}]

\bibitem{Baier:2007ix}
Baier R, Romatschke P, Son D~T, Starinets A~O and Stephanov M~A 2008
  \href{http://dx.doi.org/10.1088/1126-6708/2008/04/100}{ {\em JHEP\/} {\bf 04}
  100 } [\eprint{0712.2451}]

\bibitem{Bhattacharyya:2007vjd}
Bhattacharyya S, Hubeny V~E, Minwalla S and Rangamani M 2008
  \href{http://dx.doi.org/10.1088/1126-6708/2008/02/045}{ {\em JHEP\/} {\bf 02}
  045 } [\eprint{0712.2456}]

\bibitem{Hubeny:2011hd}
Hubeny V~E, Minwalla S and Rangamani M 2012 {The fluid/gravity correspondence}
  {\em {Theoretical Advanced Study Institute in Elementary Particle Physics}:
  {String theory and its Applications: From meV to the Planck Scale}\/} pp
  348--383 [\eprint{1107.5780}]

\bibitem{Graf:2022fve}
Graf O and Holzegel G 2023 \href{http://dx.doi.org/10.1088/1361-6382/acb0ac}{
  {\em Class. Quant. Grav.\/} {\bf 40} 045003 } [\eprint{2205.02801}]

\bibitem{Graf:2024yug}
Graf O and Holzegel G 2024  [\eprint{2410.21994}]

\bibitem{friedrich86}
Friedrich H 1986 \href{http://dx.doi.org/10.1016/0393-0440(86)90004-5}{ {\em J.
  Geom. Phys.\/} {\bf 3} 101--117 }

\bibitem{Christodoulou:1993uv}
Christodoulou D and Klainerman S 1993 {\em {The Global nonlinear stability of
  the Minkowski space}\/} ({Princeton Univ. Press})

\bibitem{Dafermos2006}
Dafermos M 2006 The black hole stability problem {\em Talk at the Newton
  Institute\/} Available at: \href{http://www-
  old.newton.ac.uk/webseminars/pg+ws/2006/gmx/1010/dafermos/}{http://www-
  old.newton.ac.uk/webseminars/pg+ws/2006/gmx/1010/dafermos/} (University of
  Cambridge)

\bibitem{DafermosHolzegel2006}
Dafermos M and Holzegel G 2006 Dynamic instability of solitons in 4+1
  dimensional gravity with negative cosmological constant {\em Seminar at
  DAMTP\/} Available at:
  \href{https://www.dpmms.cam.ac.uk/~md384/ADSinstability.pdf}{https://www.dpmms.cam.ac.uk/$\sim$md384/ADSinstability.pdf}
  (University of Cambridge)

\bibitem{Bizon:2011gg}
Bizon P and Rostworowski A 2011
  \href{http://dx.doi.org/10.1103/PhysRevLett.107.031102}{ {\em Phys. Rev.
  Lett.\/} {\bf 107} 031102 } [\eprint{1104.3702}]

\bibitem{Dias:2012tq}
Dias O~J~C, Horowitz G~T, Marolf D and Santos J~E 2012
  \href{http://dx.doi.org/10.1088/0264-9381/29/23/235019}{ {\em Class. Quant.
  Grav.\/} {\bf 29} 235019 } [\eprint{1208.5772}]

\bibitem{Buchel:2012uh}
Buchel A, Lehner L and Liebling S~L 2012
  \href{http://dx.doi.org/10.1103/PhysRevD.86.123011}{ {\em Phys. Rev.\/} {\bf
  D86} 123011 } [\eprint{1210.0890}]

\bibitem{Buchel:2013uba}
Buchel A, Liebling S~L and Lehner L 2013
  \href{http://dx.doi.org/10.1103/PhysRevD.87.123006}{ {\em Phys. Rev.\/} {\bf
  D87} 123006 } [\eprint{1304.4166}]

\bibitem{Maliborski:2013jca}
Maliborski M and Rostworowski A 2013
  \href{http://dx.doi.org/10.1103/PhysRevLett.111.051102}{ {\em Phys. Rev.
  Lett.\/} {\bf 111} 051102 } [\eprint{1303.3186}]

\bibitem{Bizon:2013xha}
Bizon P and Ja\l~$\!\!$mu\.zna J 2013
  \href{http://dx.doi.org/10.1103/PhysRevLett.111.041102}{ {\em Phys. Rev.
  Lett.\/} {\bf 111} 041102 } [\eprint{1306.0317}]

\bibitem{Maliborski:2012gx}
Maliborski M 2012 \href{http://dx.doi.org/10.1103/PhysRevLett.109.221101}{ {\em
  Phys. Rev. Lett.\/} {\bf 109} 221101 } [\eprint{1208.2934}]

\bibitem{Maliborski:2013ula}
Maliborski M and Rostworowski A 2013  [\eprint{1307.2875}]

\bibitem{Baier:2013gsa}
Baier R, Stricker S~A and Taanila O 2014
  \href{http://dx.doi.org/10.1088/0264-9381/31/2/025007}{ {\em Class. Quant.
  Grav.\/} {\bf 31} 025007 } [\eprint{1309.1629}]

\bibitem{Jalmuzna:2013rwa}
Ja\l~$\!\!$mu\.zna J 2013 \href{http://dx.doi.org/10.5506/APhysPolB.44.2603}{
  {\em Acta Phys. Polon.\/} {\bf B44} 2603--2620 } [\eprint{1311.7409}]

\bibitem{Basu:2012gg}
Basu P, Das D, Das S~R and Nishioka T 2013
  \href{http://dx.doi.org/10.1007/JHEP03(2013)146}{ {\em JHEP\/} {\bf 03} 146 }
  [\eprint{1211.7076}]

\bibitem{Gannot:2012pb}
Gannot O 2014 \href{http://dx.doi.org/10.1007/s00220-014-2002-4}{ {\em Commun.
  Math. Phys.\/} {\bf 330} 771--799 } [\eprint{1212.1907}]

\bibitem{Fodor:2013lza}
Fodor G, Forg\'acs P and Grandcl\'ement P 2014
  \href{http://dx.doi.org/10.1103/PhysRevD.89.065027}{ {\em Phys. Rev.\/} {\bf
  D89} 065027 } [\eprint{1312.7562}]

\bibitem{Friedrich:2014raa}
Friedrich H 2014 \href{http://dx.doi.org/10.1088/0264-9381/31/10/105001}{ {\em
  Class. Quant. Grav.\/} {\bf 31} 105001 } [\eprint{1401.7172}]

\bibitem{Maliborski:2014rma}
Maliborski M and Rostworowski A 2014
  \href{http://dx.doi.org/10.1103/PhysRevD.89.124006}{ {\em Phys. Rev.\/} {\bf
  D89} 124006 } [\eprint{1403.5434}]

\bibitem{Abajo-Arrastia:2014fma}
Abajo-Arrastia J, da~Silva E, Lopez E, Mas J and Serantes A 2014
  \href{http://dx.doi.org/10.1007/JHEP05(2014)126}{ {\em JHEP\/} {\bf 05} 126 }
  [\eprint{1403.2632}]

\bibitem{Balasubramanian:2014cja}
Balasubramanian V, Buchel A, Green S~R, Lehner L and Liebling S~L 2014
  \href{http://dx.doi.org/10.1103/PhysRevLett.113.071601}{ {\em Phys. Rev.
  Lett.\/} {\bf 113} 071601 } [\eprint{1403.6471}]

\bibitem{Bizon:2014bya}
Bizon P and Rostworowski A 2015
  \href{http://dx.doi.org/10.1103/PhysRevLett.115.049101}{ {\em Phys. Rev.
  Lett.\/} {\bf 115} 049101 } [\eprint{1410.2631}]

\bibitem{Balasubramanian:2015uua}
Balasubramanian V, Buchel A, Green S~R, Lehner L and Liebling S~L 2015
  \href{http://dx.doi.org/10.1103/PhysRevLett.115.049102}{ {\em Phys. Rev.
  Lett.\/} {\bf 115} 049102 } [\eprint{1506.07907}]

\bibitem{daSilva:2014zva}
da~Silva E, Lopez E, Mas J and Serantes A 2015
  \href{http://dx.doi.org/10.1007/JHEP04(2015)038}{ {\em JHEP\/} {\bf 04} 038 }
  [\eprint{1412.6002}]

\bibitem{Craps:2014vaa}
Craps B, Evnin O and Vanhoof J 2014
  \href{http://dx.doi.org/10.1007/JHEP10(2014)048}{ {\em JHEP\/} {\bf 10} 48 }
  [\eprint{1407.6273}]

\bibitem{Basu:2014sia}
Basu P, Krishnan C and Saurabh A 2015
  \href{http://dx.doi.org/10.1142/S0217751X15501286}{ {\em Int. J. Mod.
  Phys.\/} {\bf A30} 1550128 } [\eprint{1408.0624}]

\bibitem{Okawa:2014nea}
Okawa H, Cardoso V and Pani P 2014
  \href{http://dx.doi.org/10.1103/PhysRevD.90.104032}{ {\em Phys. Rev.\/} {\bf
  D90} 104032 } [\eprint{1409.0533}]

\bibitem{Deppe:2014oua}
Deppe N, Kolly A, Frey A and Kunstatter G 2015
  \href{http://dx.doi.org/10.1103/PhysRevLett.114.071102}{ {\em Phys. Rev.
  Lett.\/} {\bf 114} 071102 } [\eprint{1410.1869}]

\bibitem{Dimitrakopoulos:2014ada}
Dimitrakopoulos F~V, Freivogel B, Lippert M and Yang I~S 2015
  \href{http://dx.doi.org/10.1007/JHEP08(2015)077}{ {\em JHEP\/} {\bf 08} 077 }
  [\eprint{1410.1880}]

\bibitem{Buchel:2014xwa}
Buchel A, Green S~R, Lehner L and Liebling S~L 2015
  \href{http://dx.doi.org/10.1103/PhysRevD.91.064026}{ {\em Phys. Rev.\/} {\bf
  D91} 064026 } [\eprint{1412.4761}]

\bibitem{Craps:2014jwa}
Craps B, Evnin O and Vanhoof J 2015
  \href{http://dx.doi.org/10.1007/JHEP01(2015)108}{ {\em JHEP\/} {\bf 01} 108 }
  [\eprint{1412.3249}]

\bibitem{Basu:2015efa}
Basu P, Krishnan C and Bala~Subramanian P~N 2015
  \href{http://dx.doi.org/10.1016/j.physletb.2015.05.009}{ {\em Phys. Lett.\/}
  {\bf B746} 261--265 } [\eprint{1501.07499}]

\bibitem{Yang:2015jha}
Yang I~S 2015 \href{http://dx.doi.org/10.1103/PhysRevD.91.065011}{ {\em Phys.
  Rev.\/} {\bf D91} 065011 } [\eprint{1501.00998}]

\bibitem{Fodor:2015eia}
Fodor G, Forg\'acs P and Grandcl\'ement P 2015
  \href{http://dx.doi.org/10.1103/PhysRevD.92.025036}{ {\em Phys. Rev.\/} {\bf
  D92} 025036 } [\eprint{1503.07746}]

\bibitem{Okawa:2015xma}
Okawa H, Lopes J~C and Cardoso V 2015  [\eprint{1504.05203}]

\bibitem{Bizon:2015pfa}
Bizon P, Maliborski M and Rostworowski A 2015
  \href{http://dx.doi.org/10.1103/PhysRevLett.115.081103}{ {\em Phys. Rev.
  Lett.\/} {\bf 115} 081103 } [\eprint{1506.03519}]

\bibitem{Dimitrakopoulos:2015pwa}
Dimitrakopoulos F and Yang I~S 2015
  \href{http://dx.doi.org/10.1103/PhysRevD.92.083013}{ {\em Phys. Rev.\/} {\bf
  D92} 083013 } [\eprint{1507.02684}]

\bibitem{Green:2015dsa}
Green S~R, Maillard A, Lehner L and Liebling S~L 2015
  \href{http://dx.doi.org/10.1103/PhysRevD.92.084001}{ {\em Phys. Rev. D\/}
  {\bf 92} 084001 } [\eprint{1507.08261}]

\bibitem{Deppe:2015qsa}
Deppe N and Frey A~R 2015 \href{http://dx.doi.org/10.1007/JHEP12(2015)004}{
  {\em JHEP\/} {\bf 12} 004 } [\eprint{1508.02709}]

\bibitem{Craps:2015iia}
Craps B, Evnin O and Vanhoof J 2015
  \href{http://dx.doi.org/10.1007/JHEP10(2015)079}{ {\em JHEP\/} {\bf 10} 079 }
  [\eprint{1508.04943}]

\bibitem{Craps:2015xya}
Craps B, Evnin O, Jai-akson P and Vanhoof J 2015
  \href{http://dx.doi.org/10.1007/JHEP10(2015)080}{ {\em JHEP\/} {\bf 10} 080 }
  [\eprint{1508.05474}]

\bibitem{Evnin:2015gma}
Evnin O and Krishnan C 2015
  \href{http://dx.doi.org/10.1103/PhysRevD.91.126010}{ {\em Phys. Rev.\/} {\bf
  D91} 126010 } [\eprint{1502.03749}]

\bibitem{Menon:2015oda}
Menon D~S and Suneeta V 2016
  \href{http://dx.doi.org/10.1103/PhysRevD.93.024044}{ {\em Phys. Rev. D\/}
  {\bf 93} 024044 } [\eprint{1509.00232}]

\bibitem{Jalmuzna:2015hoa}
Jalmuzna J, Gundlach C and Chmaj T 2015
  \href{http://dx.doi.org/10.1103/PhysRevD.92.124044}{ {\em Phys. Rev.\/} {\bf
  D92} 124044 } [\eprint{1510.02592}]

\bibitem{Evnin:2015wyi}
Evnin O and Nivesvivat R 2016 \href{http://dx.doi.org/10.1007/JHEP01(2016)151}{
  {\em JHEP\/} {\bf 01} 151 } [\eprint{1512.00349}]

\bibitem{Freivogel:2015wib}
Freivogel B and Yang I~S 2016
  \href{http://dx.doi.org/10.1103/PhysRevD.93.103007}{ {\em Phys. Rev. D\/}
  {\bf 93} 103007 } [\eprint{1512.04383}]

\bibitem{Dias:2016ewl}
Dias O and Santos J~E 2016
  \href{http://dx.doi.org/10.1088/0264-9381/33/23/23LT01}{ {\em Class. Quant.
  Grav.\/} {\bf 33} 23LT01 } [\eprint{1602.03890}]

\bibitem{Evnin:2016mjx}
Evnin O and Jai-akson P 2016 \href{http://dx.doi.org/10.1007/JHEP04(2016)054}{
  {\em JHEP\/} {\bf 04} 054 } [\eprint{1602.05859}]

\bibitem{Deppe:2016gur}
Deppe N 2019 \href{http://dx.doi.org/10.1103/PhysRevD.100.124028}{ {\em Phys.
  Rev. D\/} {\bf 100} 124028 } [\eprint{1606.02712}]

\bibitem{Dimitrakopoulos:2016tss}
Dimitrakopoulos F~V, Freivogel B, Pedraza J~F and Yang I~S 2016
  \href{http://dx.doi.org/10.1103/PhysRevD.94.124008}{ {\em Phys. Rev.\/} {\bf
  D94} 124008 } [\eprint{1607.08094}]

\bibitem{Dimitrakopoulos:2016euh}
Dimitrakopoulos F~V, Freivogel B and Pedraza J~F 2018
  \href{http://dx.doi.org/10.1088/1361-6382/aac0b5}{ {\em Class. Quant.
  Grav.\/} {\bf 35} 125008 } [\eprint{1612.04758}]

\bibitem{Rostworowski:2016isb}
Rostworowski A 2017 \href{http://dx.doi.org/10.1088/1361-6382/aa71cc}{ {\em
  Class. Quant. Grav.\/} {\bf 34} 128001 } [\eprint{1612.00042}]

\bibitem{Rostworowski:2017tcx}
Rostworowski A 2017 \href{http://dx.doi.org/10.1103/PhysRevD.95.124043}{ {\em
  Phys. Rev. D\/} {\bf 95} 124043 } [\eprint{1701.07804}]

\bibitem{Moschidis:2017lcr}
Moschidis G 2017  [\eprint{1704.08685}]

\bibitem{Moschidis:2017llu}
Moschidis G 2020 \href{http://dx.doi.org/10.2140/apde.2020.13.1671}{ {\em Anal.
  Part. Diff. Eq.\/} {\bf 13} 1671--1754 } [\eprint{1704.08681}]

\bibitem{Dias:2017tjg}
Dias O~J~C and Santos J~E 2018
  \href{http://dx.doi.org/10.1088/1361-6382/aad514}{ {\em Class. Quant.
  Grav.\/} {\bf 35} 185006 } [\eprint{1705.03065}]

\bibitem{Banados:1992wn}
Banados M, Teitelboim C and Zanelli J 1992
  \href{http://dx.doi.org/10.1103/PhysRevLett.69.1849}{ {\em Phys. Rev.
  Lett.\/} {\bf 69} 1849--1851 } [\eprint{hep-th/9204099}]

\bibitem{Cardoso:2001hn}
Cardoso V and Lemos J~P~S 2001
  \href{http://dx.doi.org/10.1103/PhysRevD.63.124015}{ {\em Phys. Rev. D\/}
  {\bf 63} 124015 } [\eprint{gr-qc/0101052}]

\bibitem{Birmingham:2001hc}
Birmingham D 2001 \href{http://dx.doi.org/10.1103/PhysRevD.64.064024}{ {\em
  Phys. Rev. D\/} {\bf 64} 064024 } [\eprint{hep-th/0101194}]

\bibitem{Ishibashi:2004wx}
Ishibashi A and Wald R~M 2004
  \href{http://dx.doi.org/10.1088/0264-9381/21/12/012}{ {\em Class. Quant.
  Grav.\/} {\bf 21} 2981--3014 } [\eprint{hep-th/0402184}]

\bibitem{Astefanesei:2003rw}
Astefanesei D and Radu E 2004
  \href{http://dx.doi.org/10.1016/j.physletb.2004.03.006}{ {\em Phys. Lett.
  B\/} {\bf 587} 7--15 } [\eprint{gr-qc/0310135}]

\bibitem{Stotyn:2012ap}
Stotyn S and Mann R~B 2012
  \href{http://dx.doi.org/10.1088/1751-8113/45/37/374025}{ {\em J. Phys. A\/}
  {\bf 45} 374025 } [\eprint{1203.0214}]

\bibitem{Stotyn:2013spa}
Stotyn S, Chanona M and Mann R~B 2014
  \href{http://dx.doi.org/10.1103/PhysRevD.89.044018}{ {\em Phys. Rev. D\/}
  {\bf 89} 044018 } [\eprint{1309.2911}]

\bibitem{Iizuka:2015vsa}
Iizuka N, Ishibashi A and Maeda K 2015
  \href{http://dx.doi.org/10.1007/JHEP08(2015)112}{ {\em JHEP\/} {\bf 08} 112 }
  [\eprint{1505.00394}]

\bibitem{Ferreira:2017cta}
Ferreira H~R~C and Herdeiro C~A~R 2017
  \href{http://dx.doi.org/10.1016/j.physletb.2017.08.017}{ {\em Phys. Lett.
  B\/} {\bf 773} 129--134 } [\eprint{1707.08133}]

\bibitem{Dappiaggi:2017pbe}
Dappiaggi C, Ferreira H~R~C and Herdeiro C~A~R 2018
  \href{http://dx.doi.org/10.1016/j.physletb.2018.01.018}{ {\em Phys. Lett.
  B\/} {\bf 778} 146--154 } [\eprint{1710.08039}]

\bibitem{Gao:2023rqc}
Gao L~L, Liu Y and Lyu H~D 2024
  \href{http://dx.doi.org/10.1007/JHEP01(2024)063}{ {\em JHEP\/} {\bf 01} 063 }
  [\eprint{2310.15781}]

\bibitem{Emparan:1999wa}
Emparan R, Horowitz G~T and Myers R~C 2000
  \href{http://dx.doi.org/10.1088/1126-6708/2000/01/007}{ {\em JHEP\/} {\bf 01}
  007 } [\eprint{hep-th/9911043}]

\bibitem{Emparan:1999fd}
Emparan R, Horowitz G~T and Myers R~C 2000
  \href{http://dx.doi.org/10.1088/1126-6708/2000/01/021}{ {\em JHEP\/} {\bf 01}
  021 } [\eprint{hep-th/9912135}]

\bibitem{Cartwright:2024iwc}
Cartwright C, G{\"u}rsoy U, Pedraza J~F and Planella~Planas G 2025
  \href{http://dx.doi.org/10.1007/JHEP03(2025)039}{ {\em JHEP\/} {\bf 03} 039 }
  [\eprint{2408.08010}]

\bibitem{Cartwright:2025fay}
Cartwright C, G{\"u}rsoy U, Pedraza J~F and Svesko A 2025
  [\eprint{2501.17231}]

\bibitem{Christodoulou:2008nj}
Christodoulou D 2009 {\em The Formation of Black Holes in General Relativity\/}
  EMS Monographs in Mathematics (Z{\"u}rich: European Mathematical Society)
  ISBN 978-3-03719-068-5

\bibitem{penrose1978}
Penrose R 1978 Singularities of spacetime {\em Theoretical principles in
  astrophysics and relativity\/} ed NR~Lebovitz W~R and Vandervoort P (Chicago
  University Press) pp 217--243

\bibitem{McNamara121}
McNamara J 1978 \href{http://dx.doi.org/10.1098/rspa.1978.0191}{ {\em Proc. R.
  Soc. Lond. A\/} {\bf 364} 121--134 } ISSN 0080-4630
  \urlprefix\url{http://rspa.royalsocietypublishing.org/content/364/1716/121}

\bibitem{Ori:1991zz}
Ori A 1991 \href{http://dx.doi.org/10.1103/PhysRevLett.67.789}{ {\em Phys. Rev.
  Lett.\/} {\bf 67} 789--792 }

\bibitem{Dafermos:2003wr}
Dafermos M 2005 \href{http://dx.doi.org/10.1002/cpa.20071}{ {\em Commun. Pure
  Appl. Math.\/} {\bf 58} 0445--0504 } [\eprint{gr-qc/0307013}]

\bibitem{Franzen:2014sqa}
Franzen A~T 2016 \href{http://dx.doi.org/10.1007/s00220-015-2440-7}{ {\em
  Commun. Math. Phys.\/} {\bf 343} 601--650 } [\eprint{1407.7093}]

\bibitem{Dafermos:2017dbw}
Dafermos M and Luk J 2017  [\eprint{1710.01722}]

\bibitem{Luk:2015qja}
Luk J and Oh S~J 2017 \href{http://dx.doi.org/10.1215/00127094-3715189}{ {\em
  Duke Math. J.\/} {\bf 166} 437--493 } [\eprint{1501.04598}]

\bibitem{Dafermos:2012np}
Dafermos M 2014 \href{http://dx.doi.org/10.1007/s00220-014-2063-4}{ {\em
  Commun. Math. Phys.\/} {\bf 332} 729--757 } [\eprint{1201.1797}]

\bibitem{Dafermos:2015bzz}
Dafermos M and Shlapentokh-Rothman Y 2017
  \href{http://dx.doi.org/10.1007/s00220-016-2771-z}{ {\em Commun. Math.
  Phys.\/} {\bf 350} 985--1016 } [\eprint{1512.08260}]

\bibitem{Poisson:1990eh}
Poisson E and Israel W 1990 \href{http://dx.doi.org/10.1103/PhysRevD.41.1796}{
  {\em Phys. Rev. D\/} {\bf 41} 1796--1809 }

\bibitem{Simpson:1973ua}
Simpson M and Penrose R 1973 \href{http://dx.doi.org/10.1007/BF00792069}{ {\em
  Int. J. Theor. Phys.\/} {\bf 7} 183--197 }

\bibitem{mcnamara1978instability}
McNamara J~M 1978 {\em Proceedings of the Royal Society of London. A.
  Mathematical and Physical Sciences\/} {\bf 358} 499--517

\bibitem{chandrasekhar1982crossing}
Chandrasekhar S and Hartle J~B 1982 {\em Proceedings of the Royal Society of
  London. A. Mathematical and Physical Sciences\/} {\bf 384} 301--315

\bibitem{Hintz:2015jkj}
Hintz P and Vasy A 2017 \href{http://dx.doi.org/10.1063/1.4996575}{ {\em J.
  Math. Phys.\/} {\bf 58} 081509 } [\eprint{1512.08004}]

\bibitem{Hintz:2015koq}
Hintz P 2017 \href{http://dx.doi.org/10.4171/cmh/425}{ {\em Comment. Math.
  Helv.\/} {\bf 92} 801--837 } [\eprint{1512.08003}]

\bibitem{MR1432814}
S\'a{}~Barreto A and Zworski M 1997
  \href{http://dx.doi.org/10.4310/MRL.1997.v4.n1.a10}{ {\em Math. Res. Lett.\/}
  {\bf 4} 103--121 } ISSN 1073-2780

\bibitem{MR2426141}
Bony J~F~c and H\"afner D 2008
  \href{http://dx.doi.org/10.1007/s00220-008-0553-y}{ {\em Comm. Math. Phys.\/}
  {\bf 282} 697--719 } ISSN 0010-3616,1432-0916

\bibitem{Dyatlov:2011jd}
Dyatlov S 2012 \href{http://dx.doi.org/10.1007/s00023-012-0159-y}{ {\em Annales
  Henri Poincare\/} {\bf 13} 1101--1166 } [\eprint{1101.1260}]

\bibitem{Dyatlov:2013hba}
Dyatlov S 2015 \href{http://dx.doi.org/10.1007/s00220-014-2255-y}{ {\em Commun.
  Math. Phys.\/} {\bf 335} 1445--1485 } [\eprint{1305.1723}]

\bibitem{Hintz:2016gwb}
Hintz P and Vasy A 2018 \href{http://dx.doi.org/10.4310/ACTA.2018.v220.n1.a1}{
  {\em Acta Math.\/} {\bf 220} 1--206 } [\eprint{1606.04014}]

\bibitem{Hintz:2016jak}
Hintz P 2018 \href{http://dx.doi.org/10.1007/s40818-018-0047-y}{ {\em Annals of
  PDE\/} {\bf 4} 1--131 } [\eprint{1612.04489}]

\bibitem{Cardoso:2017soq}
Cardoso V, Costa J~a~L, Destounis K, Hintz P and Jansen A 2018
  \href{http://dx.doi.org/10.1103/PhysRevLett.120.031103}{ {\em Phys. Rev.
  Lett.\/} {\bf 120} 031103 } [\eprint{1711.10502}]

\bibitem{Cardoso:2018nvb}
Cardoso V, Costa J~L, Destounis K, Hintz P and Jansen A 2018
  \href{http://dx.doi.org/10.1103/PhysRevD.98.104007}{ {\em Phys. Rev. D\/}
  {\bf 98} 104007 } [\eprint{1808.03631}]

\bibitem{Mo:2018nnu}
Mo Y, Tian Y, Wang B, Zhang H and Zhong Z 2018
  \href{http://dx.doi.org/10.1103/PhysRevD.98.124025}{ {\em Phys. Rev. D\/}
  {\bf 98} 124025 } [\eprint{1808.03635}]

\bibitem{Dias:2018ufh}
Dias O~J~C, Reall H~S and Santos J~E 2019
  \href{http://dx.doi.org/10.1088/1361-6382/aafcf2}{ {\em Class. Quant.
  Grav.\/} {\bf 36} 045005 } [\eprint{1808.04832}]

\bibitem{Dias:2018ynt}
Dias O~J~C, Eperon F~C, Reall H~S and Santos J~E 2018
  \href{http://dx.doi.org/10.1103/PhysRevD.97.104060}{ {\em Phys. Rev. D\/}
  {\bf 97} 104060 } [\eprint{1801.09694}]

\bibitem{Casals:2020uxa}
Casals M and Marinho C~I~S 2022
  \href{http://dx.doi.org/10.1103/PhysRevD.106.044060}{ {\em Phys. Rev. D\/}
  {\bf 106} 044060 } [\eprint{2006.06483}]

\bibitem{Luna:2019olw}
Luna R, Zilh\~ao M, Cardoso V, Costa J~a~L and Nat\'ario J 2019
  \href{http://dx.doi.org/10.1103/PhysRevD.99.064014}{ {\em Phys. Rev. D\/}
  {\bf 99} 064014 } [Addendum: Phys.Rev.D 103, 104043 (2021)]
  [\eprint{1810.00886}]

\bibitem{Zhang:2019nye}
Zhang H and Zhong Z 2019  [\eprint{1910.01610}]

\bibitem{Luna:2019olwaddendum}
Luna R, Zilh\~ao M, Cardoso V, Costa J~a~L and Nat\'ario J 2021
  \href{http://dx.doi.org/10.1103/PhysRevD.103.104043}{ {\em Phys. Rev. D\/}
  {\bf 103} 104043 } [Addendum: Phys.Rev.D 103, 104043 (2021)]
  [\eprint{2012.08413}]

\bibitem{Dafermos:2018tha}
Dafermos M and Shlapentokh-Rothman Y 2018
  \href{http://dx.doi.org/10.1088/1361-6382/aadbcf}{ {\em Class. Quant.
  Grav.\/} {\bf 35} 195010 } [\eprint{1805.08764}]

\bibitem{Mellor:1989ac}
Mellor F and Moss I 1990 \href{http://dx.doi.org/10.1103/PhysRevD.41.403}{ {\em
  Phys. Rev. D\/} {\bf 41} 403 }

\bibitem{Mellor1992}
Mellor F and Moss I 1992 \href{http://dx.doi.org/10.1088/0264-9381/9/4/001}{
  {\em Class. Quantum Grav.\/} {\bf 9} L43 }

\bibitem{Brady:1998au}
Brady P~R, Moss I~G and Myers R~C 1998
  \href{http://dx.doi.org/10.1103/PhysRevLett.80.3432}{ {\em Phys. Rev.
  Lett.\/} {\bf 80} 3432--3435 } [\eprint{gr-qc/9801032}]

\bibitem{Dias:2019ery}
Dias O~J~C, Reall H~S and Santos J~E 2019
  \href{http://dx.doi.org/10.1007/JHEP12(2019)097}{ {\em JHEP\/} {\bf 12} 097 }
  [\eprint{1906.08265}]

\bibitem{Hollands:2019whz}
Hollands S, Wald R~M and Zahn J 2020
  \href{http://dx.doi.org/10.1088/1361-6382/ab8052}{ {\em Class. Quant.
  Grav.\/} {\bf 37} 115009 } [\eprint{1912.06047}]

\bibitem{Hollands:2020qpe}
Hollands S, Klein C and Zahn J 2020
  \href{http://dx.doi.org/10.1103/PhysRevD.102.085004}{ {\em Phys. Rev. D\/}
  {\bf 102} 085004 } [\eprint{2006.10991}]

\bibitem{Klein:2024sdd}
Klein C, Soltani M, Casals M and Hollands S 2024
  \href{http://dx.doi.org/10.1103/PhysRevLett.132.121501}{ {\em Phys. Rev.
  Lett.\/} {\bf 132} 121501 } [\eprint{2402.14171}]

\bibitem{Emparan:2020rnp}
Emparan R and Toma\v{s}evi\'c M 2020
  \href{http://dx.doi.org/10.1007/JHEP06(2020)038}{ {\em JHEP\/} {\bf 06} 038 }
  [\eprint{2002.02083}]

\bibitem{Emparan:2020znc}
Emparan R, Frassino A~M and Way B 2020
  \href{http://dx.doi.org/10.1007/JHEP11(2020)137}{ {\em JHEP\/} {\bf 11} 137 }
  [\eprint{2007.15999}]

\bibitem{Horowitz:1991cd}
Horowitz G~T and Strominger A 1991
  \href{http://dx.doi.org/10.1016/0550-3213(91)90440-9}{ {\em Nucl. Phys. B\/}
  {\bf 360} 197--209 }

\bibitem{Gregory:1993vy}
Gregory R and Laflamme R 1993
  \href{http://dx.doi.org/10.1103/PhysRevLett.70.2837}{ {\em Phys.Rev.Lett.\/}
  {\bf 70} 2837--2840 } [\eprint{hep-th/9301052}]

\bibitem{Gregory:1994bj}
Gregory R and Laflamme R 1994
  \href{http://dx.doi.org/10.1016/0550-3213(94)90206-2}{ {\em Nucl. Phys. B\/}
  {\bf 428} 399--434 } [\eprint{hep-th/9404071}]

\bibitem{Cardoso:2006ks}
Cardoso V and Dias O~J~C 2006
  \href{http://dx.doi.org/10.1103/PhysRevLett.96.181601}{ {\em Phys. Rev.
  Lett.\/} {\bf 96} 181601 } [\eprint{hep-th/0602017}]

\bibitem{Gubser:2001ac}
Gubser S~S 2002 \href{http://dx.doi.org/10.1088/0264-9381/19/19/303}{ {\em
  Class. Quant. Grav.\/} {\bf 19} 4825--4844 } [\eprint{hep-th/0110193}]

\bibitem{Harmark:2002tr}
Harmark T and Obers N~A 2002
  \href{http://dx.doi.org/10.1088/1126-6708/2002/05/032}{ {\em JHEP\/} {\bf 05}
  032 } [\eprint{hep-th/0204047}]

\bibitem{Kol:2002xz}
Kol B 2005 \href{http://dx.doi.org/10.1088/1126-6708/2005/10/049}{ {\em JHEP\/}
  {\bf 10} 049 } [\eprint{hep-th/0206220}]

\bibitem{Wiseman:2002zc}
Wiseman T 2003 \href{http://dx.doi.org/10.1088/0264-9381/20/6/308}{ {\em Class.
  Quant. Grav.\/} {\bf 20} 1137--1176 } [\eprint{hep-th/0209051}]

\bibitem{Kol:2003ja}
Kol B and Wiseman T 2003 \href{http://dx.doi.org/10.1088/0264-9381/20/15/315}{
  {\em Class. Quant. Grav.\/} {\bf 20} 3493--3504 } [\eprint{hep-th/0304070}]

\bibitem{Harmark:2003dg}
Harmark T and Obers N~A 2004
  \href{http://dx.doi.org/10.1088/0264-9381/21/6/026}{ {\em Class. Quant.
  Grav.\/} {\bf 21} 1709 } [\eprint{hep-th/0309116}]

\bibitem{Harmark:2003yz}
Harmark T 2004 \href{http://dx.doi.org/10.1103/PhysRevD.69.104015}{ {\em Phys.
  Rev.\/} {\bf D69} 104015 } [\eprint{hep-th/0310259}]

\bibitem{Kudoh:2003ki}
Kudoh H and Wiseman T 2004 \href{http://dx.doi.org/10.1143/PTP.111.475}{ {\em
  Prog. Theor. Phys.\/} {\bf 111} 475--507 } [\eprint{hep-th/0310104}]

\bibitem{Sorkin:2004qq}
Sorkin E 2004 \href{http://dx.doi.org/10.1103/PhysRevLett.93.031601}{ {\em
  Phys. Rev. Lett.\/} {\bf 93} 031601 } [\eprint{hep-th/0402216}]

\bibitem{Gorbonos:2004uc}
Gorbonos D and Kol B 2004
  \href{http://dx.doi.org/10.1088/1126-6708/2004/06/053}{ {\em JHEP\/} {\bf 06}
  053 } [\eprint{hep-th/0406002}]

\bibitem{Kudoh:2004hs}
Kudoh H and Wiseman T 2005
  \href{http://dx.doi.org/10.1103/PhysRevLett.94.161102}{ {\em Phys. Rev.
  Lett.\/} {\bf 94} 161102 } [\eprint{hep-th/0409111}]

\bibitem{Dias:2007hg}
Dias O~J~C, Harmark T, Myers R~C and Obers N~A 2007
  \href{http://dx.doi.org/10.1103/PhysRevD.76.104025}{ {\em Phys. Rev.\/} {\bf
  D76} 104025 } [\eprint{0706.3645}]

\bibitem{Harmark:2007md}
Harmark T, Niarchos V and Obers N~A 2007
  \href{http://dx.doi.org/10.1088/0264-9381/24/8/R01}{ {\em Class. Quant.
  Grav.\/} {\bf 24} R1--R90 } [\eprint{hep-th/0701022}]

\bibitem{Wiseman:2011by}
Wiseman T 2012 {\em {Numerical construction of static and stationary black
  holes}\/} ({Cambridge University Press}) pp 233--270 [\eprint{1107.5513}]

\bibitem{Figueras:2012xj}
Figueras P, Murata K and Reall H~S 2012
  \href{http://dx.doi.org/10.1007/JHEP11(2012)071}{ {\em JHEP\/} {\bf 11} 071 }
  [\eprint{1209.1981}]

\bibitem{Dias:2017coo}
Dias O~J, Santos J~E and Way B 2018
  \href{http://dx.doi.org/10.1007/JHEP05(2018)111}{ {\em JHEP\/} {\bf 05} 111 }
  [\eprint{1712.07663}]

\bibitem{Lehner:2010pn}
Lehner L and Pretorius F 2010
  \href{http://dx.doi.org/10.1103/PhysRevLett.105.101102}{ {\em Phys. Rev.
  Lett.\/} {\bf 105} 101102 } [\eprint{1006.5960}]

\bibitem{Emparan:2015gva}
Emparan R, Suzuki R and Tanabe K 2015
  \href{http://dx.doi.org/10.1103/PhysRevLett.115.091102}{ {\em Phys. Rev.
  Lett.\/} {\bf 115} 091102 } [\eprint{1506.06772}]

\bibitem{Figueras:2022zkg}
Figueras P, Fran\c{c}a T, Gu C and Andrade T 2023
  \href{http://dx.doi.org/10.1103/PhysRevD.107.044028}{ {\em Phys. Rev. D\/}
  {\bf 107} 044028 } [\eprint{2210.13501}]

\bibitem{Emparan:2001wn}
Emparan R and Reall H~S 2002
  \href{http://dx.doi.org/10.1103/PhysRevLett.88.101101}{ {\em Phys. Rev.
  Lett.\/} {\bf 88} 101101 } [\eprint{hep-th/0110260}]

\bibitem{Elvang:2007rd}
Elvang H and Figueras P 2007
  \href{http://dx.doi.org/10.1088/1126-6708/2007/05/050}{ {\em JHEP\/} {\bf 05}
  050 } [\eprint{hep-th/0701035}]

\bibitem{Evslin:2007fv}
Evslin J and Krishnan C 2009
  \href{http://dx.doi.org/10.1088/0264-9381/26/12/125018}{ {\em Class. Quant.
  Grav.\/} {\bf 26} 125018 } [\eprint{0706.1231}]

\bibitem{Iguchi:2007is}
Iguchi H and Mishima T 2007
  \href{http://dx.doi.org/10.1103/PhysRevD.78.069903}{ {\em Phys. Rev. D\/}
  {\bf 75} 064018 } [Erratum: Phys.Rev.D 78, 069903 (2008)]
  [\eprint{hep-th/0701043}]

\bibitem{Santos:2015iua}
Santos J~E and Way B 2015
  \href{http://dx.doi.org/10.1103/PhysRevLett.114.221101}{ {\em Phys. Rev.
  Lett.\/} {\bf 114} 221101 } [\eprint{1503.00721}]

\bibitem{Figueras:2015hkb}
Figueras P, Kunesch M and Tunyasuvunakool S 2016
  \href{http://dx.doi.org/10.1103/PhysRevLett.116.071102}{ {\em Phys. Rev.
  Lett.\/} {\bf 116} 071102 } [\eprint{1512.04532}]

\bibitem{Emparan:2008eg}
Emparan R and Reall H~S 2008 \href{http://dx.doi.org/10.12942/lrr-2008-6}{ {\em
  Living Rev. Rel.\/} {\bf 11} 6 } [\eprint{0801.3471}]

\bibitem{Dias:2022mde}
Dias O~J~C, Ishii T, Murata K, Santos J~E and Way B 2023
  \href{http://dx.doi.org/10.1007/JHEP01(2023)147}{ {\em JHEP\/} {\bf 01} 147 }
  [\eprint{2211.02672}]

\bibitem{Marolf:2004fya}
Marolf D and Cabrera~Palmer B 2004
  \href{http://dx.doi.org/10.1103/PhysRevD.70.084045}{ {\em Phys. Rev. D\/}
  {\bf 70} 084045 } [\eprint{hep-th/0404139}]

\bibitem{Cardoso:2004zz}
Cardoso V and Lemos J~P 2005
  \href{http://dx.doi.org/10.1016/j.physletb.2005.06.025}{ {\em Phys. Lett.
  B\/} {\bf 621} 219--223 } [\eprint{hep-th/0412078}]

\bibitem{Cardoso:2005vk}
Cardoso V and Yoshida S 2005
  \href{http://dx.doi.org/10.1088/1126-6708/2005/07/009}{ {\em JHEP\/} {\bf 07}
  009 } [\eprint{hep-th/0502206}]

\bibitem{Dias:2006zv}
Dias O~J 2006 \href{http://dx.doi.org/10.1103/PhysRevD.73.124035}{ {\em Phys.
  Rev. D\/} {\bf 73} 124035 } [\eprint{hep-th/0602064}]

\bibitem{Dias:2022str}
Dias O~J~C, Ishii T, Murata K, Santos J~E and Way B 2023
  \href{http://dx.doi.org/10.1007/JHEP02(2023)069}{ {\em JHEP\/} {\bf 02} 069 }
  [\eprint{2212.01400}]

\bibitem{Dias:2023nbj}
Dias O~J~C, Ishii T, Murata K, Santos J~E and Way B 2023
  \href{http://dx.doi.org/10.1007/JHEP05(2023)041}{ {\em JHEP\/} {\bf 05} 041 }
  [\eprint{2302.09085}]

\bibitem{Tangherlini:1963bw}
Tangherlini F 1963 \href{http://dx.doi.org/10.1007/BF02784569}{ {\em Nuovo
  Cim.\/} {\bf 27} 636--651 }

\bibitem{Myers:1986un}
Myers R~C and Perry M 1986
  \href{http://dx.doi.org/10.1016/0003-4916(86)90186-7}{ {\em Annals Phys.\/}
  {\bf 172} 304 }

\bibitem{Kodama:2003jz}
Kodama H and Ishibashi A 2003 \href{http://dx.doi.org/10.1143/PTP.110.701}{
  {\em Prog.Theor.Phys.\/} {\bf 110} 701--722 } [\eprint{hep-th/0305147}]

\bibitem{Kodama:2003kk}
Kodama H and Ishibashi A 2004 \href{http://dx.doi.org/10.1143/PTP.111.29}{ {\em
  Prog. Theor. Phys.\/} {\bf 111} 29--73 } [\eprint{hep-th/0308128}]

\bibitem{Durkee:2010qu}
Durkee M and Reall H~S 2011
  \href{http://dx.doi.org/10.1088/0264-9381/28/3/035011}{ {\em
  Class.Quant.Grav.\/} {\bf 28} 035011 } [\eprint{1009.0015}]

\bibitem{Godazgar:2011sn}
Godazgar M 2012 \href{http://dx.doi.org/10.1088/0264-9381/29/5/055008}{ {\em
  Class.Quant.Grav.\/} {\bf 29} 055008 } [\eprint{1110.5779}]

\bibitem{Murata:2008yx}
Murata K and Soda J 2008 \href{http://dx.doi.org/10.1143/PTP.120.561}{ {\em
  Prog.Theor.Phys.\/} {\bf 120} 561--579 } [\eprint{0803.1371}]

\bibitem{Kodama:2009bf}
Kodama H, Konoplya R and Zhidenko A 2010
  \href{http://dx.doi.org/10.1103/PhysRevD.81.044007}{ {\em Phys.Rev.\/} {\bf
  D81} 044007 } [\eprint{0904.2154}]

\bibitem{Hartnett:2013fba}
Hartnett G~S and Santos J~E 2013
  \href{http://dx.doi.org/10.1103/PhysRevD.88.041505}{ {\em Phys. Rev. D\/}
  {\bf 88} 041505 } [\eprint{1306.4318}]

\bibitem{Emparan:2014cia}
Emparan R and Tanabe K 2014
  \href{http://dx.doi.org/10.1103/PhysRevD.89.064028}{ {\em Phys. Rev.\/} {\bf
  D89} 064028 } [\eprint{1401.1957}]

\bibitem{Emparan:2003sy}
Emparan R and Myers R~C 2003
  \href{http://dx.doi.org/10.1088/1126-6708/2003/09/025}{ {\em JHEP\/} {\bf 09}
  025 } [\eprint{hep-th/0308056}]

\bibitem{Emparan:2014jca}
Emparan R, Suzuki R and Tanabe K 2014
  \href{http://dx.doi.org/10.1007/JHEP06(2014)106}{ {\em JHEP\/} {\bf 06} 106 }
  [\eprint{1402.6215}]

\bibitem{Shibata:2009ad}
Shibata M and Yoshino H 2010
  \href{http://dx.doi.org/10.1103/PhysRevD.81.021501}{ {\em Phys.Rev.\/} {\bf
  D81} 021501 } [\eprint{0912.3606}]

\bibitem{Shibata:2010wz}
Shibata M and Yoshino H 2010
  \href{http://dx.doi.org/10.1103/PhysRevD.81.104035}{ {\em Phys.Rev.\/} {\bf
  D81} 104035 } [\eprint{1004.4970}]

\bibitem{Bantilan:2019bvf}
Bantilan H, Figueras P, Kunesch M and Panosso~Macedo R 2019
  \href{http://dx.doi.org/10.1103/PhysRevD.100.086014}{ {\em Phys. Rev. D\/}
  {\bf 100} 086014 } [\eprint{1906.10696}]

\bibitem{Dias:2014cia}
Dias O~J~C, Santos J~E and Way B 2014
  \href{http://dx.doi.org/10.1007/JHEP07(2014)045}{ {\em JHEP\/} {\bf 07} 045 }
  [\eprint{1402.6345}]

\bibitem{Emparan:2014pra}
Emparan R, Figueras P and Martinez M 2014
  \href{http://dx.doi.org/10.1007/JHEP12(2014)072}{ {\em JHEP\/} {\bf 12} 072 }
  [\eprint{1410.4764}]

\bibitem{Figueras:2017zwa}
Figueras P, Kunesch M, Lehner L and Tunyasuvunakool S 2017
  \href{http://dx.doi.org/10.1103/PhysRevLett.118.151103}{ {\em Phys. Rev.
  Lett.\/} {\bf 118} 151103 } [\eprint{1702.01755}]

\bibitem{Emparan:2009cs}
Emparan R, Harmark T, Niarchos V and Obers N~A 2009
  \href{http://dx.doi.org/10.1103/PhysRevLett.102.191301}{ {\em Phys. Rev.
  Lett.\/} {\bf 102} 191301 } [\eprint{0902.0427}]

\bibitem{Emparan:2009at}
Emparan R, Harmark T, Niarchos V and Obers N~A 2010
  \href{http://dx.doi.org/10.1007/JHEP03(2010)063}{ {\em JHEP\/} {\bf 03} 063 }
  [\eprint{0910.1601}]

\bibitem{Armas:2010hz}
Armas J and Obers N~A 2011 \href{http://dx.doi.org/10.1103/PhysRevD.83.084039}{
  {\em Phys. Rev.\/} {\bf D83} 084039 } [\eprint{1012.5081}]

\bibitem{Emparan:2011br}
Emparan R 2012 {\em in Black Holes in Higher Dimensions, CUP 2012, Ed. Gary T.
  Horowitz\/} [\eprint{1106.2021}]

\bibitem{Emparan:2020inr}
Emparan R and Herzog C~P 2020
  \href{http://dx.doi.org/10.1103/RevModPhys.92.045005}{ {\em Rev. Mod.
  Phys.\/} {\bf 92} 045005 } [\eprint{2003.11394}]

\bibitem{Arvanitaki:2014wva}
Arvanitaki A, Baryakhtar M and Huang X 2015
  \href{http://dx.doi.org/10.1103/PhysRevD.91.084011}{ {\em Phys. Rev. D\/}
  {\bf 91} 084011 } [\eprint{1411.2263}]

\bibitem{Arvanitaki:2016qwi}
Arvanitaki A, Baryakhtar M, Dimopoulos S, Dubovsky S and Lasenby R 2017
  \href{http://dx.doi.org/10.1103/PhysRevD.95.043001}{ {\em Phys. Rev. D\/}
  {\bf 95} 043001 } [\eprint{1604.03958}]

\bibitem{Brito:2017wnc}
Brito R, Ghosh S, Barausse E, Berti E, Cardoso V, Dvorkin I, Klein A and Pani P
  2017 \href{http://dx.doi.org/10.1103/PhysRevLett.119.131101}{ {\em Phys. Rev.
  Lett.\/} {\bf 119} 131101 } [\eprint{1706.05097}]

\bibitem{Brito:2017zvb}
Brito R, Ghosh S, Barausse E, Berti E, Cardoso V, Dvorkin I, Klein A and Pani P
  2017 \href{http://dx.doi.org/10.1103/PhysRevD.96.064050}{ {\em Phys. Rev.
  D\/} {\bf 96} 064050 } [\eprint{1706.06311}]

\bibitem{Pani:2012vp}
Pani P, Cardoso V, Gualtieri L, Berti E and Ishibashi A 2012
  \href{http://dx.doi.org/10.1103/PhysRevLett.109.131102}{ {\em Phys. Rev.
  Lett.\/} {\bf 109} 131102 } [\eprint{1209.0465}]

\bibitem{Pani:2012bp}
Pani P, Cardoso V, Gualtieri L, Berti E and Ishibashi A 2012
  \href{http://dx.doi.org/10.1103/PhysRevD.86.104017}{ {\em Phys. Rev. D\/}
  {\bf 86} 104017 } [\eprint{1209.0773}]

\bibitem{Witek:2012tr}
Witek H, Cardoso V, Ishibashi A and Sperhake U 2013
  \href{http://dx.doi.org/10.1103/PhysRevD.87.043513}{ {\em Phys. Rev. D\/}
  {\bf 87} 043513 } [\eprint{1212.0551}]

\bibitem{Endlich:2016jgc}
Endlich S and Penco R 2017 \href{http://dx.doi.org/10.1007/JHEP05(2017)052}{
  {\em JHEP\/} {\bf 05} 052 } [\eprint{1609.06723}]

\bibitem{East:2017mrj}
East W~E 2017 \href{http://dx.doi.org/10.1103/PhysRevD.96.024004}{ {\em Phys.
  Rev. D\/} {\bf 96} 024004 } [\eprint{1705.01544}]

\bibitem{East:2017ovw}
East W~E and Pretorius F 2017
  \href{http://dx.doi.org/10.1103/PhysRevLett.119.041101}{ {\em Phys. Rev.
  Lett.\/} {\bf 119} 041101 } [\eprint{1704.04791}]

\bibitem{Baryakhtar:2017ngi}
Baryakhtar M, Lasenby R and Teo M 2017
  \href{http://dx.doi.org/10.1103/PhysRevD.96.035019}{ {\em Phys. Rev. D\/}
  {\bf 96} 035019 } [\eprint{1704.05081}]

\bibitem{East:2018glu}
East W~E 2018 \href{http://dx.doi.org/10.1103/PhysRevLett.121.131104}{ {\em
  Phys. Rev. Lett.\/} {\bf 121} 131104 } [\eprint{1807.00043}]

\bibitem{Frolov:2018ezx}
Frolov V~P, Krtou\v{s} P, Kubiz\v{n}\'ak D and Santos J~E 2018
  \href{http://dx.doi.org/10.1103/PhysRevLett.120.231103}{ {\em Phys. Rev.
  Lett.\/} {\bf 120} 231103 } [\eprint{1804.00030}]

\bibitem{Dolan:2018dqv}
Dolan S~R 2018 \href{http://dx.doi.org/10.1103/PhysRevD.98.104006}{ {\em Phys.
  Rev. D\/} {\bf 98} 104006 } [\eprint{1806.01604}]

\bibitem{Siemonsen:2019ebd}
Siemonsen N and East W~E 2020
  \href{http://dx.doi.org/10.1103/PhysRevD.101.024019}{ {\em Phys. Rev. D\/}
  {\bf 101} 024019 } [\eprint{1910.09476}]

\bibitem{Brito:2013wya}
Brito R, Cardoso V and Pani P 2013
  \href{http://dx.doi.org/10.1103/PhysRevD.88.023514}{ {\em Phys. Rev. D\/}
  {\bf 88} 023514 } [\eprint{1304.6725}]

\bibitem{Brito:2020lup}
Brito R, Grillo S and Pani P 2020
  \href{http://dx.doi.org/10.1103/PhysRevLett.124.211101}{ {\em Phys. Rev.
  Lett.\/} {\bf 124} 211101 } [\eprint{2002.04055}]

\bibitem{Dias:2023ynv}
Dias O~J~C, Lingetti G, Pani P and Santos J~E 2023
  \href{http://dx.doi.org/10.1103/PhysRevD.108.L041502}{ {\em Phys. Rev. D\/}
  {\bf 108} L041502 } [\eprint{2304.01265}]

\bibitem{East:2023nsk}
East W~E and Siemonsen N 2023
  \href{http://dx.doi.org/10.1103/PhysRevD.108.124048}{ {\em Phys. Rev. D\/}
  {\bf 108} 124048 } [\eprint{2309.05096}]

\bibitem{Brito:2014wla}
Brito R, Cardoso V and Pani P 2015
  \href{http://dx.doi.org/10.1088/0264-9381/32/13/134001}{ {\em Class. Quant.
  Grav.\/} {\bf 32} 134001 } [\eprint{1411.0686}]

\bibitem{Ficarra:2018rfu}
Ficarra G, Pani P and Witek H 2019
  \href{http://dx.doi.org/10.1103/PhysRevD.99.104019}{ {\em Phys. Rev. D\/}
  {\bf 99} 104019 } [\eprint{1812.02758}]

\bibitem{Moncrief:1980ApJ}
{Moncrief} V 1980 \href{http://dx.doi.org/10.1086/157707}{ {\em Astrophys.
  J.\/} {\bf 235} 1038--1046 }

\bibitem{Unruh:1980cg}
Unruh W~G 1981 \href{http://dx.doi.org/10.1103/PhysRevLett.46.1351}{ {\em Phys.
  Rev. Lett.\/} {\bf 46} 1351--1353 }

\bibitem{Visser:1997ux}
Visser M 1998 \href{http://dx.doi.org/10.1088/0264-9381/15/6/024}{ {\em Class.
  Quant. Grav.\/} {\bf 15} 1767--1791 } [\eprint{gr-qc/9712010}]

\bibitem{Barcelo:2005fc}
Barcelo C, Liberati S and Visser M 2005
  \href{http://dx.doi.org/10.12942/lrr-2005-12}{ {\em Living Rev. Rel.\/} {\bf
  8} 12 } [\eprint{gr-qc/0505065}]

\bibitem{Almeida:2022otk}
Almeida C~R and Jacquet M~J 2023
  \href{http://dx.doi.org/10.1140/epjh/s13129-023-00063-2}{ {\em Eur. Phys. J.
  H\/} {\bf 48} 15 } [\eprint{2212.08838}]

\bibitem{Jacquet:2020bar}
Jacquet M~J, Weinfurtner S and K\"onig F 2020
  \href{http://dx.doi.org/10.1098/rsta.2019.0239}{ {\em Phil. Trans. Roy. Soc.
  Lond. A\/} {\bf 378} 20190239 } [\eprint{2005.04027}]

\bibitem{Rousseaux:2007is}
Rousseaux G, Mathis C, Maissa P, Philbin T~G and Leonhardt U 2008
  \href{http://dx.doi.org/10.1088/1367-2630/10/5/053015}{ {\em New J. Phys.\/}
  {\bf 10} 053015 } [\eprint{0711.4767}]

\bibitem{Weinfurtner:2010nu}
Weinfurtner S, Tedford E~W, Penrice M~C~J, Unruh W~G and Lawrence G~A 2011
  \href{http://dx.doi.org/10.1103/PhysRevLett.106.021302}{ {\em Phys. Rev.
  Lett.\/} {\bf 106} 021302 } [\eprint{1008.1911}]

\bibitem{Euve:2015vml}
Euv\'e L~P, Michel F, Parentani R, Philbin T~G and Rousseaux G 2016
  \href{http://dx.doi.org/10.1103/PhysRevLett.117.121301}{ {\em Phys. Rev.
  Lett.\/} {\bf 117} 121301 } [\eprint{1511.08145}]

\bibitem{Steinhauer:2015saa}
Steinhauer J 2016 \href{http://dx.doi.org/10.1038/nphys3863}{ {\em Nature
  Phys.\/} {\bf 12} 959 } [\eprint{1510.00621}]

\bibitem{MunozdeNova:2018fxv}
Mu\~noz~de Nova J~R, Golubkov K, Kolobov V~I and Steinhauer J 2019
  \href{http://dx.doi.org/10.1038/s41586-019-1241-0}{ {\em Nature\/} {\bf 569}
  688--691 } [\eprint{1809.00913}]

\bibitem{Drori:2018ivu}
Drori J, Rosenberg Y, Bermudez D, Silberberg Y and Leonhardt U 2019
  \href{http://dx.doi.org/10.1103/PhysRevLett.122.010404}{ {\em Phys. Rev.
  Lett.\/} {\bf 122} 010404 } [\eprint{1808.09244}]

\bibitem{Kolobov:2019qfs}
Kolobov V~I, Golubkov K, Mu\~noz~de Nova J~R and Steinhauer J 2021
  \href{http://dx.doi.org/10.1038/s41567-020-01076-0}{ {\em Nature Phys.\/}
  {\bf 17} 362--367 } [\eprint{1910.09363}]

\bibitem{Shi:2021nkx}
Shi Y~H {\em et~al.\/} 2023
  \href{http://dx.doi.org/10.1038/s41467-023-39064-6}{ {\em Nature Commun.\/}
  {\bf 14} 3263 } [\eprint{2111.11092}]

\bibitem{Torres:2016iee}
Torres T, Patrick S, Coutant A, Richartz M, Tedford E~W and Weinfurtner S 2017
  \href{http://dx.doi.org/10.1038/nphys4151}{ {\em Nature Phys.\/} {\bf 13}
  833--836 } [\eprint{1612.06180}]

\bibitem{Cromb:2020ldn}
Cromb M, Gibson G~M, Toninelli E, Padgett J~M, Wright E~M and Faccio D 2020
  \href{http://dx.doi.org/10.1038/s41567-020-0944-3}{ {\em Nature Phys.\/} {\bf
  16} 1069--1073 } [\eprint{2005.03760}]

\bibitem{Braidotti:2021nhw}
Braidotti M~C, Prizia R, Maitland C, Marino F, Prain A, Starshynov I,
  Westerberg N, Wright E~M and Faccio D 2022
  \href{http://dx.doi.org/10.1103/PhysRevLett.128.013901}{ {\em Phys. Rev.
  Lett.\/} {\bf 128} 013901 } [\eprint{2109.02307}]

\bibitem{Torres:2020tzs}
Torres T, Patrick S, Richartz M and Weinfurtner S 2020
  \href{http://dx.doi.org/10.1103/PhysRevLett.125.011301}{ {\em Phys. Rev.
  Lett.\/} {\bf 125} 011301 } [\eprint{1811.07858}]

\bibitem{Smaniotto:2025hqm}
Smaniotto P, Solidoro L, \v{S}van\v{c}ara P, Patrick S, Richartz M, Barenghi
  C~F, Gregory R and Weinfurtner S 2025  [\eprint{2502.11209}]

\bibitem{Philbin:2007ji}
Philbin T~G, Kuklewicz C, Robertson S, Hill S, Konig F and Leonhardt U 2008
  \href{http://dx.doi.org/10.1126/science.1153625}{ {\em Science\/} {\bf 319}
  1367--1370 } [\eprint{0711.4796}]

\bibitem{Jaskula:2012ab}
Jaskula J~C, Partridge G~B, Bonneau M, Lopes R, Ruaudel J, Boiron D and
  Westbrook C~I 2012 \href{http://dx.doi.org/10.1103/PhysRevLett.109.220401}{
  {\em Phys. Rev. Lett.\/} {\bf 109} 220401 } [\eprint{1207.1338}]

\bibitem{Eckel:2017uqx}
Eckel S, Kumar A, Jacobson T, Spielman I~B and Campbell G~K 2018
  \href{http://dx.doi.org/10.1103/PhysRevX.8.021021}{ {\em Phys. Rev. X\/} {\bf
  8} 021021 } [\eprint{1710.05800}]

\bibitem{Patrick:2019kis}
Patrick S, Goodhew H, Gooding C and Weinfurtner S 2021
  \href{http://dx.doi.org/10.1103/PhysRevLett.126.041105}{ {\em Phys. Rev.
  Lett.\/} {\bf 126} 041105 } [\eprint{1905.03045}]

\bibitem{Steinhauer:2021fhb}
Steinhauer J, Abuzarli M, Aladjidi T, Bienaim\'e T, Piekarski C, Liu W,
  Giacobino E, Bramati A and Glorieux Q 2022
  \href{http://dx.doi.org/10.1038/s41467-022-30603-1}{ {\em Nature Commun.\/}
  {\bf 13} 2890 } [\eprint{2102.08279}]

\bibitem{Viermann:2022wgw}
Viermann C {\em et~al.\/} 2022
  \href{http://dx.doi.org/10.1038/s41586-022-05313-9}{ {\em Nature\/} {\bf 611}
  260--264 } [\eprint{2202.10399}]

\bibitem{Tajik:2022lyt}
Tajik M {\em et~al.\/} 2023 \href{http://dx.doi.org/10.1073/pnas.2301287120}{
  {\em Proc. Nat. Acad. Sci.\/} {\bf 120} e2301287120 } [\eprint{2209.09132}]

\bibitem{Gregory:2024ogi}
Gregory S~M~D, Schiattarella S, Barroso V~S, Kaiser D~I, Avgoustidis A and
  Weinfurtner S 2024  [\eprint{2410.08842}]

\bibitem{Schutzhold:2002rf}
Schutzhold R and Unruh W~G 2002
  \href{http://dx.doi.org/10.1103/PhysRevD.66.044019}{ {\em Phys. Rev. D\/}
  {\bf 66} 044019 } [\eprint{gr-qc/0205099}]

\bibitem{Volovik:1995ja}
Volovik G~E 1995  [\eprint{gr-qc/9510001}]

\bibitem{Volovik:1999fc}
Volovik G~E 1999 \href{http://dx.doi.org/10.1134/1.568079}{ {\em JETP Lett.\/}
  {\bf 69} 705--713 } [\eprint{gr-qc/9901077}]

\bibitem{Svancara:2023yrf}
\v{S}van\v{c}ara P, Smaniotto P, Solidoro L, MacDonald J~F, Patrick S, Gregory
  R, Barenghi C~F and Weinfurtner S 2024
  \href{http://dx.doi.org/10.1038/s41586-024-07176-8}{ {\em Nature\/} {\bf 628}
  66--70 } [\eprint{2308.10773}]

\bibitem{Garay:2000jj}
Garay L~J, Anglin J~R, Cirac J~I and Zoller P 2001
  \href{http://dx.doi.org/10.1103/PhysRevA.63.023611}{ {\em Phys. Rev. A\/}
  {\bf 63} 023611 } [\eprint{gr-qc/0005131}]

\bibitem{Leonhardt:2000fd}
Leonhardt U and Piwnicki P 2000
  \href{http://dx.doi.org/10.1103/PhysRevLett.84.822}{ {\em Phys. Rev. Lett.\/}
  {\bf 84} 822--825 } [\eprint{cond-mat/9906332}]

\bibitem{Marino:2008kk}
Marino F 2008 \href{http://dx.doi.org/10.1103/PhysRevA.78.063804}{ {\em Phys.
  Rev. A\/} {\bf 78} 063804 } [\eprint{0808.1624}]

\bibitem{Schutzhold:2004tv}
Schutzhold R and Unruh W~G 2005
  \href{http://dx.doi.org/10.1103/PhysRevLett.95.031301}{ {\em Phys. Rev.
  Lett.\/} {\bf 95} 031301 } [\eprint{quant-ph/0408145}]

\bibitem{Nation:2009xb}
Nation P~D, Blencowe M~P, Rimberg A~J and Buks E 2009
  \href{http://dx.doi.org/10.1103/PhysRevLett.103.087004}{ {\em Phys. Rev.
  Lett.\/} {\bf 103} 087004 } [\eprint{0904.2589}]

\bibitem{Berti:2004ju}
Berti E, Cardoso V and Lemos J~P~S 2004
  \href{http://dx.doi.org/10.1103/PhysRevD.70.124006}{ {\em Phys. Rev. D\/}
  {\bf 70} 124006 } [\eprint{gr-qc/0408099}]

\bibitem{Cardoso:2004fi}
Cardoso V, Lemos J~P~S and Yoshida S 2004
  \href{http://dx.doi.org/10.1103/PhysRevD.70.124032}{ {\em Phys. Rev. D\/}
  {\bf 70} 124032 } [\eprint{gr-qc/0410107}]

\bibitem{Torres:2019sbr}
Torres T, Patrick S, Richartz M and Weinfurtner S 2019
  \href{http://dx.doi.org/10.1088/1361-6382/ab3d48}{ {\em Class. Quant.
  Grav.\/} {\bf 36} 194002 } [\eprint{1905.00356}]

\bibitem{Barcelo:2018ynq}
Barcel\'o C 2019 \href{http://dx.doi.org/10.1038/s41567-018-0367-6}{ {\em
  Nature Phys.\/} {\bf 15} 210--213 }

\bibitem{Nakano:2004ha}
Nakano H, Kurita Y, Ogawa K and Yoo C~M 2005
  \href{http://dx.doi.org/10.1103/PhysRevD.71.084006}{ {\em Phys. Rev. D\/}
  {\bf 71} 084006 } [\eprint{gr-qc/0411041}]

\bibitem{Lepe:2004kv}
Lepe S and Saavedra J 2005
  \href{http://dx.doi.org/10.1016/j.physletb.2005.05.021}{ {\em Phys. Lett.
  B\/} {\bf 617} 174--181 } [\eprint{gr-qc/0410074}]

\bibitem{Saavedra:2005ug}
Saavedra J 2006 \href{http://dx.doi.org/10.1142/S0217732306019712}{ {\em Mod.
  Phys. Lett. A\/} {\bf 21} 1601--1608 } [\eprint{gr-qc/0508040}]

\bibitem{Xi:2007yb}
Xi P and Li X~Z 2007 \href{http://dx.doi.org/10.1142/S0218271807010687}{ {\em
  Int. J. Mod. Phys. D\/} {\bf 16} 1211--1218 } [\eprint{0709.3714}]

\bibitem{Okuzumi:2007hf}
Okuzumi S and Sakagami M~a 2007
  \href{http://dx.doi.org/10.1103/PhysRevD.76.084027}{ {\em Phys. Rev. D\/}
  {\bf 76} 084027 } [\eprint{gr-qc/0703070}]

\bibitem{Barcelo:2007ru}
Barcelo C, Cano A, Garay L~J and Jannes G 2007
  \href{http://dx.doi.org/10.1103/PhysRevD.75.084024}{ {\em Phys. Rev. D\/}
  {\bf 75} 084024 } [\eprint{gr-qc/0701173}]

\bibitem{Abdalla:2007dz}
Abdalla E, Konoplya R~A and Zhidenko A 2007
  \href{http://dx.doi.org/10.1088/0264-9381/24/23/012}{ {\em Class. Quant.
  Grav.\/} {\bf 24} 5901--5910 } [\eprint{0706.2489}]

\bibitem{Dolan:2010zza}
Dolan S~R, Oliveira L~A and Crispino L~C~B 2010
  \href{http://dx.doi.org/10.1103/PhysRevD.82.084037}{ {\em Phys. Rev. D\/}
  {\bf 82} 084037 } [\eprint{1407.3904}]

\bibitem{Dolan:2011ti}
Dolan S~R, Oliveira L~A and Crispino L~C~B 2012
  \href{http://dx.doi.org/10.1103/PhysRevD.85.044031}{ {\em Phys. Rev. D\/}
  {\bf 85} 044031 } [\eprint{1105.1795}]

\bibitem{Daghigh:2014mwa}
Daghigh R~G and Green M~D 2015
  \href{http://dx.doi.org/10.1088/0264-9381/32/9/095003}{ {\em Class. Quant.
  Grav.\/} {\bf 32} 095003 } [\eprint{1411.7066}]

\bibitem{Chaverra:2015aya}
Chaverra E, Morales M~D and Sarbach O 2015
  \href{http://dx.doi.org/10.1103/PhysRevD.91.104012}{ {\em Phys. Rev. D\/}
  {\bf 91} 104012 } [\eprint{1501.01637}]

\bibitem{Patrick:2018orp}
Patrick S, Coutant A, Richartz M and Weinfurtner S 2018
  \href{http://dx.doi.org/10.1103/PhysRevLett.121.061101}{ {\em Phys. Rev.
  Lett.\/} {\bf 121} 061101 } [\eprint{1801.08473}]

\bibitem{Assumpcao:2018bka}
Assumpcao T, Cardoso V, Ishibashi A, Richartz M and Zilhao M 2018
  \href{http://dx.doi.org/10.1103/PhysRevD.98.064036}{ {\em Phys. Rev. D\/}
  {\bf 98} 064036 } [\eprint{1806.07909}]

\bibitem{Herdeiro:2019fps}
Herdeiro C~A~R and Santos N~M 2019
  \href{http://dx.doi.org/10.1103/PhysRevD.99.084029}{ {\em Phys. Rev. D\/}
  {\bf 99} 084029 } [\eprint{1902.06748}]

\bibitem{Guo:2020blq}
Guo H, Liu H, Kuang X~M and Wang B 2020
  \href{http://dx.doi.org/10.1103/PhysRevD.102.124019}{ {\em Phys. Rev. D\/}
  {\bf 102} 124019 } [\eprint{2007.04197}]

\bibitem{Ge:2019our}
Ge X~H, Nakahara M, Sin S~J, Tian Y and Wu S~F 2019
  \href{http://dx.doi.org/10.1103/PhysRevD.99.104047}{ {\em Phys. Rev. D\/}
  {\bf 99} 104047 } [\eprint{1902.11126}]

\bibitem{Patrick:2020yyy}
Patrick S and Solidoro L 2025
  \href{http://dx.doi.org/10.1103/PhysRevD.111.104048}{ {\em Phys. Rev. D\/}
  {\bf 111} 104048 } [\eprint{2007.06671}]

\bibitem{Geelmuyden:2021sdh}
Geelmuyden A, Erne S, Patrick S, Barenghi C~F and Weinfurtner S 2022
  \href{http://dx.doi.org/10.1103/PhysRevResearch.4.023099}{ {\em Phys. Rev.
  Res.\/} {\bf 4} 023099 } [\eprint{2105.11509}]

\bibitem{Jacquet:2021scv}
Jacquet M~J, Giacomelli L, Valnais Q, Joly M, Claude F, Giacobino E, Glorieux
  Q, Carusotto I and Bramati A 2023
  \href{http://dx.doi.org/10.1103/PhysRevLett.130.111501}{ {\em Phys. Rev.
  Lett.\/} {\bf 130} 111501 } [\eprint{2110.14452}]

\bibitem{Vieira:2021ozg}
Vieira H~S, Destounis K and Kokkotas K~D 2022
  \href{http://dx.doi.org/10.1103/PhysRevD.105.045015}{ {\em Phys. Rev. D\/}
  {\bf 105} 045015 } [\eprint{2112.08711}]

\bibitem{Vieira:2021xqw}
Vieira H~S and Kokkotas K~D 2021
  \href{http://dx.doi.org/10.1103/PhysRevD.104.024035}{ {\em Phys. Rev. D\/}
  {\bf 104} 024035 } [\eprint{2104.03938}]

\bibitem{deOliveira:2023qxe}
de~Oliveira C~C, Mosna R~A and Pitelli J~a~P~M 2023
  \href{http://dx.doi.org/10.1103/PhysRevD.107.064018}{ {\em Phys. Rev. D\/}
  {\bf 107} 064018 } [\eprint{2303.05612}]

\bibitem{Burgess:2023pny}
Burgess C, Patrick S, Torres T, Gregory R and Koenig F 2024
  \href{http://dx.doi.org/10.1103/PhysRevLett.132.053802}{ {\em Phys. Rev.
  Lett.\/} {\bf 132} 053802 } [\eprint{2309.10622}]

\bibitem{Vieira:2023ylz}
Vieira H~S, Destounis K and Kokkotas K~D 2023
  \href{http://dx.doi.org/10.1103/PhysRevD.107.104038}{ {\em Phys. Rev. D\/}
  {\bf 107} 104038 } [\eprint{2301.11480}]

\bibitem{Liu:2024vde}
Liu H, Guo H and Ling R 2024
  \href{http://dx.doi.org/10.1103/PhysRevD.110.024035}{ {\em Phys. Rev. D\/}
  {\bf 110} 024035 } [\eprint{2404.04982}]

\bibitem{Matyjasek:2024uwo}
Matyjasek J, Benda K and Stafi\'nska M 2024
  \href{http://dx.doi.org/10.1103/PhysRevD.110.064083}{ {\em Phys. Rev. D\/}
  {\bf 110} 064083 } [\eprint{2408.16116}]

\bibitem{Liu:2024wch}
Liu H and Guo H 2024 \href{http://dx.doi.org/10.1103/PhysRevD.110.104058}{ {\em
  Phys. Rev. D\/} {\bf 110} 104058 } [\eprint{2409.00320}]

\bibitem{Keshet:2024hlm}
Keshet E, Shemesh I and Steinhauer J 2024  [\eprint{2407.00448}]

\bibitem{Albuquerque:2025eny}
Albuquerque S and V{\"o}lkel S~H 2025
  \href{http://dx.doi.org/10.1103/kwrg-rs71}{ {\em Phys. Rev. D\/} {\bf 111}
  124020 } [\eprint{2501.09000}]

\bibitem{Vieira:2025ljl}
Vieira H~S, Destounis K and Kokkotas K~D 2025
  \href{http://dx.doi.org/10.1103/PhysRevD.111.104025}{ {\em Phys. Rev. D\/}
  {\bf 111} 104025 } [\eprint{2502.11274}]

\bibitem{dePaula:2025fqt}
de~Paula L~T, Siqueira P~H~C, Panosso~Macedo R and Richartz M 2025
  \href{http://dx.doi.org/10.1103/PhysRevD.111.104064}{ {\em Phys. Rev. D\/}
  {\bf 111} 104064 } [\eprint{2504.00106}]

\bibitem{Torres:2017vaz}
Torres T, Coutant A, Dolan S and Weinfurtner S 2018
  \href{http://dx.doi.org/10.1017/jfm.2018.752}{ {\em J. Fluid Mech.\/} {\bf
  857} 291--311 } [\eprint{1712.04675}]

\bibitem{Falque:2023ctx}
Falque K, Delhom A, Glorieux Q, Giacobino E, Bramati A and Jacquet M~J 2025
  \href{http://dx.doi.org/10.1103/t5dh-rx6w}{ {\em Phys. Rev. Lett.\/} {\bf
  135} 023401 } [\eprint{2311.01392}]

\bibitem{Liberati:2018uev}
Liberati S, Schuster S, Tricella G and Visser M 2019
  \href{http://dx.doi.org/10.1103/PhysRevD.99.044025}{ {\em Phys. Rev. D\/}
  {\bf 99} 044025 } [\eprint{1802.04785}]

\bibitem{Solidoro:2024yxi}
Solidoro L, Patrick S, Weinfurtner S and Gregory R 2025
  \href{http://dx.doi.org/10.1103/7kjp-vrml}{ {\em Phys. Rev. Lett.\/} {\bf
  135} 051401 } [\eprint{2406.11013}]

\end{thebibliography}

\end{document}